\documentclass[pre,11pt,onecolumn,preprint,superscriptaddress]{revtex4} 
\usepackage{amssymb}
\usepackage{latexsym}
\usepackage{amsfonts}
\usepackage{graphics}
\usepackage{graphicx}
\usepackage{epsfig}
\usepackage{amsmath}
\usepackage{float}
\usepackage{verbatim}
\usepackage{caption}
\usepackage[subfigure]{graphfig}
\usepackage{subfigure}
\usepackage{diagbox}
\usepackage{makecell}
\usepackage{float}
\usepackage{soul}
\usepackage{color, xcolor}
\usepackage{color}
\begin{document}

\title{Time persistence of climate and carbon flux networks}

\author{Ting Qing}
\affiliation{School of Mathematical Sciences, Jiangsu University, Zhenjiang, 212013 Jiangsu, China}
\affiliation{Department of Physics, Bar-Ilan University, Ramat-Gan 52900, Israel}%
\author{Fan Wang}
\affiliation{Department of Physics, Bar-Ilan University, Ramat-Gan 52900, Israel}
\author{Qiuyue Li}
\affiliation{Department of Physics, Bar-Ilan University, Ramat-Gan 52900, Israel}
\author{Gaogao Dong}
\thanks{Corresponding author: gago999@126.com}
\affiliation{School of Mathematical Sciences, Jiangsu University, Zhenjiang, 212013 Jiangsu, China}

\author{Lixin Tian}
\thanks{Corresponding author: tianlx@ujs.edu.cn}
\affiliation{School of Mathematical Sciences, Jiangsu University, Zhenjiang, 212013 Jiangsu, China}
\affiliation{Research Institute of Carbon Neutralization Development, School of Mathematical Sciences, Jiangsu University, Zhenjiang 212013, China}
\affiliation{Jiangsu Province Engineering Research Center of Industrial Carbon System Analysis, School of Mathematical Sciences, Jiangsu University, Zhenjiang 212013, China}
\affiliation{Jiangsu Province Engineering Research Center of Spatial Big Data, School of Mathematical Sciences, Nanjing Normal University, Nanjing 210023, China}
\affiliation{Key Laboratory for NSLSCS, Ministry of Education, School of Mathematical Sciences, Nanjing Normal University, Nanjing 210023, China}
\author{Shlomo Havlin}
\thanks{Corresponding author: havlin@ophir.ph.biu.ac.il}
\affiliation{Department of Physics, Bar-Ilan University, Ramat-Gan 52900, Israel}

\begin{abstract}
The persistence of the global climate system is critical for assuring the sustainability of the natural ecosystem and the further development of the prosperity of socio-economics. In this paper, we develop a framework and analyze the time persistence of the yearly networks of climate and carbon flux, based on cross-correlations between sites, using daily data from China (East Asia), the contiguous United States (North America), and the Europe land region during 2000-2019. There are many studies on time persistence of single nodes, e.g., climate variables at a given location, however persistence at a network level has been rarely discussed. Here we develop a framework to study time persistence of network and we apply it to climate and carbon flux. Our framework for determining the persistence is based on analyzing the similarity between the network structures, i.e., the links of climate and carbon flux in different years of systems using the Jaccard index. Our Jaccard results reveal that the similarity of climate and carbon flux networks in different years are within the range of $0.51\pm0.09$ ($p-value<0.05$), implying that the climate and carbon flux networks studied in the Earth's climate system are generally persistent and in a steady state. Our results suggest that close to 50\% of the links appear regularly in different years. We find a very small decay in similarity when the gap between the years increases. However, we observe unique behavior of less similarity to other years in the carbon flux network of the Chinese region during the years 2004-2005 and 2015-2016. This seems to reflect China's carbon reduction policies in these specific years. Moreover, we find that the persistence of climate variables and carbon flux in the three regions decreases as the geographical distance between nodes increases. For example, for links above 500km the Jaccard is within the range of $0.45\pm0.08$ and for links above 1000km the Jaccard is within the range of $0.34\pm0.08$ ($p-value<0.05$). Analyzing the persistence and evolution of the climate and carbon flux networks, enhance our understanding of the spatial and temporal evolution of the global climate system.
\end{abstract}

\date{\today}
\maketitle

\section{Introduction}
The Earth's climate system is a persistent and a stable complex system in which all climate variables interact, and this persistence has a crucial impact on the health and sustainability of the ecological environment and species diversity \cite{salcedo2022persistence, wang2016testing, eichner2003power}. With the increasing human activities and modern technology, a significant amount of greenhouse gases, including carbon dioxide, is emitted into the Earth's atmosphere through massive burning of fossil fuels, industrial processes, and changes in land use, which is one of the important factors leading to global climate change \cite{world2017wmo}. The sustainable development of ecosystems and socioeconomic advancement depend largely on the persistence of the Earth's climate system. Therefore, a deeper understanding of the persistence of climate variables which evolve over time is crucial for understanding, predicting, and responding to climate warming. Furthermore, the persistence of the Earth's climate system during climate change has not been studied systematically.

There exist many studies on time persistence of dynamics on single nodes, e.g., climate variables at a given location \cite{koscielny1998indication}, or brain signals \cite{hardstone2012detrended}, however persistence at a network level has been rarely discussed. Here we develop a framework to study time persistence of a network and we apply it to climate and carbon flux systems. Our framework for determining the persistence is based on analyzing the similarity between the climate networks, i.e., the network links, in different years based the Jaccard index.

The primary climatic variables, including temperature, wind speed, precipitation, and geopotential height, have a significant impact on climate change in the Earth system. Wunderling et al. \cite{wunderling2023global} used a network model of four interacting climate tipping elements to investigate in different temperature scenarios, the effect of climate change. Meanwhile, based on the potential teleconnections among the tipping elements, researchers studied systematically the impact of the Amazon Rainforest Area on global climate change using the global near-surface air temperature field \cite{liu2023teleconnections}. With the increase in human activities, there has been a significant change not only in the global surface temperature but also in the sea surface temperature, which has a wide-range impact on the global climate system. Systematic studies of trends, variability, and persistence of sea surface temperature enable the analysis of the mechanisms and patterns of climate change, and thus contributing to obtain a better understanding of climate change processes \cite{bulgin2020tendencies}. Persistent extreme heat events under the influence of global climate change have severe impacts on ecosystems and societies. Rousi et al. \cite{rousi2022accelerated} identified Europe as a heatwave hotspot and found that its upward trend in heatwave occurrence over the last 42 years was related to atmospheric dynamics induced by an increase in the frequency and persistence of double jets over the Eurasian continent. Wind speed, as a crucial climatic variable, directly influences atmospheric and oceanic circulation, thereby affecting the global climate system. Ko{\c{c}}ak et al. \cite{koccak2008practical} based on autocorrelation function, conditional probabilities and wind speed duration curves methods, analyzed wind speed data in northwestern Turkey. Their results indicate that the proposed methods clearly reflect the persistence properties of the wind speed in the studied area. In addition, the climate system is a complex and giant system \cite{dai2004rainfall}, and its properties and phenomena can be analyzed and better understood from a network perspective using sophisticated methods of network science \cite{ludescher2021network}. In this approach, regions are viewed as nodes and correlations between climate variables of pairs of nodes are considered as links \cite{yamasaki2008climate}, providing a new perspective for understanding the Earth's climate system. Numerous studies have explored climate networks. For instance, constructing climate networks based on extreme winter precipitation in the United States enabled to identify regions of extreme precipitation and the climatic conditions that lead to extreme precipitation \cite{banerjee2023spatial}. In Ref. \cite{banerjee2023spatial} climate variables have been used to study the critical thresholds of climate change and provide early warning signals \cite{sunny2023predicting, lu2022early, fan2018climate, ludescher2013improved}. Moreover, integrating climate network methods have been applied to identify and measure the optimal paths that yield the interaction between pairs of remote nodes, i.e., teleconnections \cite{zhou2015teleconnection}. Furthermore, climate networks are known to generally present significant correlations when they are geographically close; while the correlations between sites at far distances are largely influenced by external and global atmospheric processes. For example, it is found the origin of a significant correlation between remote nodes is due to their correlations with atmospheric Rossby waves \cite{zhang2019significant, wang2013dominant}. These studies provide a crucial scientific basis and methods to deepen our understanding of the mechanisms and patterns behind climate change.

In 2014, the Intergovernmental Panel on Climate Change (IPCC)'s Fifth Scientific Assessment Report concluded that the global average temperature of land and ocean surfaces increased by 0.85 [0.65-1.06]$^{\circ}$C from 1880 to 2012, mainly due to the significant increase in the concentration of carbon dioxide in the atmosphere \cite{ pachauri2014climate}. While the terrestrial ecosystem and the oceanic system can effectively remove $\rm {CO_{2}}$ from the atmosphere, the carbon flux can reflect whether, at a given moment in time, a region is a carbon source area (emitting more $\rm {CO_{2}}$ than it absorbs) or a carbon sink area (removing more $\rm {CO_{2}}$ than it emits). If man-made emission sources (i.e., carbon sources) exceed the natural removal processes, it will inevitably lead to a rise in atmospheric $\rm {CO_{2}}$ concentrations. Therefore, in addition to studying climate variables such as temperature and precipitation, it is crucial to also study carbon flux for understanding changes in the Earth's climate system and the effects of global warming.

Overall, investigating the persistence of carbon flux and climate variables networks is essential for maintaining the stability of the Earth's climate system which requires significant attention. The present persistence study uses climate variables (temperature, wind speed, precipitation, and potential height) and carbon flux data in the Earth's climate system for the past 20 years. By combining network analysis of climate variables and carbon flux, we construct the corresponding complex network systems of climate and carbon flux, and other variables to investigate the persistence of the network structure of the climate system from spatial and temporal perspectives during the past 20 years. With our developed model, one can observe the persistence of climate and carbon flux network structure during the past 20 years, by using the Jaccard index. In addition, we explored the persistence of climate and carbon flux networks at different spatial distances. The results show a decreasing trend of persistence with increasing spatial distance.

\section{Methods}
\subsection{Network inference}
Here, we select three important geographical regions that have had a significant impact on global development in the past 20 years: East Asia (China (CHN)), the Americas (the Contiguous United States (USA)), and Europe (EU). The study is based on the daily average datasets of climate variables (temperature, wind speed, precipitation, and geopotential height) \cite{ERA5} and carbon flux \cite{Carbon} in the three aforementioned regions. The datasets are evenly distributed and interpolated into $2^{\circ}\times2^{\circ}$ grids, generating total of $N_A=238$ (CHN), $200$ (USA), and $385$ (EU) grid points. For more details see the SI.

The datasets of both carbon flux and climate variables contain strong unlinear trends and seasonal signals. Hence, for mitigating the effects of trends, the data has been preprocessed following Ref. \cite{ludescher2013improved}. To this end, we set $\widehat{C}_i(t')(i=1,2,\dots,N_A,t'=1,2,\dots,\widehat{L})$, $\widetilde{G}_i(t')(i=1,2,\dots,N_A,t'=1,2,\dots,\widehat{L})$, $\widetilde{T}_i(t')(i=1,2,\dots,N_A,t'=1,2,\dots,\widehat{L})$, $\widetilde{W}_i(t')(i=1,2,\dots,N_A,t'=1,2,\dots,\widehat{L})$, $\widetilde{P}_i(t')(i=1,2,\dots,N_A,t'=1,2,\dots,\widehat{L})$ as the original time series of carbon flux, geopotential height, temperature, wind speed and precipitation at grid point $i$, respectively. Here $N_A$ is the number of nodes and $\widehat{L}=365\times20$ (days) (February 29 has been removed). We removed the strong seasonal trends by de-trending the original data. We achieved this by subtracting the corresponding centered 30-day ``moving" averages from the time series of each variable, and normalizing it by the standard deviation of these 30 days \cite{leung2018synoptic, tai2010correlations}. This process helps to reduce the trends and noise impacts which is useful for evaluating the underlying cross-correlations representing the strength of the links. According to the above steps, the carbon flux, geopotential height, temperature, wind speed, and precipitation time series for each grid point are obtained and expressed respectively as $C_i(t')$, $G_i(t')$, $T_i(t')$, $W_i(t')$, and $P_i(t')$.

Next, we consider the grid sites as the nodes of the network, and the cross-correlations between the time series of different pairs of the grid sites are regarded as the links of the network. Using network science methods \cite{yamasaki2008climate}, we study the cross-correlations within the same climatic variables, thereby measuring the interactions and relations between different locations.

\begin{figure}[htbp]
    \centering
    \includegraphics[width=18em, height=12em]{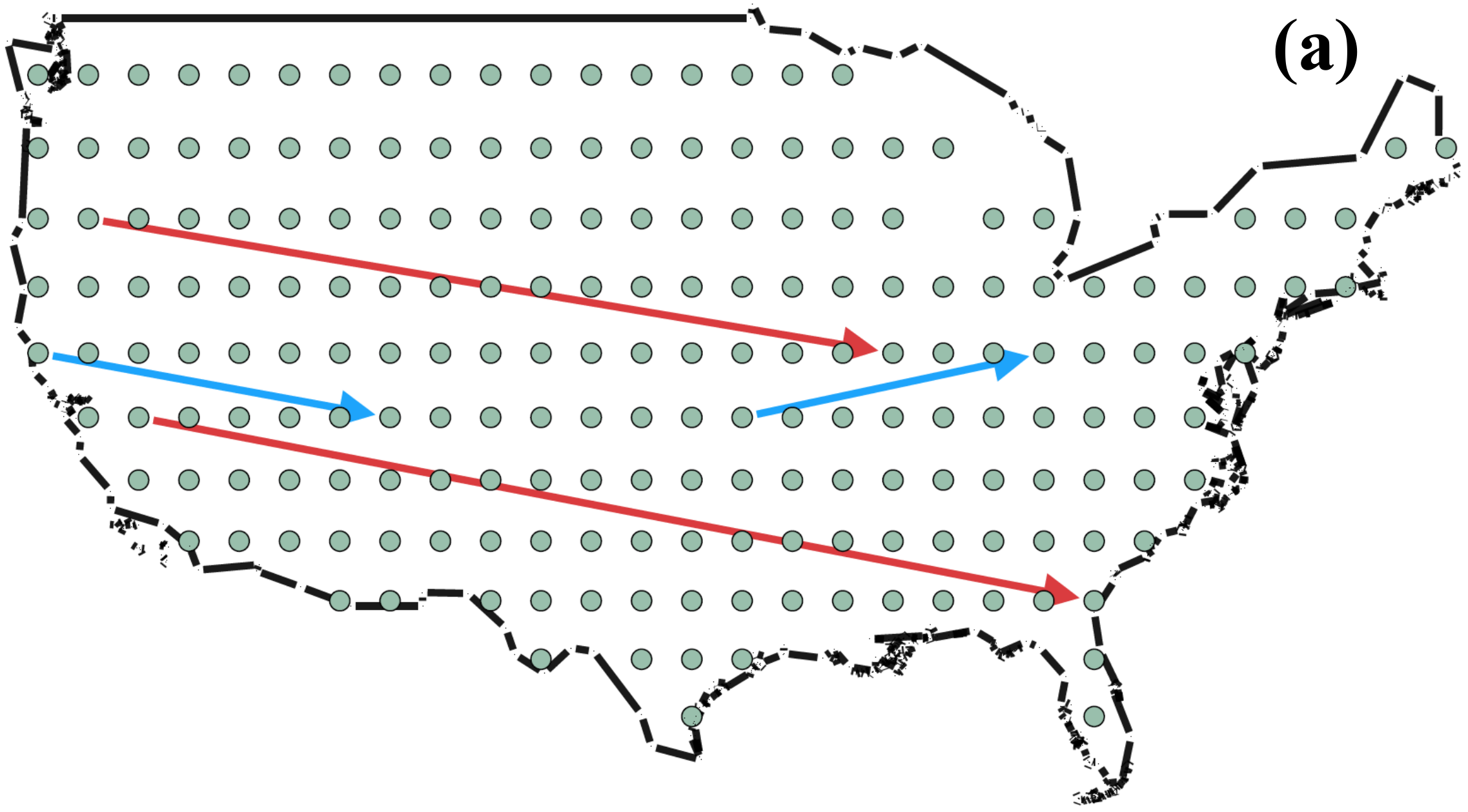}
    \includegraphics[width=14em, height=12em]{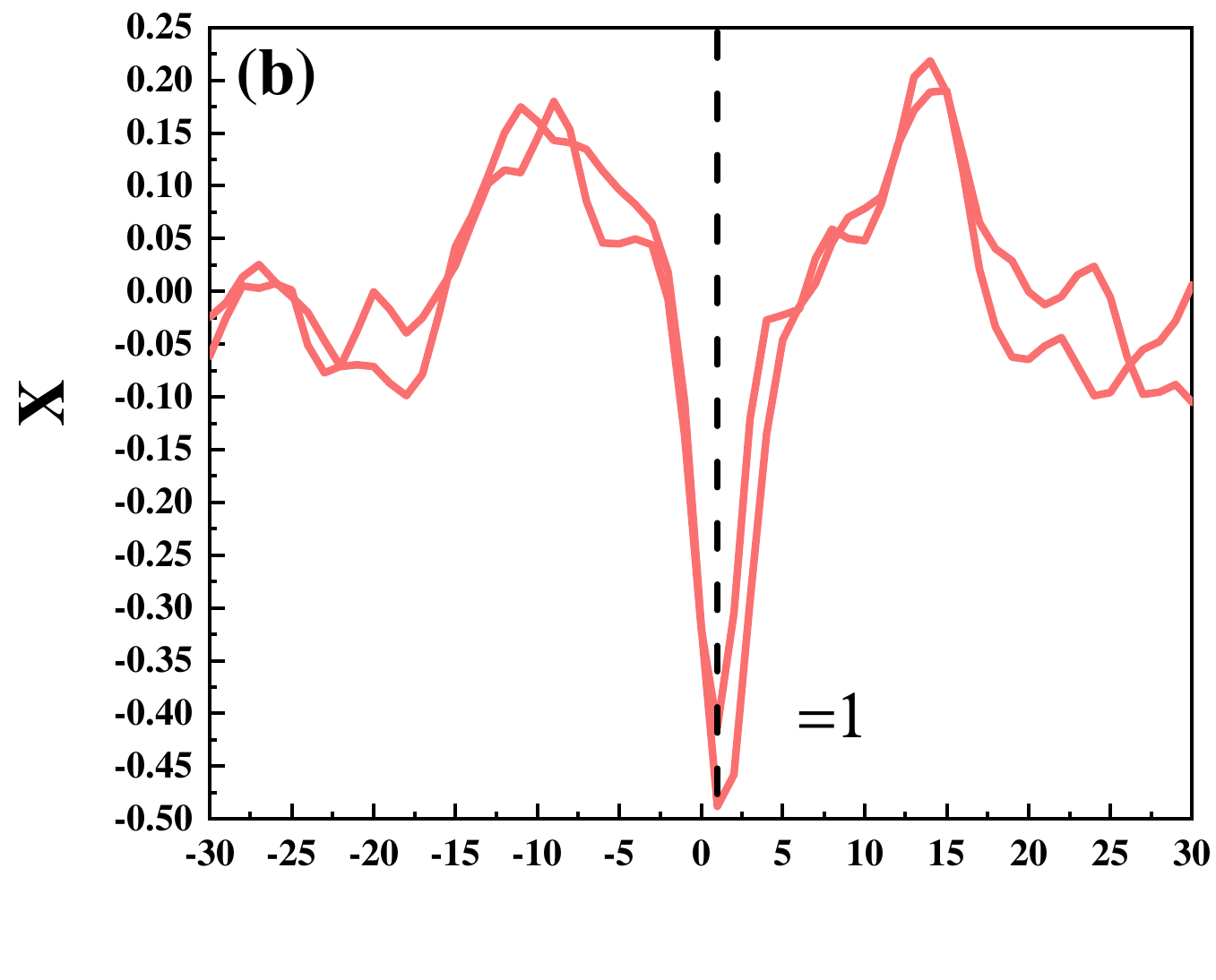}
    \includegraphics[width=14em, height=12em]{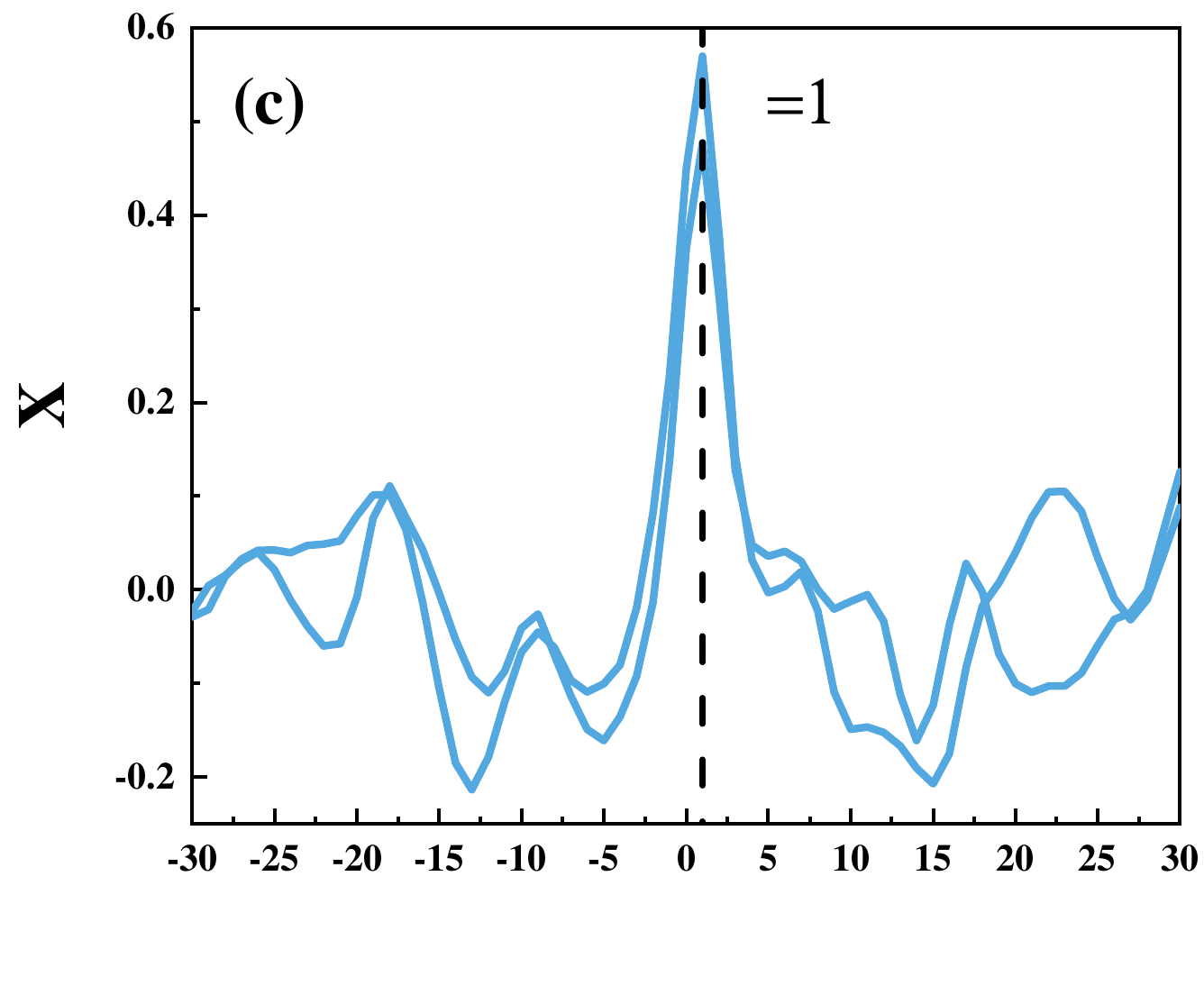}
    \caption{(a) An example for demonstration of four carbon flux network links based on 2017-2018 carbon flux data from the USA. Red arrows (direction of the link) represent negative correlated links, and blue arrows are positively correlated links. (b,c) Shown are the cross-correlation functions of the negative weights links and positive weights links (red and blue curves) for (a), respectively.}
    \label{fig1}
\end{figure}

The correlation pattern, i.e., the network structure, of the system evolves with time, and to avoid losing important information, we apply the sliding window method by dividing the entire time series into smaller fragments. By doing so, we can obtain a series of interaction networks that capture the system's evolving correlations pattern. One advantage of using the sliding window method is that the sliding windows have the properties of memory and transitivity, which are important in studying the correlation patterns of a system over time \cite{wang2016research}. Here, we consider the length of the moving window to be $L$ ($L=365\times2+90$ days), and the moving step is 365 days. Thus the time series of climate and carbon flux are divided into 18 segments of length $L$ (due to the 365 days overlap). Within the $t$th window, we compute the correlation matrix $X(t)$, in which the element $X_{i,j}$ is the cross-correlation representing the links that connect node $i$ and node $j$. First, the time-delayed cross-correlation function between the two time series ${Y_i(t)}$ and $Y_j(t)$ is calculated, see \cite{zhou2015teleconnection, ying2020rossby, gozolchiani2008pattern,guez2014influence, wang2013dominant},
\begin{equation}\label{eq1}
X_{i,j}(\tau)=\frac{\langle Y_i(t)Y_j(t+\tau)\rangle-\langle Y_i(t)\rangle \langle Y_j(t+\tau)\rangle}{\sigma_{Y_i(t)}\sigma_{Y_j(t+\tau)}},
\end{equation}
where $\langle Y_i(t)\rangle$ and $\sigma_{Y_i(t)}$ represent the average and standard deviation of $Y_i(t)$, respectively. $\tau\in[-\tau_{max},\tau_{max}]$ is the time lag, with $\tau_{max}=90$ days. The choice of $\tau_{max}$ helps to ensure a reliable estimation of the seasonal noise level in the cross-correlation. According to the time inversion symmetry, we can obtain $X_{i,j}(-\tau)=X_{i,j}(\tau)$ \cite{du2019identifying}. We identify the value of the highest peak of the absolute value of the cross-correlation function and denote the corresponding time lag of this peak to be $\tau^{\ast}$. The correlation between node $i$ and node $j$ is thus $X_{i,j}=X_{i,j}(\tau^{\ast})$. The direction of the link $X_{i,j}(\tau^{\ast})$ is from $i$ to $j$ when $\tau^{\ast}>0$ and from $j$ to $i$ when $\tau^{\ast}<0$. The direction is undefined when $\tau^{\ast}=0$. Within each window, we quantify the strength or weight of the links (correlations) using \cite{fan2017network, yamasaki2008climate}:
\begin{equation}\label{eq3}
w_{i,j}=\frac{X_{i,j}(\tau ^{\ast})-\overline{X_{i,j}(\tau)}}{\sigma_{X_{i,j}(\tau)}},
\end{equation}
where $\overline{X_{i,j}(\tau)}$, $\sigma_{X_{i,j}(\tau)}$ represent the mean and standard deviation over all $\tau$ values of the cross‐correlation function, respectively. Based on the above steps, we construct a series of time-dependent climate and carbon flux networks. A demonstration of 2017-2018 carbon flux data in the USA as an example to establish the links, is shown in Fig. {~\ref{fig1}}. In Figs. {~\ref{fig1}}(b) and {~\ref{fig1}}(c), one can see that the peak point of the correlation coefficient of its links corresponds to $\tau^{\ast}=1$, which implies that the time delay between these locations of carbon flux is 1 day from west to east.

\subsection{Significant tests}
\begin{figure}[htpb]
    \centering
    \includegraphics[width=9em, height=7.5em]{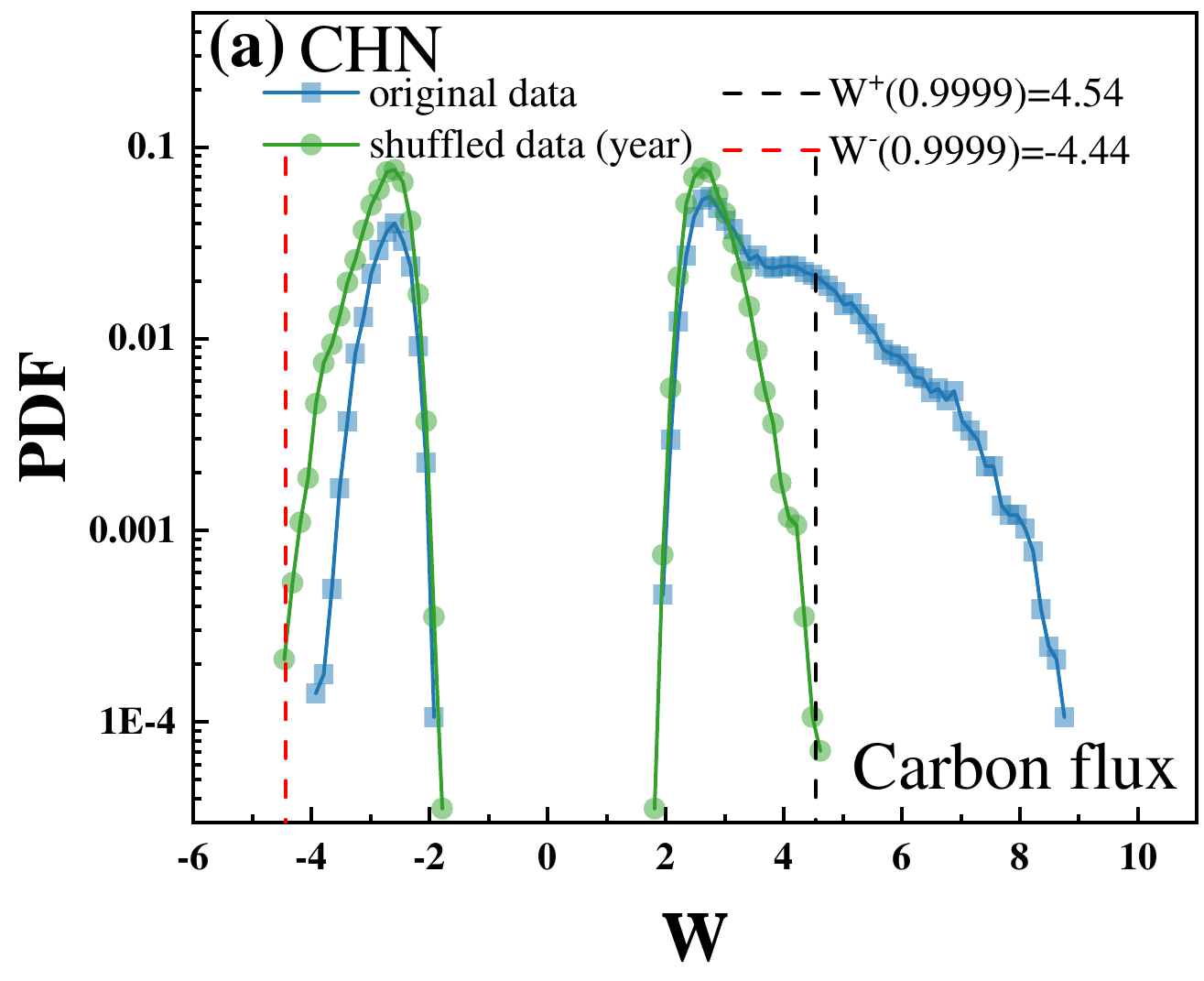}
    \includegraphics[width=9em, height=7.5em]{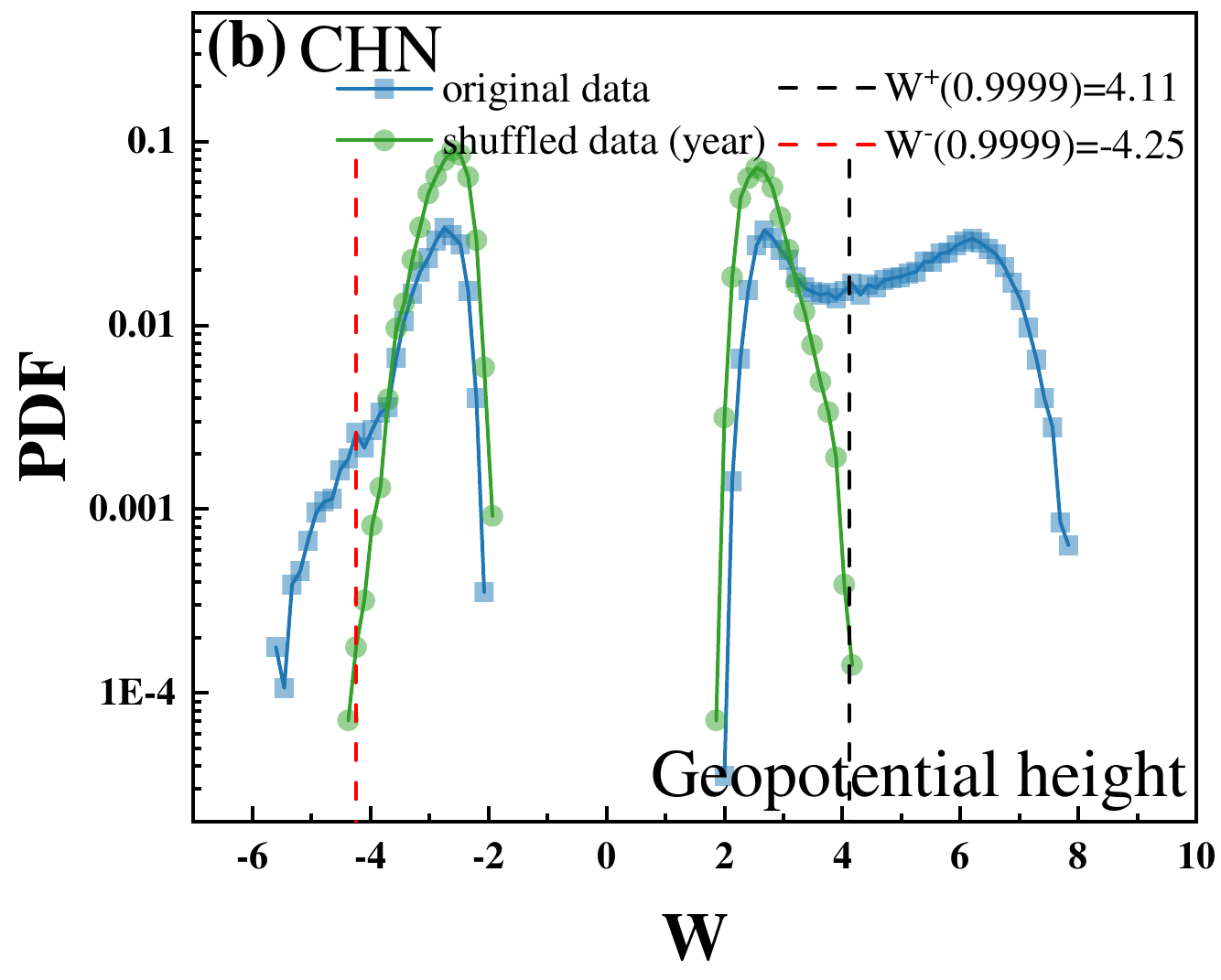}
    \includegraphics[width=9em, height=7.5em]{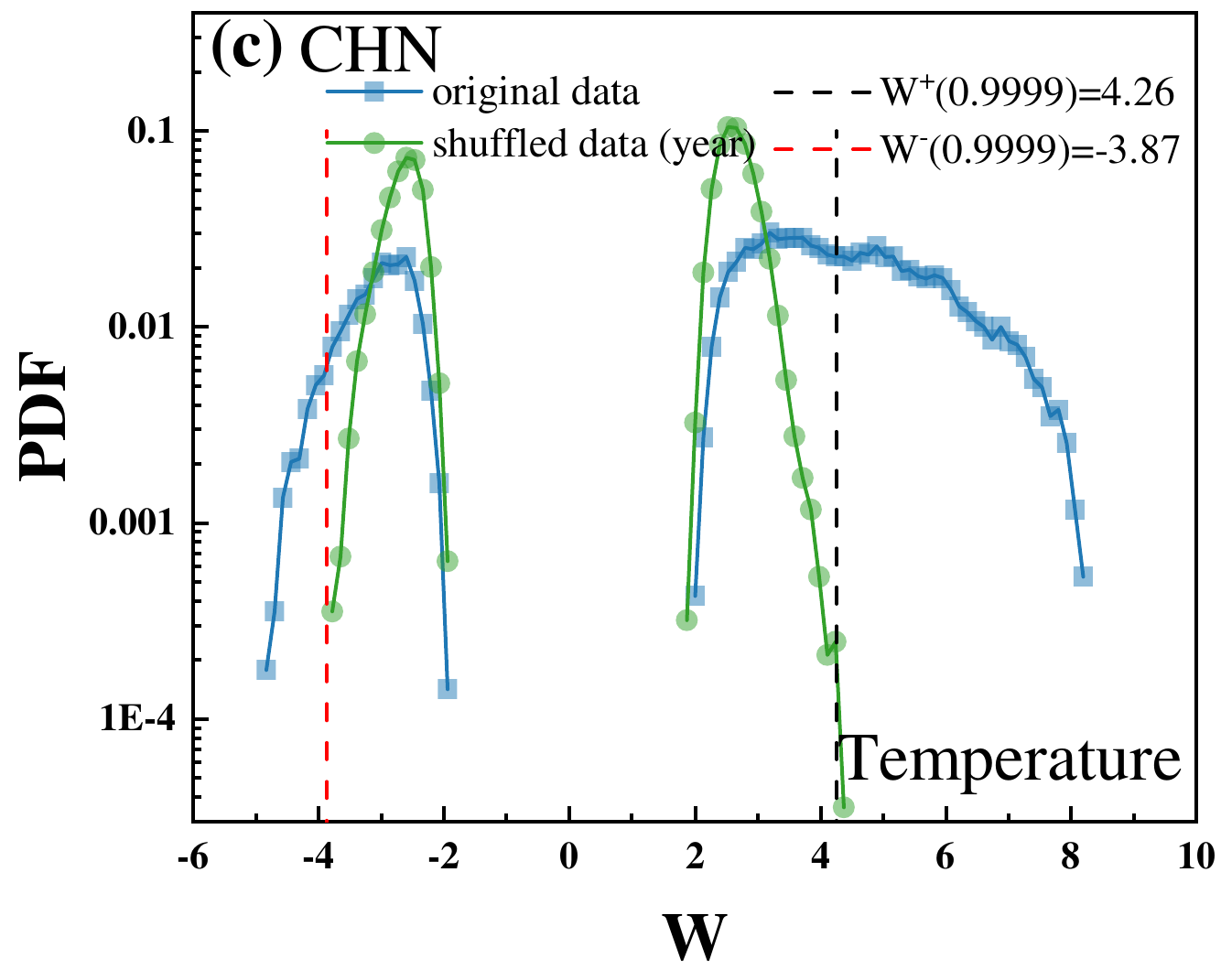}
    \includegraphics[width=9em, height=7.5em]{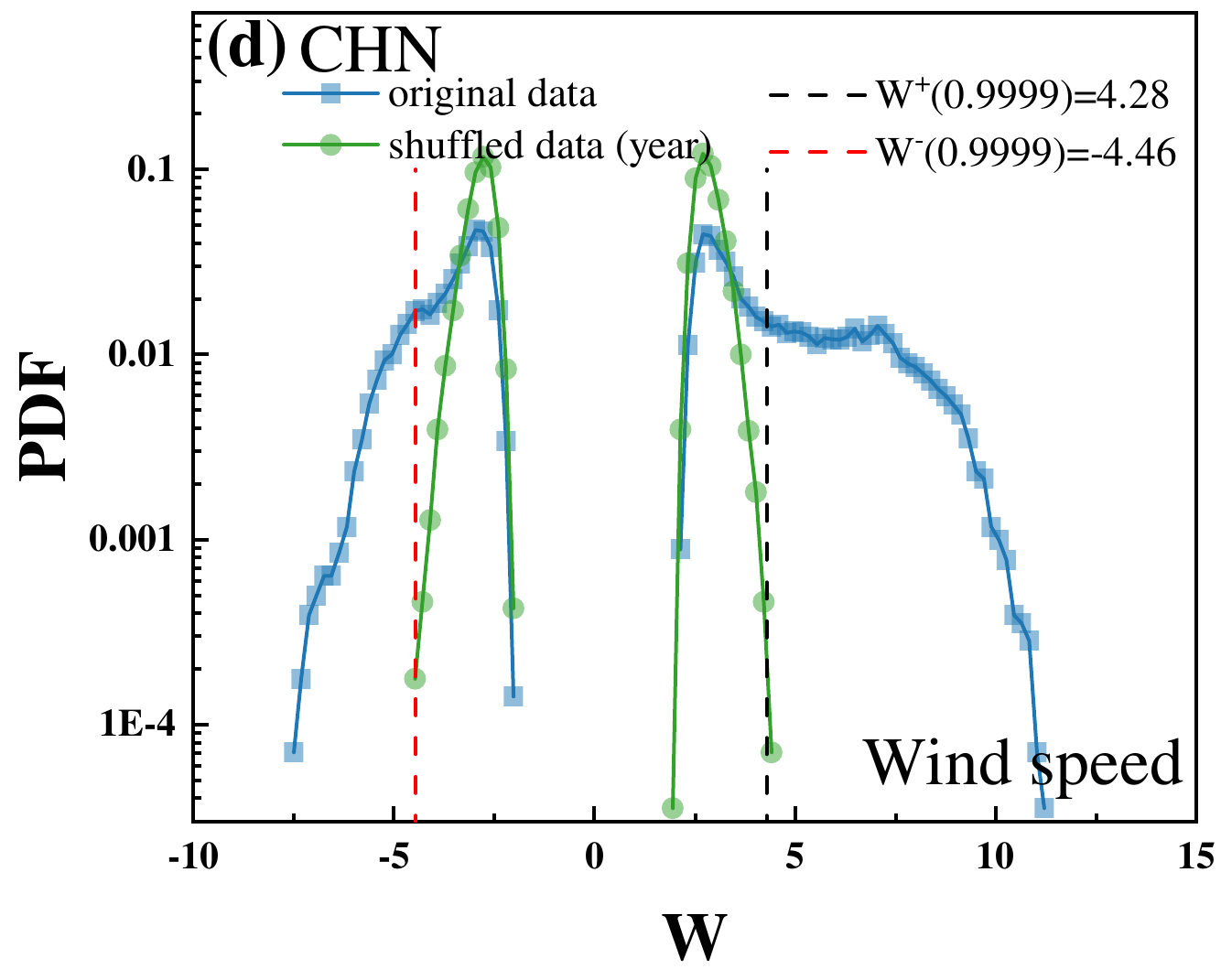}
    \includegraphics[width=9em, height=7.5em]{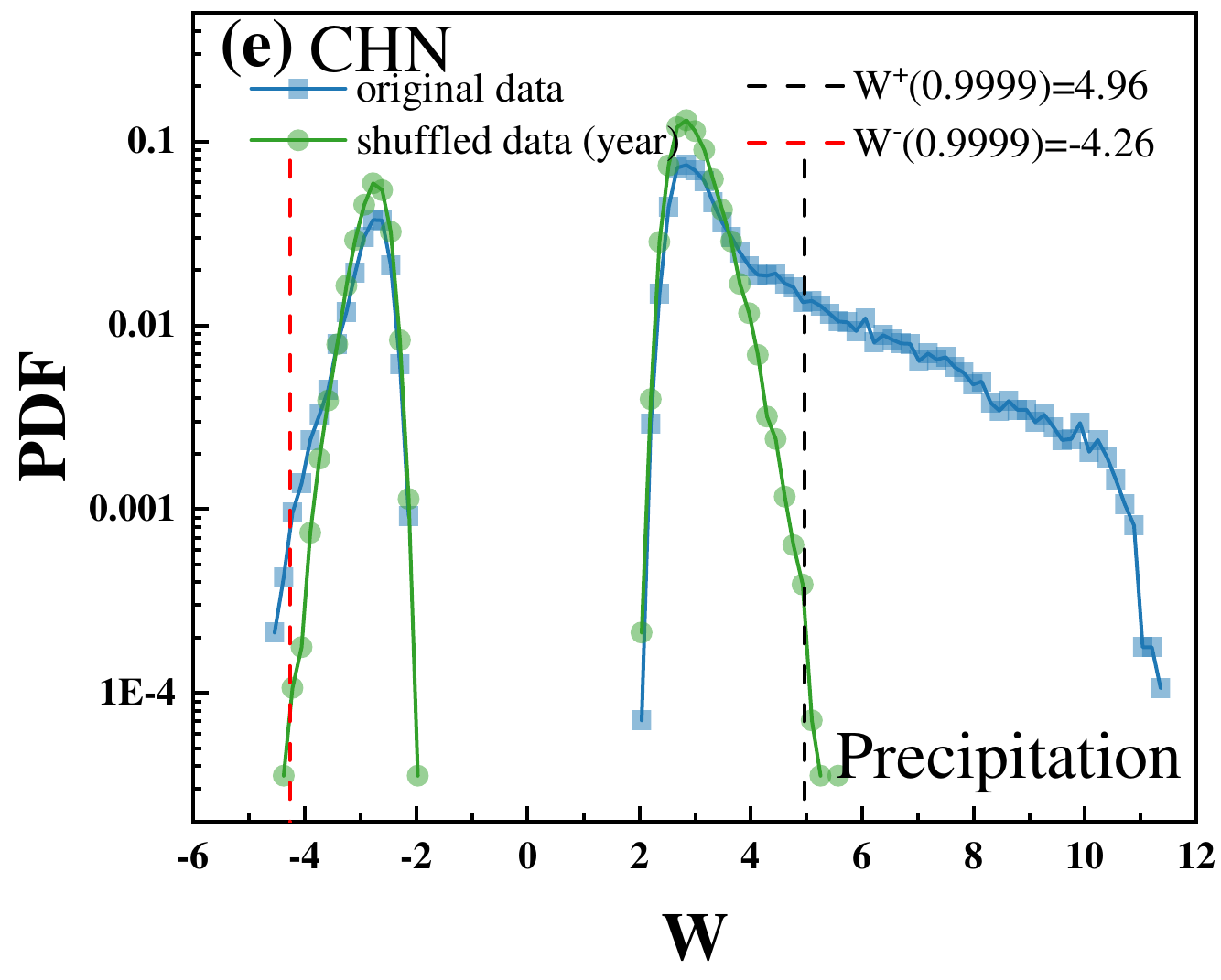}
    \includegraphics[width=9em, height=7.5em]{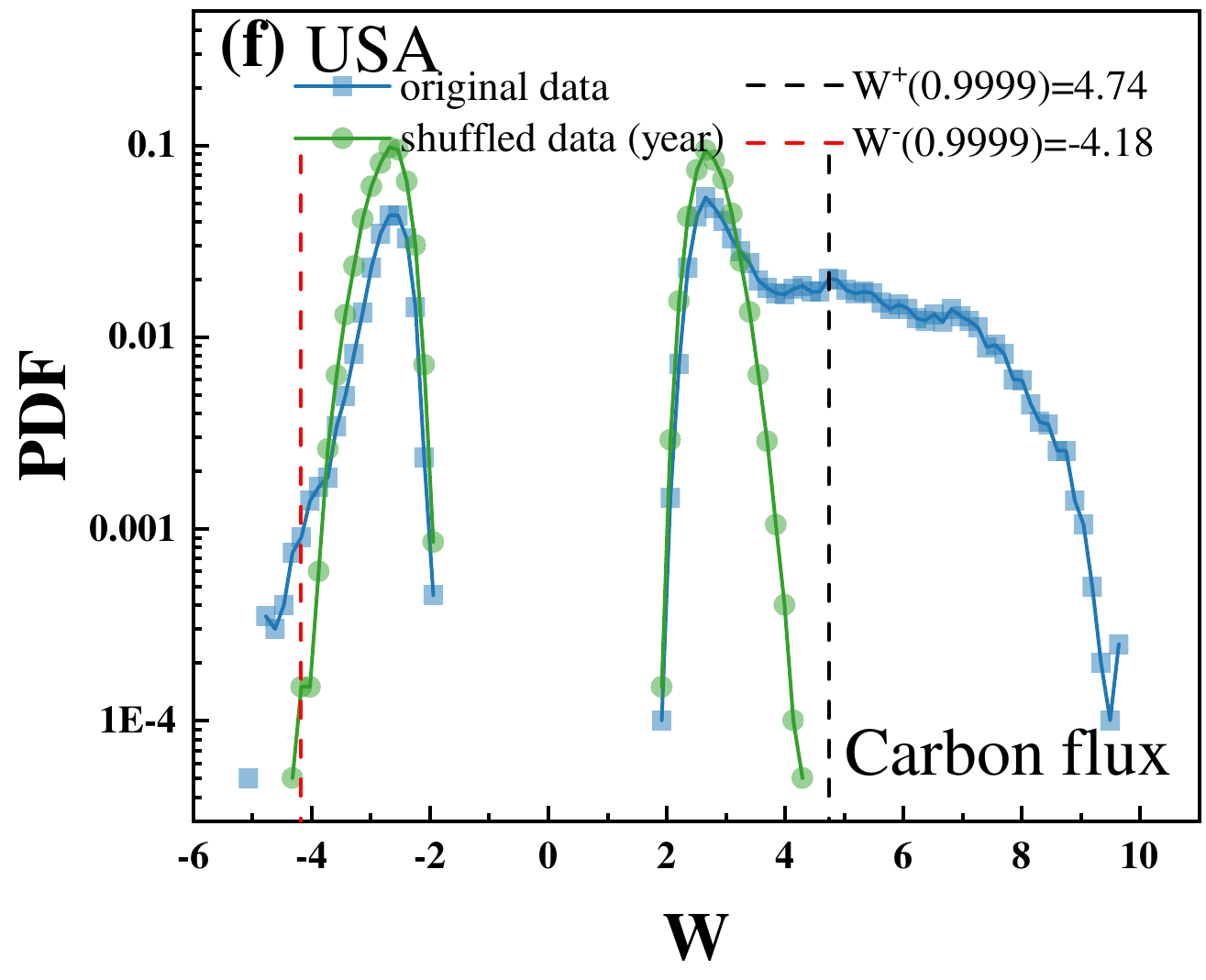}
    \includegraphics[width=9em, height=7.5em]{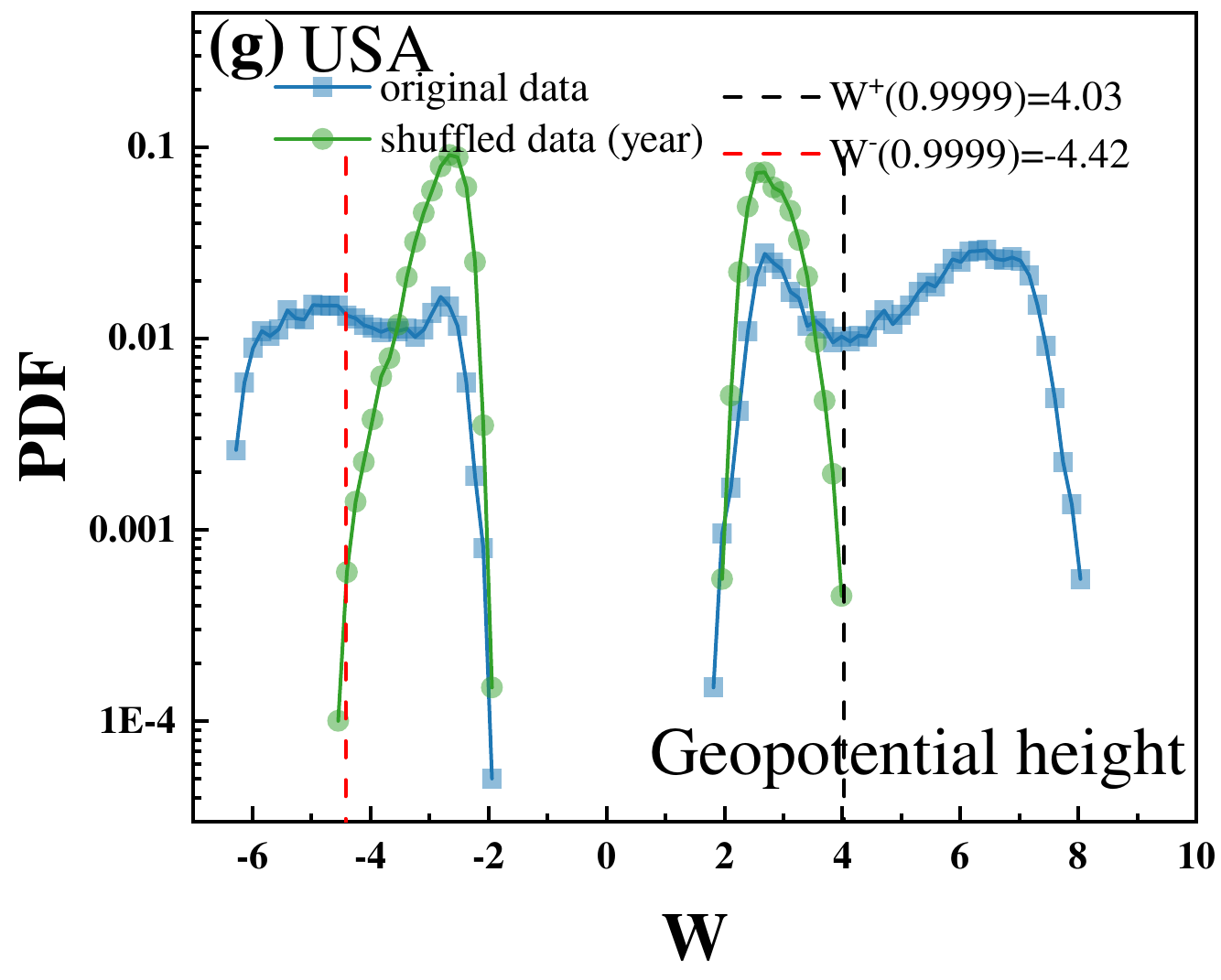}
    \includegraphics[width=9em, height=7.5em]{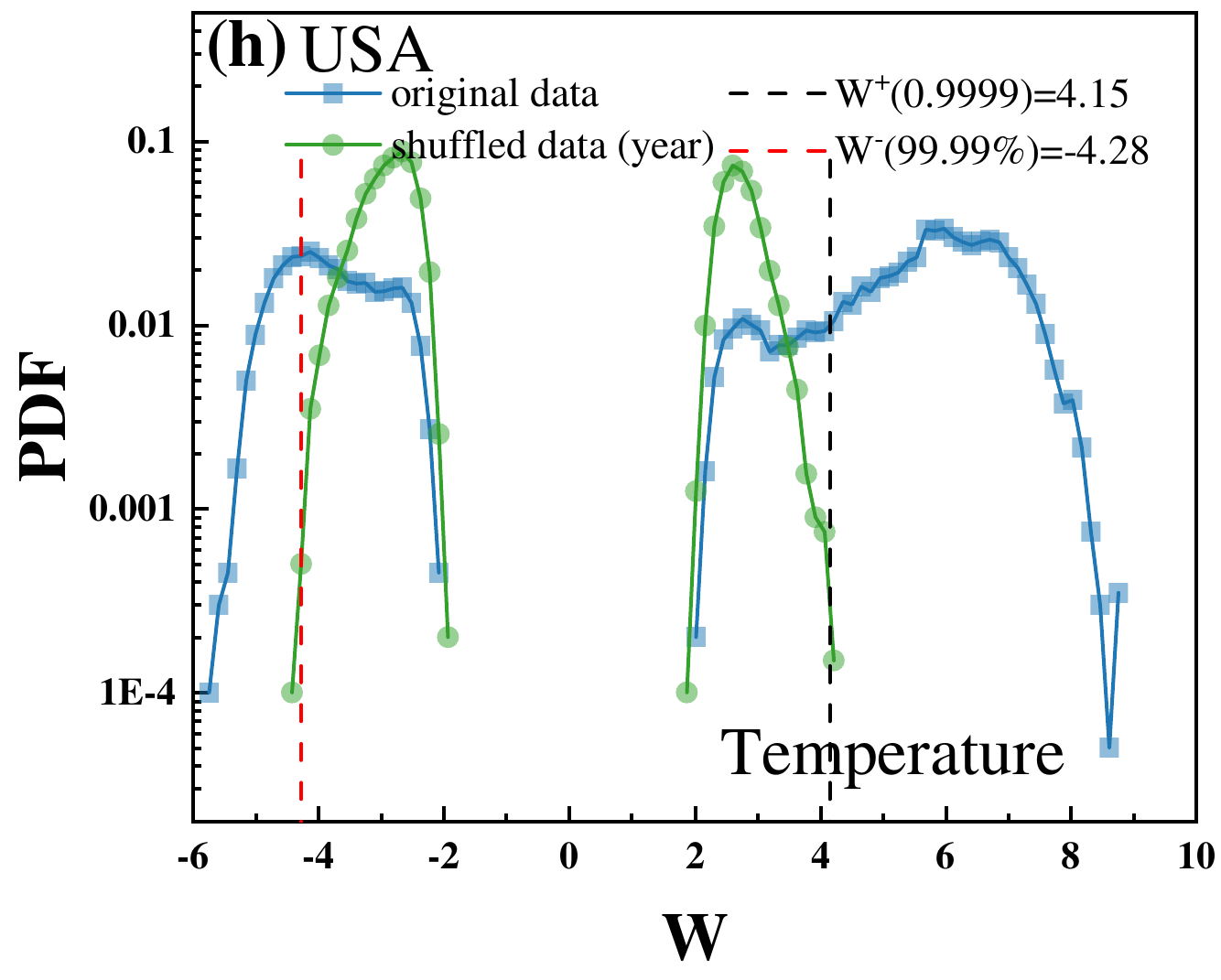}
    \includegraphics[width=9em, height=7.5em]{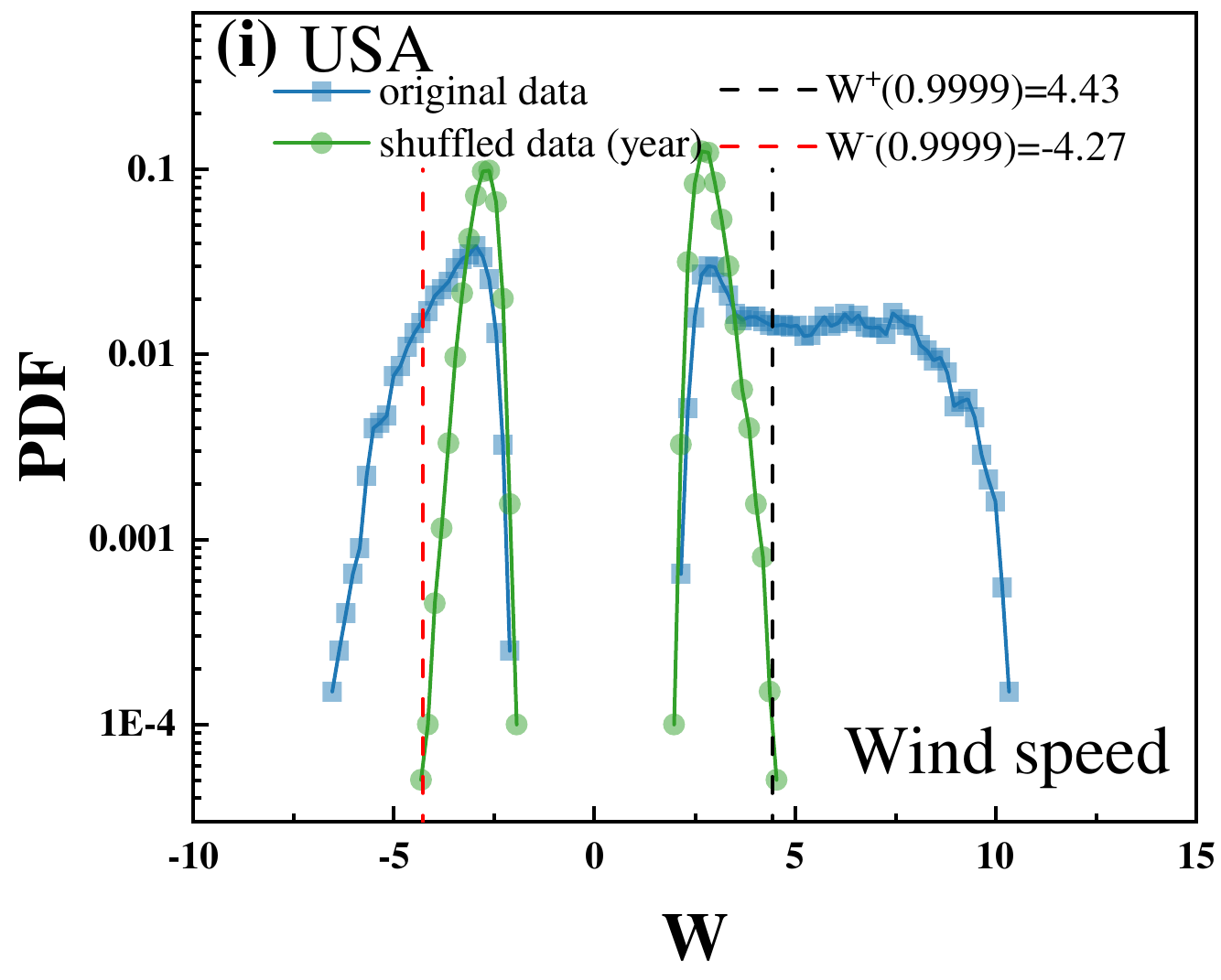}
    \includegraphics[width=9em, height=7.5em]{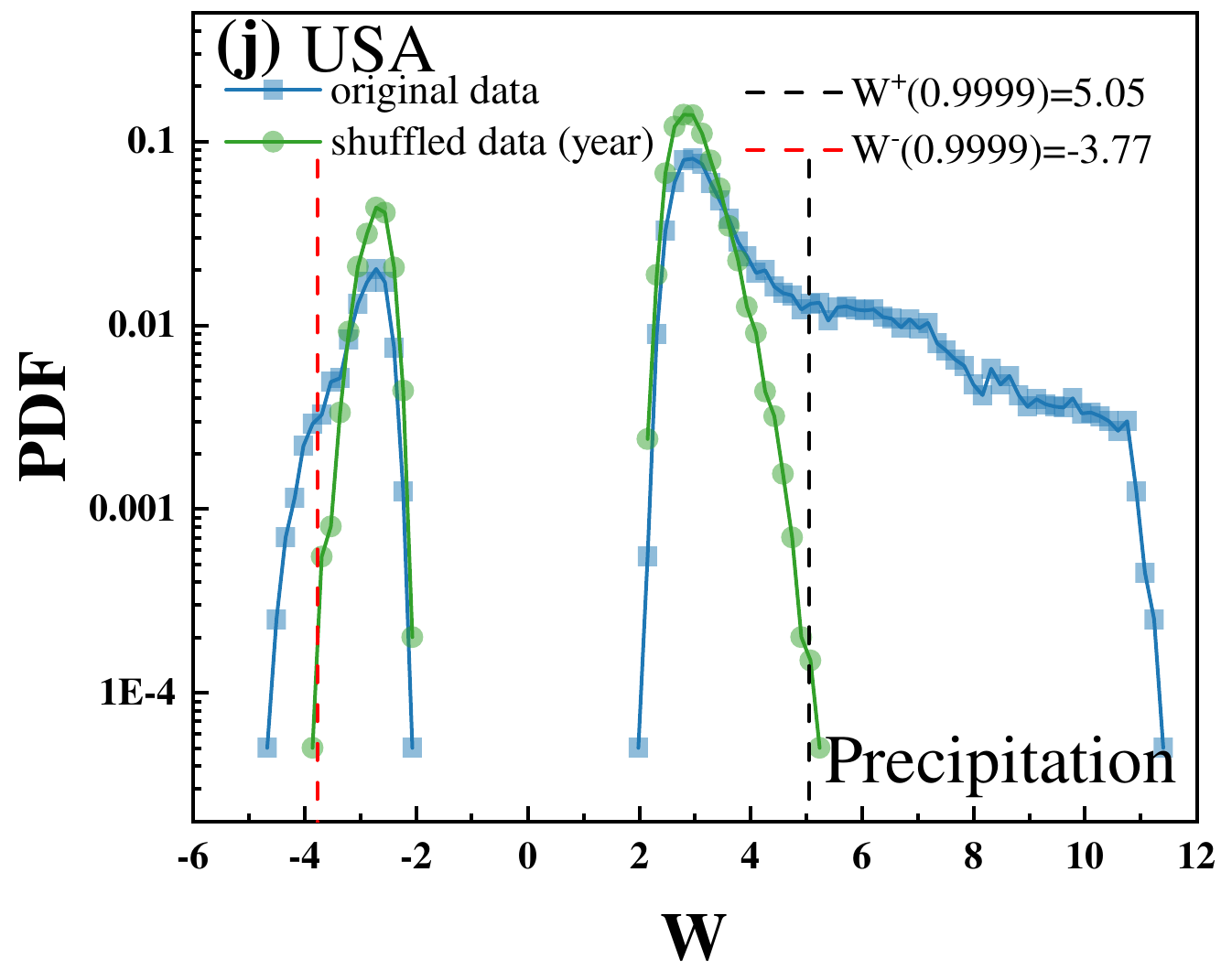}
    \includegraphics[width=9em, height=7.5em]{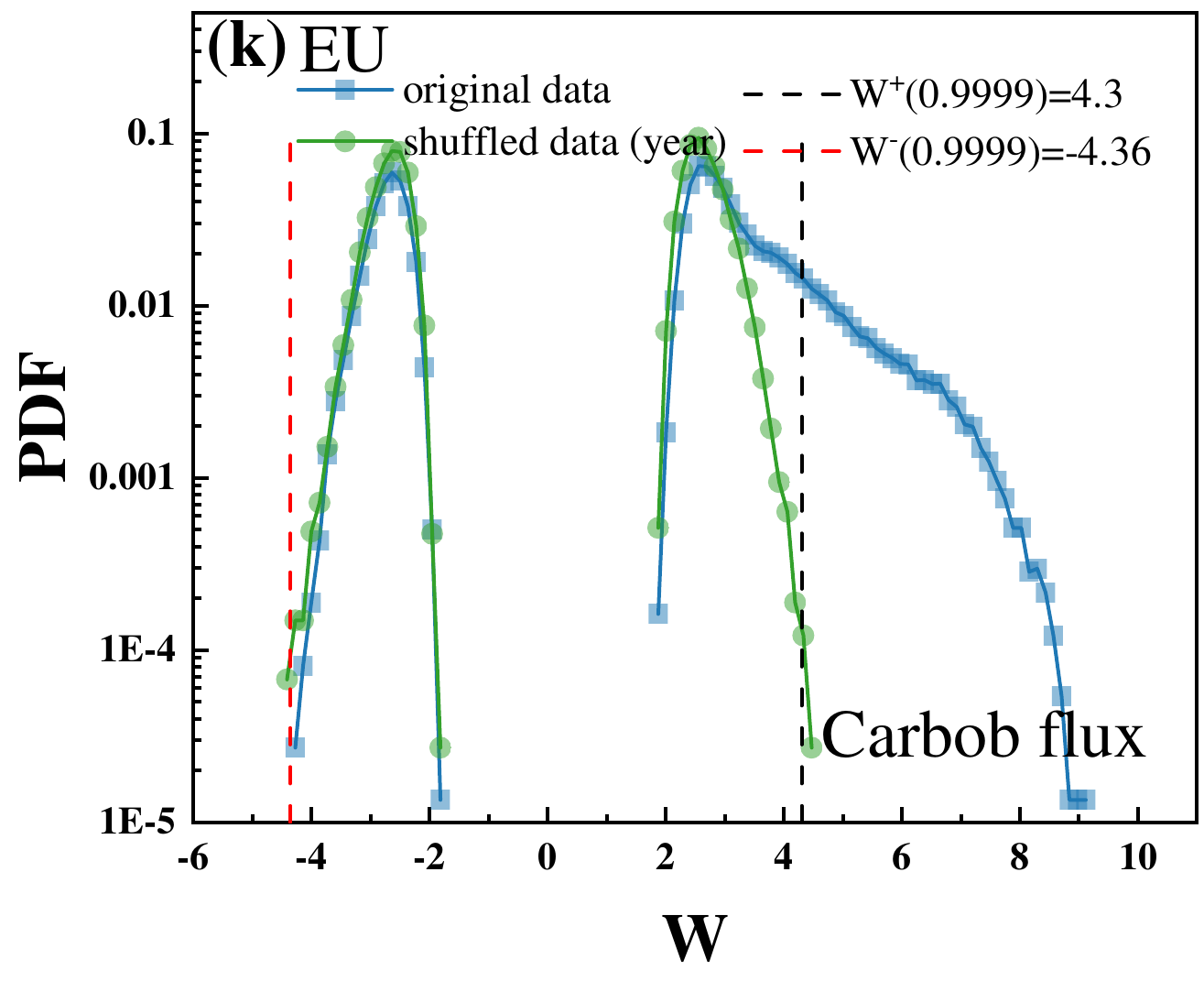}
    \includegraphics[width=9em, height=7.5em]{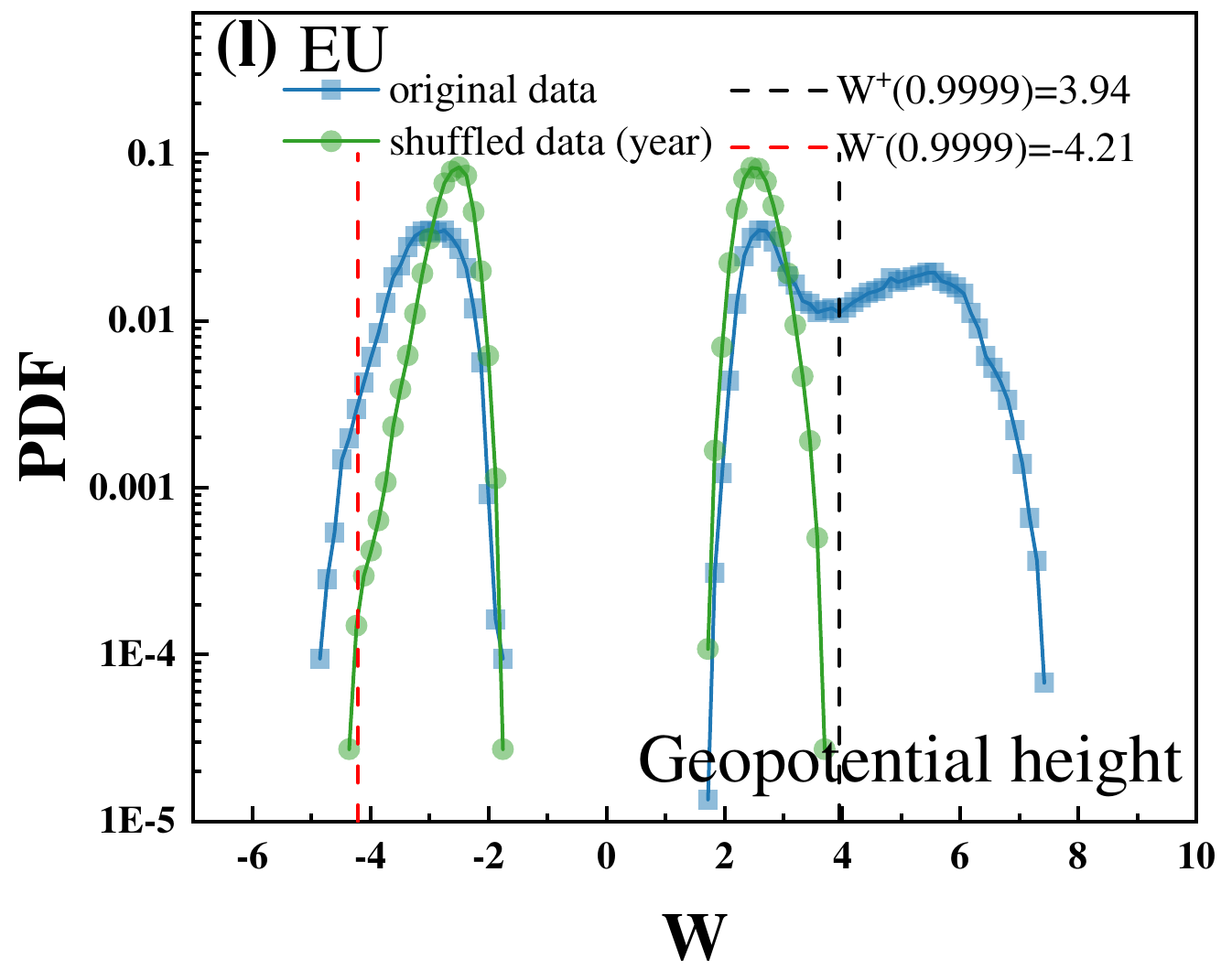}
    \includegraphics[width=9em, height=7.5em]{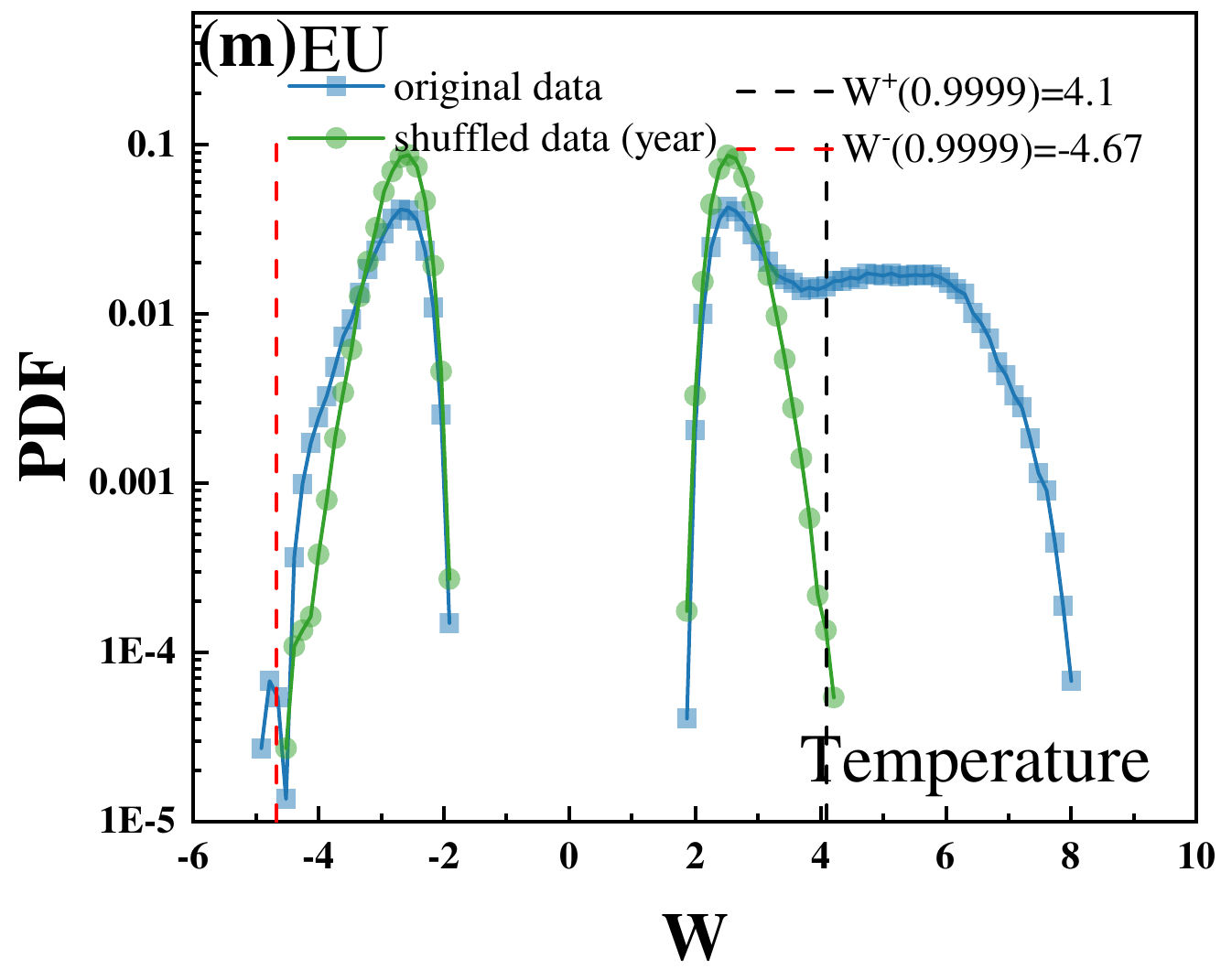}
    \includegraphics[width=9em, height=7.5em]{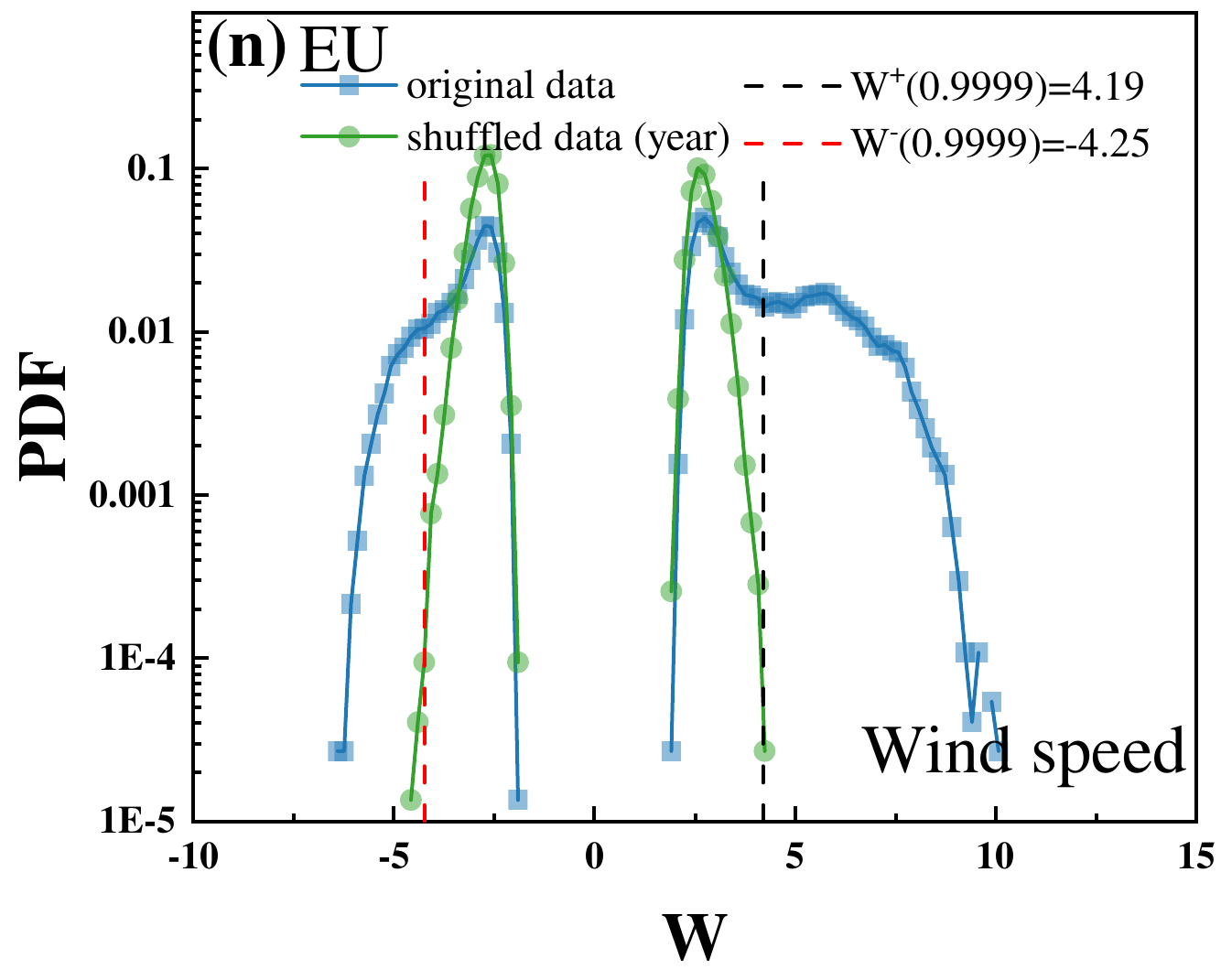}
    \includegraphics[width=9em, height=7.5em]{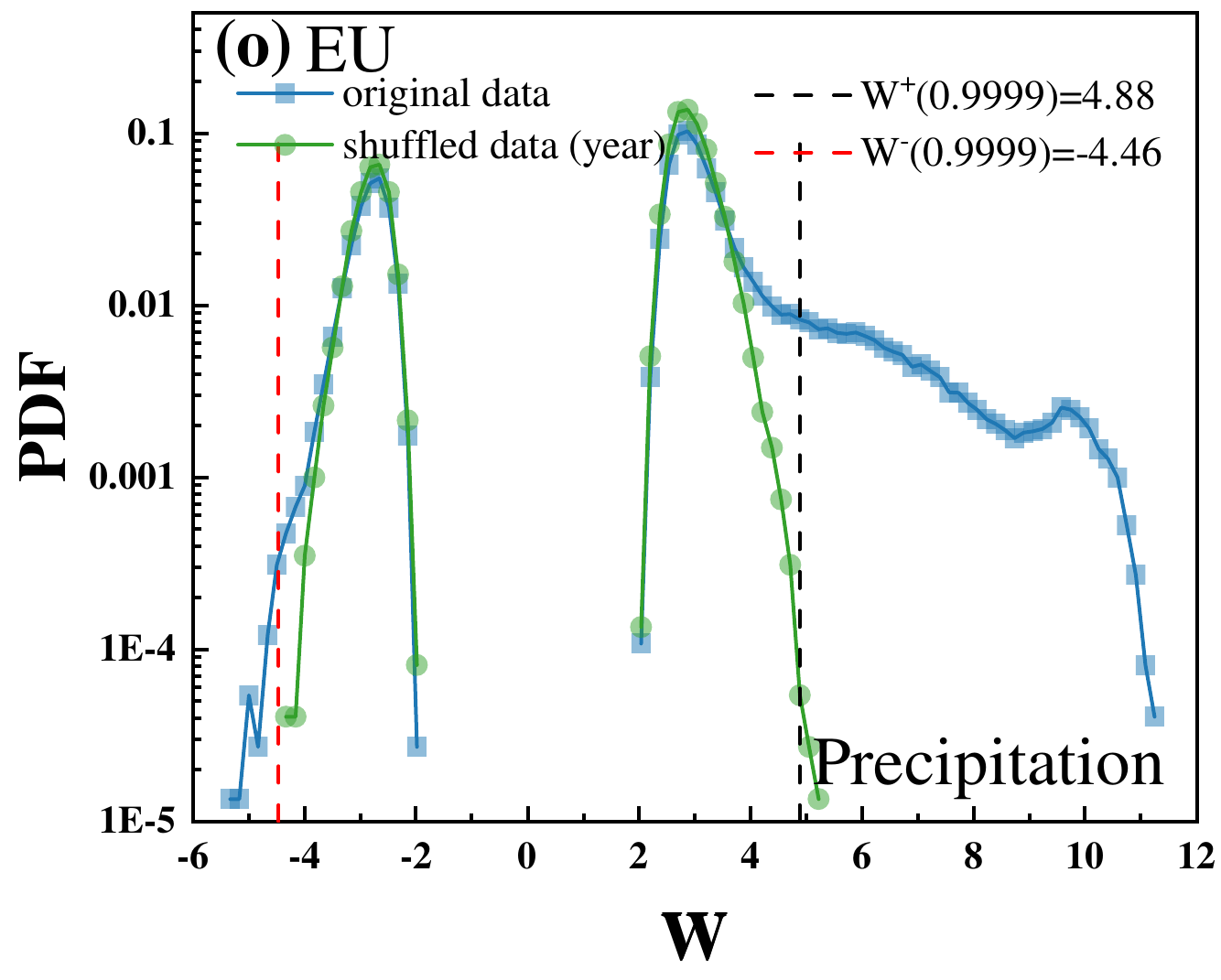}
    \caption{Probability distribution function (PDF) of link weights for the real time window data and for the random shuffled data for the years 2017-2018. The blue color points is the PDF of link weight of the real records, the green color is the PDF of the link weight for the shuffled data, and the dashed line is the threshold of link weight above which the chance that the links obtained from the real data is less than 0.01\% being random and false.}
    \label{fig2}
\end{figure}

In this study, a shuffling procedure is used to test the statistical significance of the calculated link weights. For each pair of climate variables nodes $i$ and $j$, we calculate the correlation coefficients between the $t$th window for node $i$ and the $(t-2)$th window of node $j$ as the shuffled procedure, and obtain the link strengths based on shuffled correlation coefficients (SI Fig. S1 illustrates that the interval years ($(t-Interval)$th time window) as a function of the threshold in a shuffling procedure presents a relatively stable trend. It is worth noting that we consider the length of the moving window to be $L$, which further verifies the reliability of the $(t-2)$th time window in the shuffling procedure, and ensures that there is no overlap of the window data at the $(t-2)$th time window with the $t$th). In this shuffling approach, we preserve the distribution of values and the natural auto-correlations in each year for each shuffled record. If the original link weights are significantly higher than that of the moved control, we regard it as a real link; otherwise, it is suspected to be a spurious link. Since positive weights ($w_{i,j}^{+}$) and negative weights ($w_{i,j}^{-}$) have different properties, (of positive correlations and negative correlations respectively) critical threshold $\theta_{+}(\theta_{-})$ is set to be such that a correlation value is regarded to be real when it has a probability below 0.01\% to be random. In  Fig. ~\ref{fig2}, the dashed lines represent location of the thresholds where the real cross-correlations values appear to have probability of at least 100 times than that of random. Thus, we obtain the weighted adjacency matrix $W(t)$ of the climate variables links in the $t$th time window. As an example, we present the probability distribution function of link weights for the original and shuffled data for the 2017-2018 time window in Fig.~\ref{fig2}. It is seen from Fig. ~\ref{fig2} that the threshold appears to be mostly for $W$ above close to 4,  that is, above this value of $W$, the chance of a link to be false is less than 0.01\%. Note that the fraction of links as a function of threshold also supports the reliability of this threshold since only above this threshold the fraction of links highly increases. (See SI Fig. S17). Note also, that $W=4$ means that the peak value of the shifted cross-correlation function is above 4 standard deviations of all correlation values (For W=4 as the threshold, the variation in the fraction of significant links with year is shown in SI Fig. S18). Note that similar results are seen in other years, see SI Figs. S2-S16.

\begin{figure}
    \centering
    \includegraphics[width=9em, height=7.5em]{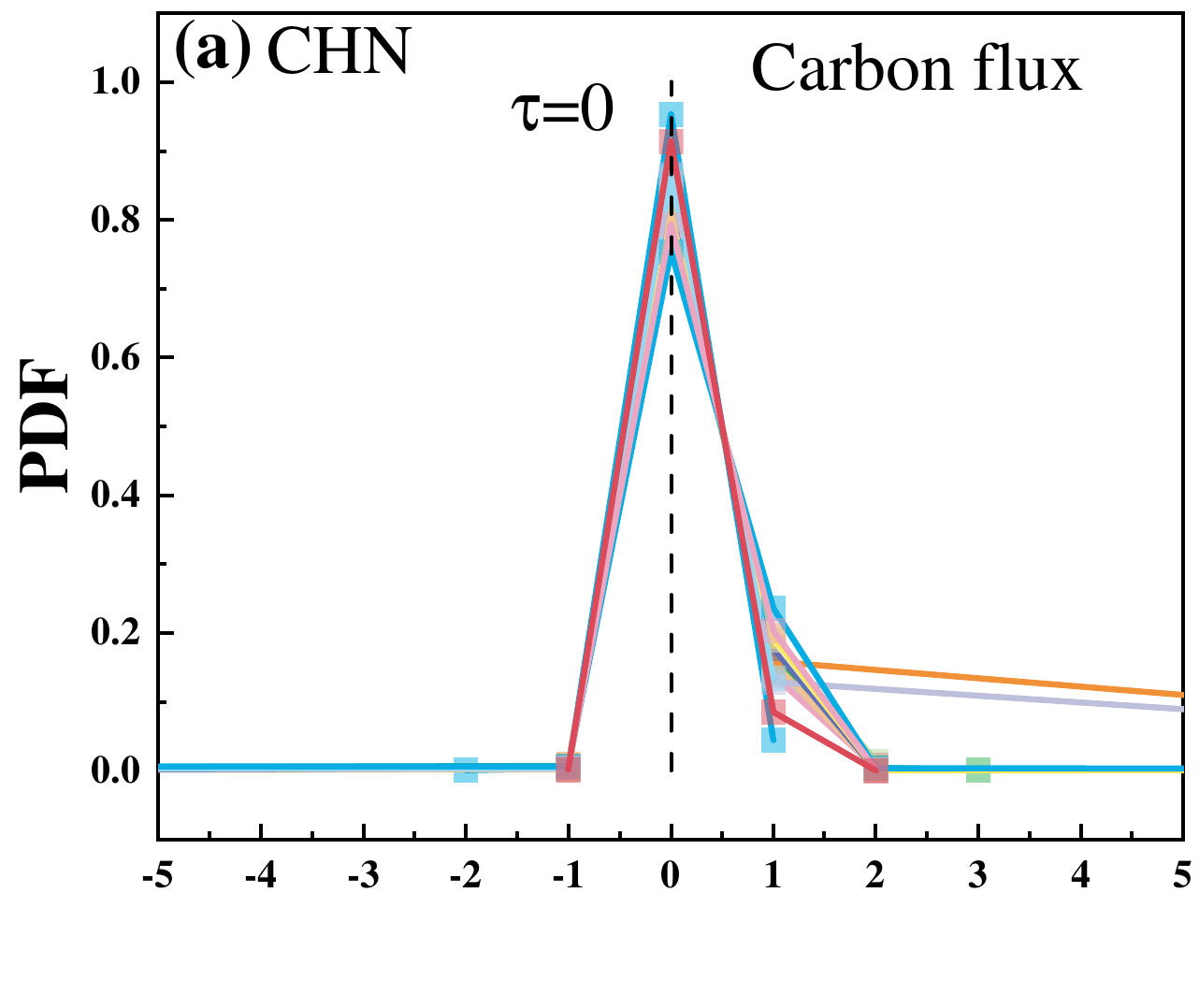}
    \includegraphics[width=9em, height=7.5em]{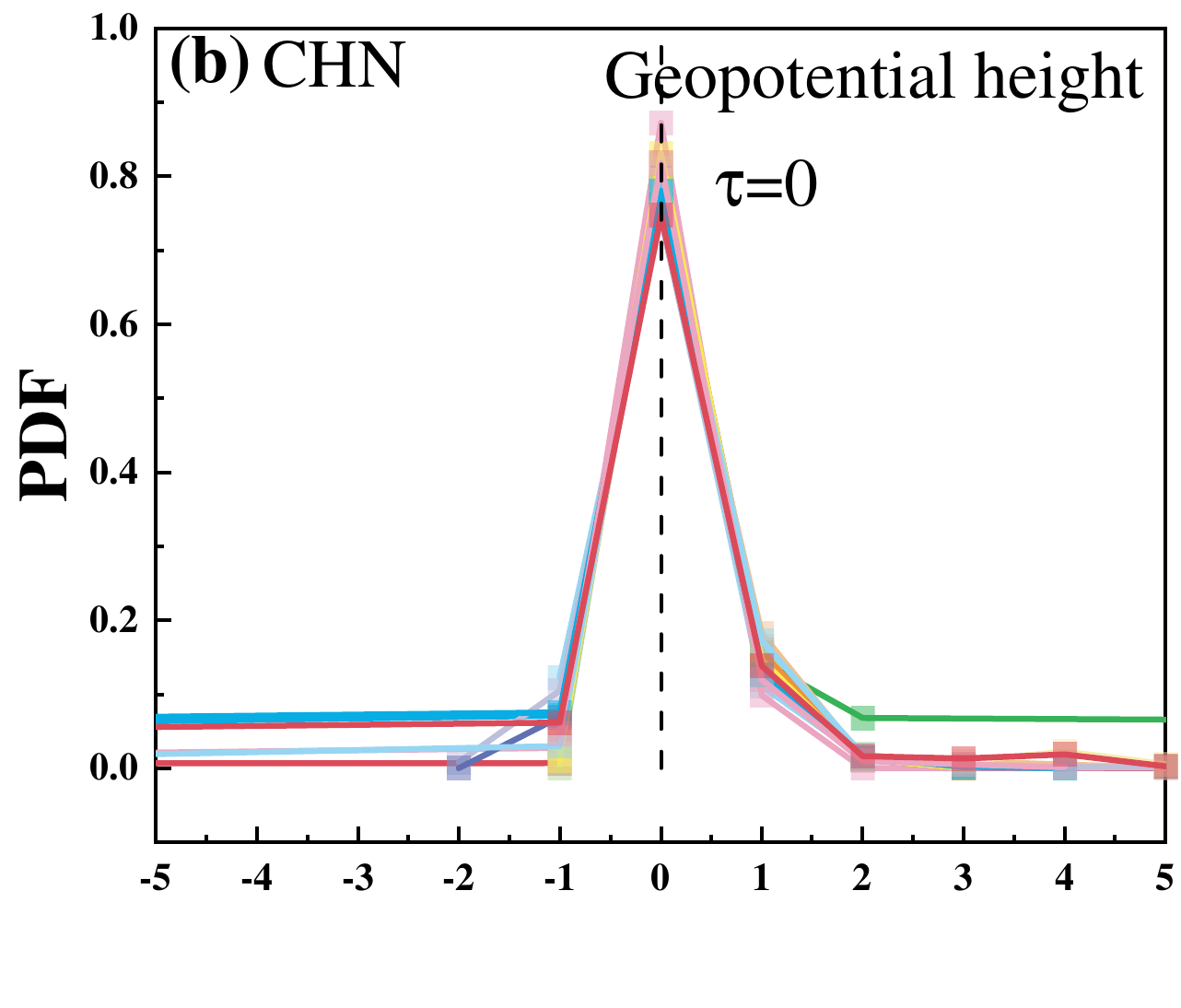}
    \includegraphics[width=9em, height=7.5em]{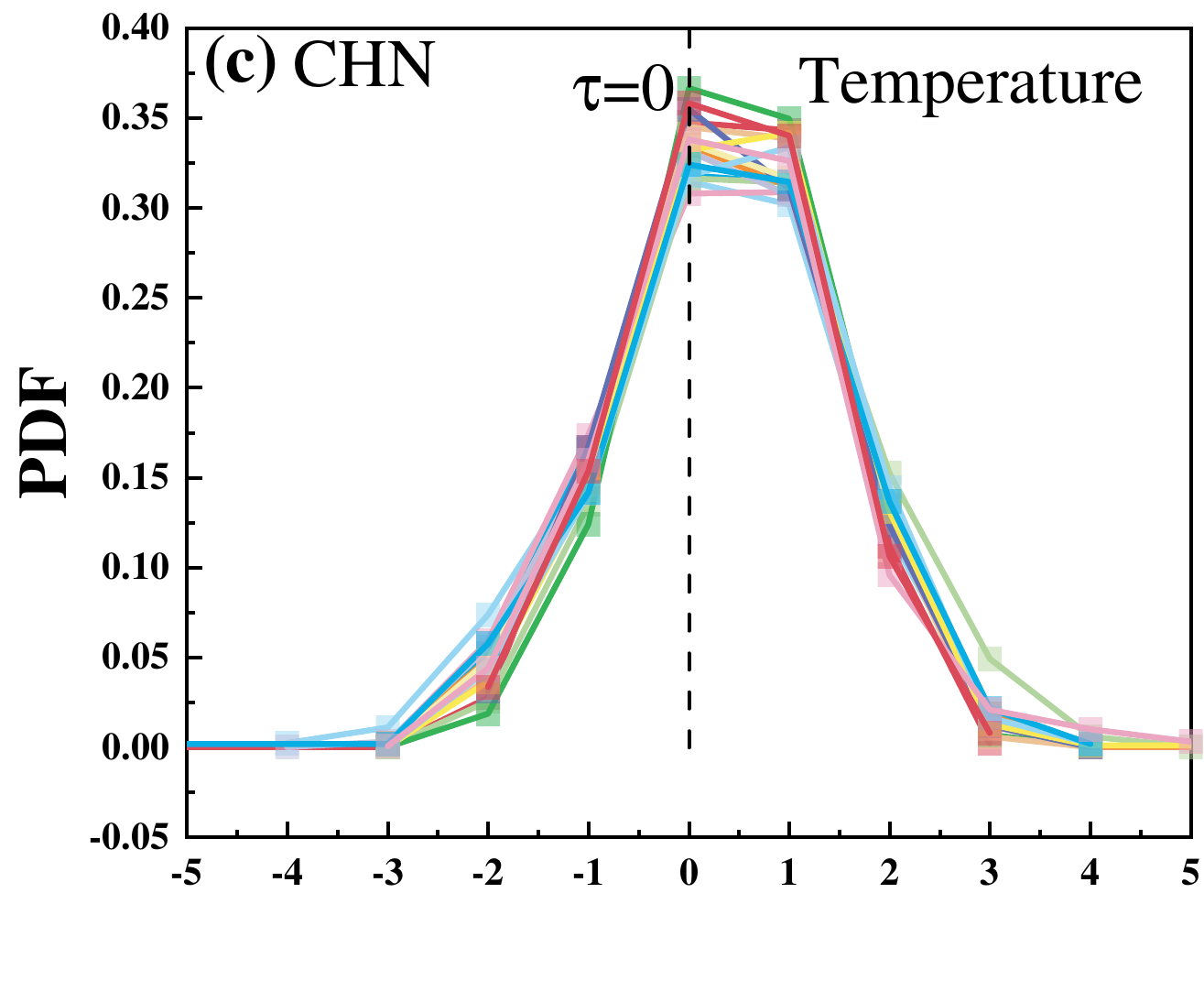}
    \includegraphics[width=9em, height=7.5em]{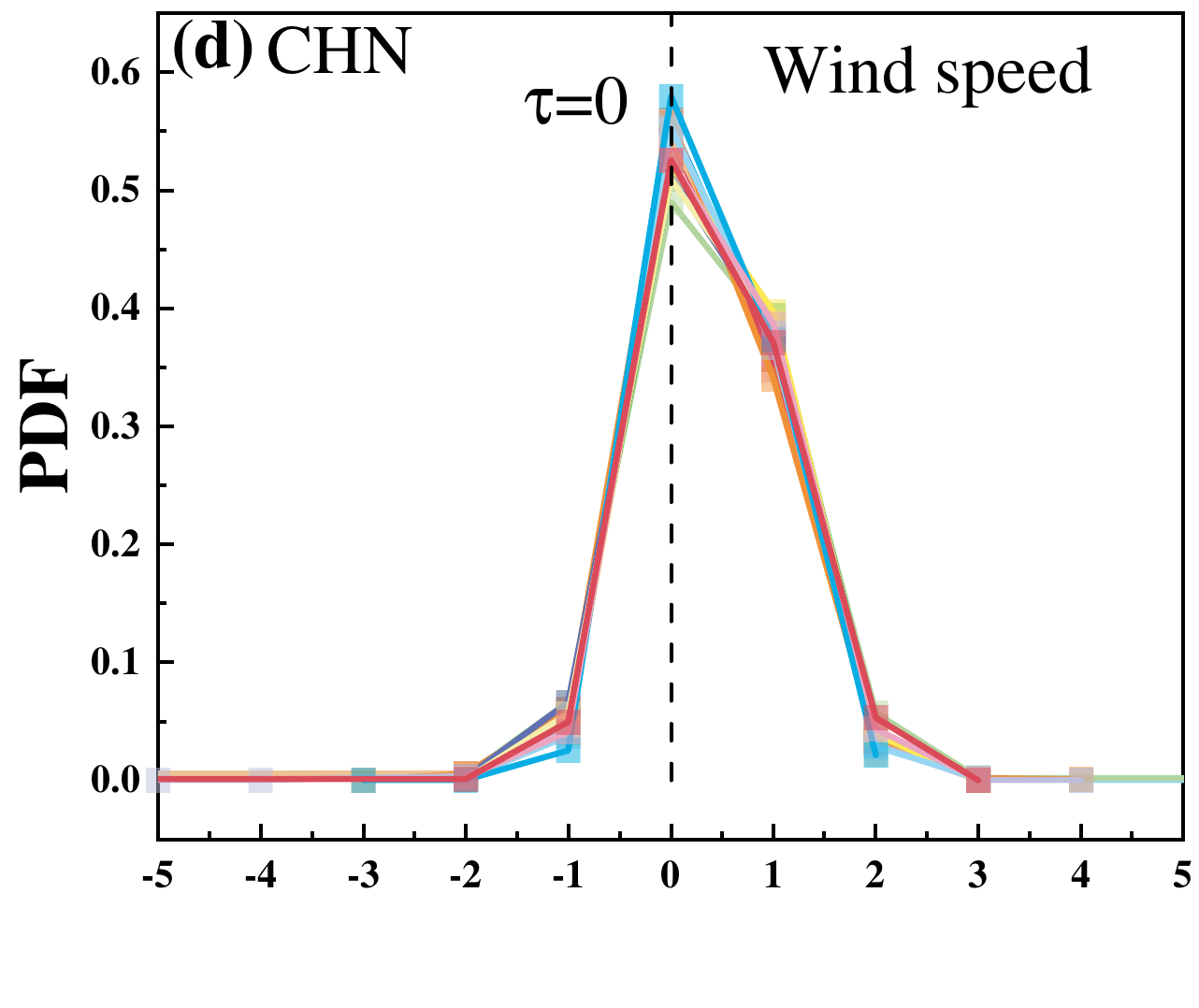}
    \includegraphics[width=9em, height=7.5em]{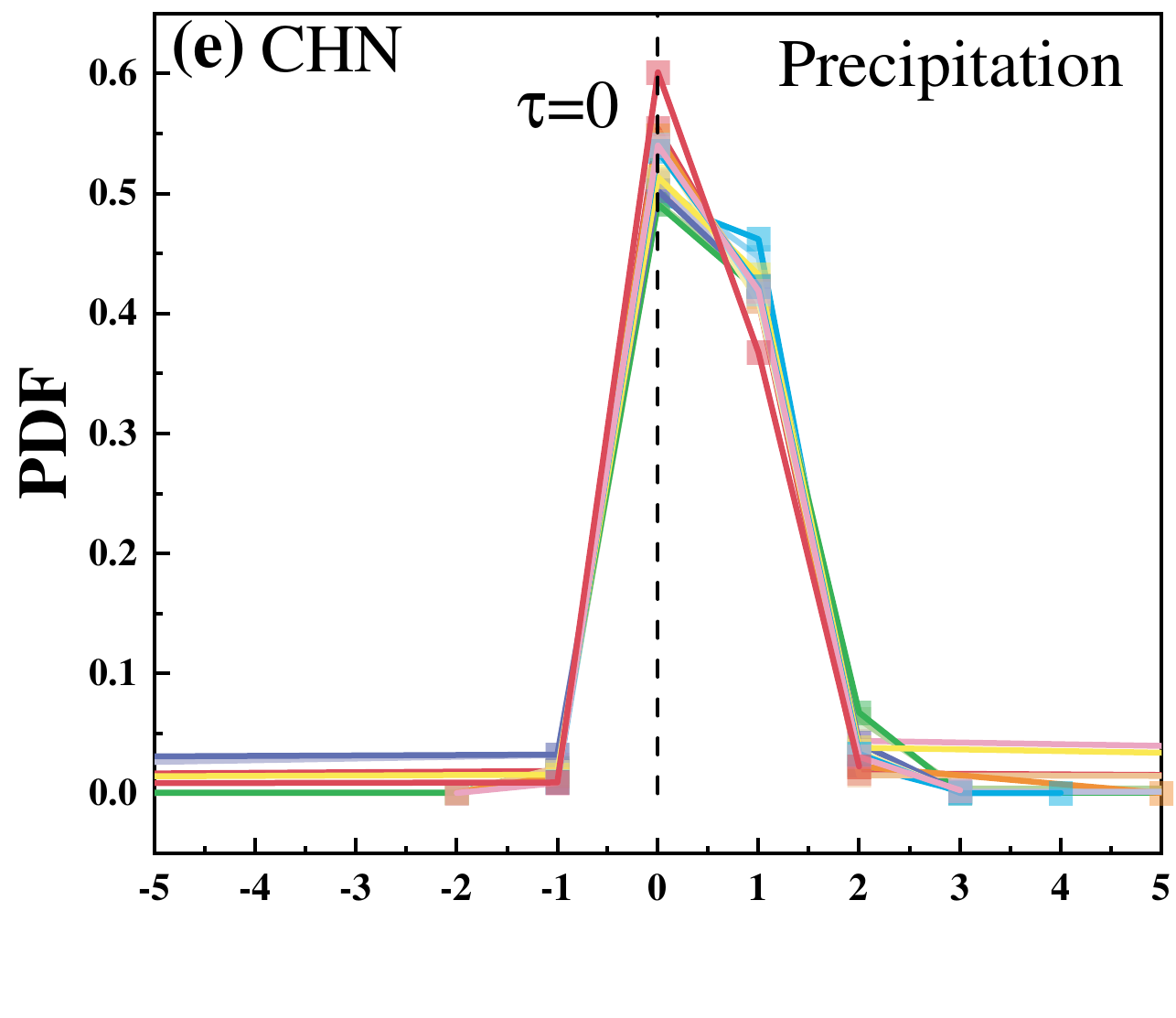}
    \includegraphics[width=9em, height=7.5em]{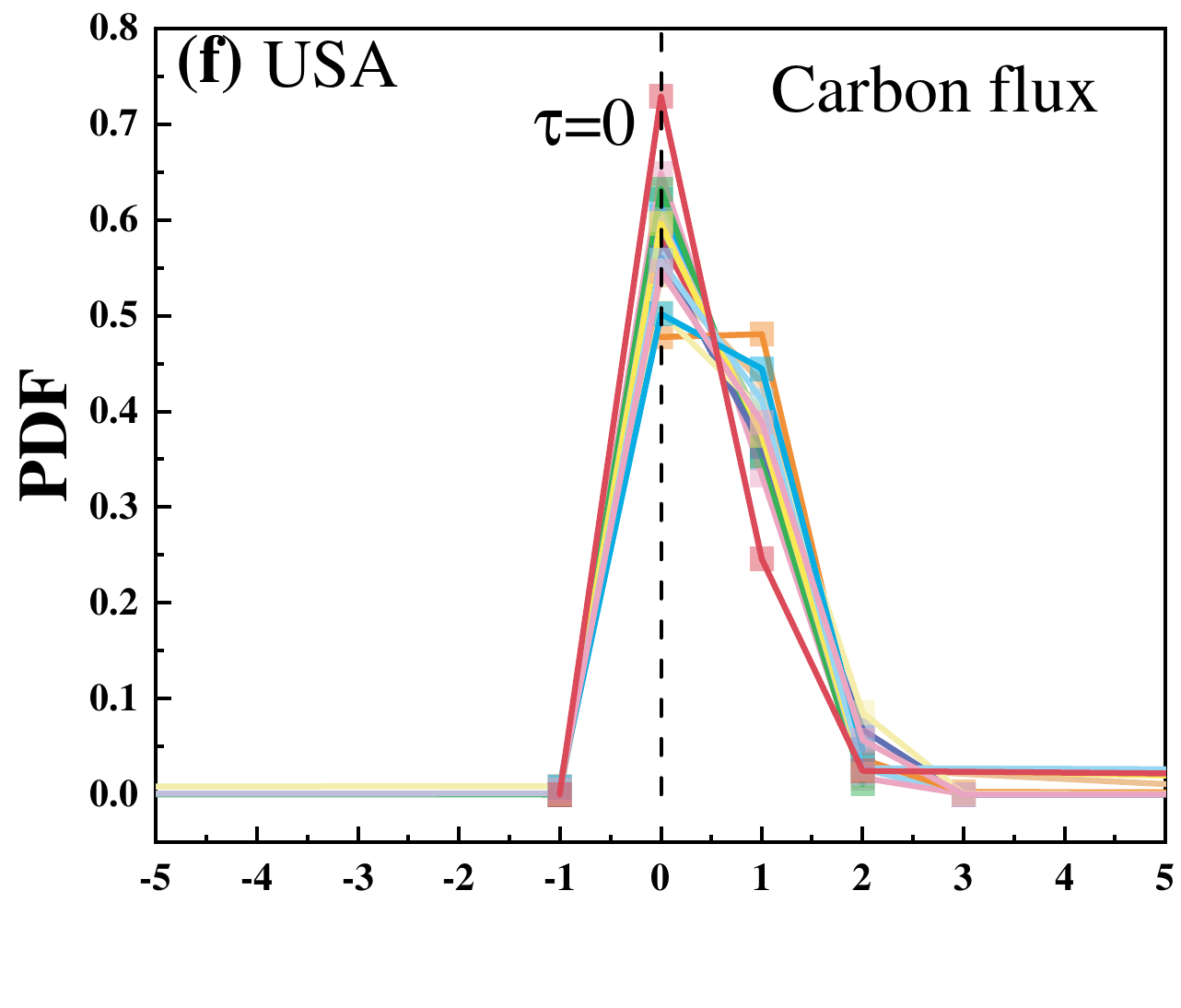}
    \includegraphics[width=9em, height=7.5em]{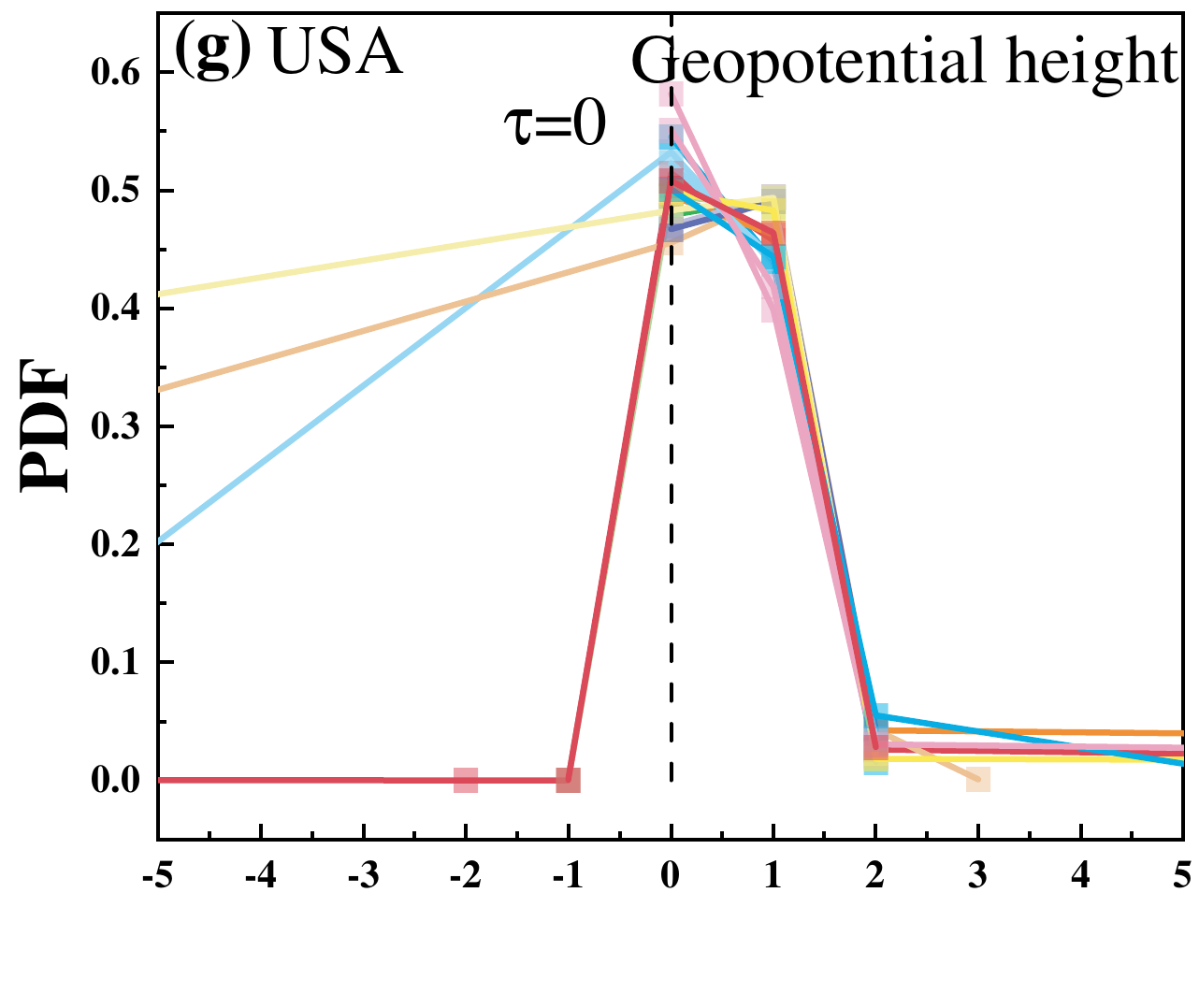}
    \includegraphics[width=9em, height=7.5em]{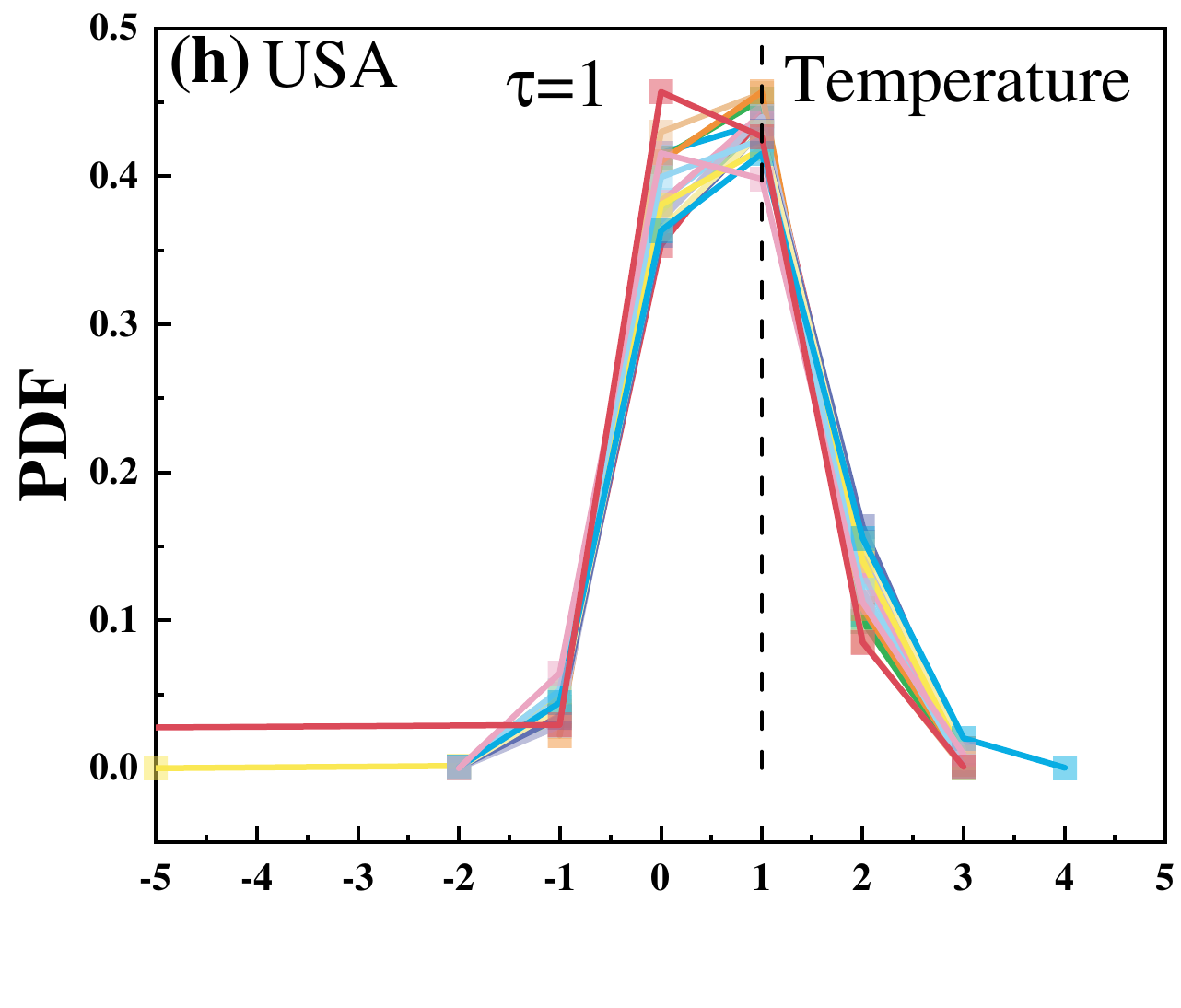}
    \includegraphics[width=9em, height=7.5em]{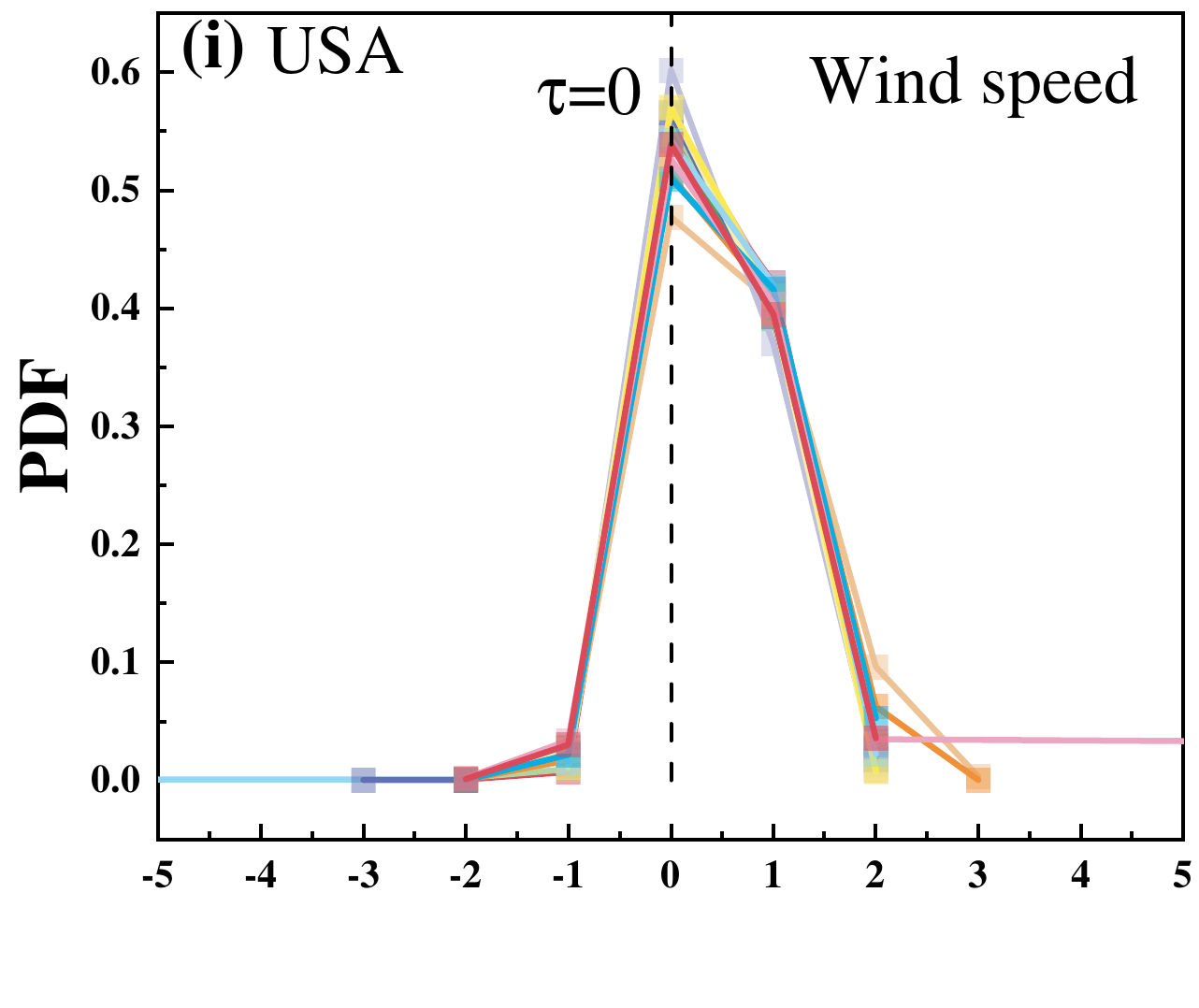}
    \includegraphics[width=9em, height=7.5em]{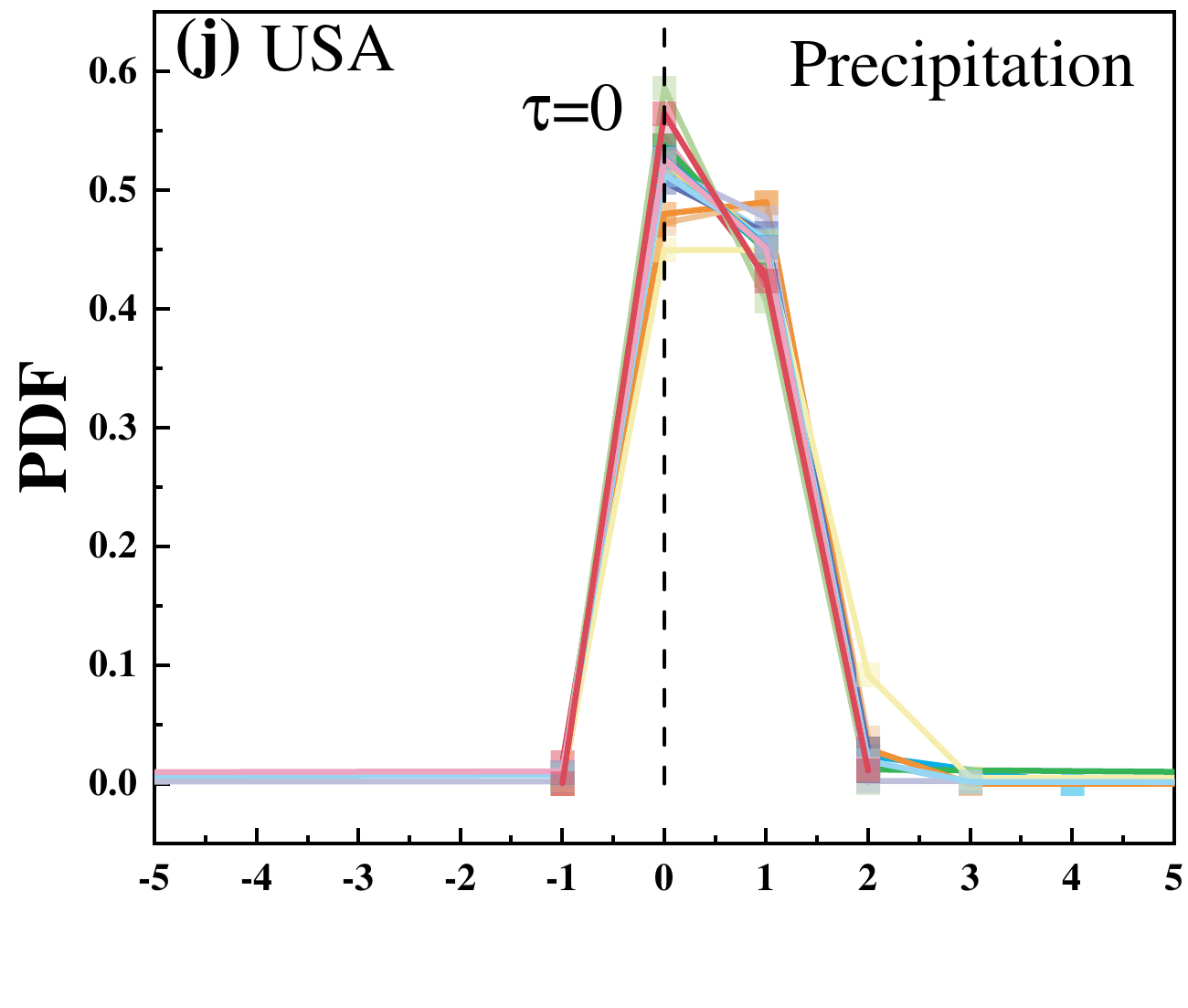}
    \includegraphics[width=9em, height=7.5em]{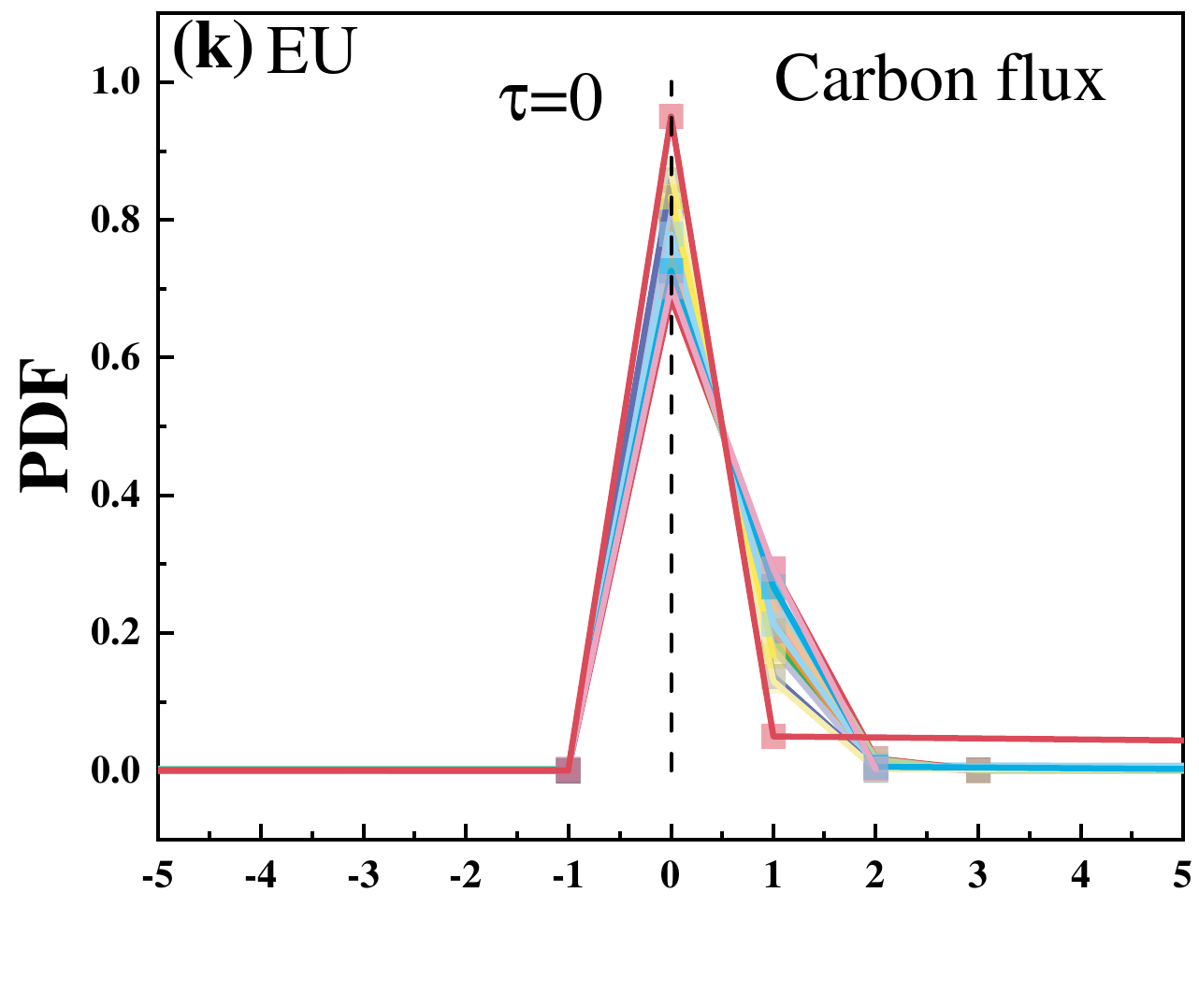}
    \includegraphics[width=9em, height=7.5em]{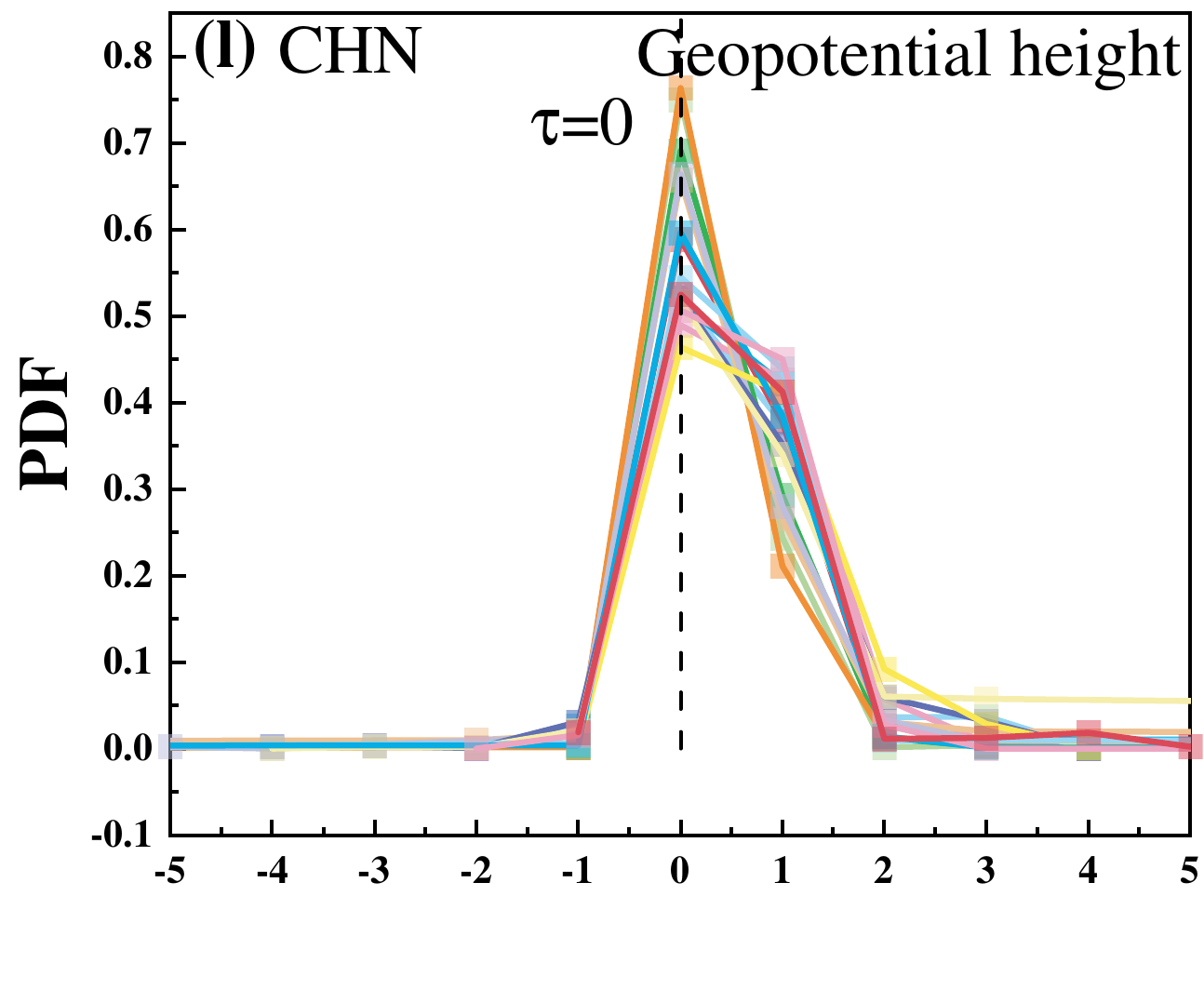}
    \includegraphics[width=9em, height=7.5em]{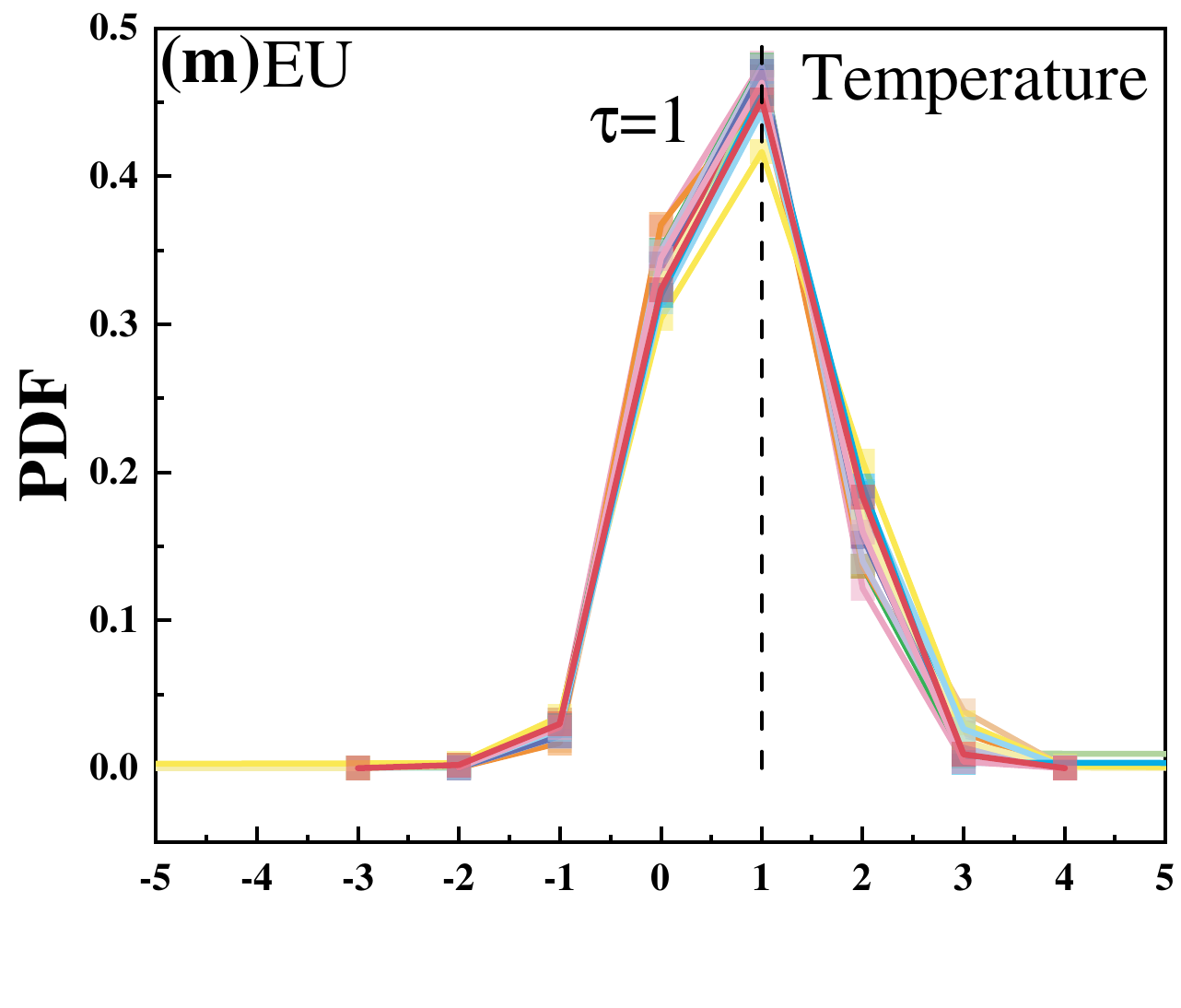}
    \includegraphics[width=9em, height=7.5em]{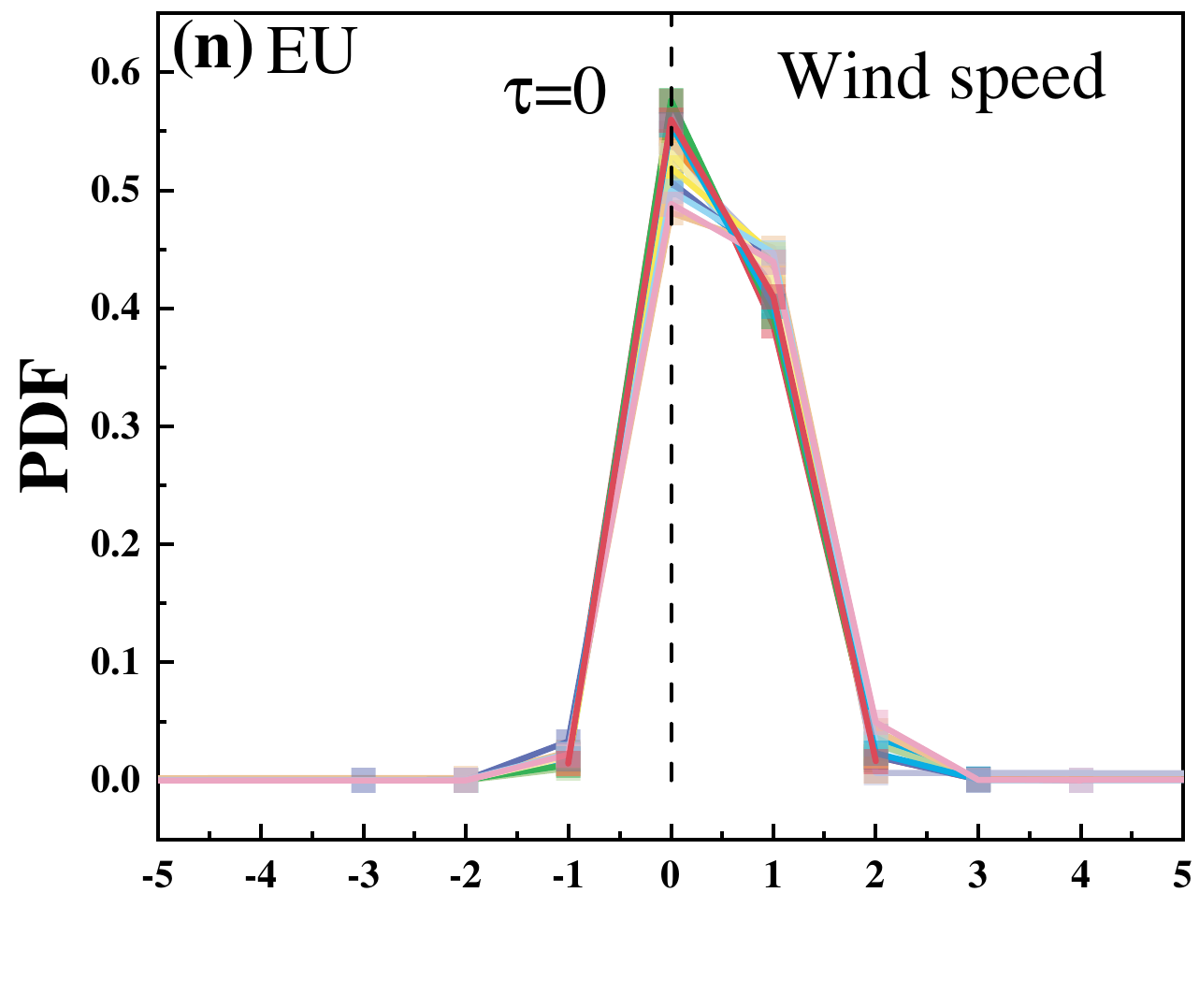}
    \includegraphics[width=9em, height=7.5em]{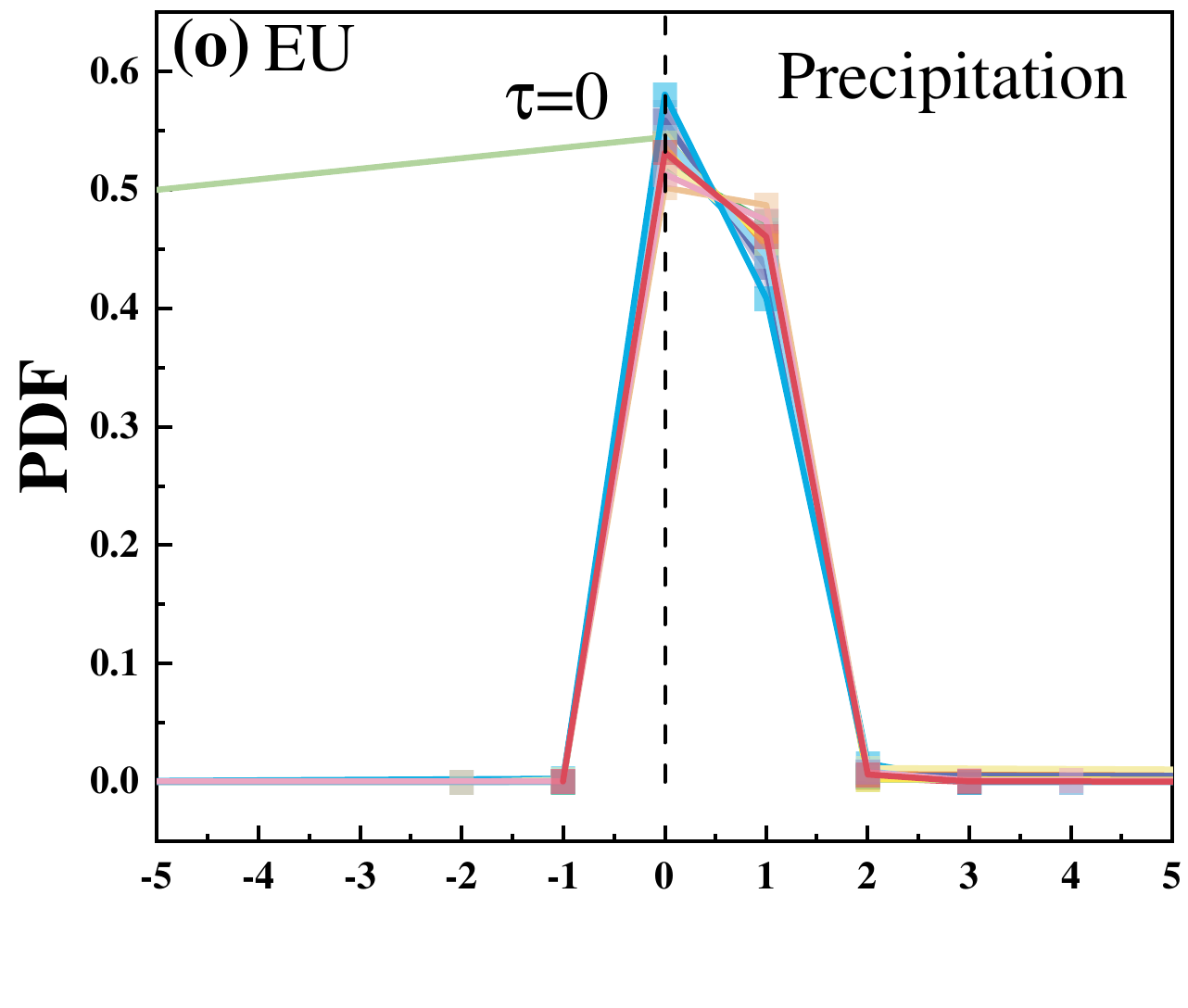}
    \caption{Probability distribution functions (PDF) of the time lags of the real links in different time series. Different colors represent the time lag probability distribution of the network for different years.}
    \label{fig3}
\end{figure}

In addition, we analyze the probability distribution functions of time lags of the links between different locations for the studied climate variables and carbon flux, as shown in Fig. ~\ref{fig3}. It can be seen that the maximum value of the time lags of most climate variables and carbon flux is 0, 1 or -1, i.e., $\tau^{\ast}=$0, 1 or -1 day. This indicates that the time delay between most locations of climate variables and carbon flux is within 1 day from west to east (The number and fraction of significant links between west to east and east to west at time lag $\tau^{\ast}=\pm 1$ are shown in SI Table I-IX). The size of these time lag and the direction of the network structure may be related to atmospheric circulation patterns, ocean surface temperature distribution, and other climatic phenomena. For increasing geographical distance (as shown in SI Figs. S19 and S20), the time lag increases, but the maximum value does not exceed 1 or -1.

\begin{figure}
    \centering
    \includegraphics[width=9em, height=7.5em]{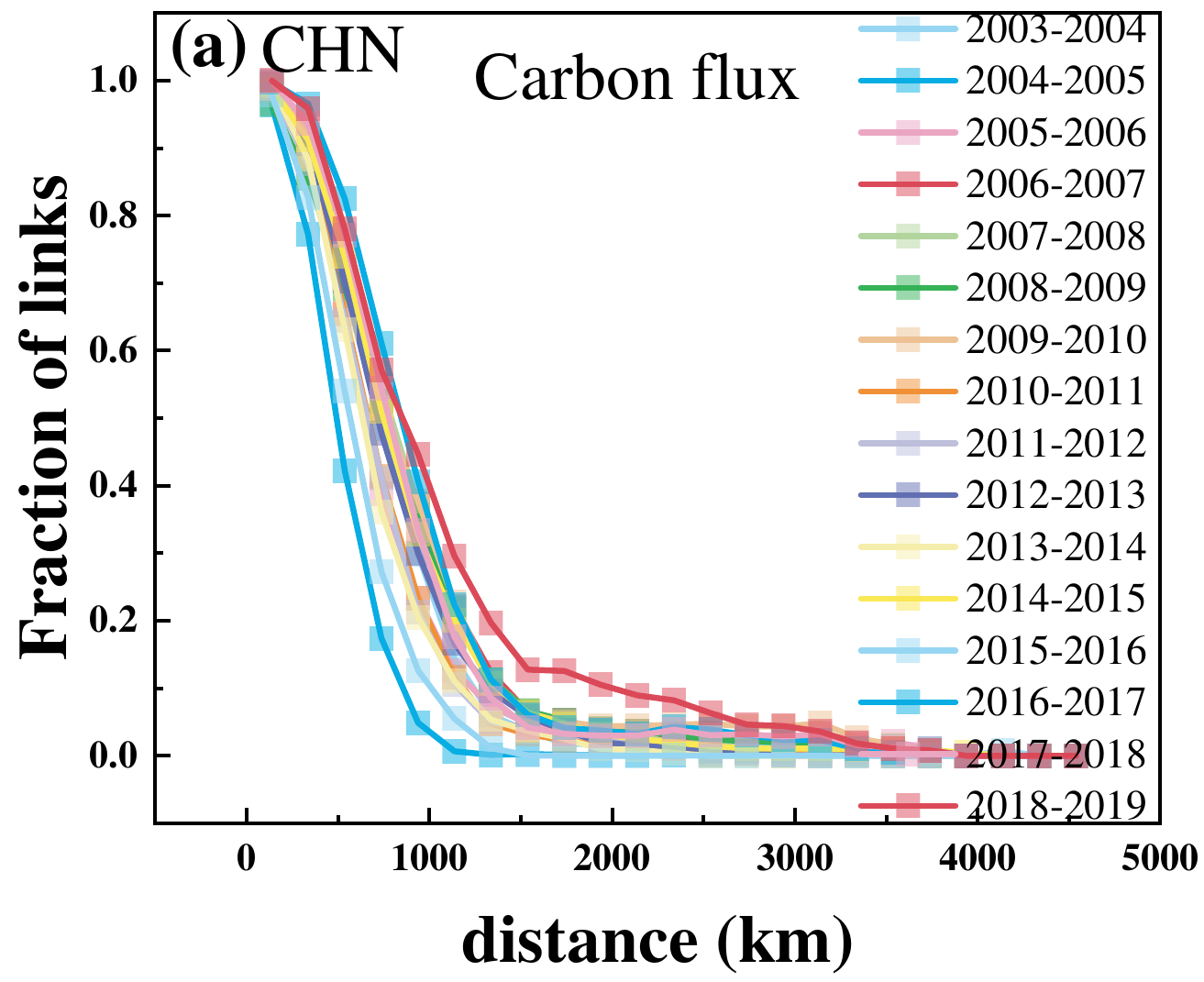}
    \includegraphics[width=9em, height=7.5em]{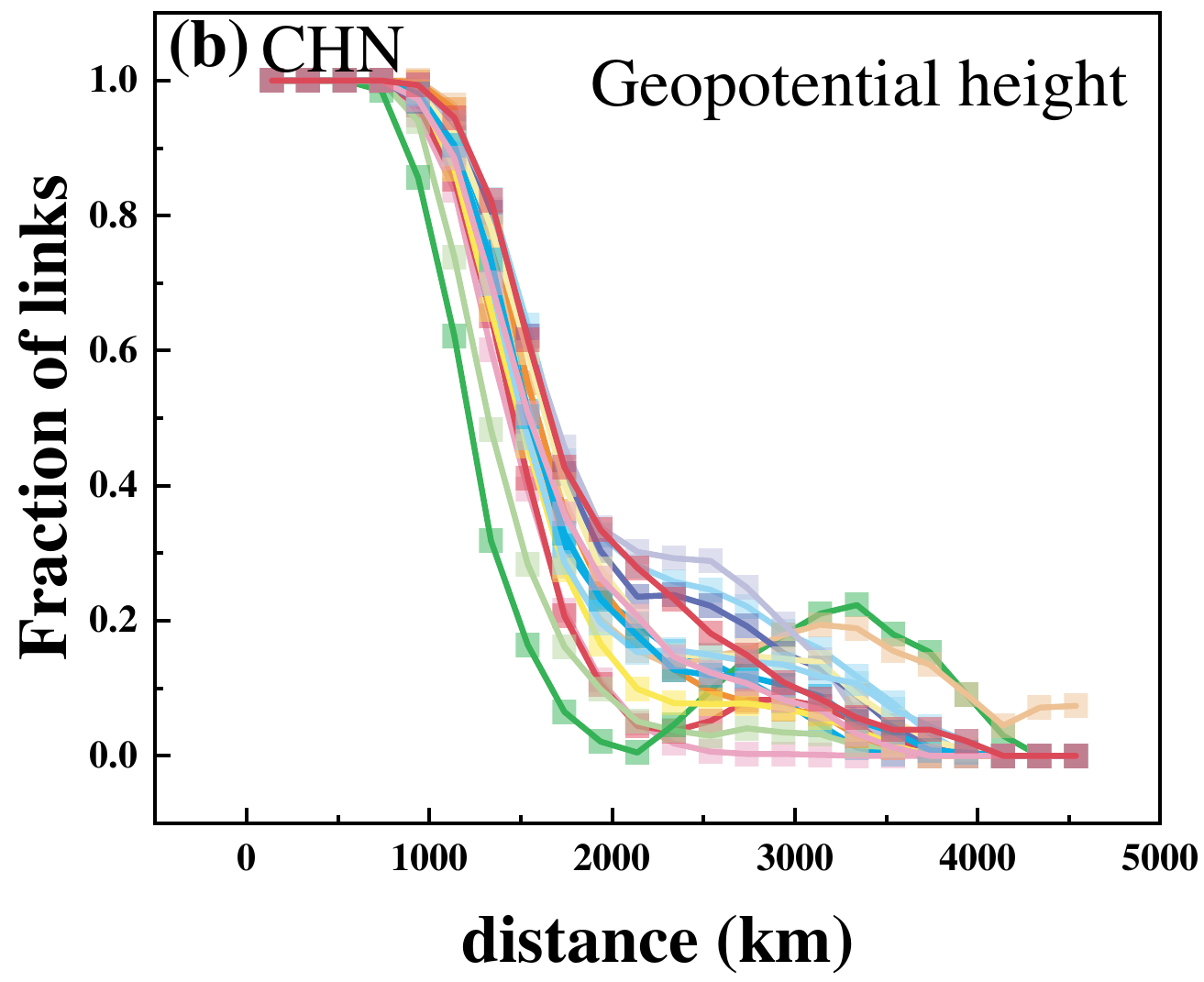}
    \includegraphics[width=9em, height=7.5em]{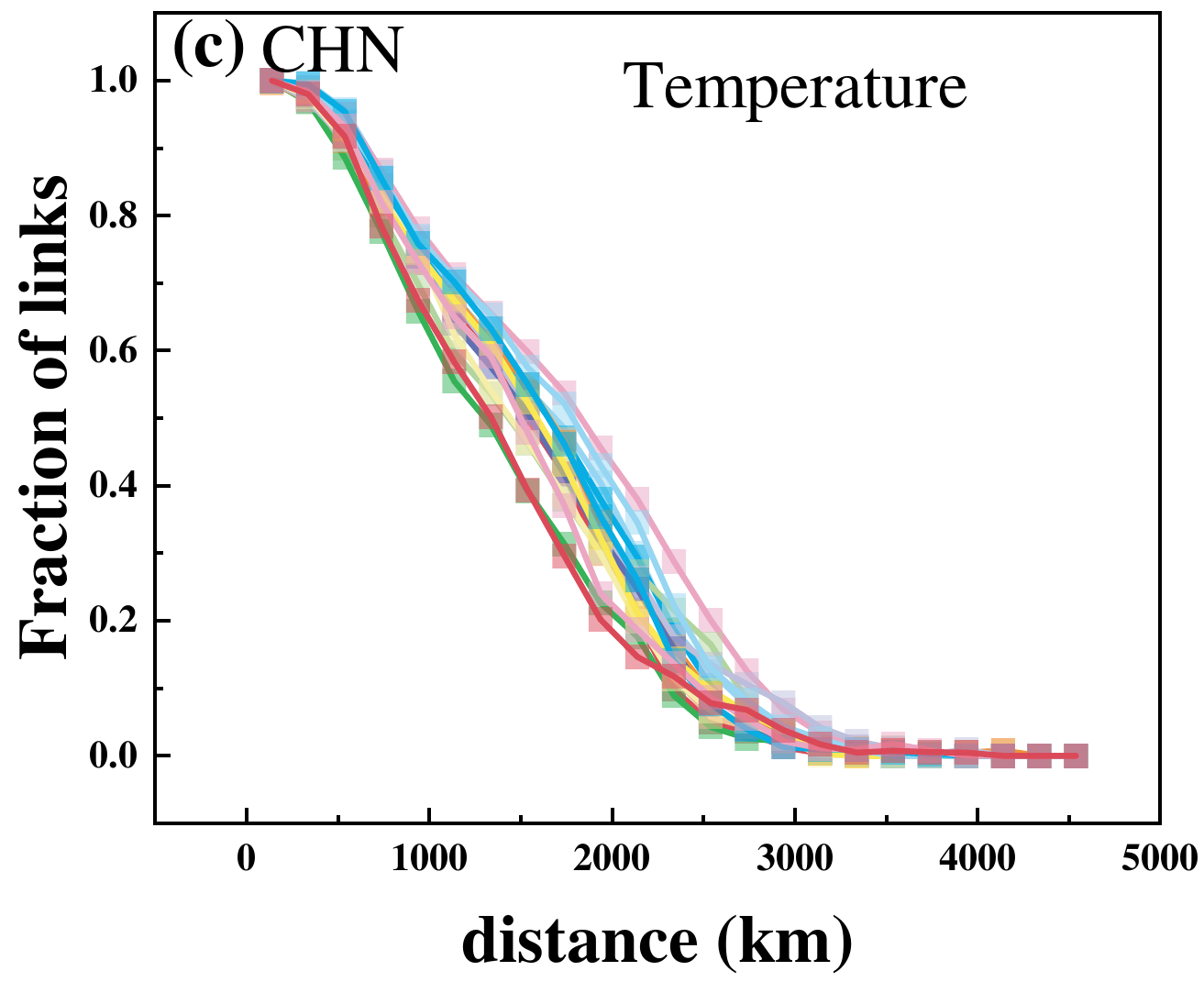}
    \includegraphics[width=9em, height=7.5em]{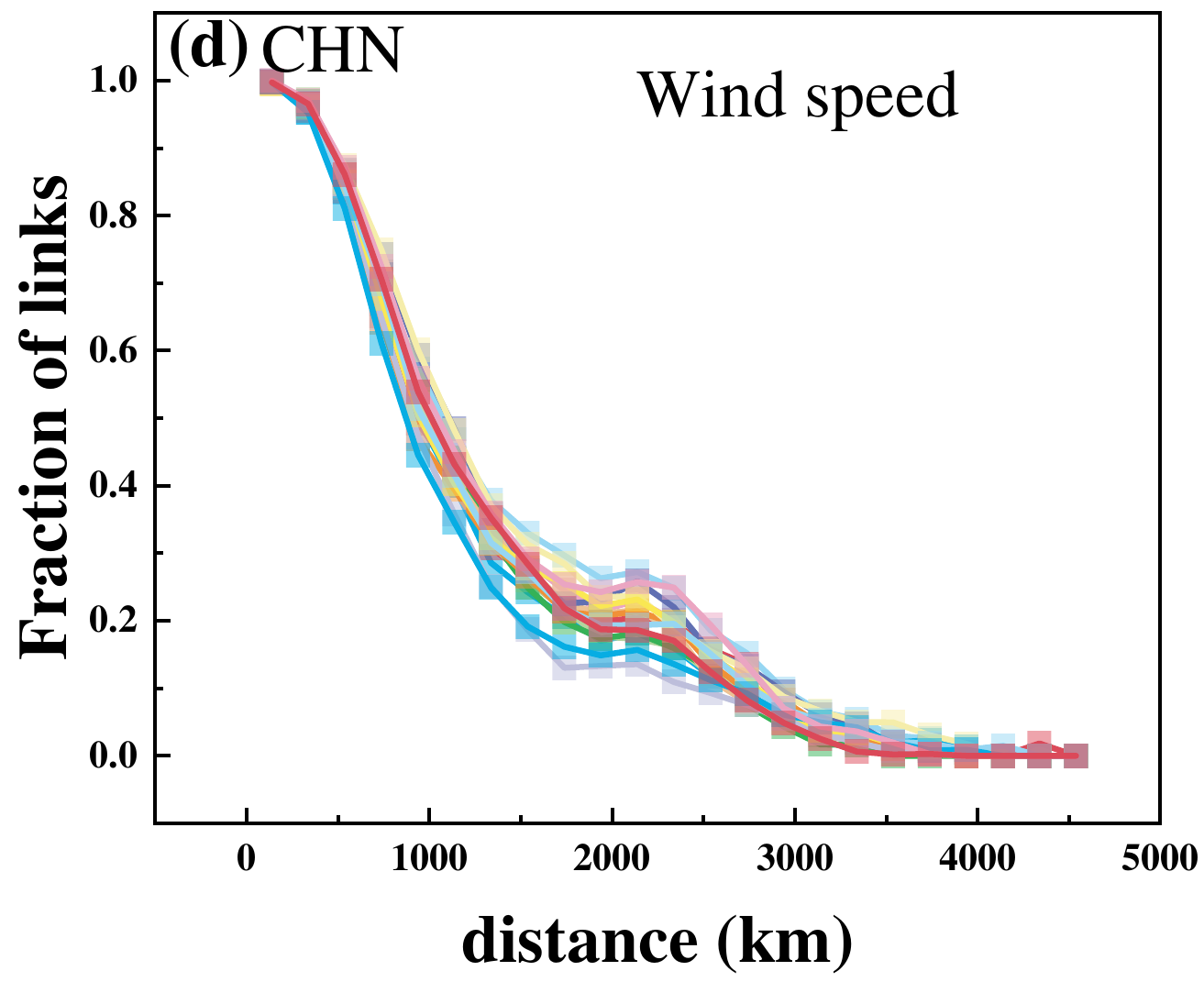}
    \includegraphics[width=9em, height=7.5em]{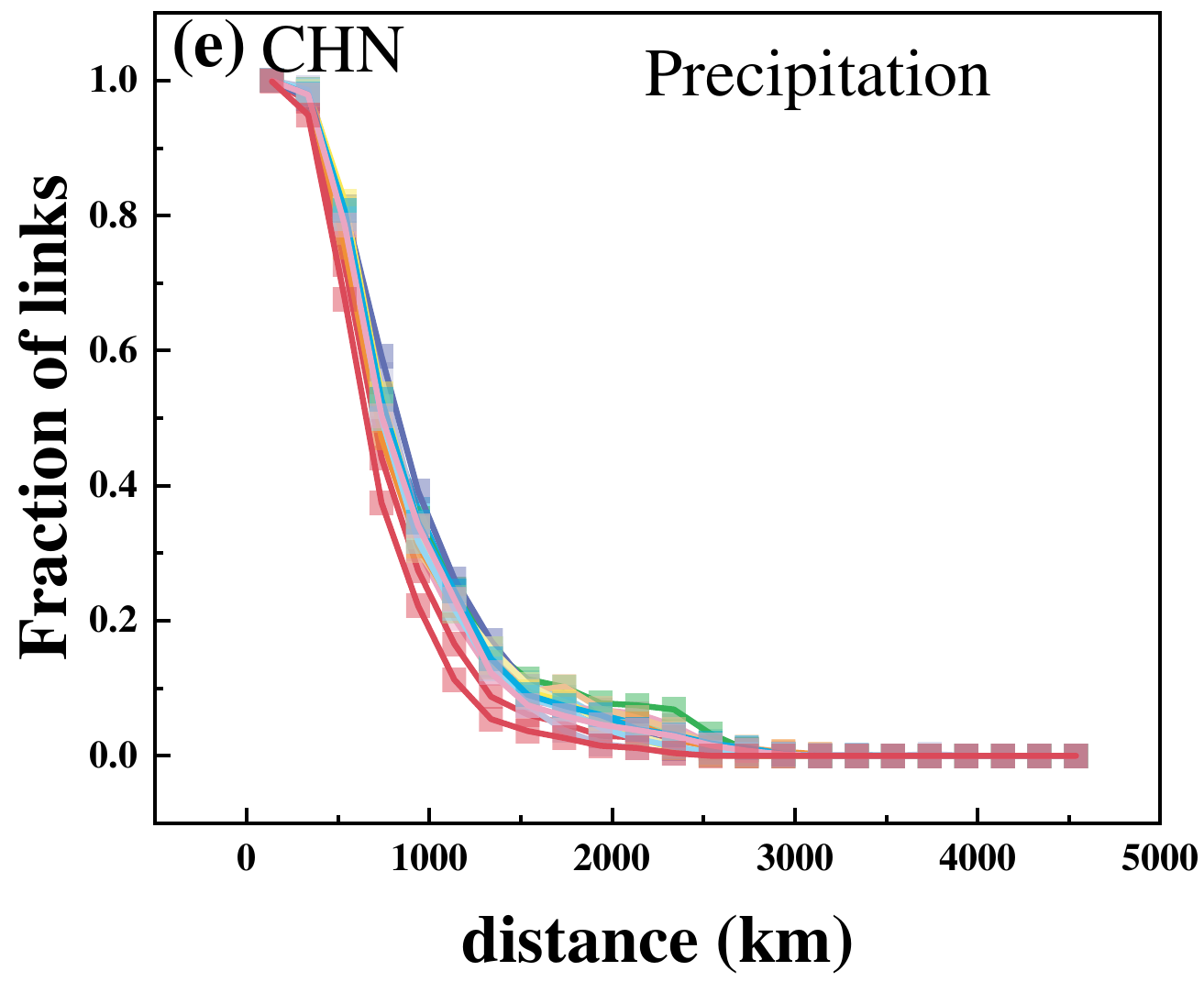}
    \includegraphics[width=9em, height=7.5em]{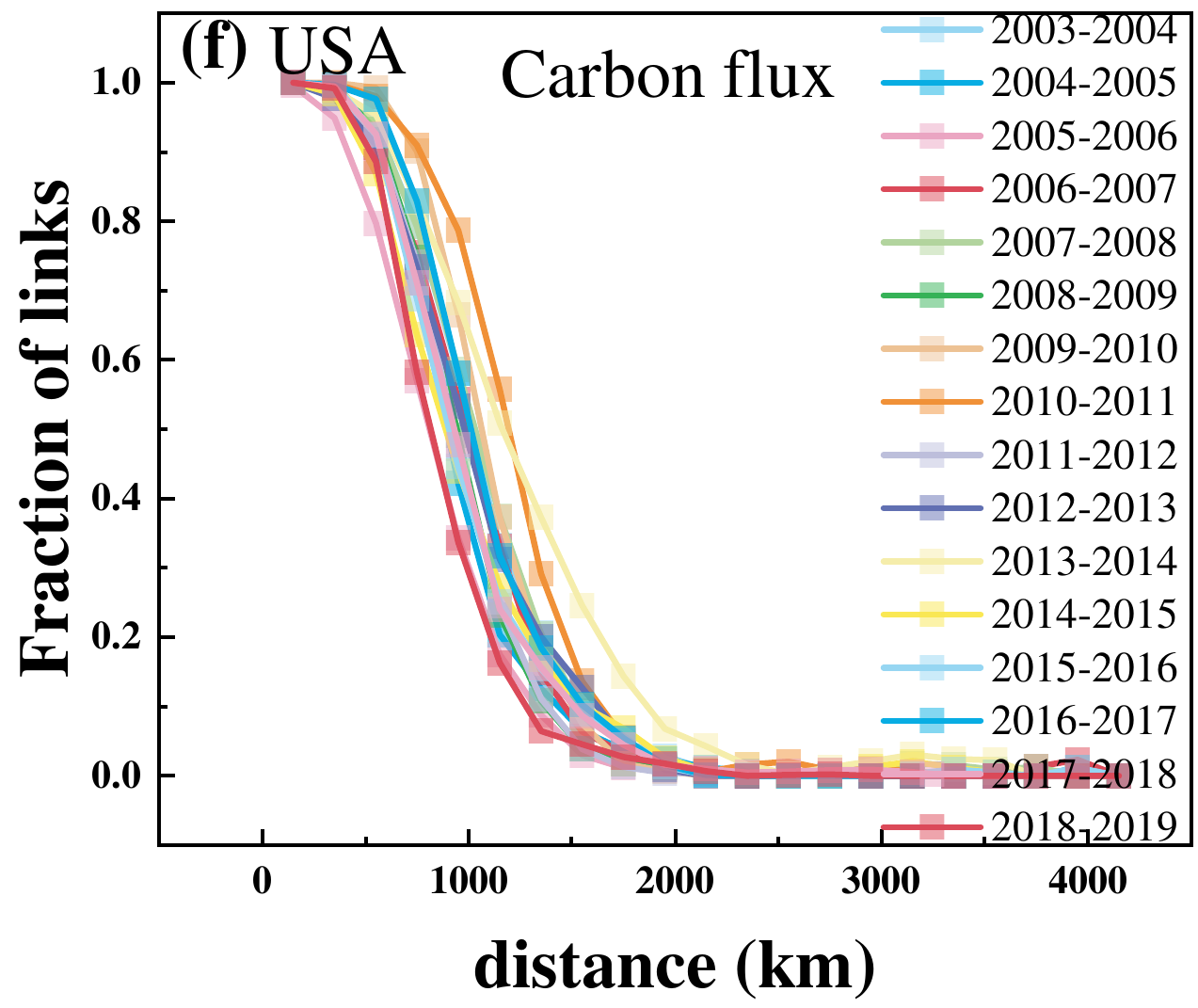}
    \includegraphics[width=9em, height=7.5em]{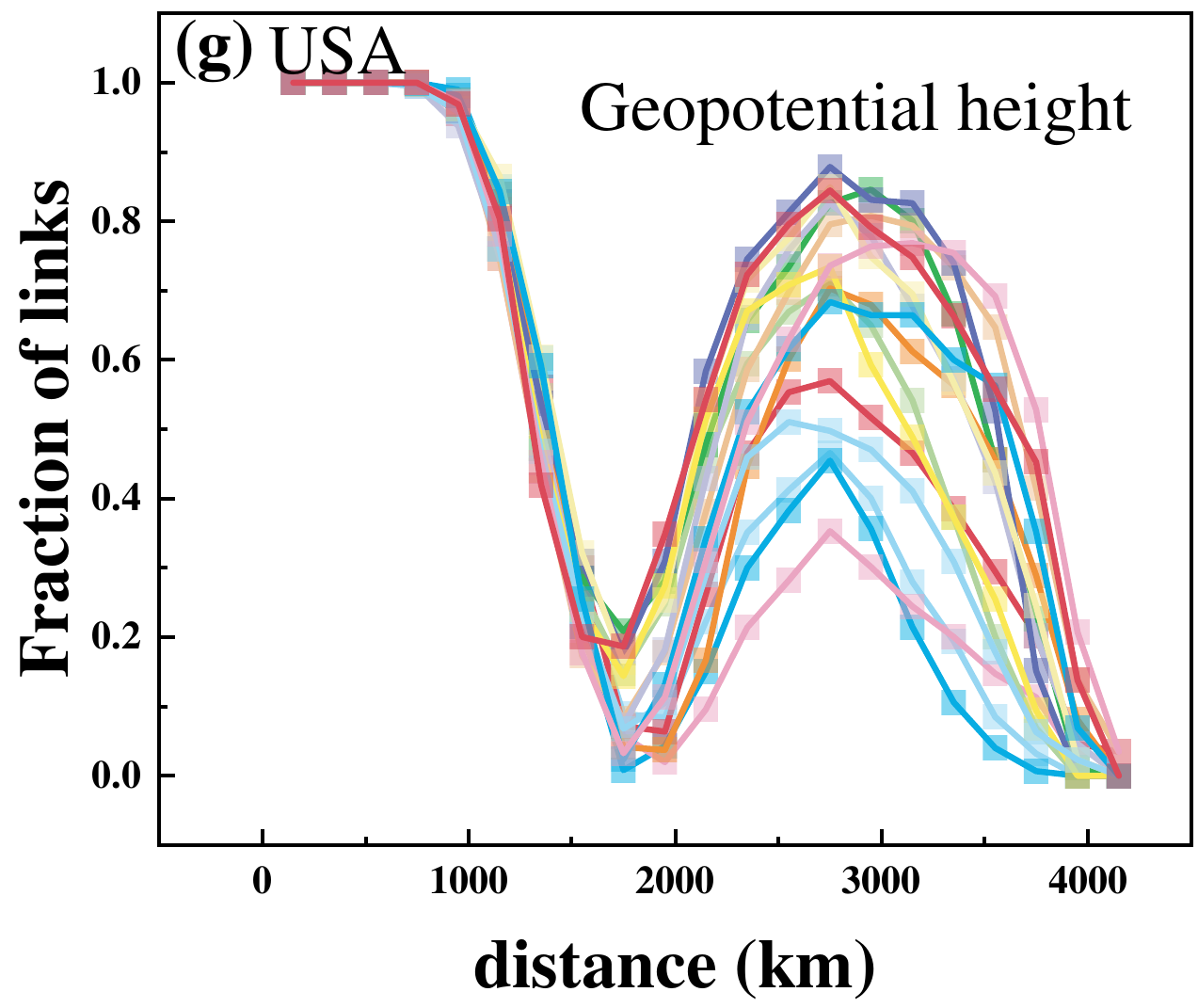}
    \includegraphics[width=9em, height=7.5em]{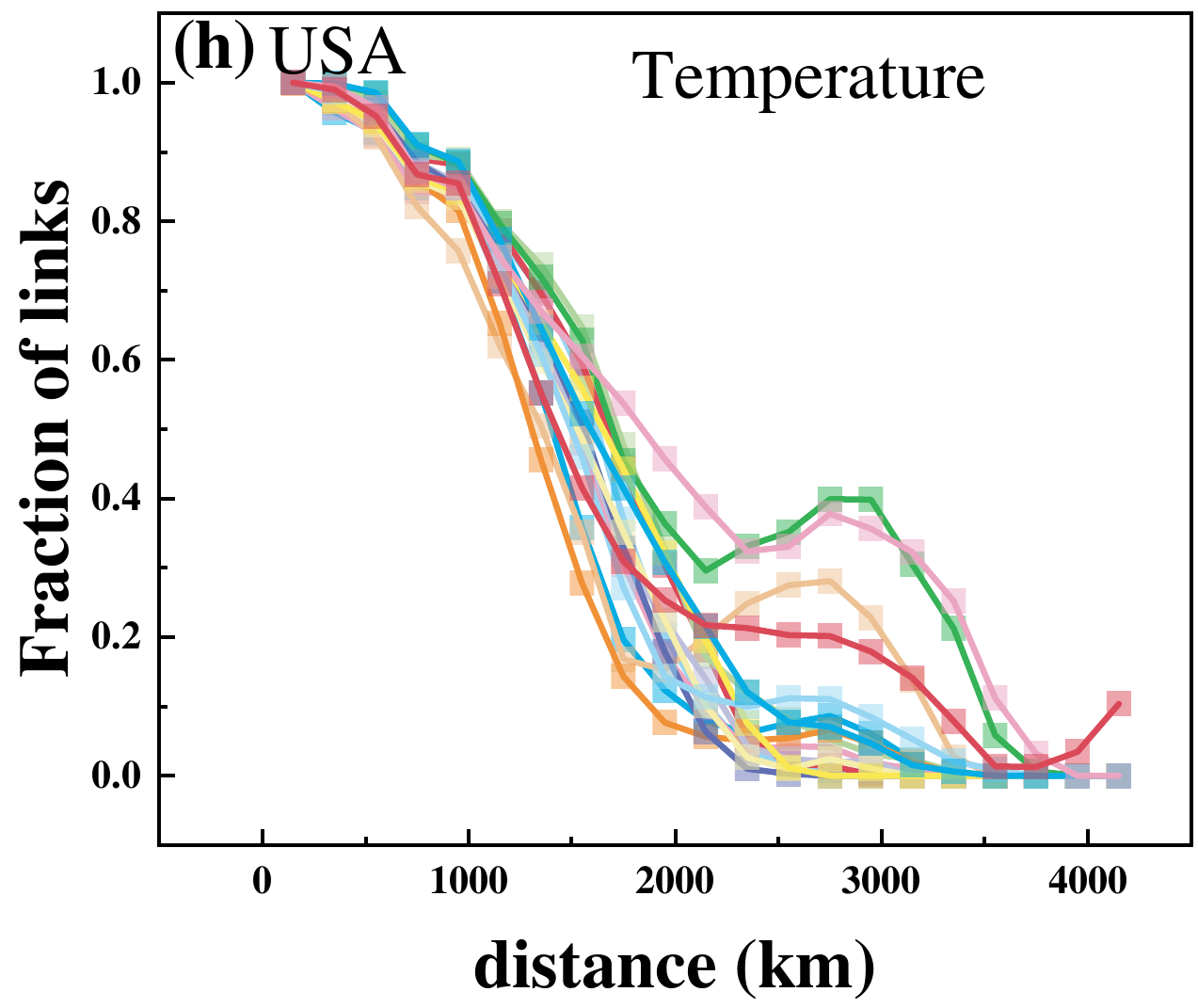}
    \includegraphics[width=9em, height=7.5em]{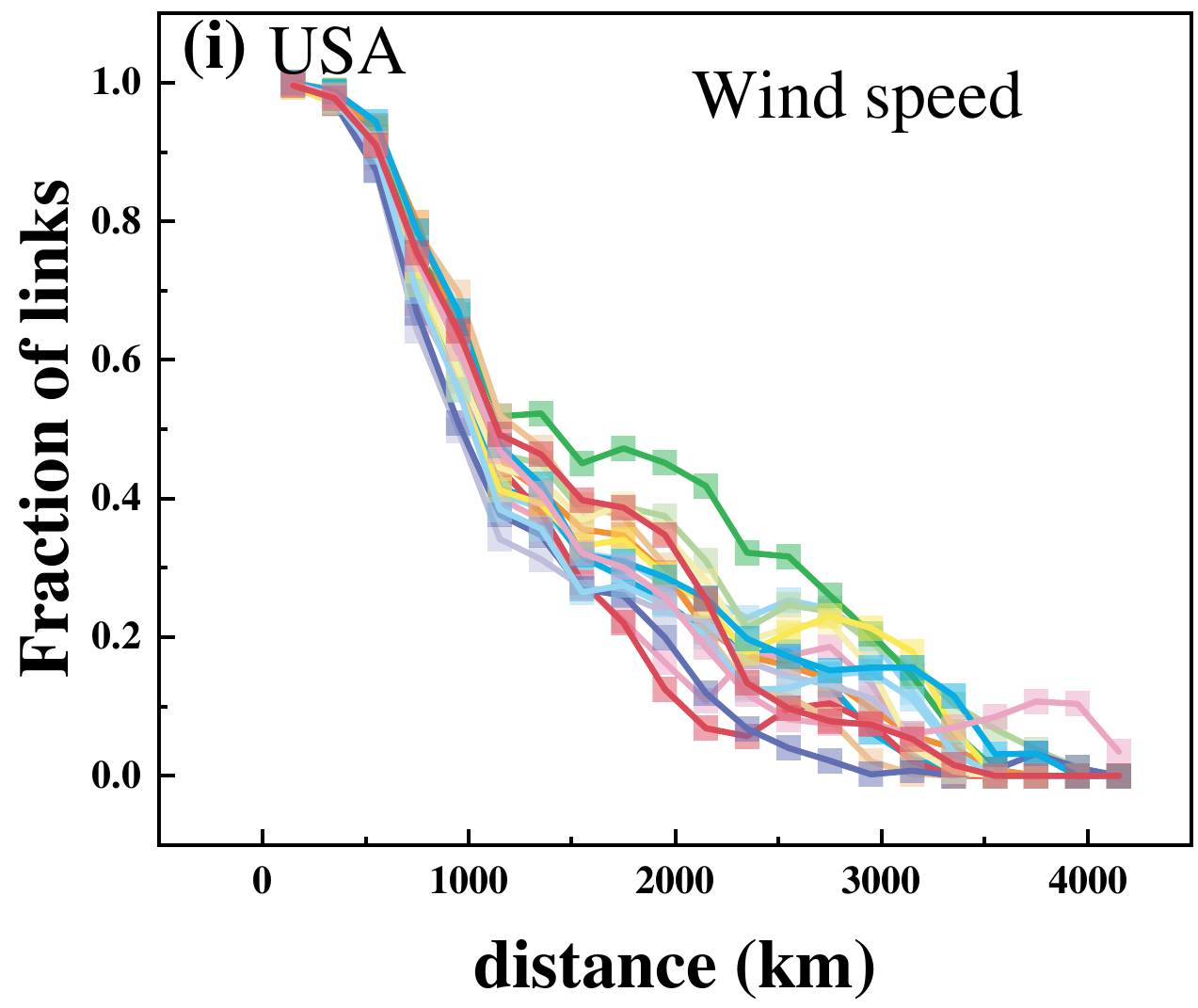}
    \includegraphics[width=9em, height=7.5em]{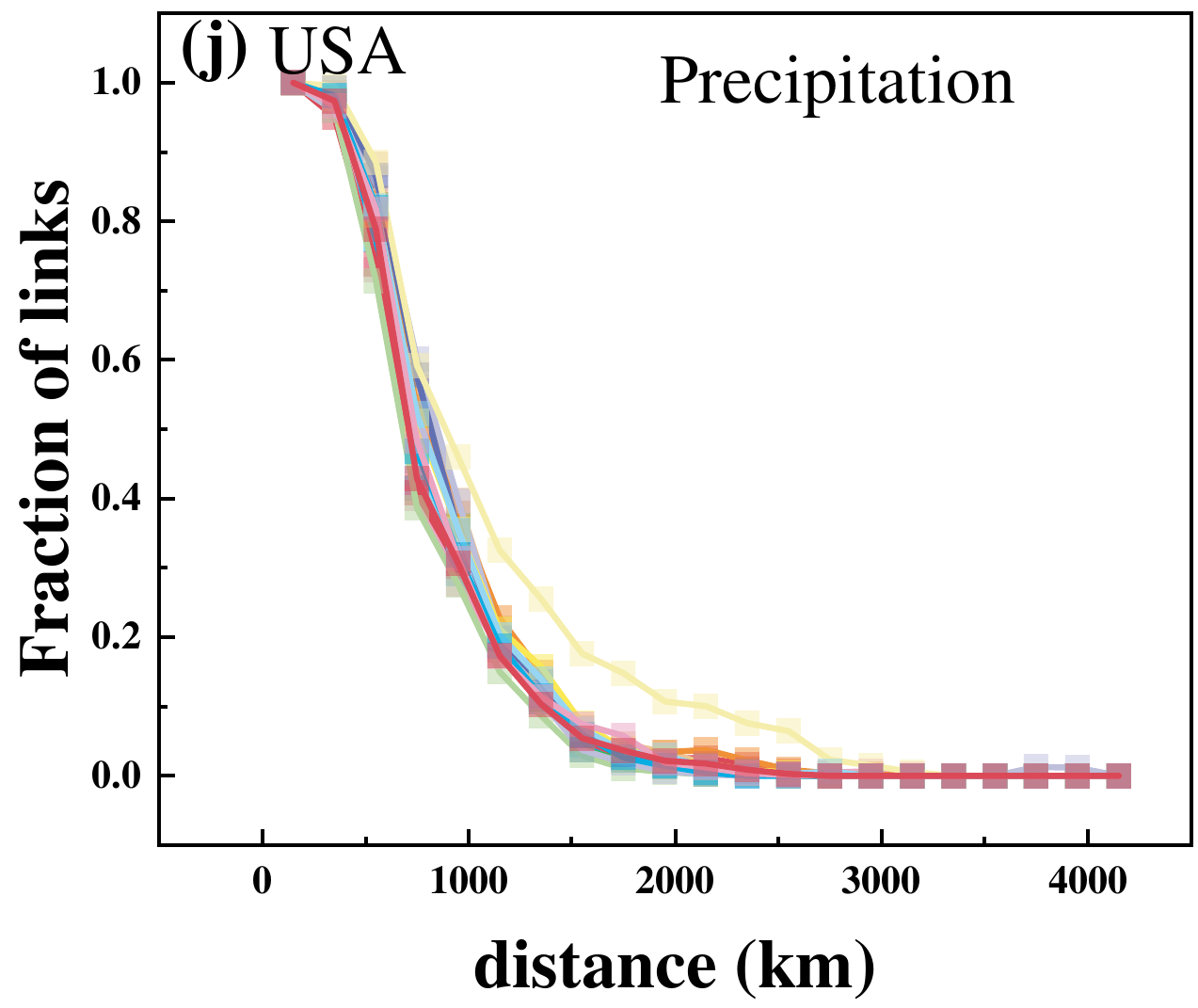}
    \includegraphics[width=9em, height=7.5em]{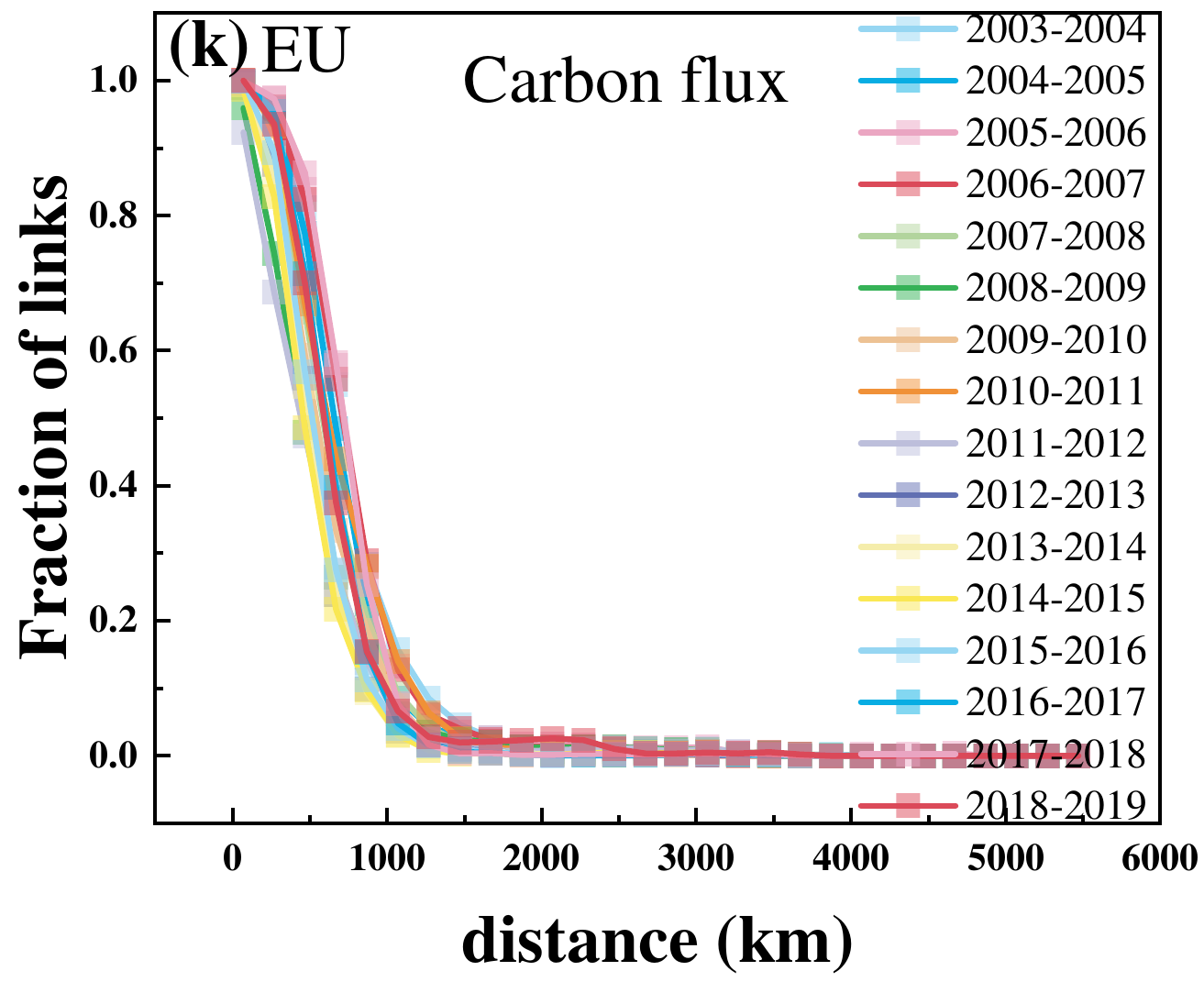}
    \includegraphics[width=9em, height=7.5em]{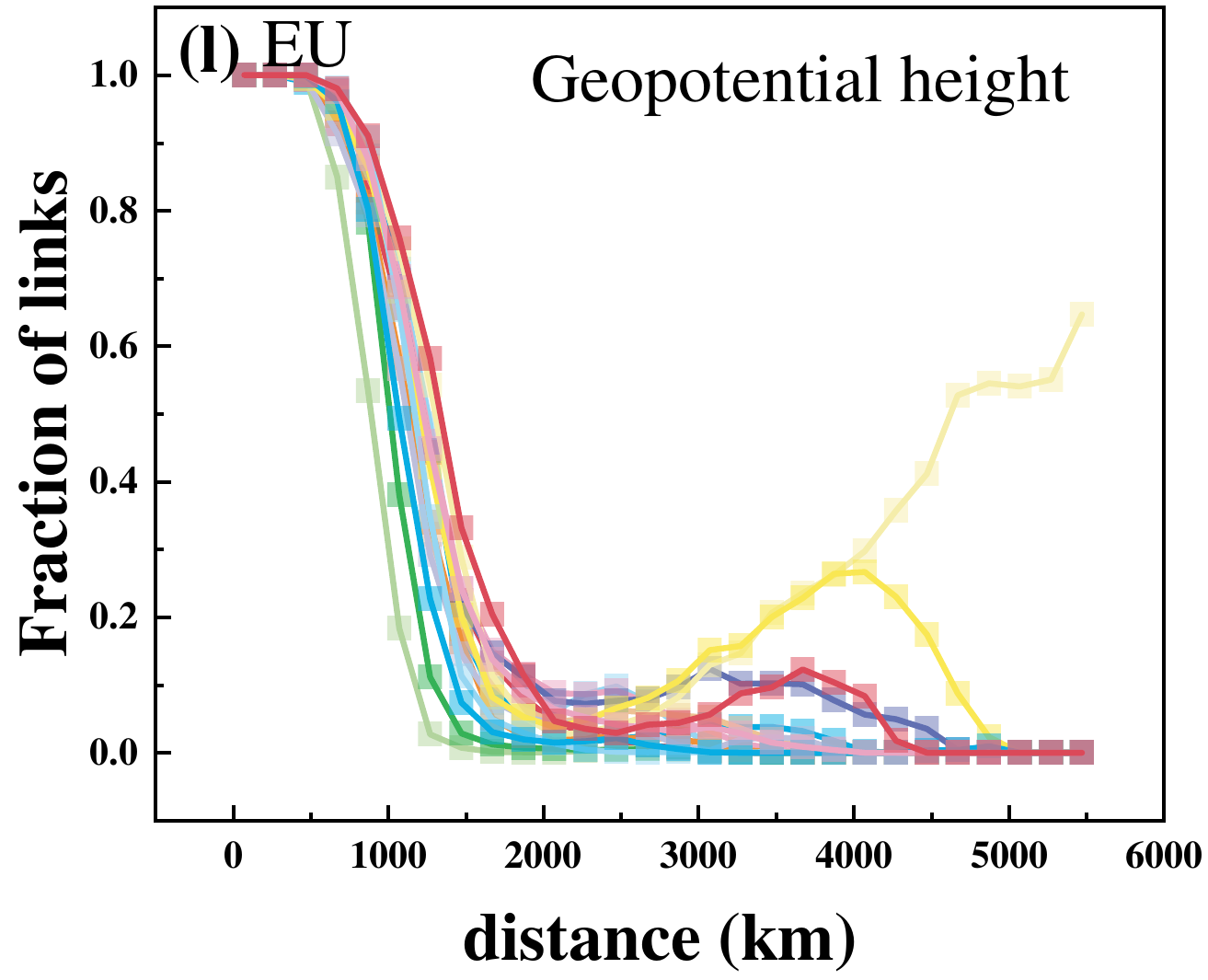}
    \includegraphics[width=9em, height=7.5em]{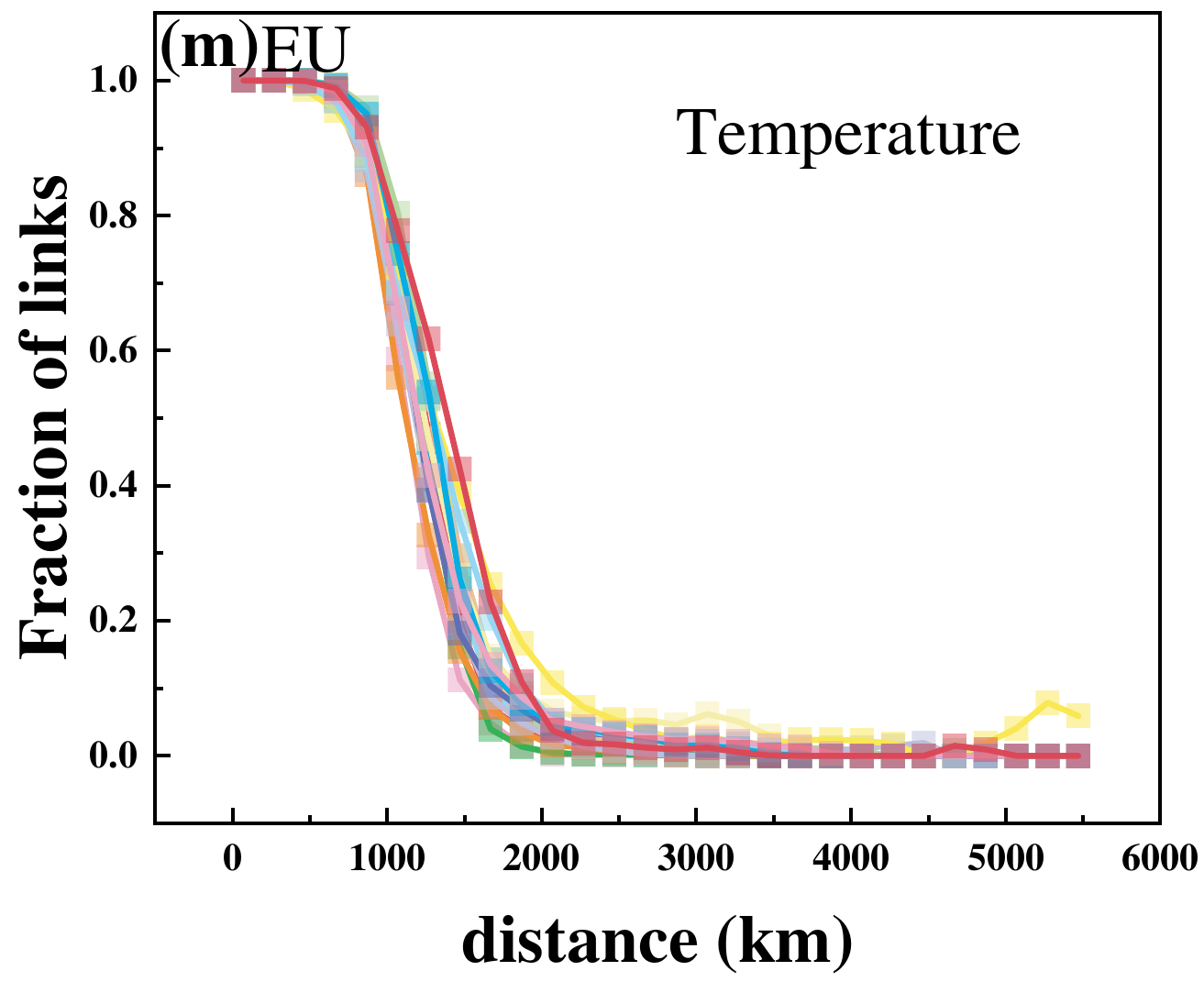}
    \includegraphics[width=9em, height=7.5em]{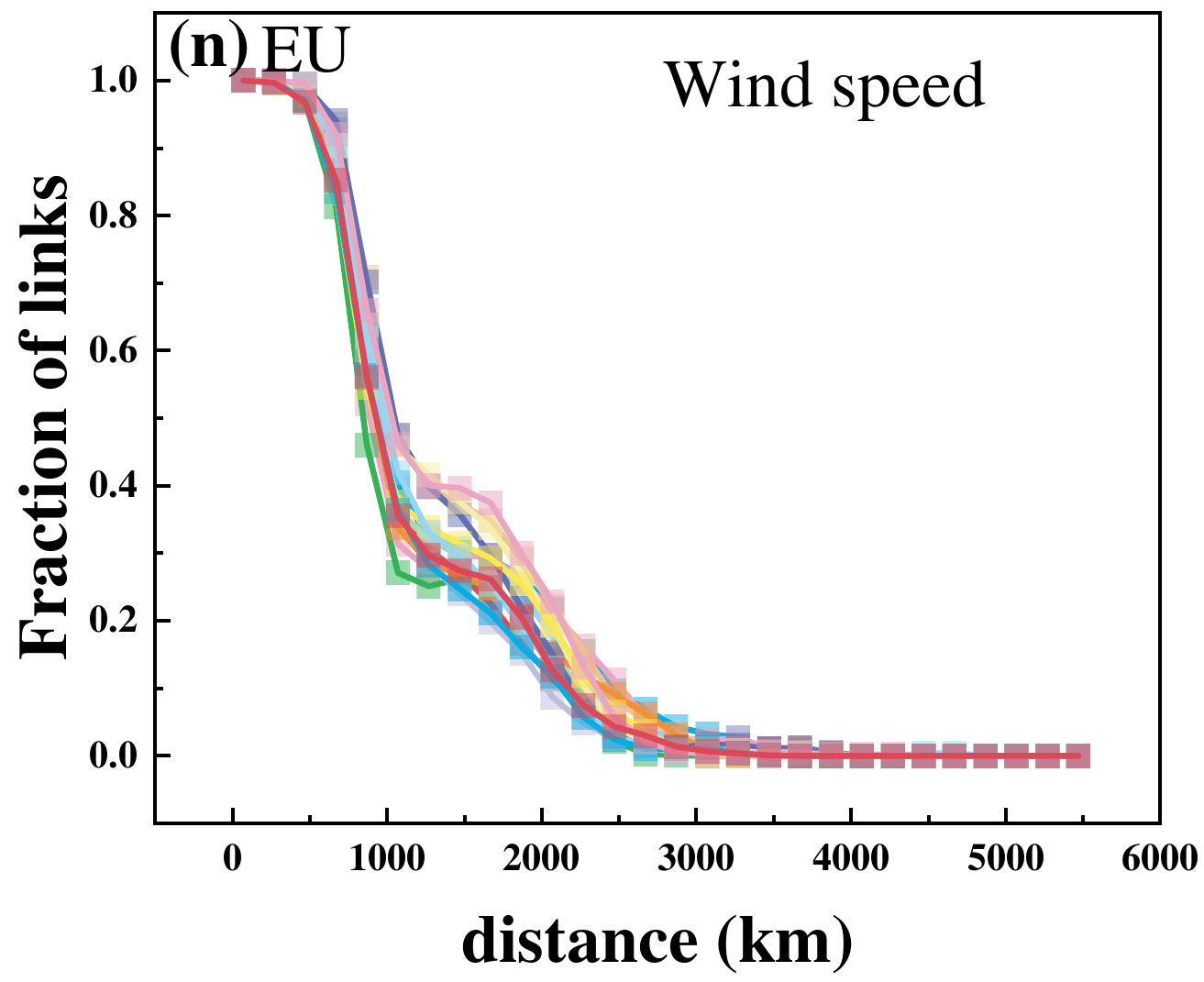}
    \includegraphics[width=9em, height=7.5em]{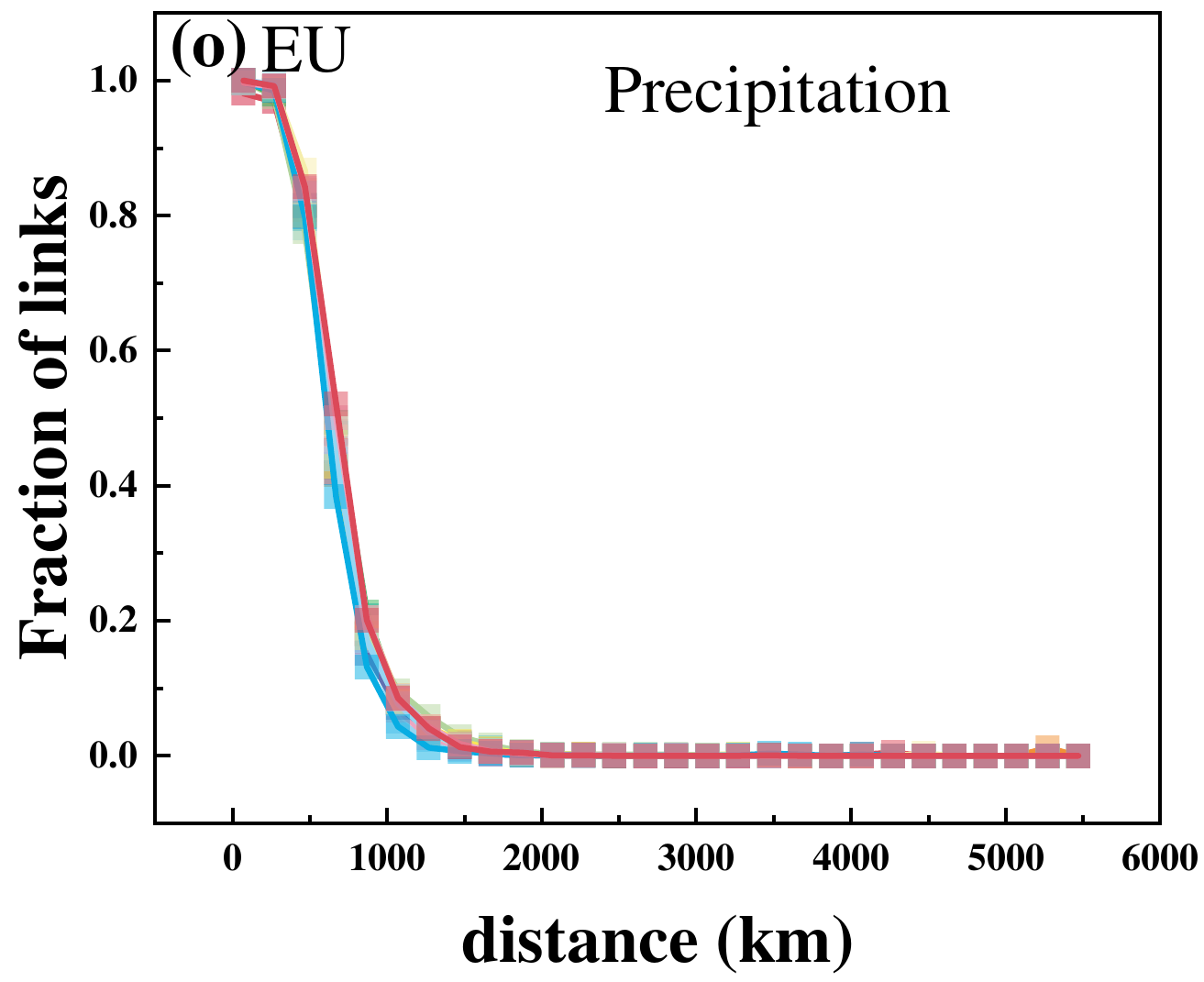}
    \caption{Fraction of significant links in the network as a function of geographic distance between the sites for different years.}
    \label{fig4}
\end{figure}

Next, we study how the correlation strength change with distance. Here, the geographic distance is calculated based on the latitude and longitude of the two nodes.  Given that the geographic distance dataset is discrete, we partition the range of the dataset into uniform subintervals with a length of 200km, which is approximately the distance between two adjacent nodes. By doing so, each pair of nodes can be assigned to a specific subinterval based on their geographic distance, such that $dis _{i,j} \in (dis_n, dis_{n+1}]$. Here, $dis_{i,j}$ denotes the geographic distance between node $i$ and node $j$, and $dis_n$ and $dis_{n+1}$ represent the $n$th and $(n+1)$th distance intervals, respectively. Fig. ~\ref{fig4} shows the fraction of links of climate and carbon flux with geographic distance in the different time series of $dis \in (dis_n,dis_{n+1}]$. The results show in general, that the closer the nodes are geographically, there is many more significant links. This is expected since winds or other transport processes move the climate and carbon from place to place, and the further the places are, the similarity between locations which is represented by links becomes weaker. However, we can also see strong outliers. In many cases, there is a bump at few thousands kilometer. This suggests that not only transport is the mechanism but also a third party could play a significant role such as the Rossby waves \cite{ying2020rossby, wang2013dominant, riboldi2022circumglobal}.

\section{Persistence analysis}

After identifying the significant links in the correlation network of climate and carbon flux constructed above (the number of significant links for each of climate and carbon flux networks for different years is given in SI Fig. S21), we wish to measure here the persistence  (similarity) of the spatial network during the years. To this end, we analyze their temporal persistence during the recent 20 years using Jaccard similarity coefficients \cite{real1996probabilistic}. Here the Jaccard similarity is defined as the ratio between the intersection of the links in both networks in different years and the union number of links in both networks. The number of link intersections/unions for each two networks is shown in SI Figs. S22 and S23. Based on the definition of the Jaccard similarity coefficient, one can conclude that a higher Jaccard index in the networks indicates a greater structural similarity between the networks in these years. This, in turn, signifies a stronger persistence in the spatial and temporal evolution of the network system. Therefore, Jaccard similarity coefficients is chosen in this study to measure the structural persistence of the climate networks and of the carbon flux network. We present the actual Jaccard coefficient results in SI Figs. S24-S26, demonstrating significant persistence during the years in the carbon flux and climate networks. However, when the geographic distance is above $1000 km$, as shown in Fig. S26, it becomes evident that the links of the carbon flux and climate networks in the regions of China and Europe exhibit much less persistence.  Note also that the persistence of precipitation is lower compared to climate, in particular in the Europe.

\begin{figure}[t]
    \centering
    \includegraphics[width=9em, height=7.5em]{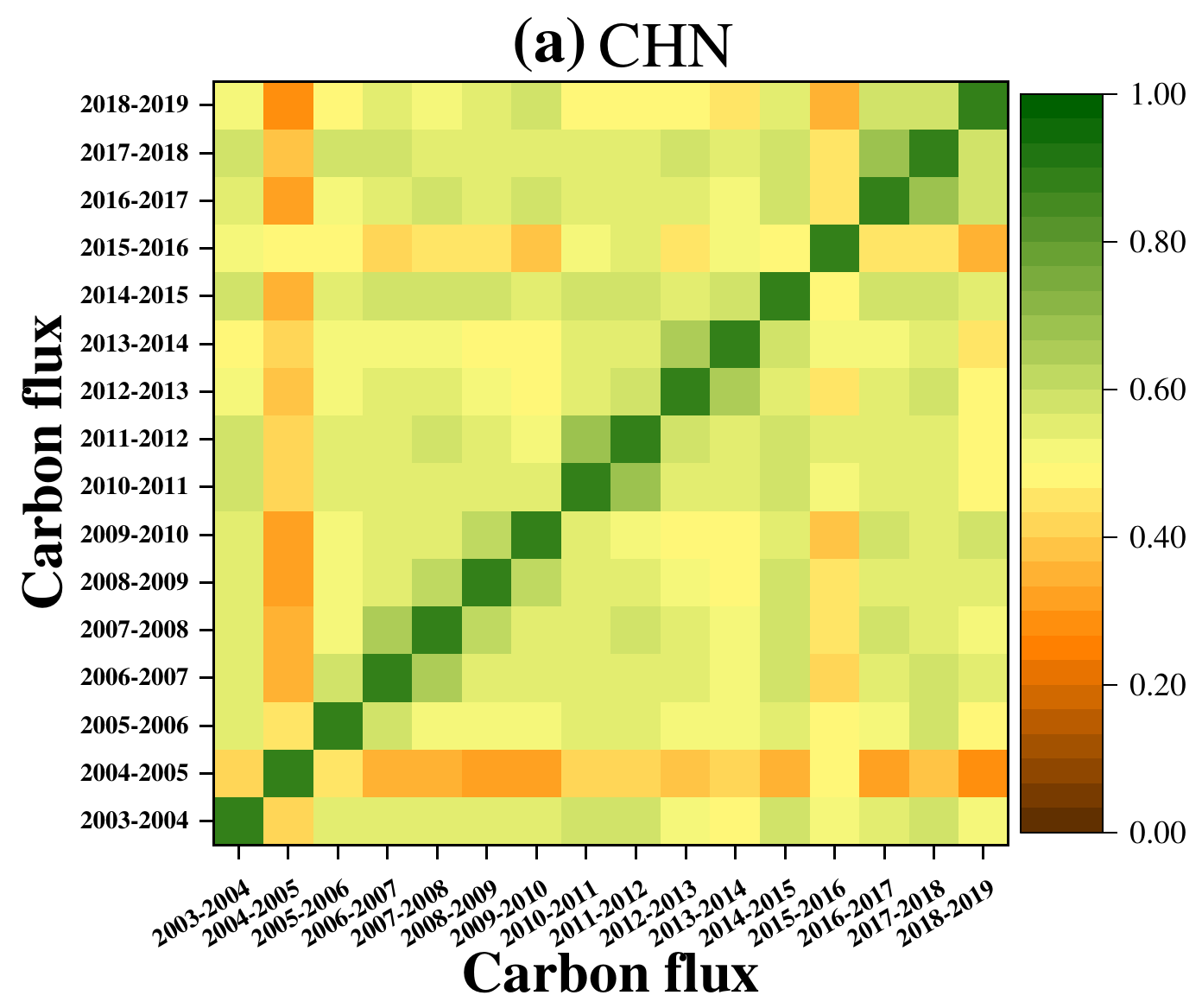}
    \includegraphics[width=9em, height=7.5em]{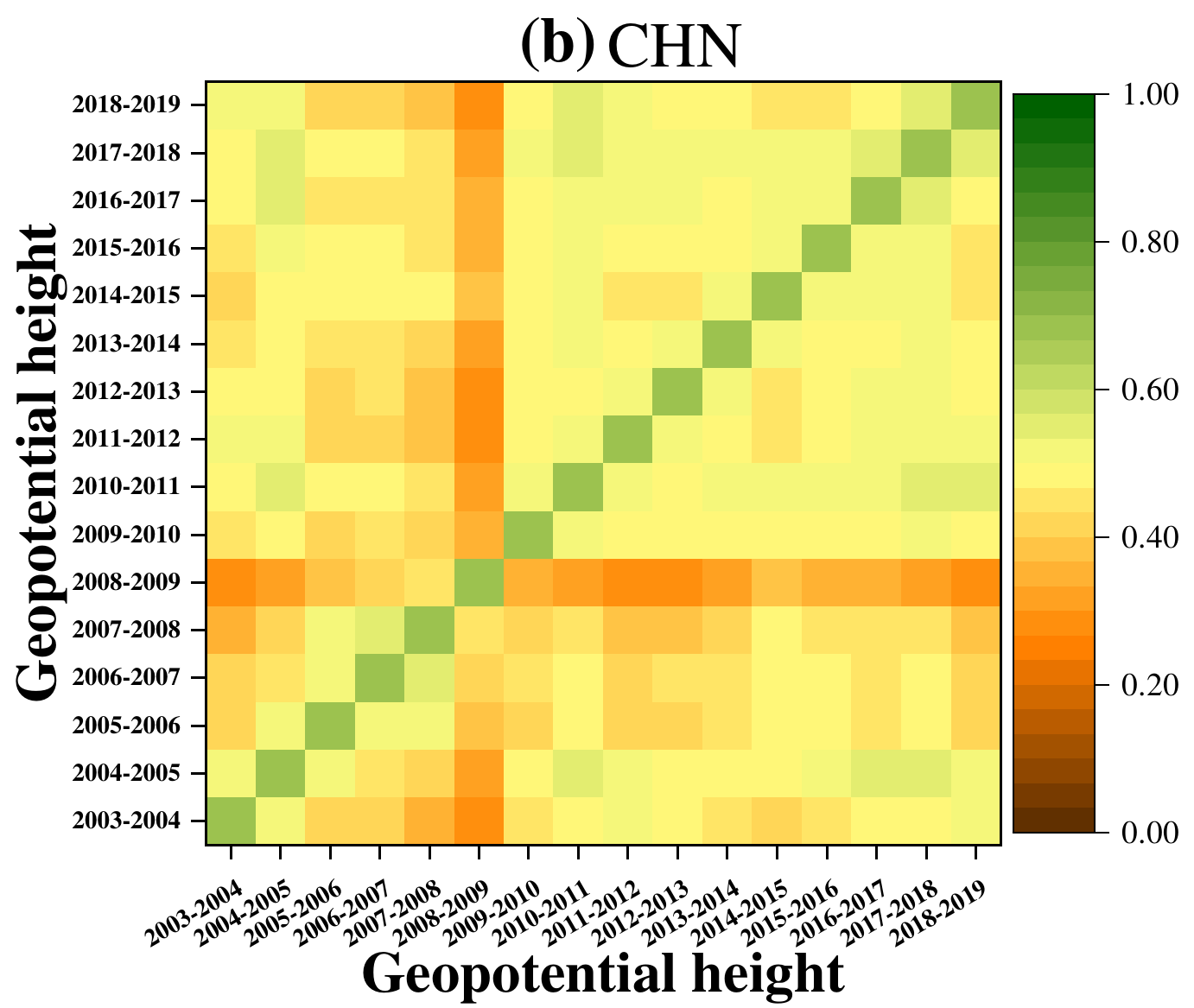}
    \includegraphics[width=9em, height=7.5em]{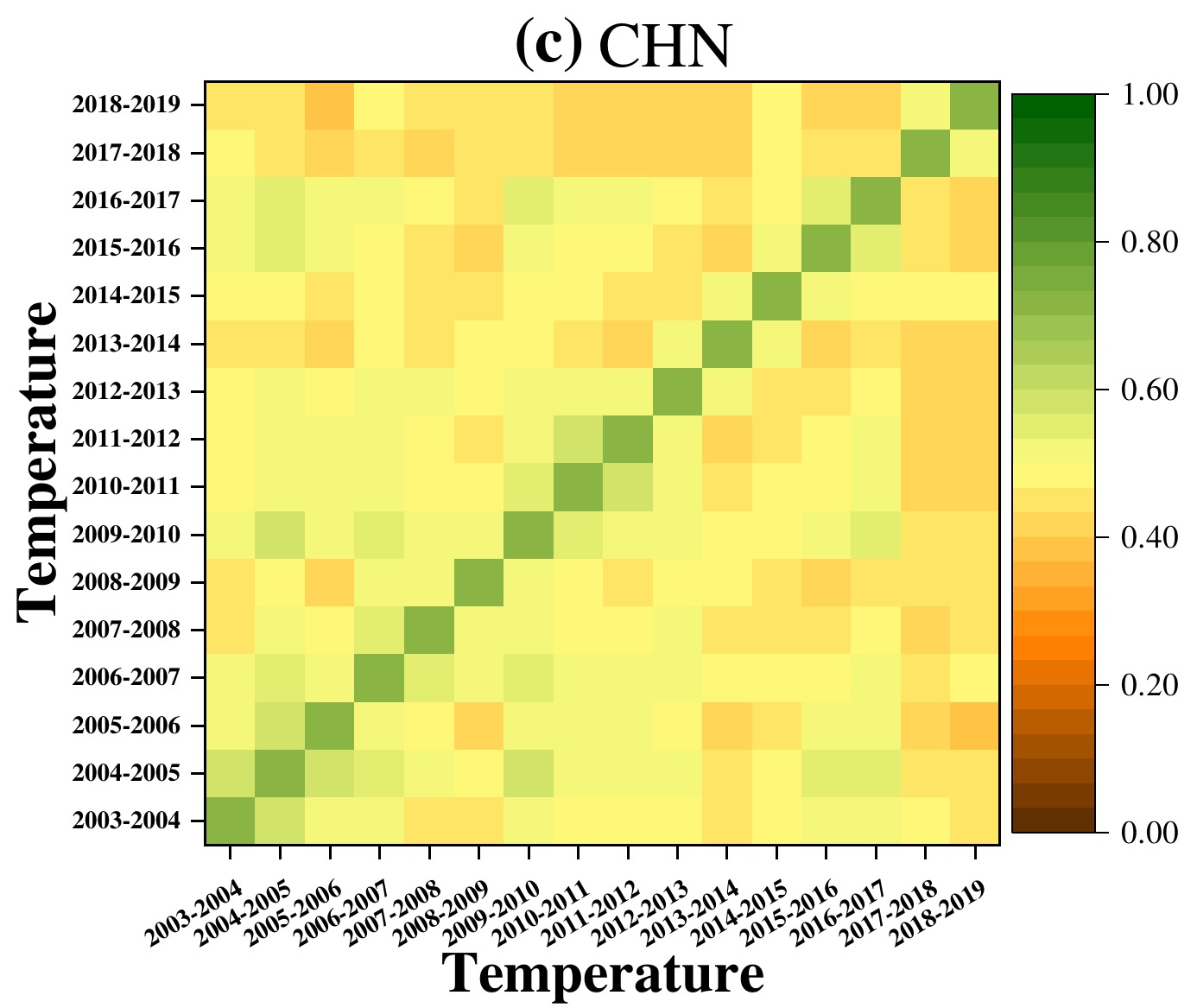}
    \includegraphics[width=9em, height=7.5em]{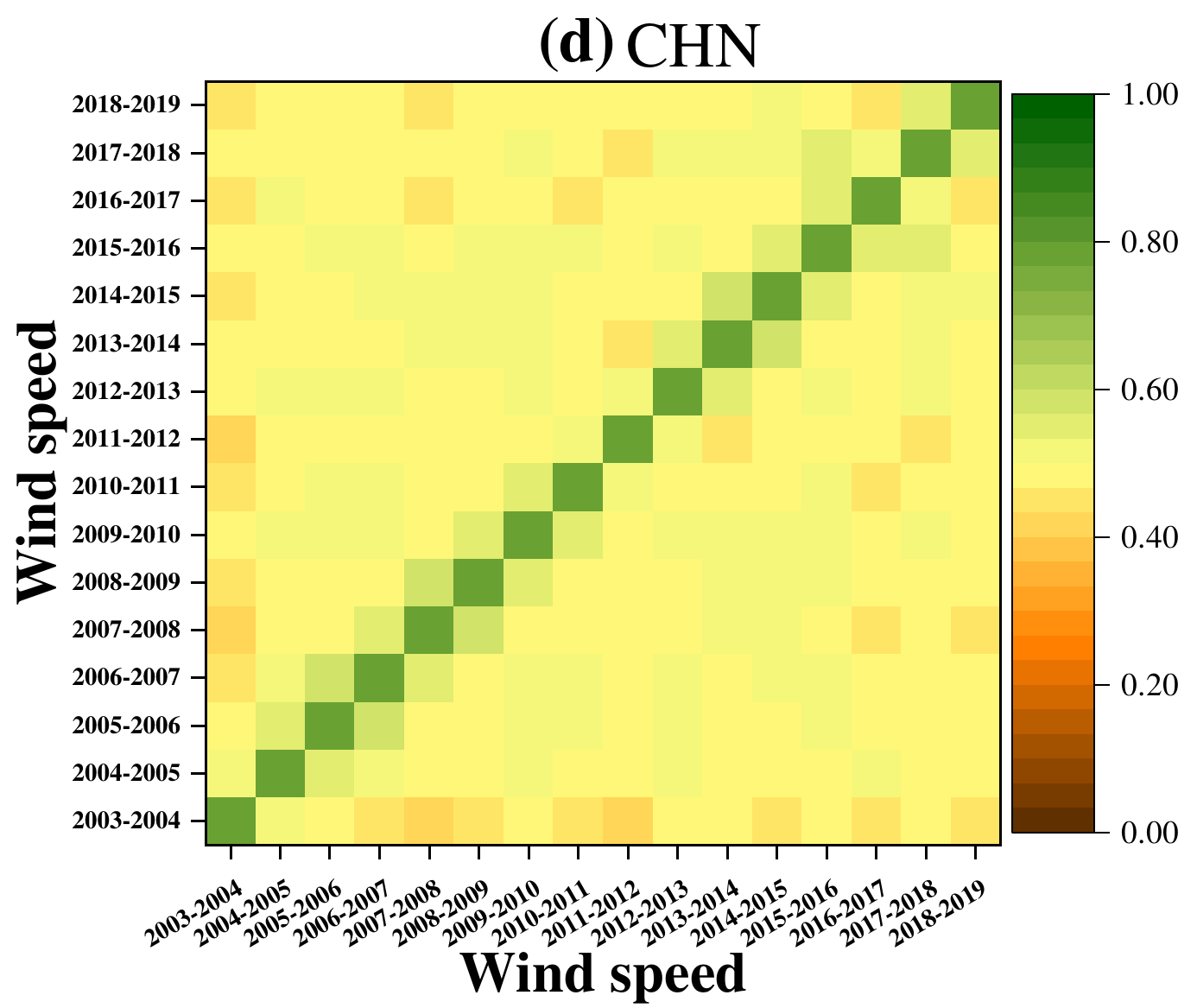}
    \includegraphics[width=9em, height=7.5em]{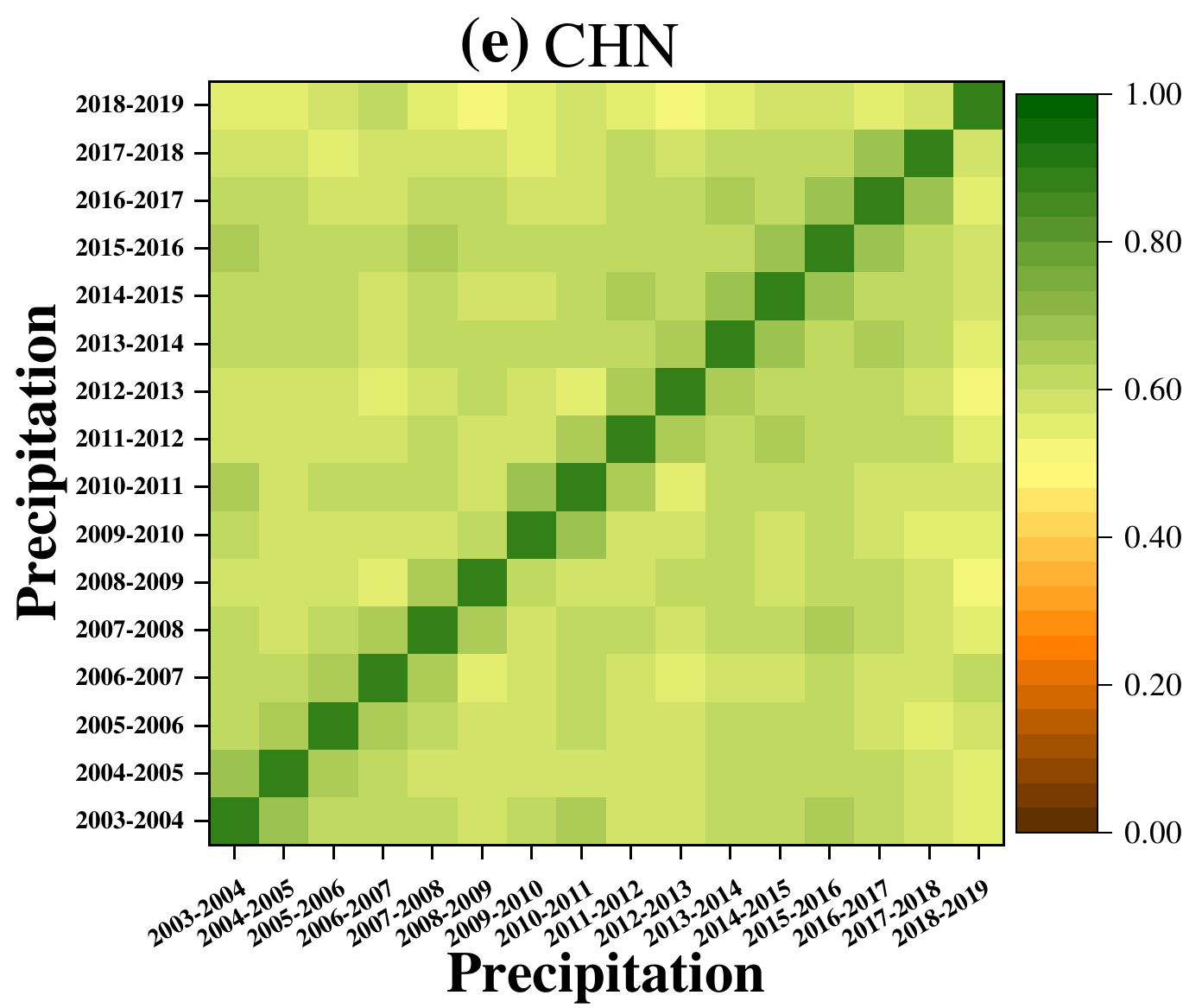}
    \includegraphics[width=9em, height=7.5em]{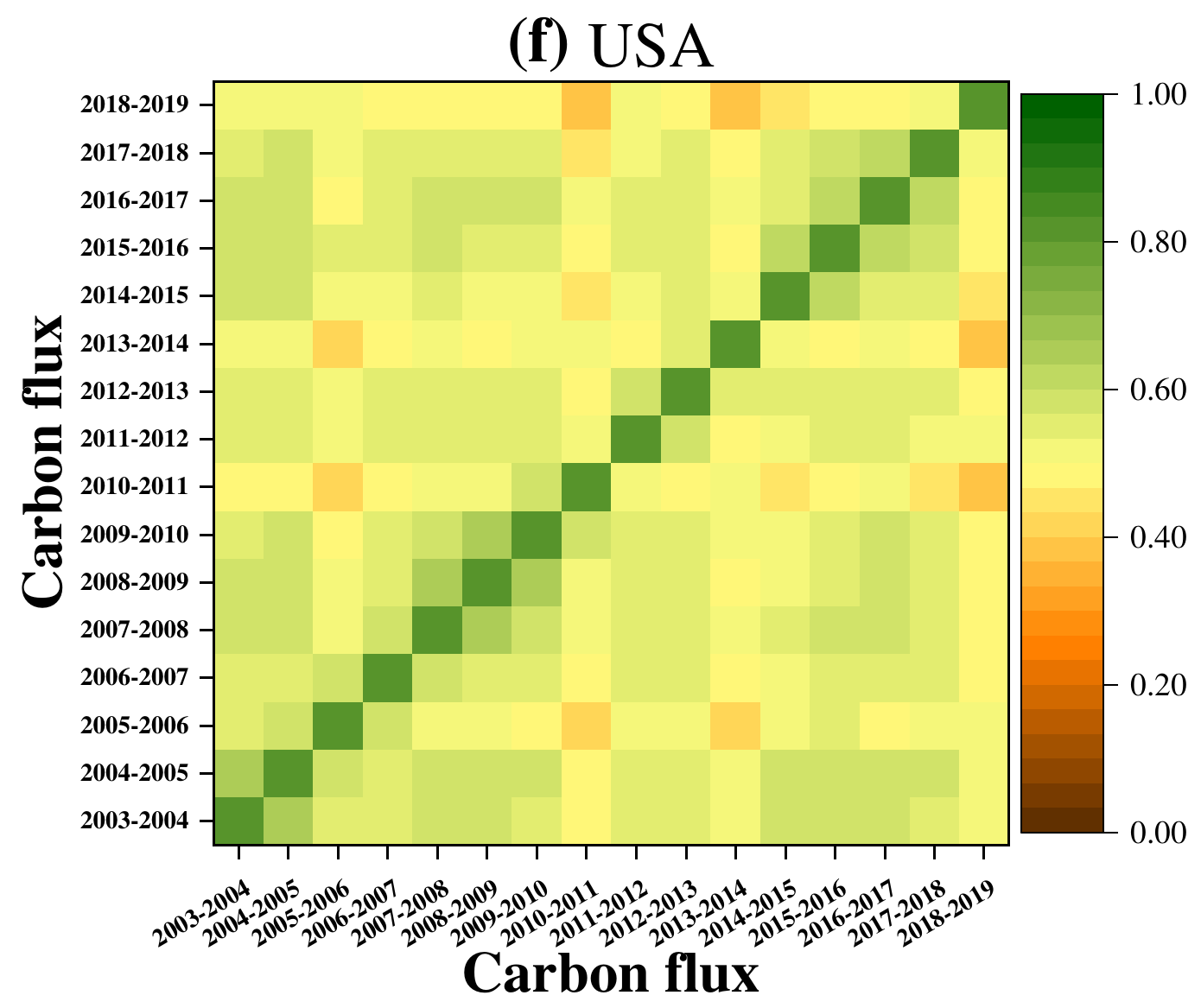}
    \includegraphics[width=9em, height=7.5em]{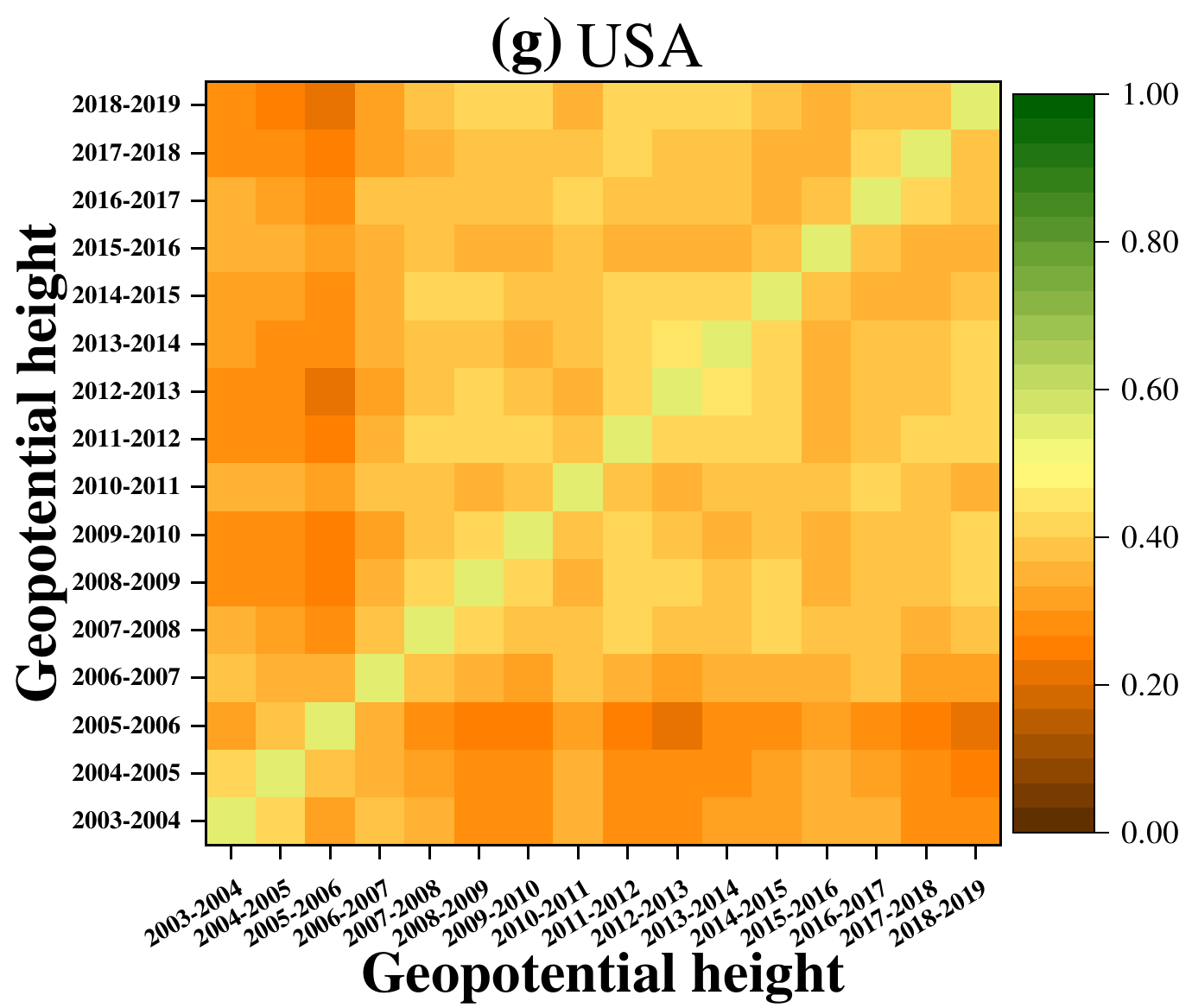}
    \includegraphics[width=9em, height=7.5em]{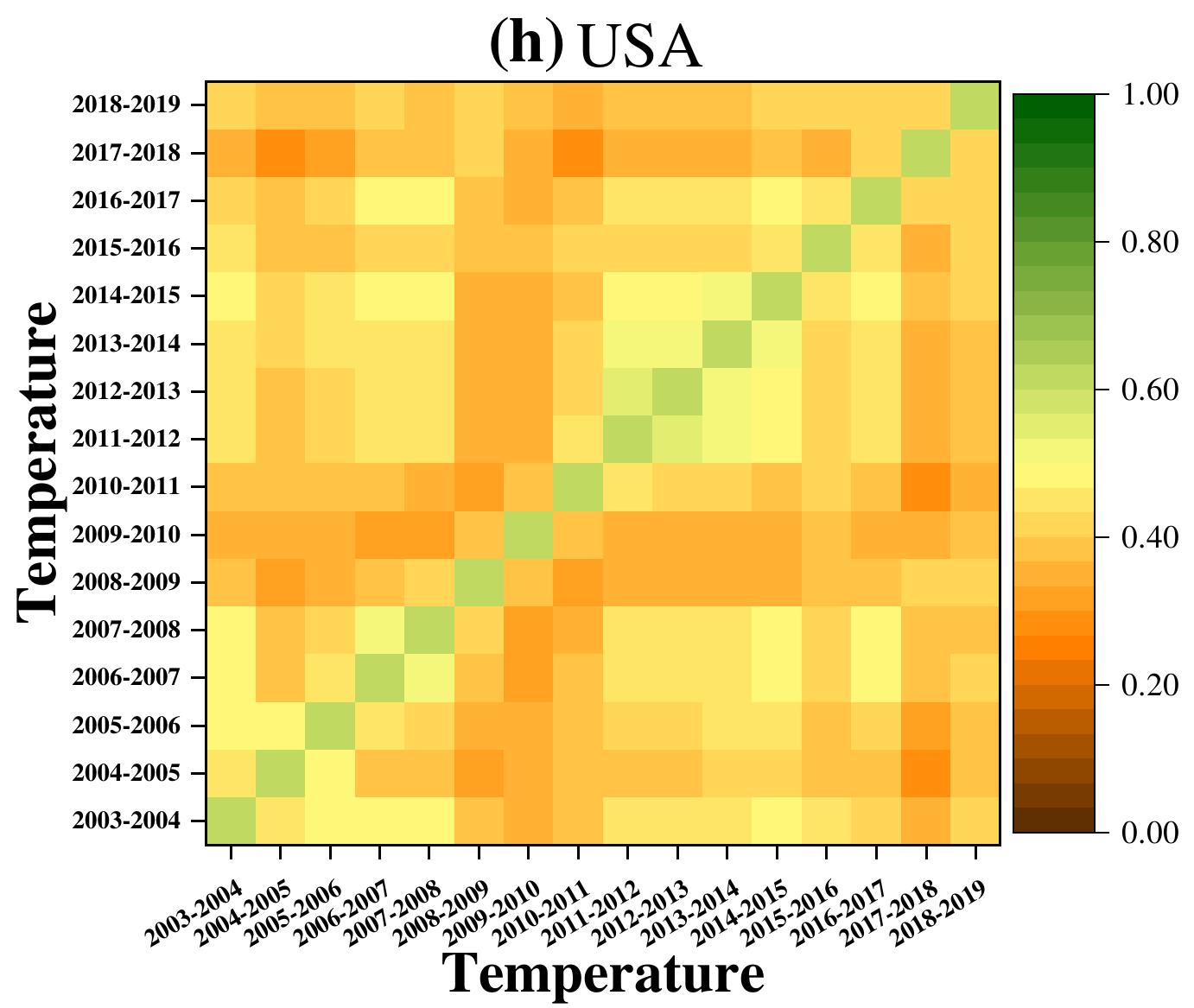}
    \includegraphics[width=9em, height=7.5em]{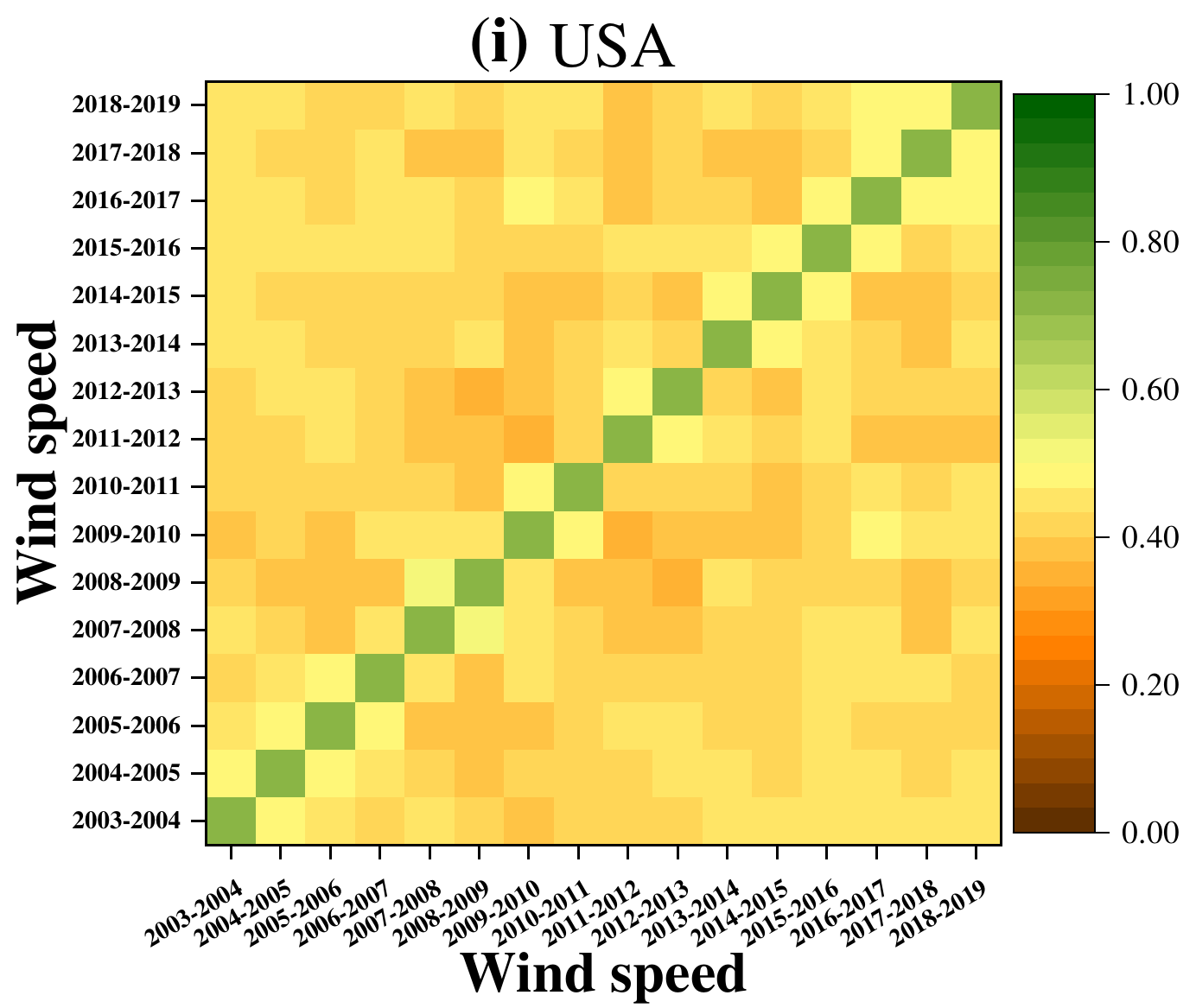}
    \includegraphics[width=9em, height=7.5em]{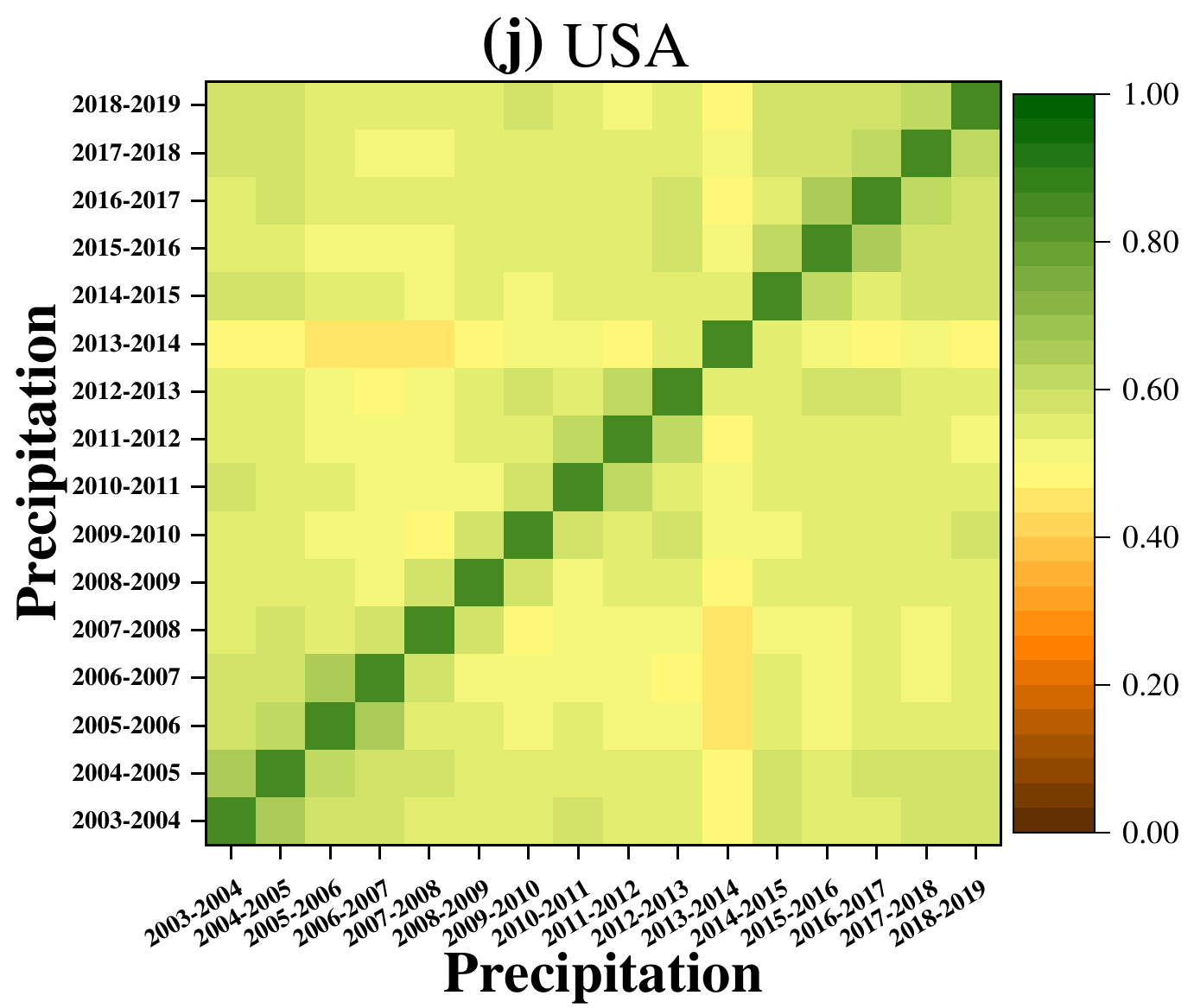}
    \includegraphics[width=9em, height=7.5em]{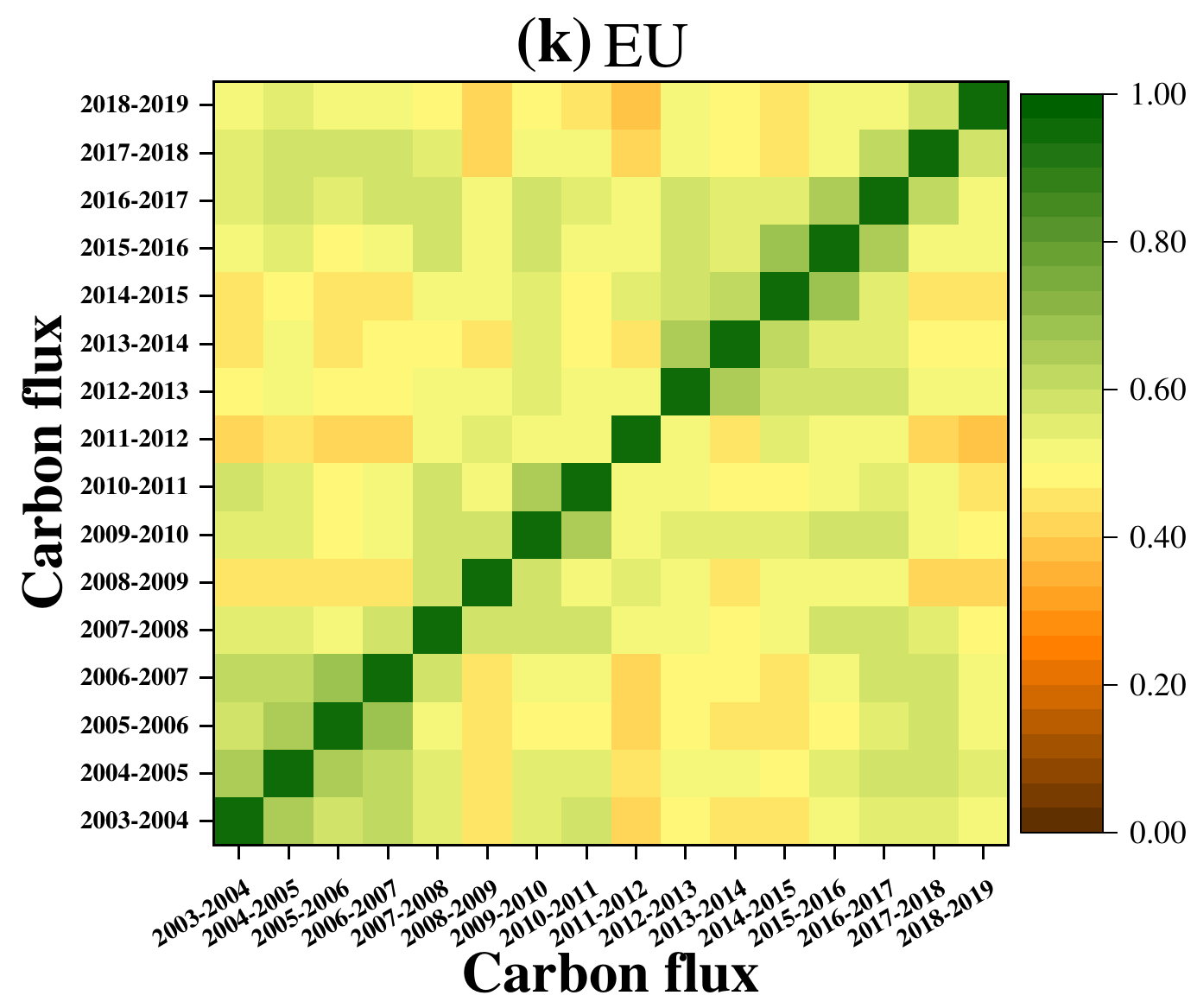}
    \includegraphics[width=9em, height=7.5em]{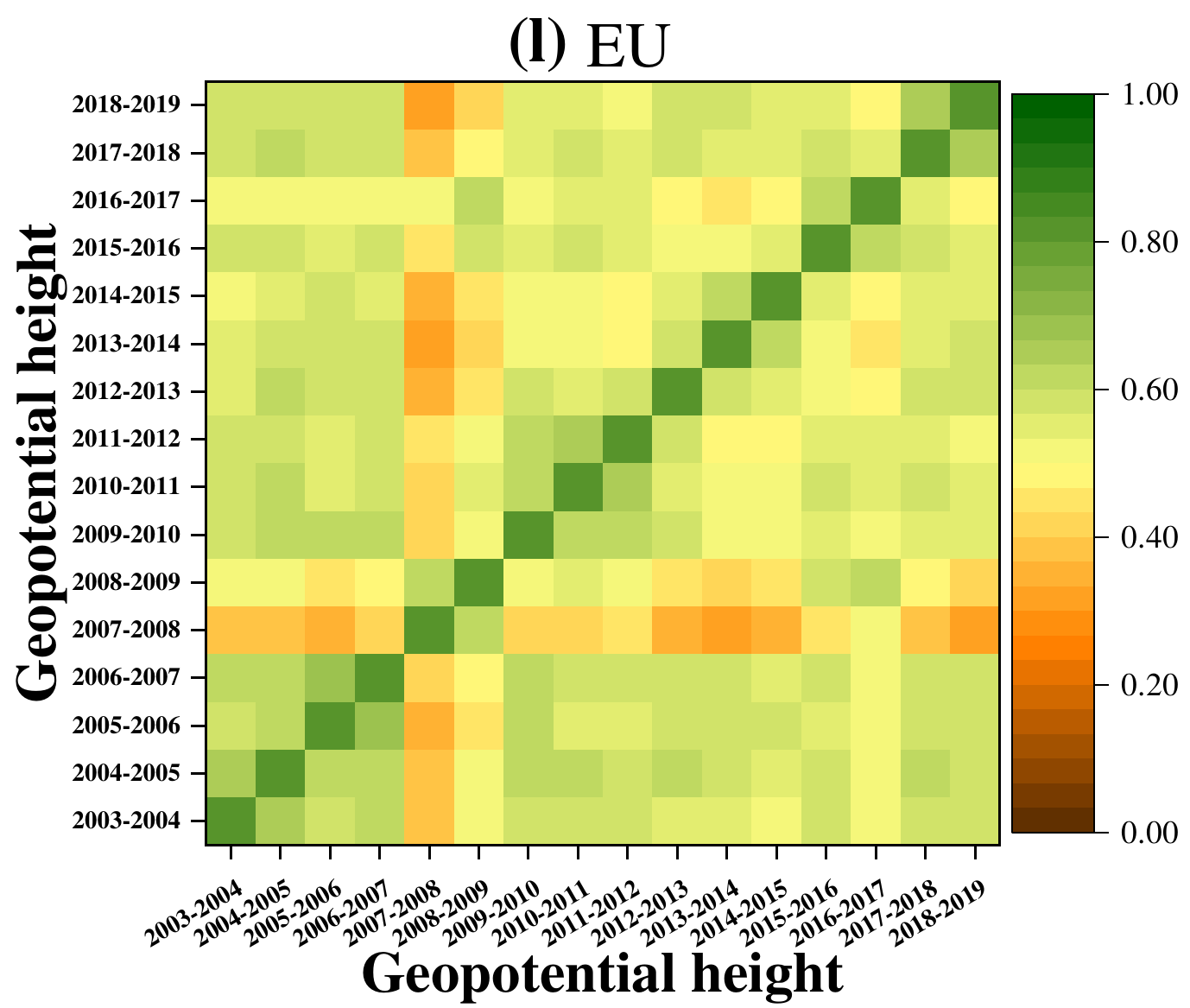}
    \includegraphics[width=9em, height=7.5em]{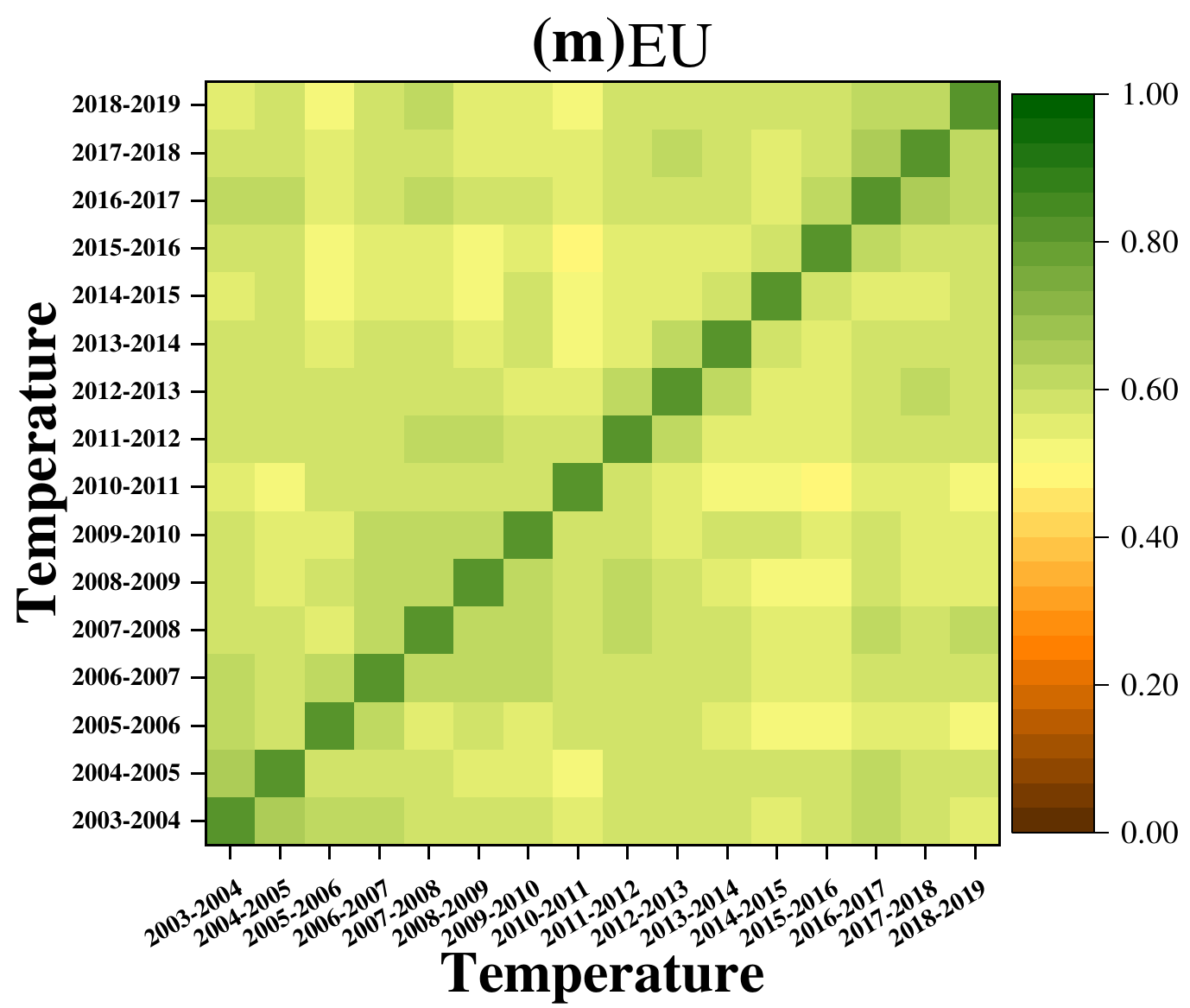}
    \includegraphics[width=9em, height=7.5em]{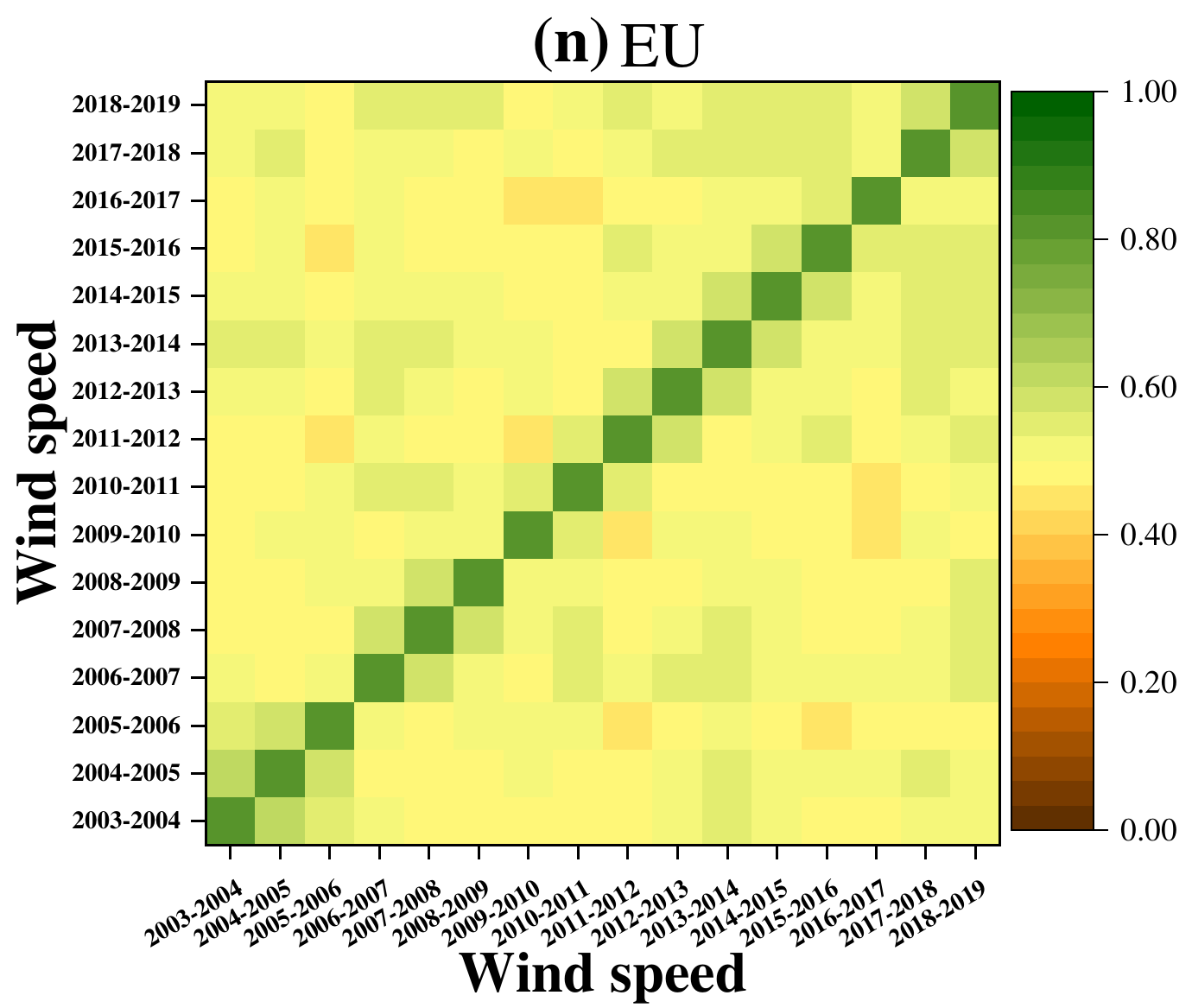}
    \includegraphics[width=9em, height=7.5em]{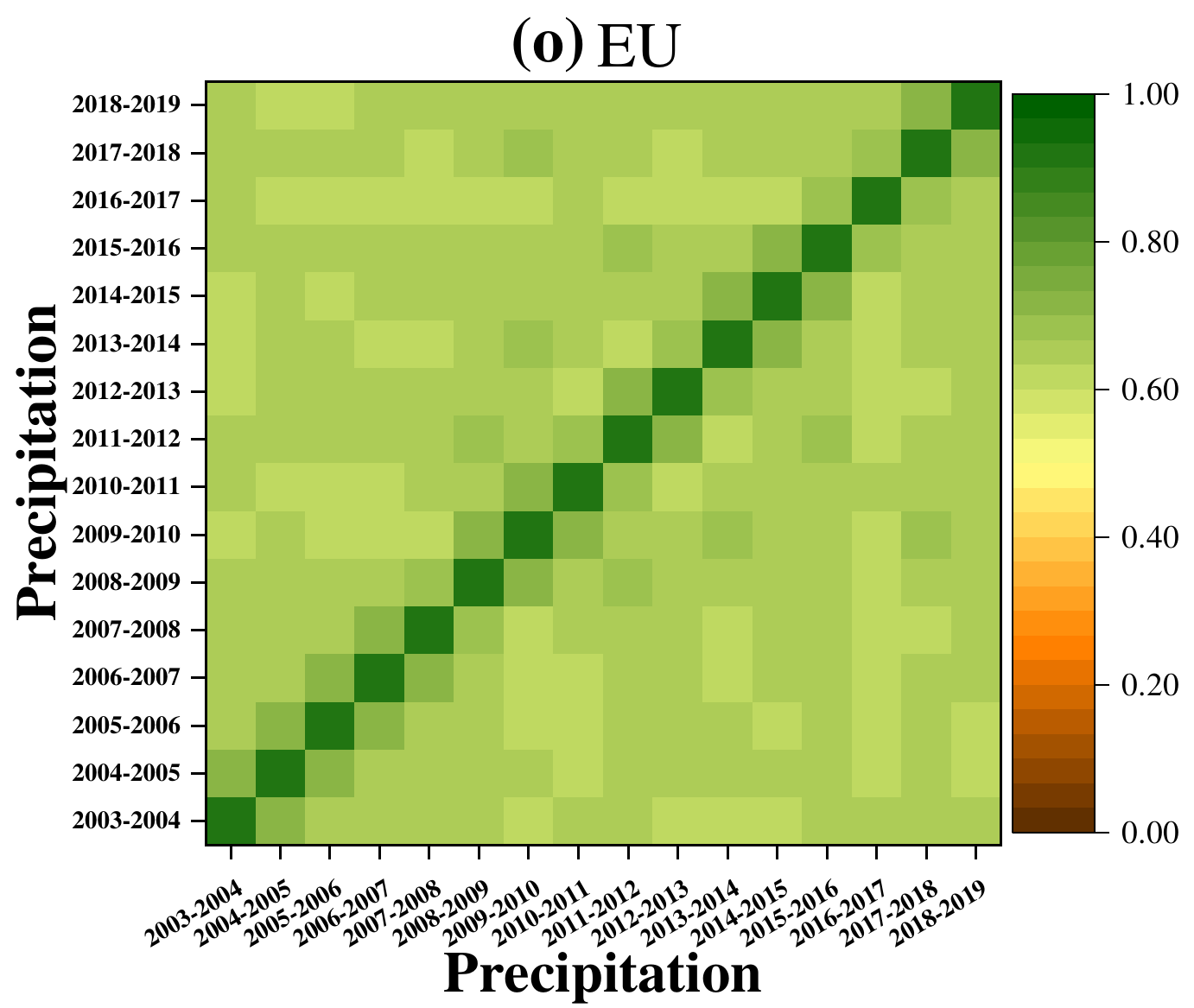}
    \caption{The effective Jaccard similarity coefficient matrix for links in two networks of different years for each of the climate variables. Each matrix element represents the difference between the actual Jaccard similarity coefficient and the corresponding average values obtained from the controlled random case.}
    \label{fig5}
\end{figure}

To test the significance of the actual Jaccard similarity coefficients $J_{ij}$ between networks $i$ and $j$ in the climate and carbon flux, we randomly chose the same number of edges for networks $i$ and $j$ and generate a pair of new networks and measure $J^{ran}_{ij}$. The controlled process is iterated 500 times, and the Jaccard coefficient in the controlled scenario ($J^{ran}$) is determined as the average value of $J^{ran}_{ij}$ computed across these 500 iterations (see SI Figs. S27-S29). Next, we examined the value of the actual Jaccard coefficient ($J_{ij}$) with respect to the values of the Jaccard coefficient in the controlled random case ($J^{ran}$) for determining the significance (t-test). The t-test reveals that the actual Jaccard coefficients are significantly higher and different ($p<0.05$) from the Jaccard coefficients in the controlled random case. Thus, the average of the Jaccard similarity coefficients in the controlled random case ($\overline{J^{ran}}$) is subtracted from the actual Jaccard similarity coefficients to obtain the corrected effective Jaccard similarity coefficient (E-Jaccard), i.e., $E\mbox{-}Jaccard_{ij}$ defined ---$J_{ij}-\overline{J^{ran}}$ (see SI TABLE X for the average and standard deviation of the actual and controlled Jaccard similarity coefficients, $p<0.05$). Fig. ~\ref{fig5} displays the effective Jaccard similarity coefficient matrix for all year pairs based on the carbon flux and climate network links. One can observe that the effective Jaccard coefficients are in the range of $0.51\pm0.09$ ($p<0.05$) for different carbon flux and climate variables in different regions. This implies that the climate variables and carbon flux studied in the Earth's climate system are generally persistent and maintain a steady state. It is notable that the similarity between a given year and the years after does not decay systematically --but only when averaged (See Fig. S38). This suggests further that there is on average a small decay but the persistent network of links are quite stable. However, note that, for the carbon flux network in China, unique low values of Jaccard have been observed in two years, 2004-2005 and 2015-2016. This could be related to the fact that in 2004, China became the world's largest emitter of $\rm {CO_{2}}$, although forests in the country absorbed more than 8\% of national emissions during that year \cite{WRI2023}. The following year, 2005, $\rm {CO_{2}}$ emissions reached a new record high. Also, in 2015, China's carbon emissions declined for the first time, marking a turning point in the country's carbon emissions \cite{NDRC2023}. These results indicate that China has been actively pursuing the goals of the ``Paris Agreement". In addition, Figs. ~\ref{fig5}(b) and ~\ref{fig5}(l), the geopotential height networks in China and Europe, under the general persistence, also show unique low Jaccards in the 2008-2009 and 2007-2008 time windows, respectively. This finding, can be probably attributed to the largest snow extent in Eurasia in January 2008, and the unusually warm temperatures resulted in the smallest spring snow extent in Eurasia in March and April 2008, leading to severe winter weather and unique behavior of geopotential heights in Eurasia \cite{world2009wmo, fink2009european}. Furthermore, the Jaccard similarity coefficients of the links of each two networks in carbon flux and climate variables are also analyzed for links having geographic distances above $500km$ and $1000km$ (see SI Figs. S33 and S34). For links above $500km$ the Jaccard values are within the range of $0.45\pm0.08$, and for links above $1000km$ the Jaccard values are within the range of $0.34\pm0.11$ ($p<0.05$). Thus, the similarity represented by the Jaccard values decreases as the geographical distance between nodes increase. Moreover, we analyze the average of Jaccard similarity coefficients based on the effective Jaccard similarity coefficient matrix, considering in between intervals ranging from one to five years (i.e., averaging over each of the first five diagonal columns below the central dark green column of Fig. ~\ref{fig5}) (see SI Figs. S38). One can observe, in general, a gradually decreasing trend in the average of Jaccard as the interval years increase. Note that the decrease is significant at the interval of two years, followed by a slow decreasing trend subsequently. This phenomenon can be attributed to the presence of a 365-day overlap in the data for a one-year interval, resulting in an increased similarity in the network structure, which in turn leads to a higher average of Jaccard value. Thus, when the interval of two years, the decrease in the average value of Jaccard reaches the most significant extent. Nevertheless, one can see that the similarity of the climate variables decreases with time, which is reflected in the gradual decrease in persistence. In addition, we notice that the trends of climatic variables in different regions at different geographic distances remain generally consistent. Notably, within the China region, there is a noteworthy slow increase in the persistence of temperature and wind speed after the interval of 3 years. This change in trend may reflect climate dynamics processes or climate variability patterns specific to China.

\begin{figure}[h]
    \centering
    \includegraphics[width=45em, height=25em]{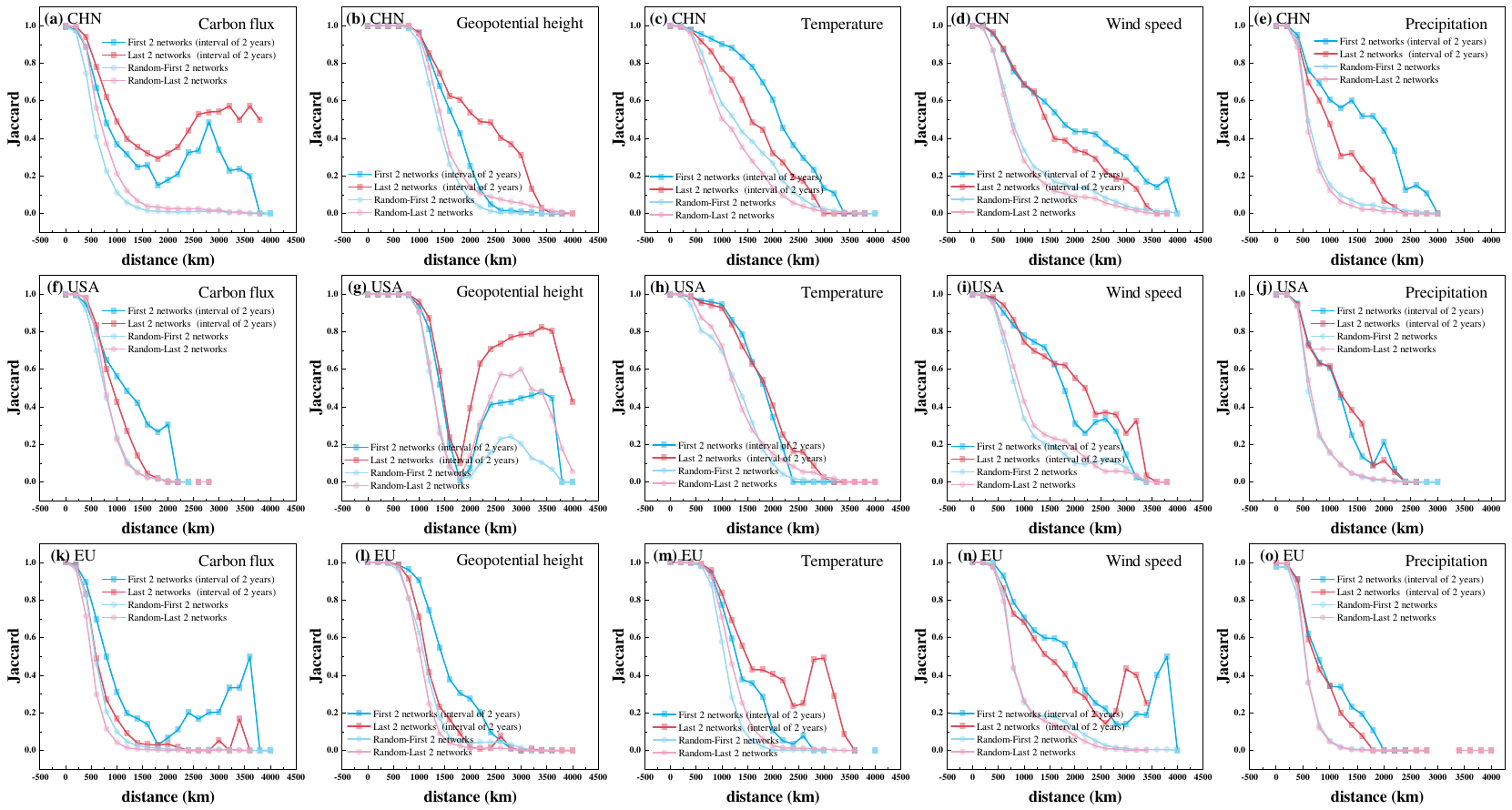}
    \caption{Variation of Jaccard similarity coefficients for two yearly networks at different geographical distances.}
    \label{fig6}
\end{figure}

In addition, this paper extends the Jaccard similarity coefficient by analyzing the variation of Jaccard similarity coefficients with years within each of the two networks for different geographical distances. To test the significance of our results, we analyze the network structure shuffled at least 500 times to obtain the Jaccard similarity coefficients at different geographic distances in the controlled scenario. The findings, depicted in Fig. ~\ref{fig6}, reveal that in the controlled scenario, the Jaccard similarity coefficient at different geographic distances is consistently lower than the corresponding actual Jaccard similarity coefficient as shown in Fig. ~\ref{fig6}. Furthermore, in contrast to the carbon flux network, the Jaccard index of the other climate networks exhibits similarity for the two networks at different times. It is also seen that the shorter geographically links are significantly, more similar over the years. The results are consistent with those in Fig. ~\ref{fig4}, with increasing geographic distance, the number of significant links decreases, i.e., there are fewer connections between more distant geographic distances. Moreover, the Jaccard value is also smaller, i.e., there is less climatic similarity between longer geographic distances. Furthermore, Fig. ~\ref{fig6}(a, g) shows some anomalies of the large number of links at geographic distances between 1500-3500 km, indicating the possible influence of Rossby waves under the effects of air transport \cite{zhang2019significant, ying2020rossby}

\section{Conclusions}
In this manuscript we developed a framework to study the time persistence of climate networks. While persistence of single records such as climate, heartbeat and DNA sequences have been extensively studied \cite{bunde2013fractals} the persistence of a network as a whole has been rarely addressed. The steady state is a crucial feature of the climate and carbon flux system, playing an essential role for understanding climate changes in the Earth system. However, the persistence and steady state of the climate and carbon flux systems are not yet fully understood. In this study, we construct climate and carbon networks where nodes are spatial locations and links are derived based on evaluating the crosscorrelations between climate variables and carbon flux in different locations. This framework enables to test the similarity between the networks in different years in order to evaluate the persistence of the climate and carbon systems.

We investigate the similarity of climate and carbon flux networks in different years for three land regions, namely East Asia (China), North America (contiguous United States), and Europe, using daily data of climate and carbon flux during 2000-2019. We focus on the spatio-temporal perspective to explore the evolution of the similarity of the system network structure. The results indicate that the similarity of the climate and carbon flux networks between different years represented by the Jaccard measure are in the range of $0.51\pm0.08$ ($p<0.05$). That is close to 50\% of the links remain the same during the years, indicating that the correlation network structure of the system is generally persistent and in a stable state. However, the carbon flux network in China exhibits unique behavior in specific two years, 2004-2005 and 2015-2016, which is probably attributed to China's rapid development and active participation in the ``Paris Agreement". On the other hand, the study reveals that the persistence of the network structure of climate and carbon flux networks tends to significantly decrease as the geographical distance increases. Finally, our study provides a new methodological and theoretical framework for analyzing the network persistence of the association structure for carbon flux and the climate system, which could help to understand the structural steady states of the Earth's climate system that could provide new insights for forecasting extreme events in the climate system.
\section{Acknowledgements}
This research is supported by grants from the National Natural Science Foundation of China (Grant No. 72174091, 62373169), National Key Research and Development Program of the Ministry of Science and Technology of China (Grant No.2020YFA0608601), Major Projects of the National Social Science Foundation of China (Grant No. 22\&ZD136), Science and technology innovation project of Carbon Peaking and Carbon Neutrality of Jiangsu Province of China (Grant No. BE2022612),  National Statistical Science Research Project (Grant No. 2022LZ03), Special Project of Emergency Management Institute of Jiangsu University (Grant No. KY-A-08), Israel Science Foundation (Grant No. 189/19), Binational Israel-China Science Foundation (Grant No. 3132/19), Science Minister-Smart Mobility (Grant No. 1001706769), European Union’s Horizon 534 2020 research and innovation programme (DIT4Tram, Grant Agreement 953783), and the Jiangsu Postgraduate Research and Innovation Plan (Grant No. KYCX22$\_$3662). Ting Qing acknowledges the support from the program of the China Scholarship Council.

\bibliographystyle{unsrt}

\end{document}


\title{Supporting information for ``Time persistence of climate and carbon flux networks"}

\author{Ting Qing}
\affiliation{School of Mathematical Sciences, Jiangsu University, Zhenjiang, 212013 Jiangsu, China}%
\affiliation{Department of Physics, Bar-Ilan University, Ramat-Gan 52900, Israel}%
\author{Fan Wang}%
\affiliation{Department of Physics, Bar-Ilan University, Ramat-Gan 52900, Israel}%
\author{Qiuyue Li}%
\affiliation{Department of Physics, Bar-Ilan University, Ramat-Gan 52900, Israel}%

\author{Gaogao Dong}%
\thanks{Corresponding author: gago999@126.com}
\affiliation{School of Mathematical Sciences, Jiangsu University, Zhenjiang, 212013 Jiangsu, China}%
\author{Lixin Tian}%
\thanks{Corresponding author: tianlx@ujs.edu.cn}
\affiliation{School of Mathematical Sciences, Jiangsu University, Zhenjiang, 212013 Jiangsu, China}%
\affiliation{Research Institute of Carbon Neutralization Development, School of Mathematical Sciences, Jiangsu University, Zhenjiang 212013, China}%
\affiliation{Jiangsu Province Engineering Research Center of Industrial Carbon System Analysis, School of Mathematical Sciences, Jiangsu University, Zhenjiang 212013, China}%
\affiliation{Jiangsu Province Engineering Research Center of Spatial Big Data, School of Mathematical Sciences, Nanjing Normal University, Nanjing 210023, China}%
\affiliation{Key Laboratory for NSLSCS, Ministry of Education, School of Mathematical Sciences, Nanjing Normal University, Nanjing 210023, China}%
\author{Shlomo Havlin }%
\thanks{Corresponding author: havlin@ophir.ph.biu.ac.il}
\affiliation{Department of Physics, Bar-Ilan University, Ramat-Gan 52900, Israel}%

\maketitle
The supplementary information presents the data used in the paper and lists the figures supplemented in the paper, as shown in Figs. S1-S23.
\section{Data}
\subsection{Carbon flux}
This paper uses CarbonTracker, an open-source product of the NOAA Earth System Research Laboratory, which uses data from the NOAA ESRL Greenhouse Gas Observing Network and partner agencies. CarbonTracker is a $\rm {CO_{2}}$ measurement and modeling system developed by NOAA to track the global sources and sinks of $\rm {CO_{2}}$. The carbon flux dataset is composed of four components, namely fossil fuel emissions, land biosphere NEE excluding fires, wildfire emissions, and air-sea exchange. These components are cautioned by summing flux components to get total surface CO2 exchange with the atmosphere. The global surface flux has a spatial resolution of $1^{\circ}\times1^{\circ}$, and a temporal resolution of three-hourly. For this study, we use the CarbonTracker version CT2019B\cite{Carbon} dataset covering the period between 2000 and 2019.
\subsection{Climate variables}
In this study, the European Centre for Medium-Range Weather Forecasts (ECMWF) reanalysis v5 (ERA5)\cite{ERA5} dataset, which is derived from satellite and ground-based observations, is interpolated to obtain the gridded data. The selected climate variables are Geopotential height, Temperature, Wind speed, and Precipitation (including rain and snow), with a spatial resolution of $0.25^{\circ}\times0.25^{\circ}$ and a temporal resolution of one hour. Similar to the carbon flux data, we utilize the ERA5 dataset covering the period between 2000 and 2019. Geopotential height (500hpa) is measured in $m^2s^{-2}$ and represents the gravitational potential energy per unit mass at a given location relative to the mean sea level. Temperature (1000hpa at the surface) is measured in $K$ and represents the temperature in the atmosphere. Wind speed (1000hpa at the surface) is measured in $ms^{-1}$ and represents the eastward component of the wind, which is the horizontal velocity of the air moving eastward, while a negative value indicates that the air is moving westward. Precipitation is measured in $m$ and represents the cumulative liquid and frozen ice, including rain and snow, falling to the Earth's surface (1000hpa), and is the sum of widespread and convective precipitation.

\section{Supplementary Figures}

\begin{center}
\includegraphics[width=40em, height=21em]{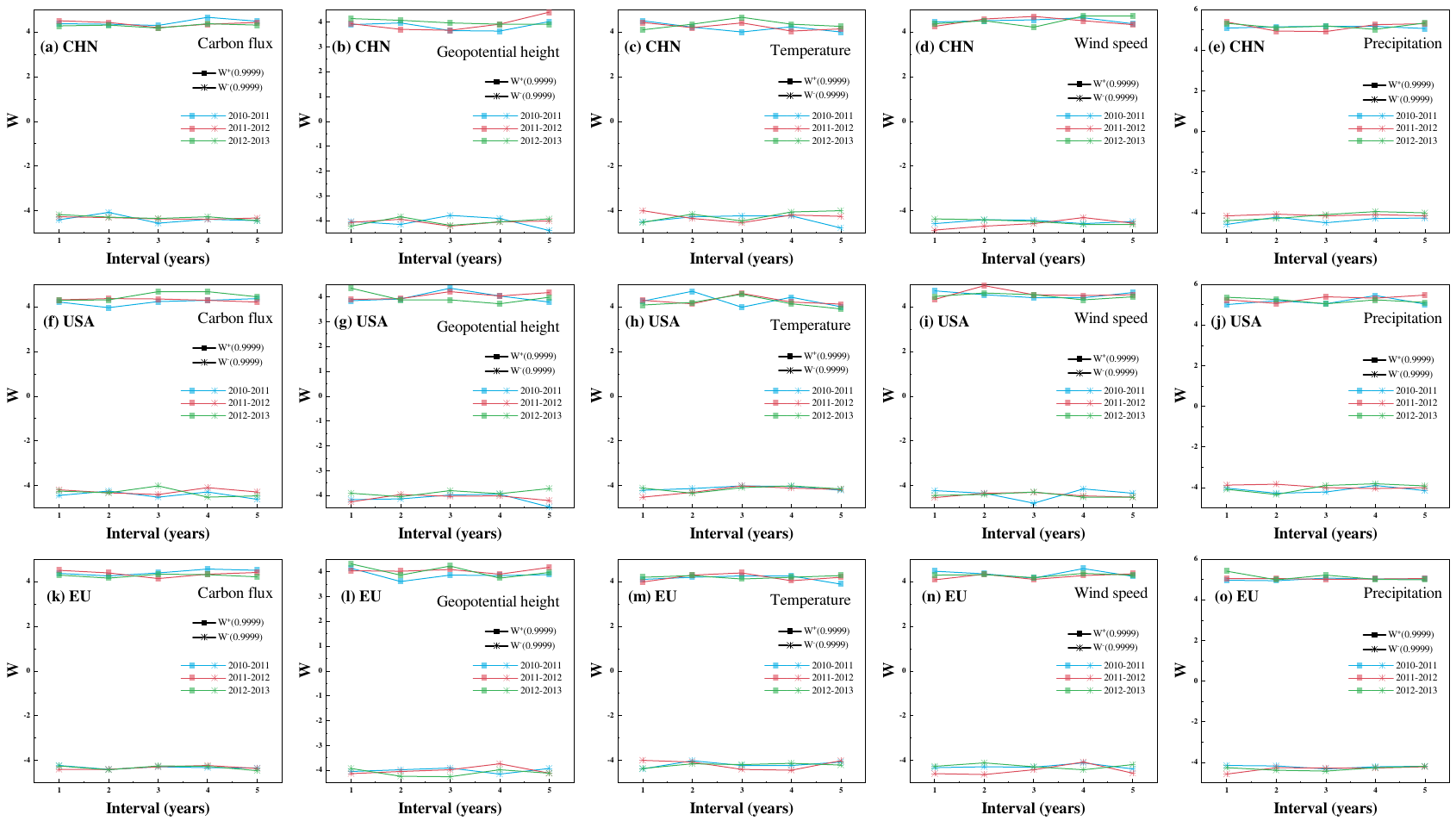}
\end{center}

\begin{center}
\noindent {\small {\bf Fig. S1} Function of the interval years ($(t-Interval)$th time window) and the $W$(threshold) in shuffling procedure.}
\end{center}

\begin{center}
\includegraphics[width=8.5em, height=7em]{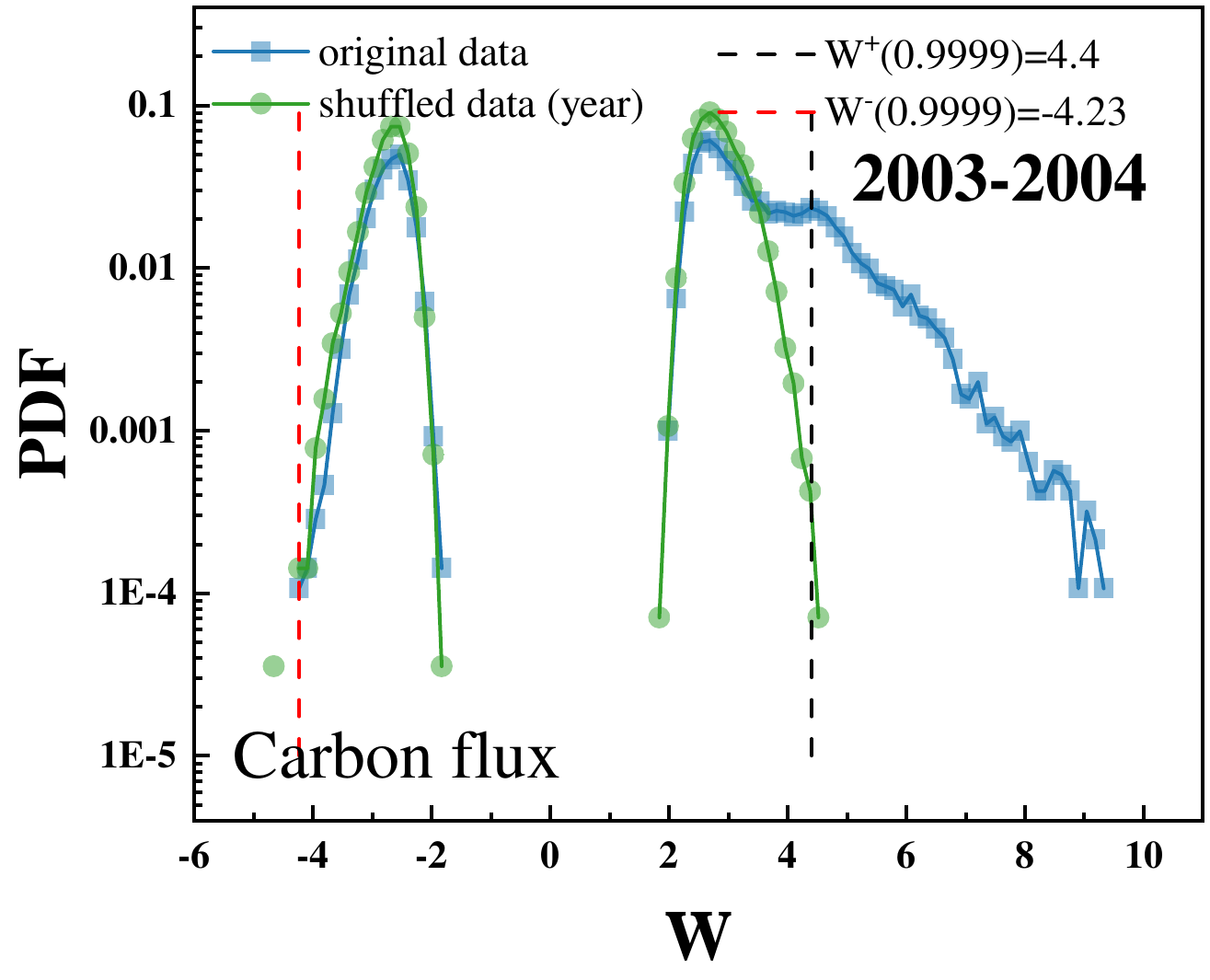}
\includegraphics[width=8.5em, height=7em]{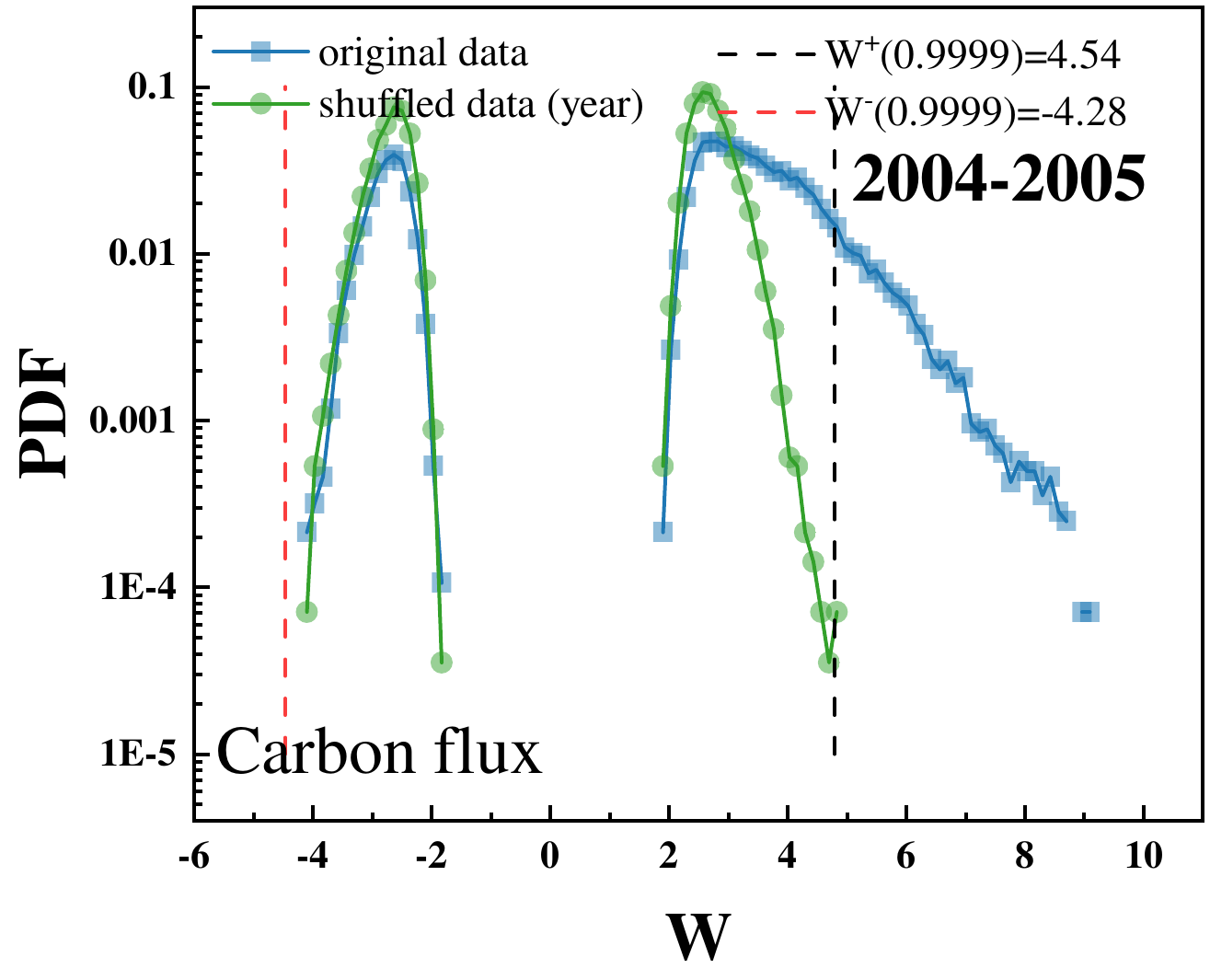}
\includegraphics[width=8.5em, height=7em]{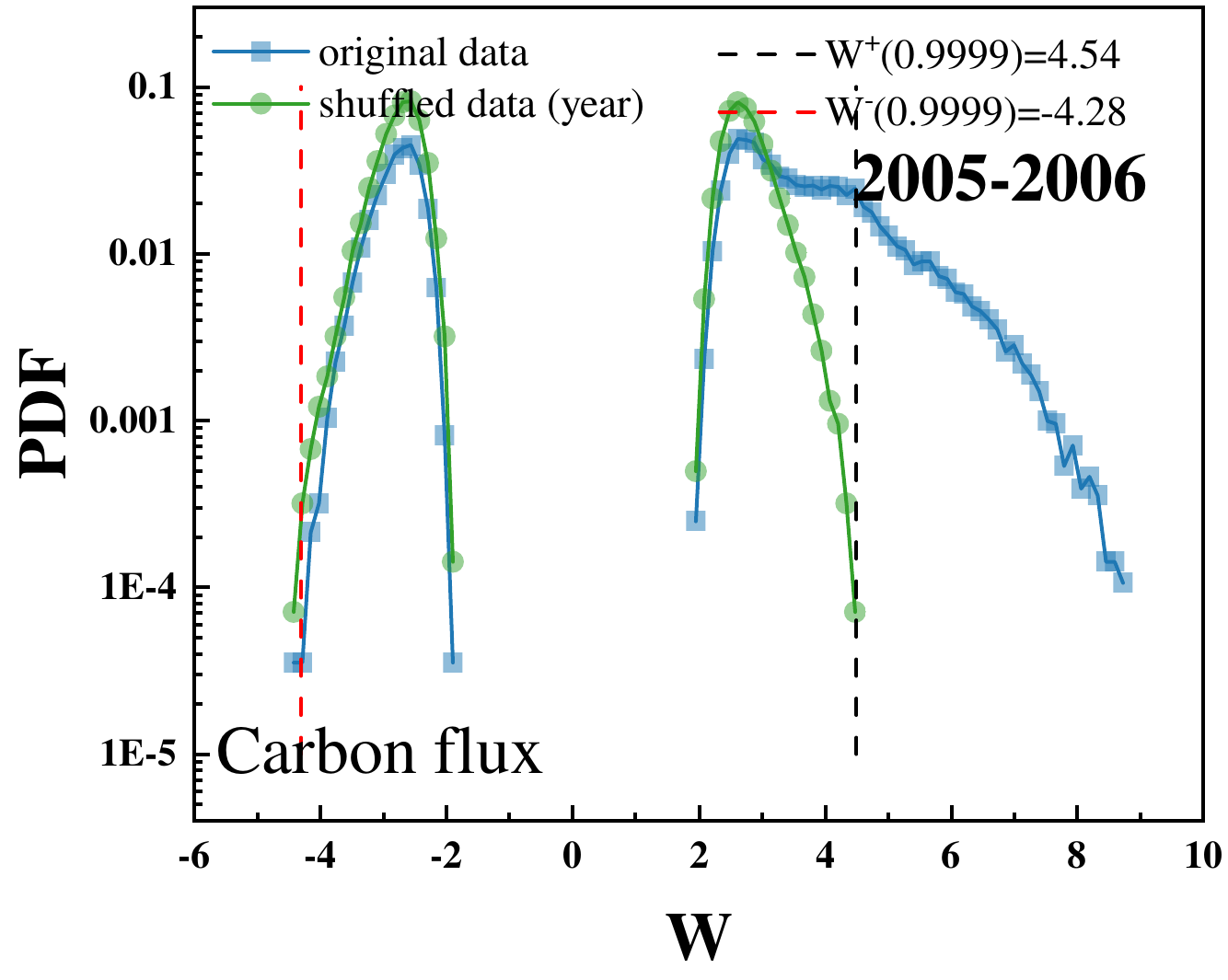}
\includegraphics[width=8.5em, height=7em]{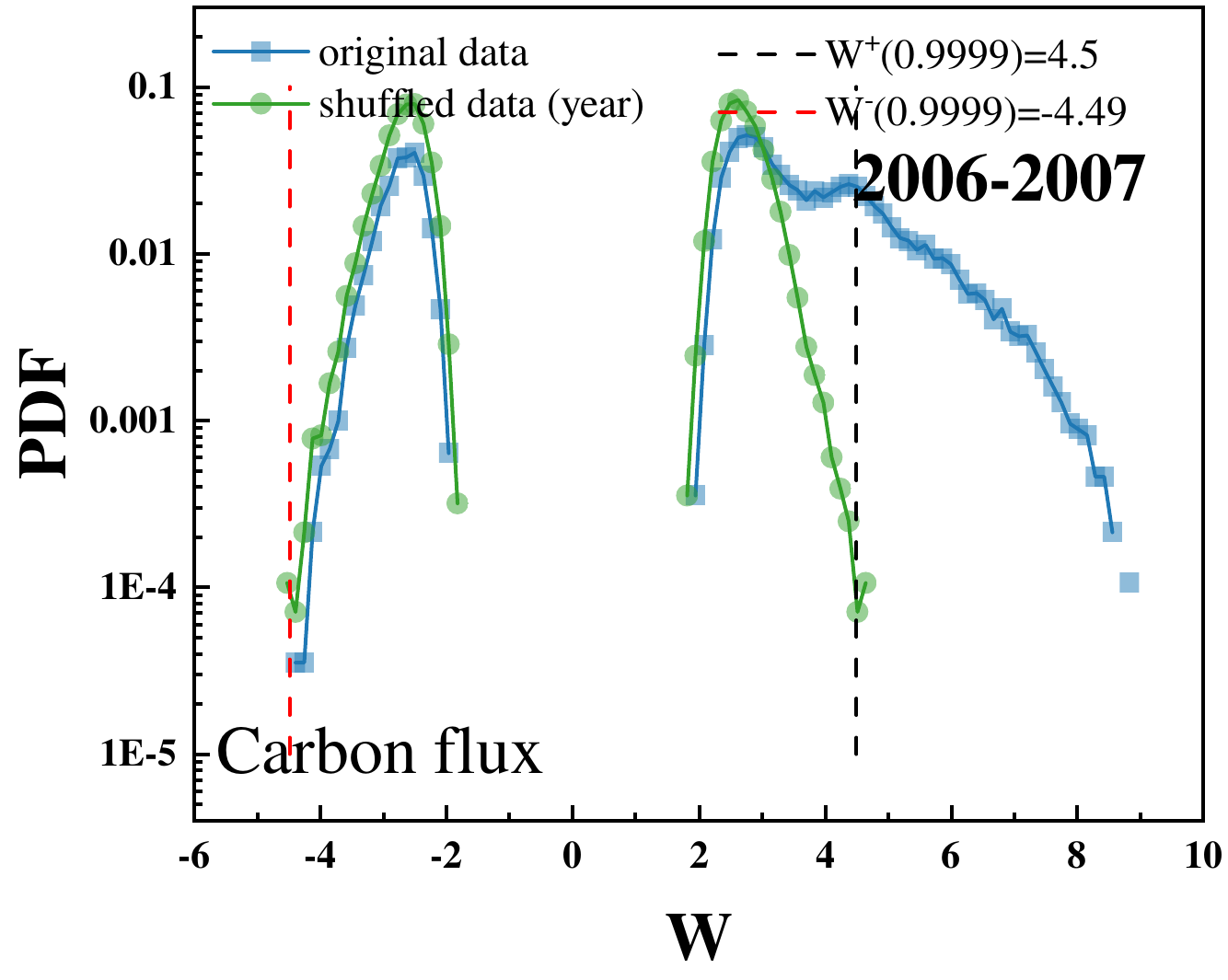}
\includegraphics[width=8.5em, height=7em]{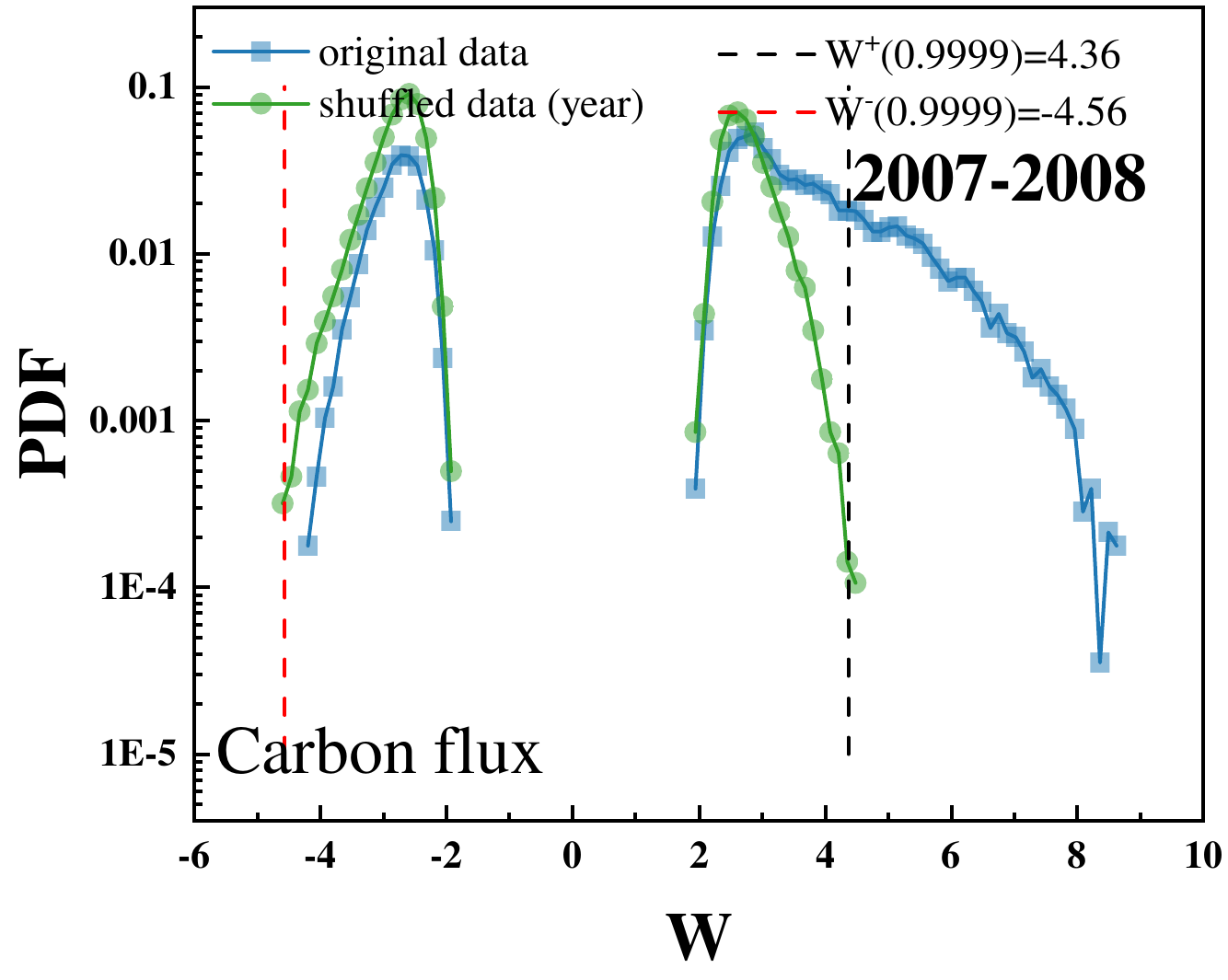}
\includegraphics[width=8.5em, height=7em]{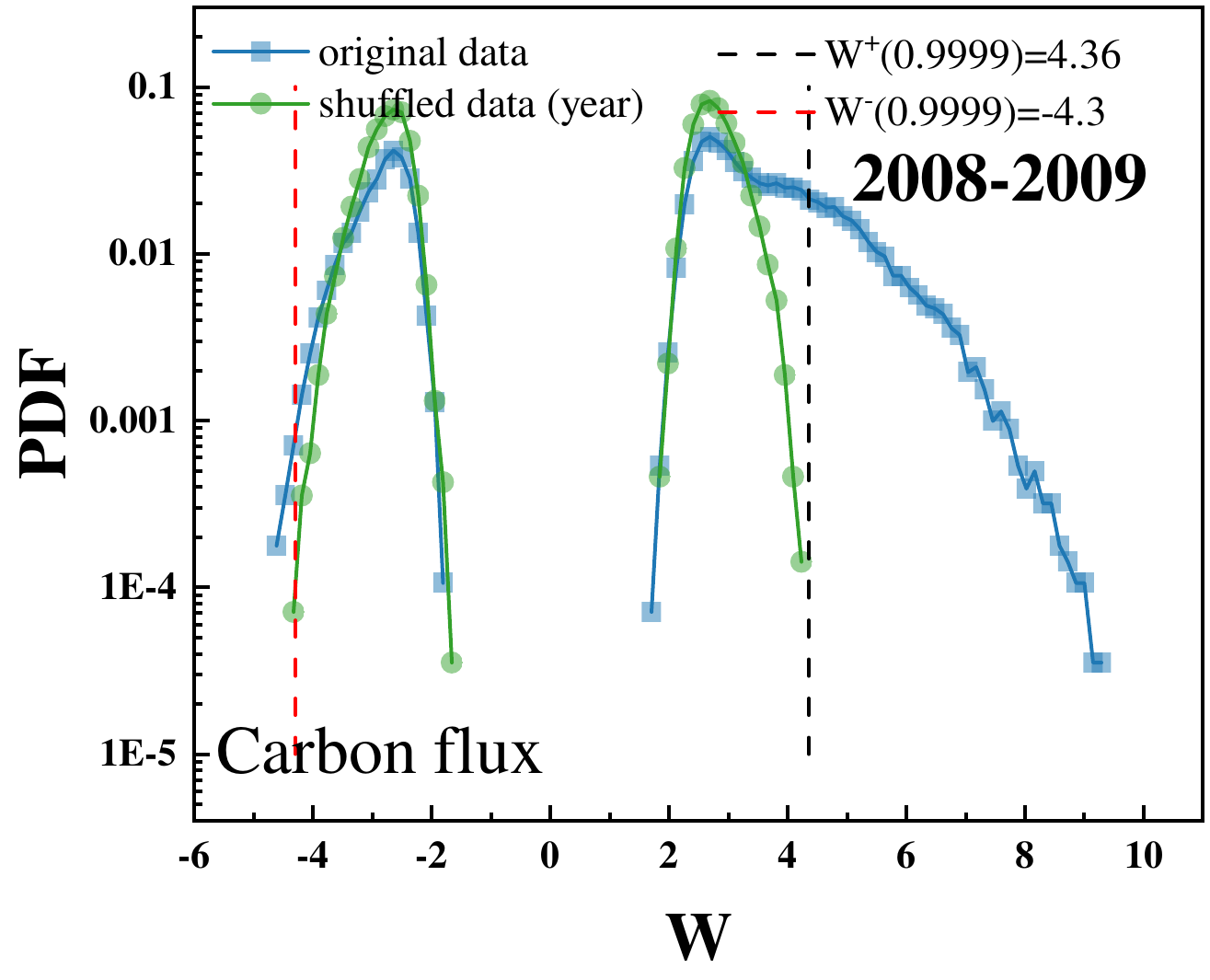}
\includegraphics[width=8.5em, height=7em]{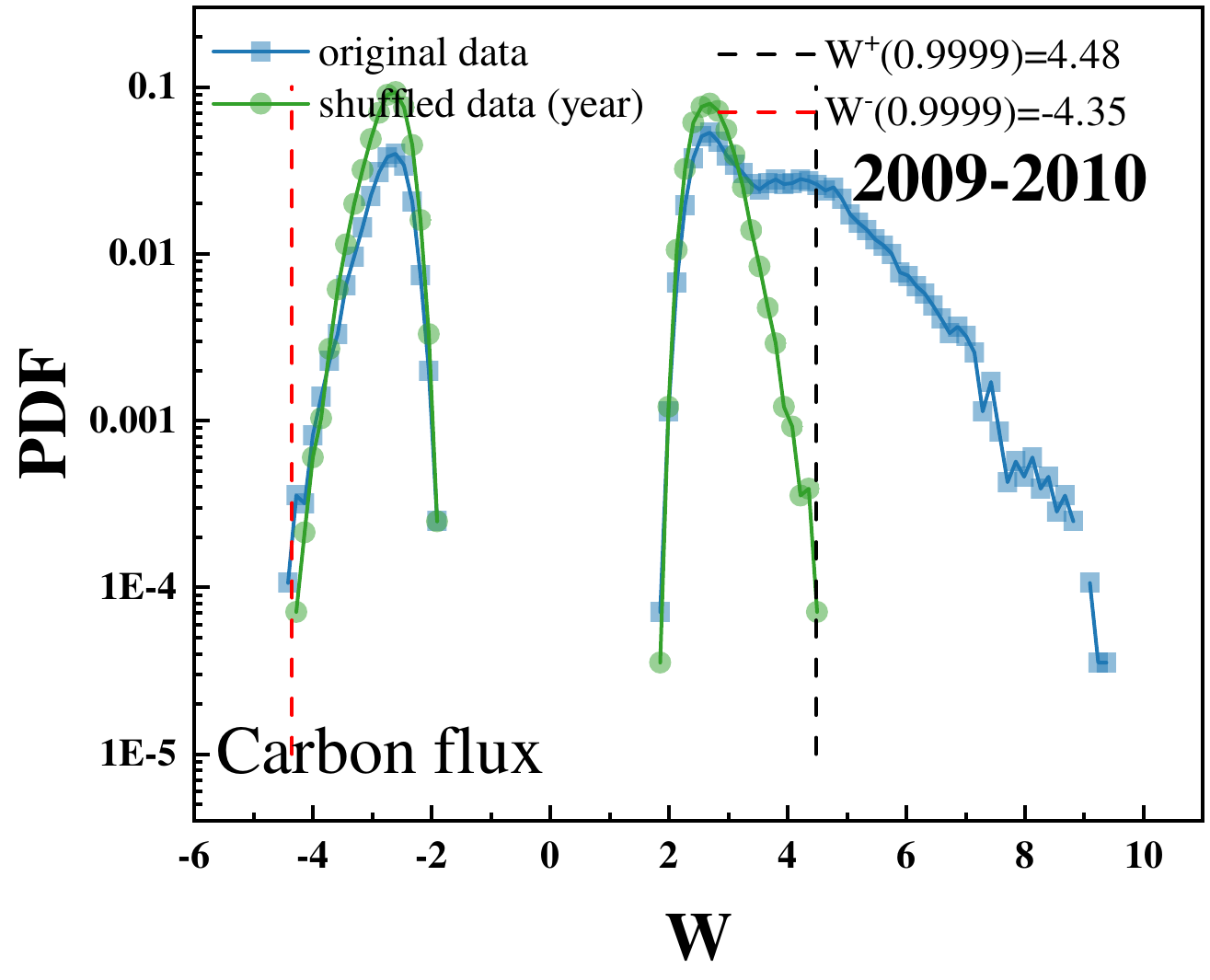}
\includegraphics[width=8.5em, height=7em]{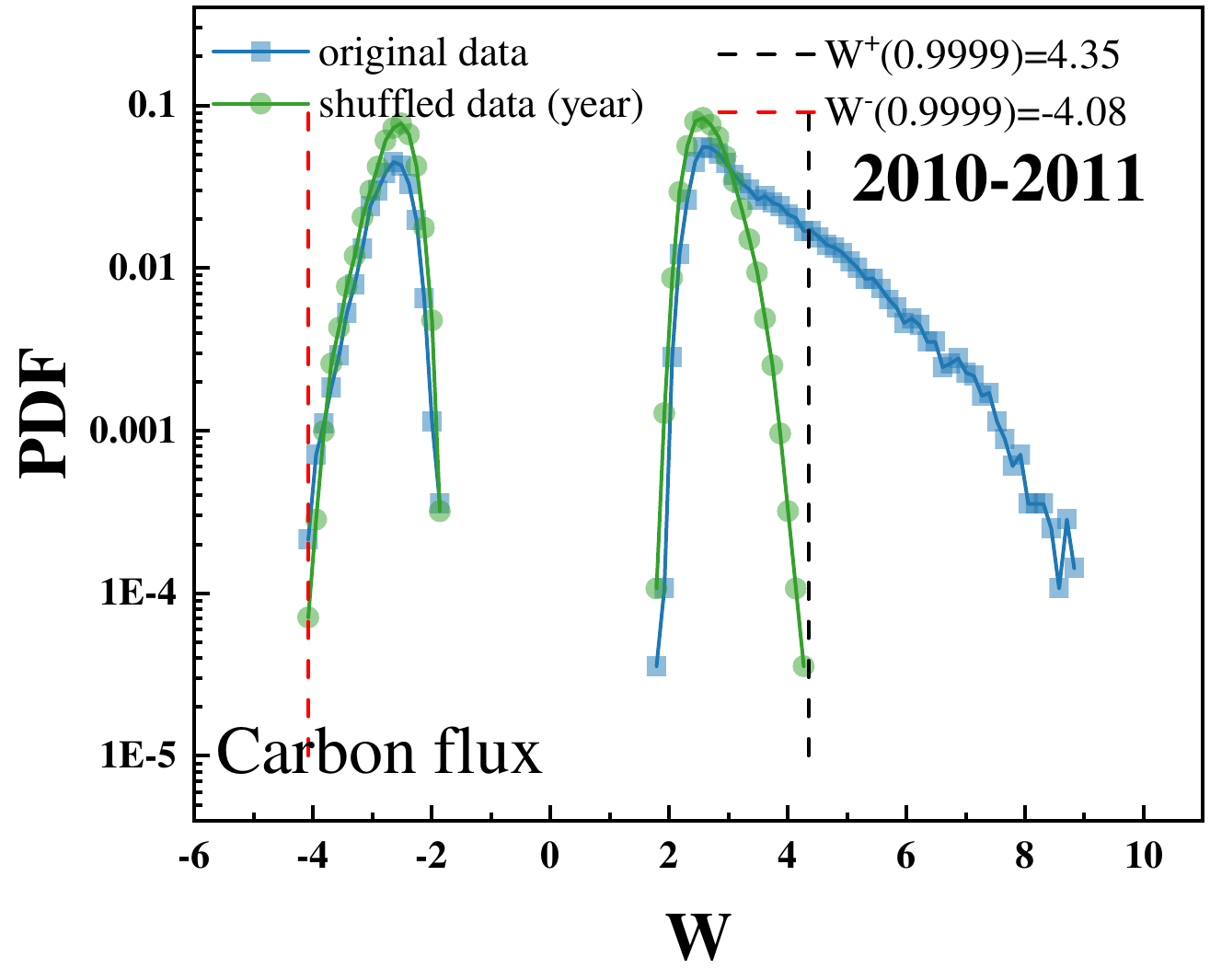}
\includegraphics[width=8.5em, height=7em]{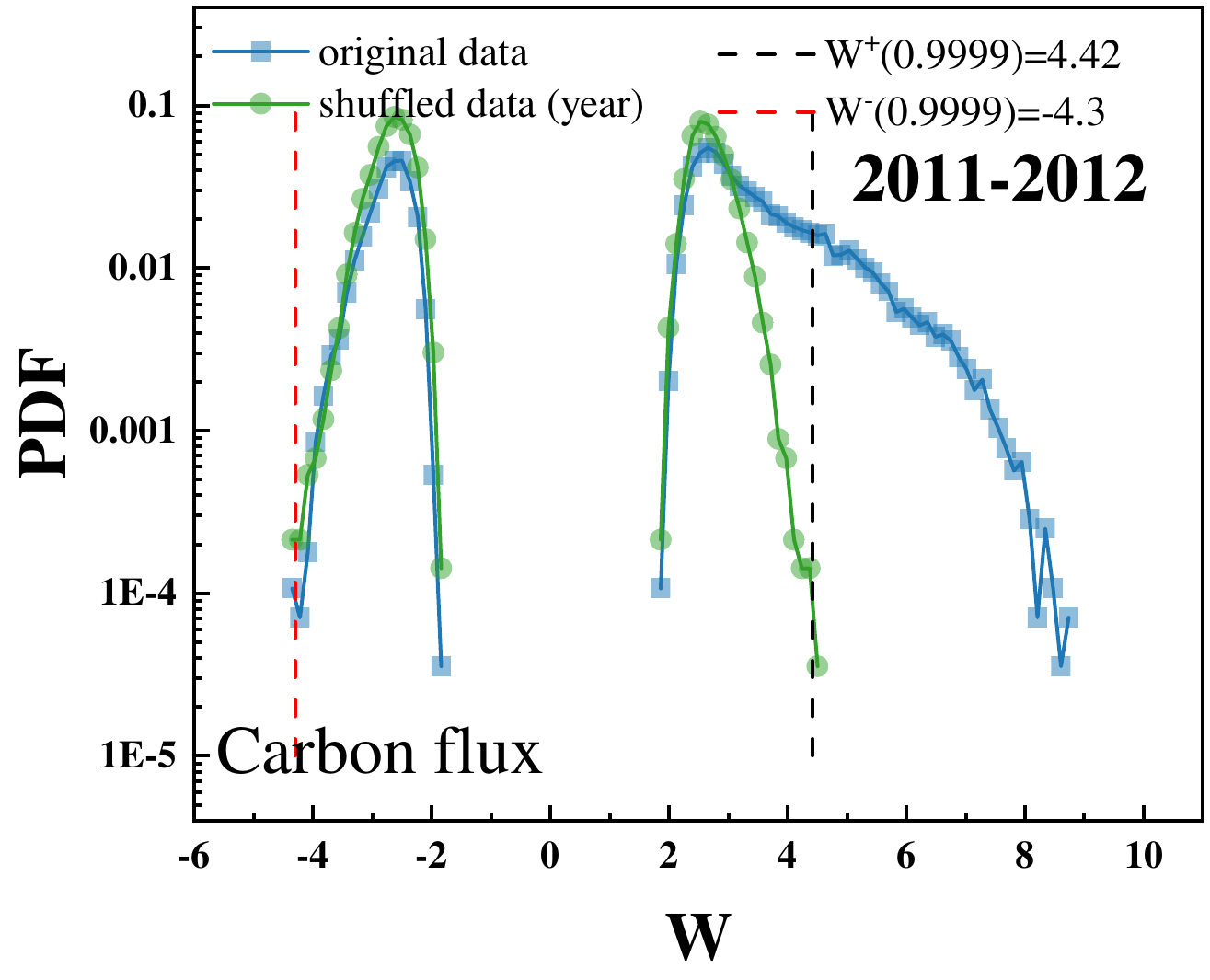}
\includegraphics[width=8.5em, height=7em]{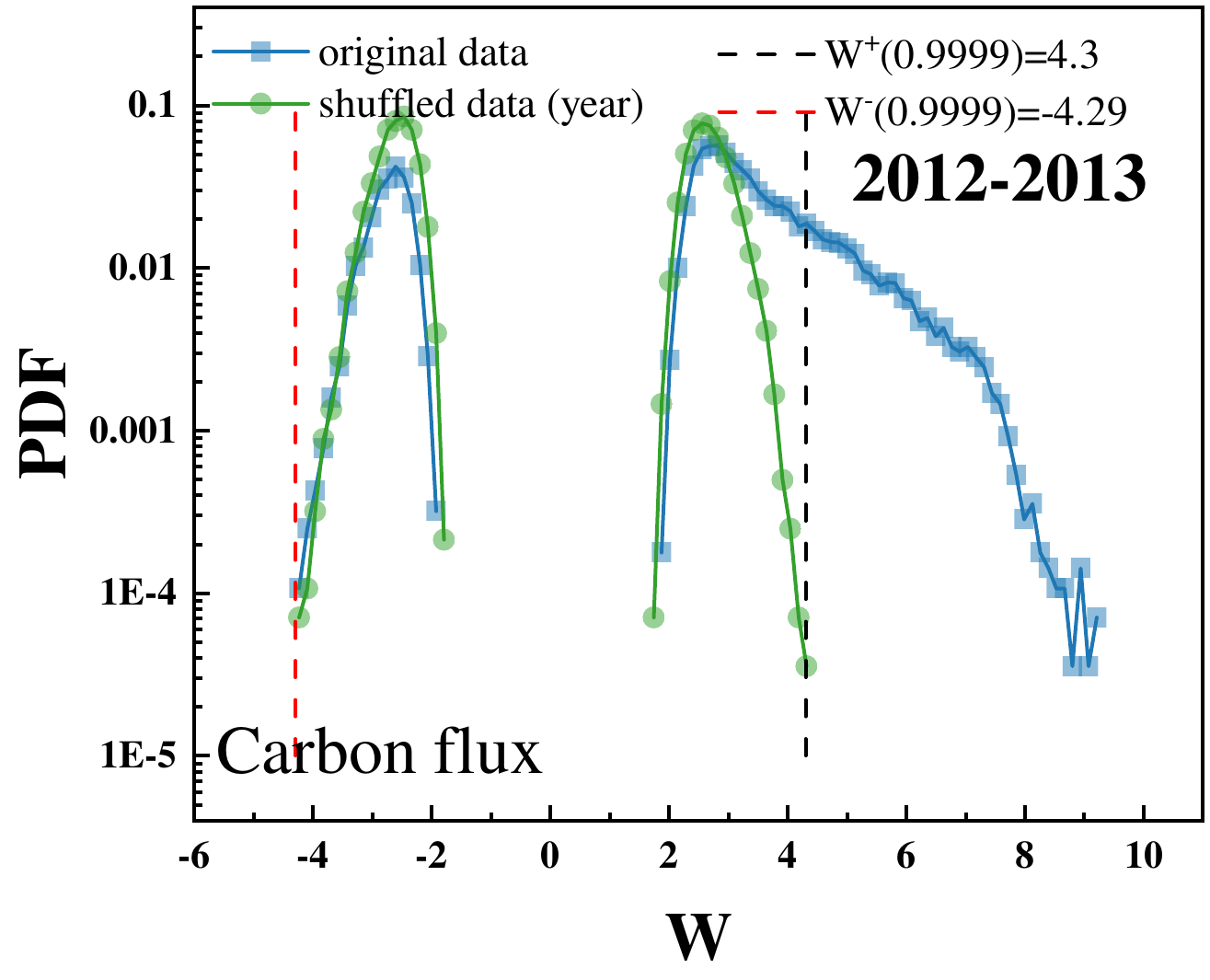}
\includegraphics[width=8.5em, height=7em]{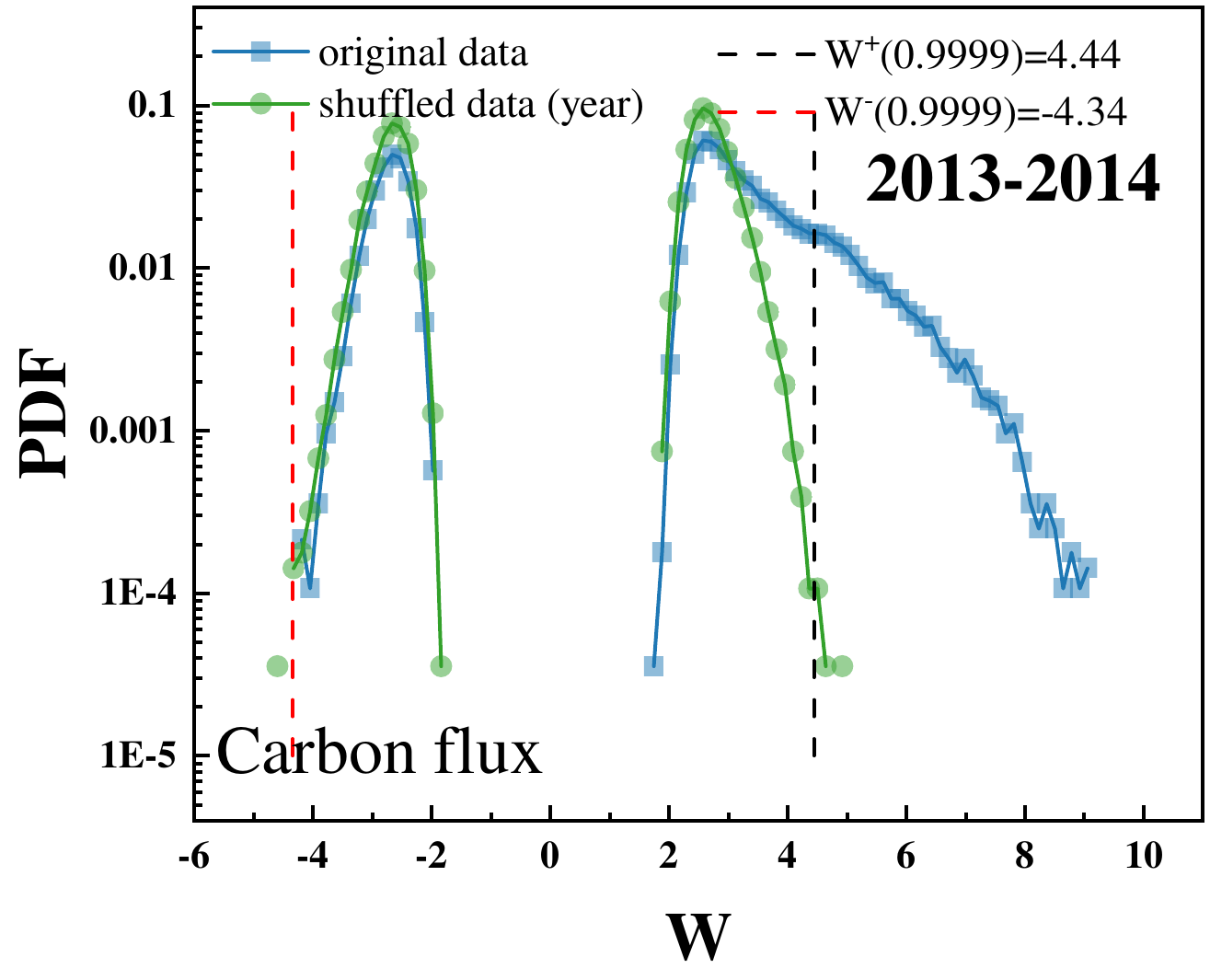}
\includegraphics[width=8.5em, height=7em]{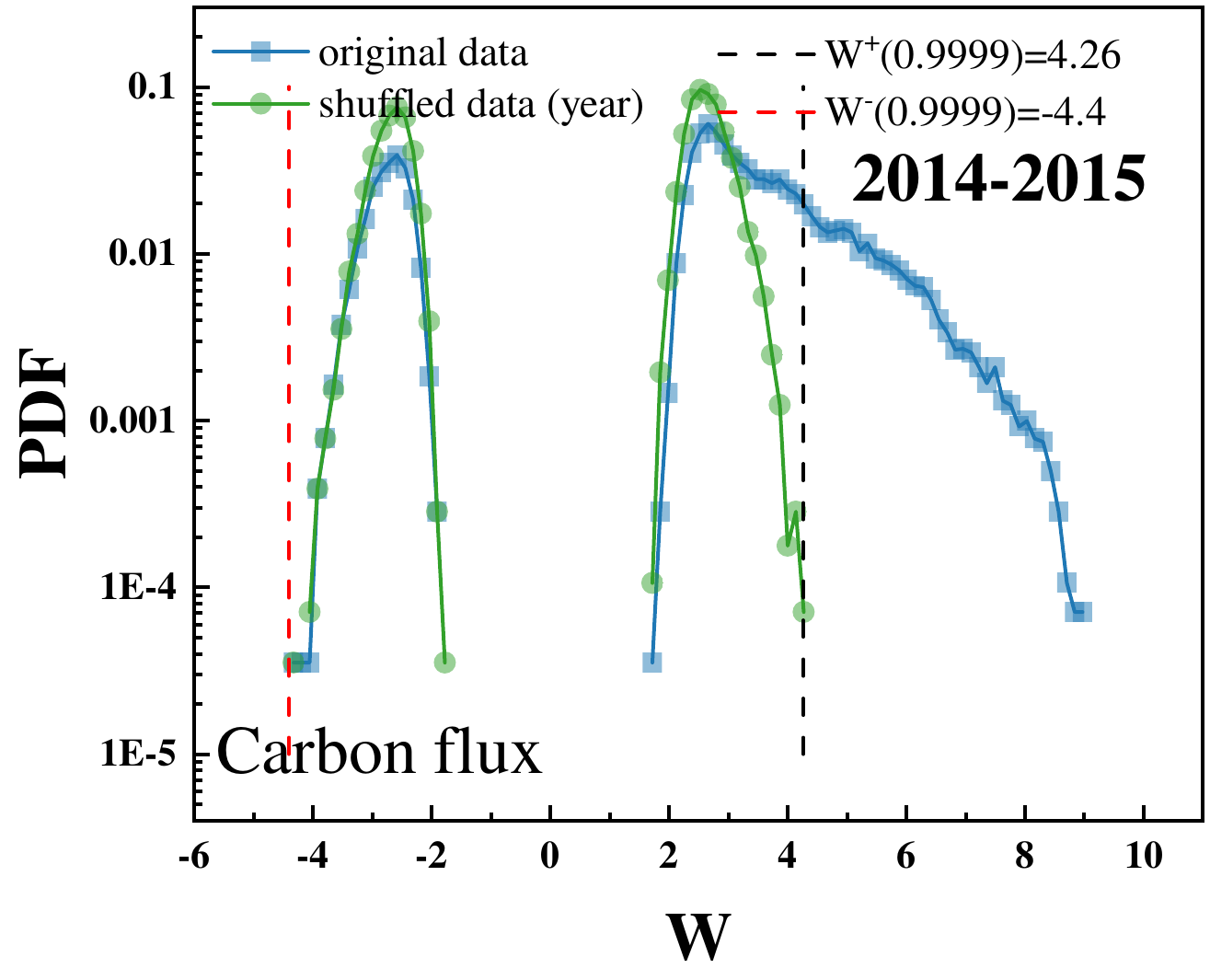}
\includegraphics[width=8.5em, height=7em]{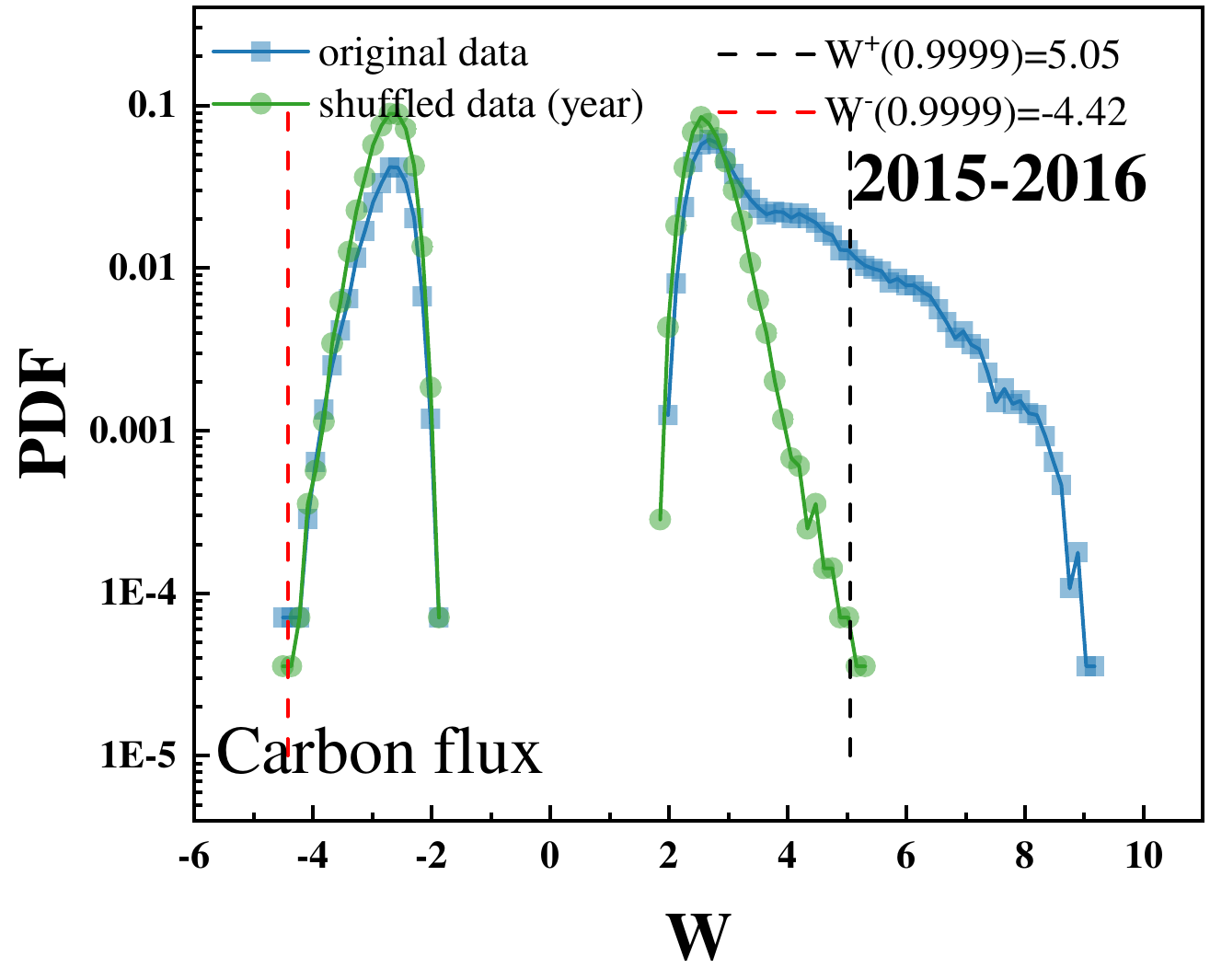}
\includegraphics[width=8.5em, height=7em]{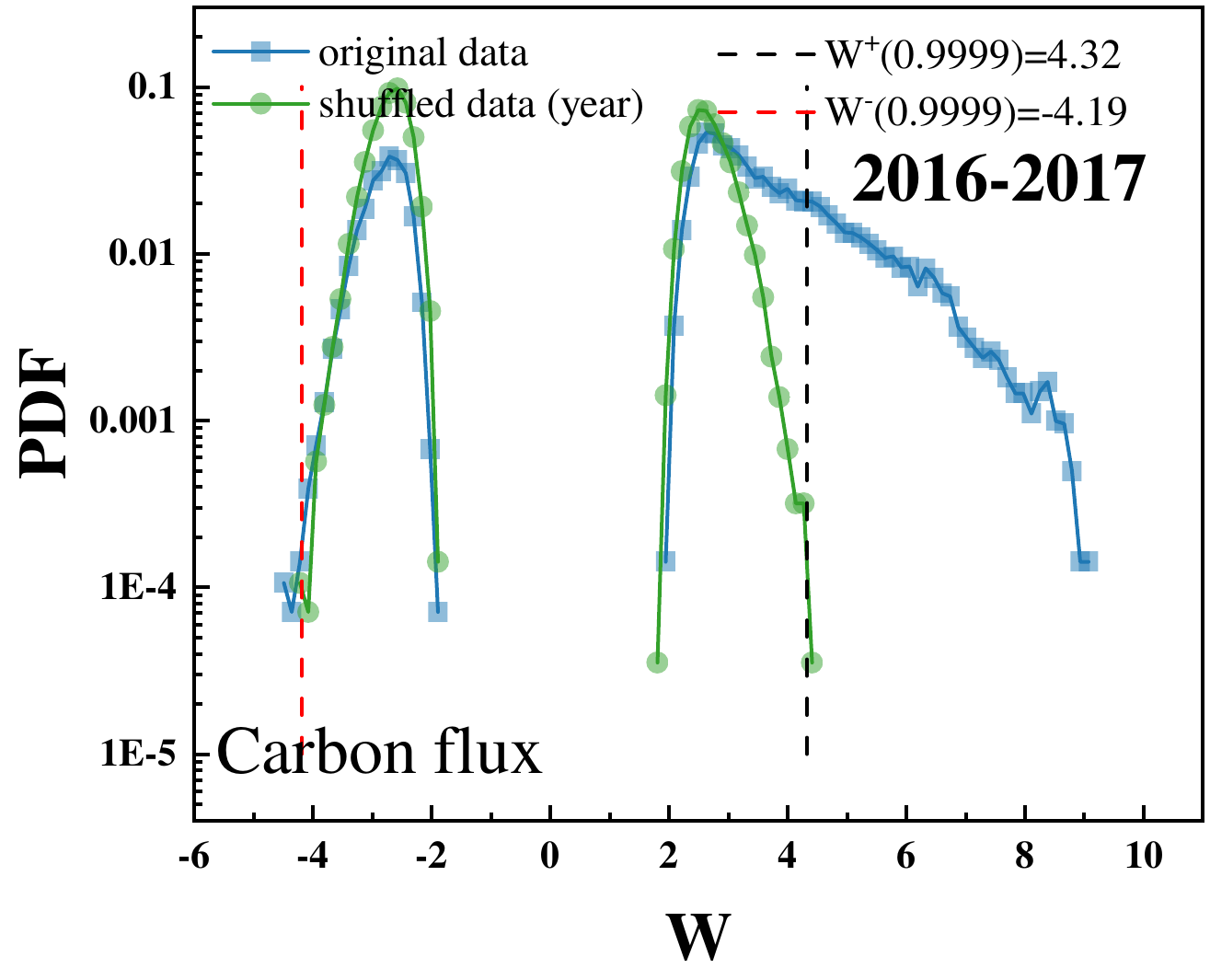}
\includegraphics[width=8.5em, height=7em]{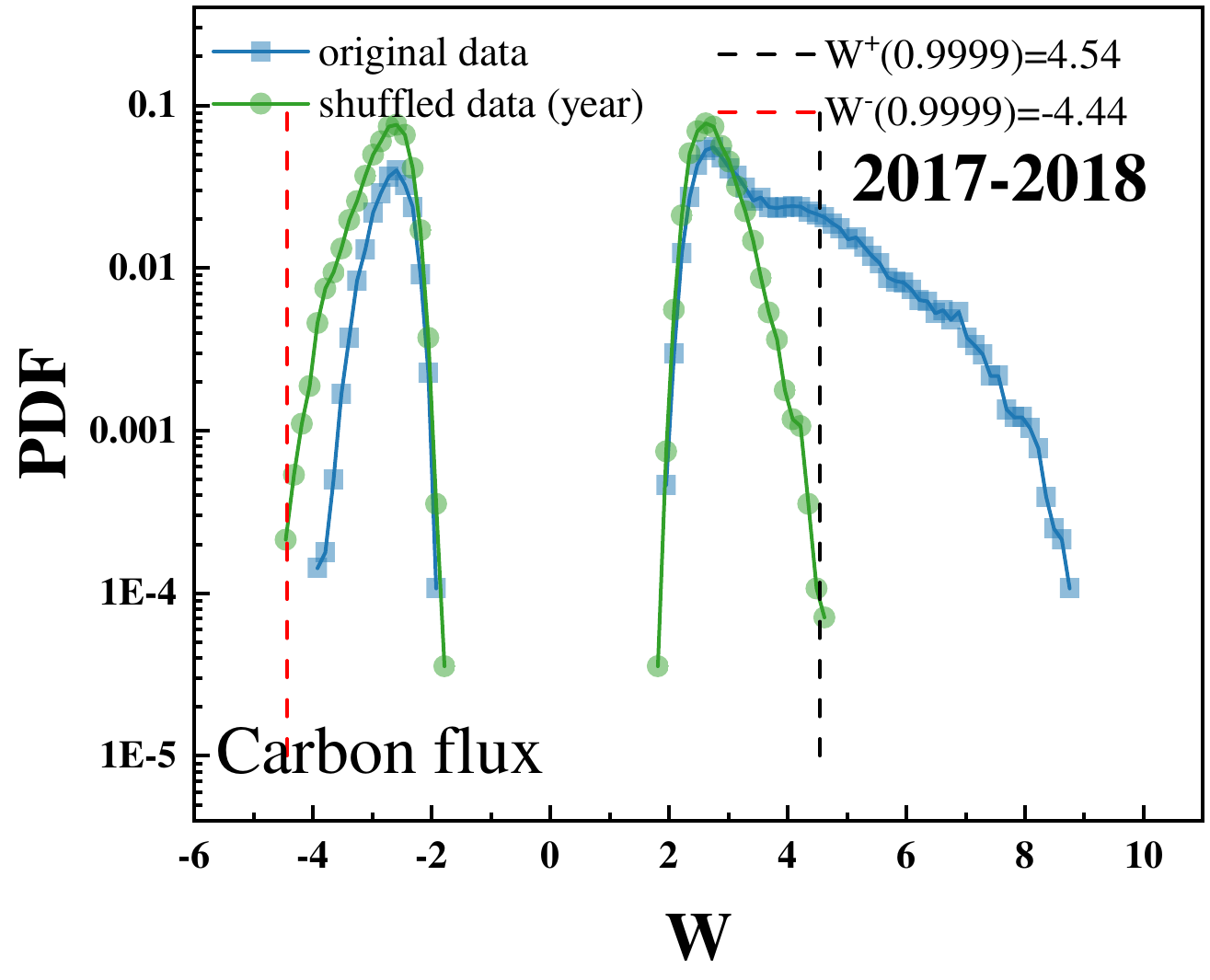}
\includegraphics[width=8.5em, height=7em]{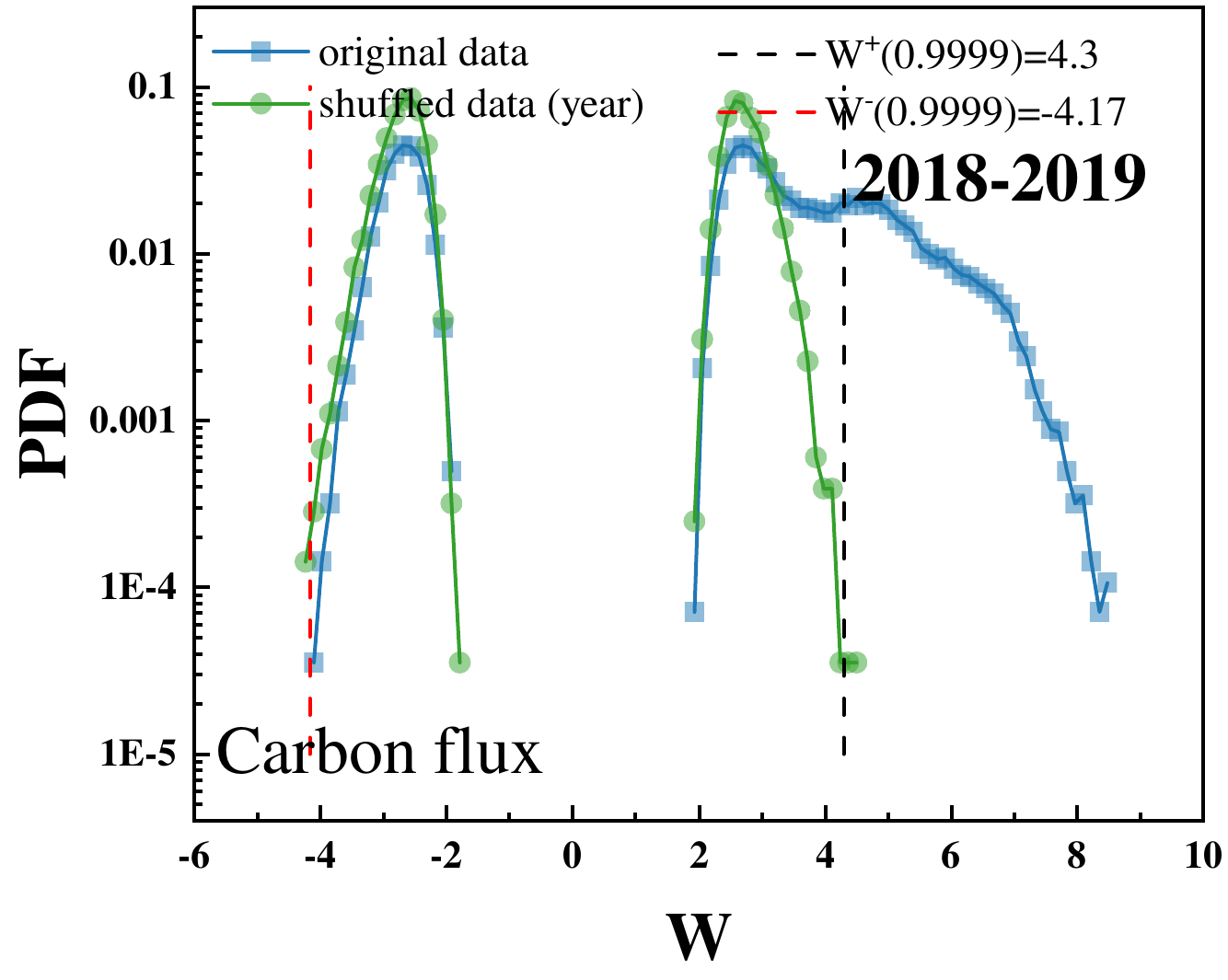}
\end{center}

\begin{center}
\noindent {\small {\bf Fig. S2} Probability distribution function (PDF) of link weights for the original data and shuffled data of carbon flux in China. }
\end{center}

\begin{center}
\includegraphics[width=8.5em, height=7em]{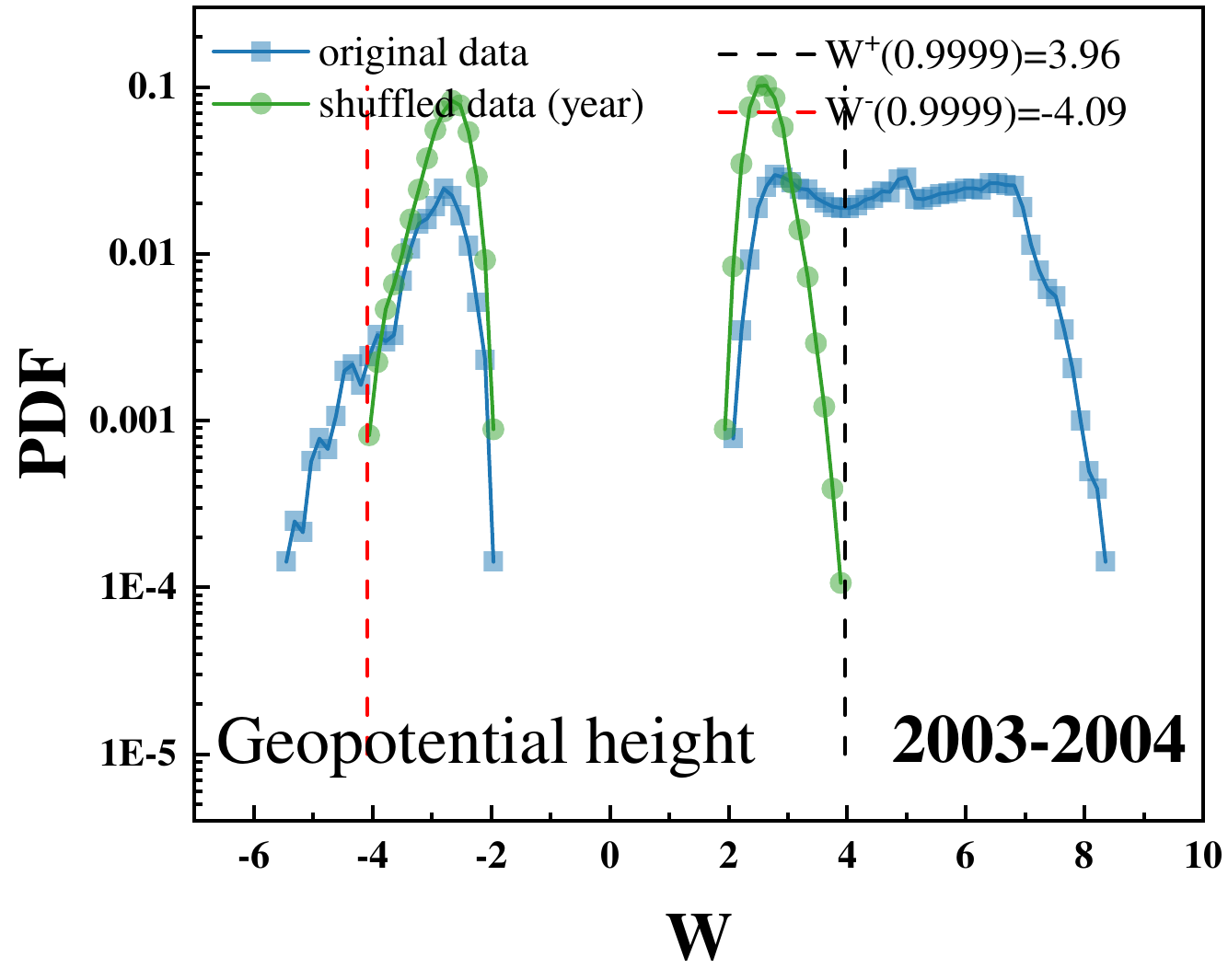}
\includegraphics[width=8.5em, height=7em]{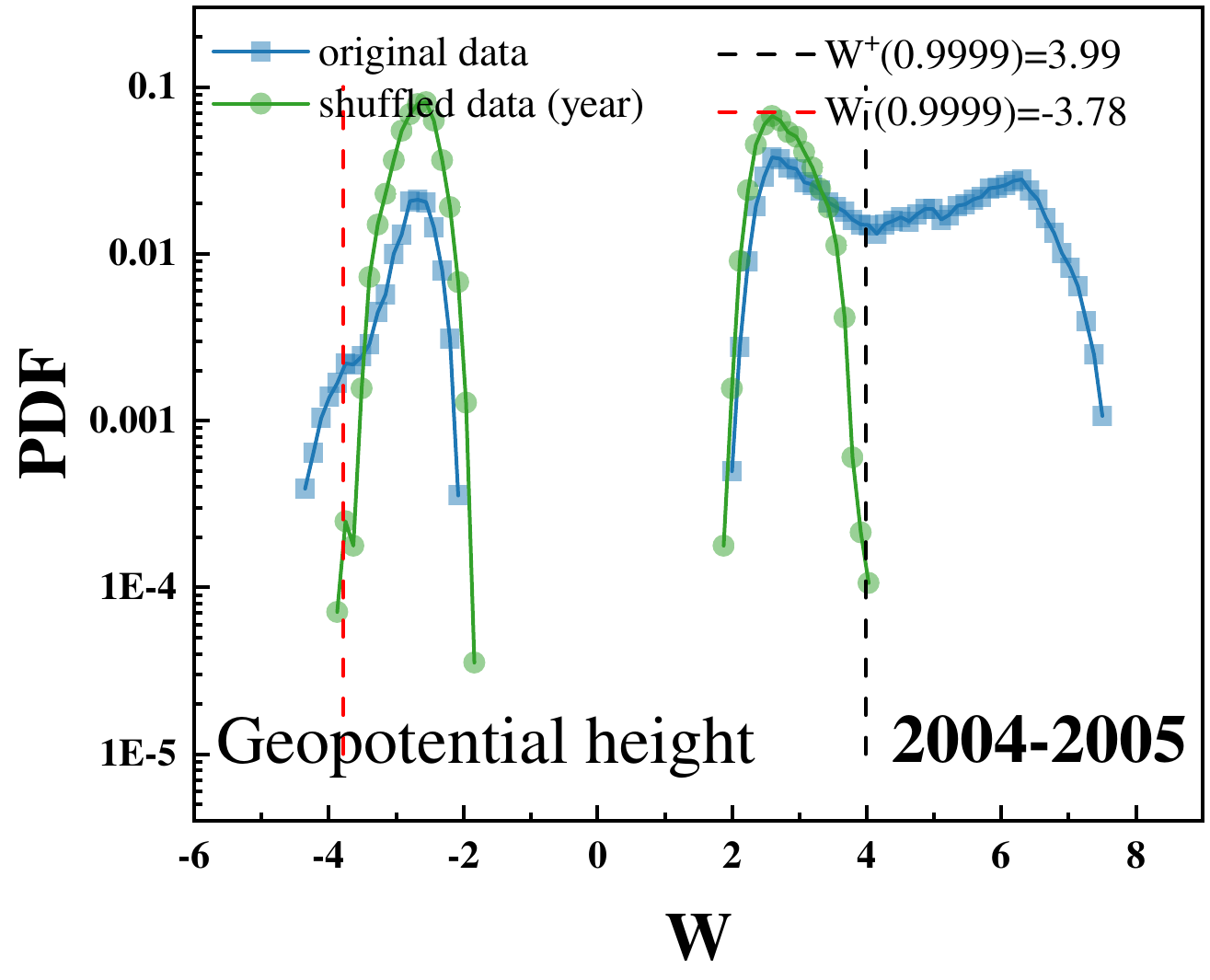}
\includegraphics[width=8.5em, height=7em]{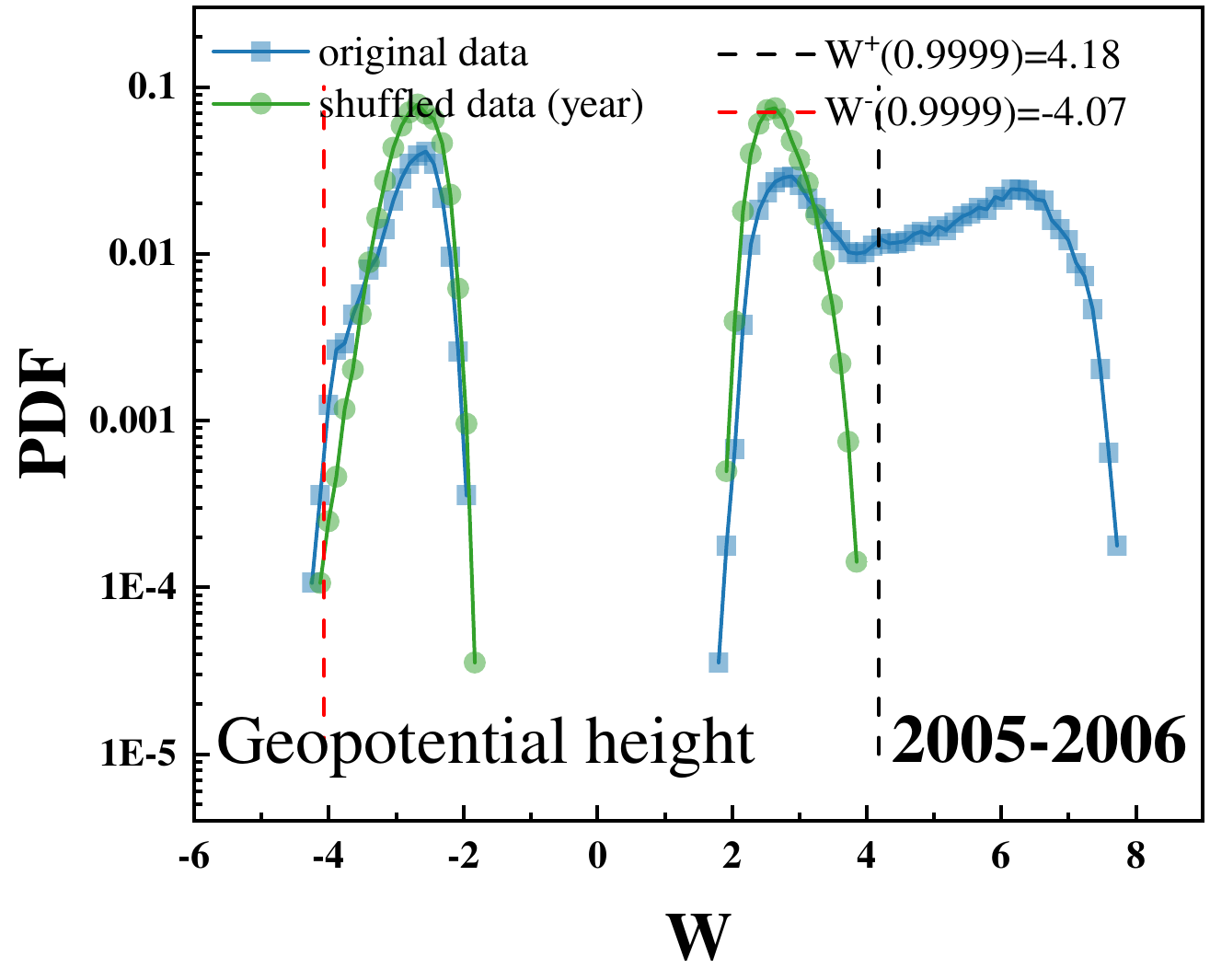}
\includegraphics[width=8.5em, height=7em]{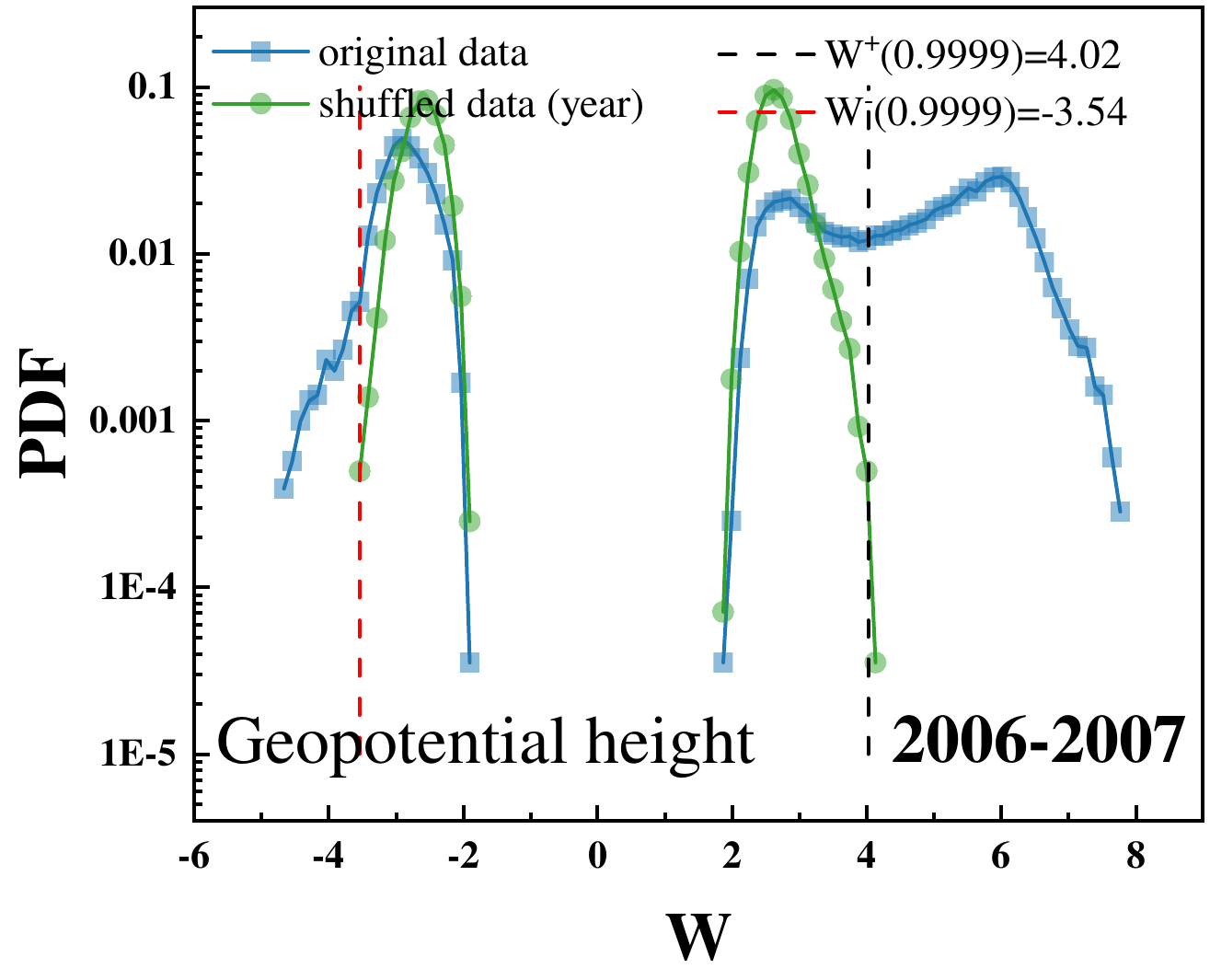}
\includegraphics[width=8.5em, height=7em]{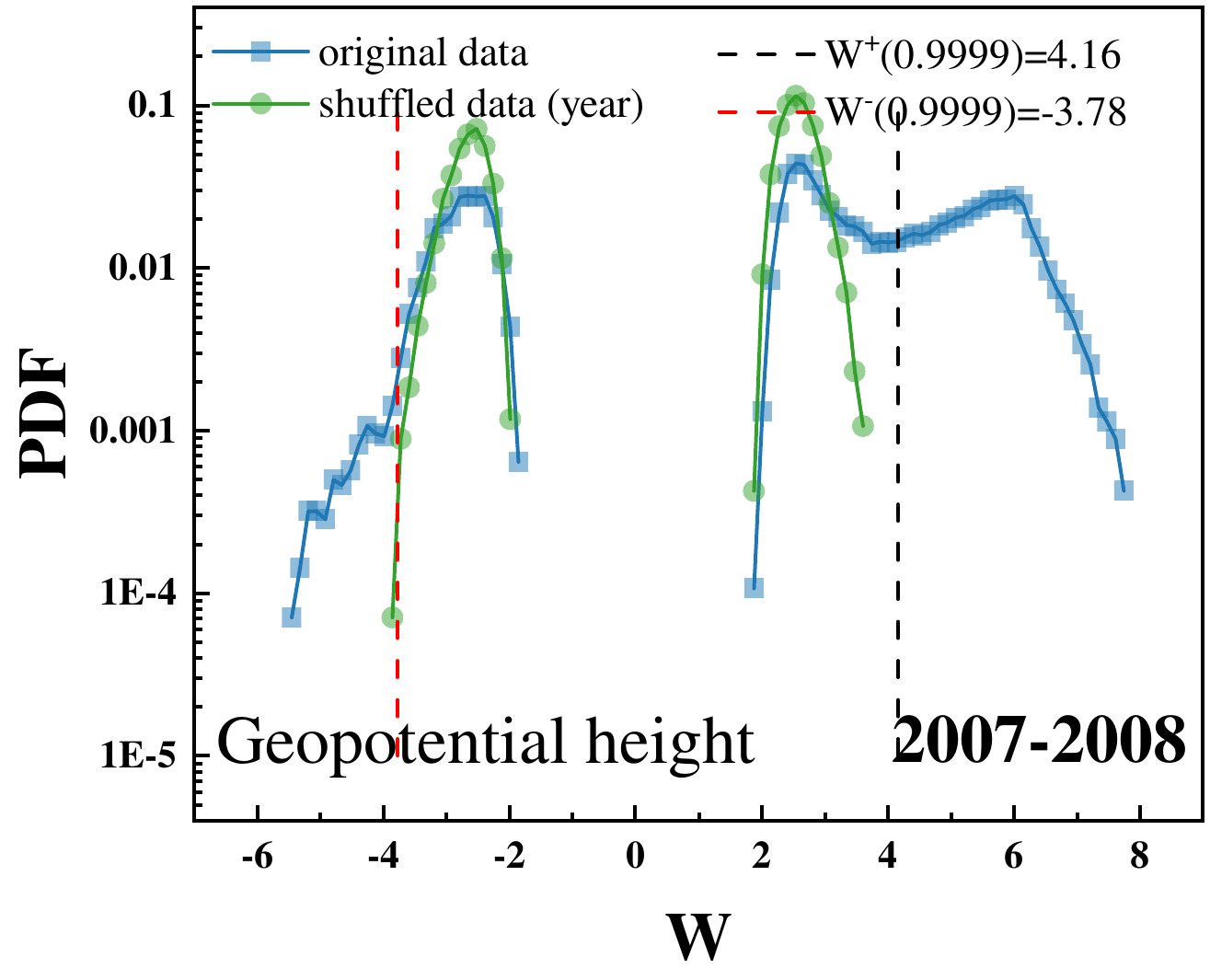}
\includegraphics[width=8.5em, height=7em]{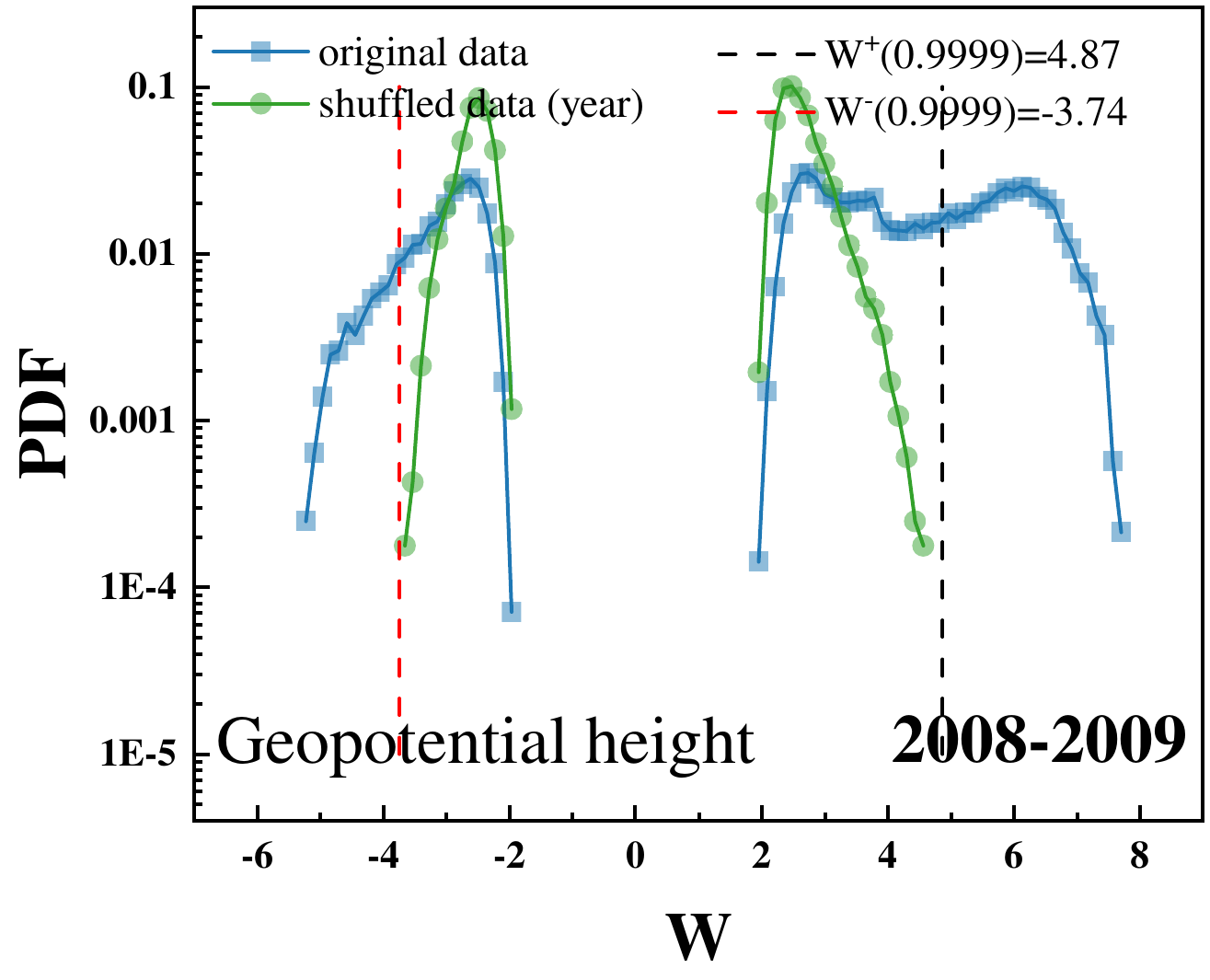}
\includegraphics[width=8.5em, height=7em]{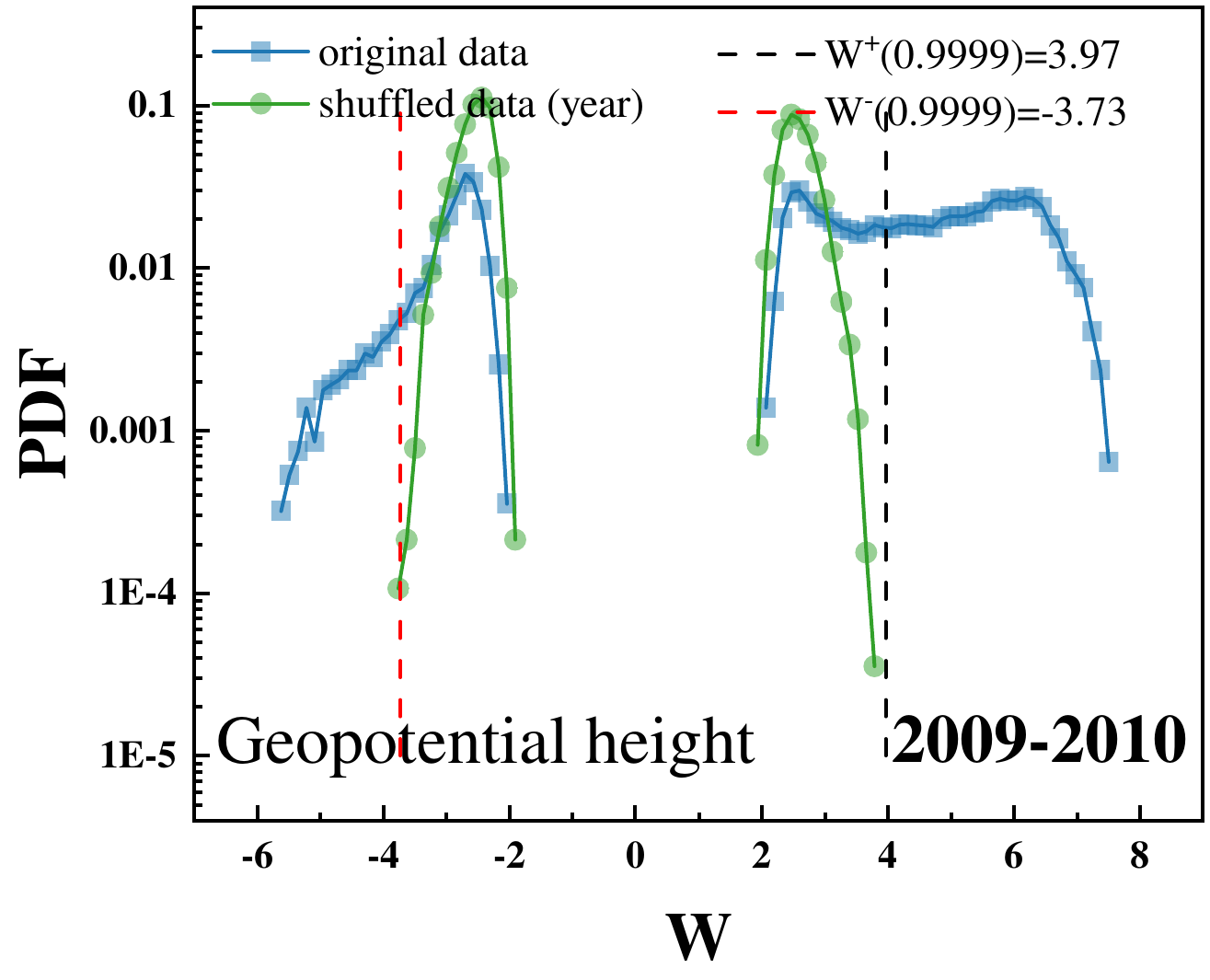}
\includegraphics[width=8.5em, height=7em]{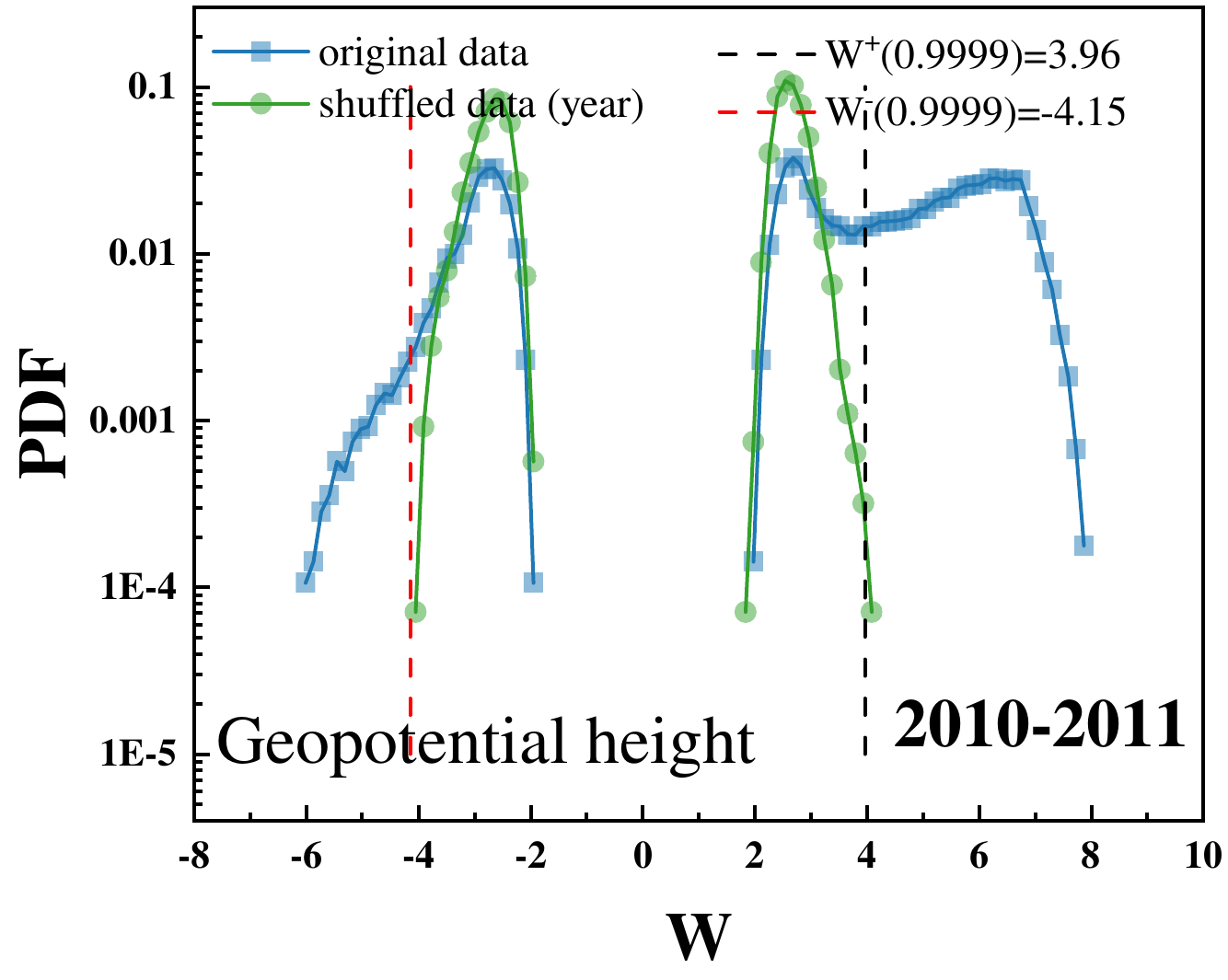}
\includegraphics[width=8.5em, height=7em]{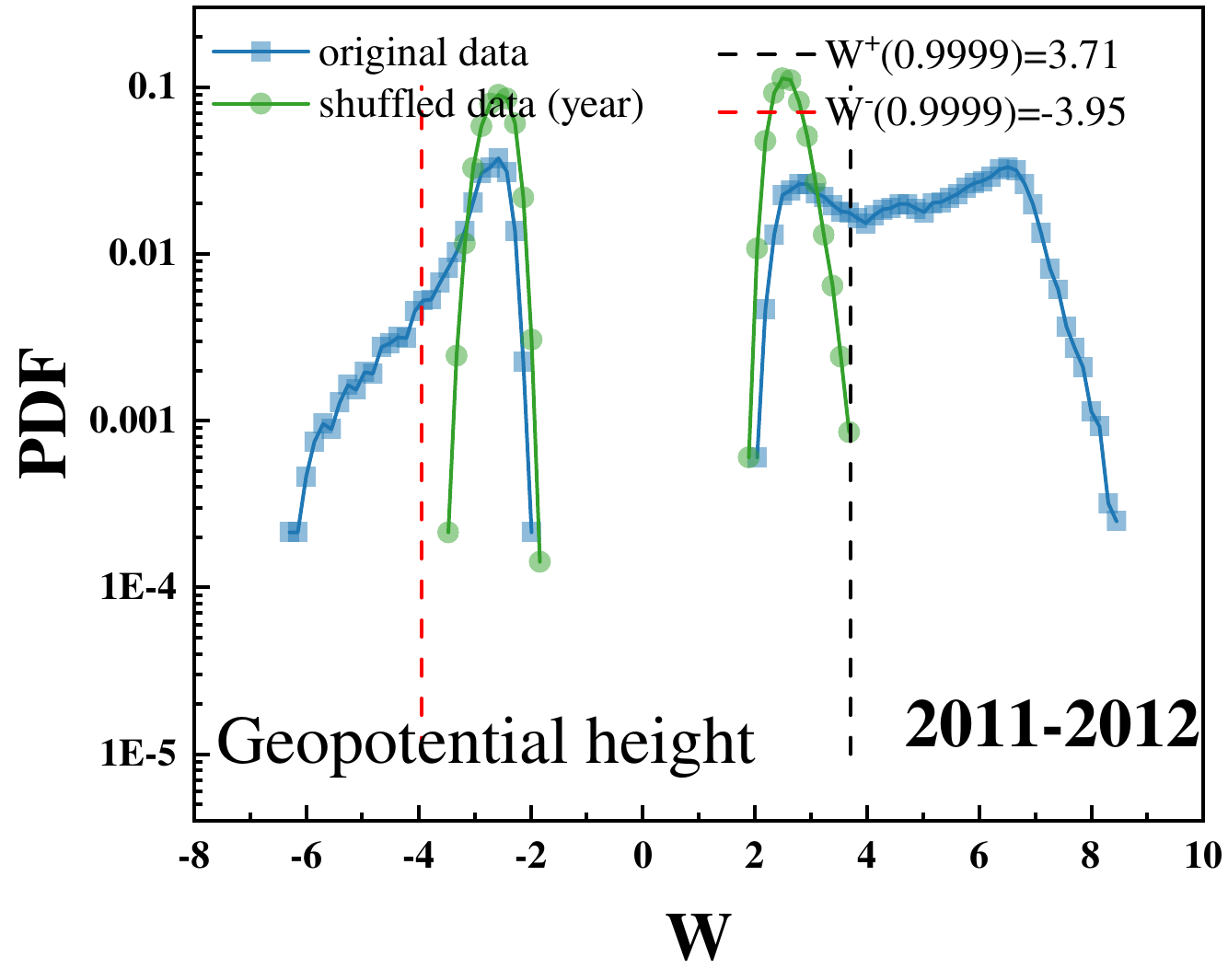}
\includegraphics[width=8.5em, height=7em]{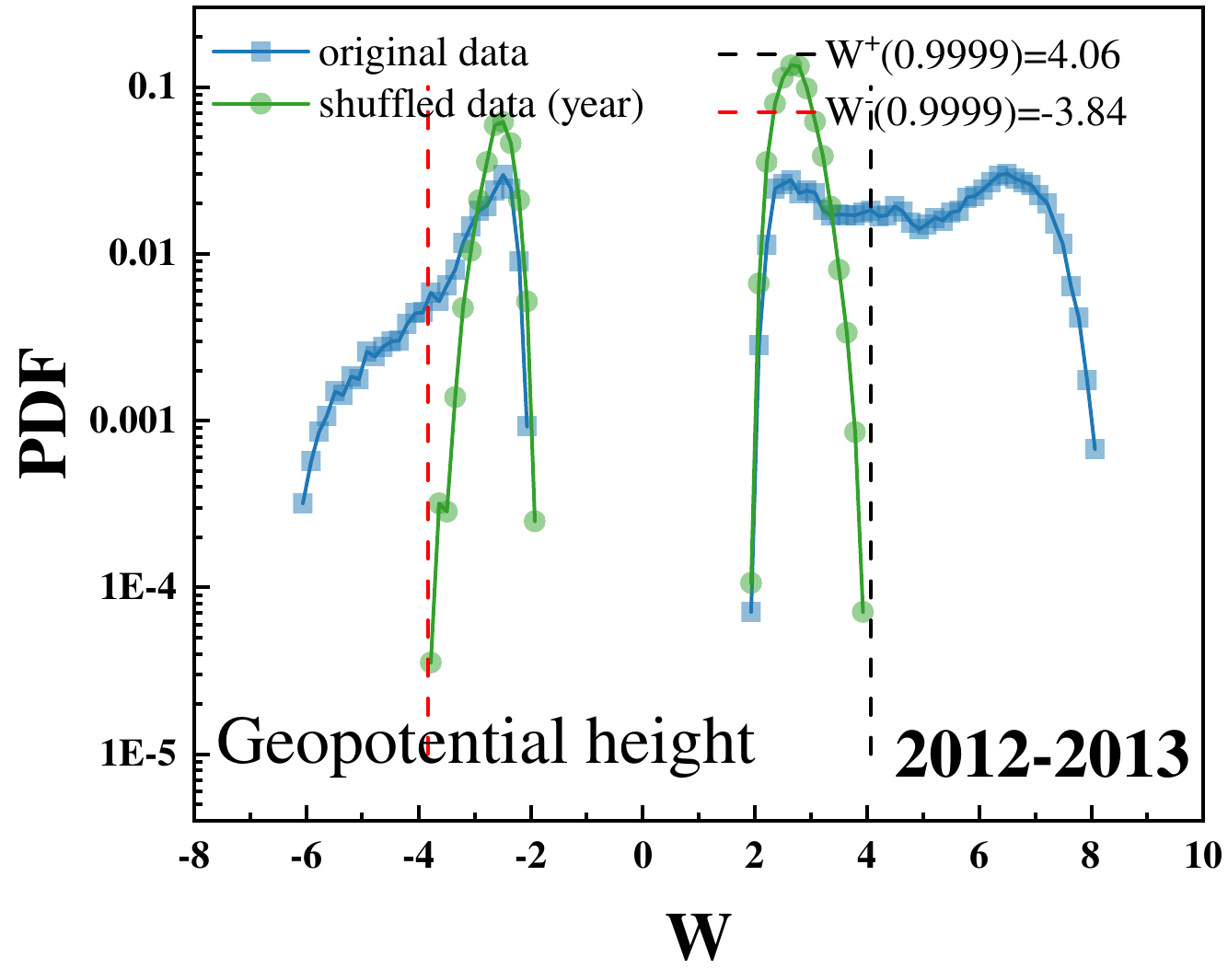}
\includegraphics[width=8.5em, height=7em]{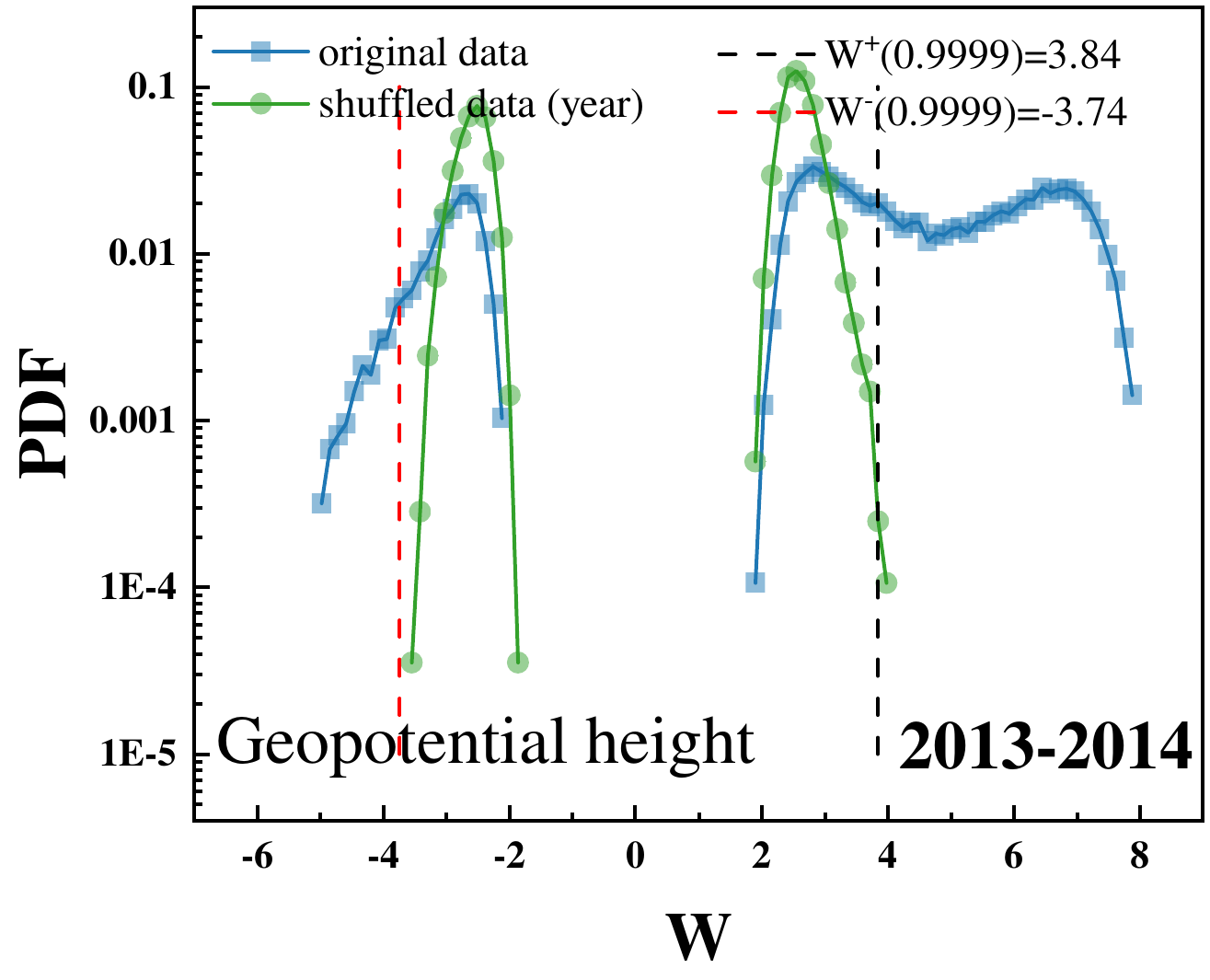}
\includegraphics[width=8.5em, height=7em]{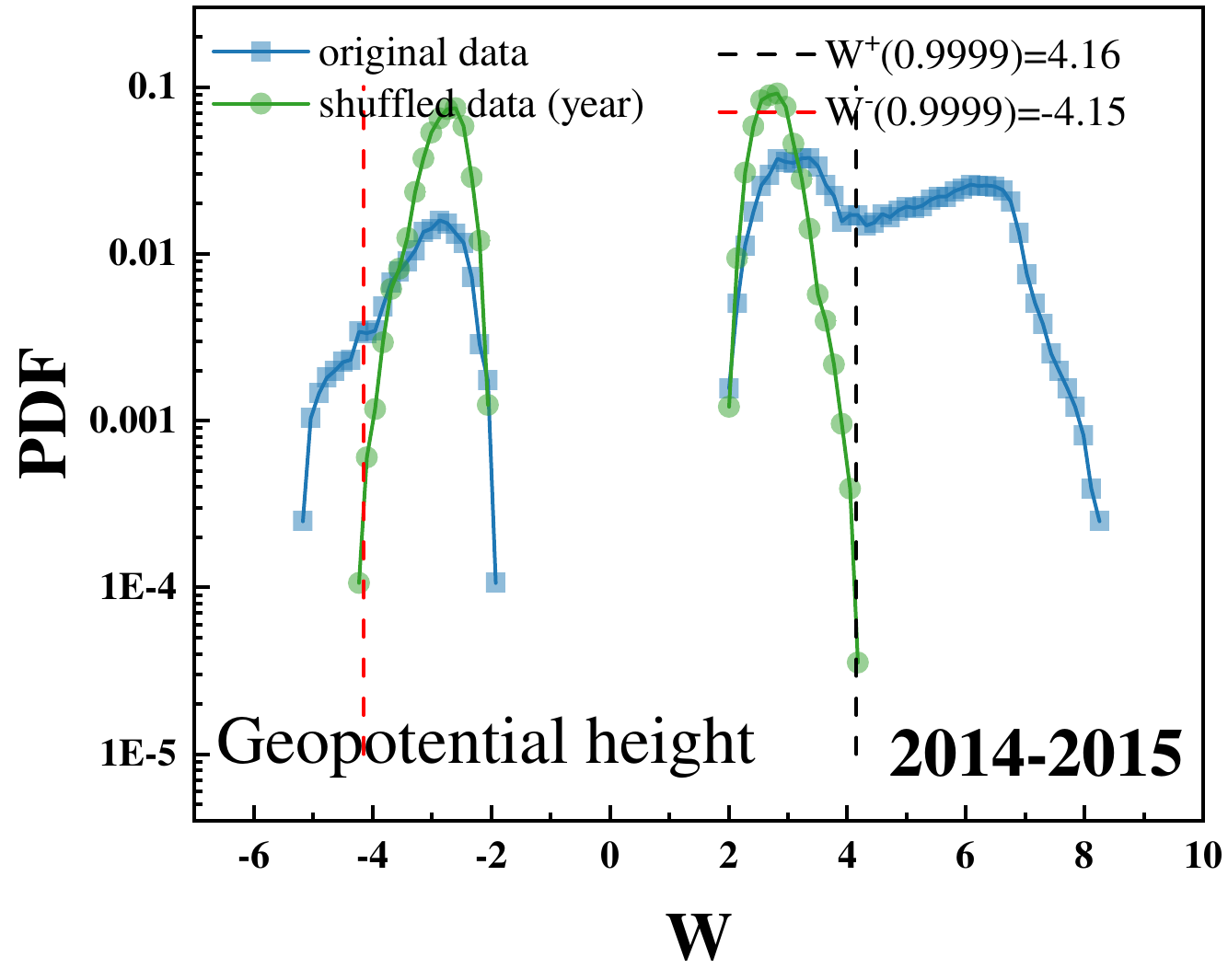}
\includegraphics[width=8.5em, height=7em]{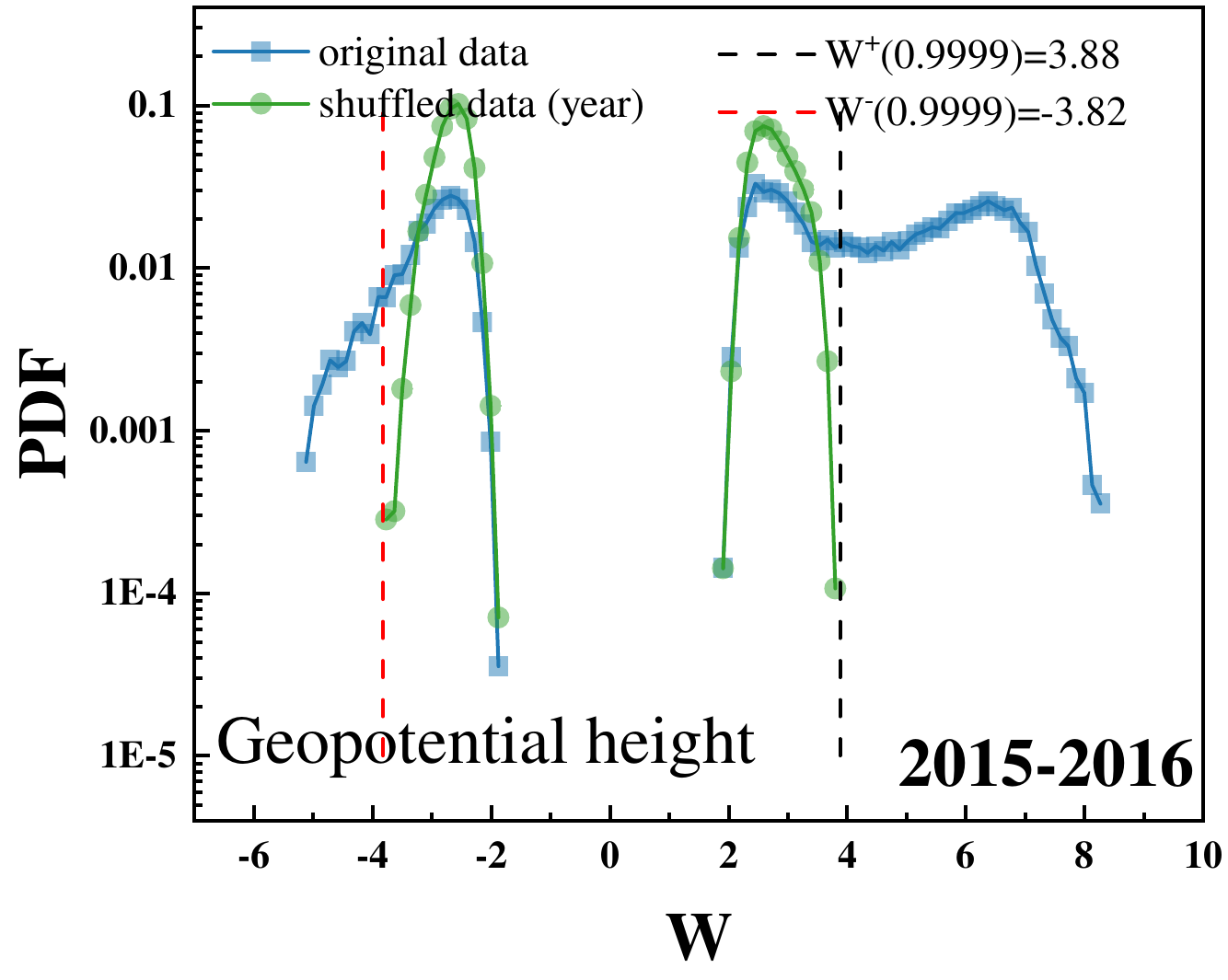}
\includegraphics[width=8.5em, height=7em]{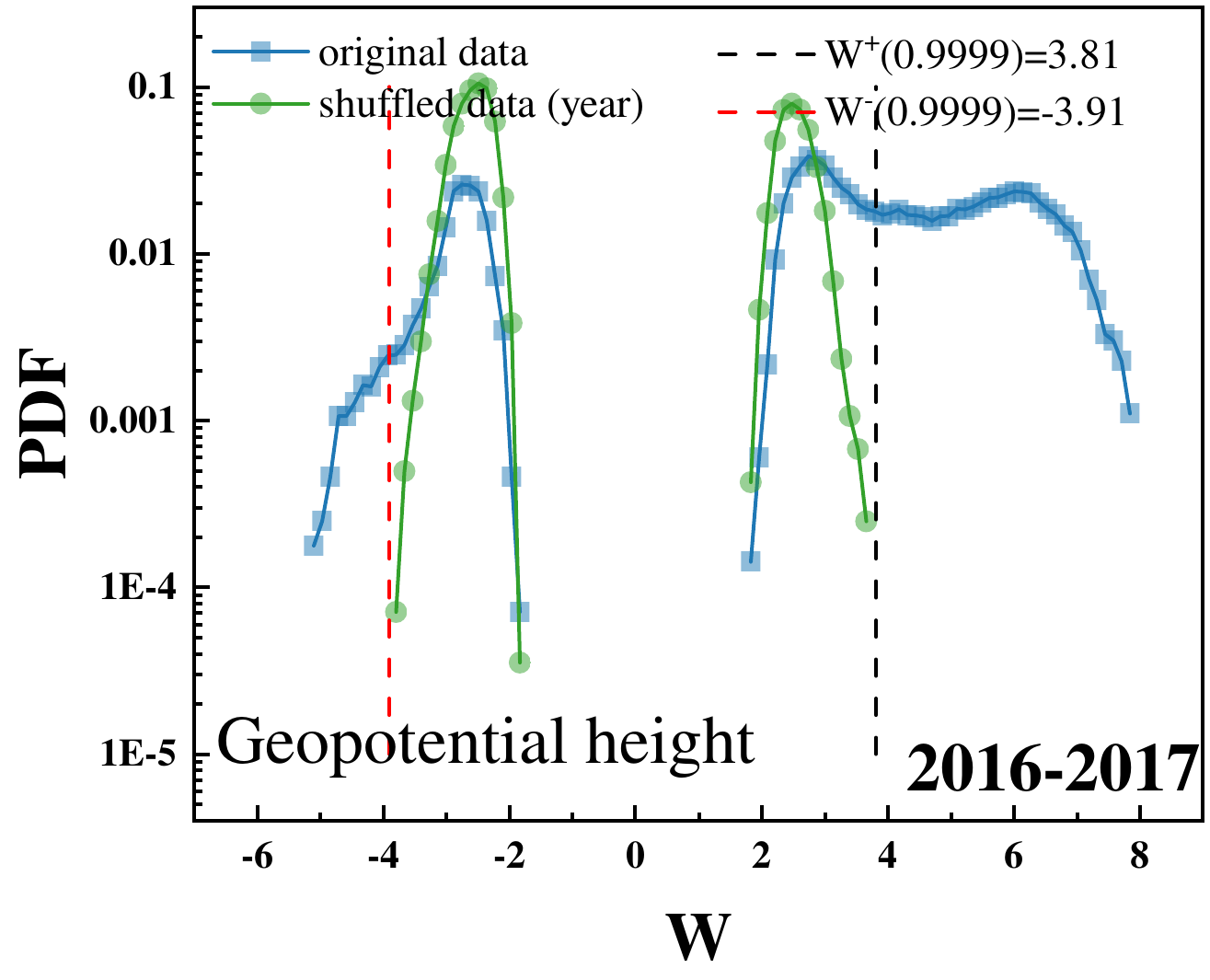}
\includegraphics[width=8.5em, height=7em]{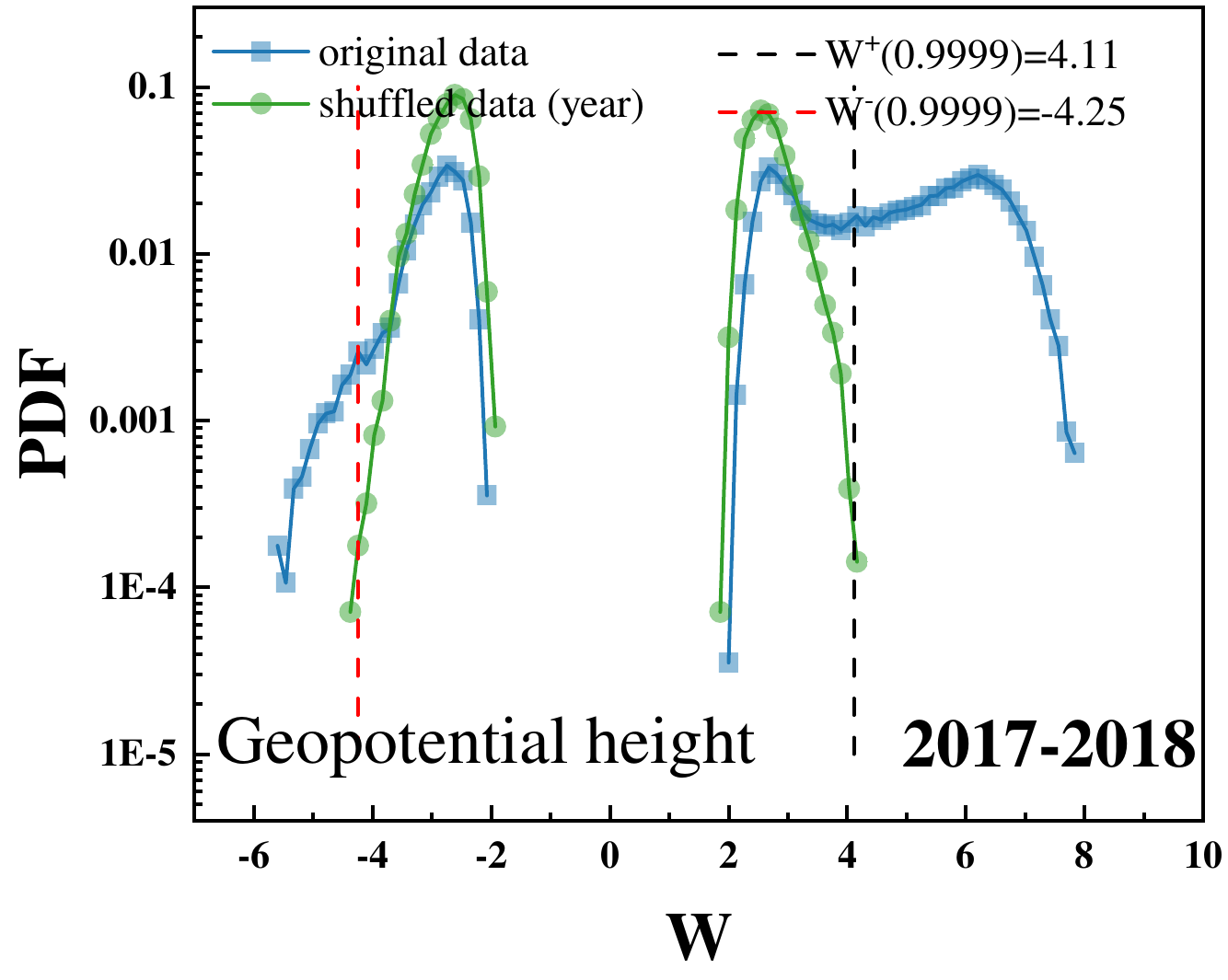}
\includegraphics[width=8.5em, height=7em]{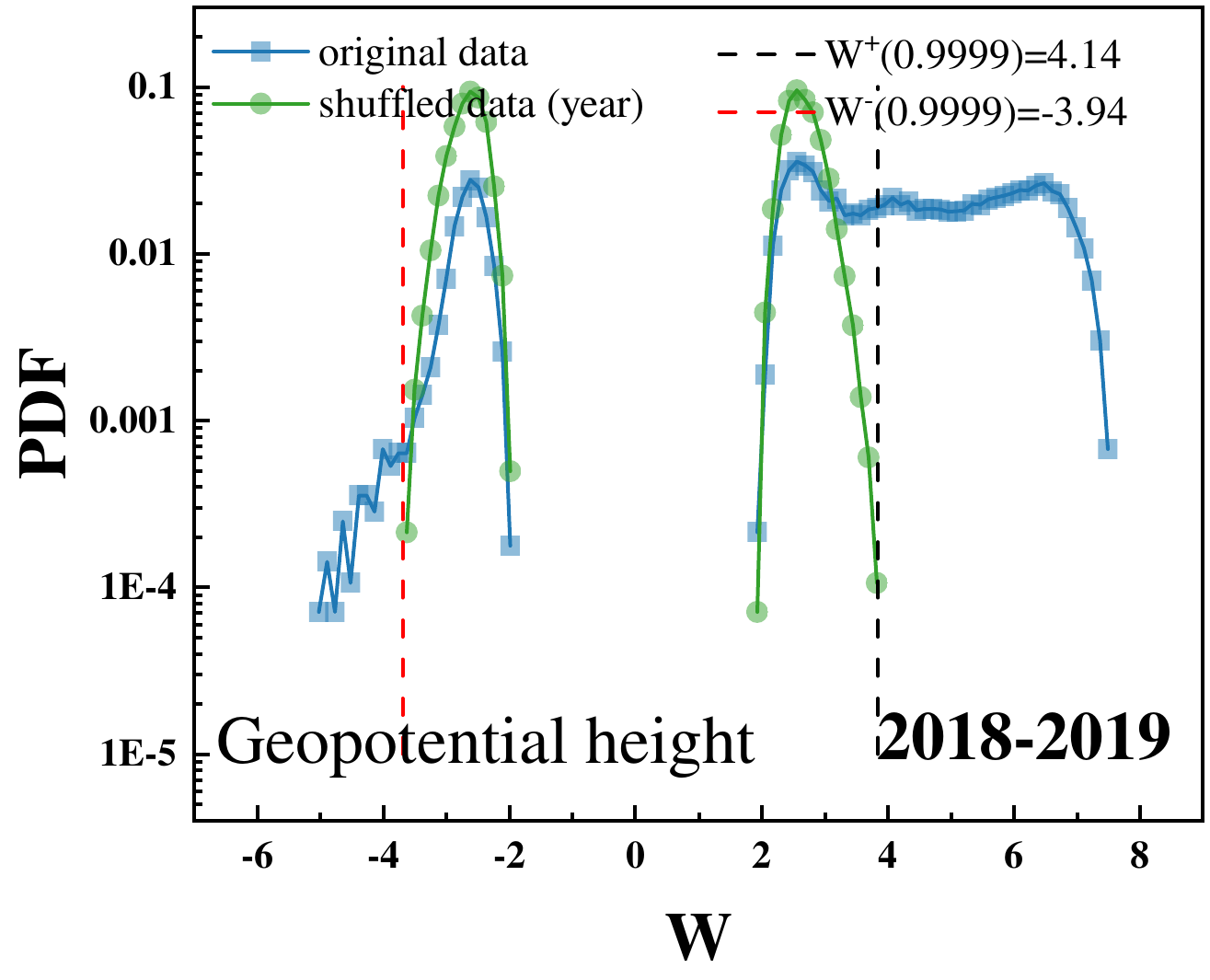}
\end{center}

\begin{center}
\noindent {\small {\bf Fig. S3} Probability distribution function (PDF) of link weights for the original data and shuffled data of Geopotential height in China. }
\end{center}

\begin{center}
\includegraphics[width=8.5em, height=7em]{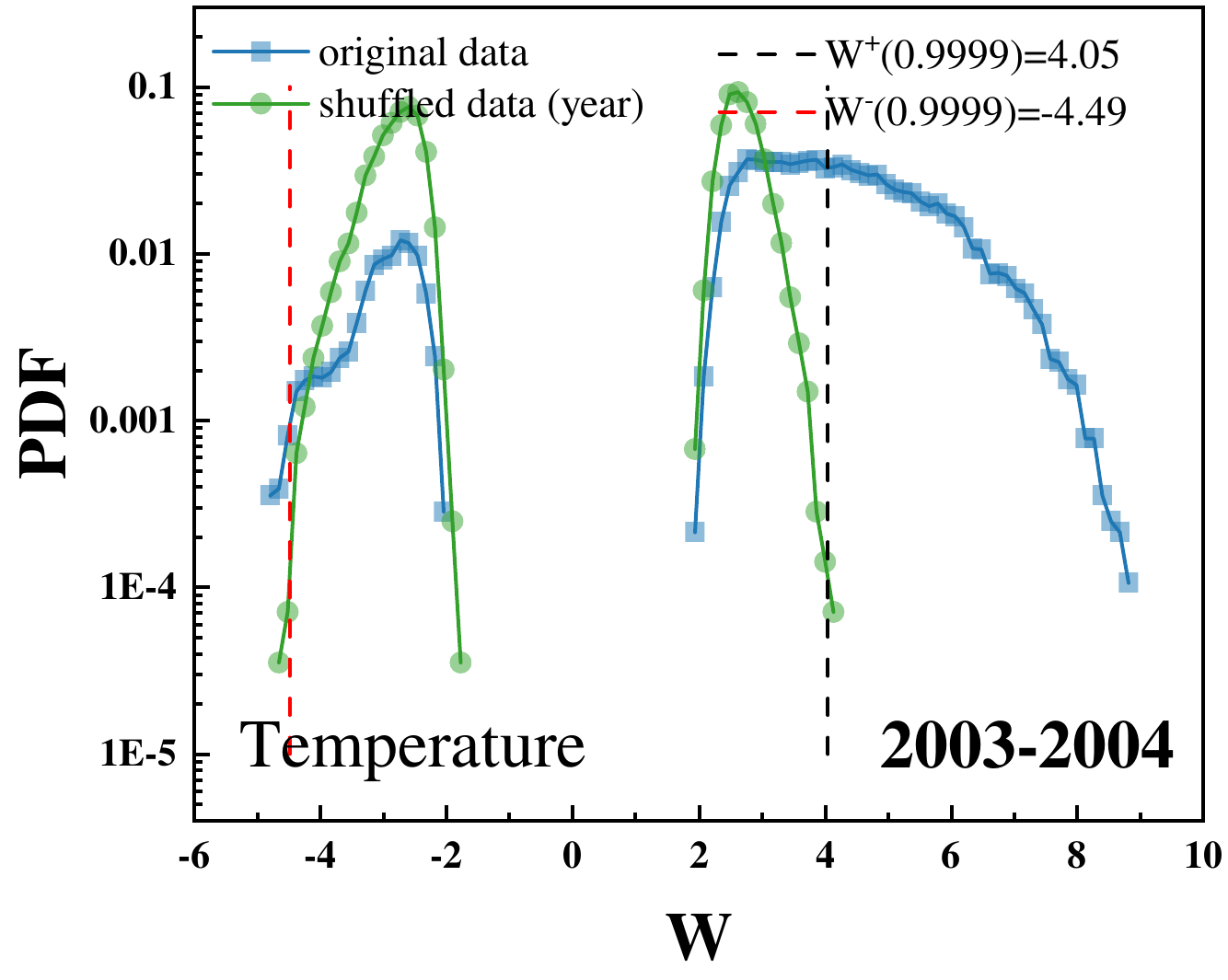}
\includegraphics[width=8.5em, height=7em]{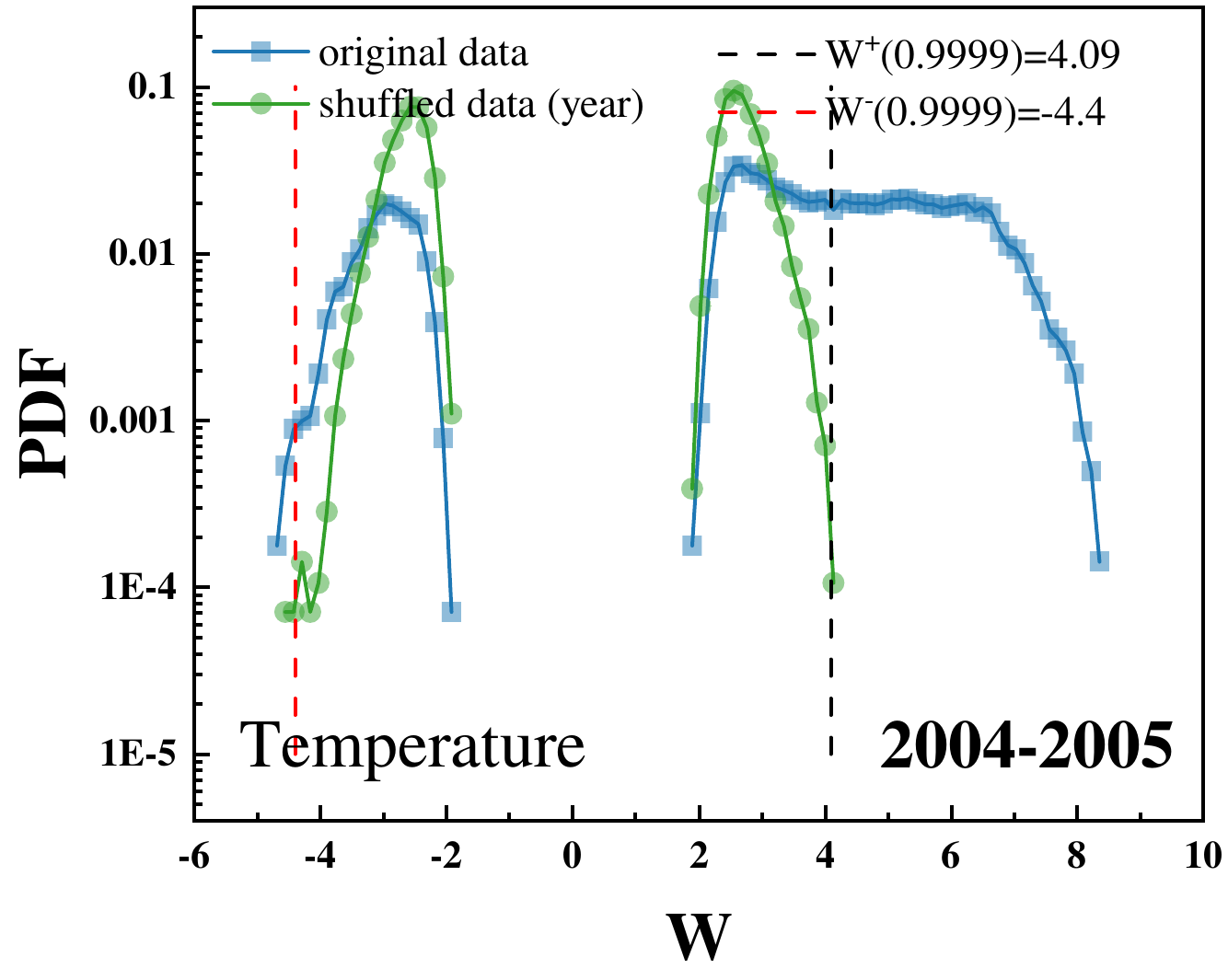}
\includegraphics[width=8.5em, height=7em]{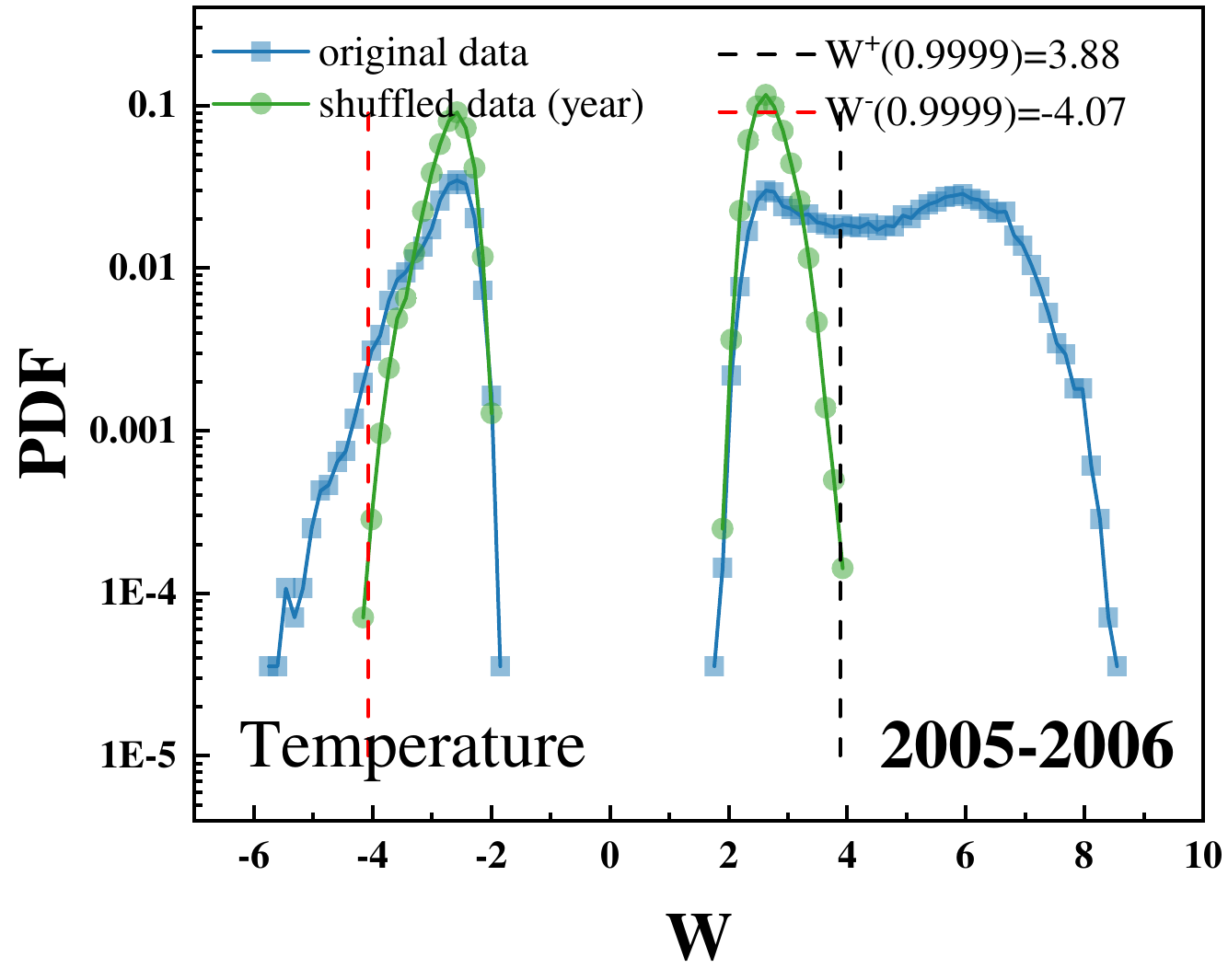}
\includegraphics[width=8.5em, height=7em]{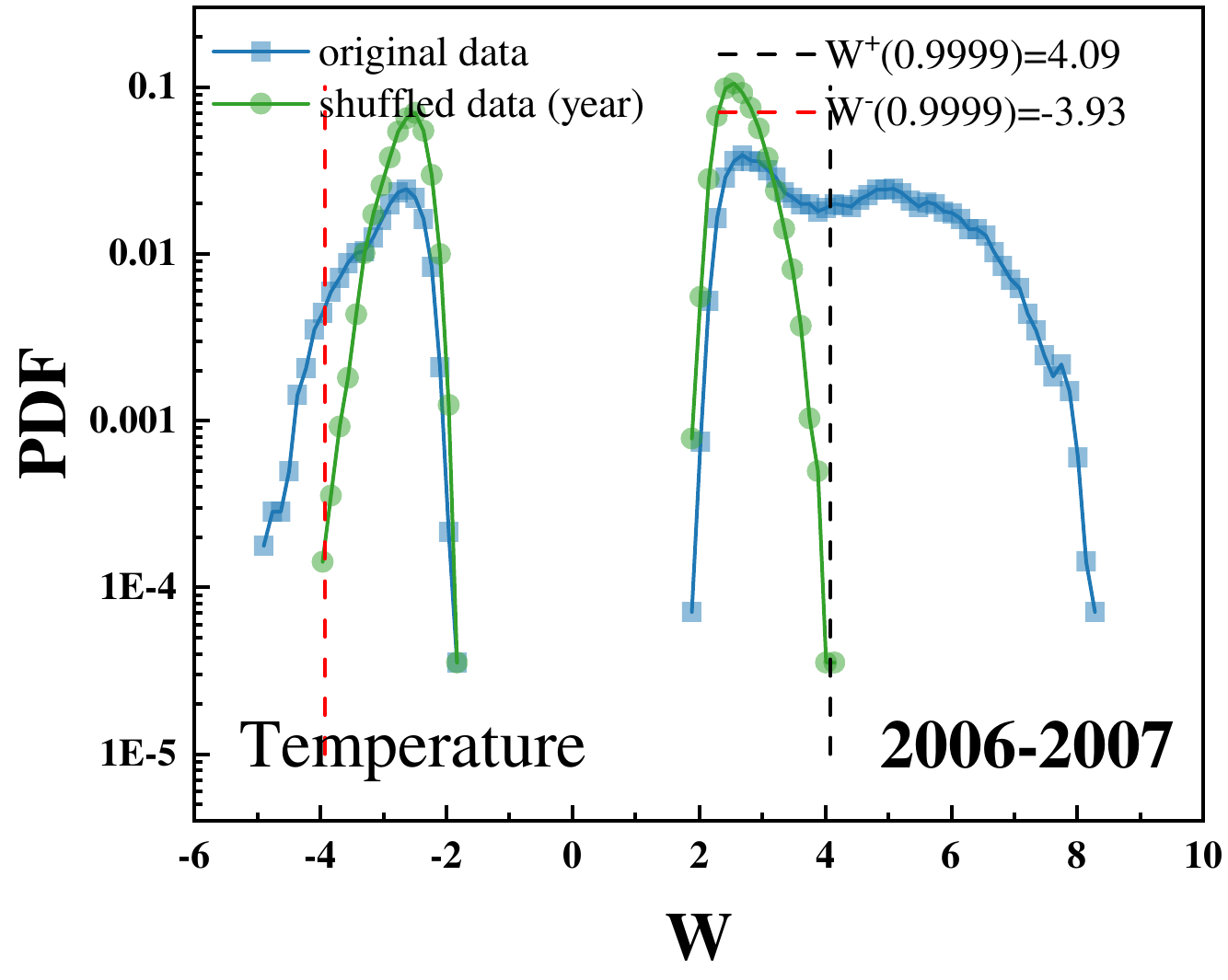}
\includegraphics[width=8.5em, height=7em]{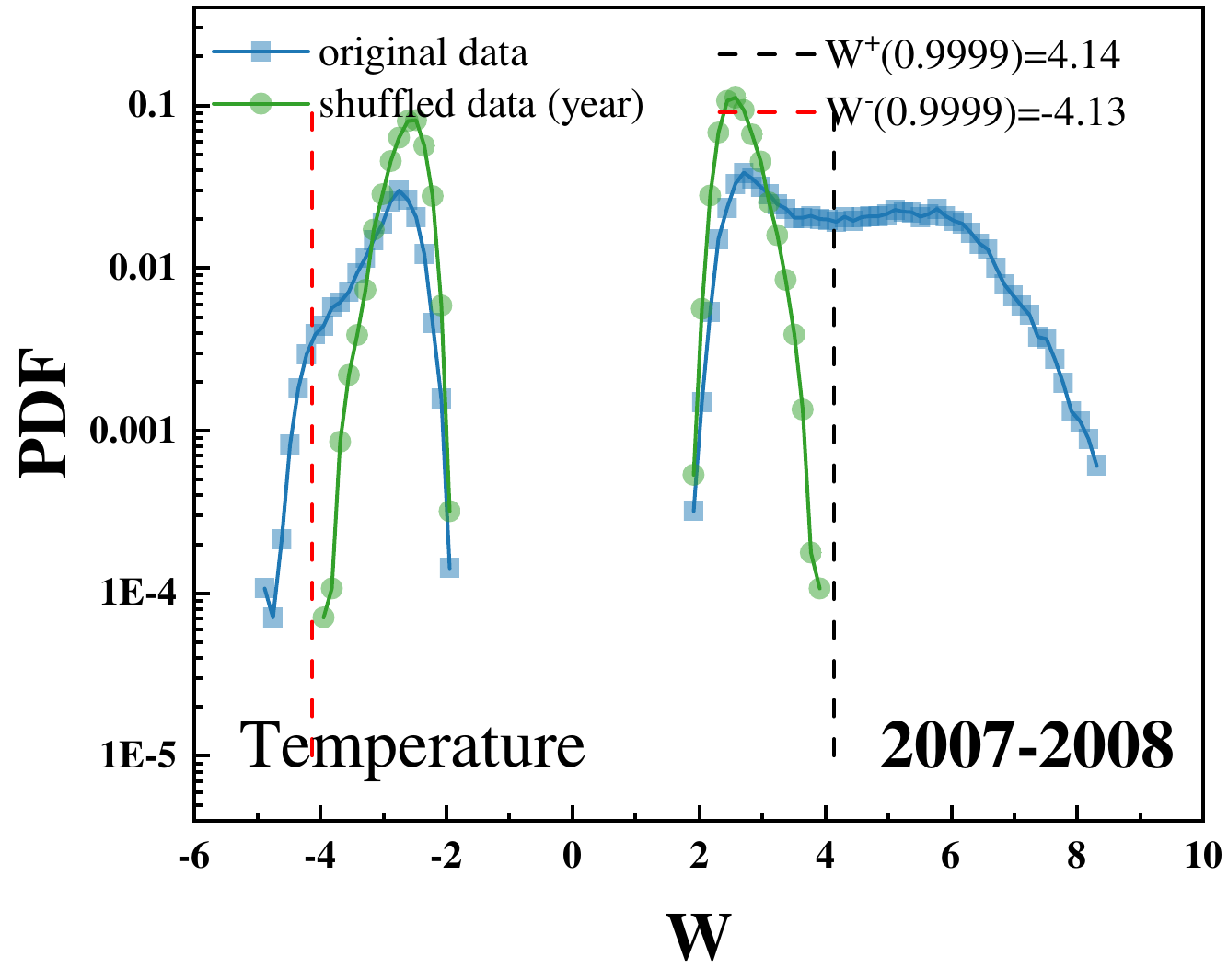}
\includegraphics[width=8.5em, height=7em]{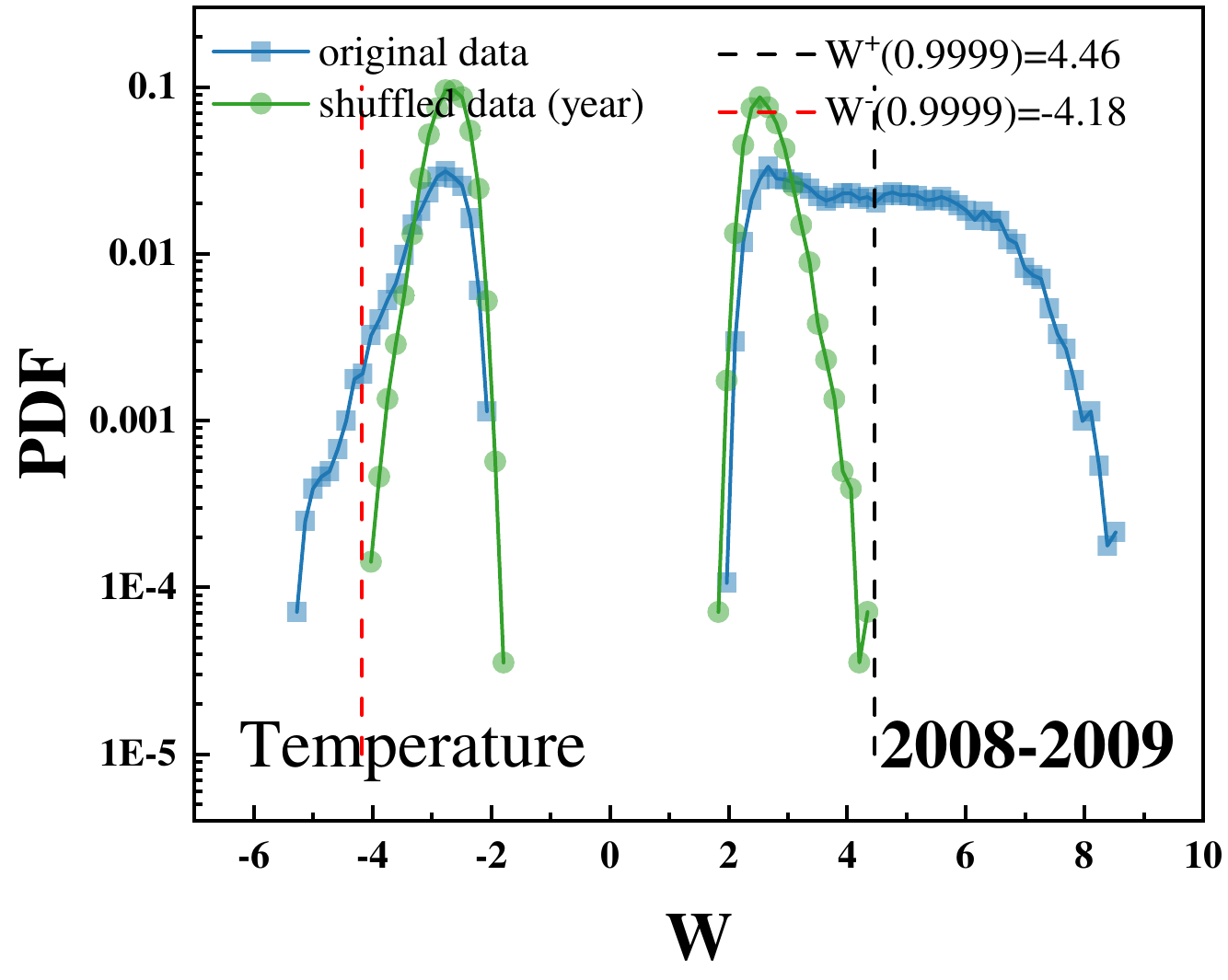}
\includegraphics[width=8.5em, height=7em]{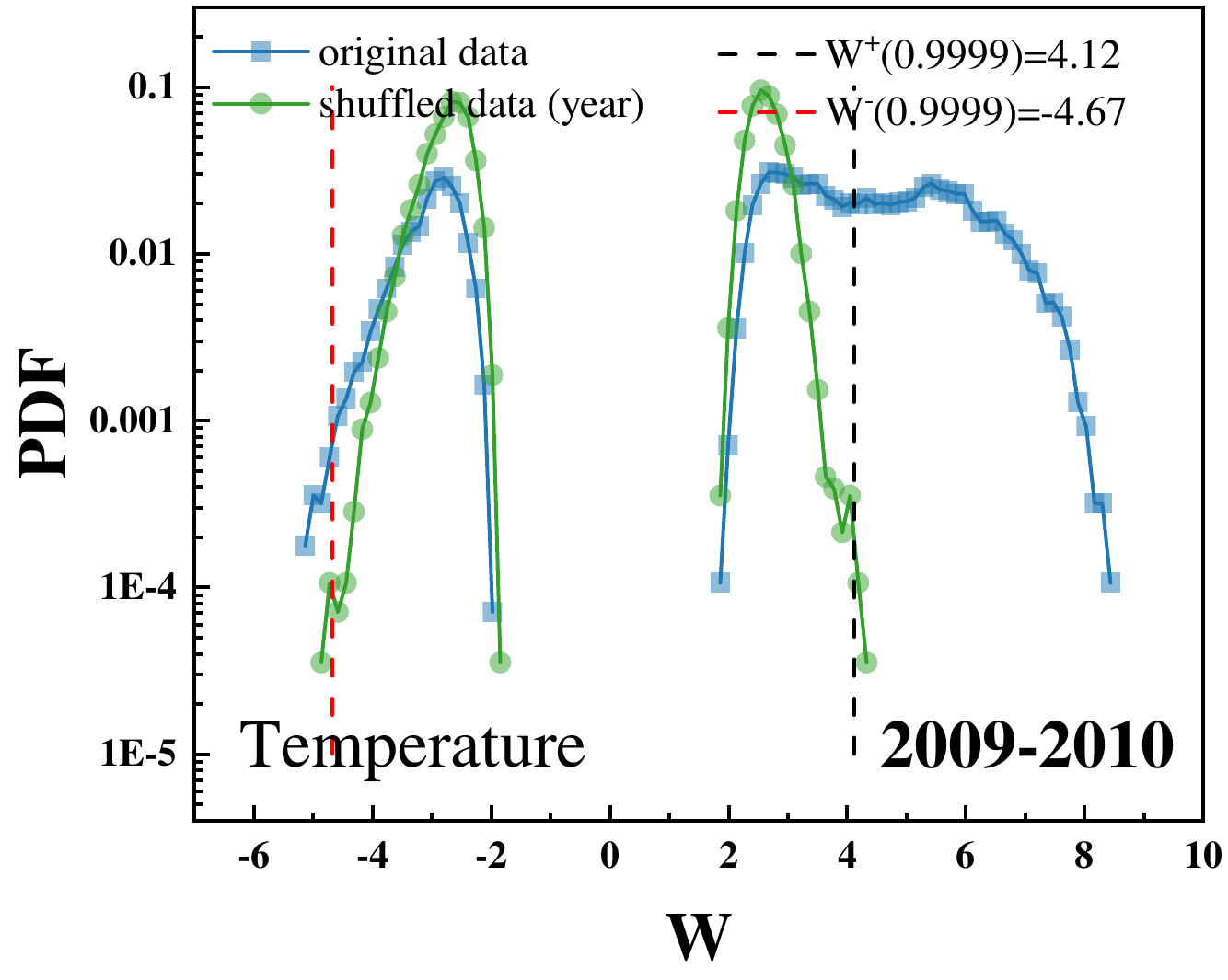}
\includegraphics[width=8.5em, height=7em]{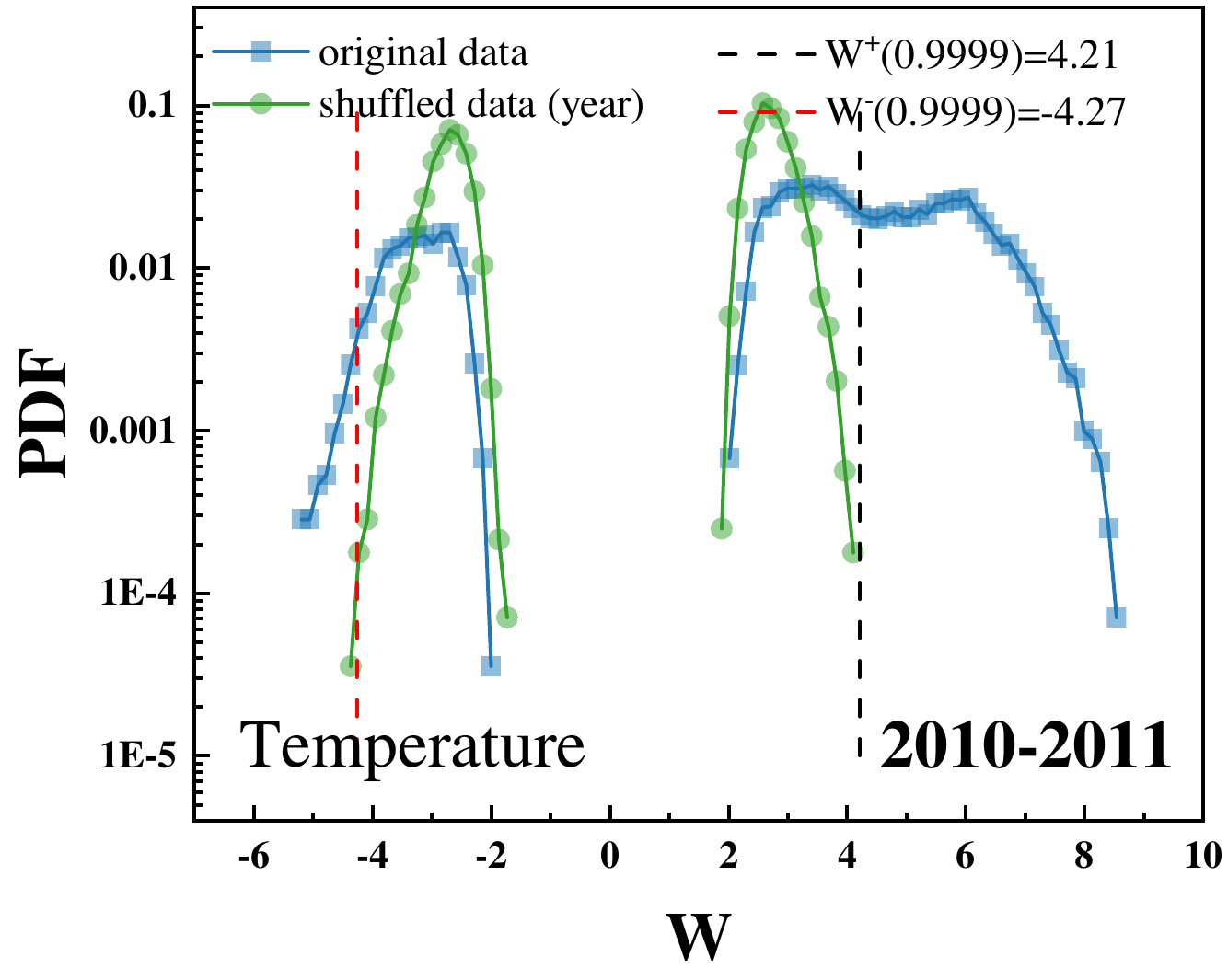}
\includegraphics[width=8.5em, height=7em]{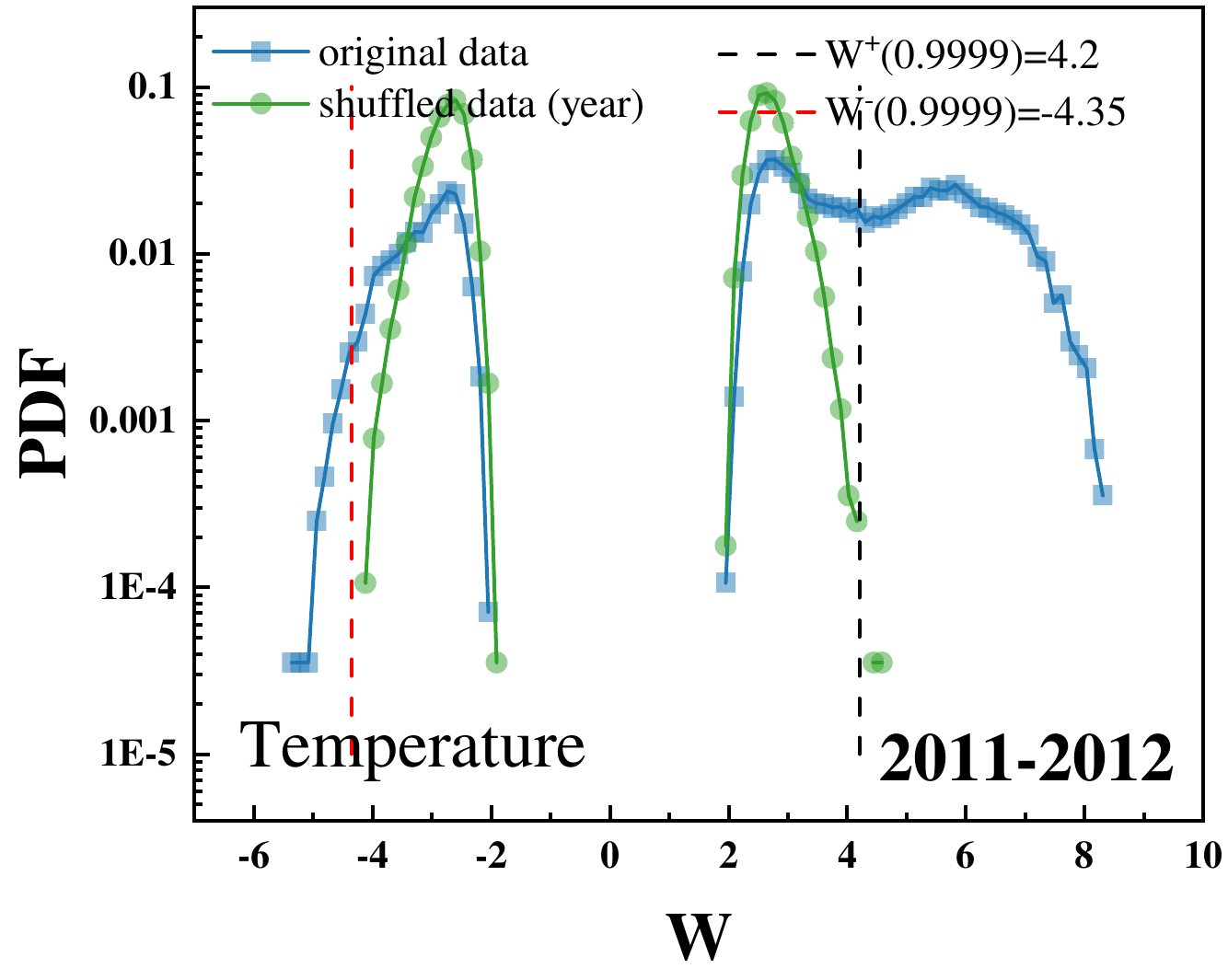}
\includegraphics[width=8.5em, height=7em]{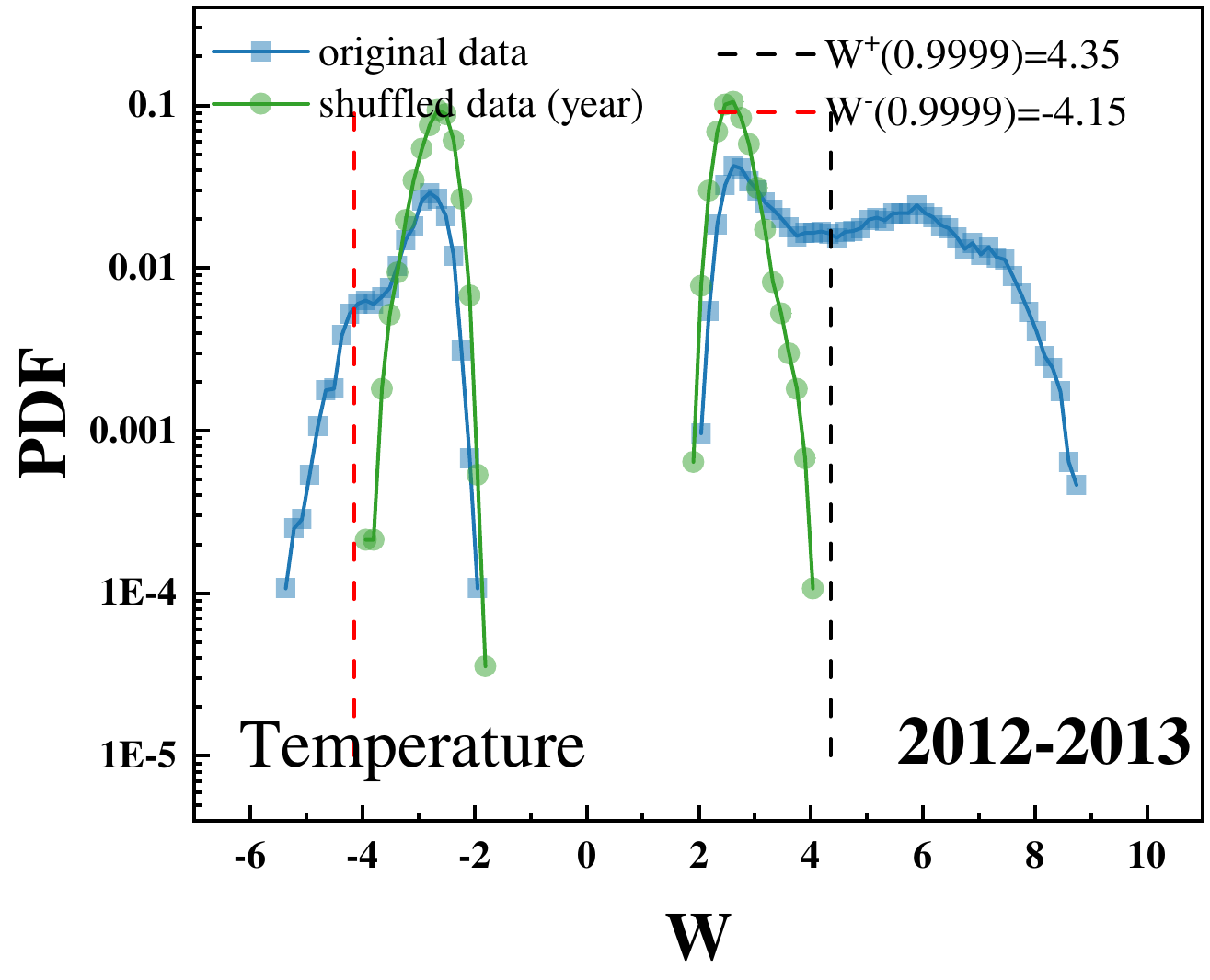}
\includegraphics[width=8.5em, height=7em]{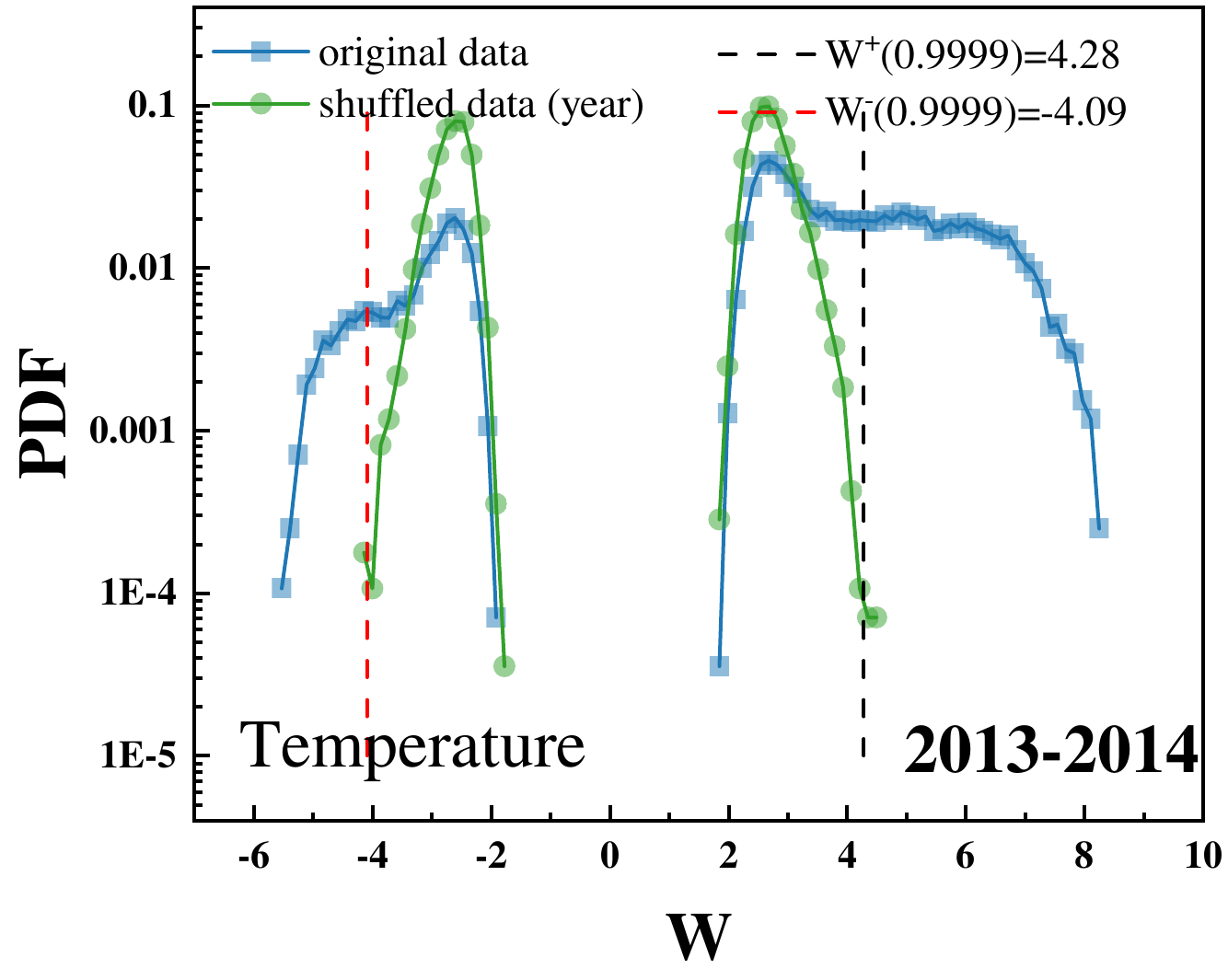}
\includegraphics[width=8.5em, height=7em]{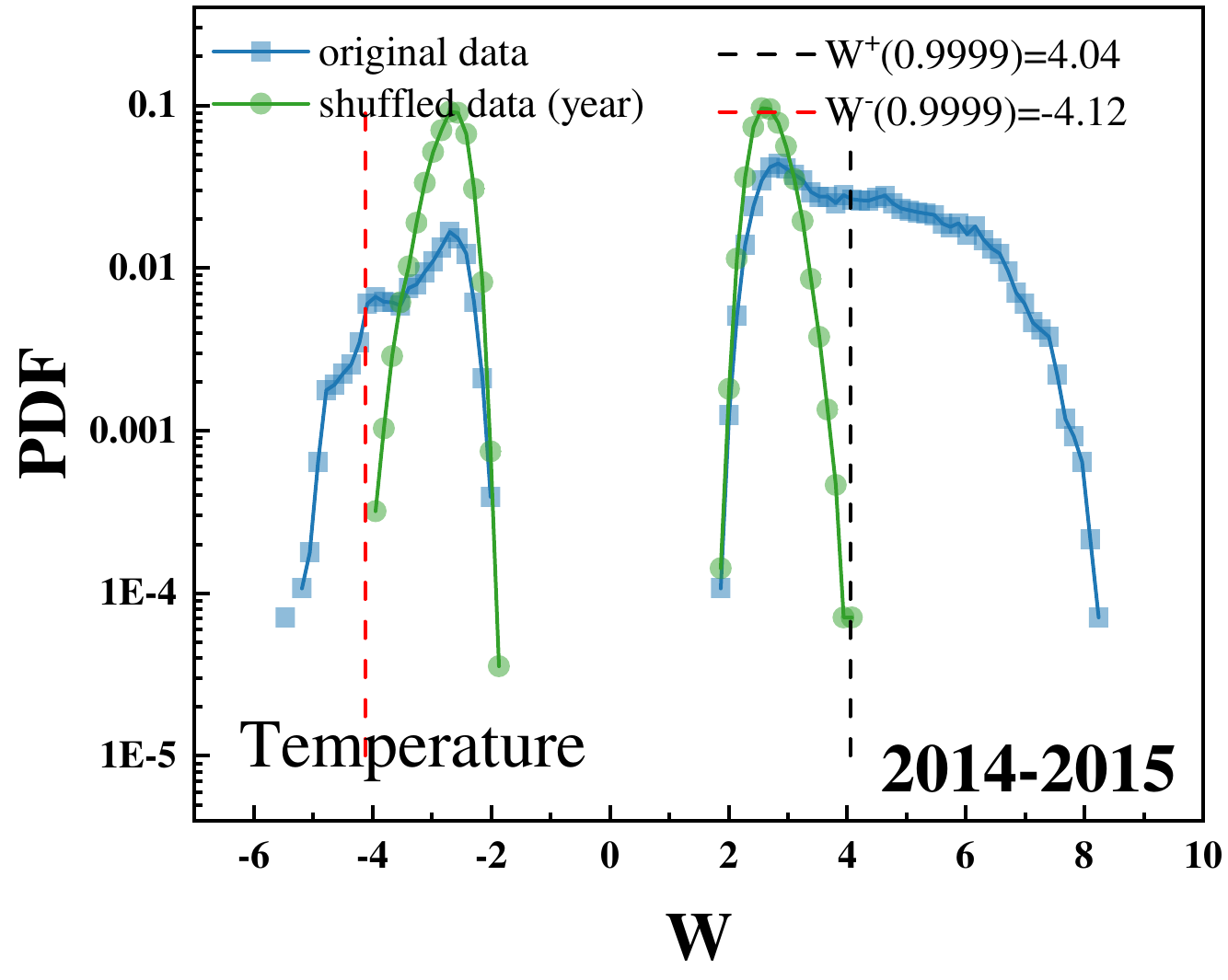}
\includegraphics[width=8.5em, height=7em]{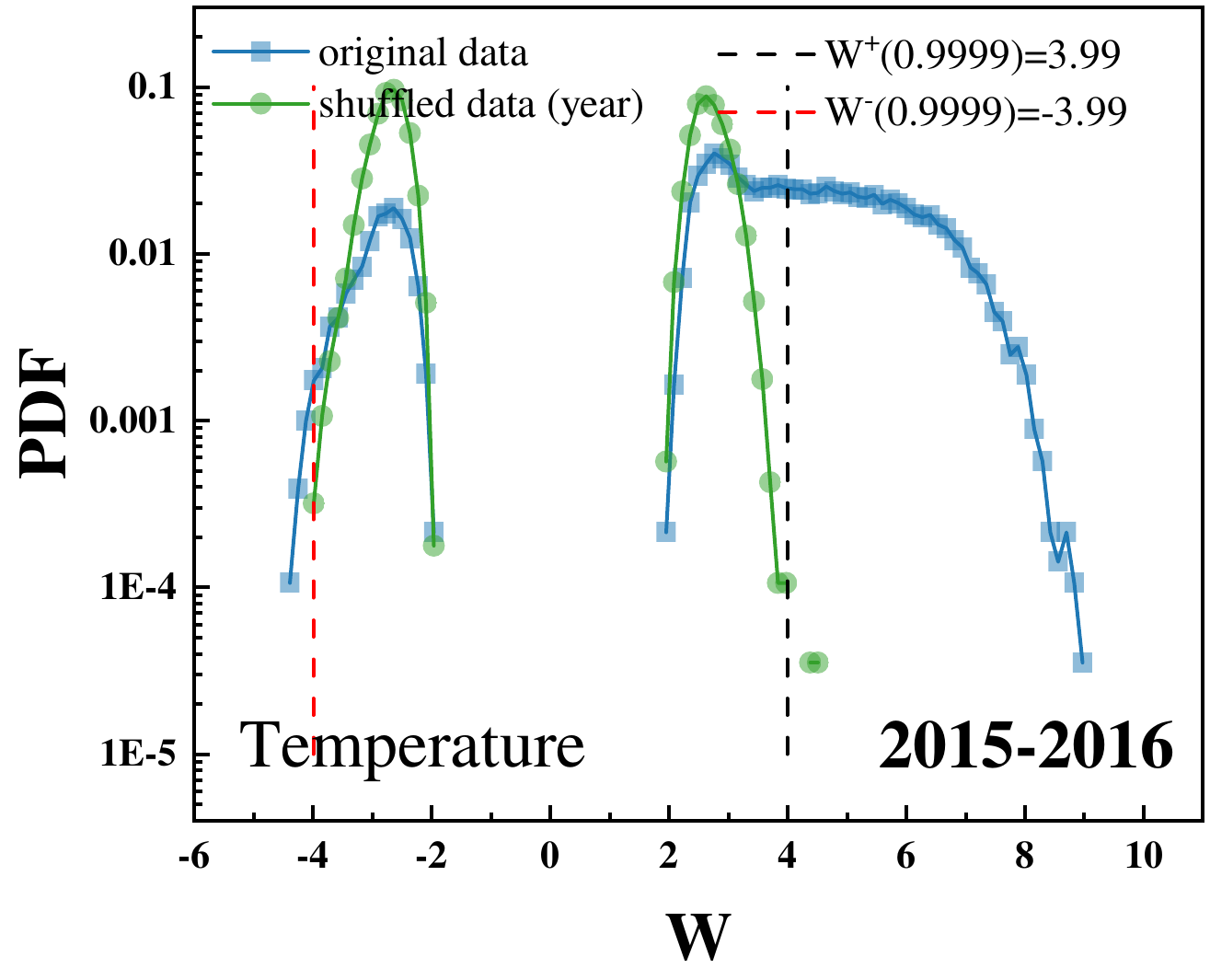}
\includegraphics[width=8.5em, height=7em]{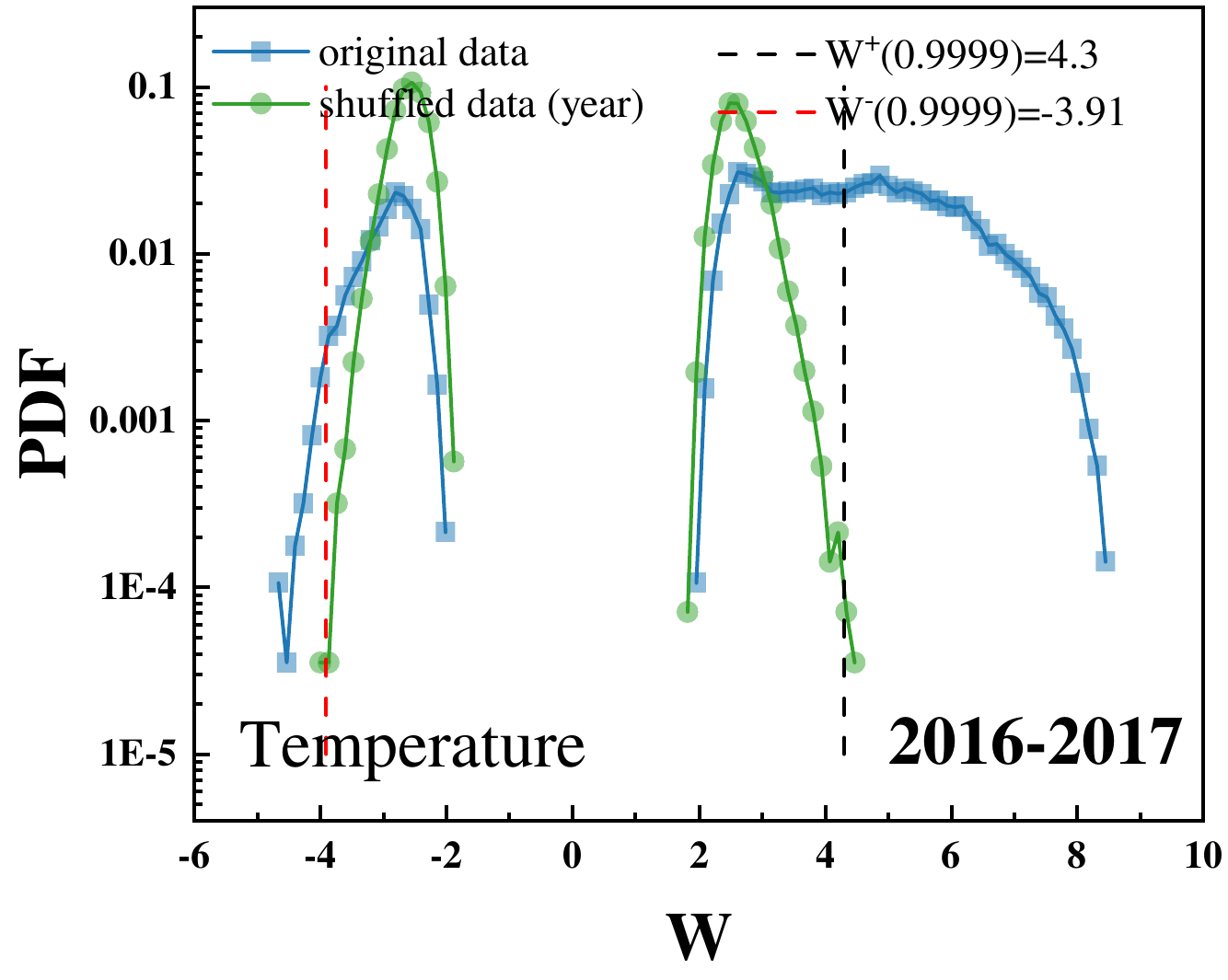}
\includegraphics[width=8.5em, height=7em]{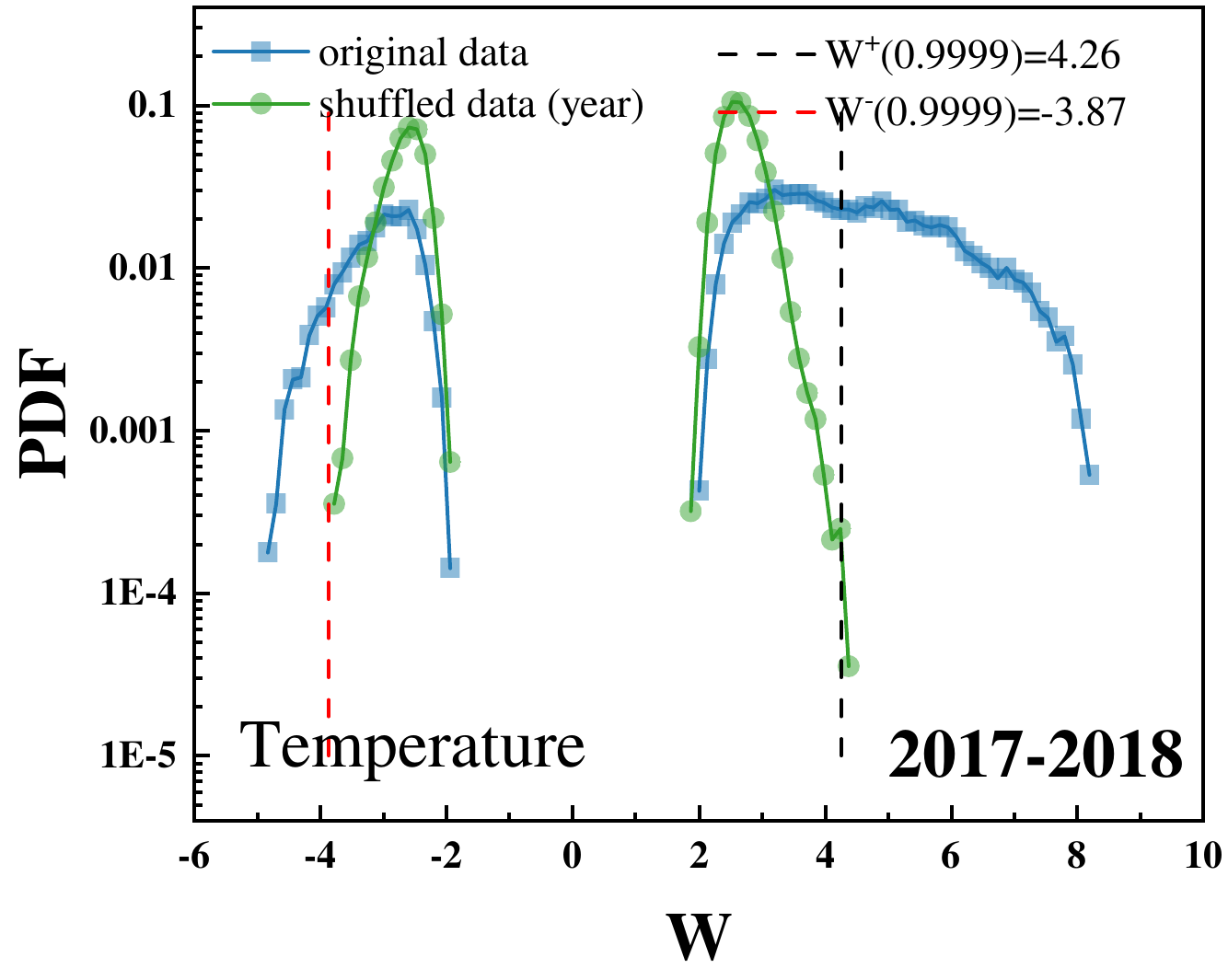}
\includegraphics[width=8.5em, height=7em]{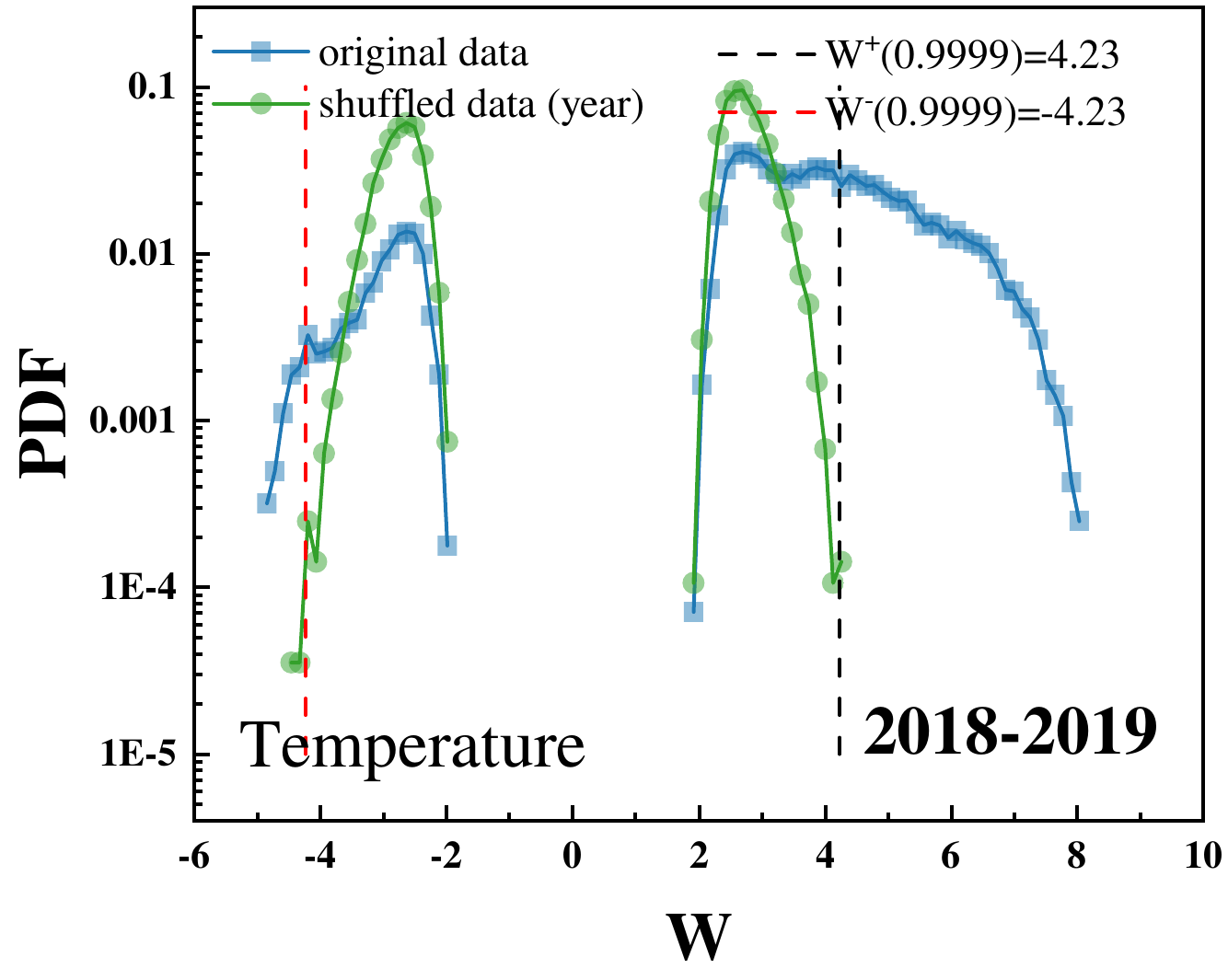}
\end{center}

\begin{center}
\noindent {\small {\bf Fig. S4} Probability distribution function (PDF) of link weights for the original data and shuffled data of temperature in China. }
\end{center}

\begin{center}
\includegraphics[width=8.5em, height=7em]{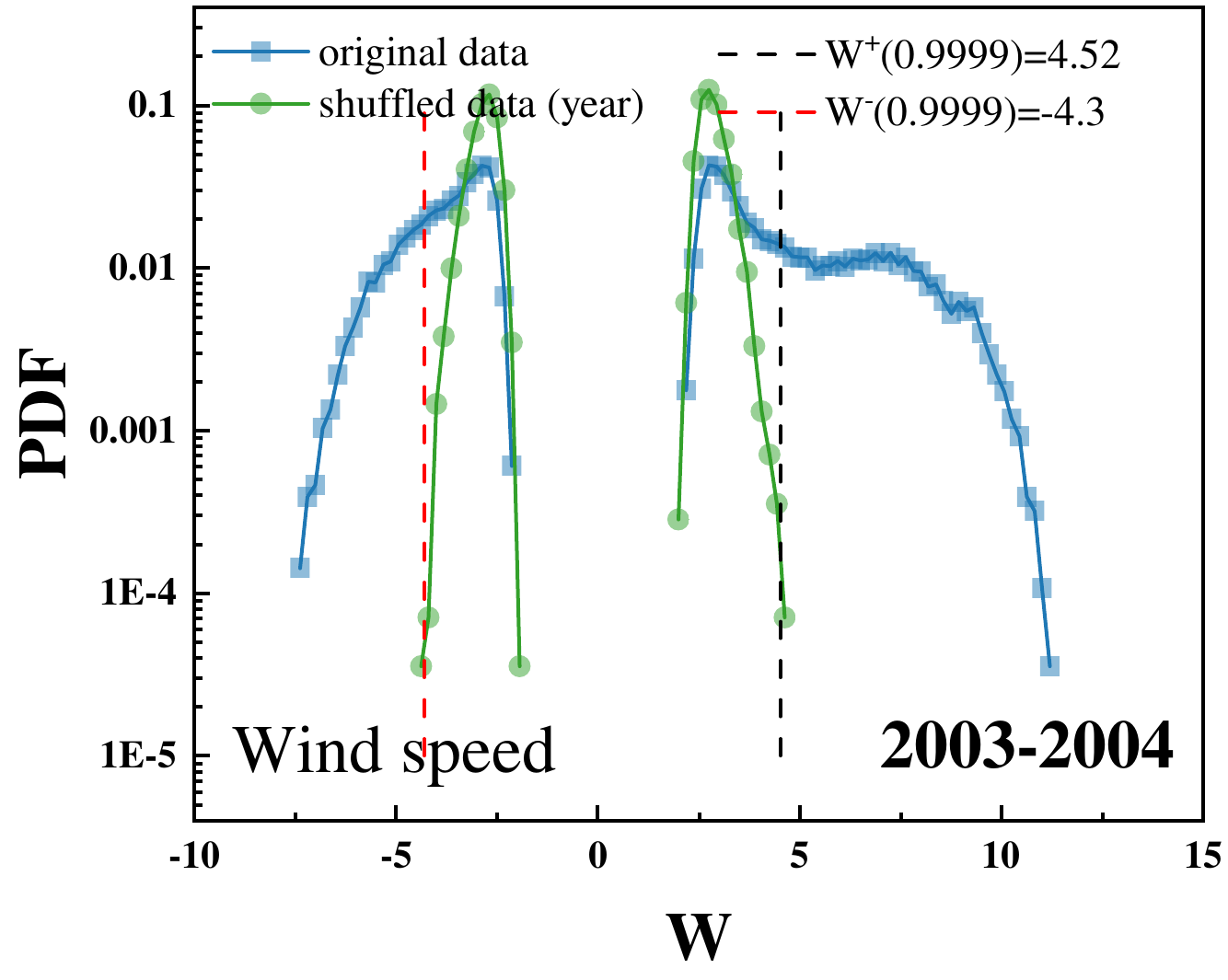}
\includegraphics[width=8.5em, height=7em]{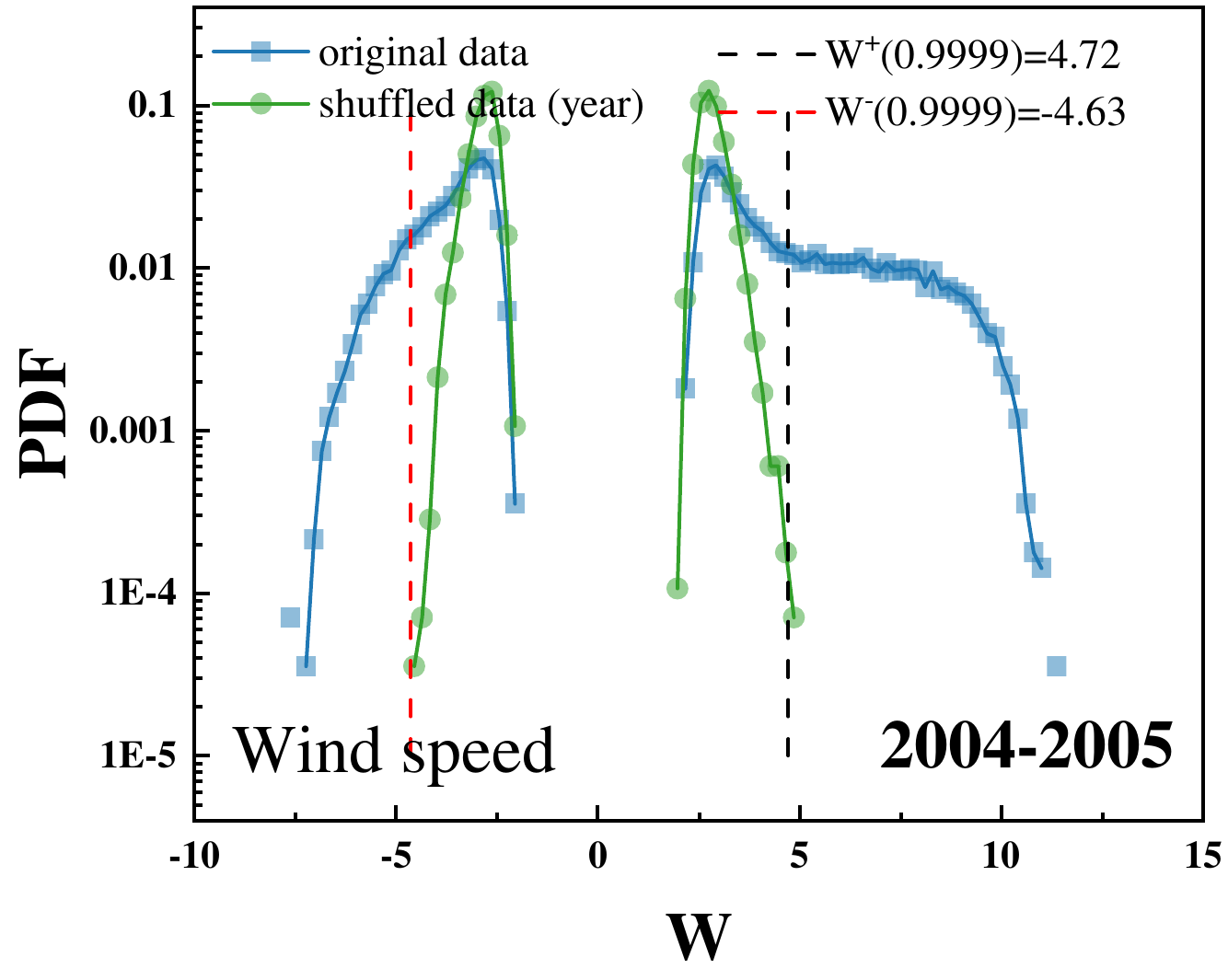}
\includegraphics[width=8.5em, height=7em]{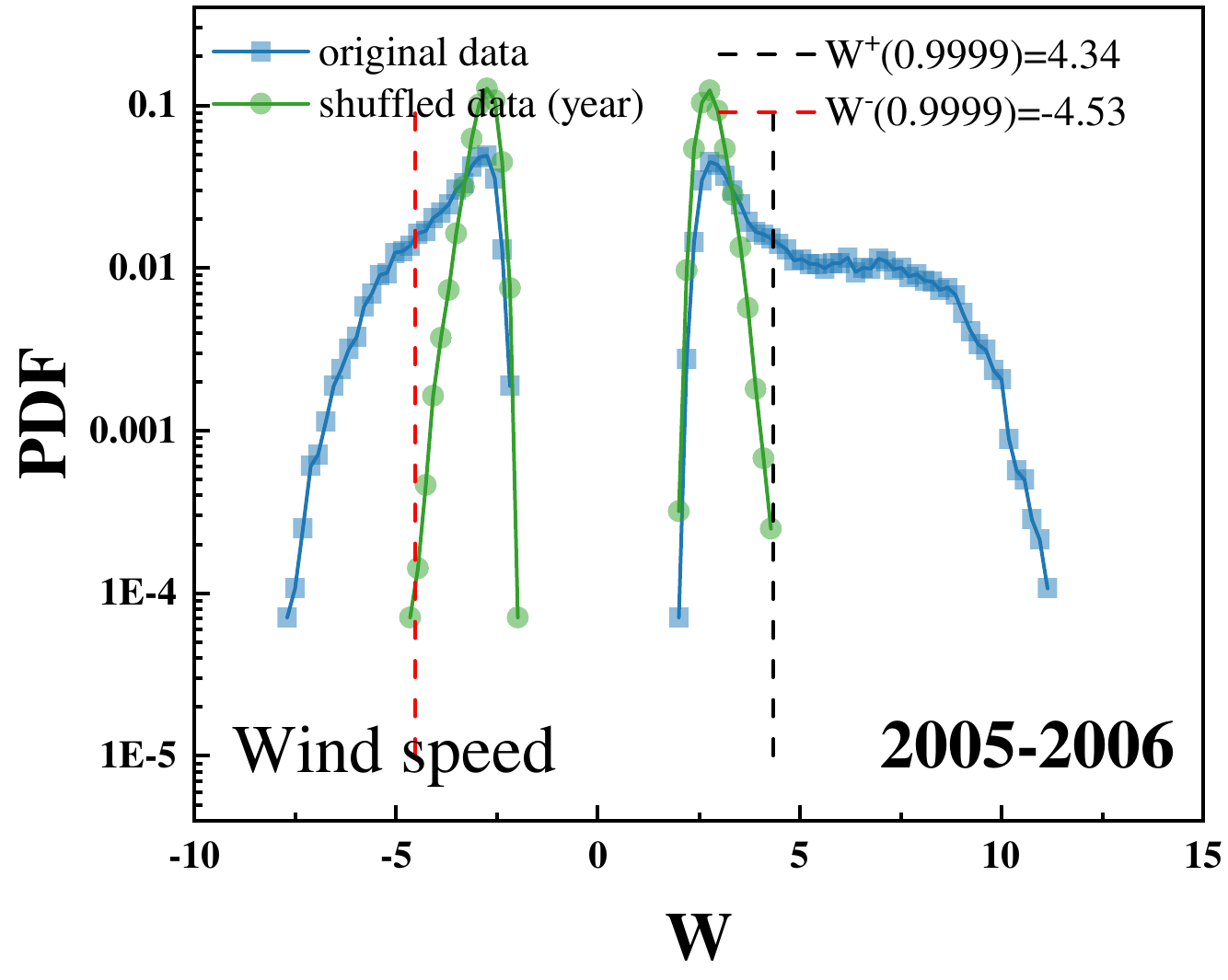}
\includegraphics[width=8.5em, height=7em]{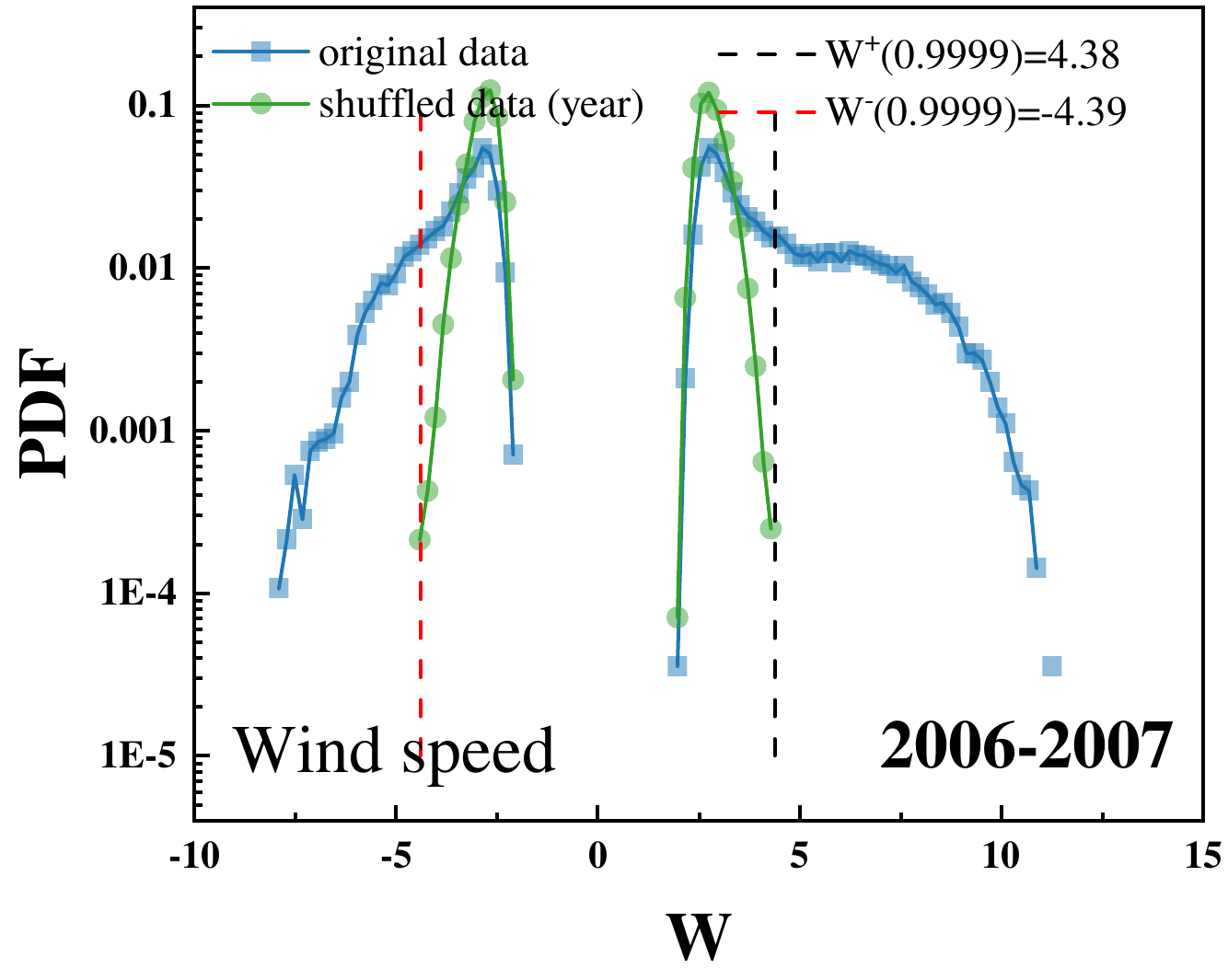}
\includegraphics[width=8.5em, height=7em]{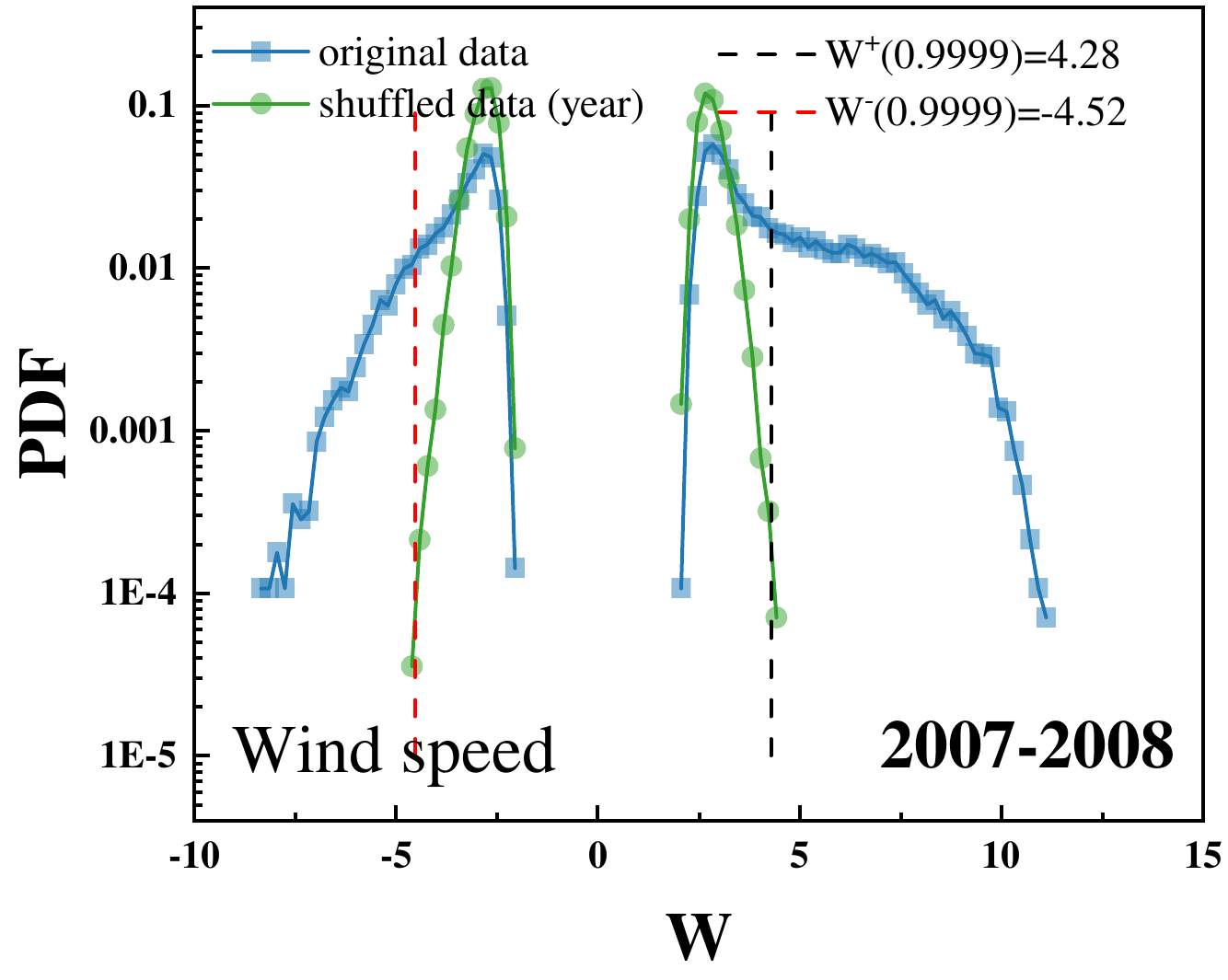}
\includegraphics[width=8.5em, height=7em]{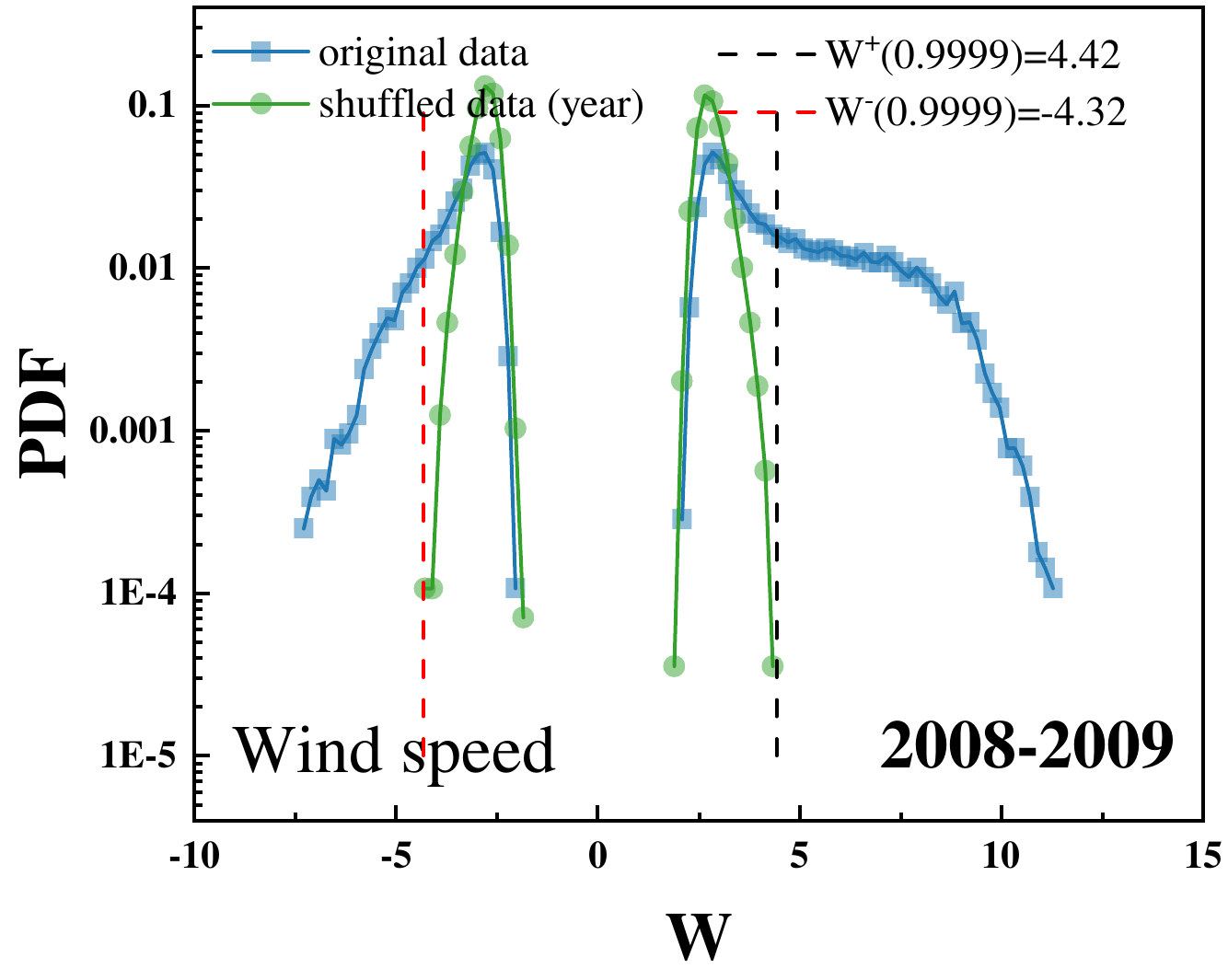}
\includegraphics[width=8.5em, height=7em]{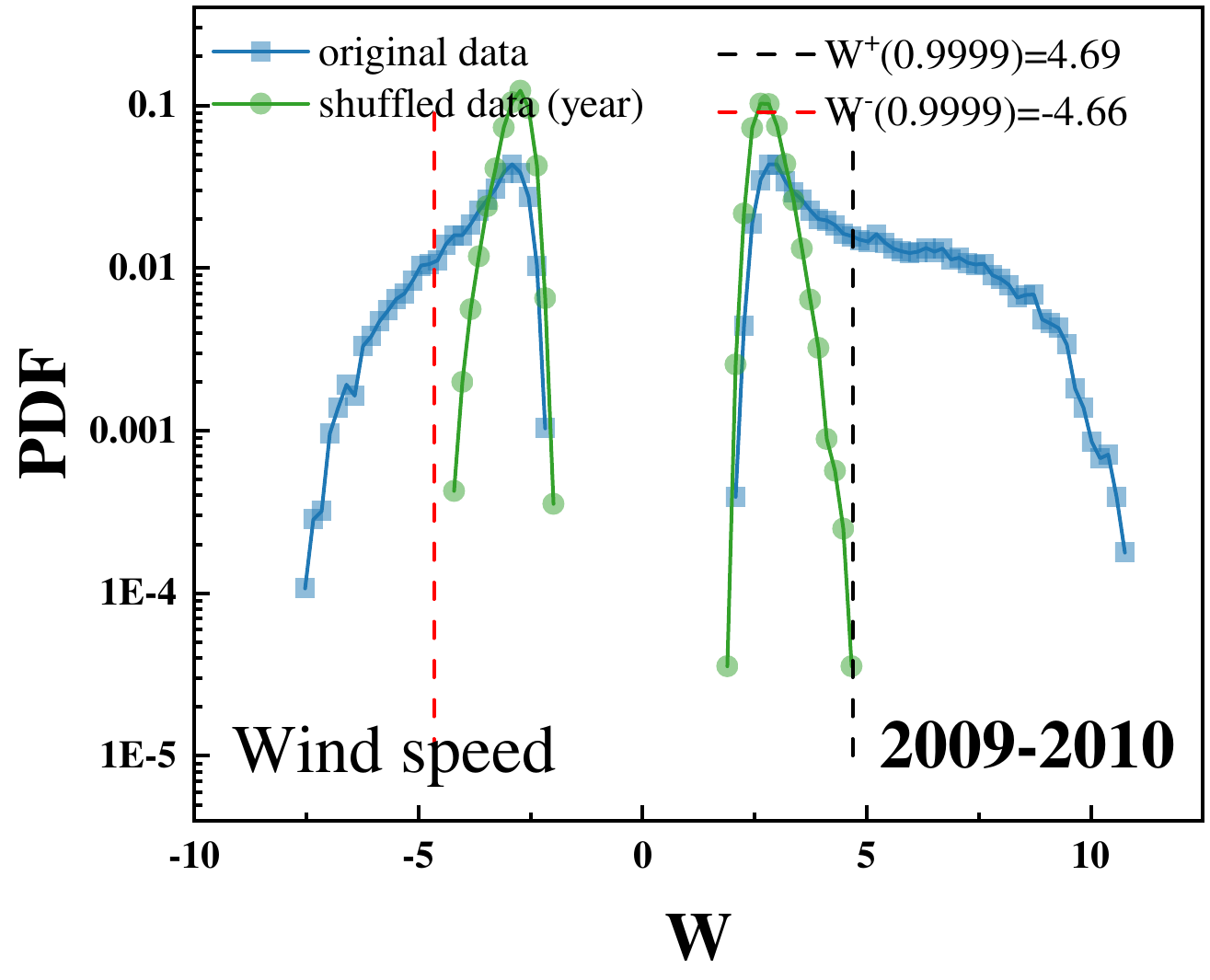}
\includegraphics[width=8.5em, height=7em]{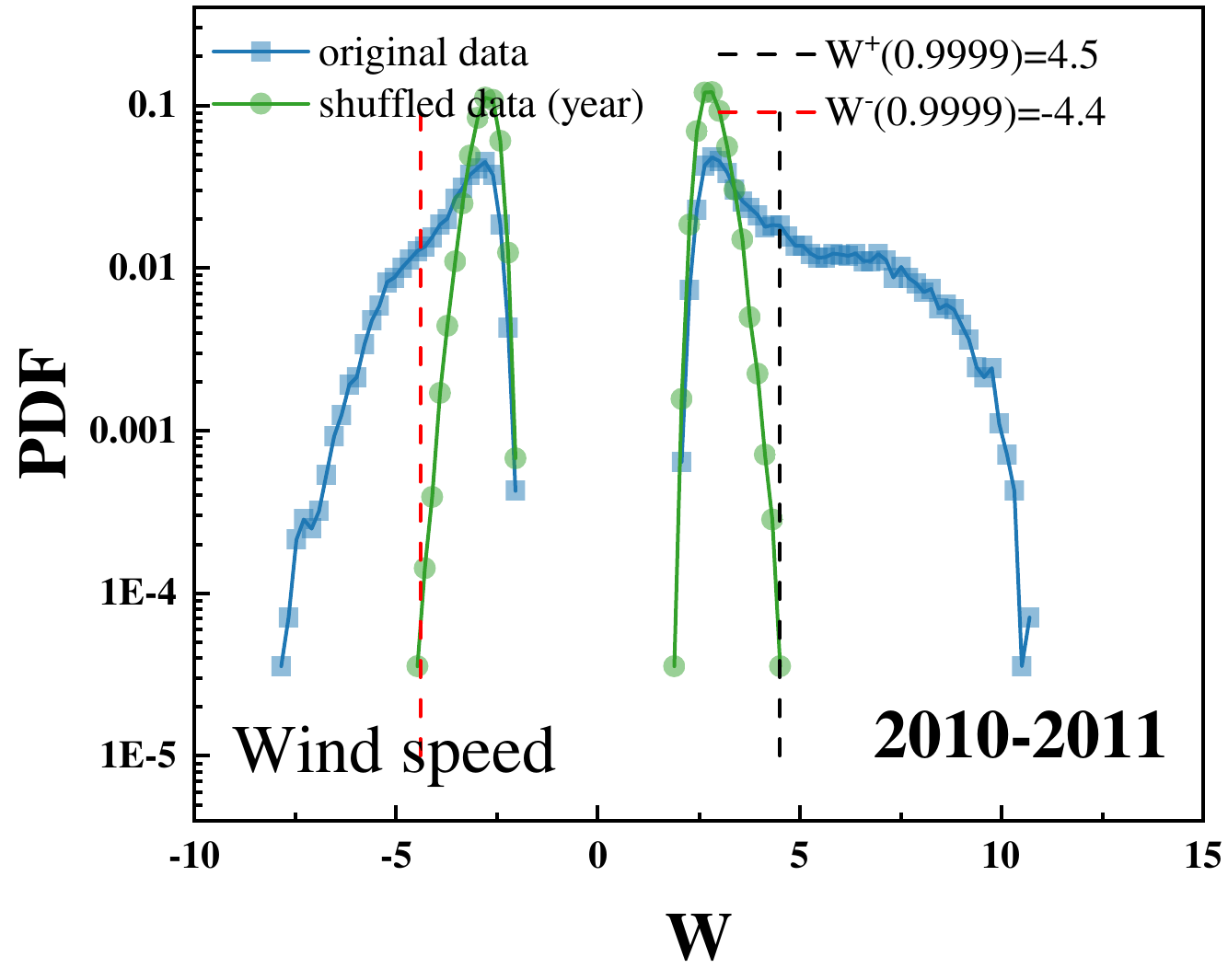}
\includegraphics[width=8.5em, height=7em]{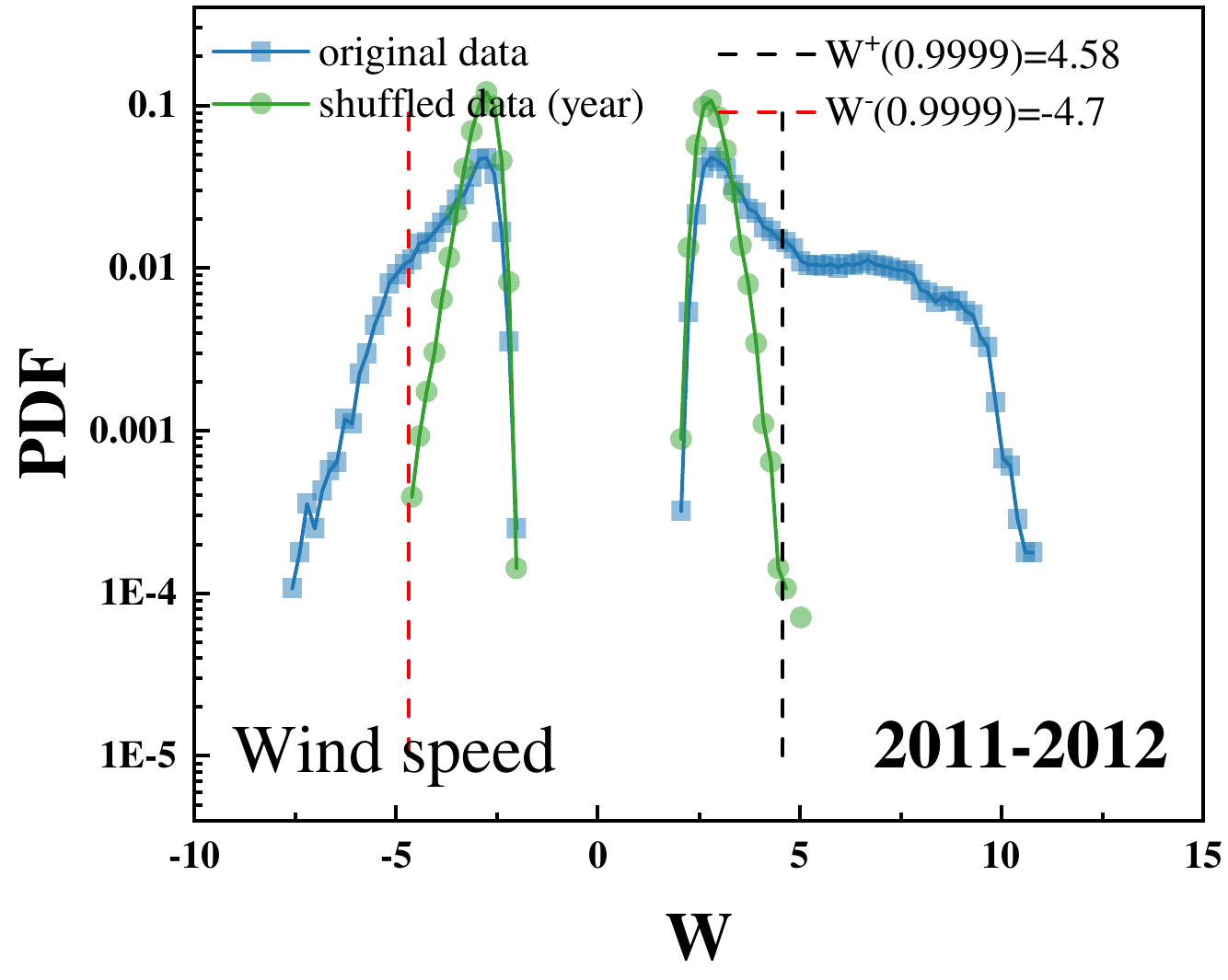}
\includegraphics[width=8.5em, height=7em]{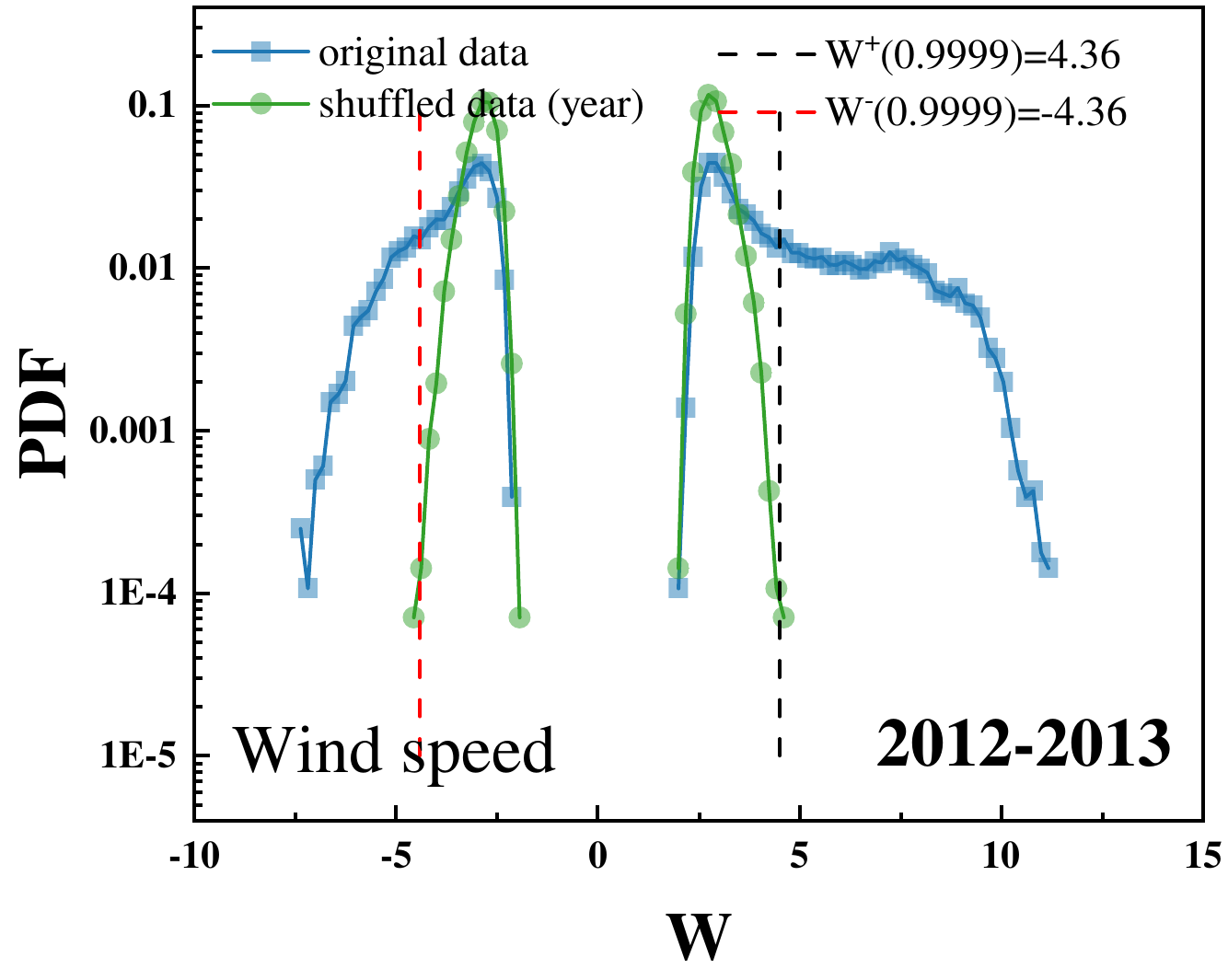}
\includegraphics[width=8.5em, height=7em]{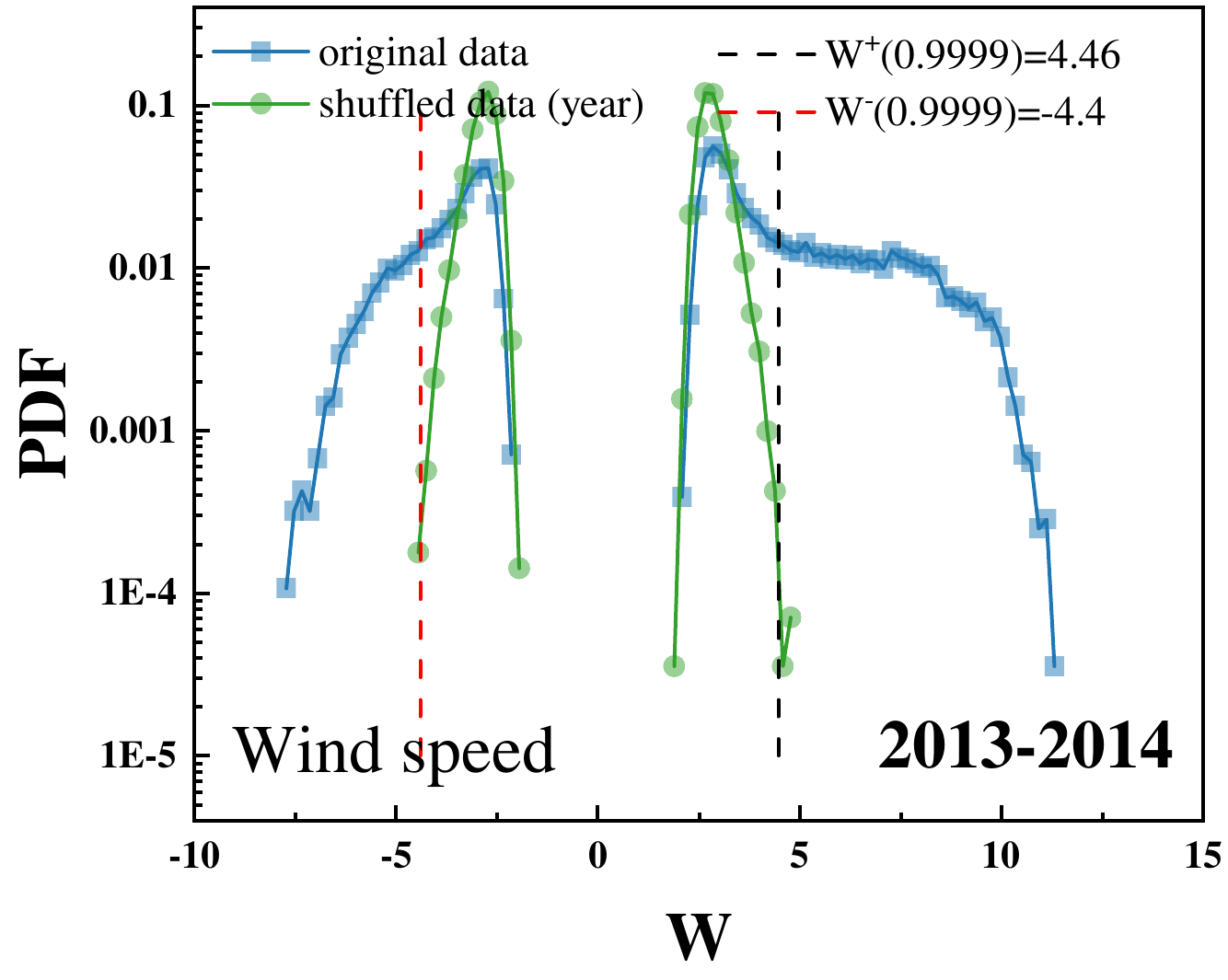}
\includegraphics[width=8.5em, height=7em]{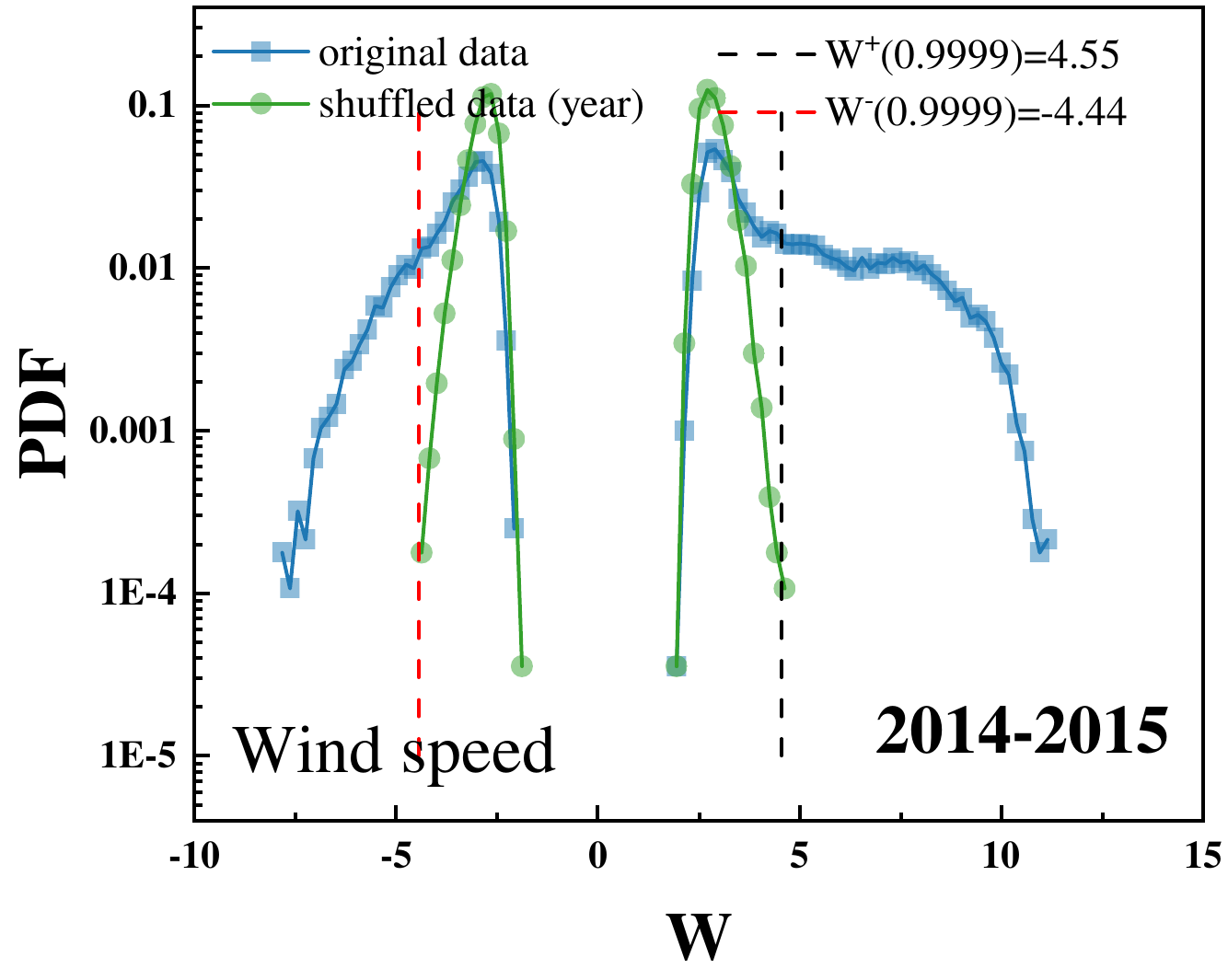}
\includegraphics[width=8.5em, height=7em]{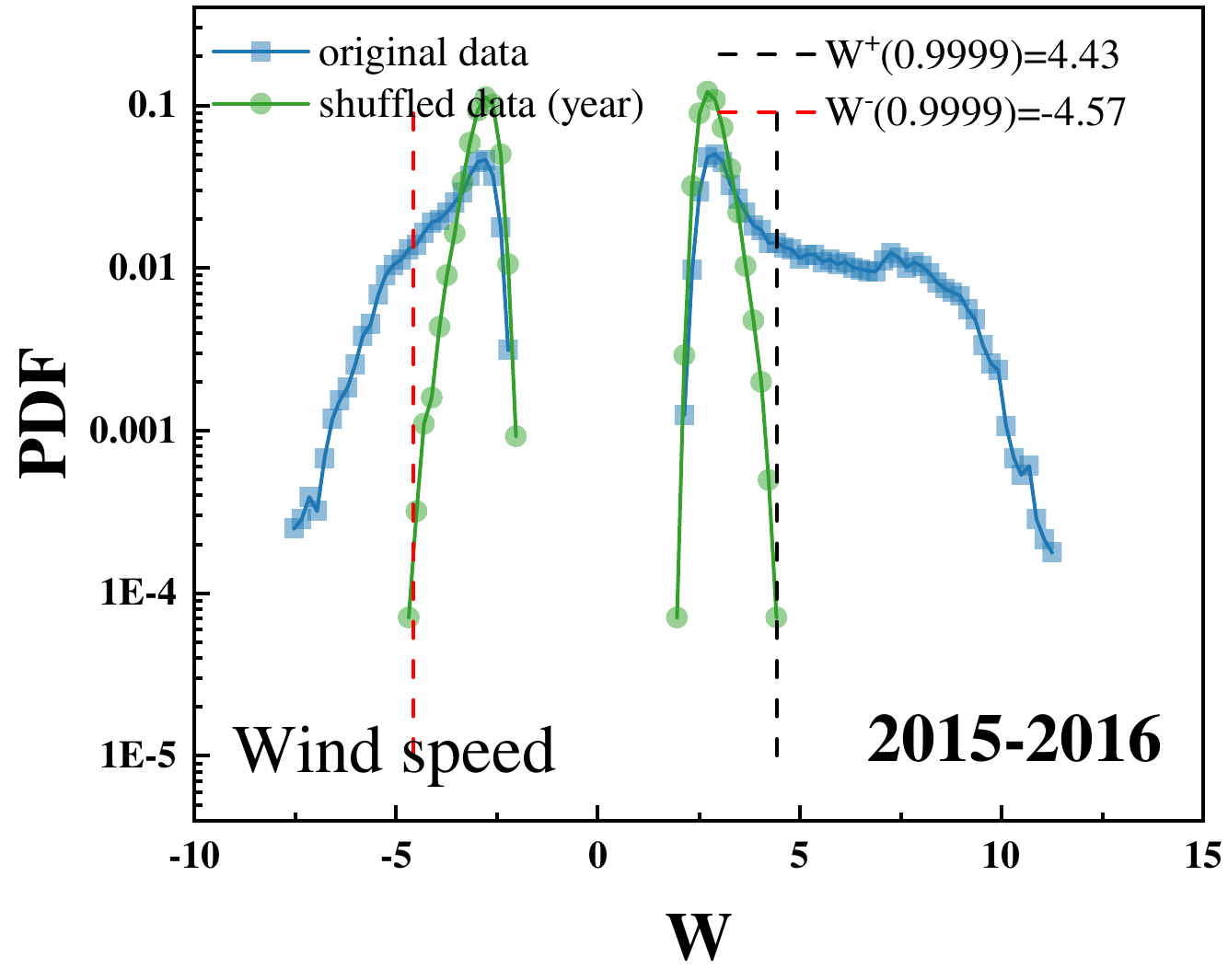}
\includegraphics[width=8.5em, height=7em]{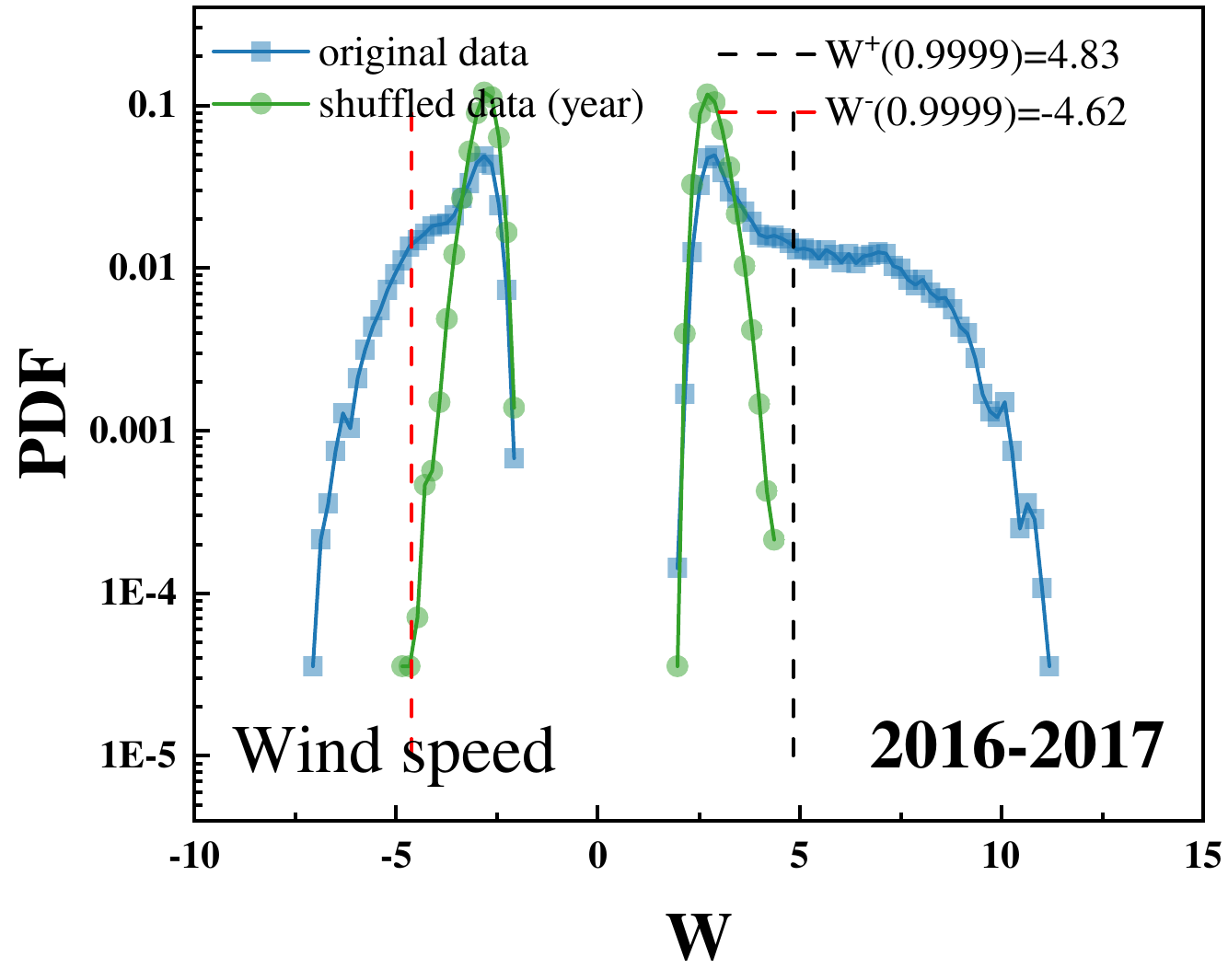}
\includegraphics[width=8.5em, height=7em]{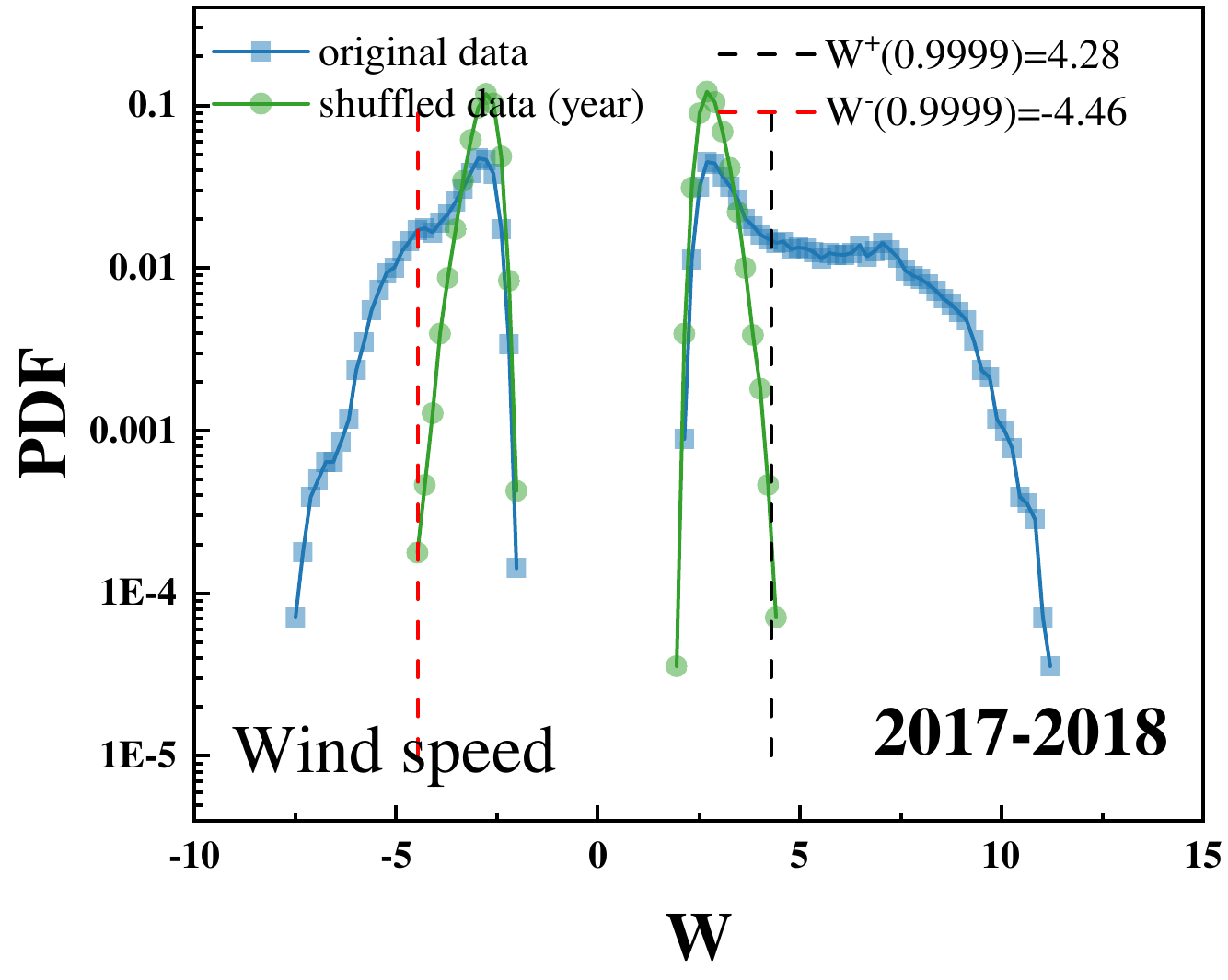}
\includegraphics[width=8.5em, height=7em]{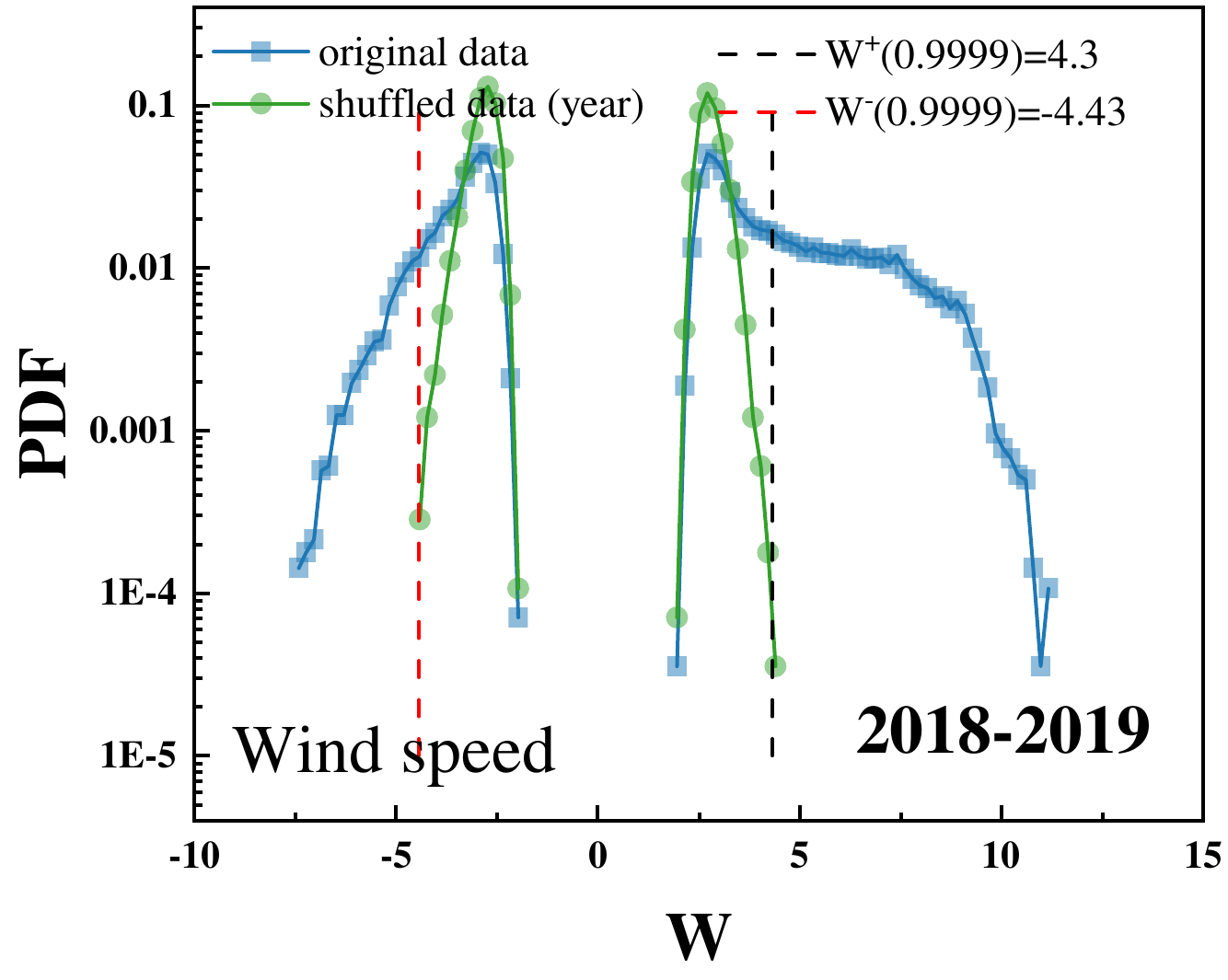}
\end{center}

\begin{center}
\noindent {\small {\bf Fig. S5} Probability distribution function (PDF) of link weights for the original data and shuffled data of wind speed in China. }
\end{center}

\begin{center}
\includegraphics[width=8.5em, height=7em]{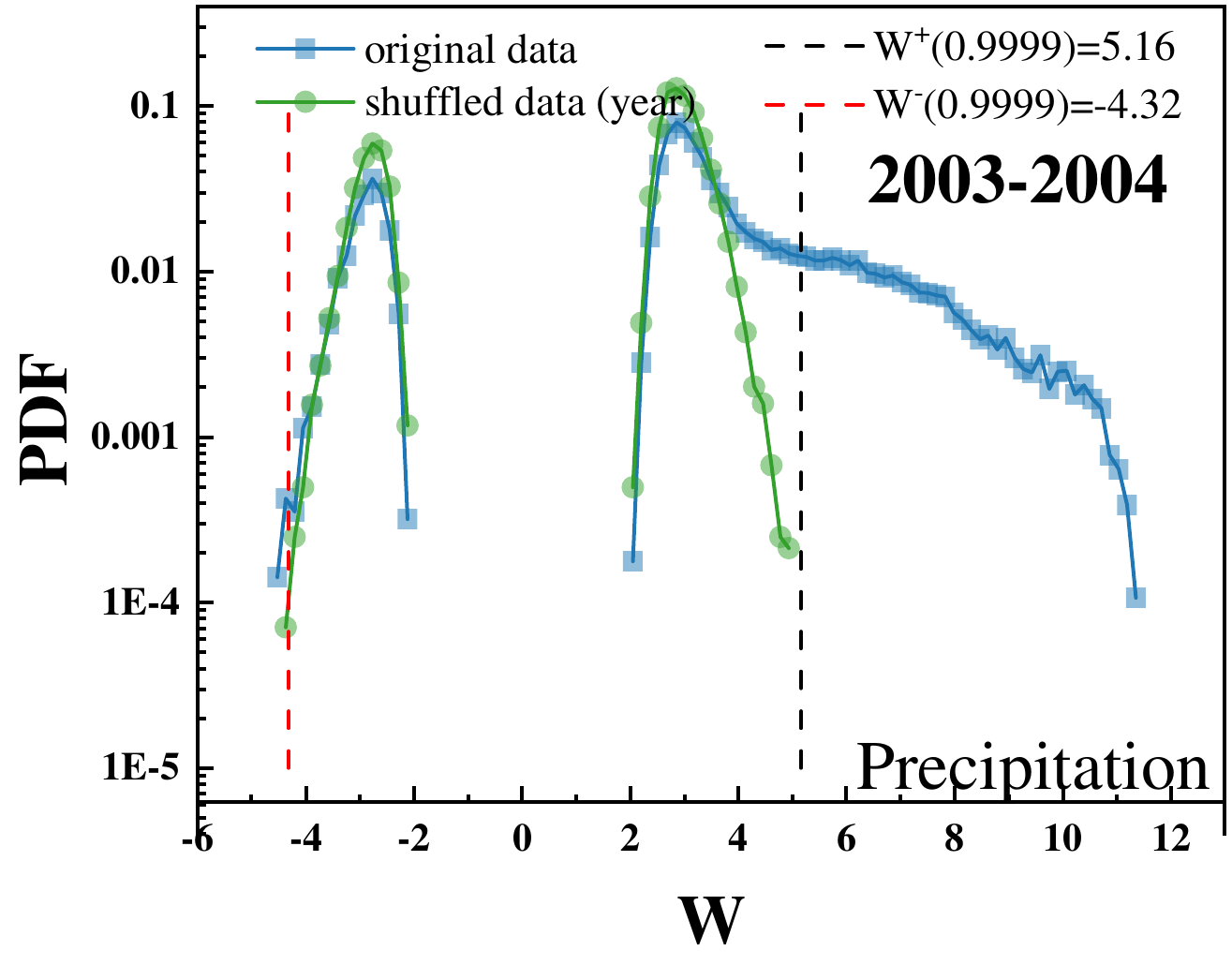}
\includegraphics[width=8.5em, height=7em]{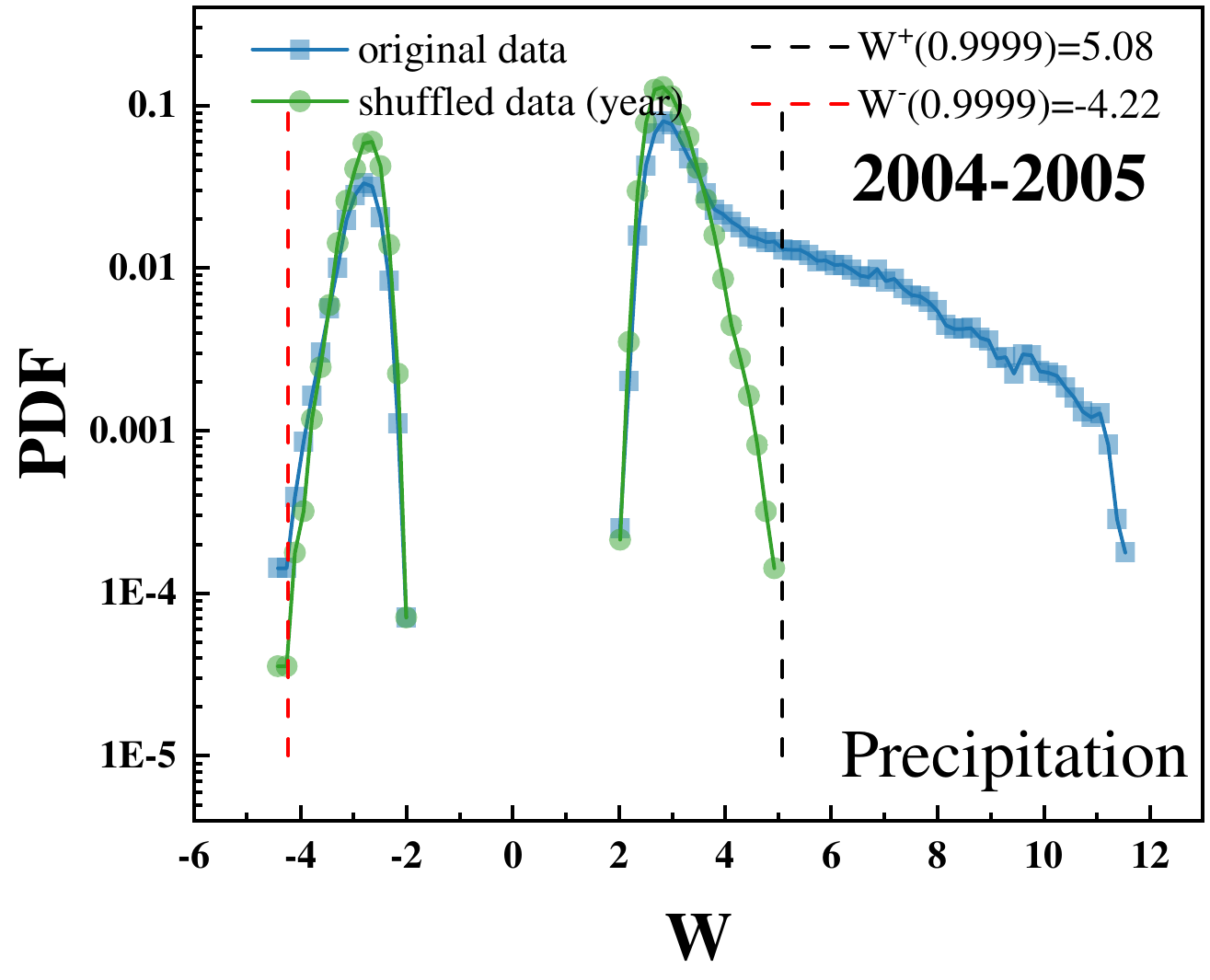}
\includegraphics[width=8.5em, height=7em]{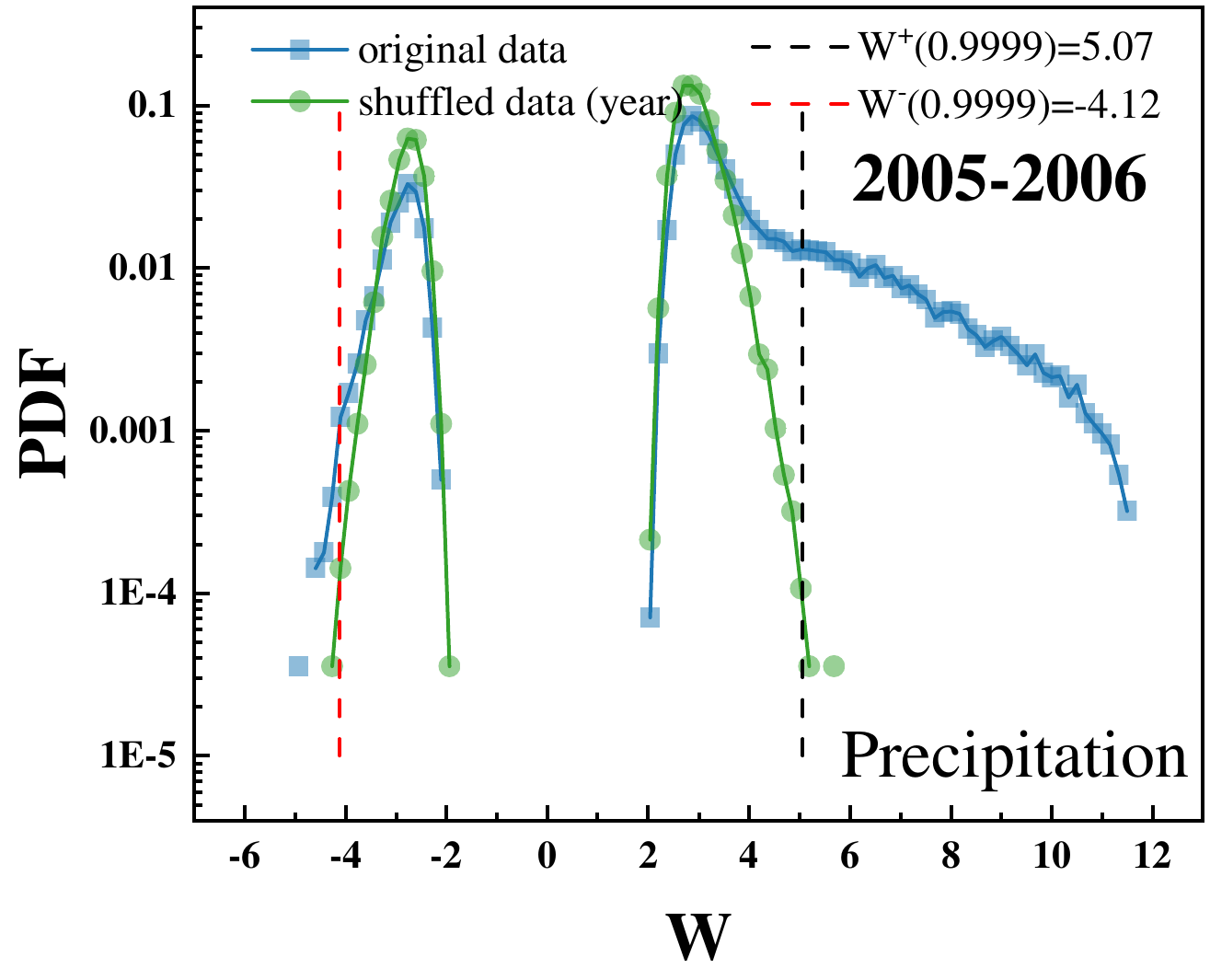}
\includegraphics[width=8.5em, height=7em]{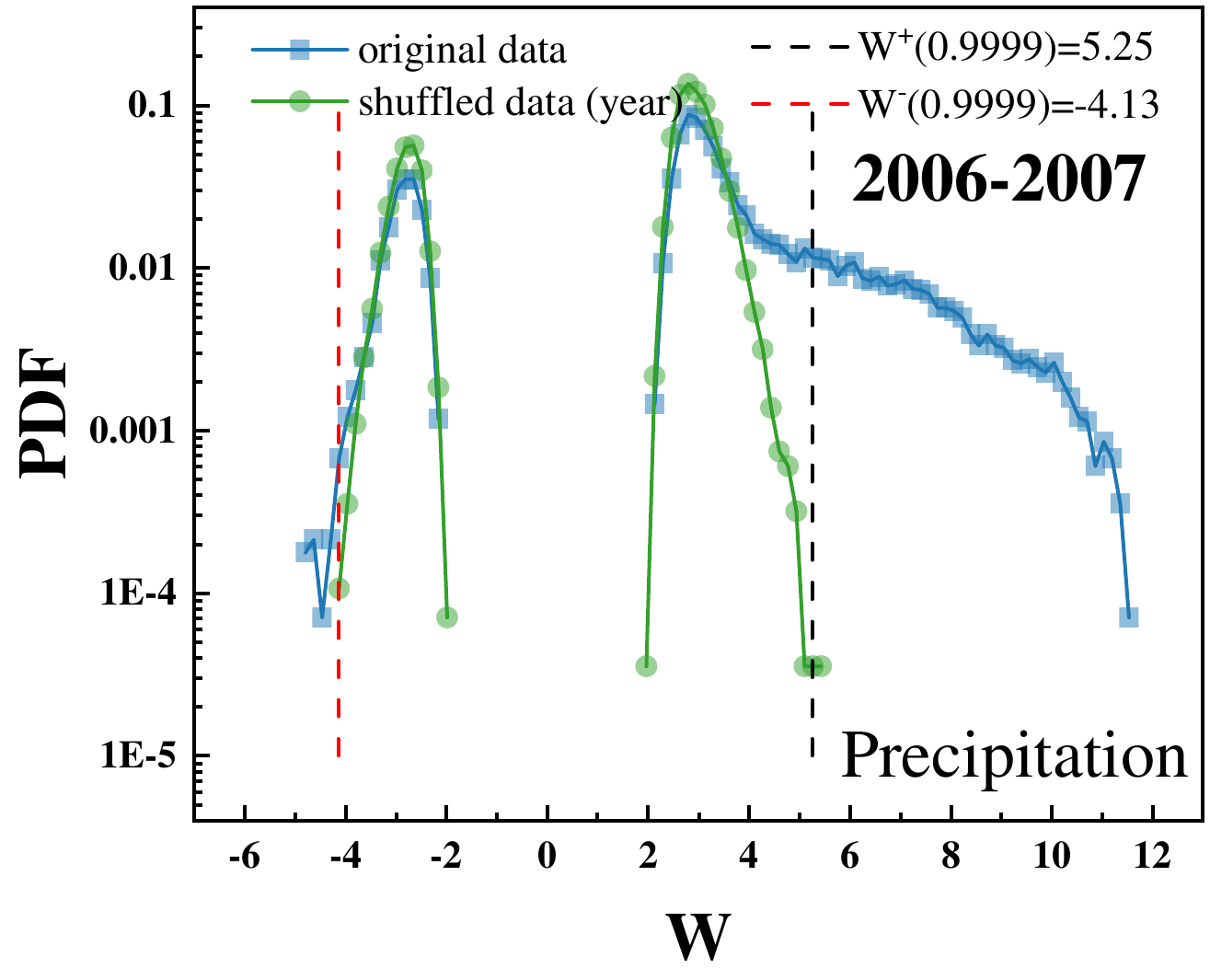}
\includegraphics[width=8.5em, height=7em]{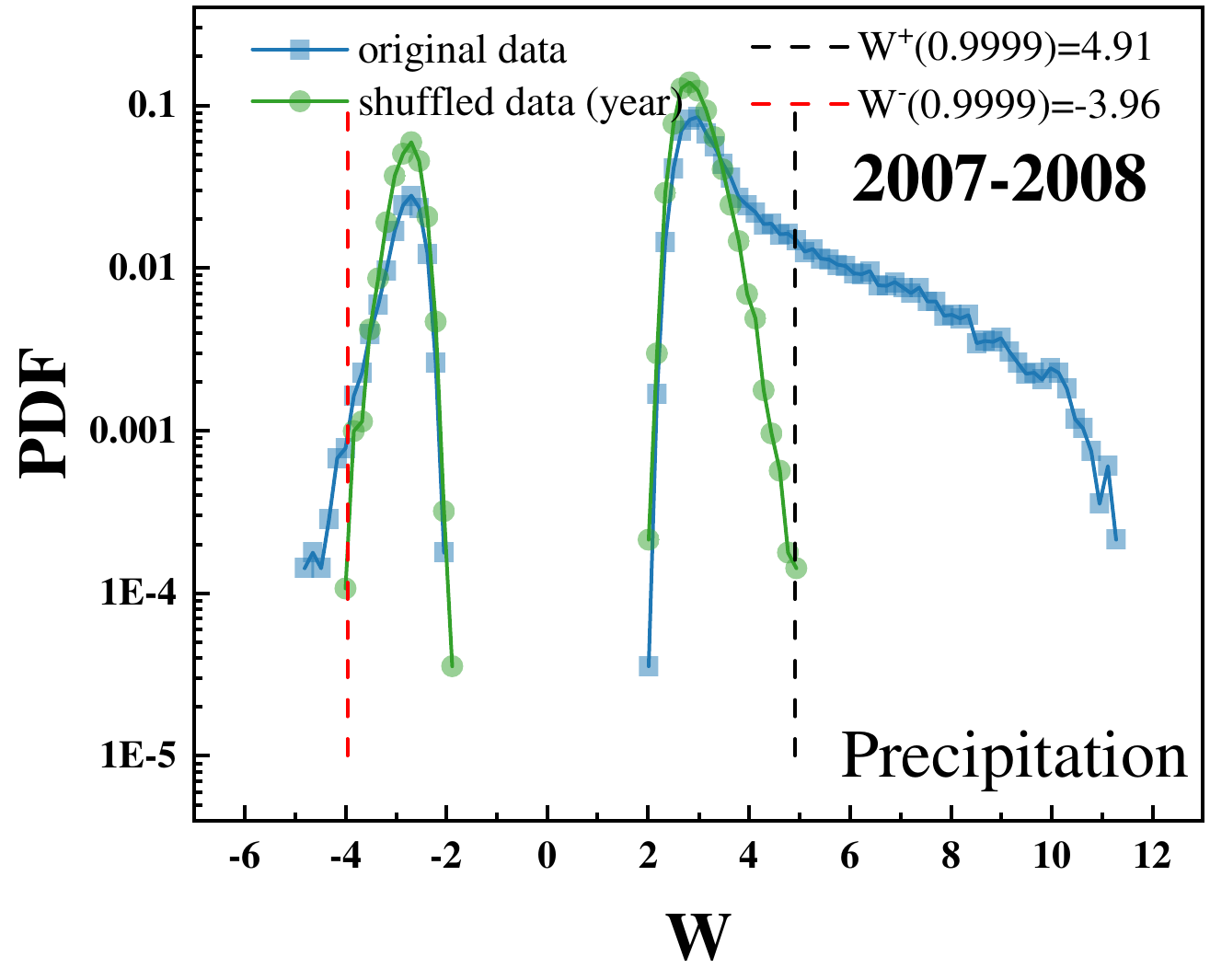}
\includegraphics[width=8.5em, height=7em]{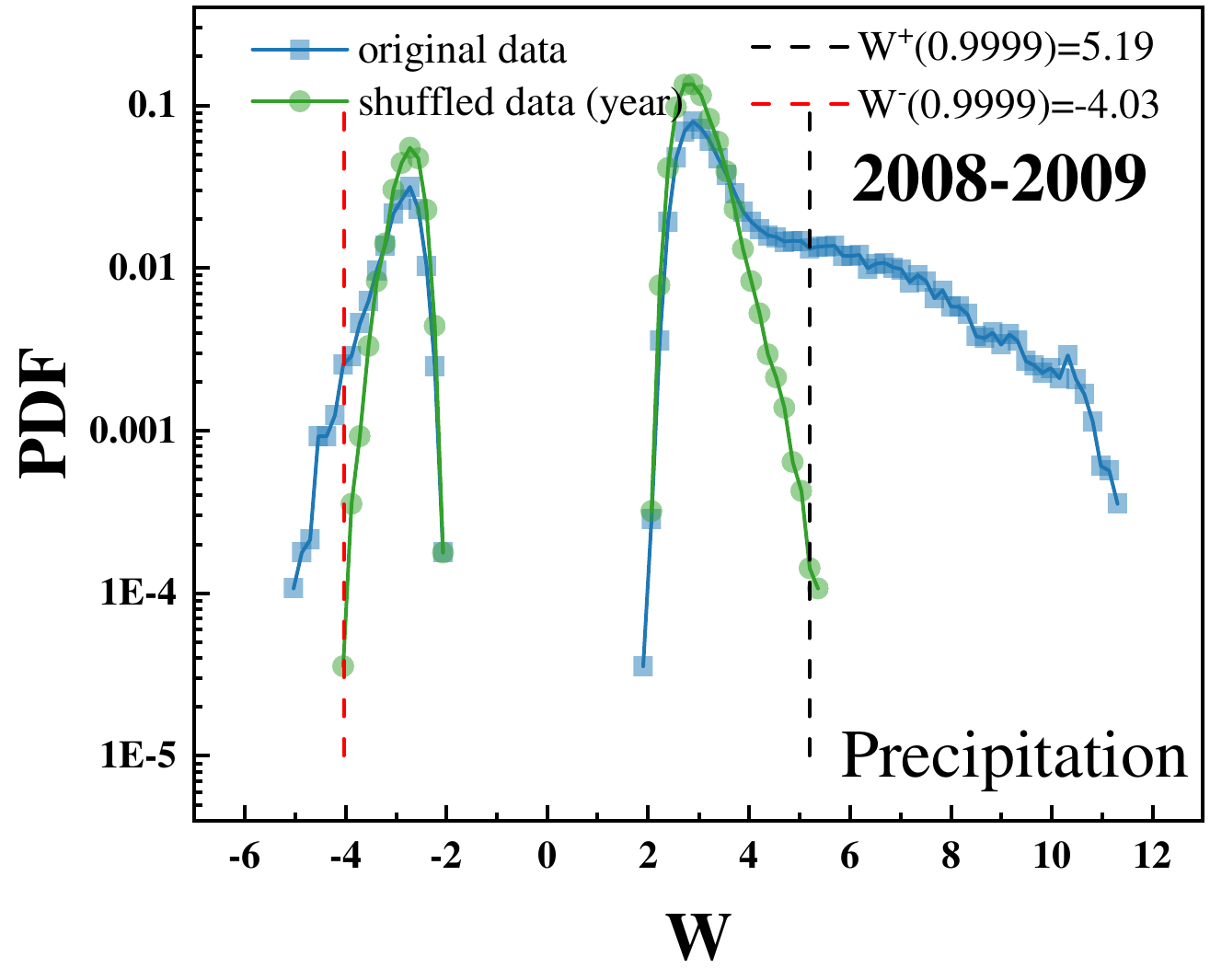}
\includegraphics[width=8.5em, height=7em]{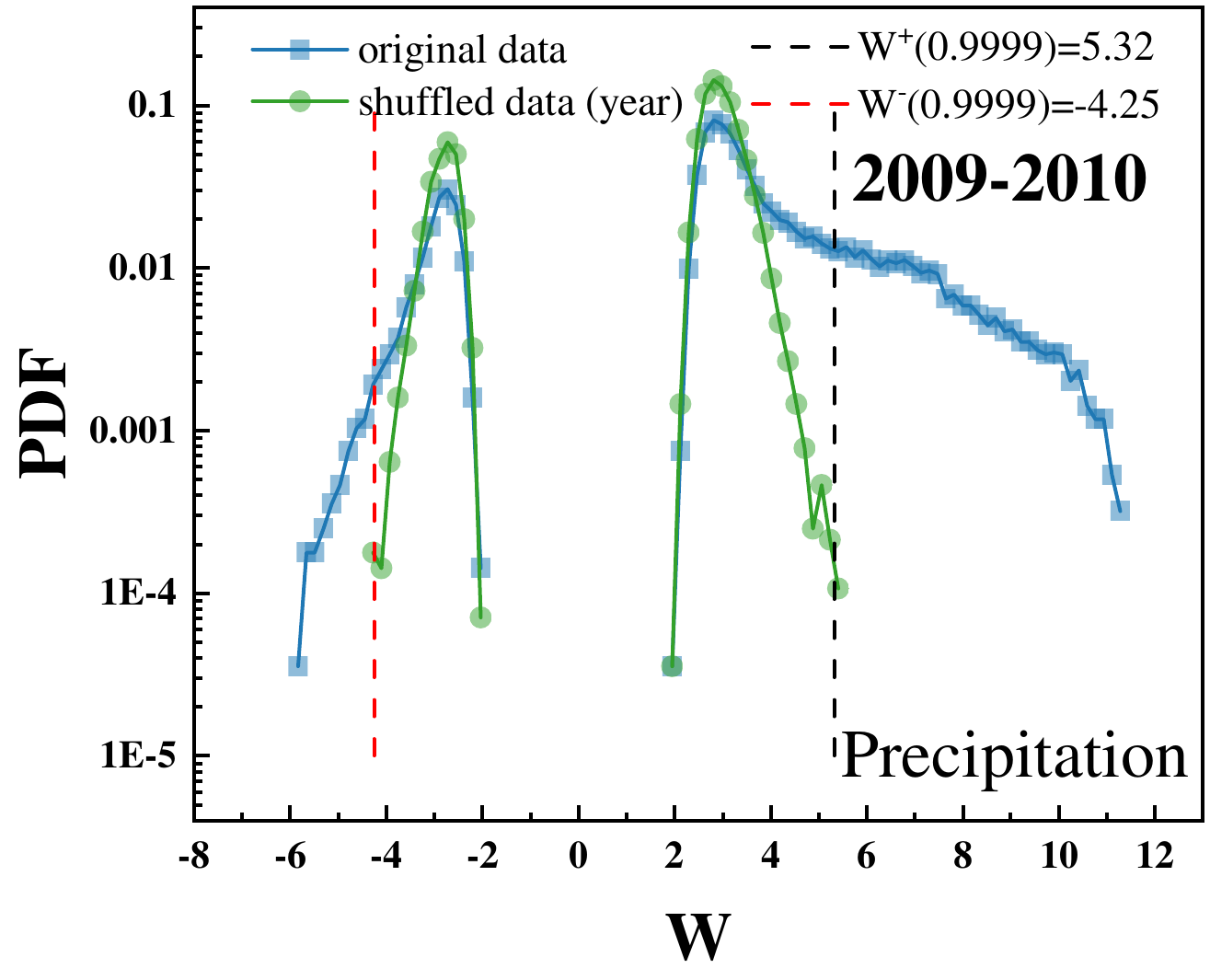}
\includegraphics[width=8.5em, height=7em]{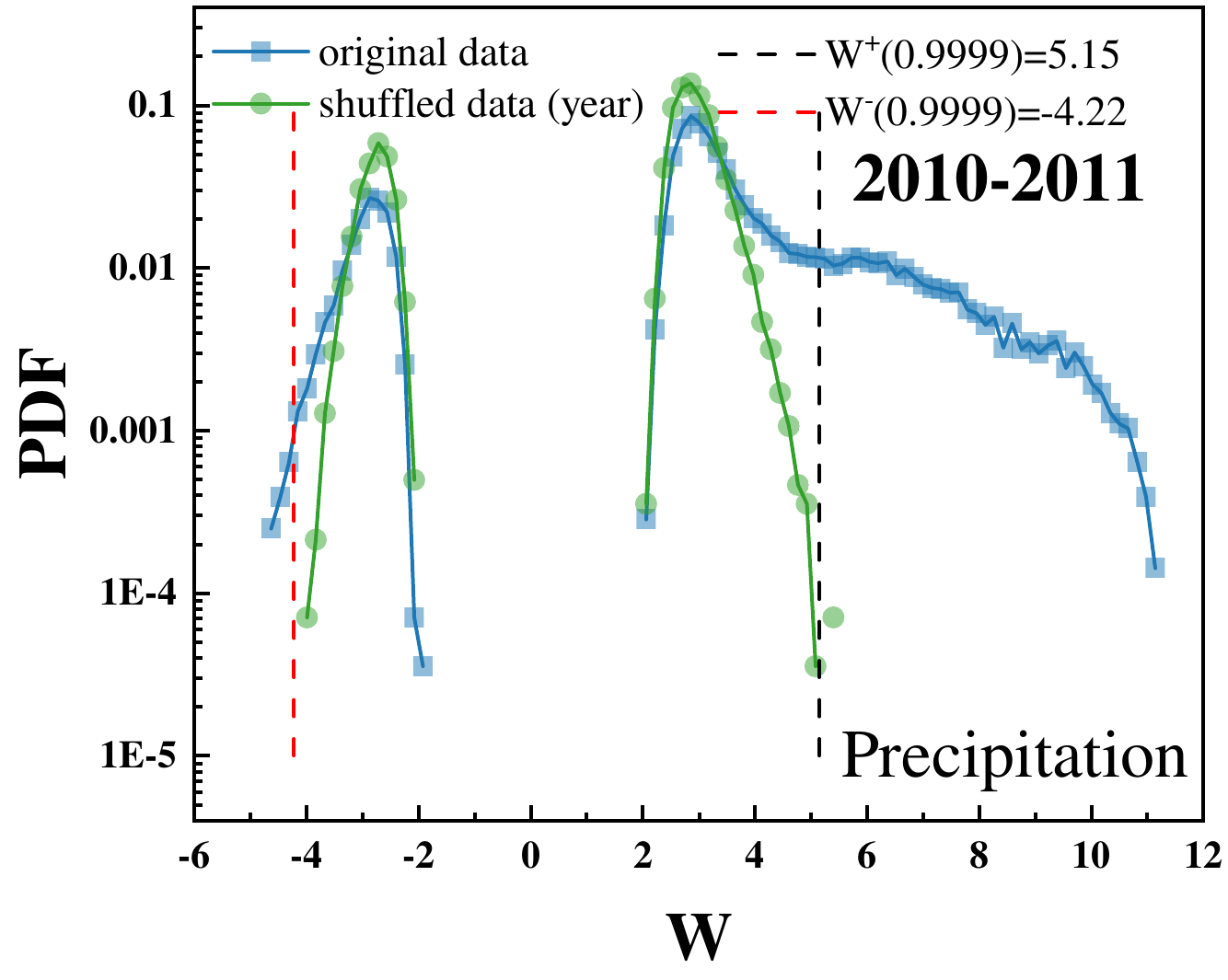}
\includegraphics[width=8.5em, height=7em]{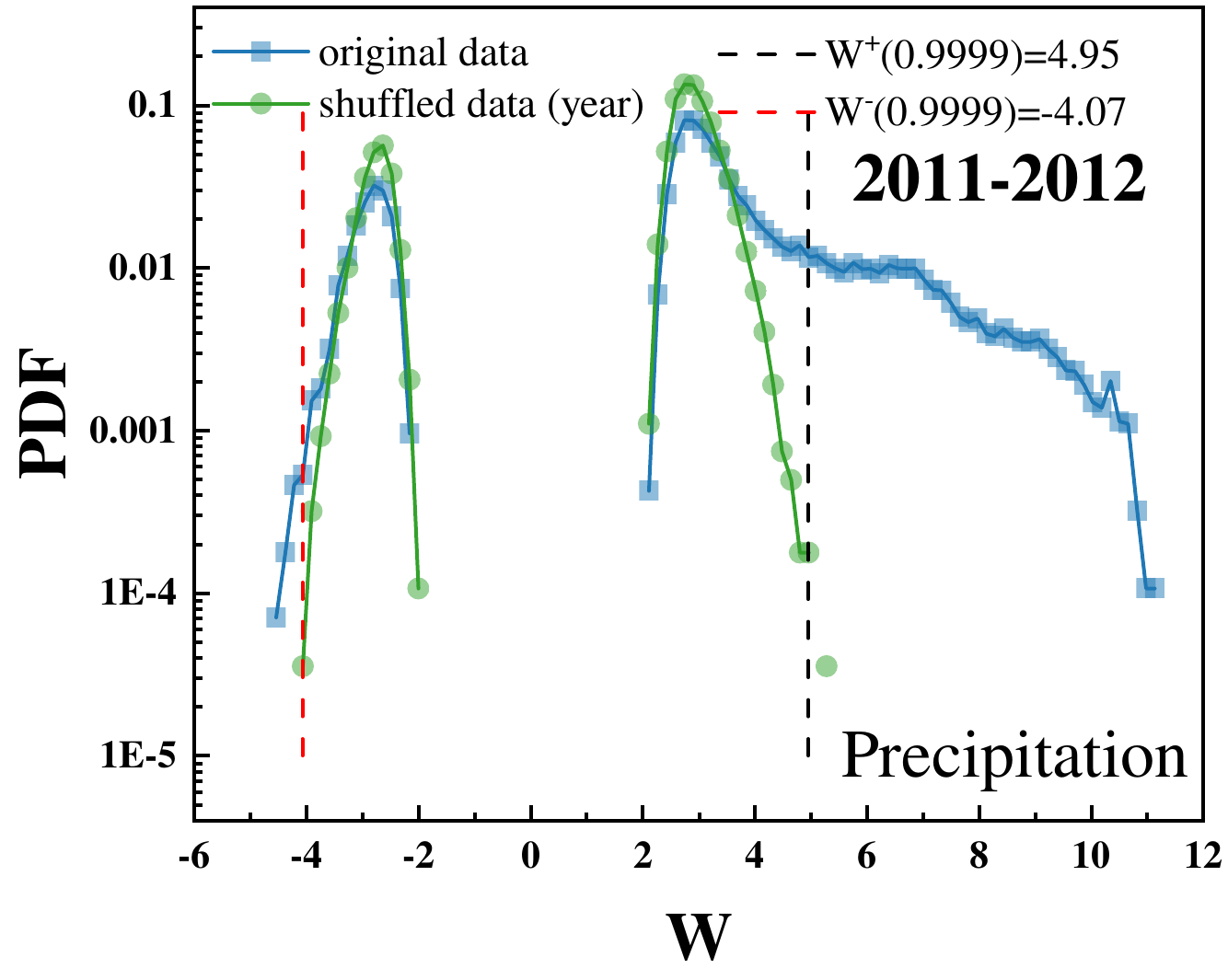}
\includegraphics[width=8.5em, height=7em]{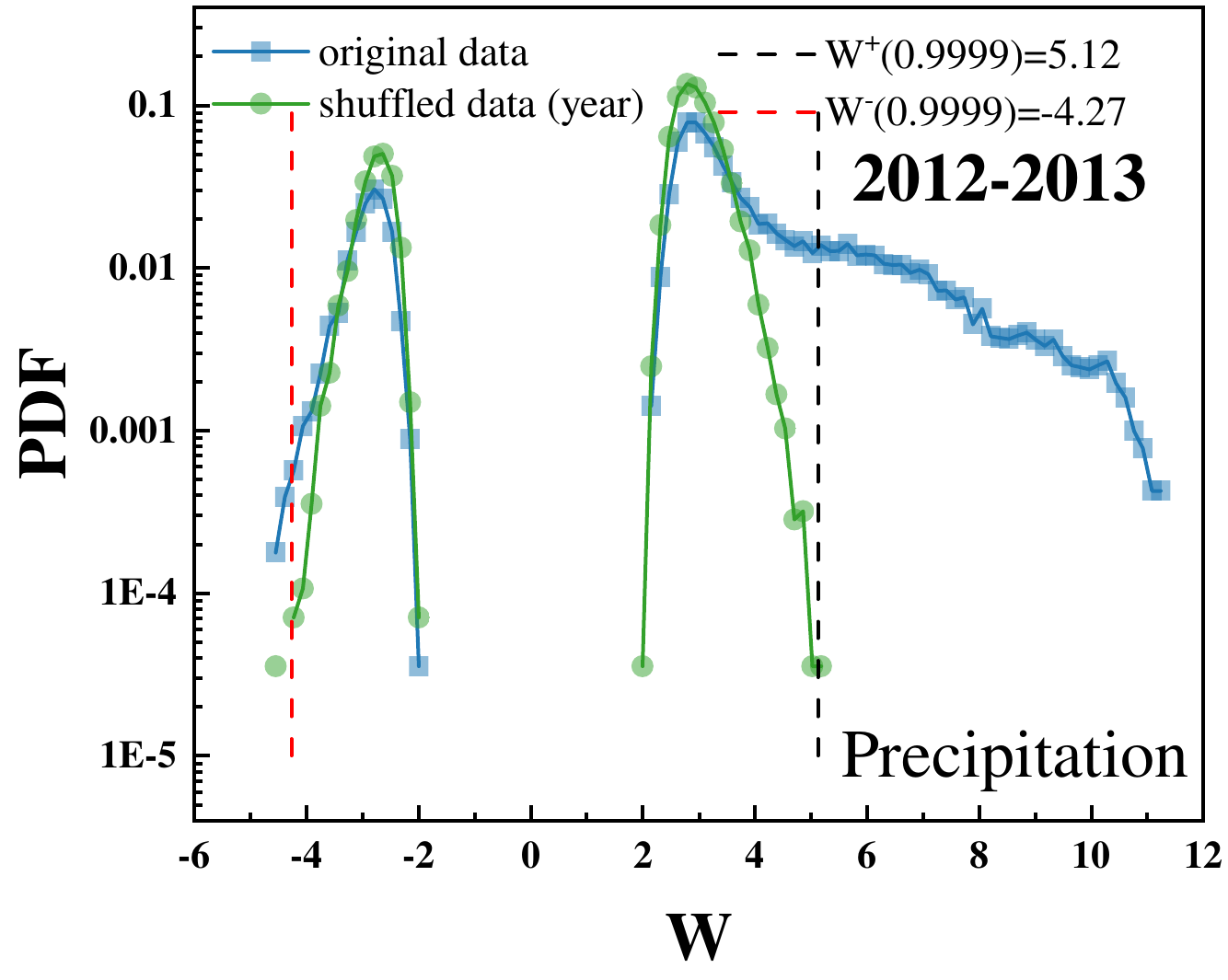}
\includegraphics[width=8.5em, height=7em]{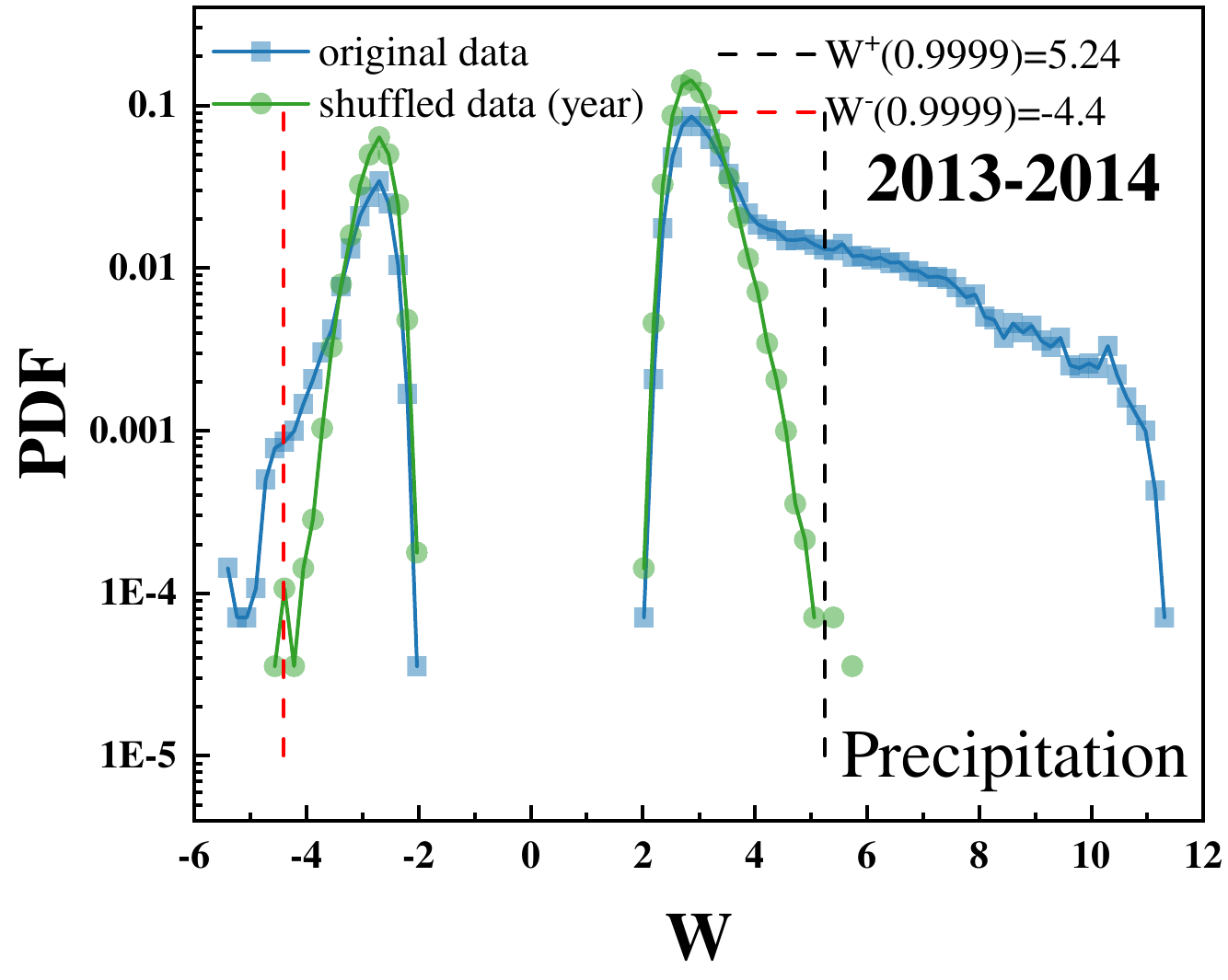}
\includegraphics[width=8.5em, height=7em]{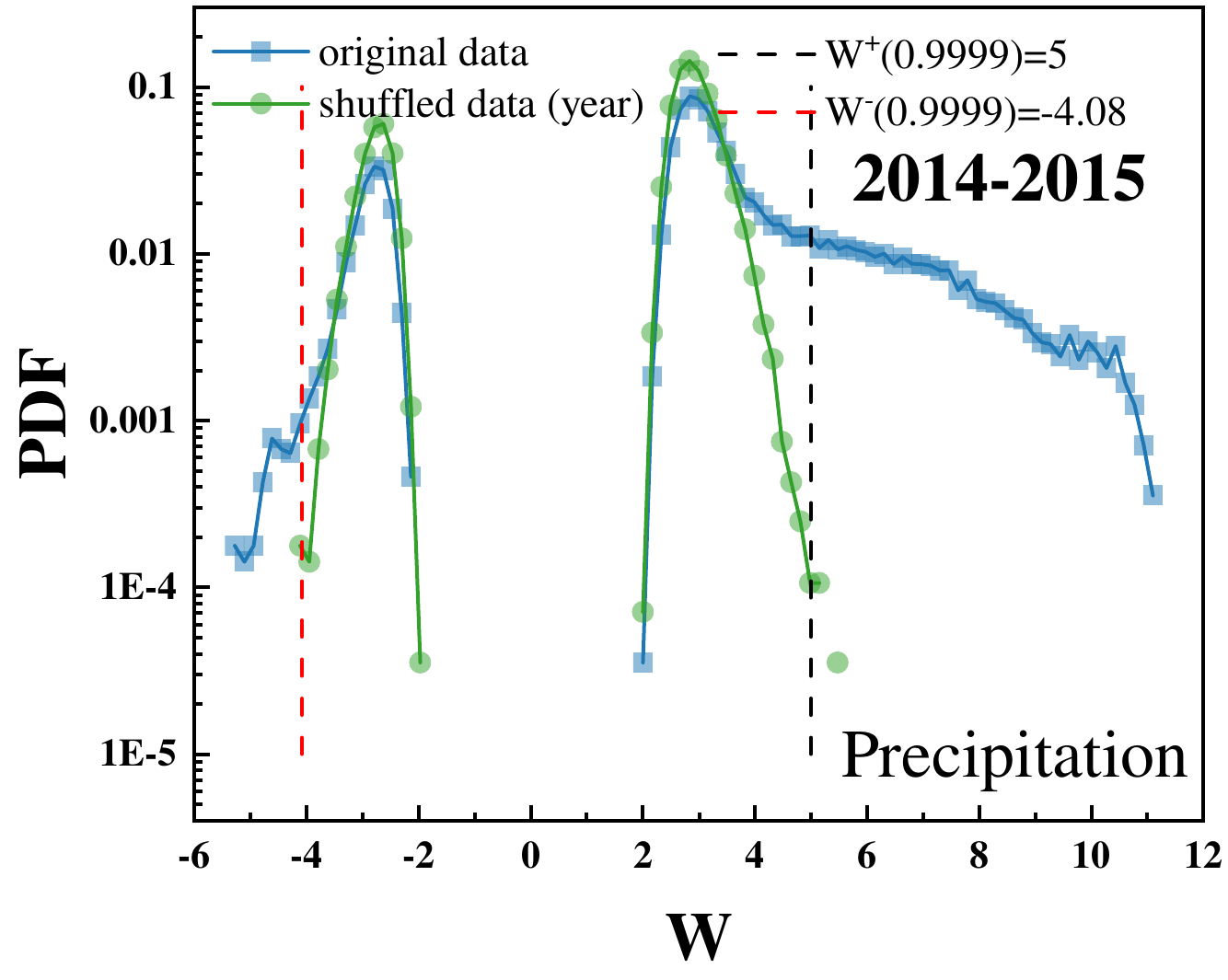}
\includegraphics[width=8.5em, height=7em]{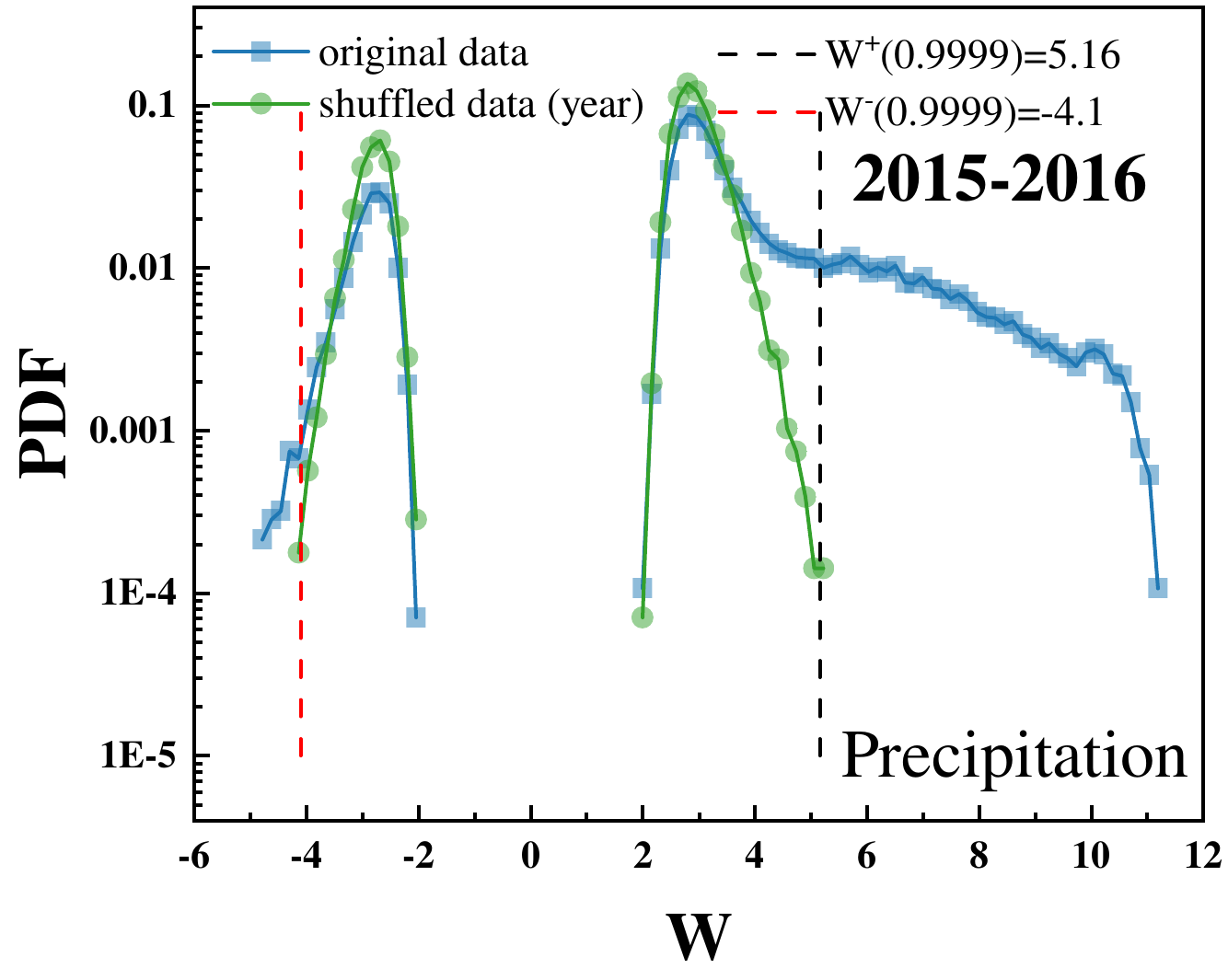}
\includegraphics[width=8.5em, height=7em]{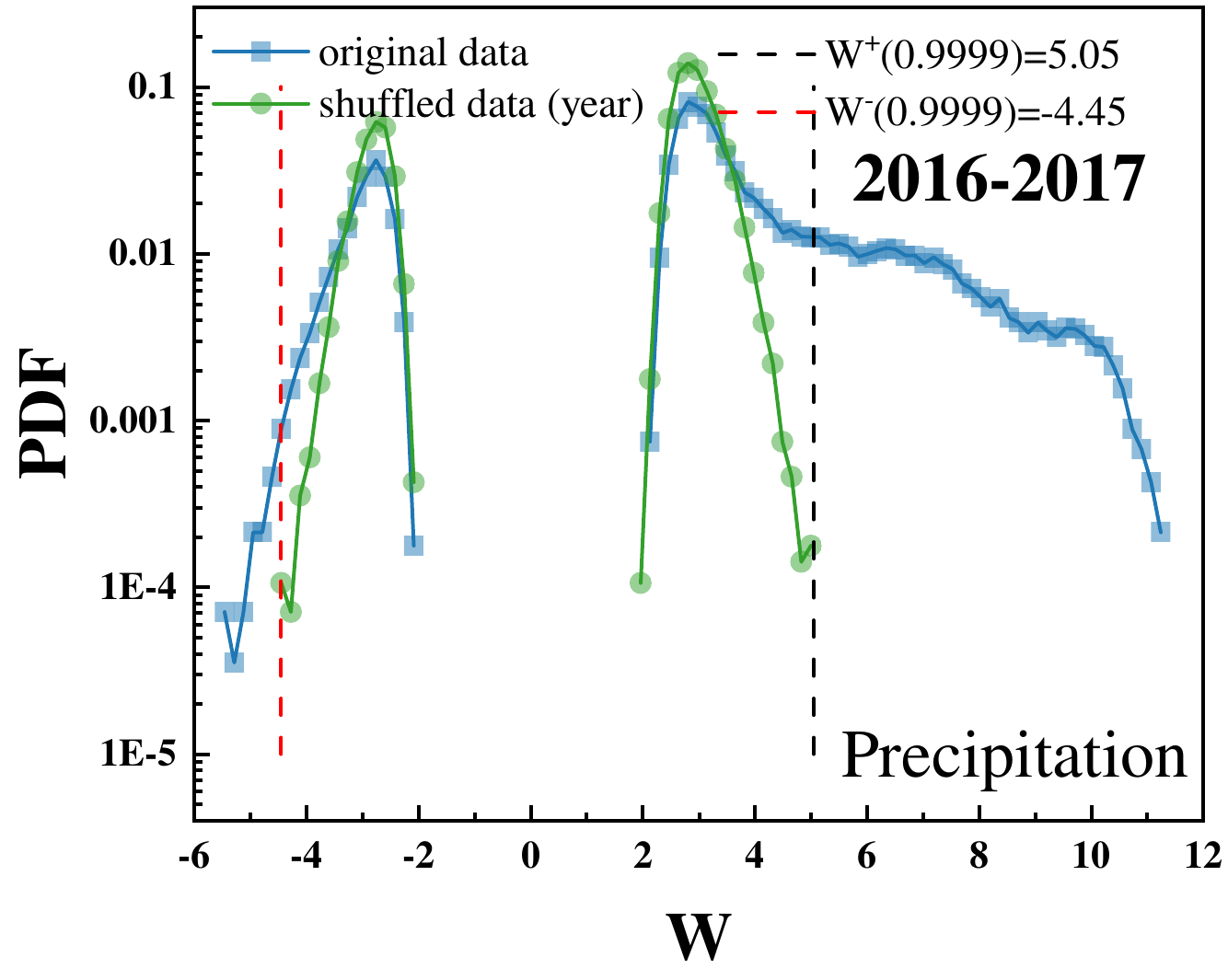}
\includegraphics[width=8.5em, height=7em]{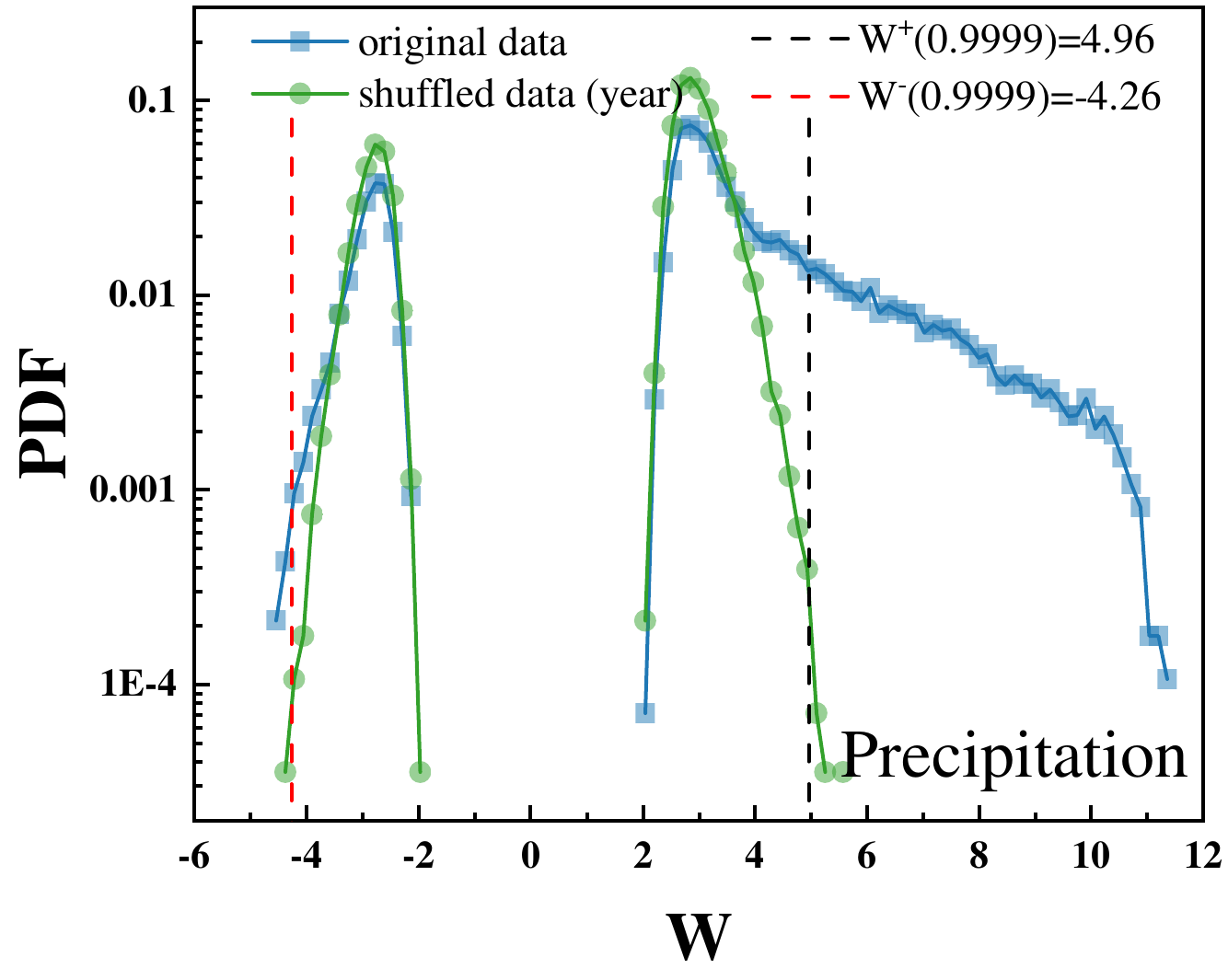}
\includegraphics[width=8.5em, height=7em]{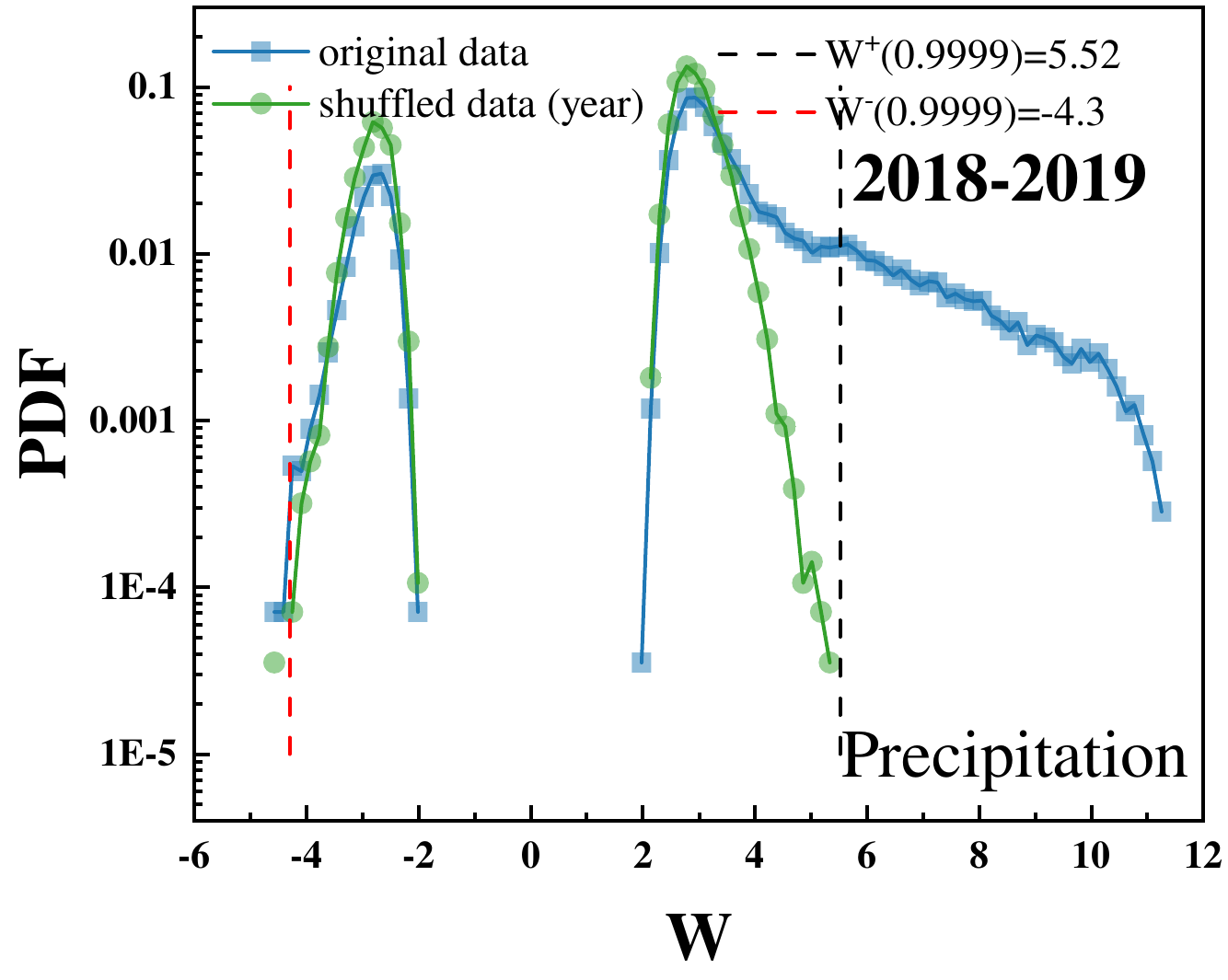}
\end{center}

\begin{center}
\noindent {\small {\bf Fig. S6} Probability distribution function (PDF) of link weights for the original data and shuffled data of precipitation in China. }
\end{center}

\begin{center}
\includegraphics[width=8.5em, height=7em]{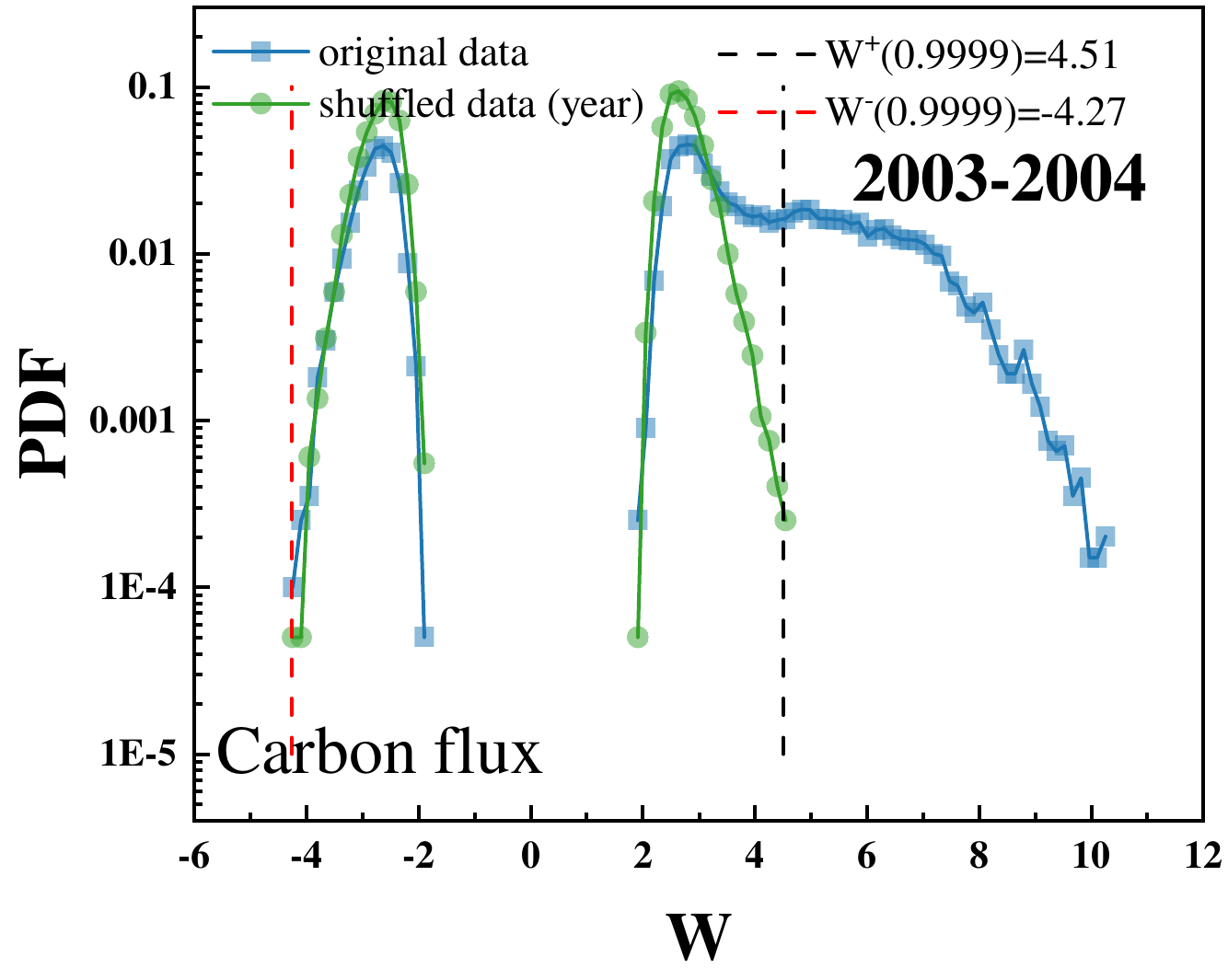}
\includegraphics[width=8.5em, height=7em]{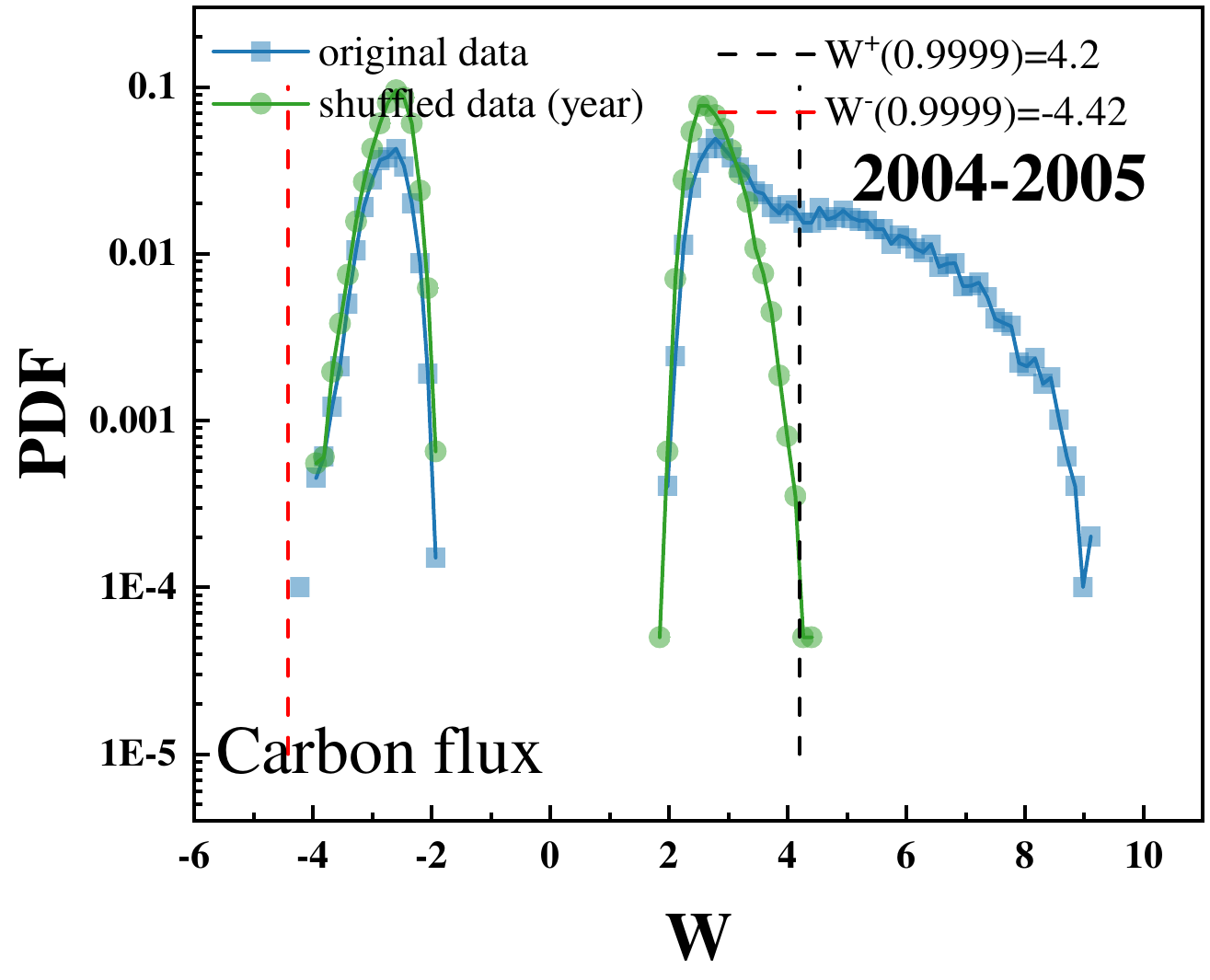}
\includegraphics[width=8.5em, height=7em]{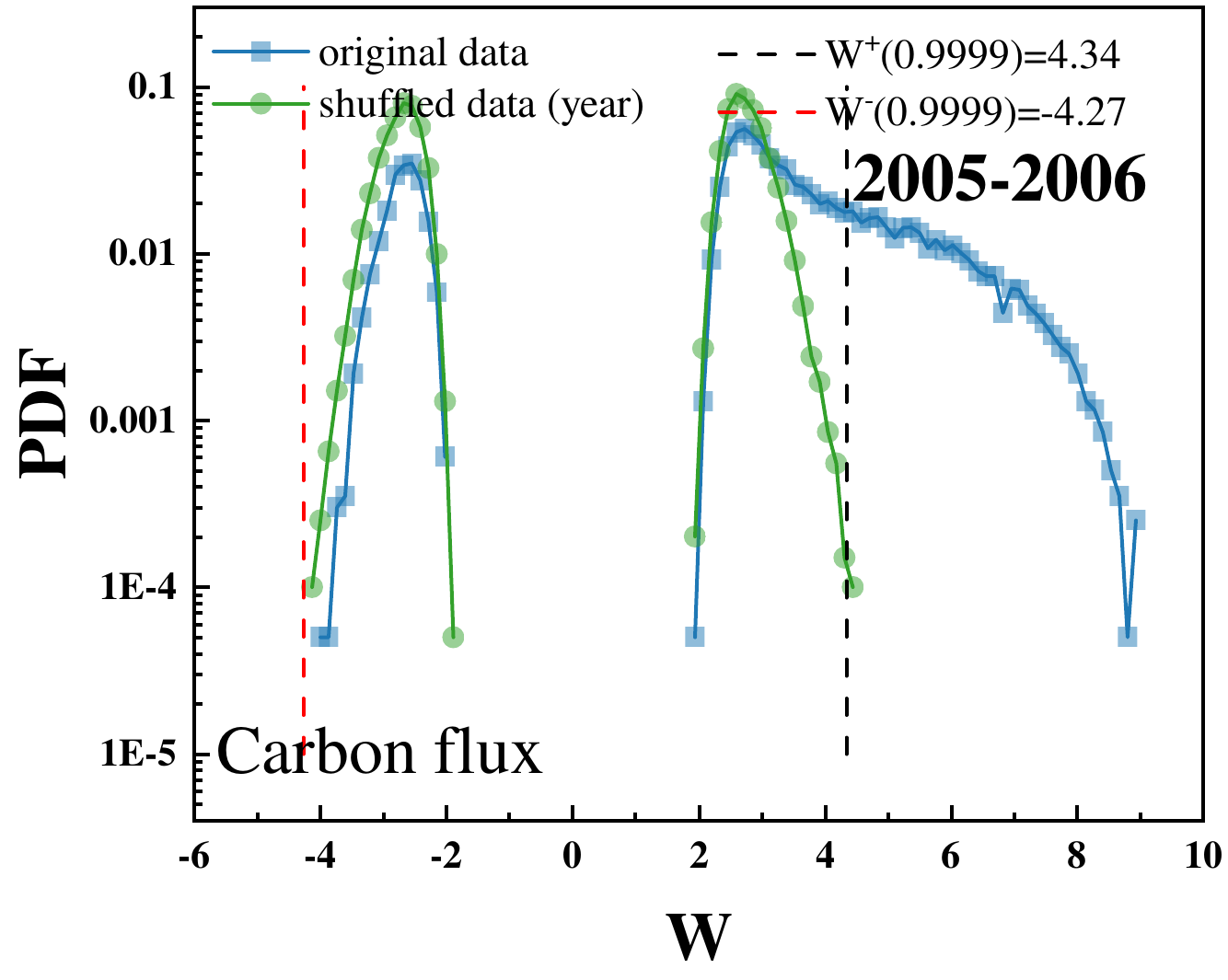}
\includegraphics[width=8.5em, height=7em]{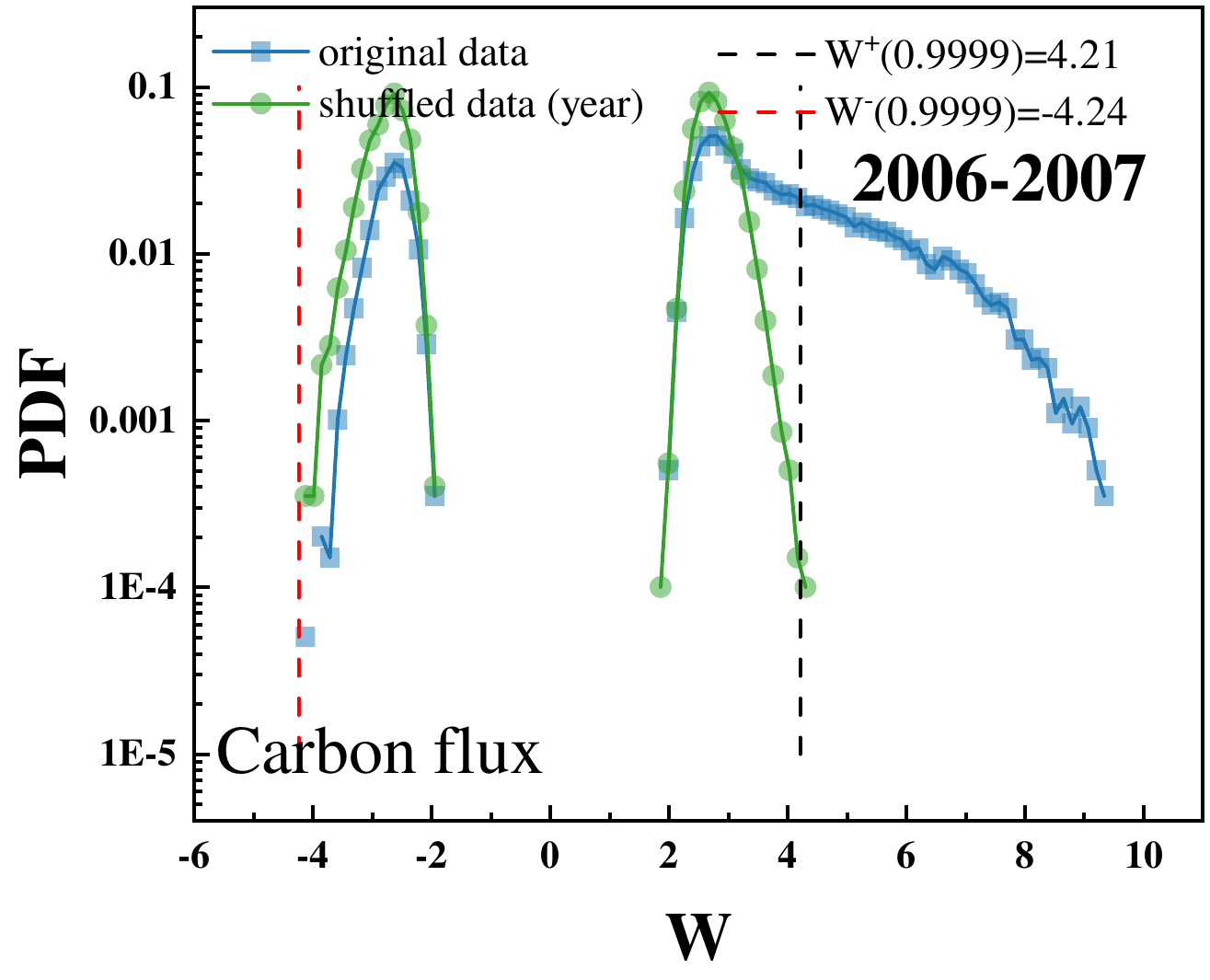}
\includegraphics[width=8.5em, height=7em]{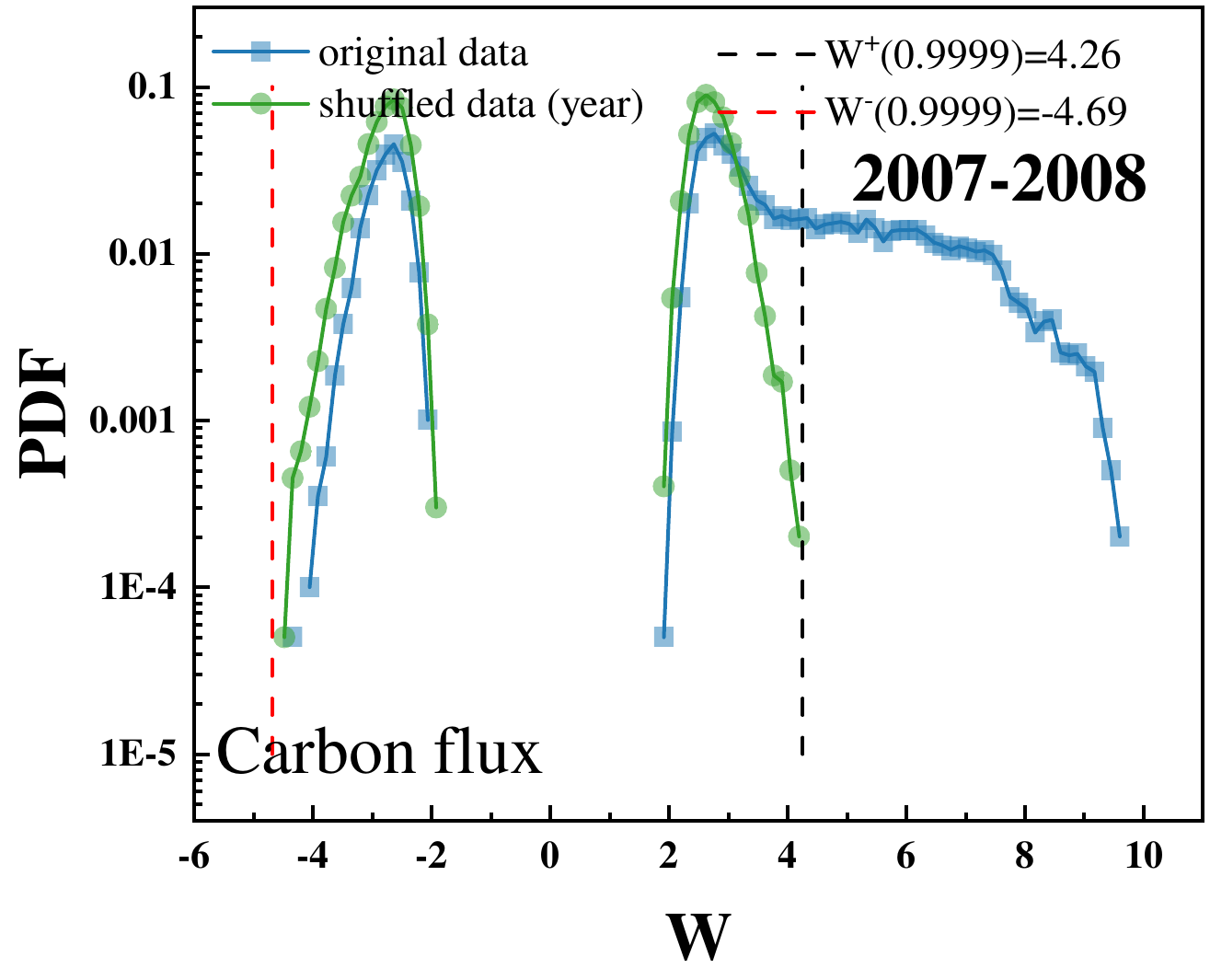}
\includegraphics[width=8.5em, height=7em]{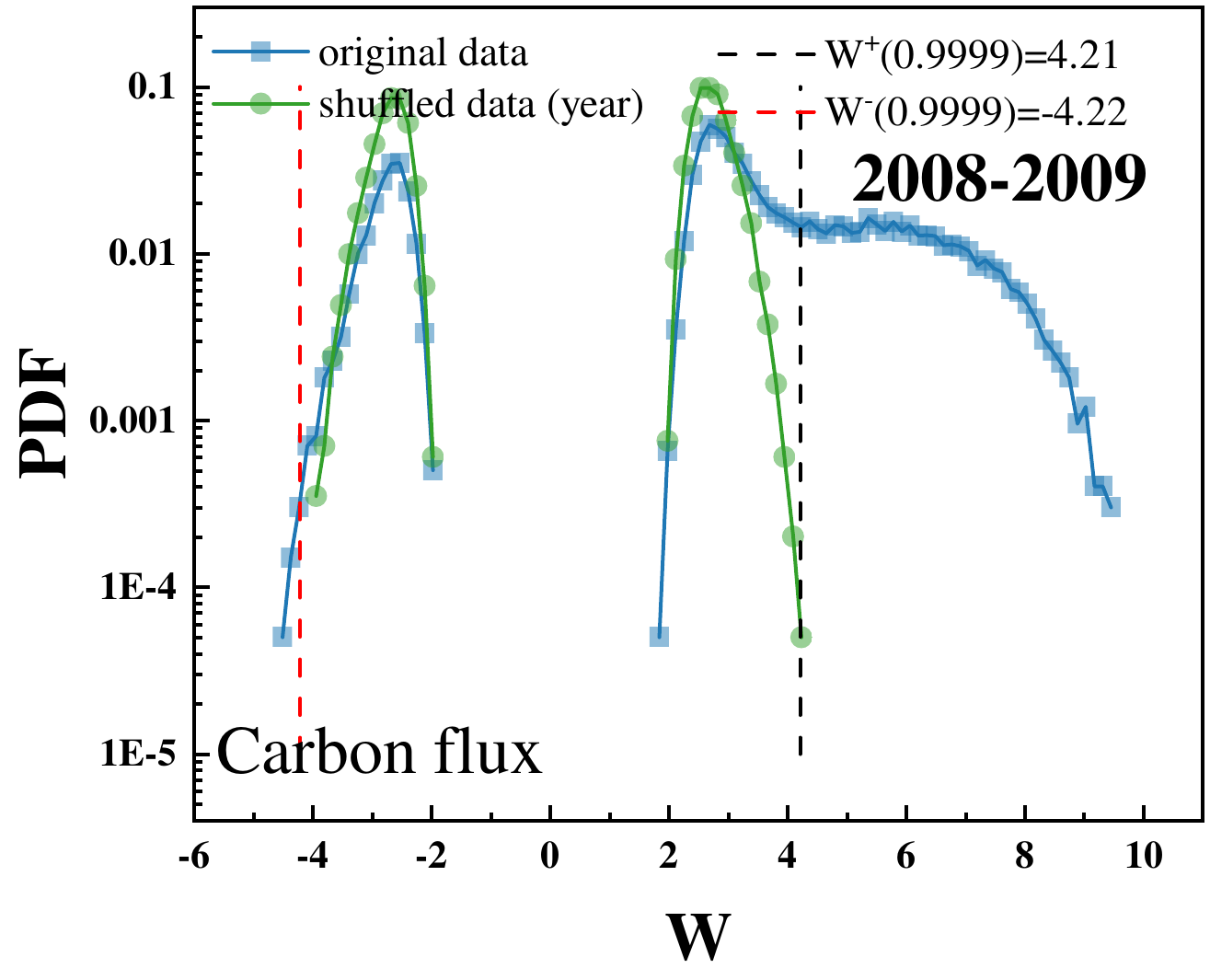}
\includegraphics[width=8.5em, height=7em]{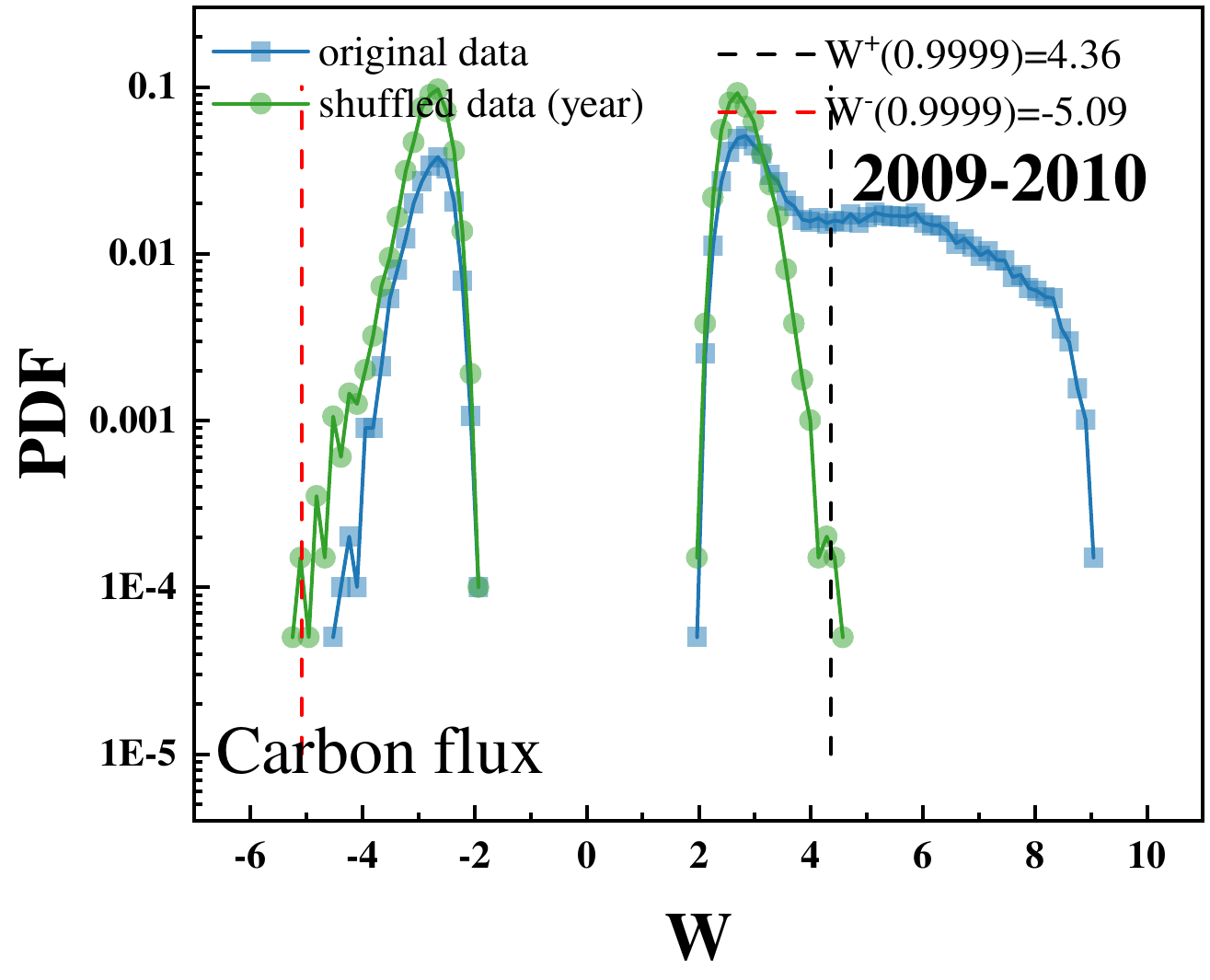}
\includegraphics[width=8.5em, height=7em]{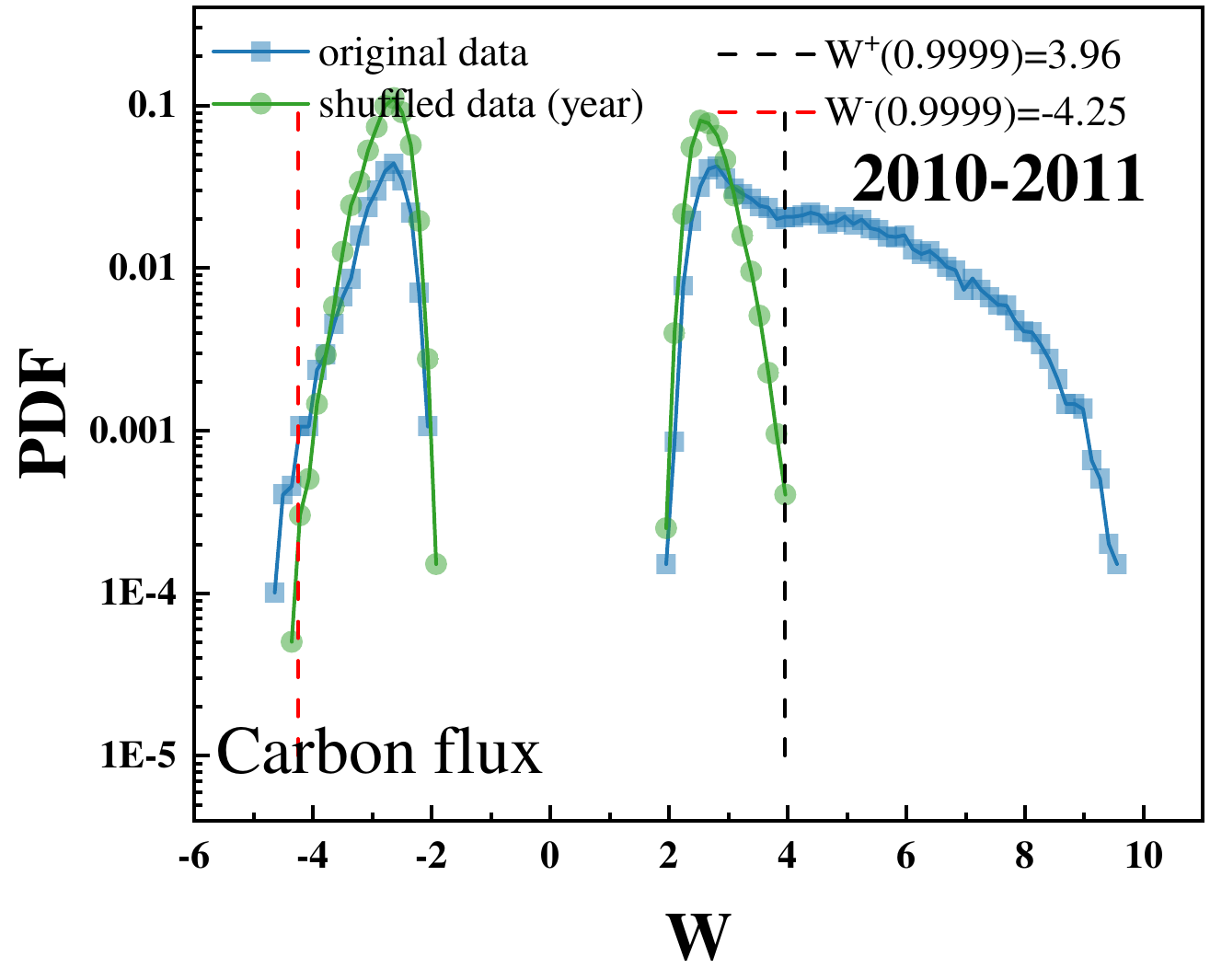}
\includegraphics[width=8.5em, height=7em]{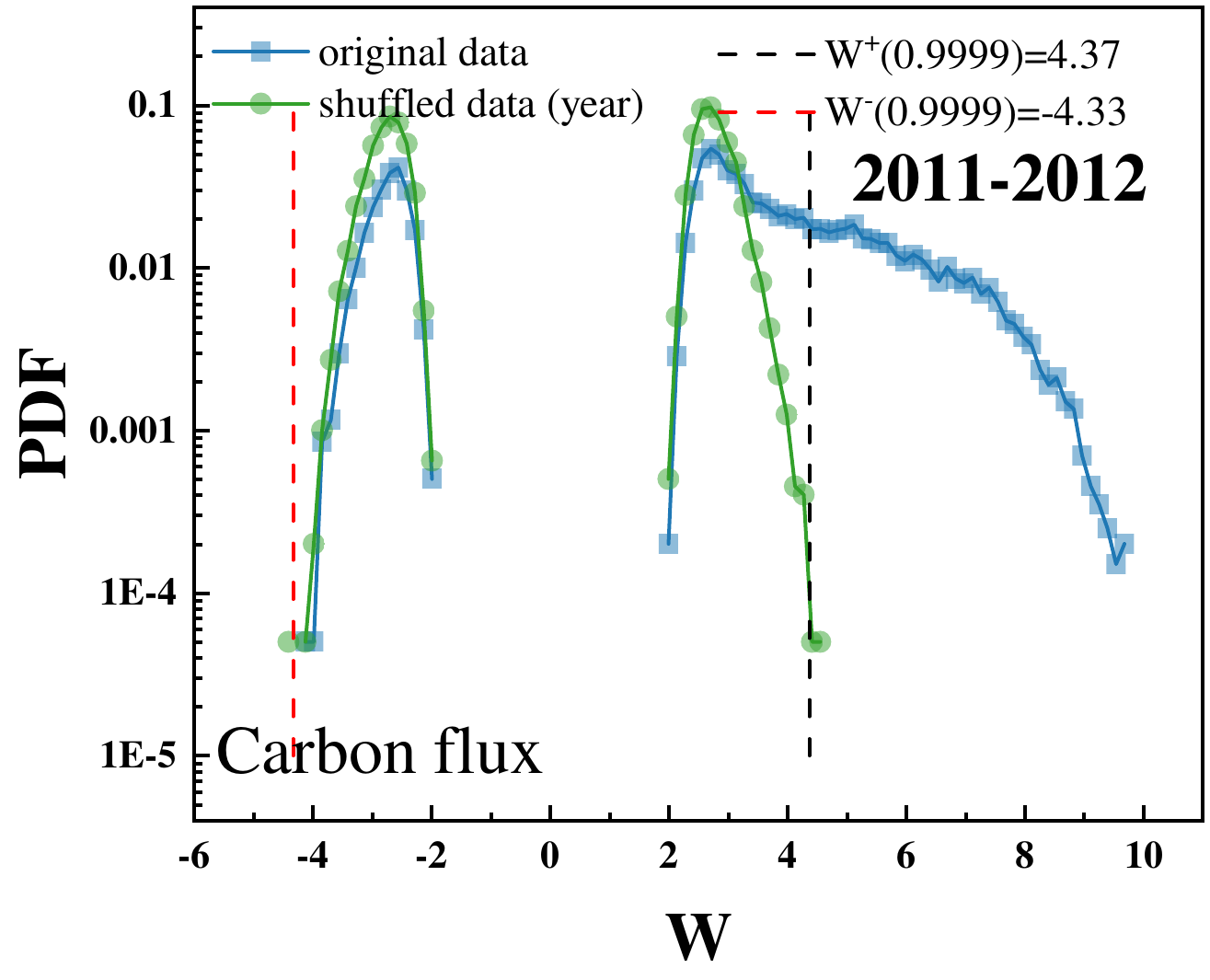}
\includegraphics[width=8.5em, height=7em]{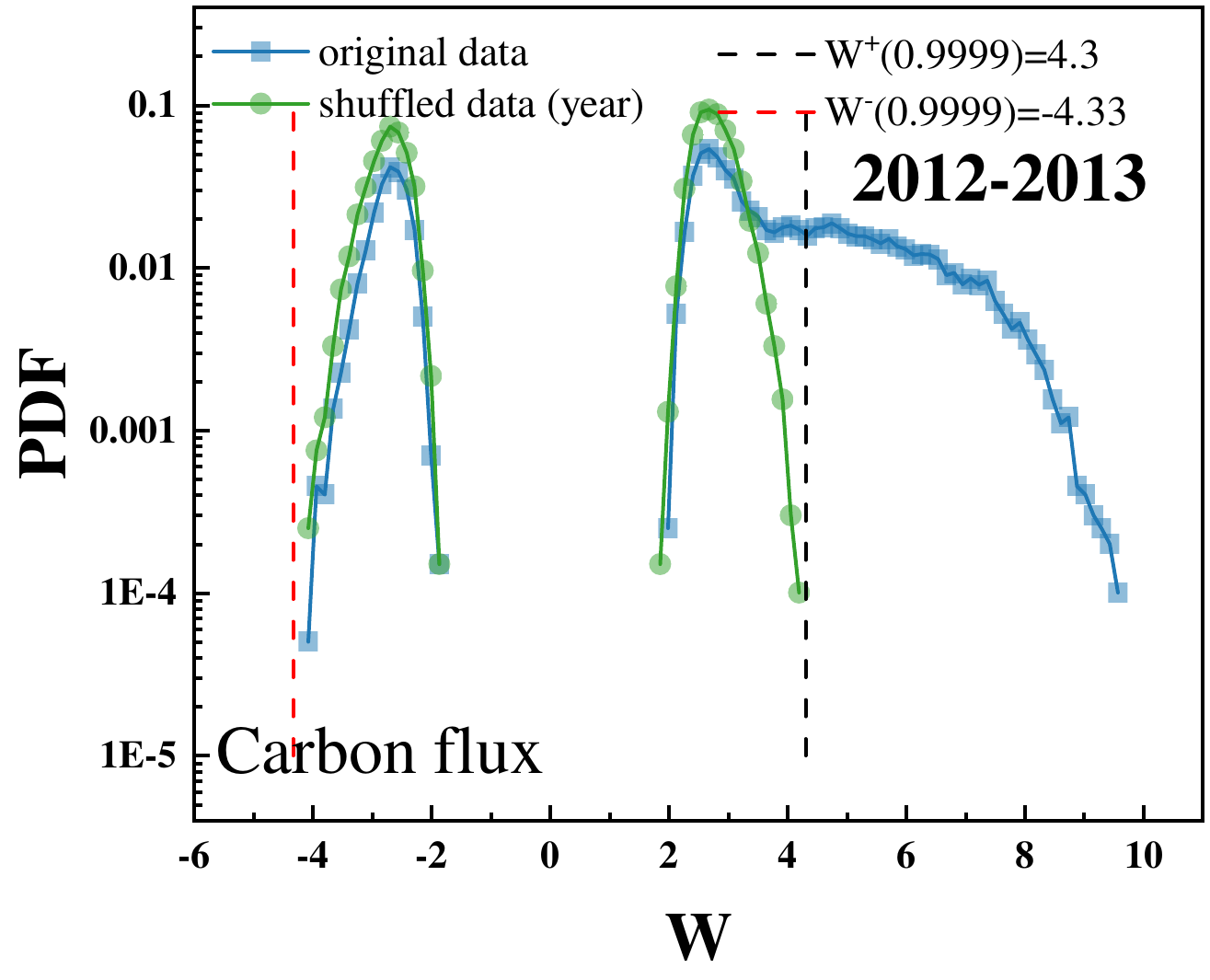}
\includegraphics[width=8.5em, height=7em]{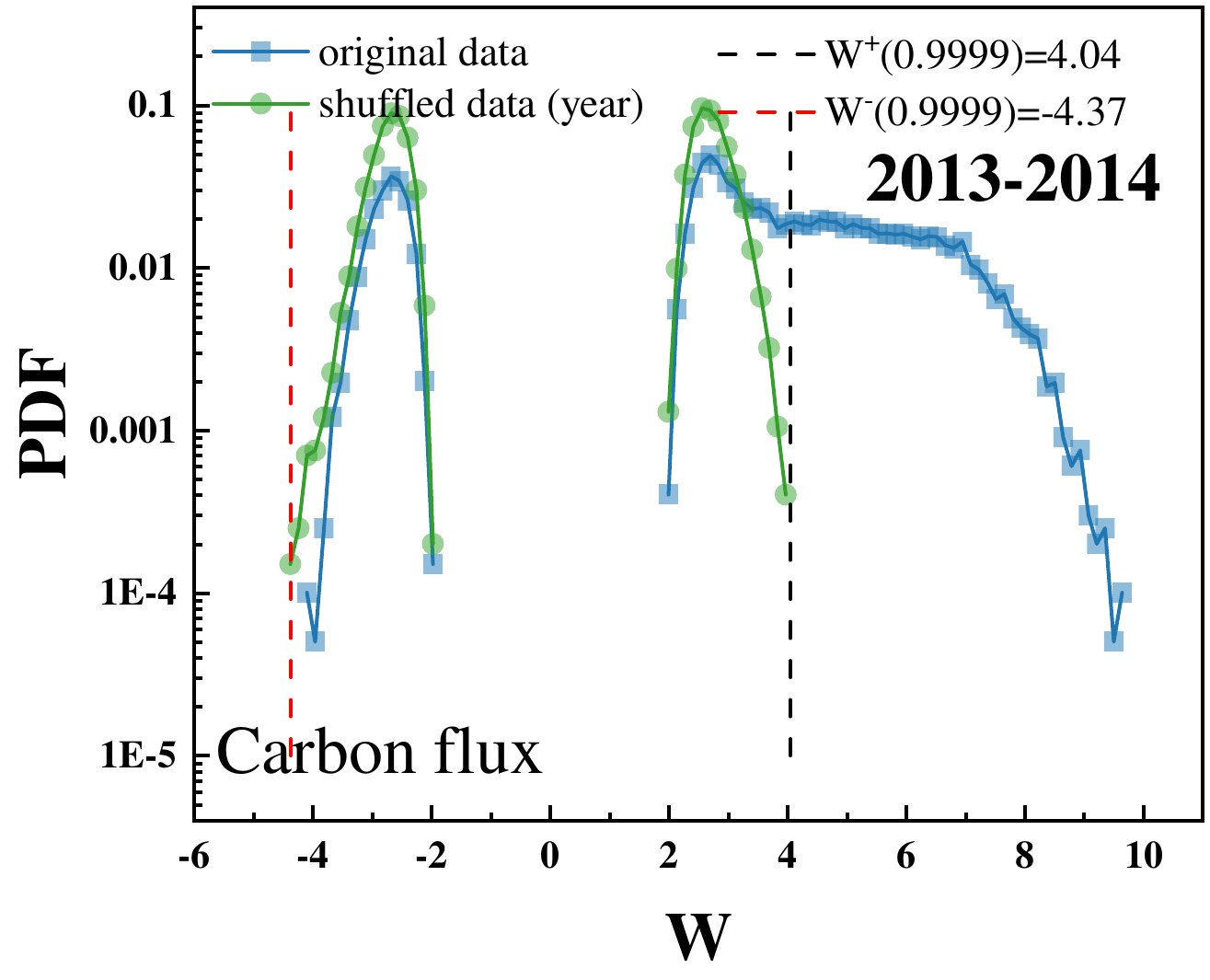}
\includegraphics[width=8.5em, height=7em]{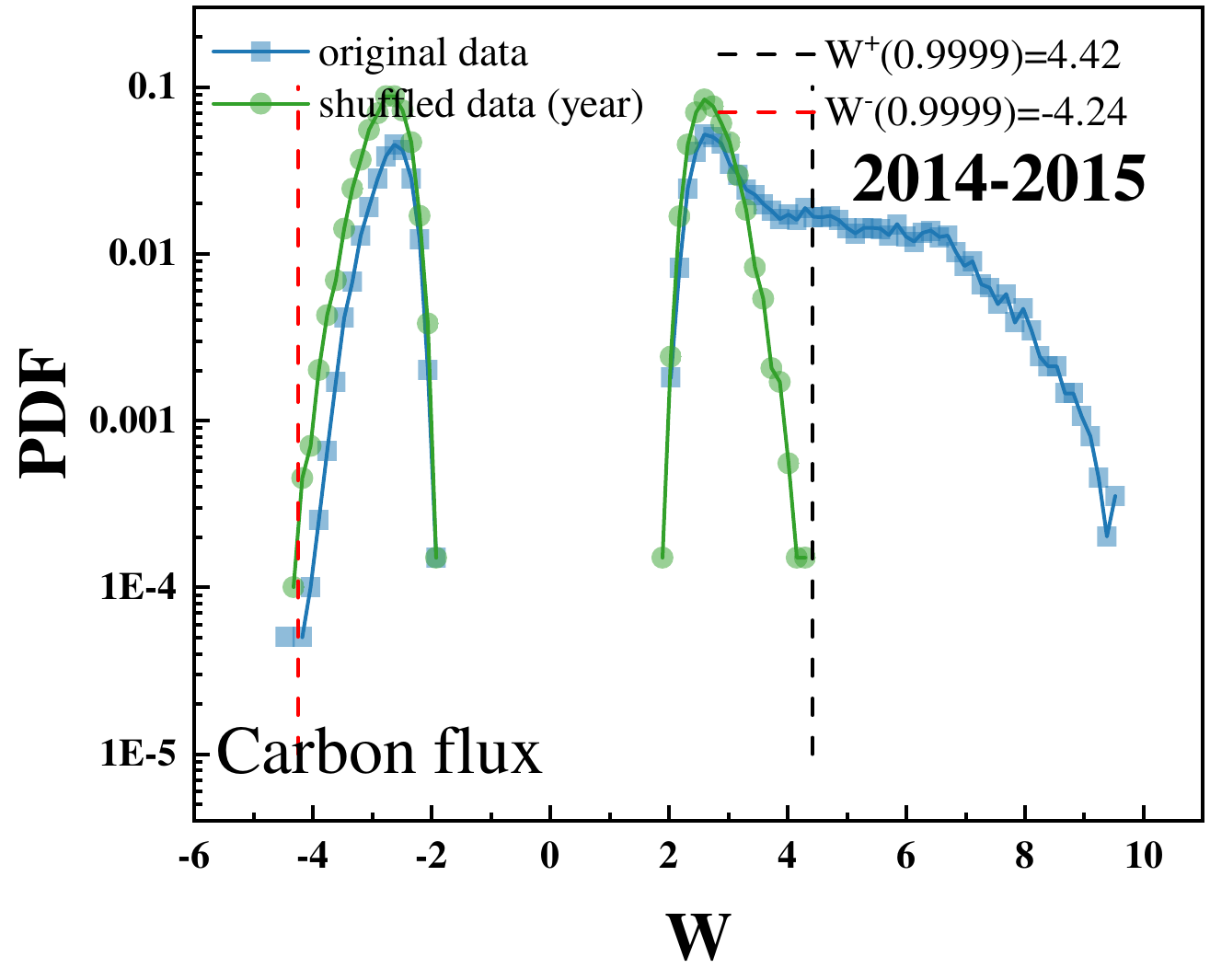}
\includegraphics[width=8.5em, height=7em]{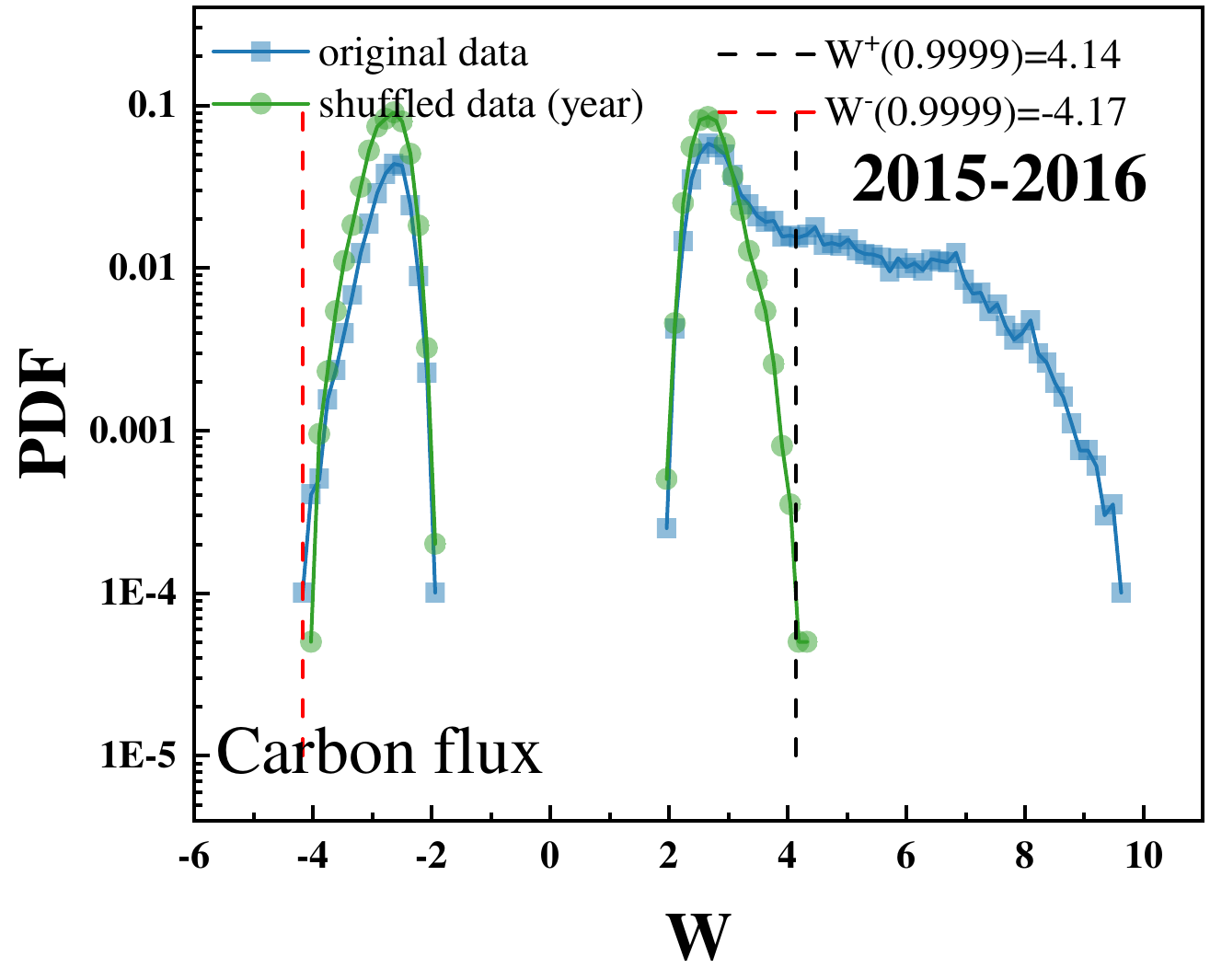}
\includegraphics[width=8.5em, height=7em]{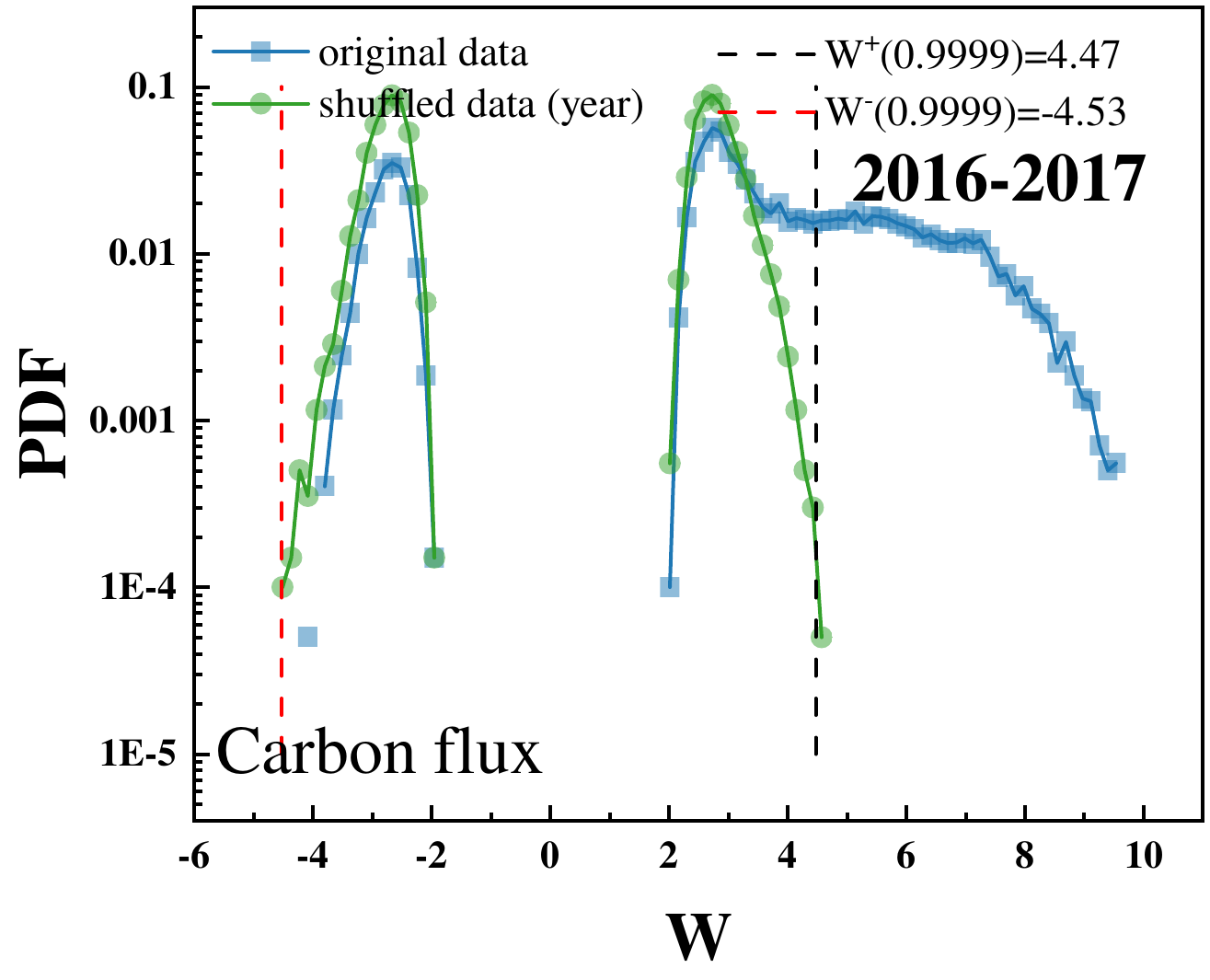}
\includegraphics[width=8.5em, height=7em]{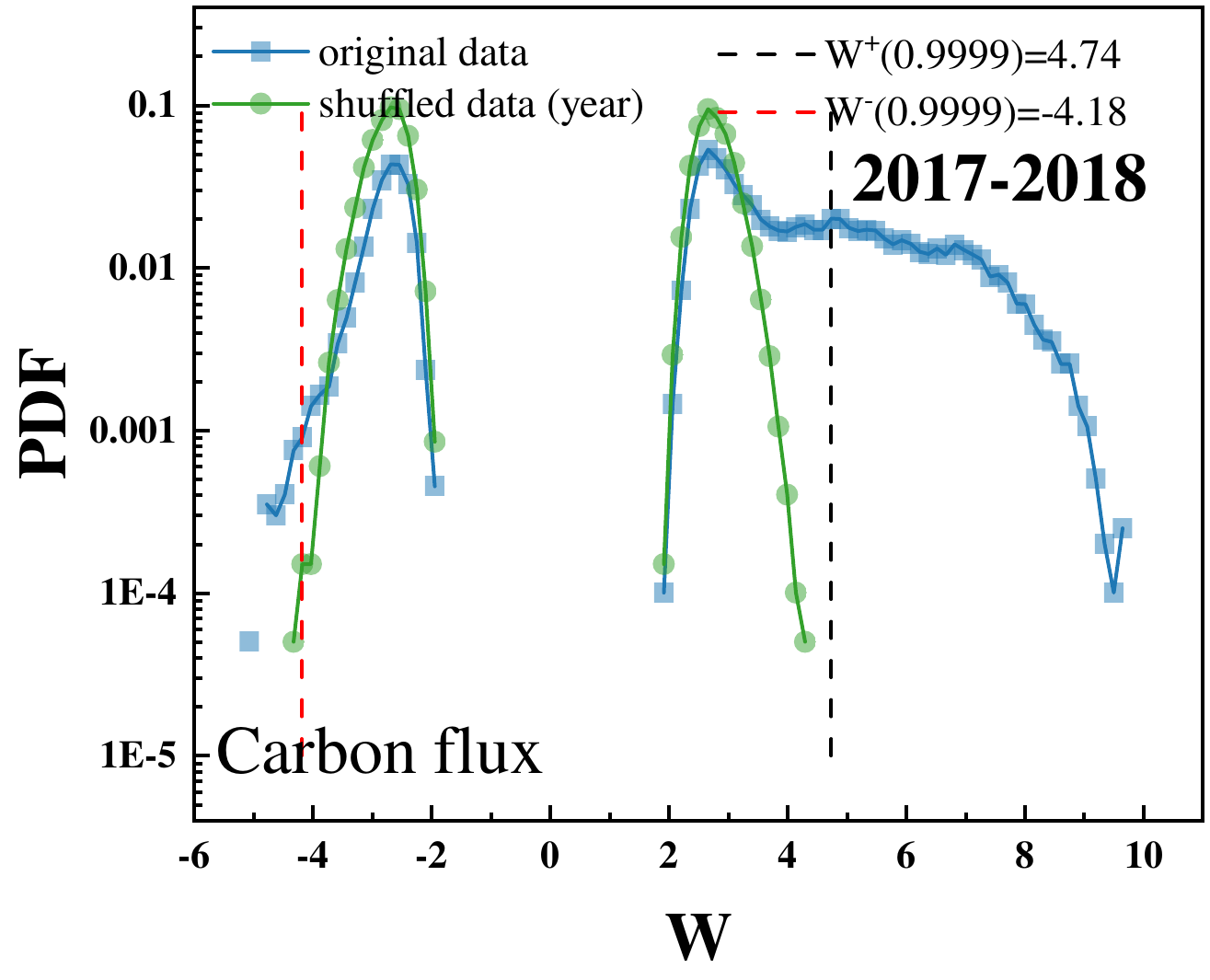}
\includegraphics[width=8.5em, height=7em]{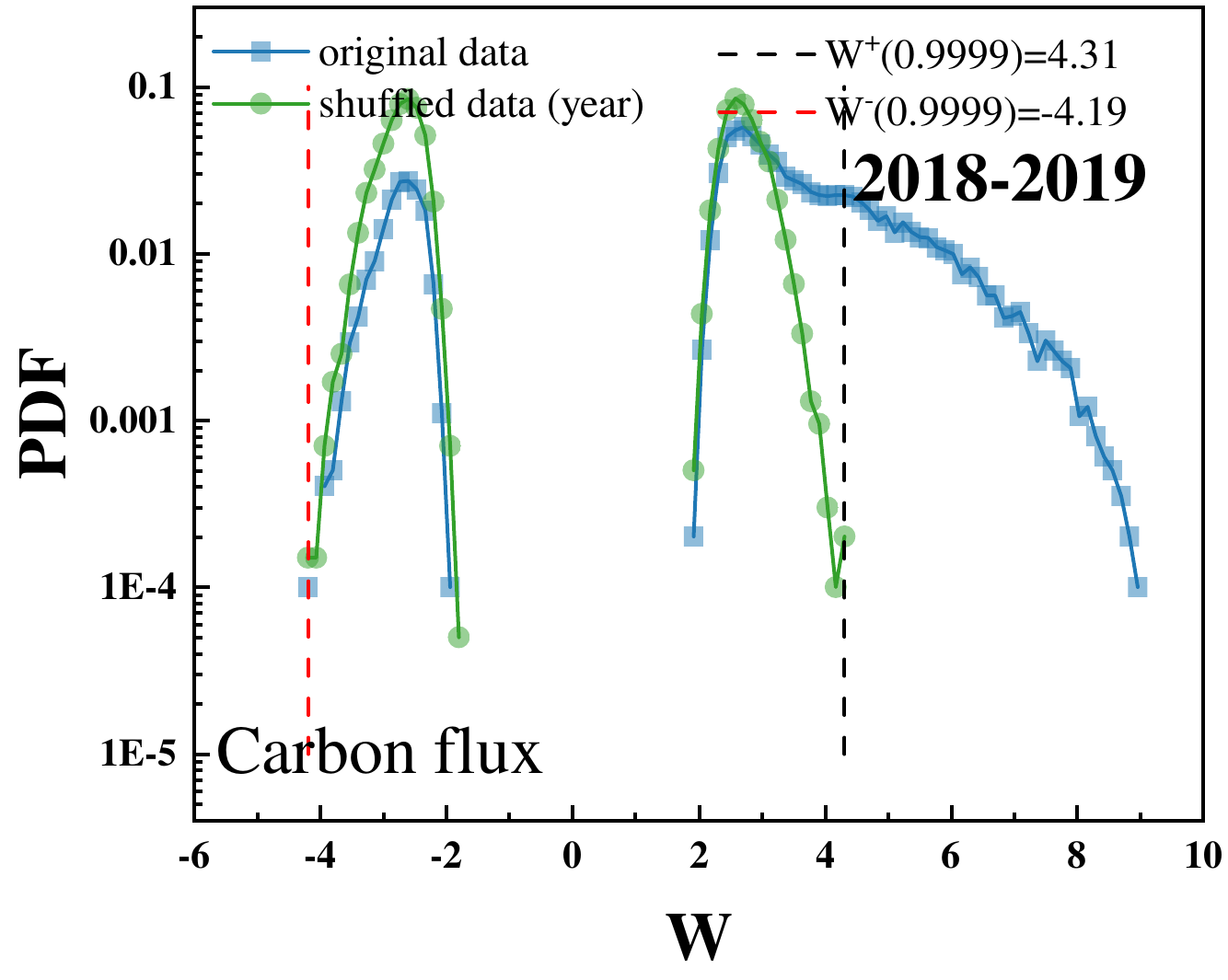}
\end{center}

\begin{center}
\noindent {\small {\bf Fig. S7} Probability distribution function (PDF) of link weights for the original data and shuffled data of carbon flux in the Contiguous United States. }
\end{center}

\begin{center}
\includegraphics[width=8.5em, height=7em]{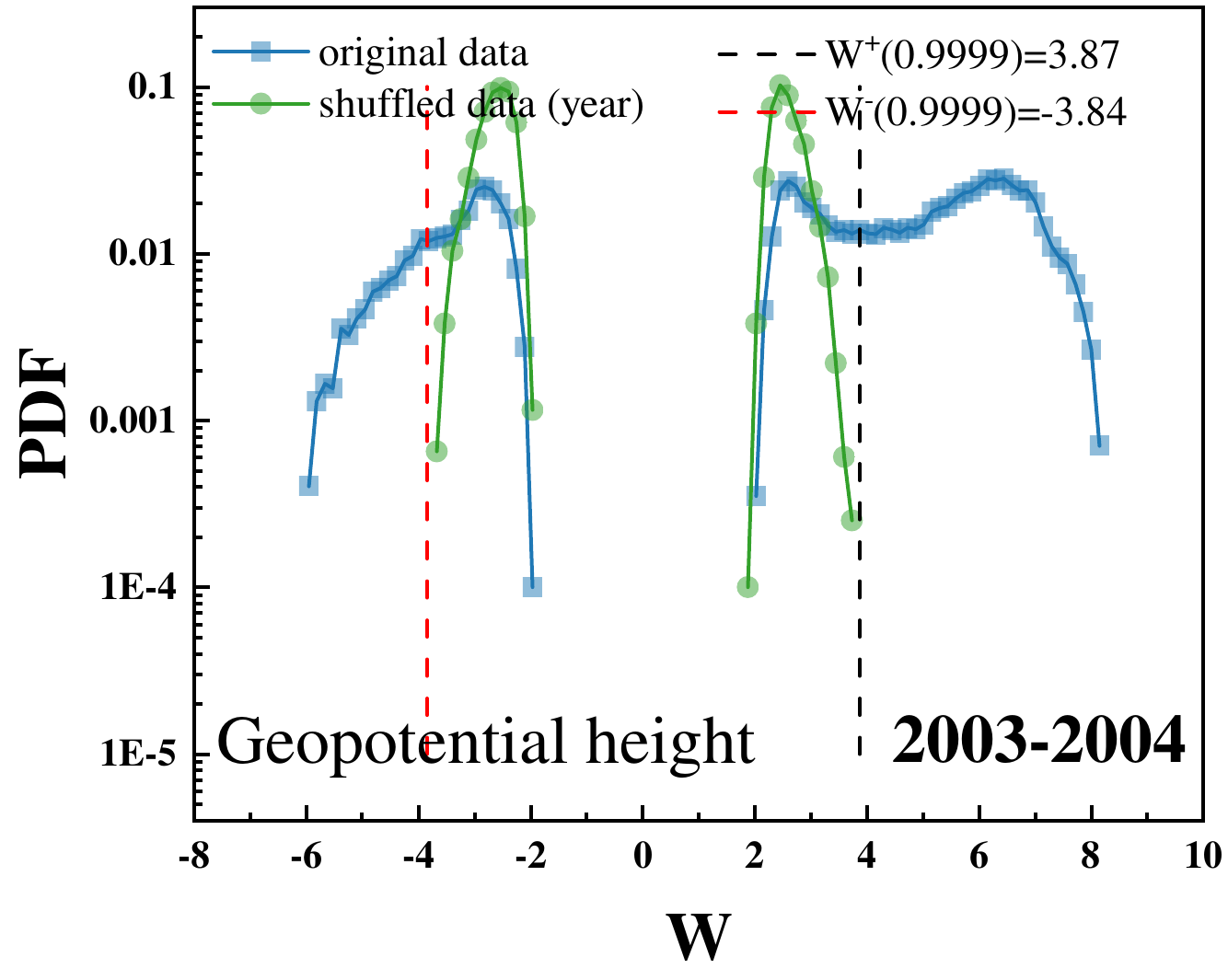}
\includegraphics[width=8.5em, height=7em]{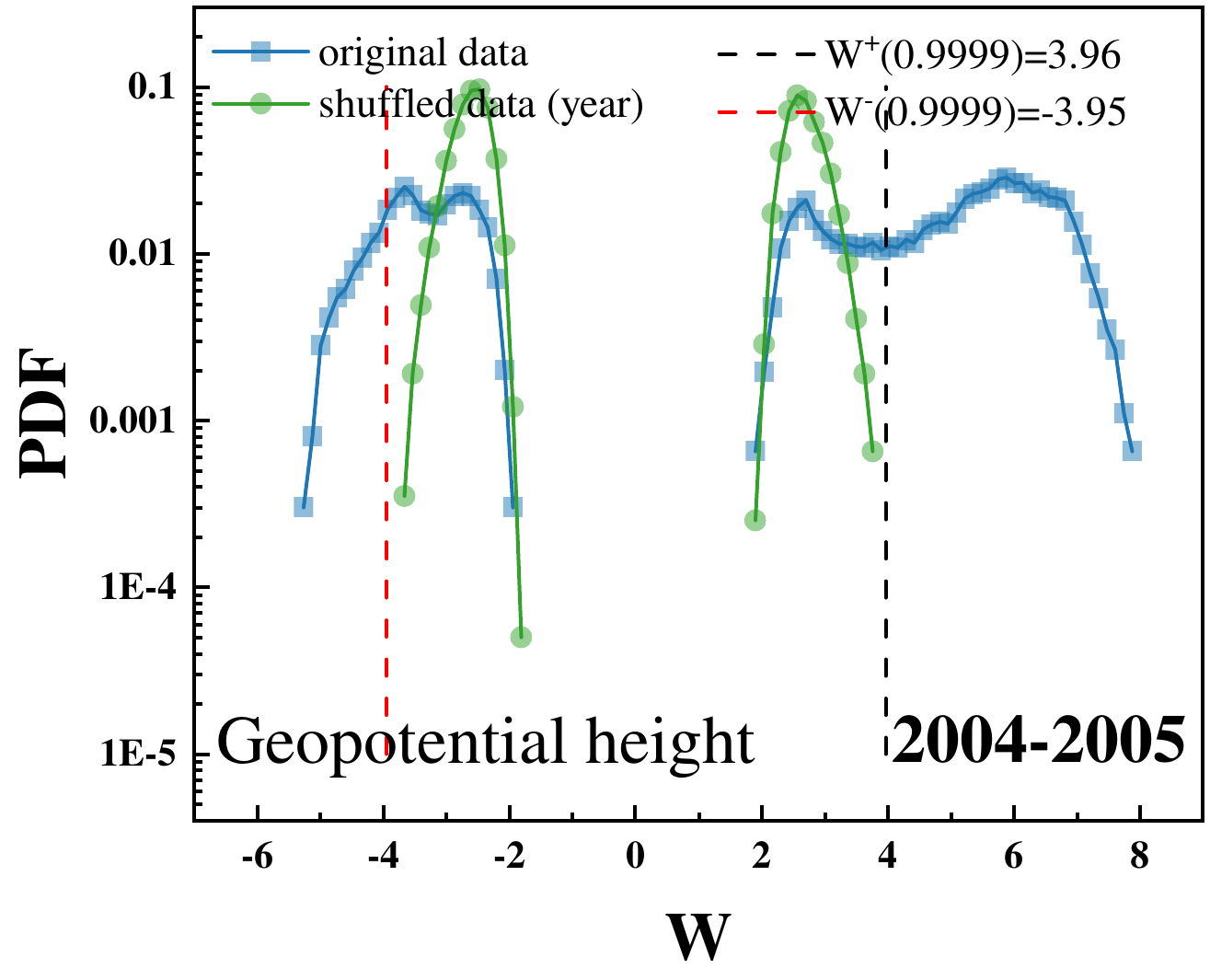}
\includegraphics[width=8.5em, height=7em]{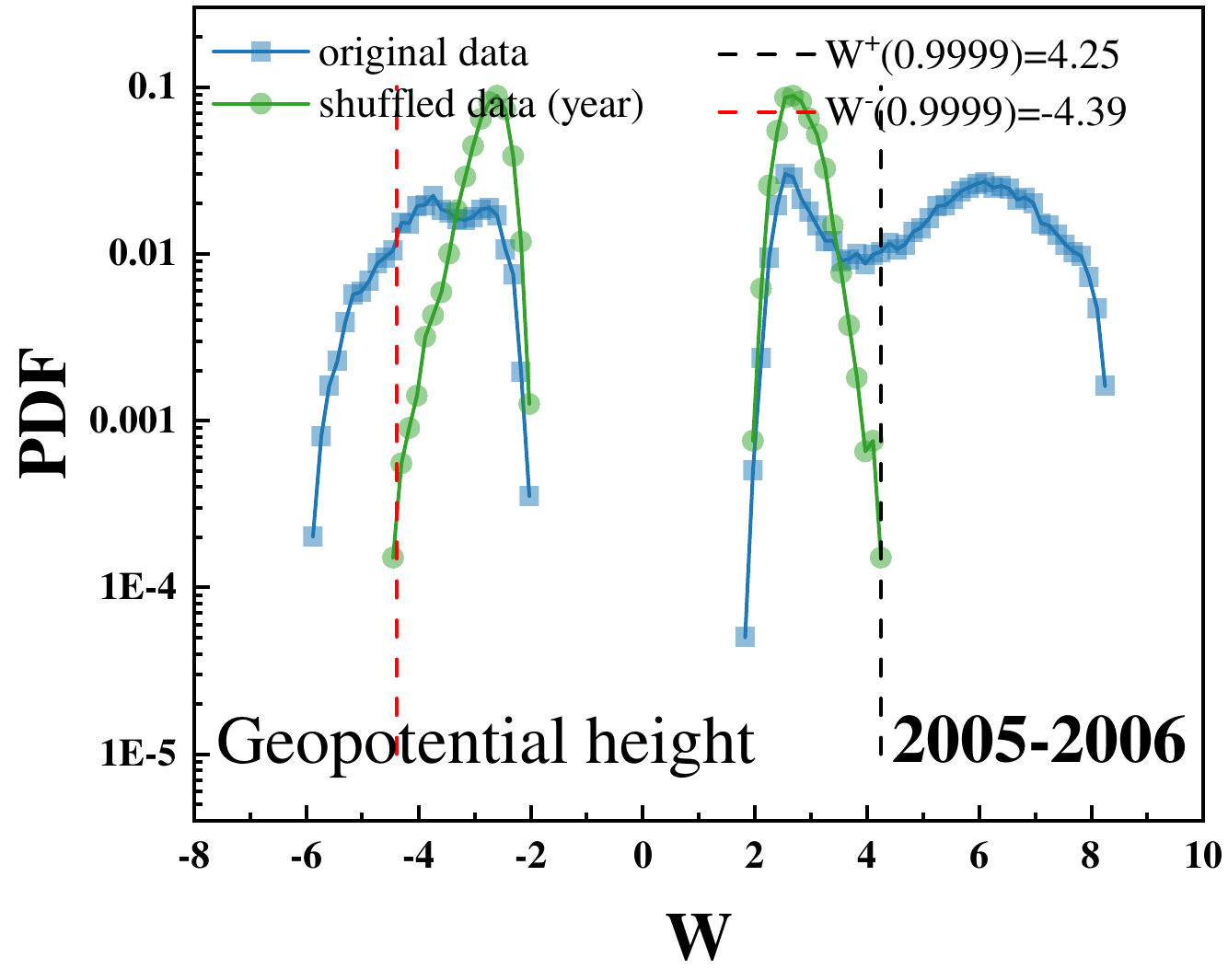}
\includegraphics[width=8.5em, height=7em]{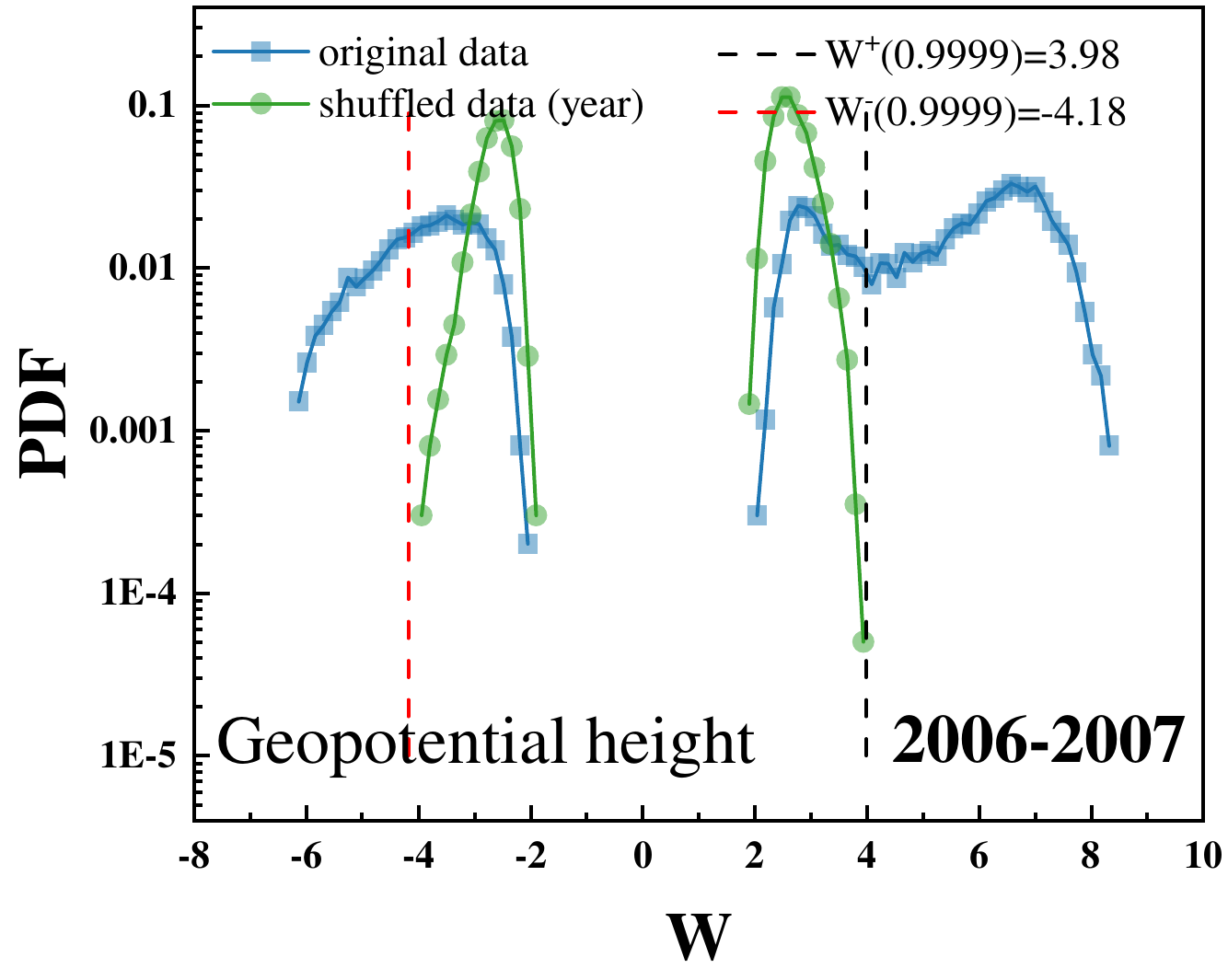}
\includegraphics[width=8.5em, height=7em]{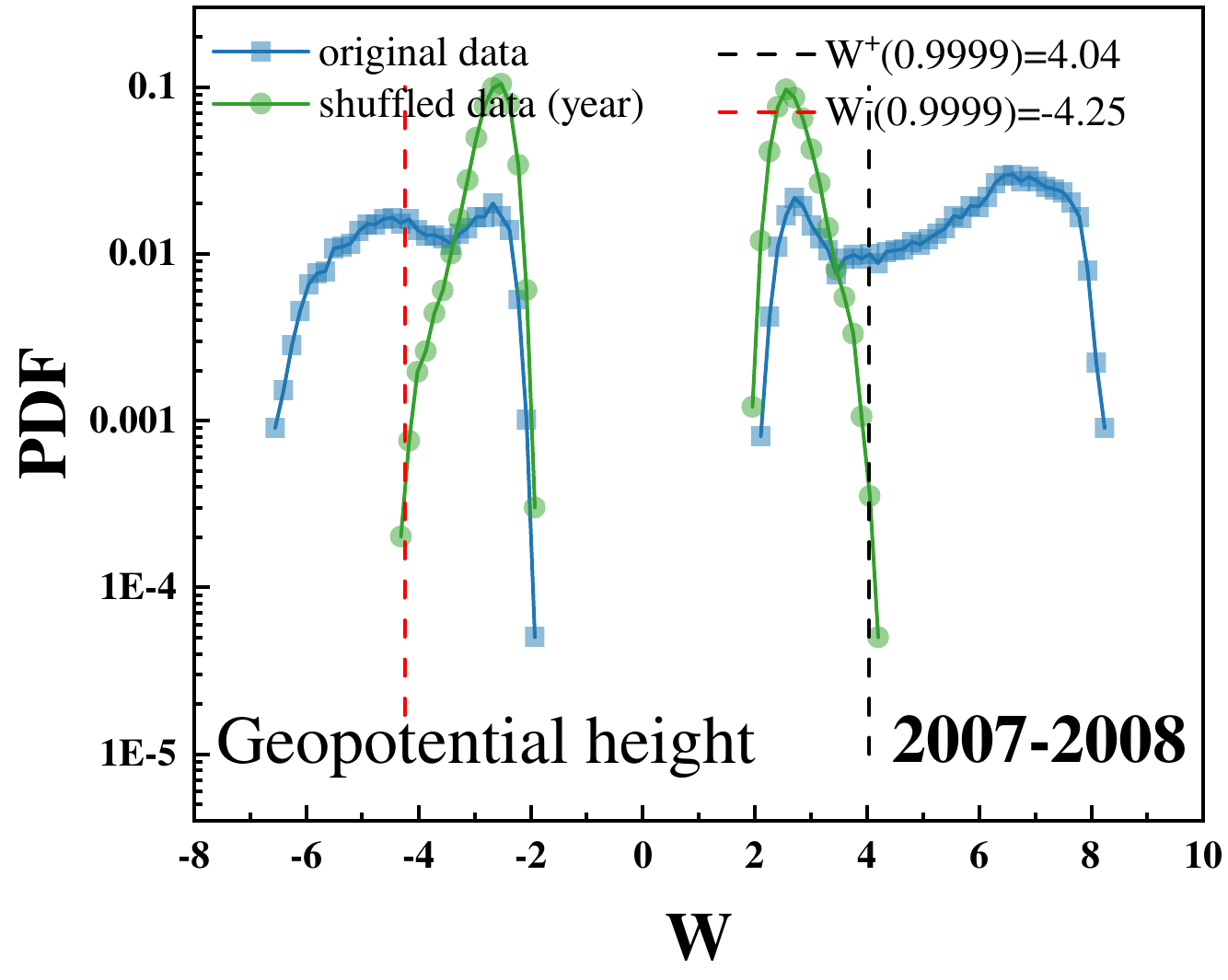}
\includegraphics[width=8.5em, height=7em]{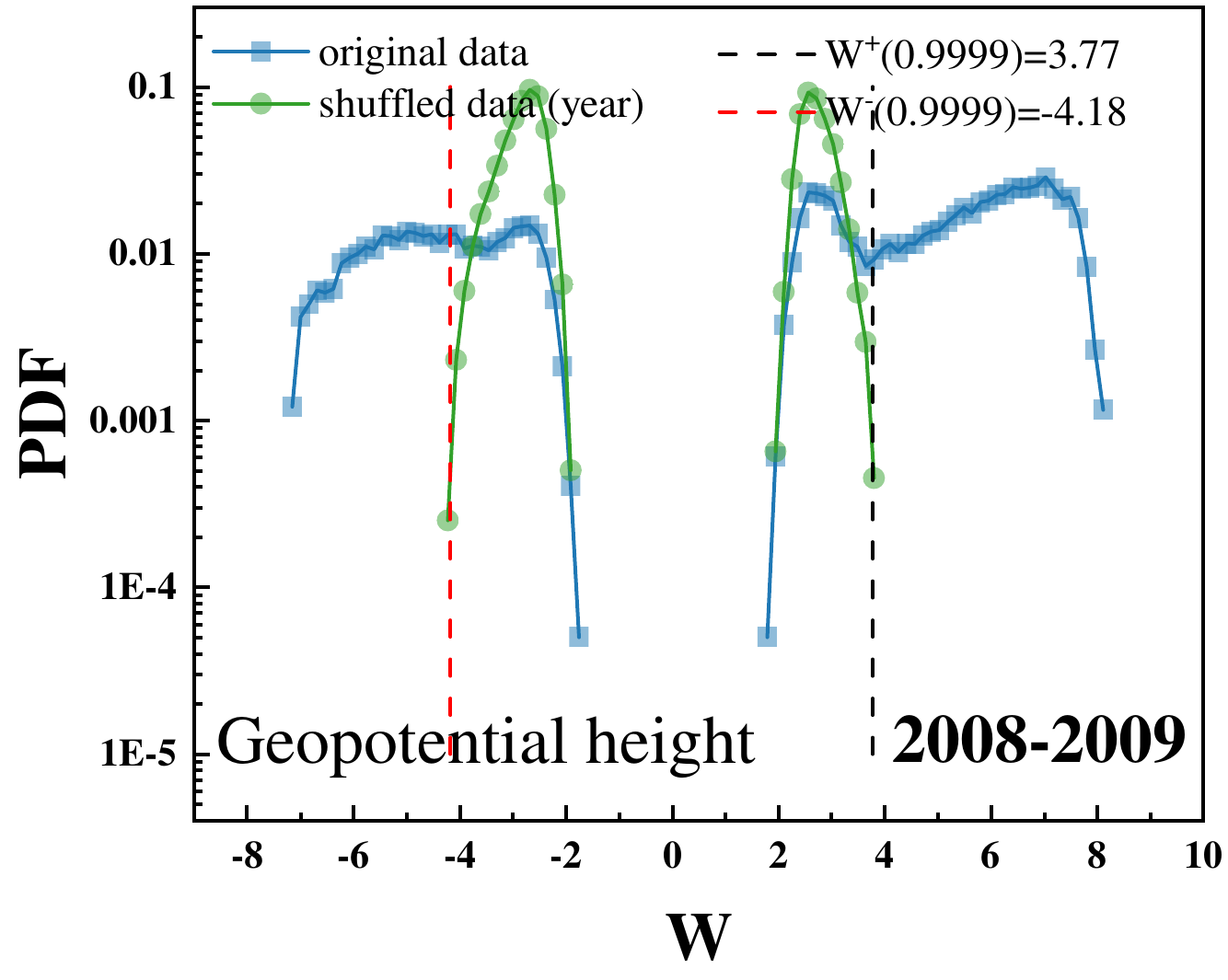}
\includegraphics[width=8.5em, height=7em]{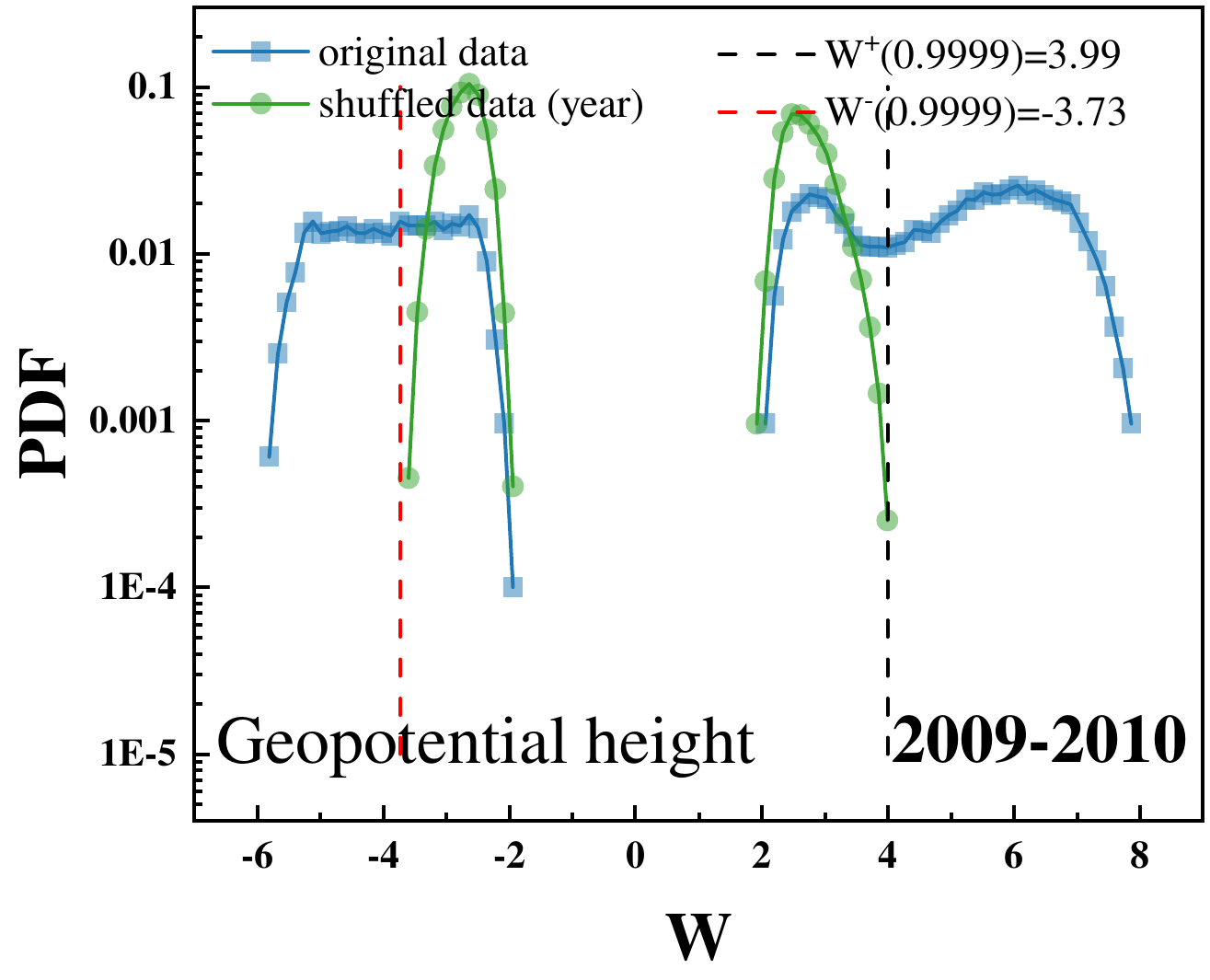}
\includegraphics[width=8.5em, height=7em]{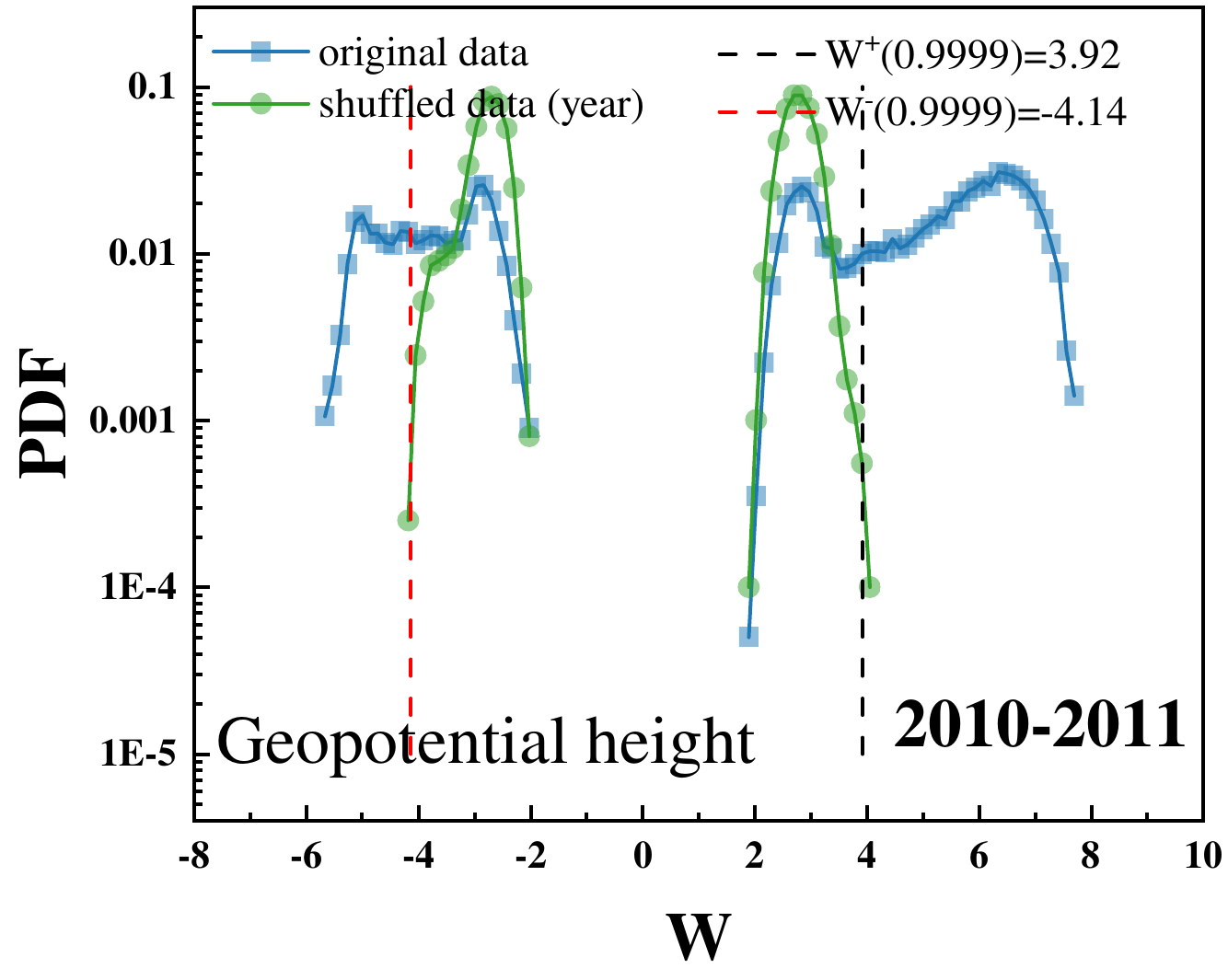}
\includegraphics[width=8.5em, height=7em]{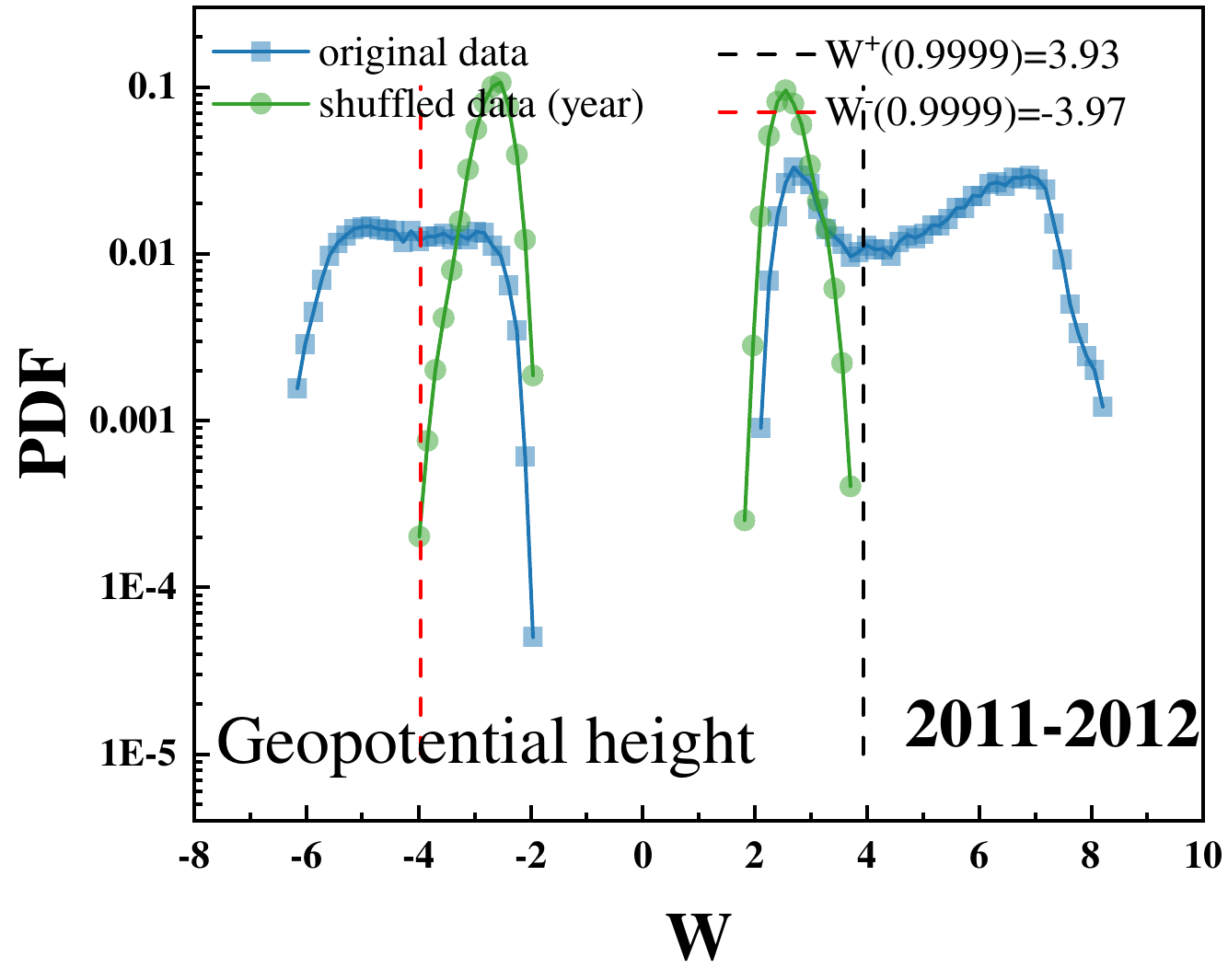}
\includegraphics[width=8.5em, height=7em]{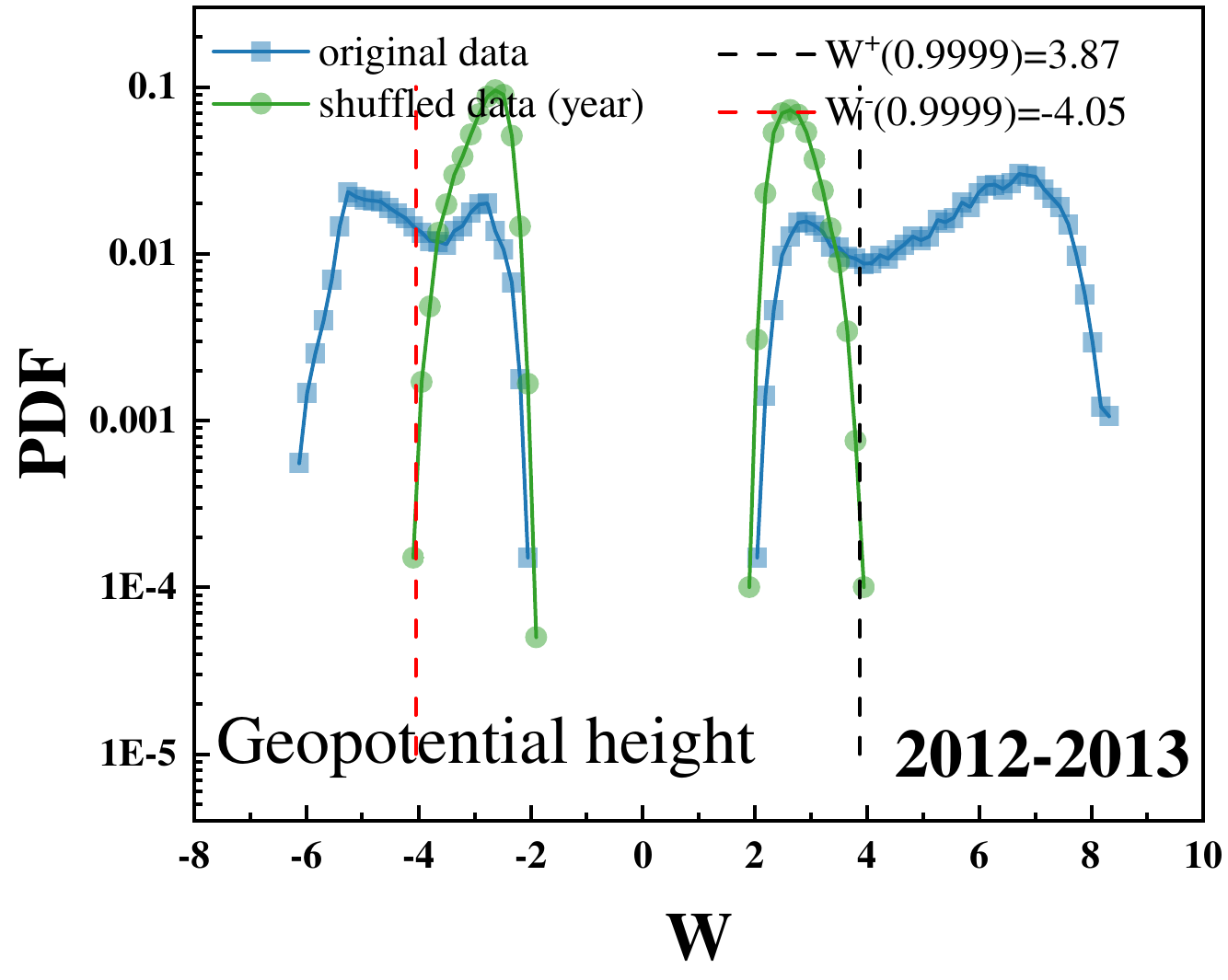}
\includegraphics[width=8.5em, height=7em]{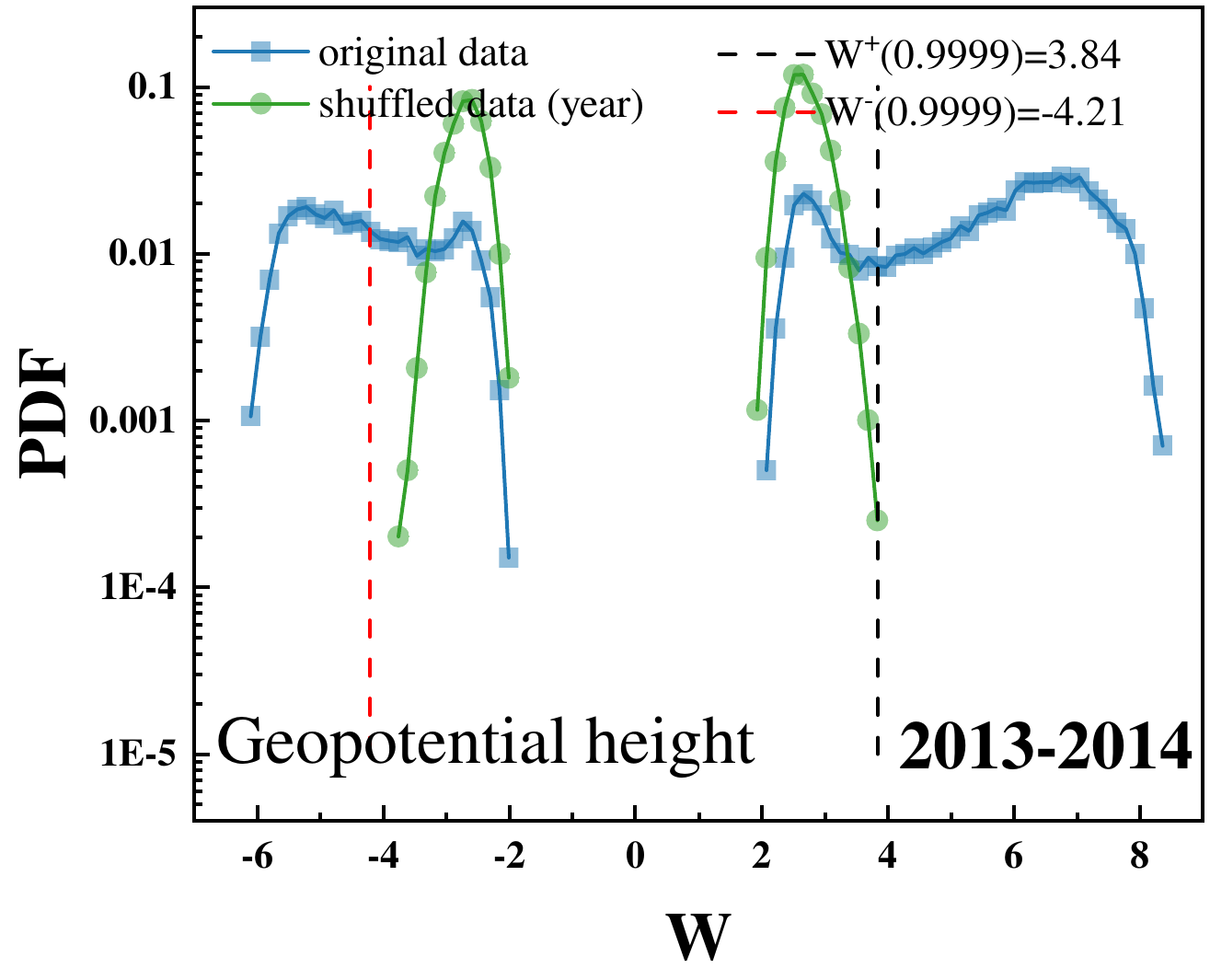}
\includegraphics[width=8.5em, height=7em]{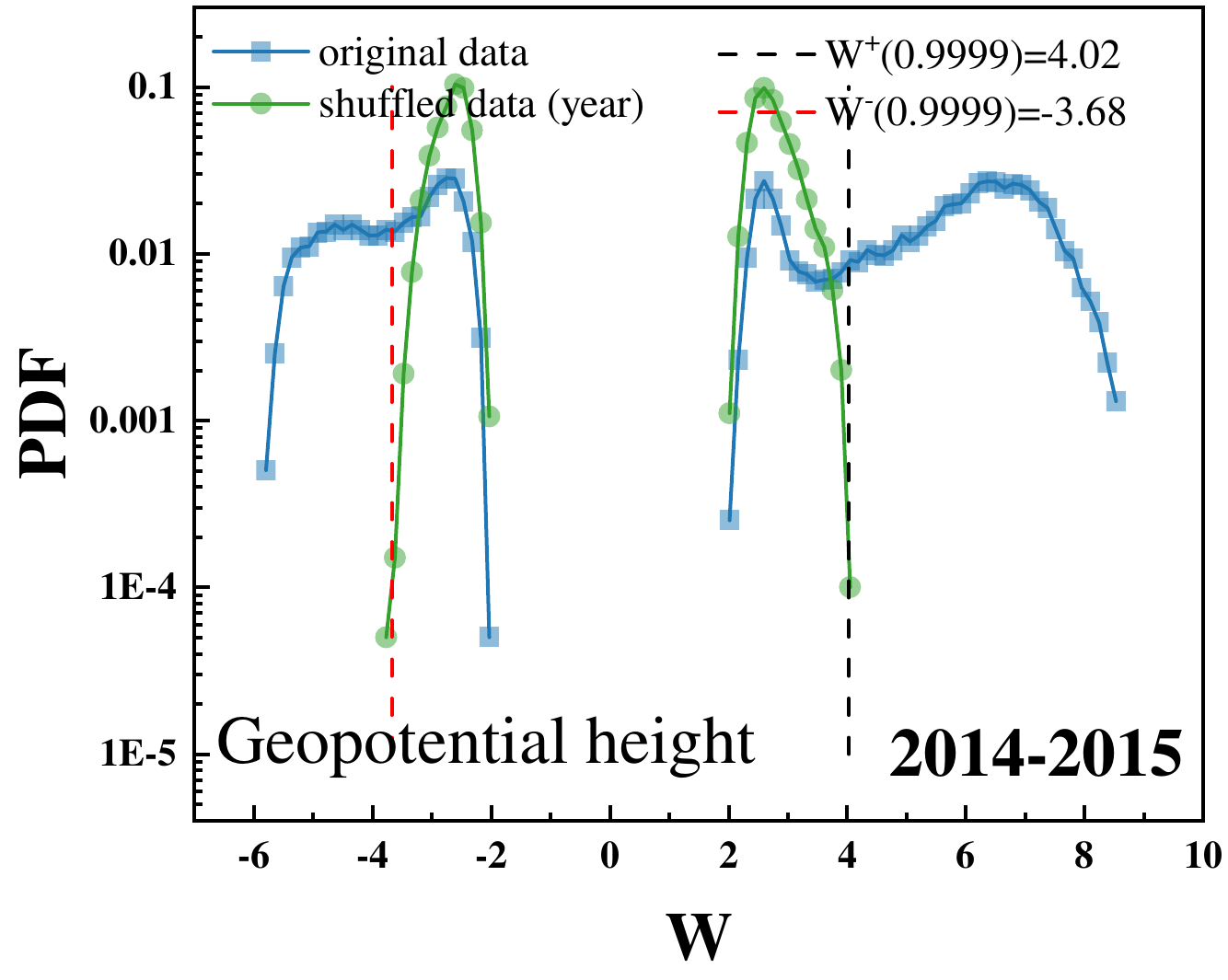}
\includegraphics[width=8.5em, height=7em]{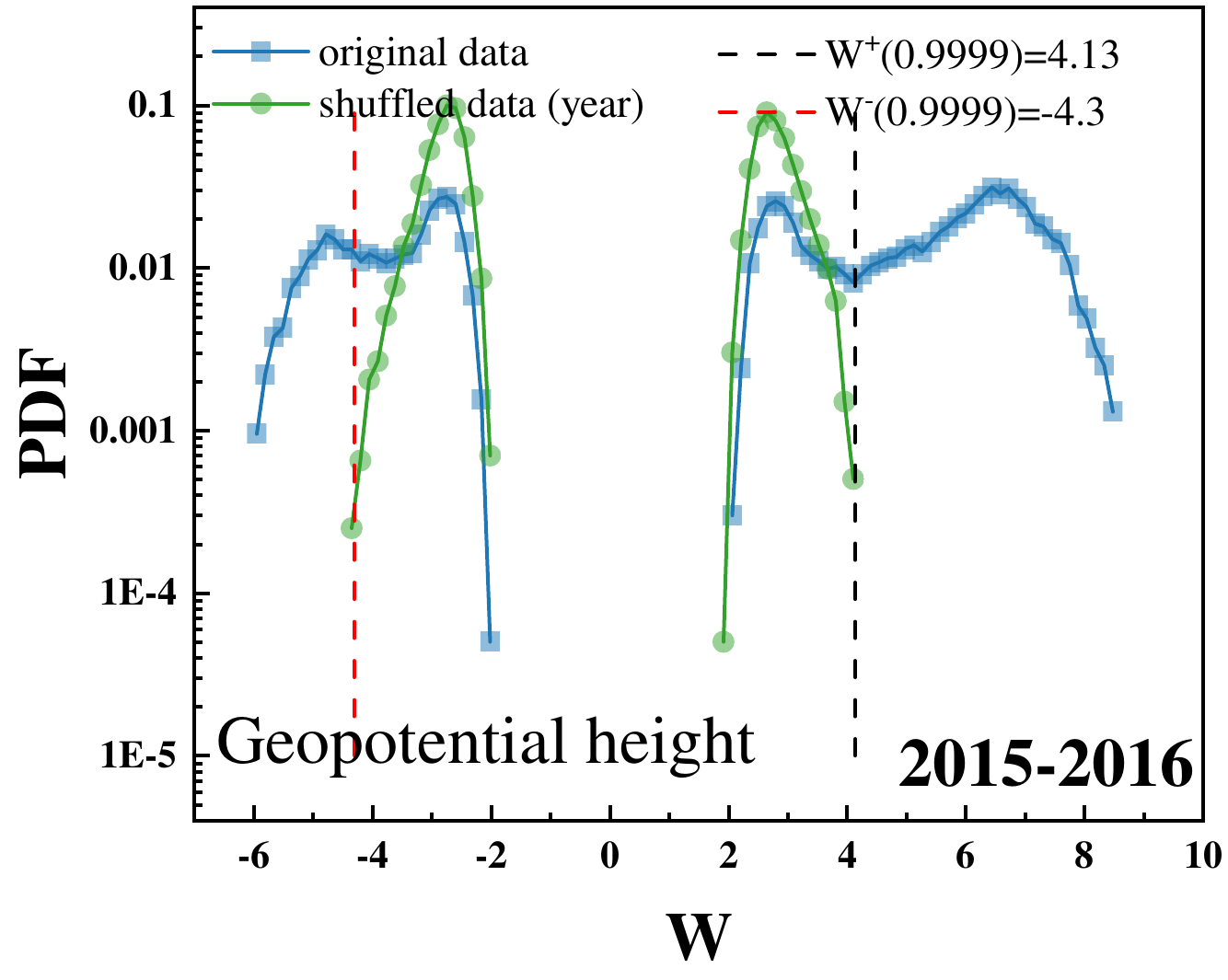}
\includegraphics[width=8.5em, height=7em]{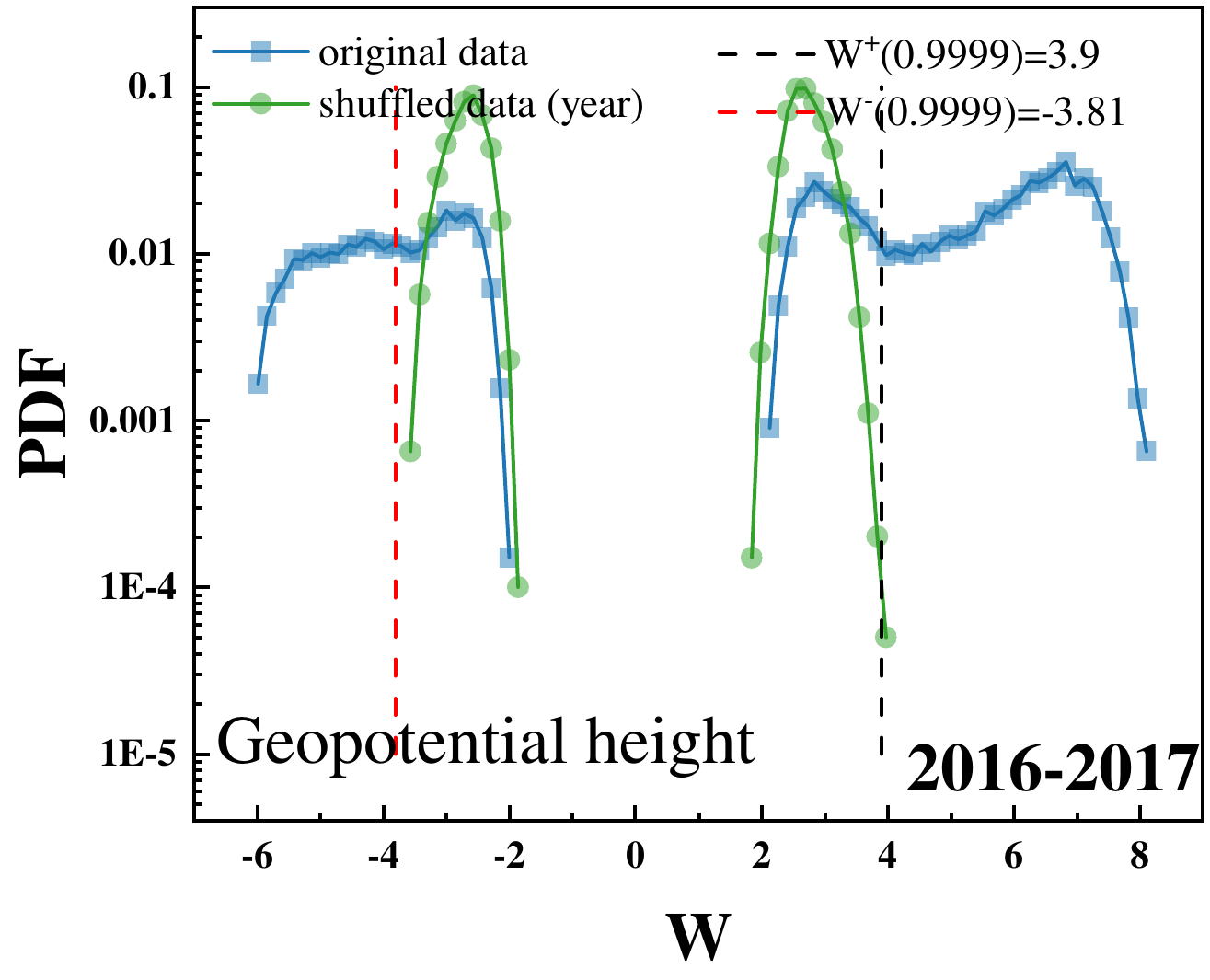}
\includegraphics[width=8.5em, height=7em]{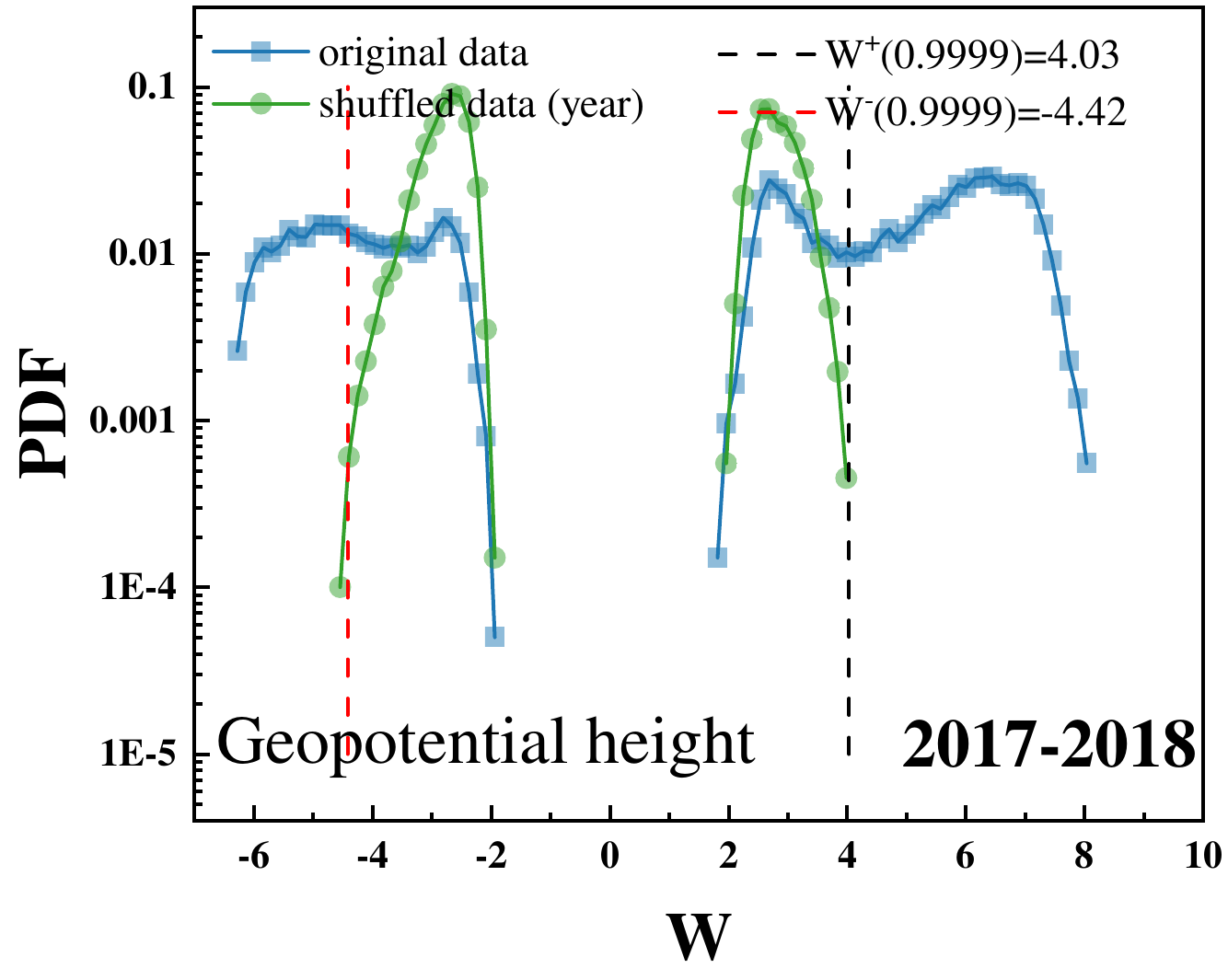}
\includegraphics[width=8.5em, height=7em]{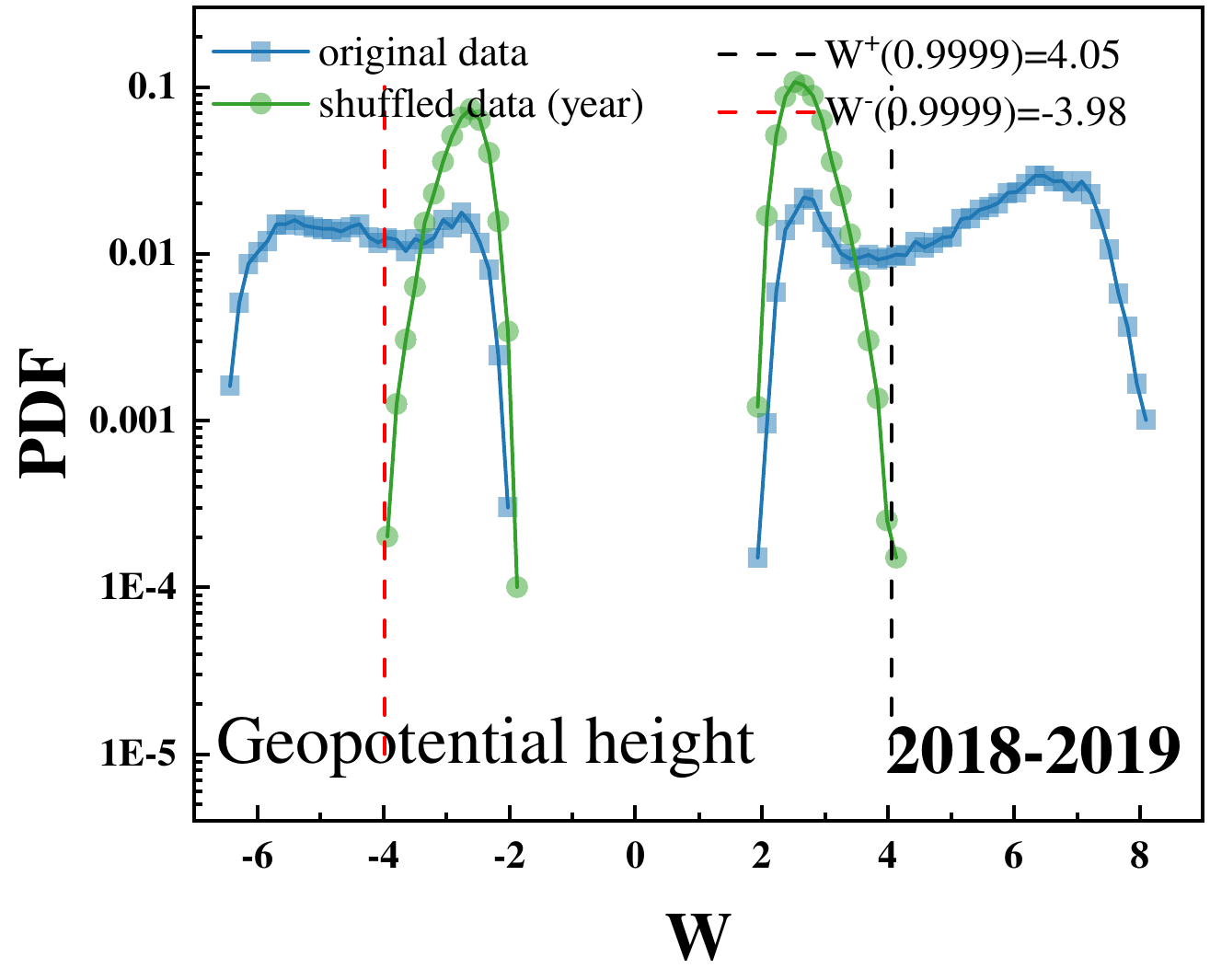}
\end{center}

\begin{center}
\noindent {\small {\bf Fig. S8} Probability distribution function (PDF) of link weights for the original data and shuffled data of Geopotential height in the Contiguous United States. }
\end{center}

\begin{center}
\includegraphics[width=8.5em, height=7em]{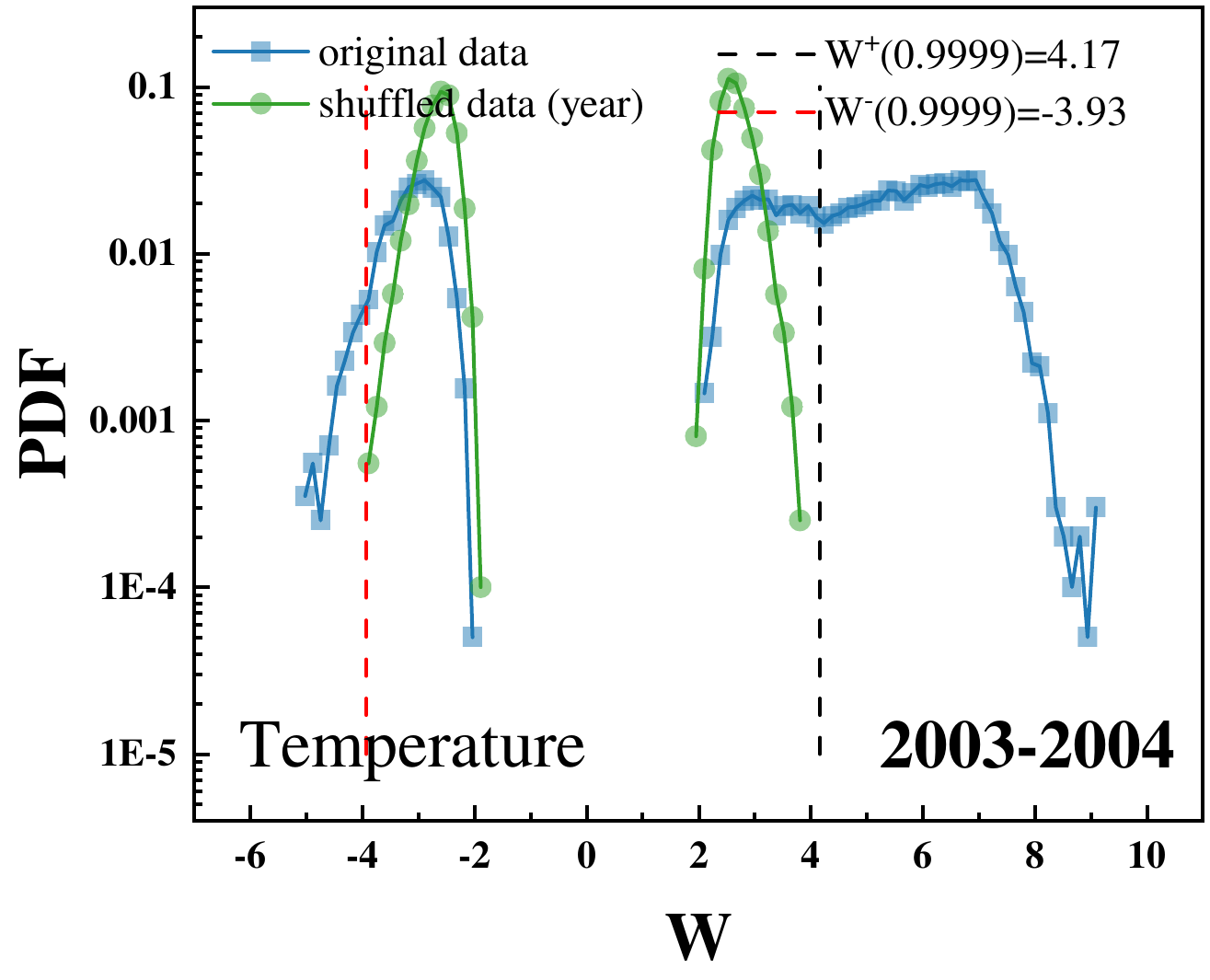}
\includegraphics[width=8.5em, height=7em]{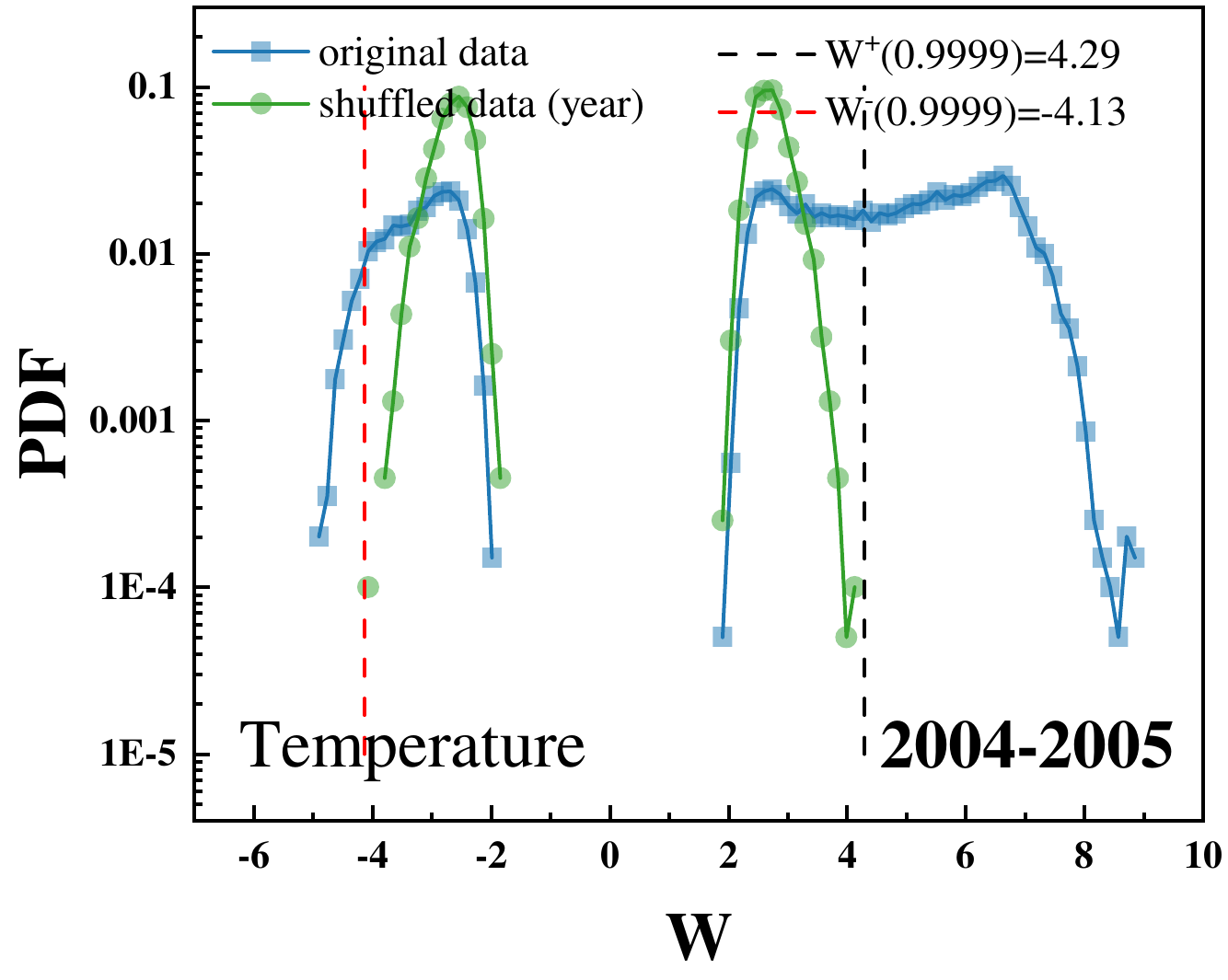}
\includegraphics[width=8.5em, height=7em]{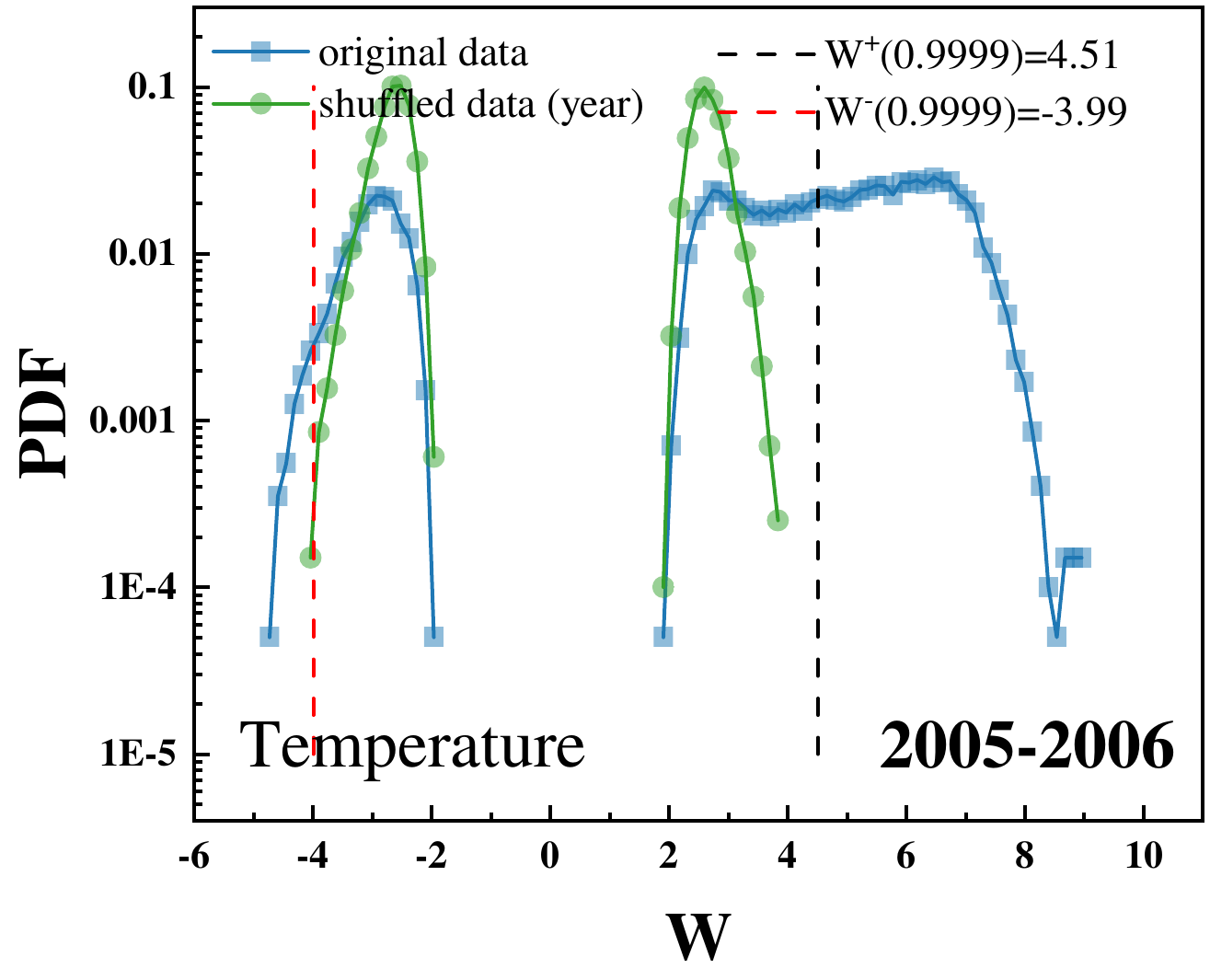}
\includegraphics[width=8.5em, height=7em]{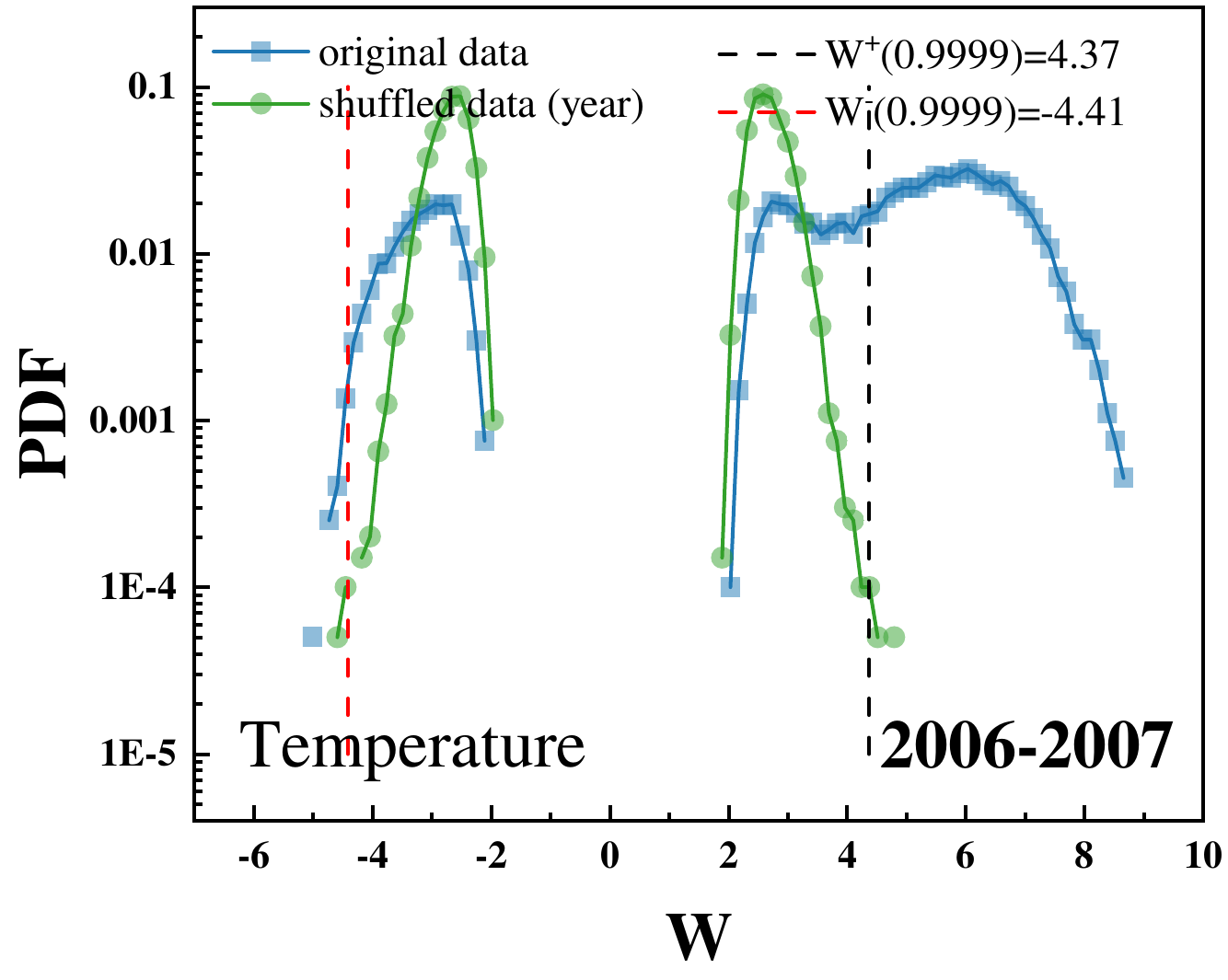}
\includegraphics[width=8.5em, height=7em]{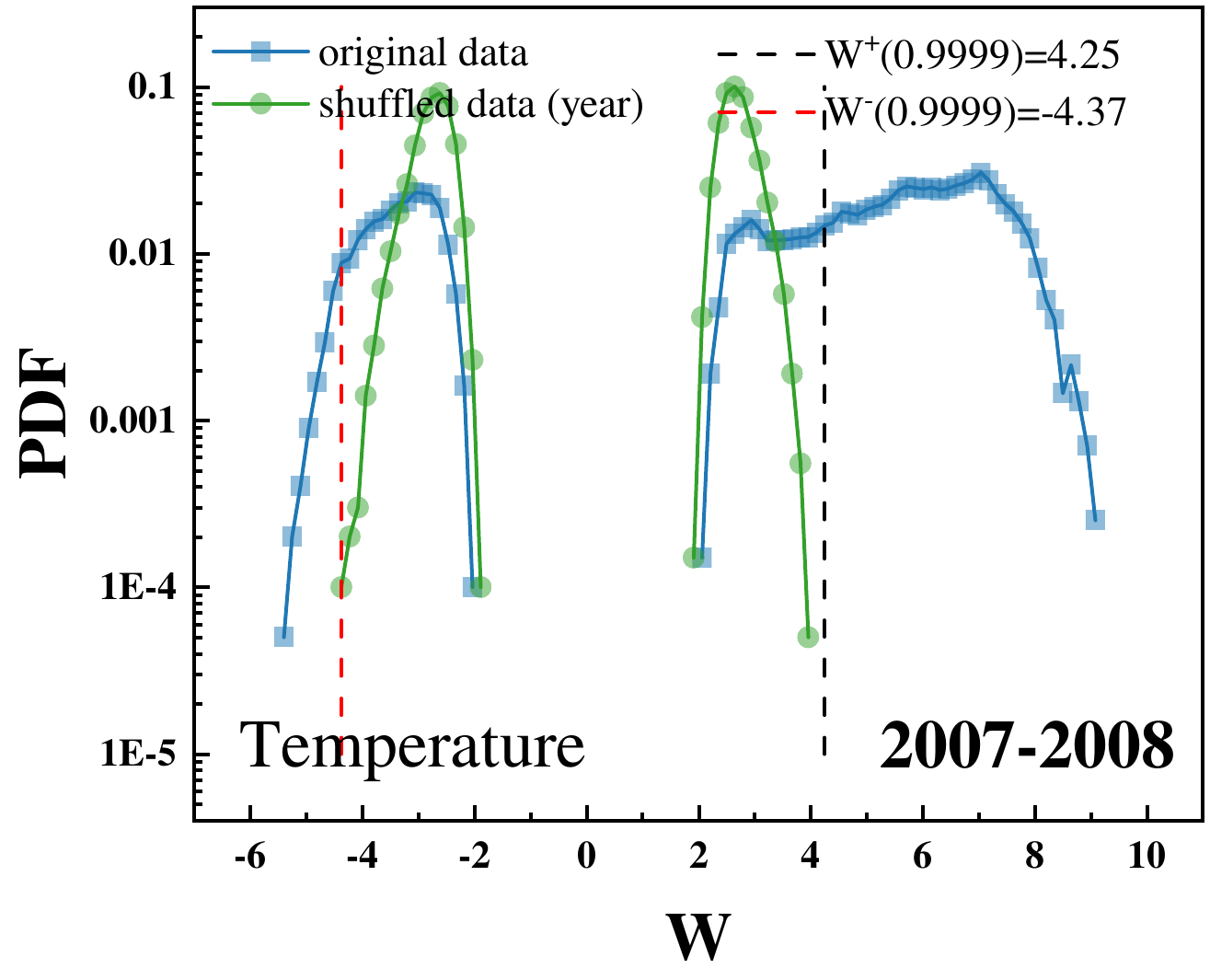}
\includegraphics[width=8.5em, height=7em]{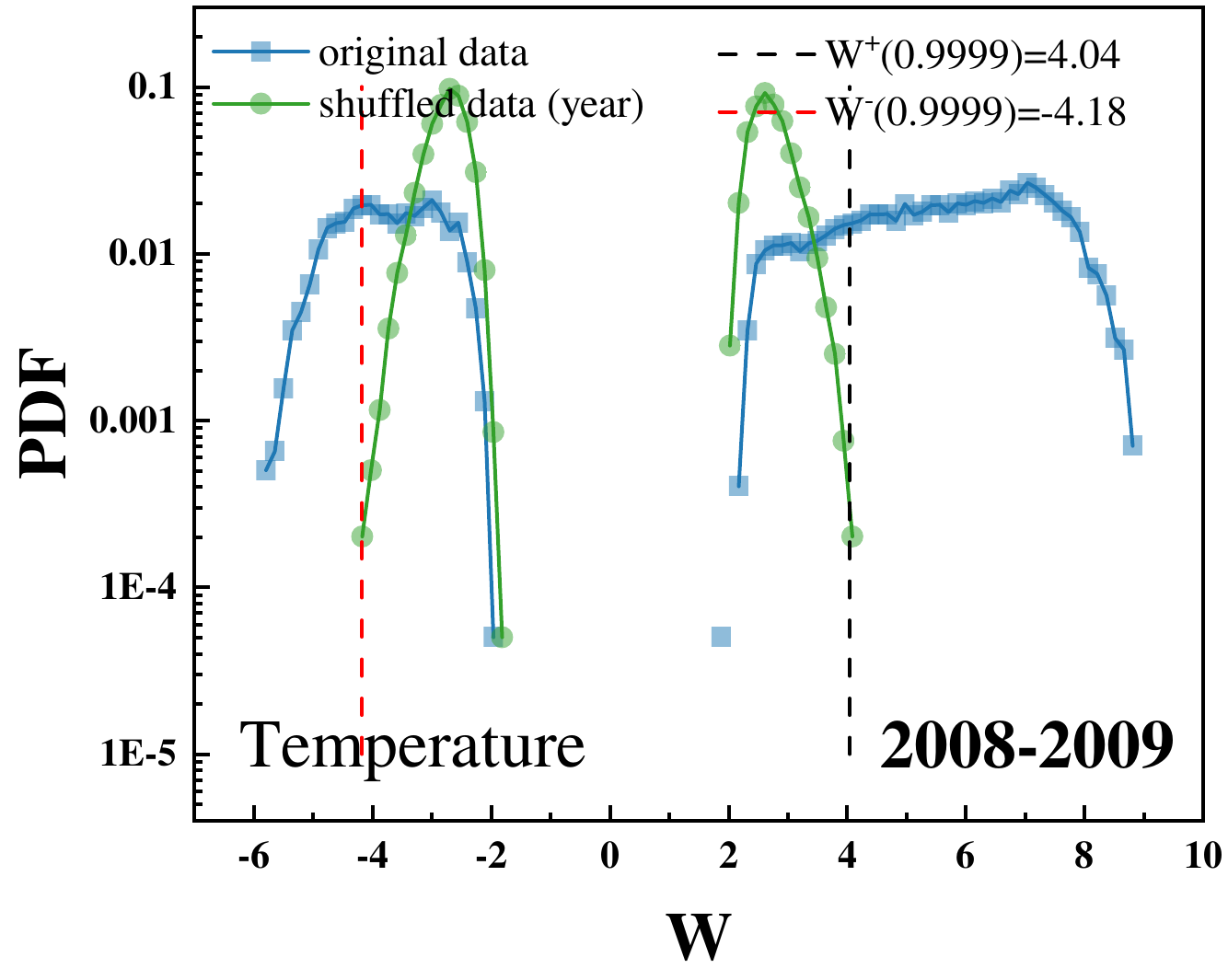}
\includegraphics[width=8.5em, height=7em]{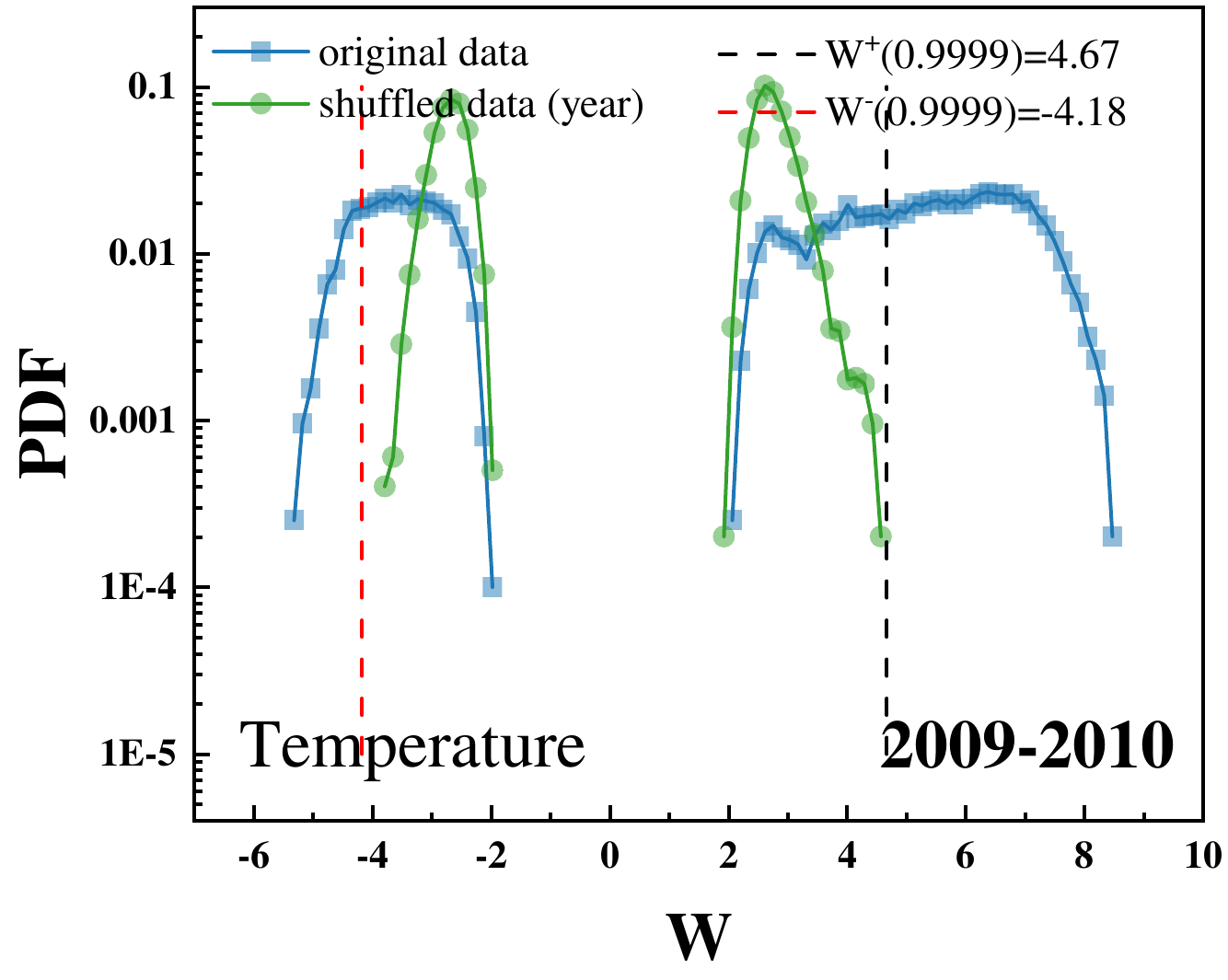}
\includegraphics[width=8.5em, height=7em]{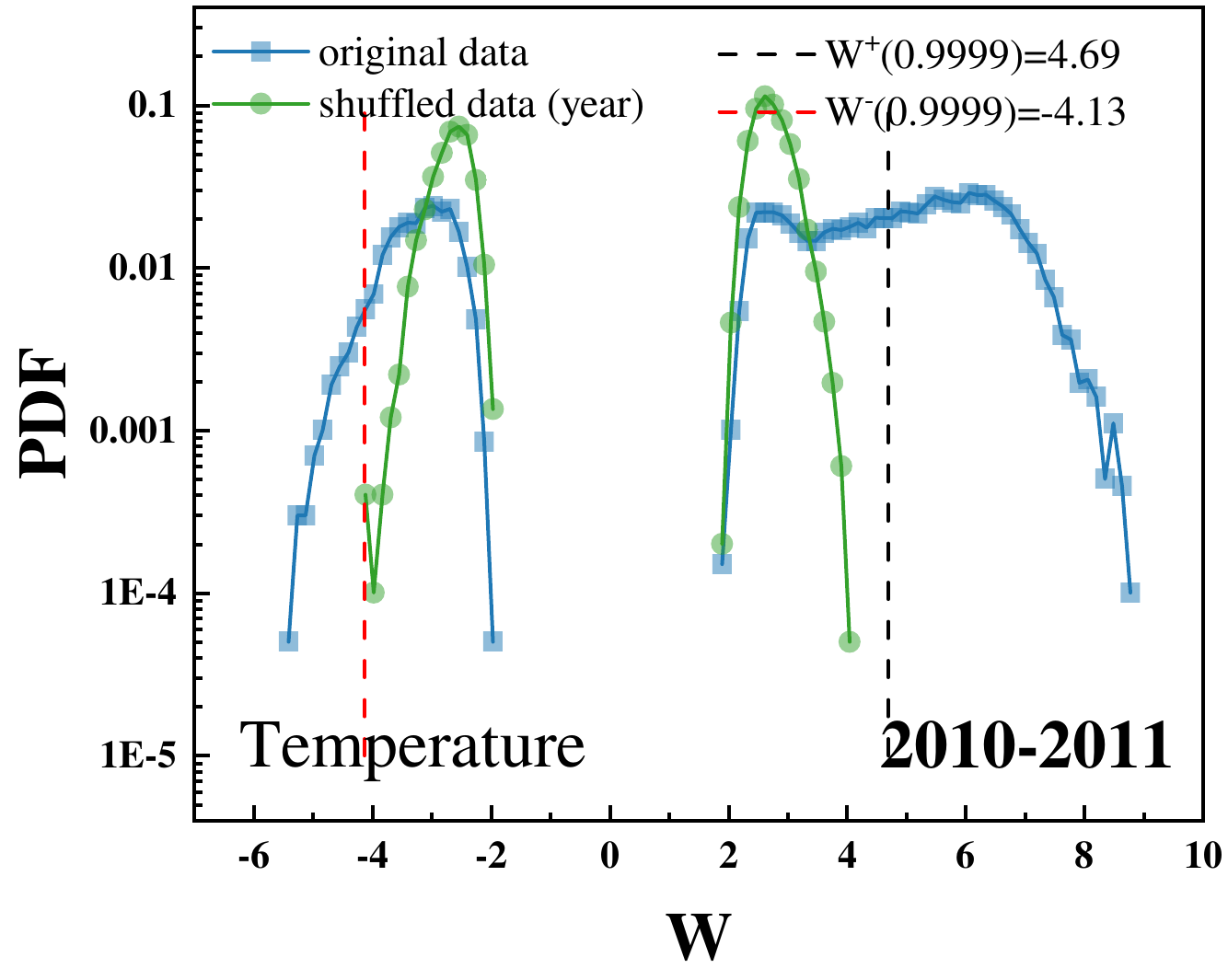}
\includegraphics[width=8.5em, height=7em]{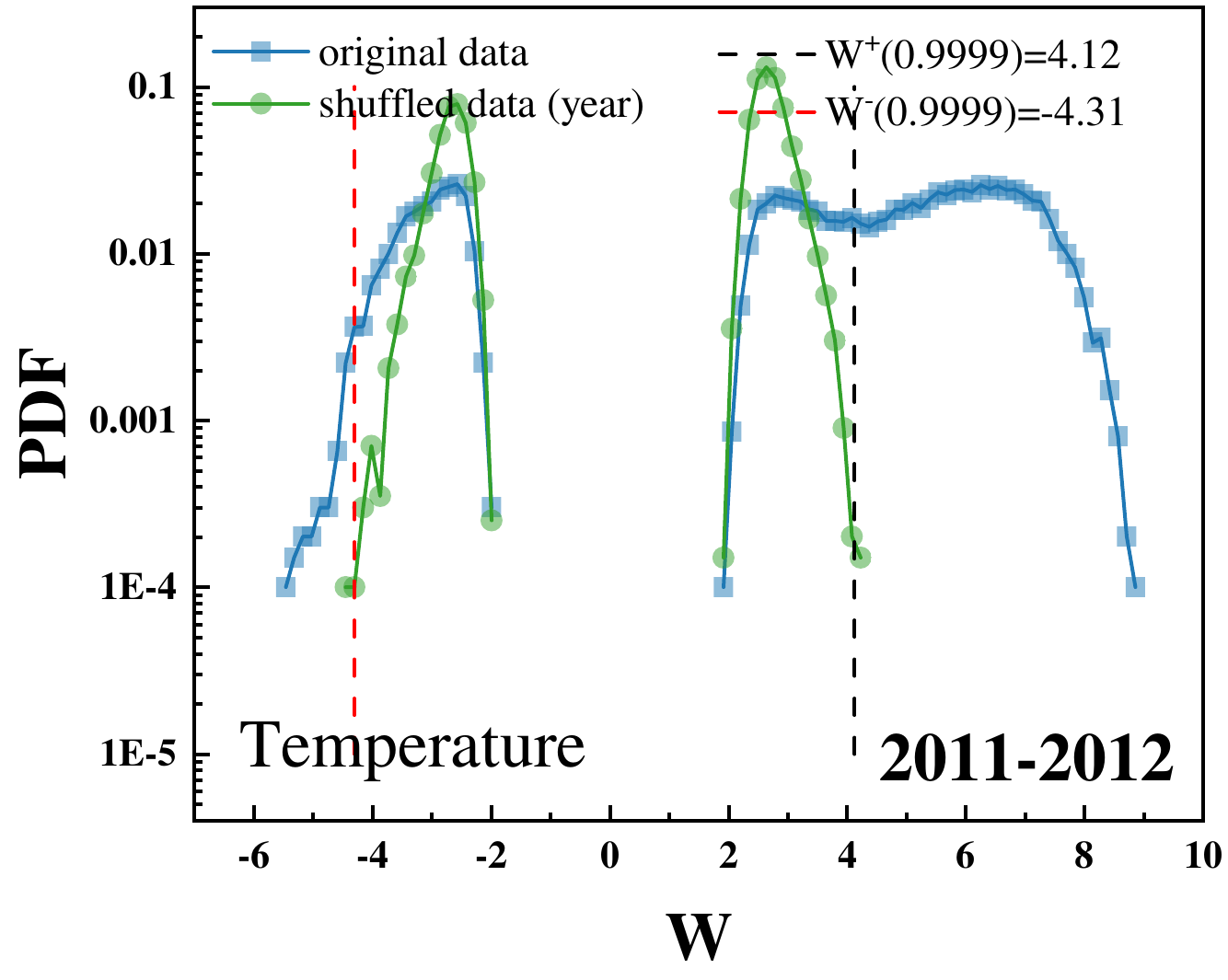}
\includegraphics[width=8.5em, height=7em]{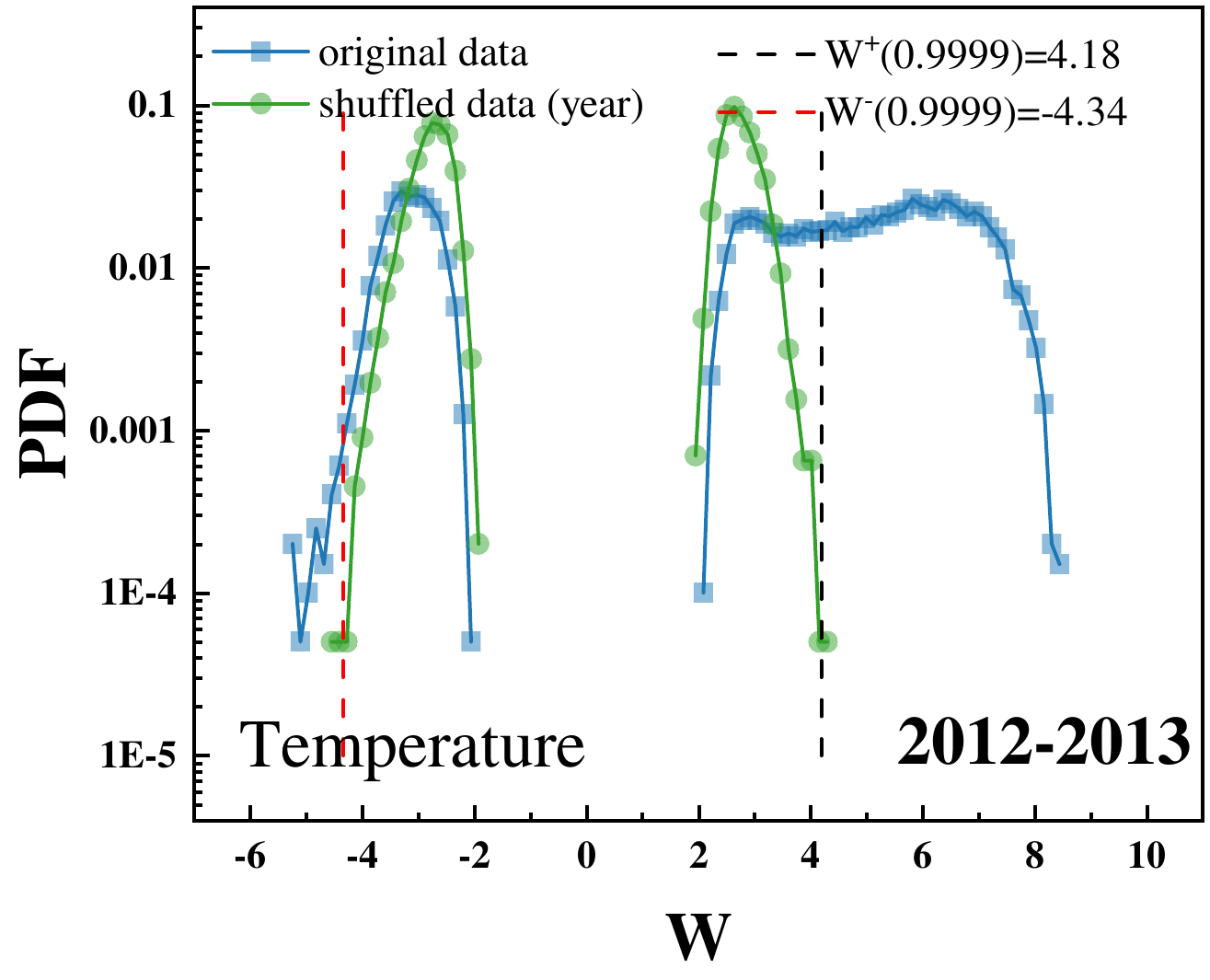}
\includegraphics[width=8.5em, height=7em]{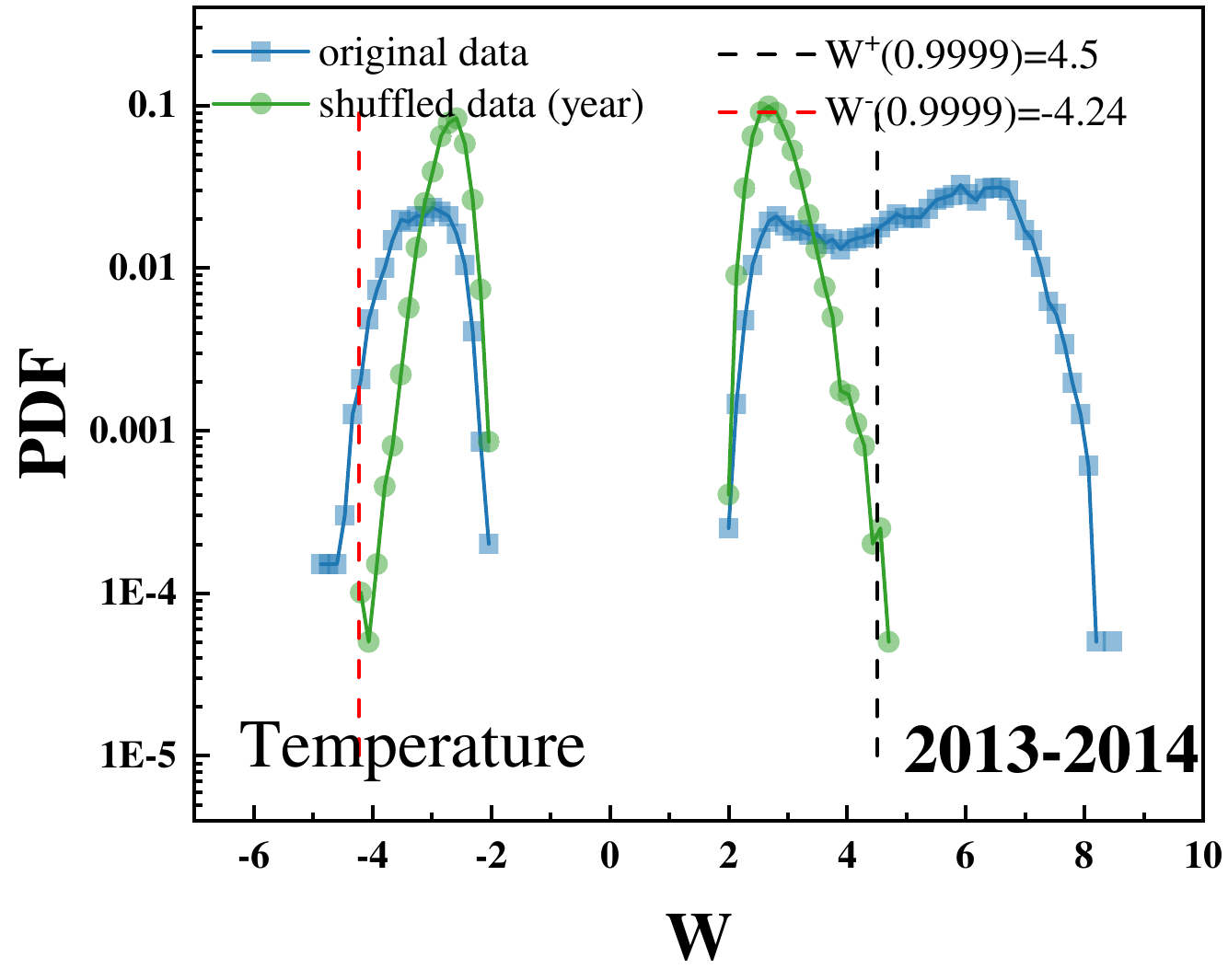}
\includegraphics[width=8.5em, height=7em]{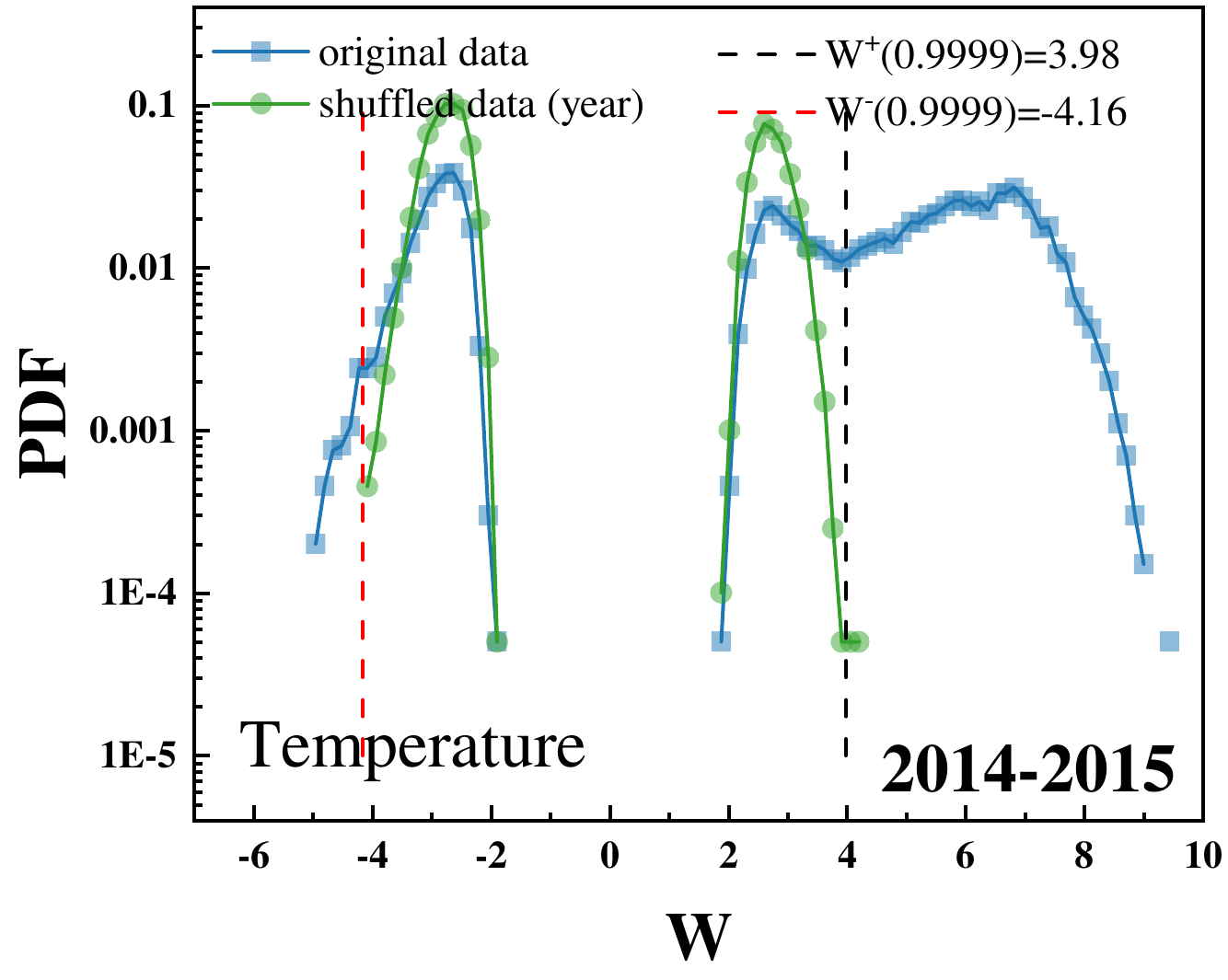}
\includegraphics[width=8.5em, height=7em]{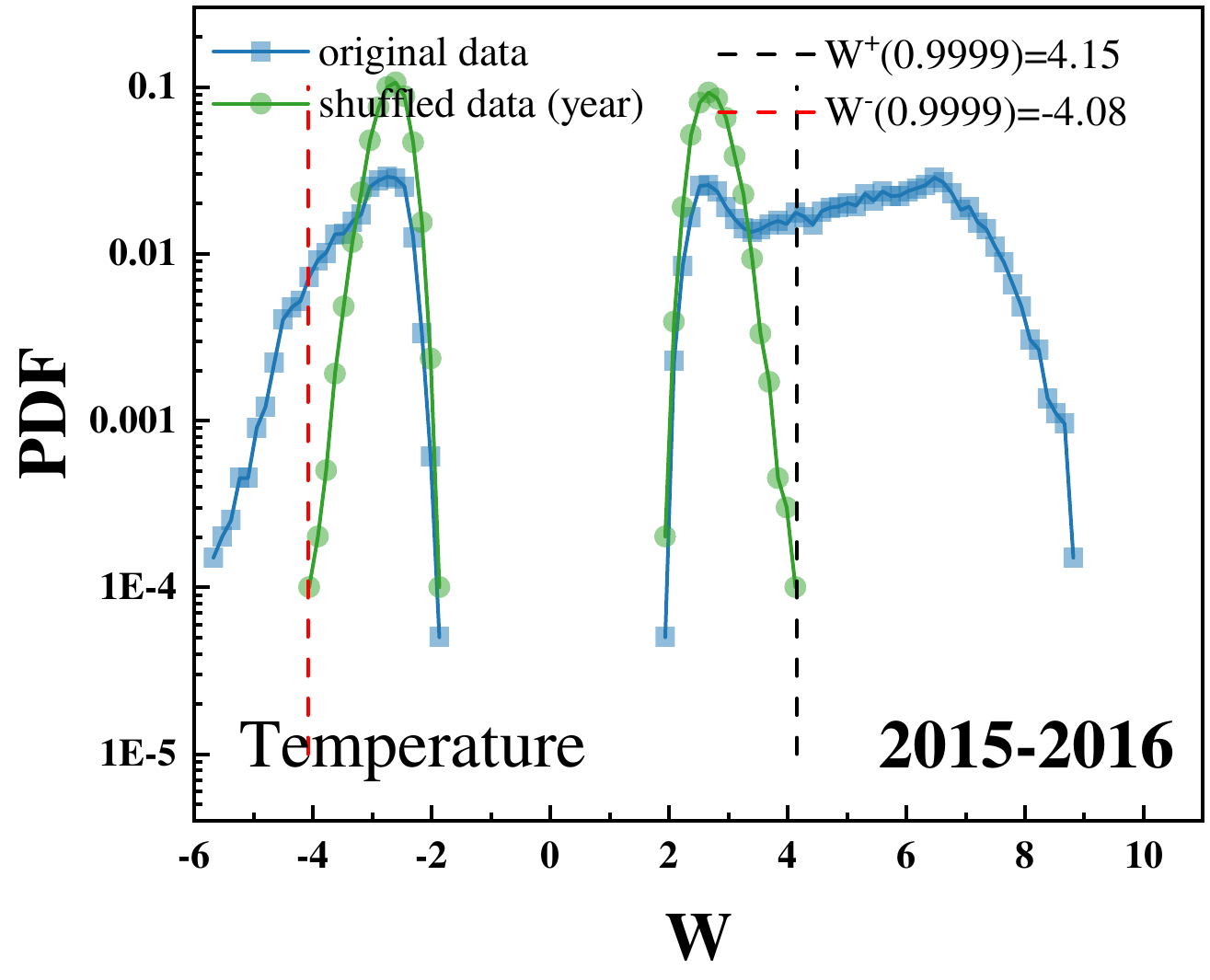}
\includegraphics[width=8.5em, height=7em]{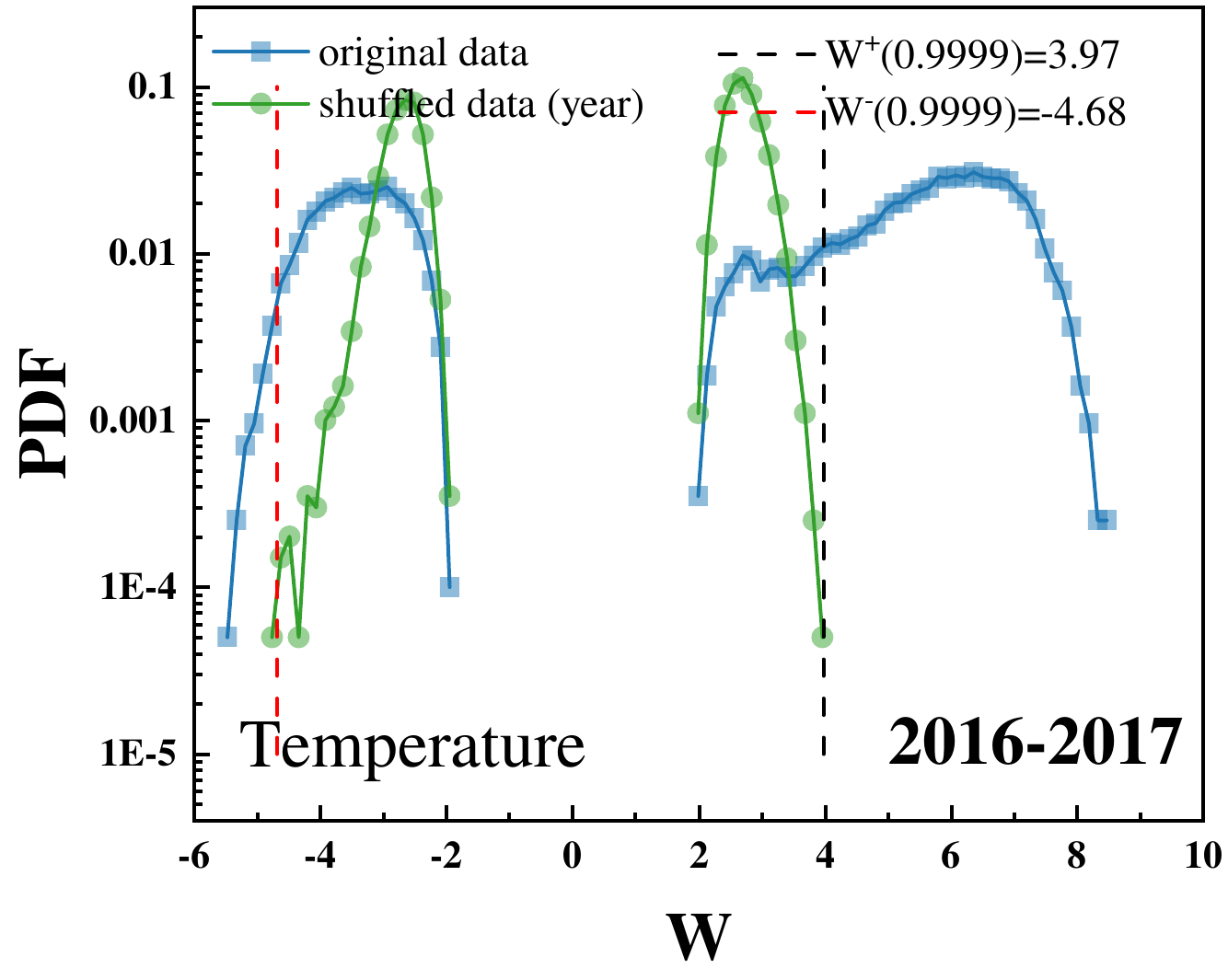}
\includegraphics[width=8.5em, height=7em]{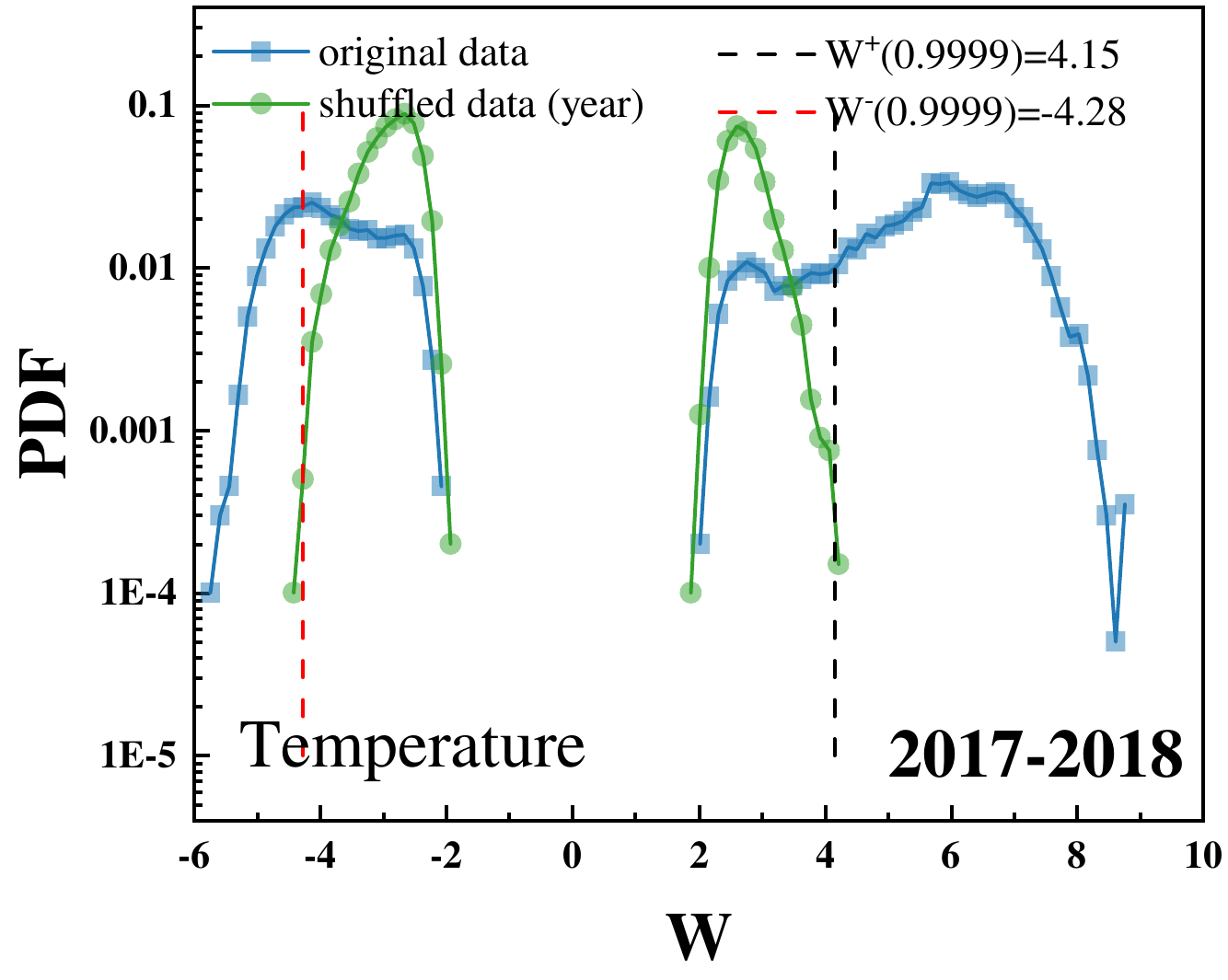}
\includegraphics[width=8.5em, height=7em]{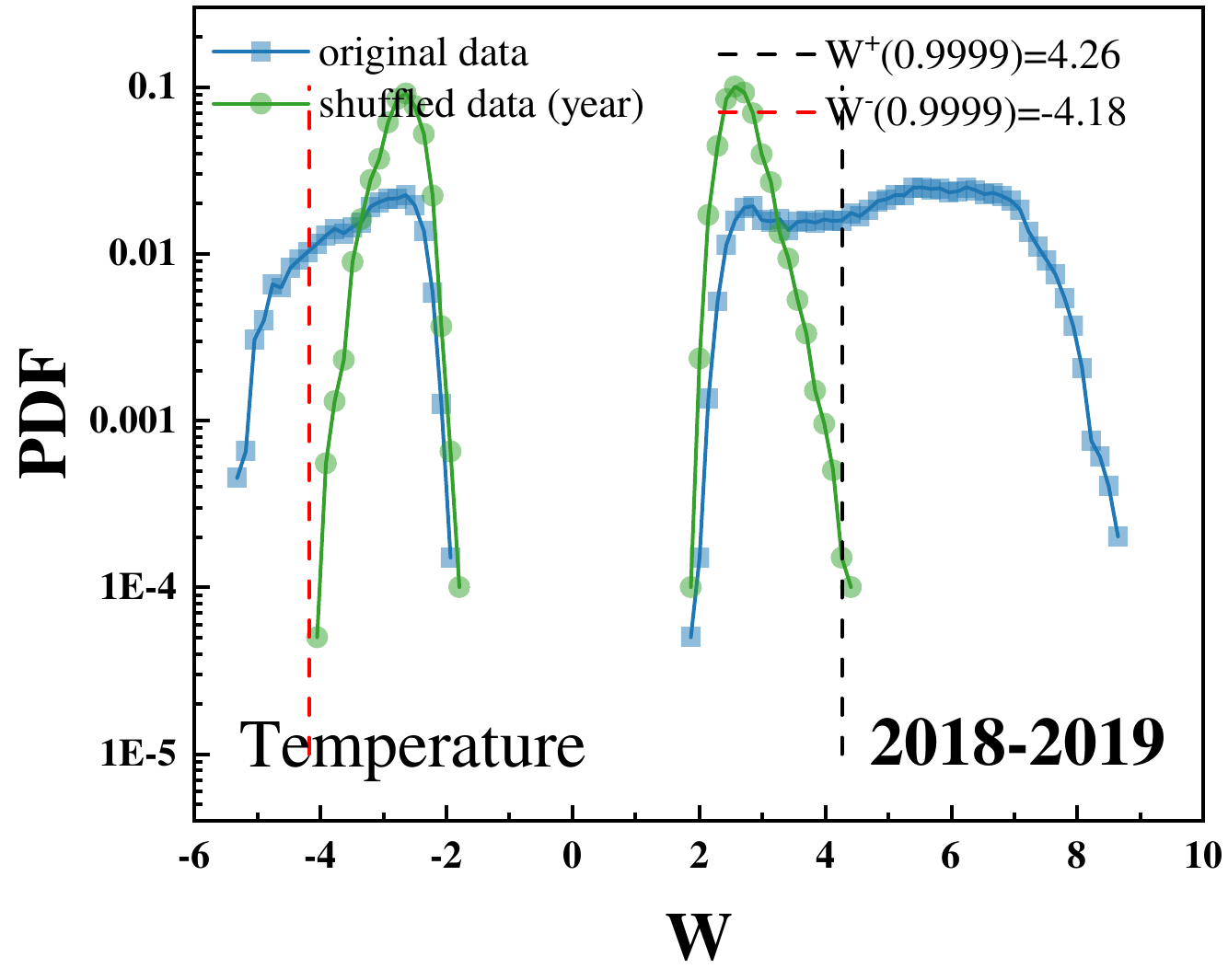}
\end{center}

\begin{center}
\noindent {\small {\bf Fig. S9} Probability distribution function (PDF) of link weights for the original data and shuffled data of temperature in the Contiguous United States. }
\end{center}

\begin{center}
\includegraphics[width=8.5em, height=7em]{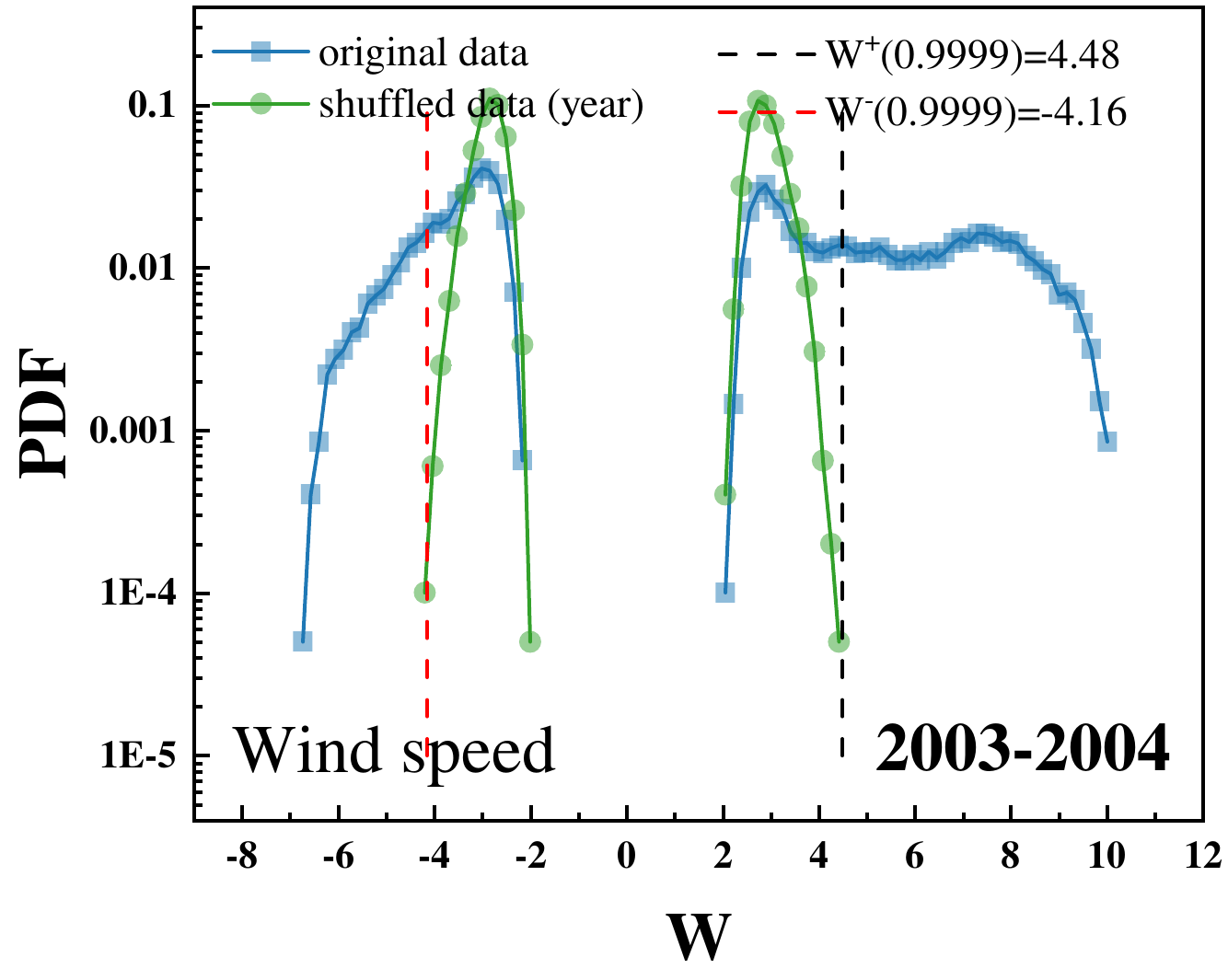}
\includegraphics[width=8.5em, height=7em]{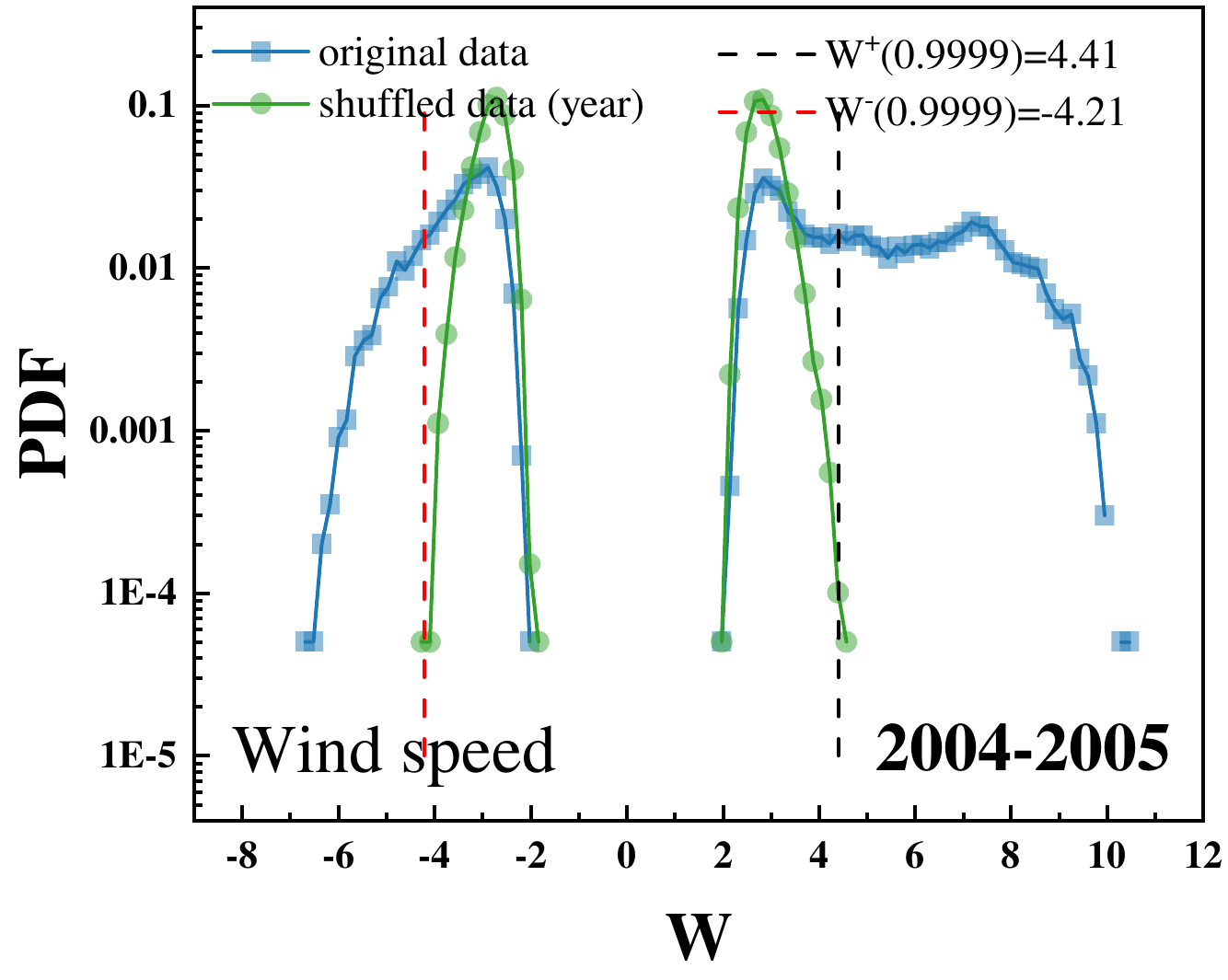}
\includegraphics[width=8.5em, height=7em]{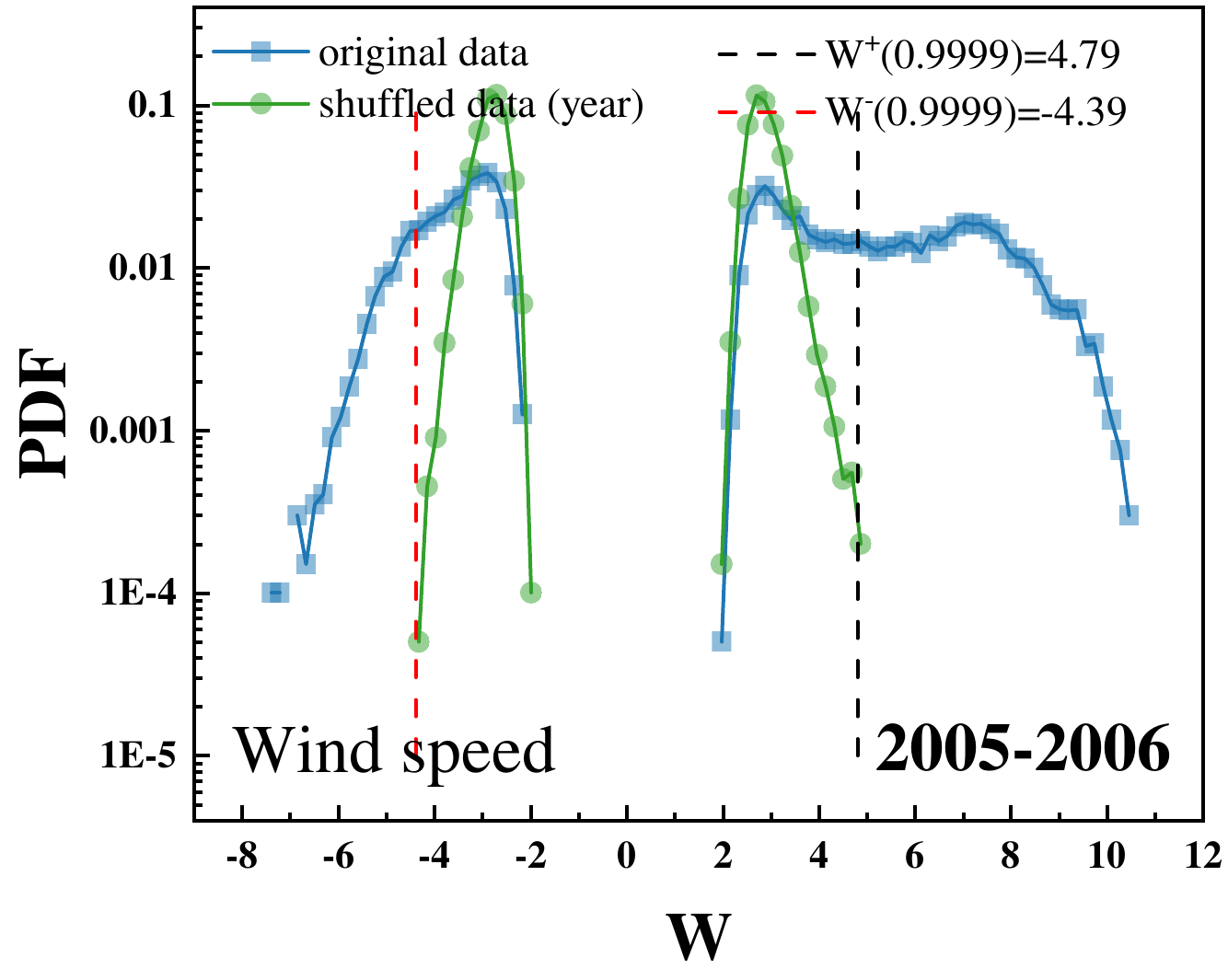}
\includegraphics[width=8.5em, height=7em]{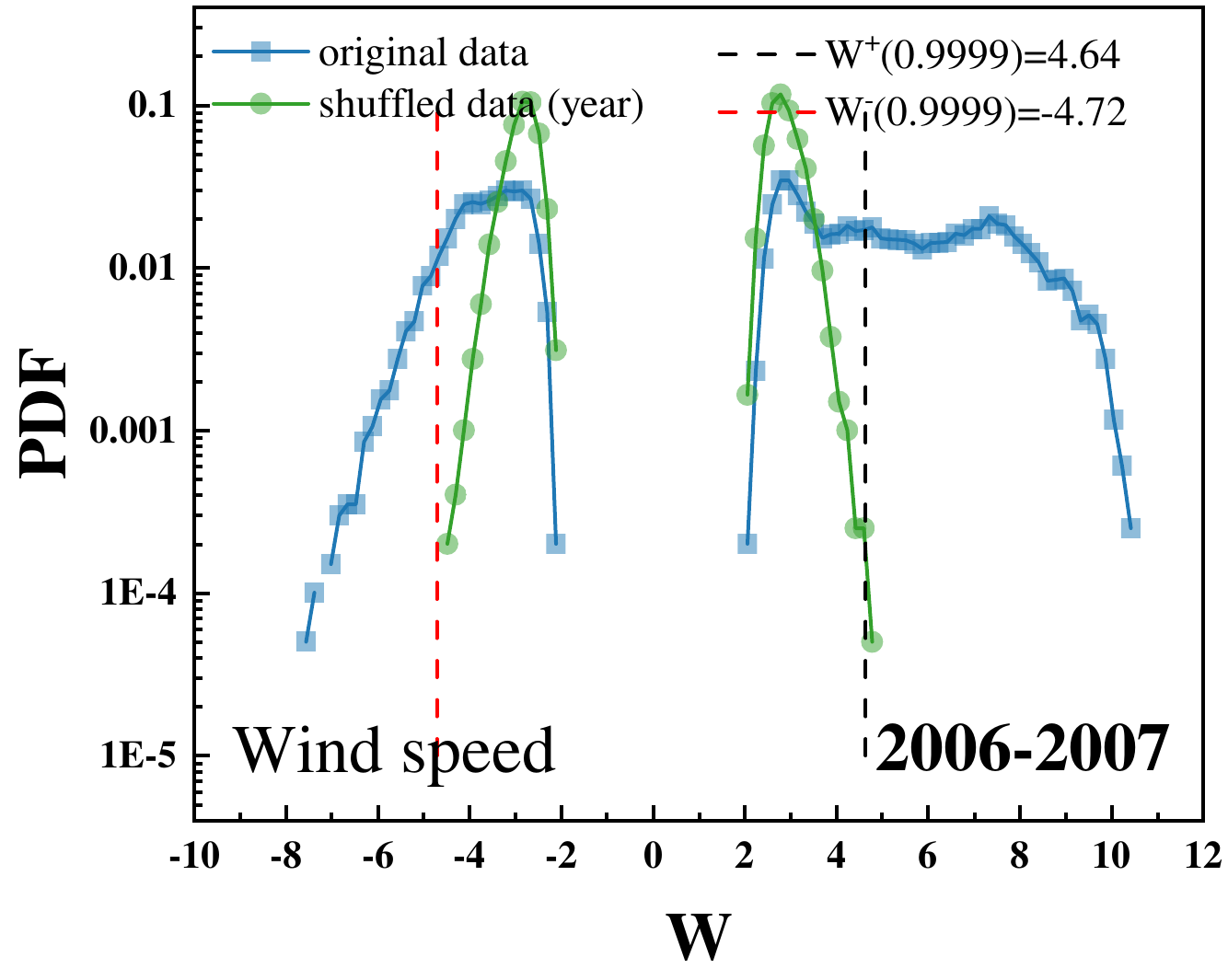}
\includegraphics[width=8.5em, height=7em]{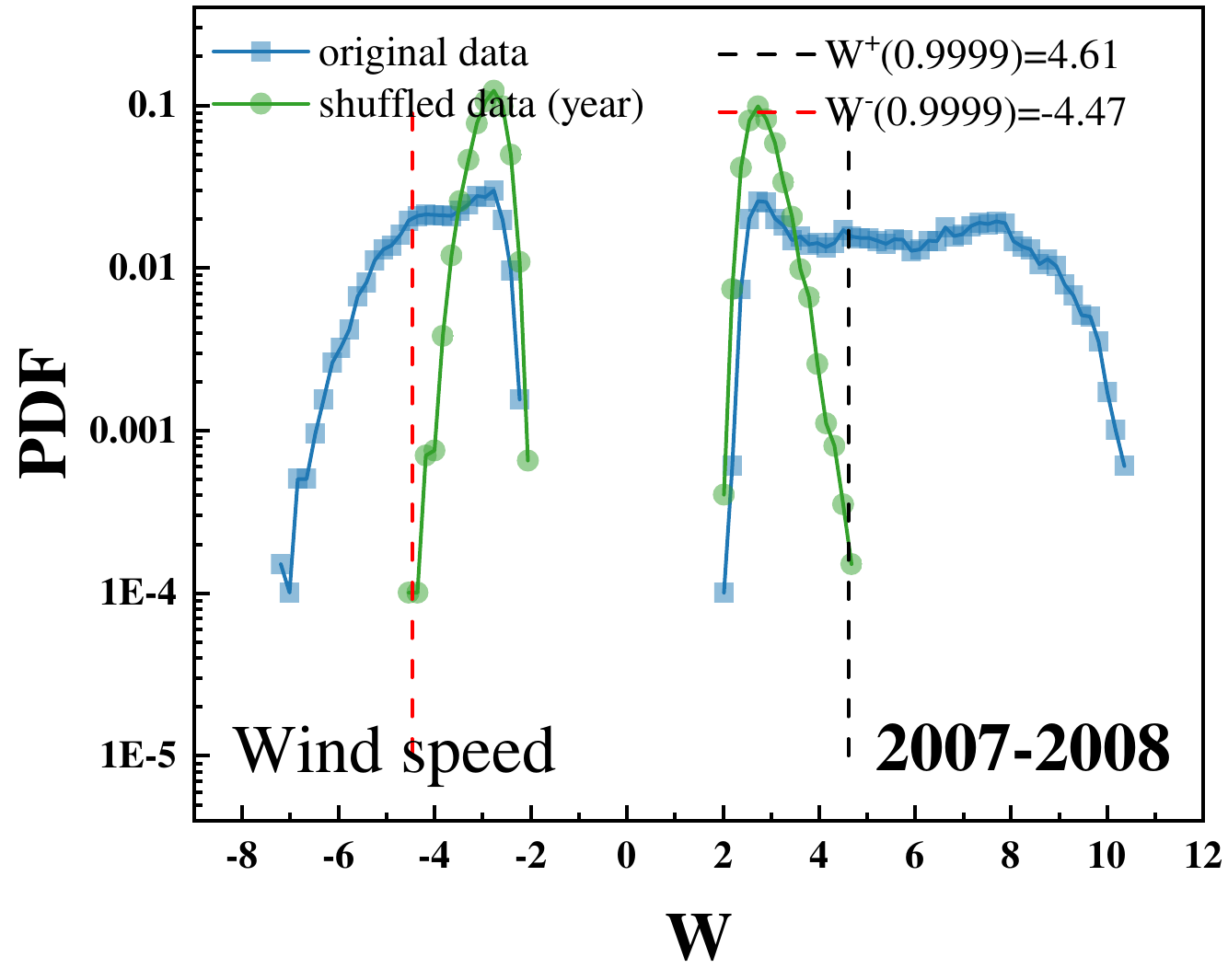}
\includegraphics[width=8.5em, height=7em]{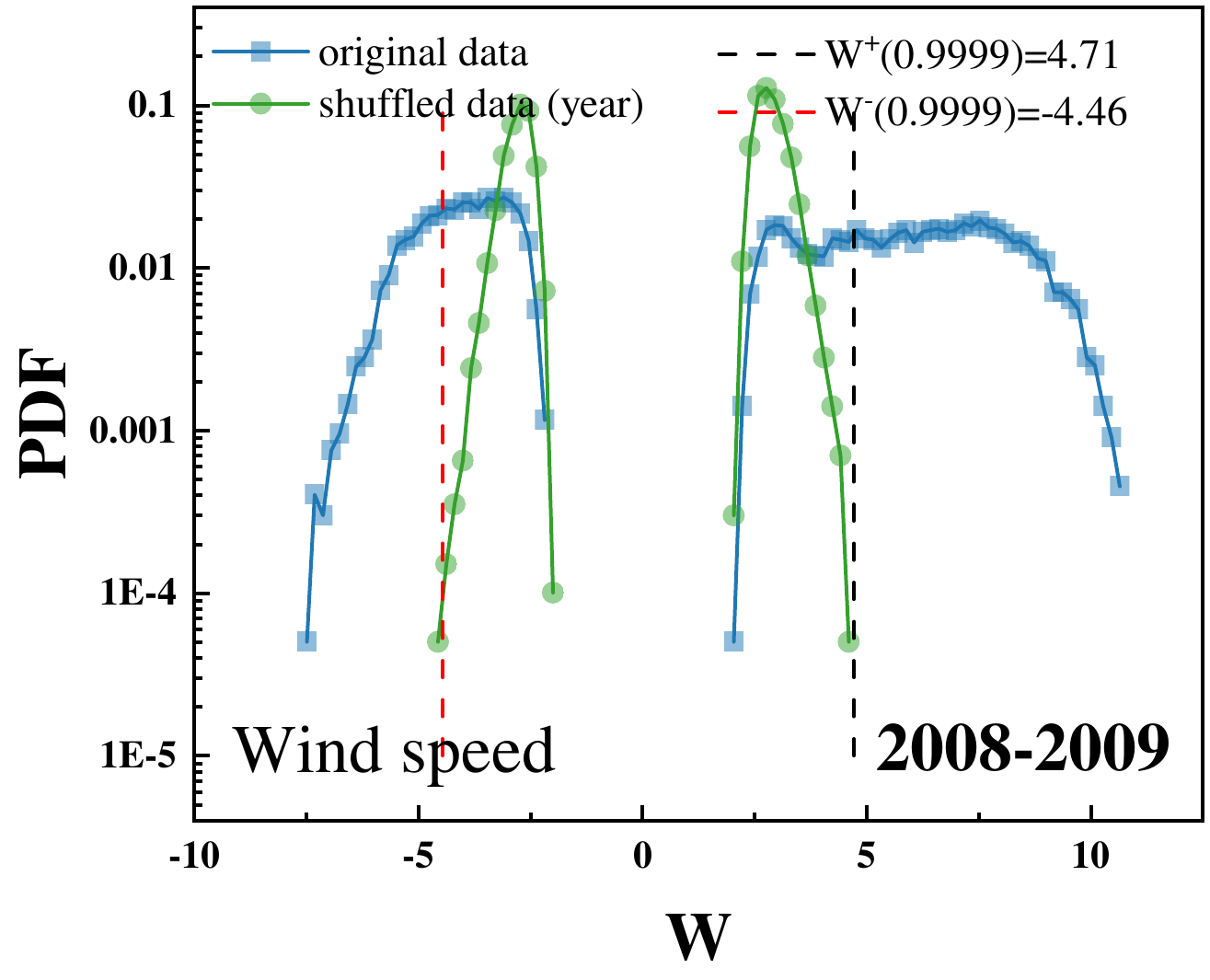}
\includegraphics[width=8.5em, height=7em]{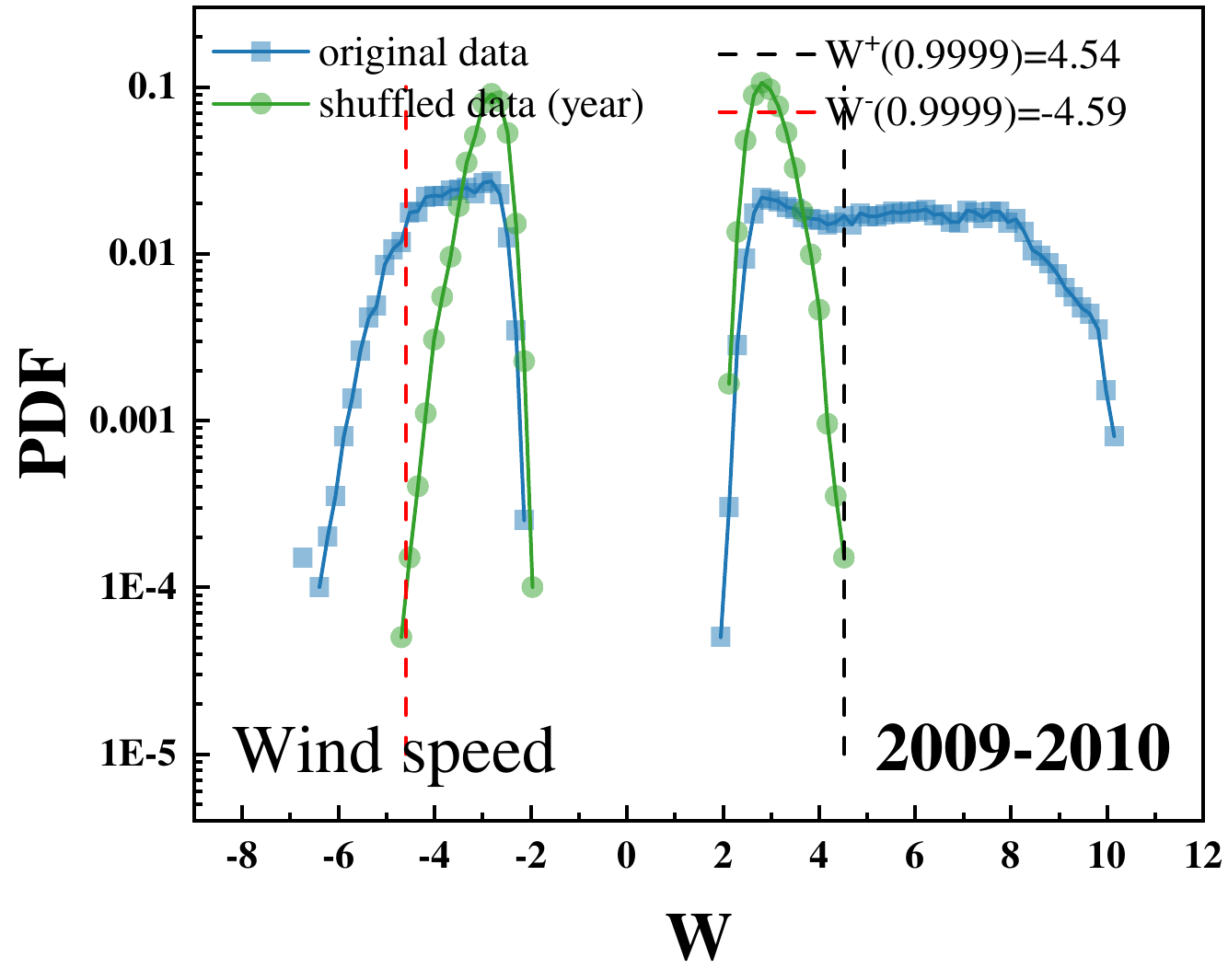}
\includegraphics[width=8.5em, height=7em]{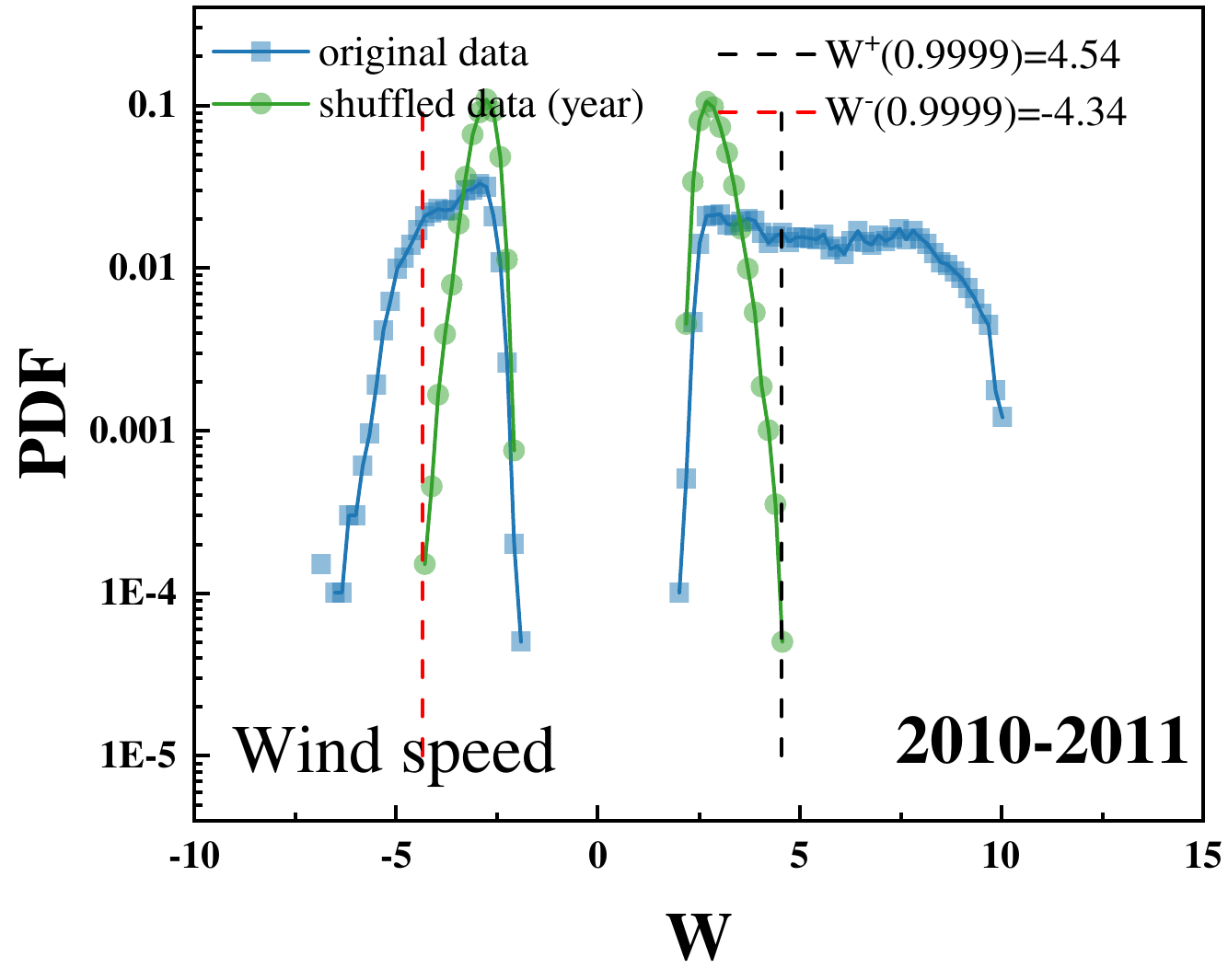}
\includegraphics[width=8.5em, height=7em]{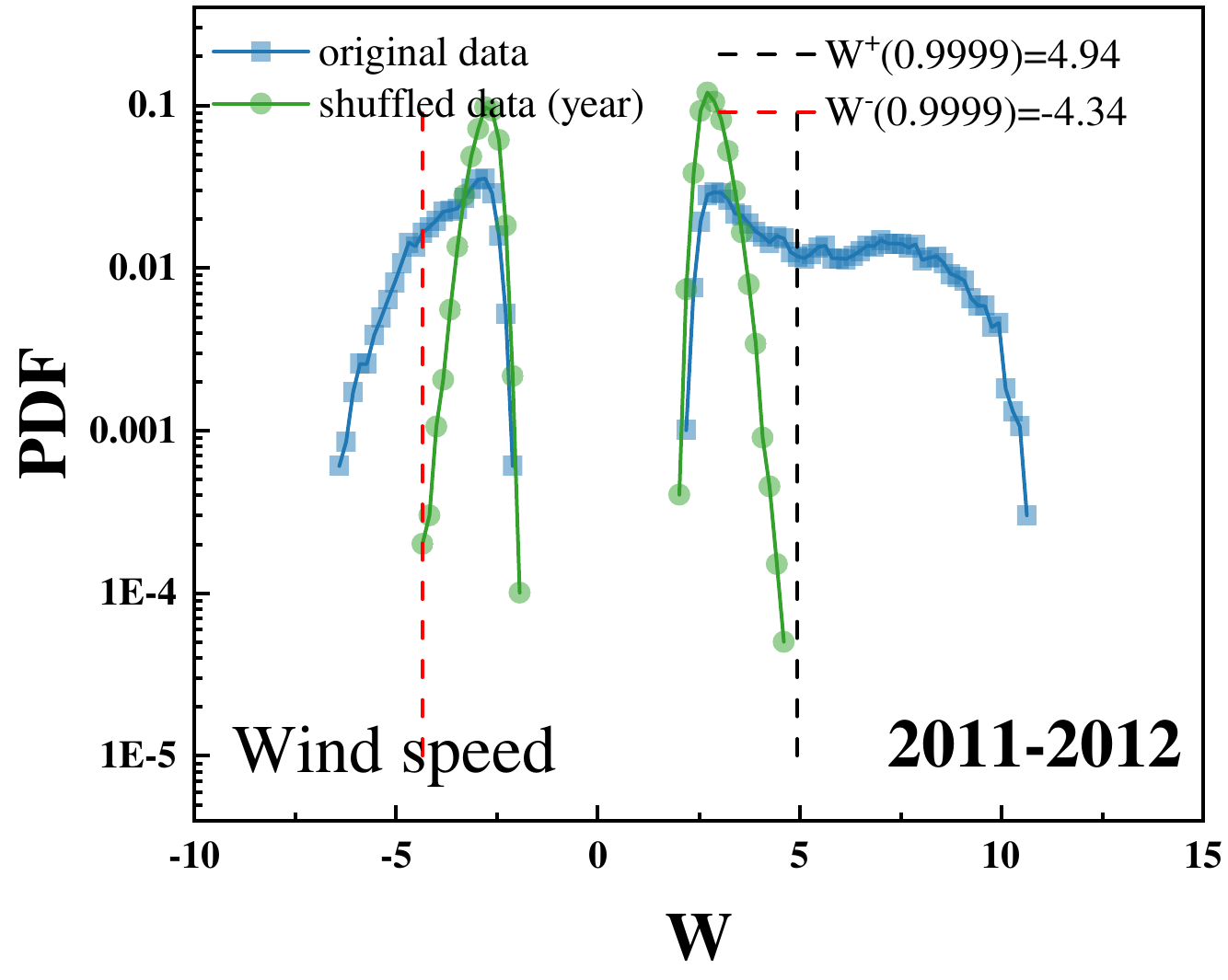}
\includegraphics[width=8.5em, height=7em]{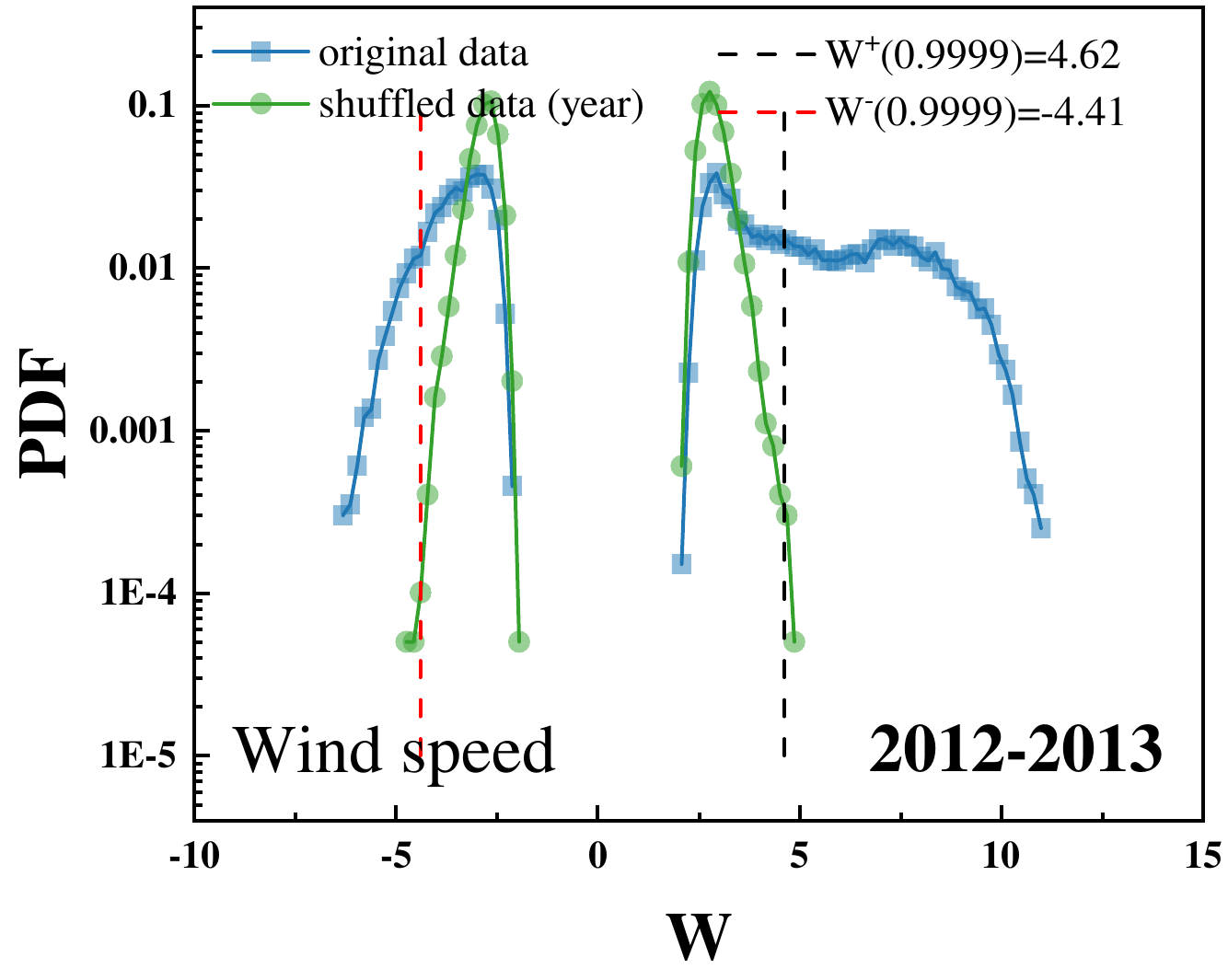}
\includegraphics[width=8.5em, height=7em]{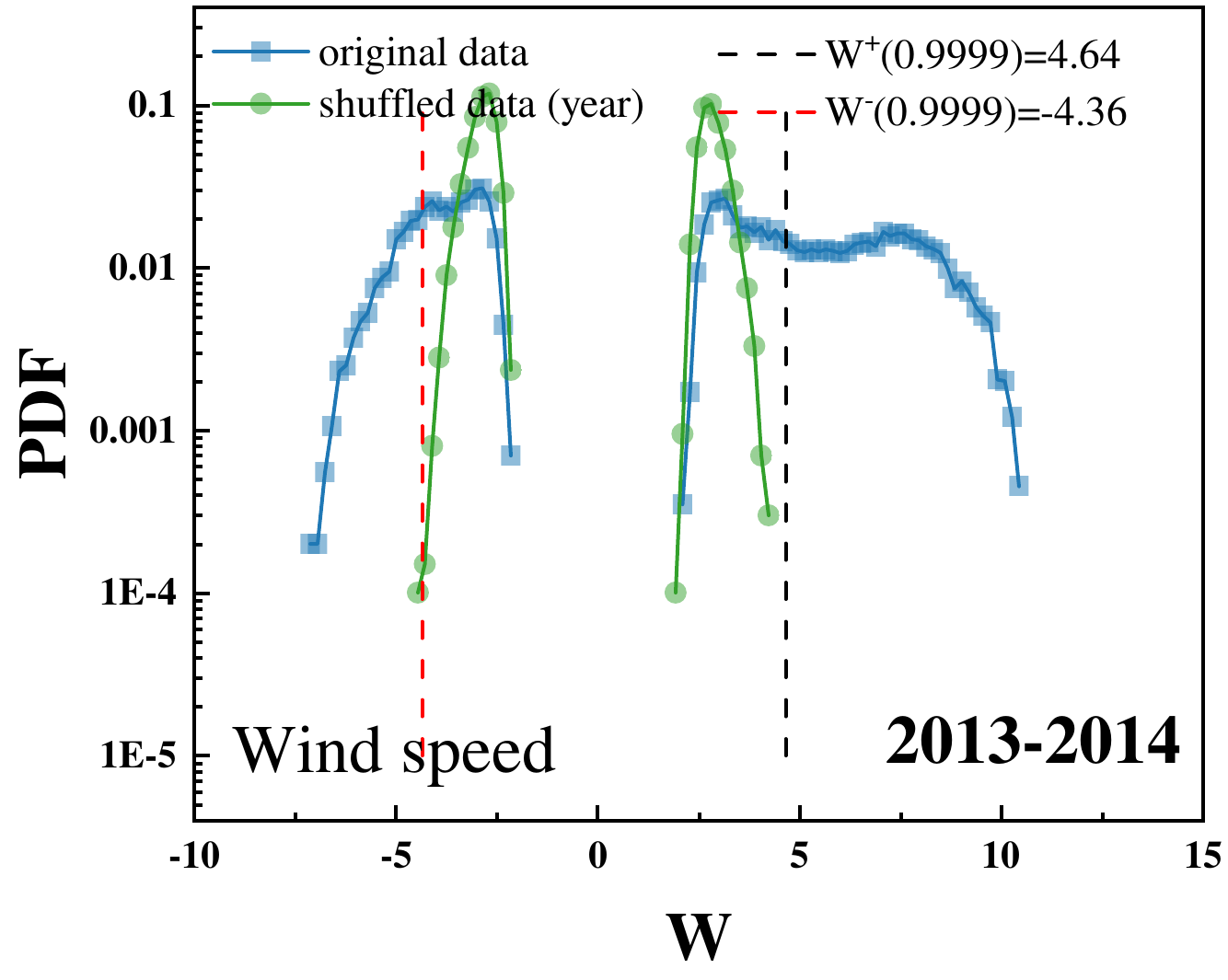}
\includegraphics[width=8.5em, height=7em]{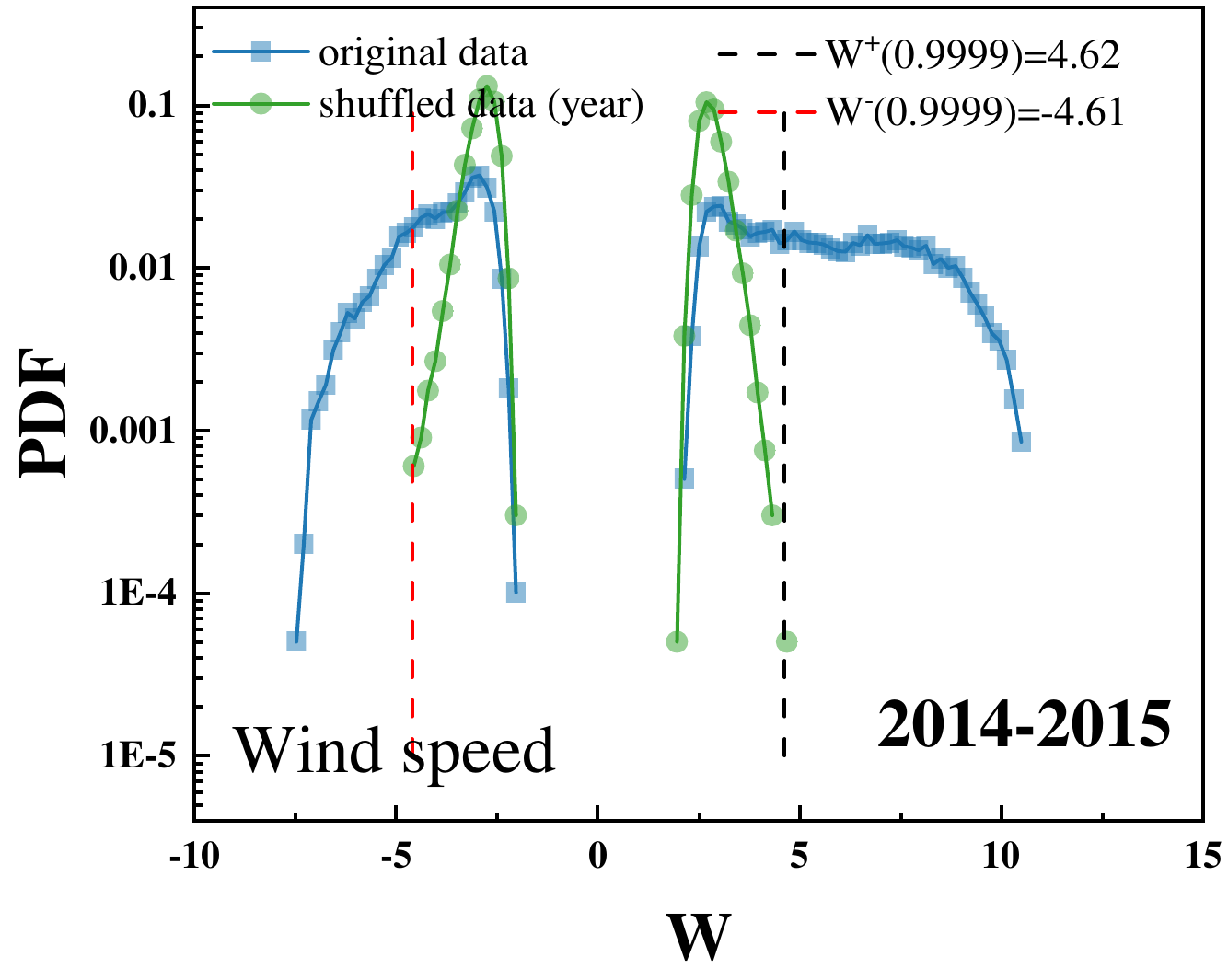}
\includegraphics[width=8.5em, height=7em]{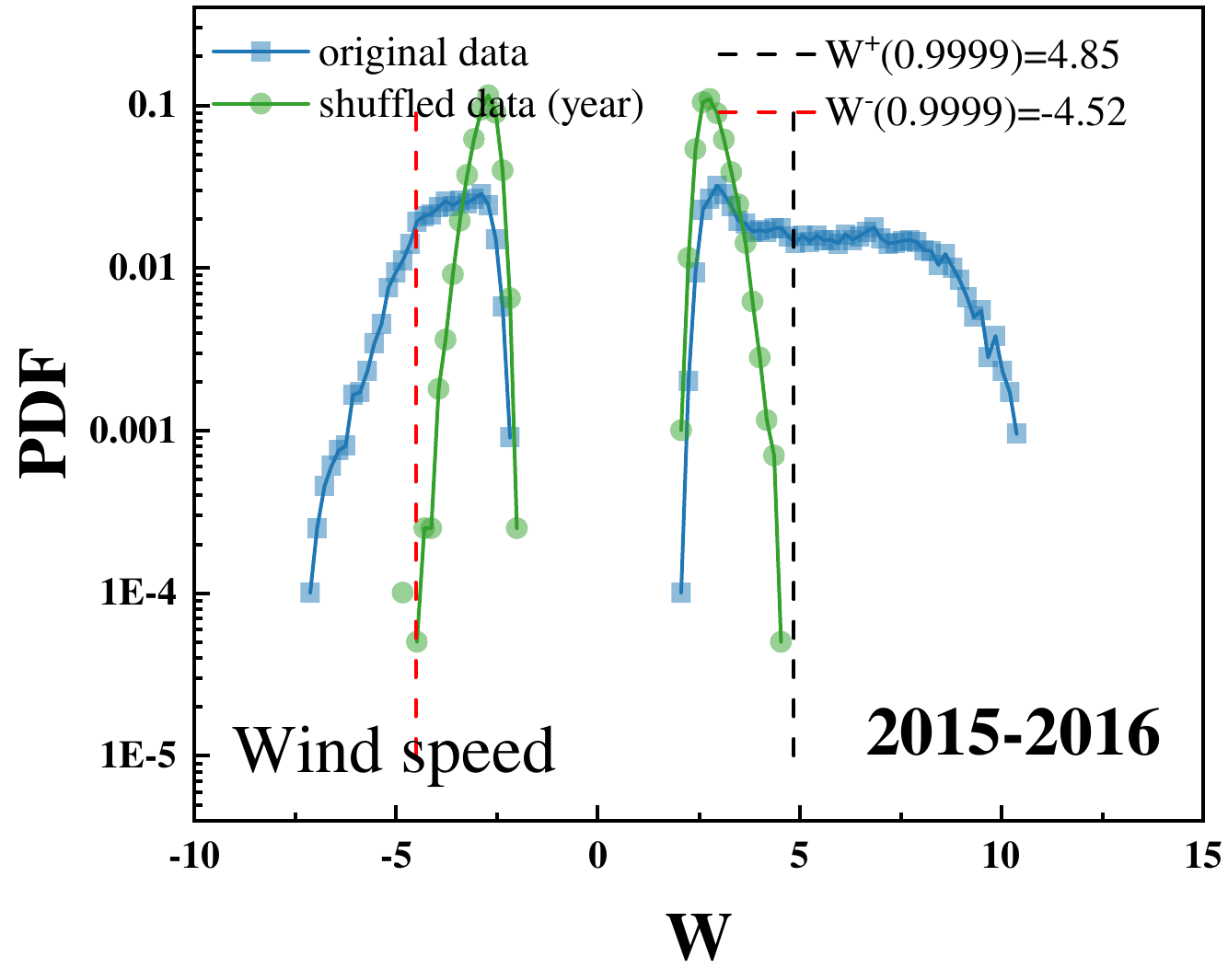}
\includegraphics[width=8.5em, height=7em]{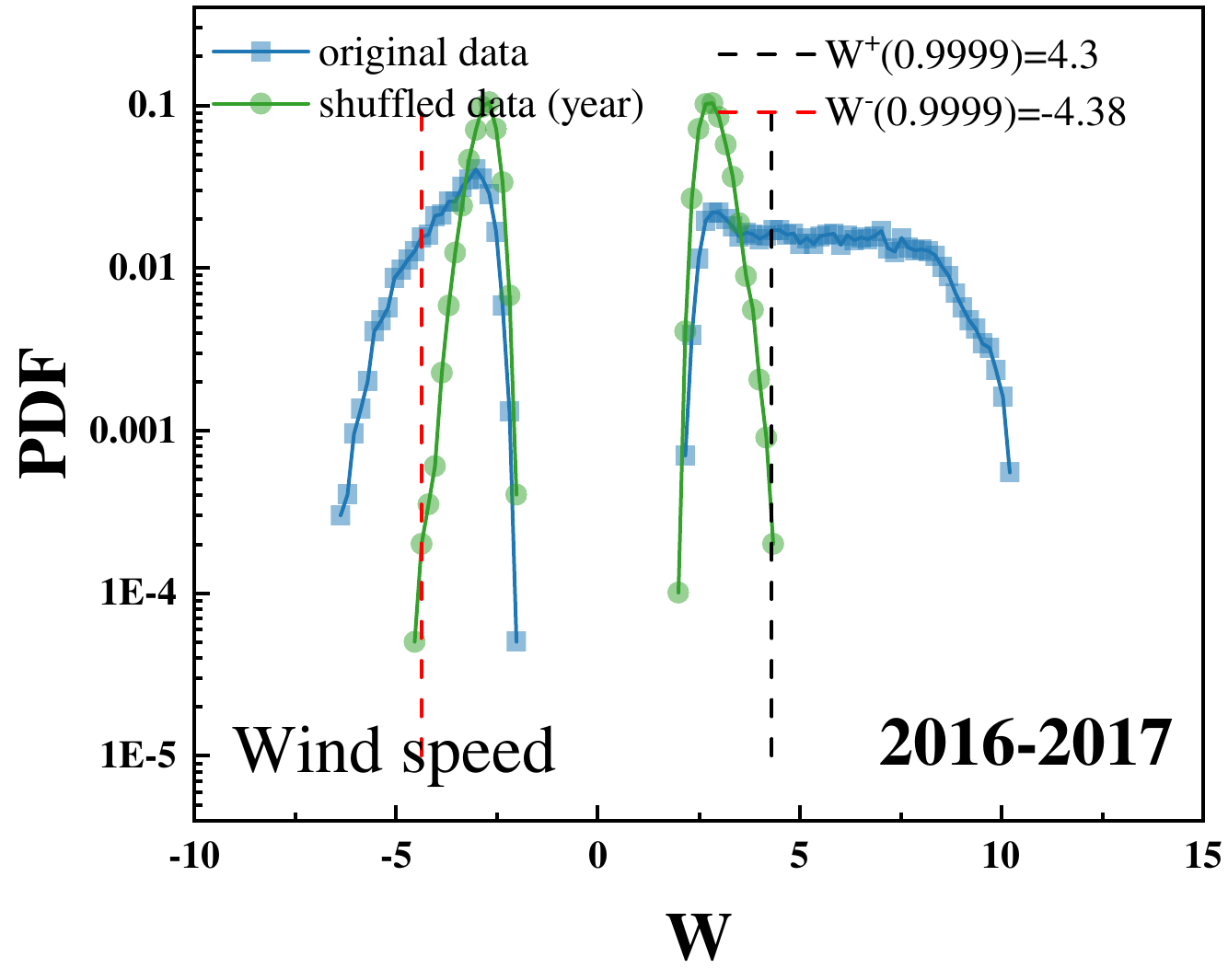}
\includegraphics[width=8.5em, height=7em]{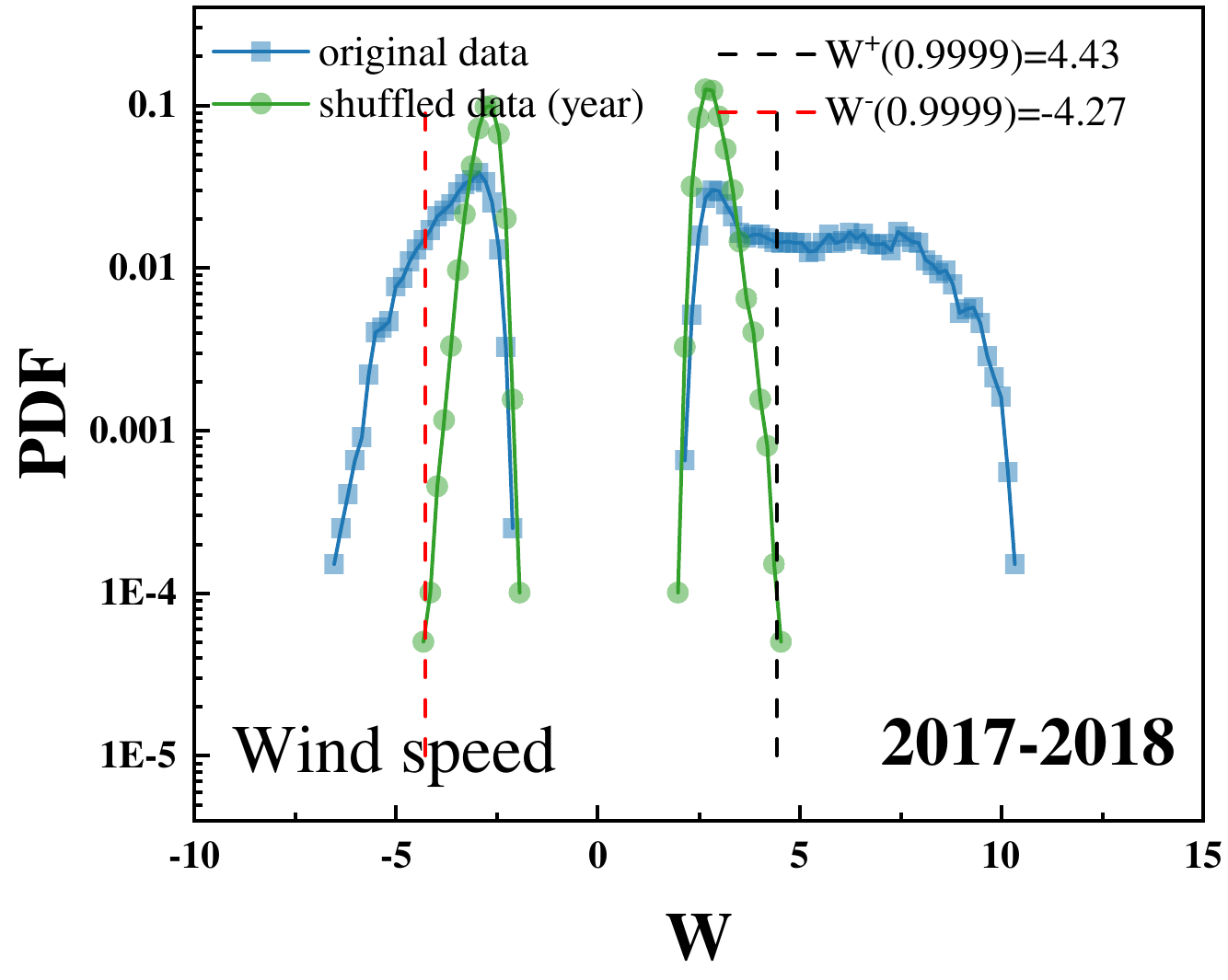}
\includegraphics[width=8.5em, height=7em]{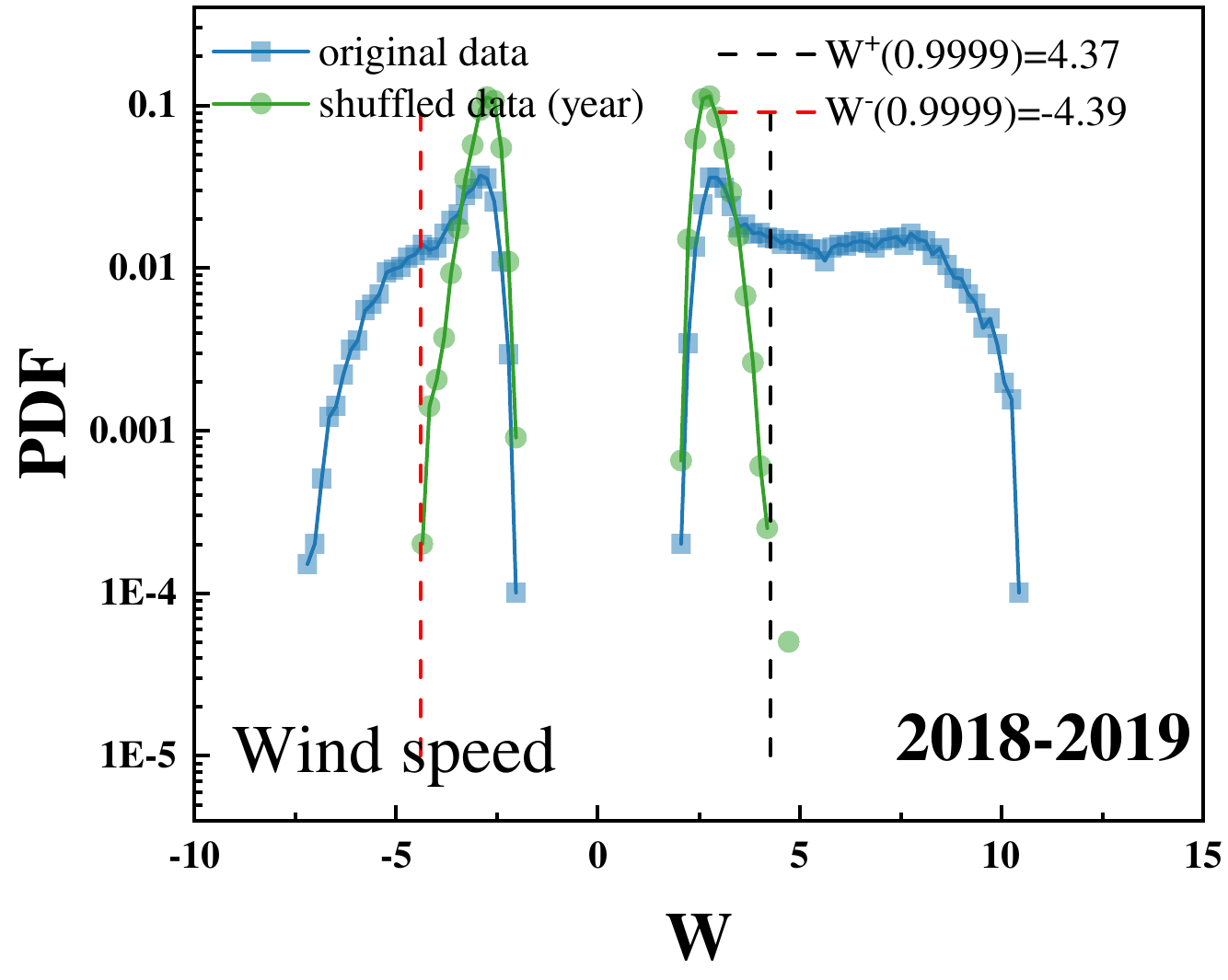}
\end{center}

\begin{center}
\noindent {\small {\bf Fig. S10} Probability distribution function (PDF) of link weights for the original data and shuffled data of wind speed in the Contiguous United States. }
\end{center}

\begin{center}
\includegraphics[width=8.5em, height=7em]{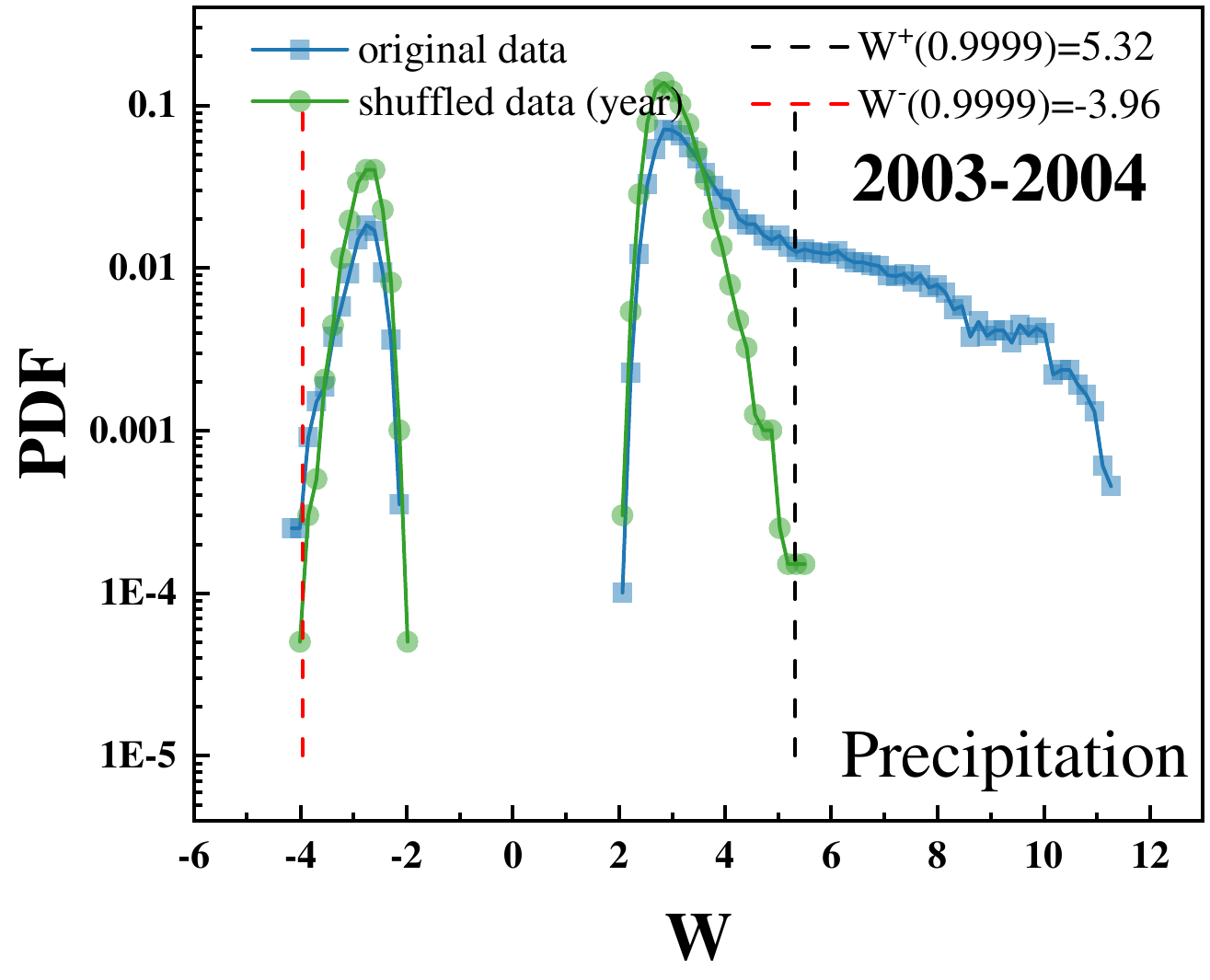}
\includegraphics[width=8.5em, height=7em]{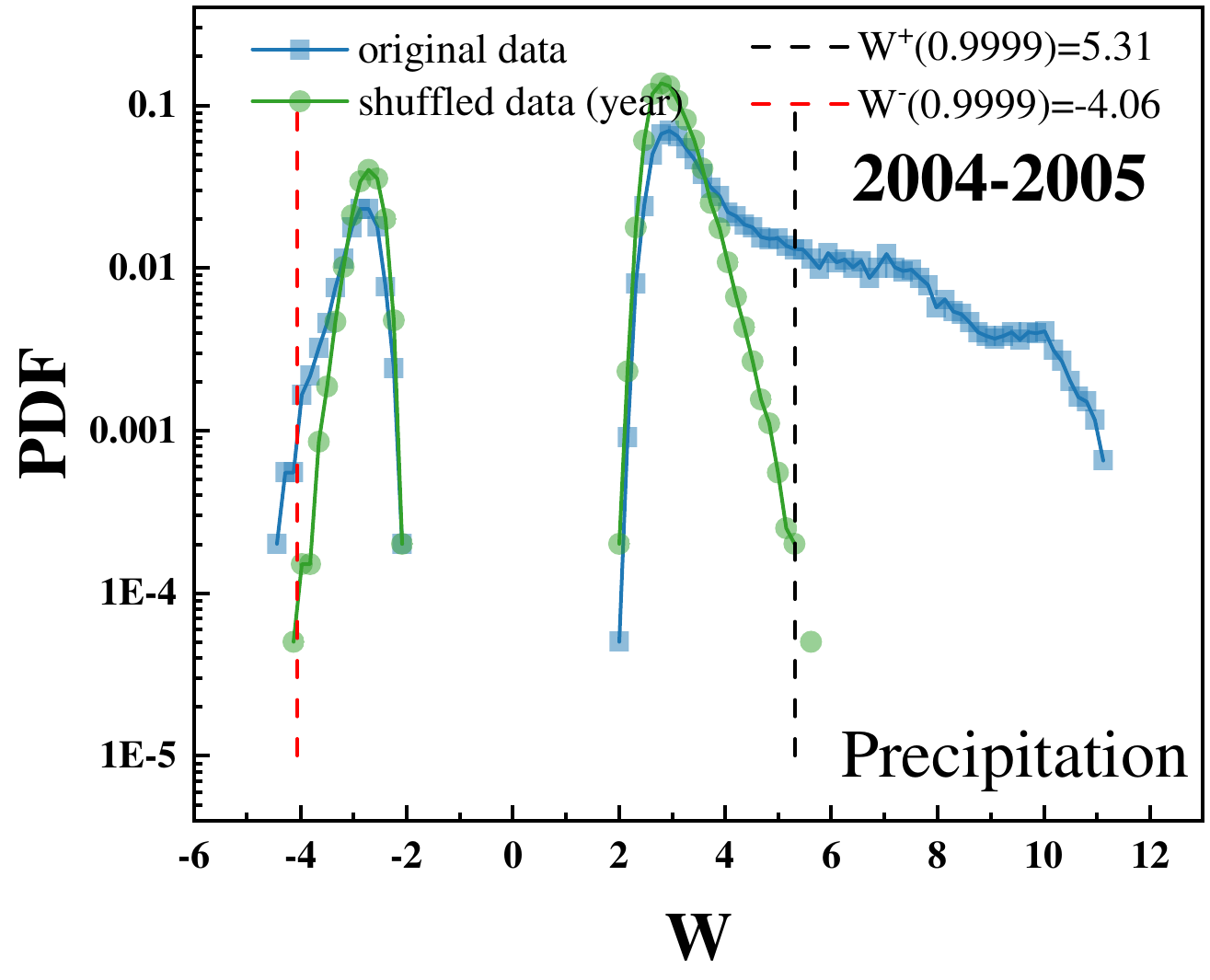}
\includegraphics[width=8.5em, height=7em]{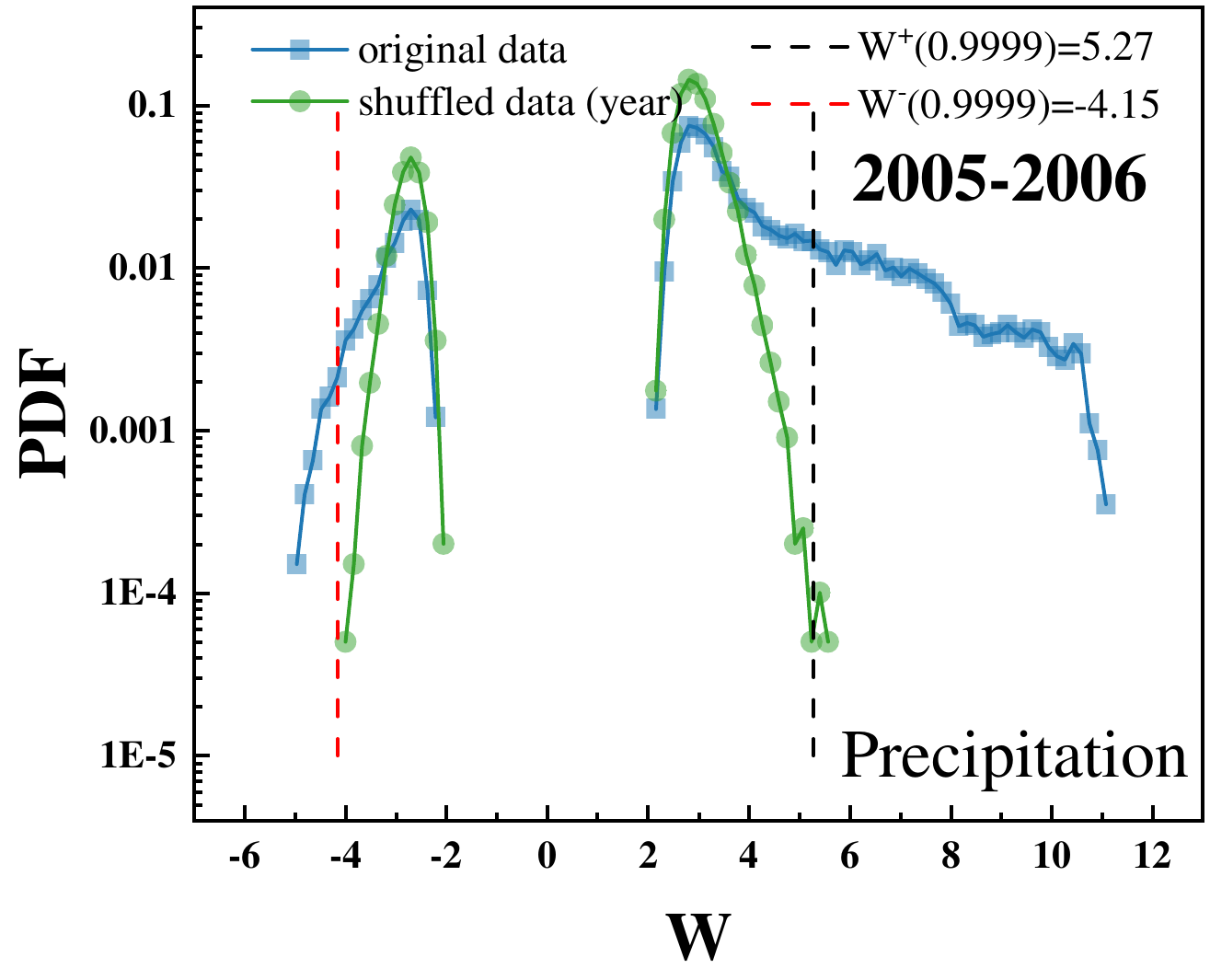}
\includegraphics[width=8.5em, height=7em]{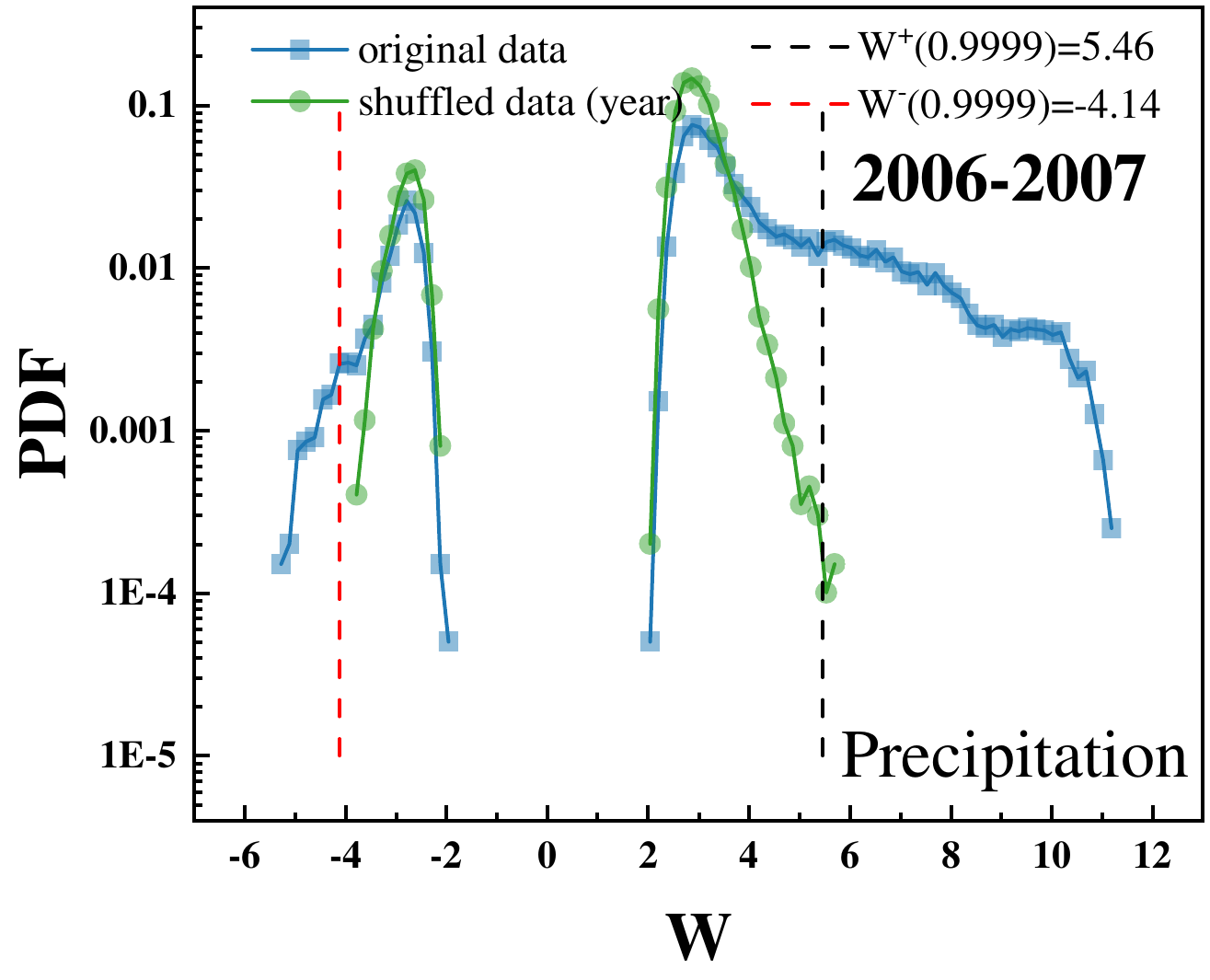}
\includegraphics[width=8.5em, height=7em]{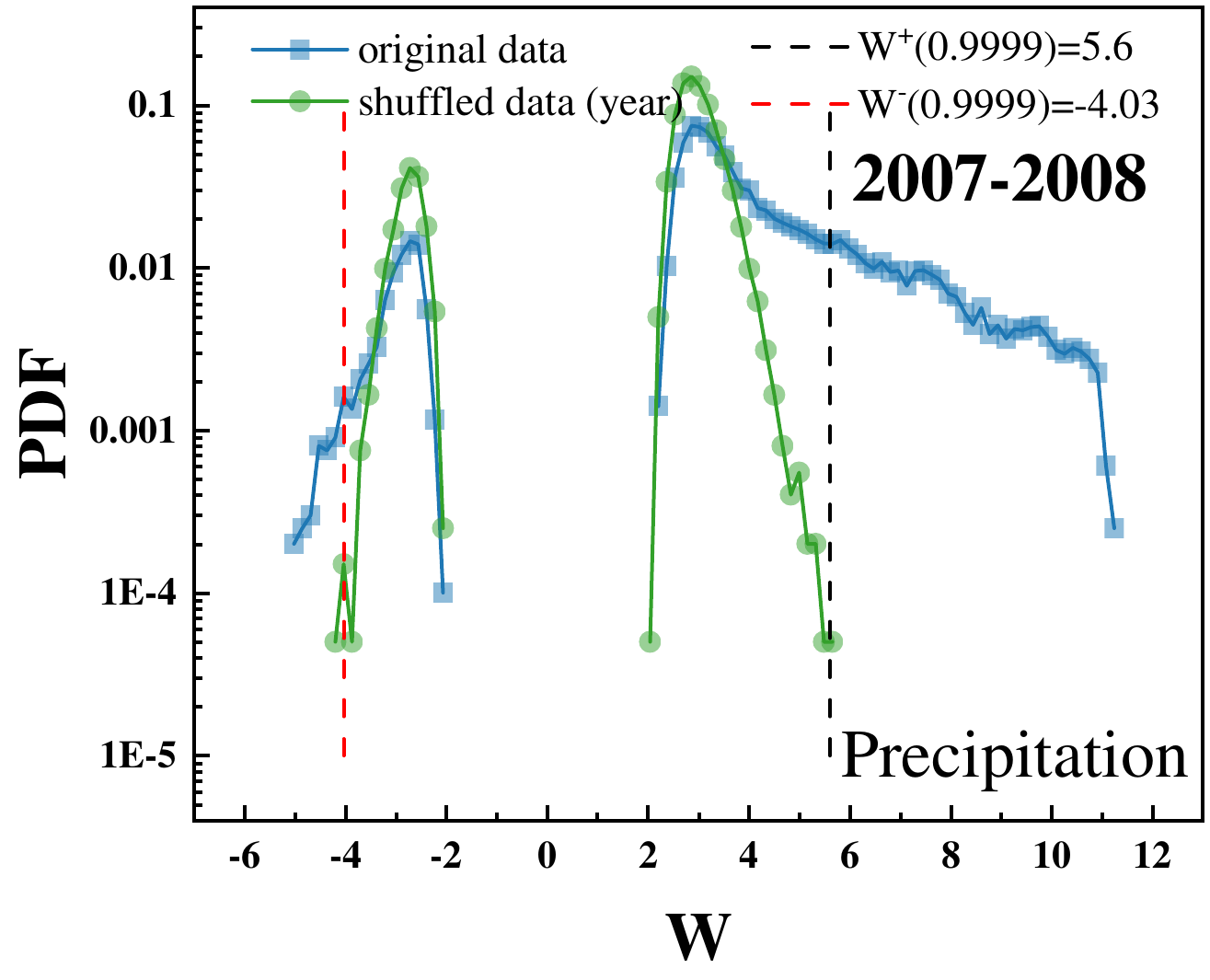}
\includegraphics[width=8.5em, height=7em]{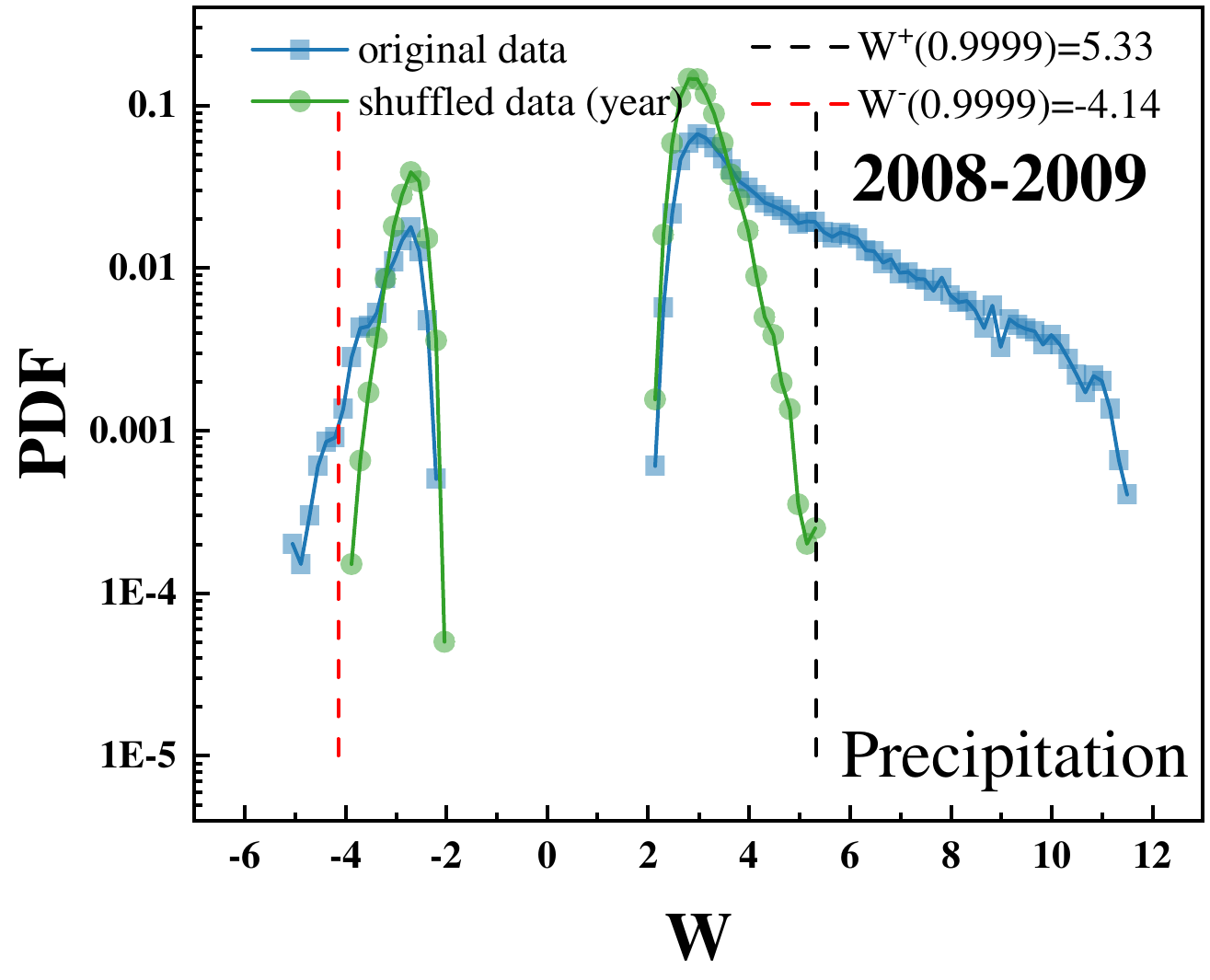}
\includegraphics[width=8.5em, height=7em]{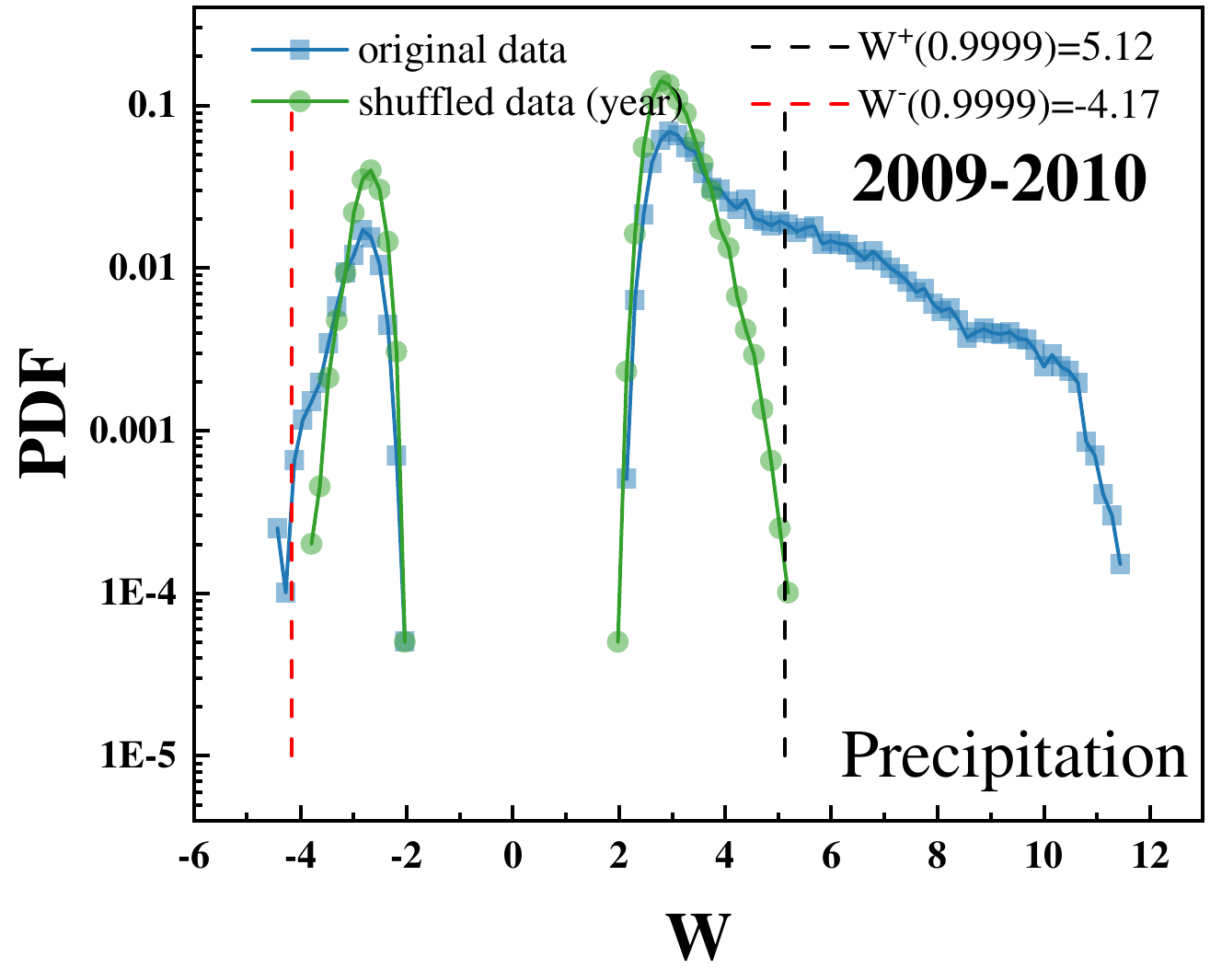}
\includegraphics[width=8.5em, height=7em]{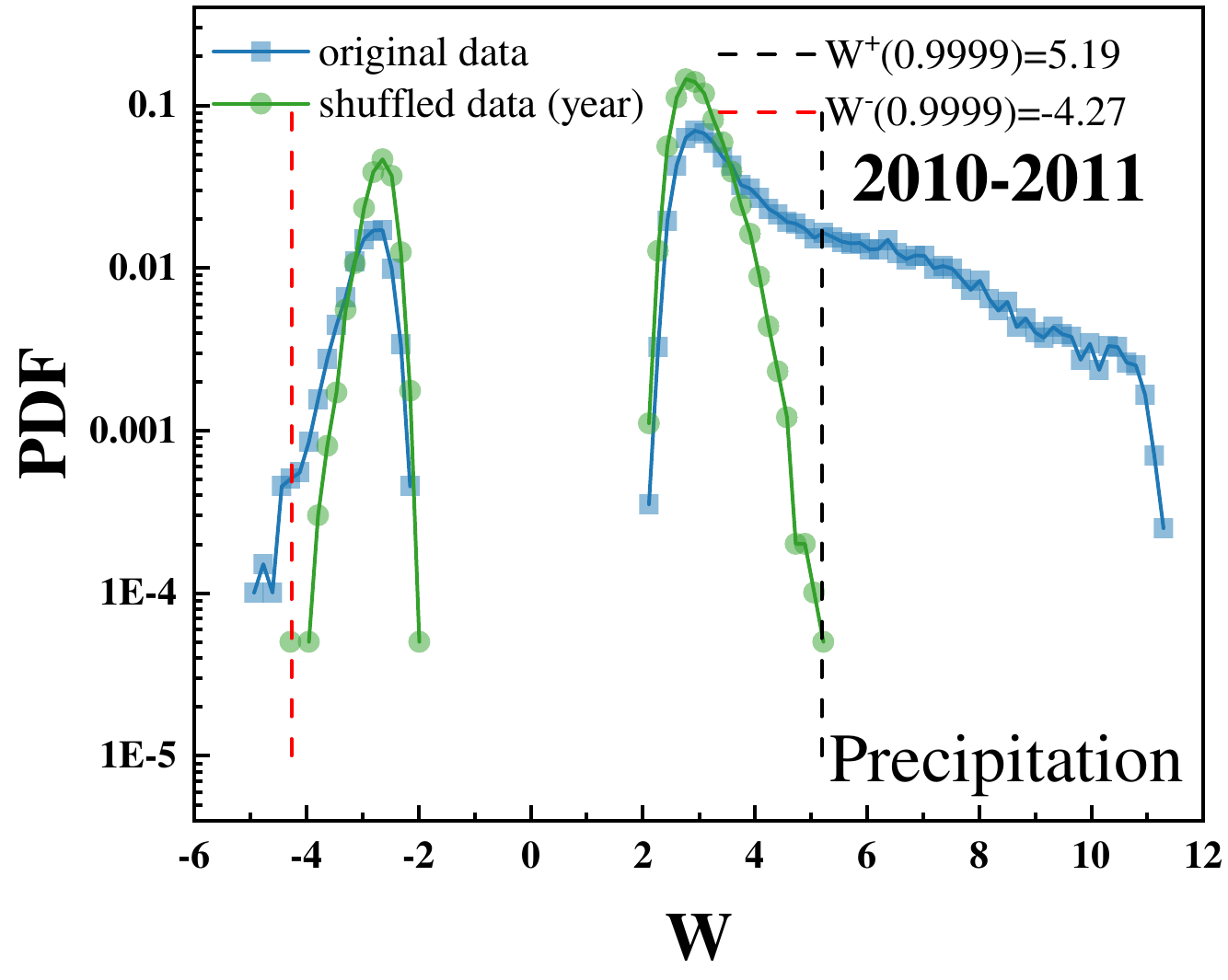}
\includegraphics[width=8.5em, height=7em]{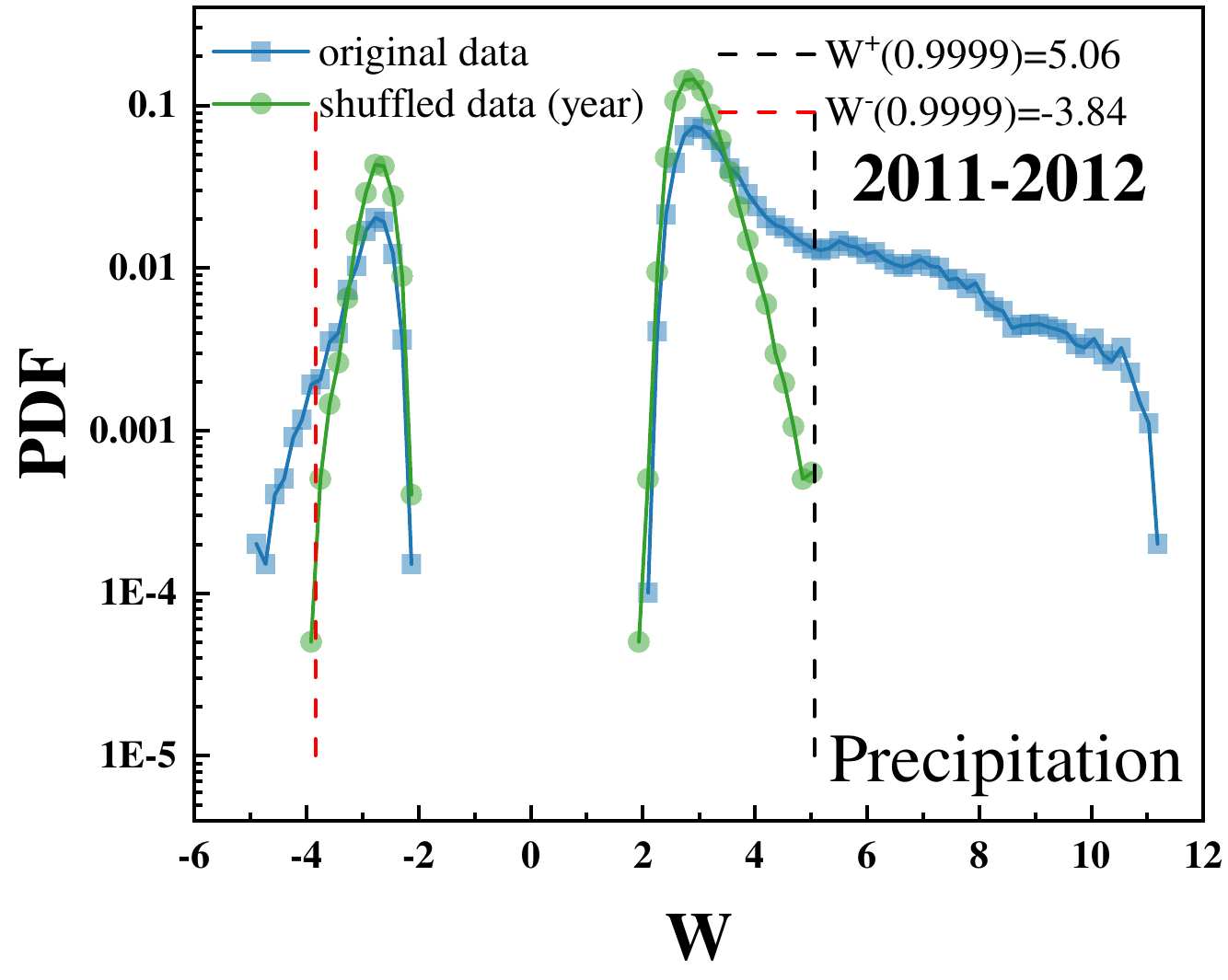}
\includegraphics[width=8.5em, height=7em]{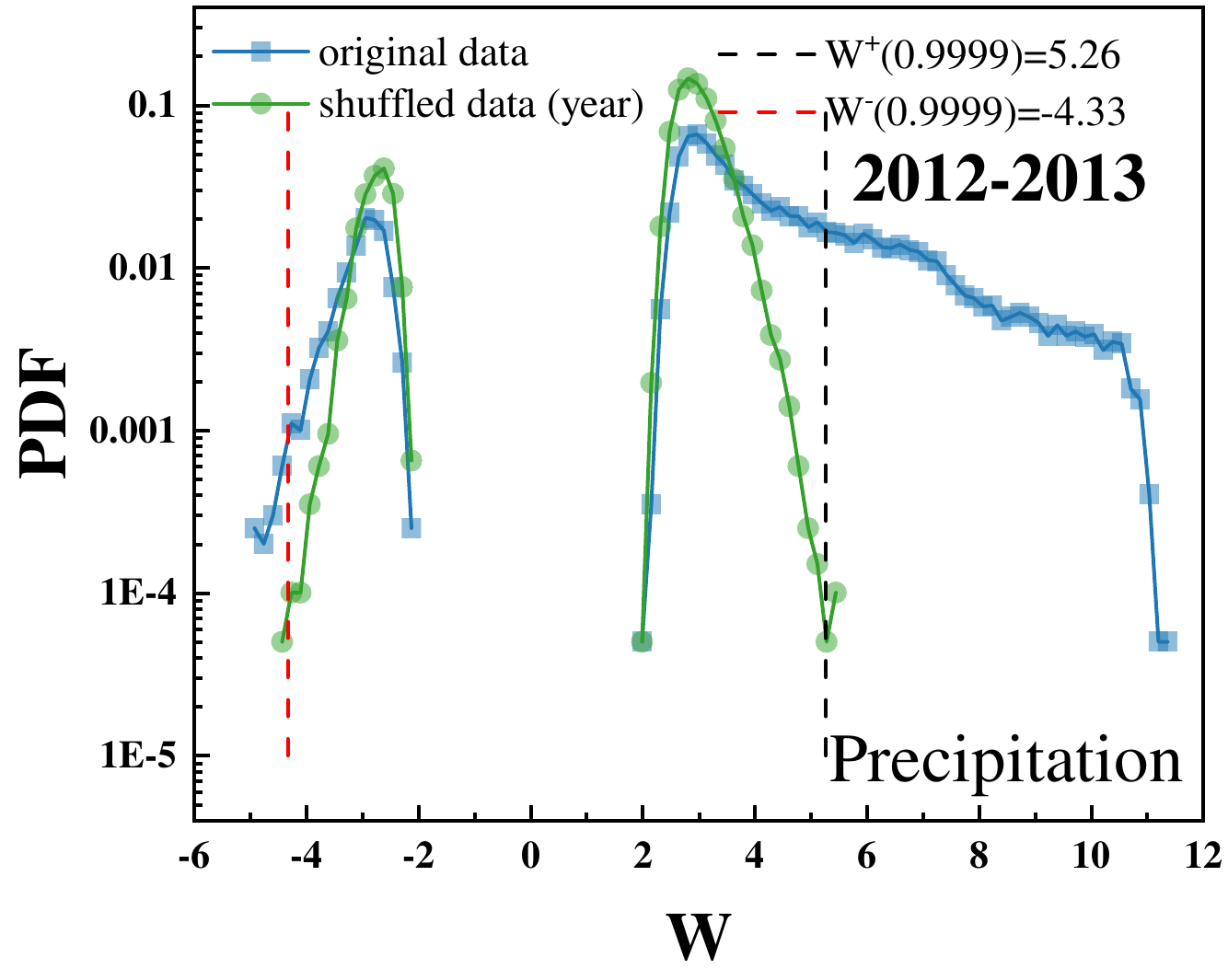}
\includegraphics[width=8.5em, height=7em]{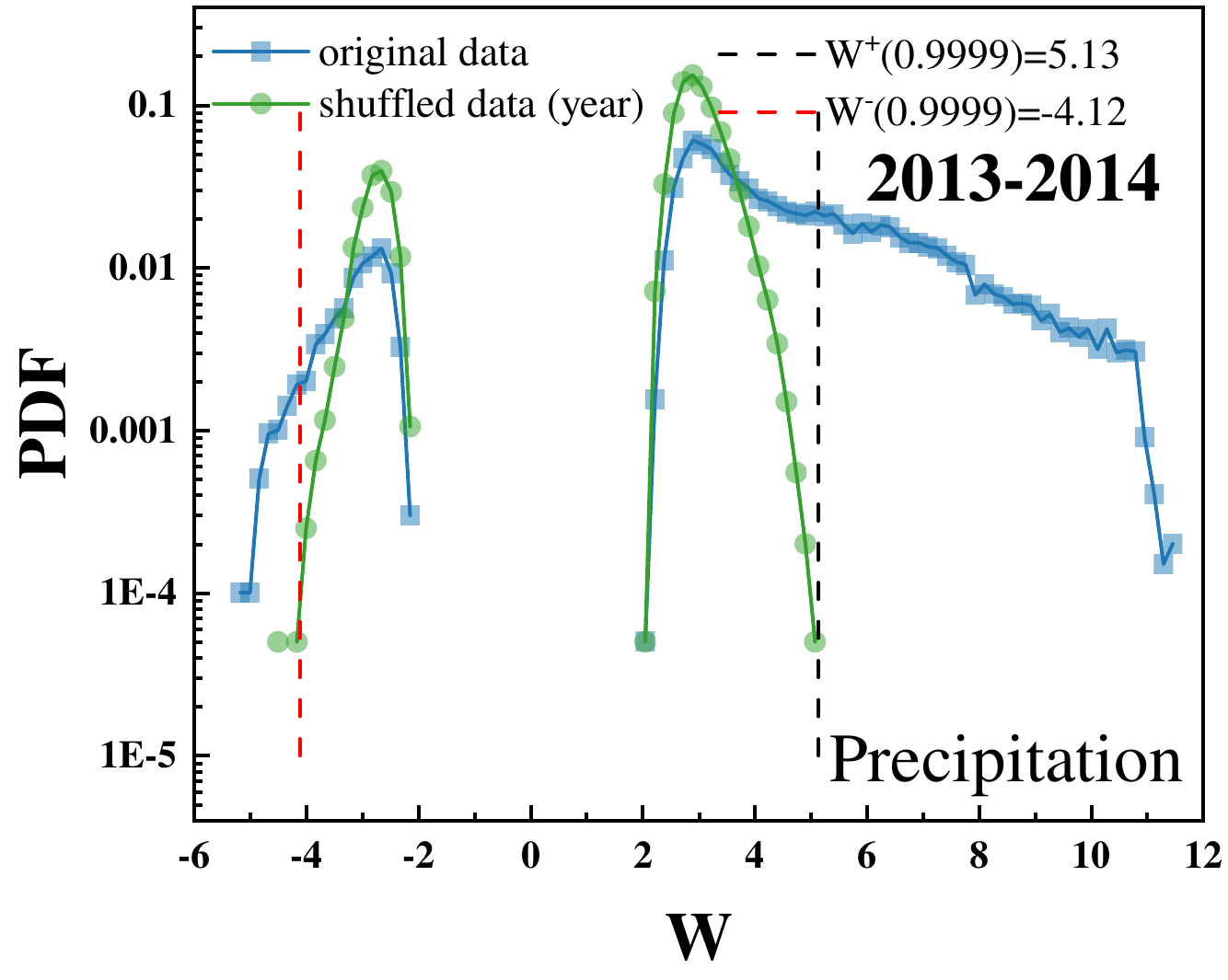}
\includegraphics[width=8.5em, height=7em]{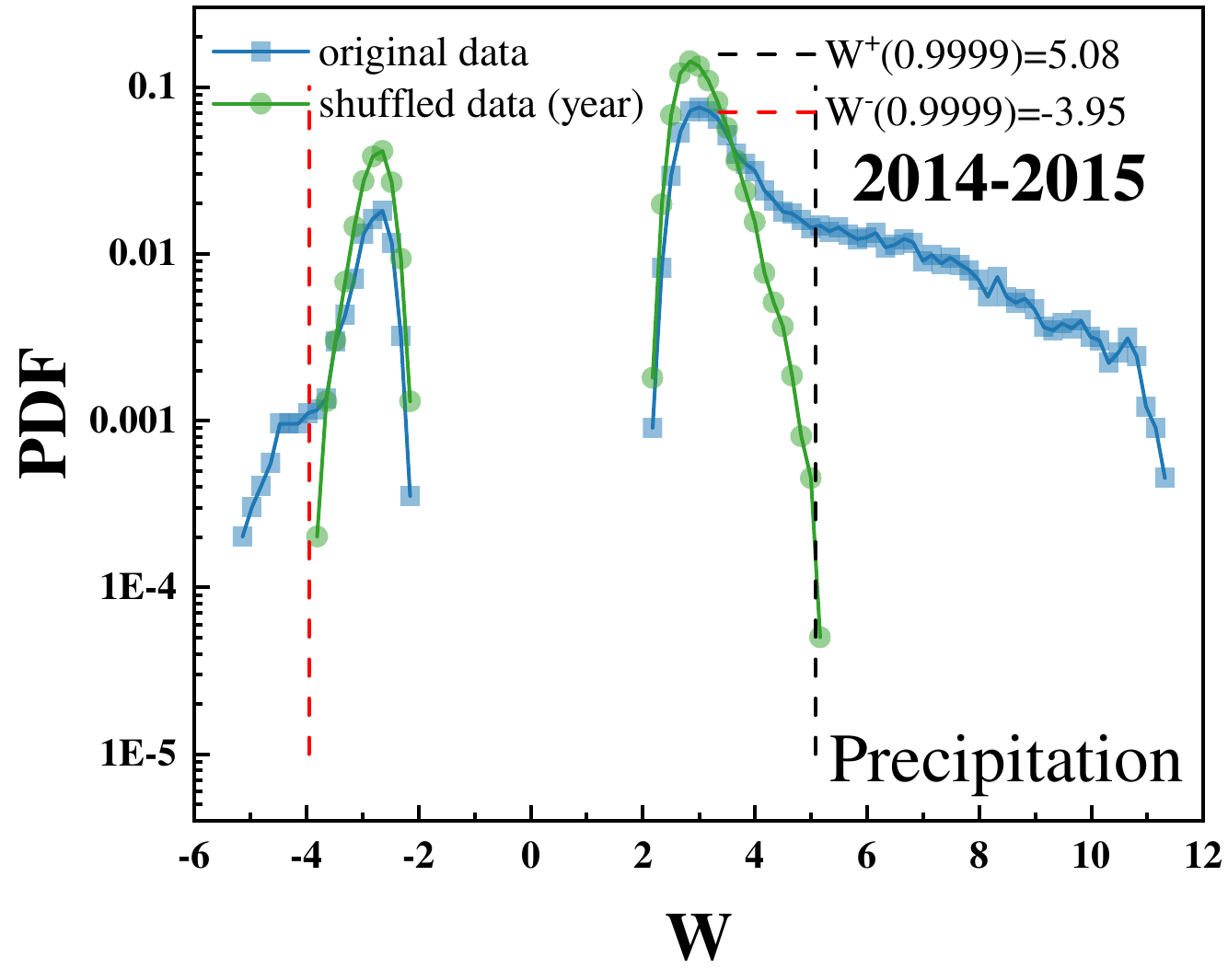}
\includegraphics[width=8.5em, height=7em]{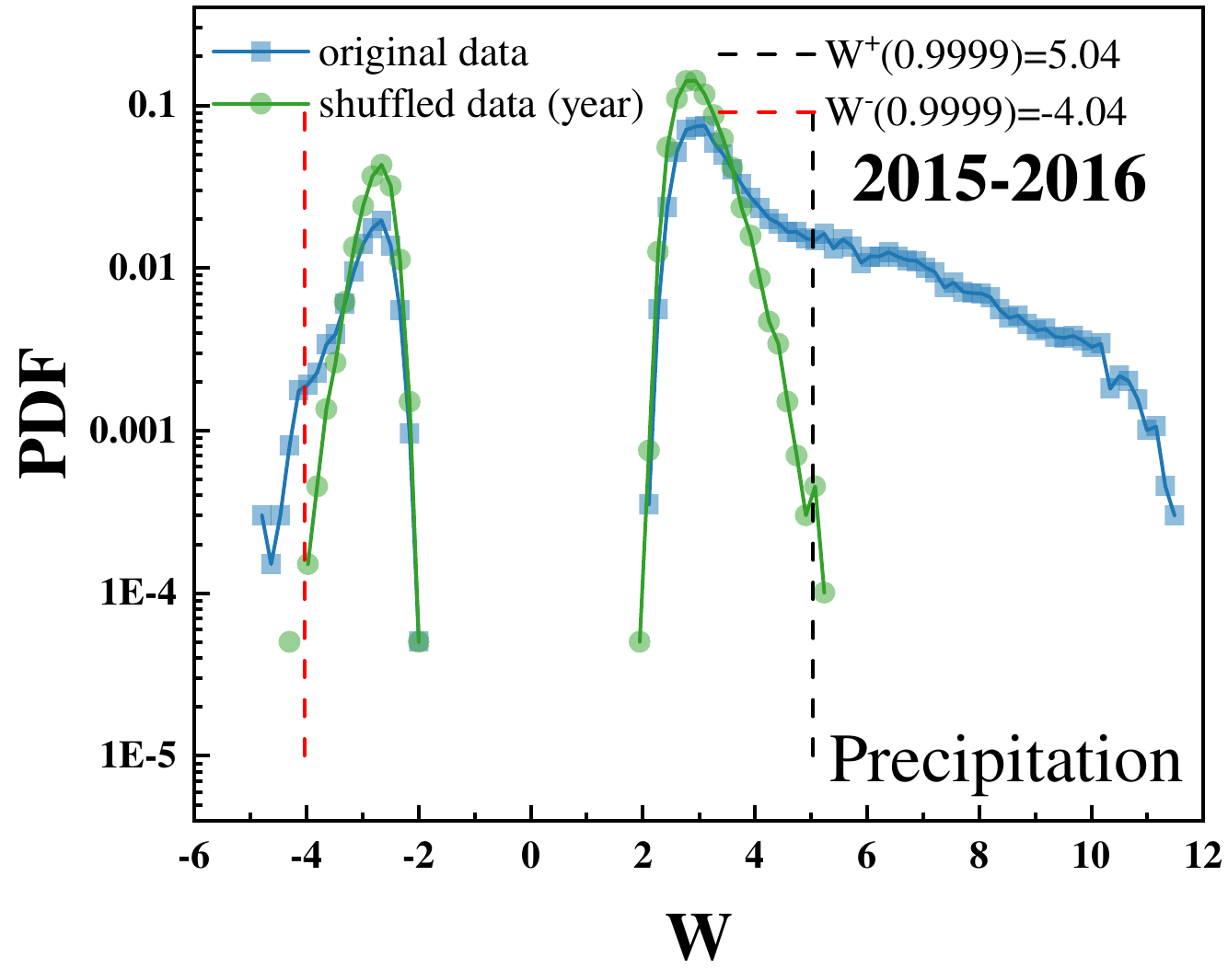}
\includegraphics[width=8.5em, height=7em]{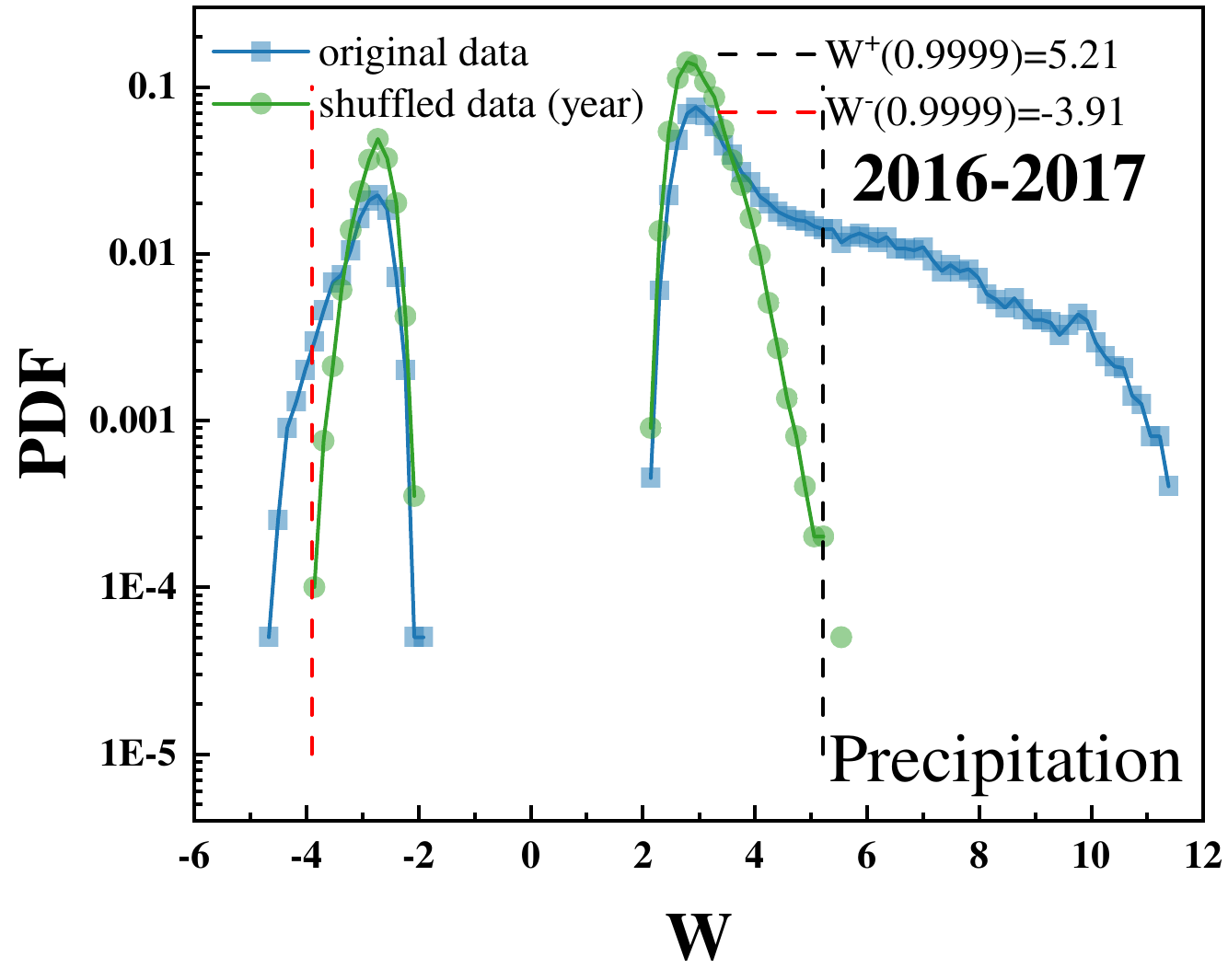}
\includegraphics[width=8.5em, height=7em]{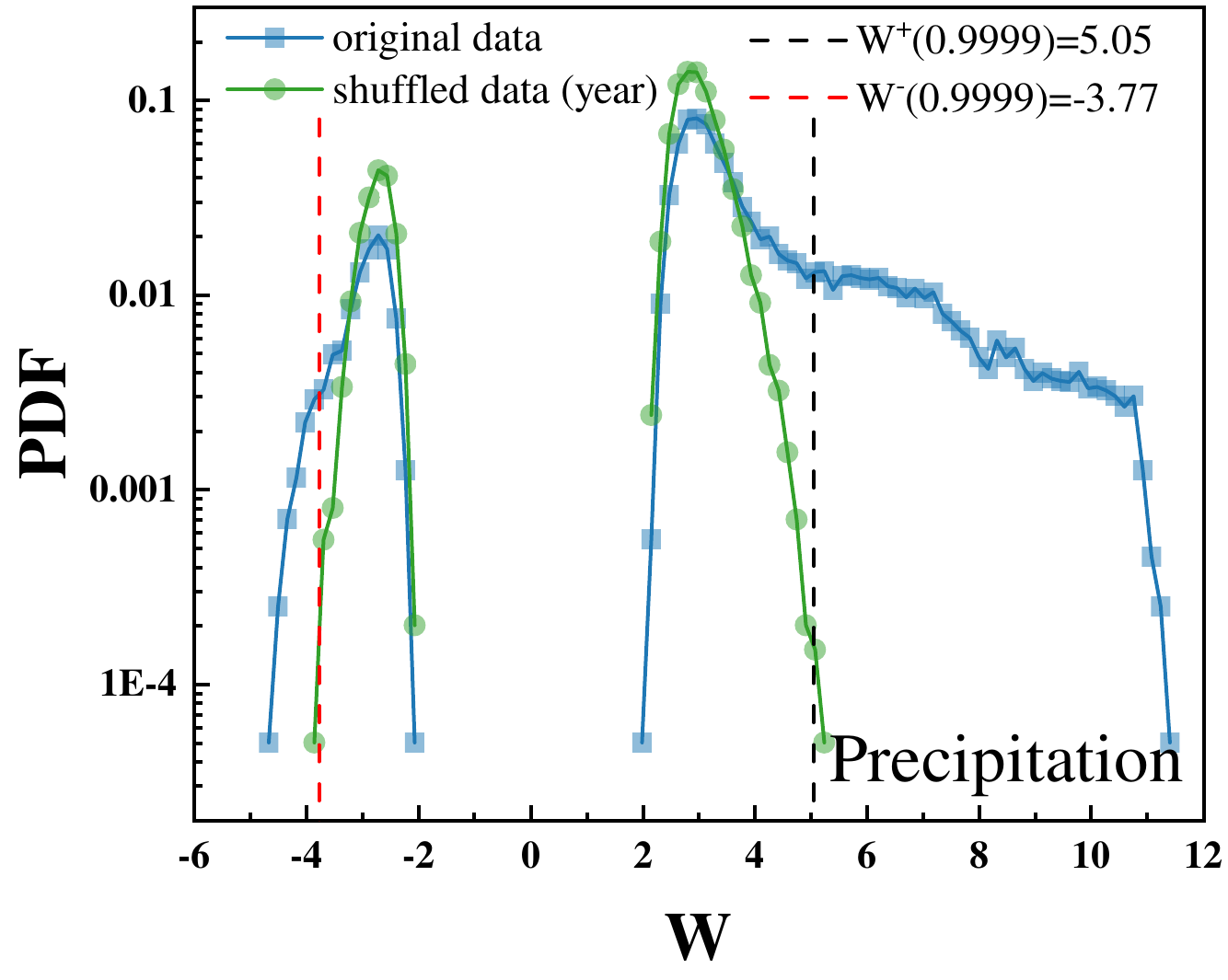}
\includegraphics[width=8.5em, height=7em]{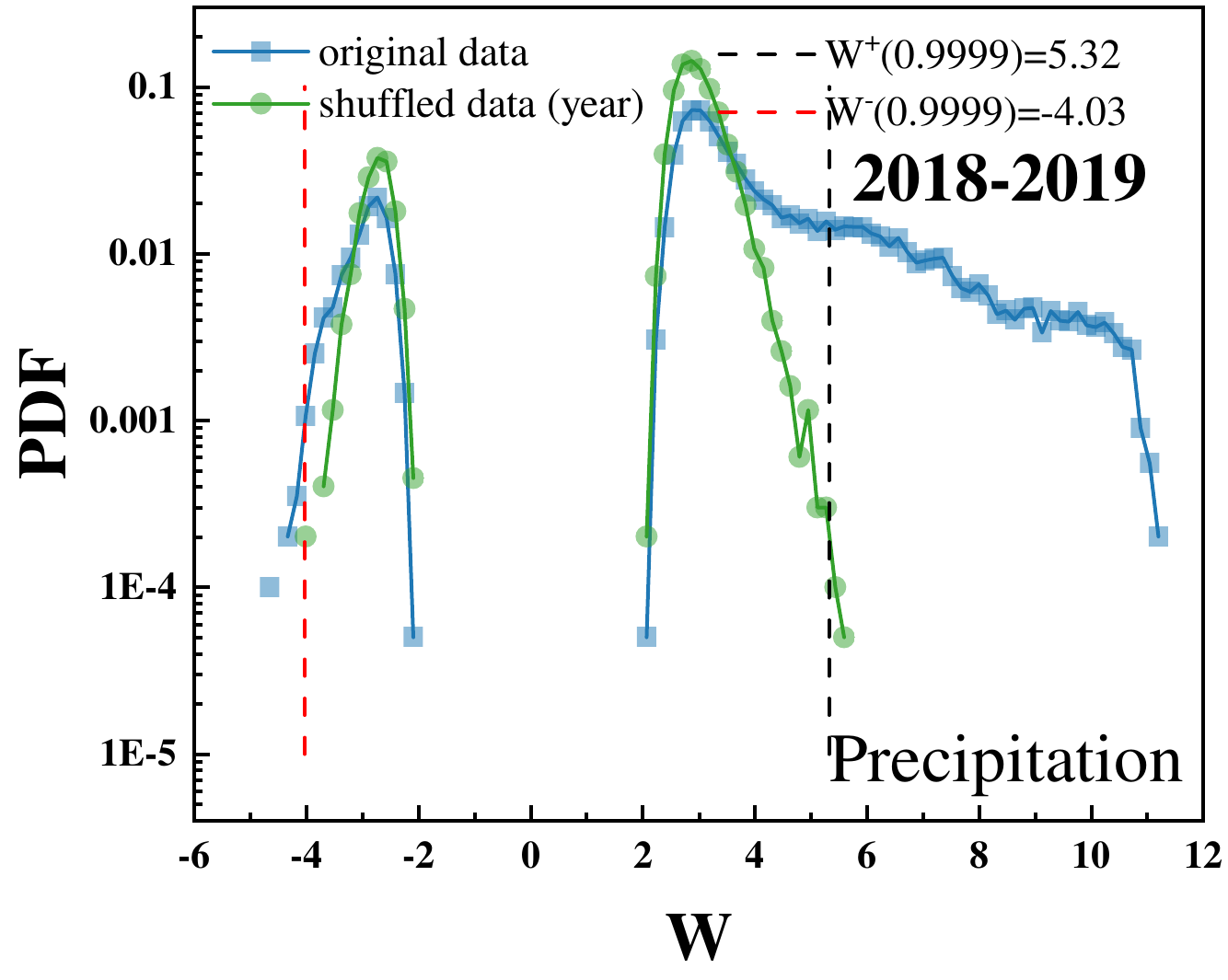}
\end{center}

\begin{center}
\noindent {\small {\bf Fig. S11} Probability distribution function (PDF) of link weights for the original data and shuffled data of precipitation in the Contiguous United States. }
\end{center}

\begin{center}
\includegraphics[width=8.5em, height=7em]{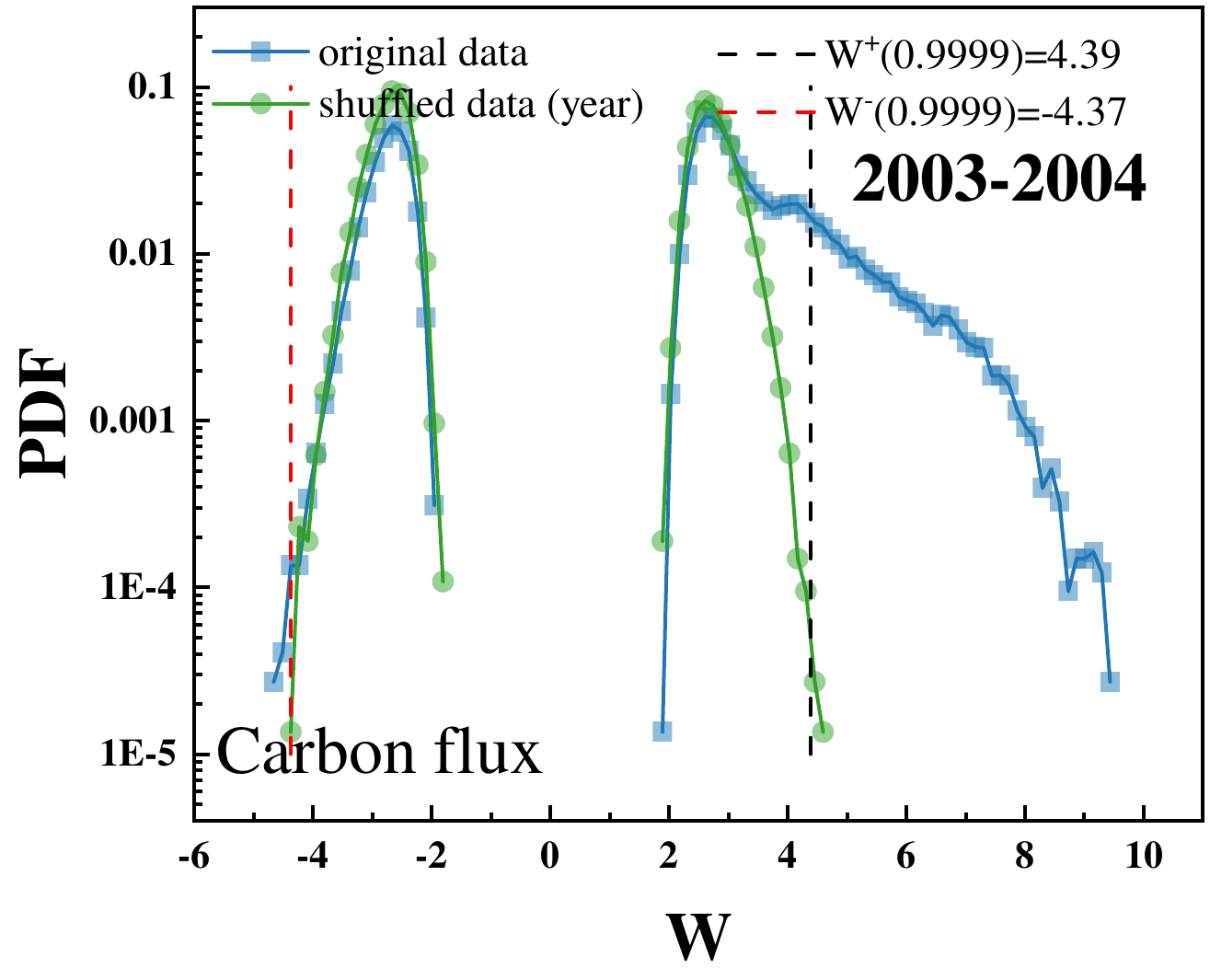}
\includegraphics[width=8.5em, height=7em]{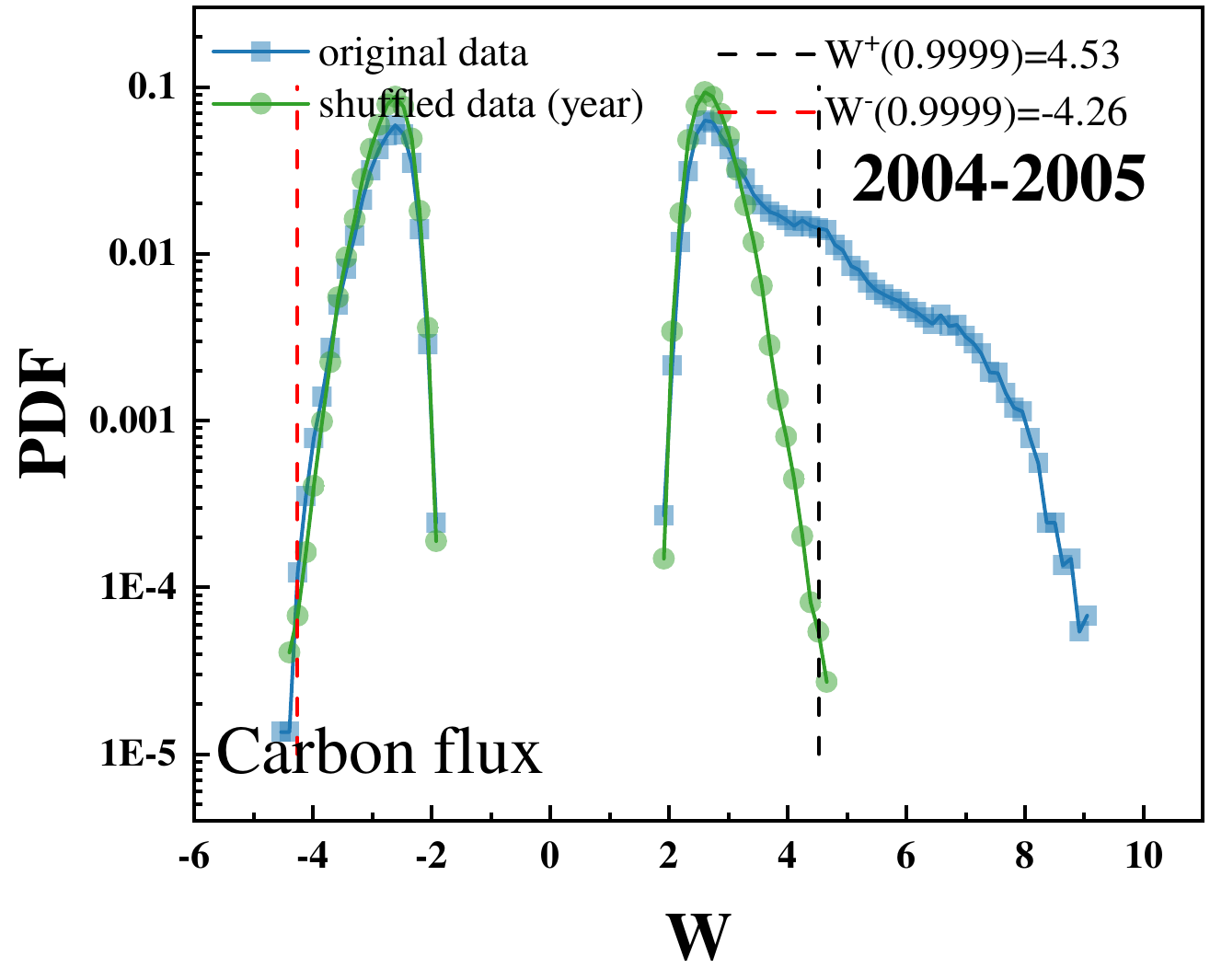}
\includegraphics[width=8.5em, height=7em]{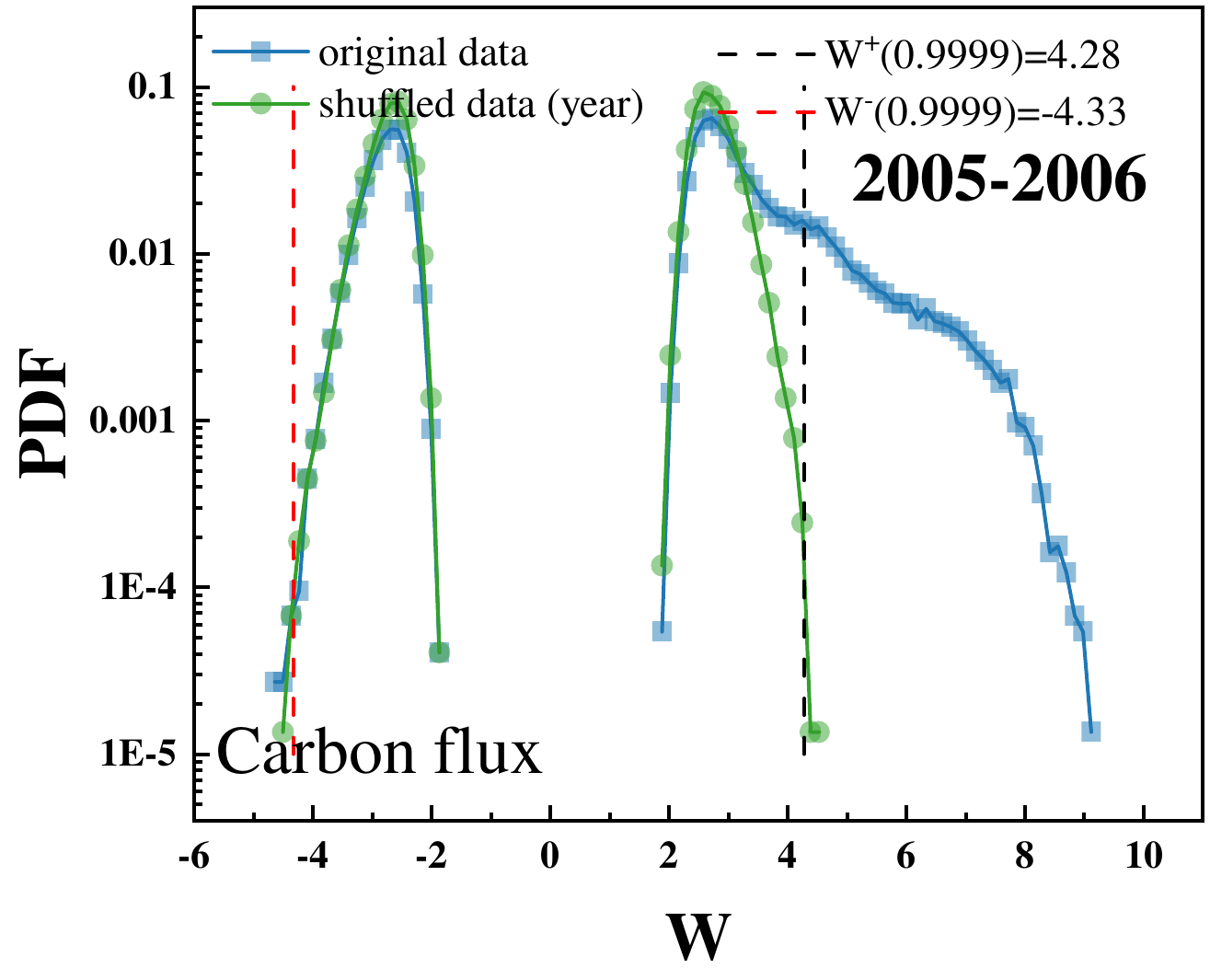}
\includegraphics[width=8.5em, height=7em]{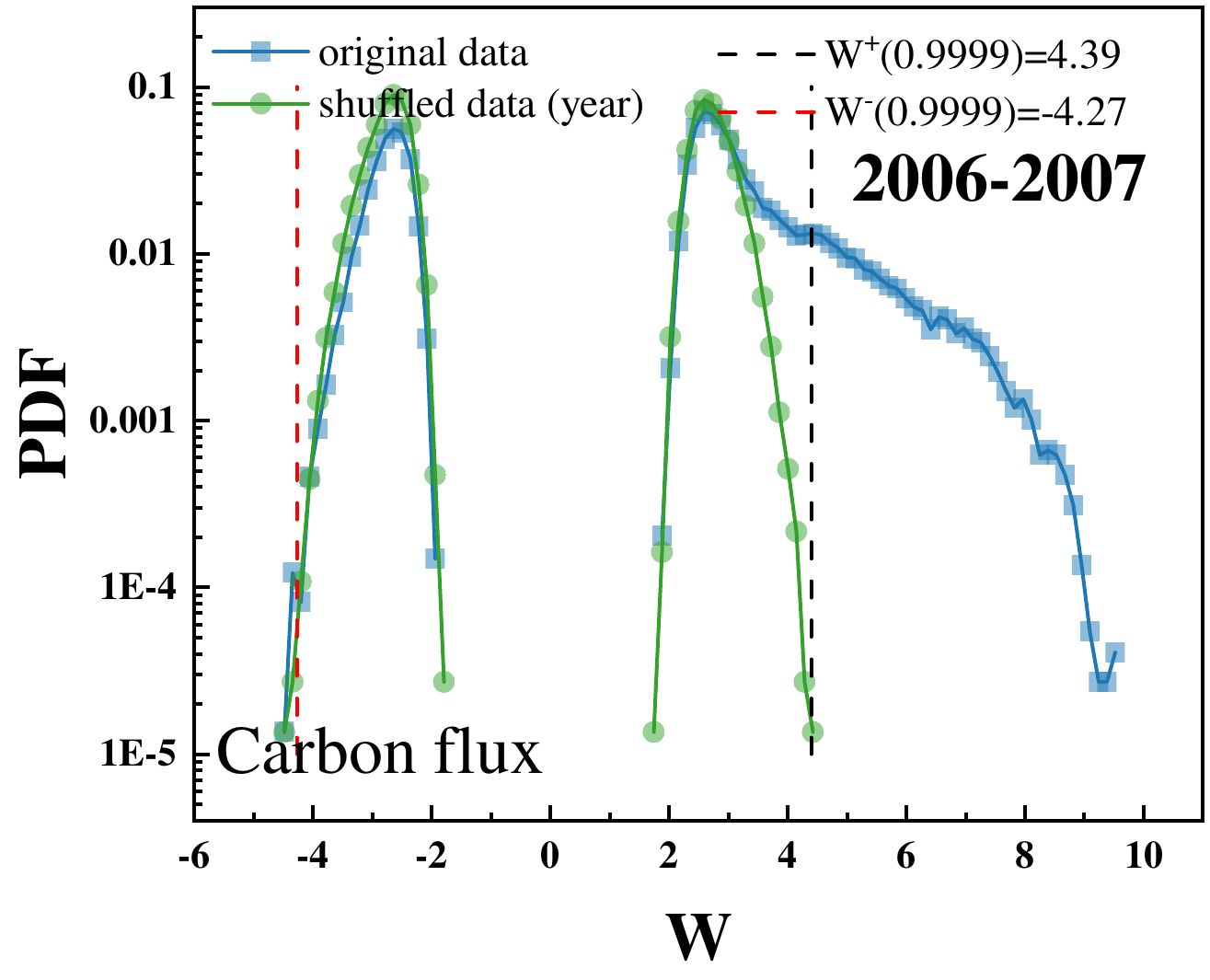}
\includegraphics[width=8.5em, height=7em]{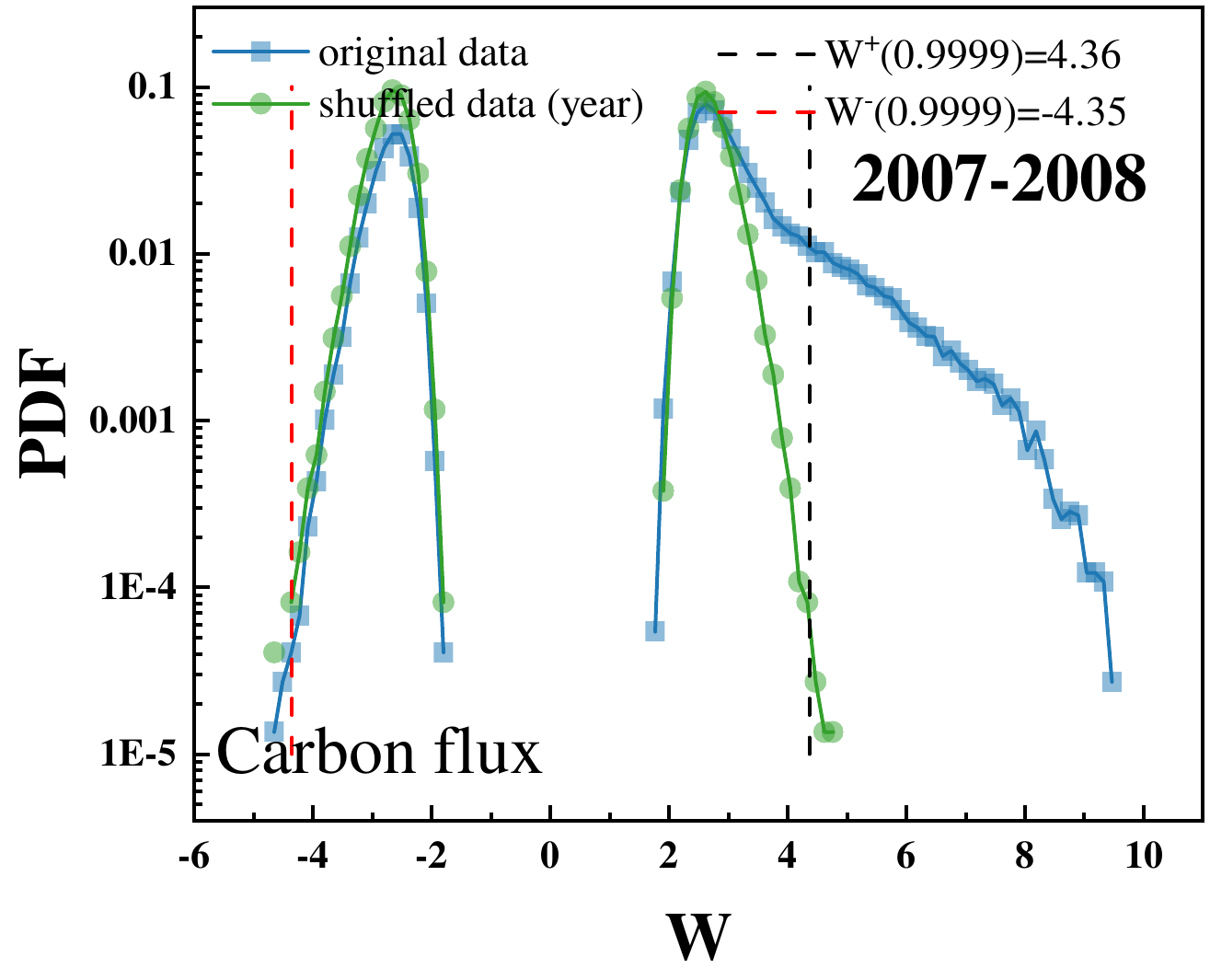}
\includegraphics[width=8.5em, height=7em]{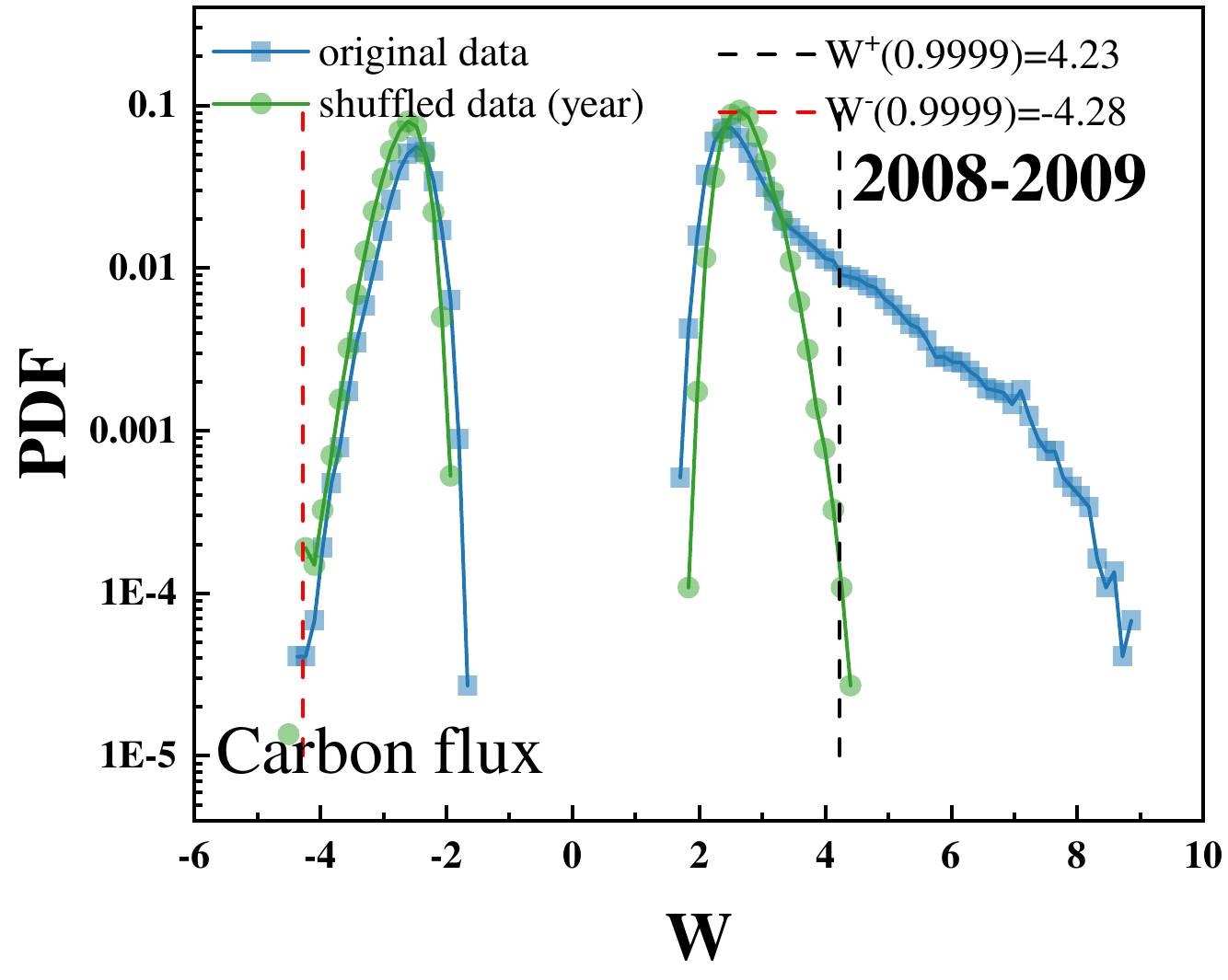}
\includegraphics[width=8.5em, height=7em]{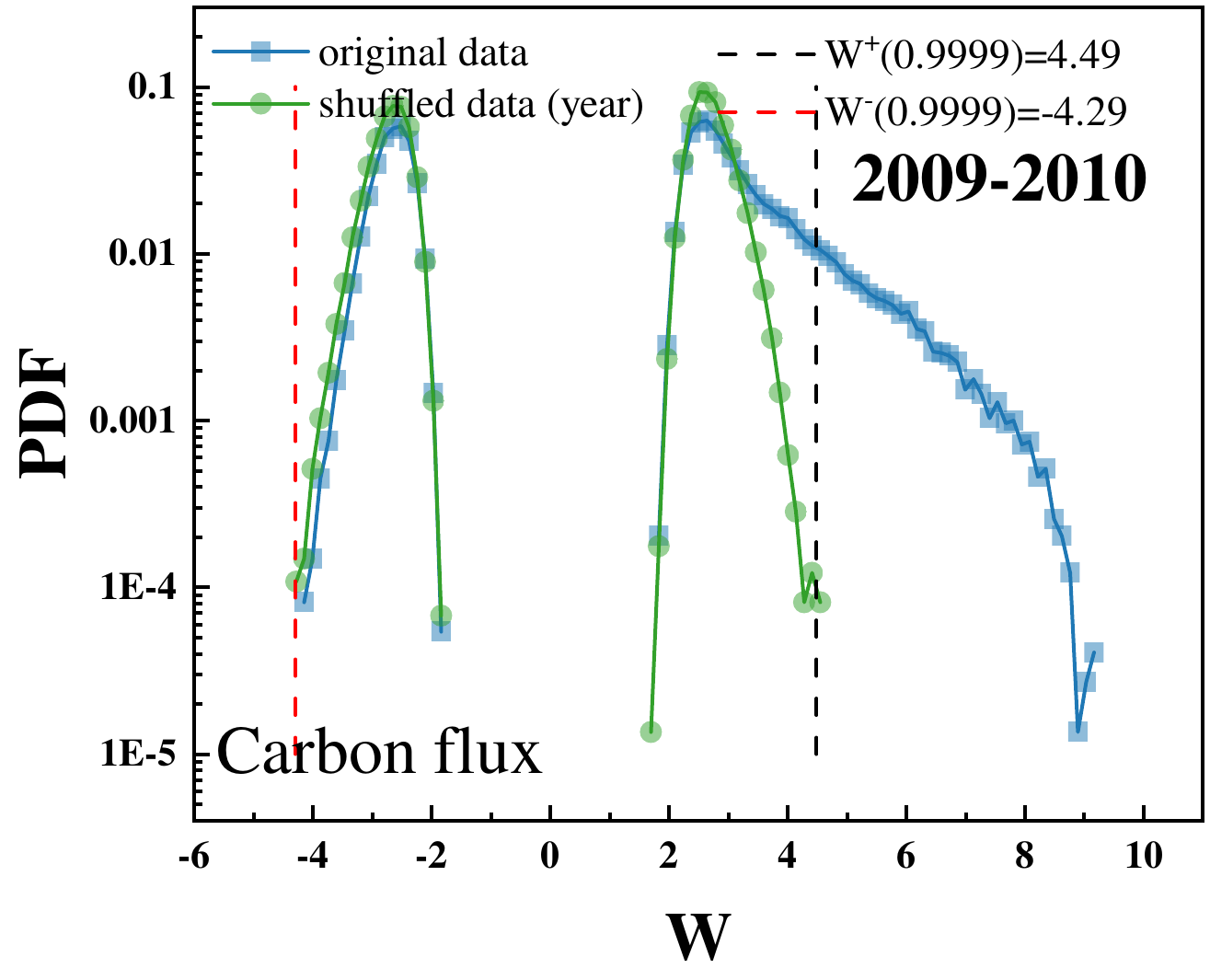}
\includegraphics[width=8.5em, height=7em]{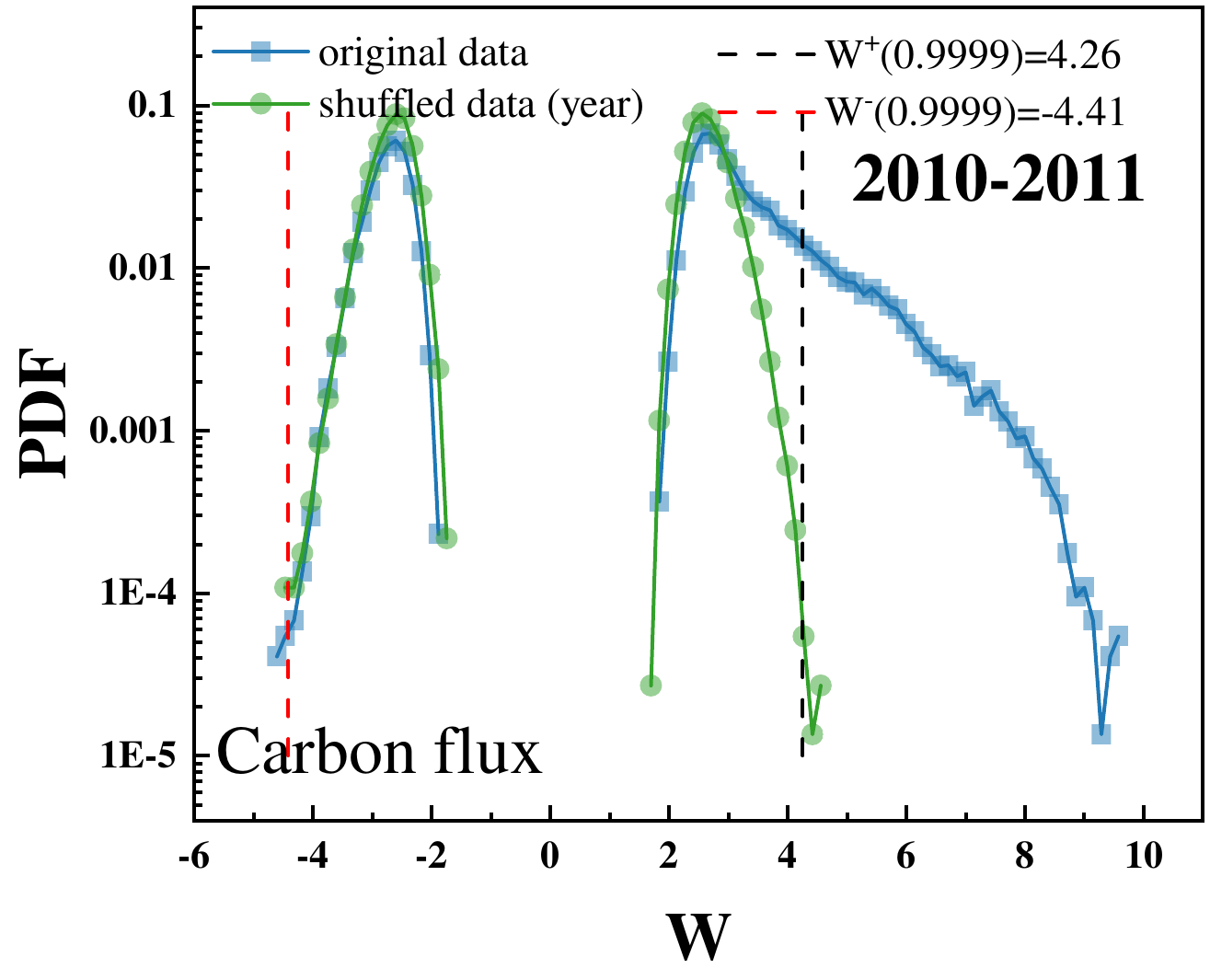}
\includegraphics[width=8.5em, height=7em]{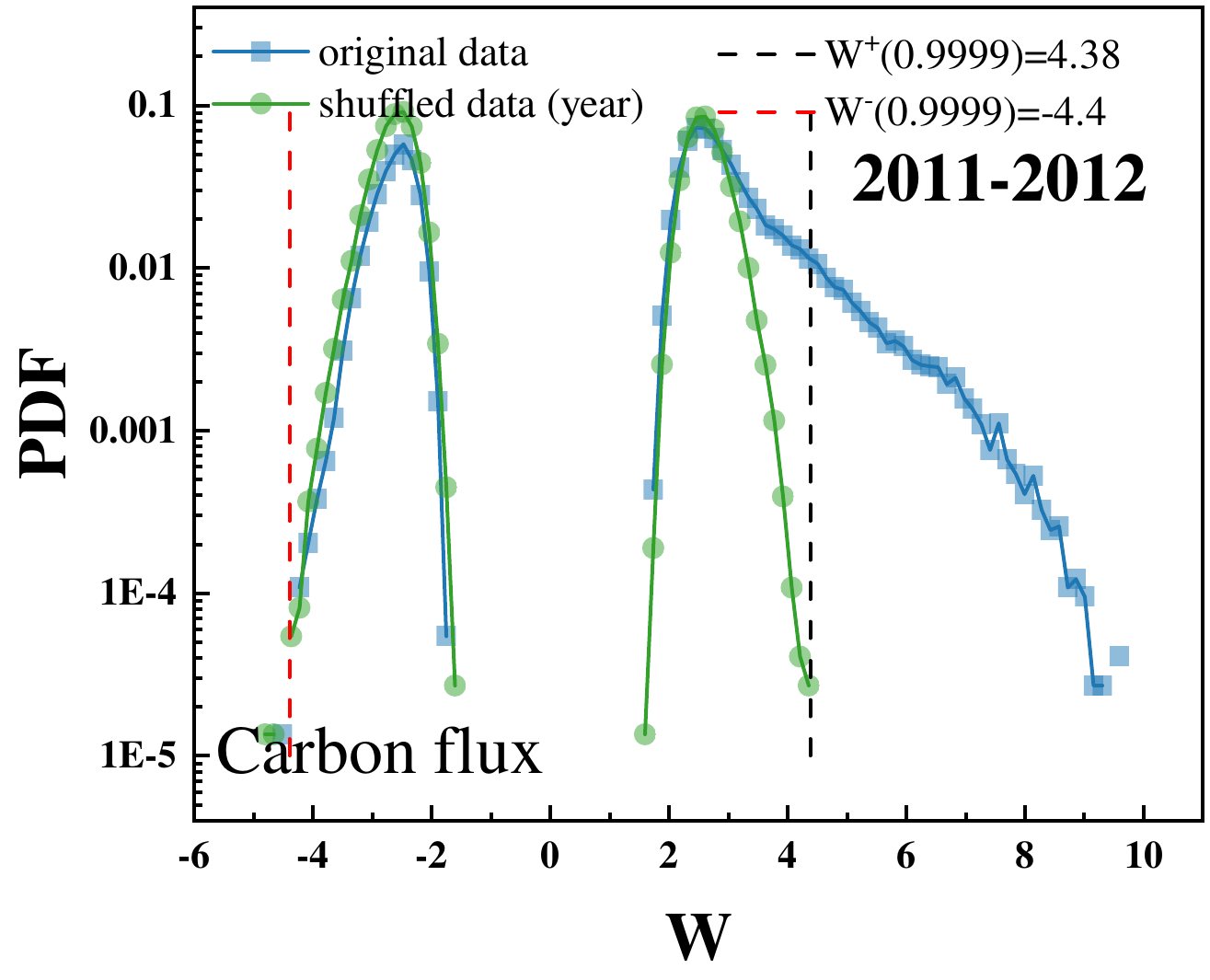}
\includegraphics[width=8.5em, height=7em]{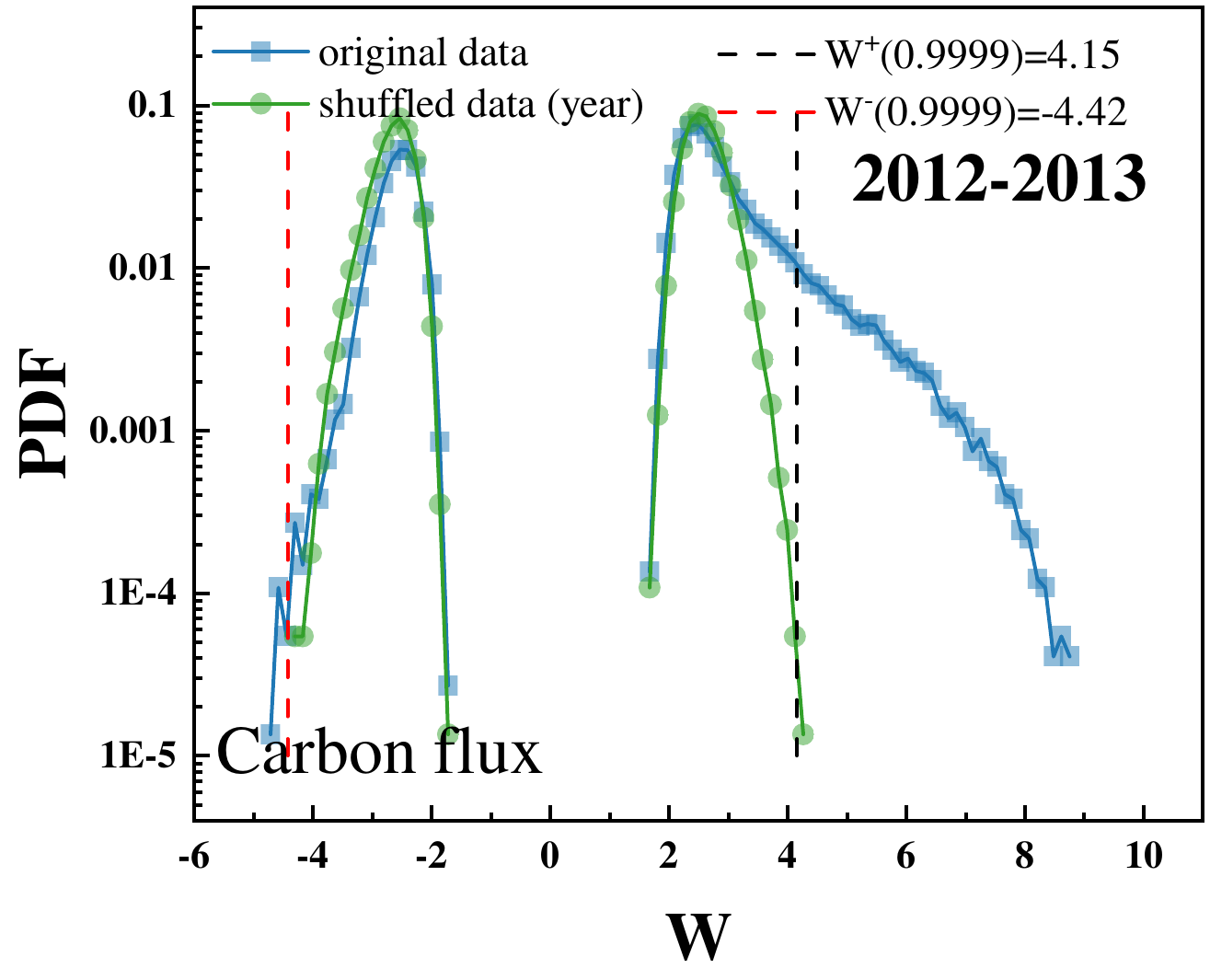}
\includegraphics[width=8.5em, height=7em]{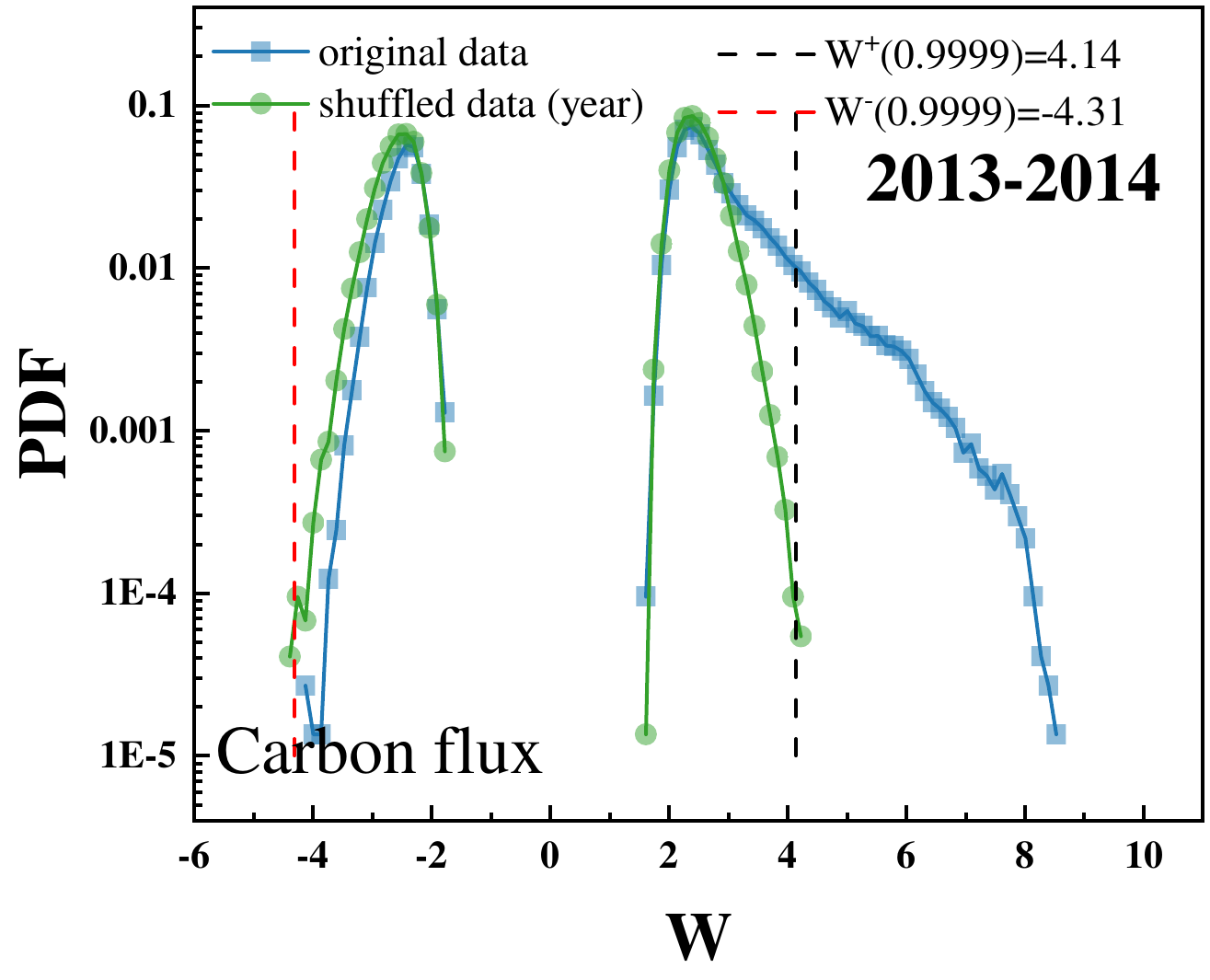}
\includegraphics[width=8.5em, height=7em]{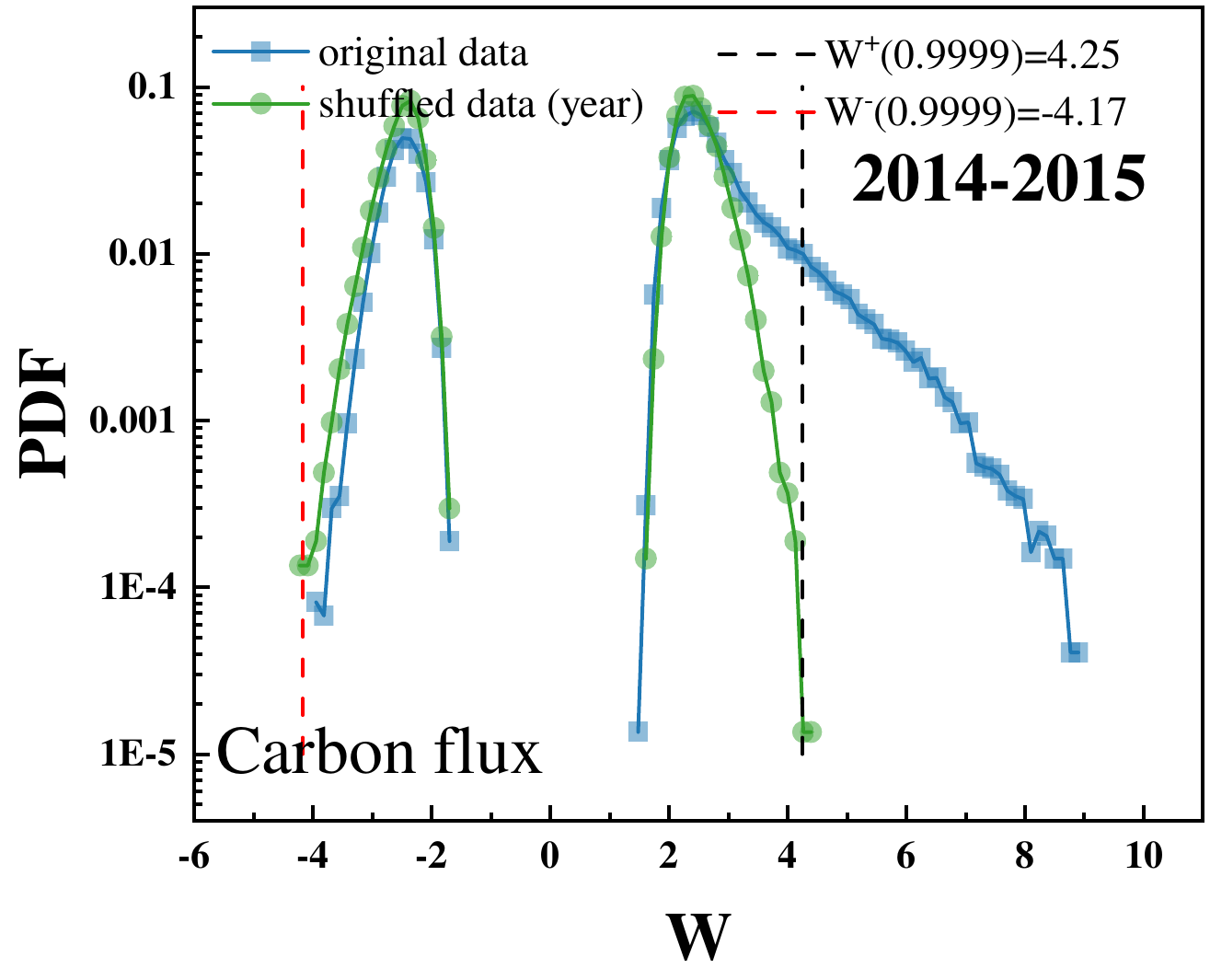}
\includegraphics[width=8.5em, height=7em]{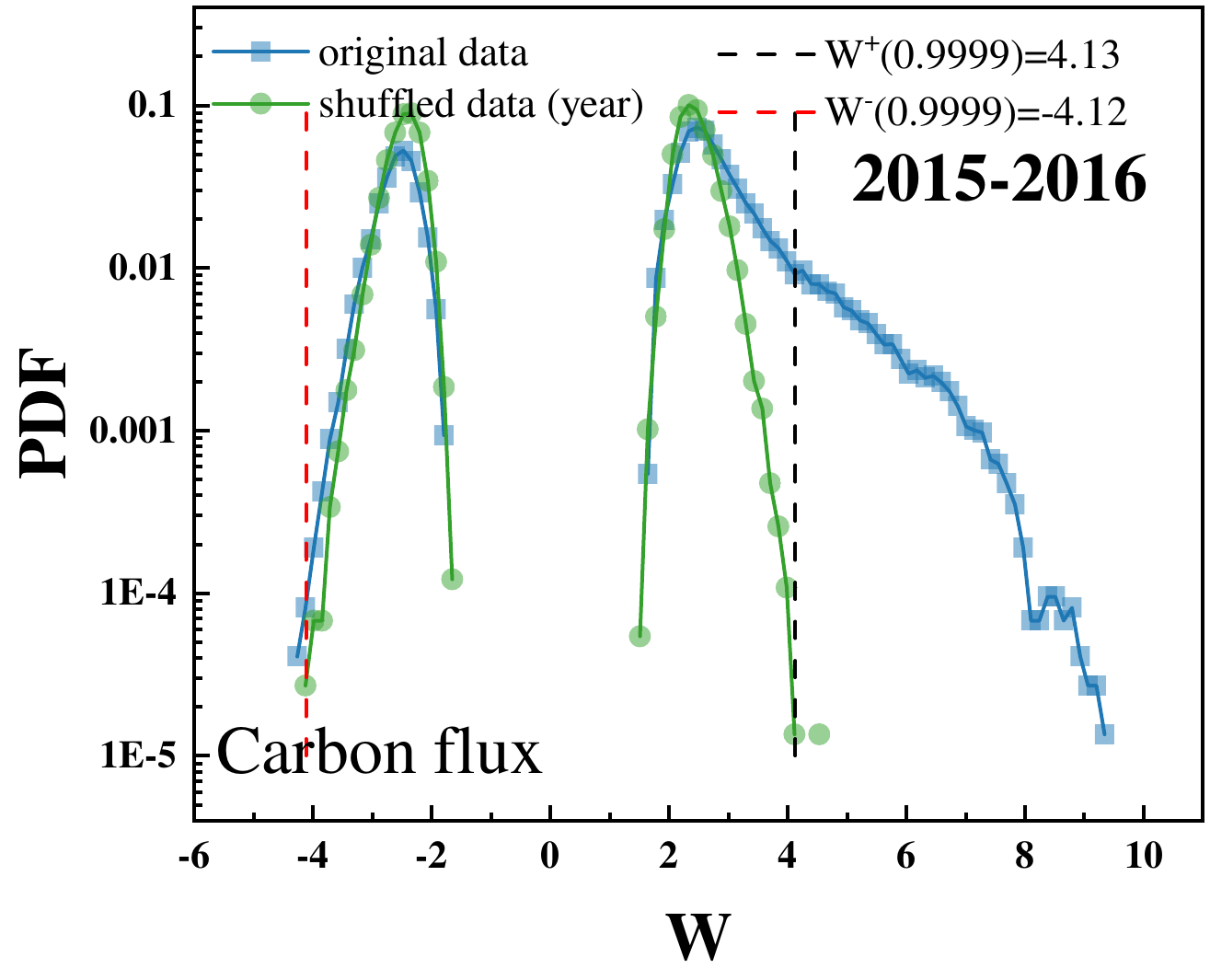}
\includegraphics[width=8.5em, height=7em]{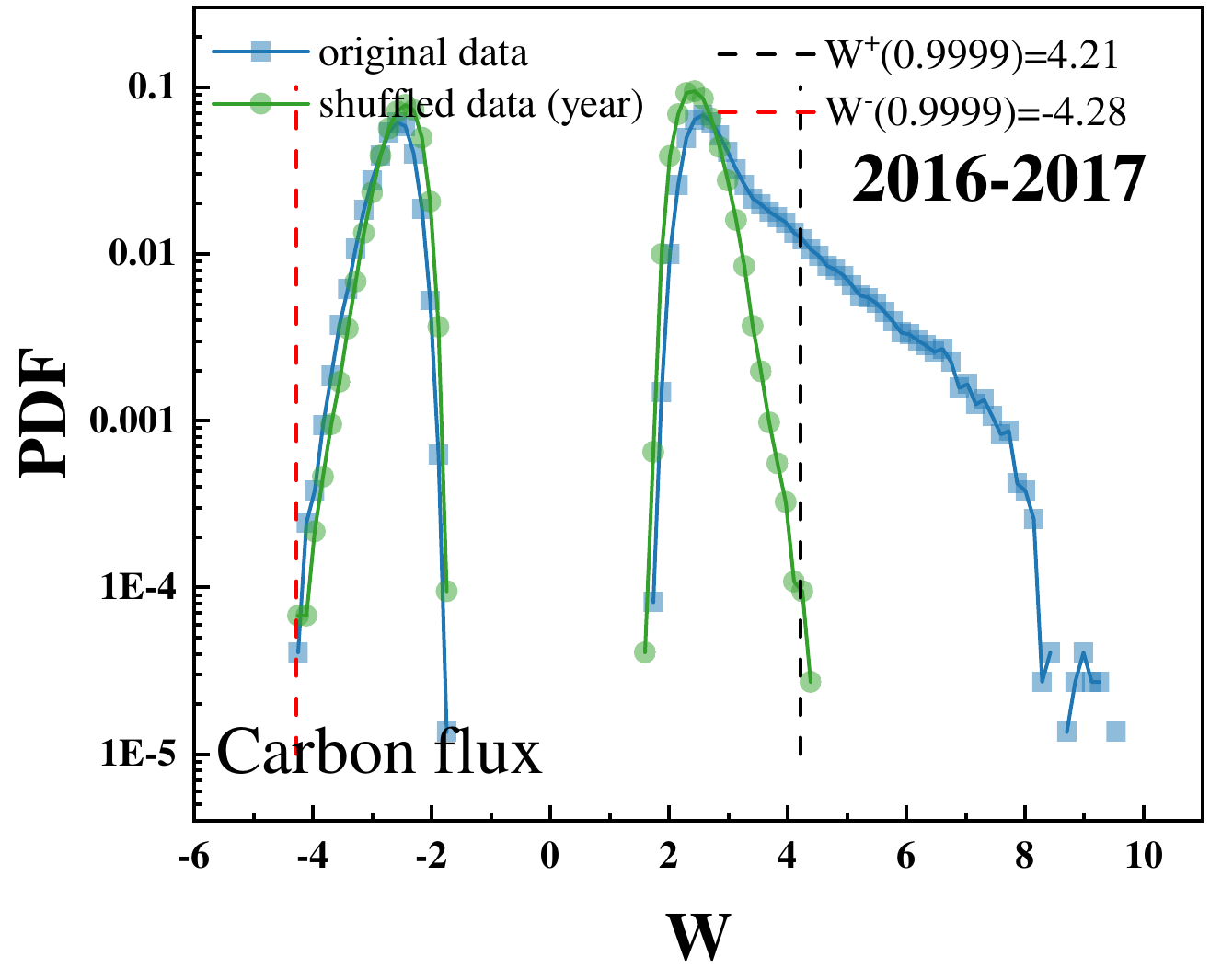}
\includegraphics[width=8.5em, height=7em]{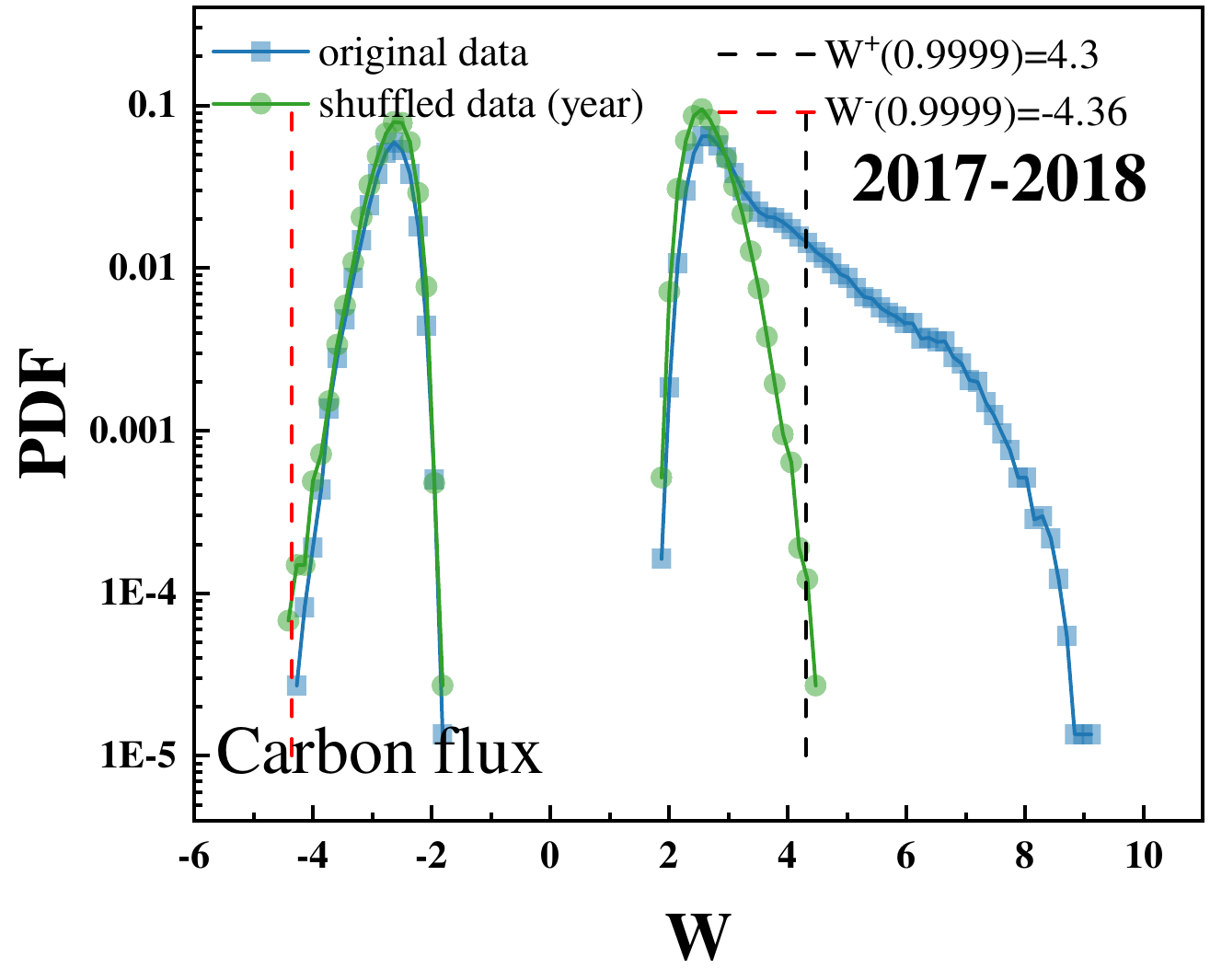}
\includegraphics[width=8.5em, height=7em]{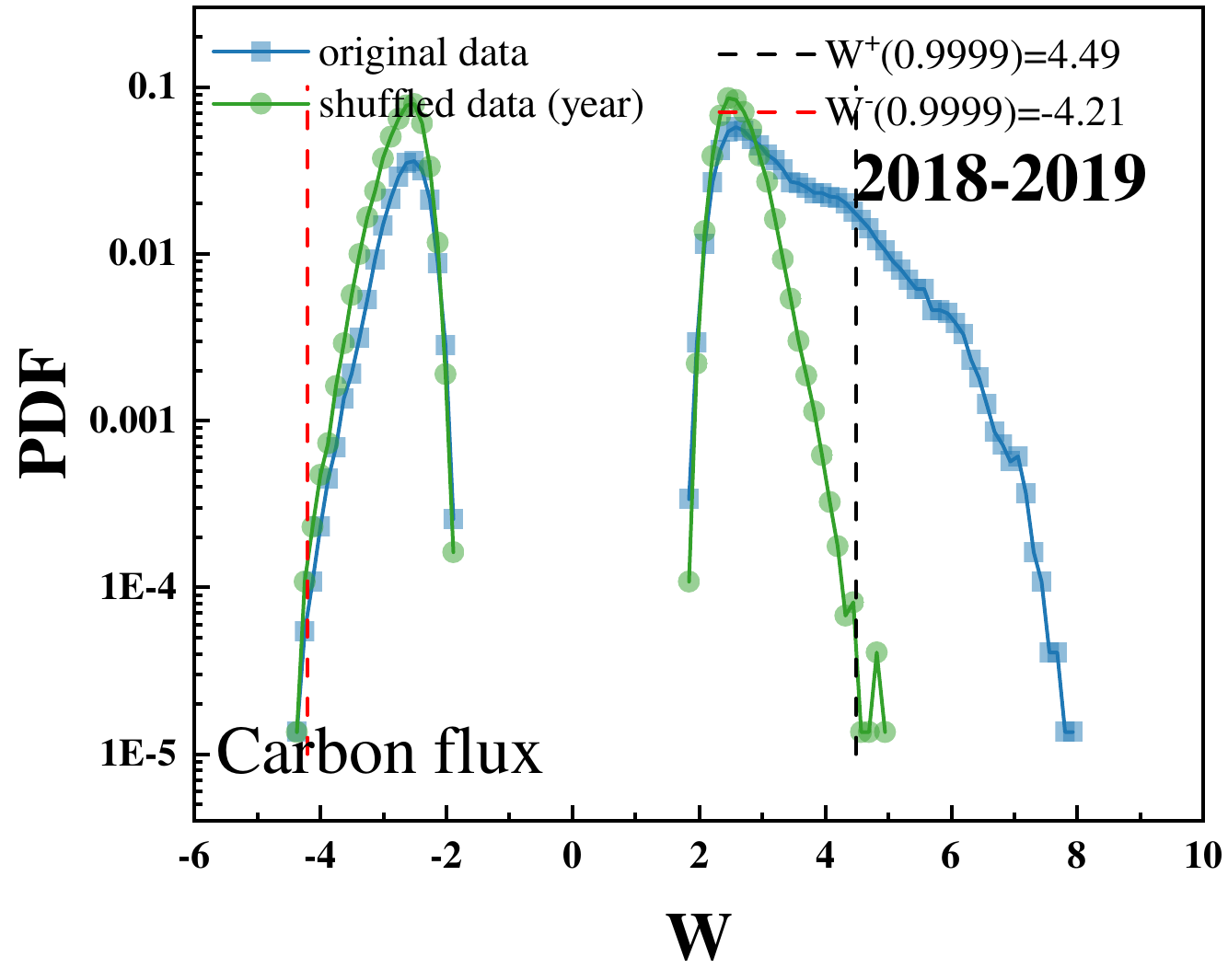}
\end{center}

\begin{center}
\noindent {\small {\bf Fig. S12} Probability distribution function (PDF) of link weights for the original data and shuffled data of carbon flux in Europe. }
\end{center}

\begin{center}
\includegraphics[width=8.5em, height=7em]{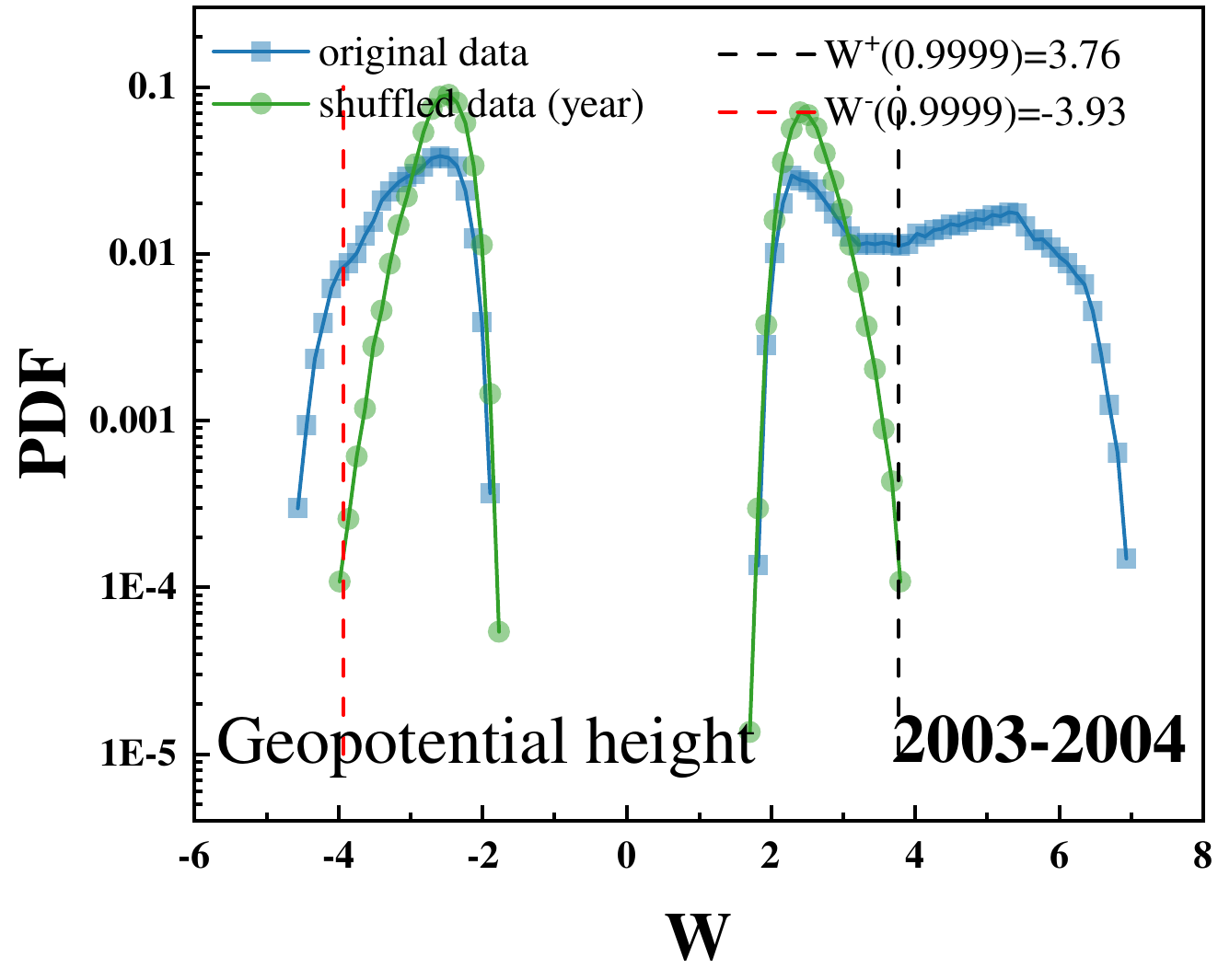}
\includegraphics[width=8.5em, height=7em]{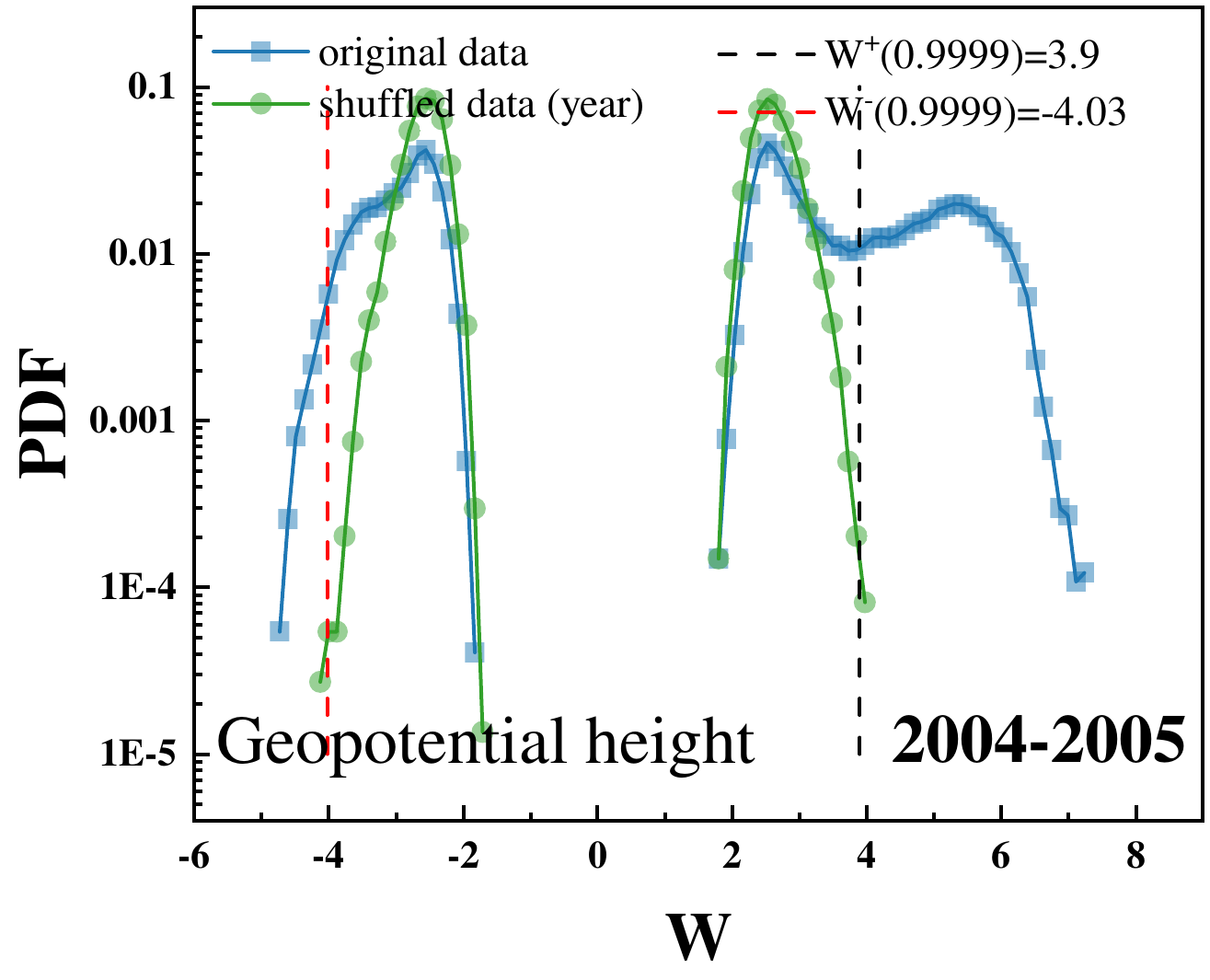}
\includegraphics[width=8.5em, height=7em]{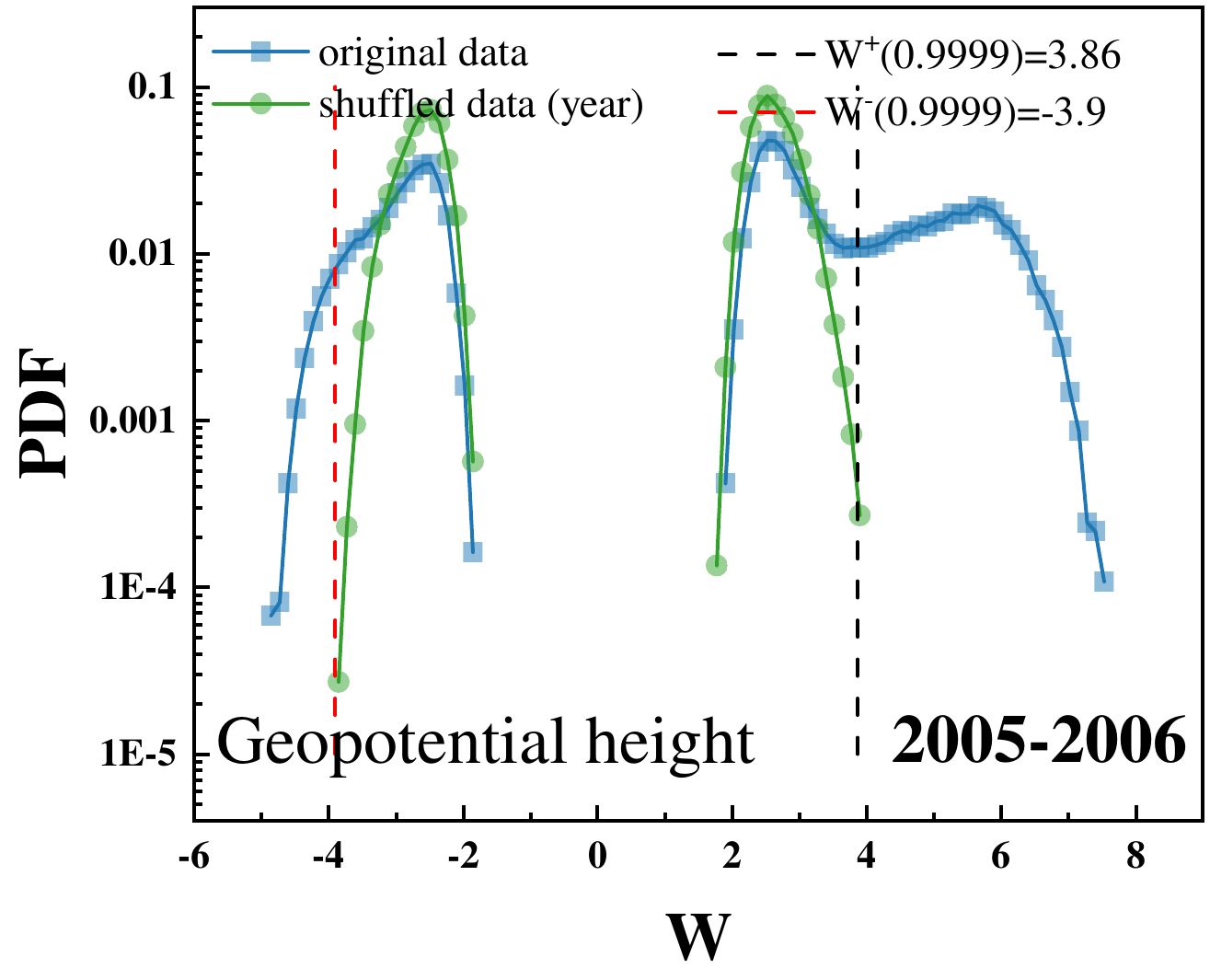}
\includegraphics[width=8.5em, height=7em]{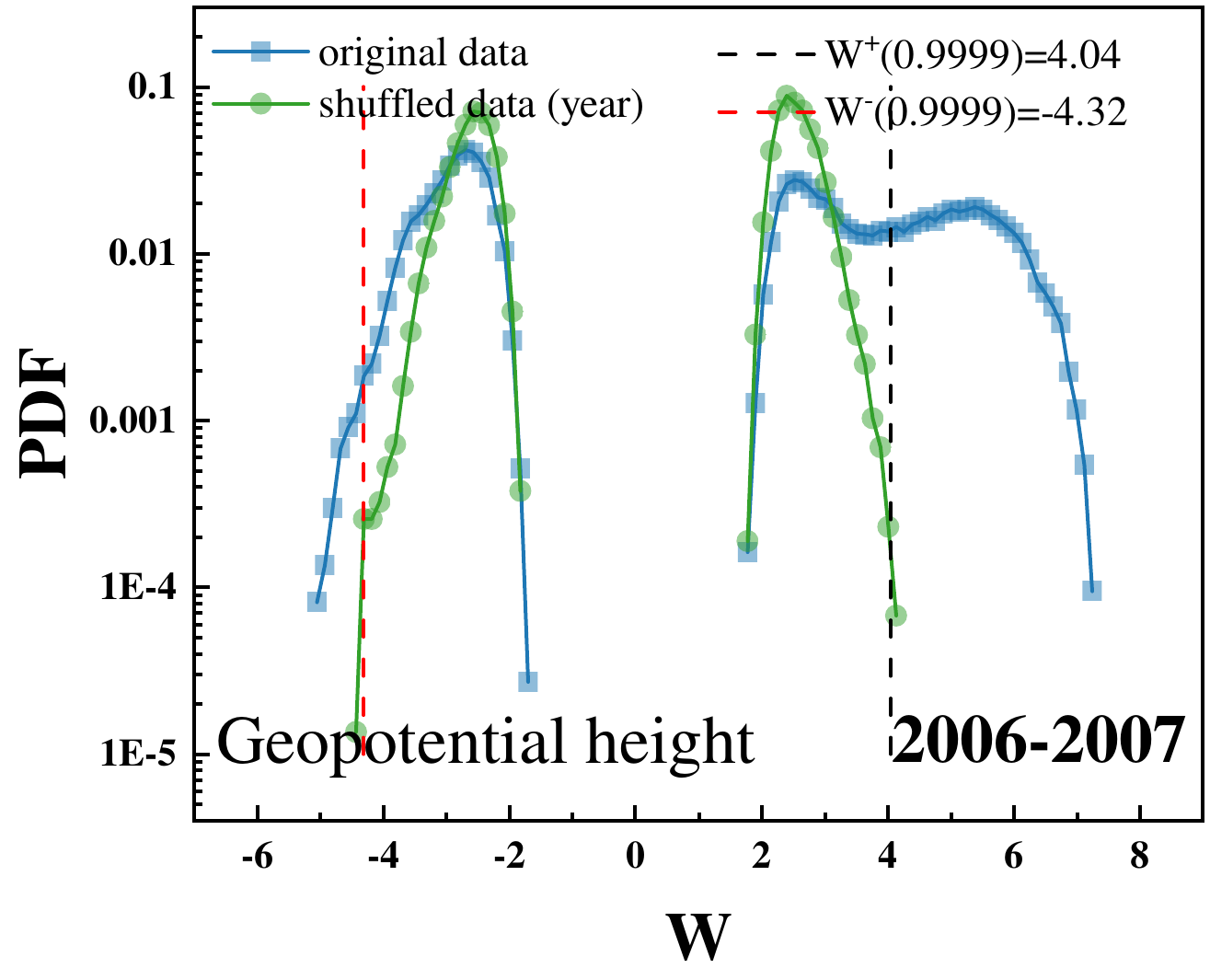}
\includegraphics[width=8.5em, height=7em]{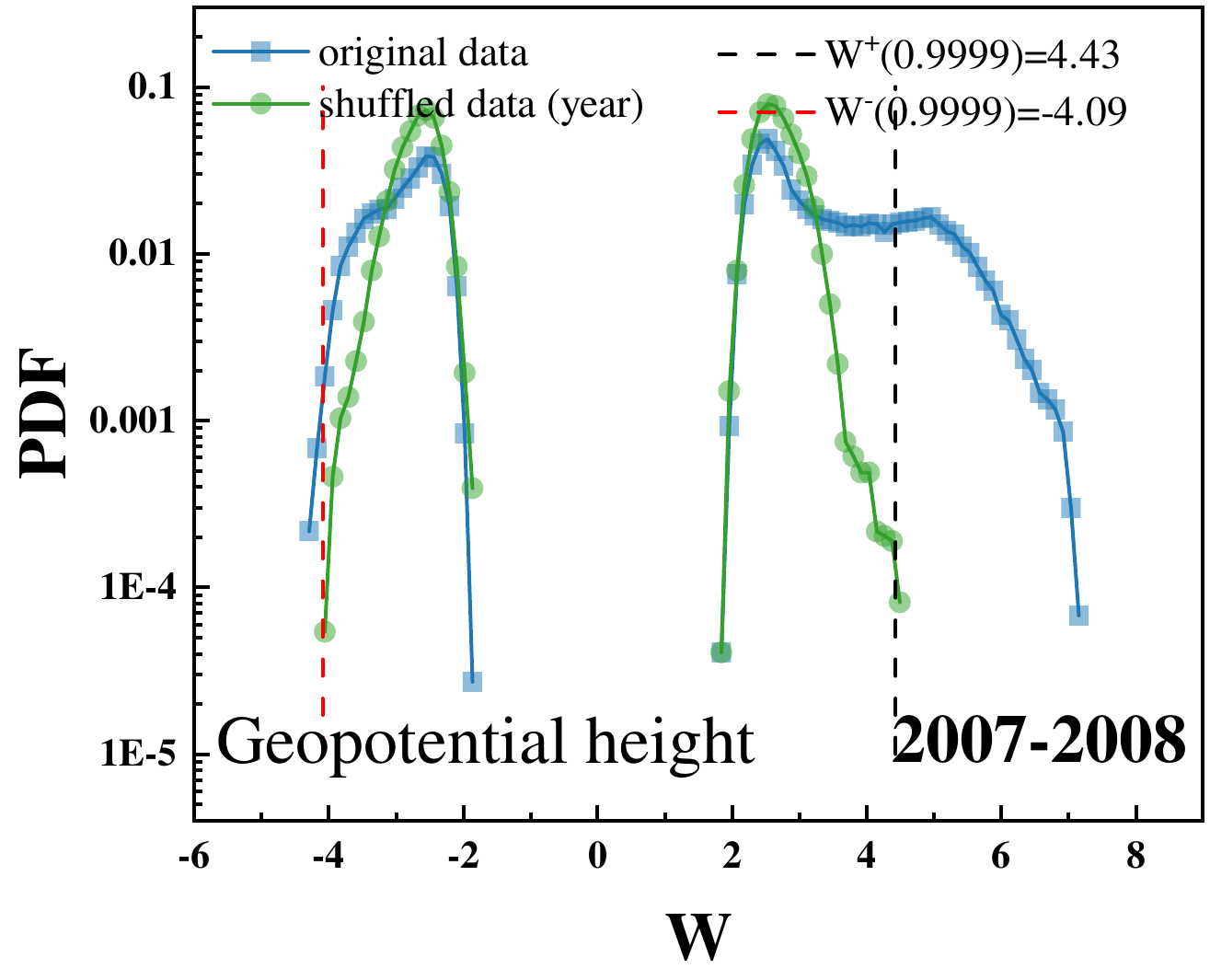}
\includegraphics[width=8.5em, height=7em]{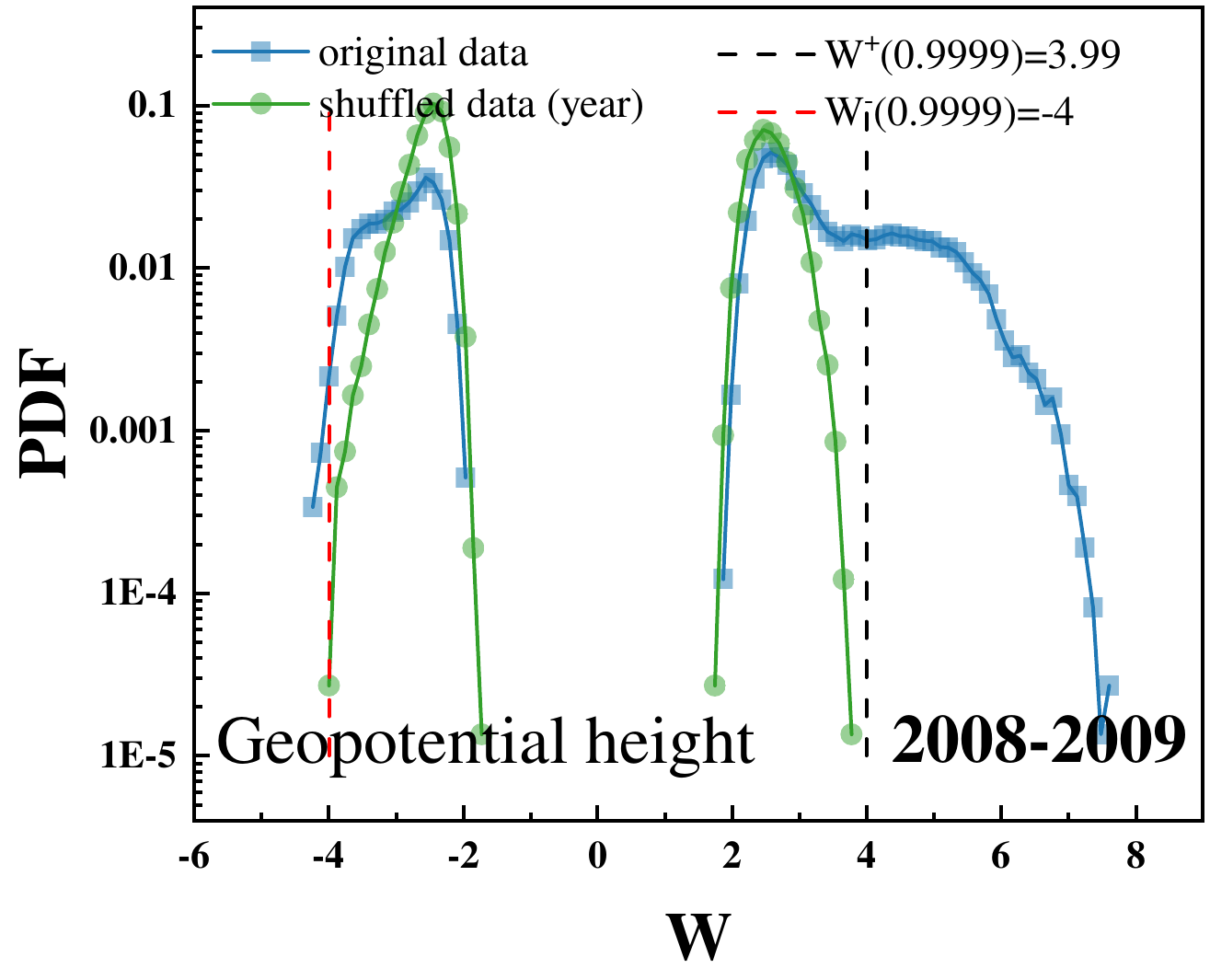}
\includegraphics[width=8.5em, height=7em]{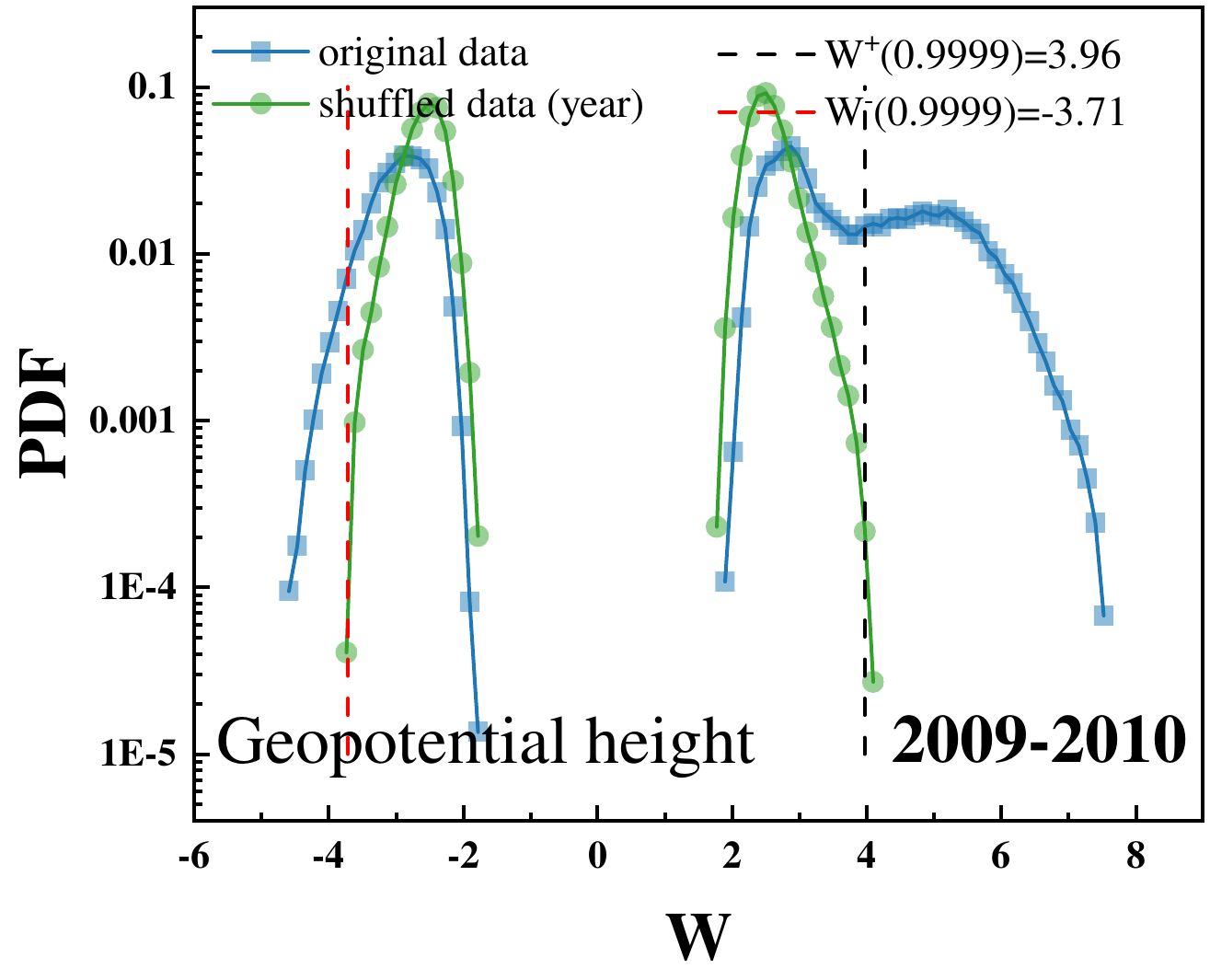}
\includegraphics[width=8.5em, height=7em]{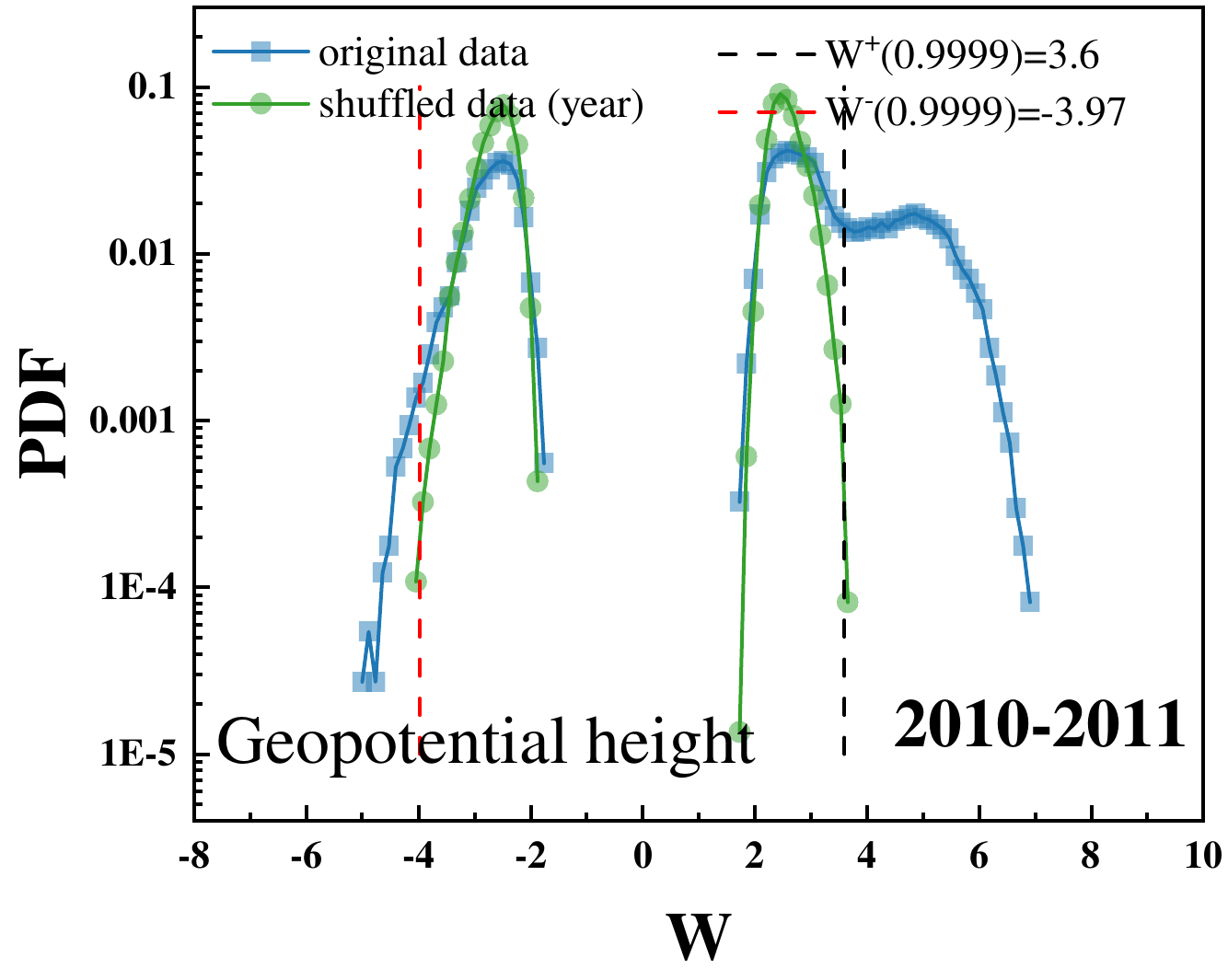}
\includegraphics[width=8.5em, height=7em]{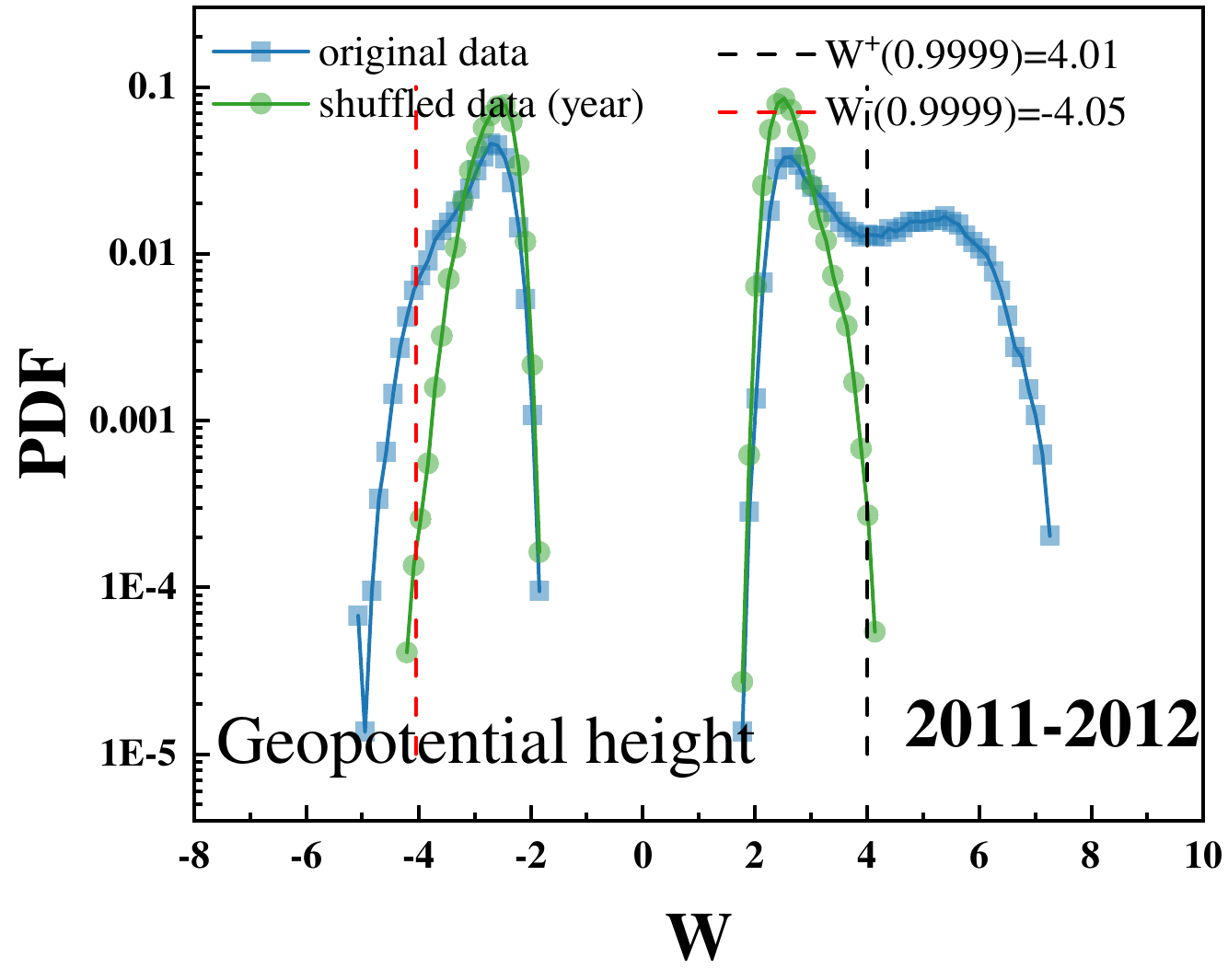}
\includegraphics[width=8.5em, height=7em]{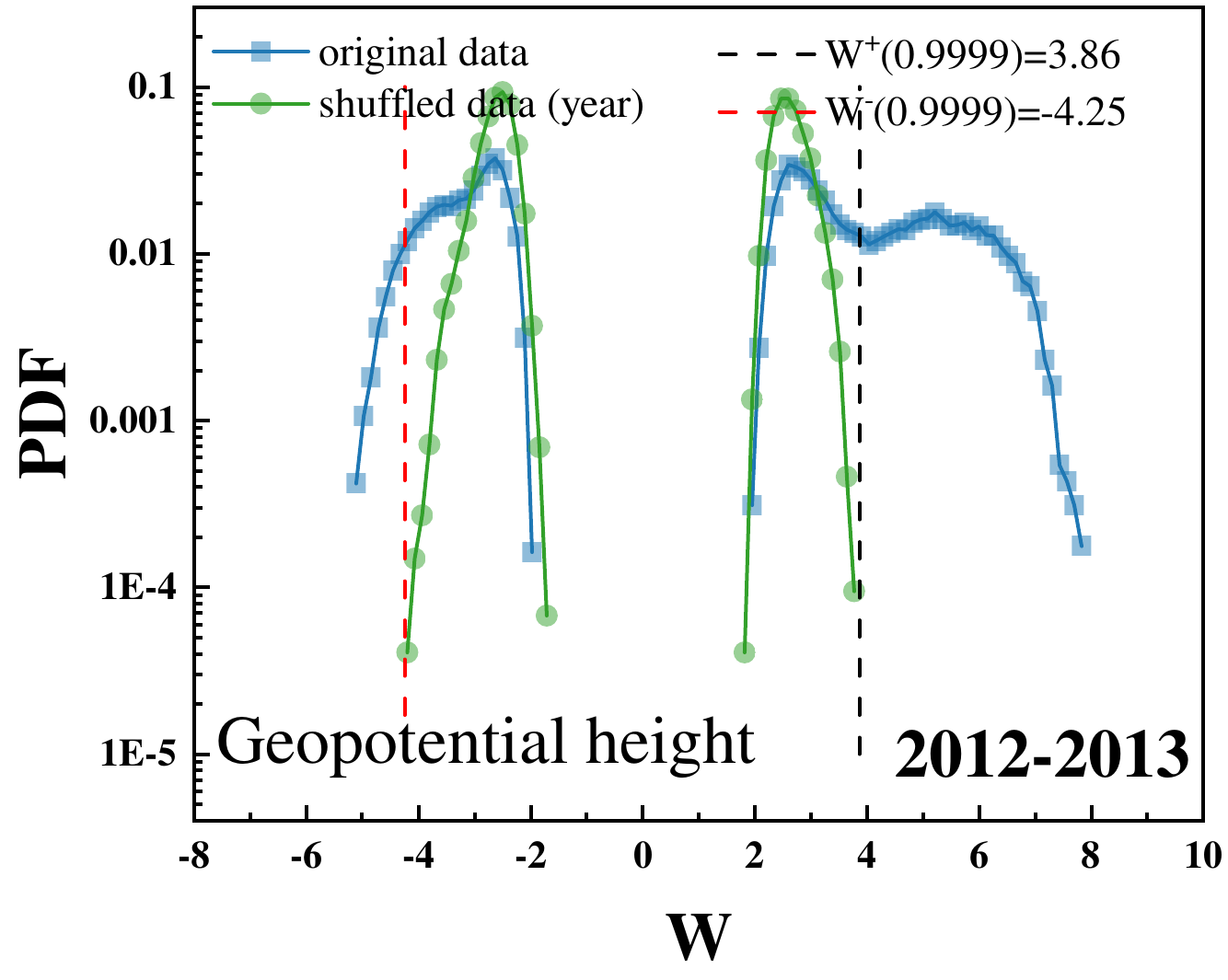}
\includegraphics[width=8.5em, height=7em]{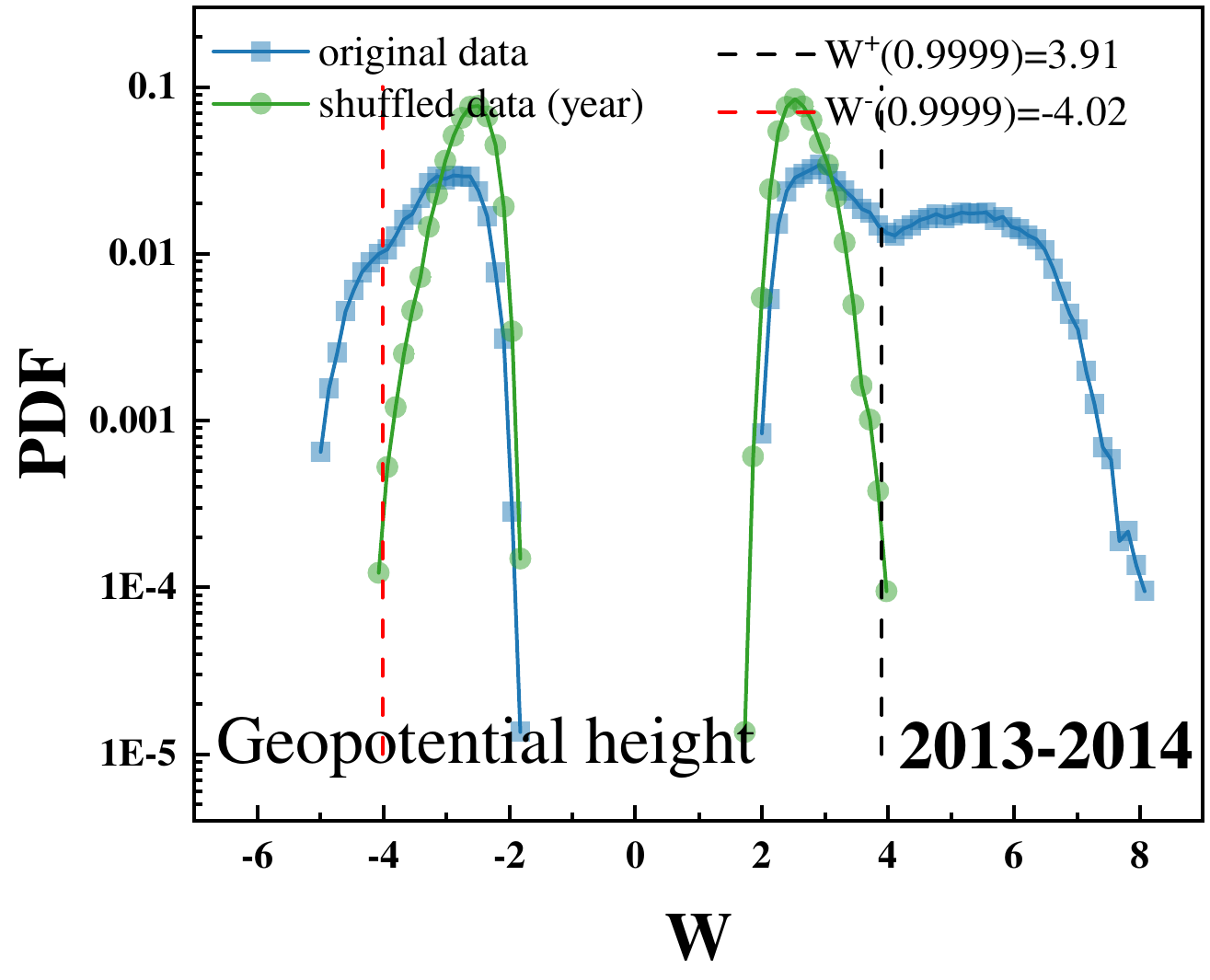}
\includegraphics[width=8.5em, height=7em]{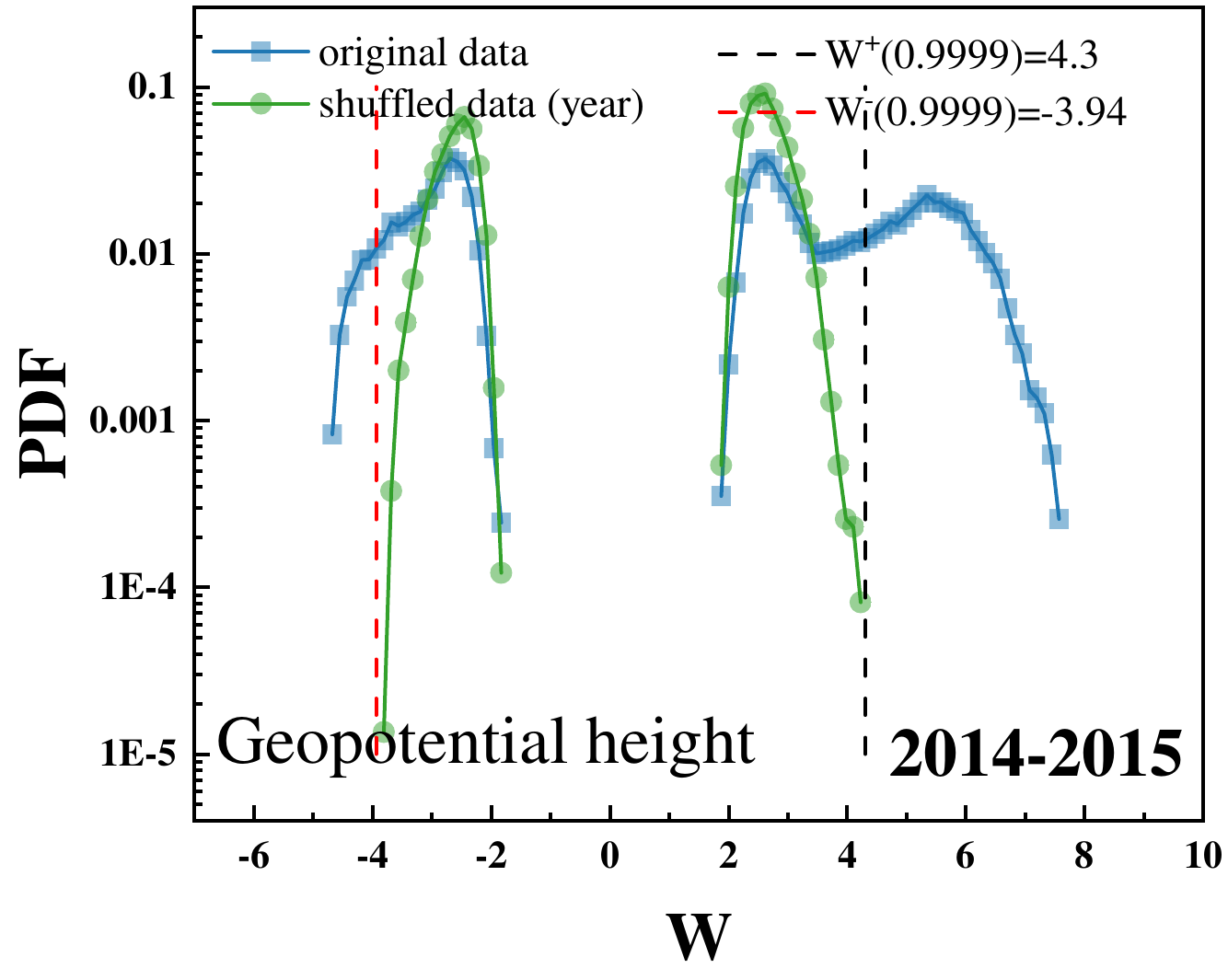}
\includegraphics[width=8.5em, height=7em]{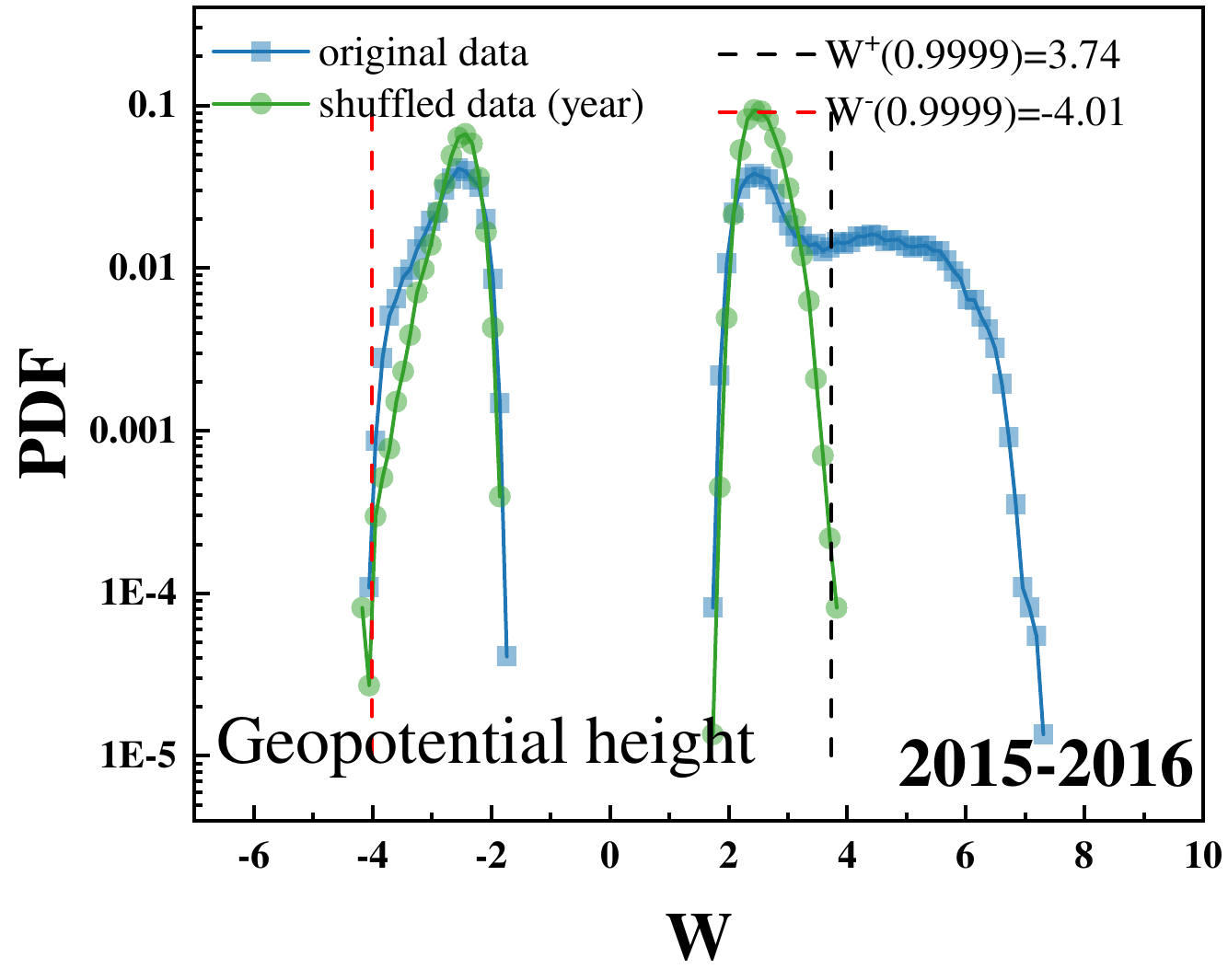}
\includegraphics[width=8.5em, height=7em]{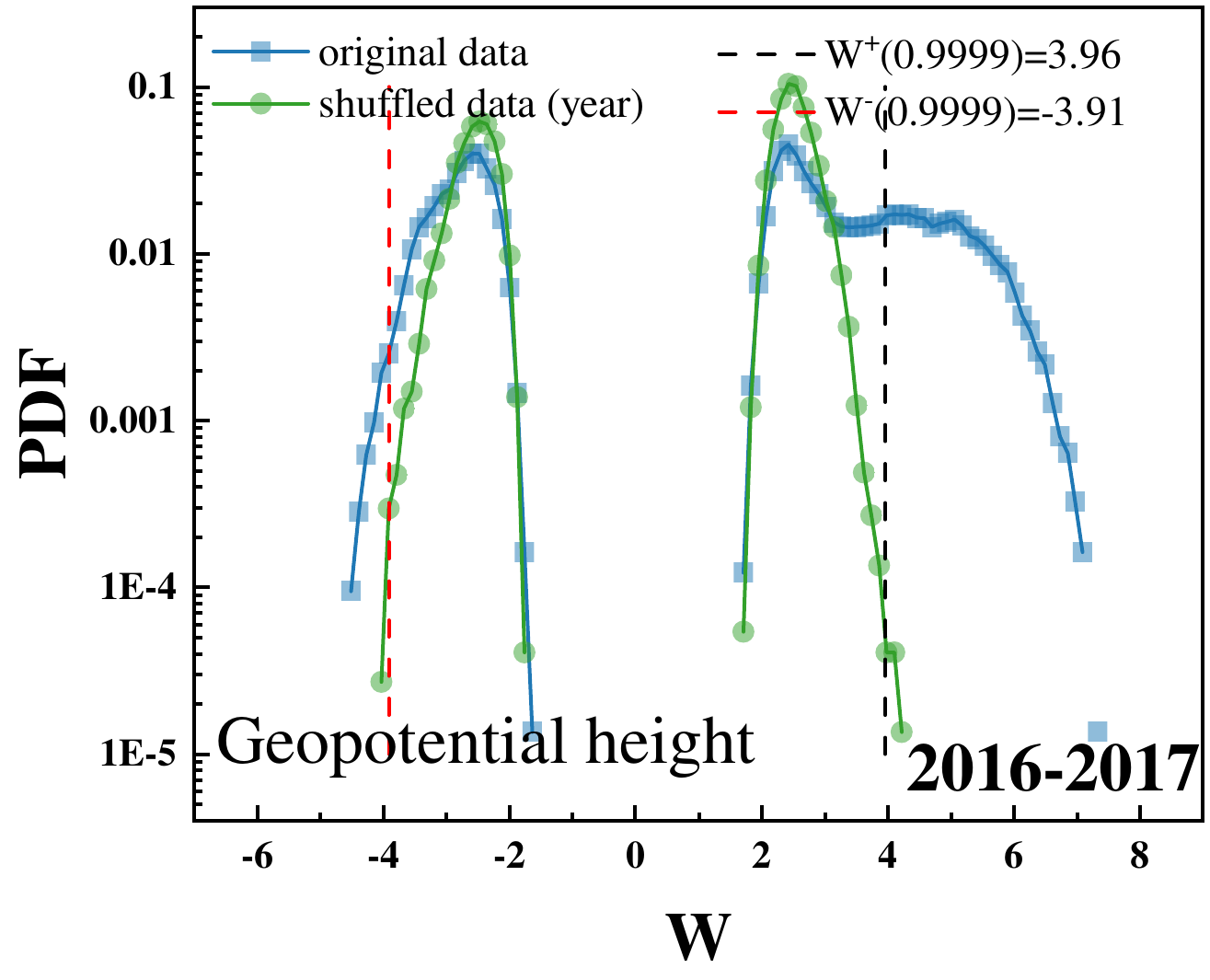}
\includegraphics[width=8.5em, height=7em]{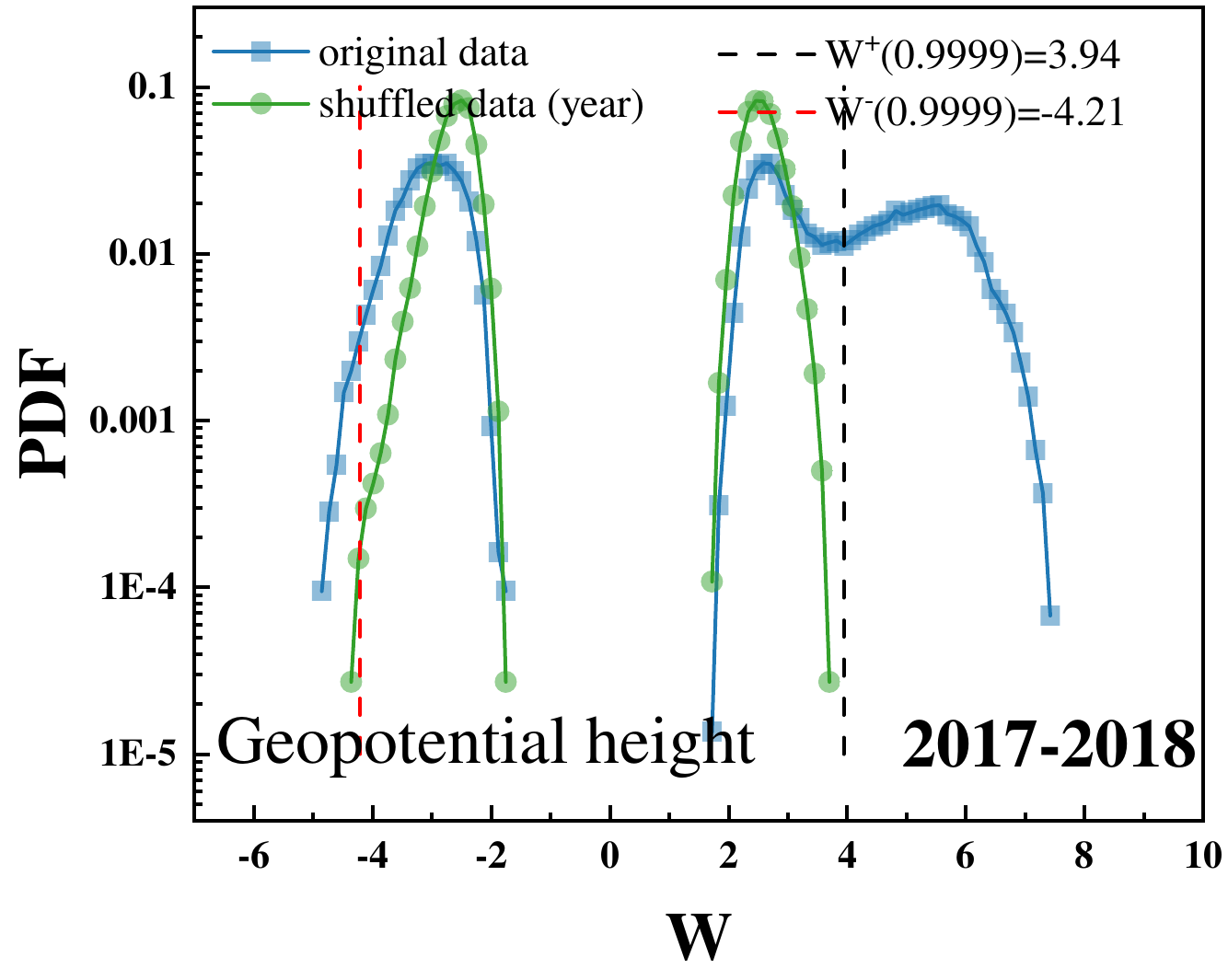}
\includegraphics[width=8.5em, height=7em]{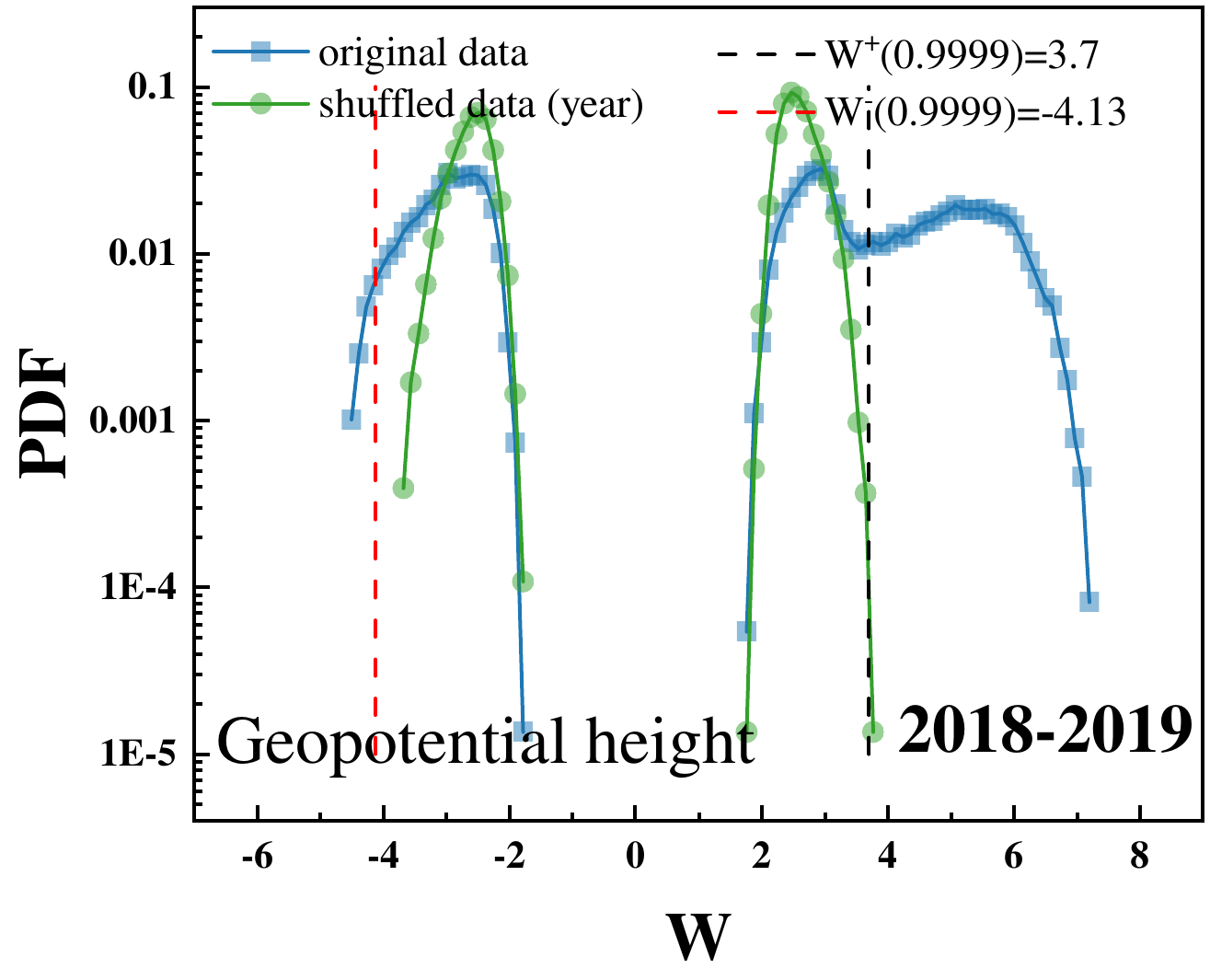}
\end{center}

\begin{center}
\noindent {\small {\bf Fig. S13} Probability distribution function (PDF) of link weights for the original data and shuffled data of Geopotential height in Europe. }
\end{center}

\begin{center}
\includegraphics[width=8.5em, height=7em]{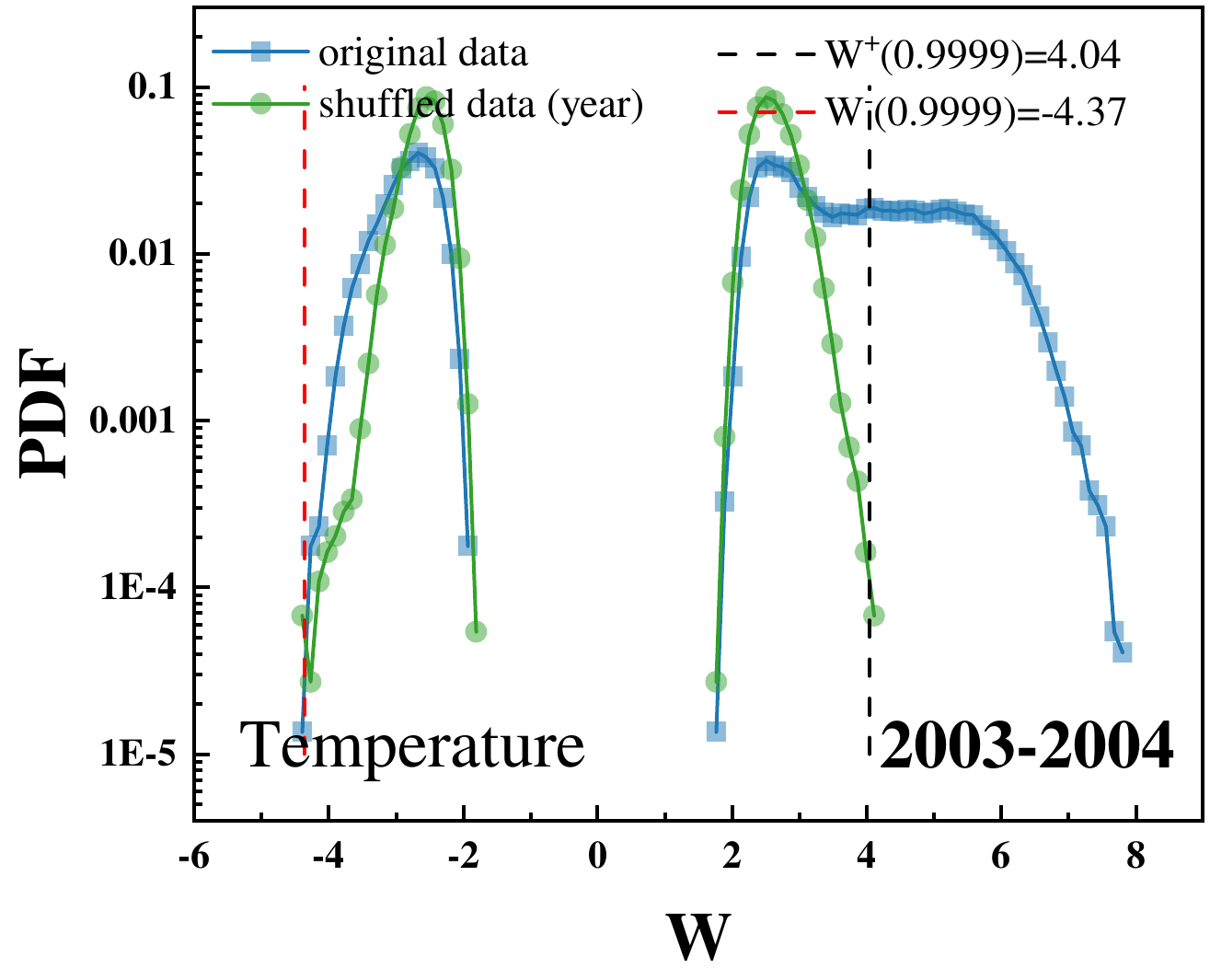}
\includegraphics[width=8.5em, height=7em]{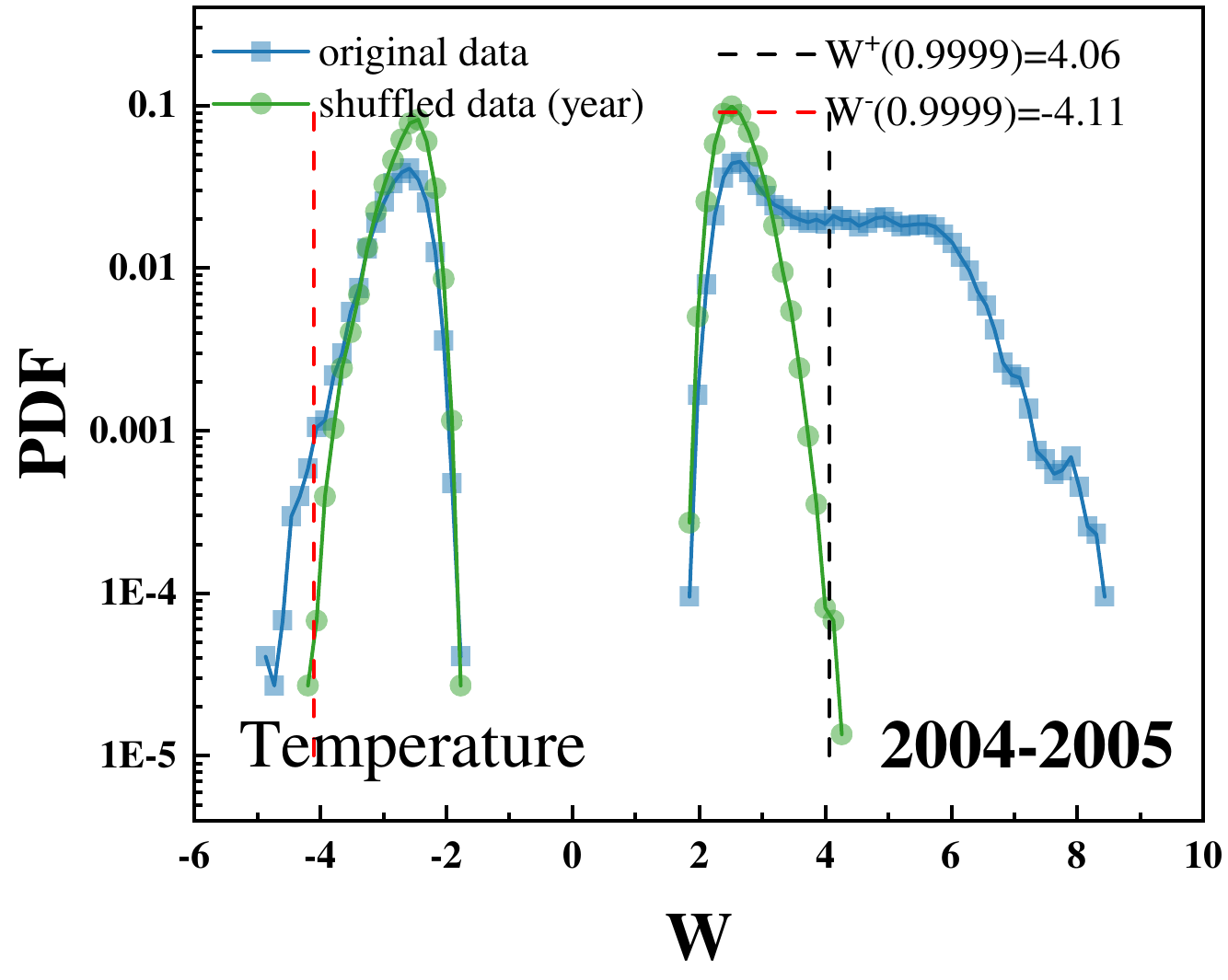}
\includegraphics[width=8.5em, height=7em]{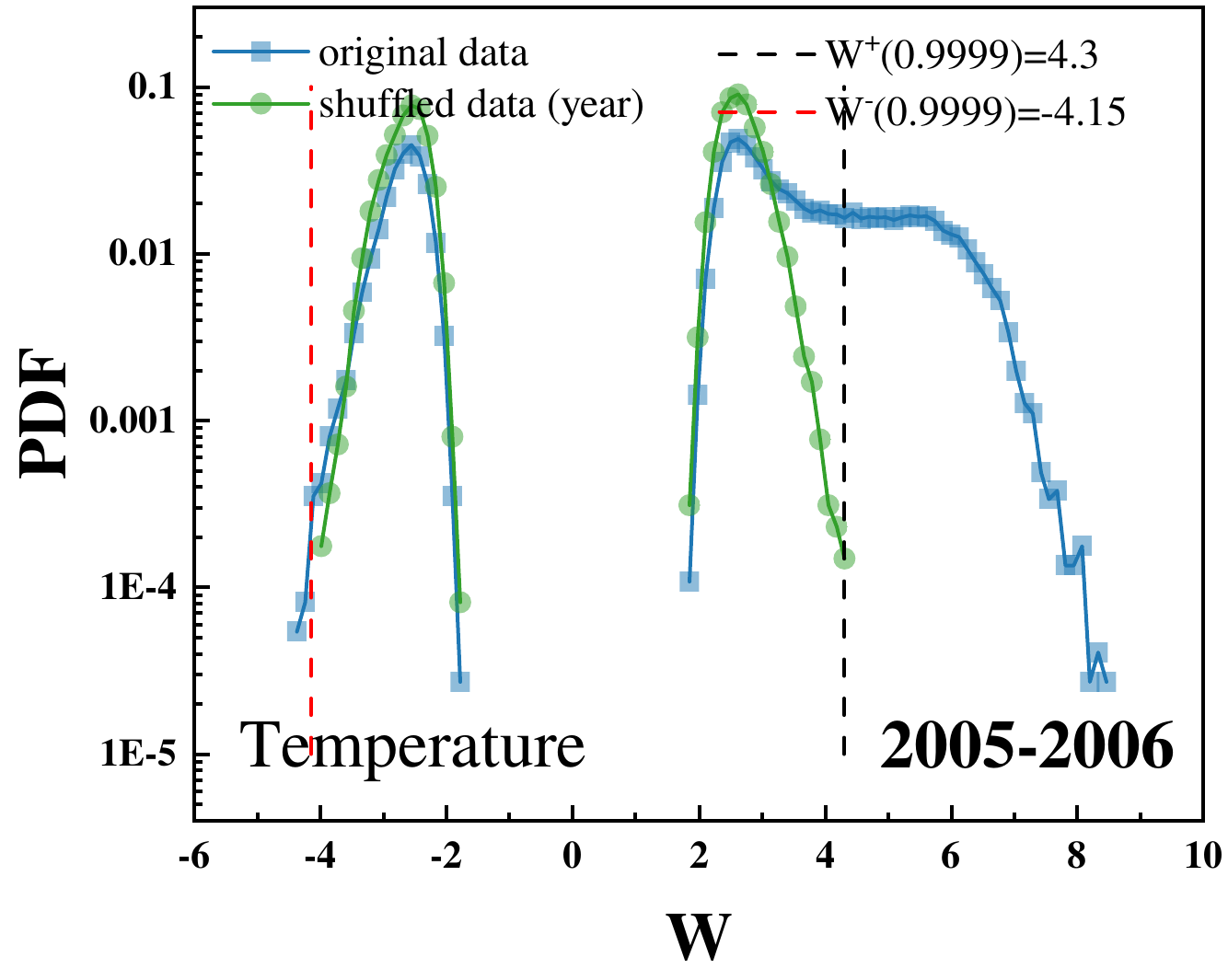}
\includegraphics[width=8.5em, height=7em]{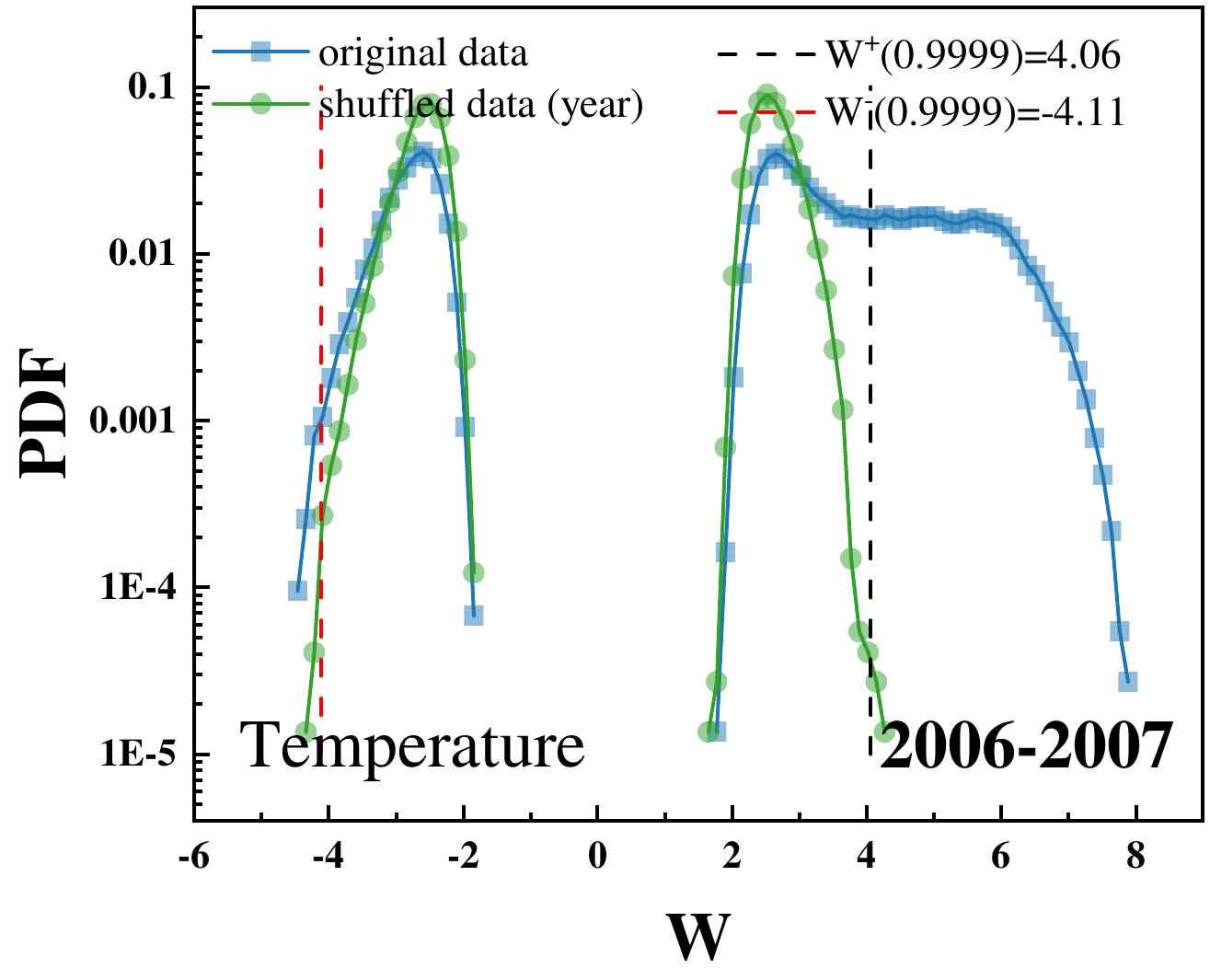}
\includegraphics[width=8.5em, height=7em]{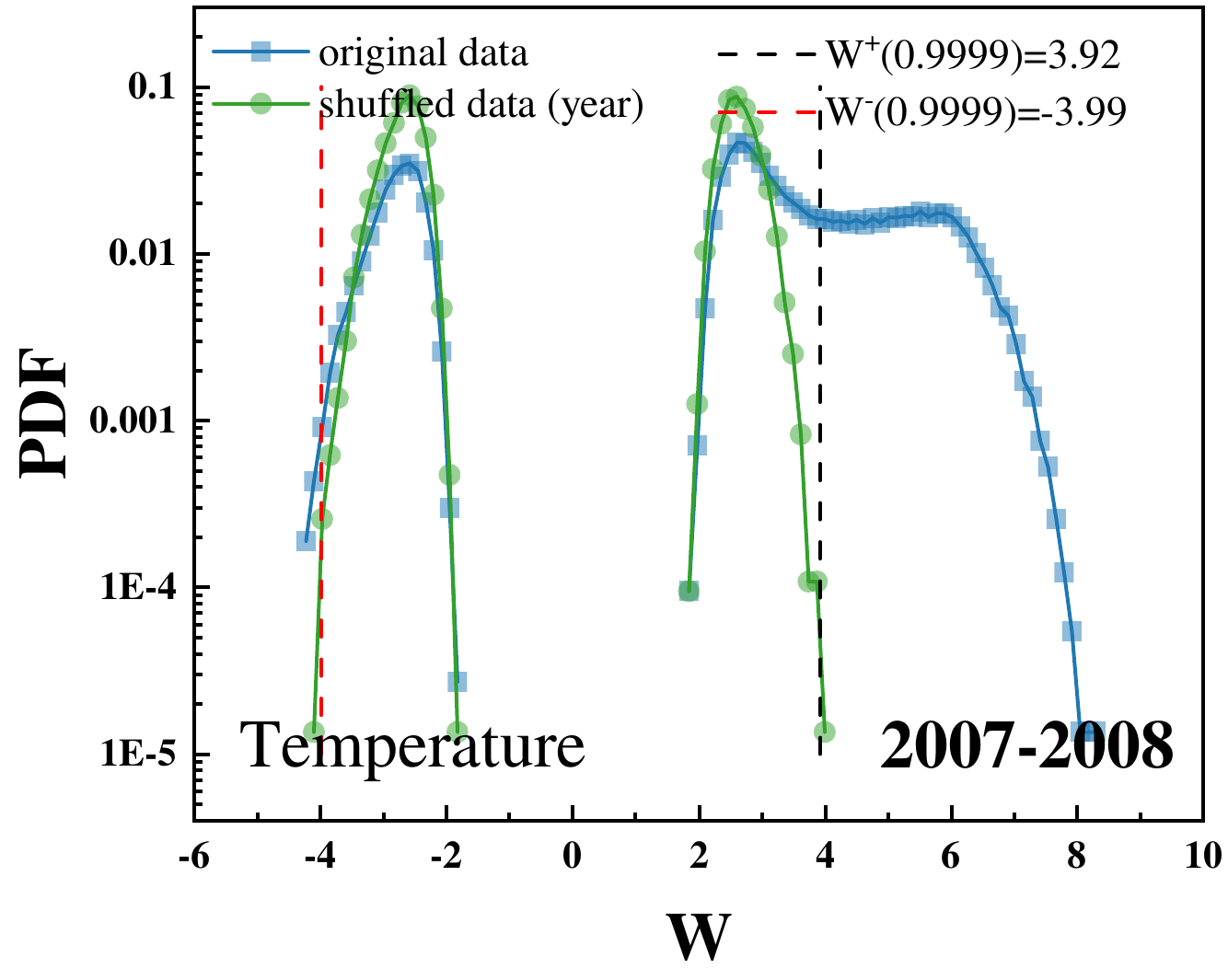}
\includegraphics[width=8.5em, height=7em]{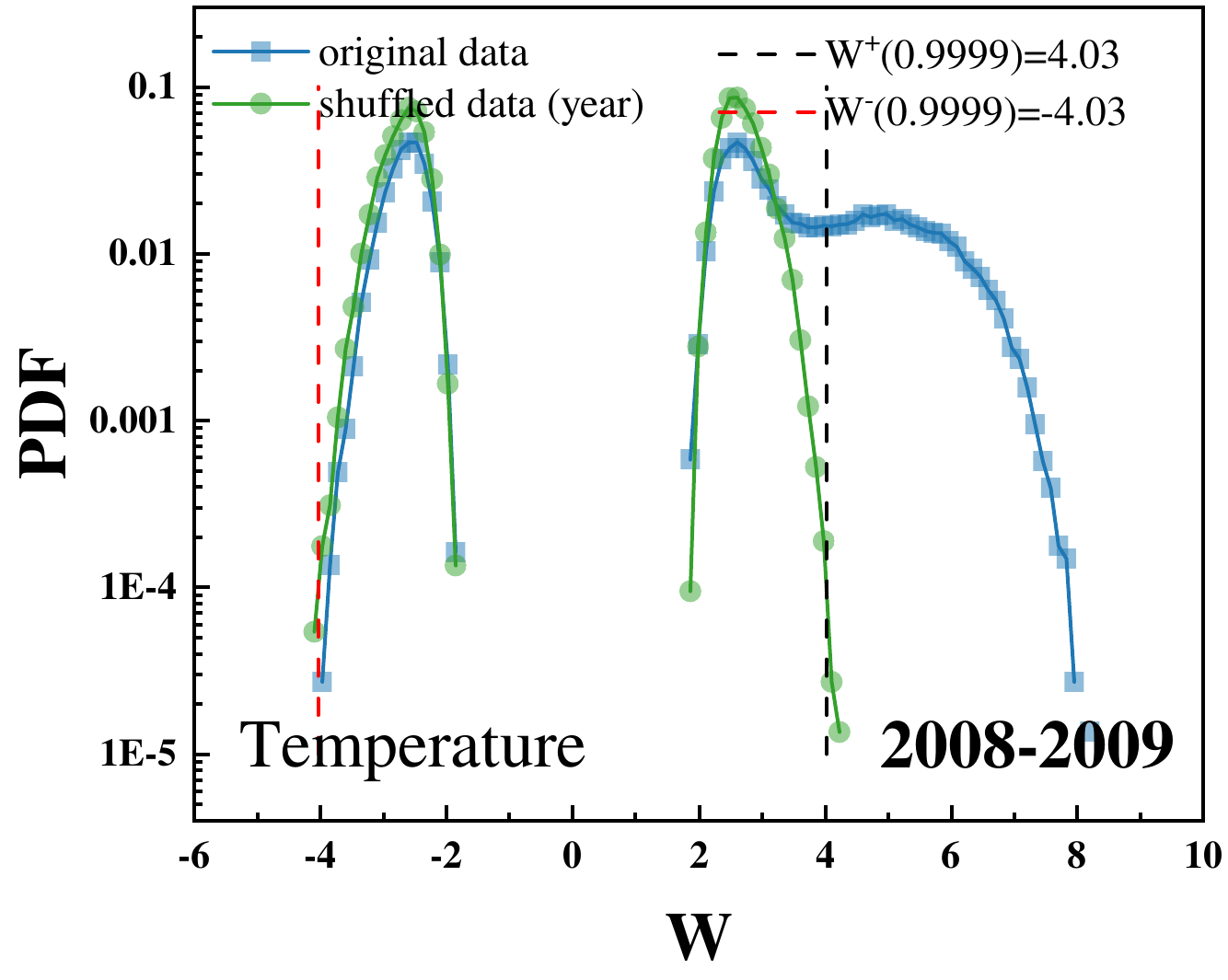}
\includegraphics[width=8.5em, height=7em]{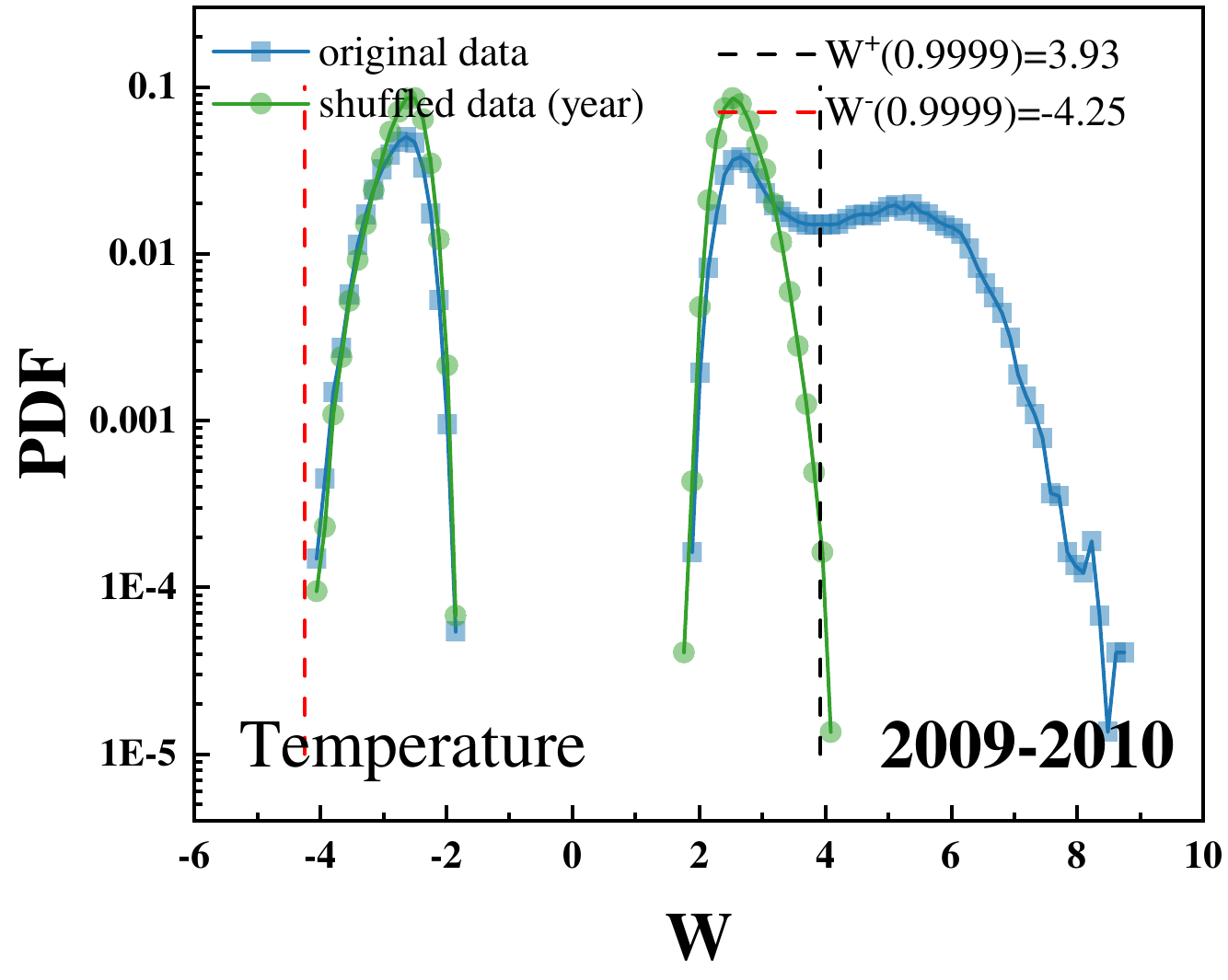}
\includegraphics[width=8.5em, height=7em]{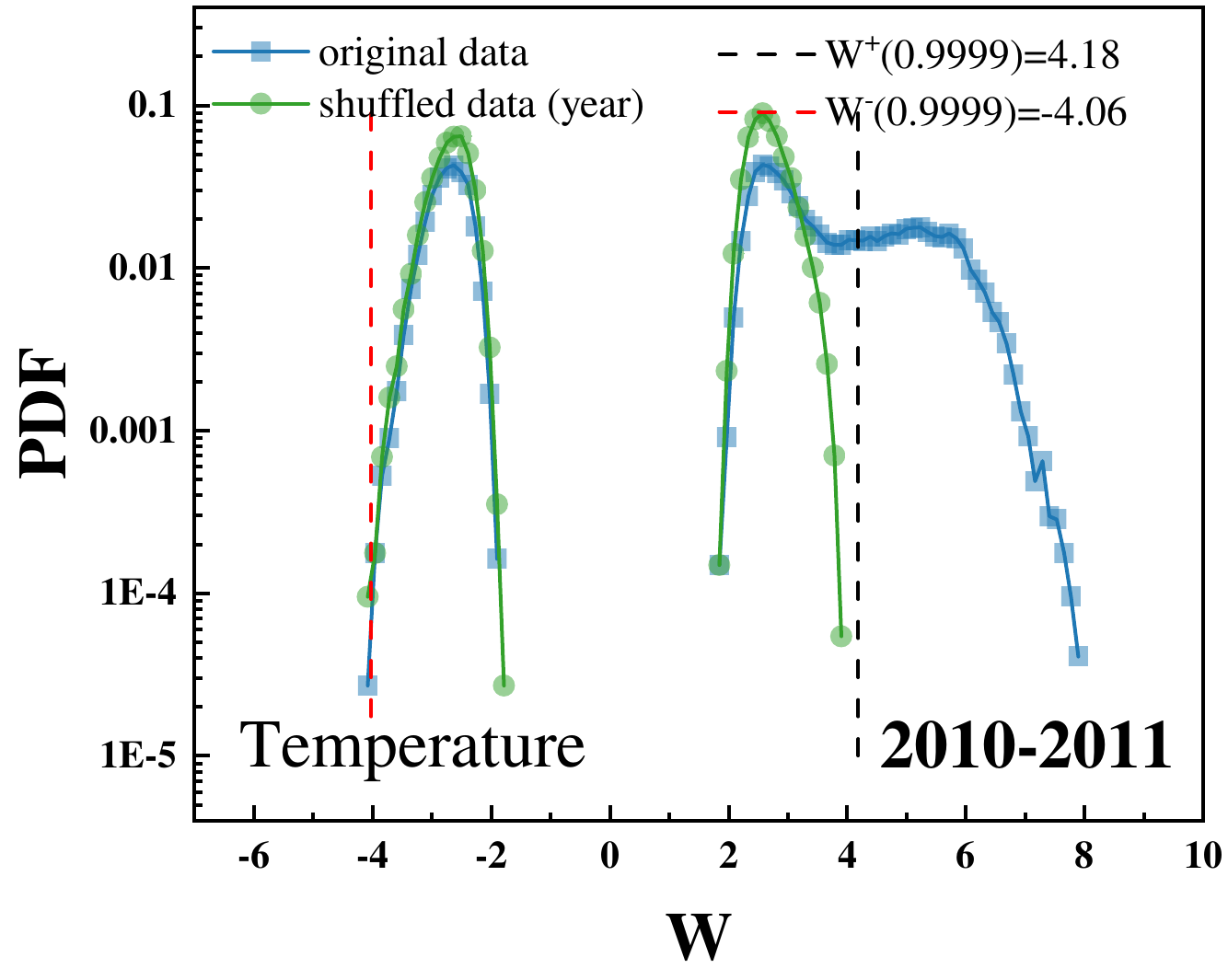}
\includegraphics[width=8.5em, height=7em]{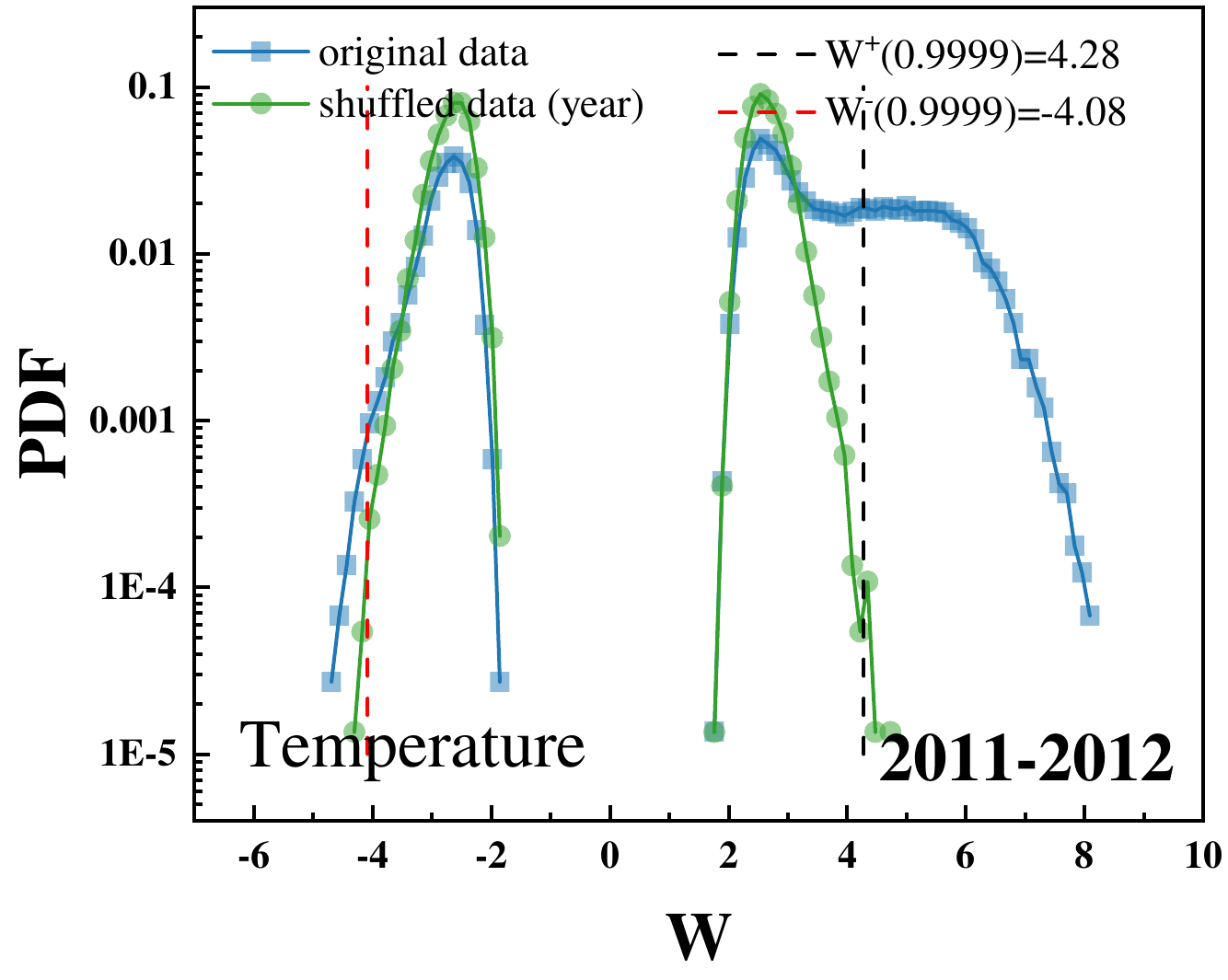}
\includegraphics[width=8.5em, height=7em]{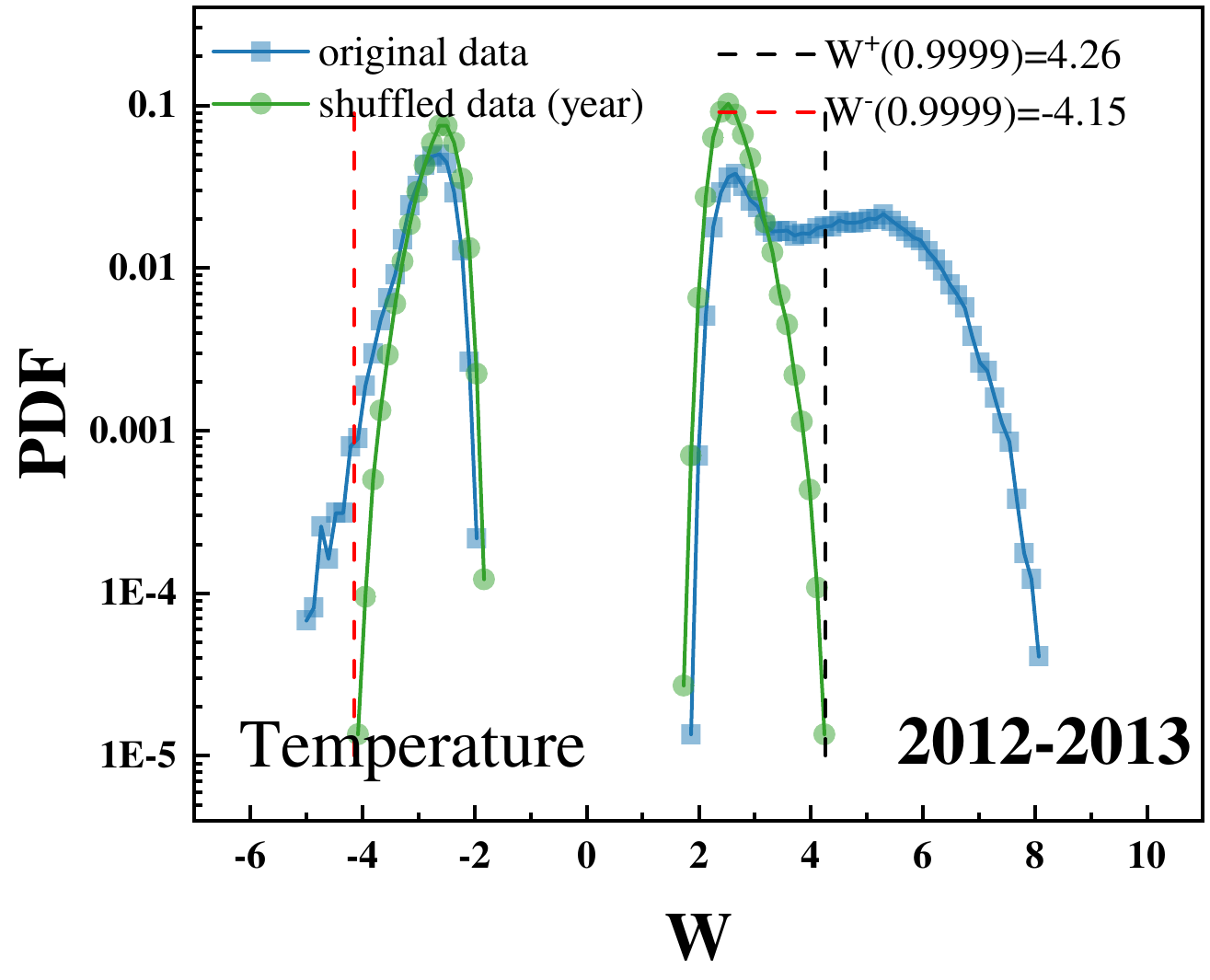}
\includegraphics[width=8.5em, height=7em]{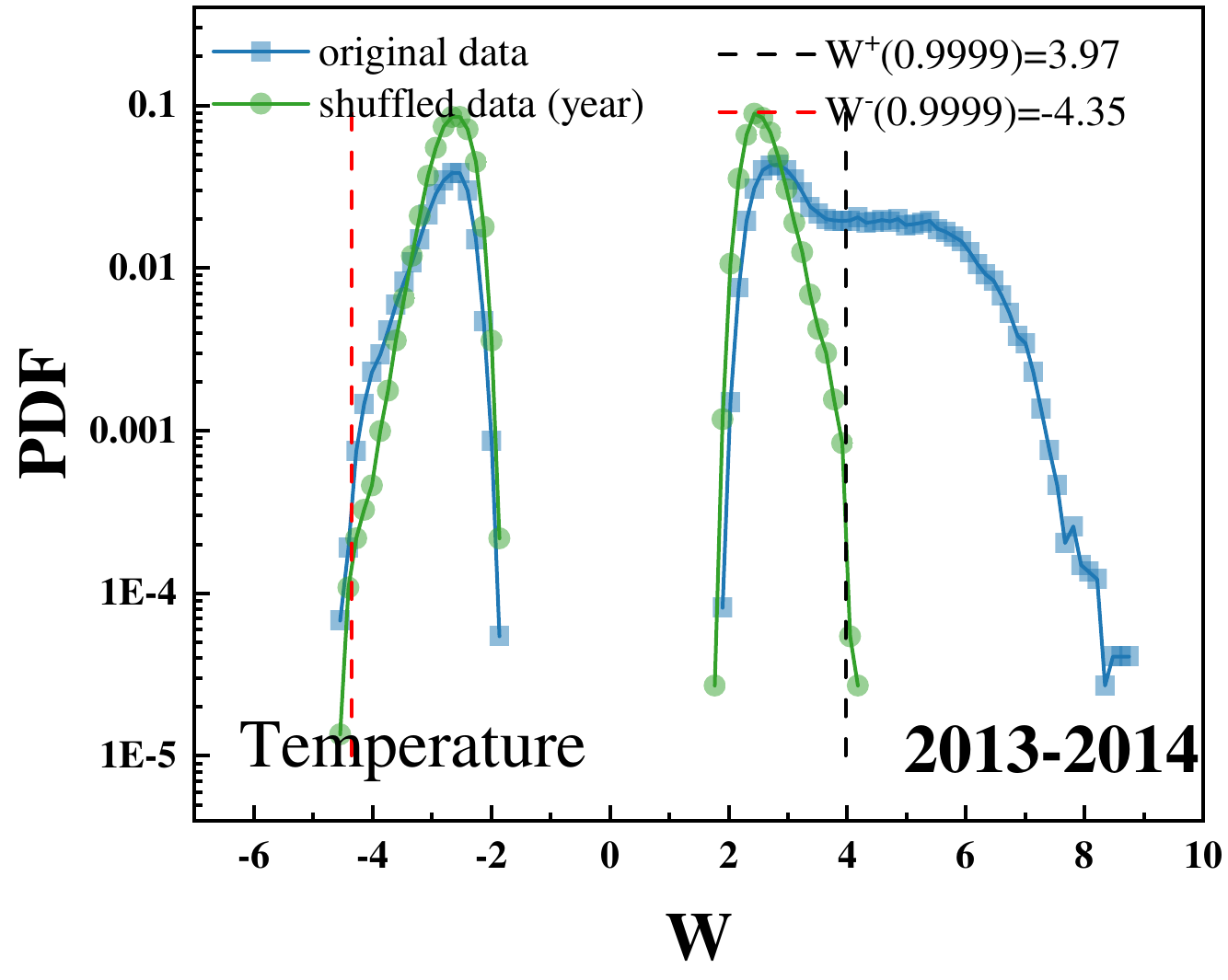}
\includegraphics[width=8.5em, height=7em]{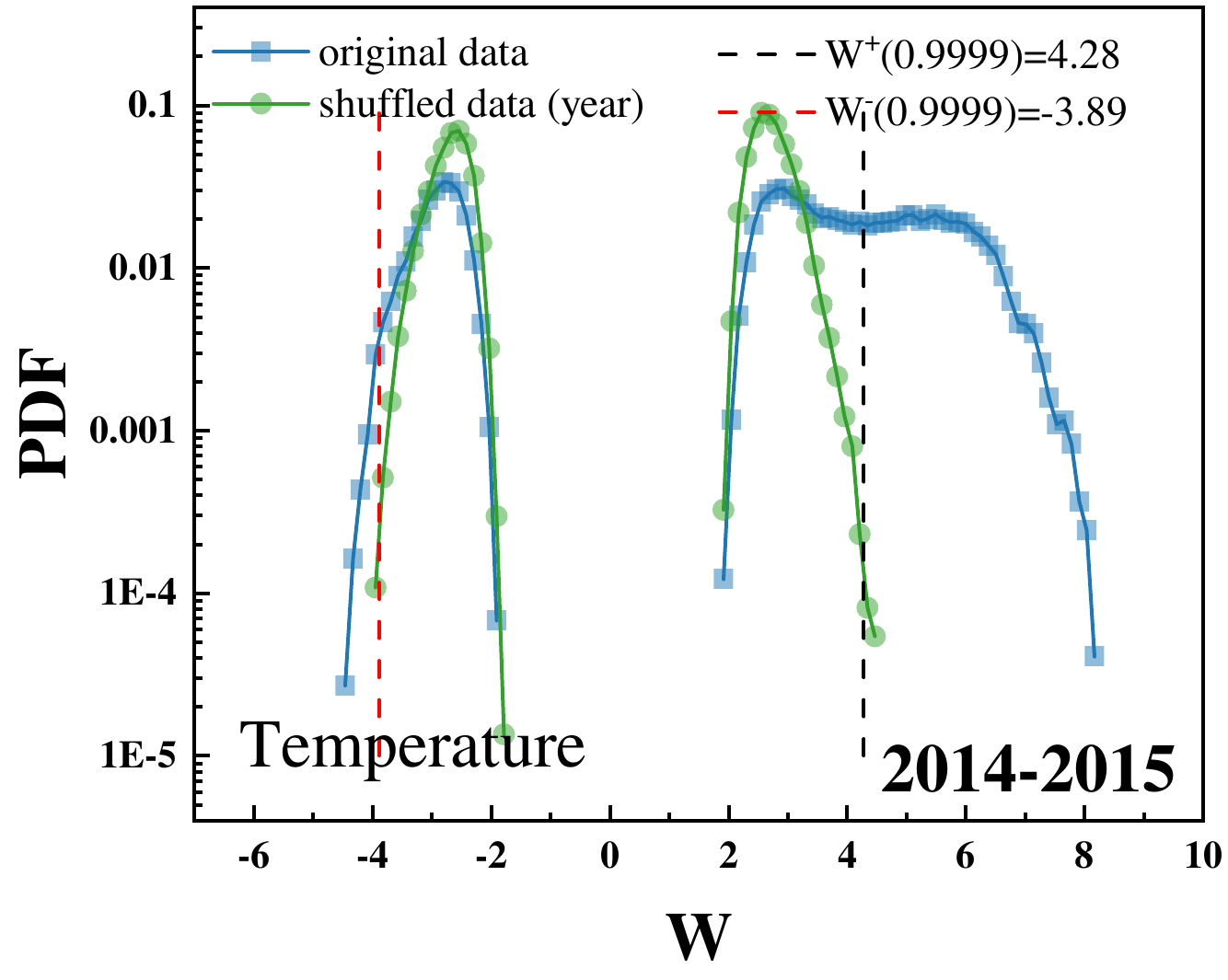}
\includegraphics[width=8.5em, height=7em]{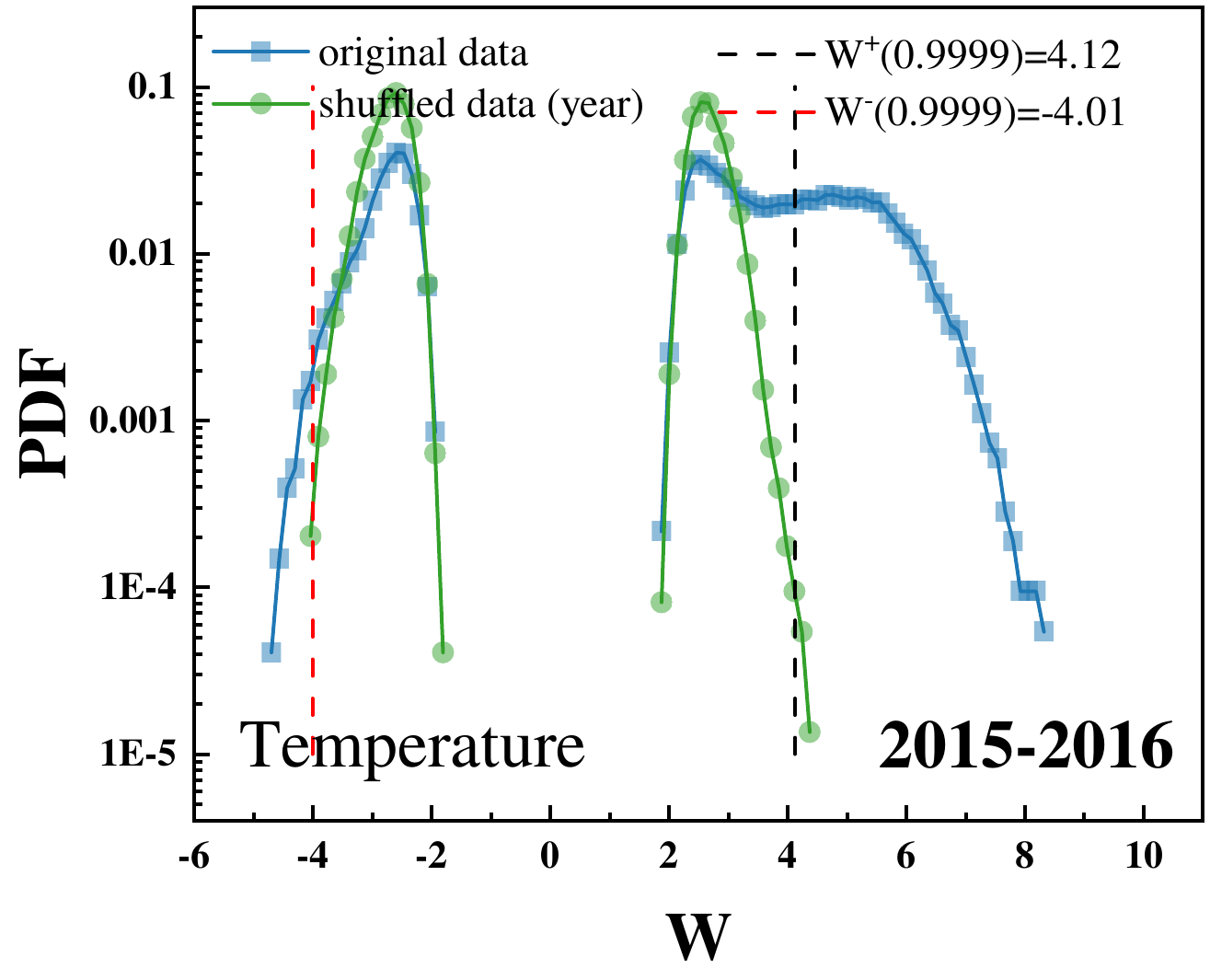}
\includegraphics[width=8.5em, height=7em]{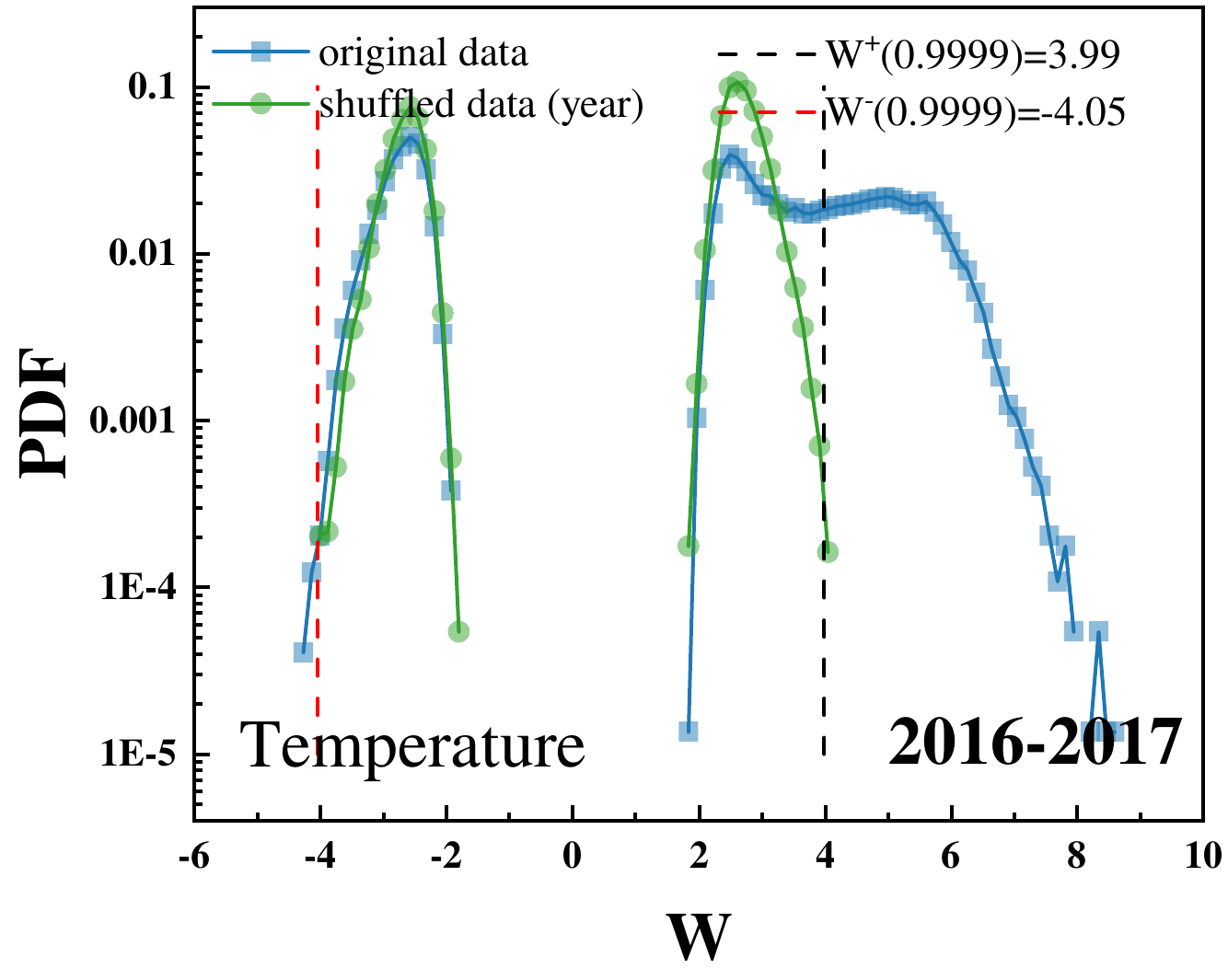}
\includegraphics[width=8.5em, height=7em]{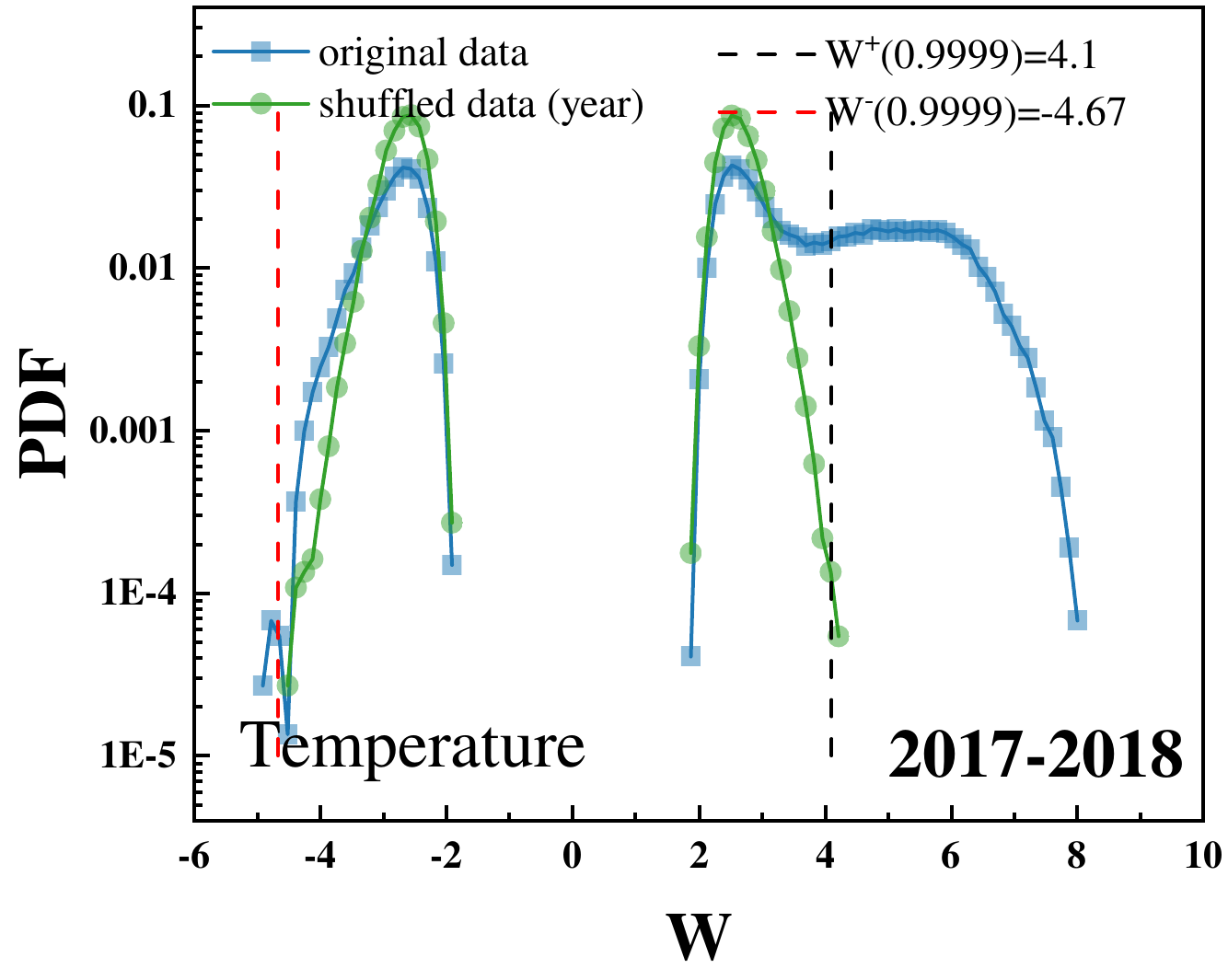}
\includegraphics[width=8.5em, height=7em]{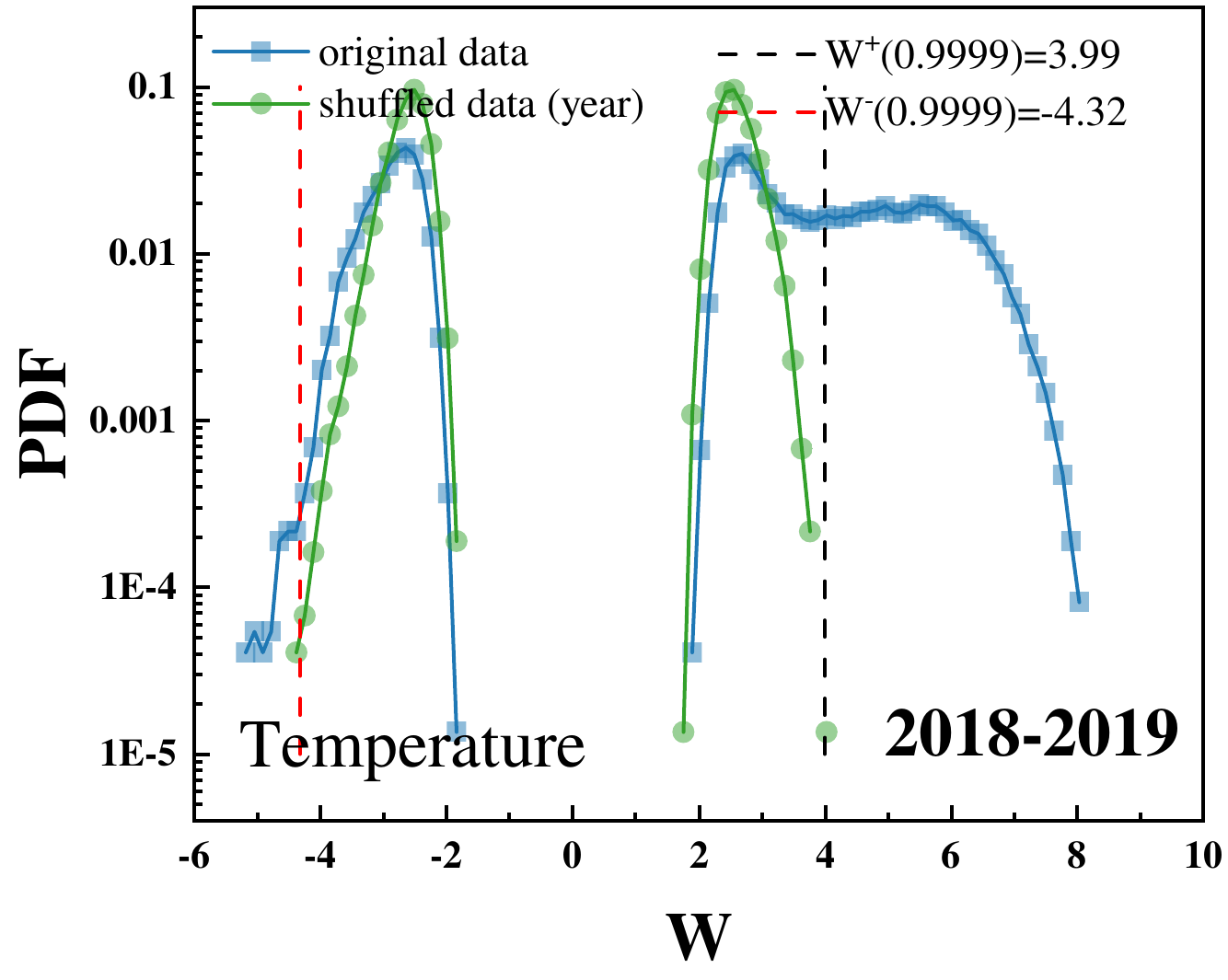}
\end{center}

\begin{center}
\noindent {\small {\bf Fig. S14} Probability distribution function (PDF) of link weights for the original data and shuffled data of temperature in Europe. }
\end{center}

\begin{center}
\includegraphics[width=8.5em, height=7em]{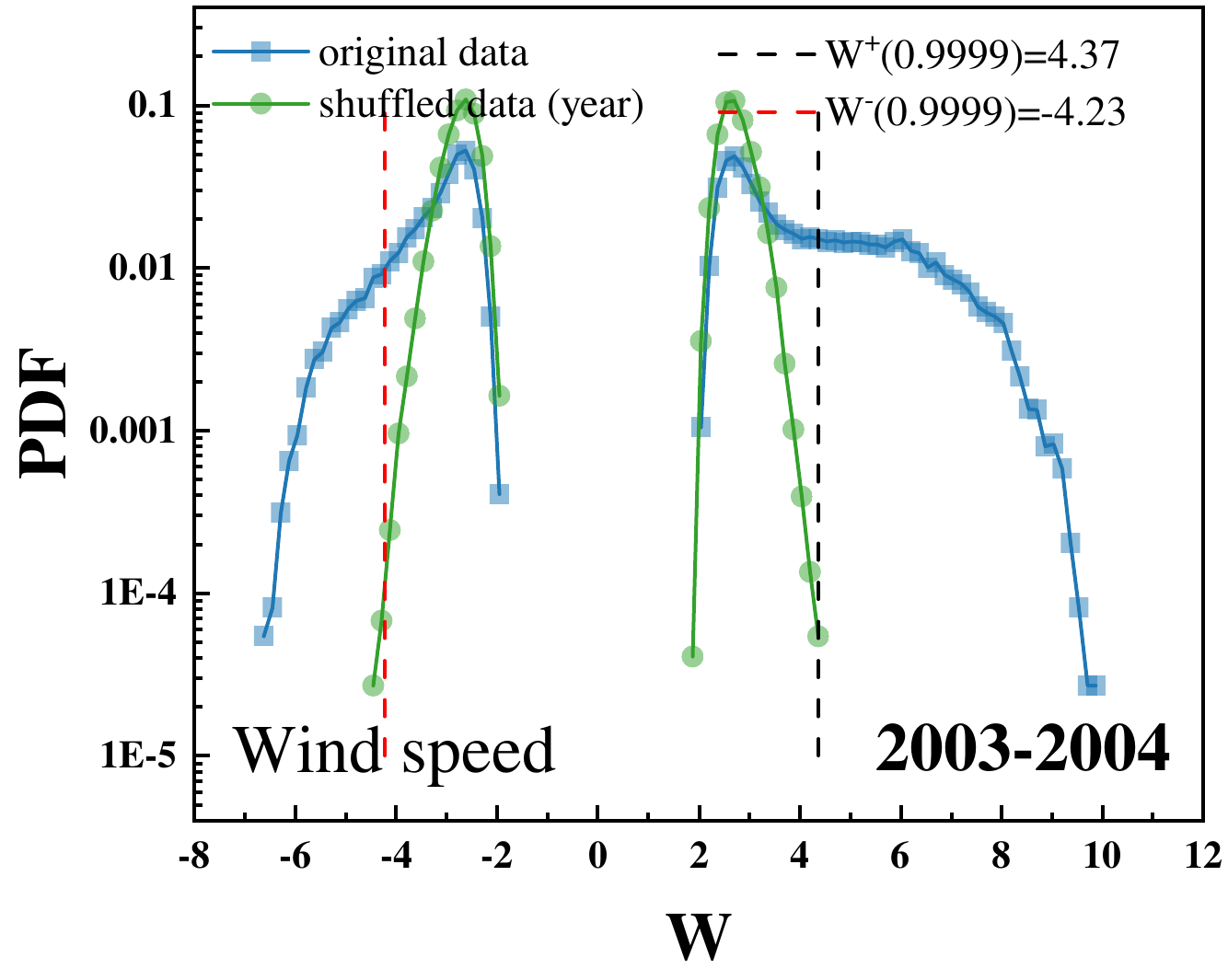}
\includegraphics[width=8.5em, height=7em]{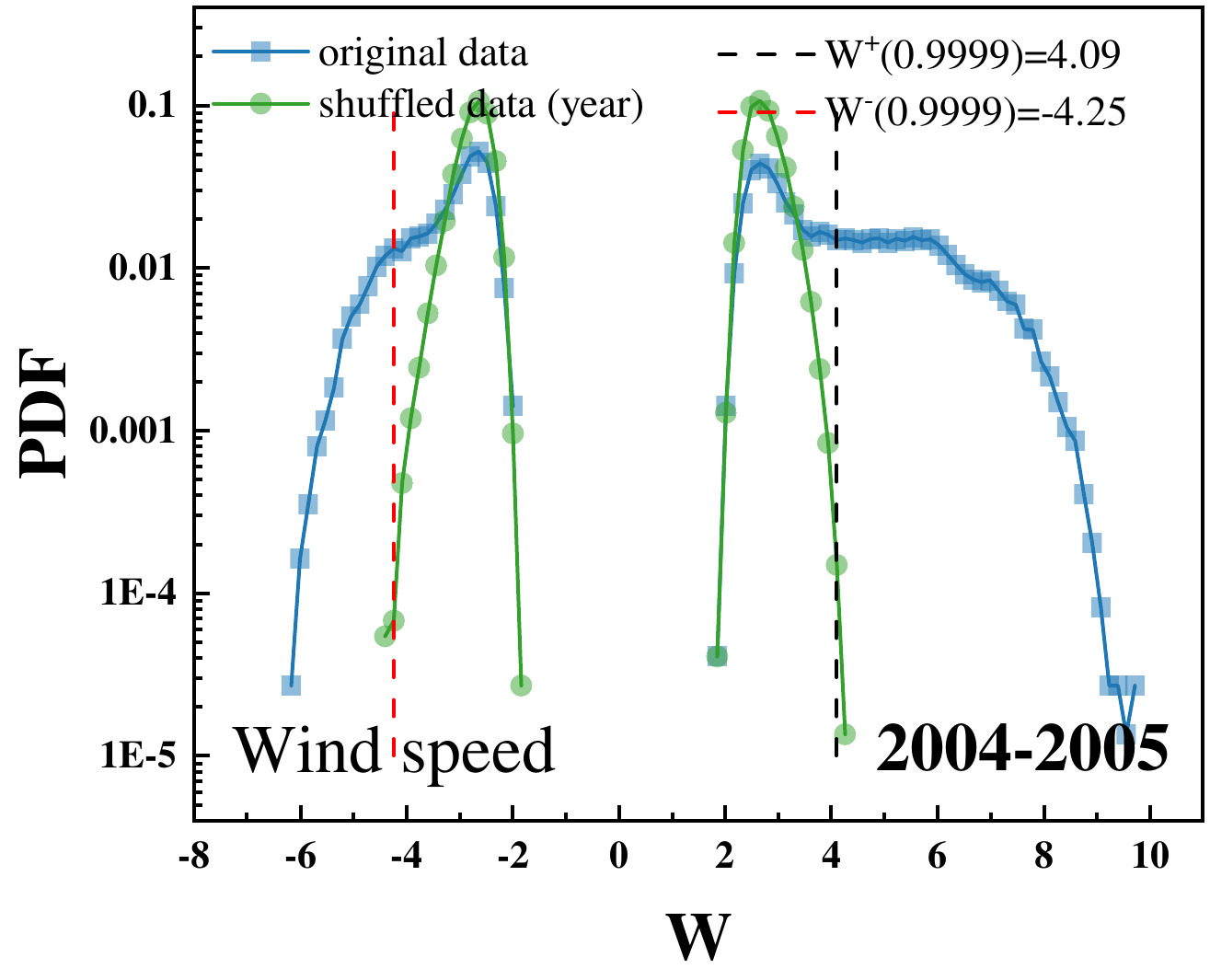}
\includegraphics[width=8.5em, height=7em]{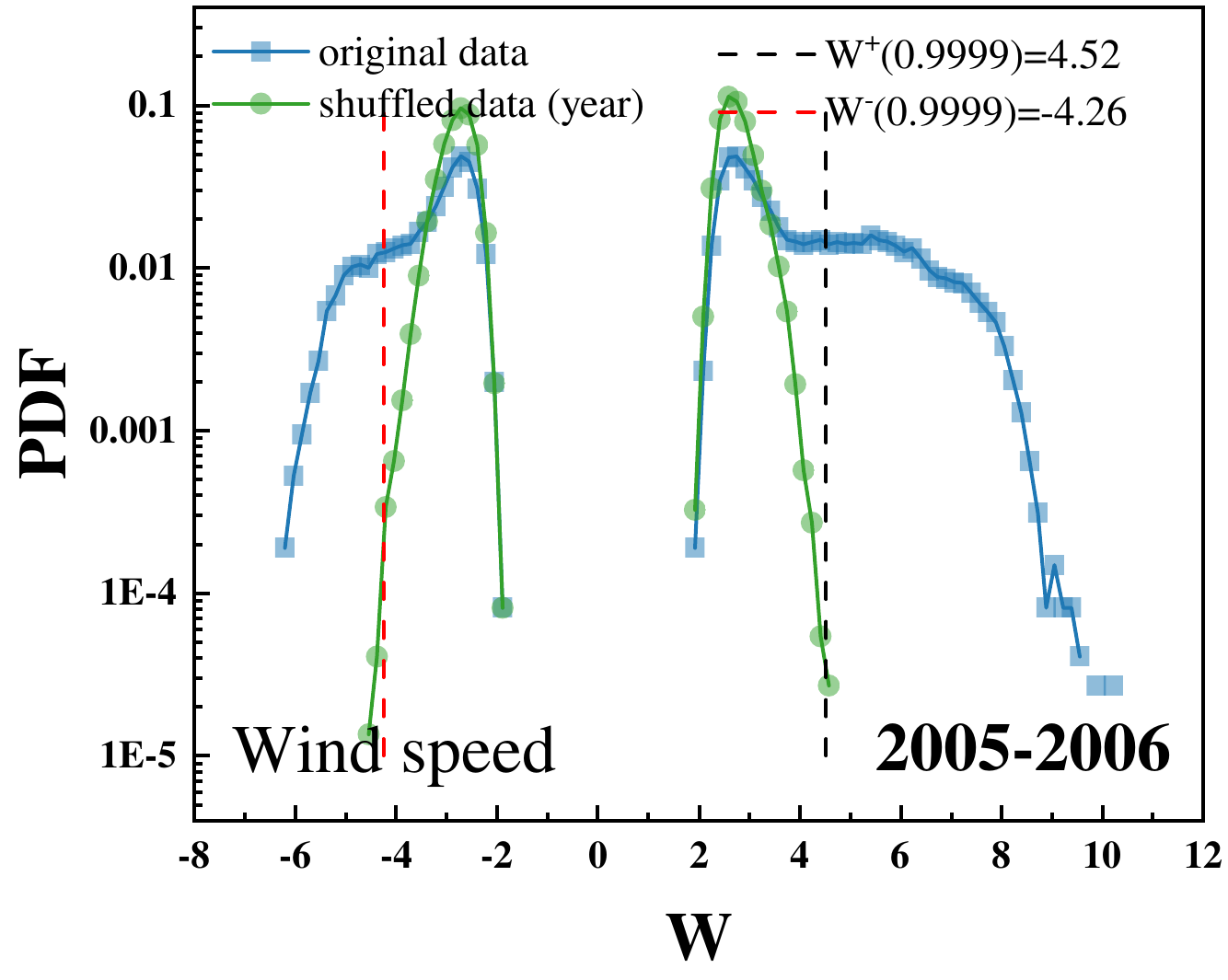}
\includegraphics[width=8.5em, height=7em]{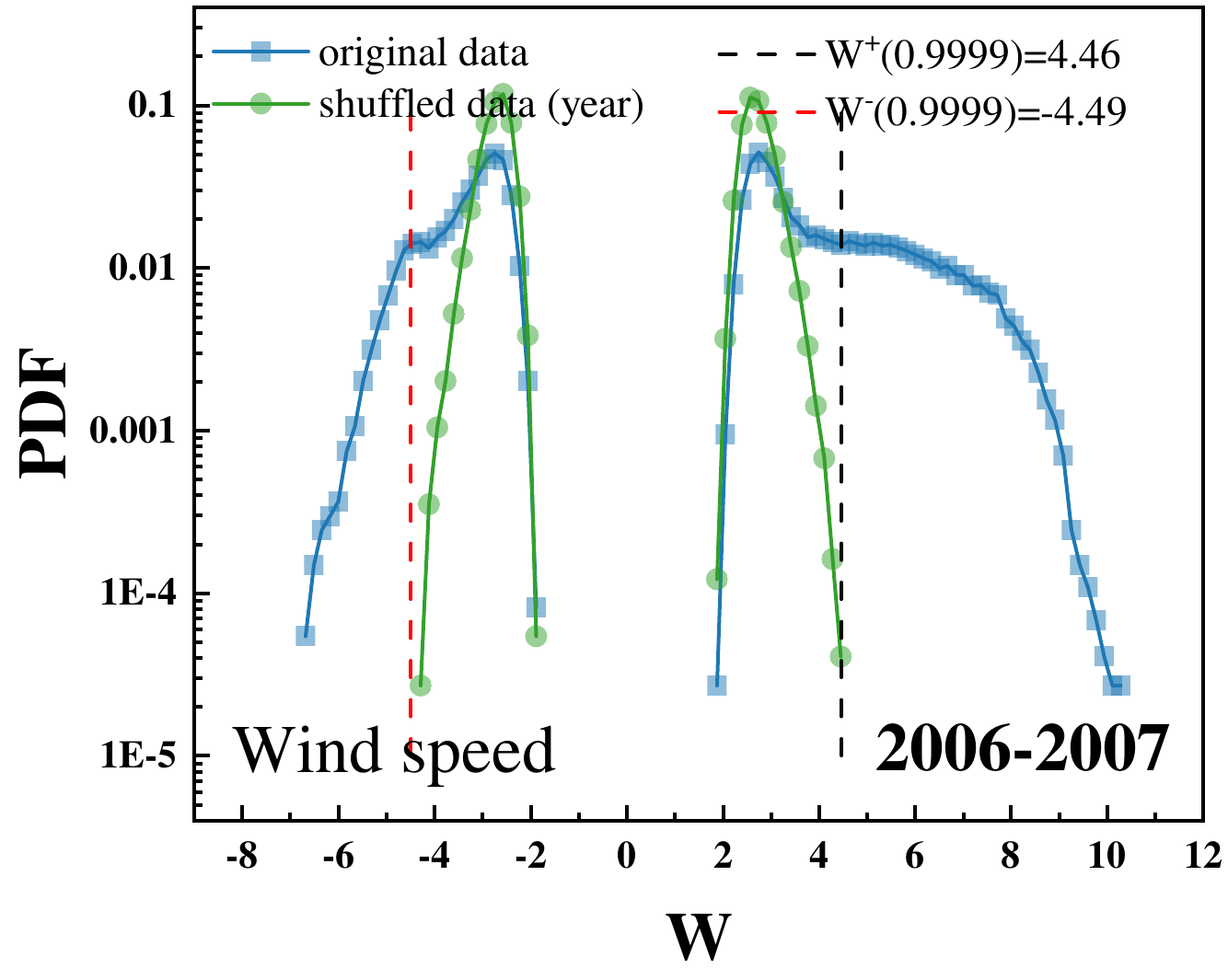}
\includegraphics[width=8.5em, height=7em]{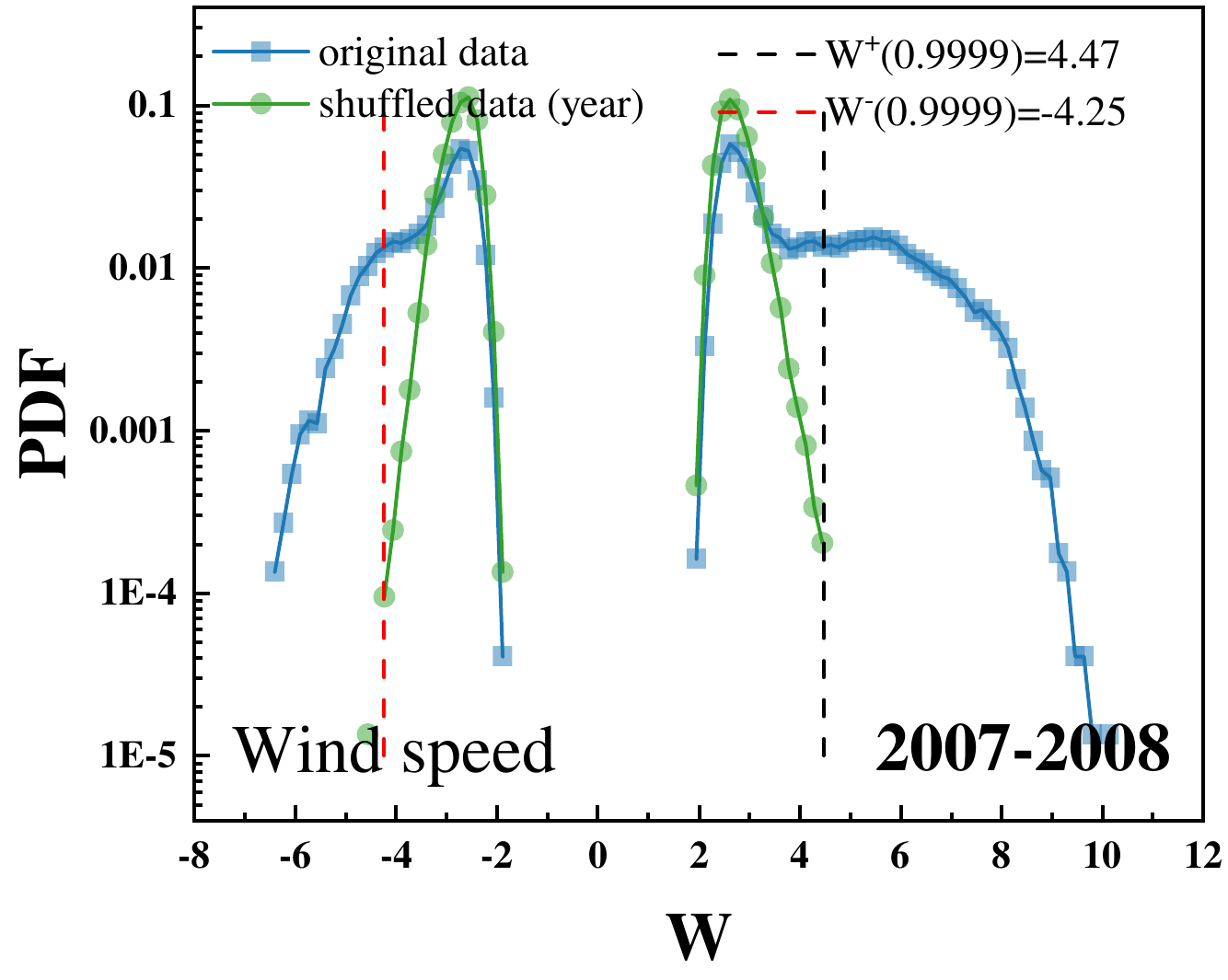}
\includegraphics[width=8.5em, height=7em]{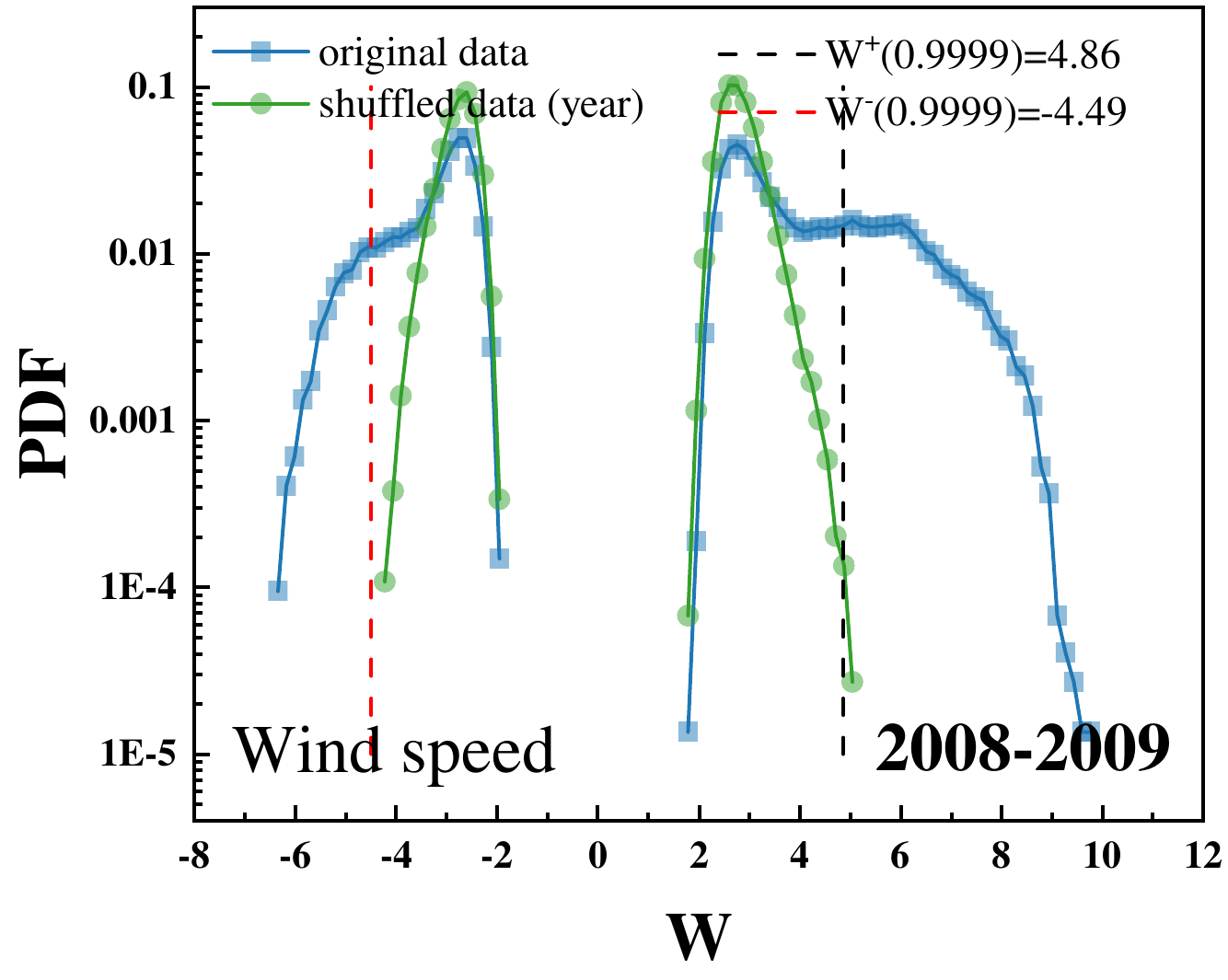}
\includegraphics[width=8.5em, height=7em]{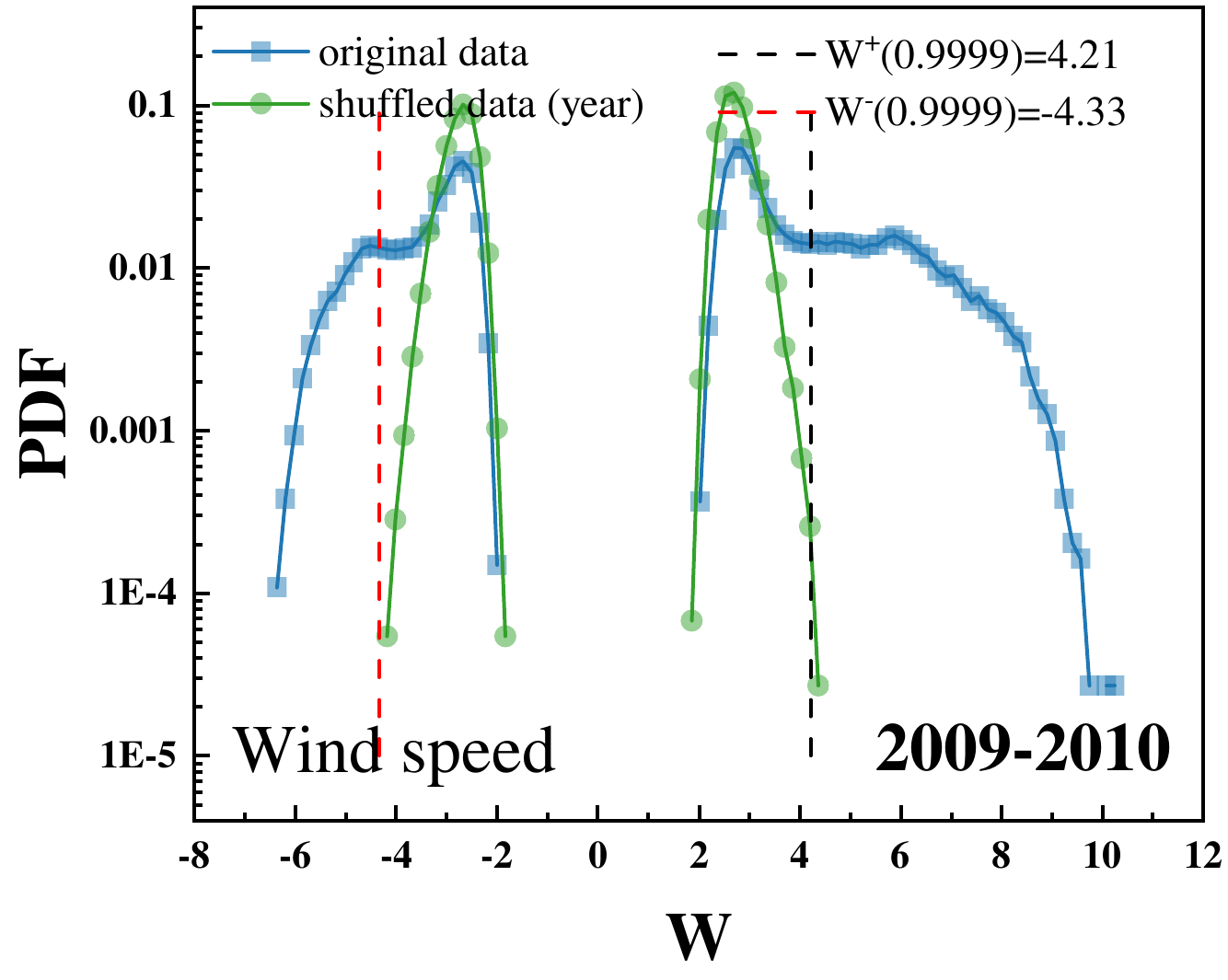}
\includegraphics[width=8.5em, height=7em]{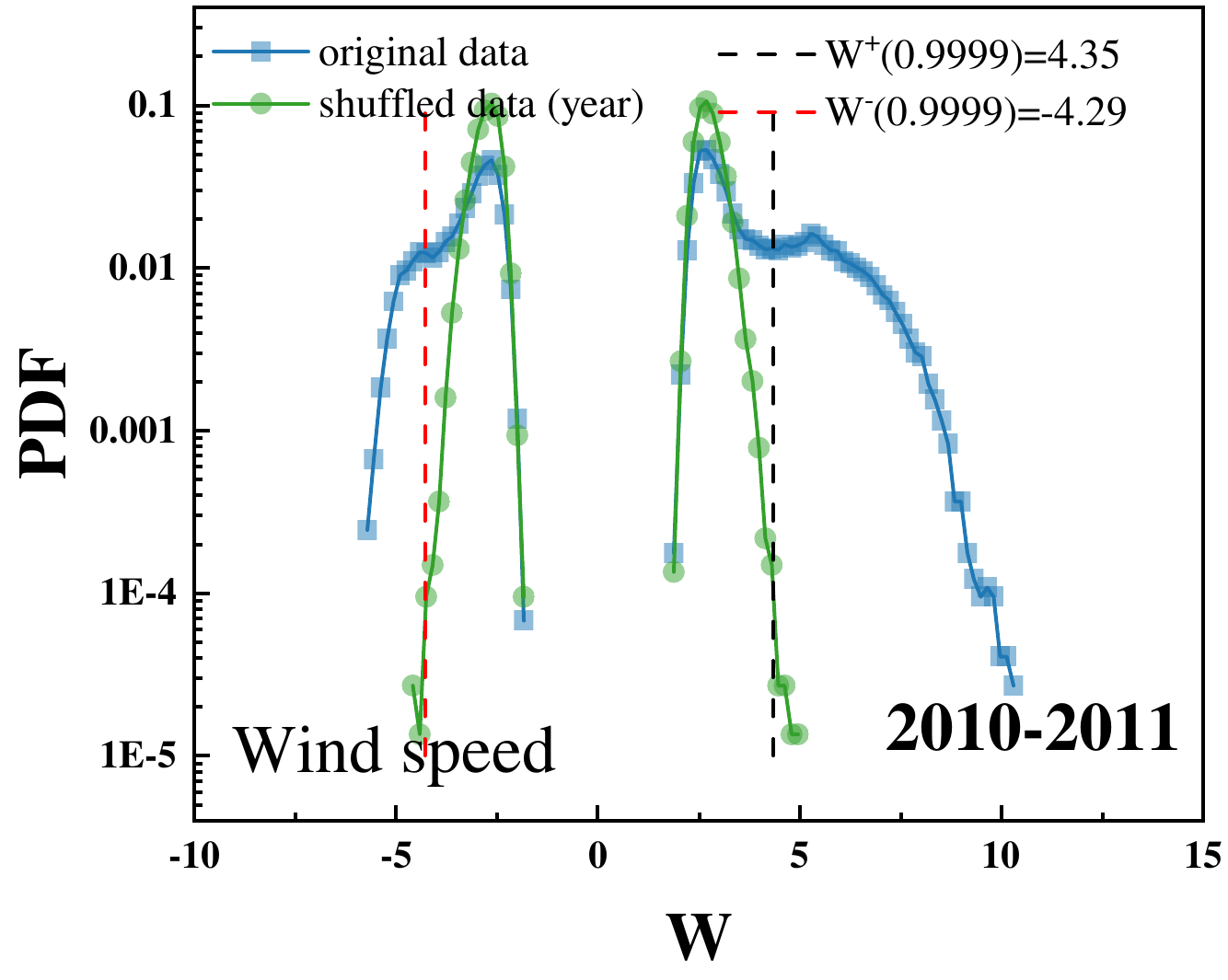}
\includegraphics[width=8.5em, height=7em]{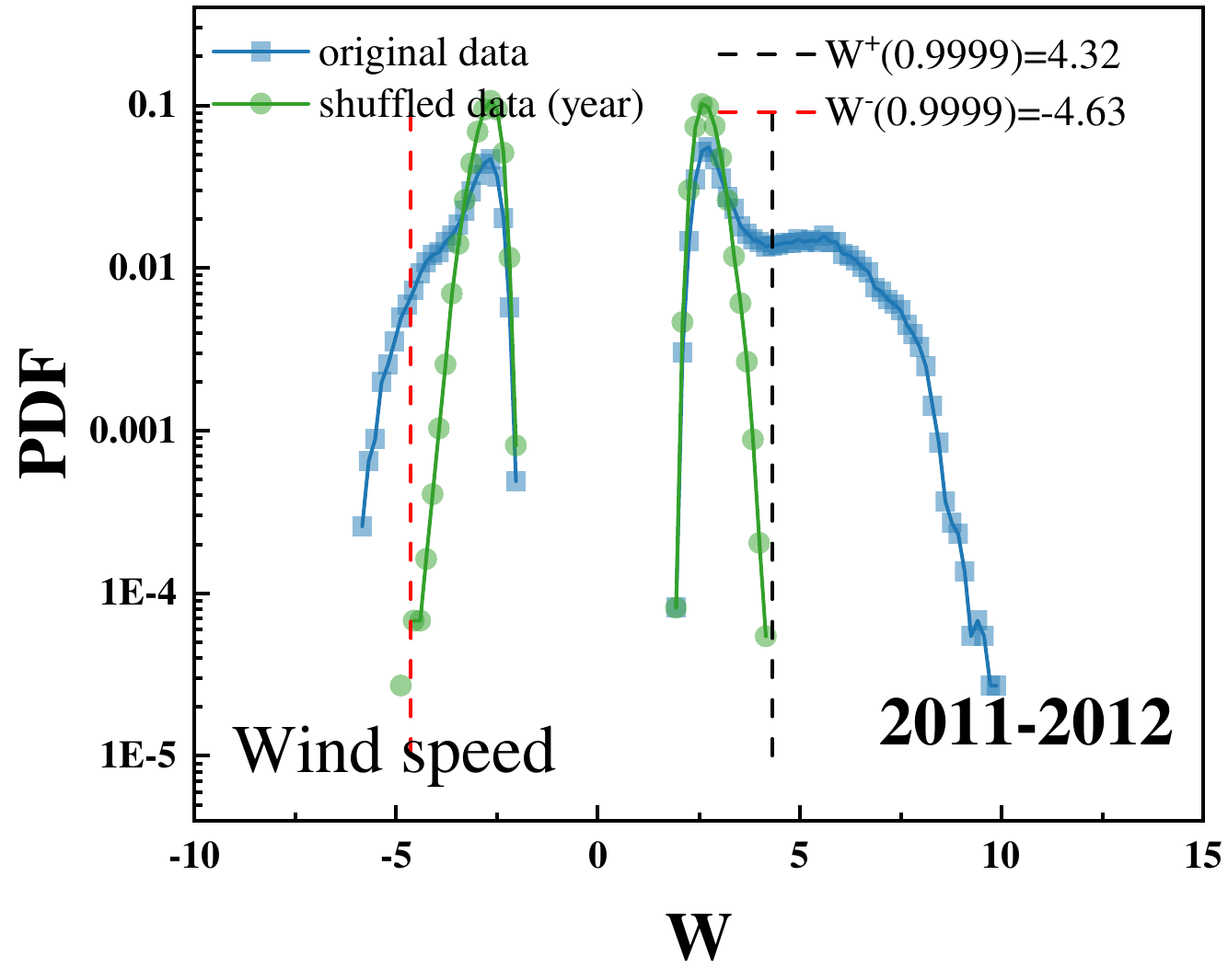}
\includegraphics[width=8.5em, height=7em]{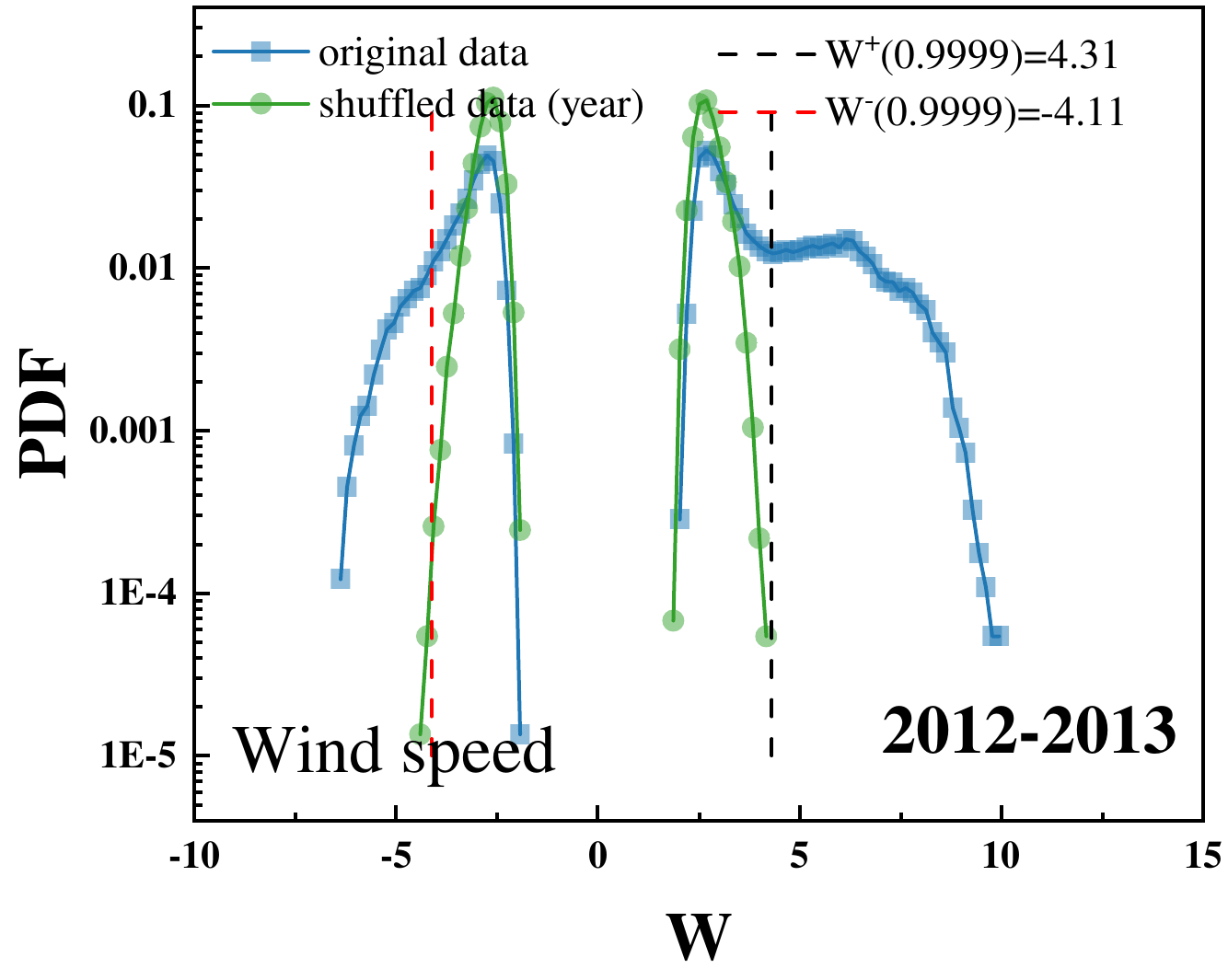}
\includegraphics[width=8.5em, height=7em]{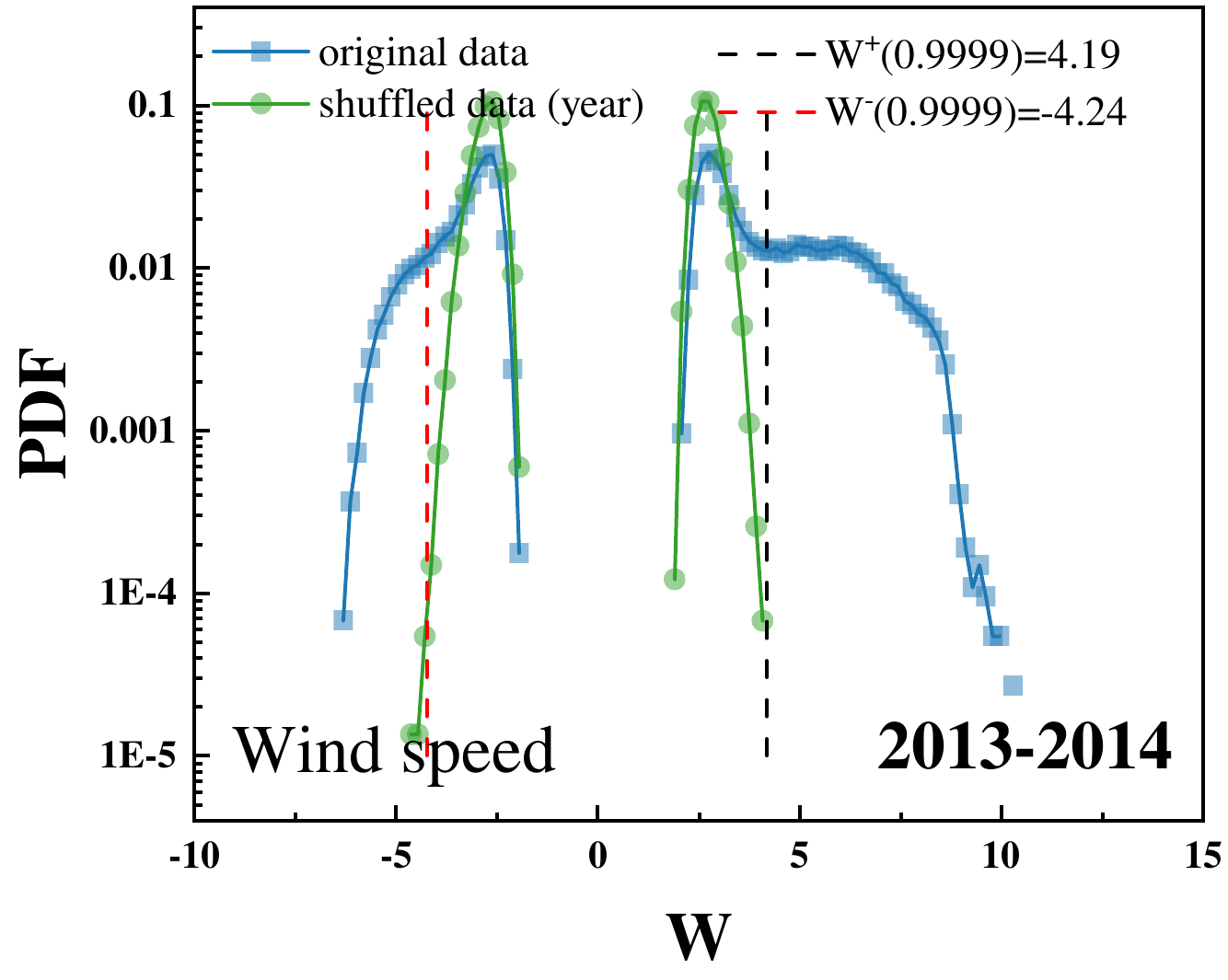}
\includegraphics[width=8.5em, height=7em]{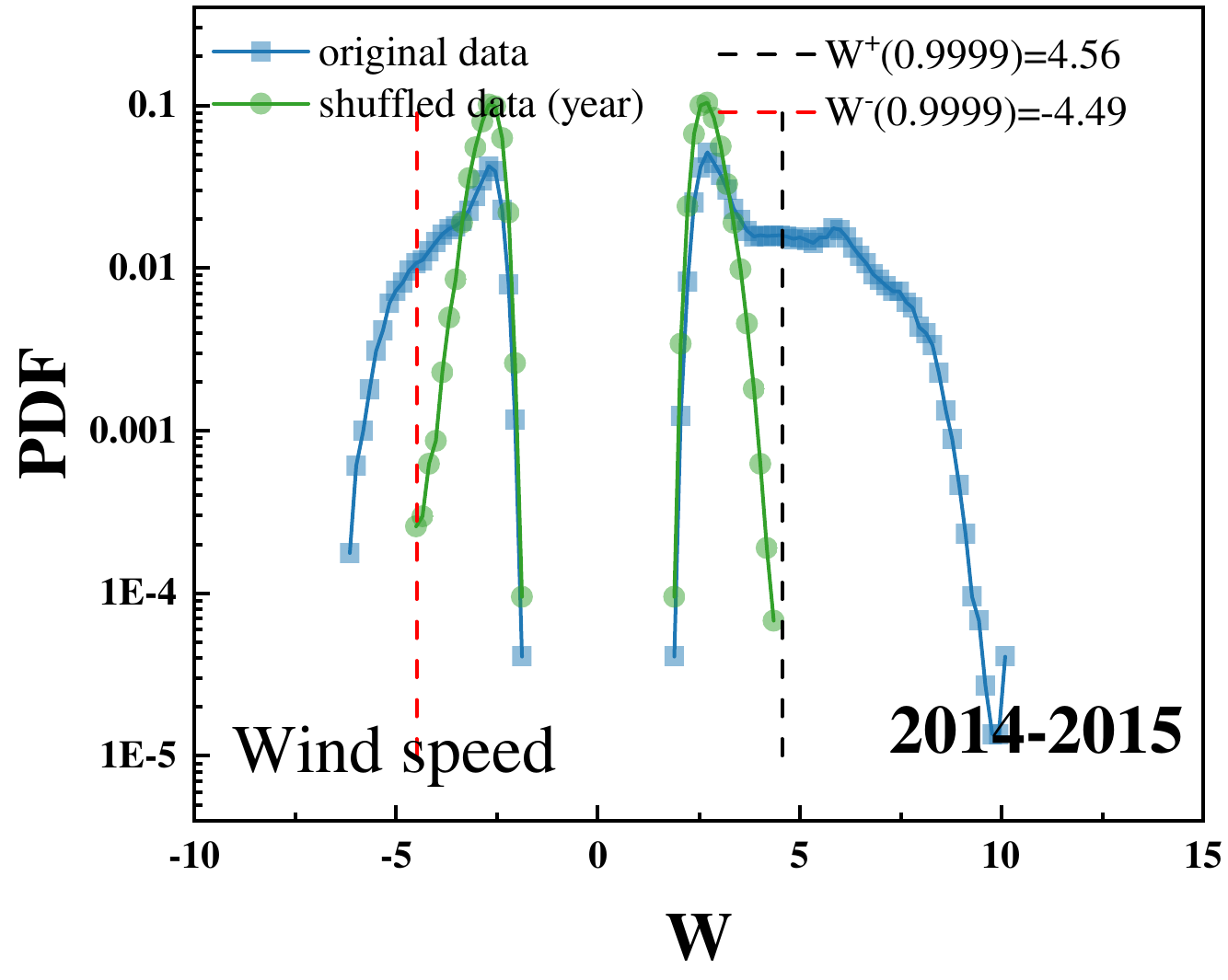}
\includegraphics[width=8.5em, height=7em]{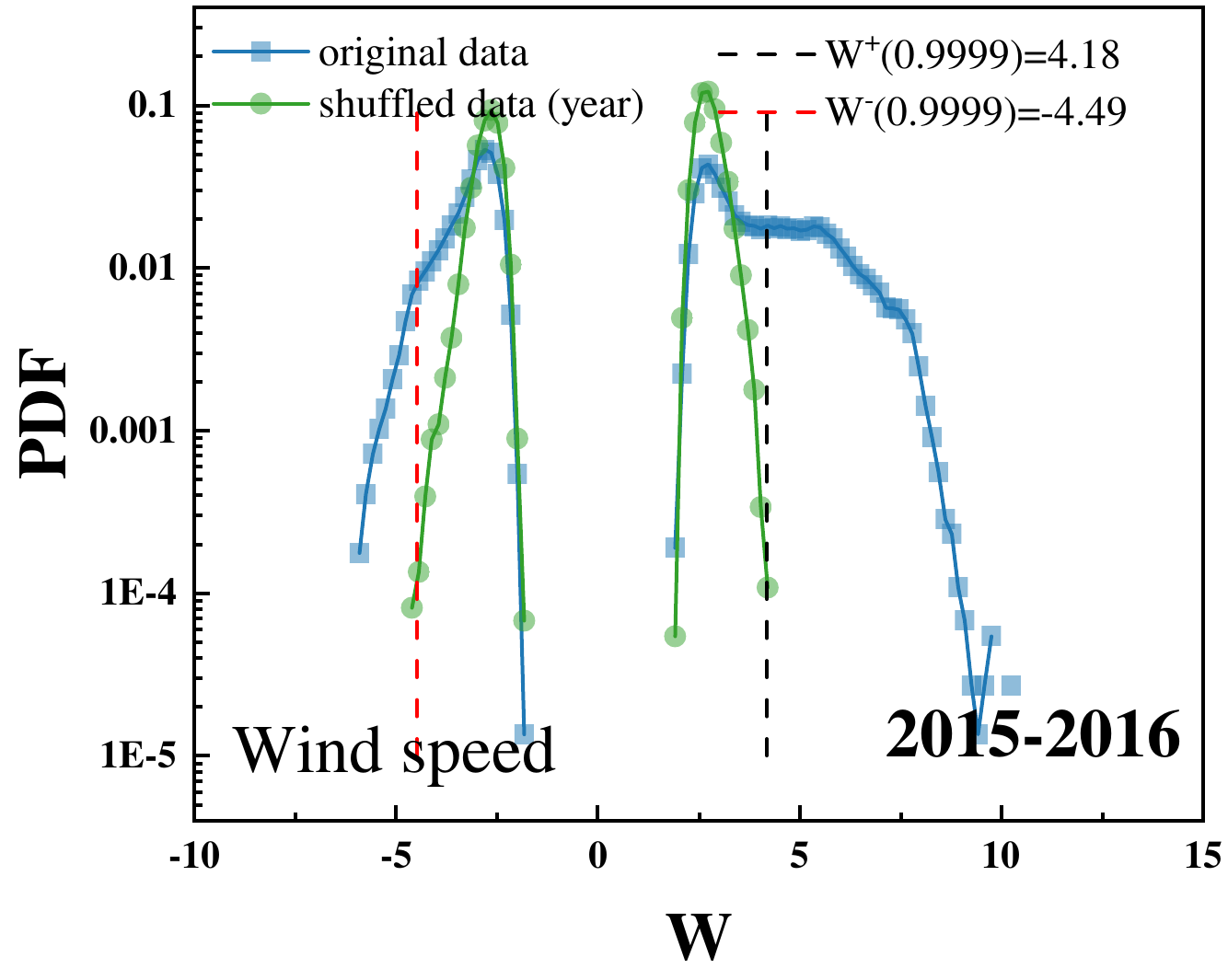}
\includegraphics[width=8.5em, height=7em]{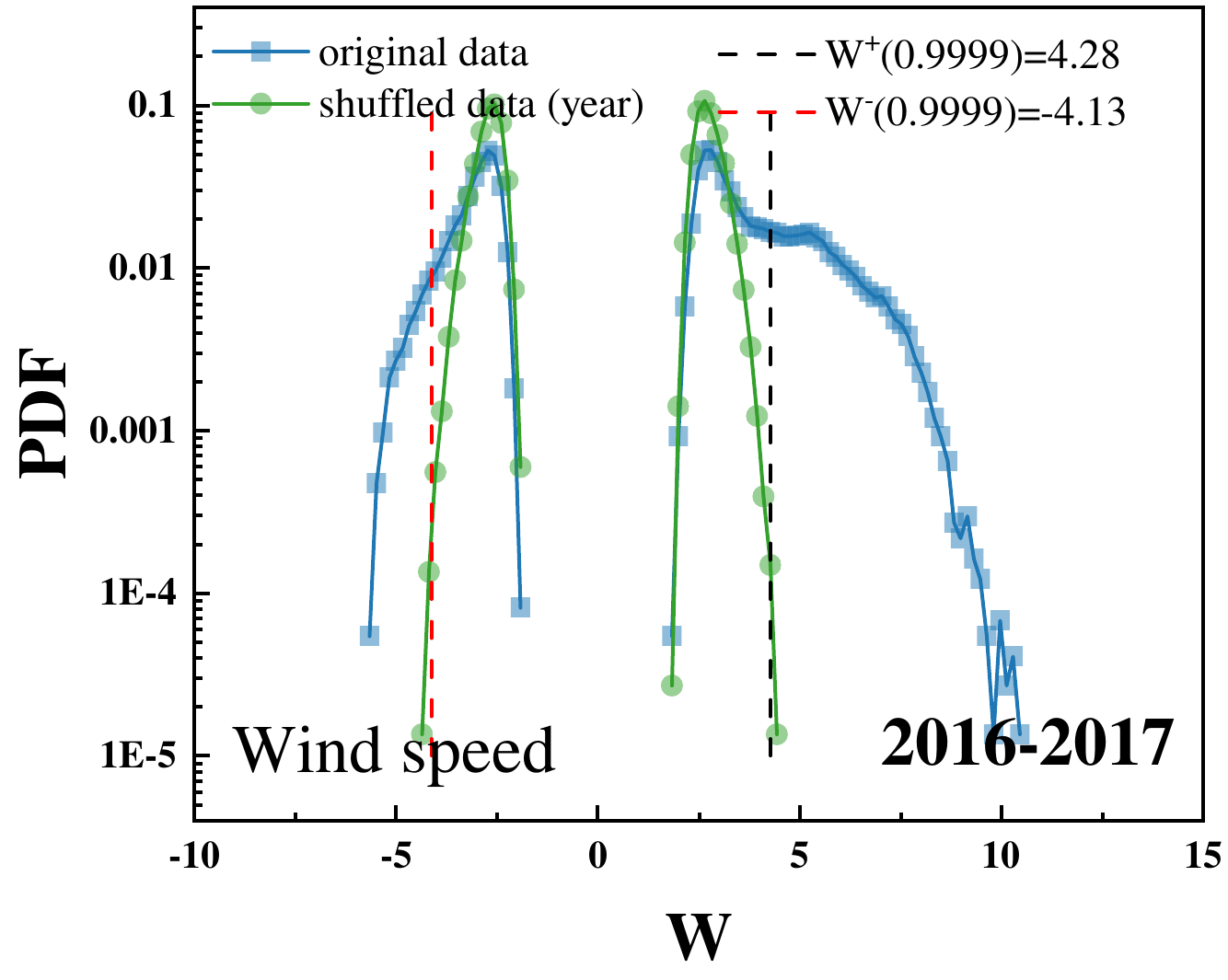}
\includegraphics[width=8.5em, height=7em]{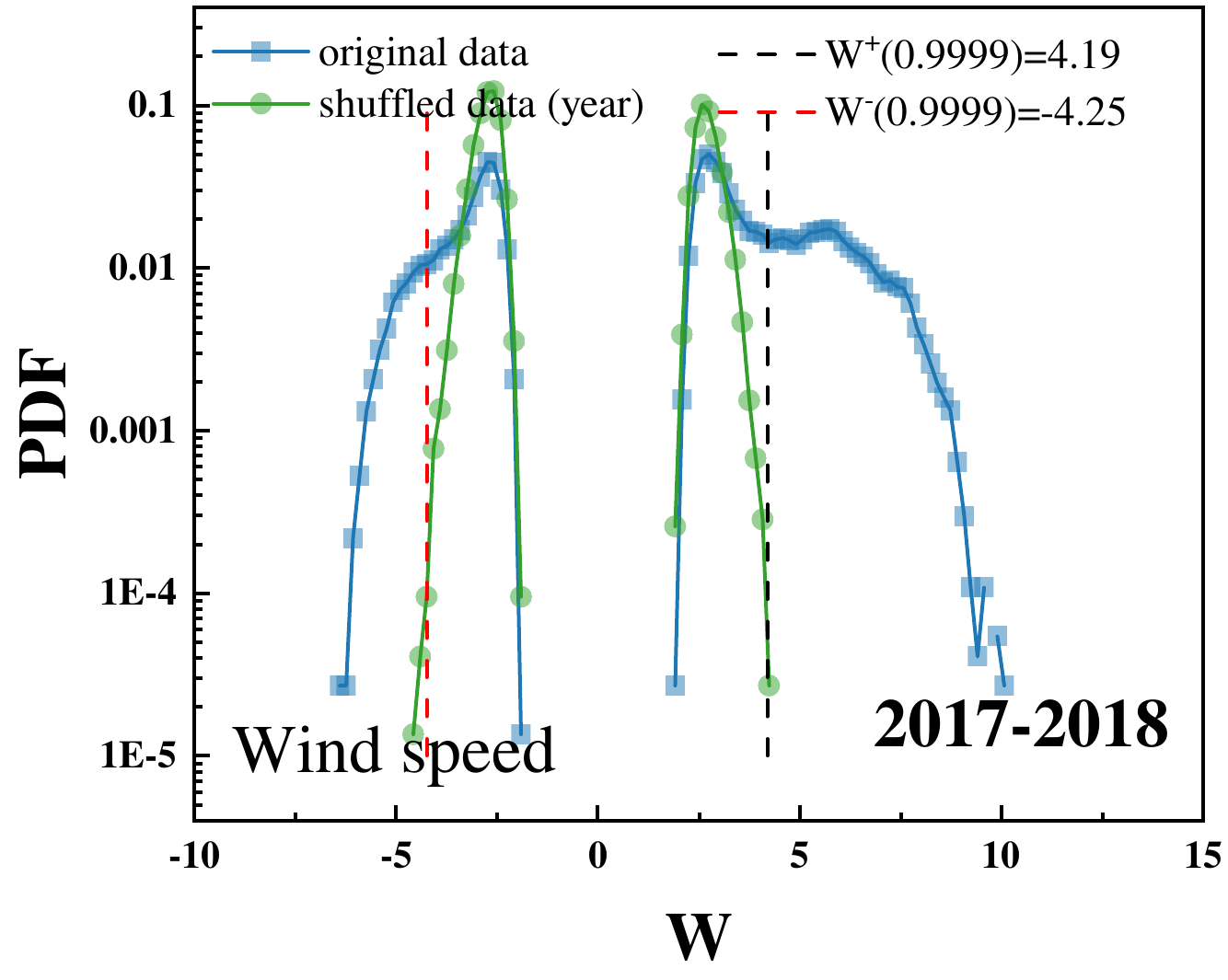}
\includegraphics[width=8.5em, height=7em]{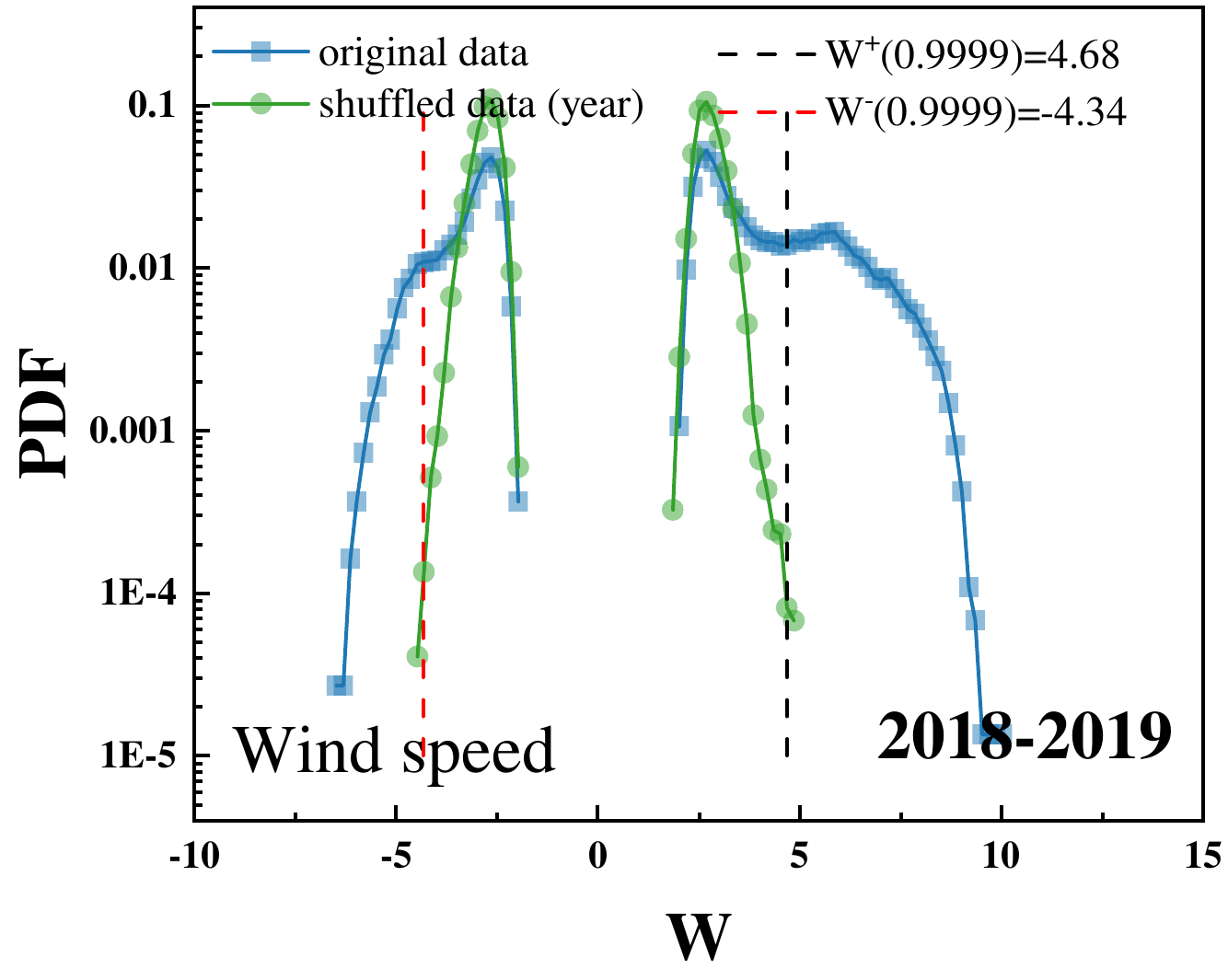}
\end{center}

\begin{center}
\noindent {\small {\bf Fig. S15} Probability distribution function (PDF) of link weights for the original data and shuffled data of wind speed in Europe. }
\end{center}

\begin{center}
\includegraphics[width=8.5em, height=7em]{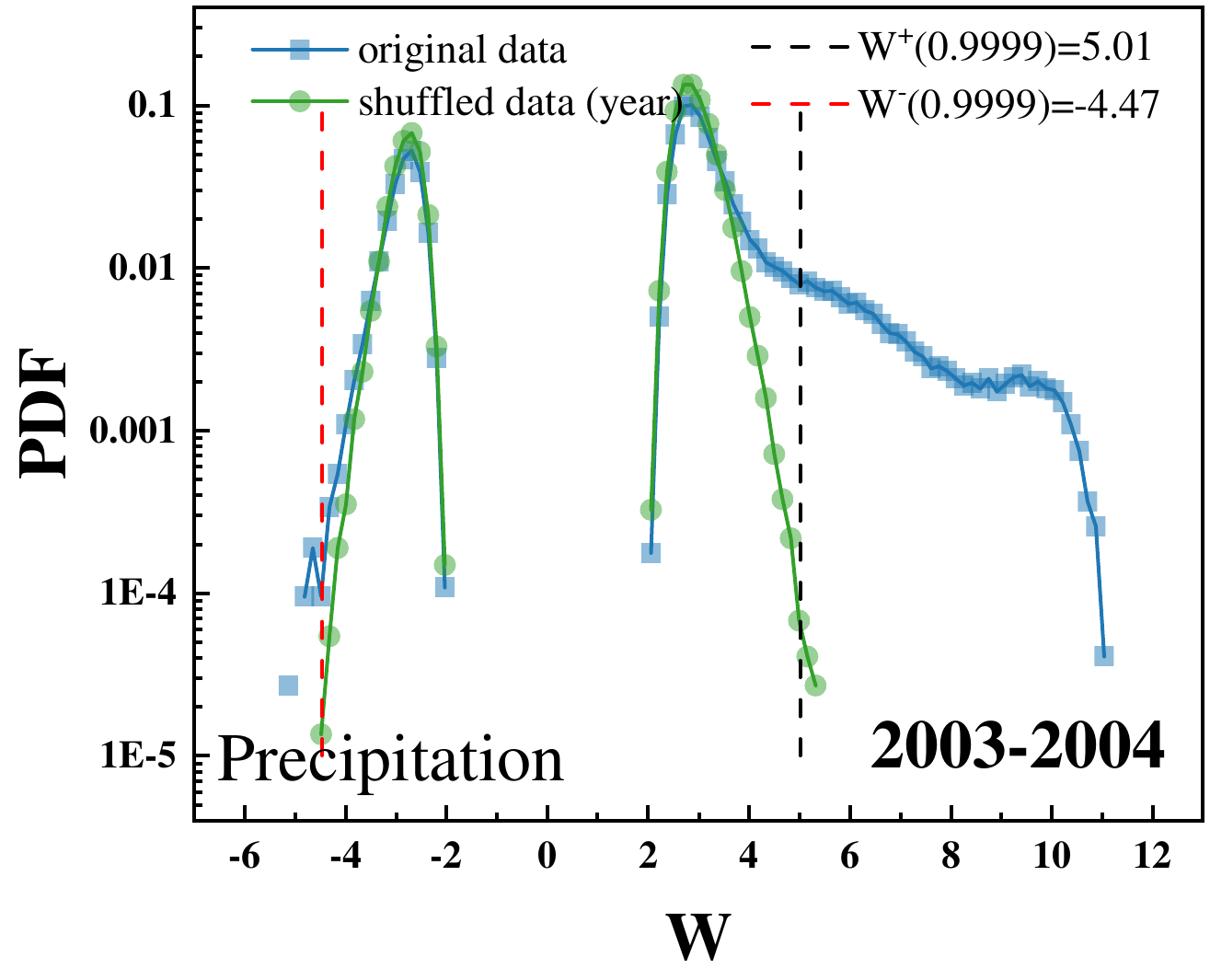}
\includegraphics[width=8.5em, height=7em]{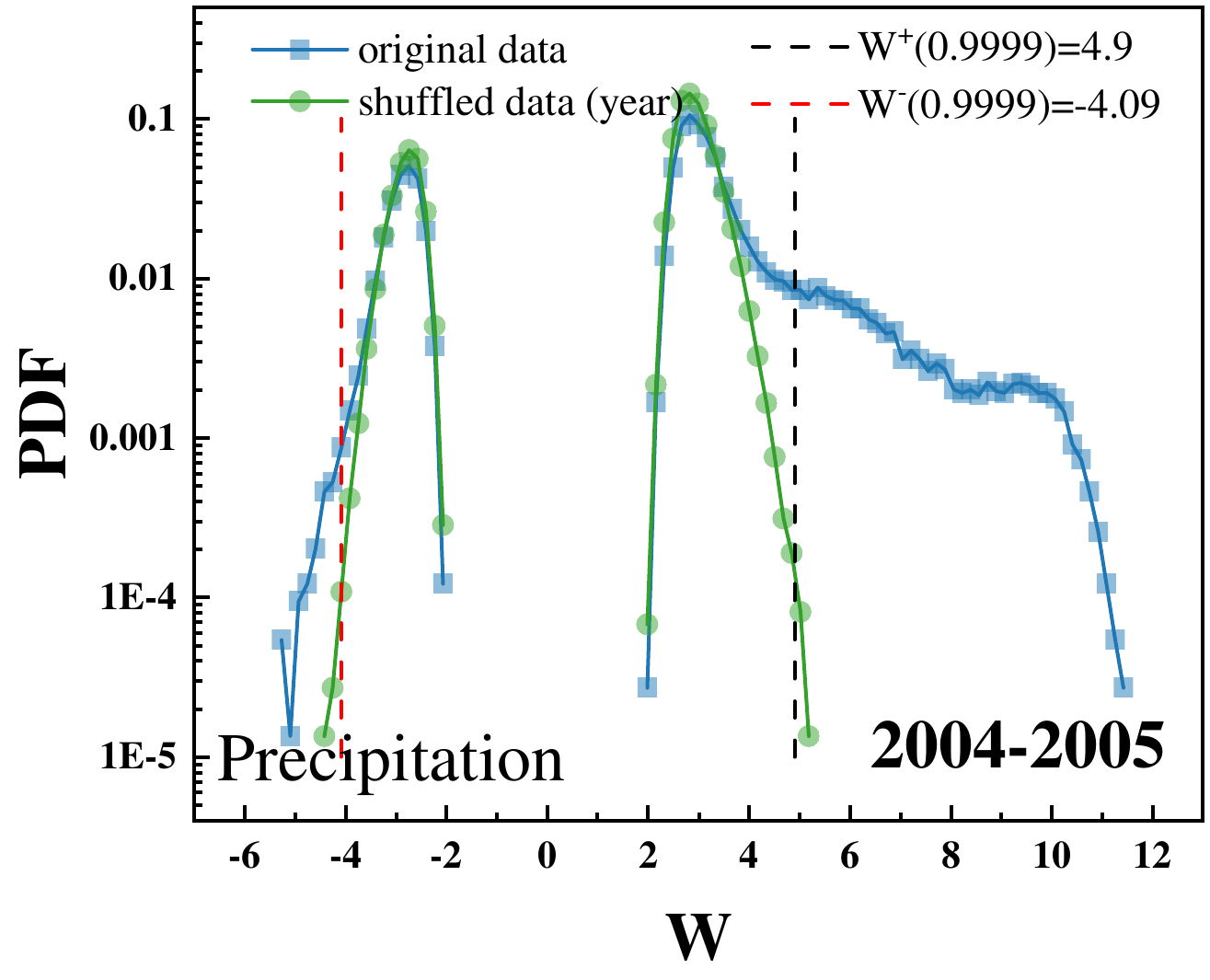}
\includegraphics[width=8.5em, height=7em]{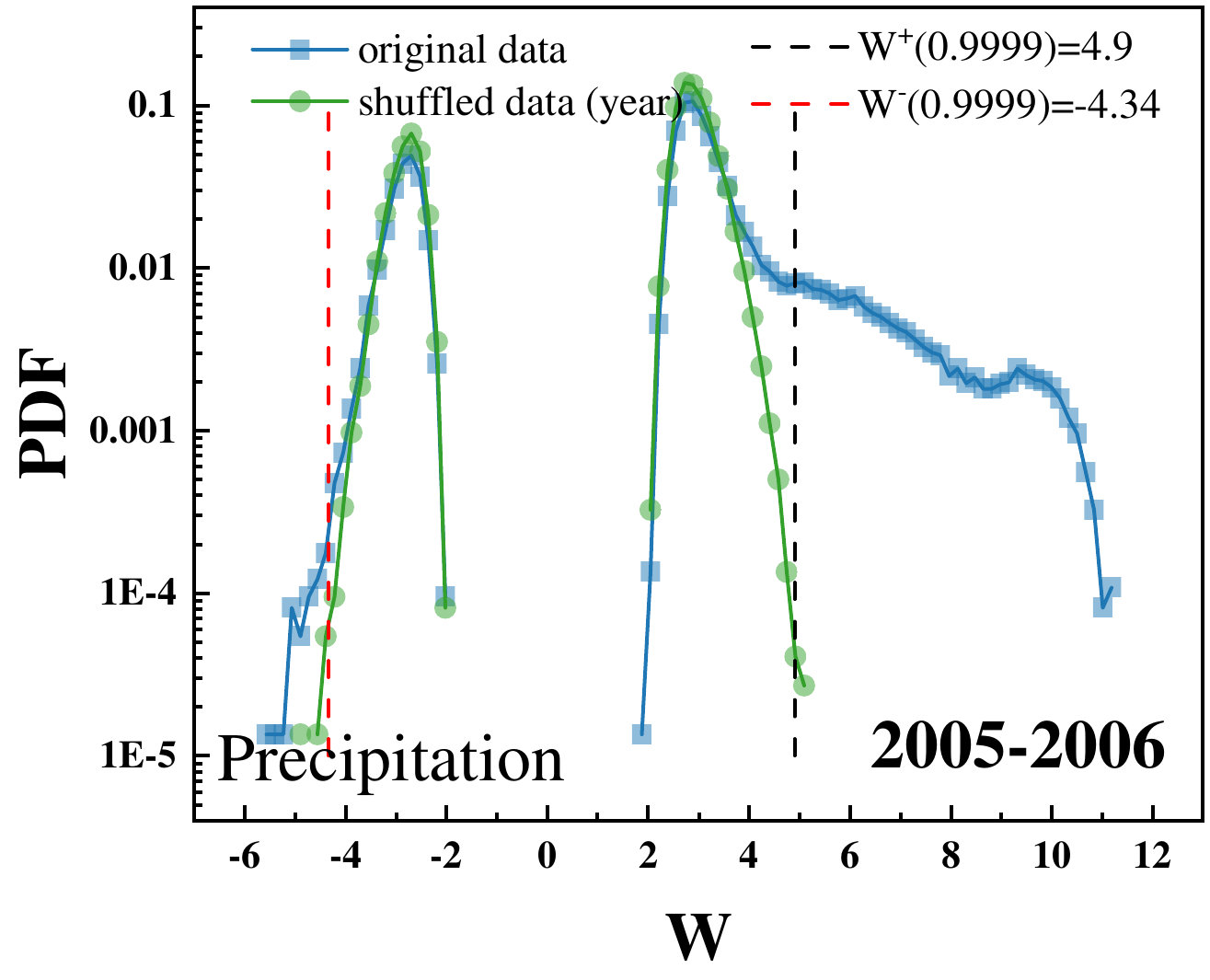}
\includegraphics[width=8.5em, height=7em]{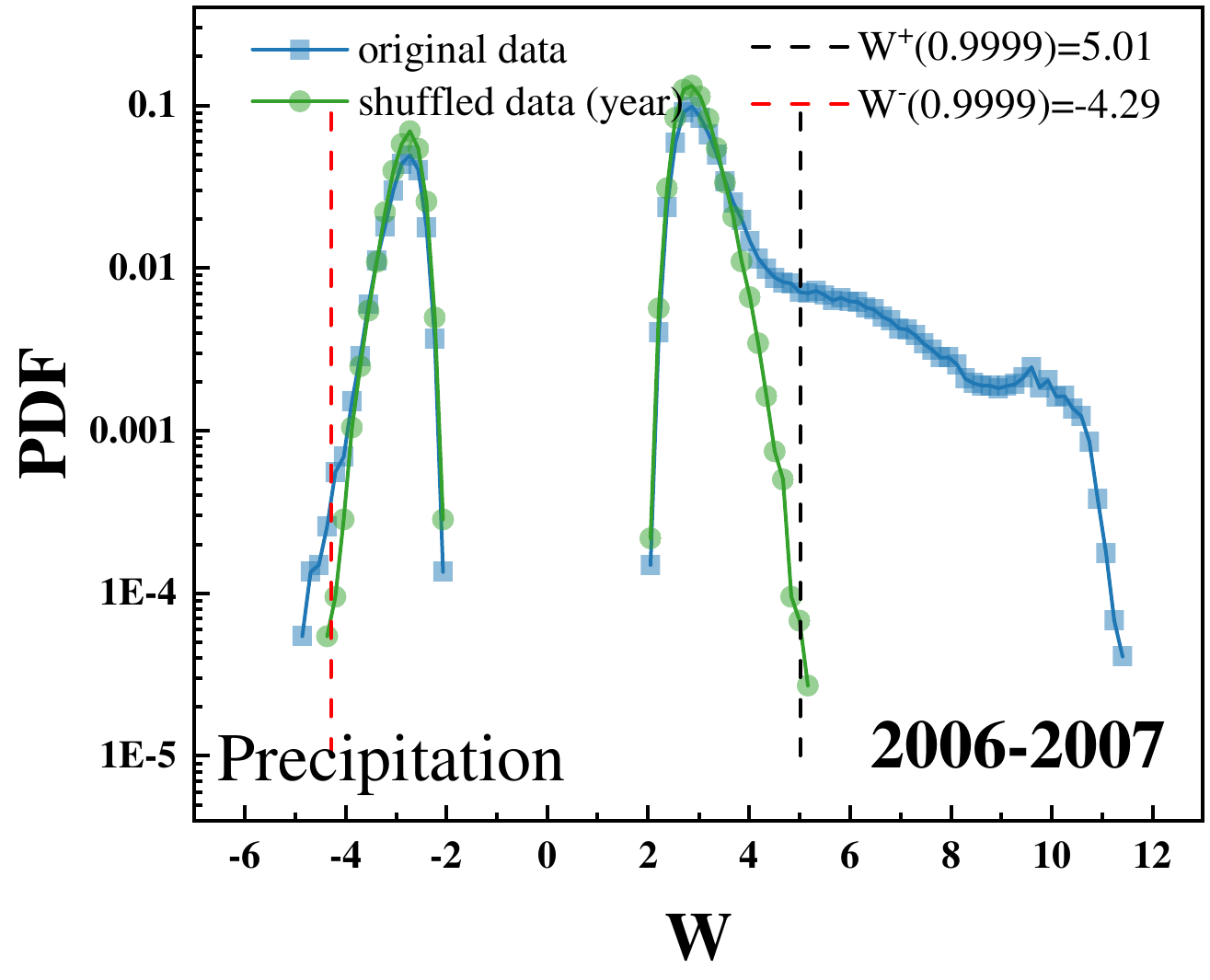}
\includegraphics[width=8.5em, height=7em]{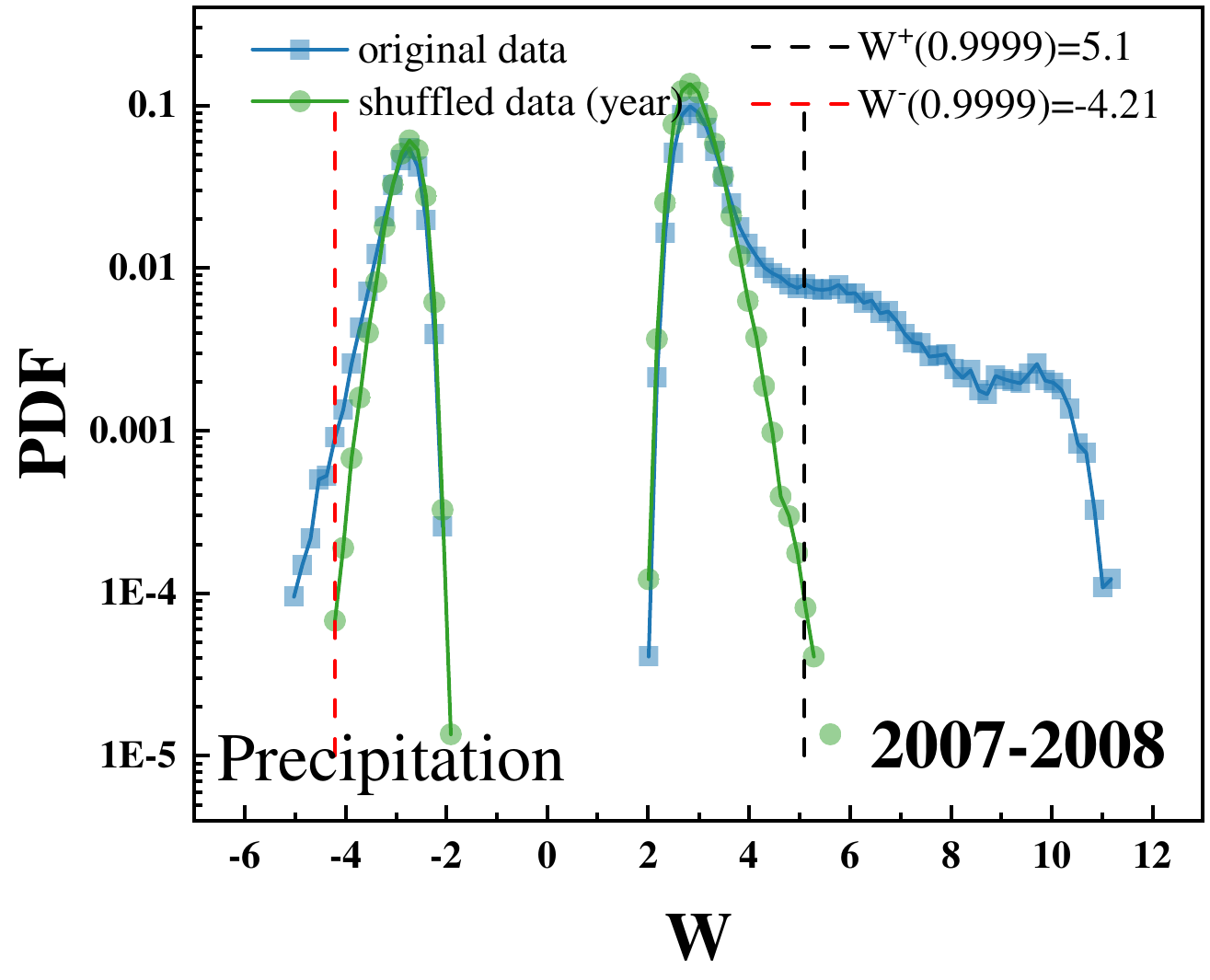}
\includegraphics[width=8.5em, height=7em]{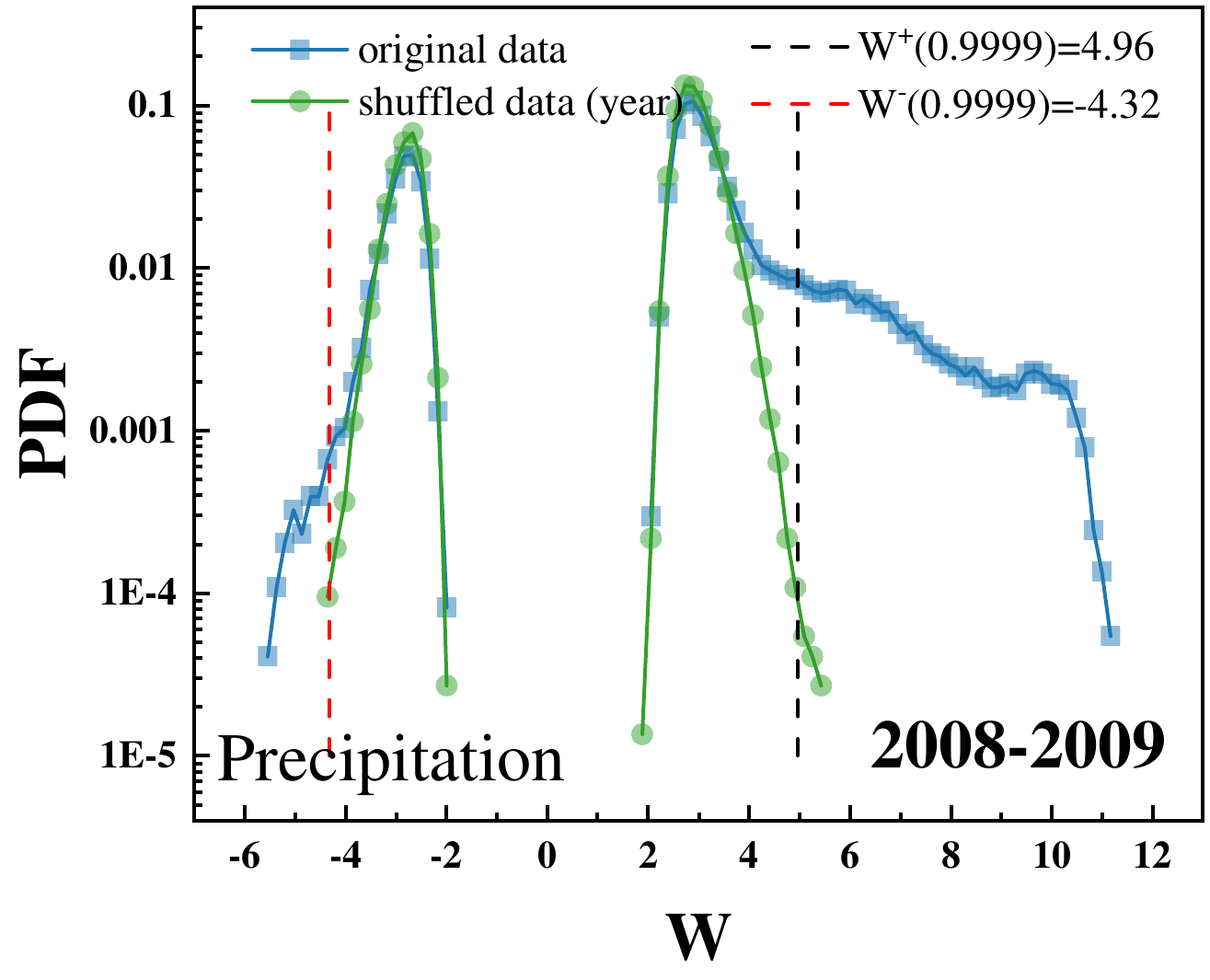}
\includegraphics[width=8.5em, height=7em]{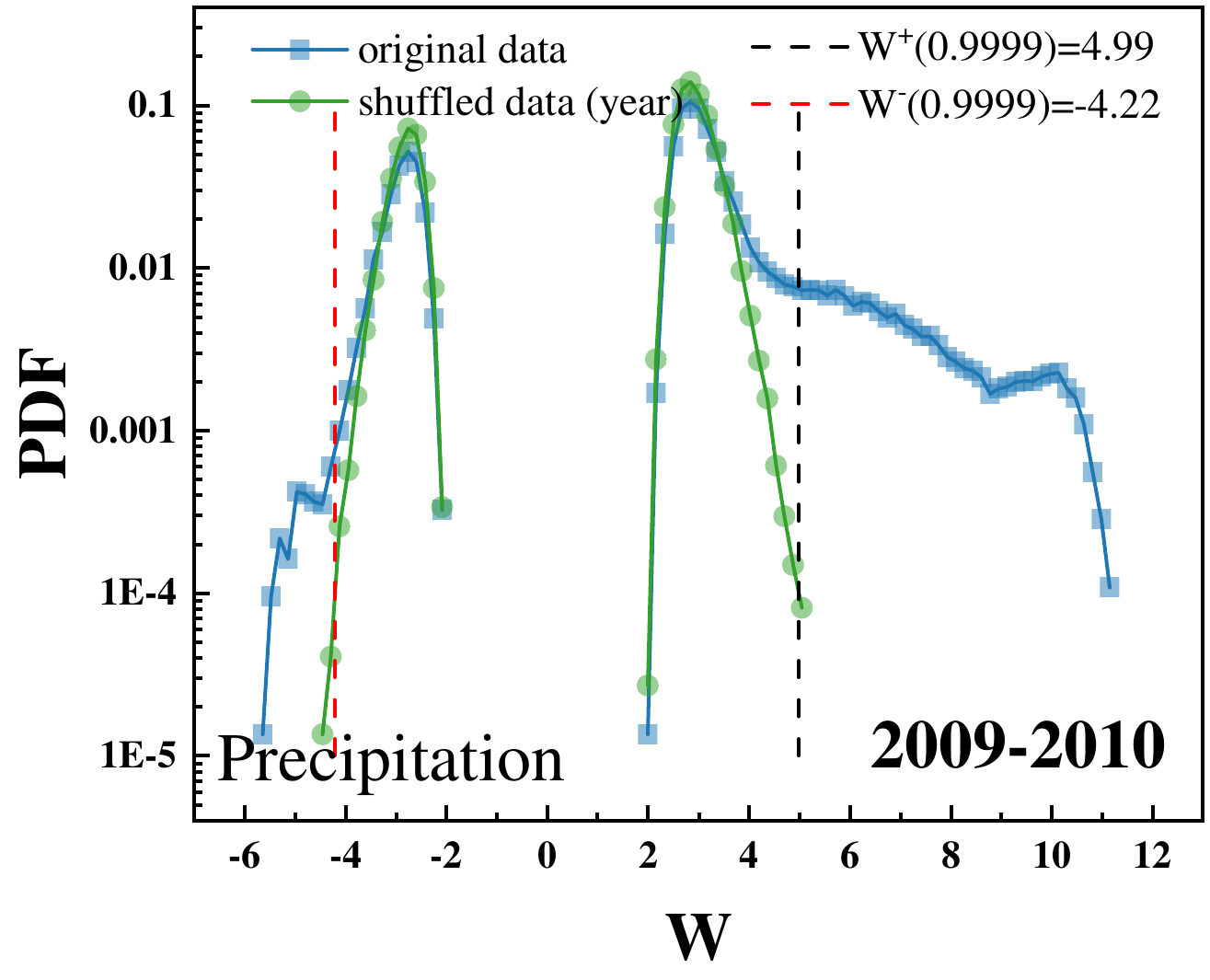}
\includegraphics[width=8.5em, height=7em]{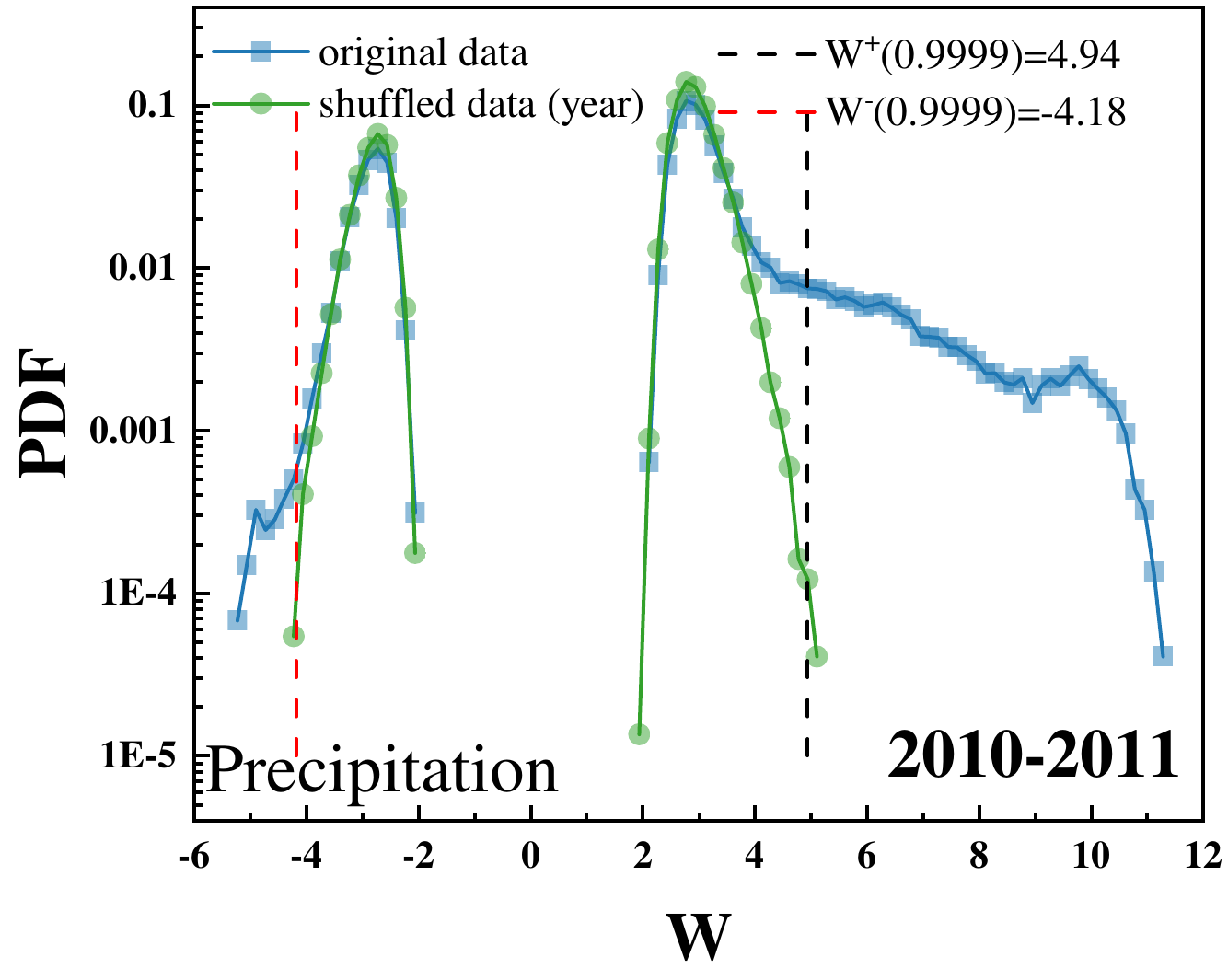}
\includegraphics[width=8.5em, height=7em]{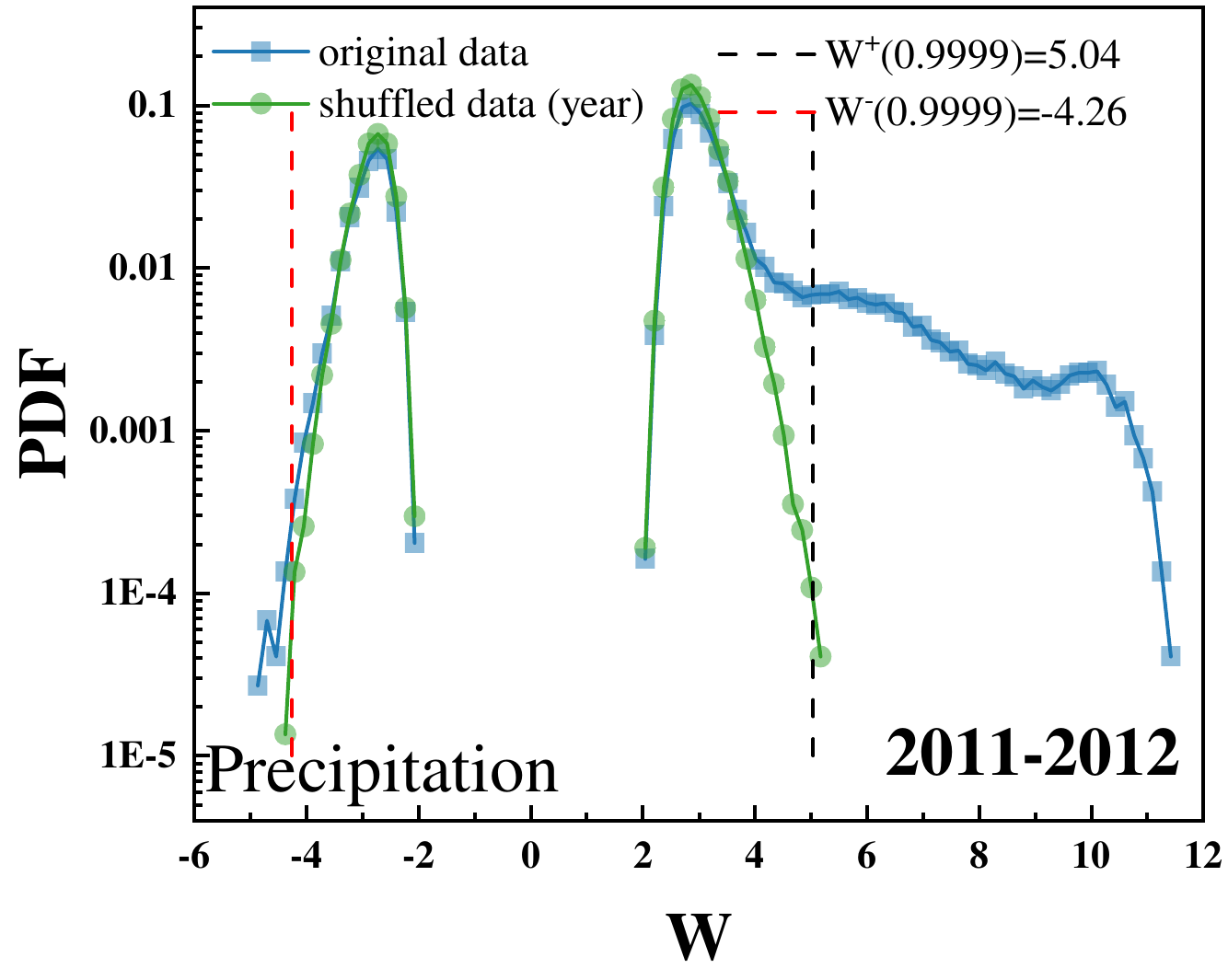}
\includegraphics[width=8.5em, height=7em]{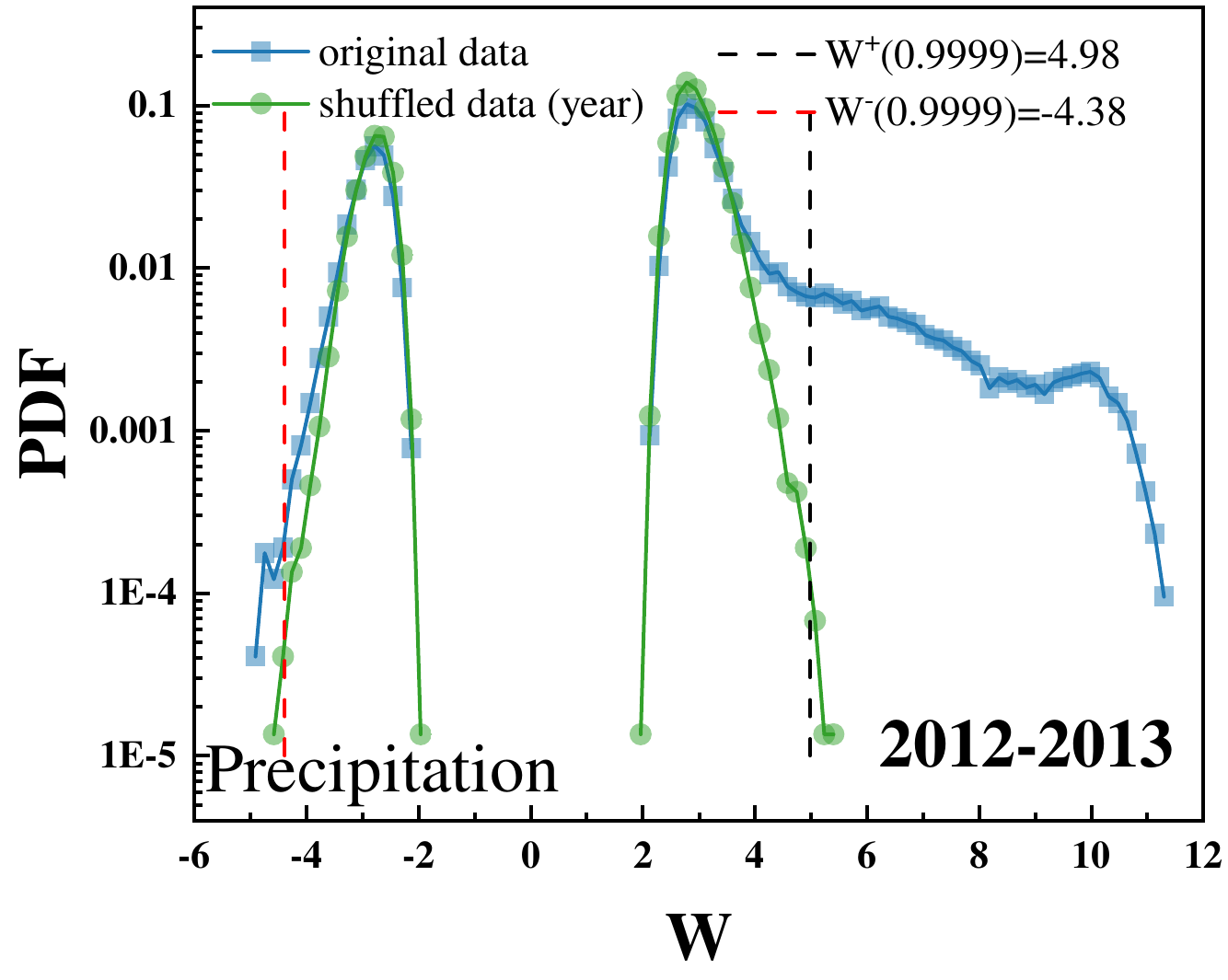}
\includegraphics[width=8.5em, height=7em]{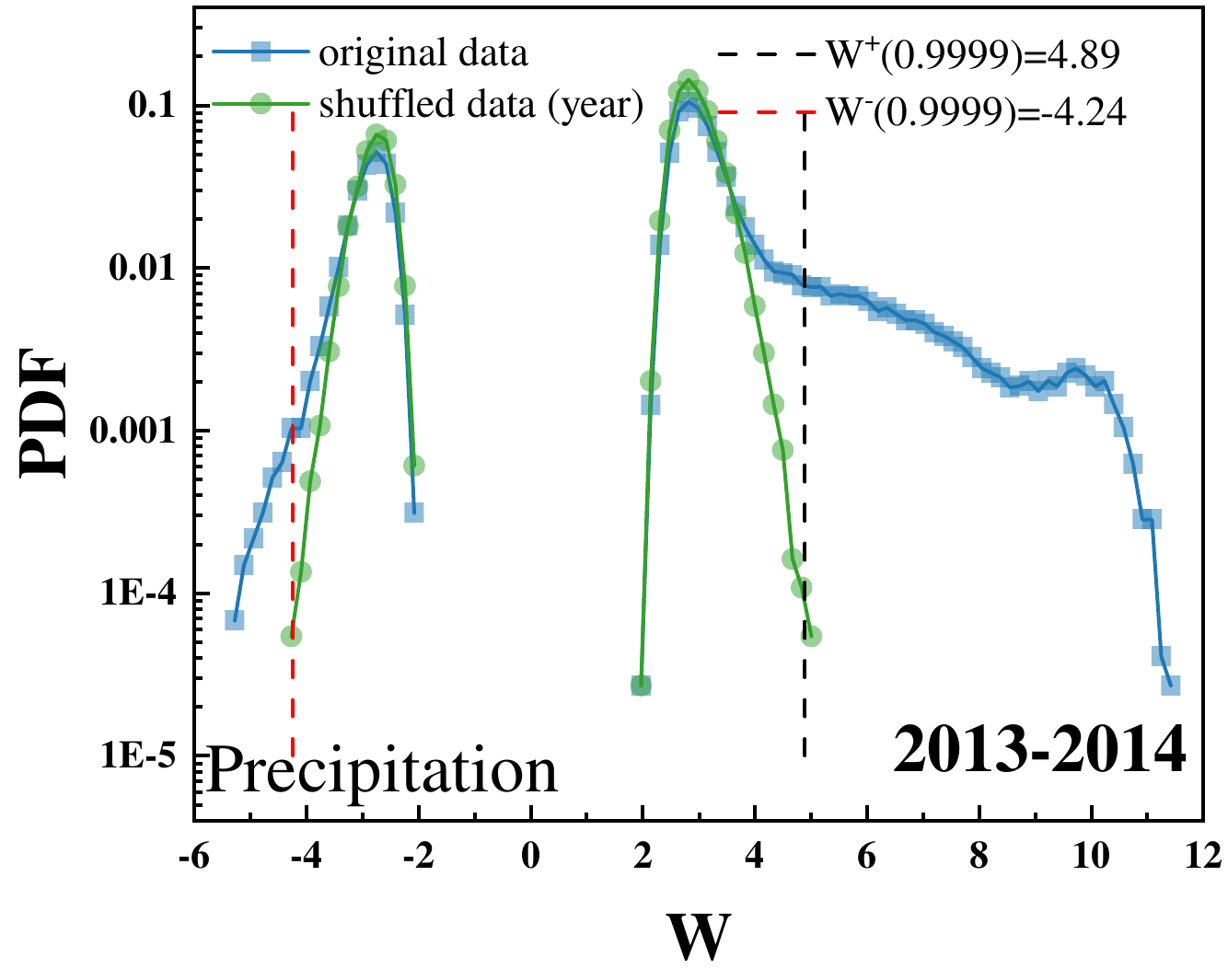}
\includegraphics[width=8.5em, height=7em]{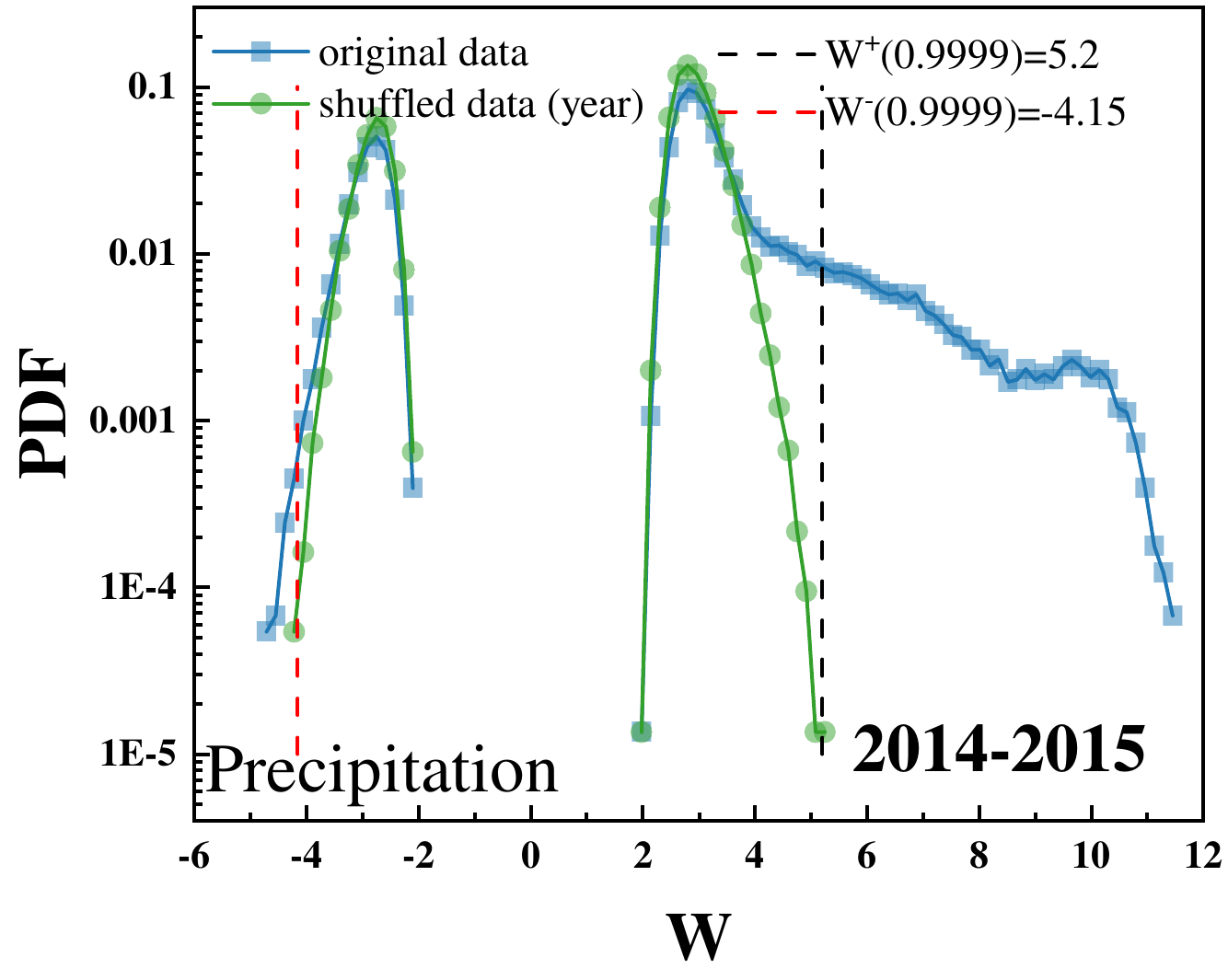}
\includegraphics[width=8.5em, height=7em]{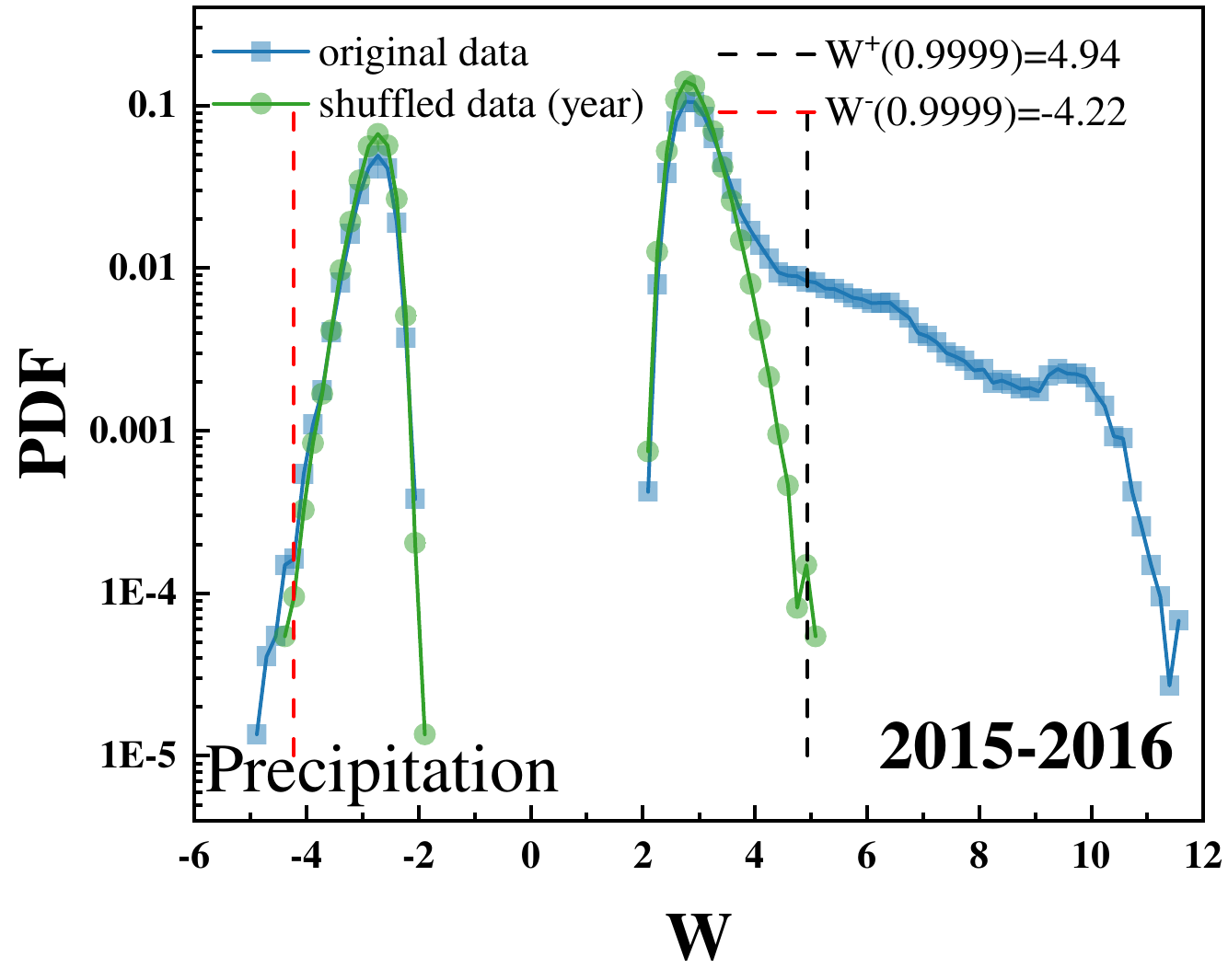}
\includegraphics[width=8.5em, height=7em]{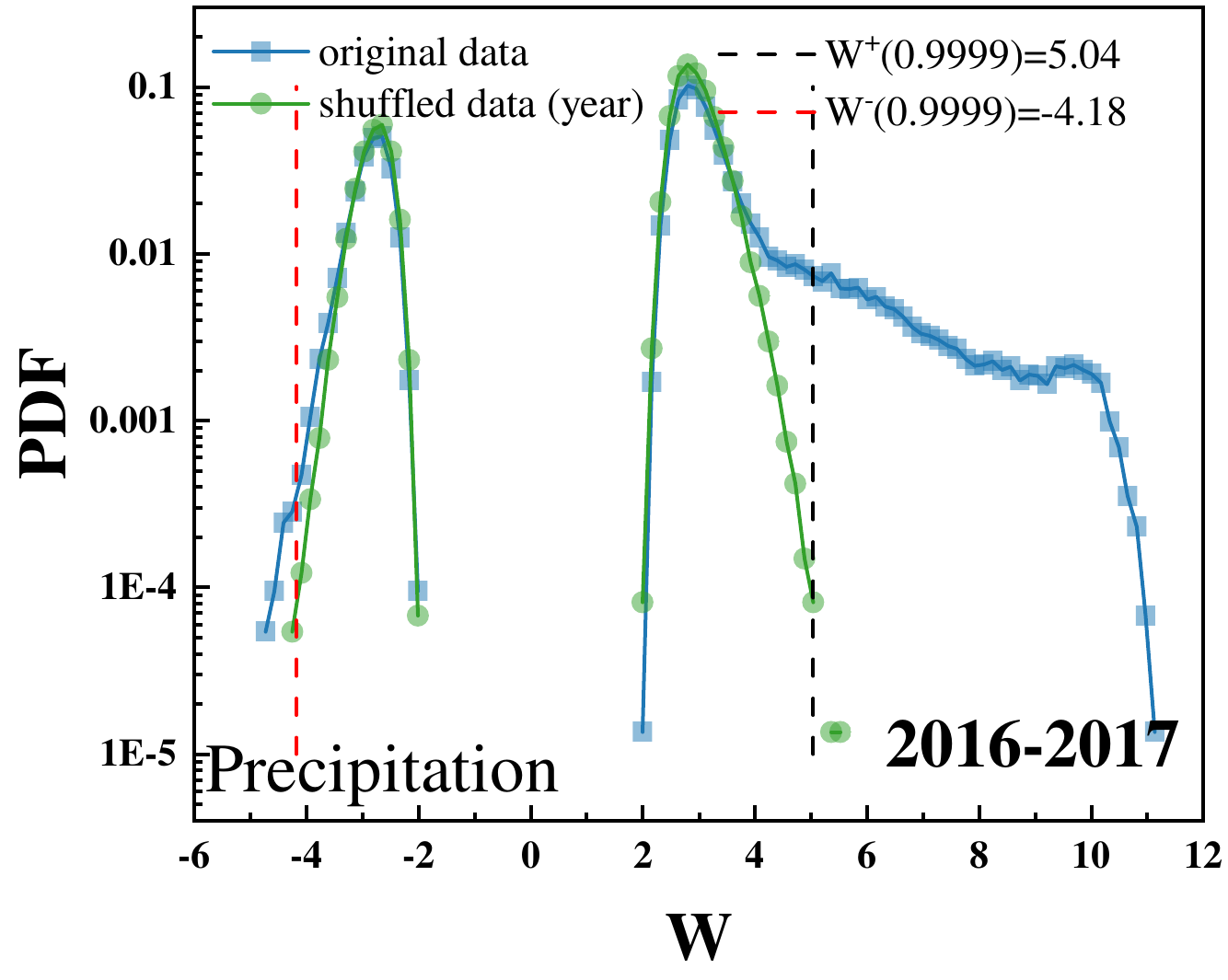}
\includegraphics[width=8.5em, height=7em]{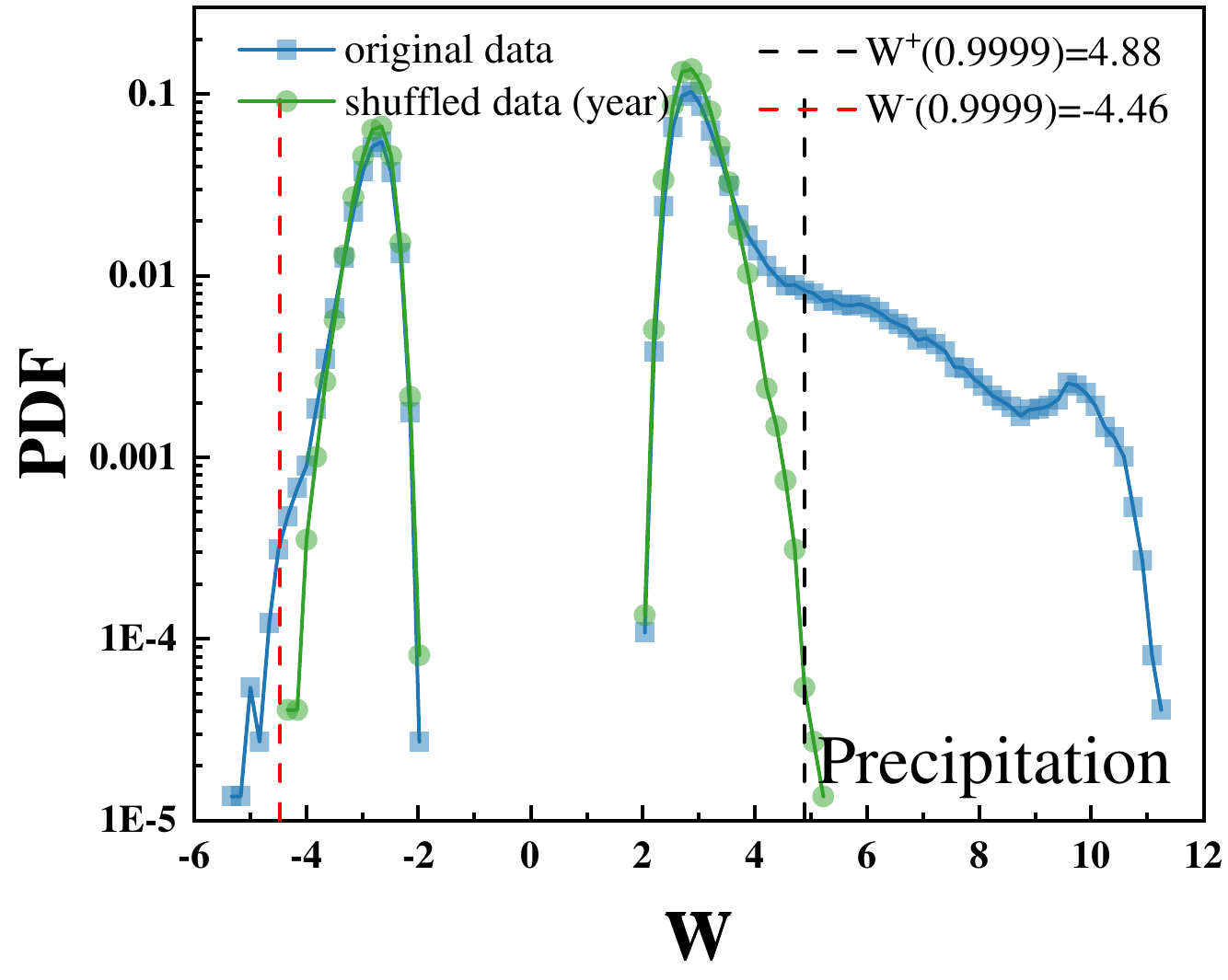}
\includegraphics[width=8.5em, height=7em]{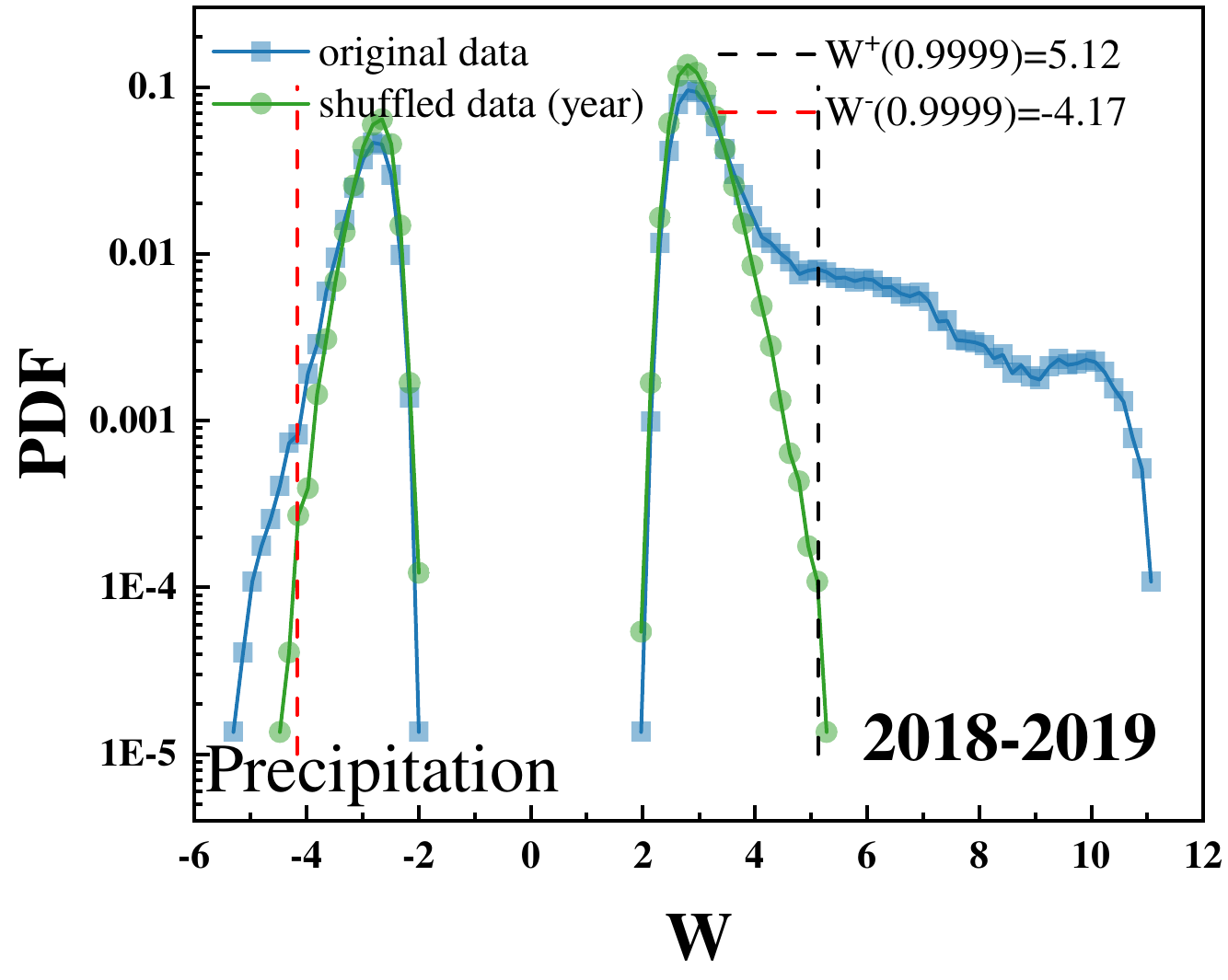}
\end{center}

\begin{center}
\noindent {\small {\bf Fig. S16} Probability distribution function (PDF) of link weights for the original data and shuffled data of precipitation in Europe. }
\end{center}

\begin{center}
\includegraphics[width=8em, height=7em]{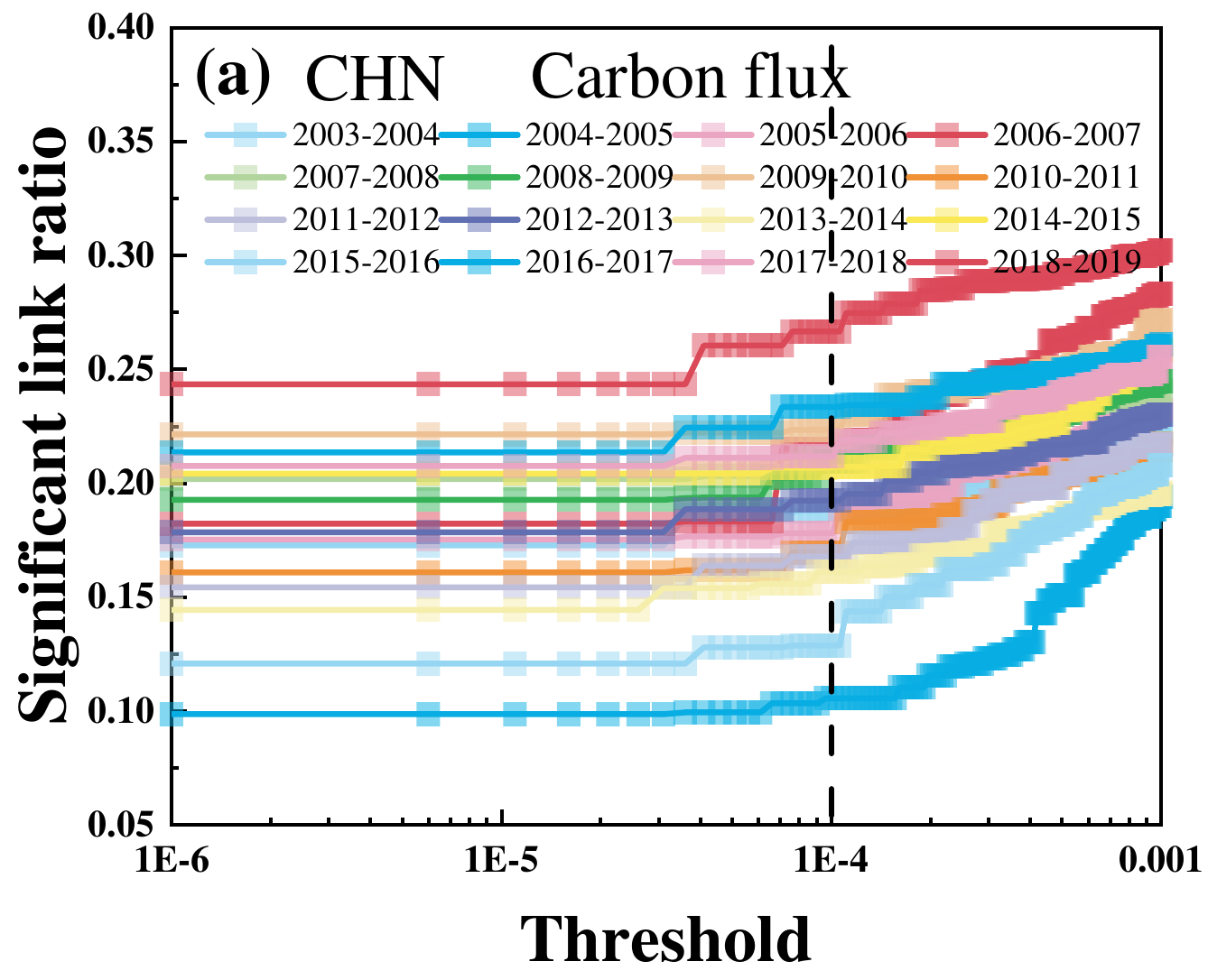}
\includegraphics[width=8em, height=7em]{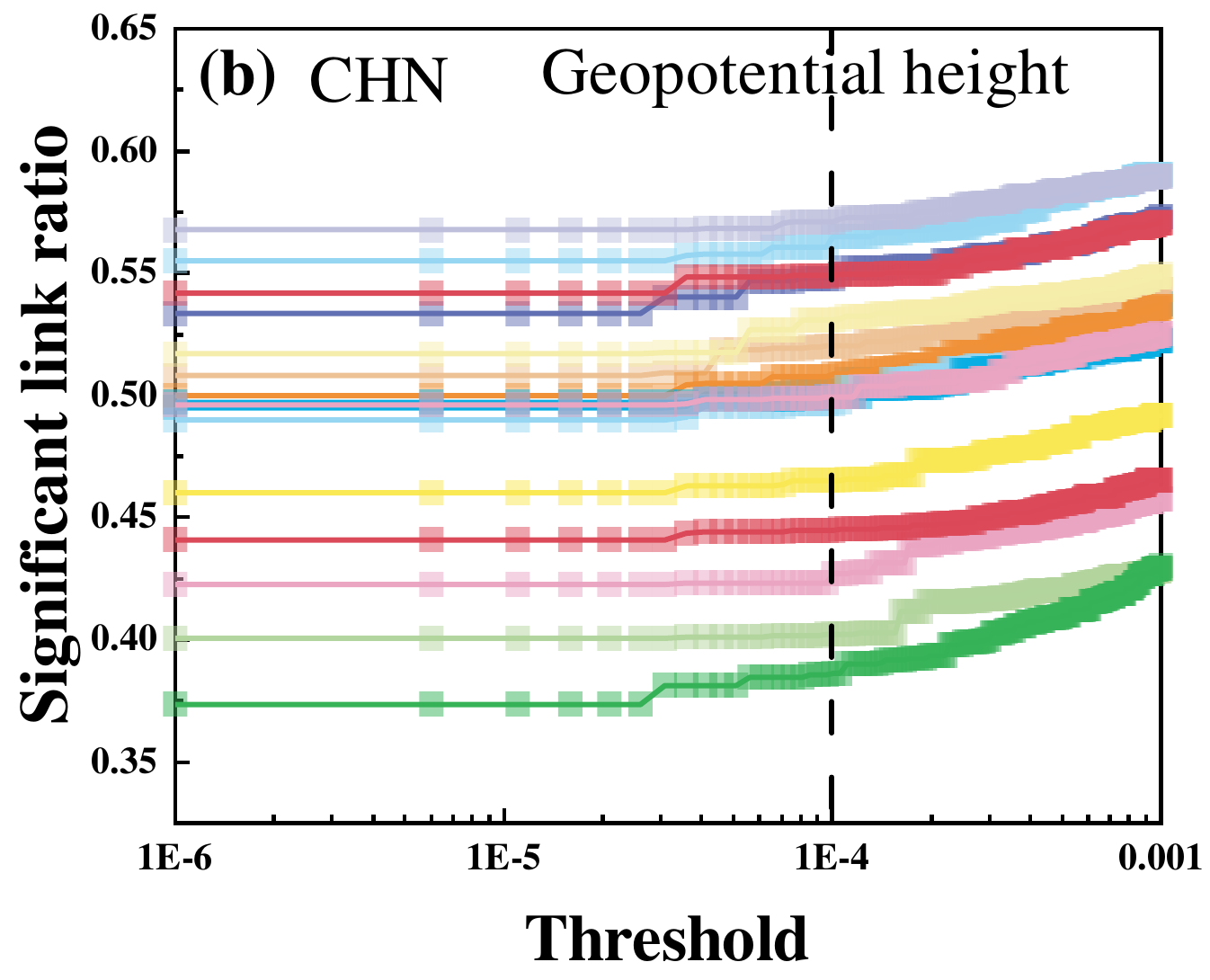}
\includegraphics[width=8em, height=7em]{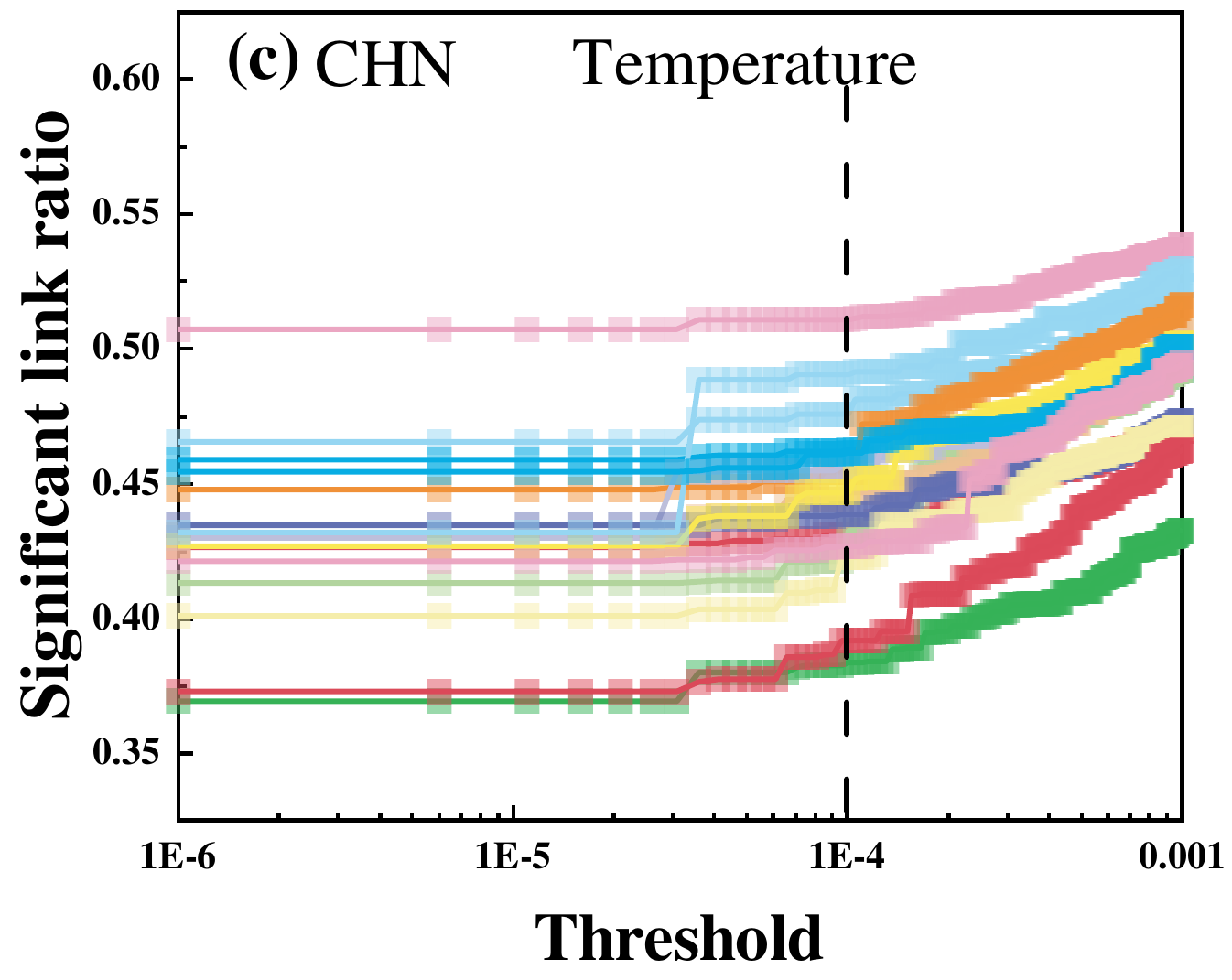}
\includegraphics[width=8em, height=7em]{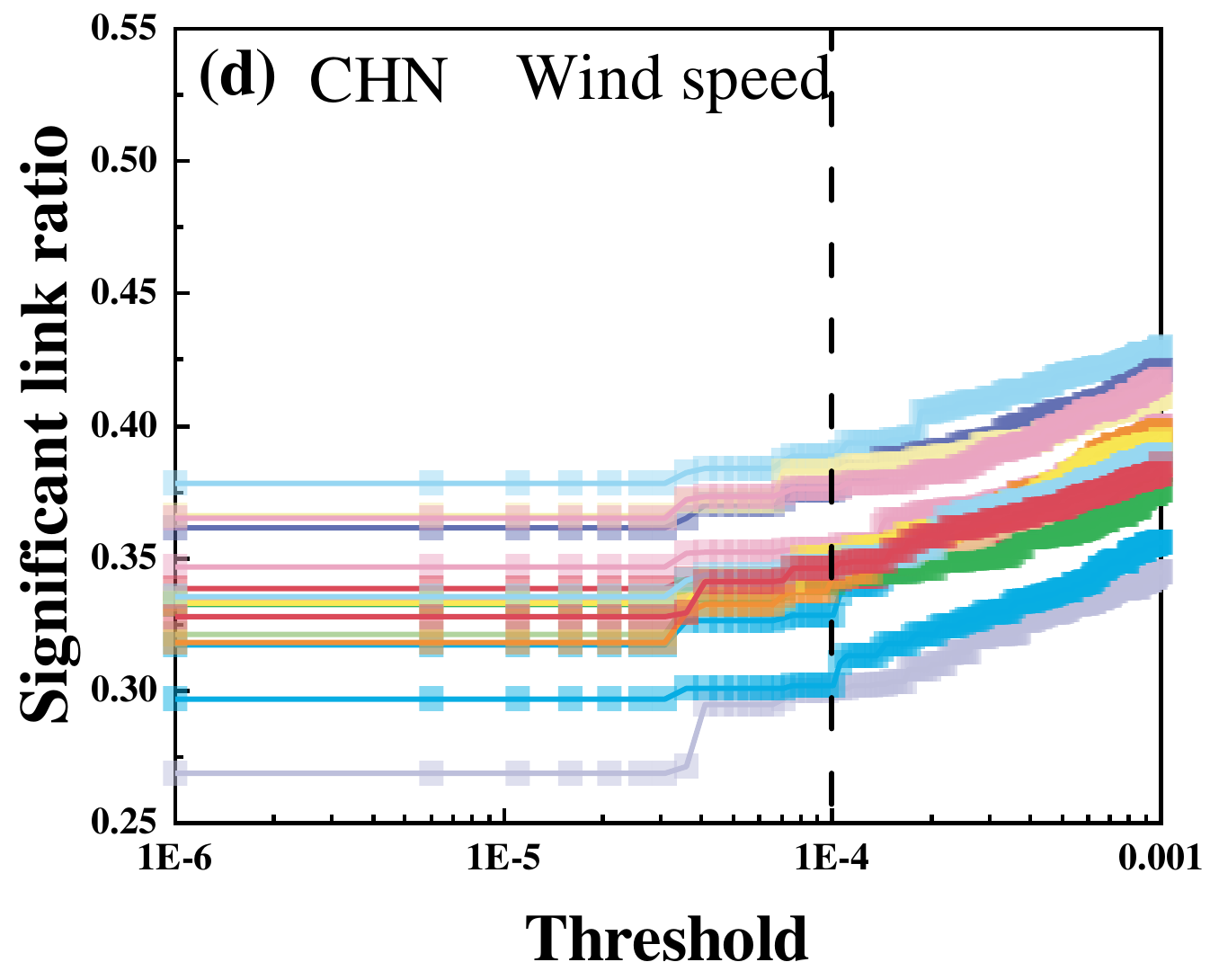}
\includegraphics[width=8em, height=7em]{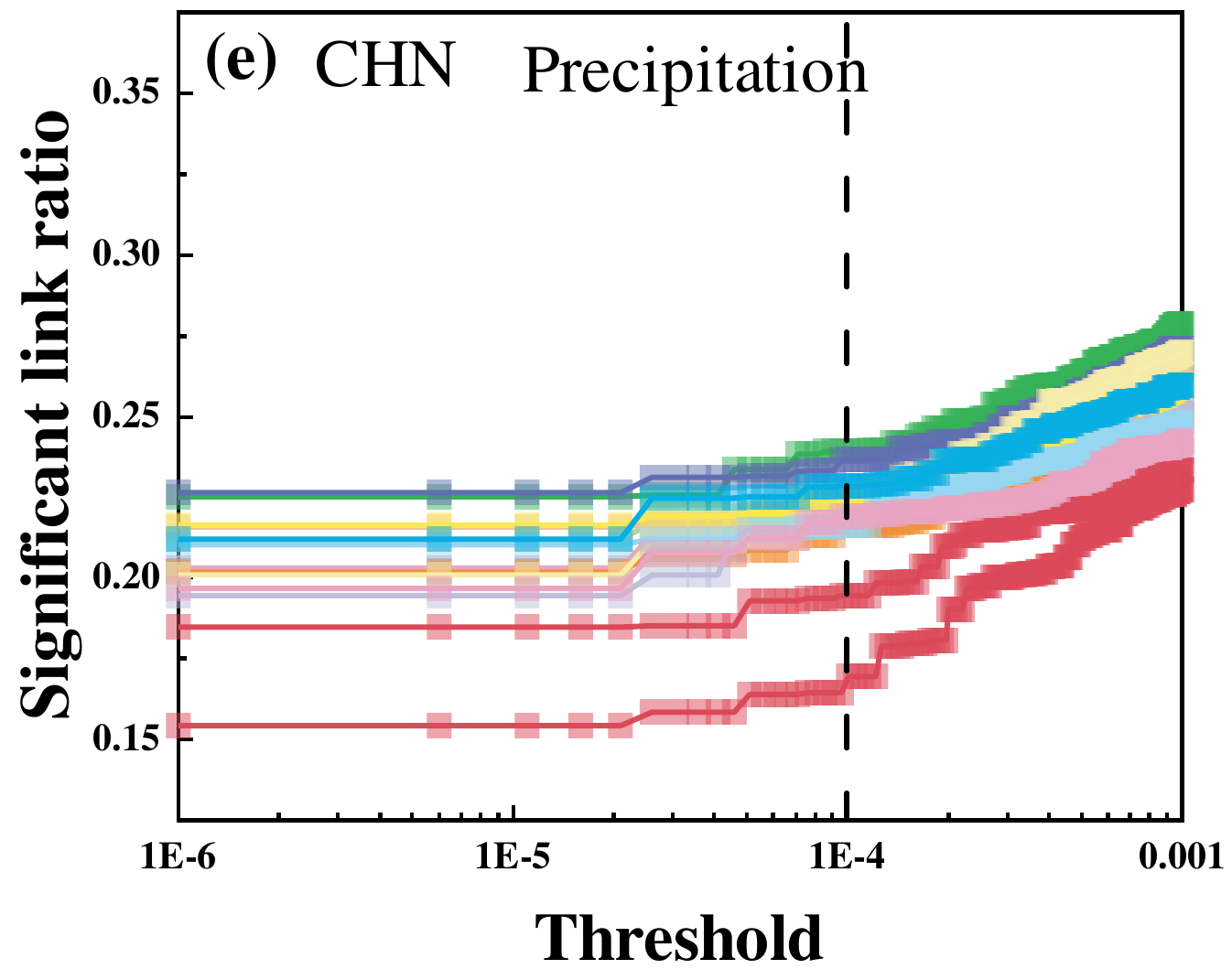}
\includegraphics[width=8em, height=7em]{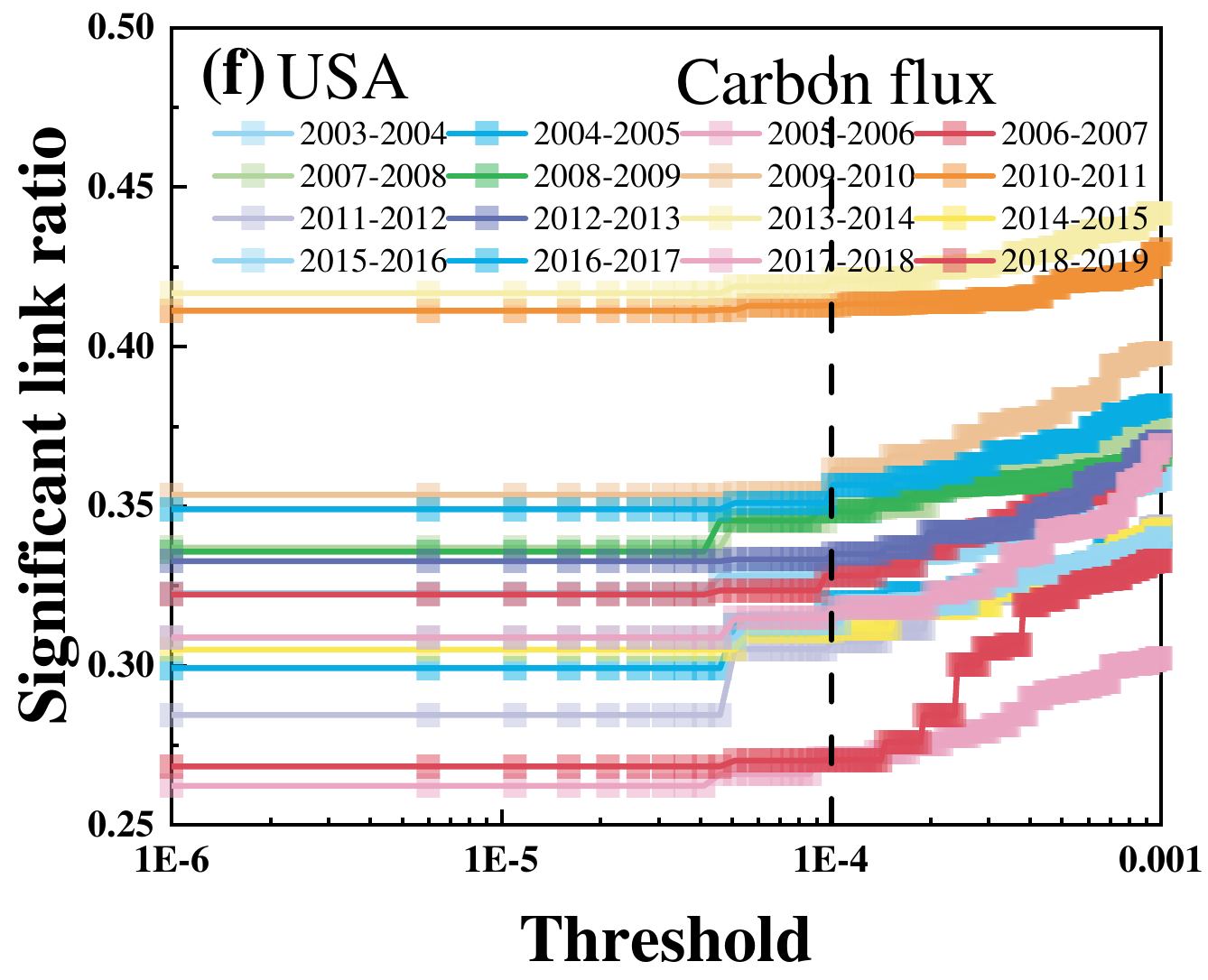}
\includegraphics[width=8em, height=7em]{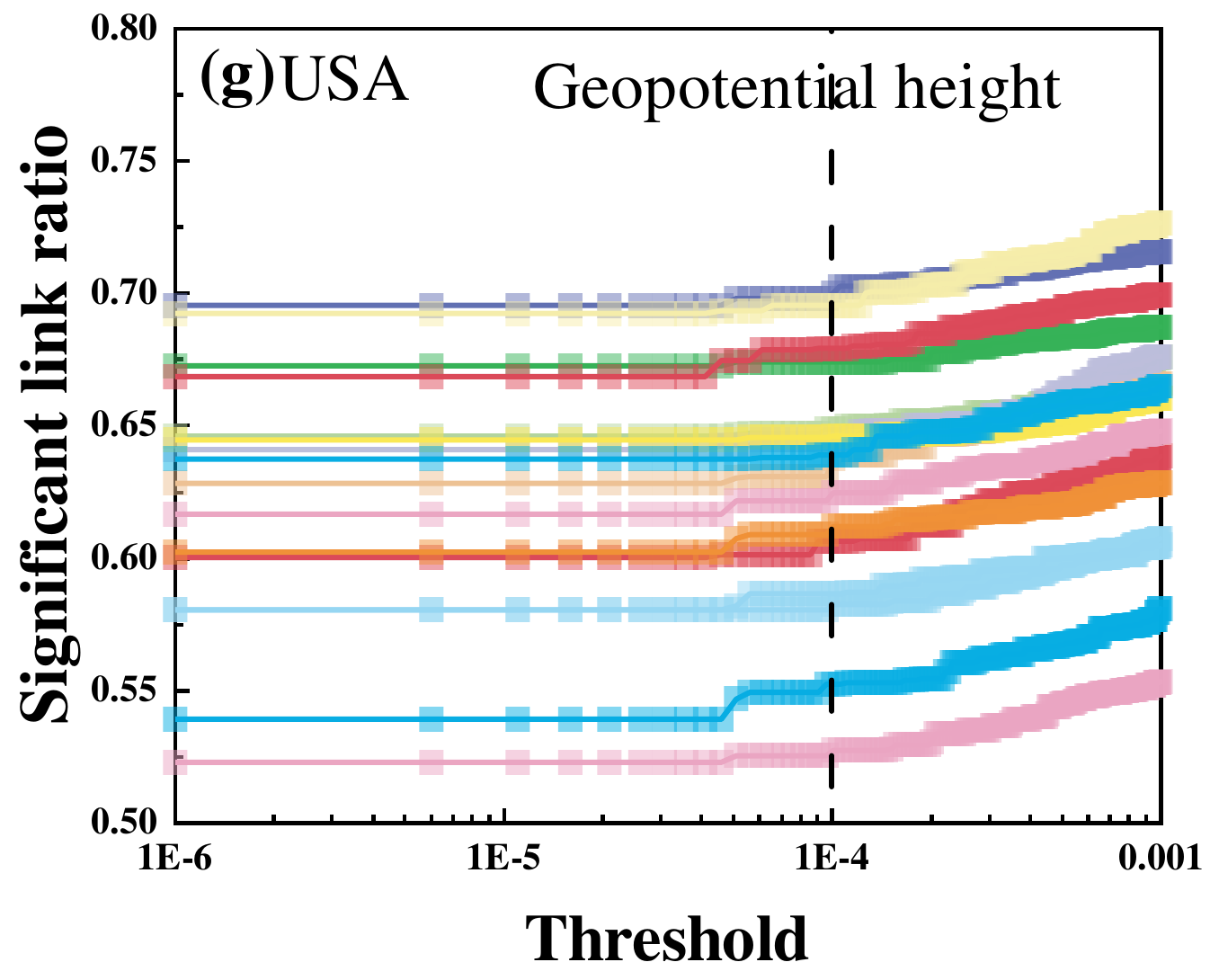}
\includegraphics[width=8em, height=7em]{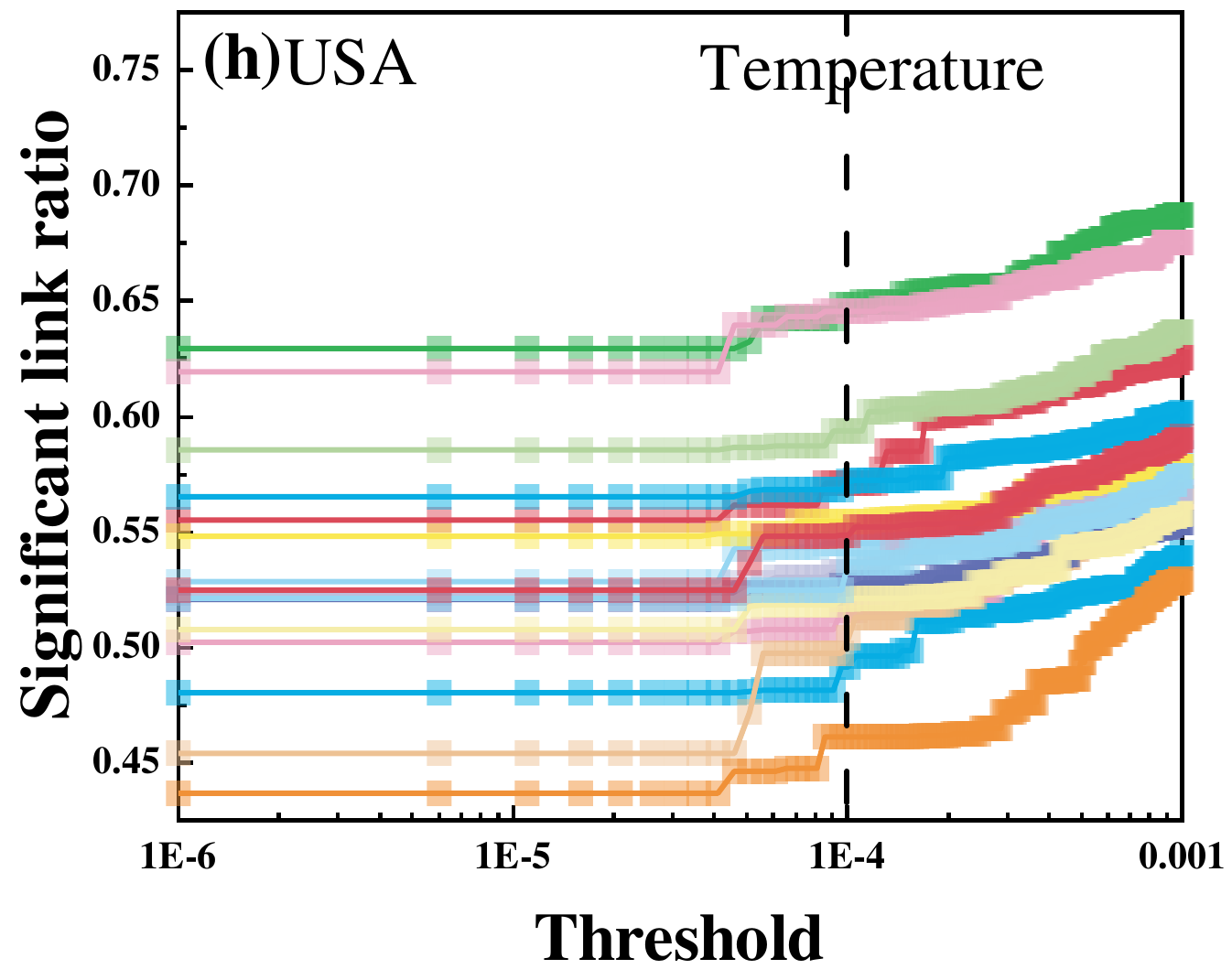}
\includegraphics[width=8em, height=7em]{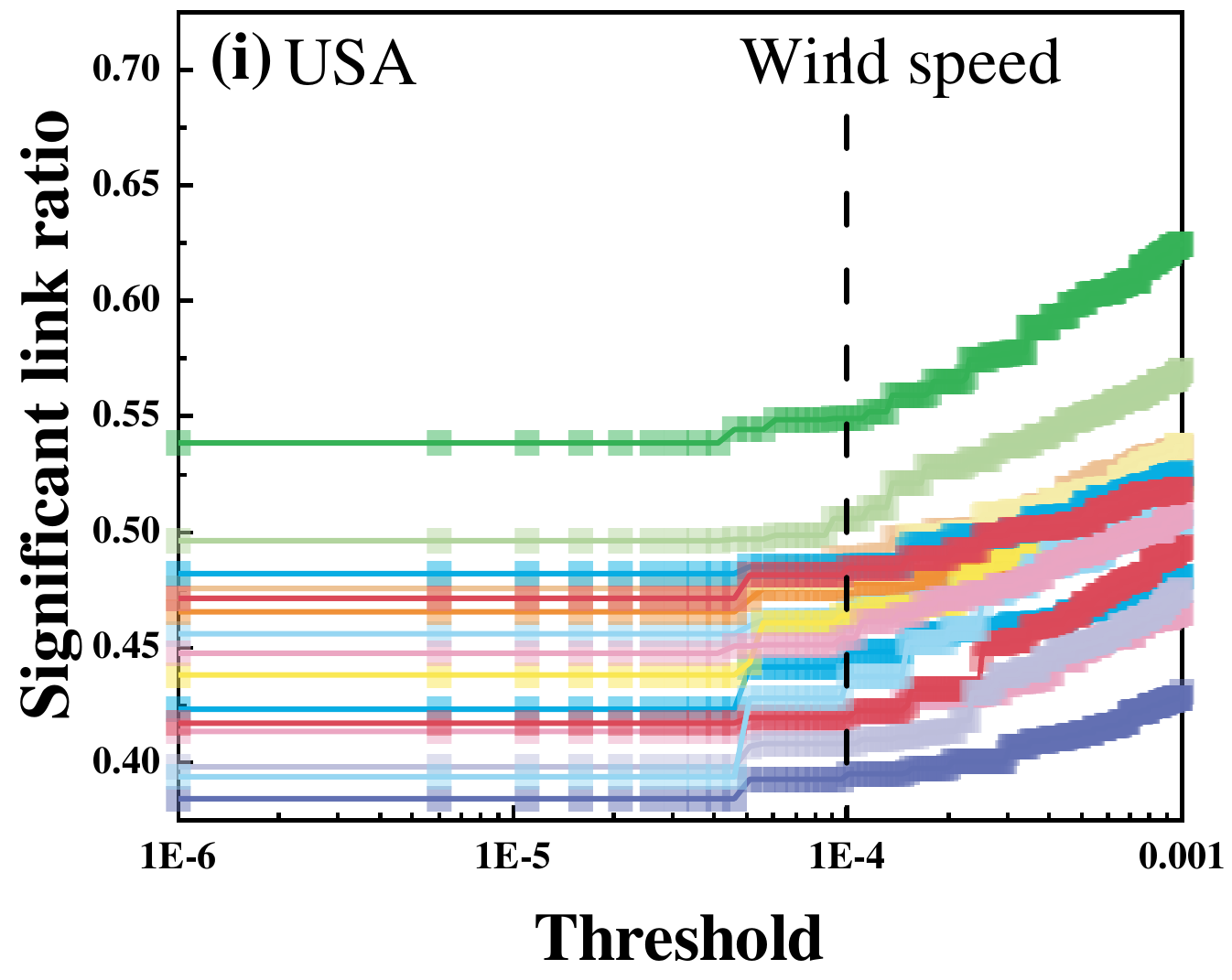}
\includegraphics[width=8em, height=7em]{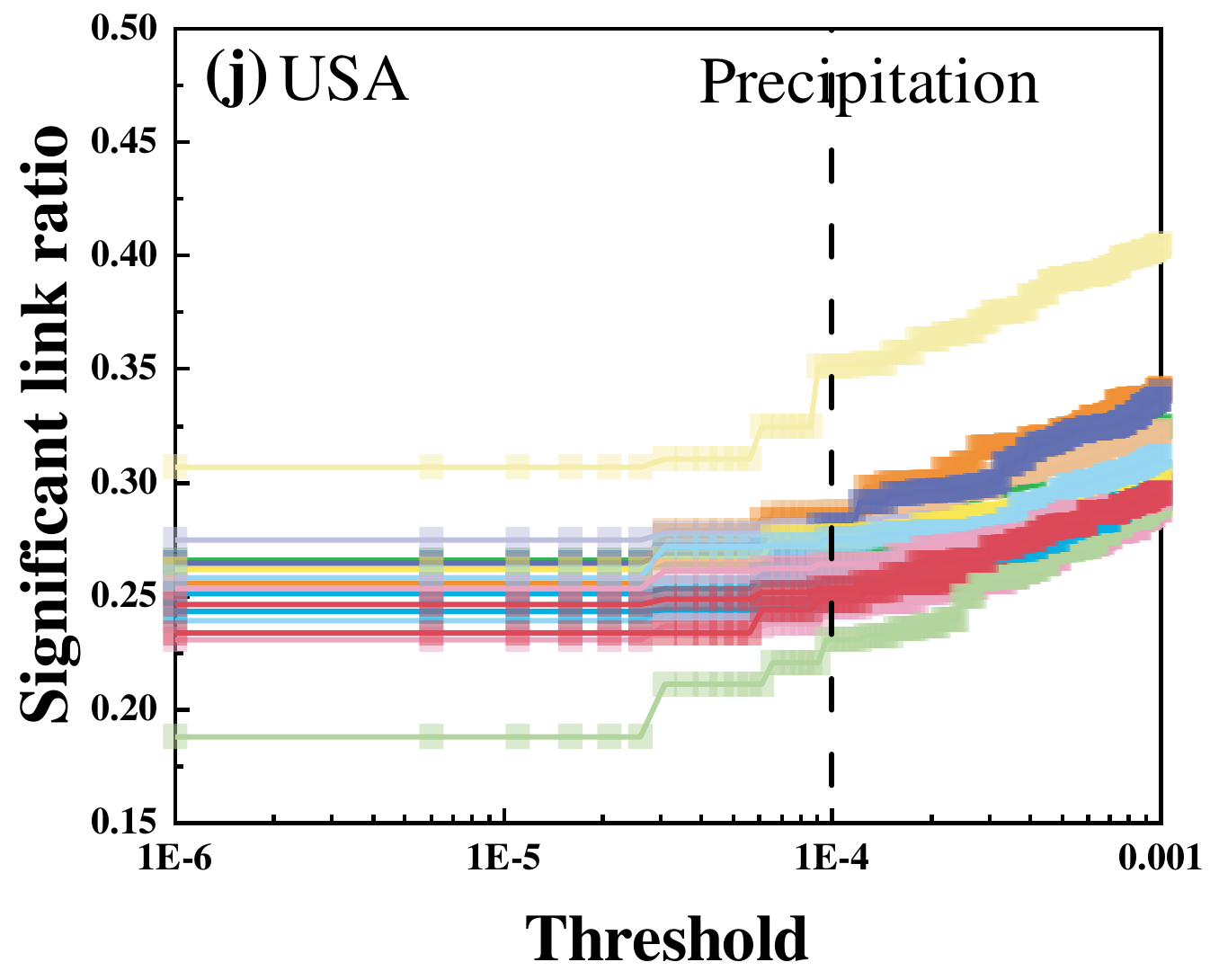}
\includegraphics[width=8em, height=7em]{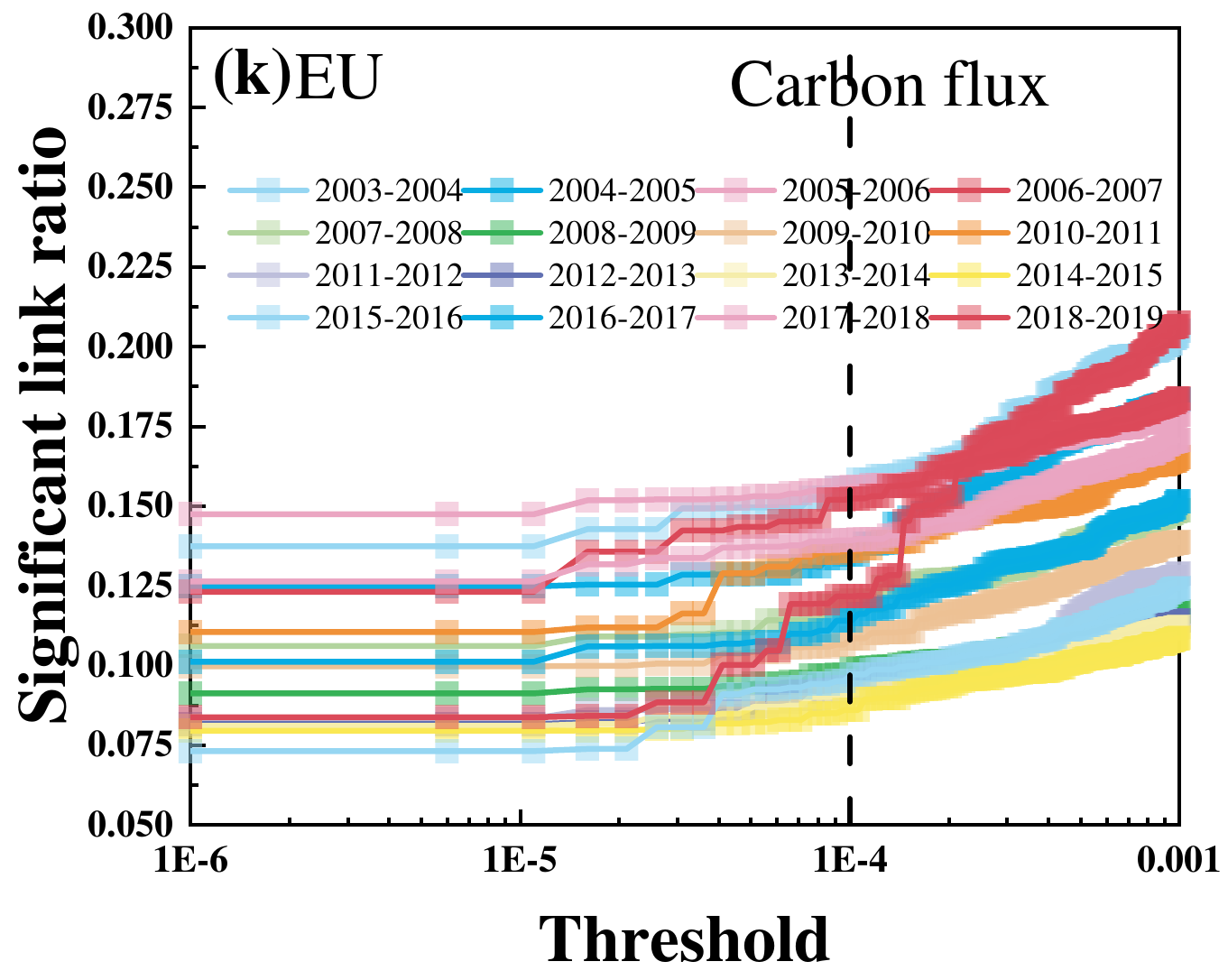}
\includegraphics[width=8em, height=7em]{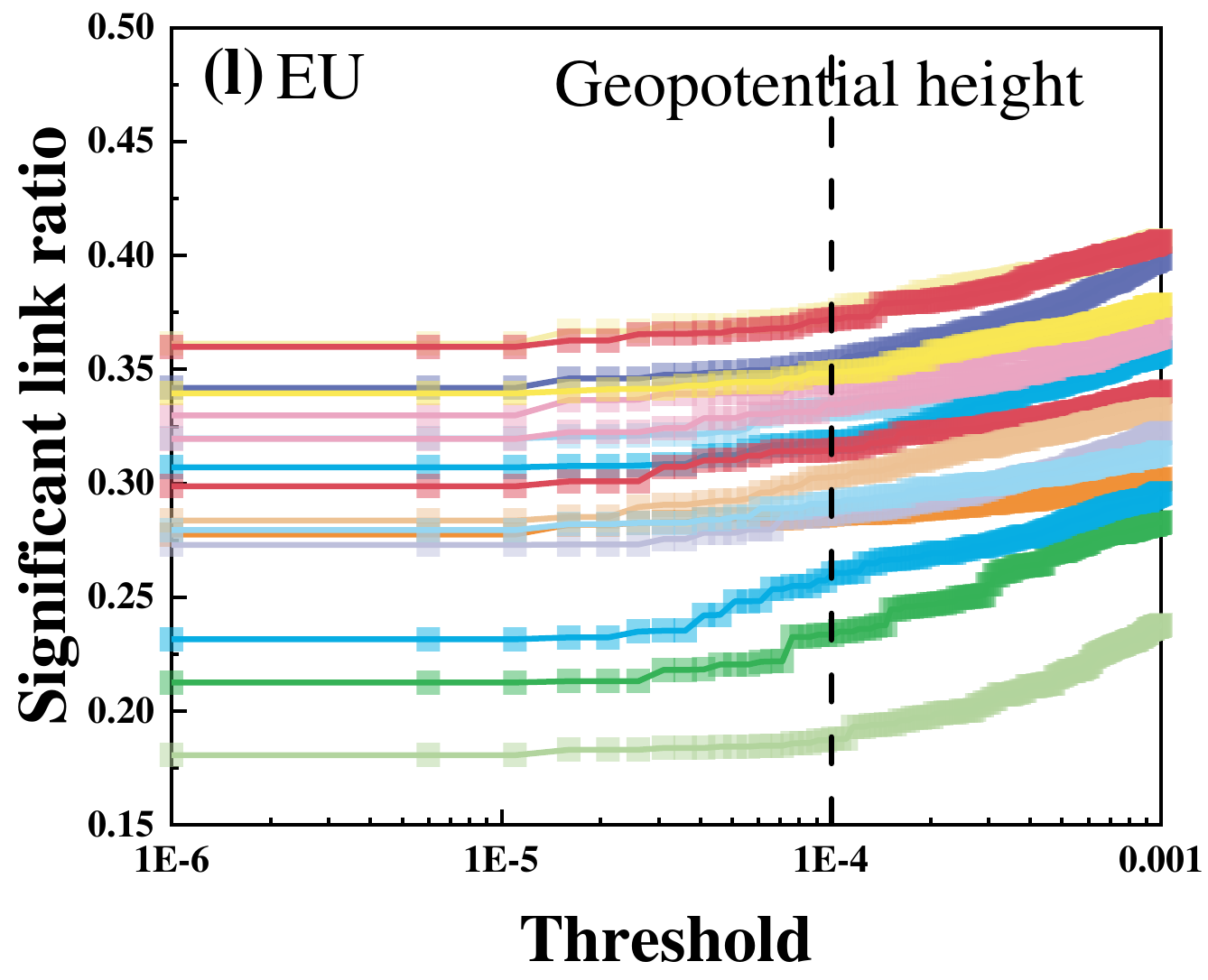}
\includegraphics[width=8em, height=7em]{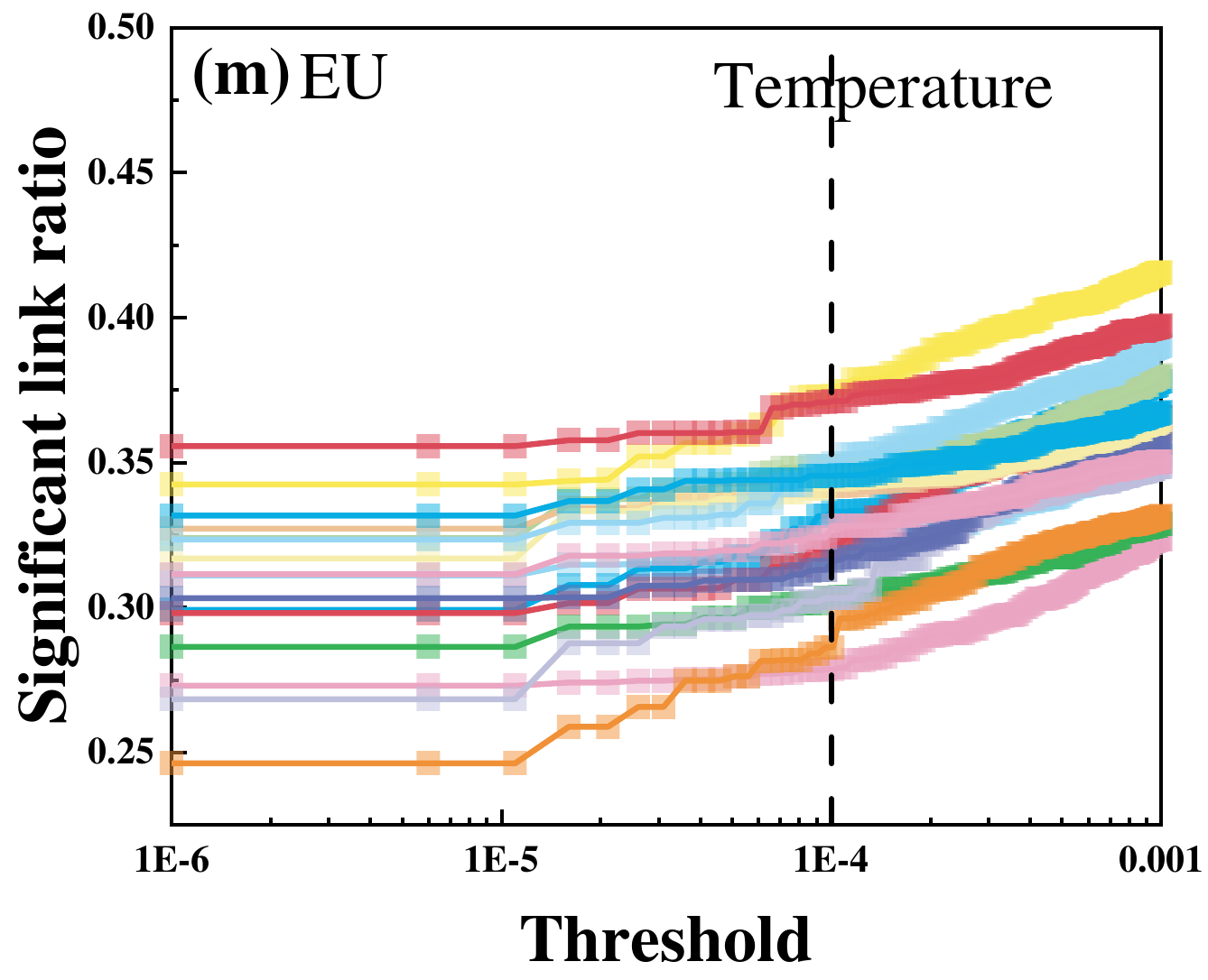}
\includegraphics[width=8em, height=7em]{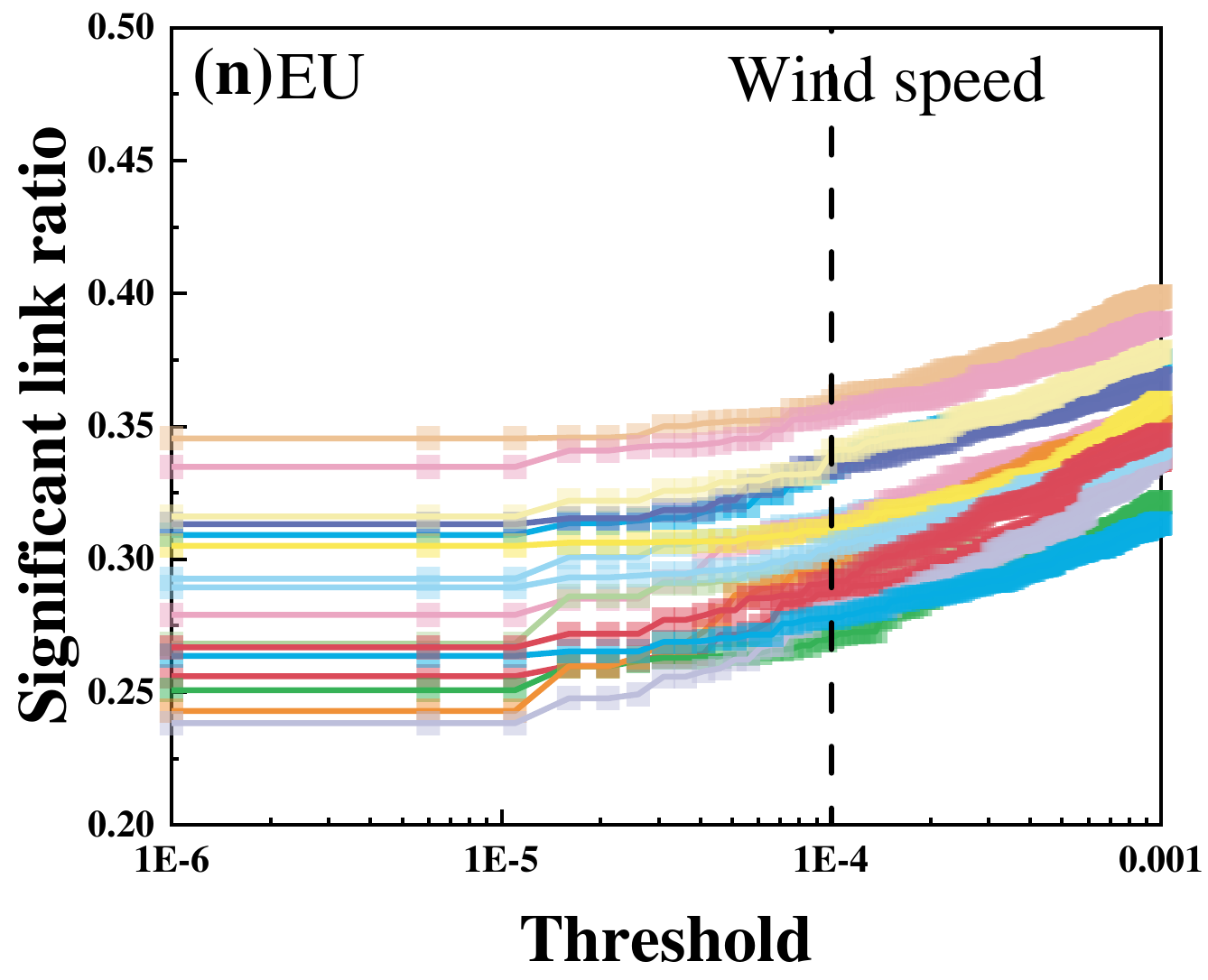}
\includegraphics[width=8em, height=7em]{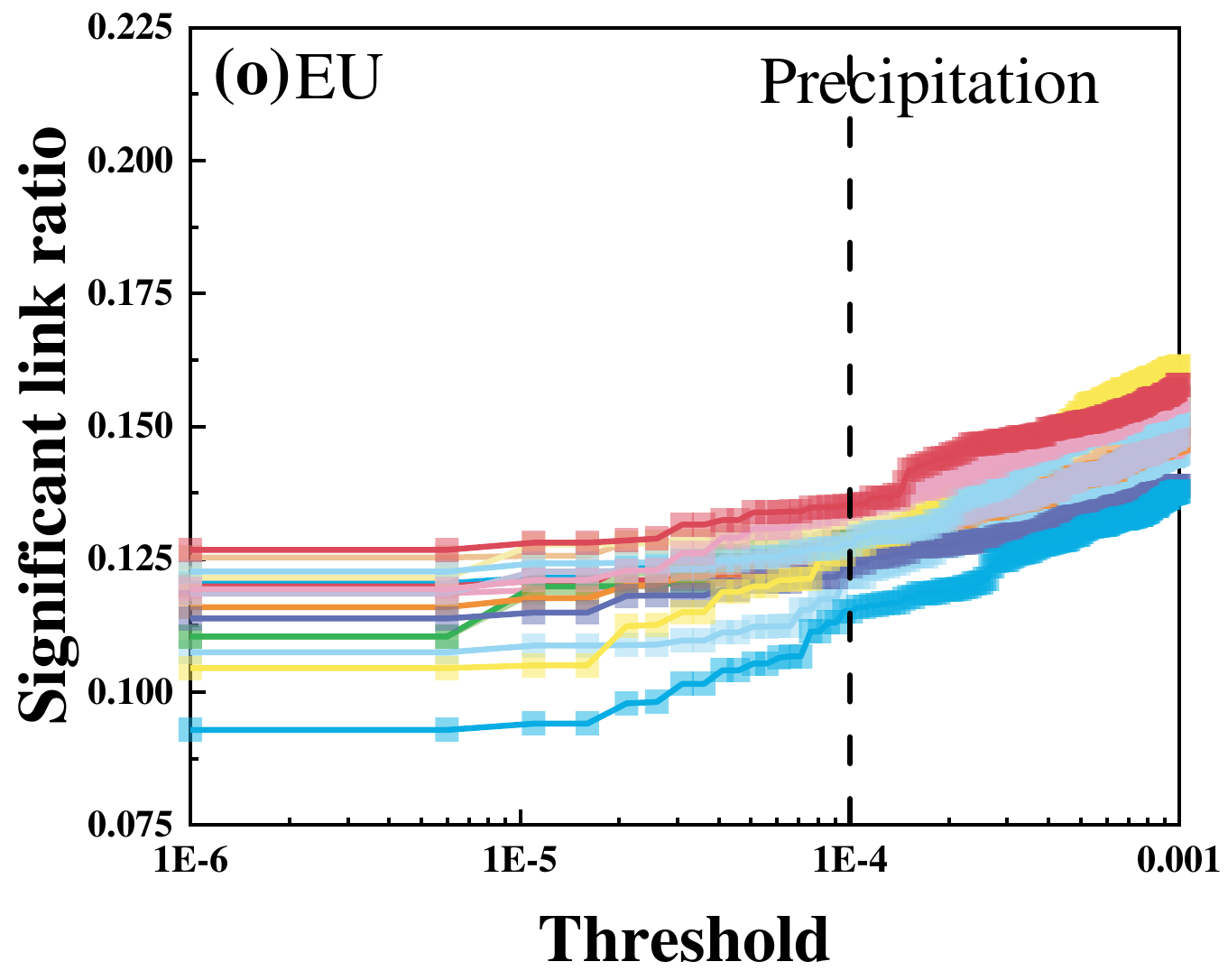}
\end{center}

\begin{center}
\noindent {\small {\bf Fig. S17} Significant link ratio as a function of threshold. The meaning of the horizontal coordinate is the value greater than $W(1-threshold)$ as the threshold for filtering significant links, and the meaning of the dashed line is the chosen threshold in the main text, i.e., W(0.9999).}
\end{center}

\begin{center}
\includegraphics[width=8em, height=7em]{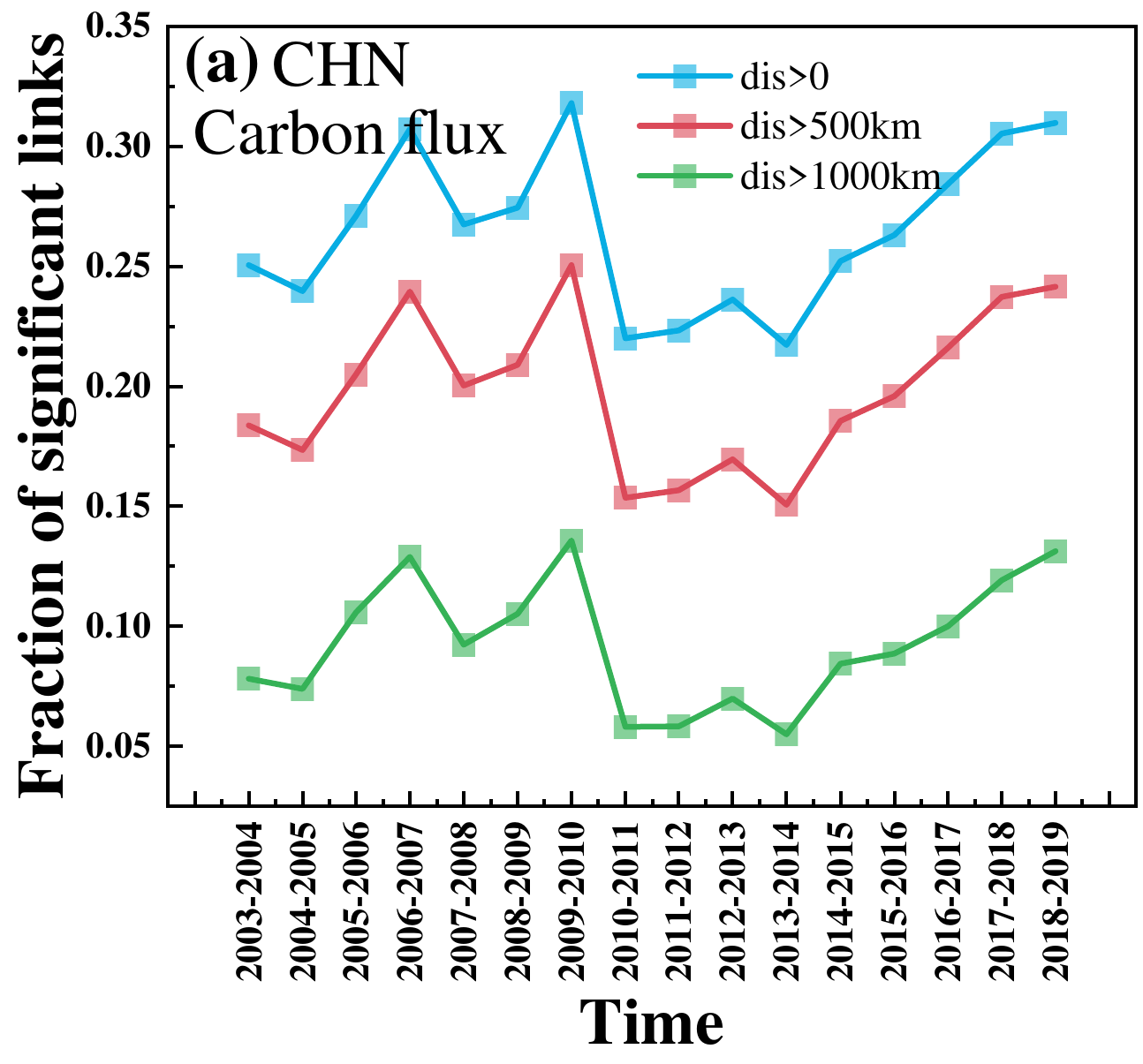}
\includegraphics[width=8em, height=7em]{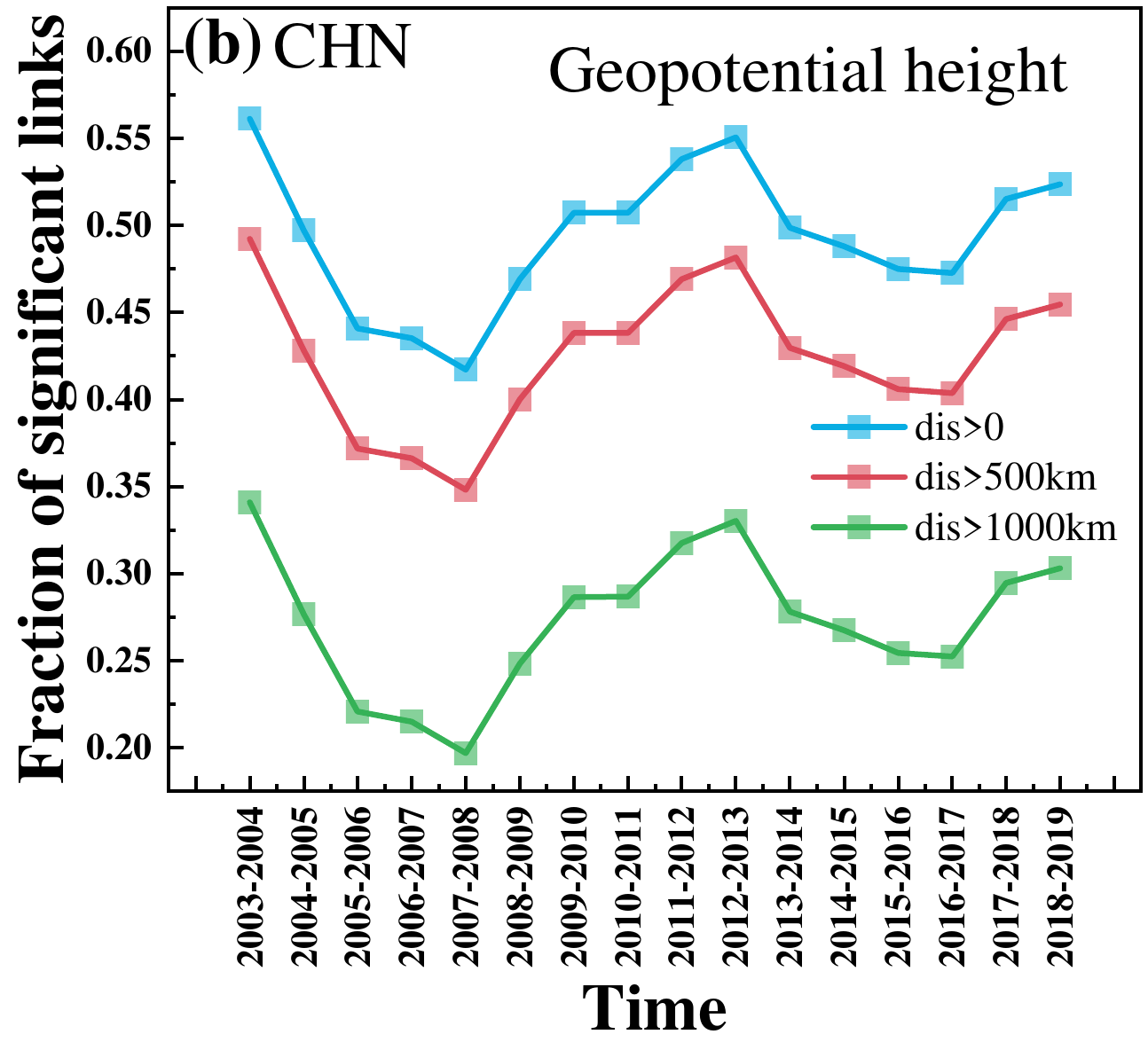}
\includegraphics[width=8em, height=7em]{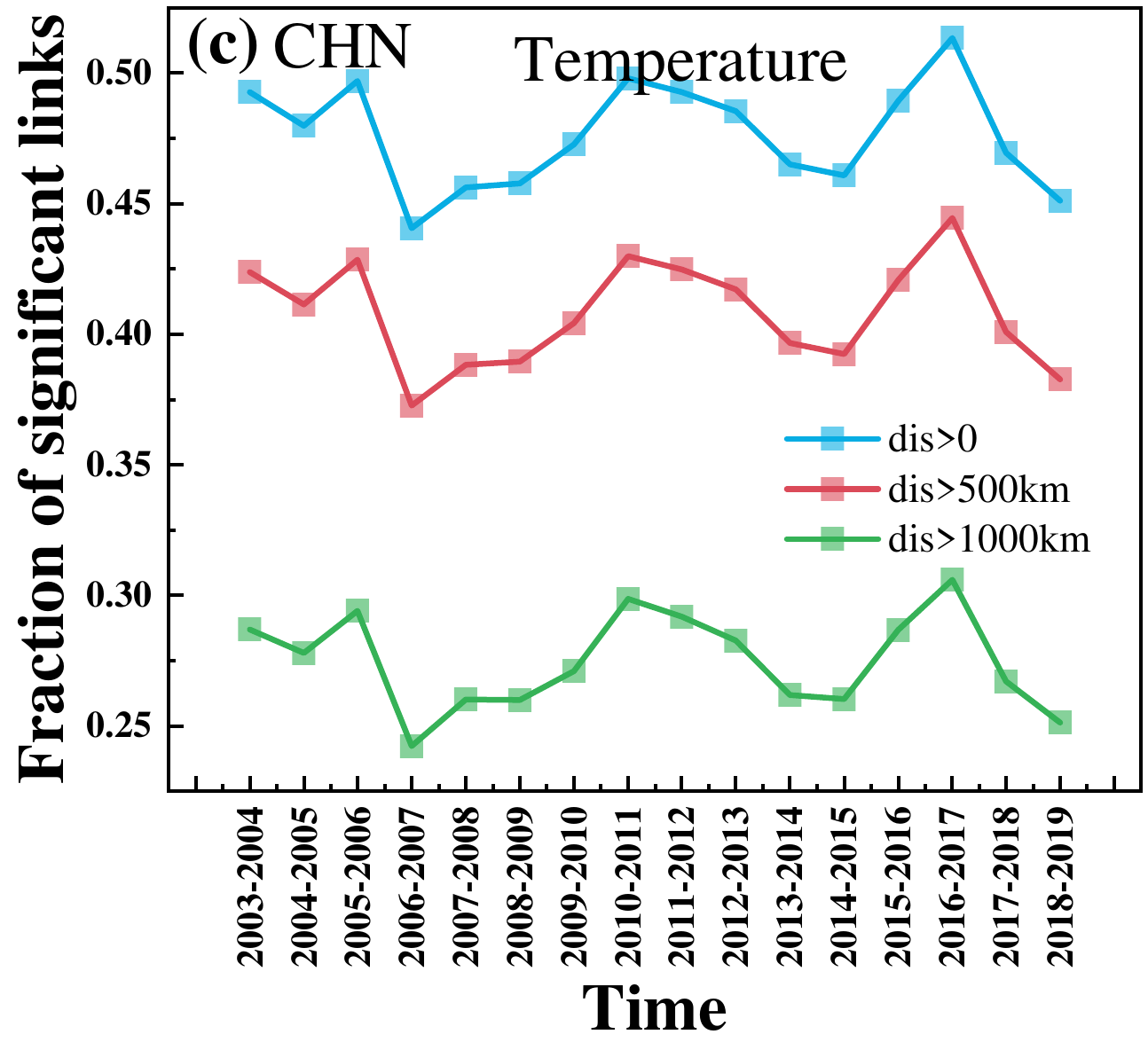}
\includegraphics[width=8em, height=7em]{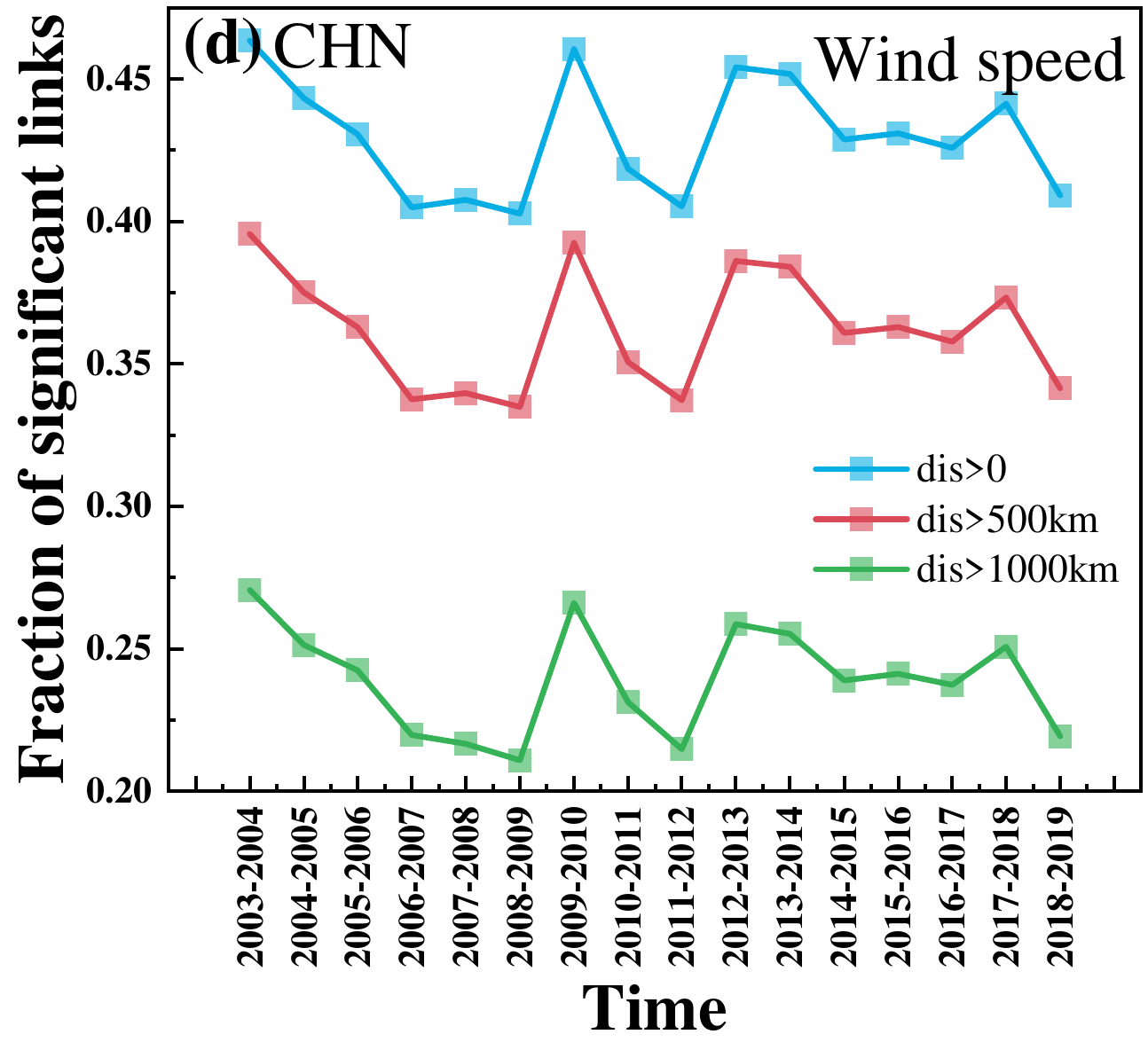}
\includegraphics[width=8em, height=7em]{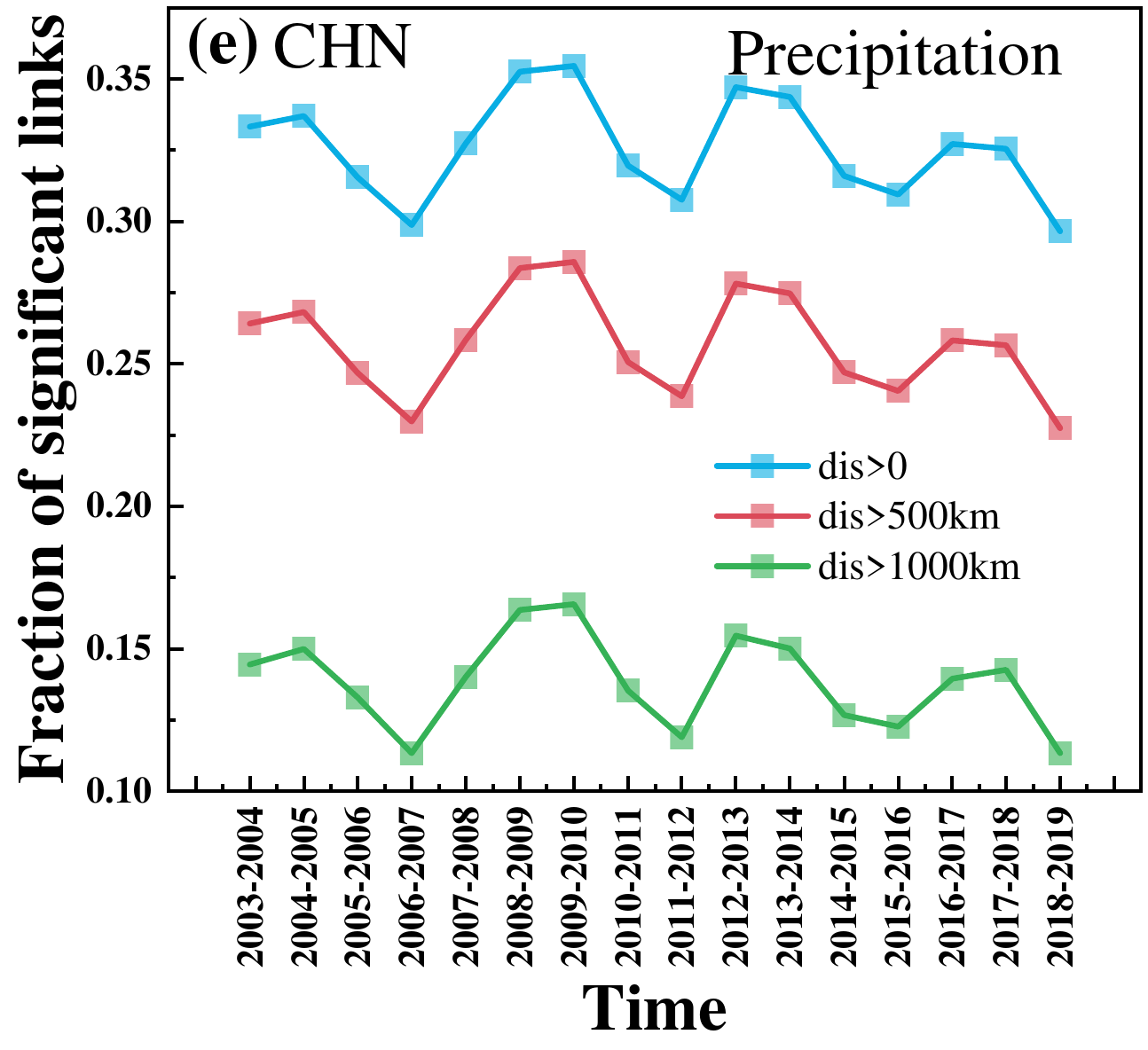}
\includegraphics[width=8em, height=7em]{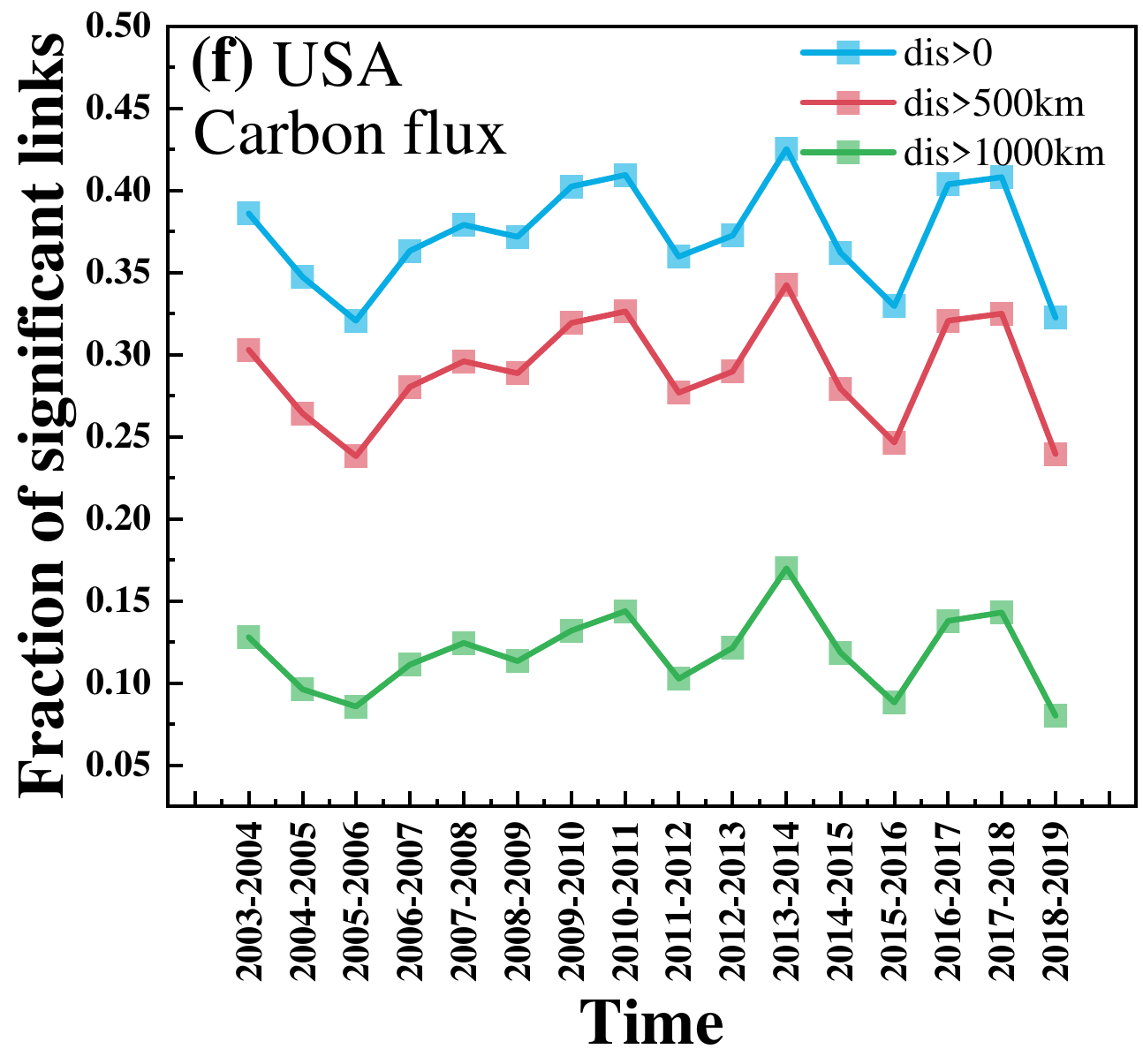}
\includegraphics[width=8em, height=7em]{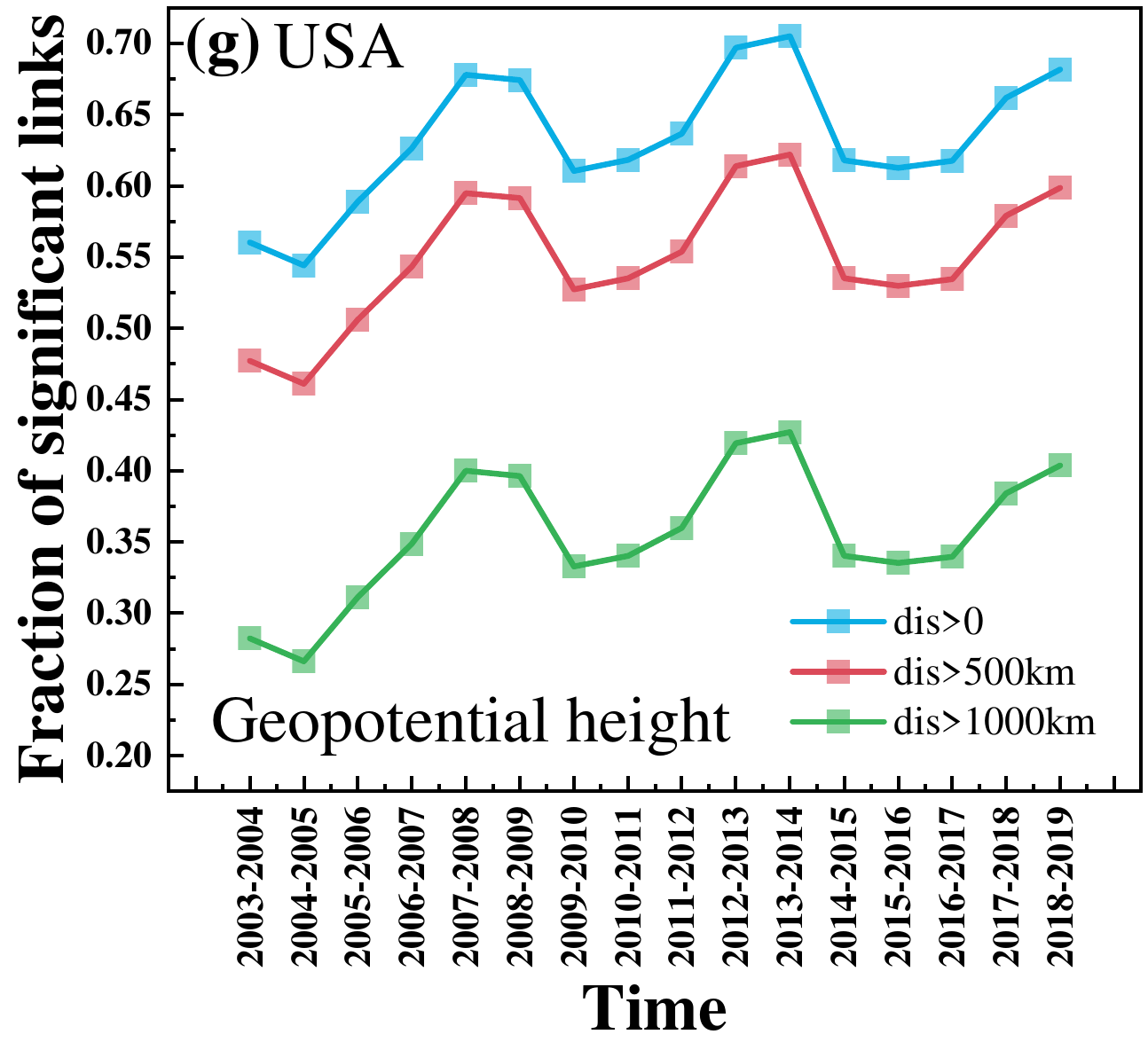}
\includegraphics[width=8em, height=7em]{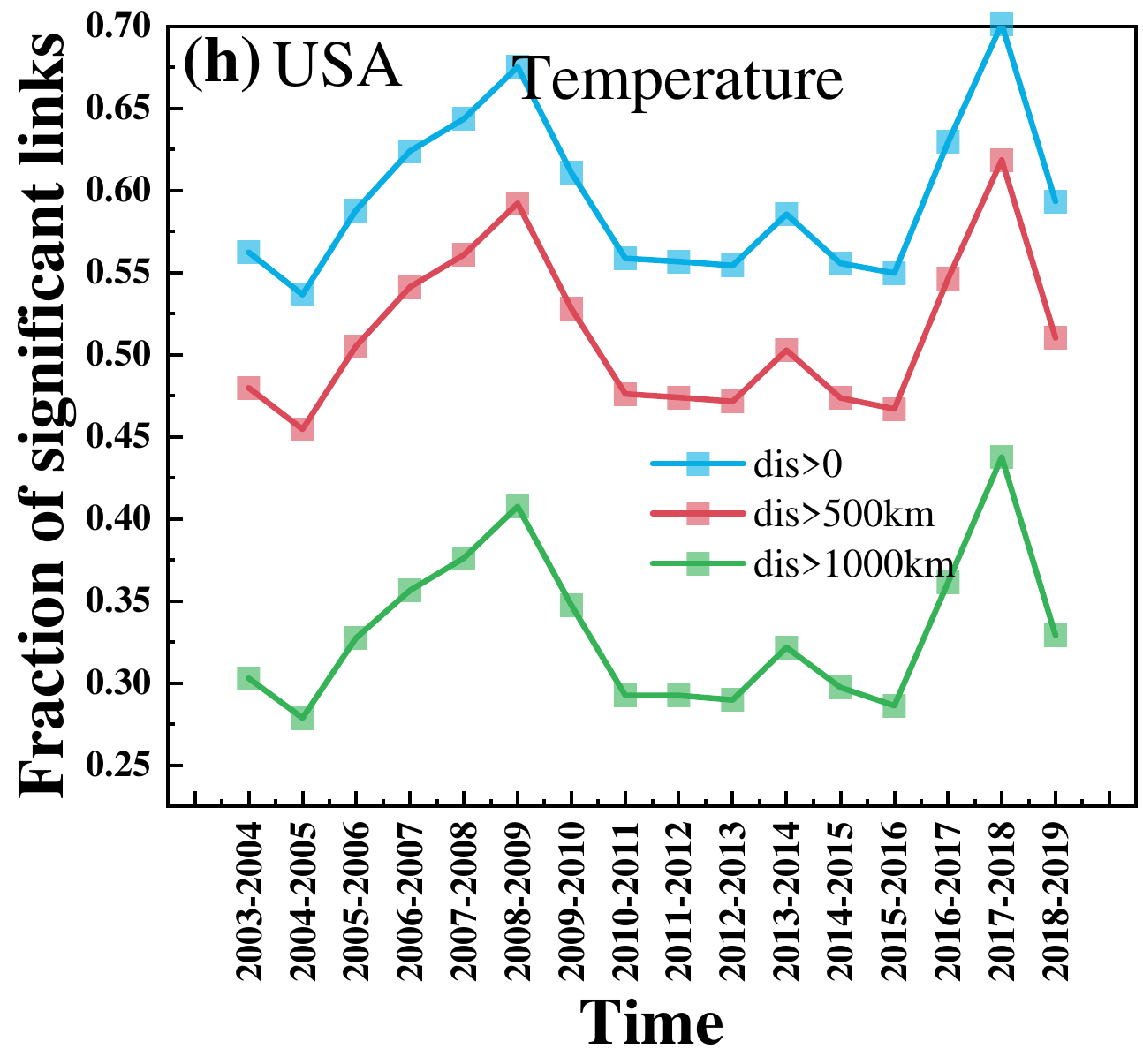}
\includegraphics[width=8em, height=7em]{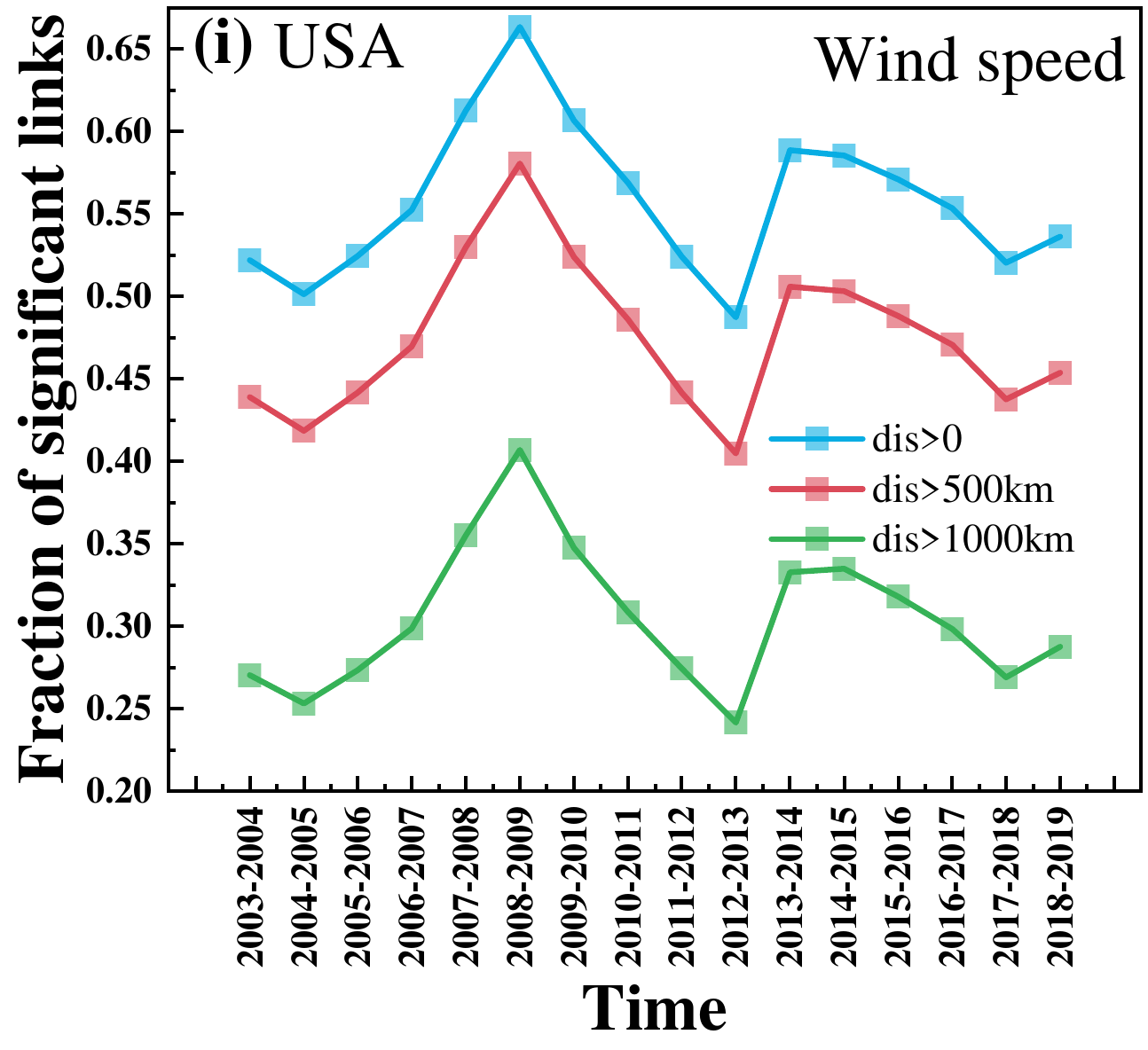}
\includegraphics[width=8em, height=7em]{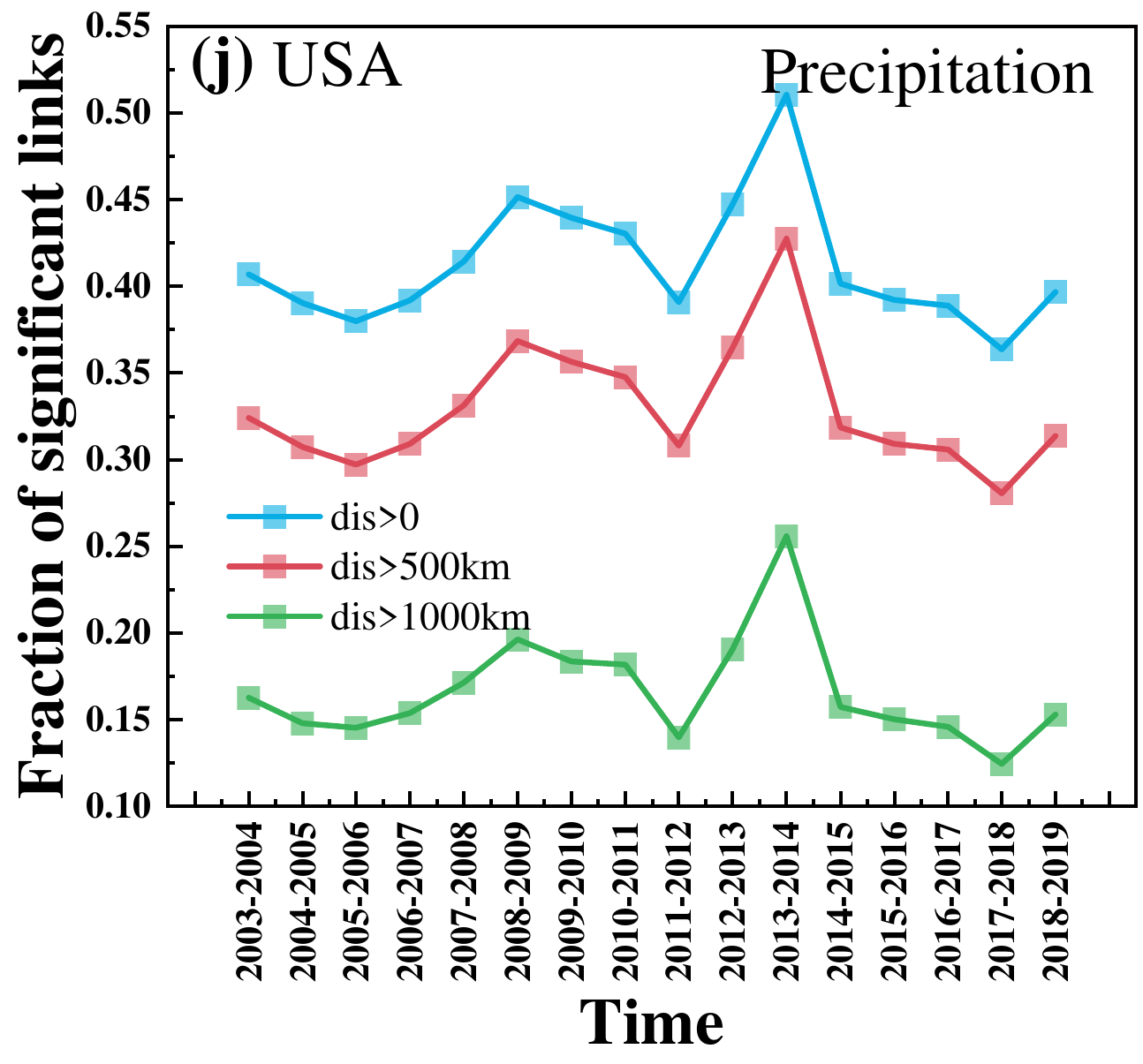}
\includegraphics[width=8em, height=7em]{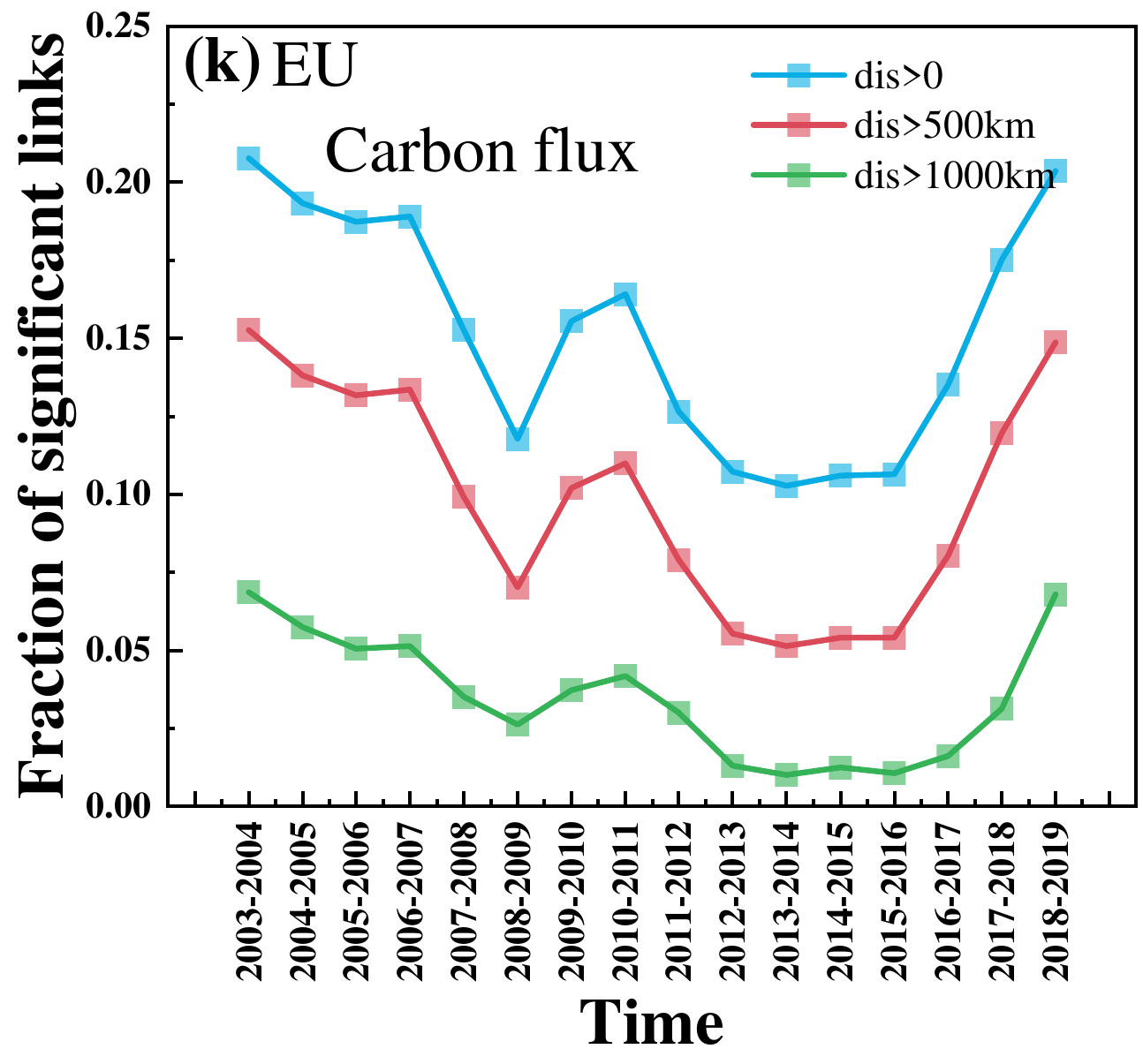}
\includegraphics[width=8em, height=7em]{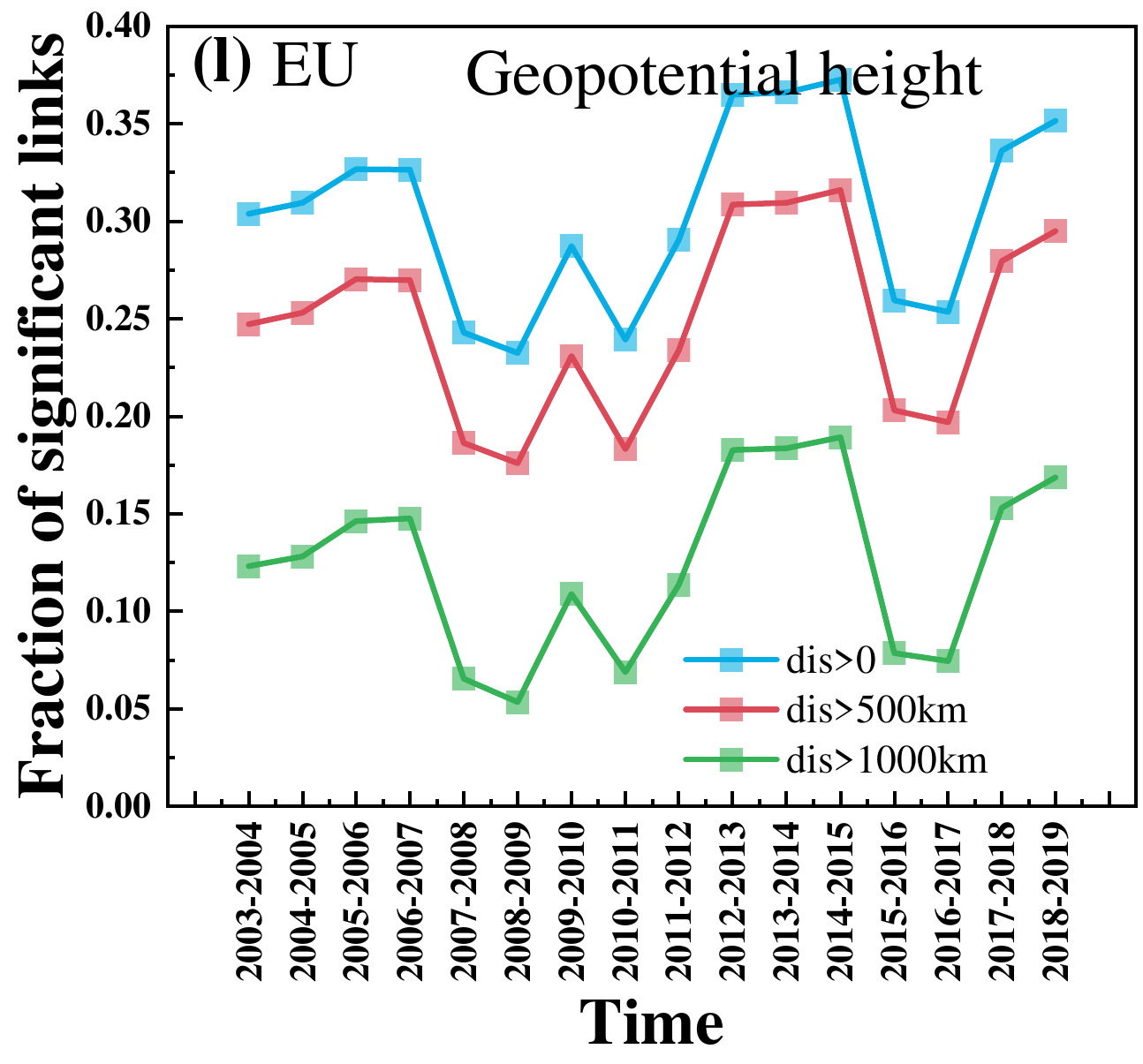}
\includegraphics[width=8em, height=7em]{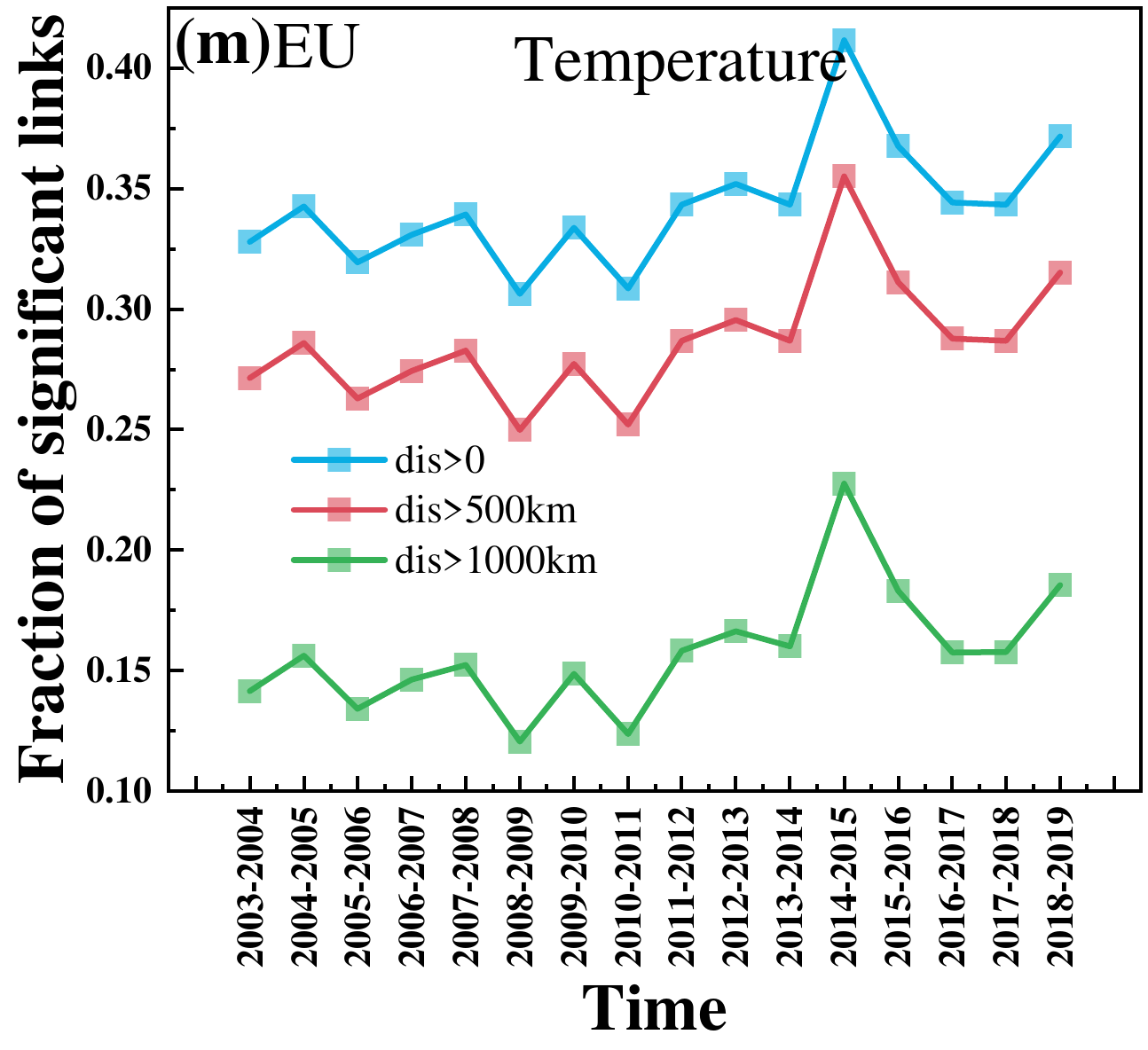}
\includegraphics[width=8em, height=7em]{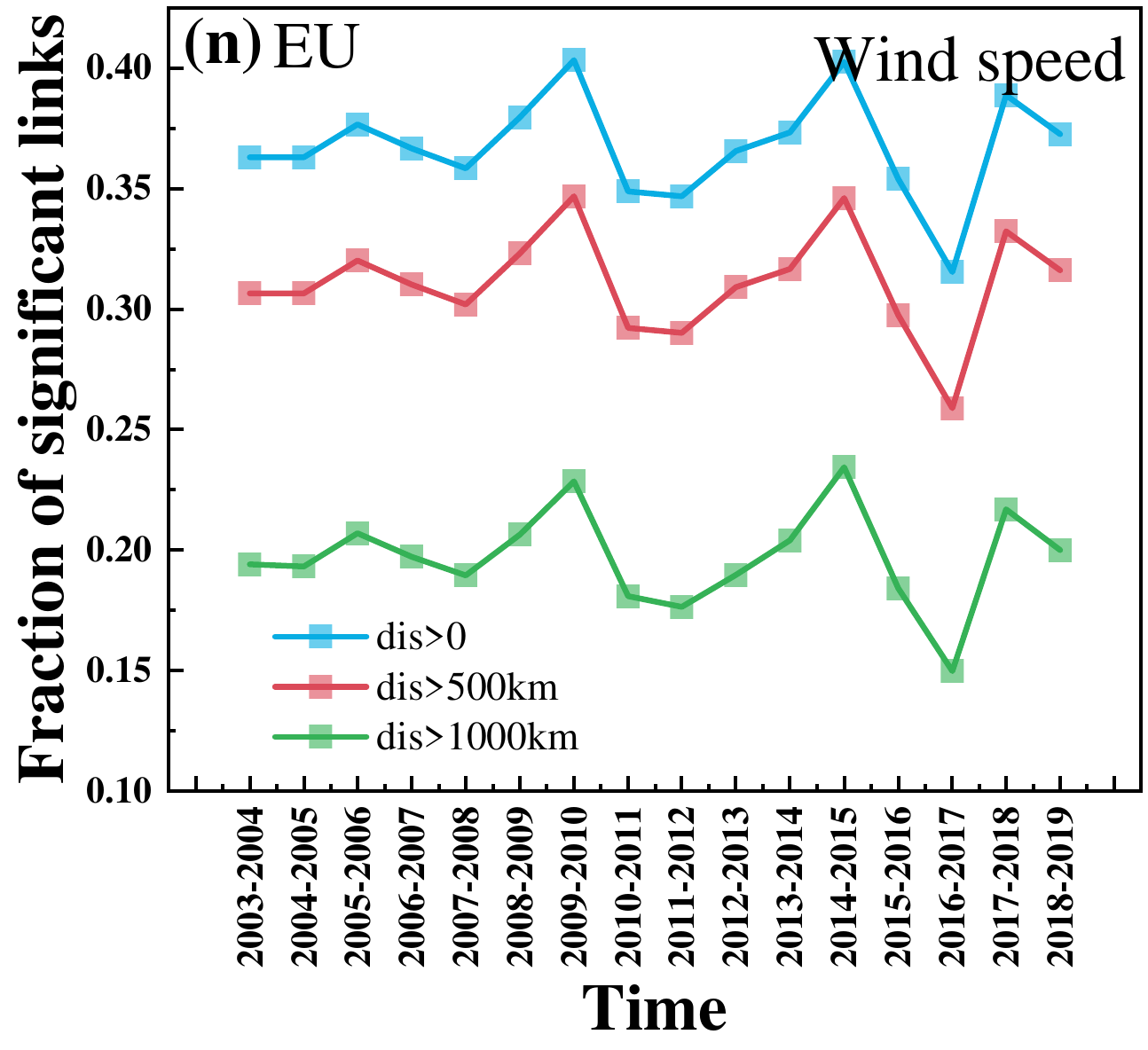}
\includegraphics[width=8em, height=7em]{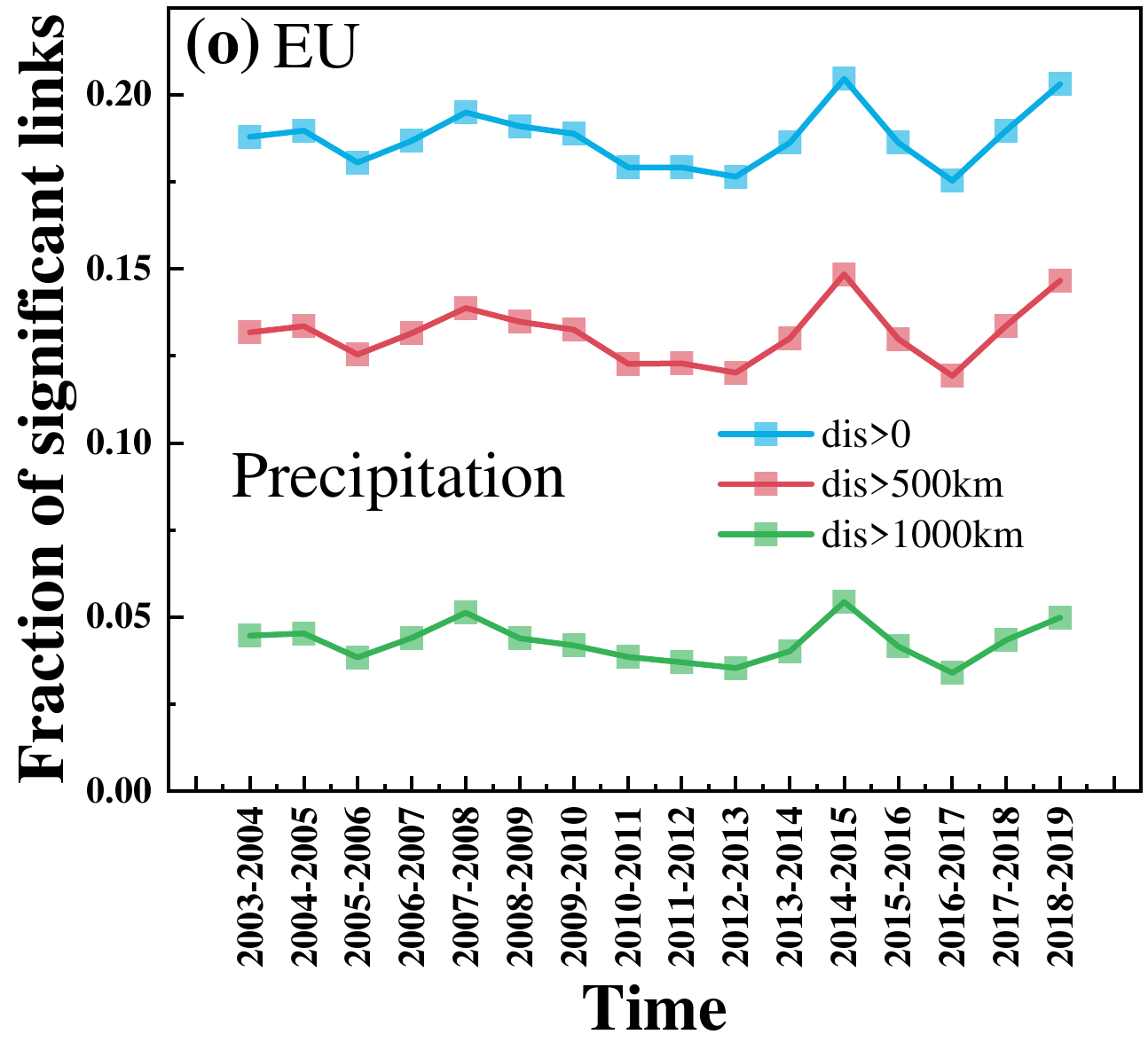}
\end{center}

\begin{center}
\noindent {\small {\bf Fig. S18} The number of significant links as a function of time in different networks at the threshold $W=4$.}
\end{center}

\begin{table}
\scriptsize
  \centering
  \caption{In CHN, the number of significant links (N) and the fraction of links (F) with time lag $\tau^{\ast}=\pm 1$ from West to East (W $\to$ E, $\tau^{\ast}=1$) and from East to West (E $\to$ W, $\tau^{\ast}=-1$) at $dis>0$.}
\begin{tabular}{|c|c|c|c|c|c|c|c|c|c|c|c|c|c|c|c|c|c|c|c|c|}

    \hline
    \multirow{3}[6]{*}{} & \multicolumn{4}{c|}{Carbon flux} & \multicolumn{4}{c|}{Geopotential height} & \multicolumn{4}{c|}{Temperature} & \multicolumn{4}{c|}{Wind speed} & \multicolumn{4}{c|}{Precipitation} \\
\cline{2-21}          & \multicolumn{2}{c|}{W$\to$ E} & \multicolumn{2}{c|}{E$\to$ W} & \multicolumn{2}{c|}{W$\to$ E} & \multicolumn{2}{c|}{E$\to$ W} & \multicolumn{2}{c|}{W$\to$ E} & \multicolumn{2}{c|}{E$\to$ W} & \multicolumn{2}{c|}{W$\to$ E} & \multicolumn{2}{c|}{E$\to$ W} & \multicolumn{2}{c|}{W$\to$ E} & \multicolumn{2}{c|}{E$\to$ W} \\
\cline{2-21}          & N     & F     & N     & F     & N     & F     & N     & F     & N     & F     & N     & F     & N     & F     & N     & F     & N     & F     & N     & F \\
    \hline
    2003-2004 & 822   & 0.15  & 7     & 0.001  & 1767  & 0.11  & 1952  & 0.12  & 4051  & 0.30  & 1811  & 0.13  & 4154  & 0.38  & 547   & 0.05  & 2848  & 0.45  & 31    & 0.005  \\
    \hline
    2004-2005 & 131   & 0.04  & 6     & 0.002  & 1814  & 0.13  & 1002  & 0.07  & 4096  & 0.31  & 1816  & 0.14  & 3386  & 0.37  & 319   & 0.03  & 2972  & 0.46  & 61    & 0.009  \\
    \hline
    2005-2006 & 694   & 0.14  & 9     & 0.002  & 1180  & 0.10  & 326   & 0.03  & 4446  & 0.31  & 2137  & 0.15  & 3458  & 0.35  & 481   & 0.05  & 2578  & 0.42  & 90    & 0.015  \\
    \hline
    2006-2007 & 1077  & 0.18  & 16    & 0.003  & 2012  & 0.16  & 84    & 0.01  & 4167  & 0.34  & 1674  & 0.14  & 3435  & 0.36  & 509   & 0.05  & 2250  & 0.41  & 58    & 0.011  \\
    \hline
    2007-2008 & 1131  & 0.19  & 12    & 0.002  & 1873  & 0.17  & 1     & 0.00  & 3830  & 0.31  & 1312  & 0.11  & 3675  & 0.38  & 571   & 0.06  & 2698  & 0.43  & 59    & 0.009  \\
    \hline
    2008-2009 & 1011  & 0.17  & 18    & 0.003  & 1578  & 0.14  & 0     & 0.00  & 3769  & 0.35  & 1031  & 0.10  & 3777  & 0.39  & 411   & 0.04  & 2825  & 0.42  & 89    & 0.013  \\
    \hline
    2009-2010 & 988   & 0.16  & 2     & 0.000  & 2675  & 0.18  & 383   & 0.03  & 4264  & 0.34  & 1492  & 0.12  & 3621  & 0.37  & 464   & 0.05  & 2665  & 0.42  & 88    & 0.014  \\
    \hline
    2010-2011 & 756   & 0.16  & 18    & 0.004  & 2242  & 0.16  & 513   & 0.04  & 3989  & 0.31  & 1571  & 0.12  & 3232  & 0.34  & 516   & 0.05  & 2523  & 0.41  & 53    & 0.009  \\
    \hline
    2011-2012 & 608   & 0.13  & 11    & 0.002  & 2054  & 0.13  & 1677  & 0.10  & 3957  & 0.31  & 1723  & 0.13  & 3096  & 0.37  & 318   & 0.04  & 2611  & 0.42  & 125   & 0.020  \\
    \hline
    2012-2013 & 960   & 0.18  & 10    & 0.002  & 2234  & 0.14  & 1070  & 0.07  & 3840  & 0.31  & 1741  & 0.14  & 4003  & 0.38  & 625   & 0.06  & 2820  & 0.42  & 137   & 0.021  \\
    \hline
    2013-2014 & 864   & 0.19  & 2     & 0.000  & 1890  & 0.13  & 264   & 0.02  & 3763  & 0.32  & 1510  & 0.13  & 4173  & 0.39  & 521   & 0.05  & 2608  & 0.41  & 71    & 0.011  \\
    \hline
    2014-2015 & 1143  & 0.20  & 14    & 0.002  & 1921  & 0.15  & 127   & 0.01  & 4301  & 0.34  & 1478  & 0.12  & 3929  & 0.40  & 323   & 0.03  & 2749  & 0.43  & 55    & 0.009  \\
    \hline
    2015-2016 & 488   & 0.13  & 3     & 0.001  & 2417  & 0.17  & 424   & 0.03  & 4615  & 0.33  & 1610  & 0.12  & 3701  & 0.38  & 309   & 0.03  & 2540  & 0.42  & 31    & 0.005  \\
    \hline
    2016-2017 & 1552  & 0.24  & 15    & 0.002  & 1781  & 0.13  & 1058  & 0.08  & 4089  & 0.31  & 1525  & 0.12  & 3172  & 0.37  & 174   & 0.02  & 2719  & 0.42  & 28    & 0.004  \\
    \hline
    2017-2018 & 1211  & 0.20  & 8     & 0.001  & 1685  & 0.12  & 630   & 0.04  & 3922  & 0.33  & 1599  & 0.13  & 4091  & 0.39  & 368   & 0.03  & 2586  & 0.42  & 19    & 0.003  \\
    \hline
    2018-2019 & 639   & 0.08  & 7     & 0.001  & 2149  & 0.14  & 923   & 0.06  & 3755  & 0.34  & 1373  & 0.12  & 3618  & 0.37  & 395   & 0.04  & 1757  & 0.37  & 17    & 0.004  \\
    \hline
  
\end{tabular}%
\label{tab:addlabel}
\end{table}

\begin{table}
\scriptsize
  \centering
\caption{In USA, the number of significant links (N) and the fraction of links (F) with time lag $\tau^{\ast}=\pm 1$ from West to East (W $\to$ E, $\tau^{\ast}=1$) and from East to West (E $\to$ W, $\tau^{\ast}=-1$) at $dis>0$.}
\begin{tabular}{|c|c|c|c|c|c|c|c|c|c|c|c|c|c|c|c|c|c|c|c|c|}

    \hline
    \multirow{3}[6]{*}{} & \multicolumn{4}{c|}{Carbon flux} & \multicolumn{4}{c|}{Geopotential height} & \multicolumn{4}{c|}{Temperature} & \multicolumn{4}{c|}{Wind speed} & \multicolumn{4}{c|}{Precipitation} \\
\cline{2-21}          & \multicolumn{2}{c|}{W$\to$ E} & \multicolumn{2}{c|}{E$\to$ W} & \multicolumn{2}{c|}{W$\to$ E} & \multicolumn{2}{c|}{E$\to$ W} & \multicolumn{2}{c|}{W$\to$ E} & \multicolumn{2}{c|}{E$\to$ W} & \multicolumn{2}{c|}{W$\to$ E} & \multicolumn{2}{c|}{E$\to$ W} & \multicolumn{2}{c|}{W$\to$ E} & \multicolumn{2}{c|}{E$\to$ W} \\
\cline{2-21}          & N     & F     & N     & F     & N     & F     & N     & F     & N     & F     & N     & F     & N     & F     & N     & F     & N     & F     & N     & F \\
    \hline
    2003-2004 & 2338  & 0.36  & 0     & 0.000  & 5105  & 0.44  & 0     & 0.000  & 4478  & 0.41  & 546   & 0.05  & 3704  & 0.40  & 154   & 0.02  & 2373  & 0.46  & 3     & 0.001  \\
    \hline
    2004-2005 & 2309  & 0.36  & 0     & 0.000  & 4836  & 0.44  & 0     & 0.000  & 4070  & 0.42  & 406   & 0.04  & 3654  & 0.41  & 185   & 0.02  & 2218  & 0.45  & 13    & 0.003  \\
    \hline
    2005-2006 & 1851  & 0.34  & 0     & 0.000  & 4173  & 0.40  & 0     & 0.000  & 4324  & 0.42  & 380   & 0.04  & 3481  & 0.42  & 105   & 0.01  & 2092  & 0.43  & 46    & 0.009  \\
    \hline
    2006-2007 & 2520  & 0.39  & 0     & 0.000  & 5522  & 0.46  & 1     & 0.000  & 4787  & 0.42  & 435   & 0.04  & 3515  & 0.42  & 59    & 0.01  & 2115  & 0.43  & 87    & 0.018  \\
    \hline
    2007-2008 & 2679  & 0.39  & 0     & 0.000  & 6265  & 0.49  & 1     & 0.000  & 4860  & 0.41  & 394   & 0.03  & 4080  & 0.41  & 95    & 0.01  & 1864  & 0.41  & 31    & 0.007  \\
    \hline
    2008-2009 & 2661  & 0.38  & 0     & 0.000  & 6623  & 0.49  & 0     & 0.000  & 5631  & 0.44  & 453   & 0.04  & 4271  & 0.39  & 301   & 0.03  & 2397  & 0.45  & 12    & 0.002  \\
    \hline
    2009-2010 & 3075  & 0.43  & 0     & 0.000  & 6271  & 0.49  & 0     & 0.000  & 4372  & 0.44  & 281   & 0.03  & 3922  & 0.40  & 155   & 0.02  & 2794  & 0.49  & 1     & 0.000  \\
    \hline
    2010-2011 & 3950  & 0.48  & 0     & 0.000  & 5627  & 0.46  & 0     & 0.000  & 4024  & 0.44  & 205   & 0.02  & 3861  & 0.41  & 146   & 0.02  & 2769  & 0.49  & 1     & 0.000  \\
    \hline
    2011-2012 & 2307  & 0.38  & 1     & 0.000  & 6348  & 0.49  & 0     & 0.000  & 4466  & 0.42  & 305   & 0.03  & 2997  & 0.37  & 120   & 0.01  & 2647  & 0.48  & 11    & 0.002  \\
    \hline
    2012-2013 & 2423  & 0.36  & 16    & 0.002  & 6832  & 0.49  & 0     & 0.000  & 4297  & 0.41  & 362   & 0.03  & 3080  & 0.39  & 178   & 0.02  & 2594  & 0.46  & 0     & 0.000  \\
    \hline
    2013-2014 & 3342  & 0.40  & 38    & 0.005  & 6833  & 0.49  & 0     & 0.000  & 4233  & 0.41  & 419   & 0.04  & 3775  & 0.39  & 242   & 0.03  & 3131  & 0.45  & 21    & 0.003  \\
    \hline
    2014-2015 & 2322  & 0.38  & 9     & 0.001  & 6210  & 0.48  & 0     & 0.000  & 4376  & 0.40  & 576   & 0.05  & 3773  & 0.41  & 97    & 0.01  & 2511  & 0.46  & 44    & 0.008  \\
    \hline
    2015-2016 & 2569  & 0.41  & 5     & 0.001  & 5117  & 0.44  & 0     & 0.000  & 4232  & 0.40  & 552   & 0.05  & 3578  & 0.42  & 130   & 0.02  & 2507  & 0.46  & 40    & 0.007  \\
    \hline
    2016-2017 & 3104  & 0.44  & 16    & 0.002  & 5642  & 0.44  & 0     & 0.000  & 4498  & 0.40  & 505   & 0.04  & 4009  & 0.42  & 198   & 0.02  & 2344  & 0.45  & 48    & 0.009  \\
    \hline
    2017-2018 & 2459  & 0.39  & 10    & 0.002  & 5198  & 0.42  & 0     & 0.000  & 4884  & 0.38  & 826   & 0.06  & 3636  & 0.40  & 299   & 0.03  & 2356  & 0.45  & 54    & 0.010  \\
    \hline
    2018-2019 & 1322  & 0.25  & 0     & 0.000  & 6270  & 0.46  & 3     & 0.000  & 4475  & 0.41  & 322   & 0.03  & 3795  & 0.39  & 283   & 0.03  & 2136  & 0.42  & 3     & 0.001  \\
    \hline
  
\end{tabular}%
\label{tab:addlabel}
\end{table}

\begin{table}
\scriptsize
  \centering
\caption{In EU, the number of significant links (N) and the fraction of links (F) with time lag $\tau^{\ast}=\pm 1$ from West to East (W $\to$ E, $\tau^{\ast}=1$) and from East to West (E $\to$ W, $\tau^{\ast}=-1$) at $dis>0$.}
\begin{tabular}{|c|c|c|c|c|c|c|c|c|c|c|c|c|c|c|c|c|c|c|c|c|}

    \hline
    \multirow{3}[6]{*}{} & \multicolumn{4}{c|}{Carbon flux} & \multicolumn{4}{c|}{Geopotential height} & \multicolumn{4}{c|}{Temperature} & \multicolumn{4}{c|}{Wind speed} & \multicolumn{4}{c|}{Precipitation} \\
\cline{2-21}          & \multicolumn{2}{c|}{W$\to$ E} & \multicolumn{2}{c|}{E$\to$ W} & \multicolumn{2}{c|}{W$\to$ E} & \multicolumn{2}{c|}{E$\to$ W} & \multicolumn{2}{c|}{W$\to$ E} & \multicolumn{2}{c|}{E$\to$ W} & \multicolumn{2}{c|}{W$\to$ E} & \multicolumn{2}{c|}{E$\to$ W} & \multicolumn{2}{c|}{W$\to$ E} & \multicolumn{2}{c|}{E$\to$ W} \\
\cline{2-21}          & N     & F     & N     & F     & N     & F     & N     & F     & N     & F     & N     & F     & N     & F     & N     & F     & N     & F     & N     & F \\
    \hline
    2003-2004 & 2905  & 0.25  & 0     & 0.000  & 9006  & 0.37  & 735   & 0.03  & 10283 & 0.43  & 779   & 0.03  & 9602  & 0.41  & 282   & 0.01  & 3906  & 0.43  & 1     & 0.000  \\
    \hline
    2004-2005 & 1941  & 0.19  & 9     & 0.001  & 9717  & 0.41  & 619   & 0.03  & 10624 & 0.43  & 733   & 0.03  & 10862 & 0.44  & 279   & 0.01  & 4325  & 0.45  & 28    & 0.003  \\
    \hline
    2005-2006 & 2666  & 0.23  & 4     & 0.000  & 10680 & 0.42  & 516   & 0.02  & 9414  & 0.46  & 624   & 0.03  & 9558  & 0.41  & 165   & 0.01  & 4364  & 0.46  & 8     & 0.001  \\
    \hline
    2006-2007 & 3298  & 0.29  & 3     & 0.000  & 8679  & 0.37  & 369   & 0.02  & 10653 & 0.45  & 781   & 0.03  & 8125  & 0.38  & 242   & 0.01  & 4282  & 0.45  & 4     & 0.000  \\
    \hline
    2007-2008 & 2348  & 0.26  & 16    & 0.002  & 3351  & 0.24  & 13    & 0.00  & 11478 & 0.45  & 733   & 0.03  & 8823  & 0.40  & 114   & 0.01  & 4335  & 0.45  & 0     & 0.000  \\
    \hline
    2008-2009 & 1360  & 0.19  & 1     & 0.000  & 5038  & 0.29  & 10    & 0.00  & 10255 & 0.46  & 608   & 0.03  & 7828  & 0.39  & 151   & 0.01  & 4630  & 0.47  & 0     & 0.000  \\
    \hline
    2009-2010 & 1904  & 0.24  & 0     & 0.000  & 5894  & 0.26  & 378   & 0.02  & 10988 & 0.43  & 472   & 0.02  & 11990 & 0.45  & 413   & 0.02  & 4779  & 0.49  & 0     & 0.000  \\
    \hline
    2010-2011 & 2049  & 0.20  & 1     & 0.000  & 4455  & 0.21  & 32    & 0.00  & 9331  & 0.44  & 364   & 0.02  & 9462  & 0.43  & 262   & 0.01  & 4252  & 0.45  & 3     & 0.000  \\
    \hline
    2011-2012 & 1201  & 0.18  & 0     & 0.000  & 5866  & 0.28  & 297   & 0.01  & 10272 & 0.46  & 484   & 0.02  & 9130  & 0.44  & 225   & 0.01  & 4008  & 0.43  & 5     & 0.001  \\
    \hline
    2012-2013 & 945   & 0.14  & 0     & 0.000  & 9015  & 0.35  & 803   & 0.03  & 10613 & 0.45  & 510   & 0.02  & 10813 & 0.44  & 553   & 0.02  & 3965  & 0.43  & 3     & 0.000  \\
    \hline
    2013-2014 & 876   & 0.13  & 0     & 0.000  & 9279  & 0.34  & 579   & 0.02  & 10996 & 0.43  & 758   & 0.03  & 10720 & 0.43  & 325   & 0.01  & 4508  & 0.46  & 18    & 0.002  \\
    \hline
    2014-2015 & 1061  & 0.17  & 0     & 0.000  & 10491 & 0.41  & 89    & 0.00  & 10998 & 0.40  & 966   & 0.04  & 10244 & 0.44  & 212   & 0.01  & 4292  & 0.45  & 2     & 0.000  \\
    \hline
    2015-2016 & 1530  & 0.21  & 0     & 0.000  & 9401  & 0.44  & 116   & 0.01  & 10987 & 0.43  & 675   & 0.03  & 10055 & 0.45  & 336   & 0.01  & 4184  & 0.44  & 7     & 0.001  \\
    \hline
    2016-2017 & 2251  & 0.27  & 3     & 0.000  & 7353  & 0.38  & 74    & 0.00  & 11174 & 0.44  & 744   & 0.03  & 8255  & 0.40  & 238   & 0.01  & 3473  & 0.41  & 1     & 0.000  \\
    \hline
    2017-2018 & 3013  & 0.29  & 8     & 0.001  & 10966 & 0.44  & 381   & 0.02  & 10756 & 0.44  & 671   & 0.03  & 11457 & 0.44  & 413   & 0.02  & 4721  & 0.47  & 3     & 0.000  \\
    \hline
    2018-2019 & 444   & 0.05  & 0     & 0.000  & 11157 & 0.41  & 516   & 0.02  & 11867 & 0.43  & 820   & 0.03  & 8813  & 0.41  & 192   & 0.01  & 4594  & 0.46  & 6     & 0.001  \\
    \hline
  
\end{tabular}%
\label{tab:addlabel}
\end{table}

\begin{table}
\scriptsize
  \centering
   \caption{In CHN, the number of significant links (N) and the fraction of links (F) with time lag $\tau^{\ast}=\pm 1$ from West to East (W $\to$ E, $\tau^{\ast}=1$) and from East to West (E $\to$ W, $\tau^{\ast}=-1$) at $dis>500km$.}
\begin{tabular}{|c|c|c|c|c|c|c|c|c|c|c|c|c|c|c|c|c|c|c|c|c|}

    \hline
    \multirow{3}[6]{*}{} & \multicolumn{4}{c|}{Carbon flux} & \multicolumn{4}{c|}{Geopotential height} & \multicolumn{4}{c|}{Temperature} & \multicolumn{4}{c|}{Wind speed} & \multicolumn{4}{c|}{Precipitation} \\
\cline{2-21}          & \multicolumn{2}{c|}{W$\to$ E} & \multicolumn{2}{c|}{E$\to$ W} & \multicolumn{2}{c|}{W$\to$ E} & \multicolumn{2}{c|}{E$\to$ W} & \multicolumn{2}{c|}{W$\to$ E} & \multicolumn{2}{c|}{E$\to$ W} & \multicolumn{2}{c|}{W$\to$ E} & \multicolumn{2}{c|}{E$\to$ W} & \multicolumn{2}{c|}{W$\to$ E} & \multicolumn{2}{c|}{E$\to$ W} \\
\cline{2-21}          & N     & F     & N     & F     & N     & F     & N     & F     & N     & F     & N     & F     & N     & F     & N     & F     & N     & F     & N     & F \\
    \hline
    2003-2004 & 817   & 0.15  & 7     & 0.001  & 1767  & 0.11  & 1952  & 0.12  & 3899  & 0.29  & 1726  & 0.13  & 4111  & 0.38  & 531   & 0.05  & 2785  & 0.44  & 20    & 0.003  \\
    \hline
    2004-2005 & 131   & 0.04  & 6     & 0.002  & 1814  & 0.13  & 1002  & 0.07  & 3945  & 0.30  & 1749  & 0.13  & 3347  & 0.36  & 310   & 0.03  & 2912  & 0.45  & 45    & 0.007  \\
    \hline
    2005-2006 & 689   & 0.14  & 7     & 0.001  & 1173  & 0.10  & 326   & 0.03  & 4288  & 0.30  & 2048  & 0.14  & 3422  & 0.34  & 468   & 0.05  & 2528  & 0.41  & 61    & 0.010  \\
    \hline
    2006-2007 & 1071  & 0.18  & 13    & 0.002  & 2006  & 0.16  & 84    & 0.01  & 4016  & 0.33  & 1577  & 0.13  & 3407  & 0.35  & 493   & 0.05  & 2181  & 0.40  & 34    & 0.006  \\
    \hline
    2007-2008 & 1123  & 0.19  & 12    & 0.002  & 1873  & 0.17  & 1     & 0.00  & 3673  & 0.30  & 1235  & 0.10  & 3637  & 0.38  & 551   & 0.06  & 2624  & 0.41  & 45    & 0.007  \\
    \hline
    2008-2009 & 1006  & 0.17  & 18    & 0.003  & 1578  & 0.14  & 0     & 0.00  & 3610  & 0.33  & 975   & 0.09  & 3741  & 0.39  & 399   & 0.04  & 2752  & 0.41  & 77    & 0.011  \\
    \hline
    2009-2010 & 979   & 0.16  & 2     & 0.000  & 2669  & 0.18  & 383   & 0.03  & 4126  & 0.33  & 1435  & 0.11  & 3587  & 0.37  & 453   & 0.05  & 2603  & 0.41  & 75    & 0.012  \\
    \hline
    2010-2011 & 754   & 0.16  & 17    & 0.004  & 2241  & 0.16  & 513   & 0.04  & 3823  & 0.30  & 1470  & 0.11  & 3208  & 0.34  & 497   & 0.05  & 2473  & 0.41  & 41    & 0.007  \\
    \hline
    2011-2012 & 607   & 0.13  & 10    & 0.002  & 2052  & 0.13  & 1677  & 0.10  & 3794  & 0.29  & 1637  & 0.13  & 3068  & 0.36  & 304   & 0.04  & 2534  & 0.41  & 108   & 0.017  \\
    \hline
    2012-2013 & 954   & 0.18  & 10    & 0.002  & 2230  & 0.14  & 1070  & 0.07  & 3712  & 0.30  & 1671  & 0.14  & 3971  & 0.37  & 611   & 0.06  & 2766  & 0.41  & 123   & 0.018  \\
    \hline
    2013-2014 & 855   & 0.19  & 2     & 0.000  & 1886  & 0.13  & 264   & 0.02  & 3594  & 0.30  & 1430  & 0.12  & 4128  & 0.38  & 507   & 0.05  & 2556  & 0.40  & 55    & 0.009  \\
    \hline
    2014-2015 & 1137  & 0.19  & 14    & 0.002  & 1916  & 0.15  & 127   & 0.01  & 4121  & 0.33  & 1397  & 0.11  & 3890  & 0.39  & 315   & 0.03  & 2681  & 0.42  & 42    & 0.007  \\
    \hline
    2015-2016 & 488   & 0.13  & 3     & 0.001  & 2417  & 0.17  & 424   & 0.03  & 4452  & 0.32  & 1544  & 0.11  & 3662  & 0.37  & 300   & 0.03  & 2481  & 0.41  & 22    & 0.004  \\
    \hline
    2016-2017 & 1549  & 0.24  & 15    & 0.002  & 1779  & 0.13  & 1058  & 0.08  & 3929  & 0.30  & 1449  & 0.11  & 3138  & 0.37  & 166   & 0.02  & 2658  & 0.41  & 21    & 0.003  \\
    \hline
    2017-2018 & 1207  & 0.20  & 8     & 0.001  & 1678  & 0.12  & 630   & 0.04  & 3767  & 0.31  & 1528  & 0.13  & 4052  & 0.38  & 352   & 0.03  & 2534  & 0.41  & 14    & 0.002  \\
    \hline
    2018-2019 & 638   & 0.08  & 6     & 0.001  & 2143  & 0.14  & 923   & 0.06  & 3585  & 0.32  & 1306  & 0.12  & 3588  & 0.37  & 384   & 0.04  & 1707  & 0.36  & 14    & 0.003  \\
    \hline
  
\end{tabular}%
\label{tab:addlabel}
\end{table}

\begin{table}
\scriptsize
  \centering
\caption{In USA, the number of significant links (N) and the fraction of links (F) with time lag $\tau^{\ast}=\pm 1$ from West to East (W $\to$ E, $\tau^{\ast}=1$) and from East to West (E $\to$ W, $\tau^{\ast}=-1$) at $dis>500km$.}
\begin{tabular}{|c|c|c|c|c|c|c|c|c|c|c|c|c|c|c|c|c|c|c|c|c|}

    \hline
    \multirow{3}[6]{*}{} & \multicolumn{4}{c|}{Carbon flux} & \multicolumn{4}{c|}{Geopotential height} & \multicolumn{4}{c|}{Temperature} & \multicolumn{4}{c|}{Wind speed} & \multicolumn{4}{c|}{Precipitation} \\
\cline{2-21}          & \multicolumn{2}{c|}{W$\to$ E} & \multicolumn{2}{c|}{E$\to$ W} & \multicolumn{2}{c|}{W$\to$ E} & \multicolumn{2}{c|}{E$\to$ W} & \multicolumn{2}{c|}{W$\to$ E} & \multicolumn{2}{c|}{E$\to$ W} & \multicolumn{2}{c|}{W$\to$ E} & \multicolumn{2}{c|}{E$\to$ W} & \multicolumn{2}{c|}{W$\to$ E} & \multicolumn{2}{c|}{E$\to$ W} \\
\cline{2-21}          & N     & F     & N     & F     & N     & F     & N     & F     & N     & F     & N     & F     & N     & F     & N     & F     & N     & F     & N     & F \\
    \hline
    2003-2004 & 2338  & 0.36  & 0     & 0.000  & 5105  & 0.44  & 0     & 0.000  & 4425  & 0.41  & 546   & 0.05  & 3700  & 0.40  & 147   & 0.02  & 2340  & 0.46  & 3     & 0.001  \\
    \hline
    2004-2005 & 2309  & 0.36  & 0     & 0.000  & 4836  & 0.44  & 0     & 0.000  & 4013  & 0.41  & 406   & 0.04  & 3650  & 0.41  & 179   & 0.02  & 2188  & 0.44  & 13    & 0.003  \\
    \hline
    2005-2006 & 1851  & 0.34  & 0     & 0.000  & 4173  & 0.40  & 0     & 0.000  & 4271  & 0.42  & 380   & 0.04  & 3480  & 0.42  & 103   & 0.01  & 2080  & 0.43  & 46    & 0.009  \\
    \hline
    2006-2007 & 2520  & 0.39  & 0     & 0.000  & 5522  & 0.46  & 1     & 0.000  & 4720  & 0.42  & 435   & 0.04  & 3513  & 0.42  & 52    & 0.01  & 2094  & 0.42  & 87    & 0.018  \\
    \hline
    2007-2008 & 2679  & 0.39  & 0     & 0.000  & 6265  & 0.49  & 1     & 0.000  & 4804  & 0.41  & 394   & 0.03  & 4079  & 0.41  & 82    & 0.01  & 1839  & 0.40  & 31    & 0.007  \\
    \hline
    2008-2009 & 2655  & 0.38  & 0     & 0.000  & 6623  & 0.49  & 0     & 0.000  & 5597  & 0.43  & 453   & 0.04  & 4269  & 0.39  & 292   & 0.03  & 2360  & 0.44  & 12    & 0.002  \\
    \hline
    2009-2010 & 3068  & 0.43  & 0     & 0.000  & 6271  & 0.49  & 0     & 0.000  & 4329  & 0.43  & 281   & 0.03  & 3911  & 0.40  & 149   & 0.02  & 2761  & 0.48  & 1     & 0.000  \\
    \hline
    2010-2011 & 3949  & 0.48  & 0     & 0.000  & 5627  & 0.46  & 0     & 0.000  & 3963  & 0.43  & 205   & 0.02  & 3854  & 0.41  & 143   & 0.02  & 2738  & 0.48  & 1     & 0.000  \\
    \hline
    2011-2012 & 2307  & 0.38  & 1     & 0.000  & 6348  & 0.49  & 0     & 0.000  & 4419  & 0.42  & 305   & 0.03  & 2994  & 0.37  & 119   & 0.01  & 2607  & 0.47  & 11    & 0.002  \\
    \hline
    2012-2013 & 2418  & 0.36  & 16    & 0.002  & 6832  & 0.49  & 0     & 0.000  & 4256  & 0.41  & 362   & 0.03  & 3077  & 0.39  & 175   & 0.02  & 2554  & 0.46  & 0     & 0.000  \\
    \hline
    2013-2014 & 3341  & 0.40  & 38    & 0.005  & 6833  & 0.49  & 0     & 0.000  & 4155  & 0.40  & 419   & 0.04  & 3773  & 0.39  & 240   & 0.03  & 3091  & 0.44  & 21    & 0.003  \\
    \hline
    2014-2015 & 2322  & 0.38  & 9     & 0.001  & 6210  & 0.48  & 0     & 0.000  & 4302  & 0.39  & 576   & 0.05  & 3771  & 0.41  & 95    & 0.01  & 2477  & 0.45  & 44    & 0.008  \\
    \hline
    2015-2016 & 2568  & 0.41  & 5     & 0.001  & 5117  & 0.44  & 0     & 0.000  & 4156  & 0.40  & 544   & 0.05  & 3576  & 0.42  & 128   & 0.02  & 2476  & 0.45  & 40    & 0.007  \\
    \hline
    2016-2017 & 3095  & 0.44  & 16    & 0.002  & 5642  & 0.44  & 0     & 0.000  & 4417  & 0.39  & 505   & 0.04  & 4007  & 0.42  & 193   & 0.02  & 2300  & 0.44  & 48    & 0.009  \\
    \hline
    2017-2018 & 2450  & 0.39  & 10    & 0.002  & 5198  & 0.42  & 0     & 0.000  & 4836  & 0.38  & 826   & 0.06  & 3633  & 0.40  & 292   & 0.03  & 2339  & 0.45  & 54    & 0.010  \\
    \hline
    2018-2019 & 1320  & 0.25  & 0     & 0.000  & 6270  & 0.46  & 3     & 0.000  & 4449  & 0.41  & 321   & 0.03  & 3794  & 0.39  & 281   & 0.03  & 2118  & 0.42  & 3     & 0.001  \\
    \hline
  
\end{tabular}%
\label{tab:addlabel}
\end{table}

\begin{table}
\scriptsize
  \centering
\caption{In EU, the number of significant links (N) and the fraction of links (F) with time lag $\tau^{\ast}=\pm 1$ from West to East (W $\to$ E, $\tau^{\ast}=1$) and from East to West (E $\to$ W, $\tau^{\ast}=-1$) at $dis>500km$.}
\begin{tabular}{|c|c|c|c|c|c|c|c|c|c|c|c|c|c|c|c|c|c|c|c|c|}

    \hline
    \multirow{3}[6]{*}{} & \multicolumn{4}{c|}{Carbon flux} & \multicolumn{4}{c|}{Geopotential height} & \multicolumn{4}{c|}{Temperature} & \multicolumn{4}{c|}{Wind speed} & \multicolumn{4}{c|}{Precipitation} \\
\cline{2-21}          & \multicolumn{2}{c|}{W$\to$ E} & \multicolumn{2}{c|}{E$\to$ W} & \multicolumn{2}{c|}{W$\to$ E} & \multicolumn{2}{c|}{E$\to$ W} & \multicolumn{2}{c|}{W$\to$ E} & \multicolumn{2}{c|}{E$\to$ W} & \multicolumn{2}{c|}{W$\to$ E} & \multicolumn{2}{c|}{E$\to$ W} & \multicolumn{2}{c|}{W$\to$ E} & \multicolumn{2}{c|}{E$\to$ W} \\
\cline{2-21}          & N     & F     & N     & F     & N     & F     & N     & F     & N     & F     & N     & F     & N     & F     & N     & F     & N     & F     & N     & F \\
    \hline
    2003-2004 & 2894  & 0.25  & 0     & 0.000  & 8762  & 0.36  & 735   & 0.03  & 9638  & 0.41  & 777   & 0.03  & 9289  & 0.40  & 282   & 0.01  & 3373  & 0.38  & 1     & 0.000  \\
    \hline
    2004-2005 & 1919  & 0.19  & 9     & 0.001  & 9473  & 0.40  & 619   & 0.03  & 9921  & 0.40  & 732   & 0.03  & 10468 & 0.42  & 278   & 0.01  & 3712  & 0.38  & 28    & 0.003  \\
    \hline
    2005-2006 & 2628  & 0.23  & 4     & 0.000  & 10466 & 0.41  & 516   & 0.02  & 8746  & 0.42  & 623   & 0.03  & 9149  & 0.40  & 164   & 0.01  & 3742  & 0.40  & 8     & 0.001  \\
    \hline
    2006-2007 & 3272  & 0.29  & 3     & 0.000  & 8483  & 0.36  & 369   & 0.02  & 10040 & 0.42  & 779   & 0.03  & 7806  & 0.37  & 241   & 0.01  & 3735  & 0.40  & 4     & 0.000  \\
    \hline
    2007-2008 & 2340  & 0.26  & 16    & 0.002  & 3161  & 0.23  & 13    & 0.00  & 10868 & 0.42  & 732   & 0.03  & 8475  & 0.39  & 113   & 0.01  & 3807  & 0.39  & 0     & 0.000  \\
    \hline
    2008-2009 & 1352  & 0.18  & 1     & 0.000  & 4841  & 0.28  & 10    & 0.00  & 9654  & 0.43  & 608   & 0.03  & 7344  & 0.37  & 150   & 0.01  & 3992  & 0.41  & 0     & 0.000  \\
    \hline
    2009-2010 & 1891  & 0.24  & 0     & 0.000  & 5738  & 0.26  & 378   & 0.02  & 10467 & 0.41  & 472   & 0.02  & 11537 & 0.43  & 412   & 0.02  & 4127  & 0.42  & 0     & 0.000  \\
    \hline
    2010-2011 & 2030  & 0.20  & 1     & 0.000  & 4356  & 0.21  & 32    & 0.00  & 8738  & 0.41  & 364   & 0.02  & 9065  & 0.41  & 261   & 0.01  & 3685  & 0.39  & 3     & 0.000  \\
    \hline
    2011-2012 & 1182  & 0.17  & 0     & 0.000  & 5743  & 0.27  & 297   & 0.01  & 9626  & 0.43  & 482   & 0.02  & 8783  & 0.43  & 224   & 0.01  & 3439  & 0.37  & 5     & 0.001  \\
    \hline
    2012-2013 & 934   & 0.14  & 0     & 0.000  & 8807  & 0.34  & 803   & 0.03  & 9916  & 0.42  & 510   & 0.02  & 10407 & 0.42  & 553   & 0.02  & 3345  & 0.37  & 3     & 0.000  \\
    \hline
    2013-2014 & 860   & 0.13  & 0     & 0.000  & 8986  & 0.32  & 579   & 0.02  & 10344 & 0.41  & 757   & 0.03  & 10310 & 0.41  & 325   & 0.01  & 3873  & 0.39  & 18    & 0.002  \\
    \hline
    2014-2015 & 1046  & 0.16  & 0     & 0.000  & 10245 & 0.40  & 89    & 0.00  & 10383 & 0.38  & 964   & 0.03  & 9919  & 0.43  & 212   & 0.01  & 3669  & 0.39  & 2     & 0.000  \\
    \hline
    2015-2016 & 1510  & 0.21  & 0     & 0.000  & 9155  & 0.42  & 116   & 0.01  & 10329 & 0.40  & 673   & 0.03  & 9640  & 0.43  & 336   & 0.01  & 3543  & 0.38  & 7     & 0.001  \\
    \hline
    2016-2017 & 2223  & 0.26  & 3     & 0.000  & 7077  & 0.37  & 74    & 0.00  & 10518 & 0.41  & 743   & 0.03  & 7853  & 0.38  & 238   & 0.01  & 2877  & 0.34  & 1     & 0.000  \\
    \hline
    2017-2018 & 2979  & 0.29  & 8     & 0.001  & 10786 & 0.44  & 381   & 0.02  & 10226 & 0.42  & 671   & 0.03  & 11102 & 0.42  & 412   & 0.02  & 4133  & 0.42  & 3     & 0.000  \\
    \hline
    2018-2019 & 441   & 0.05  & 0     & 0.000  & 11021 & 0.40  & 516   & 0.02  & 11357 & 0.41  & 820   & 0.03  & 8477  & 0.39  & 192   & 0.01  & 3988  & 0.40  & 6     & 0.001  \\
    \hline
  
\end{tabular}%
\label{tab:addlabel}
\end{table}

\begin{table}
\scriptsize
  \centering
  \caption{In CHN, the number of significant links (N) and the fraction of links (F) with time lag $\tau^{\ast}=\pm 1$ from West to East (W $\to$ E, $\tau^{\ast}=1$) and from East to West (E $\to$ W, $\tau^{\ast}=-1$) at $dis>1000km$.}
\begin{tabular}{|c|c|c|c|c|c|c|c|c|c|c|c|c|c|c|c|c|c|c|c|c|}

    \hline
    \multirow{3}[6]{*}{} & \multicolumn{4}{c|}{Carbon flux} & \multicolumn{4}{c|}{Geopotential height} & \multicolumn{4}{c|}{Temperature} & \multicolumn{4}{c|}{Wind speed} & \multicolumn{4}{c|}{Precipitation} \\
\cline{2-21}          & \multicolumn{2}{c|}{W$\to$ E} & \multicolumn{2}{c|}{E$\to$ W} & \multicolumn{2}{c|}{W$\to$ E} & \multicolumn{2}{c|}{E$\to$ W} & \multicolumn{2}{c|}{W$\to$ E} & \multicolumn{2}{c|}{E$\to$ W} & \multicolumn{2}{c|}{W$\to$ E} & \multicolumn{2}{c|}{E$\to$ W} & \multicolumn{2}{c|}{W$\to$ E} & \multicolumn{2}{c|}{E$\to$ W} \\
\cline{2-21}          & N     & F     & N     & F     & N     & F     & N     & F     & N     & F     & N     & F     & N     & F     & N     & F     & N     & F     & N     & F \\
    \hline
    2003-2004 & 299   & 0.06  & 0     & 0.000  & 1191  & 0.07  & 1952  & 0.12  & 2463  & 0.18  & 1135  & 0.08  & 2818  & 0.26  & 372   & 0.03  & 1549  & 0.24  & 0     & 0.000  \\
    \hline
    2004-2005 & 21    & 0.01  & 0     & 0.000  & 1151  & 0.08  & 1002  & 0.07  & 2459  & 0.19  & 1218  & 0.09  & 2213  & 0.24  & 196   & 0.02  & 1568  & 0.24  & 10    & 0.002  \\
    \hline
    2005-2006 & 270   & 0.05  & 0     & 0.000  & 670   & 0.06  & 326   & 0.03  & 2846  & 0.20  & 1442  & 0.10  & 2484  & 0.25  & 312   & 0.03  & 1354  & 0.22  & 4     & 0.001  \\
    \hline
    2006-2007 & 474   & 0.08  & 3     & 0.000  & 1434  & 0.11  & 84    & 0.01  & 2613  & 0.21  & 1040  & 0.09  & 2466  & 0.26  & 319   & 0.03  & 1060  & 0.19  & 1     & 0.000  \\
    \hline
    2007-2008 & 449   & 0.08  & 4     & 0.001  & 1138  & 0.10  & 1     & 0.00  & 2222  & 0.18  & 771   & 0.06  & 2353  & 0.24  & 396   & 0.04  & 1229  & 0.19  & 12    & 0.002  \\
    \hline
    2008-2009 & 447   & 0.08  & 2     & 0.000  & 1074  & 0.10  & 0     & 0.00  & 2174  & 0.20  & 574   & 0.05  & 2483  & 0.26  & 273   & 0.03  & 1494  & 0.22  & 40    & 0.006  \\
    \hline
    2009-2010 & 492   & 0.08  & 0     & 0.000  & 2105  & 0.14  & 383   & 0.03  & 2720  & 0.22  & 931   & 0.07  & 2517  & 0.26  & 288   & 0.03  & 1549  & 0.24  & 41    & 0.006  \\
    \hline
    2010-2011 & 290   & 0.06  & 0     & 0.000  & 1680  & 0.12  & 513   & 0.04  & 2475  & 0.19  & 886   & 0.07  & 2301  & 0.24  & 295   & 0.03  & 1403  & 0.23  & 18    & 0.003  \\
    \hline
    2011-2012 & 209   & 0.04  & 0     & 0.000  & 1620  & 0.10  & 1677  & 0.10  & 2426  & 0.19  & 1057  & 0.08  & 2064  & 0.24  & 194   & 0.02  & 1155  & 0.19  & 22    & 0.004  \\
    \hline
    2012-2013 & 438   & 0.08  & 4     & 0.001  & 1805  & 0.12  & 1070  & 0.07  & 2356  & 0.19  & 1181  & 0.10  & 2779  & 0.26  & 422   & 0.04  & 1401  & 0.21  & 48    & 0.007  \\
    \hline
    2013-2014 & 393   & 0.09  & 0     & 0.000  & 1390  & 0.09  & 264   & 0.02  & 2135  & 0.18  & 909   & 0.08  & 2932  & 0.27  & 303   & 0.03  & 1363  & 0.21  & 7     & 0.001  \\
    \hline
    2014-2015 & 590   & 0.10  & 7     & 0.001  & 1369  & 0.10  & 127   & 0.01  & 2596  & 0.21  & 913   & 0.07  & 2833  & 0.29  & 170   & 0.02  & 1292  & 0.20  & 5     & 0.001  \\
    \hline
    2015-2016 & 181   & 0.05  & 0     & 0.000  & 1879  & 0.13  & 424   & 0.03  & 2950  & 0.21  & 1066  & 0.08  & 2610  & 0.27  & 173   & 0.02  & 1260  & 0.21  & 2     & 0.000  \\
    \hline
    2016-2017 & 680   & 0.10  & 0     & 0.000  & 1301  & 0.09  & 1058  & 0.08  & 2471  & 0.19  & 922   & 0.07  & 2140  & 0.25  & 91    & 0.01  & 1420  & 0.22  & 0     & 0.000  \\
    \hline
    2017-2018 & 520   & 0.09  & 0     & 0.000  & 1215  & 0.09  & 630   & 0.04  & 2345  & 0.19  & 1026  & 0.09  & 2896  & 0.27  & 225   & 0.02  & 1321  & 0.21  & 1     & 0.000  \\
    \hline
    2018-2019 & 332   & 0.04  & 5     & 0.001  & 1562  & 0.10  & 923   & 0.06  & 2177  & 0.20  & 795   & 0.07  & 2417  & 0.25  & 255   & 0.03  & 634   & 0.13  & 0     & 0.000  \\
    \hline
  
\end{tabular}%
\label{tab:addlabel}
\end{table}

\begin{table}
\scriptsize
  \centering
\caption{In USA, the number of significant links (N) and the fraction of links (F) with time lag $\tau^{\ast}=\pm 1$ from West to East (W $\to$ E, $\tau^{\ast}=1$) and from East to West (E $\to$ W, $\tau^{\ast}=-1$) at $dis>1000km$.}
\begin{tabular}{|c|c|c|c|c|c|c|c|c|c|c|c|c|c|c|c|c|c|c|c|c|}

    \hline
    \multirow{3}[6]{*}{} & \multicolumn{4}{c|}{Carbon flux} & \multicolumn{4}{c|}{Geopotential height} & \multicolumn{4}{c|}{Temperature} & \multicolumn{4}{c|}{Wind speed} & \multicolumn{4}{c|}{Precipitation} \\
\cline{2-21}          & \multicolumn{2}{c|}{W$\to$ E} & \multicolumn{2}{c|}{E$\to$ W} & \multicolumn{2}{c|}{W$\to$ E} & \multicolumn{2}{c|}{E$\to$ W} & \multicolumn{2}{c|}{W$\to$ E} & \multicolumn{2}{c|}{E$\to$ W} & \multicolumn{2}{c|}{W$\to$ E} & \multicolumn{2}{c|}{E$\to$ W} & \multicolumn{2}{c|}{W$\to$ E} & \multicolumn{2}{c|}{E$\to$ W} \\
\cline{2-21}          & N     & F     & N     & F     & N     & F     & N     & F     & N     & F     & N     & F     & N     & F     & N     & F     & N     & F     & N     & F \\
    \hline
    2003-2004 & 1184  & 0.18  & 0     & 0.000  & 4129  & 0.35  & 0     & 0.000  & 2671  & 0.25  & 424   & 0.04  & 2502  & 0.27  & 116   & 0.01  & 981   & 0.19  & 3     & 0.001  \\
    \hline
    2004-2005 & 1113  & 0.17  & 0     & 0.000  & 3816  & 0.35  & 0     & 0.000  & 2261  & 0.23  & 300   & 0.03  & 2436  & 0.27  & 160   & 0.02  & 902   & 0.18  & 11    & 0.002  \\
    \hline
    2005-2006 & 838   & 0.16  & 0     & 0.000  & 3166  & 0.30  & 0     & 0.000  & 2512  & 0.24  & 280   & 0.03  & 2339  & 0.28  & 79    & 0.01  & 920   & 0.19  & 40    & 0.008  \\
    \hline
    2006-2007 & 1216  & 0.19  & 0     & 0.000  & 4458  & 0.37  & 1     & 0.000  & 2774  & 0.24  & 340   & 0.03  & 2262  & 0.27  & 43    & 0.01  & 903   & 0.18  & 77    & 0.016  \\
    \hline
    2007-2008 & 1464  & 0.21  & 0     & 0.000  & 5190  & 0.40  & 1     & 0.000  & 2897  & 0.25  & 307   & 0.03  & 2929  & 0.29  & 65    & 0.01  & 759   & 0.17  & 29    & 0.006  \\
    \hline
    2008-2009 & 1332  & 0.19  & 0     & 0.000  & 5521  & 0.41  & 0     & 0.000  & 3700  & 0.29  & 390   & 0.03  & 3052  & 0.28  & 271   & 0.02  & 1004  & 0.19  & 12    & 0.002  \\
    \hline
    2009-2010 & 1417  & 0.20  & 0     & 0.000  & 5029  & 0.40  & 0     & 0.000  & 2460  & 0.25  & 232   & 0.02  & 2400  & 0.25  & 126   & 0.01  & 1110  & 0.19  & 1     & 0.000  \\
    \hline
    2010-2011 & 2159  & 0.26  & 0     & 0.000  & 4589  & 0.38  & 0     & 0.000  & 2020  & 0.22  & 150   & 0.02  & 2358  & 0.25  & 136   & 0.01  & 1226  & 0.22  & 1     & 0.000  \\
    \hline
    2011-2012 & 941   & 0.15  & 1     & 0.000  & 5271  & 0.41  & 0     & 0.000  & 2483  & 0.23  & 252   & 0.02  & 1921  & 0.24  & 107   & 0.01  & 994   & 0.18  & 11    & 0.002  \\
    \hline
    2012-2013 & 1032  & 0.15  & 16    & 0.002  & 5667  & 0.41  & 0     & 0.000  & 2286  & 0.22  & 284   & 0.03  & 1981  & 0.25  & 139   & 0.02  & 1026  & 0.18  & 0     & 0.000  \\
    \hline
    2013-2014 & 1942  & 0.23  & 38    & 0.005  & 5813  & 0.42  & 0     & 0.000  & 2268  & 0.22  & 320   & 0.03  & 2543  & 0.27  & 213   & 0.02  & 1641  & 0.23  & 13    & 0.002  \\
    \hline
    2014-2015 & 1244  & 0.20  & 9     & 0.001  & 5282  & 0.41  & 0     & 0.000  & 2521  & 0.23  & 450   & 0.04  & 2585  & 0.28  & 64    & 0.01  & 1167  & 0.21  & 25    & 0.005  \\
    \hline
    2015-2016 & 1117  & 0.18  & 5     & 0.001  & 4076  & 0.35  & 0     & 0.000  & 2300  & 0.22  & 393   & 0.04  & 2319  & 0.27  & 105   & 0.01  & 1090  & 0.20  & 28    & 0.005  \\
    \hline
    2016-2017 & 1309  & 0.19  & 13    & 0.002  & 4718  & 0.37  & 0     & 0.000  & 2403  & 0.21  & 372   & 0.03  & 2657  & 0.28  & 162   & 0.02  & 924   & 0.18  & 43    & 0.008  \\
    \hline
    2017-2018 & 965   & 0.15  & 9     & 0.001  & 4519  & 0.36  & 0     & 0.000  & 2991  & 0.23  & 744   & 0.06  & 2313  & 0.26  & 251   & 0.03  & 963   & 0.18  & 40    & 0.008  \\
    \hline
    2018-2019 & 512   & 0.10  & 0     & 0.000  & 5407  & 0.40  & 3     & 0.000  & 2737  & 0.25  & 236   & 0.02  & 2495  & 0.26  & 238   & 0.02  & 1022  & 0.20  & 3     & 0.001  \\
    \hline
  
\end{tabular}%
\label{tab:addlabel}
\end{table}

\begin{table}
\scriptsize
  \centering
\caption{In EU, the number of significant links (N) and the fraction of links (F) with time lag $\tau^{\ast}=\pm 1$ from West to East (W $\to$ E, $\tau^{\ast}=1$) and from East to West (E $\to$ W, $\tau^{\ast}=-1$) at $dis>1000km$.}
\begin{tabular}{|c|c|c|c|c|c|c|c|c|c|c|c|c|c|c|c|c|c|c|c|c|}

    \hline
    \multirow{3}[6]{*}{} & \multicolumn{4}{c|}{Carbon flux} & \multicolumn{4}{c|}{Geopotential height} & \multicolumn{4}{c|}{Temperature} & \multicolumn{4}{c|}{Wind speed} & \multicolumn{4}{c|}{Precipitation} \\
\cline{2-21}          & \multicolumn{2}{c|}{W$\to$ E} & \multicolumn{2}{c|}{E$\to$ W} & \multicolumn{2}{c|}{W$\to$ E} & \multicolumn{2}{c|}{E$\to$ W} & \multicolumn{2}{c|}{W$\to$ E} & \multicolumn{2}{c|}{E$\to$ W} & \multicolumn{2}{c|}{W$\to$ E} & \multicolumn{2}{c|}{E$\to$ W} & \multicolumn{2}{c|}{W$\to$ E} & \multicolumn{2}{c|}{E$\to$ W} \\
\cline{2-21}          & N     & F     & N     & F     & N     & F     & N     & F     & N     & F     & N     & F     & N     & F     & N     & F     & N     & F     & N     & F \\
    \hline
    2003-2004 & 1120  & 0.10  & 0     & 0.000  & 5234  & 0.21  & 692   & 0.03  & 4505  & 0.19  & 518   & 0.02  & 6413  & 0.28  & 187   & 0.01  & 678   & 0.08  & 1     & 0.000  \\
    \hline
    2004-2005 & 363   & 0.04  & 7     & 0.001  & 5813  & 0.25  & 611   & 0.03  & 4649  & 0.19  & 533   & 0.02  & 7278  & 0.29  & 122   & 0.00  & 999   & 0.10  & 3     & 0.000  \\
    \hline
    2005-2006 & 637   & 0.06  & 2     & 0.000  & 7019  & 0.28  & 516   & 0.02  & 3579  & 0.17  & 374   & 0.02  & 6224  & 0.27  & 60    & 0.00  & 855   & 0.09  & 0     & 0.000  \\
    \hline
    2006-2007 & 1045  & 0.09  & 3     & 0.000  & 5585  & 0.24  & 369   & 0.02  & 4854  & 0.20  & 589   & 0.02  & 4998  & 0.24  & 154   & 0.01  & 1028  & 0.11  & 2     & 0.000  \\
    \hline
    2007-2008 & 767   & 0.09  & 11    & 0.001  & 887   & 0.06  & 13    & 0.00  & 5554  & 0.22  & 569   & 0.02  & 5634  & 0.26  & 43    & 0.00  & 1311  & 0.14  & 0     & 0.000  \\
    \hline
    2008-2009 & 427   & 0.06  & 1     & 0.000  & 1939  & 0.11  & 10    & 0.00  & 4241  & 0.19  & 447   & 0.02  & 4703  & 0.24  & 27    & 0.00  & 1046  & 0.11  & 0     & 0.000  \\
    \hline
    2009-2010 & 589   & 0.07  & 0     & 0.000  & 3329  & 0.15  & 378   & 0.02  & 5172  & 0.20  & 312   & 0.01  & 7883  & 0.30  & 212   & 0.01  & 1038  & 0.11  & 0     & 0.000  \\
    \hline
    2010-2011 & 804   & 0.08  & 0     & 0.000  & 2169  & 0.10  & 32    & 0.00  & 3814  & 0.18  & 177   & 0.01  & 5513  & 0.25  & 120   & 0.01  & 1010  & 0.11  & 0     & 0.000  \\
    \hline
    2011-2012 & 252   & 0.04  & 0     & 0.000  & 3340  & 0.16  & 297   & 0.01  & 4515  & 0.20  & 332   & 0.01  & 5213  & 0.25  & 94    & 0.00  & 976   & 0.10  & 0     & 0.000  \\
    \hline
    2012-2013 & 186   & 0.03  & 0     & 0.000  & 5549  & 0.21  & 794   & 0.03  & 4820  & 0.21  & 386   & 0.02  & 6734  & 0.27  & 317   & 0.01  & 853   & 0.09  & 0     & 0.000  \\
    \hline
    2013-2014 & 170   & 0.03  & 0     & 0.000  & 5930  & 0.21  & 579   & 0.02  & 5418  & 0.21  & 507   & 0.02  & 7063  & 0.28  & 148   & 0.01  & 877   & 0.09  & 7     & 0.001  \\
    \hline
    2014-2015 & 205   & 0.03  & 0     & 0.000  & 6511  & 0.25  & 89    & 0.00  & 5615  & 0.20  & 602   & 0.02  & 7110  & 0.31  & 82    & 0.00  & 1053  & 0.11  & 1     & 0.000  \\
    \hline
    2015-2016 & 282   & 0.04  & 0     & 0.000  & 5271  & 0.24  & 116   & 0.01  & 5144  & 0.20  & 506   & 0.02  & 6309  & 0.28  & 139   & 0.01  & 931   & 0.10  & 0     & 0.000  \\
    \hline
    2016-2017 & 390   & 0.05  & 0     & 0.000  & 3688  & 0.19  & 74    & 0.00  & 5207  & 0.20  & 563   & 0.02  & 4868  & 0.24  & 107   & 0.01  & 484   & 0.06  & 0     & 0.000  \\
    \hline
    2017-2018 & 657   & 0.06  & 5     & 0.000  & 7106  & 0.29  & 381   & 0.02  & 5025  & 0.21  & 417   & 0.02  & 7764  & 0.30  & 289   & 0.01  & 1027  & 0.10  & 0     & 0.000  \\
    \hline
    2018-2019 & 71    & 0.01  & 0     & 0.000  & 7848  & 0.29  & 516   & 0.02  & 6167  & 0.22  & 643   & 0.02  & 5473  & 0.25  & 122   & 0.01  & 930   & 0.09  & 2     & 0.000  \\
    \hline
  
\end{tabular}%
\label{tab:addlabel}
\end{table} 

\begin{center}
\includegraphics[width=8em, height=7em]{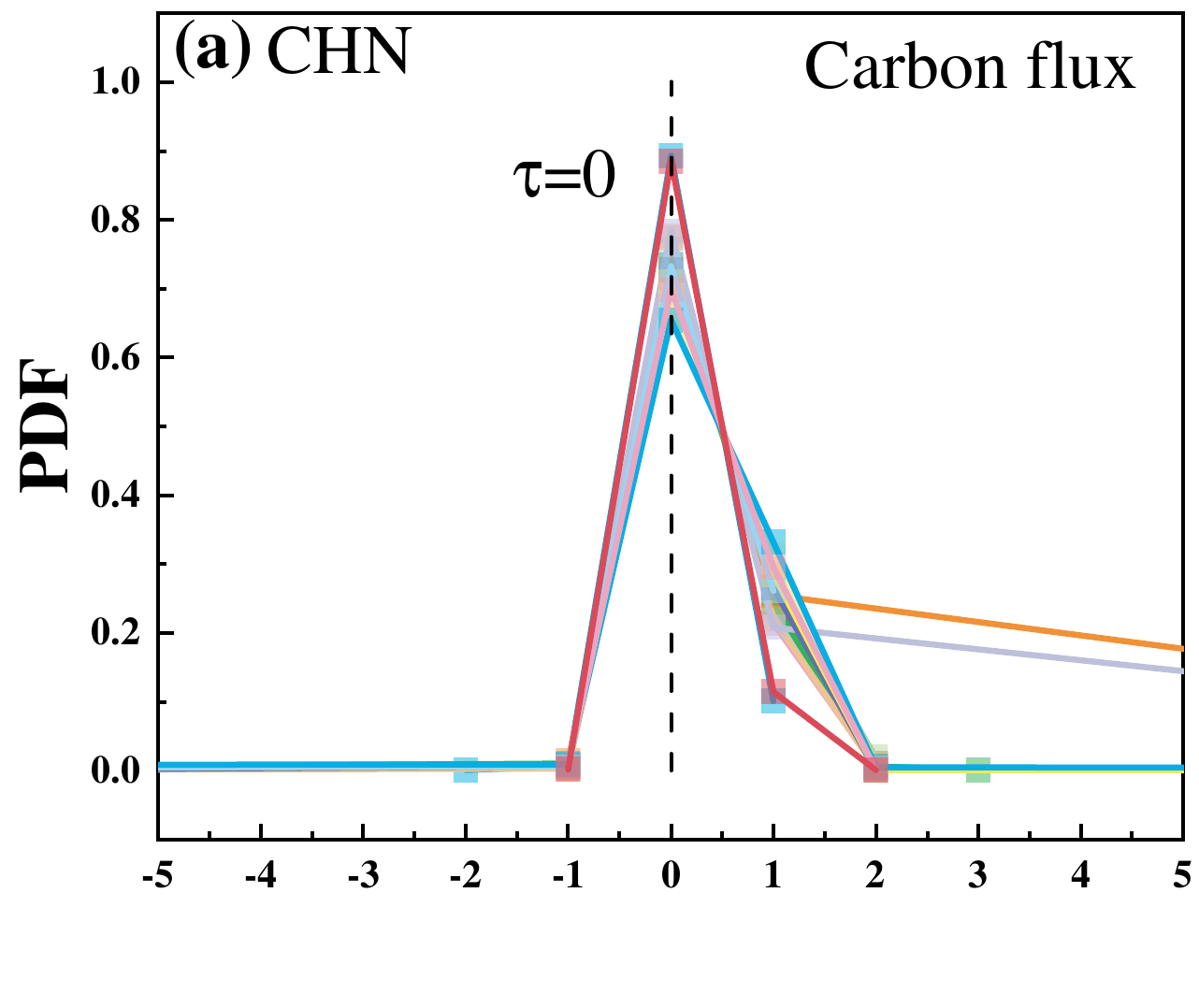}
    \includegraphics[width=8em, height=7em]{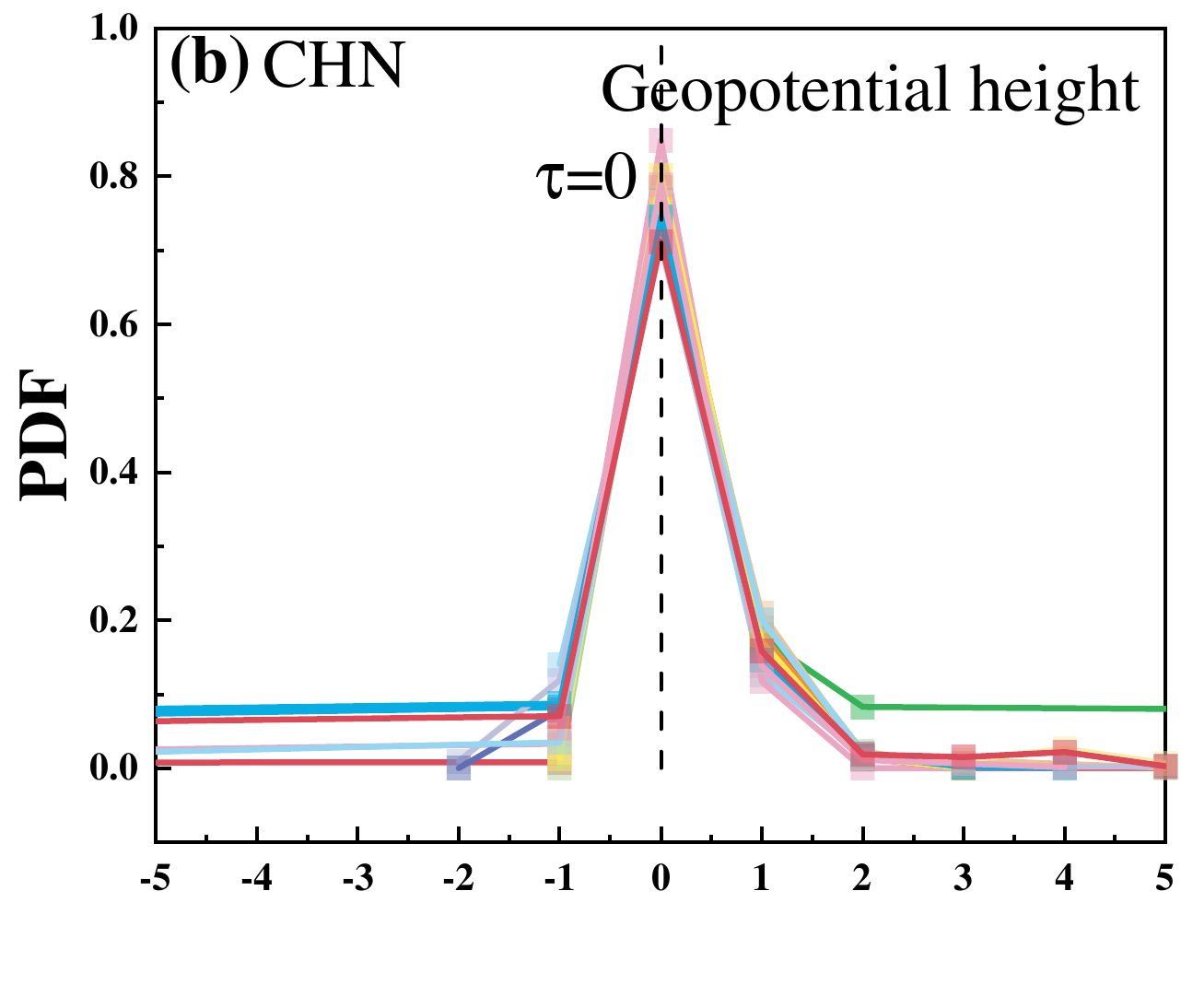}
    \includegraphics[width=8em, height=7em]{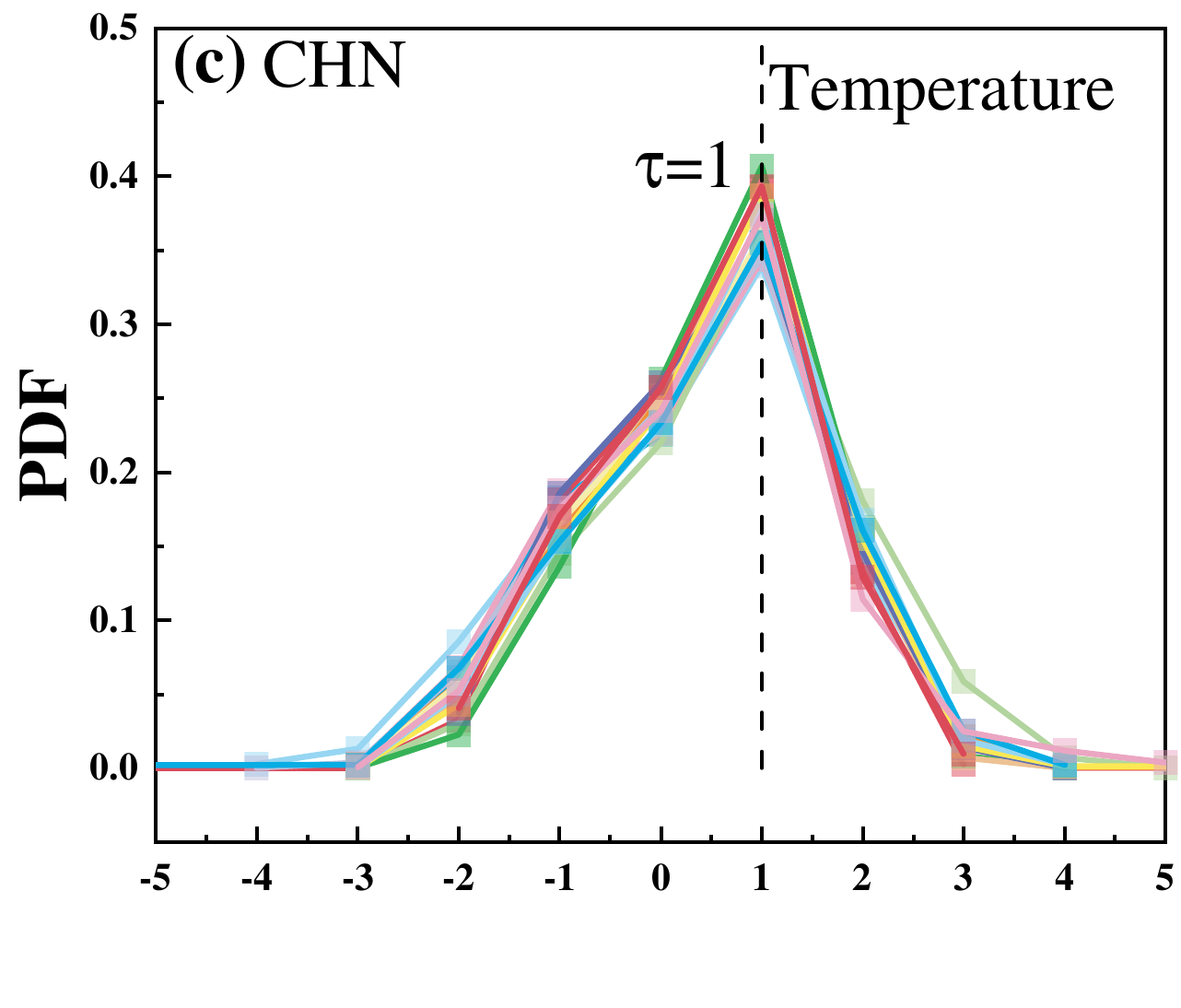}
    \includegraphics[width=8em, height=7em]{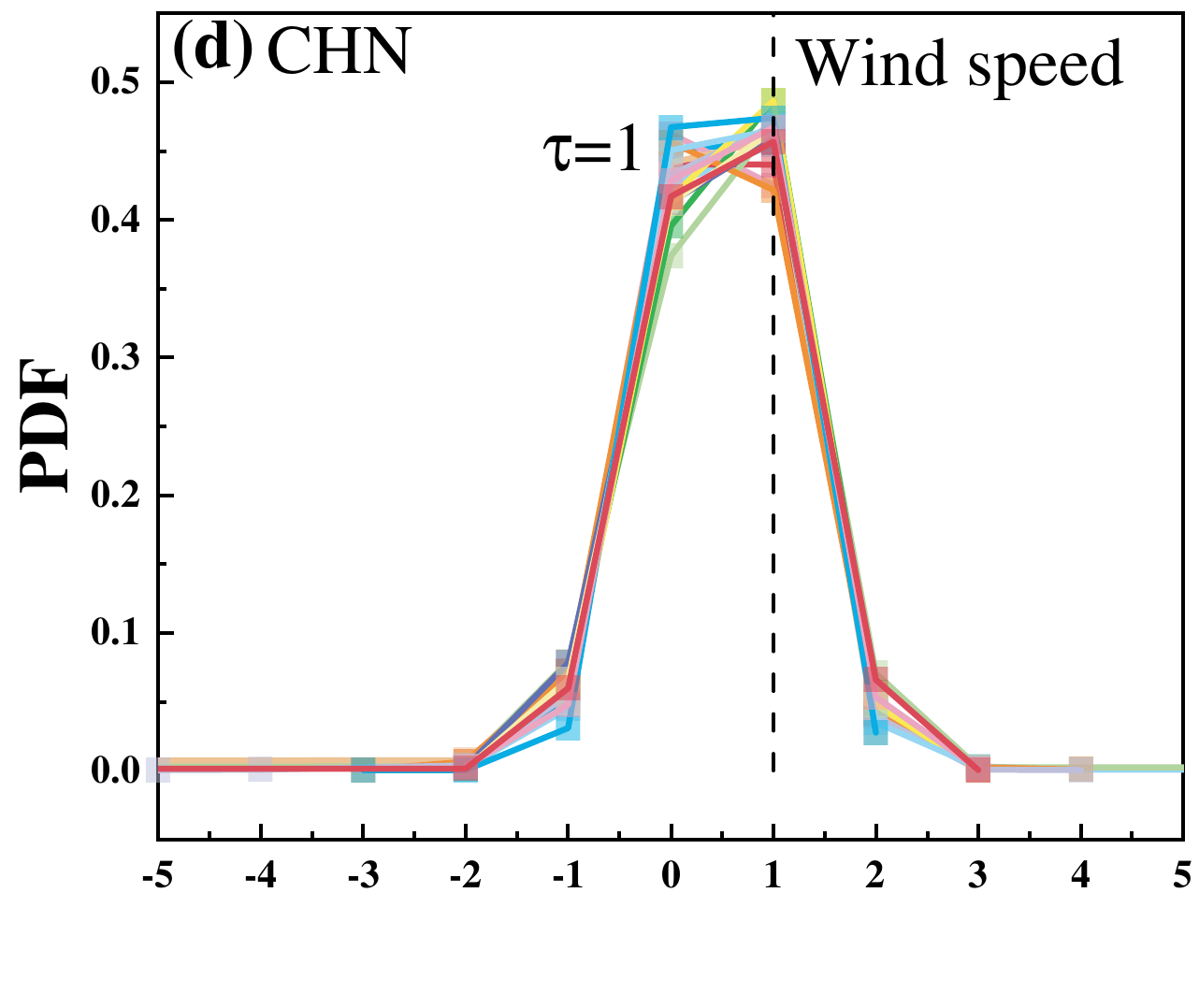}
    \includegraphics[width=8em, height=7em]{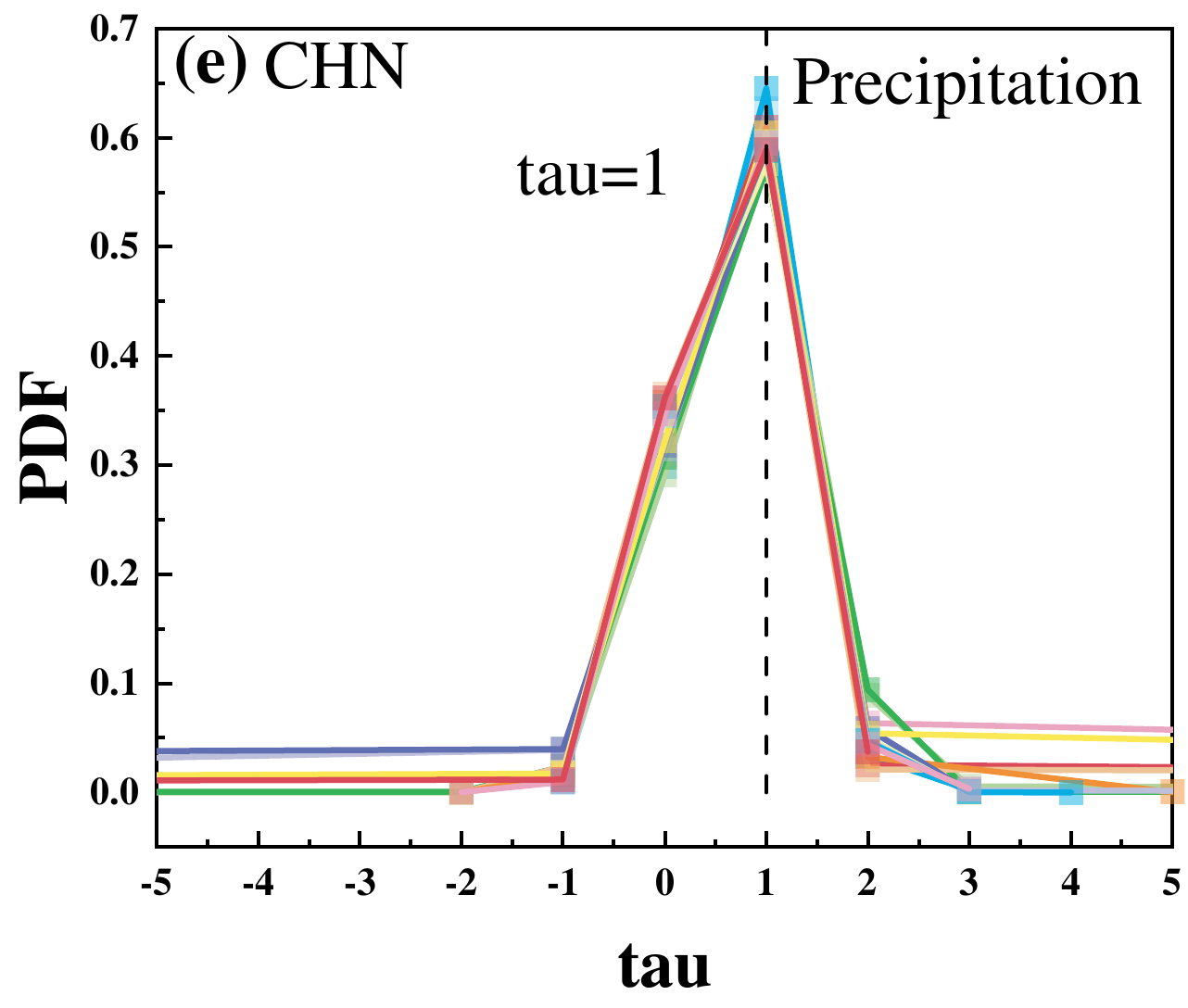}
    \includegraphics[width=8em, height=7em]{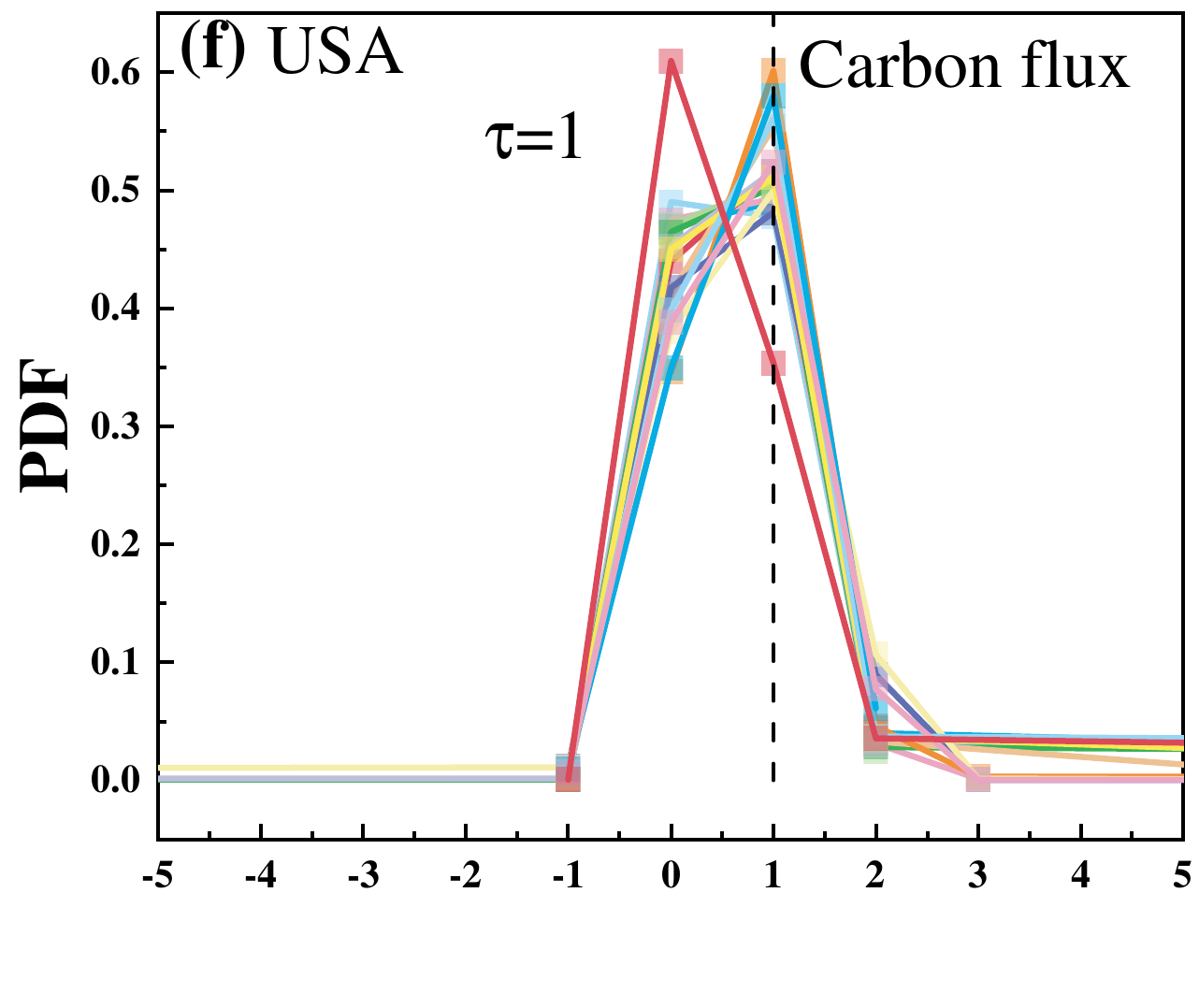}
    \includegraphics[width=8em, height=7em]{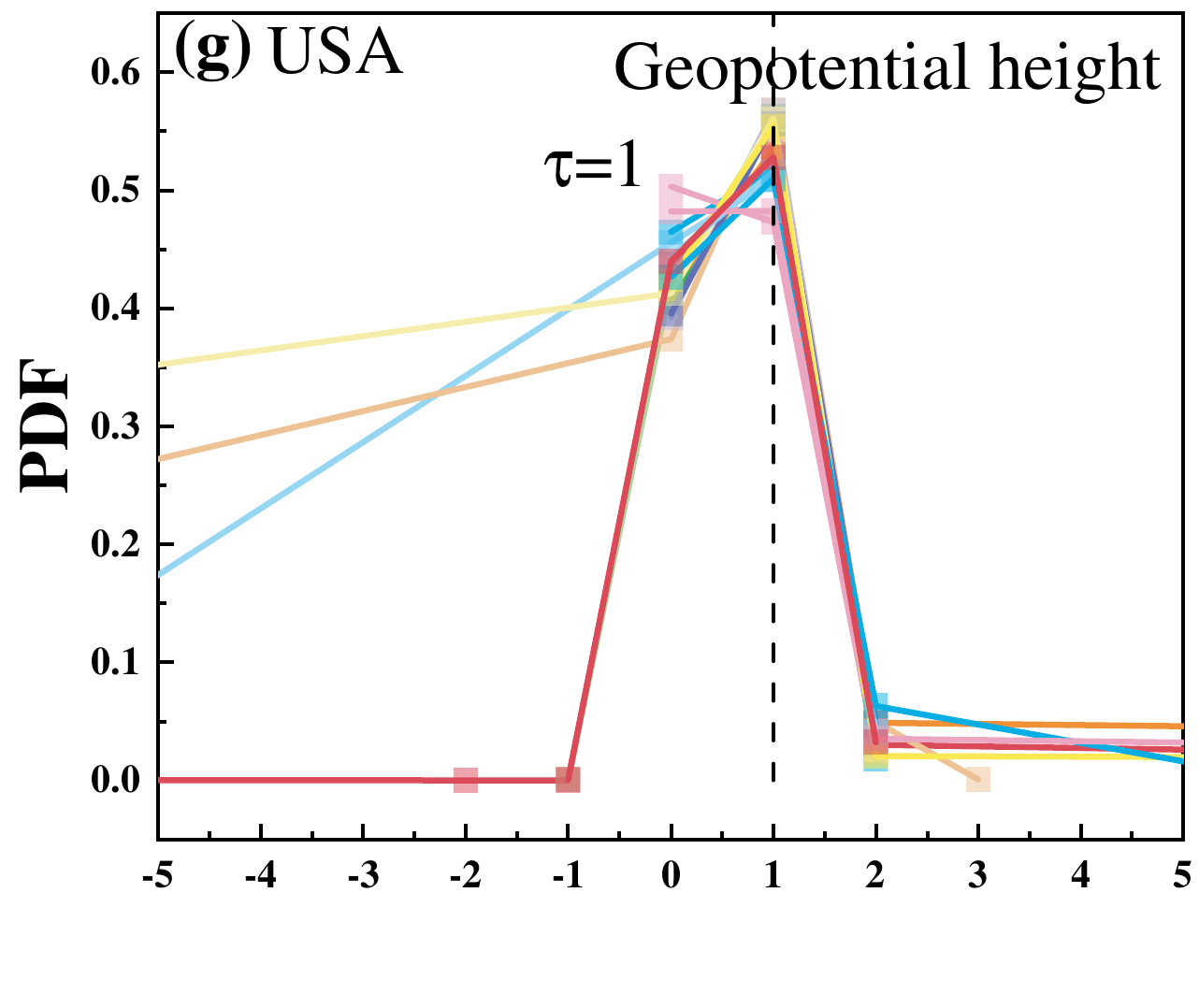}
    \includegraphics[width=8em, height=7em]{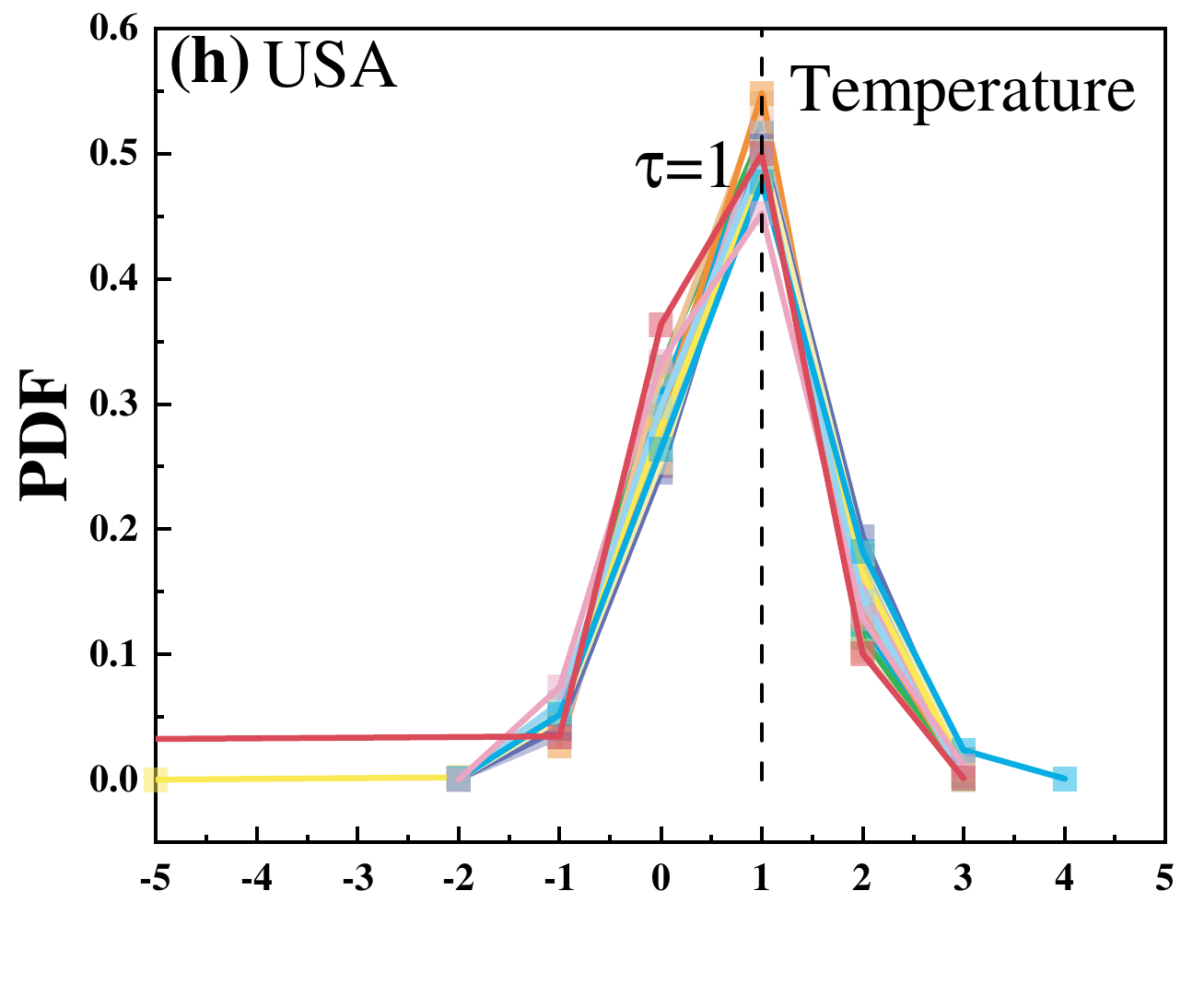}
    \includegraphics[width=8em, height=7em]{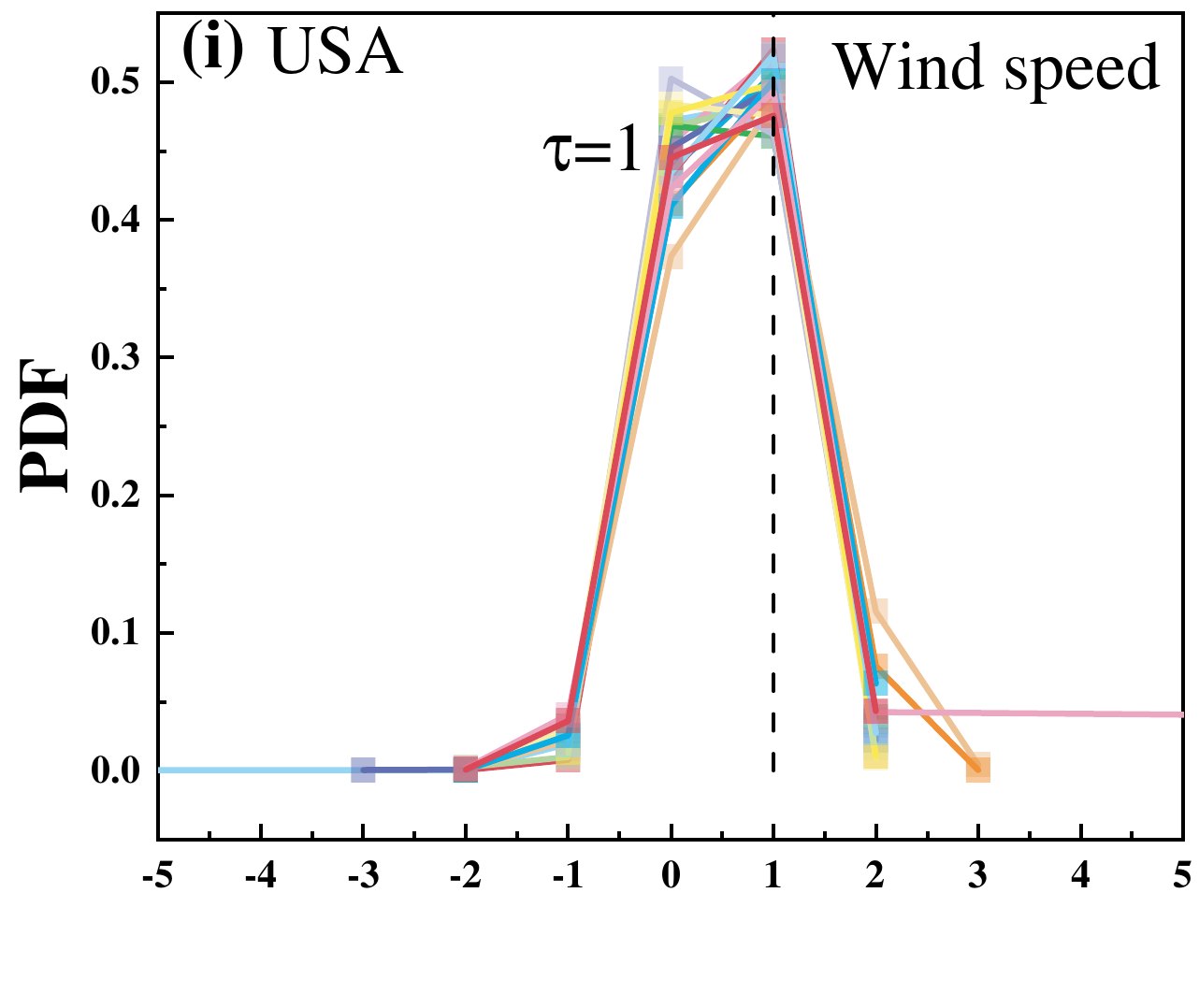}
    \includegraphics[width=8em, height=7em]{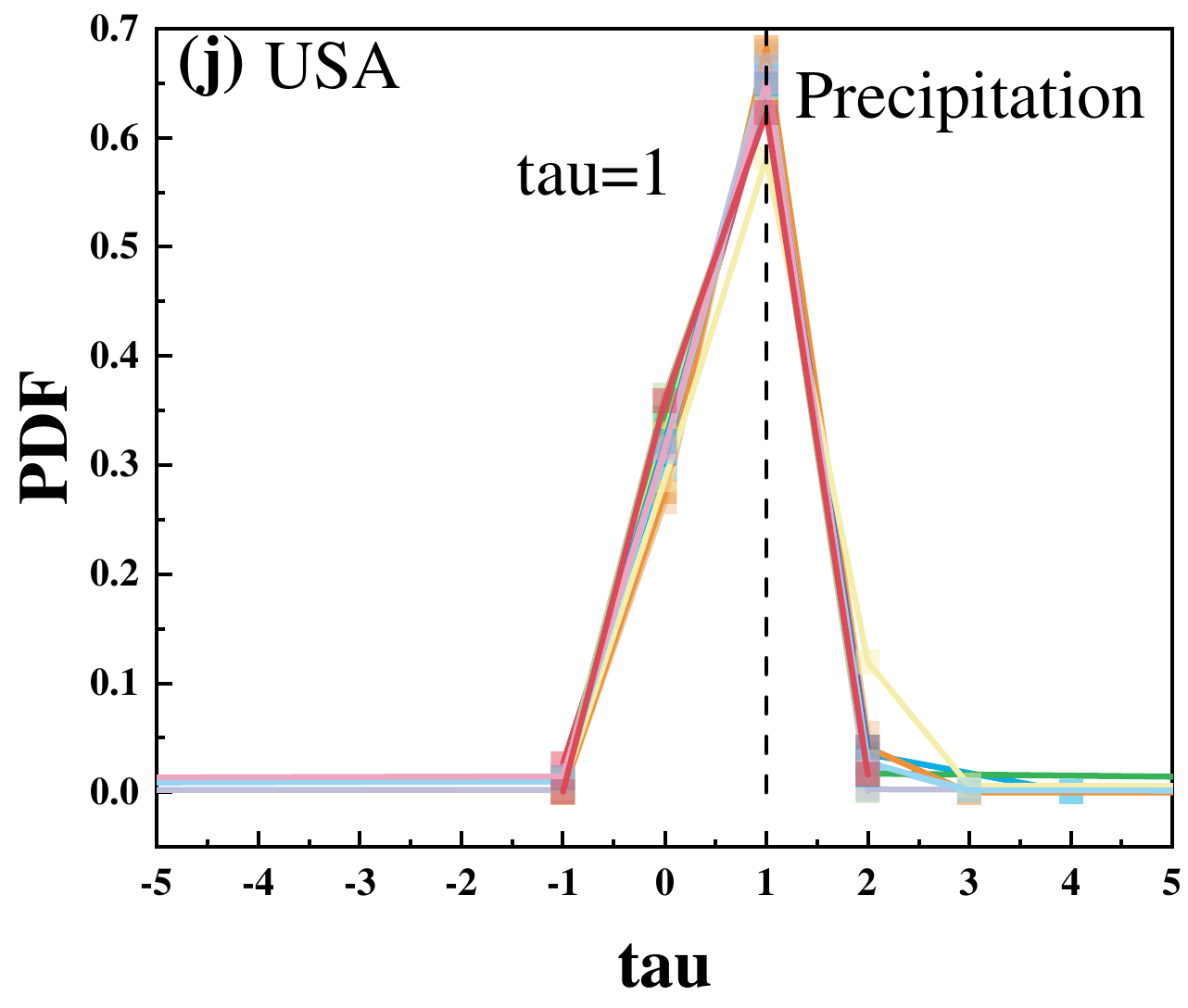}
    \includegraphics[width=8em, height=7em]{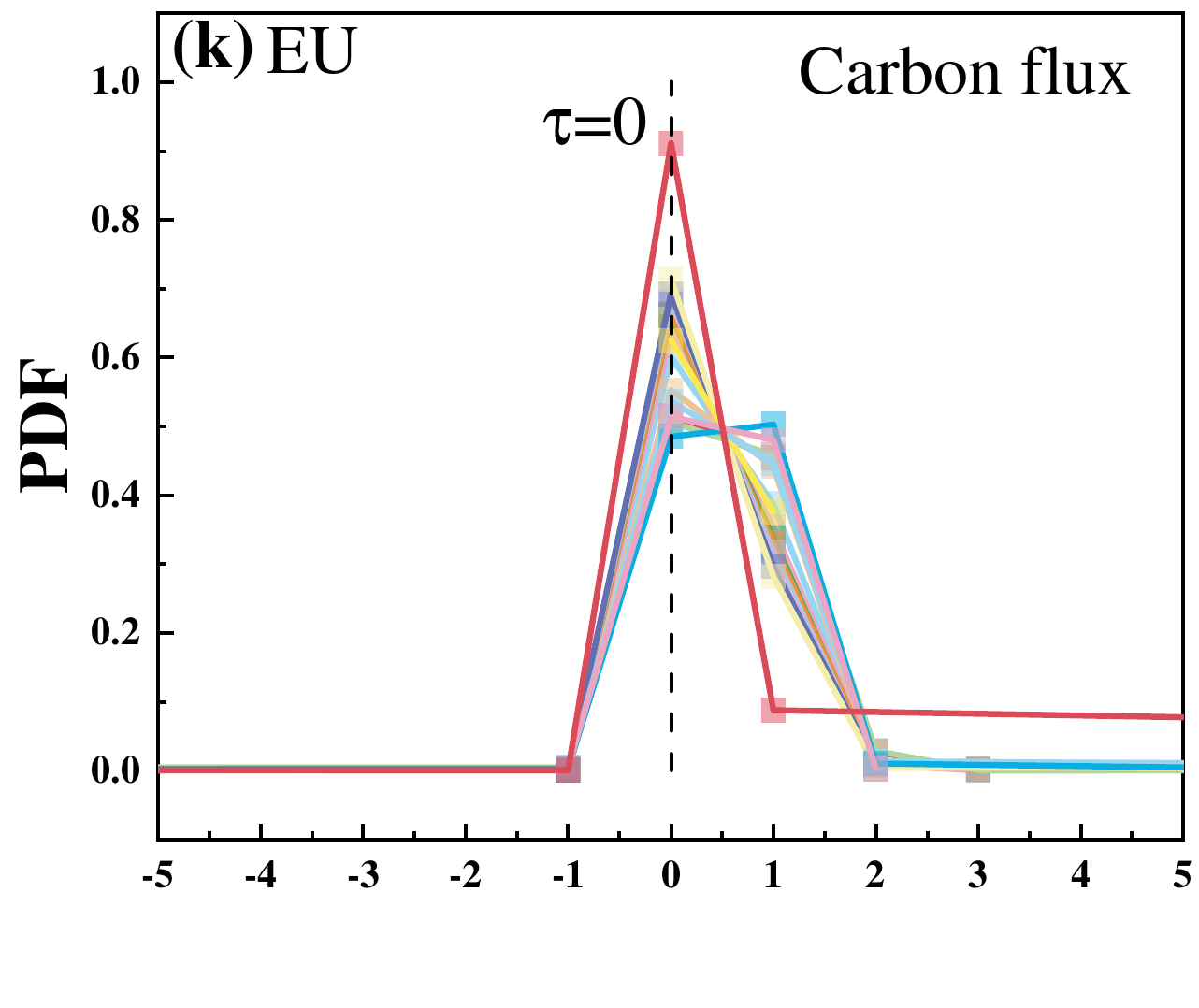}
    \includegraphics[width=8em, height=7em]{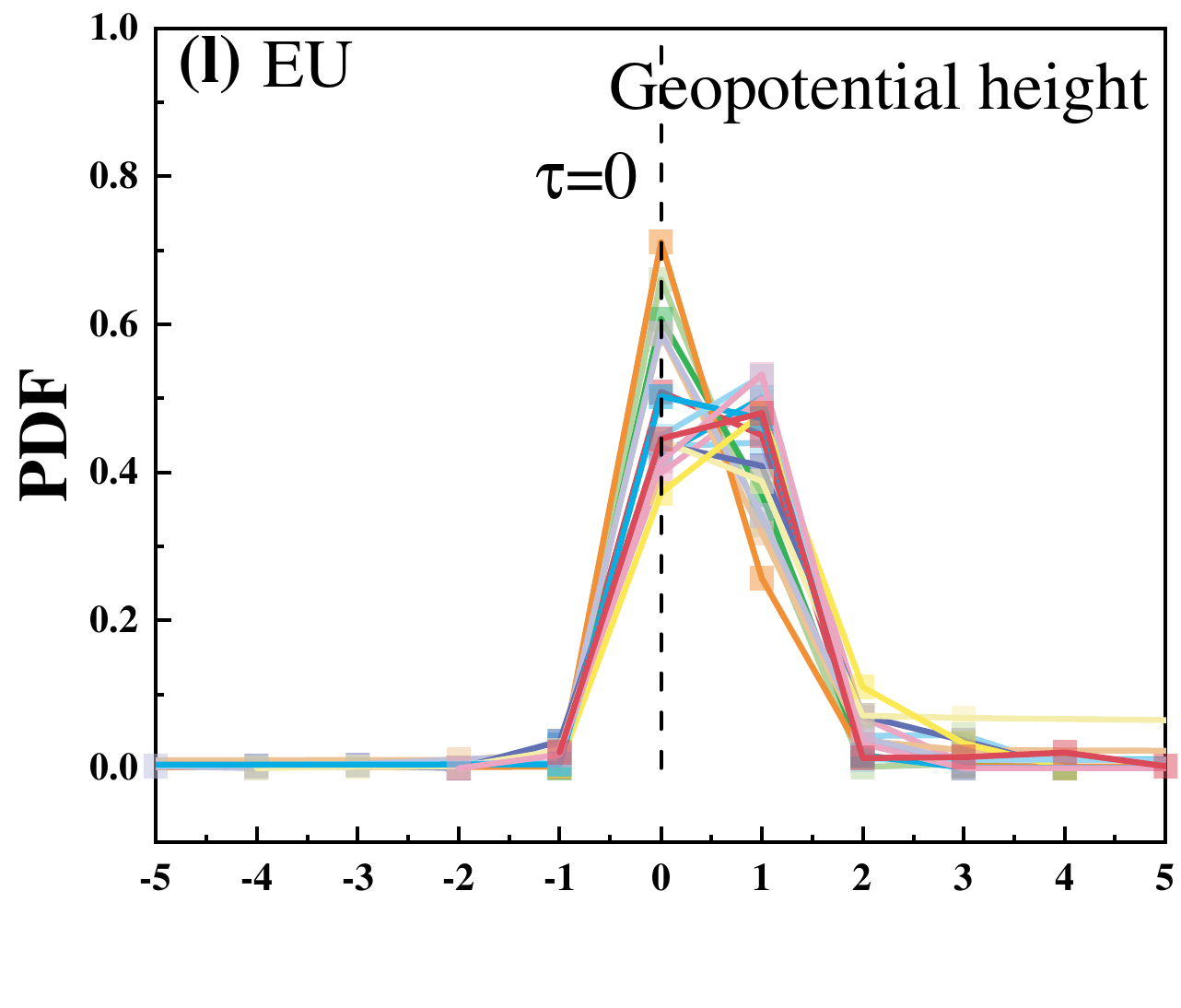}
    \includegraphics[width=8em, height=7em]{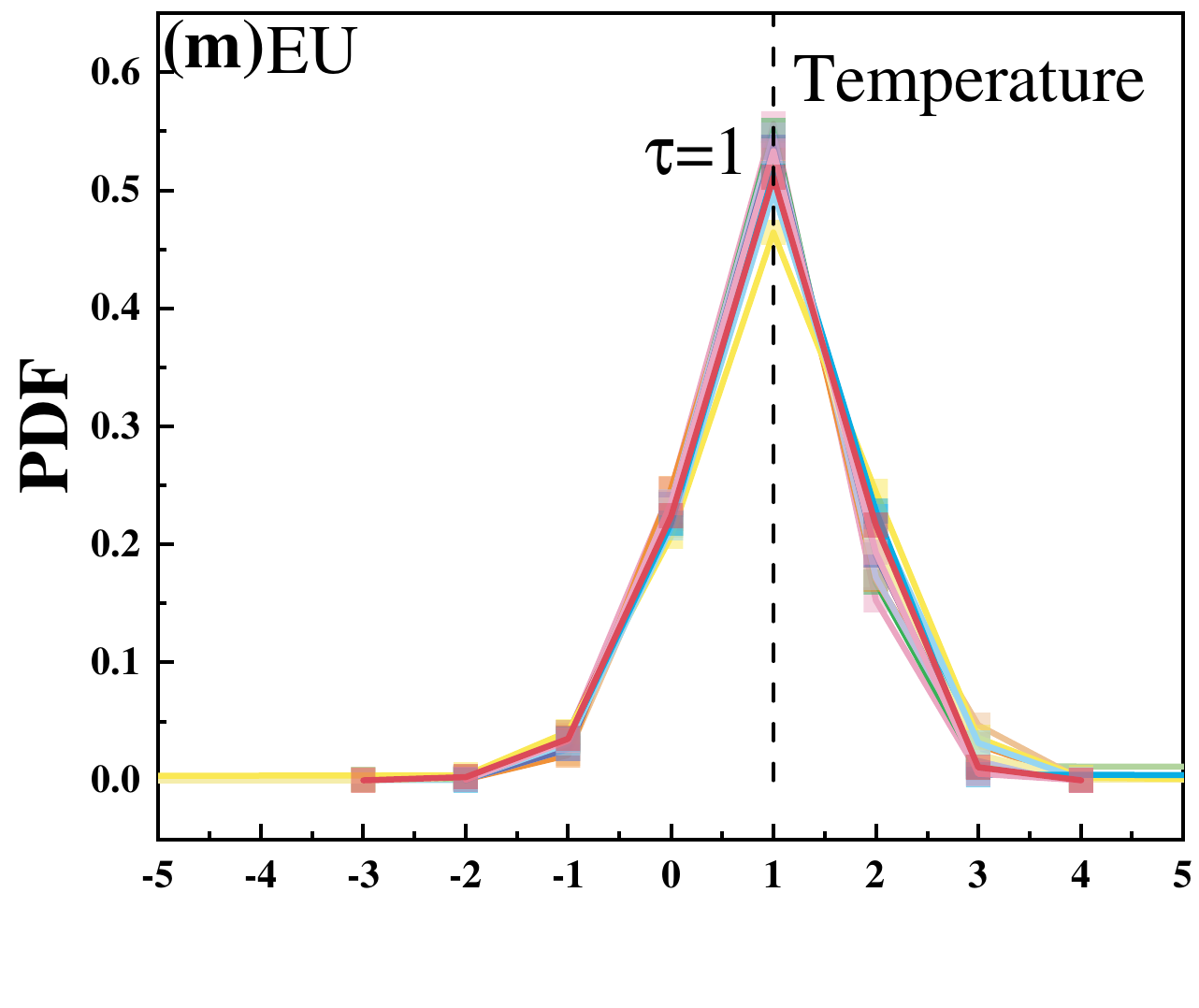}
    \includegraphics[width=8em, height=7em]{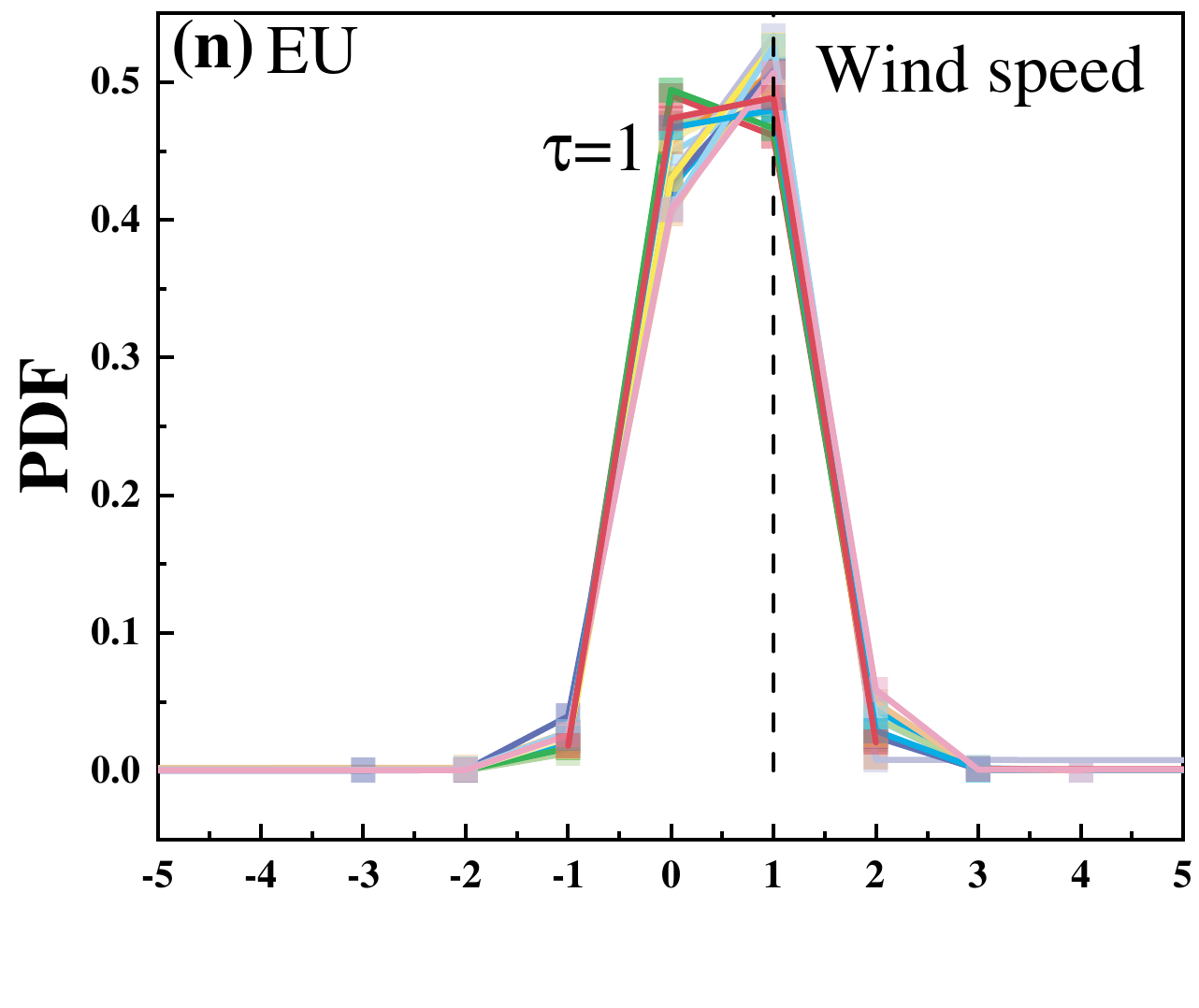}
    \includegraphics[width=8em, height=7em]{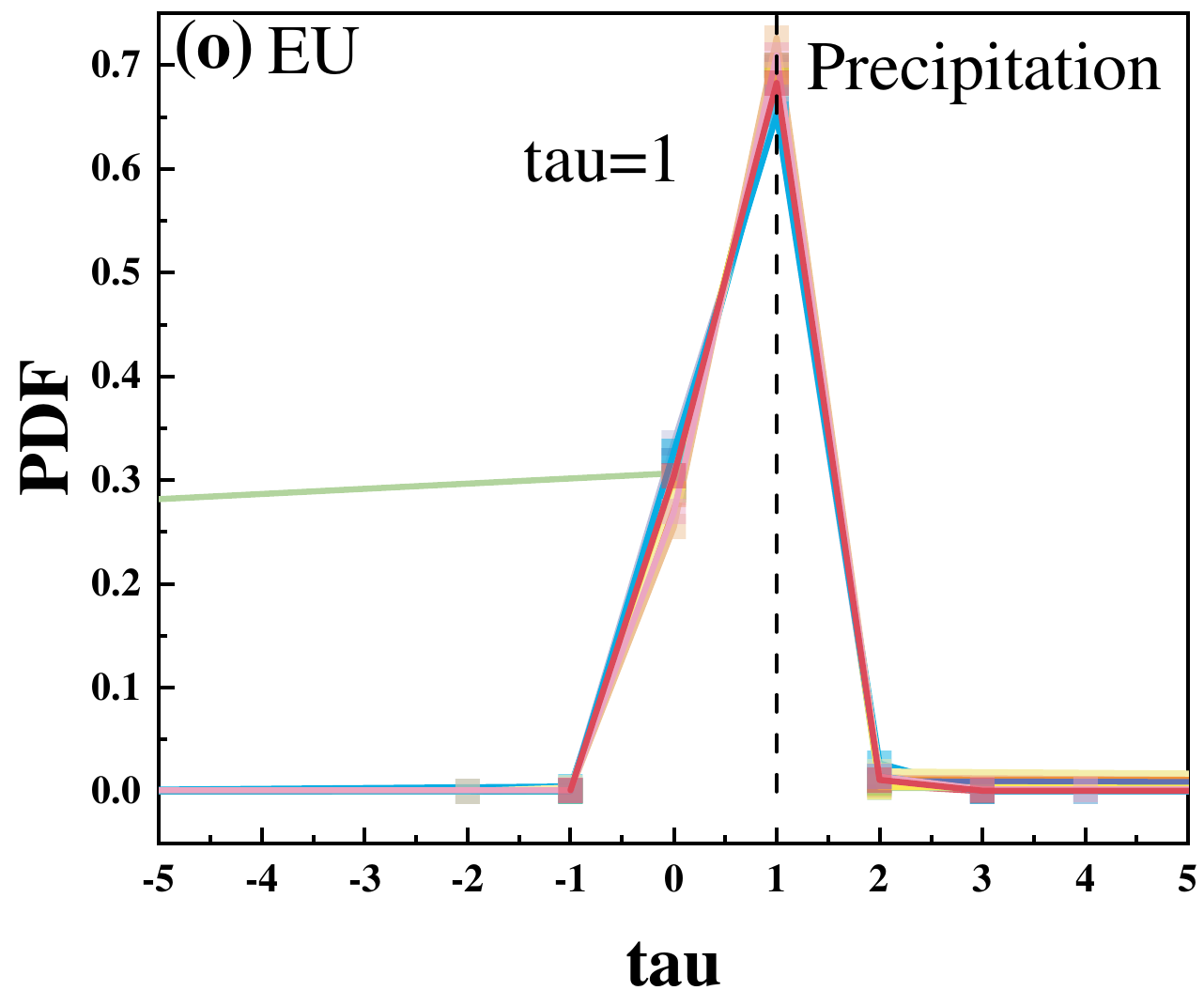}
\end{center}

\begin{center}
\noindent {\small {\bf Fig. S19} For lengths above 500$km$, probability distribution functions (PDF) of time lags in different time series. Different colors represent the time lag probability distribution of the network for different years. }
\end{center}

\begin{center}
\includegraphics[width=8em, height=7em]{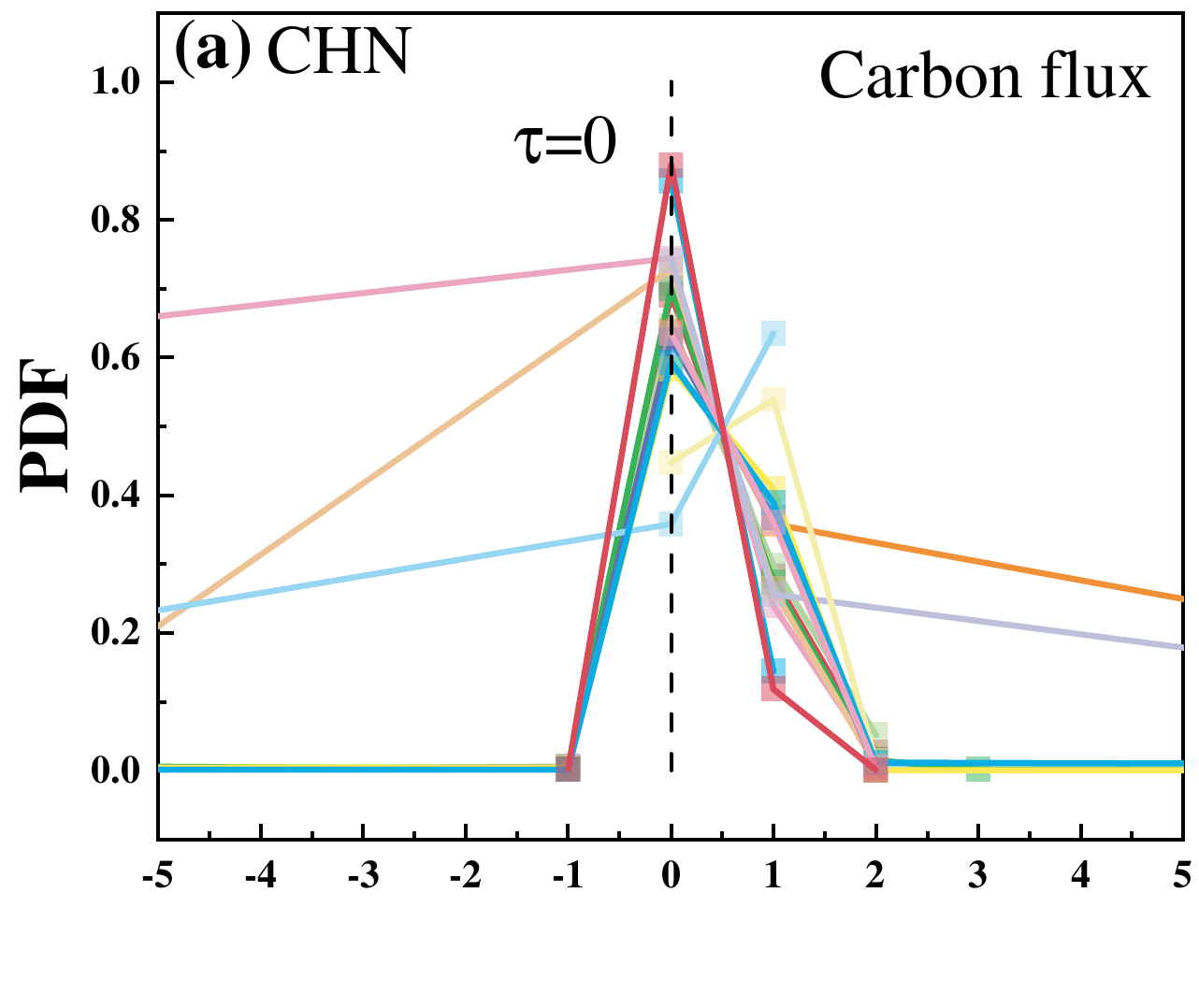}
    \includegraphics[width=8em, height=7em]{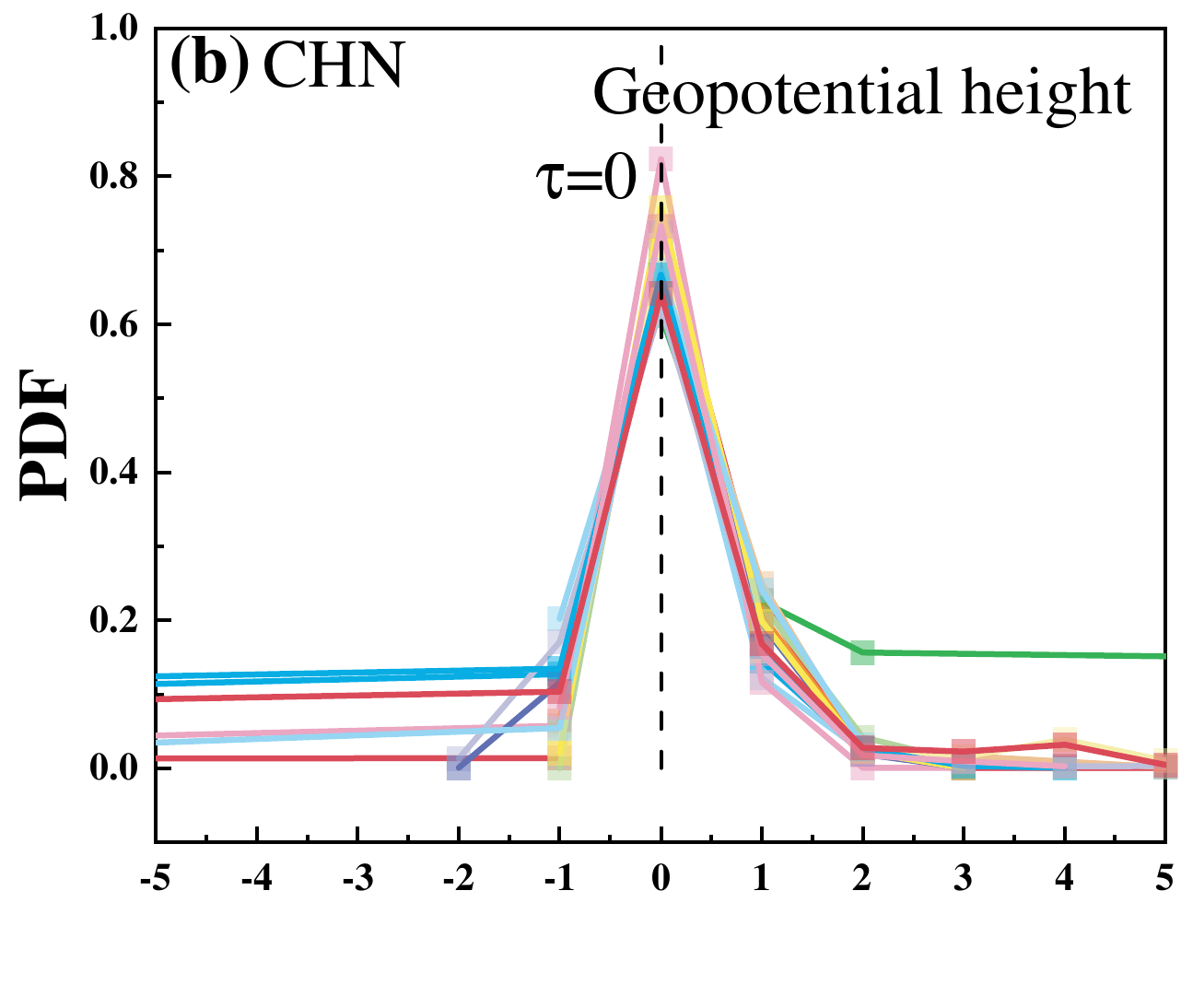}
    \includegraphics[width=8em, height=7em]{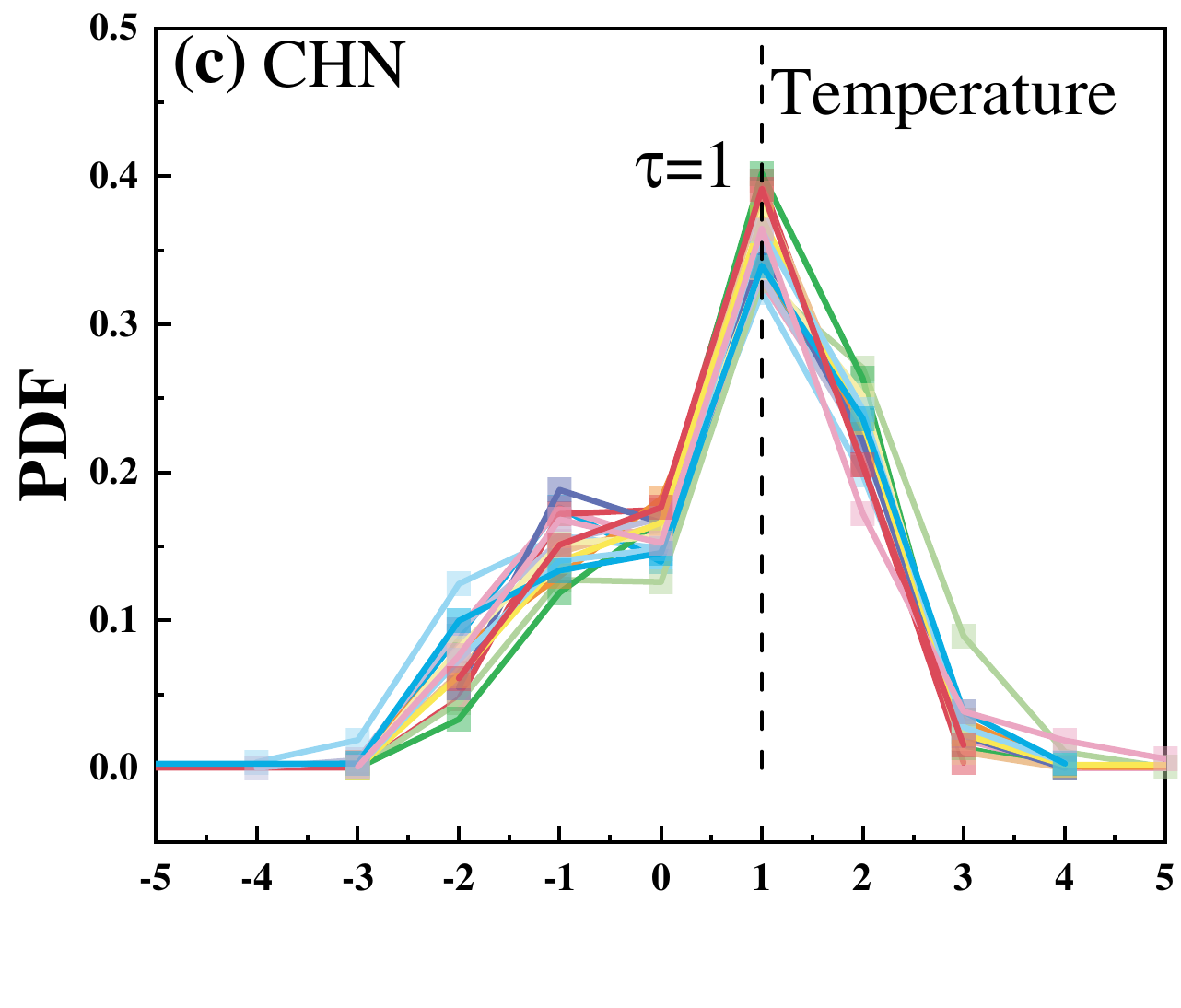}
    \includegraphics[width=8em, height=7em]{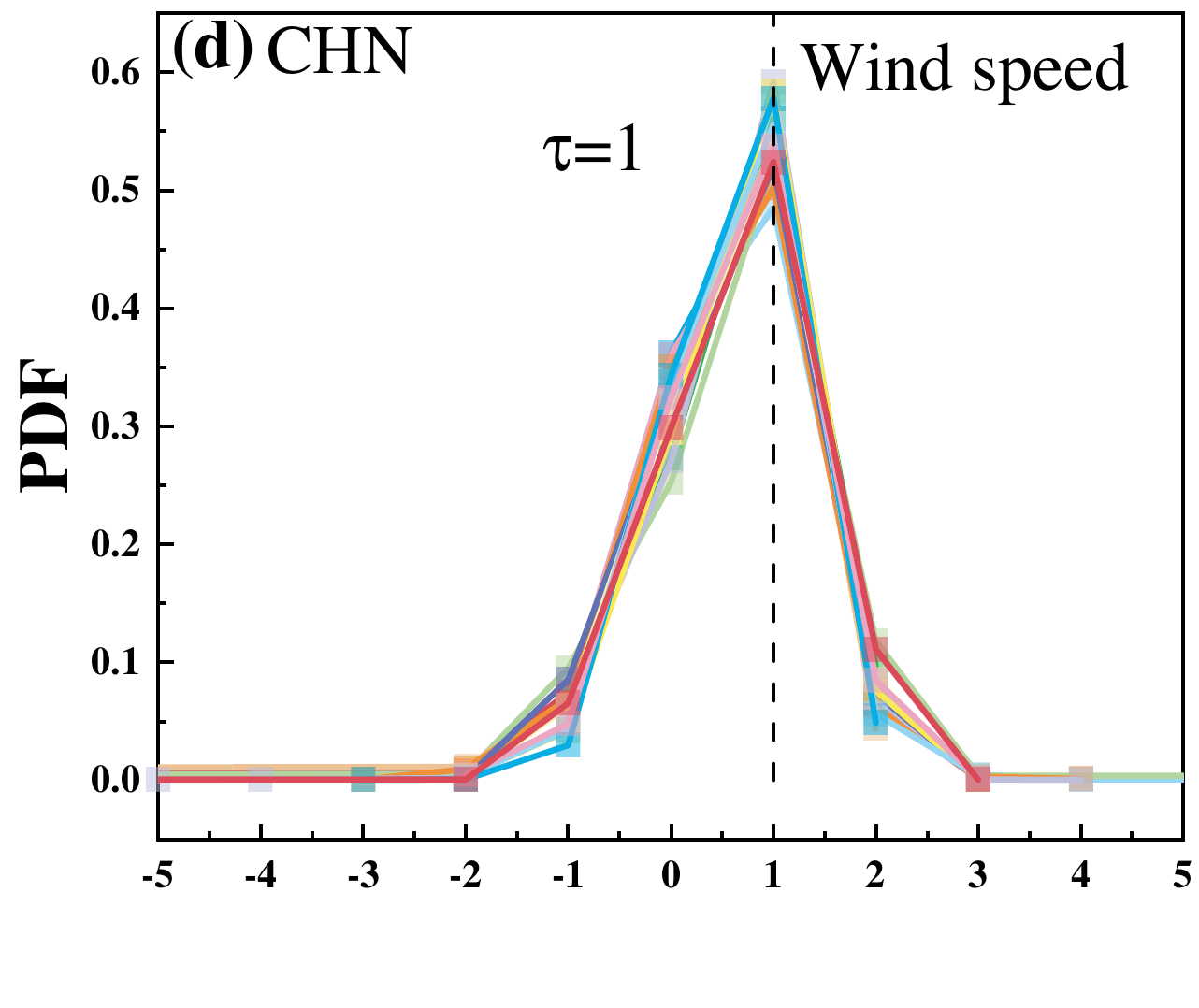}
    \includegraphics[width=8em, height=7em]{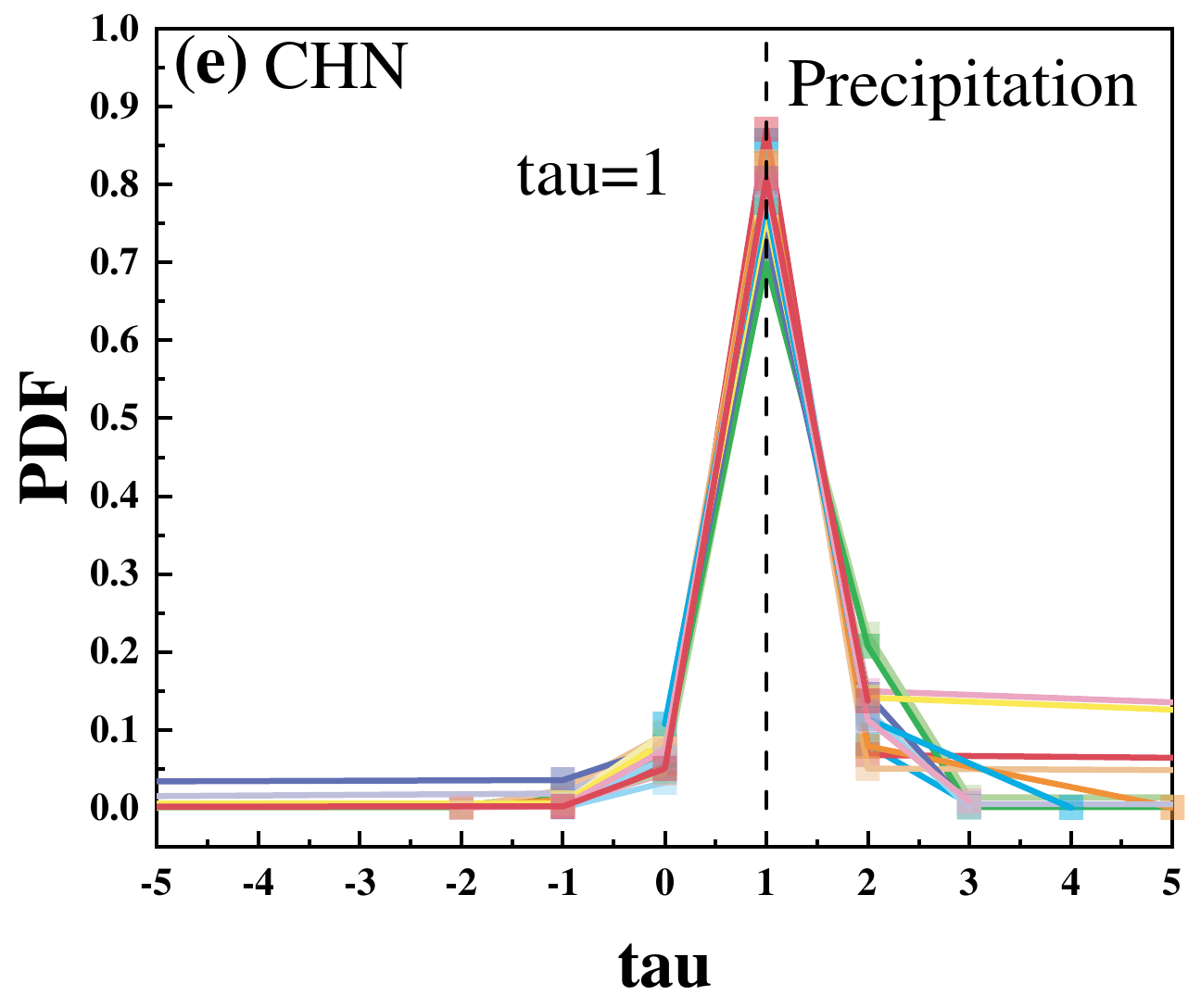}
    \includegraphics[width=8em, height=7em]{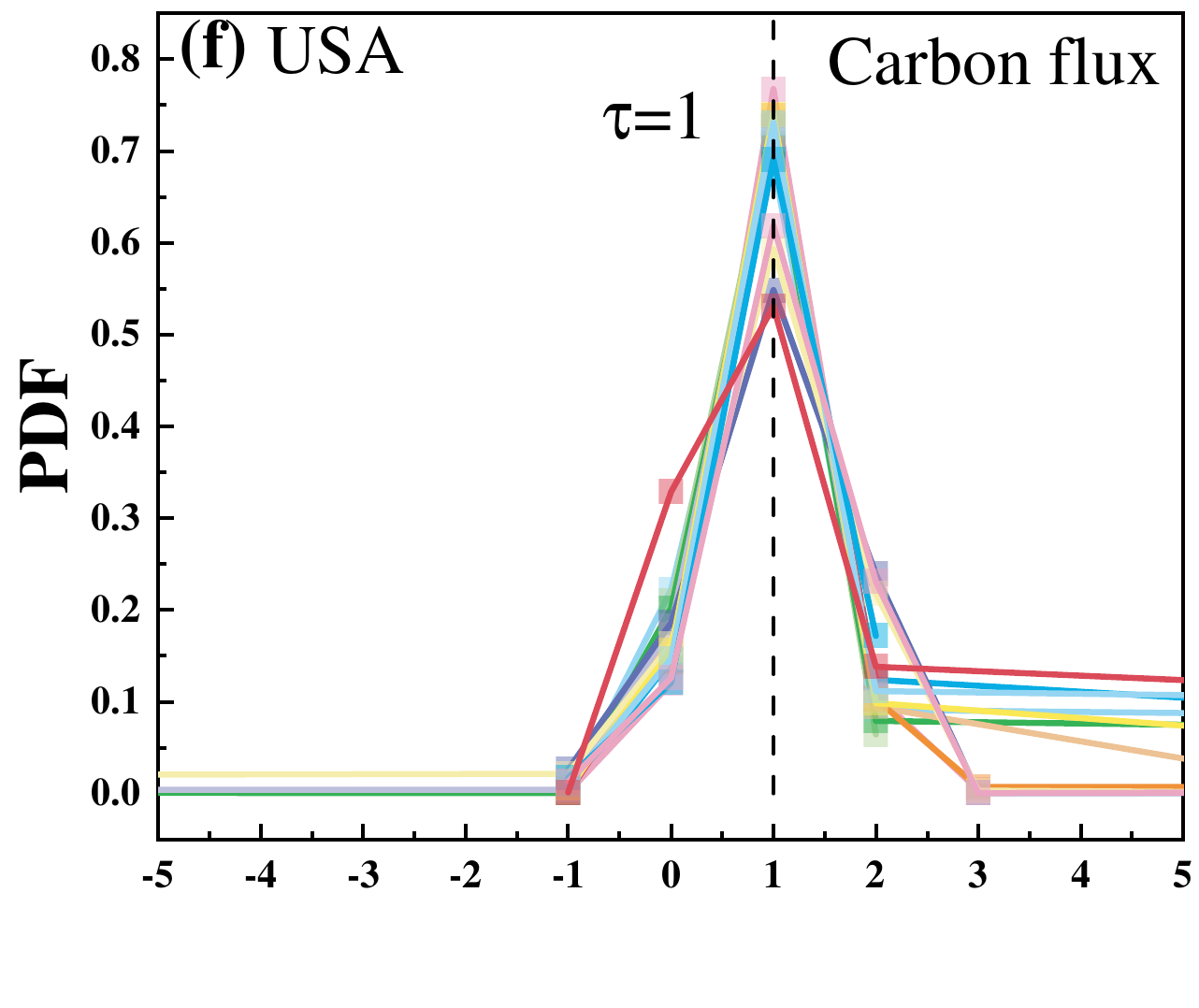}
    \includegraphics[width=8em, height=7em]{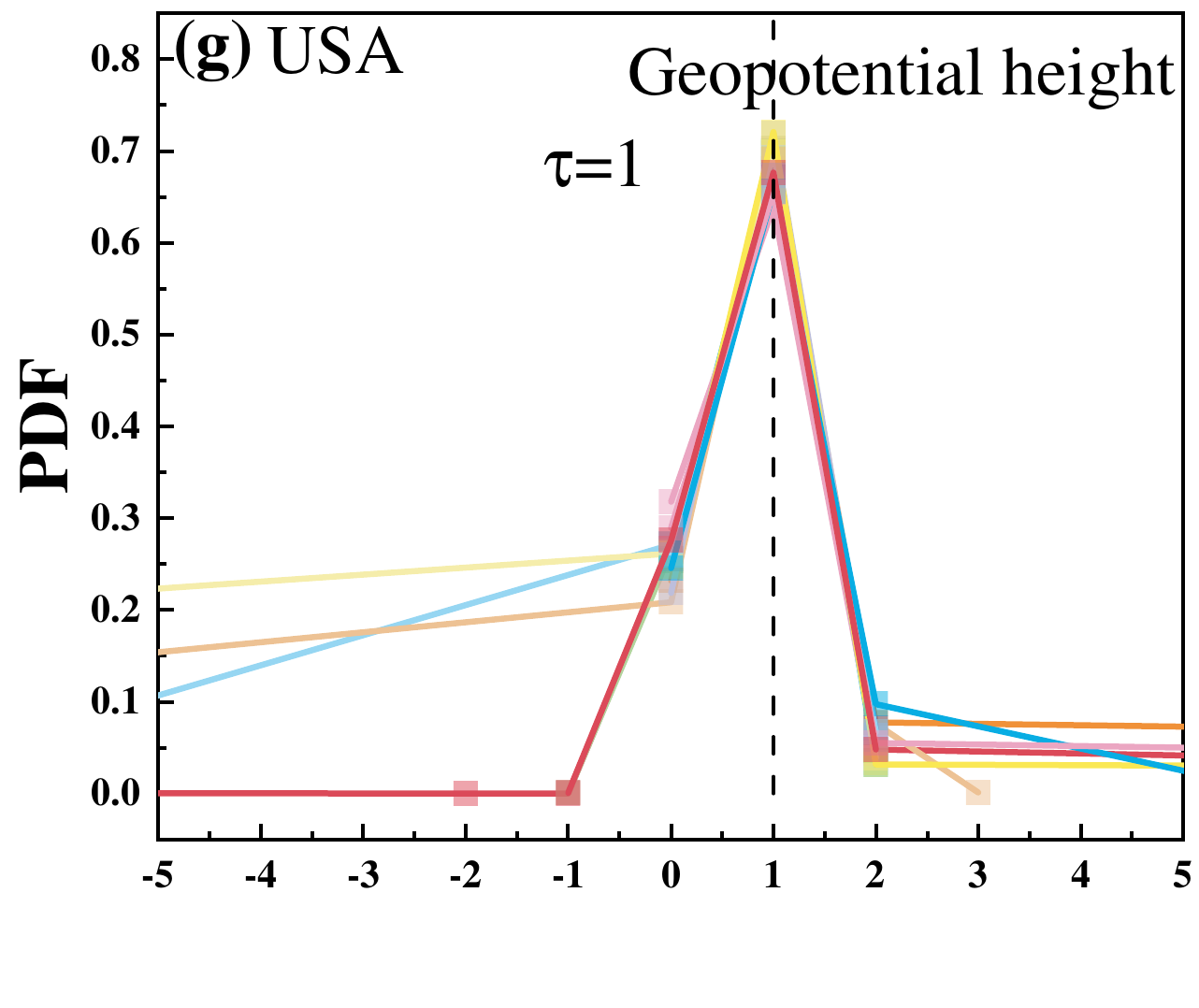}
    \includegraphics[width=8em, height=7em]{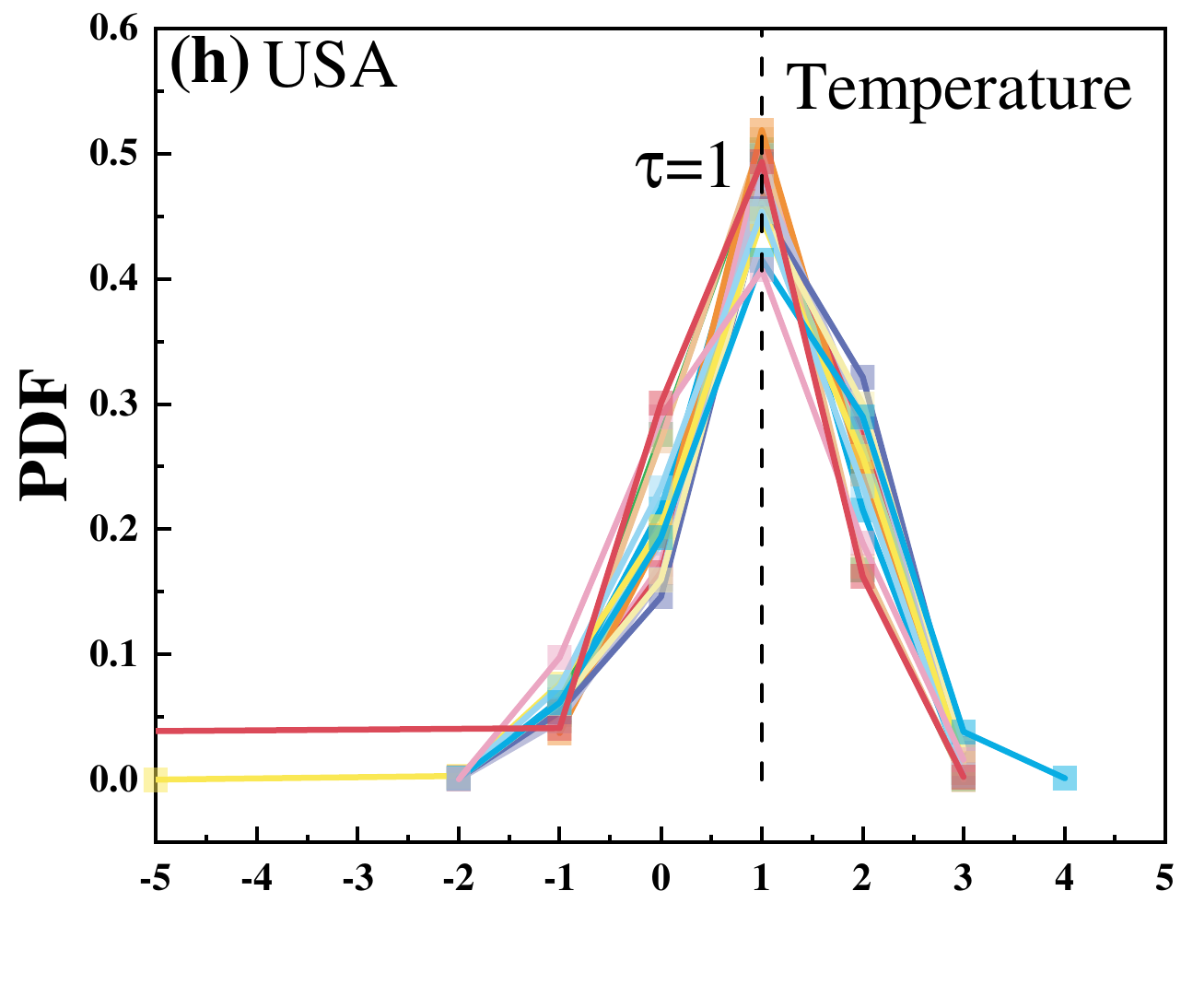}
    \includegraphics[width=8em, height=7em]{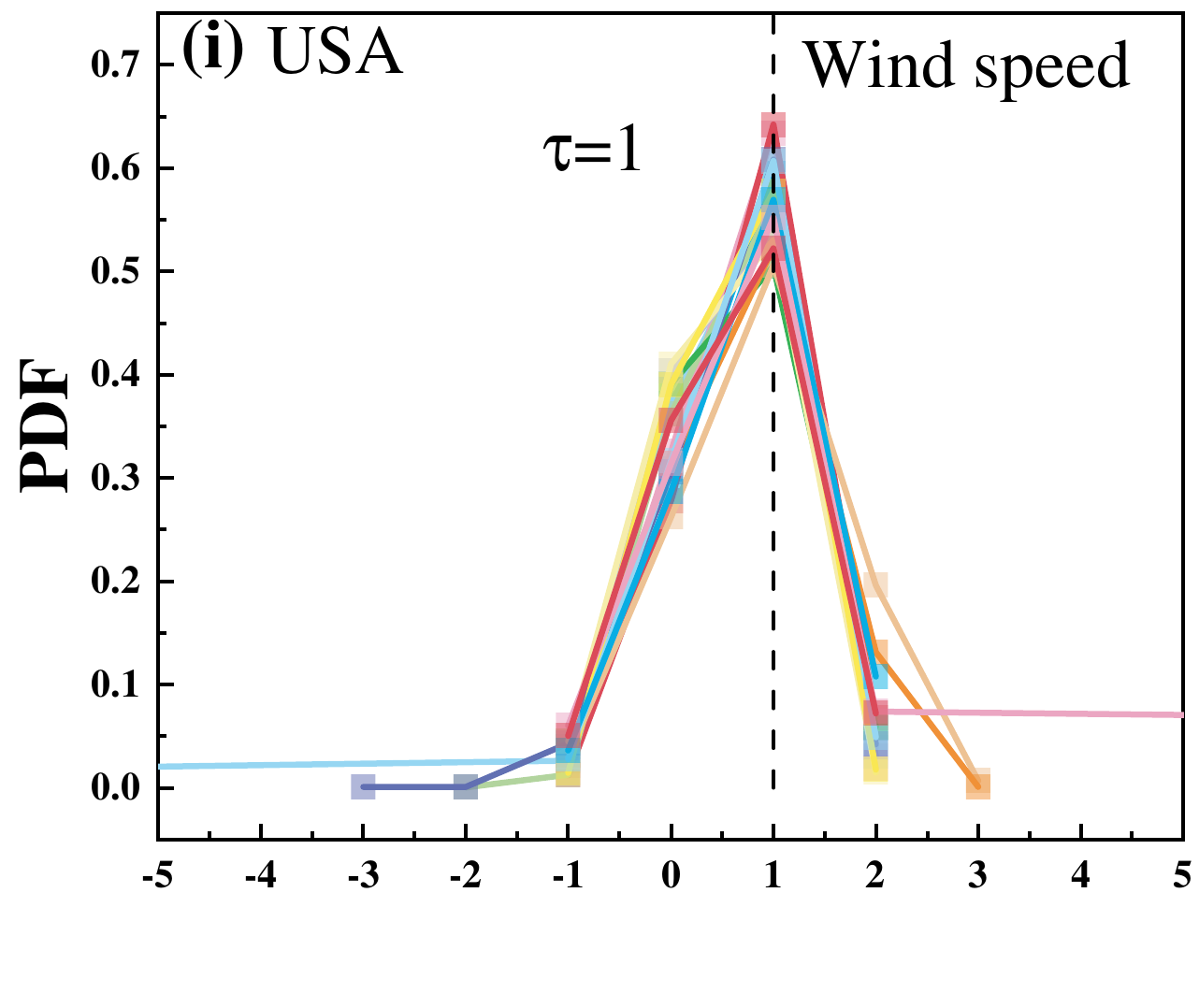}
    \includegraphics[width=8em, height=7em]{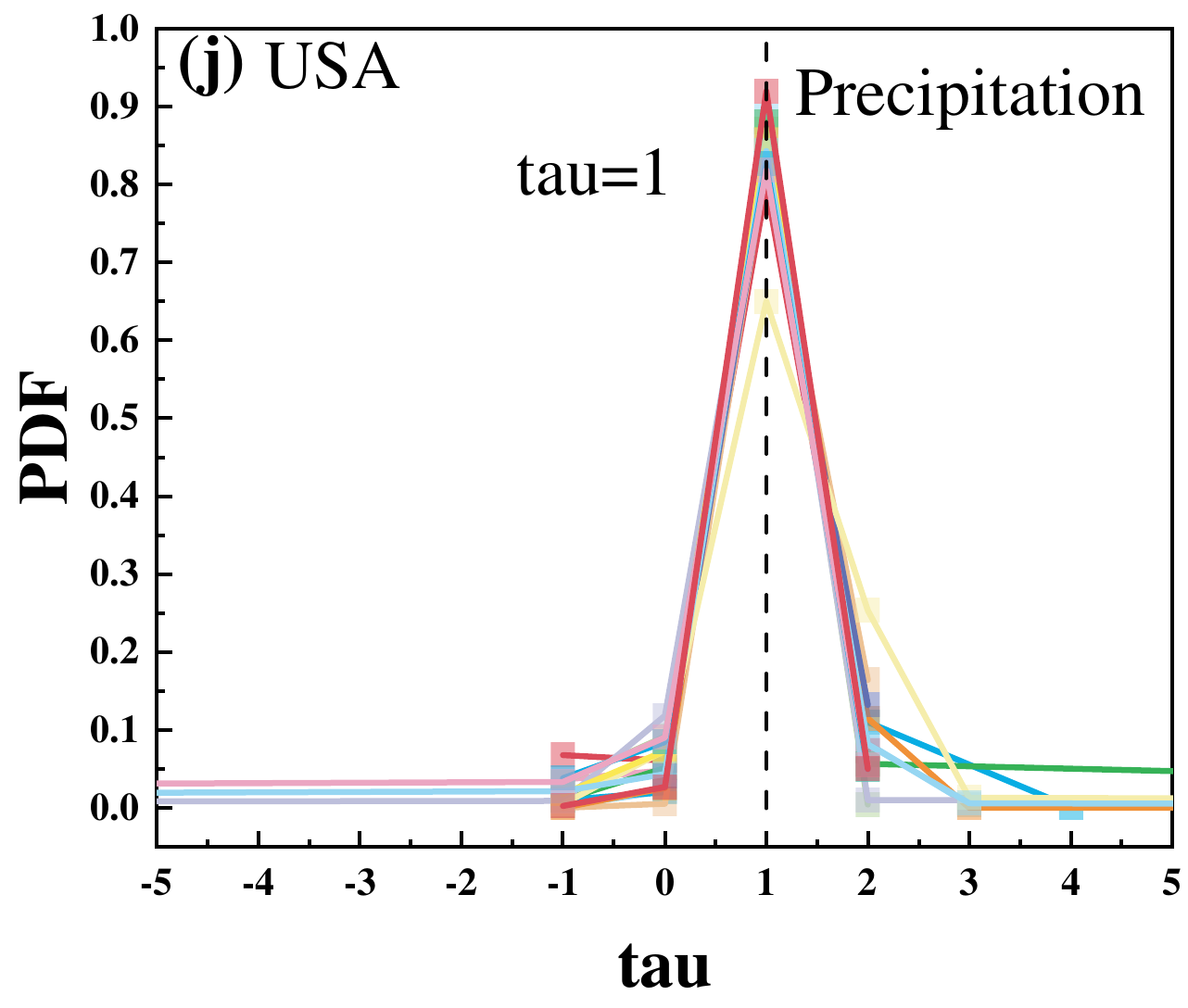}
    \includegraphics[width=8em, height=7em]{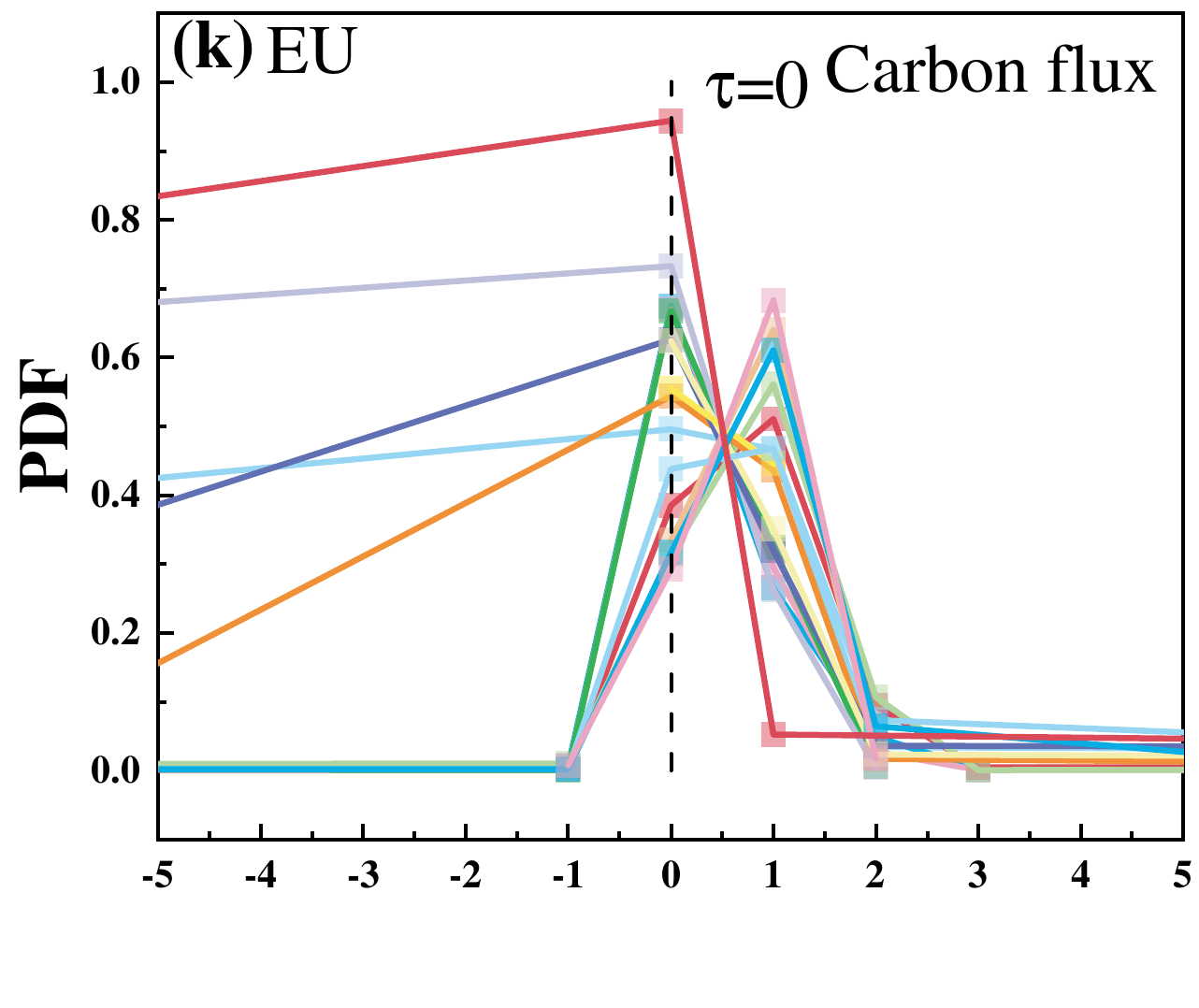}
    \includegraphics[width=8em, height=7em]{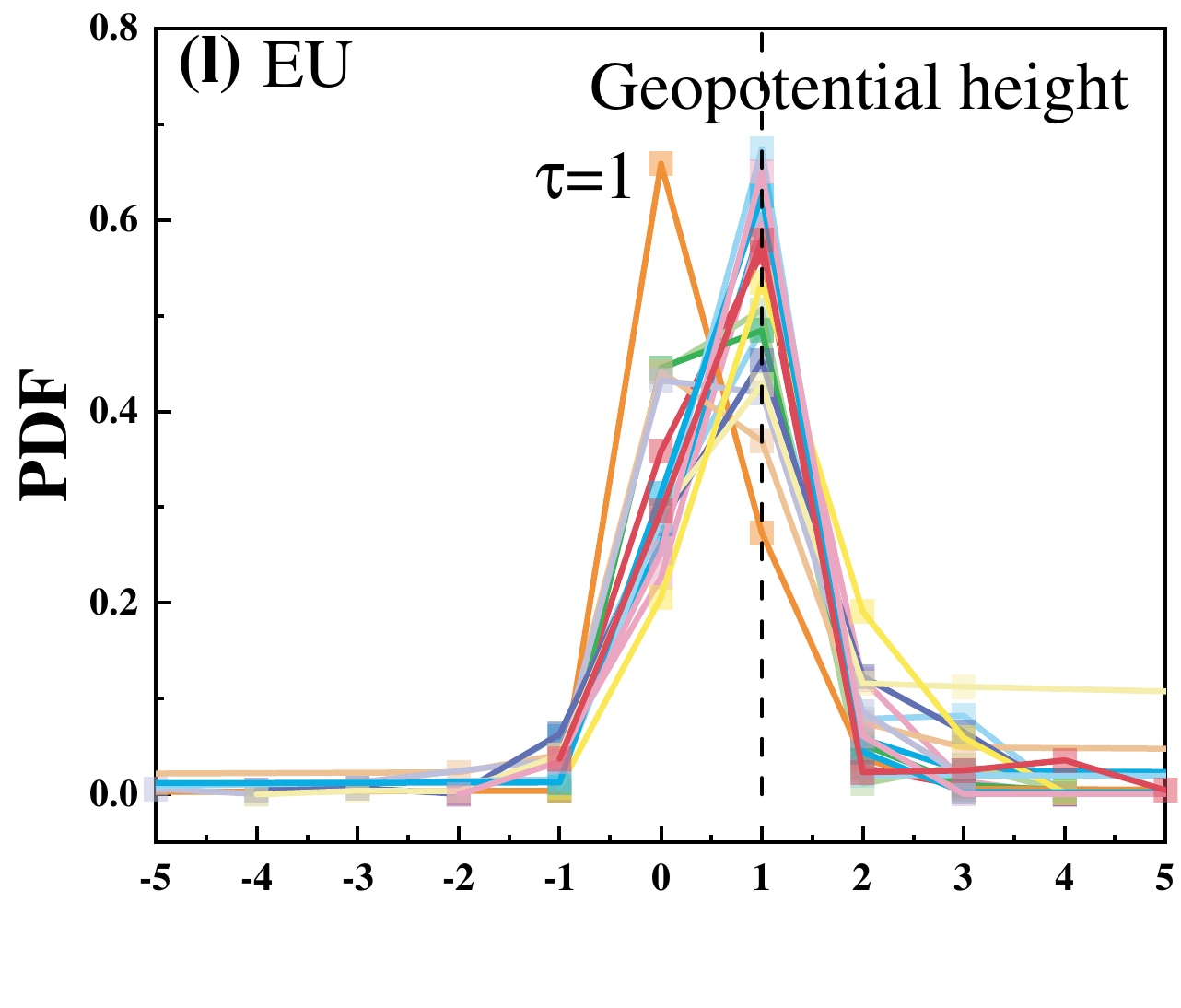}
    \includegraphics[width=8em, height=7em]{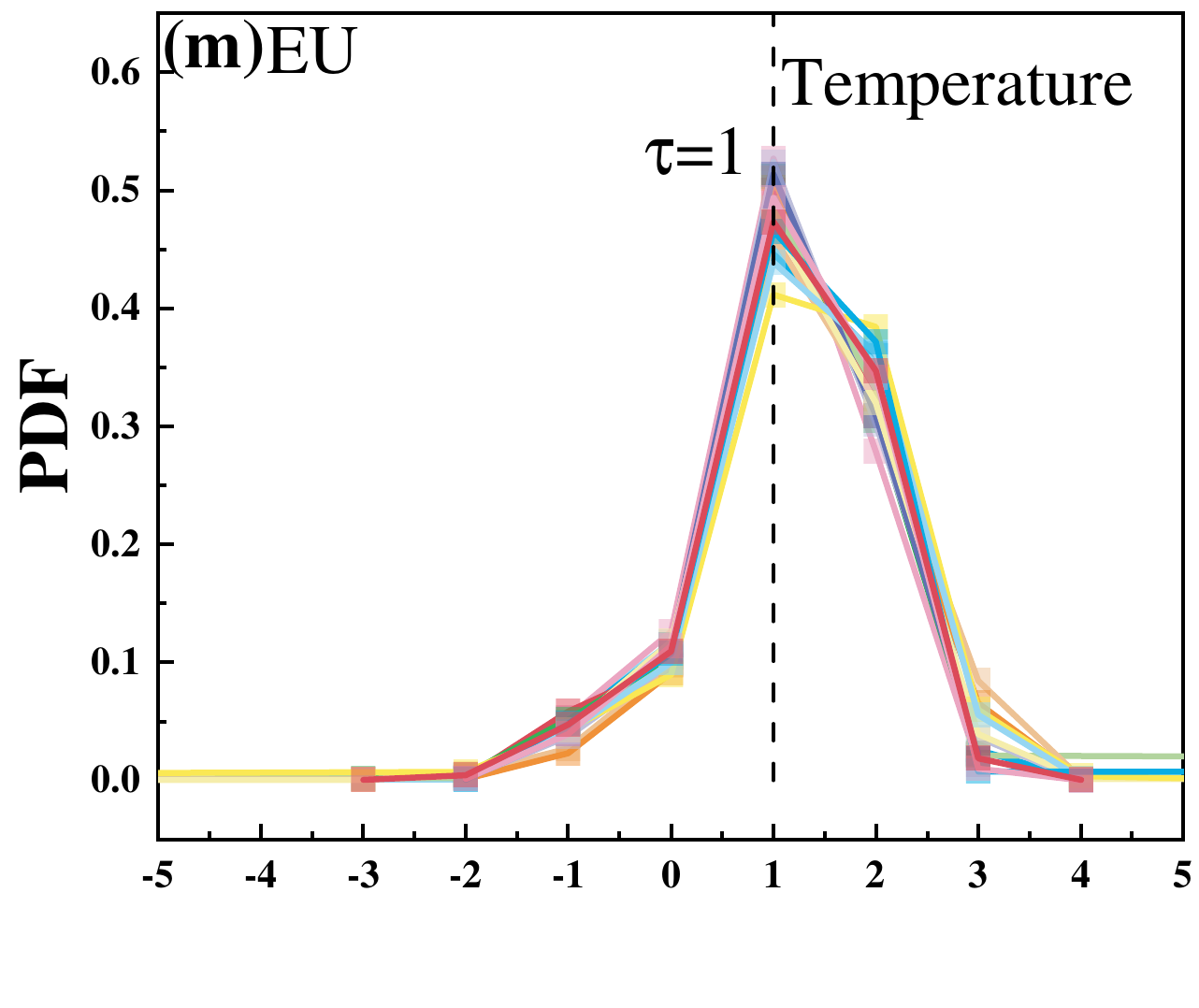}
    \includegraphics[width=8em, height=7em]{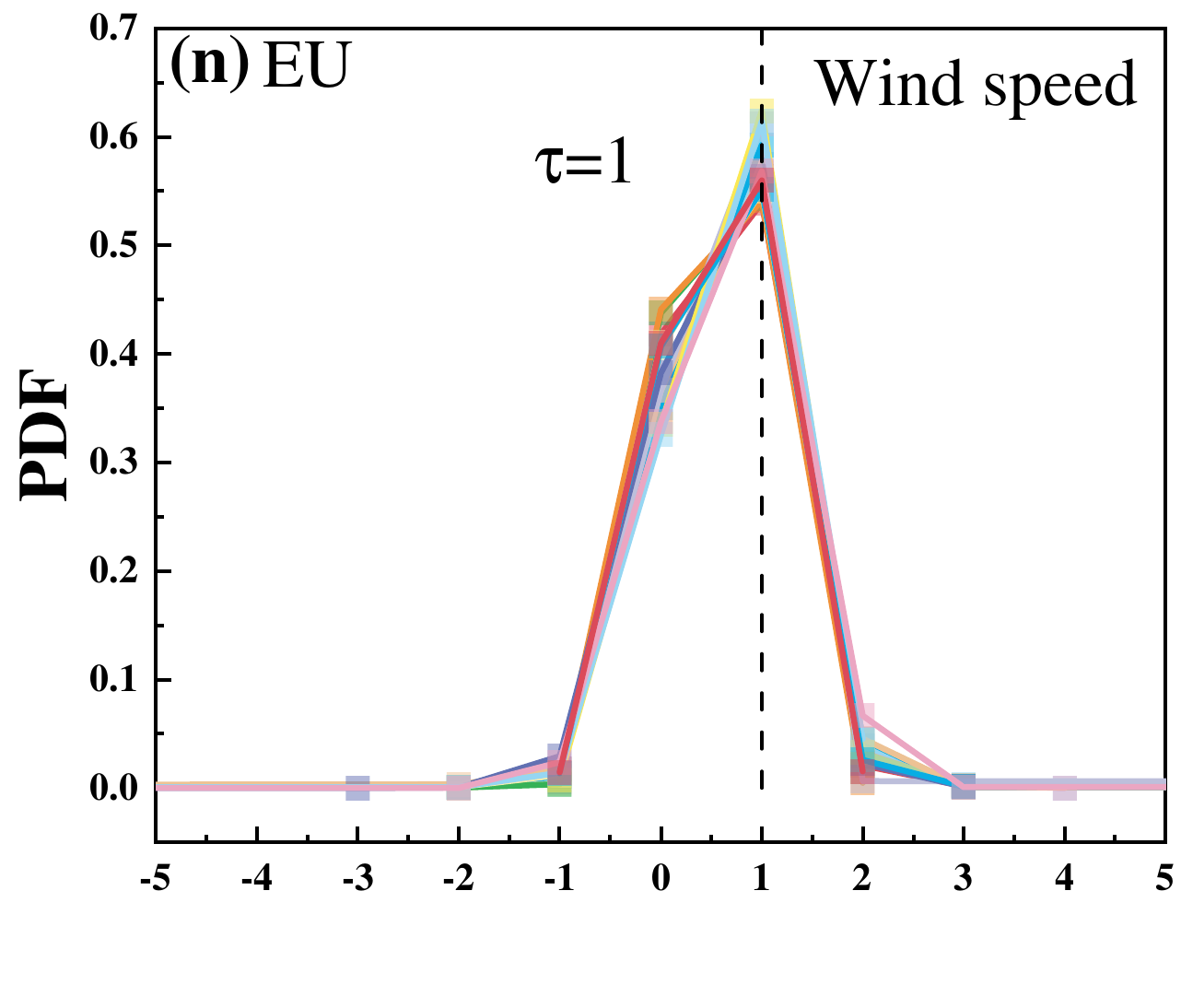}
    \includegraphics[width=8em, height=7em]{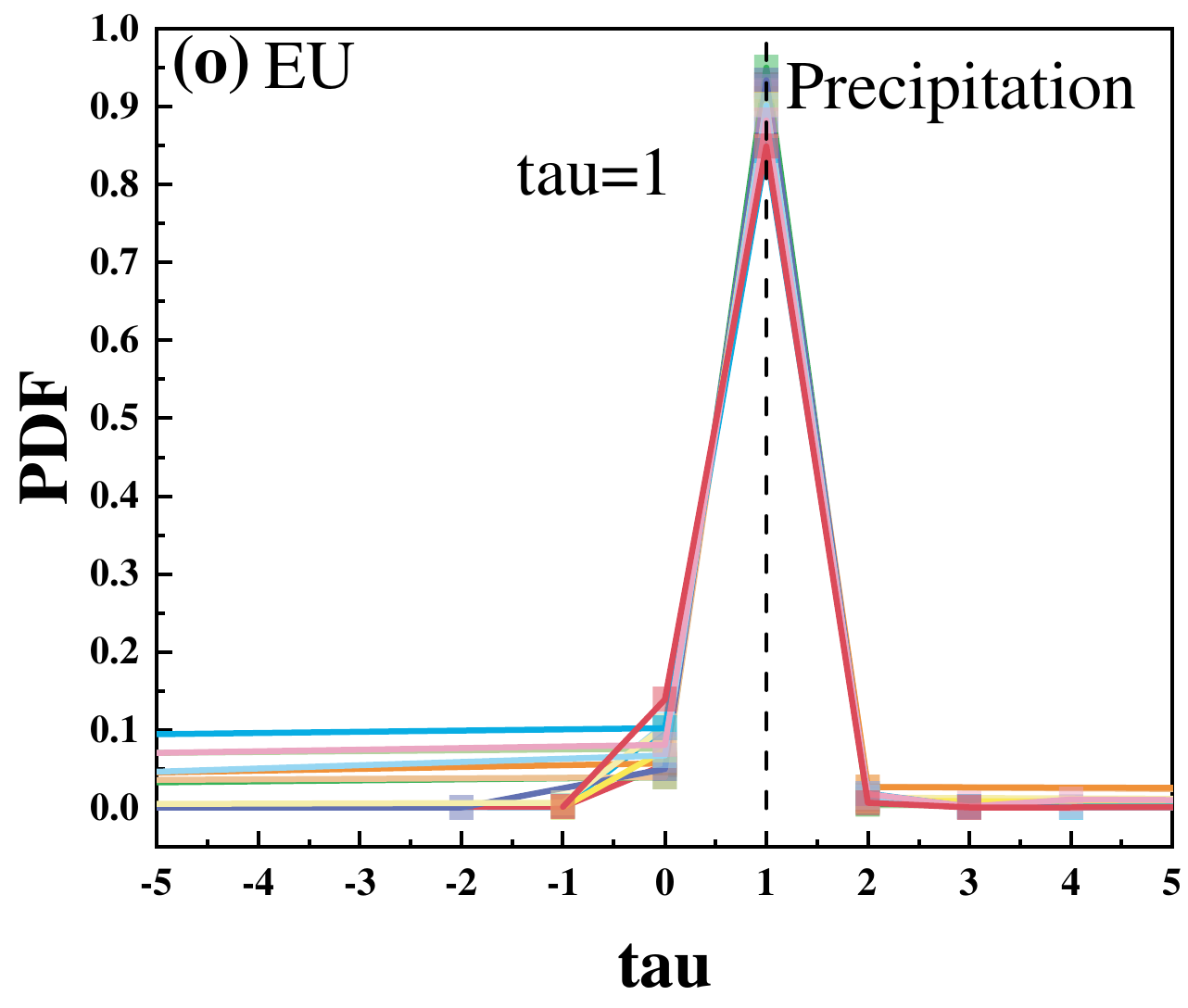}
\end{center}

\begin{center}
\noindent {\small {\bf Fig. S20} For lengths above 1000$km$, probability distribution functions (PDF) of time lags in different time series. Different colors represent the time lag probability distribution of the network for different years. }
\end{center}

\begin{center}
\includegraphics[width=8em, height=7em]{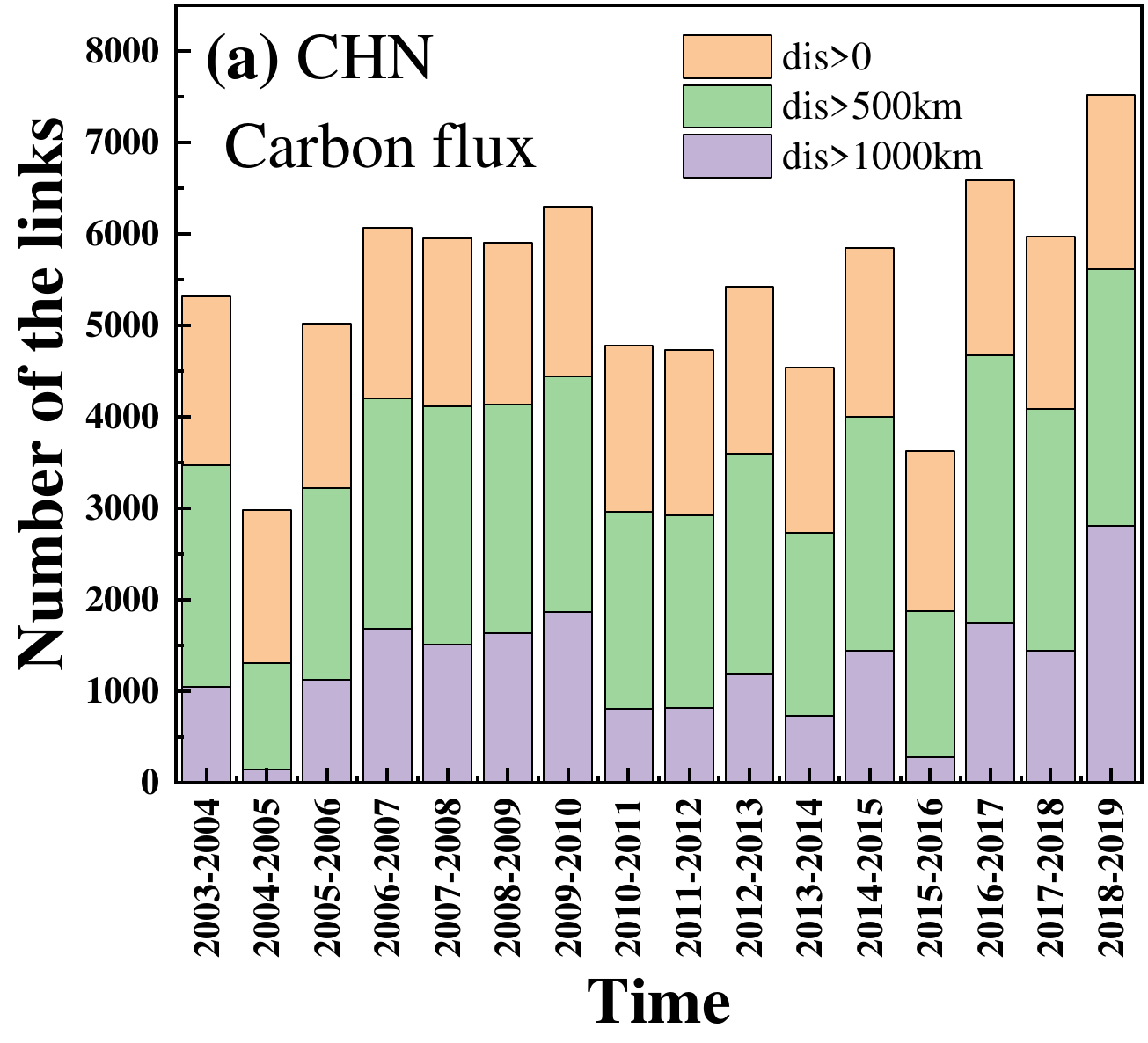}
\includegraphics[width=8em, height=7em]{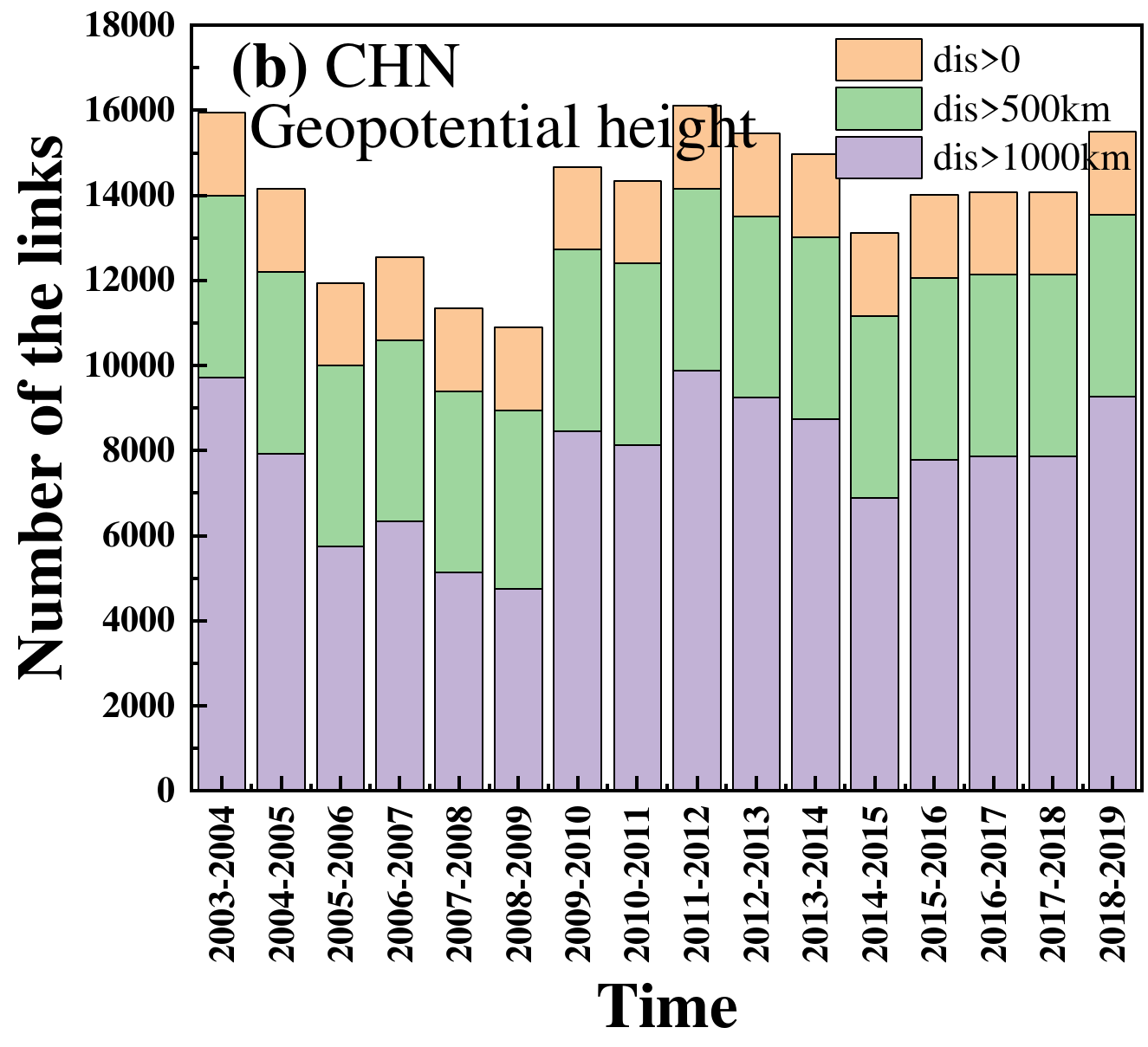}
\includegraphics[width=8em, height=7em]{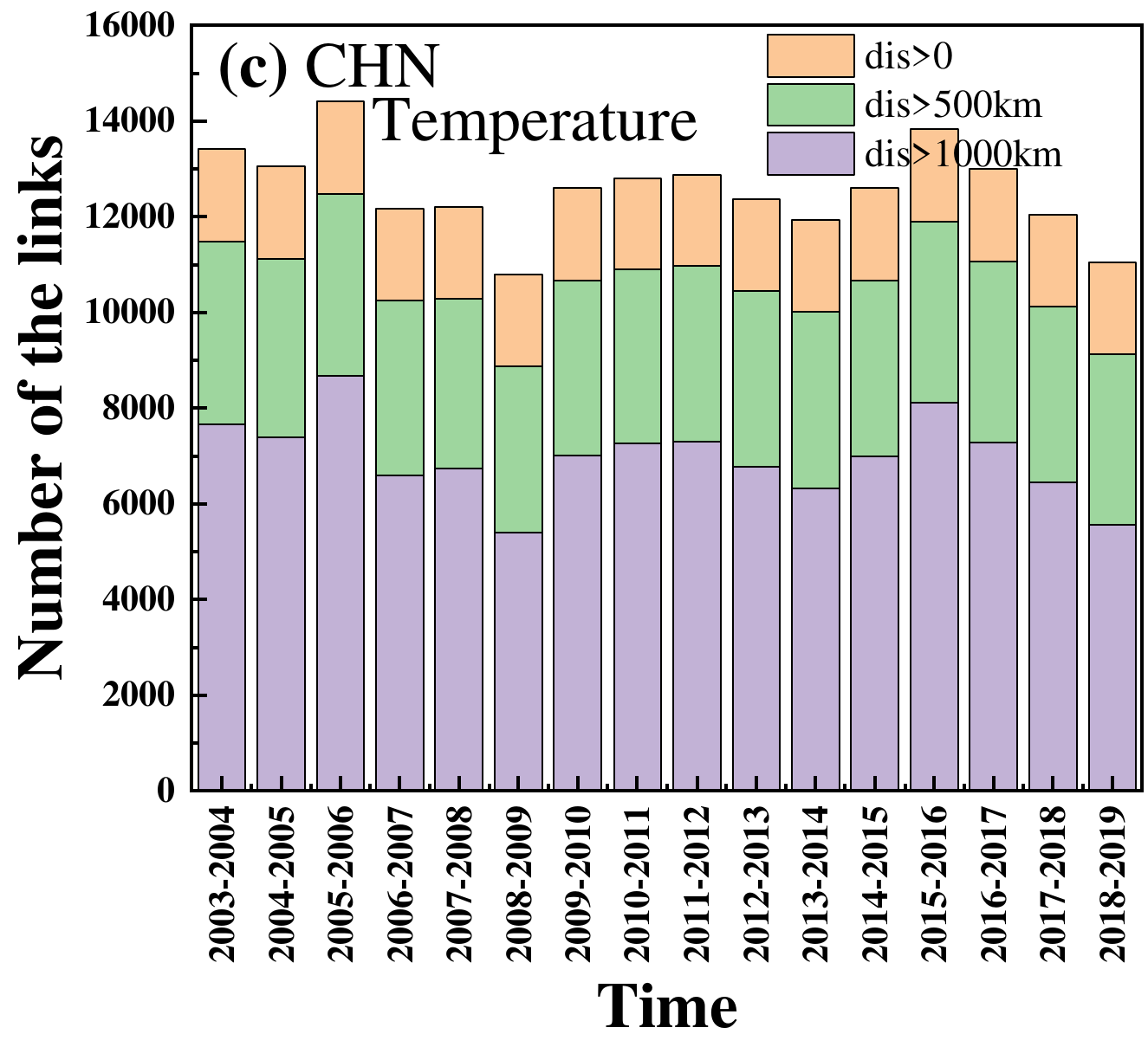}
\includegraphics[width=8em, height=7em]{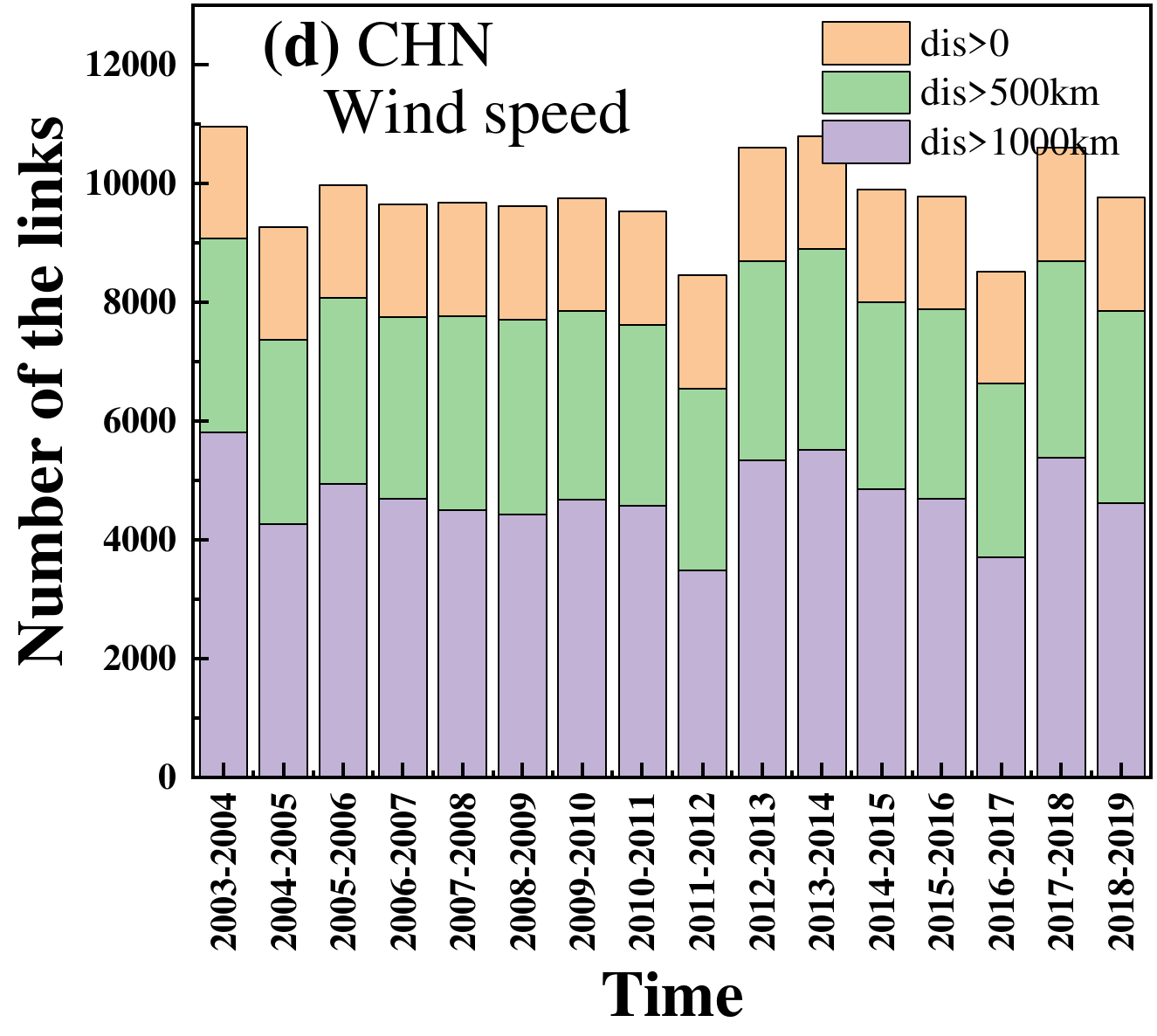}
\includegraphics[width=8em, height=7em]{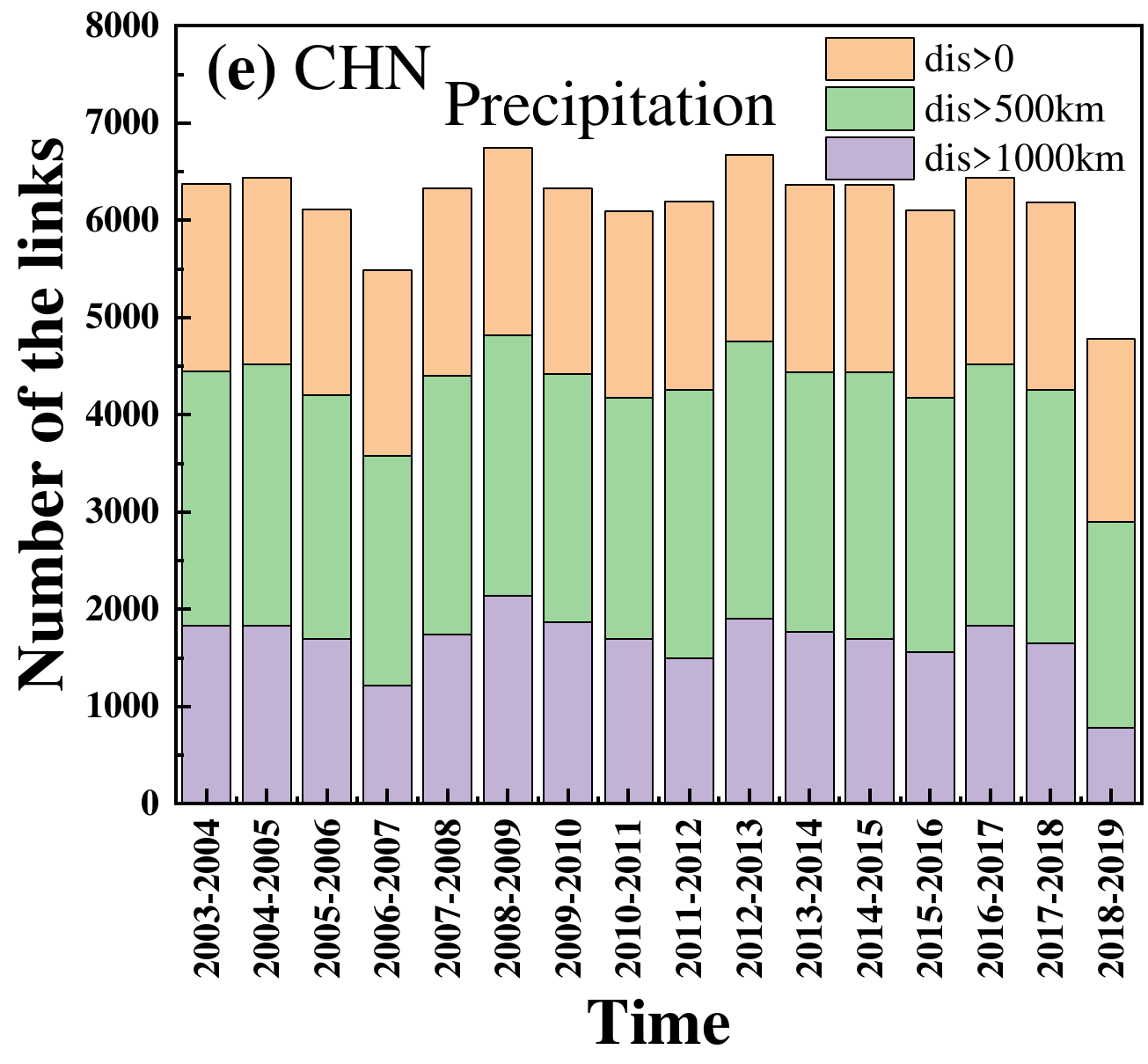}
\includegraphics[width=8em, height=7em]{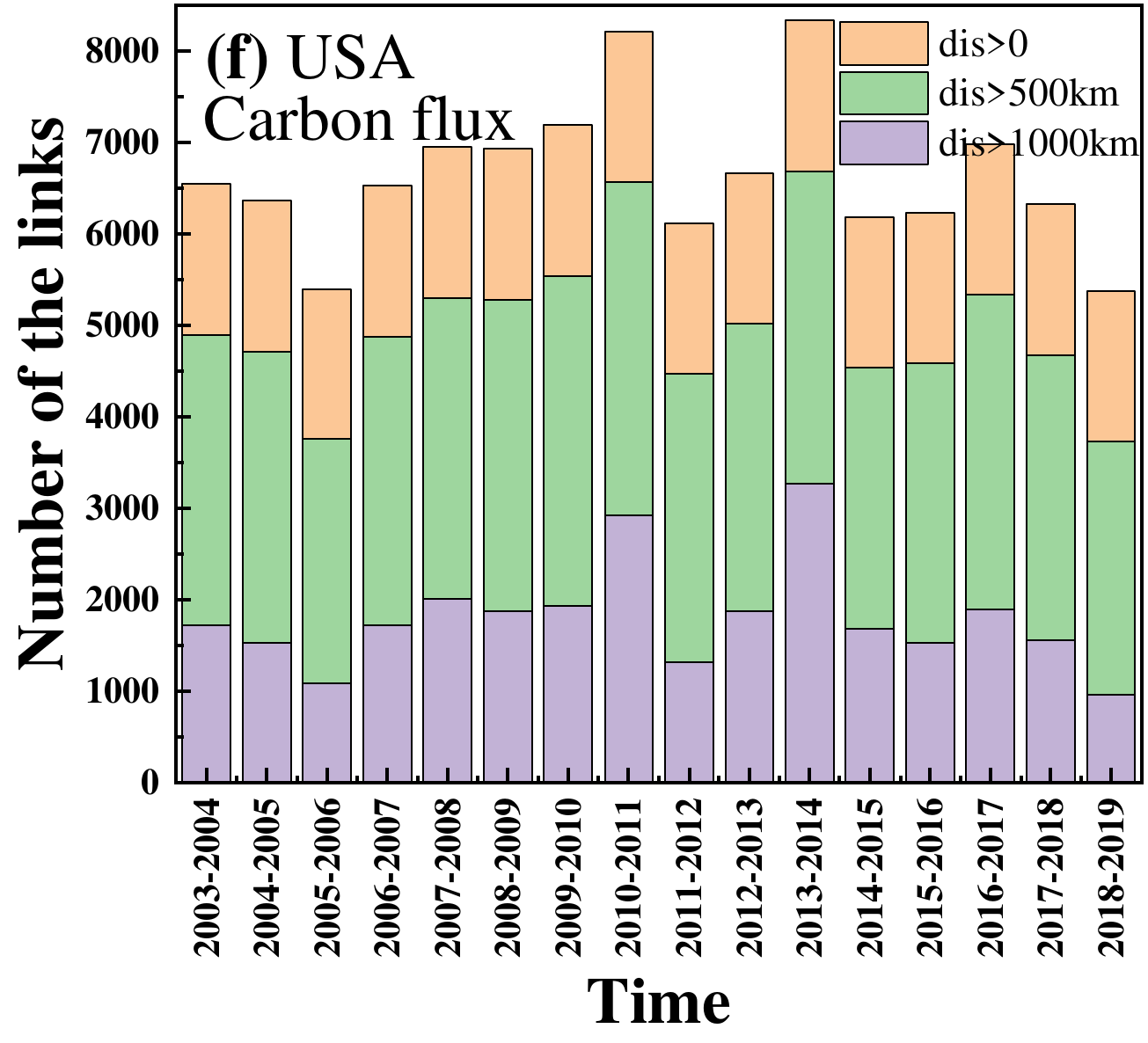}
\includegraphics[width=8em, height=7em]{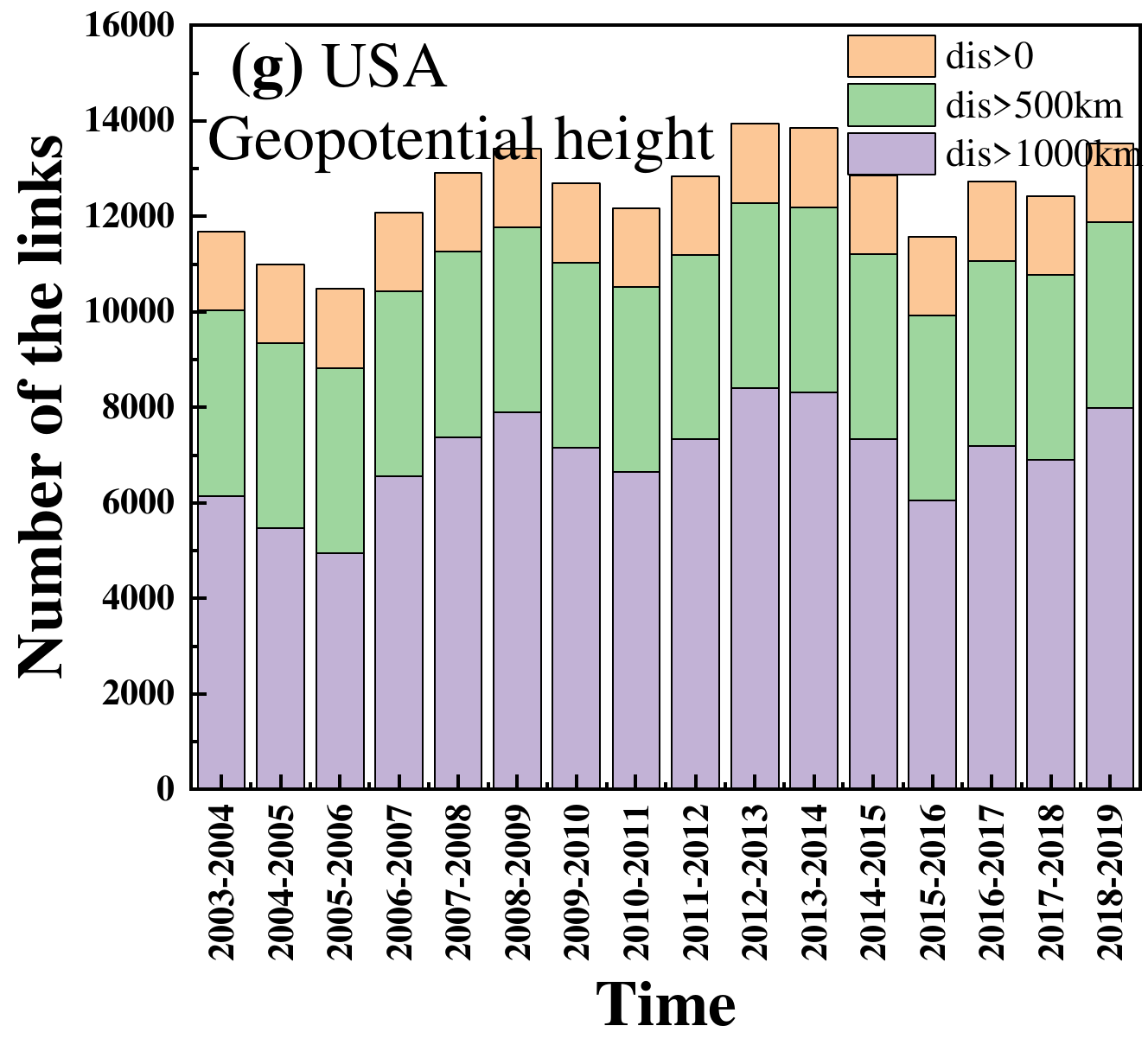}
\includegraphics[width=8em, height=7em]{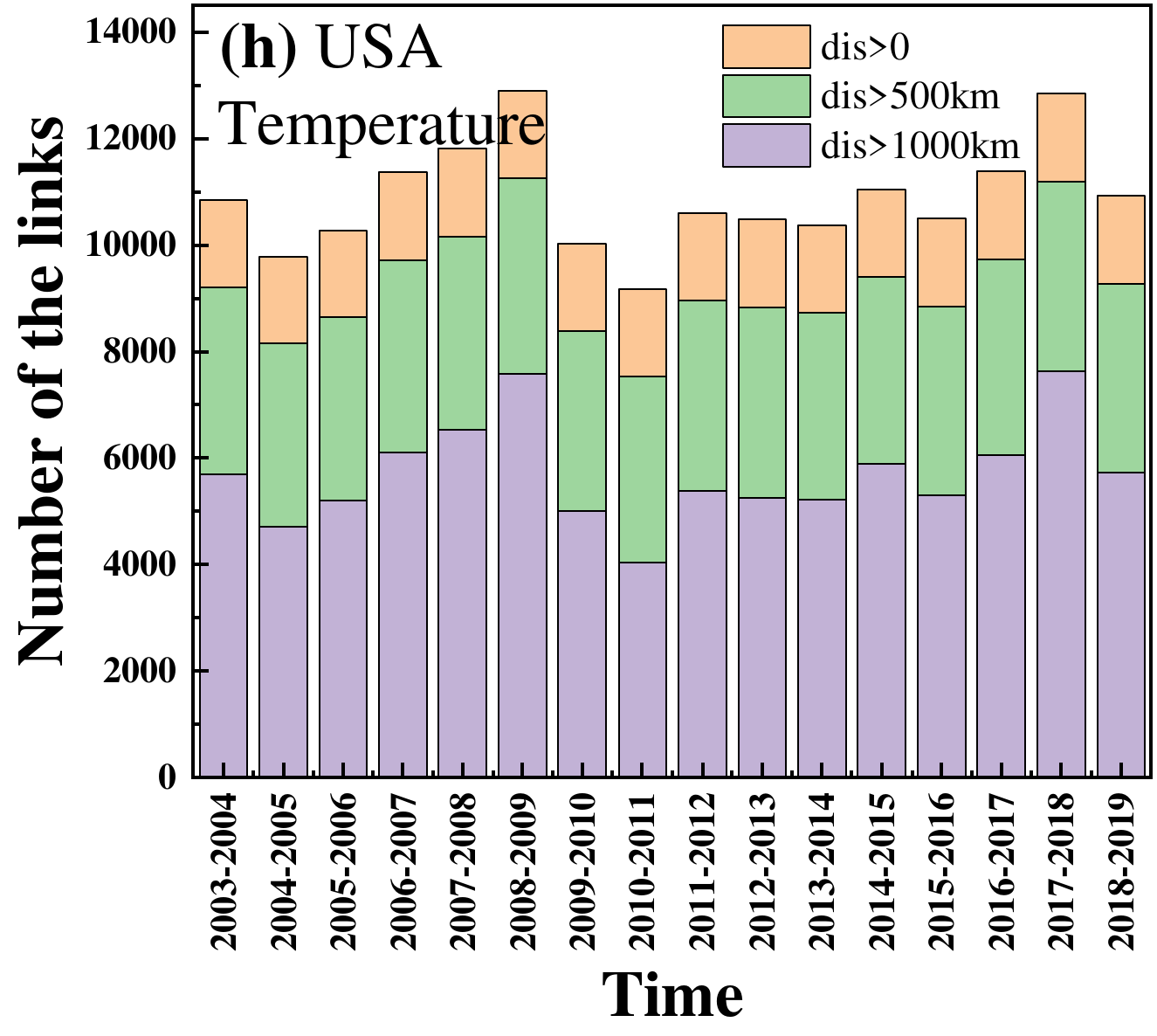}
\includegraphics[width=8em, height=7em]{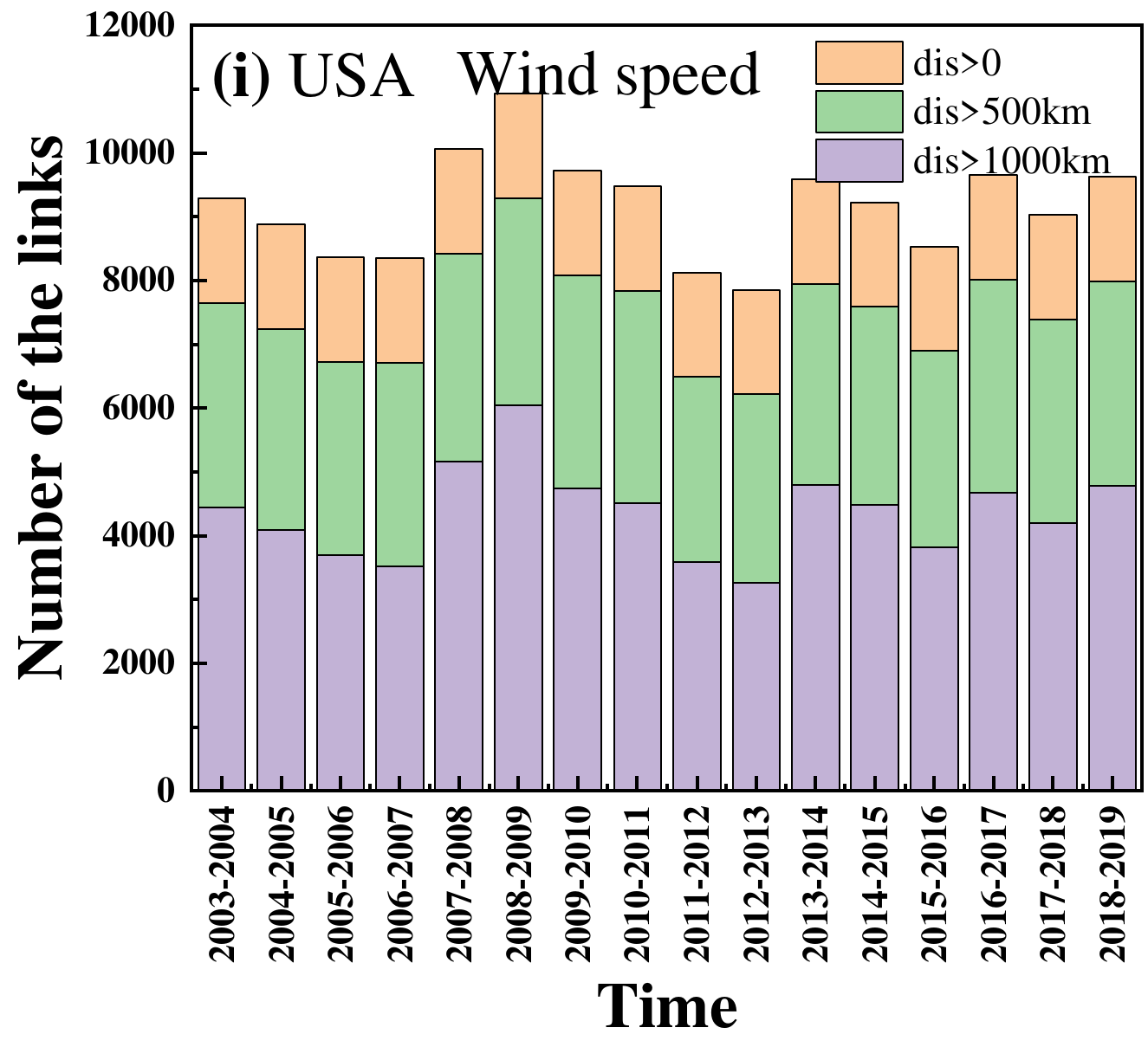}
\includegraphics[width=8em, height=7em]{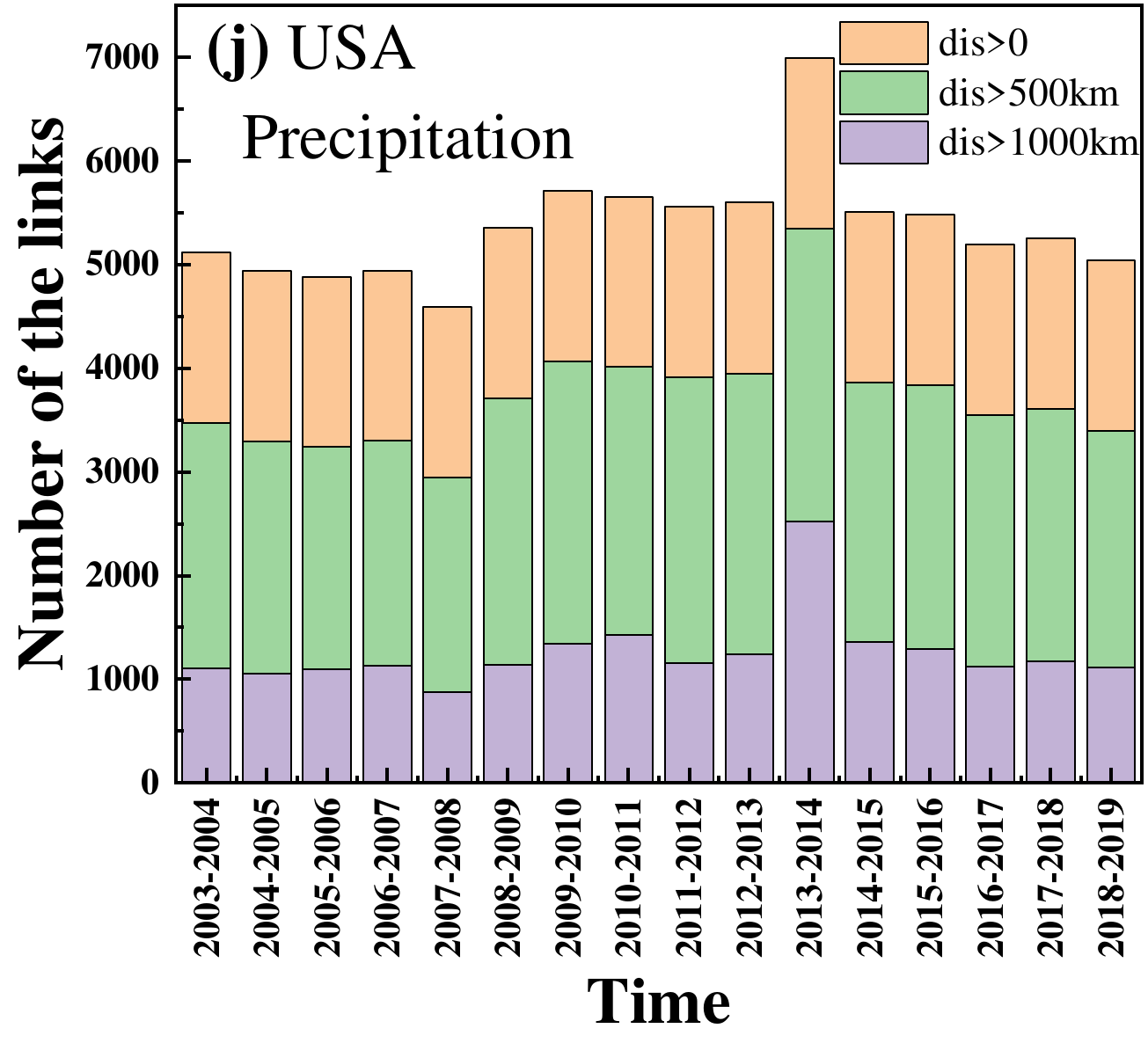}
\includegraphics[width=8em, height=7em]{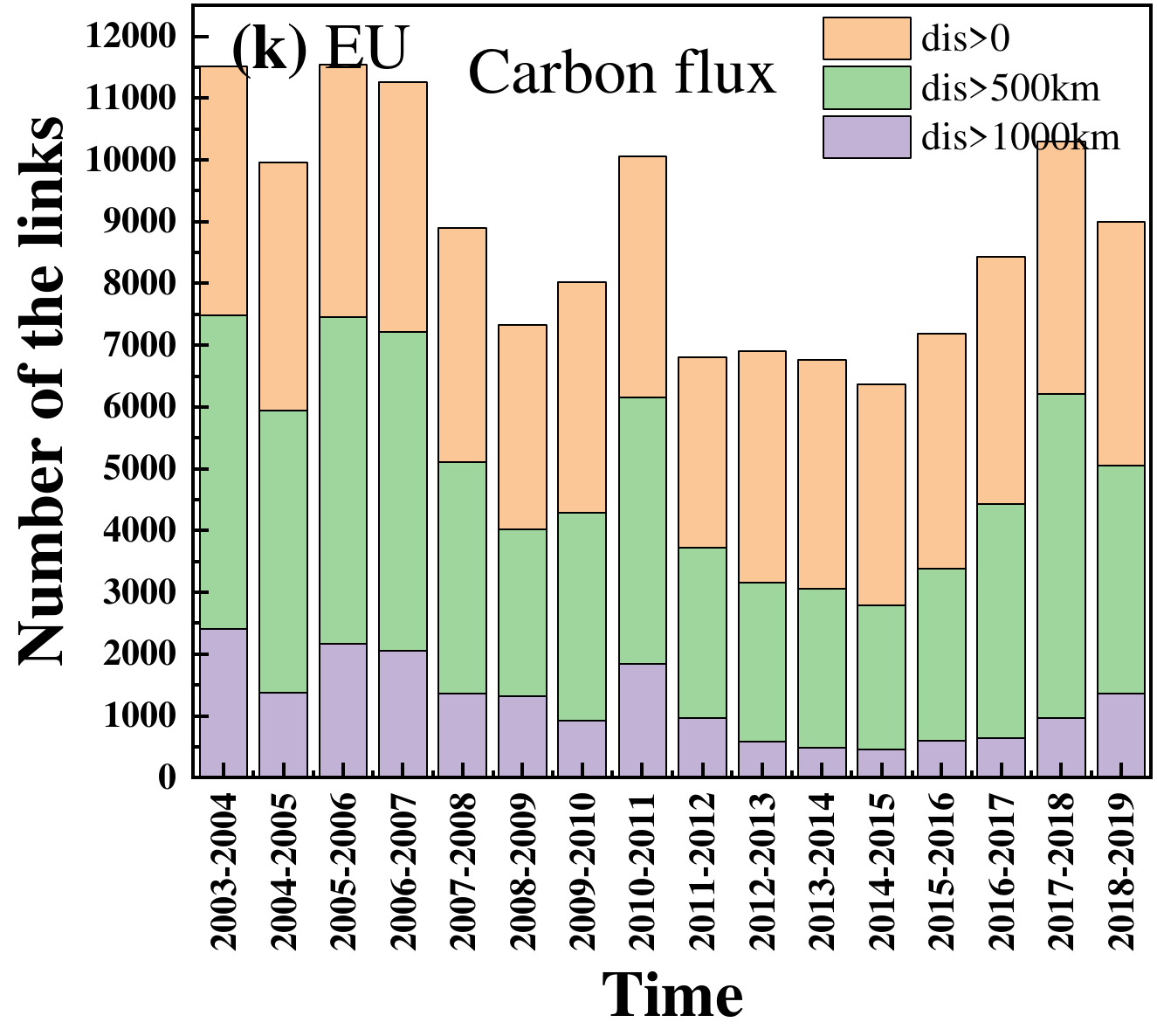}
\includegraphics[width=8em, height=7em]{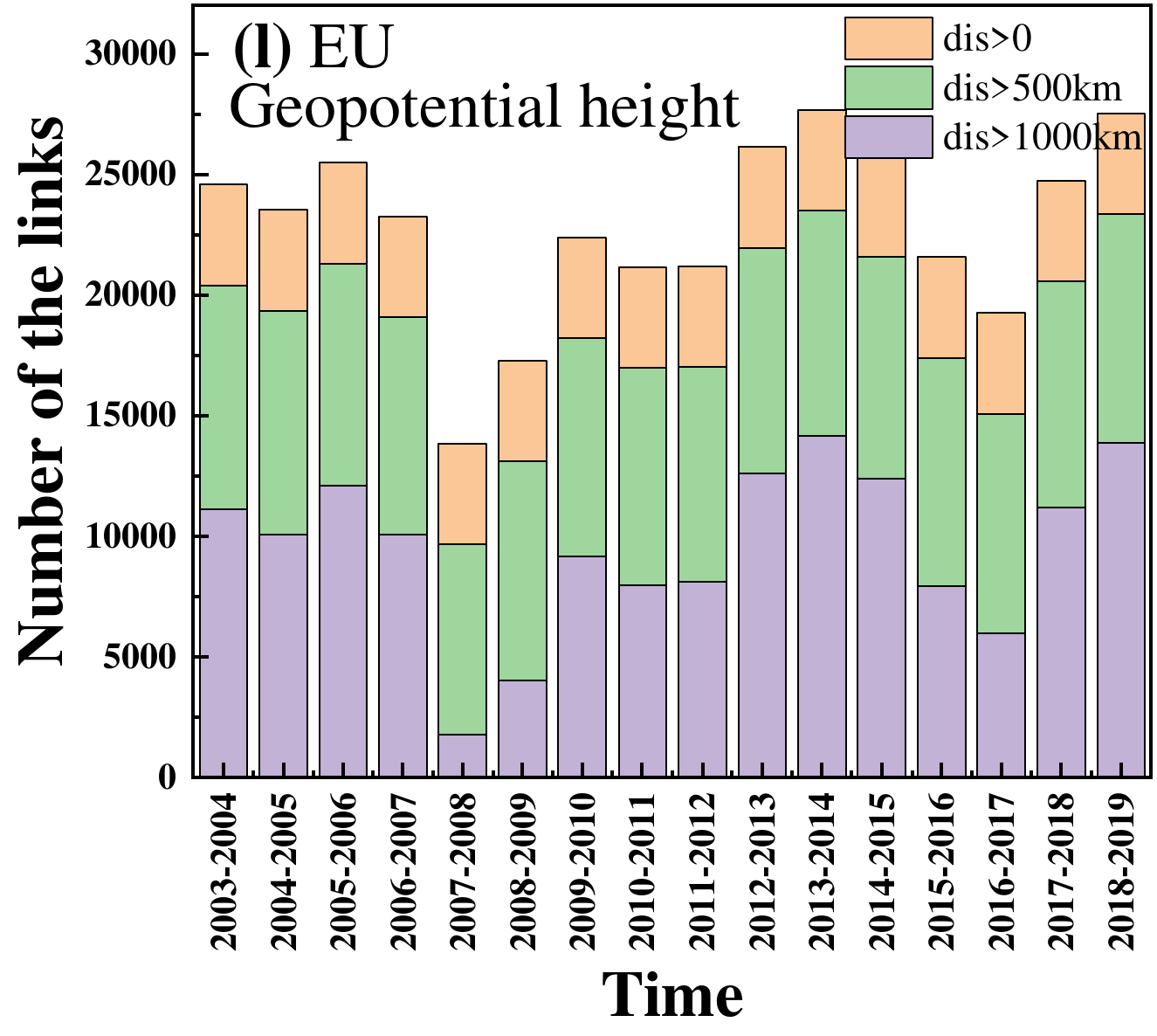}
\includegraphics[width=8em, height=7em]{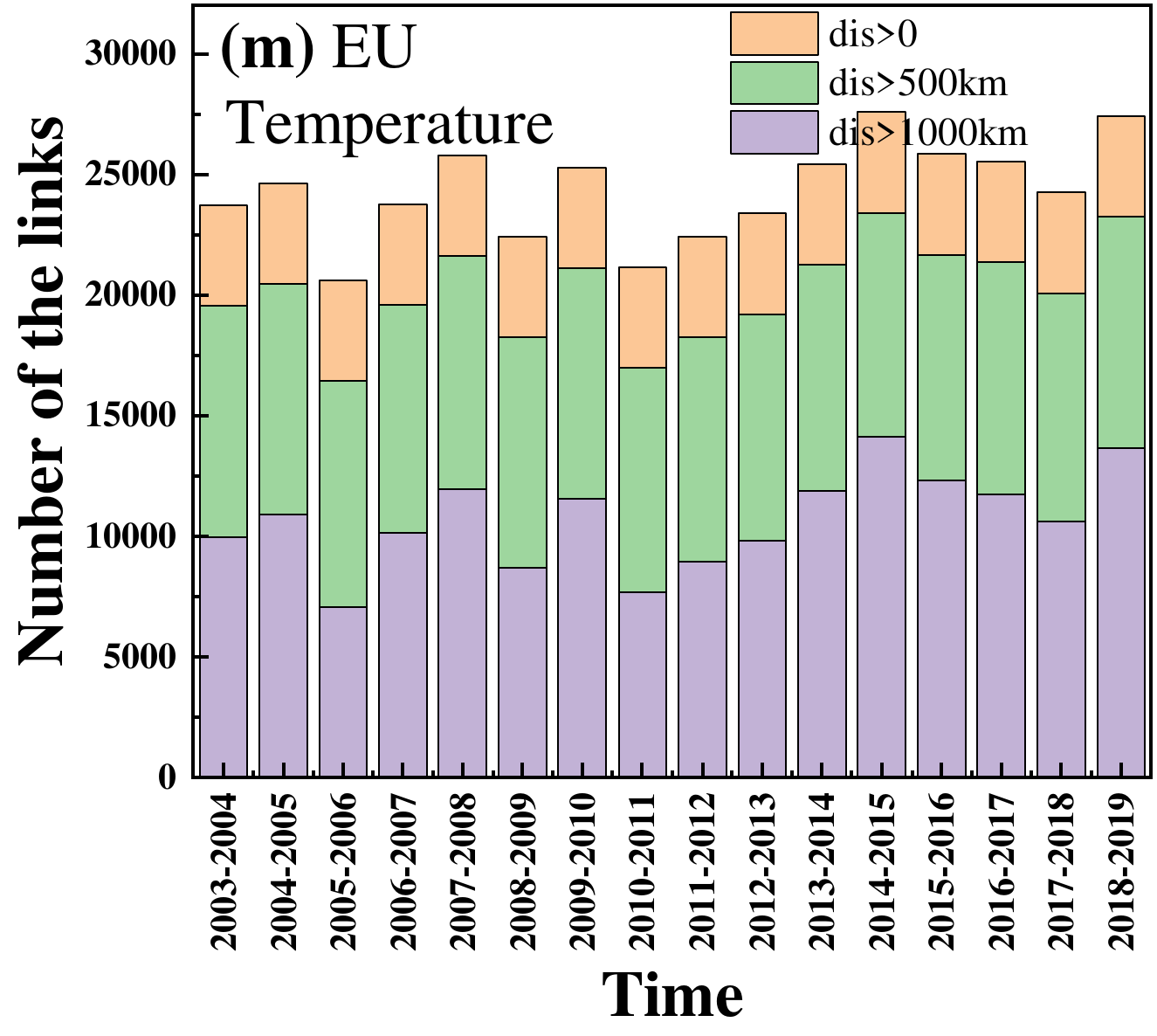}
\includegraphics[width=8em, height=7em]{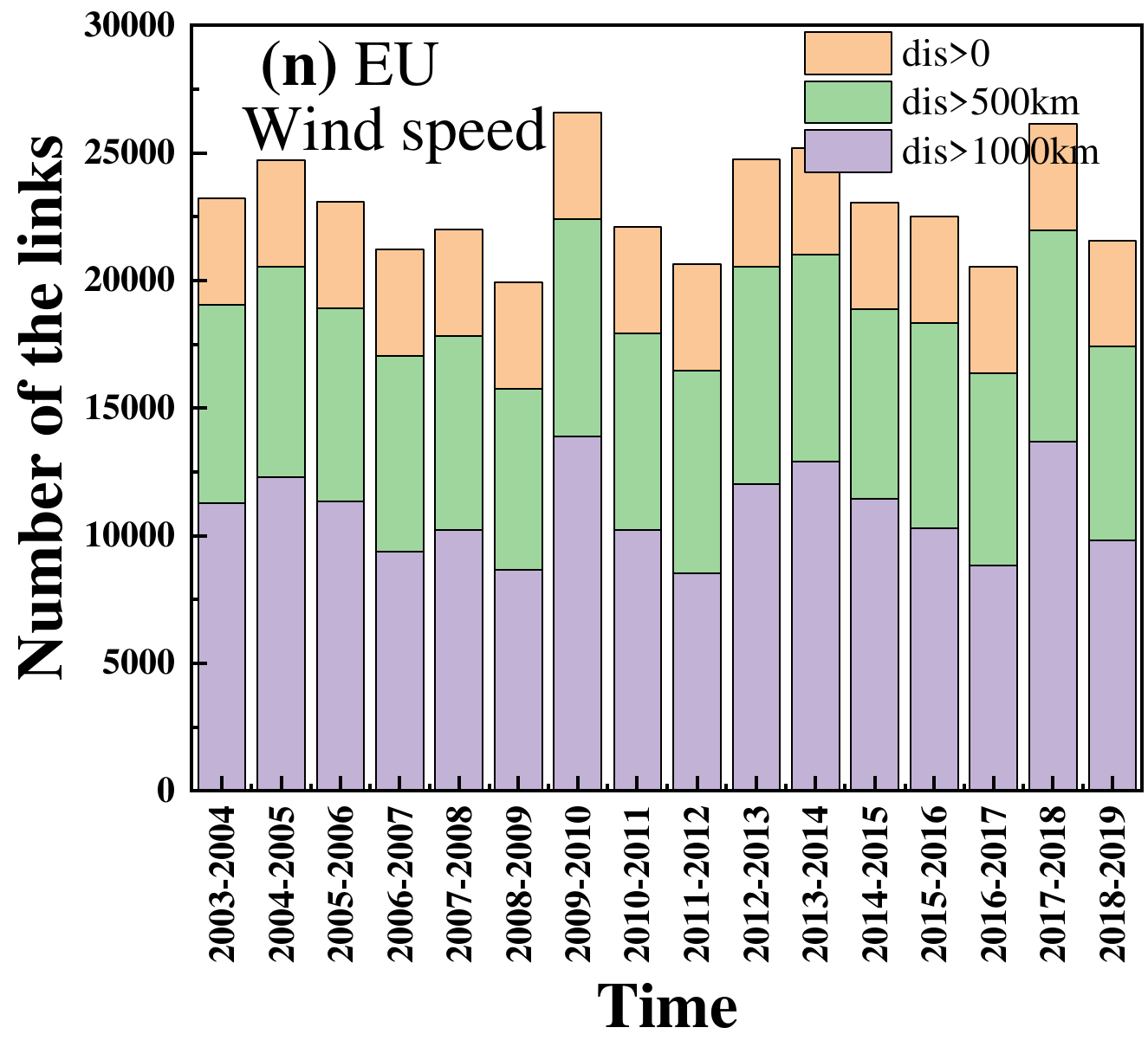}
\includegraphics[width=8em, height=7em]{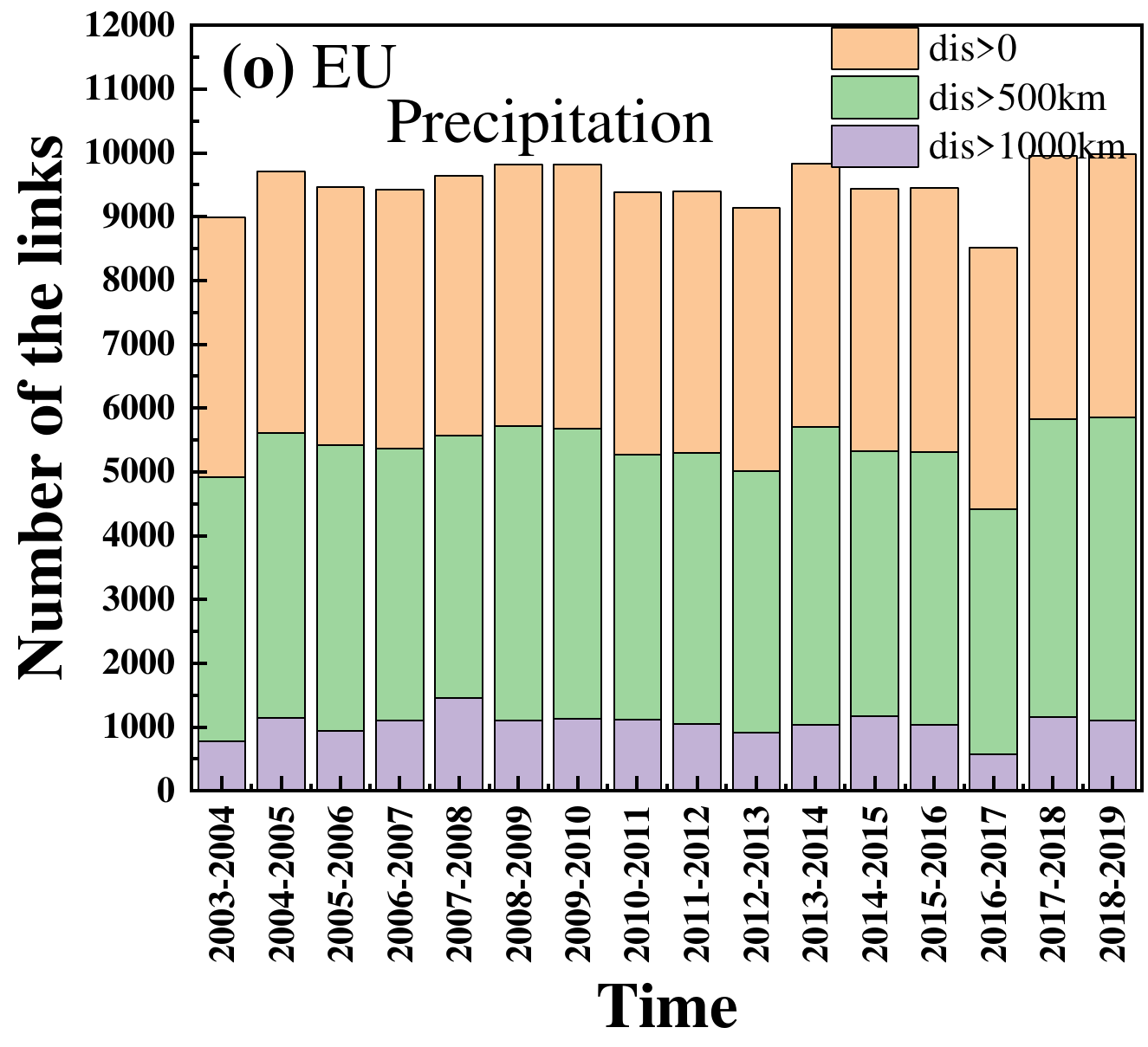}
\end{center}

\begin{center}
\noindent {\small {\bf Fig. S21} The number of significant links as a function of time in different networks.}
\end{center}

\begin{center}
\includegraphics[width=8em, height=7em]{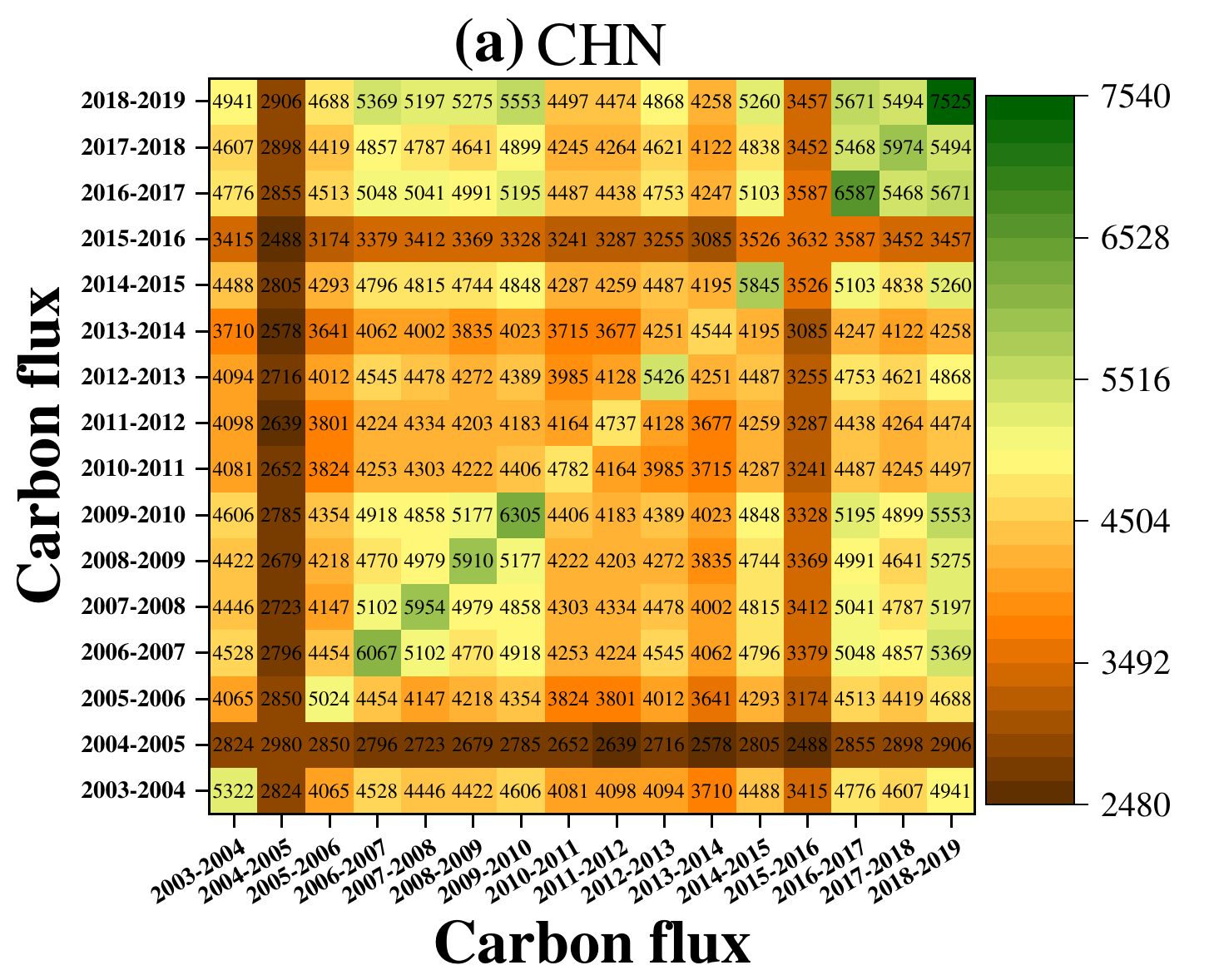}
\includegraphics[width=8em, height=7em]{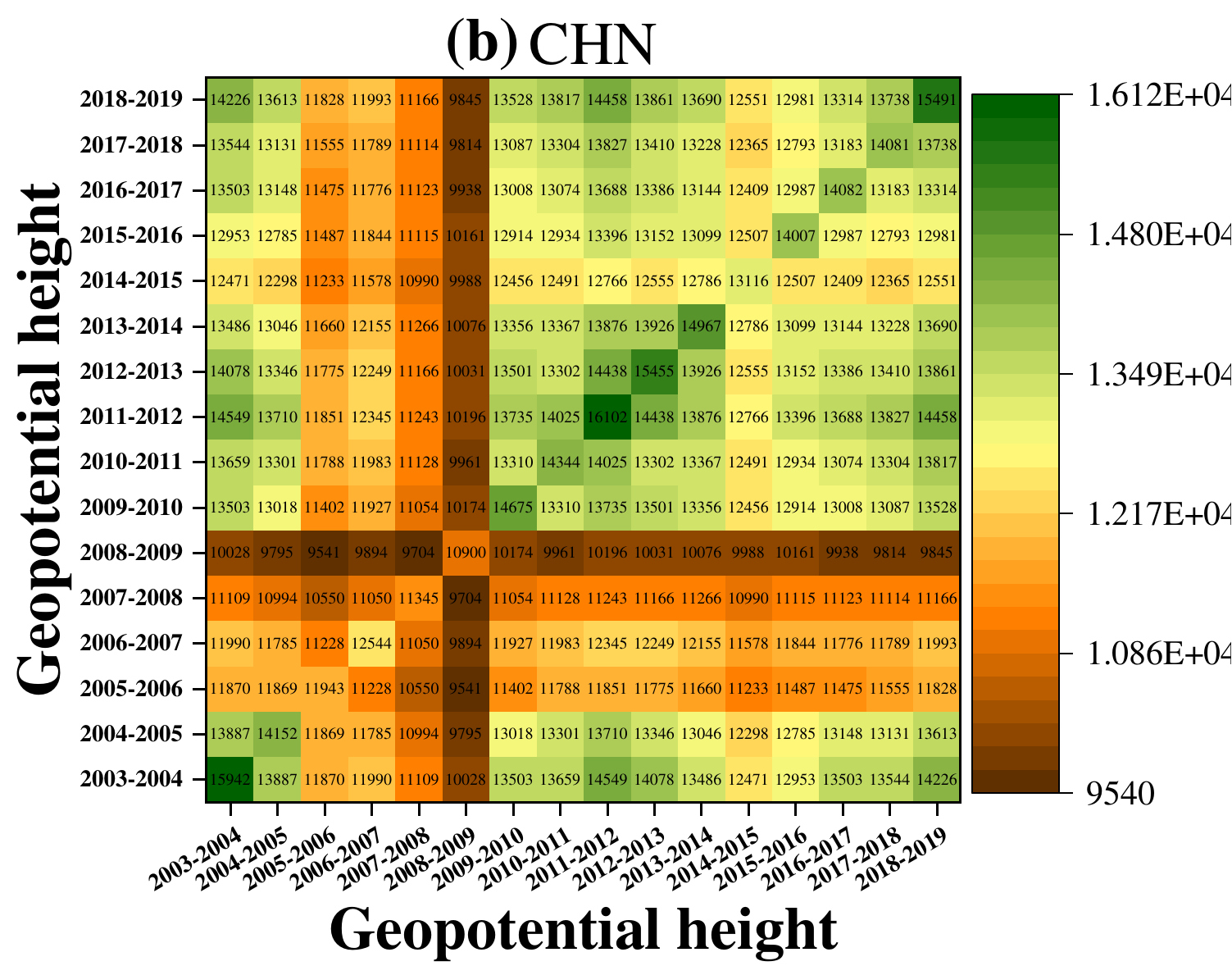}
\includegraphics[width=8em, height=7em]{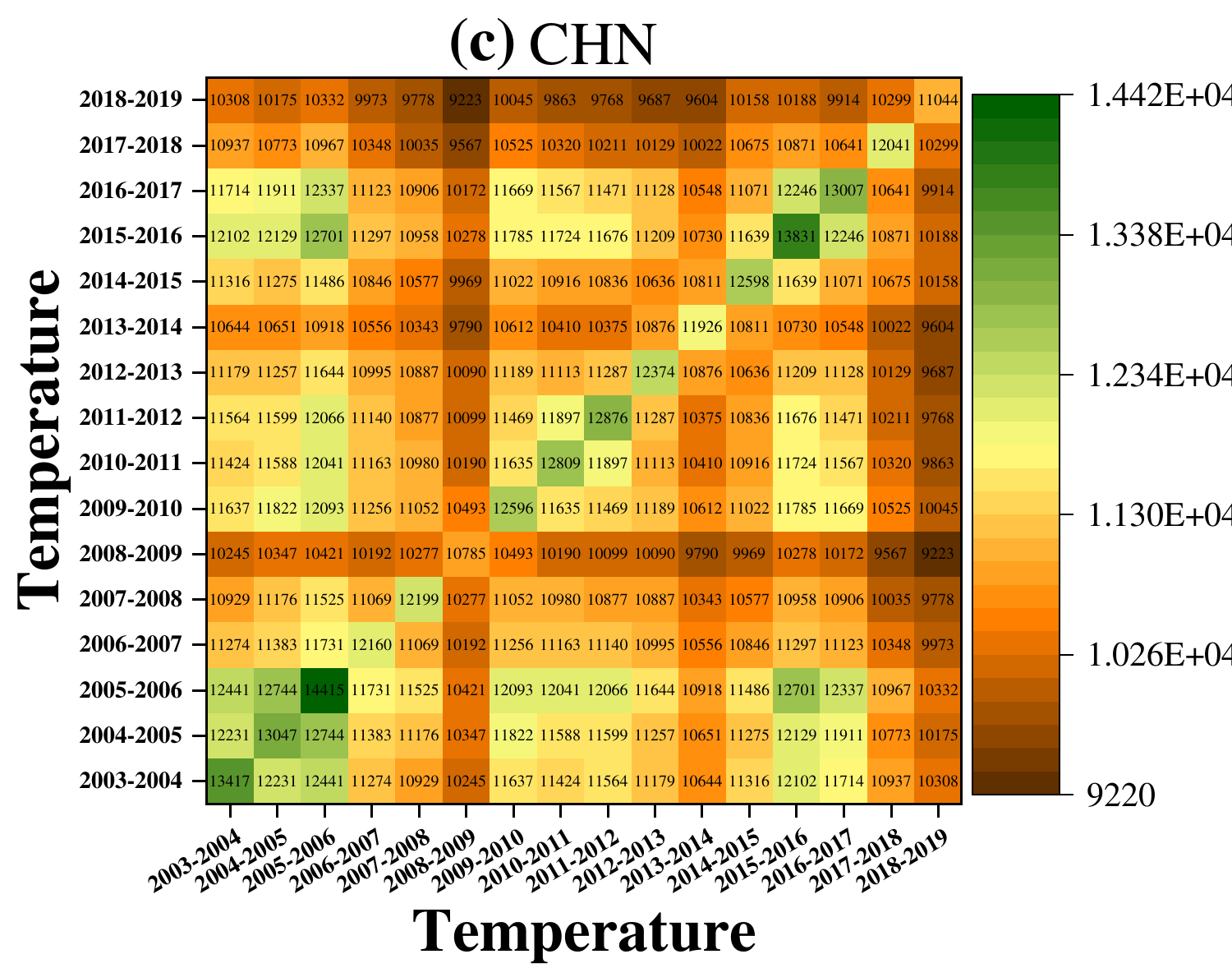}
\includegraphics[width=8em, height=7em]{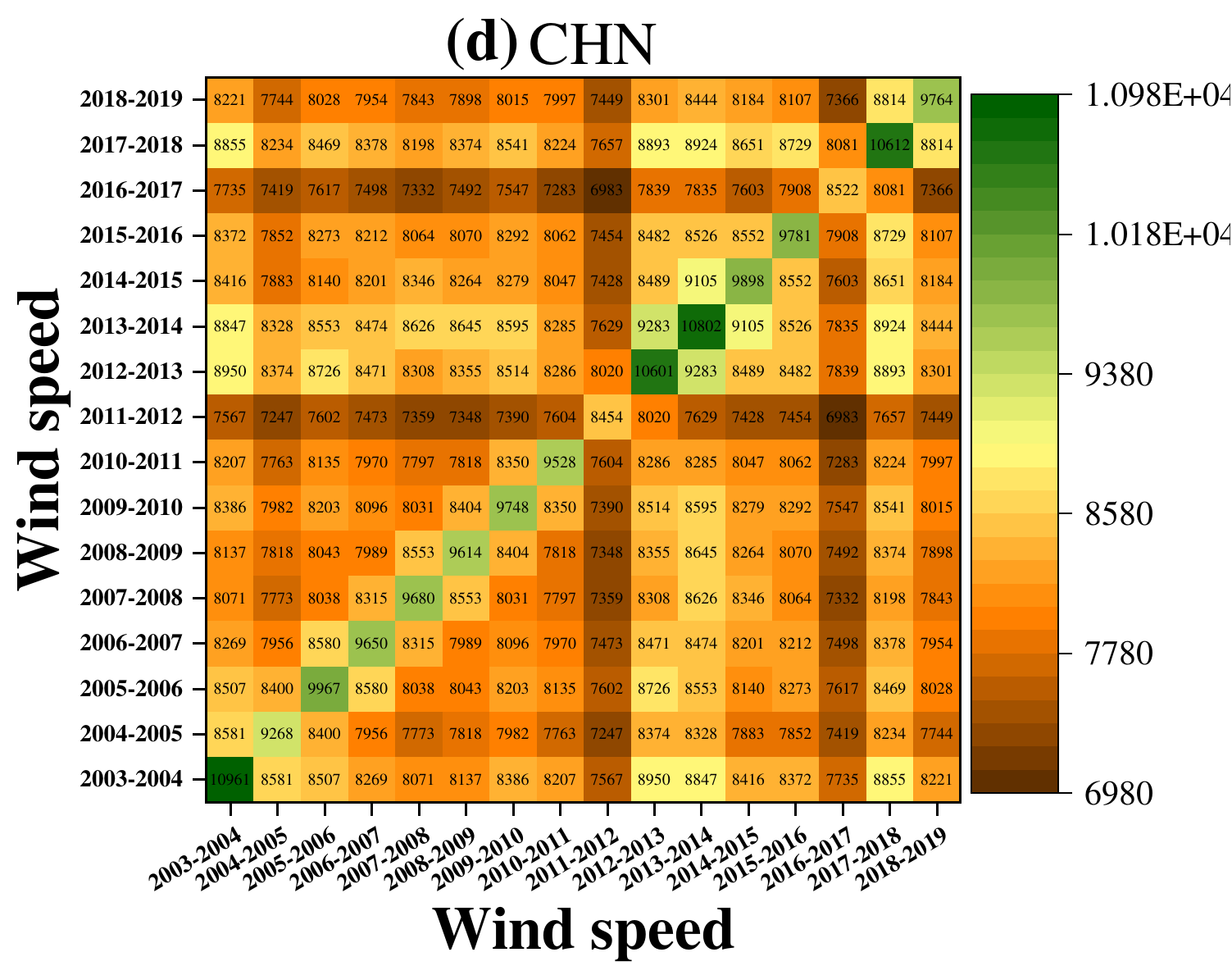}
\includegraphics[width=8em, height=7em]{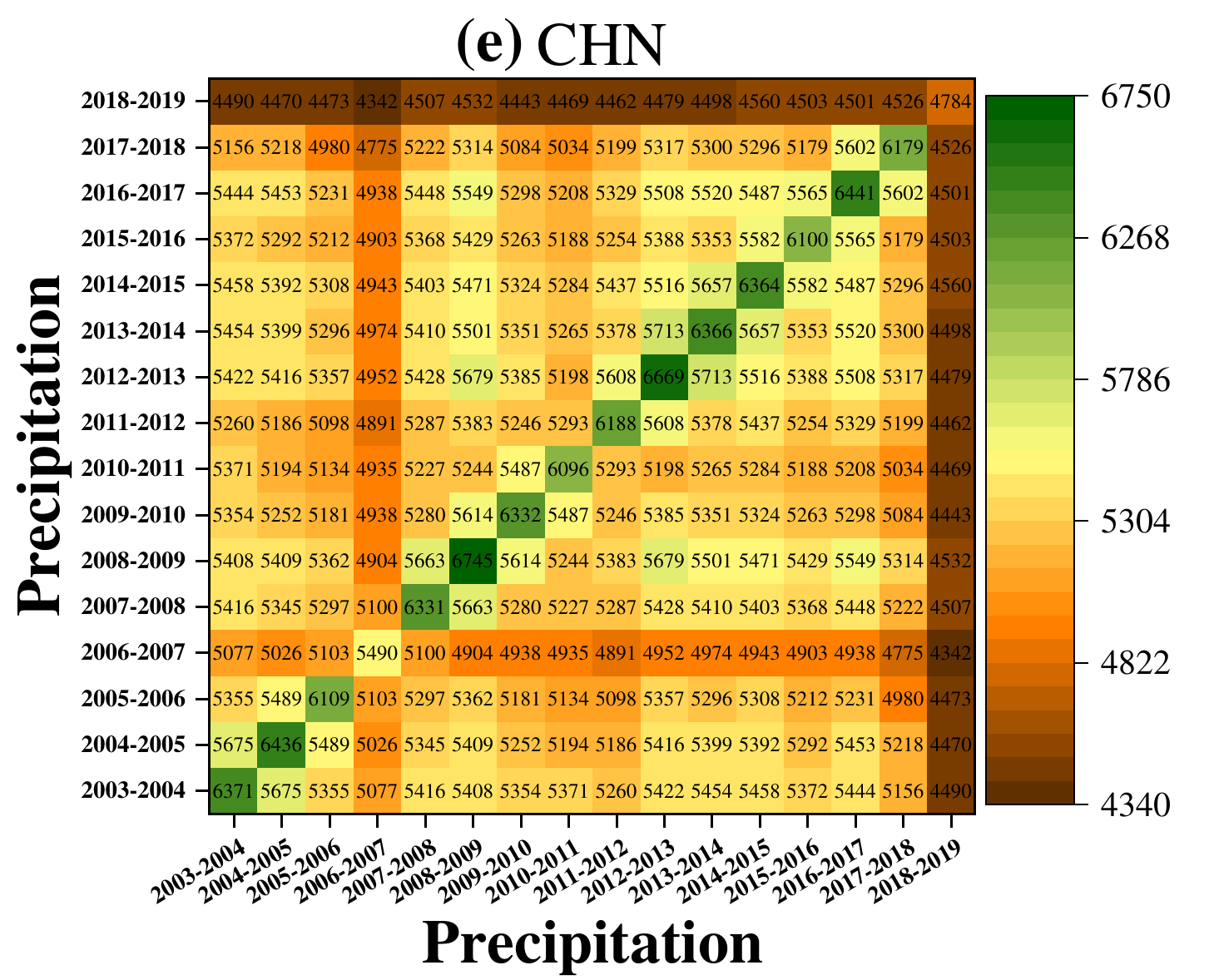}
\includegraphics[width=8em, height=7em]{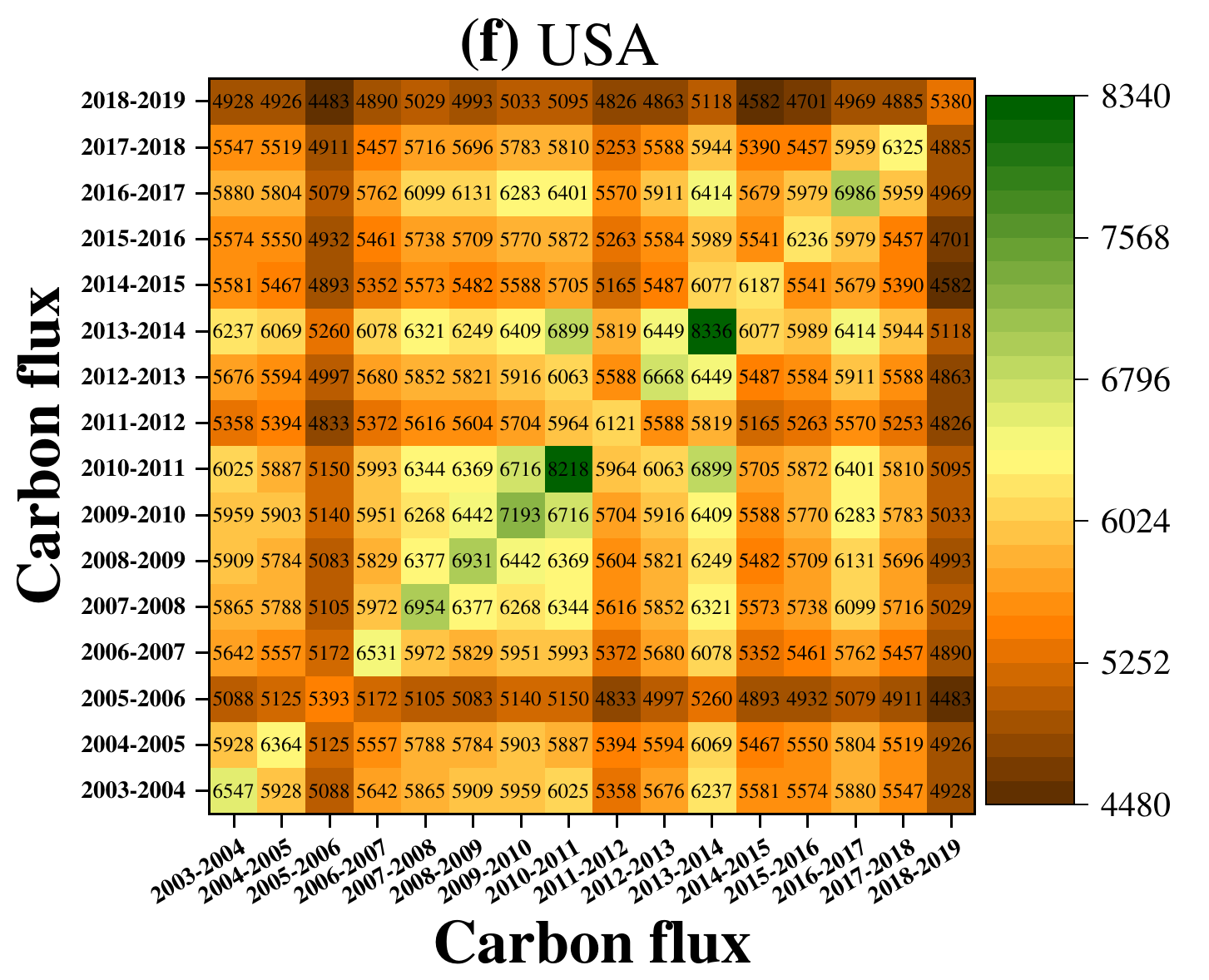}
\includegraphics[width=8em, height=7em]{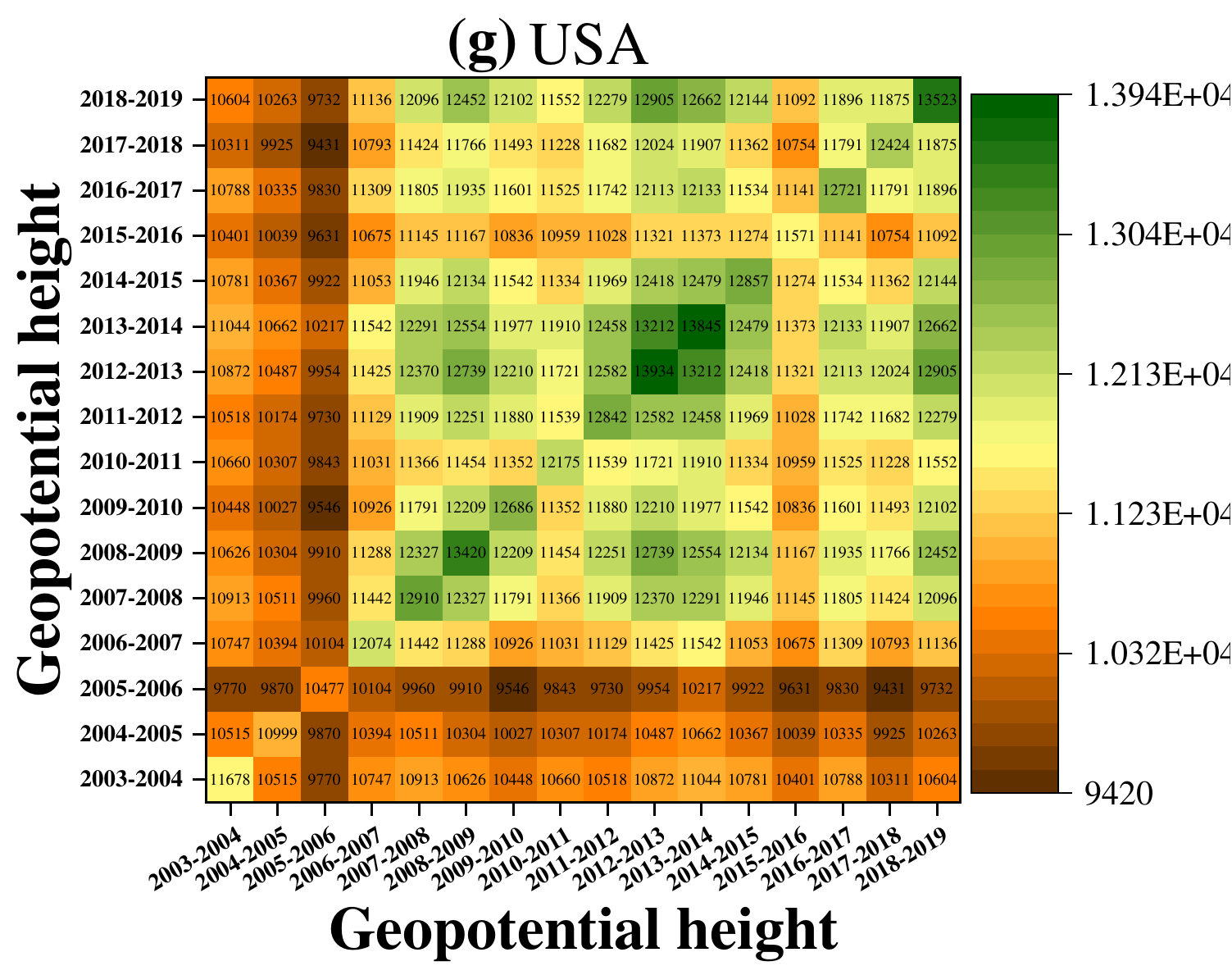}
\includegraphics[width=8em, height=7em]{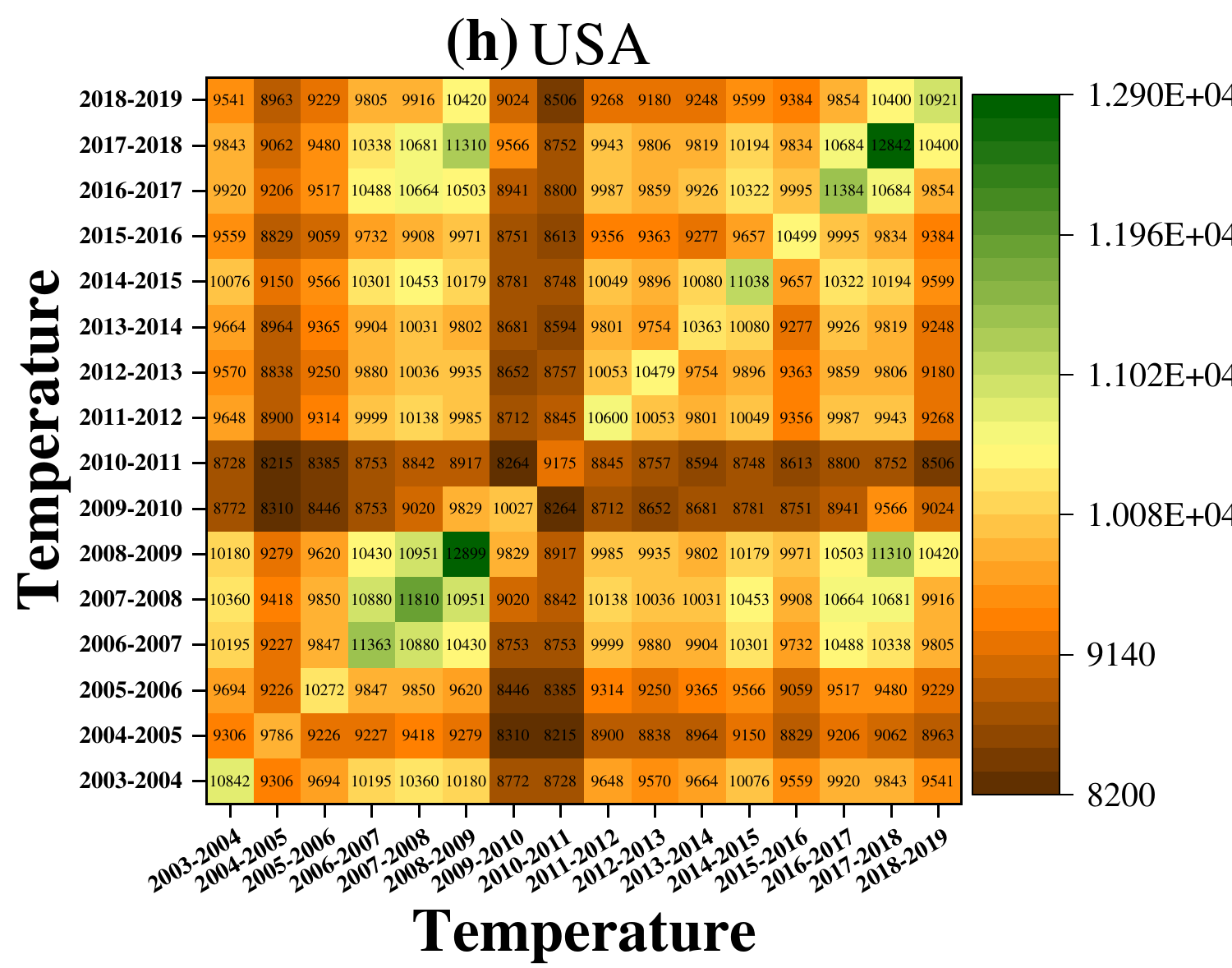}
\includegraphics[width=8em, height=7em]{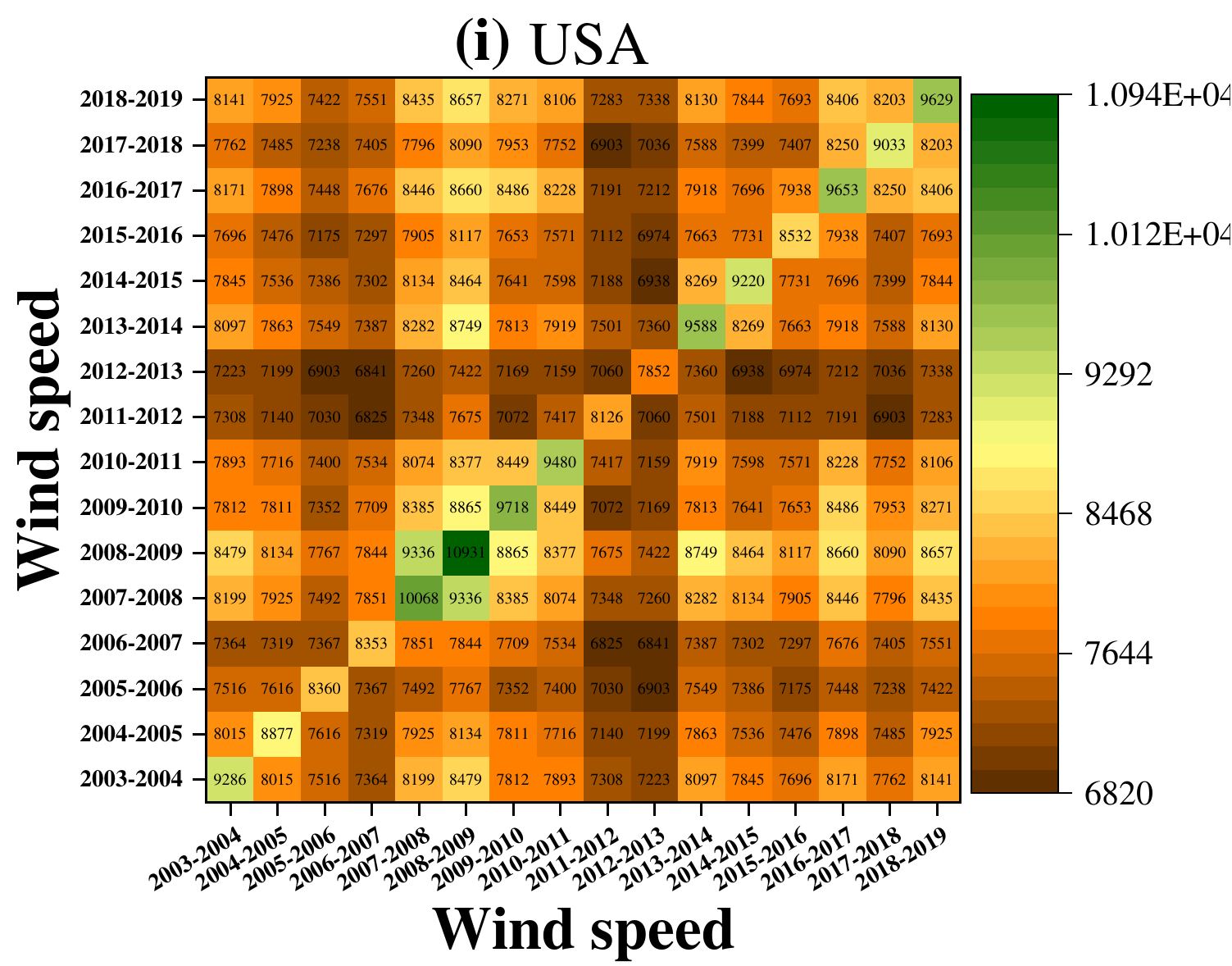}
\includegraphics[width=8em, height=7em]{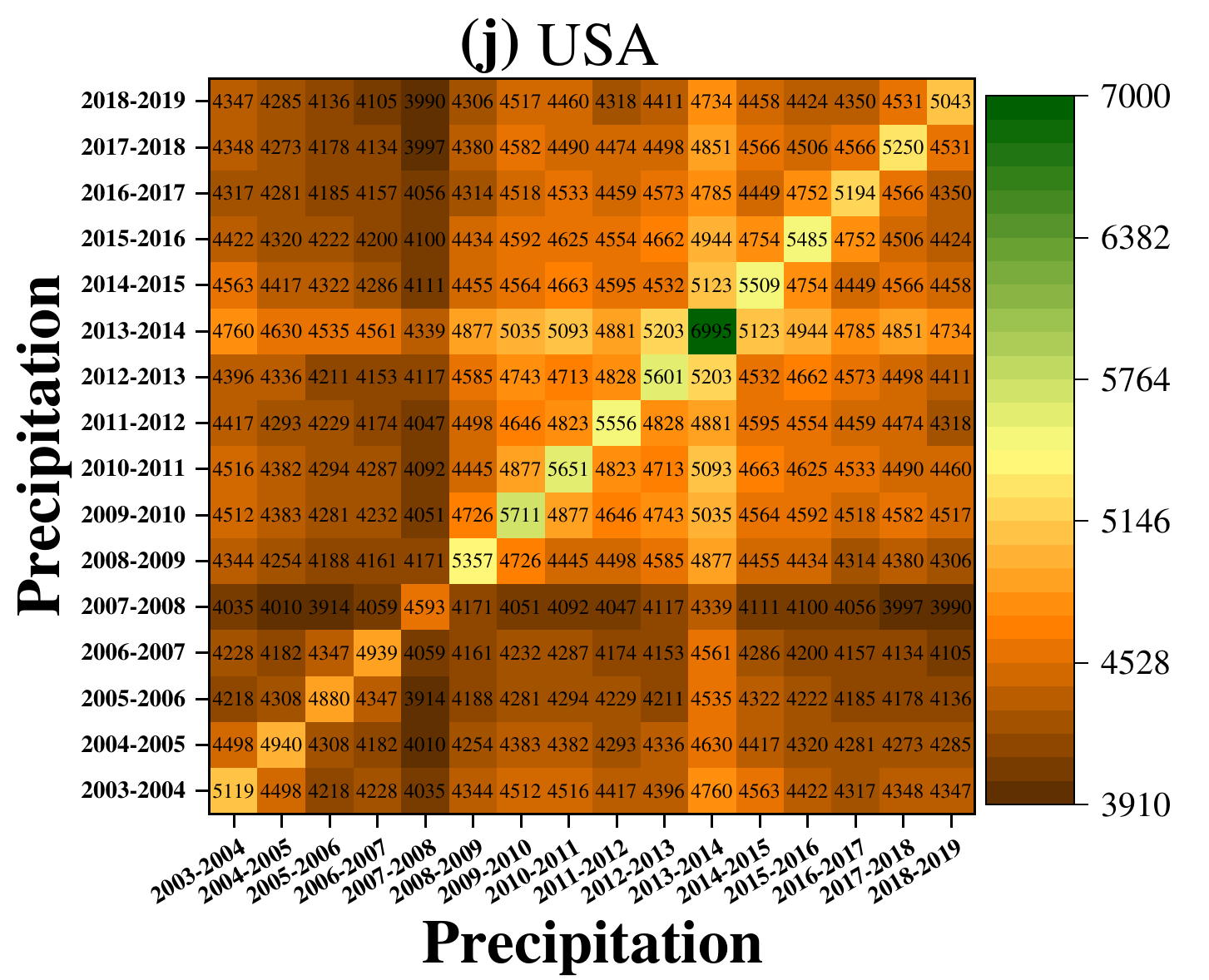}
\includegraphics[width=8em, height=7em]{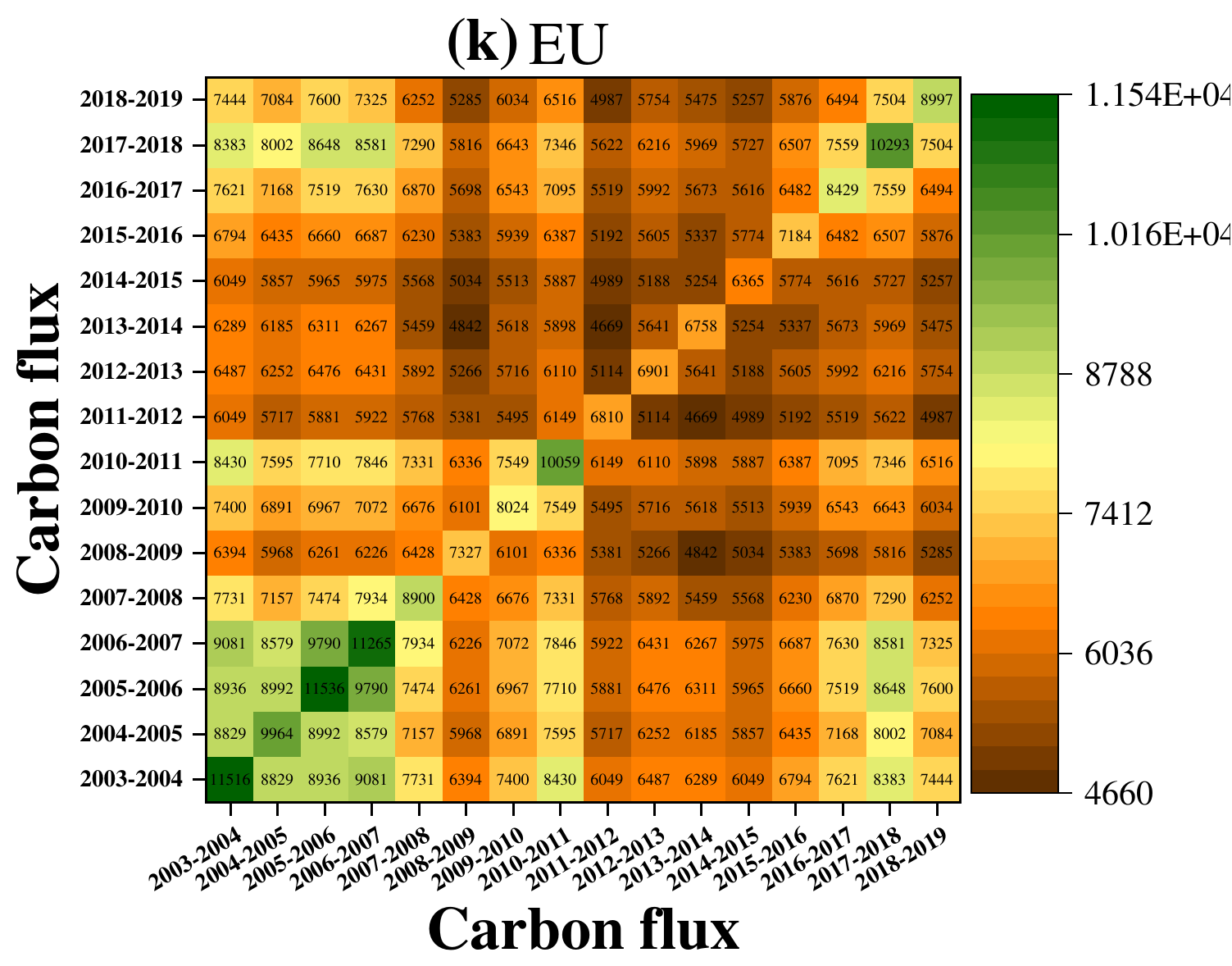}
\includegraphics[width=8em, height=7em]{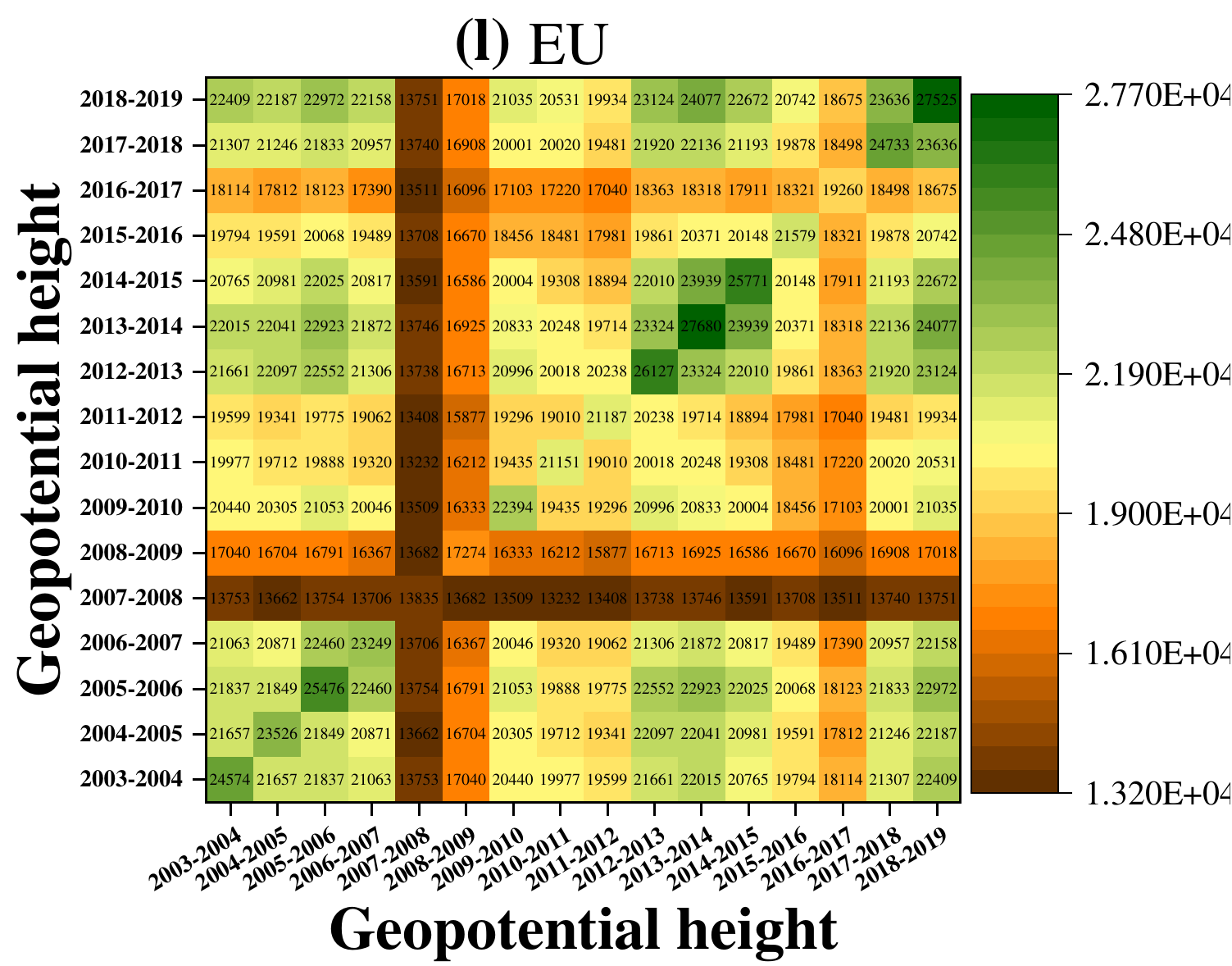}
\includegraphics[width=8em, height=7em]{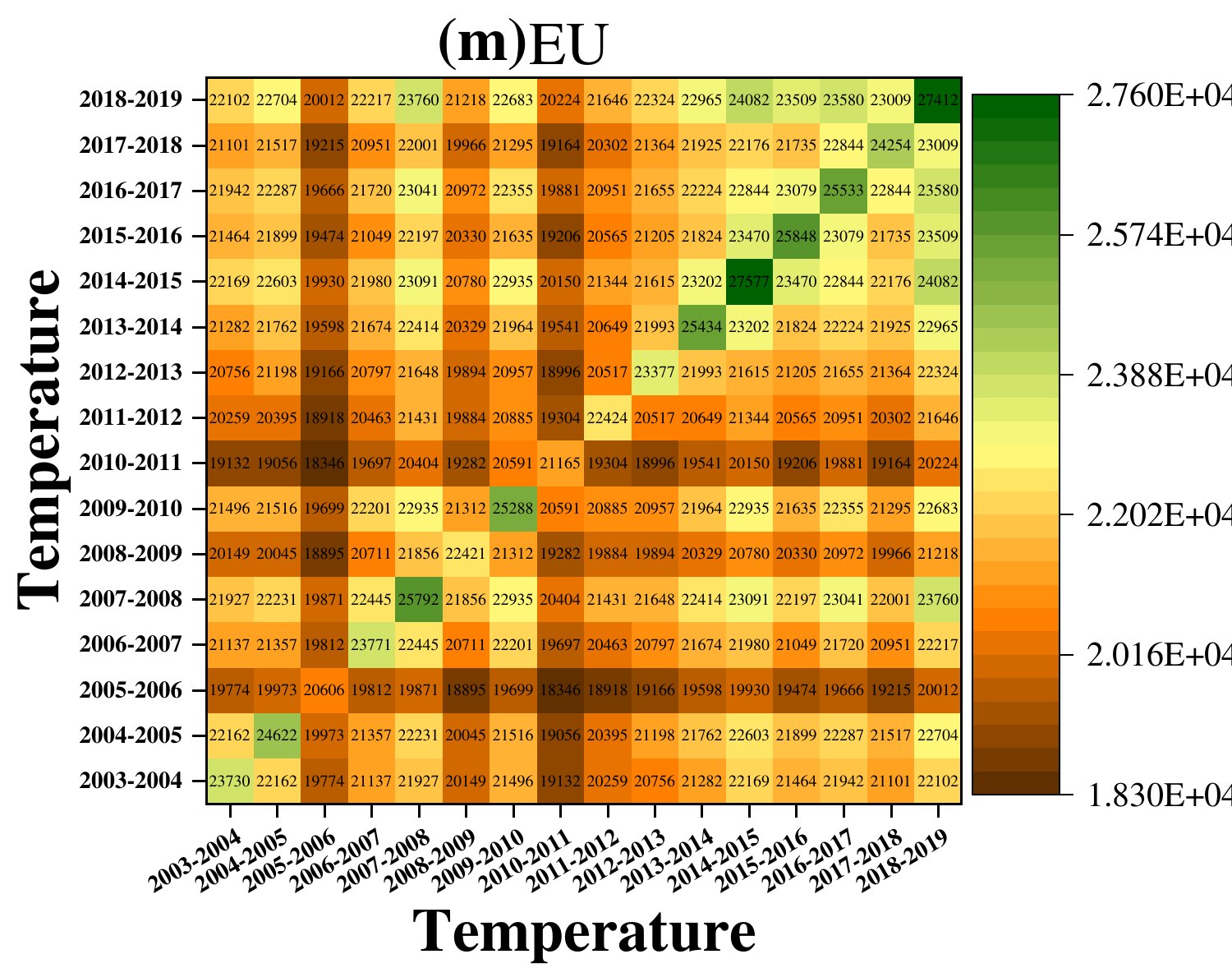}
\includegraphics[width=8em, height=7em]{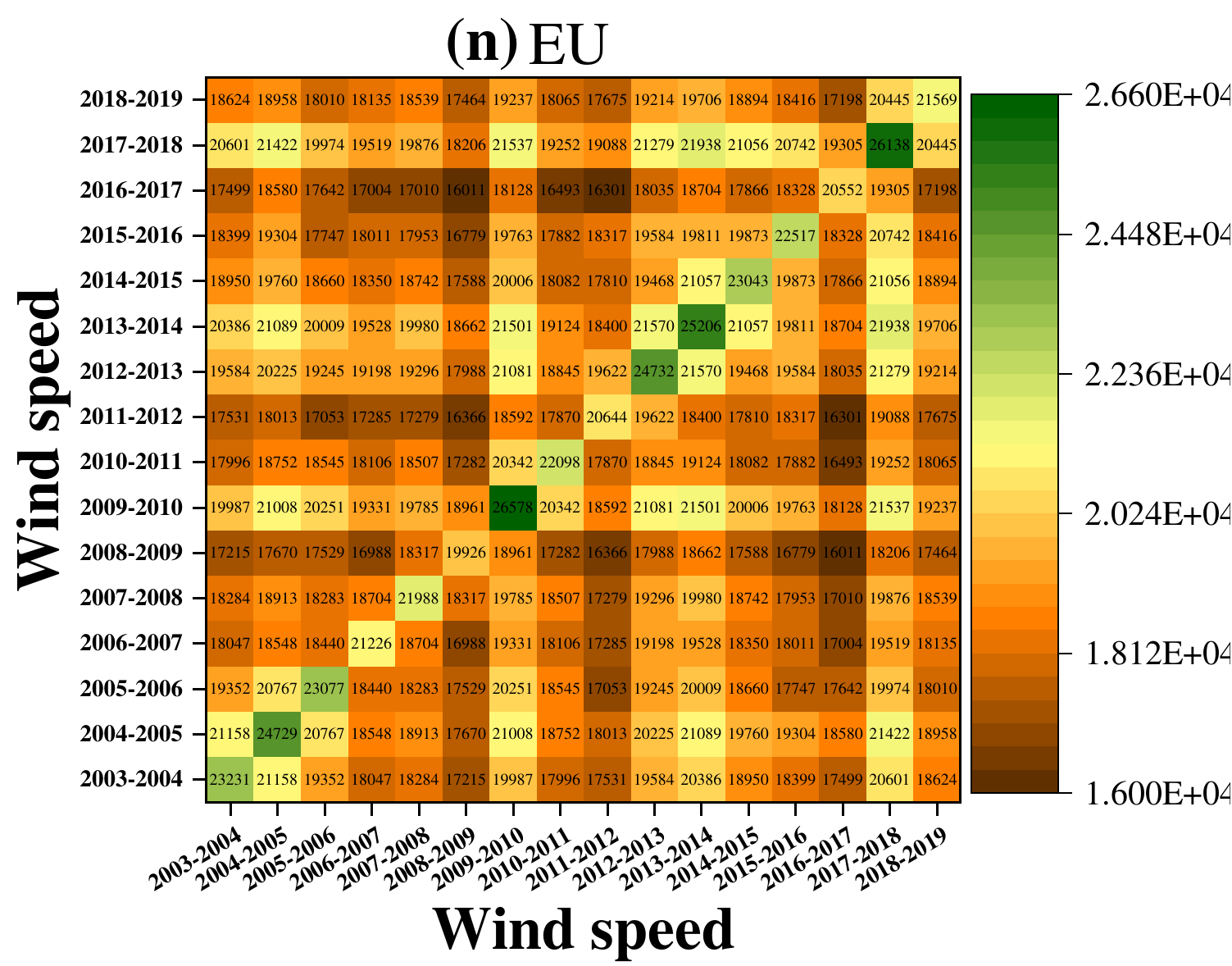}
\includegraphics[width=8em, height=7em]{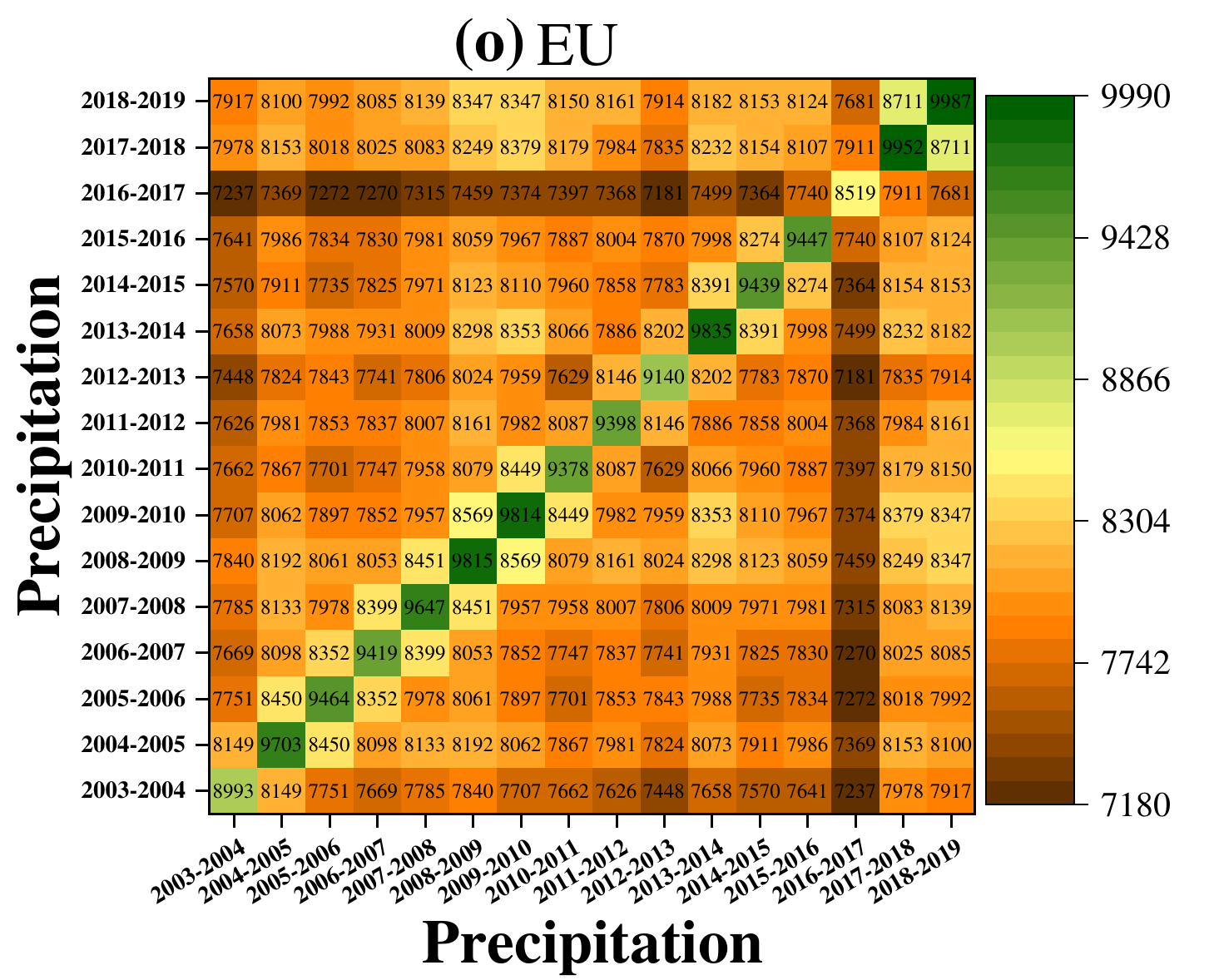}
\end{center}

\begin{center}
\noindent {\small {\bf Fig. S22} The matrix of the intersection of the links of each two yearly networks for different climate variables. Each matrix element represents the number of link intersections in both networks.}
\end{center}

\begin{center}
\includegraphics[width=8em, height=7em]{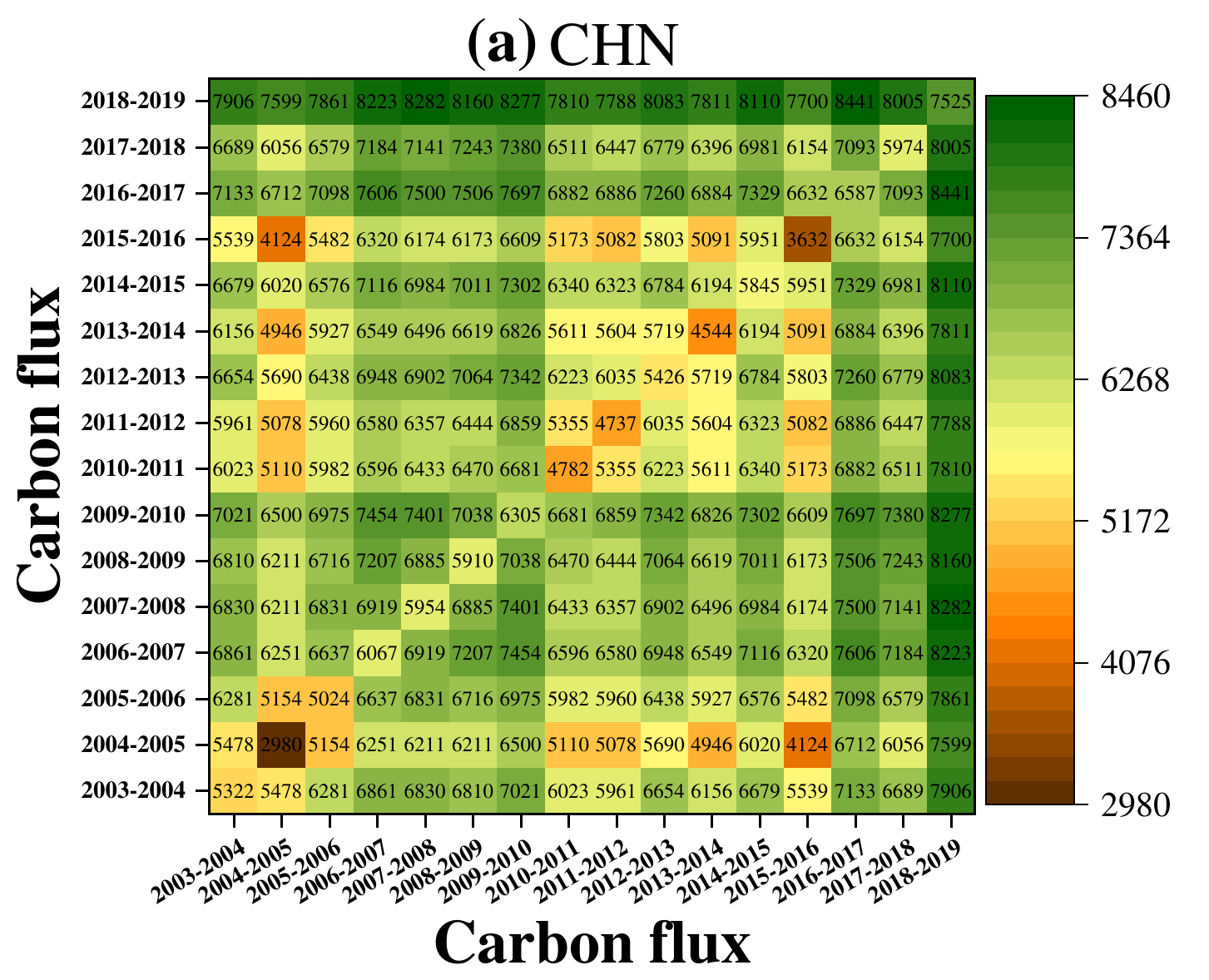}
\includegraphics[width=8em, height=7em]{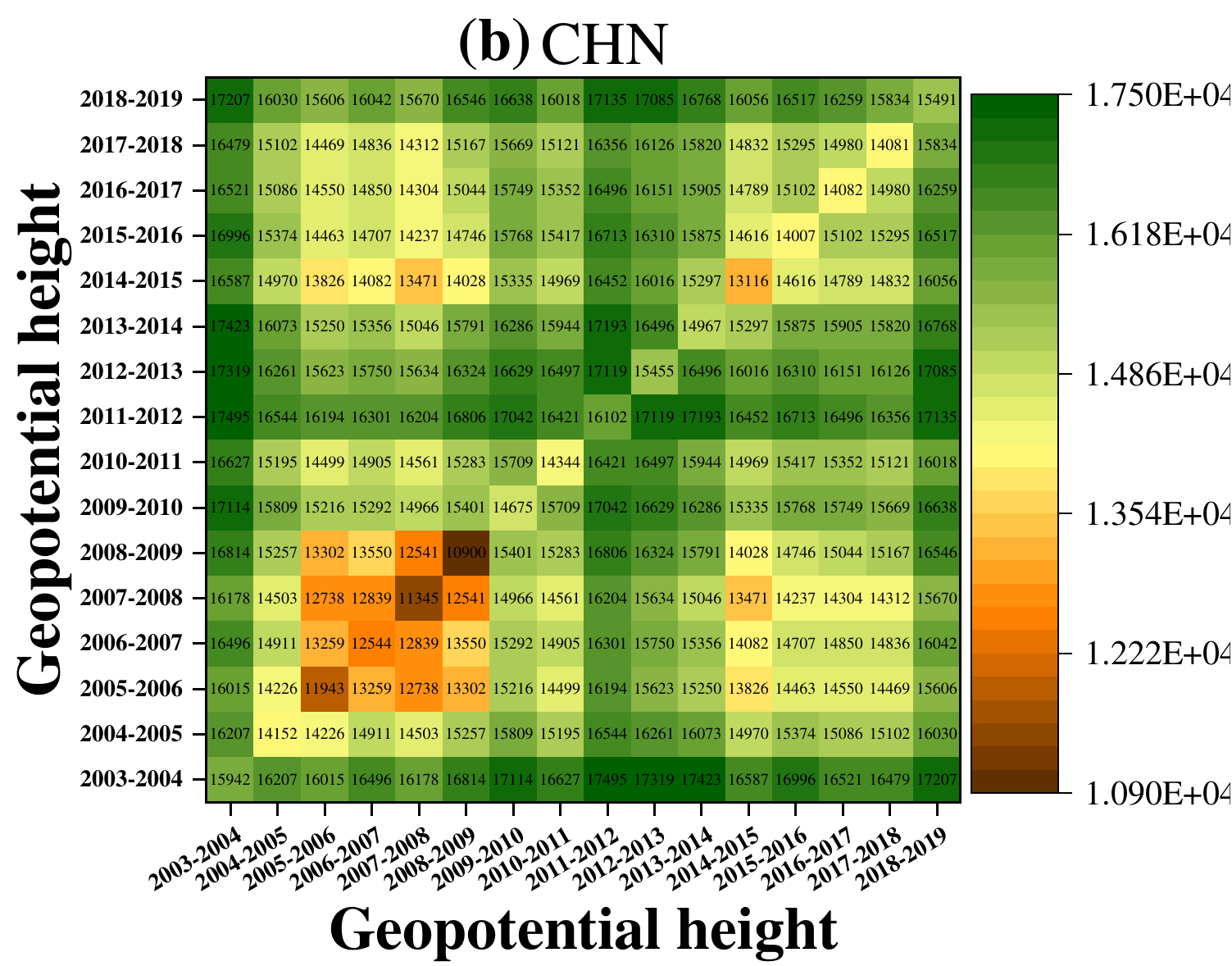}
\includegraphics[width=8em, height=7em]{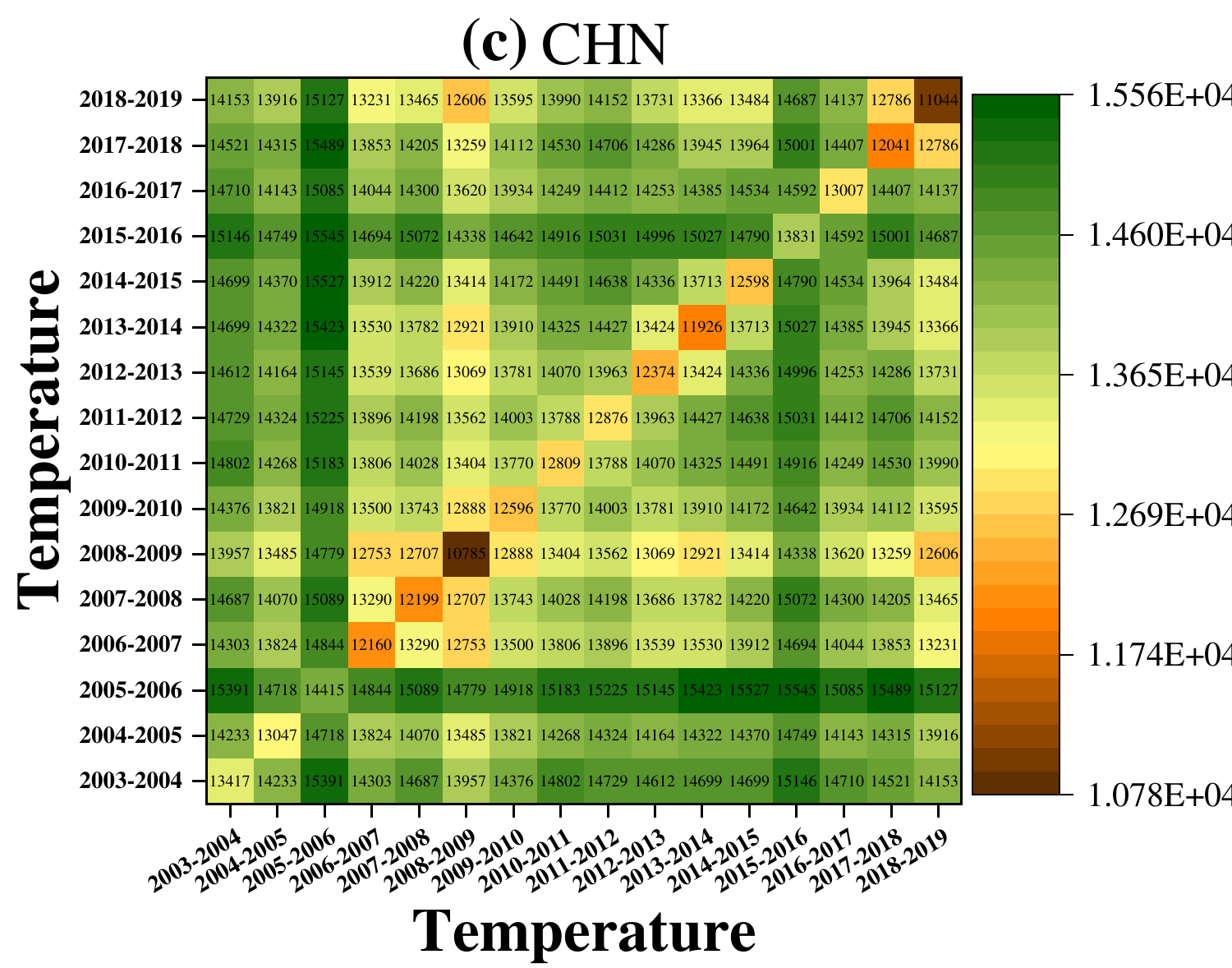}
\includegraphics[width=8em, height=7em]{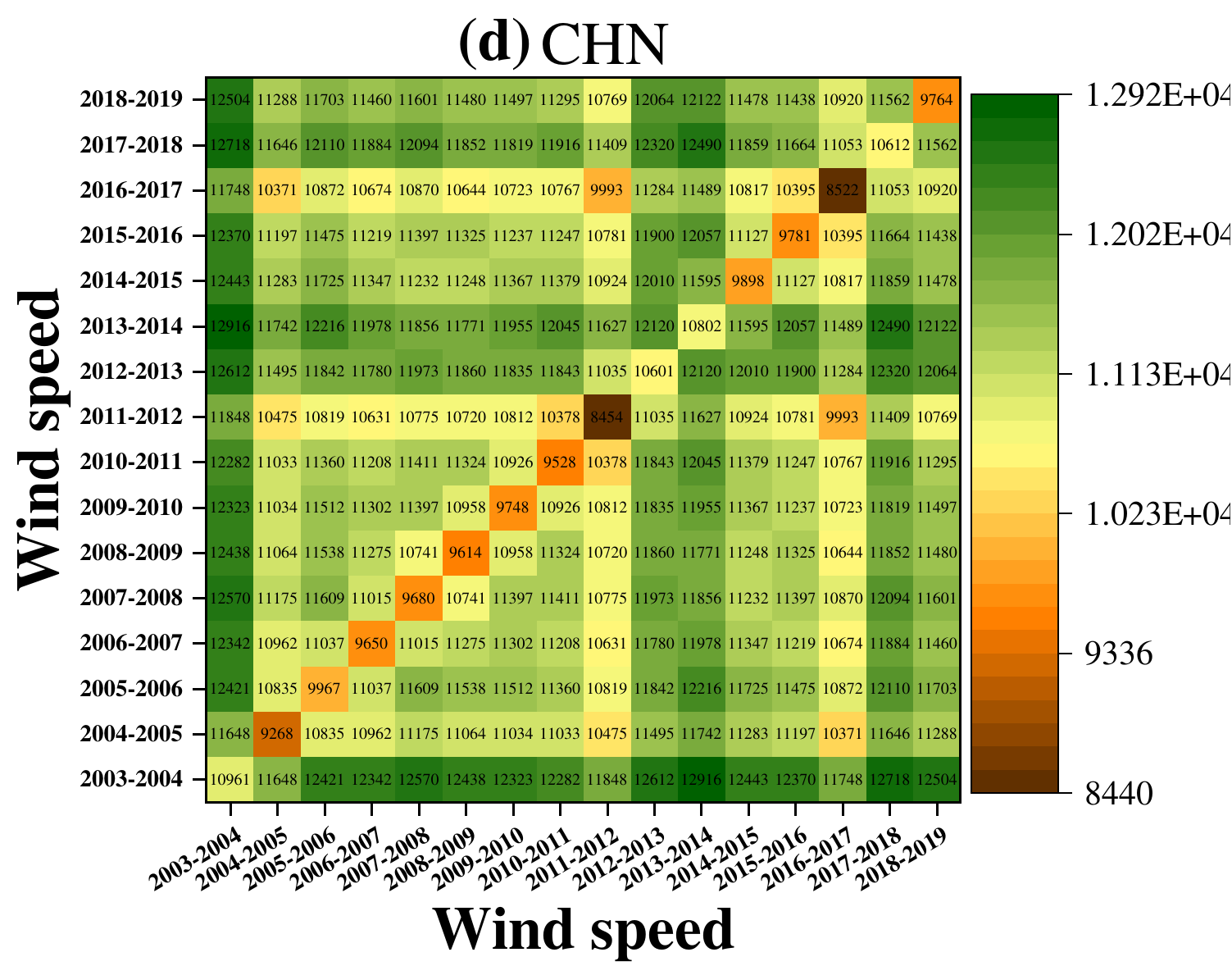}
\includegraphics[width=8em, height=7em]{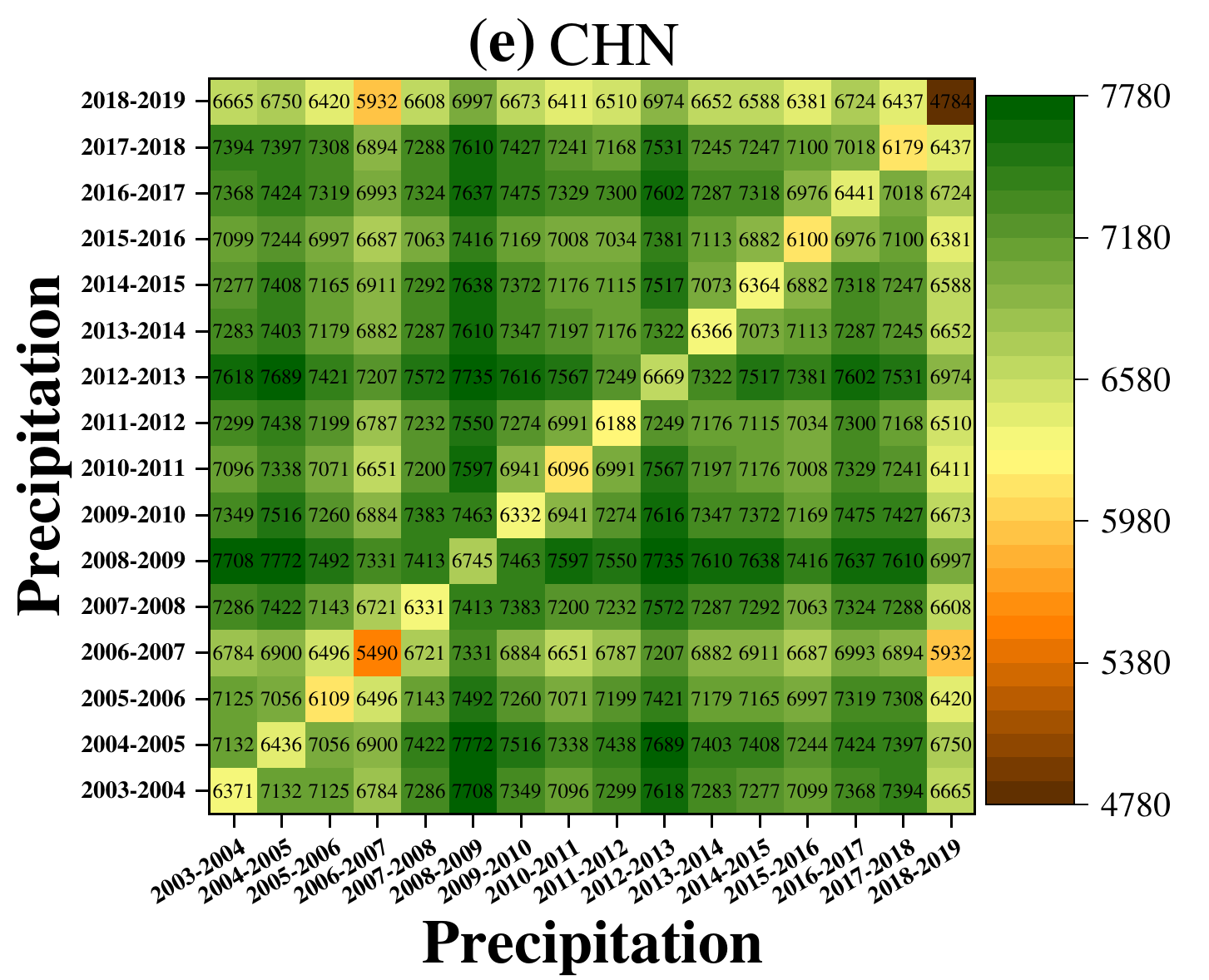}
\includegraphics[width=8em, height=7em]{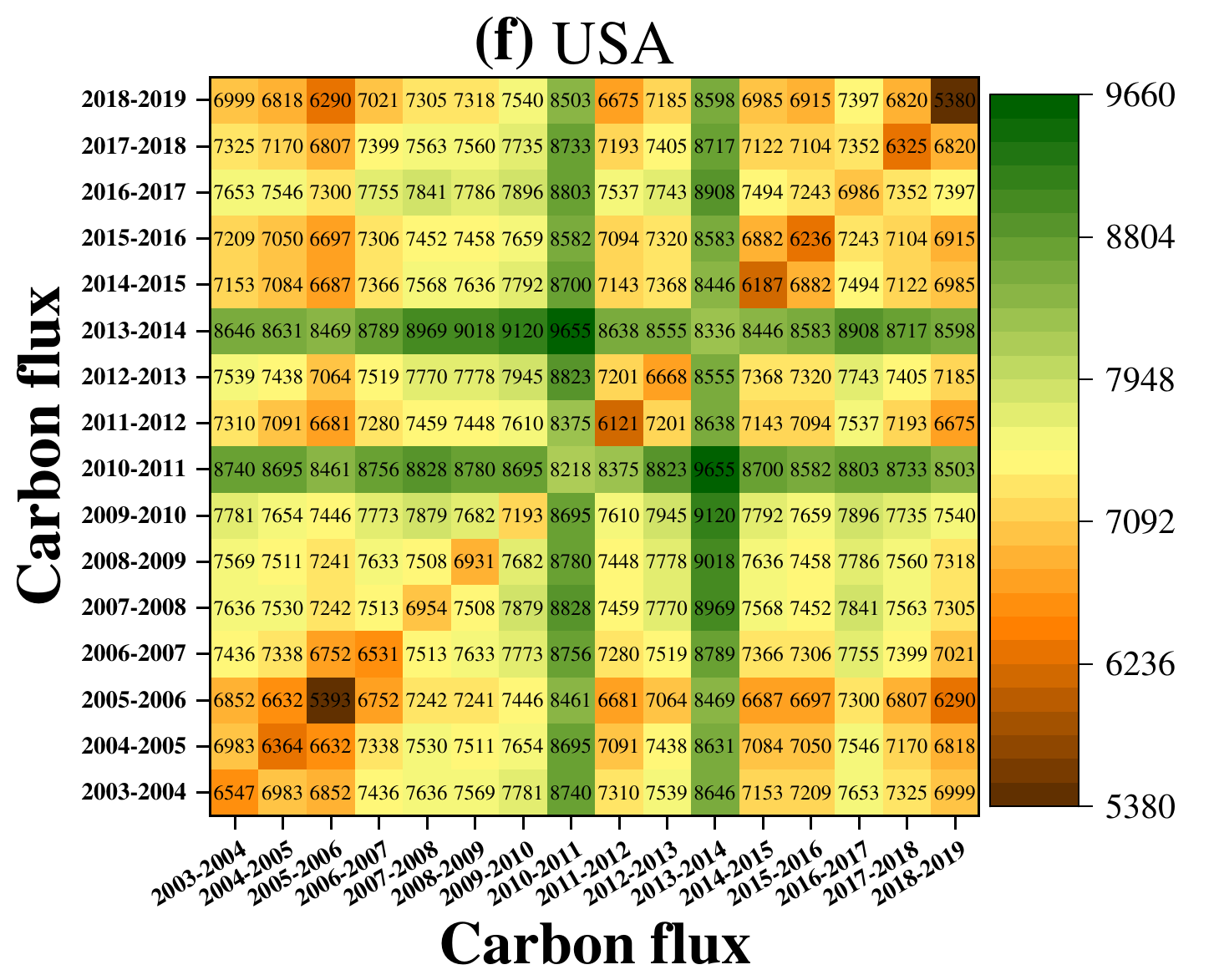}
\includegraphics[width=8em, height=7em]{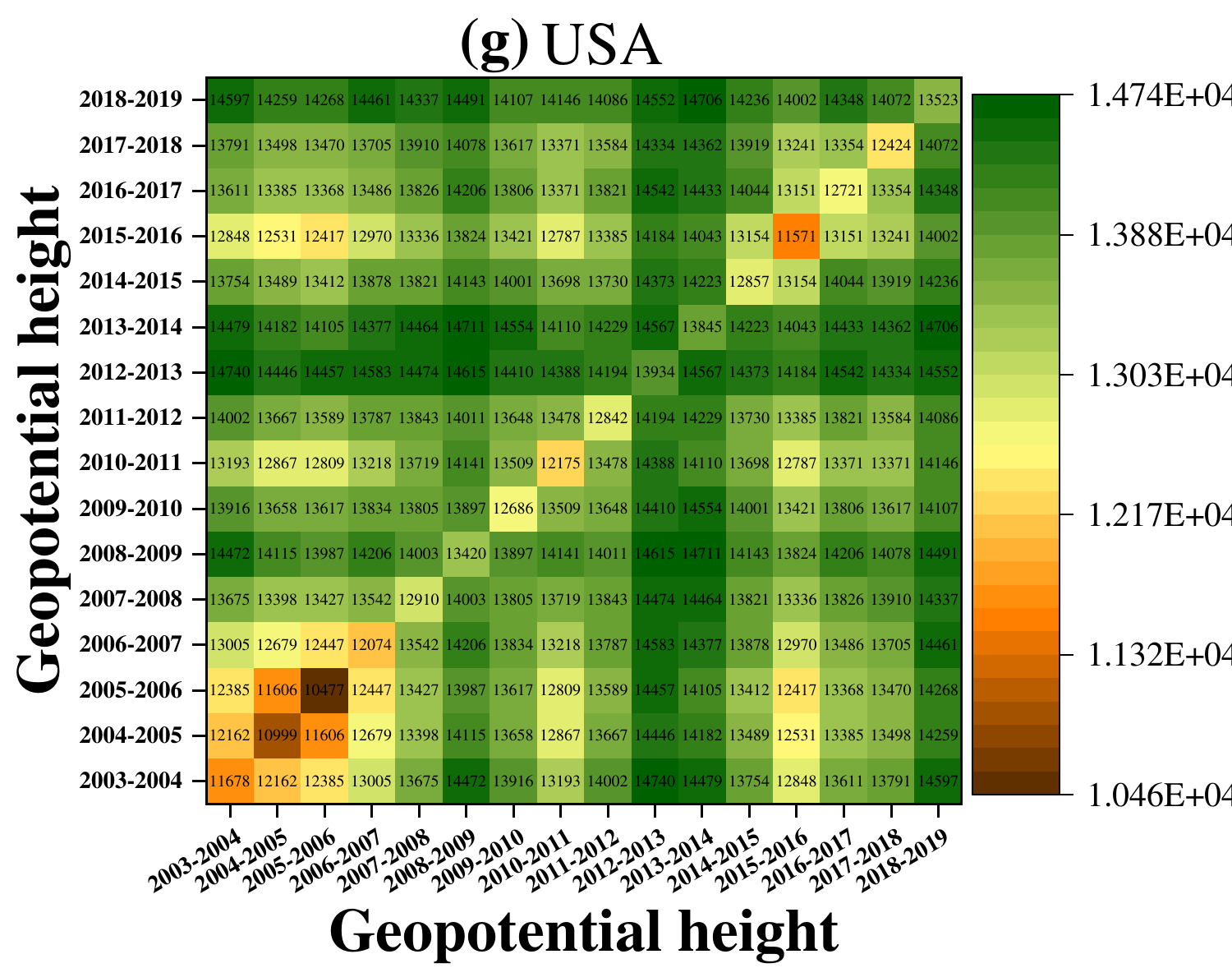}
\includegraphics[width=8em, height=7em]{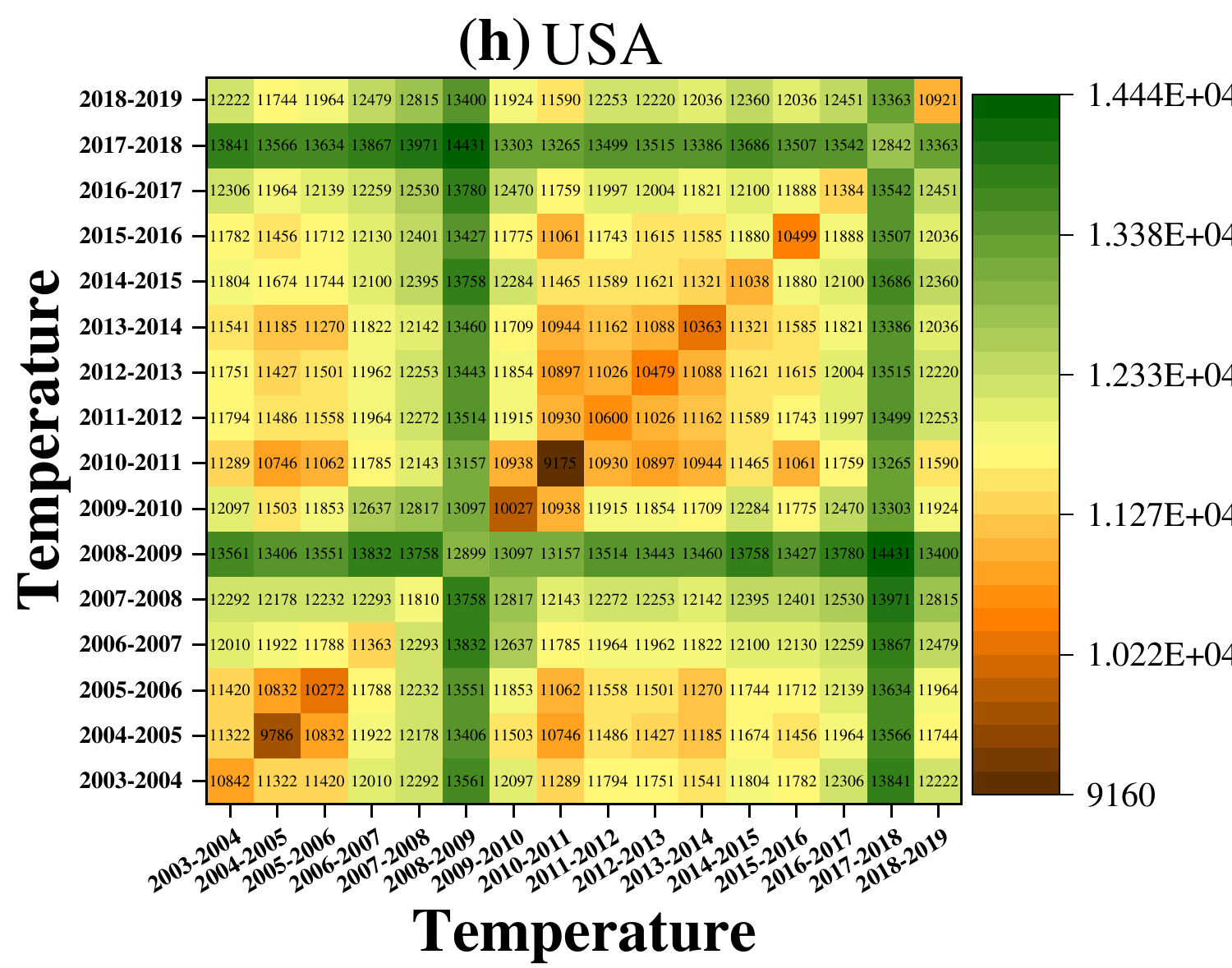}
\includegraphics[width=8em, height=7em]{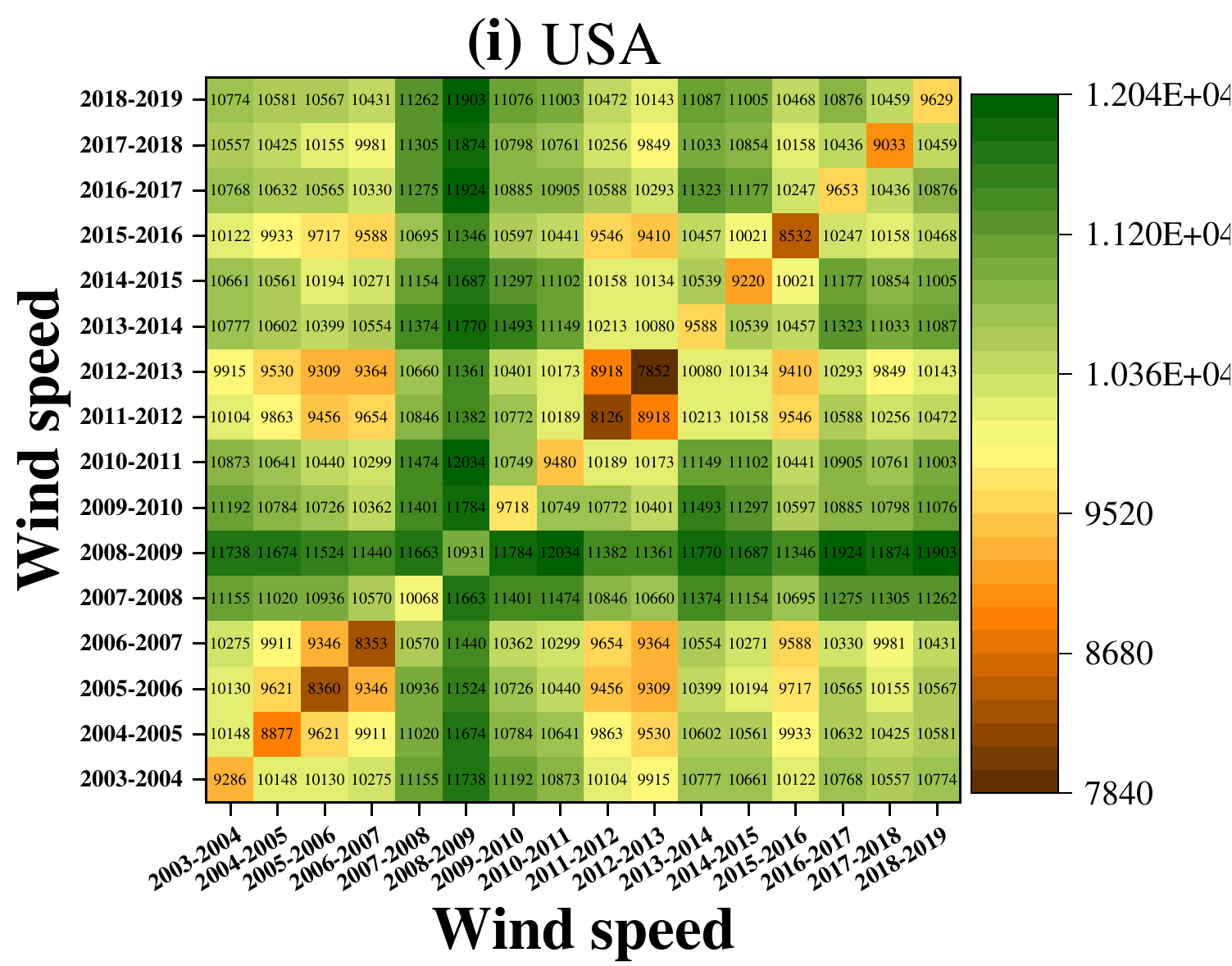}
\includegraphics[width=8em, height=7em]{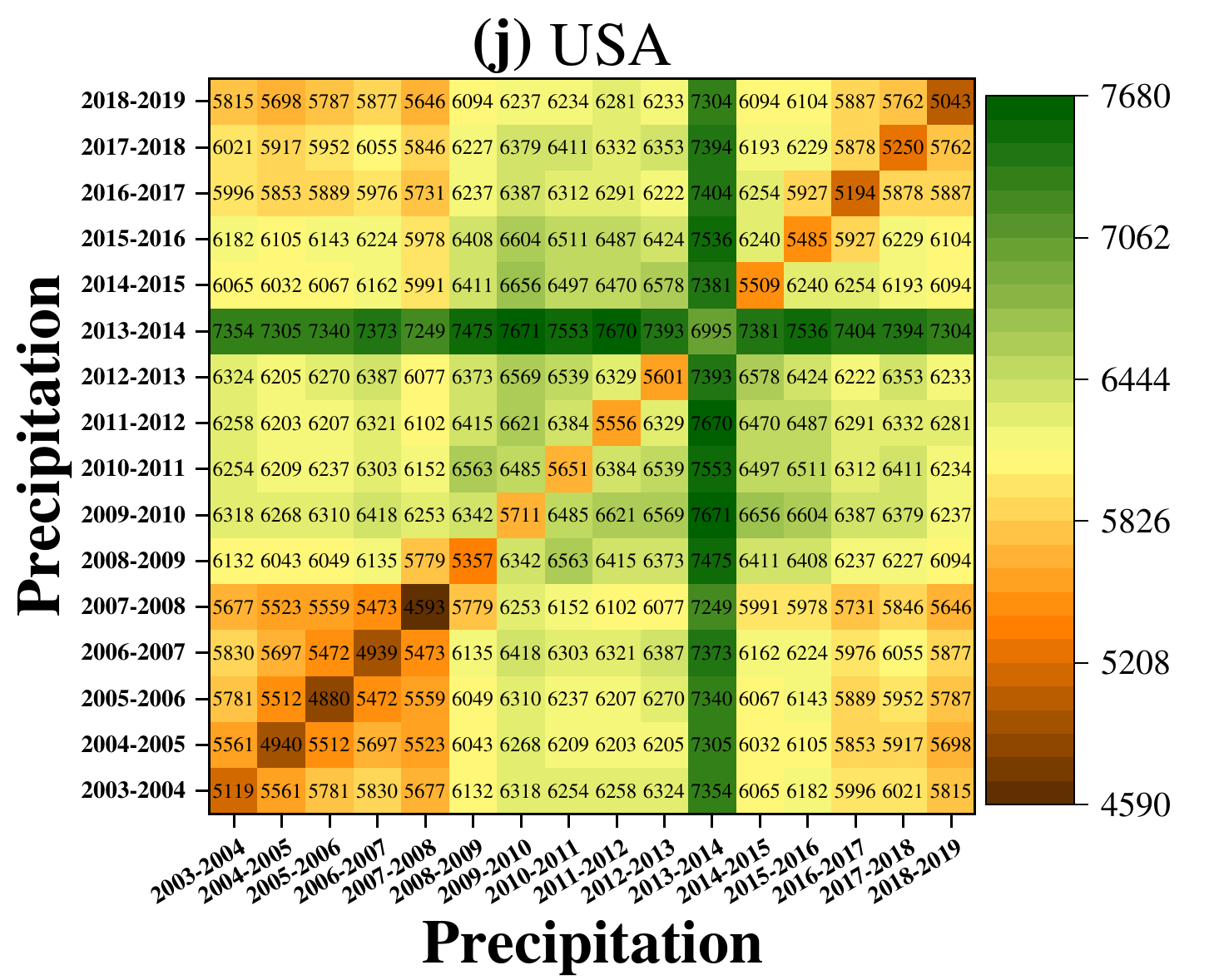}
\includegraphics[width=8em, height=7em]{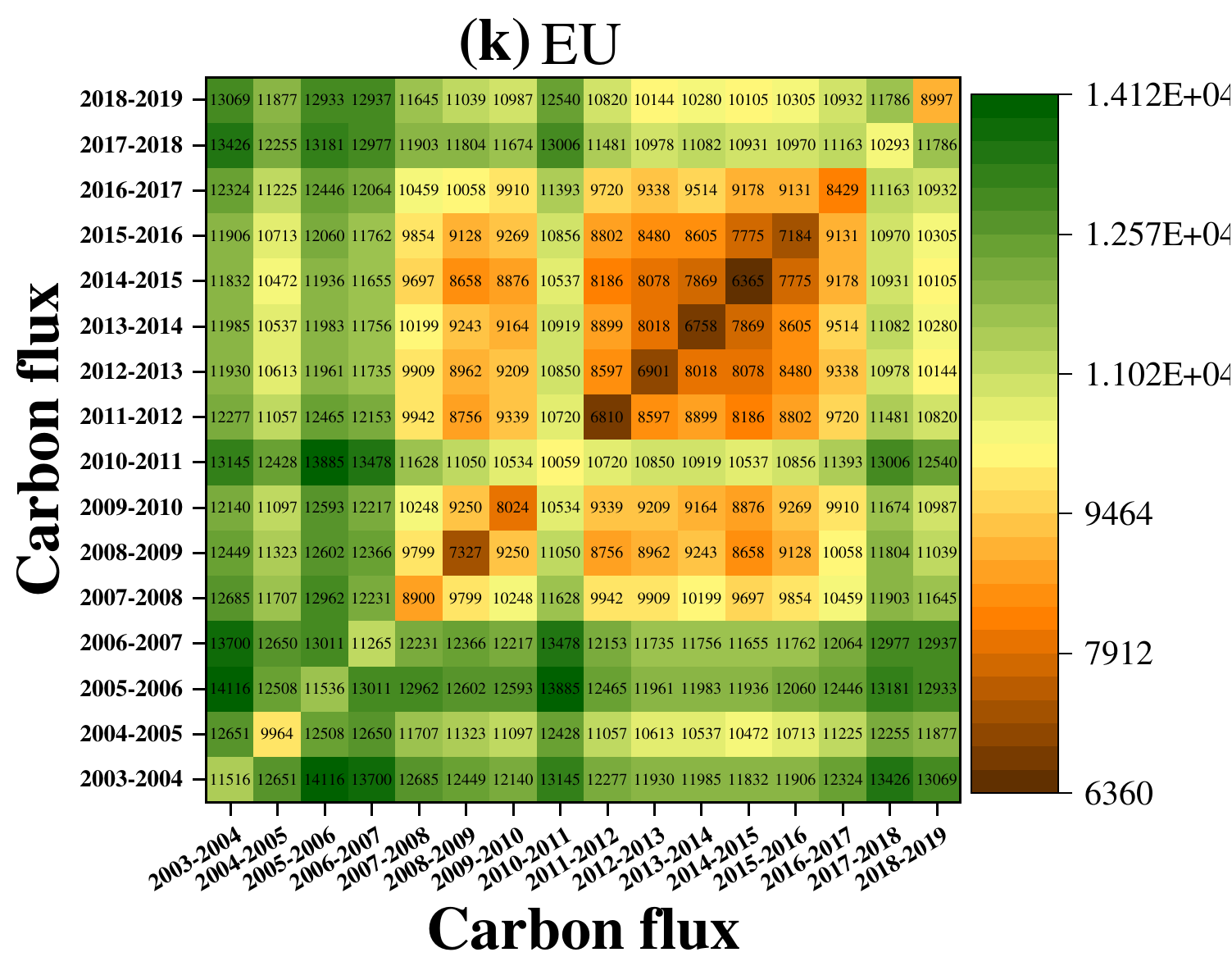}
\includegraphics[width=8em, height=7em]{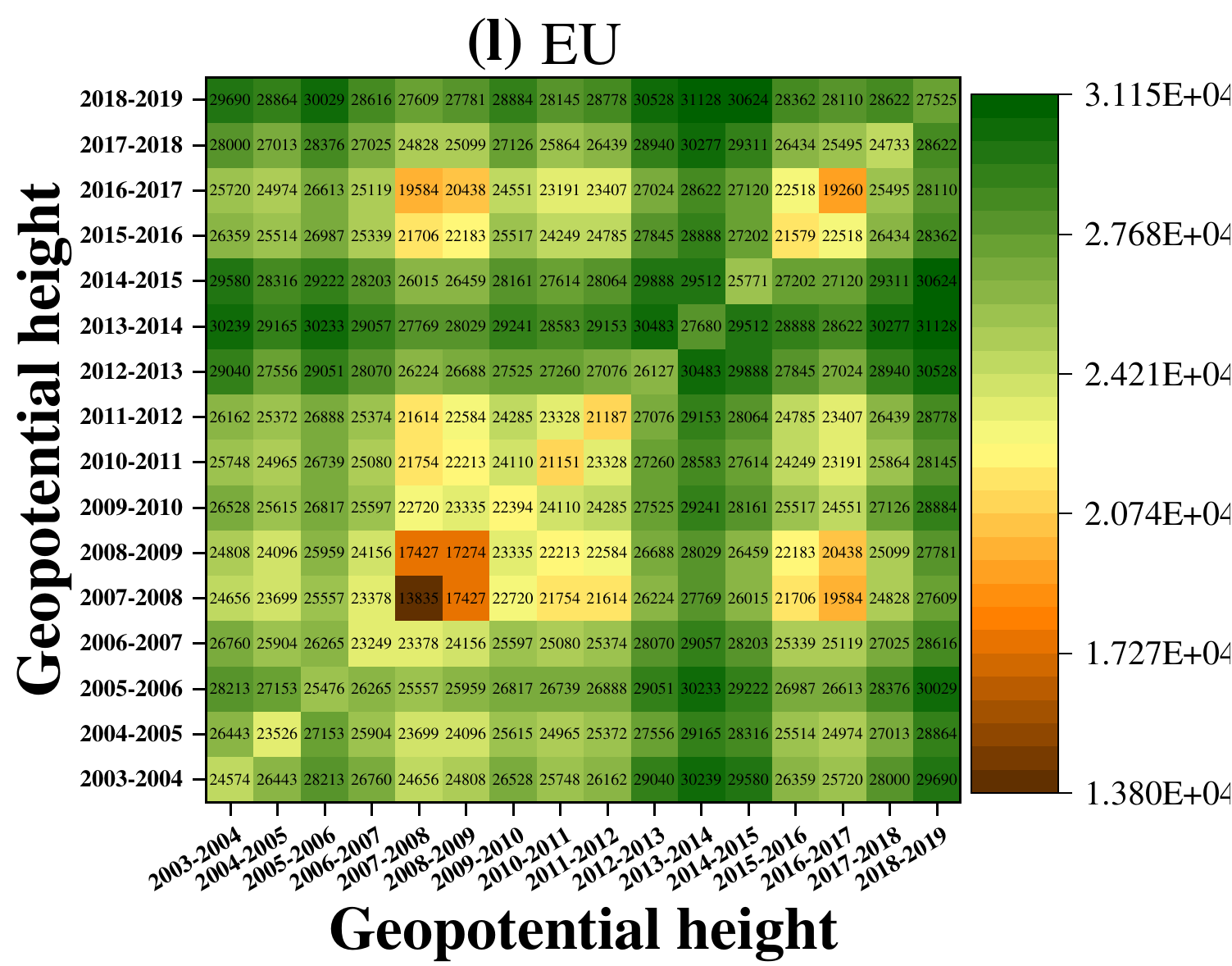}
\includegraphics[width=8em, height=7em]{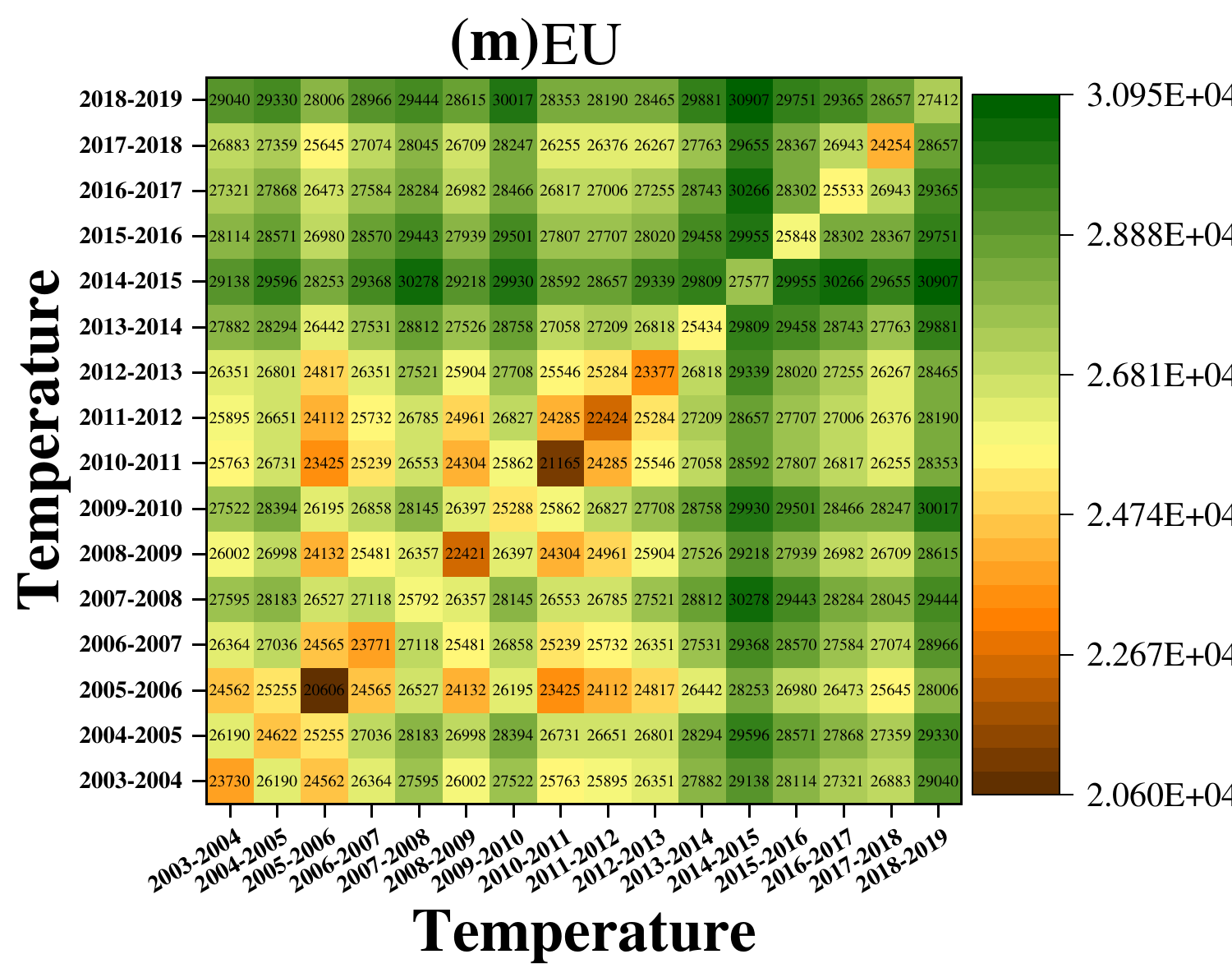}
\includegraphics[width=8em, height=7em]{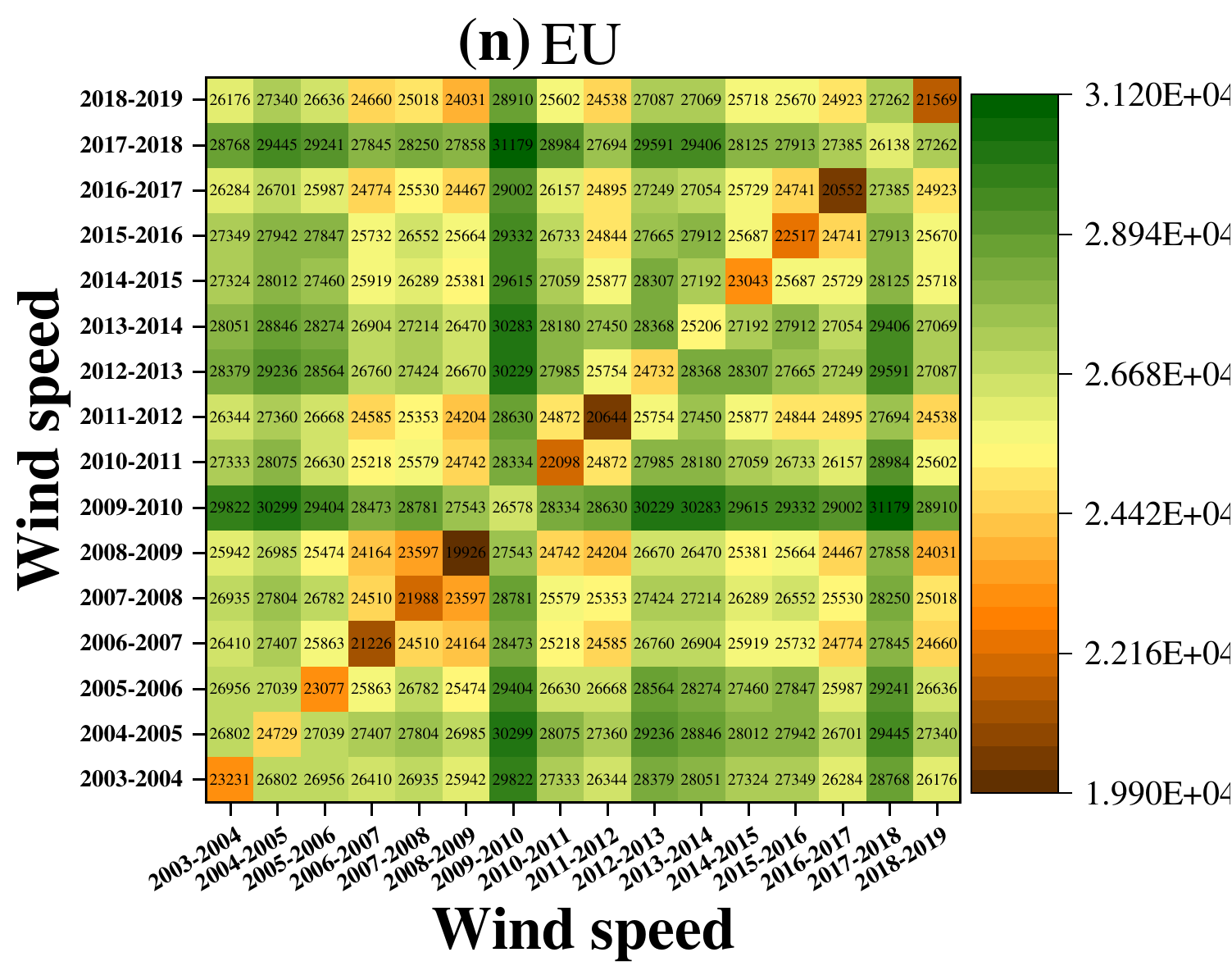}
\includegraphics[width=8em, height=7em]{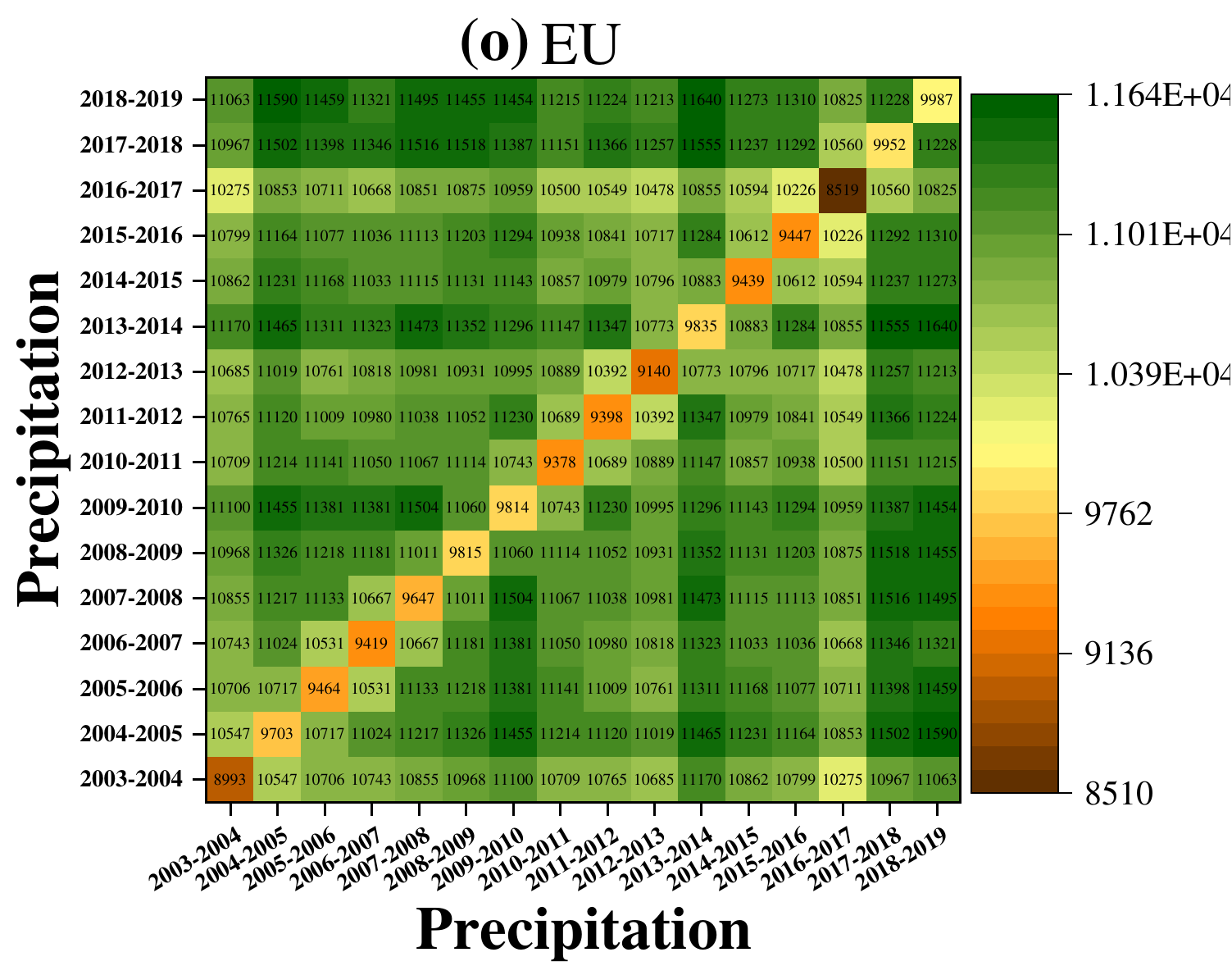}
\end{center}

\begin{center}
\noindent {\small {\bf Fig. S23} The matrix of the union of the links of each two yearly networks for different climate variables. Each matrix element represents the number of link unions in both networks.}
\end{center}

\begin{center}
\includegraphics[width=8em, height=7em]{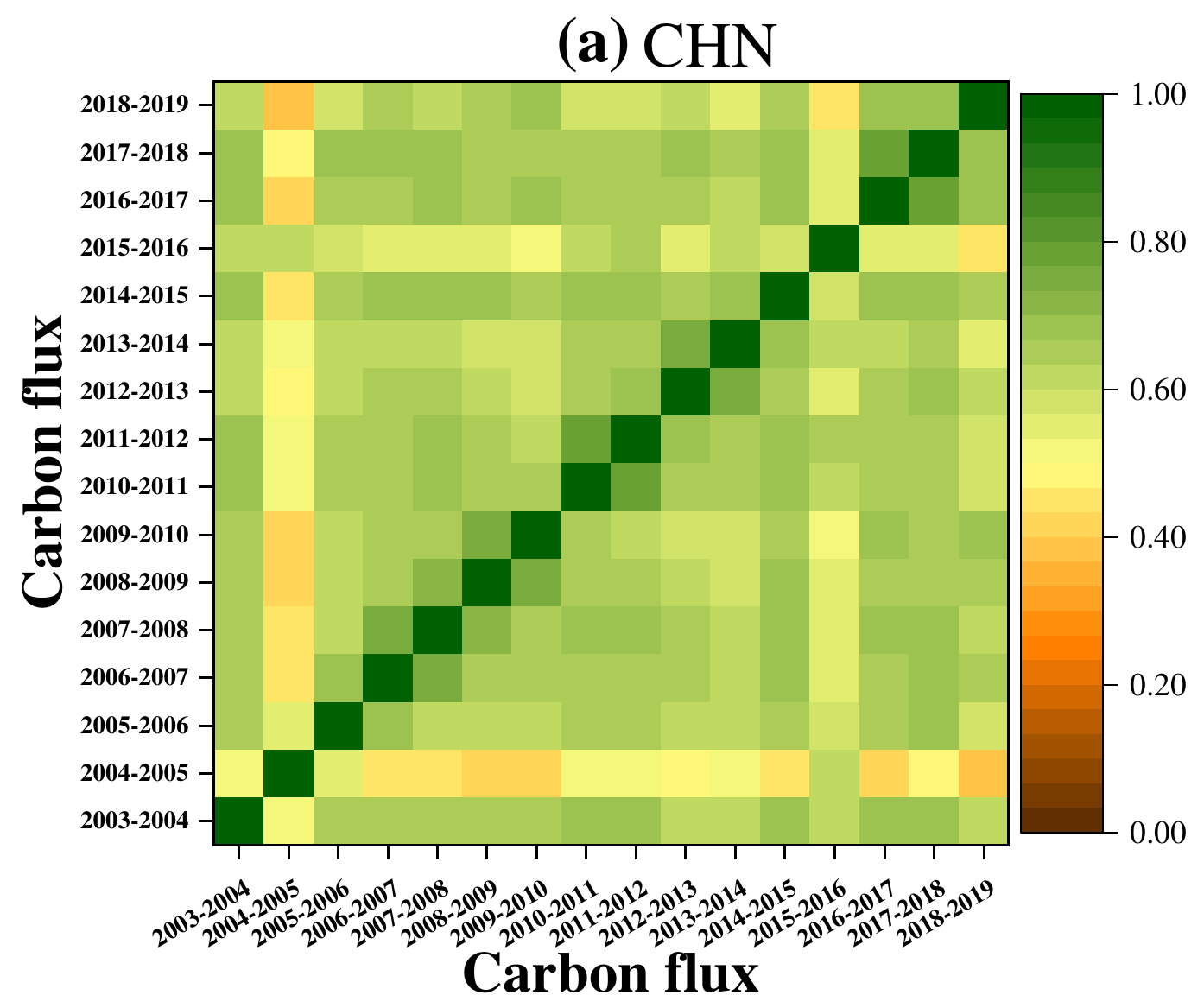}
\includegraphics[width=8em, height=7em]{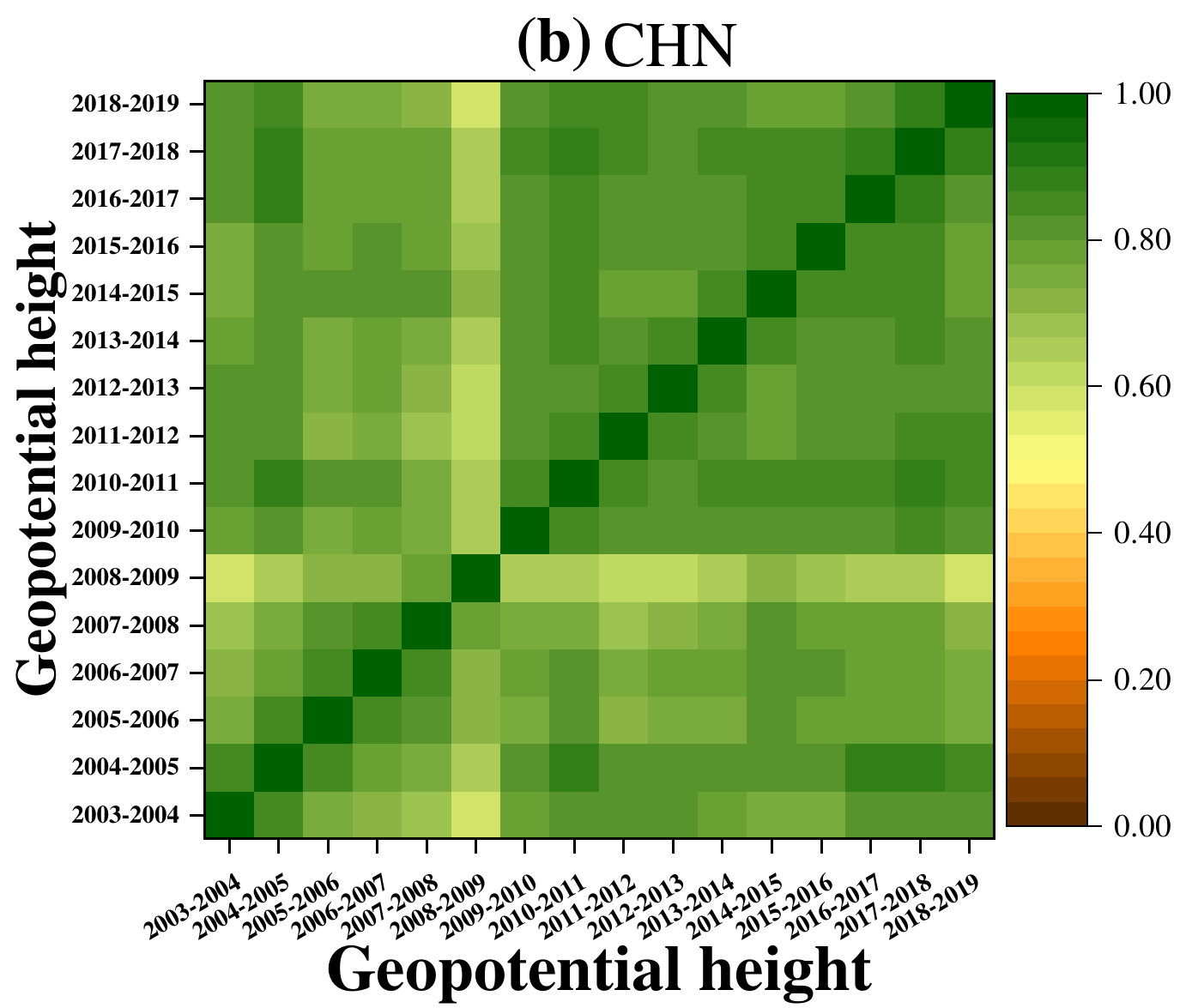}
\includegraphics[width=8em, height=7em]{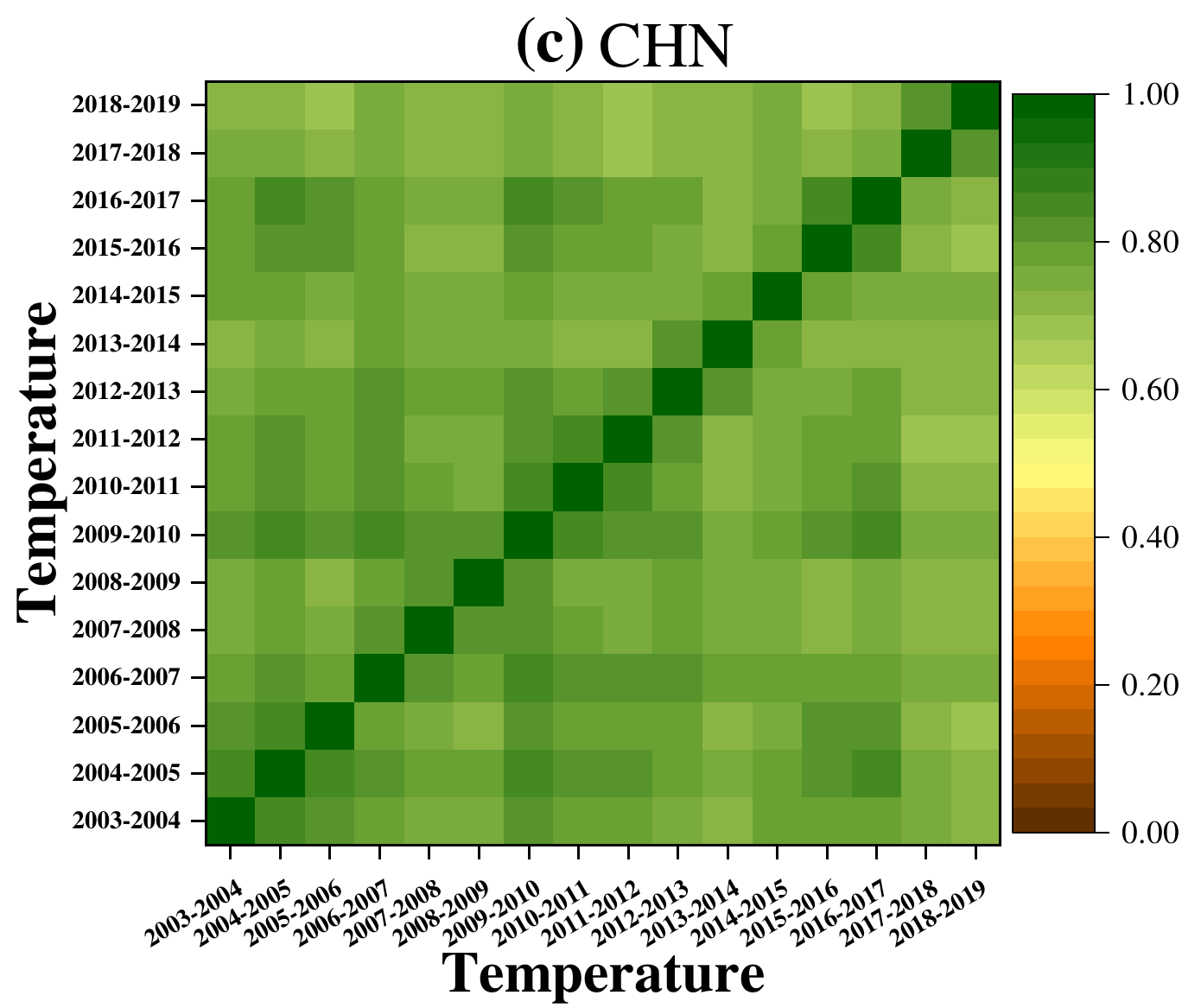}
\includegraphics[width=8em, height=7em]{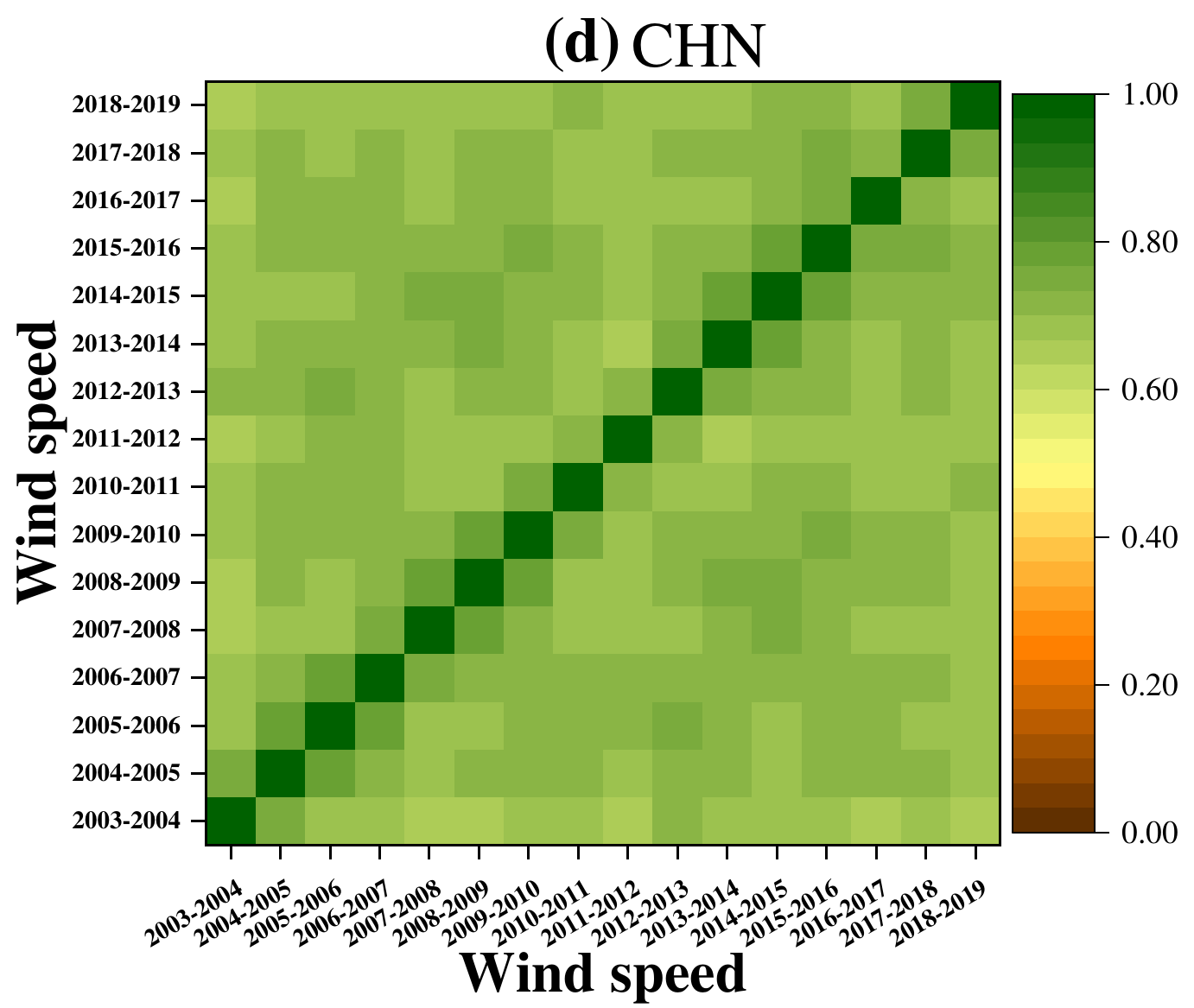}
\includegraphics[width=8em, height=7em]{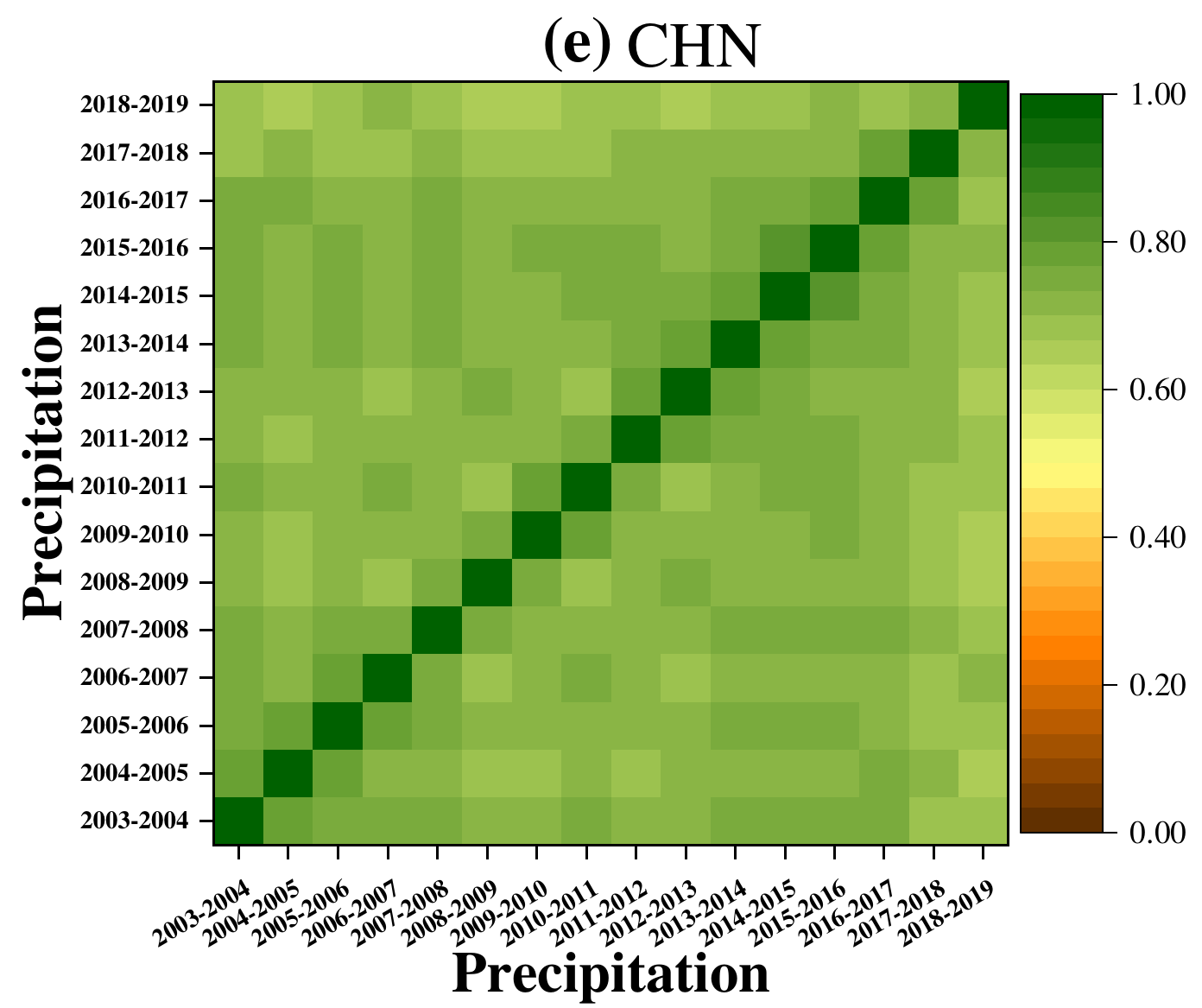}
\includegraphics[width=8em, height=7em]{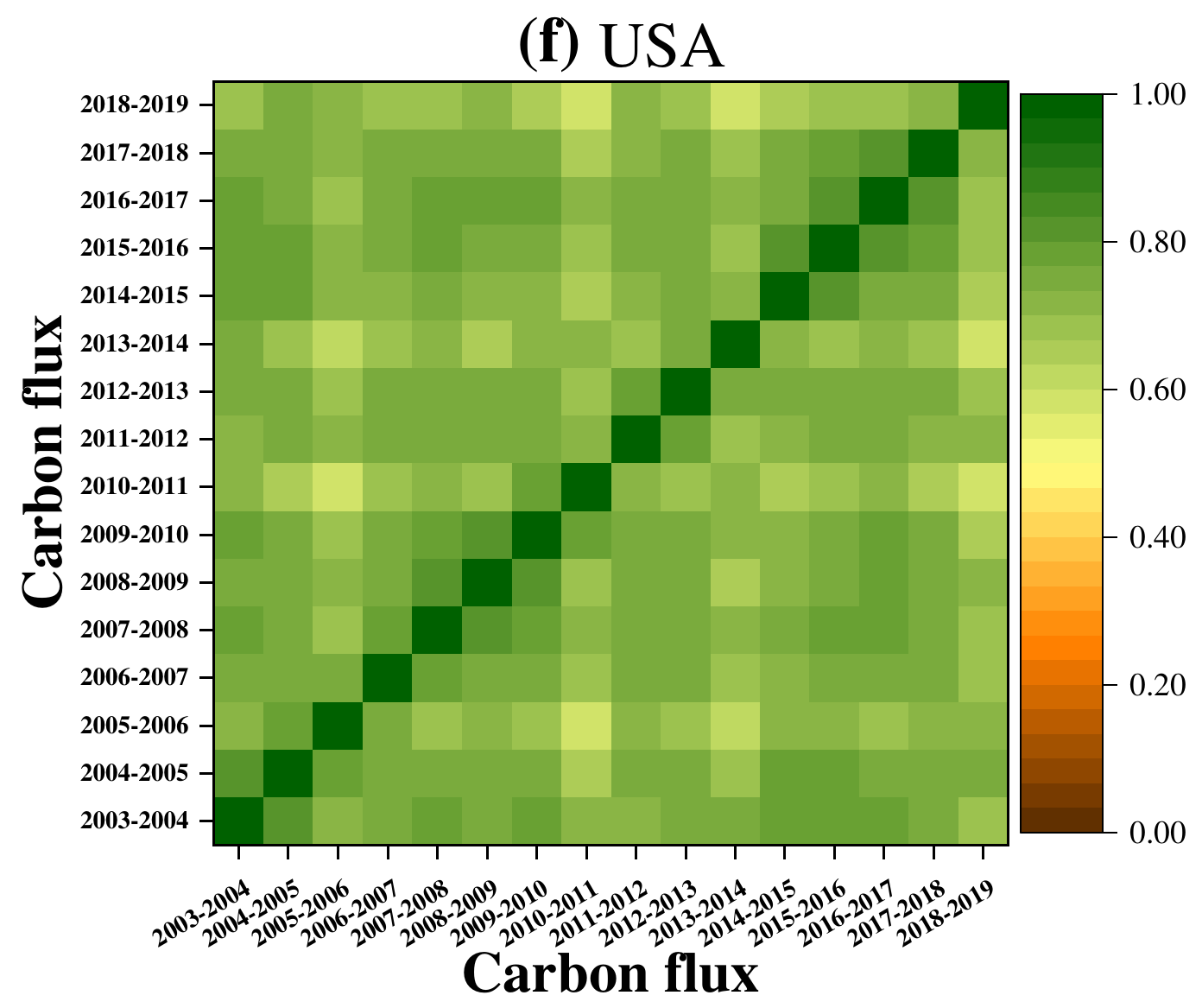}
\includegraphics[width=8em, height=7em]{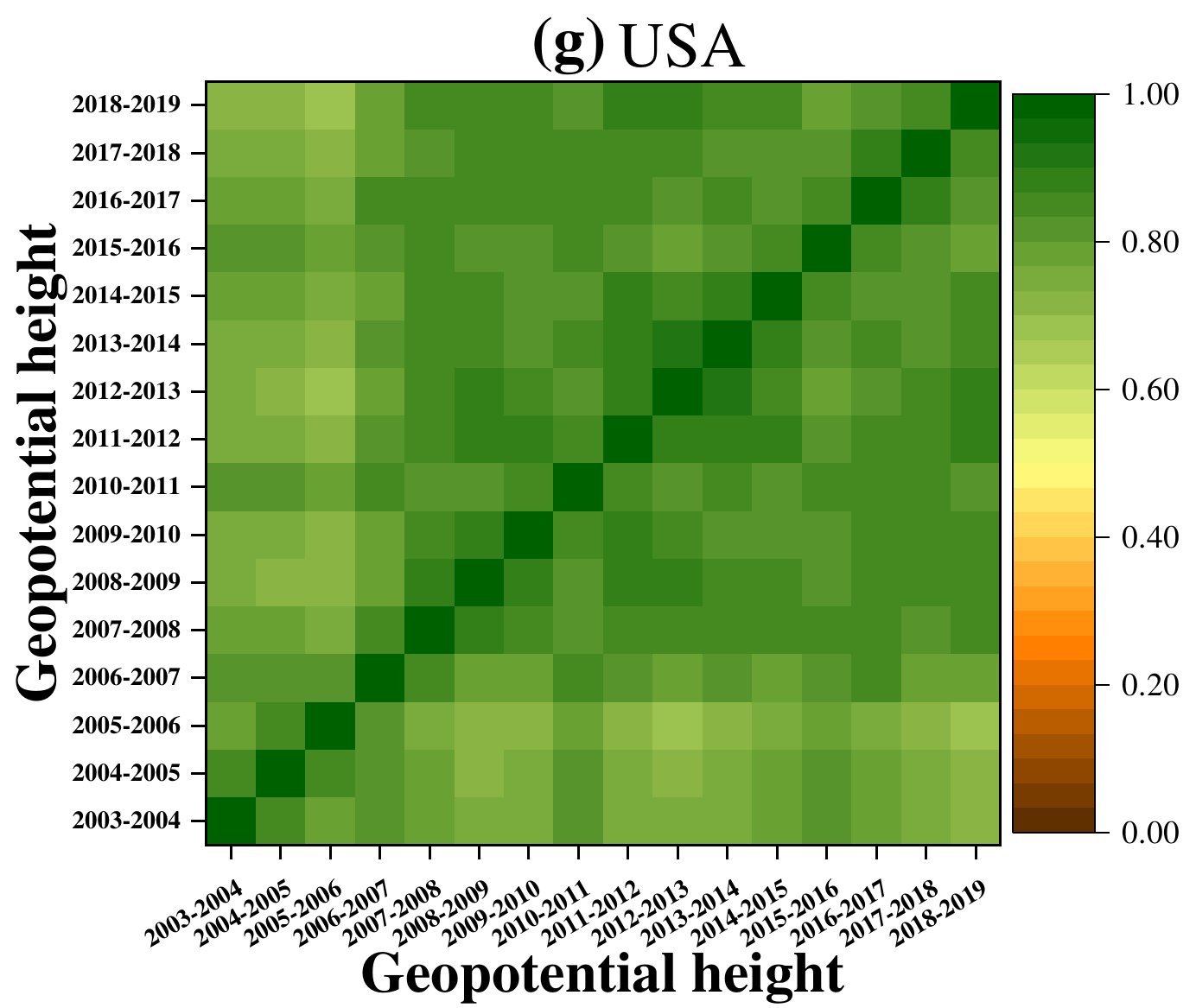}
\includegraphics[width=8em, height=7em]{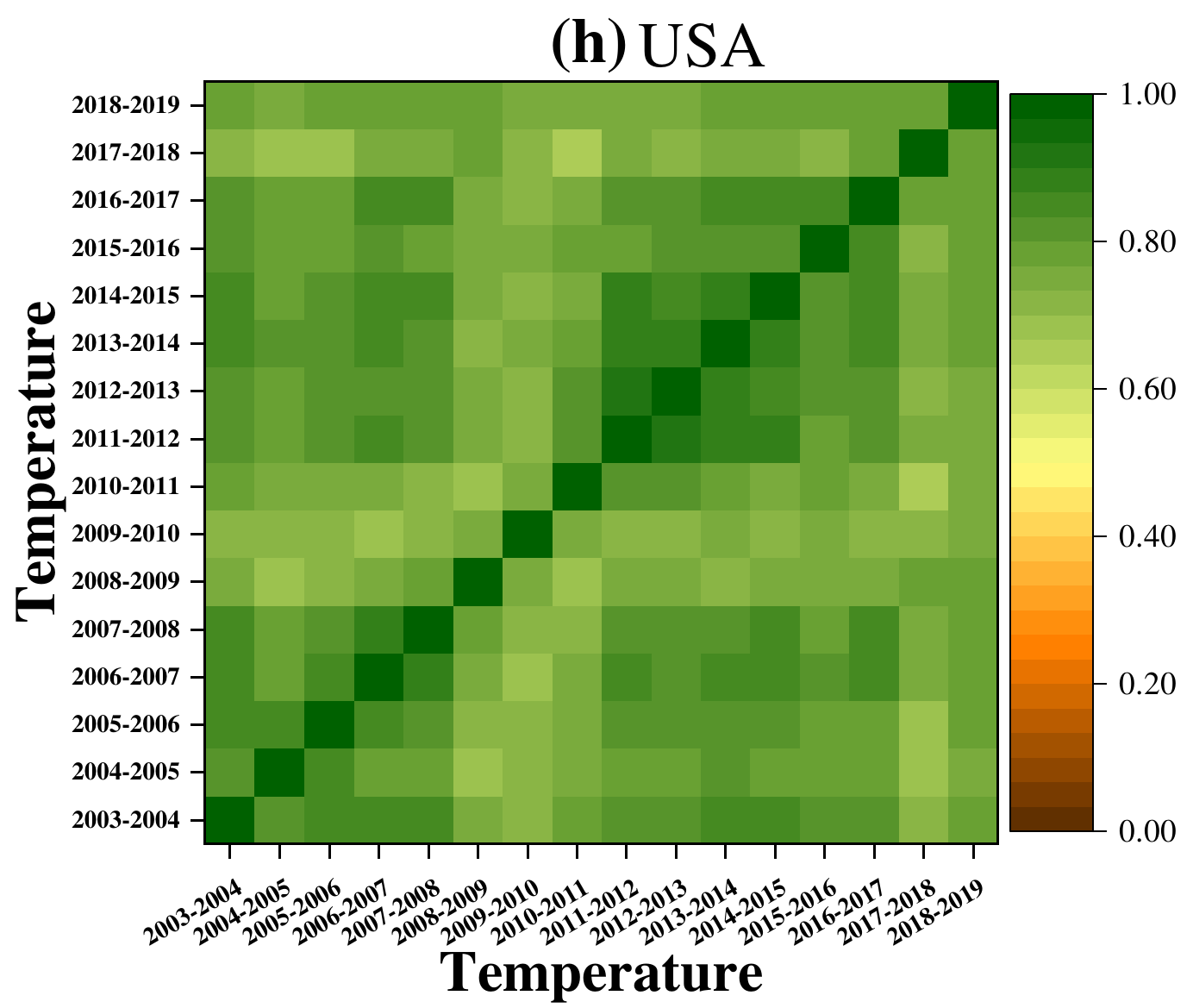}
\includegraphics[width=8em, height=7em]{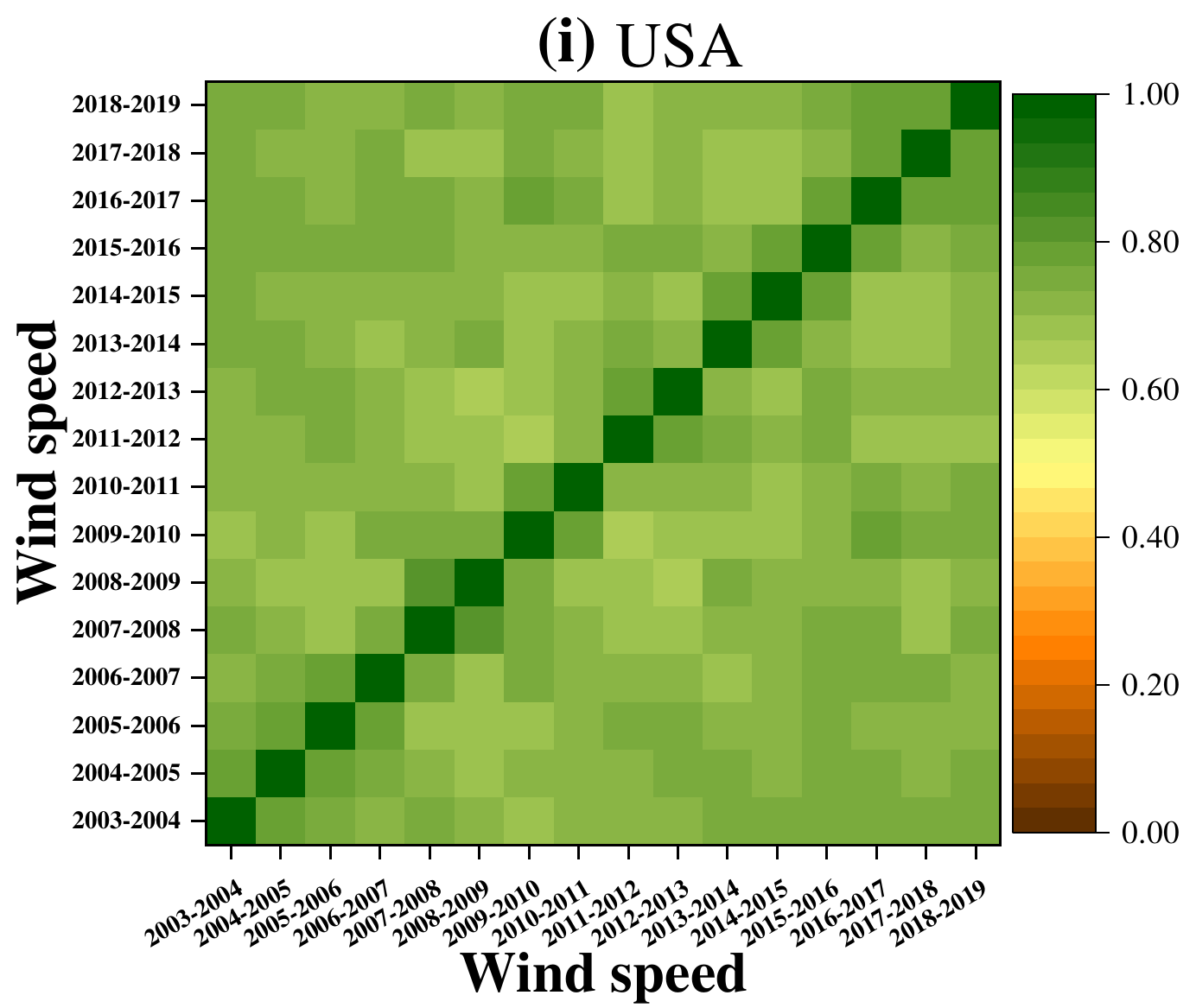}
\includegraphics[width=8em, height=7em]{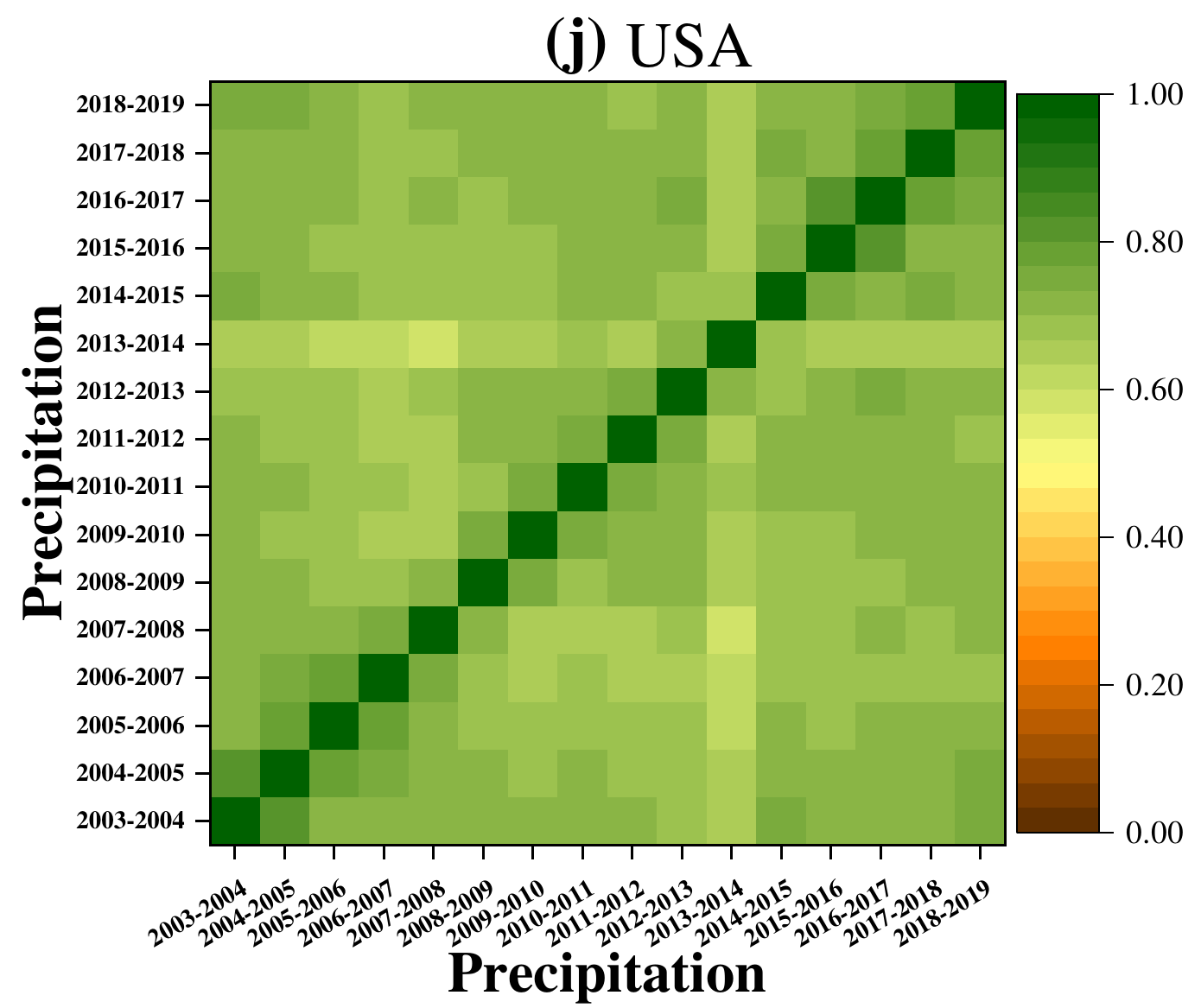}
\includegraphics[width=8em, height=7em]{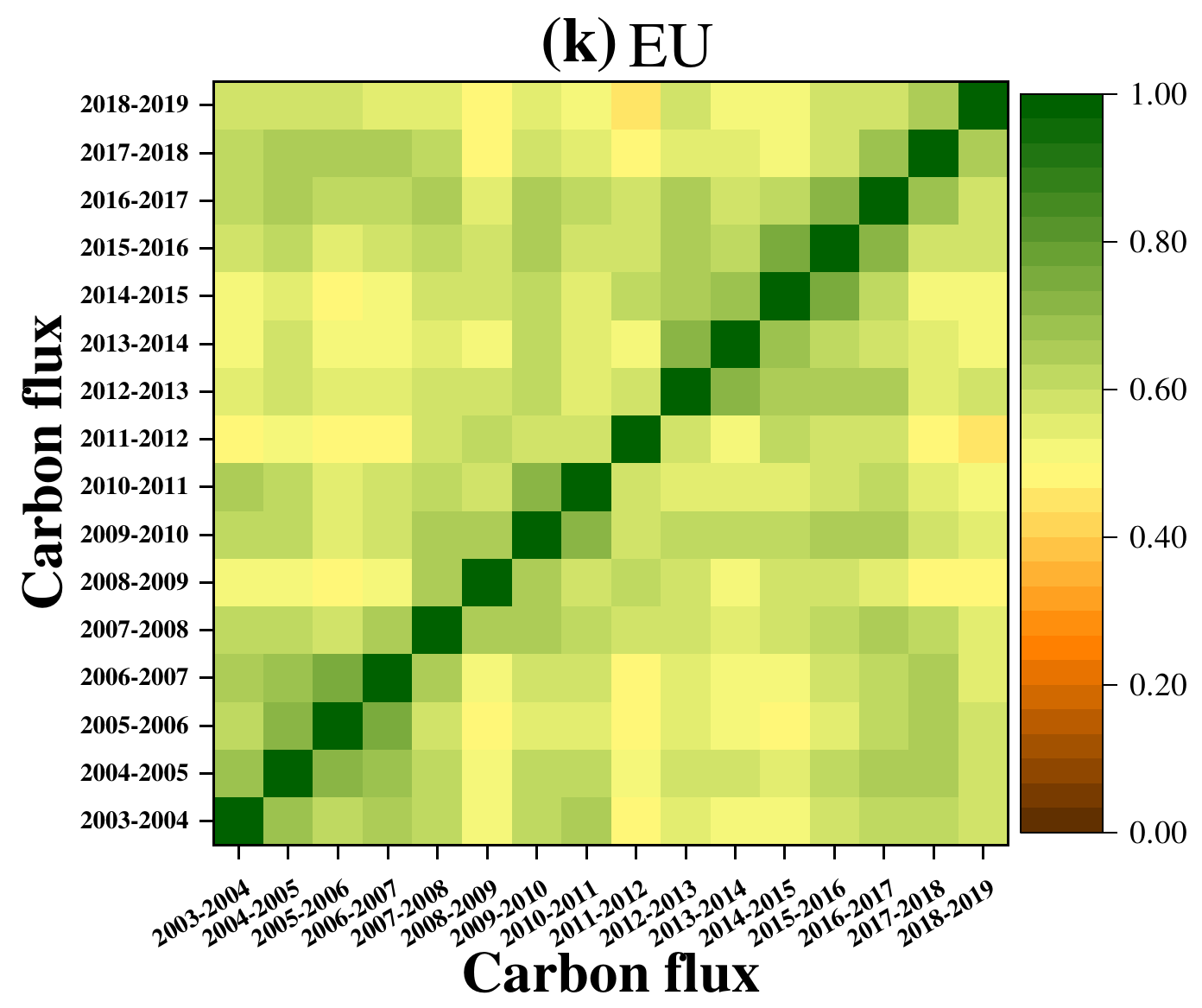}
\includegraphics[width=8em, height=7em]{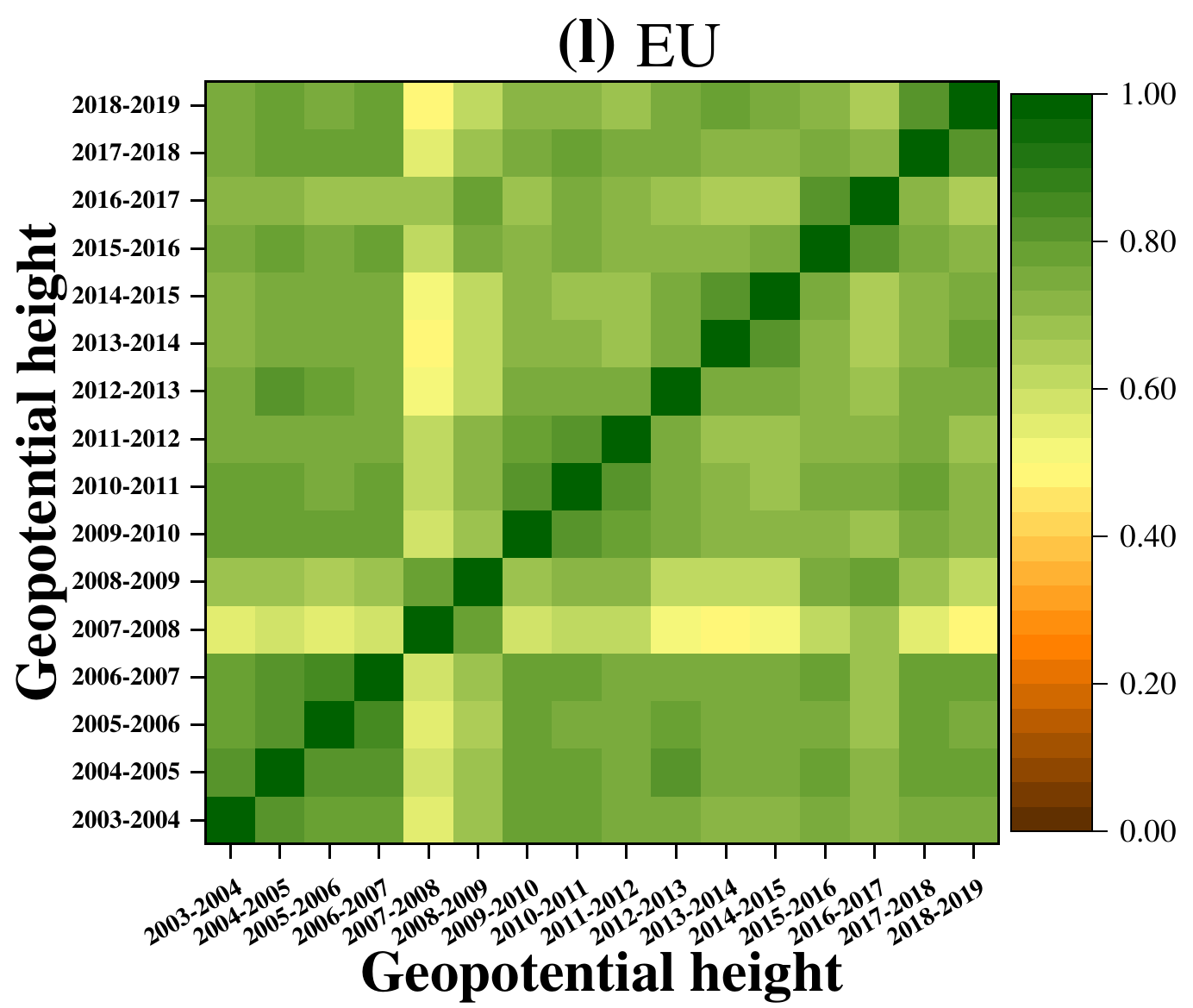}
\includegraphics[width=8em, height=7em]{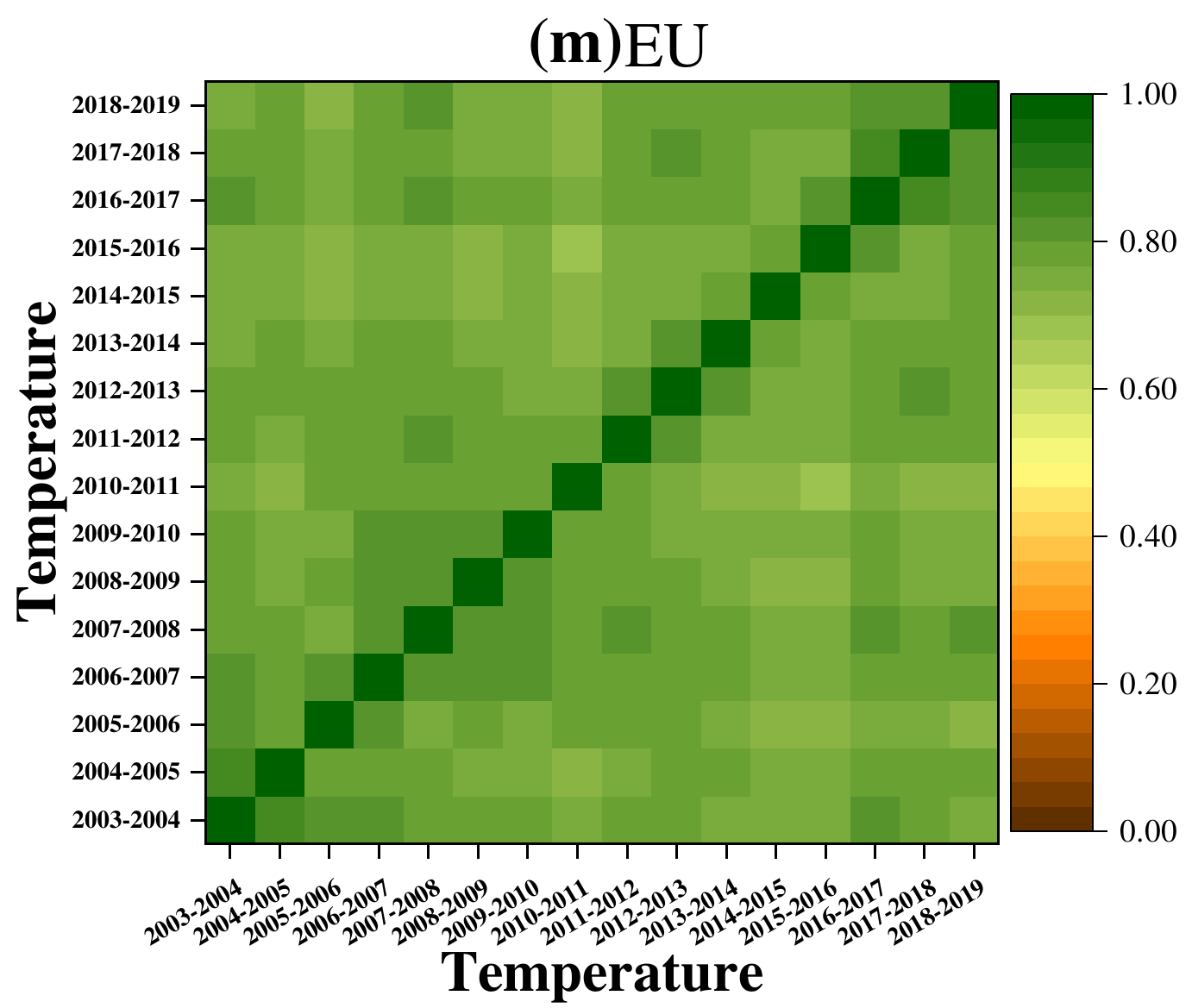}
\includegraphics[width=8em, height=7em]{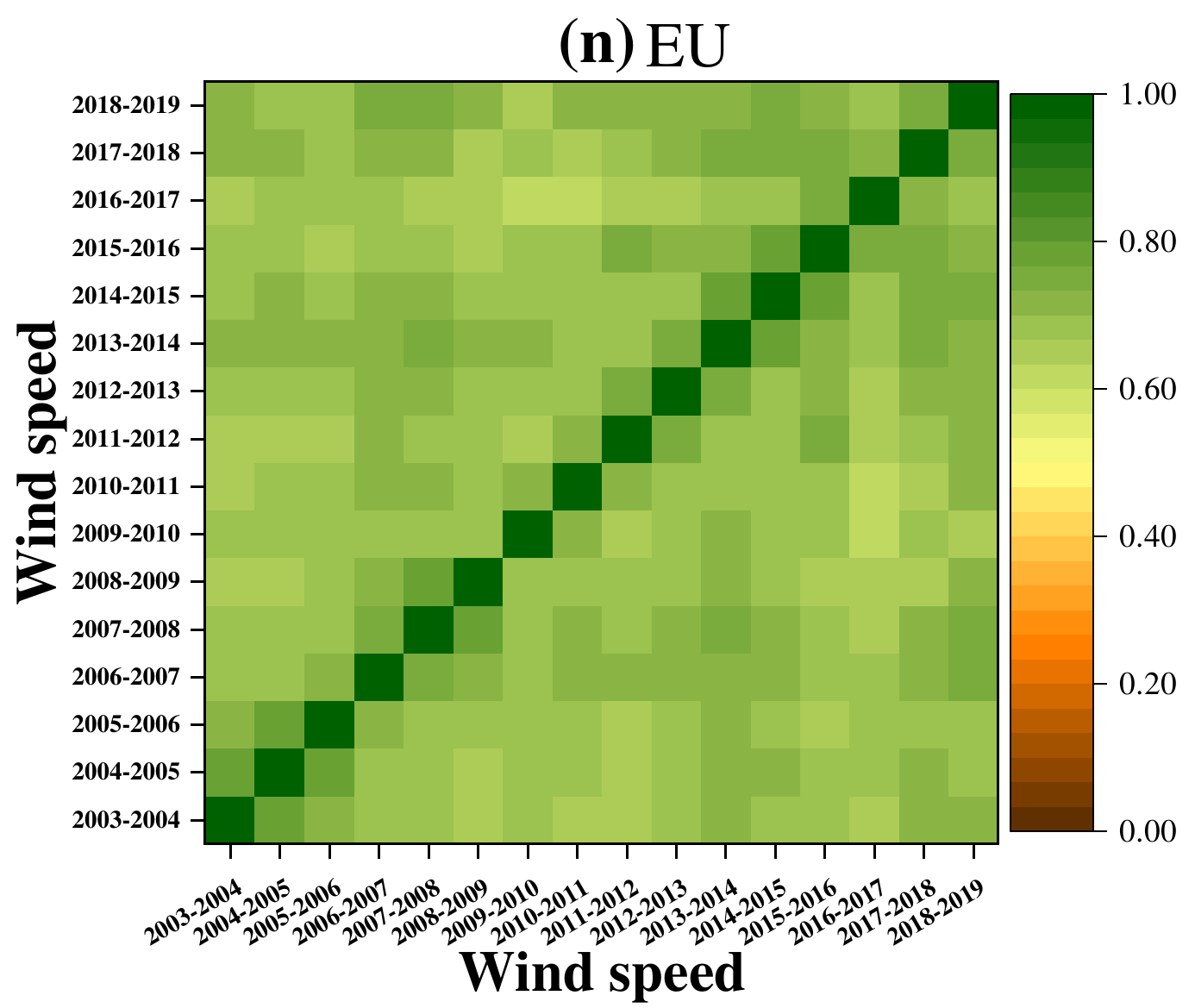}
\includegraphics[width=8em, height=7em]{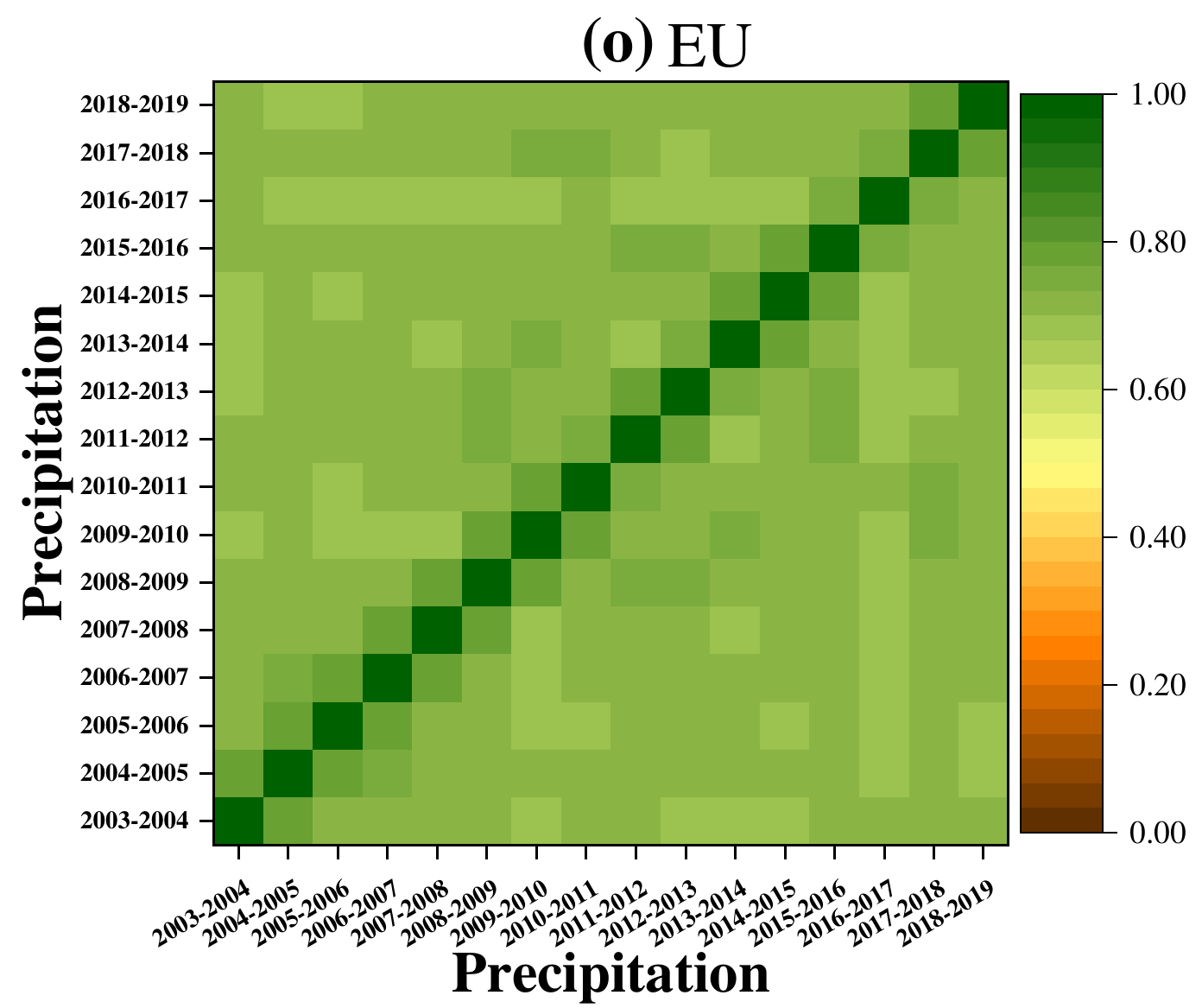}
\end{center}

\begin{center}
\noindent {\small {\bf Fig. S24} Actual Jaccard similarity coefficient matrix of links between two networks of different years for each of the climate variables.}
\end{center}

\begin{center}
\includegraphics[width=8em, height=7em]{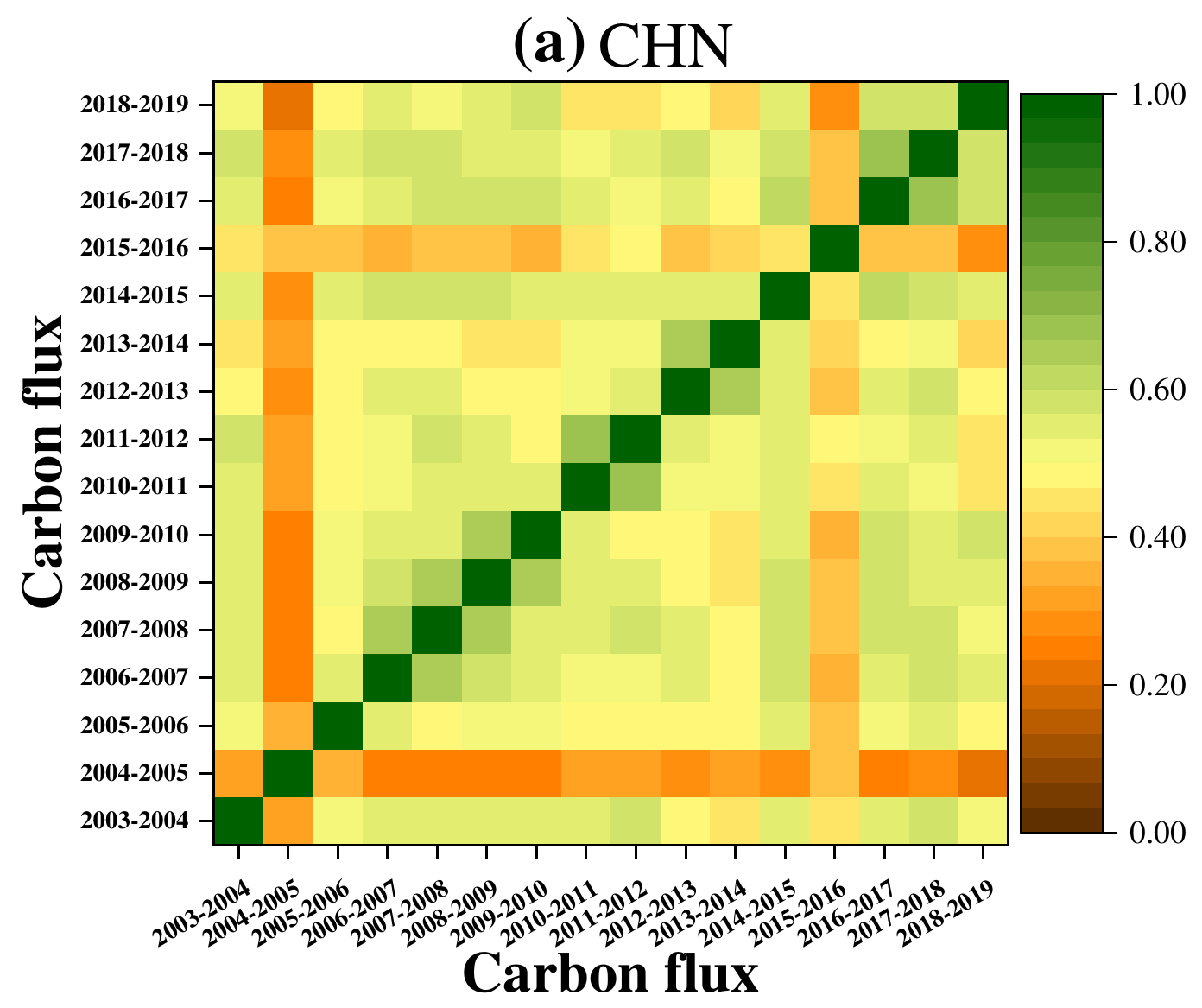}
\includegraphics[width=8em, height=7em]{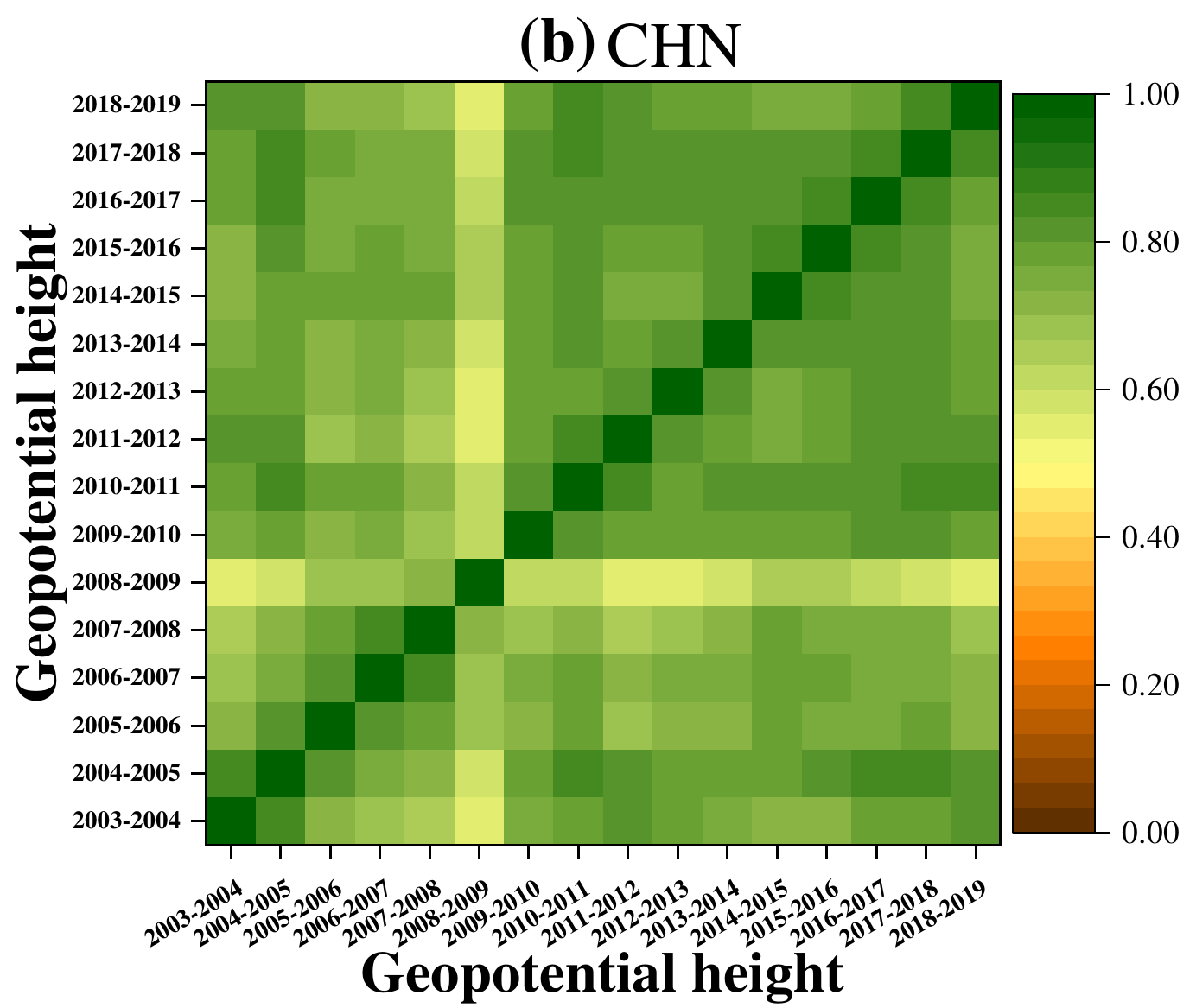}
\includegraphics[width=8em, height=7em]{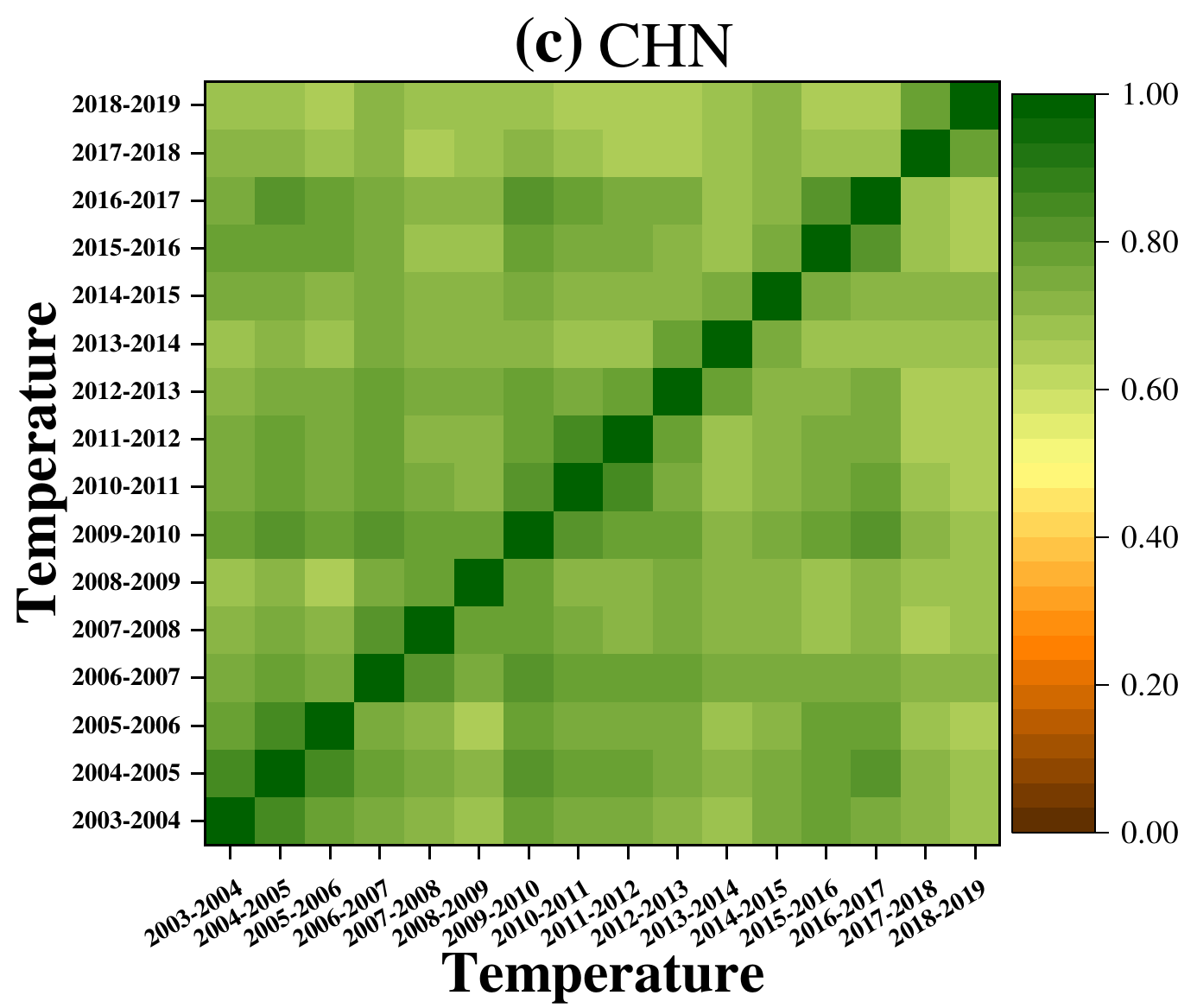}
\includegraphics[width=8em, height=7em]{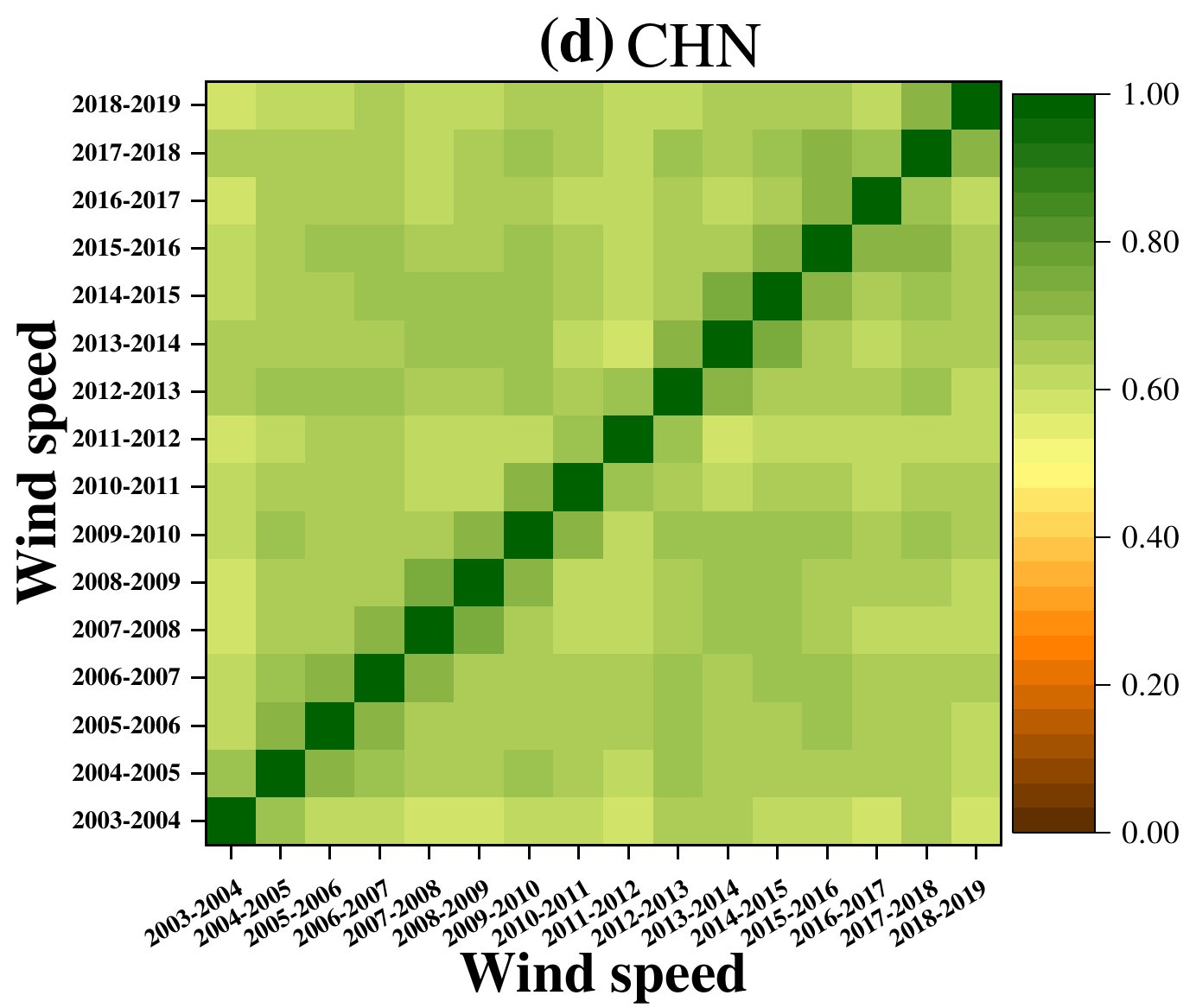}
\includegraphics[width=8em, height=7em]{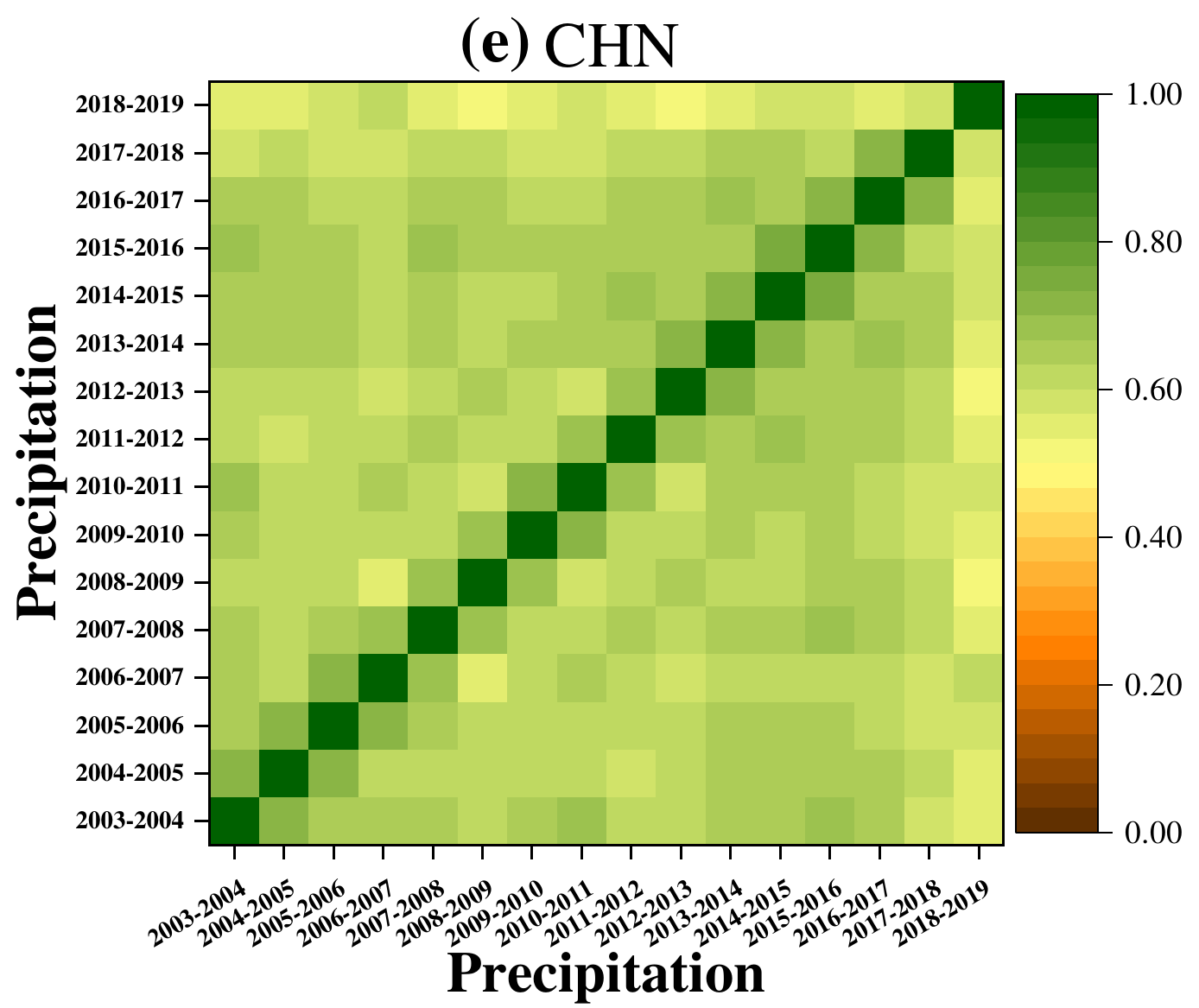}
\includegraphics[width=8em, height=7em]{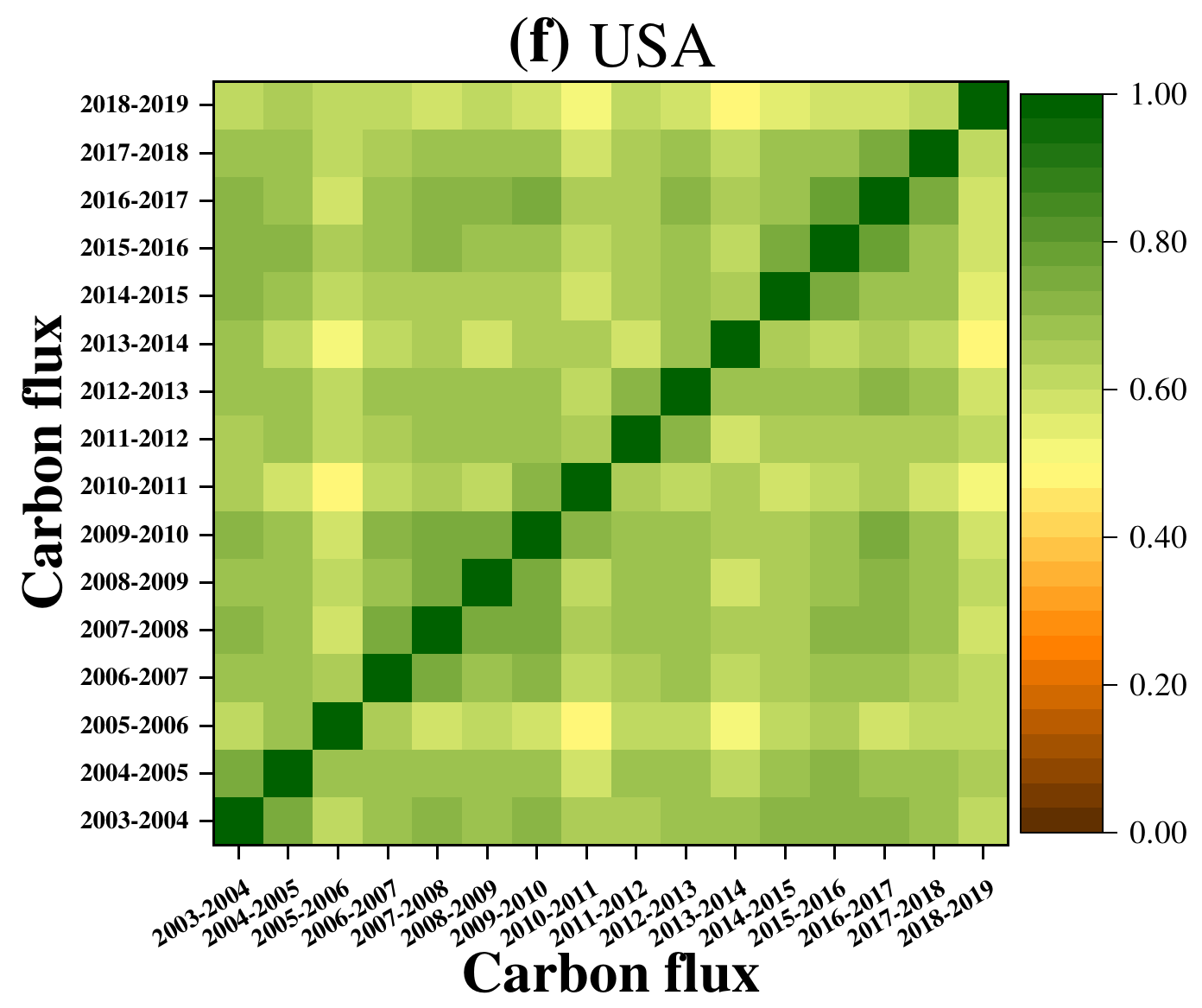}
\includegraphics[width=8em, height=7em]{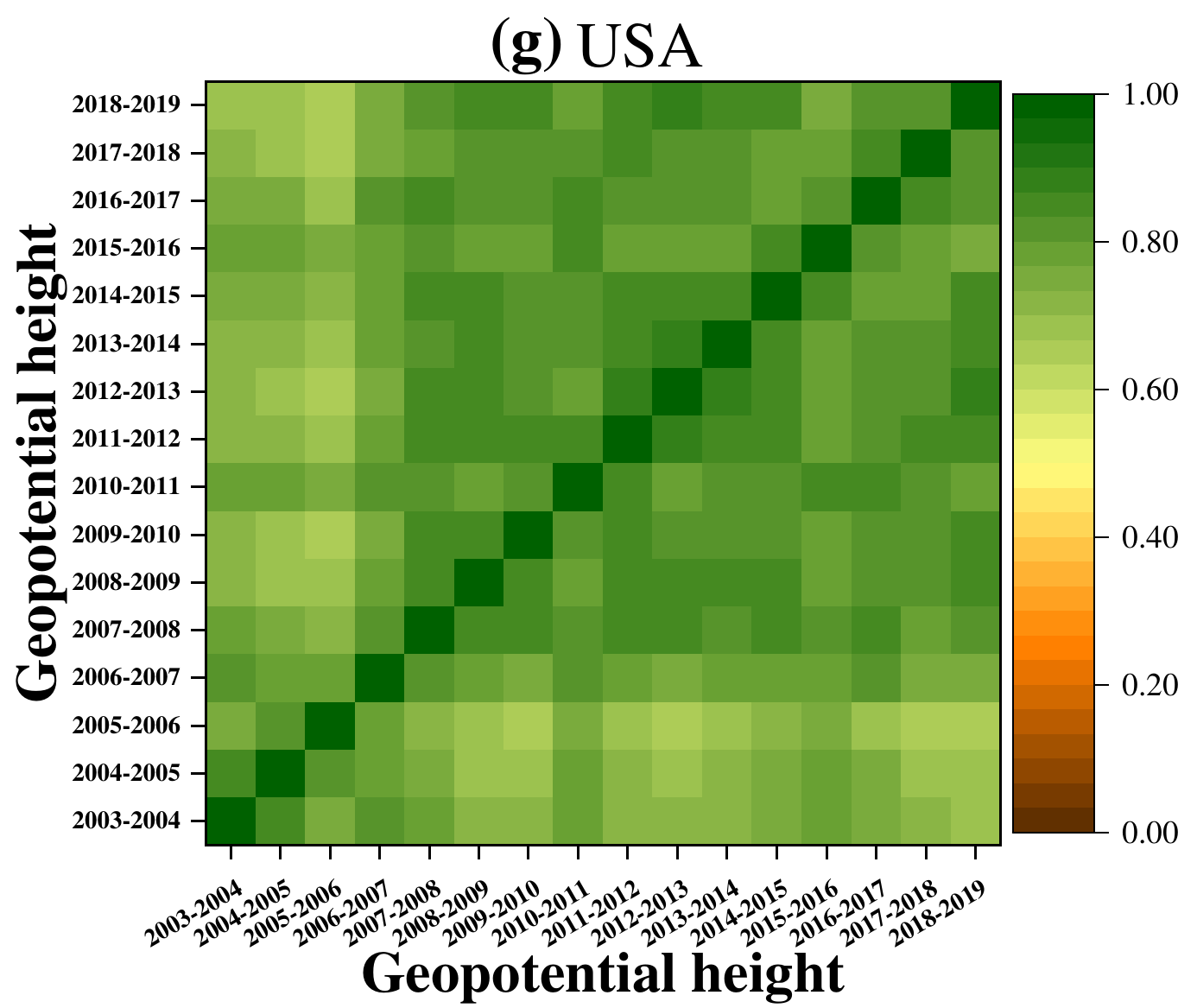}
\includegraphics[width=8em, height=7em]{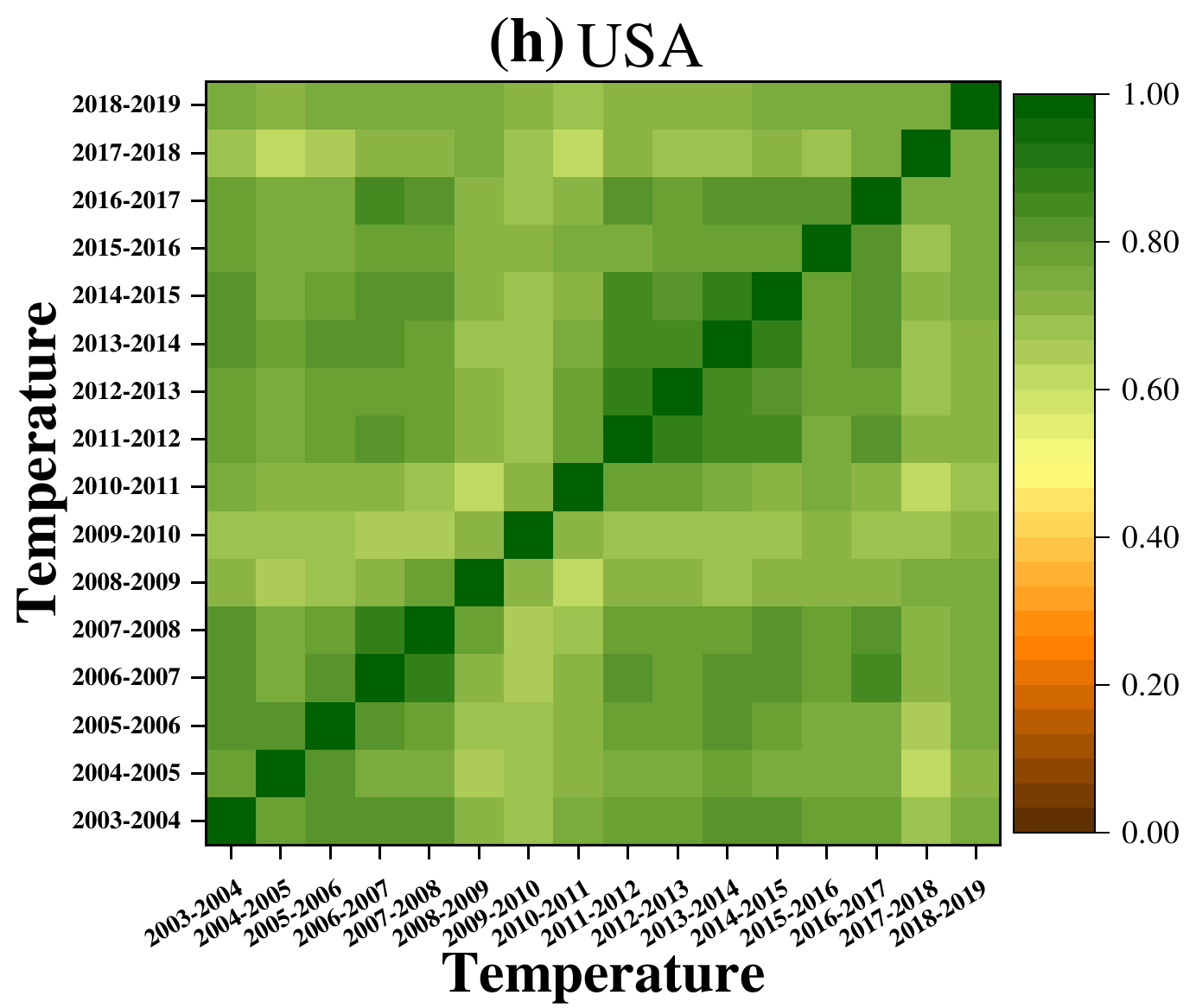}
\includegraphics[width=8em, height=7em]{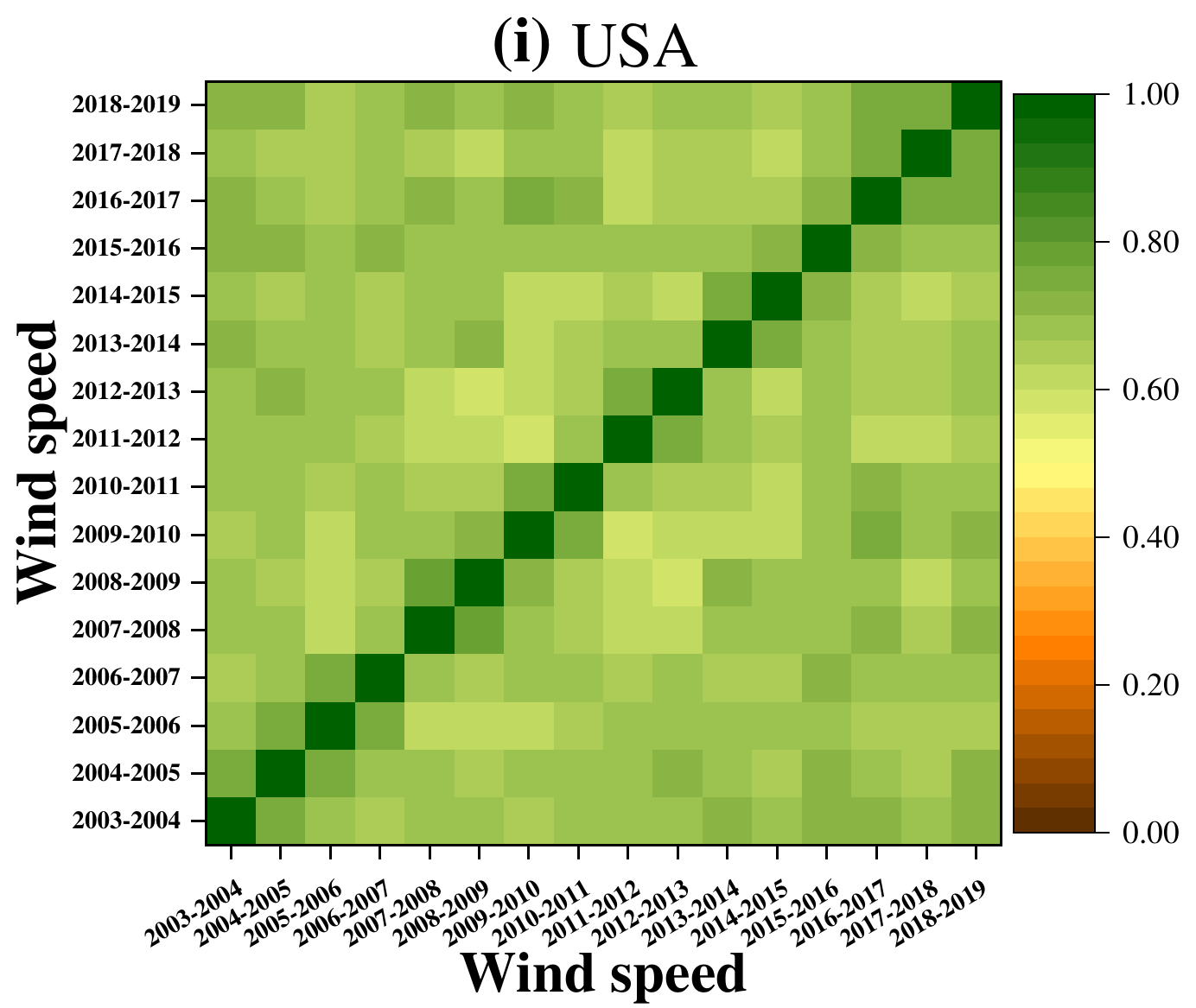}
\includegraphics[width=8em, height=7em]{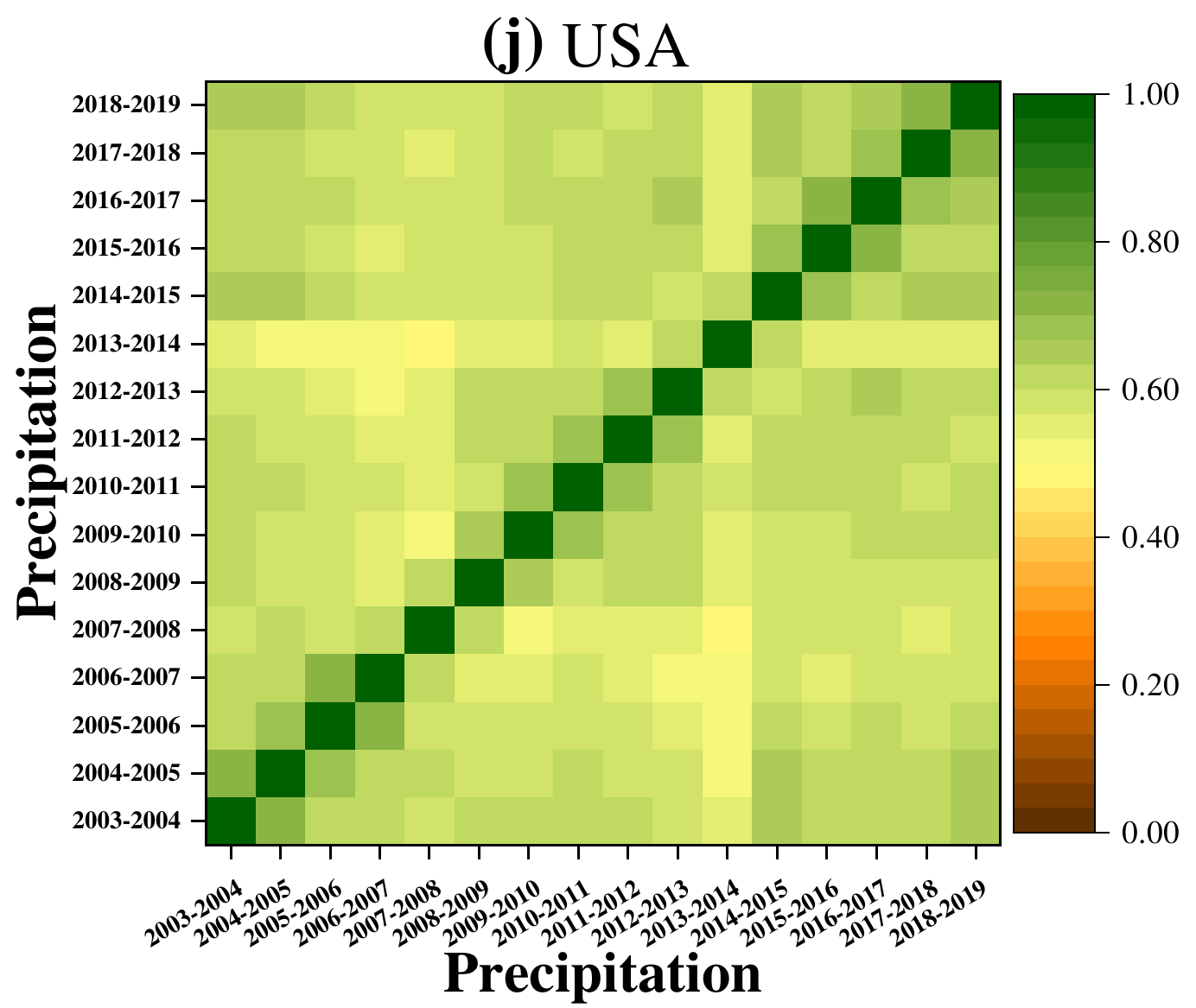}
\includegraphics[width=8em, height=7em]{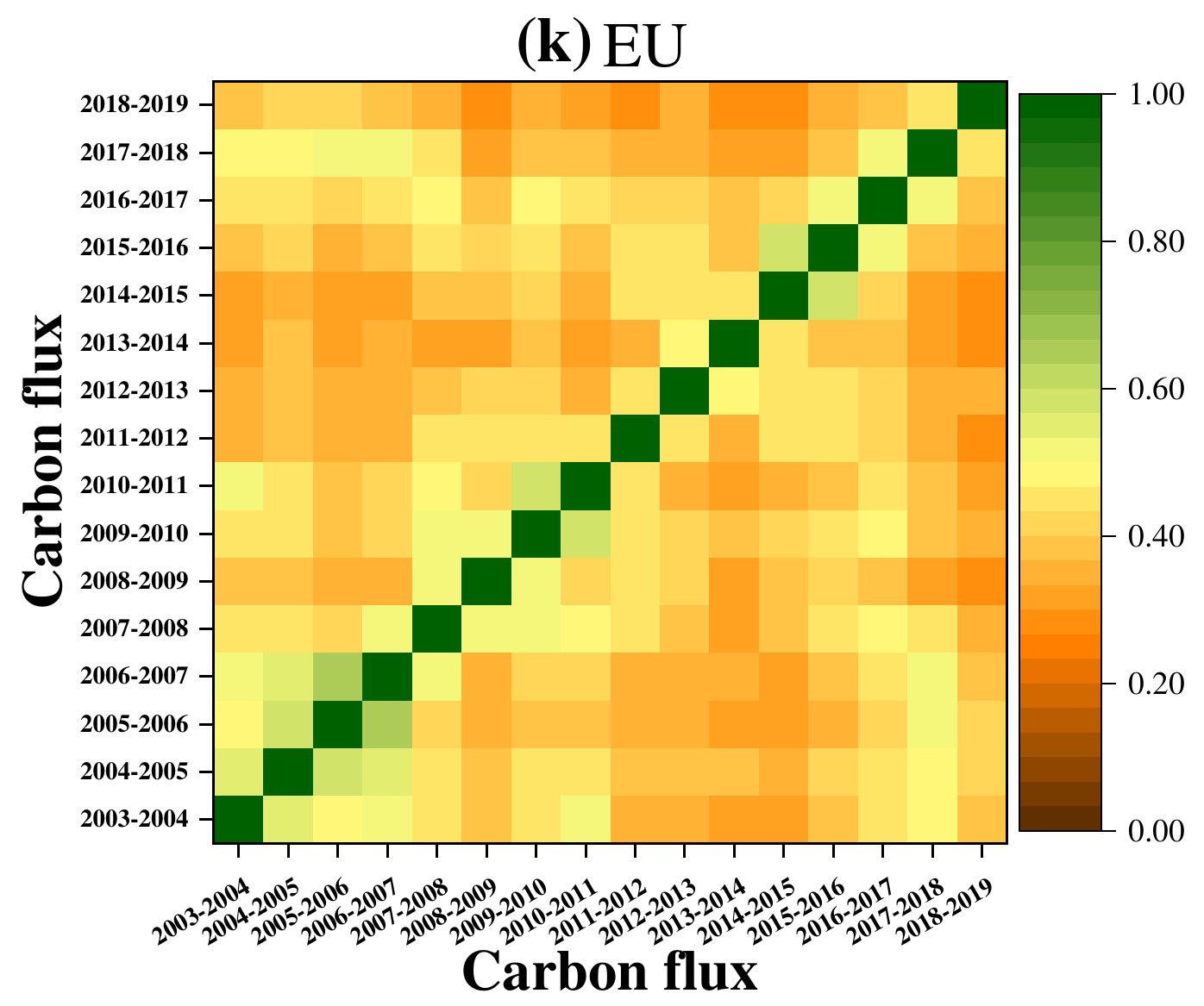}
\includegraphics[width=8em, height=7em]{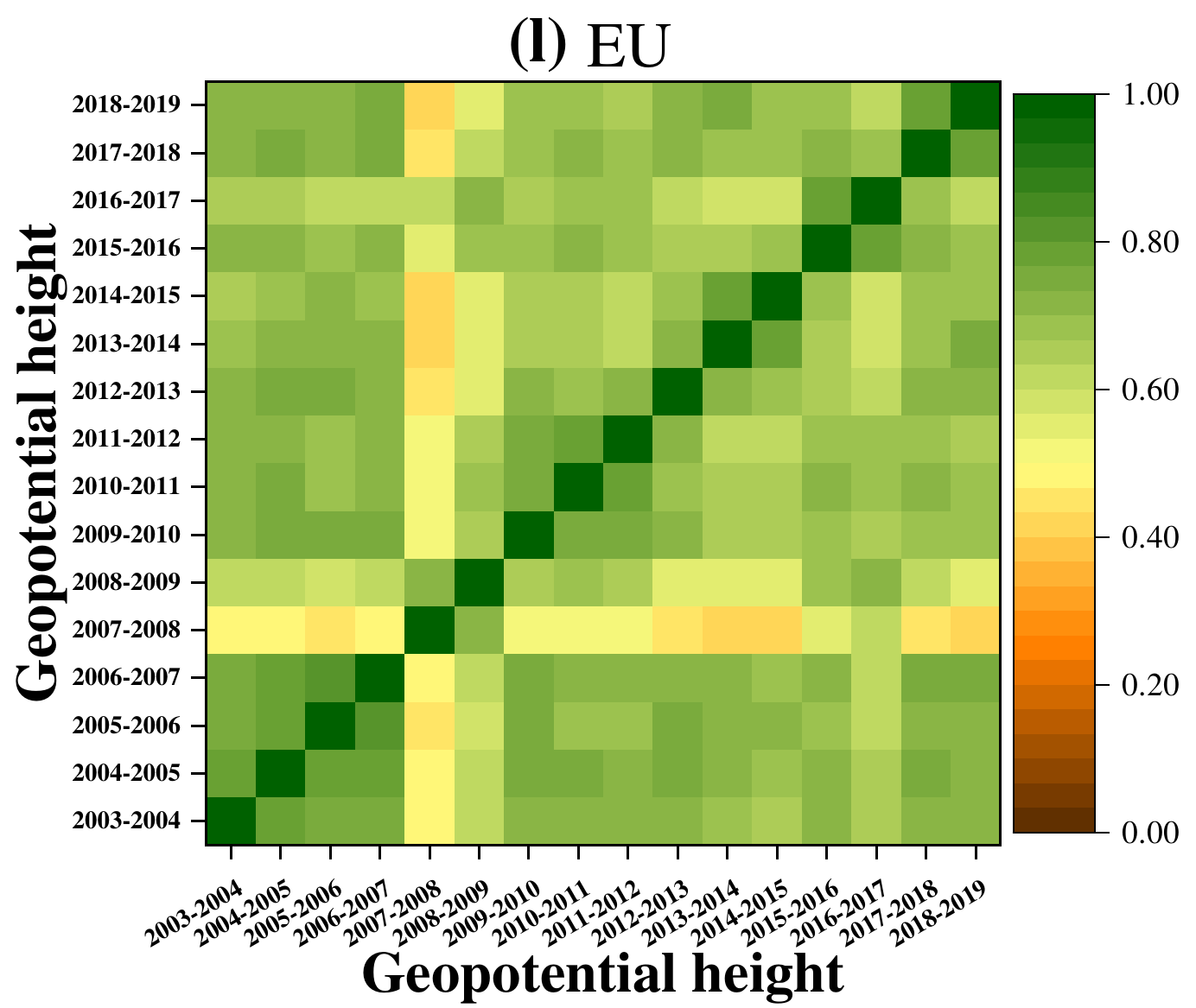}
\includegraphics[width=8em, height=7em]{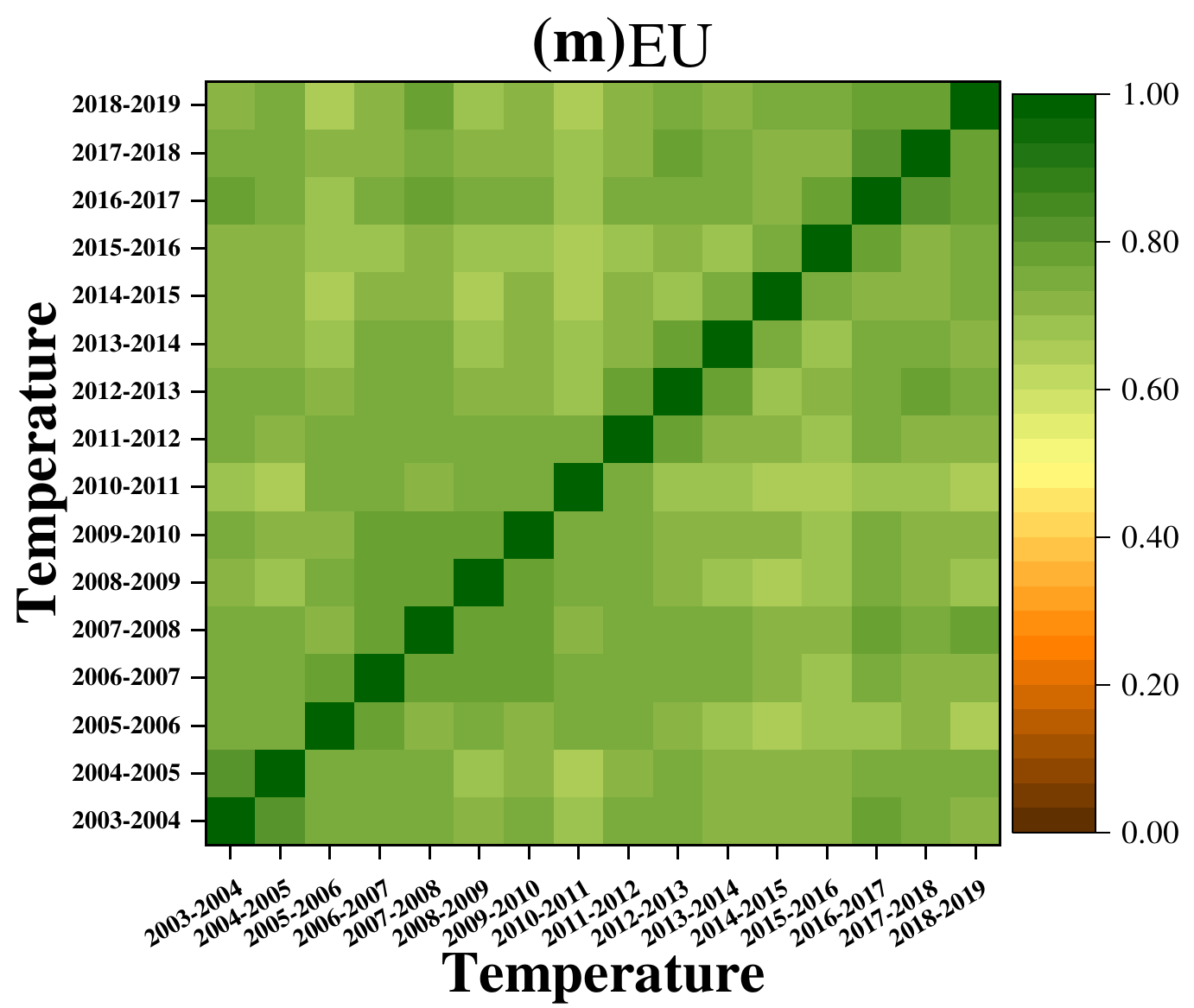}
\includegraphics[width=8em, height=7em]{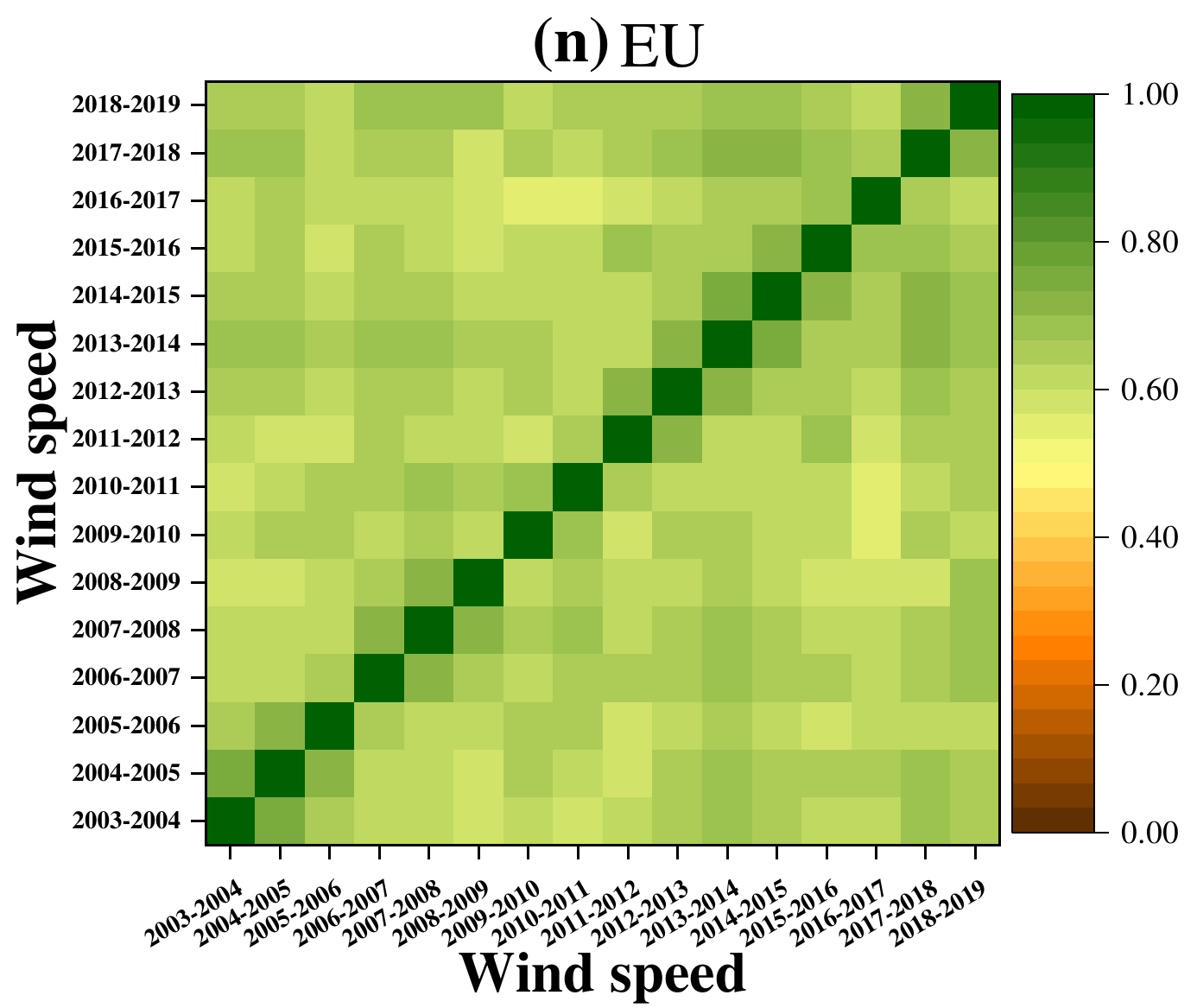}
\includegraphics[width=8em, height=7em]{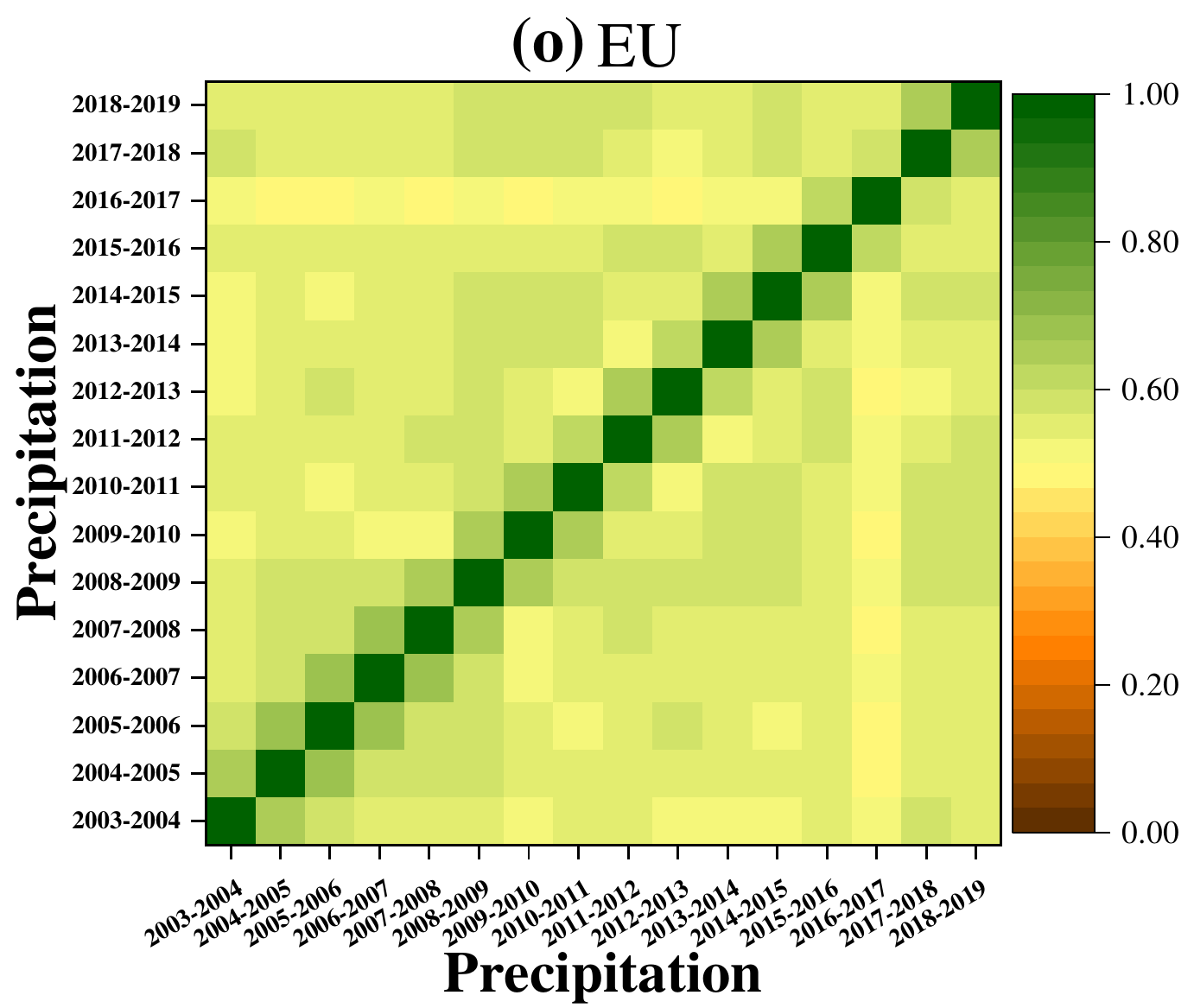}
\end{center}

\begin{center}
\noindent {\small {\bf Fig. S25} The actual Jaccard similarity coefficient matrix of links of lengths above $500km$ in two networks of different years for each of the climate variables.}
\end{center}

\begin{center}
\includegraphics[width=8em, height=7em]{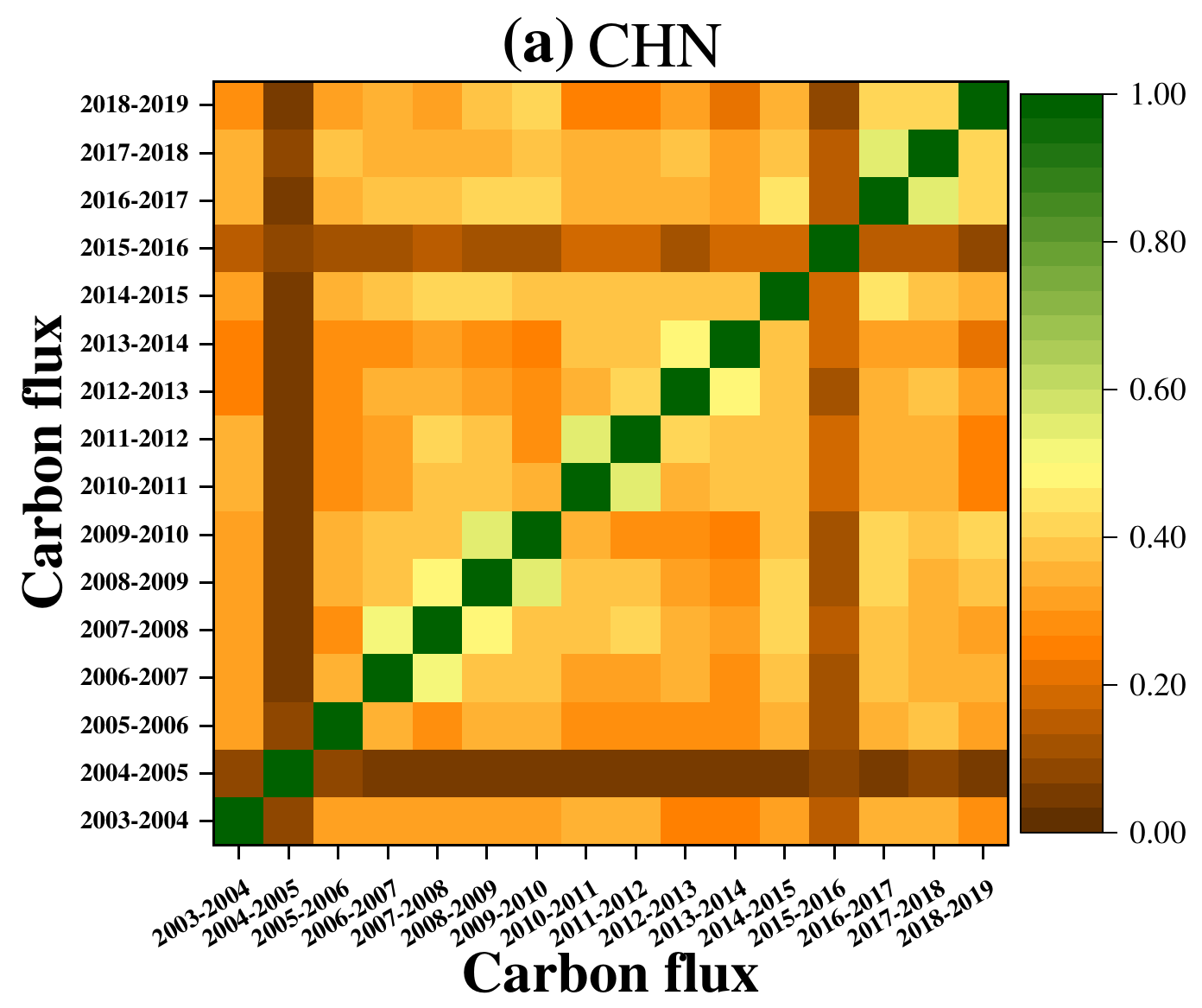}
\includegraphics[width=8em, height=7em]{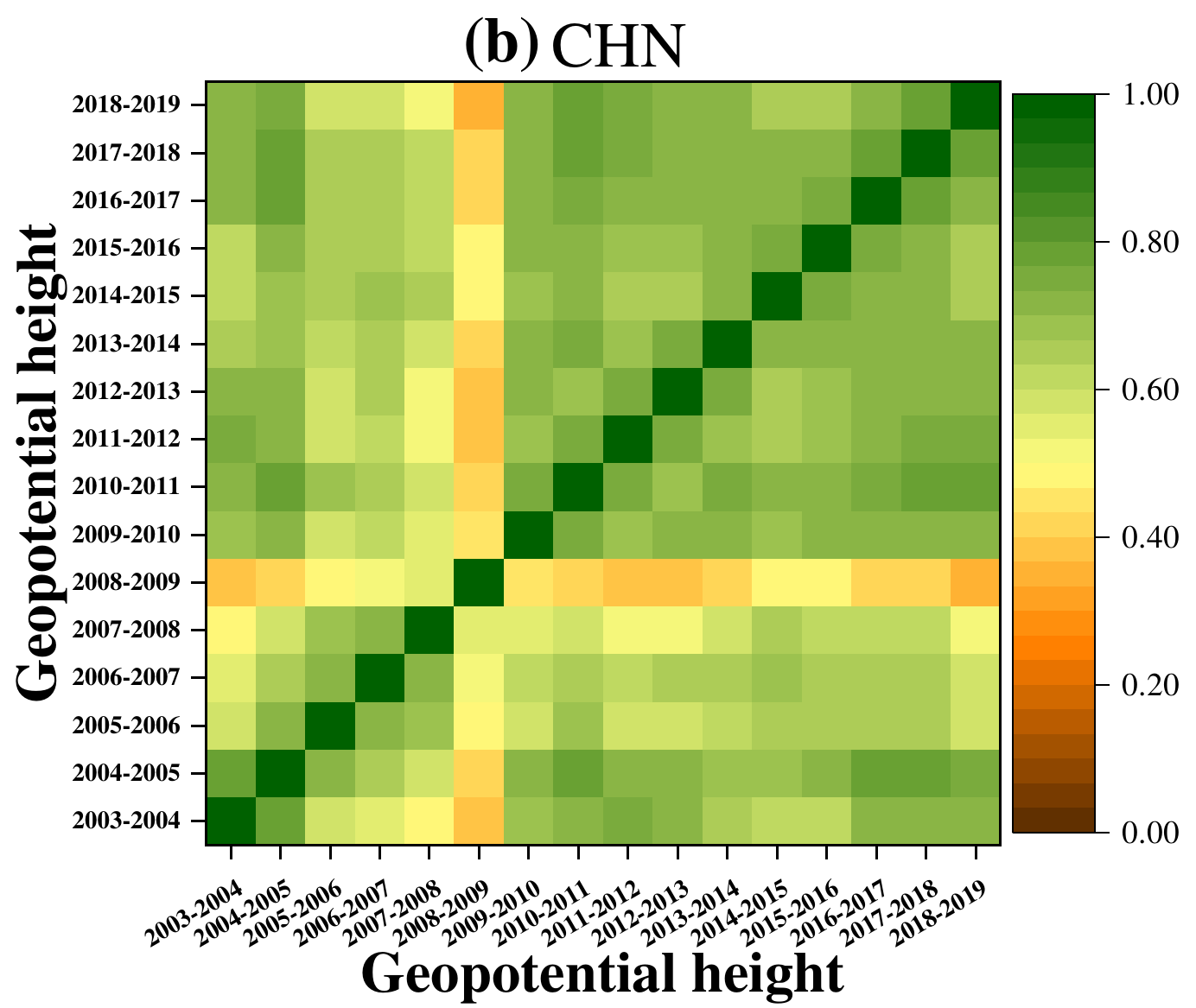}
\includegraphics[width=8em, height=7em]{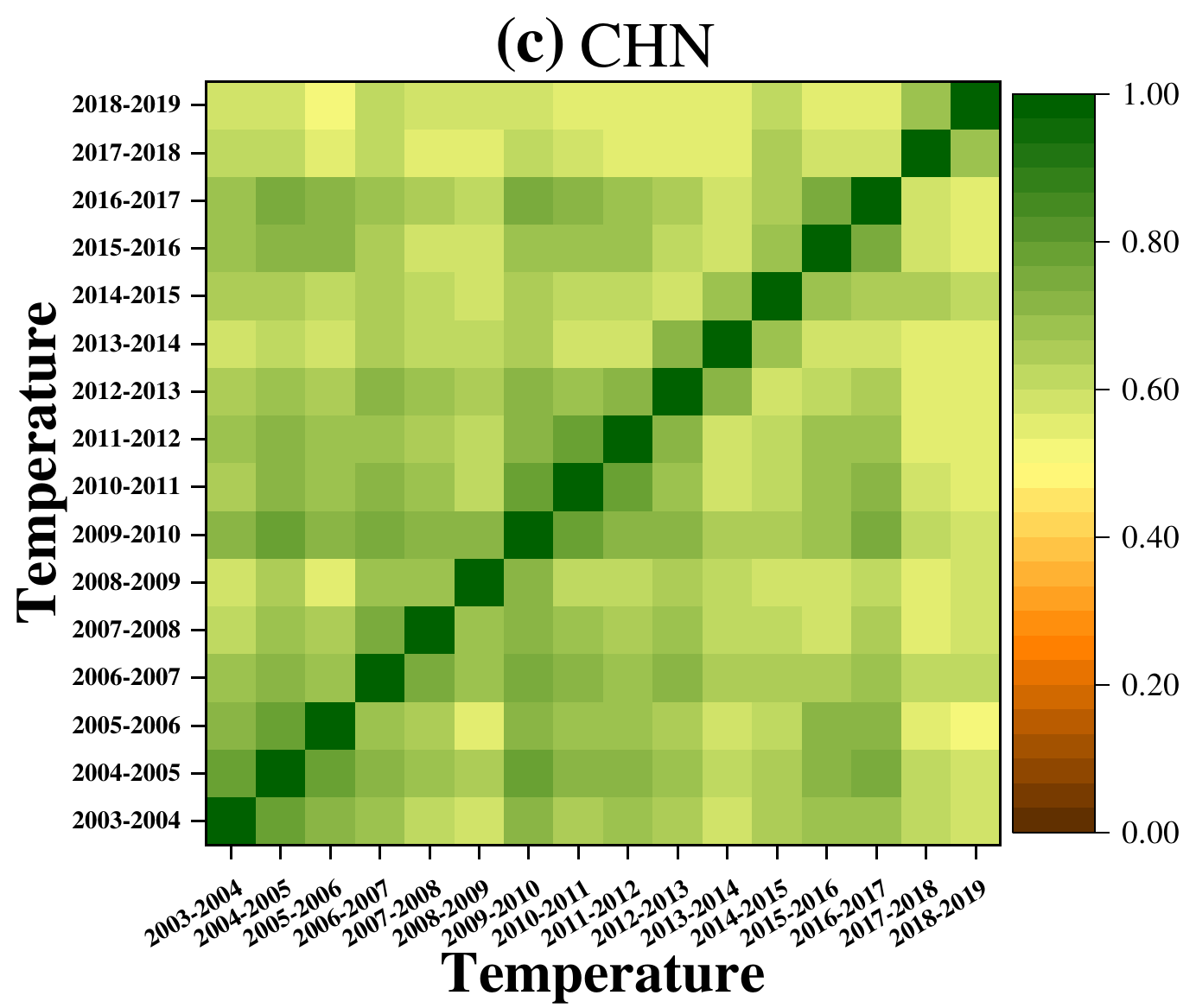}
\includegraphics[width=8em, height=7em]{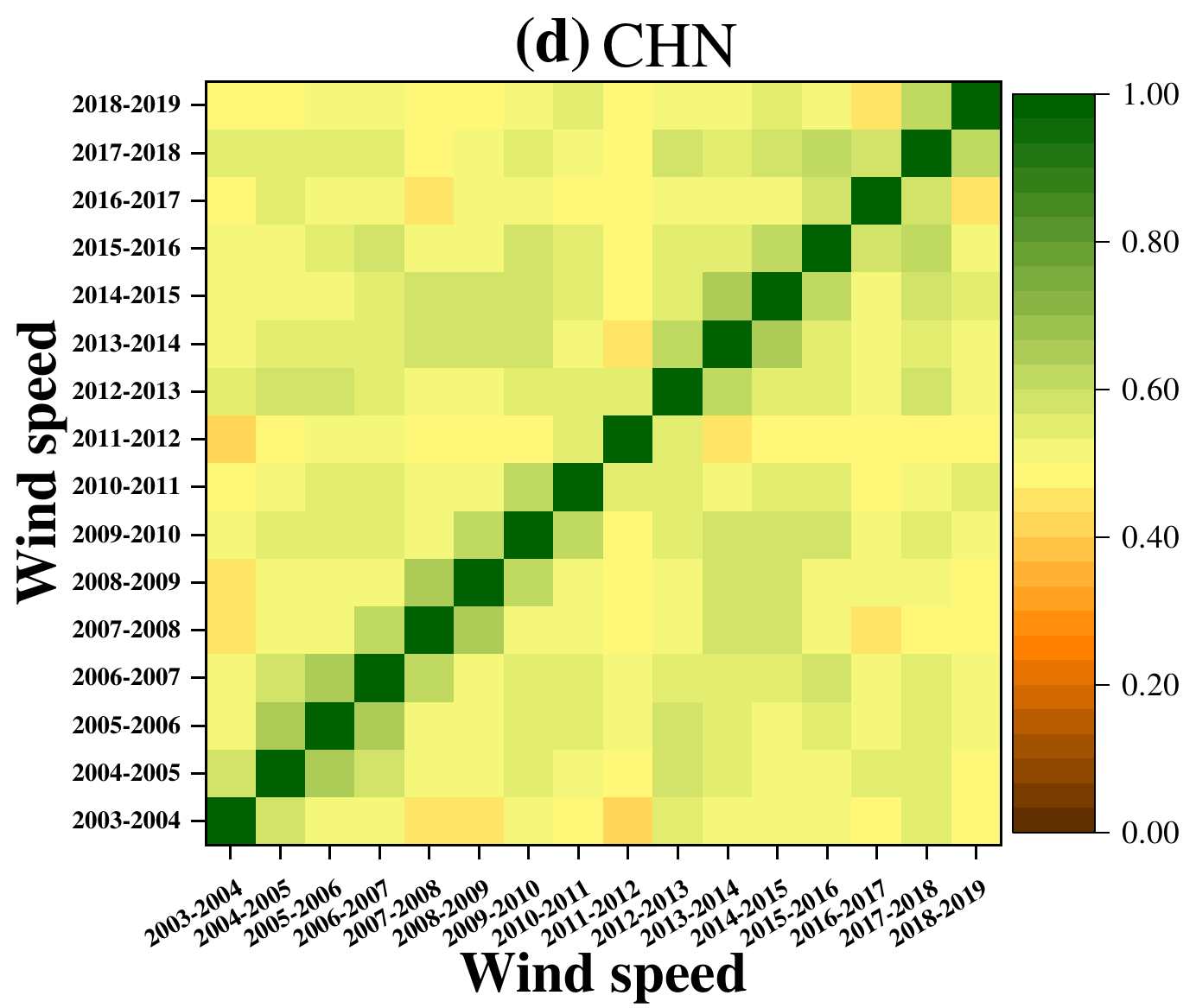}
\includegraphics[width=8em, height=7em]{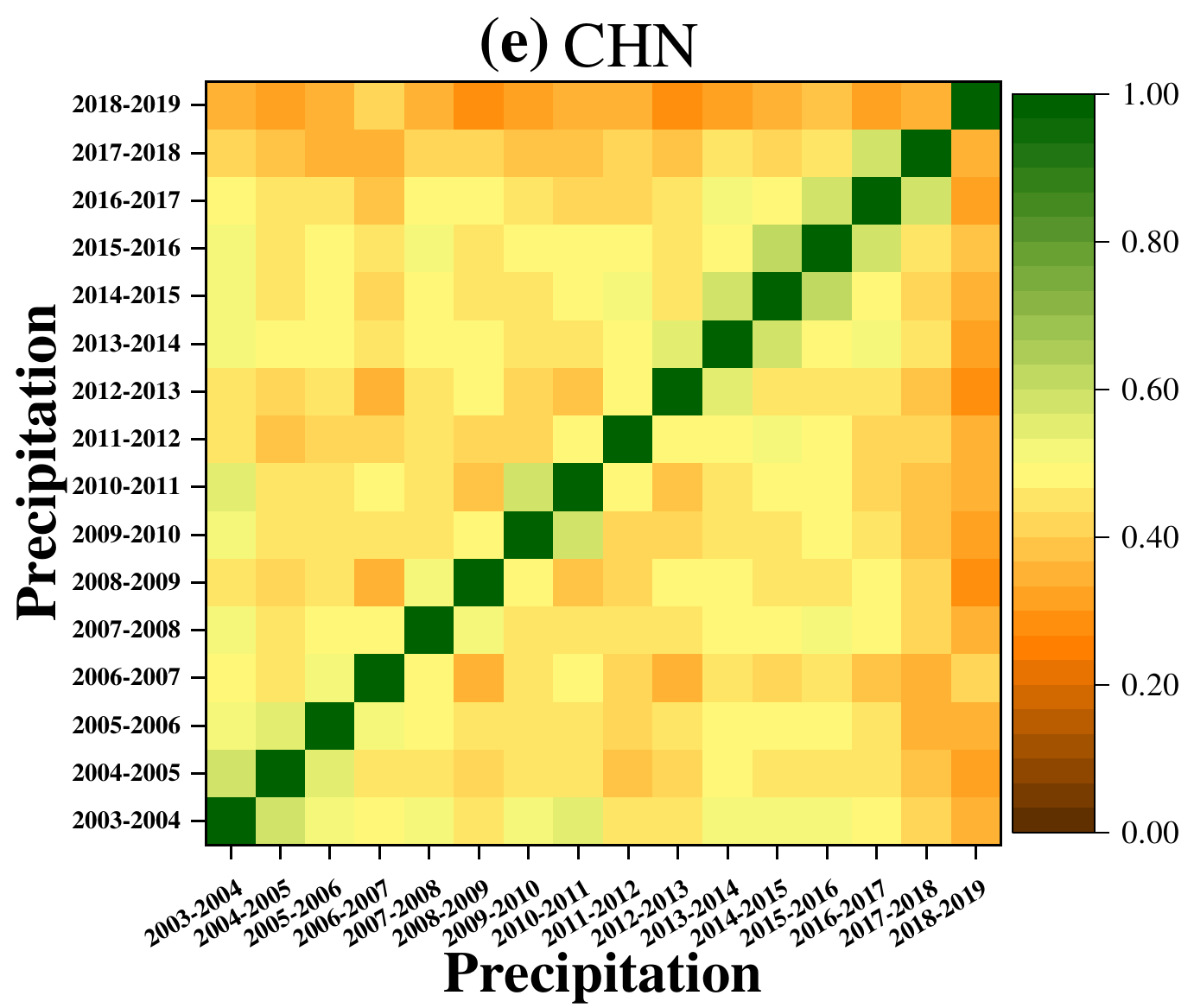}
\includegraphics[width=8em, height=7em]{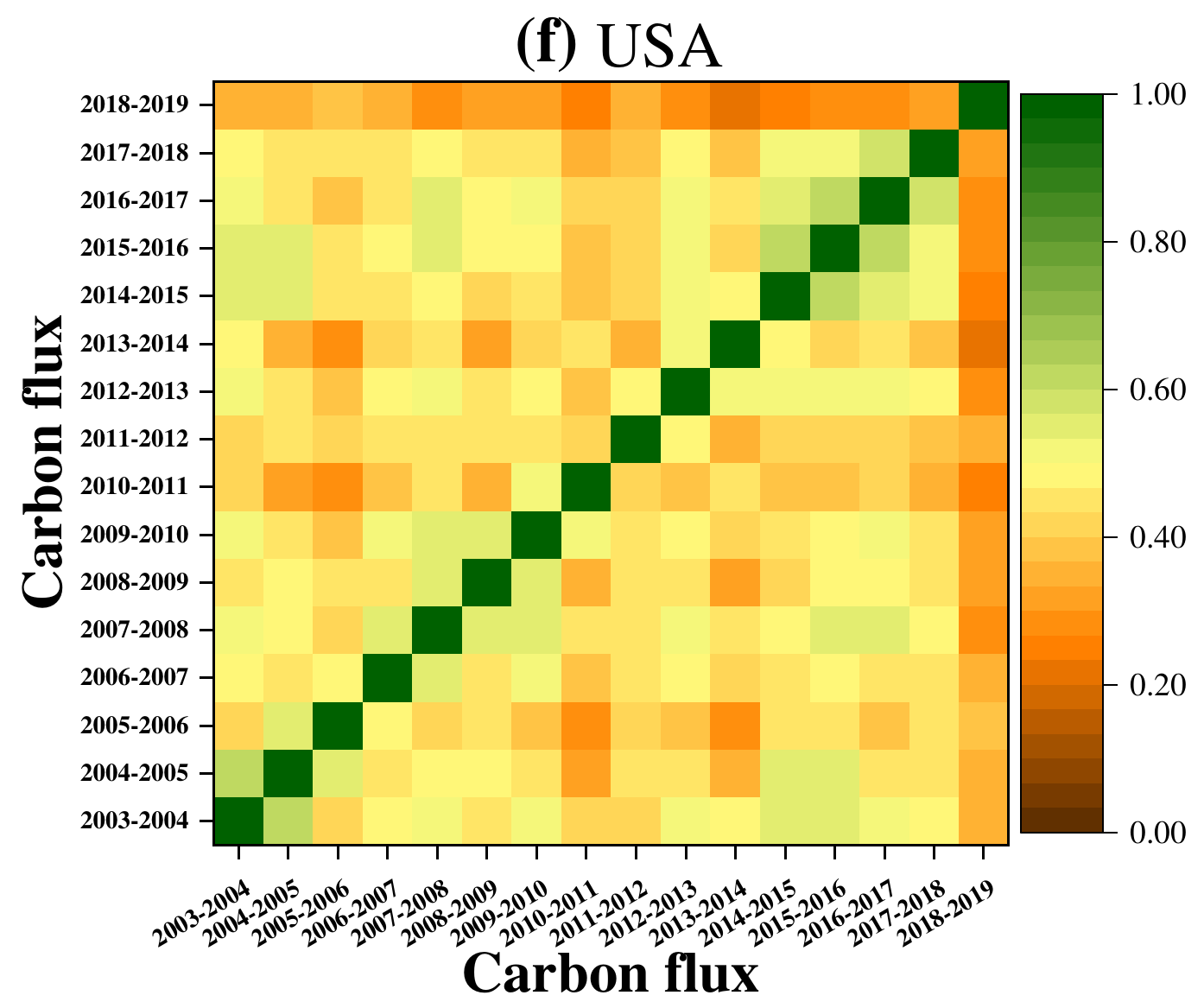}
\includegraphics[width=8em, height=7em]{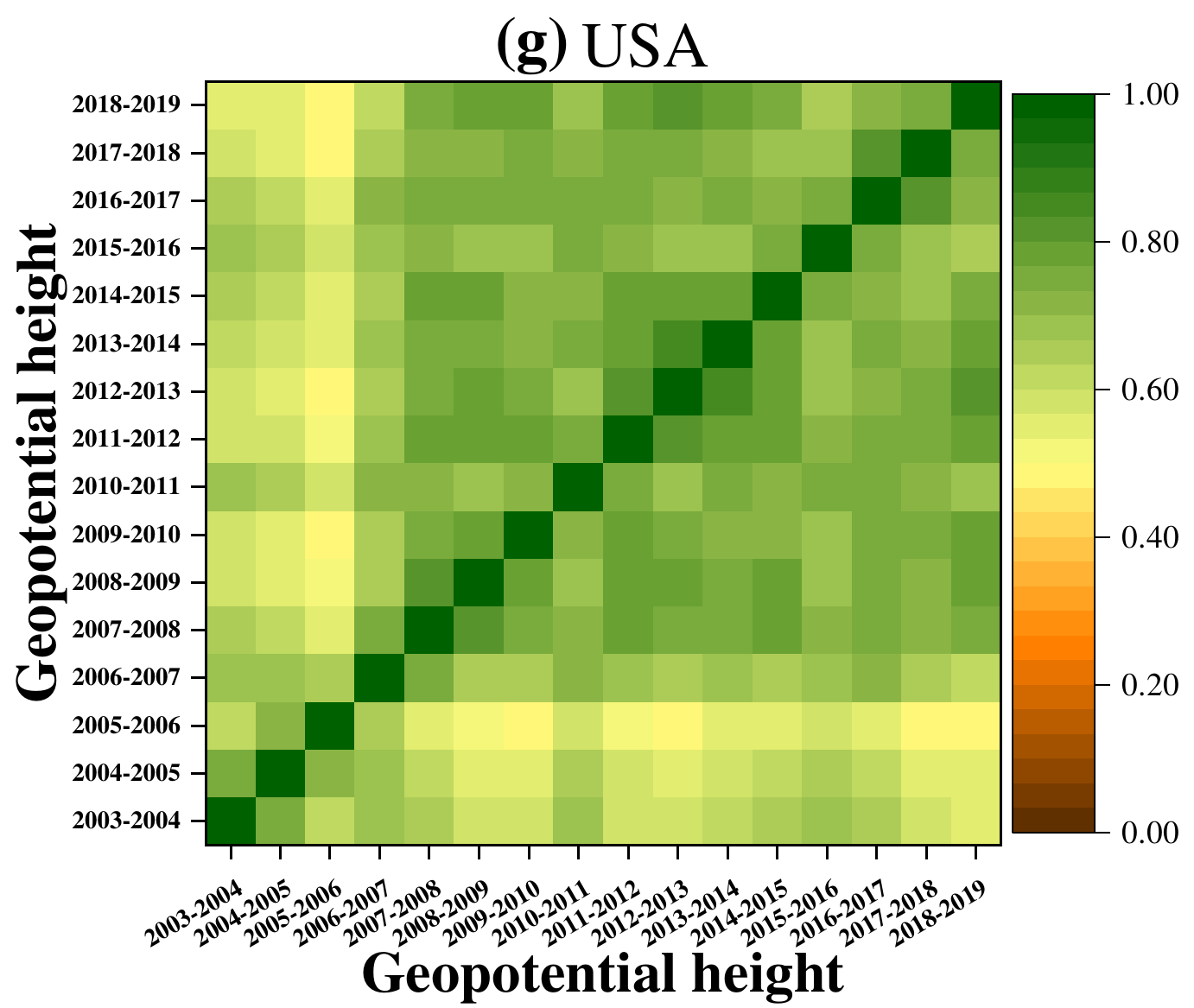}
\includegraphics[width=8em, height=7em]{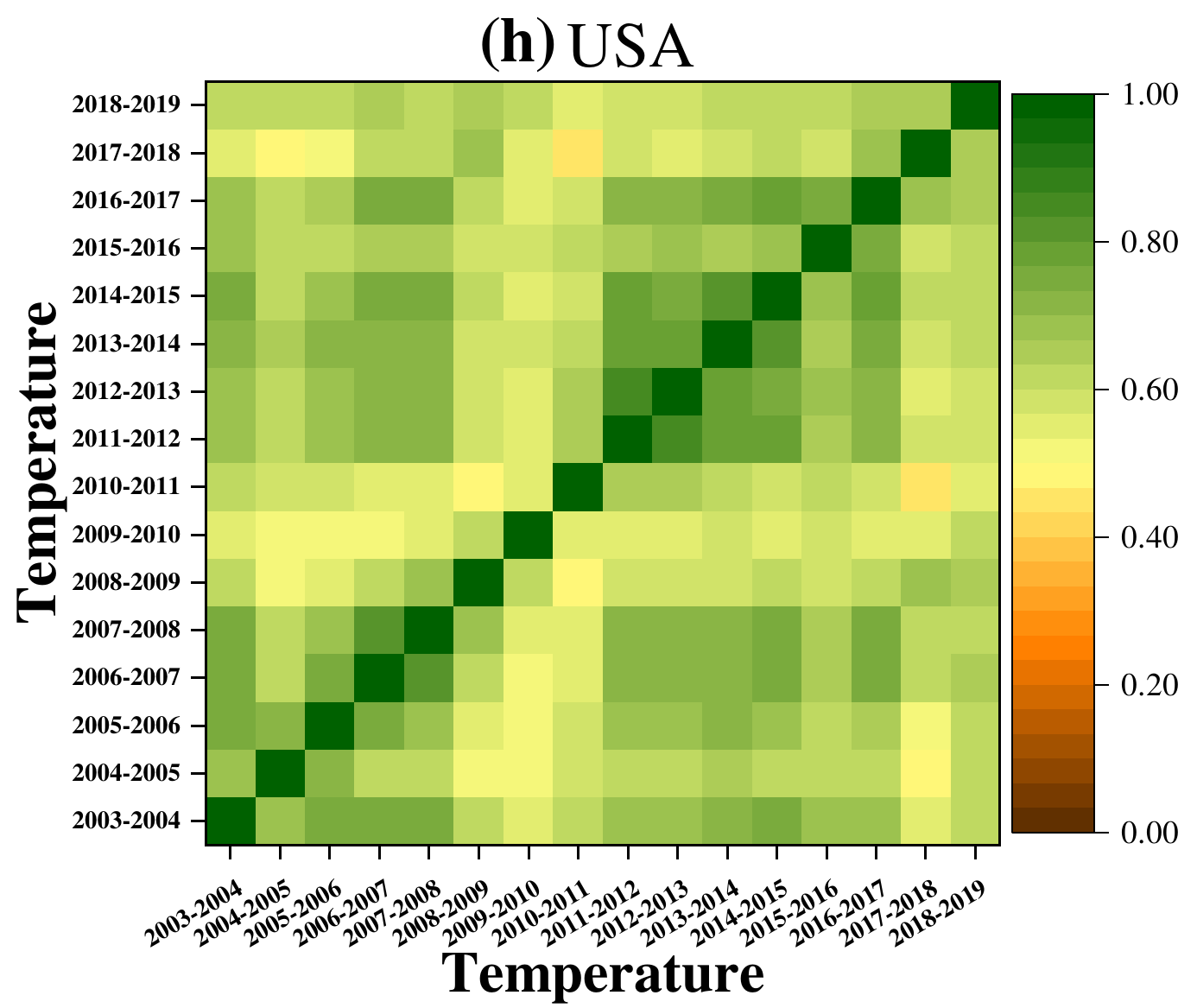}
\includegraphics[width=8em, height=7em]{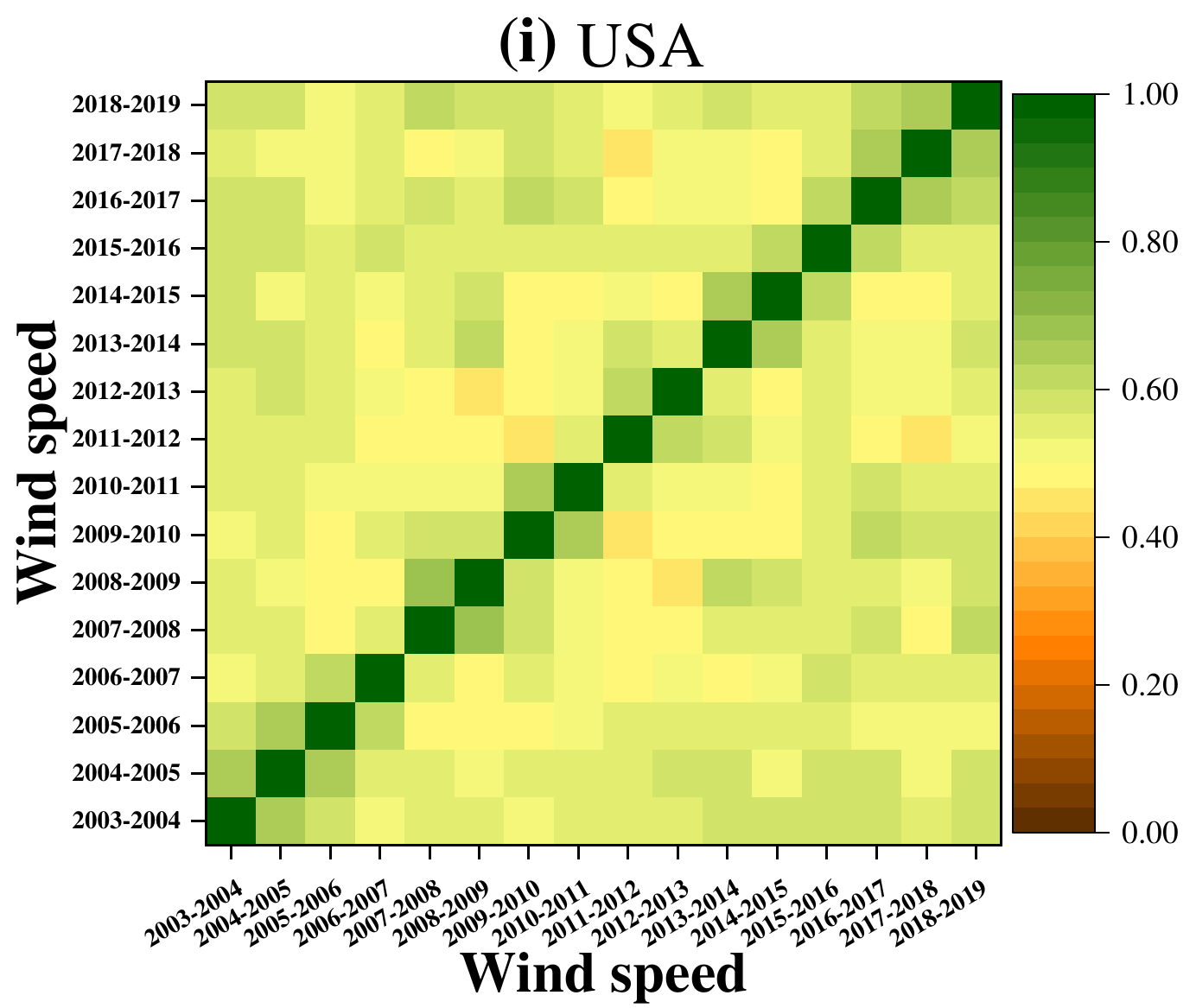}
\includegraphics[width=8em, height=7em]{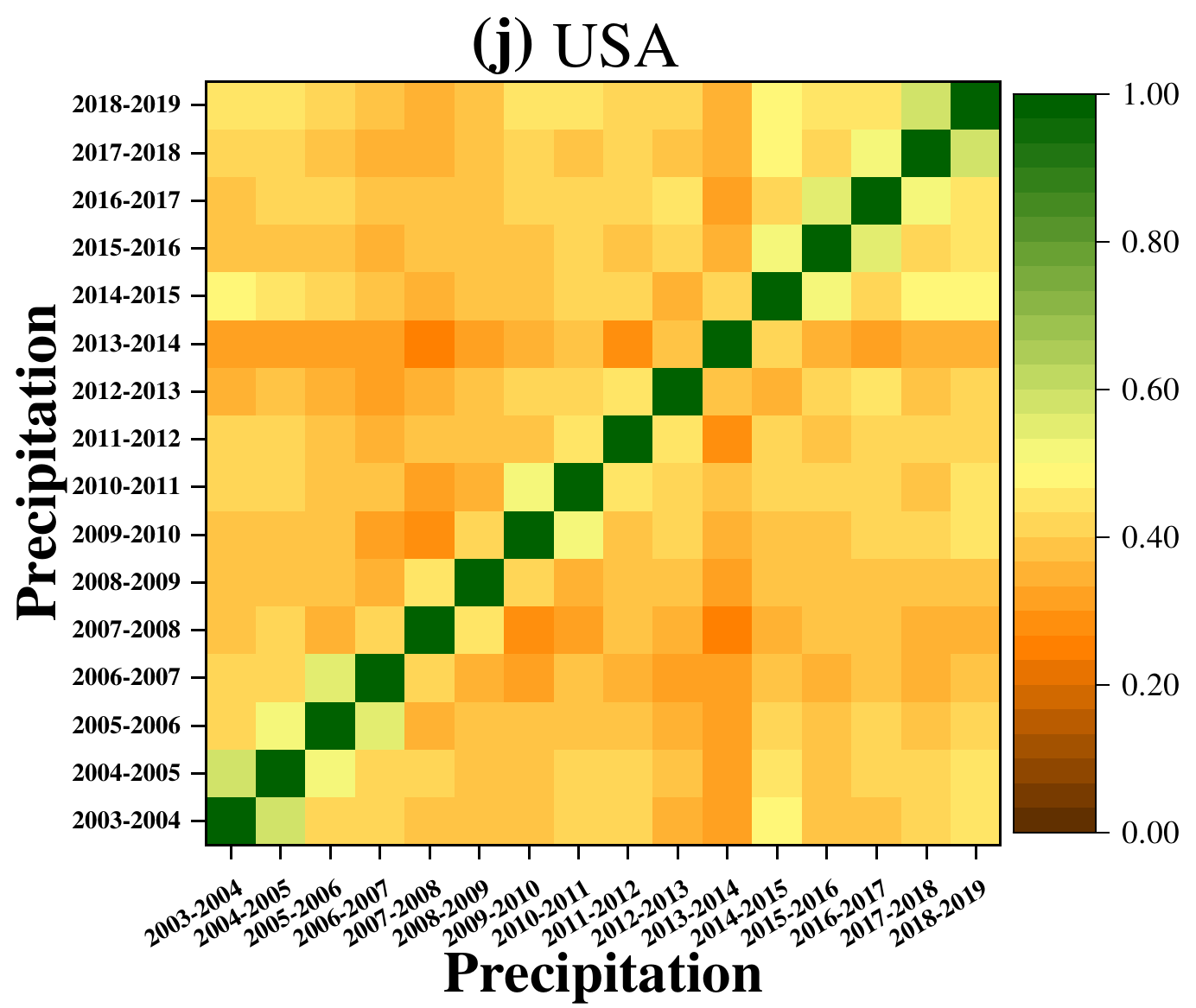}
\includegraphics[width=8em, height=7em]{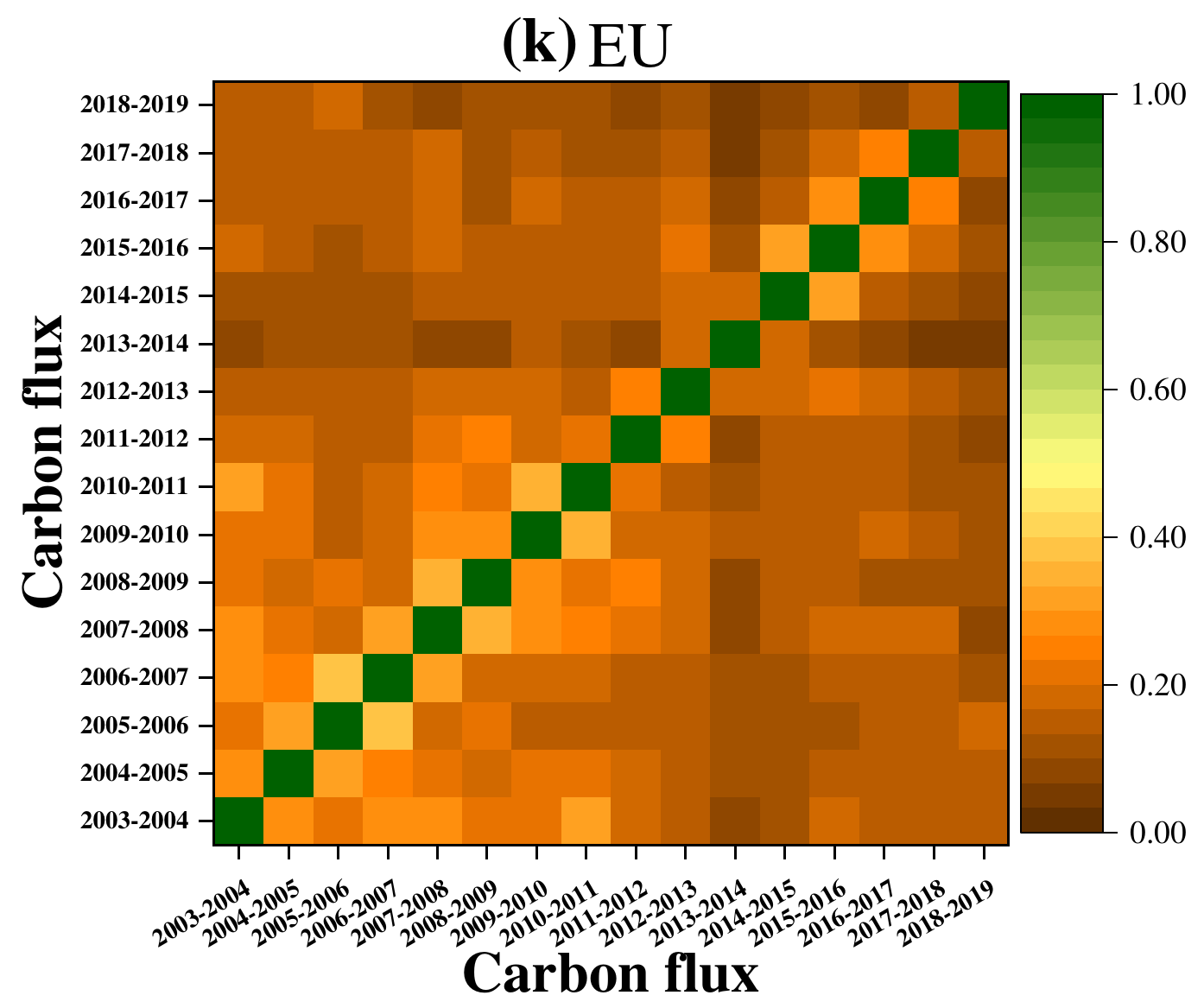}
\includegraphics[width=8em, height=7em]{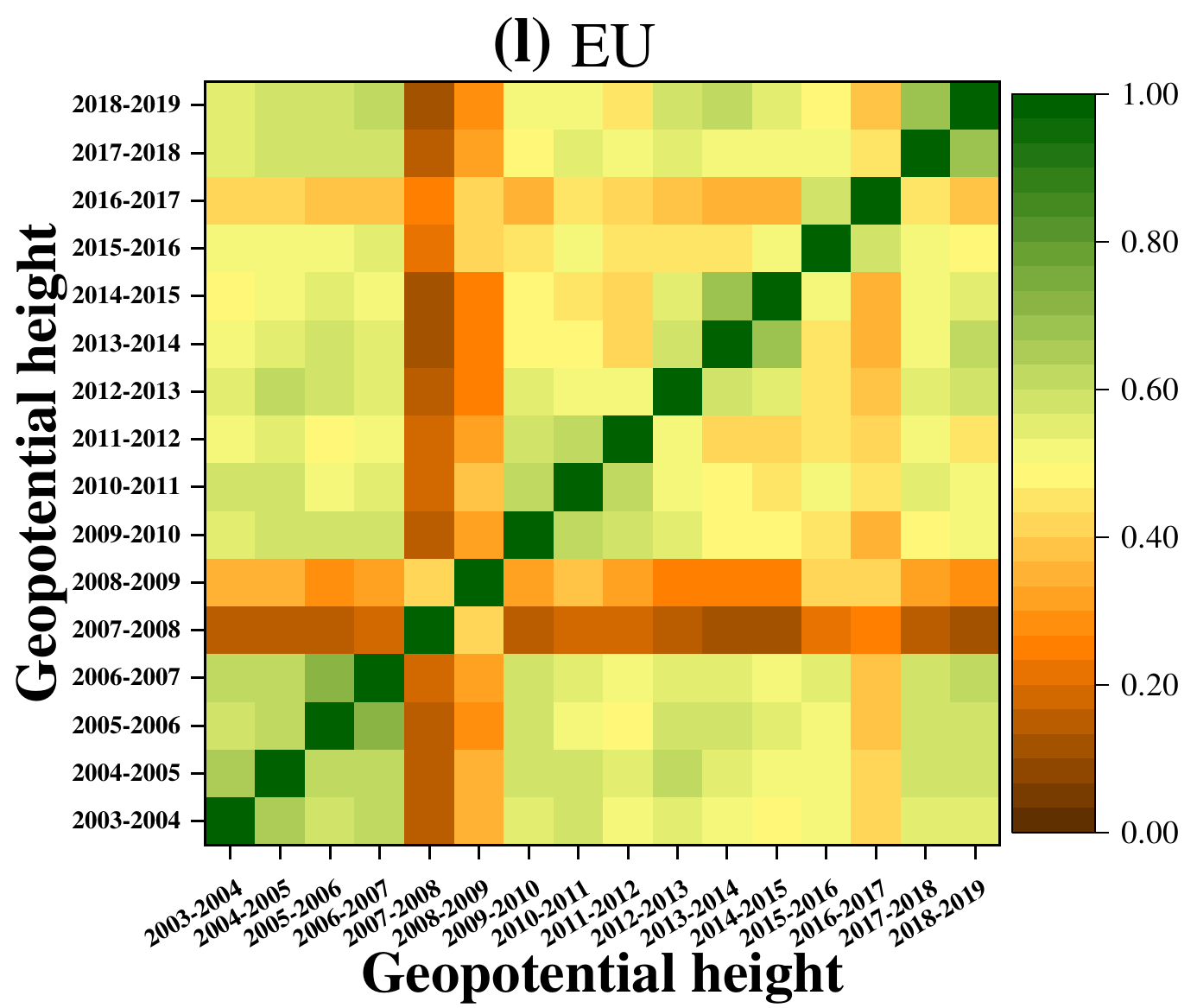}
\includegraphics[width=8em, height=7em]{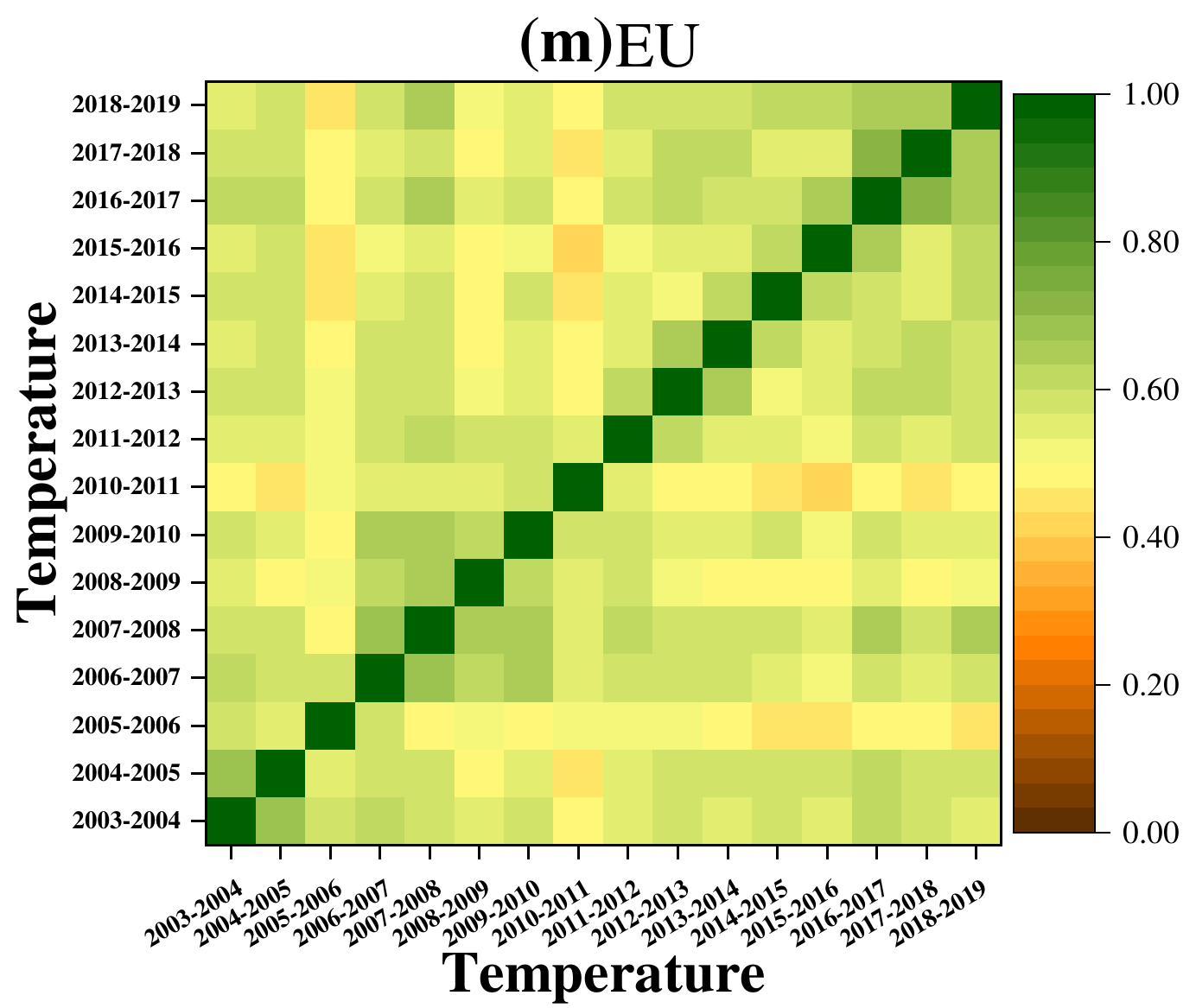}
\includegraphics[width=8em, height=7em]{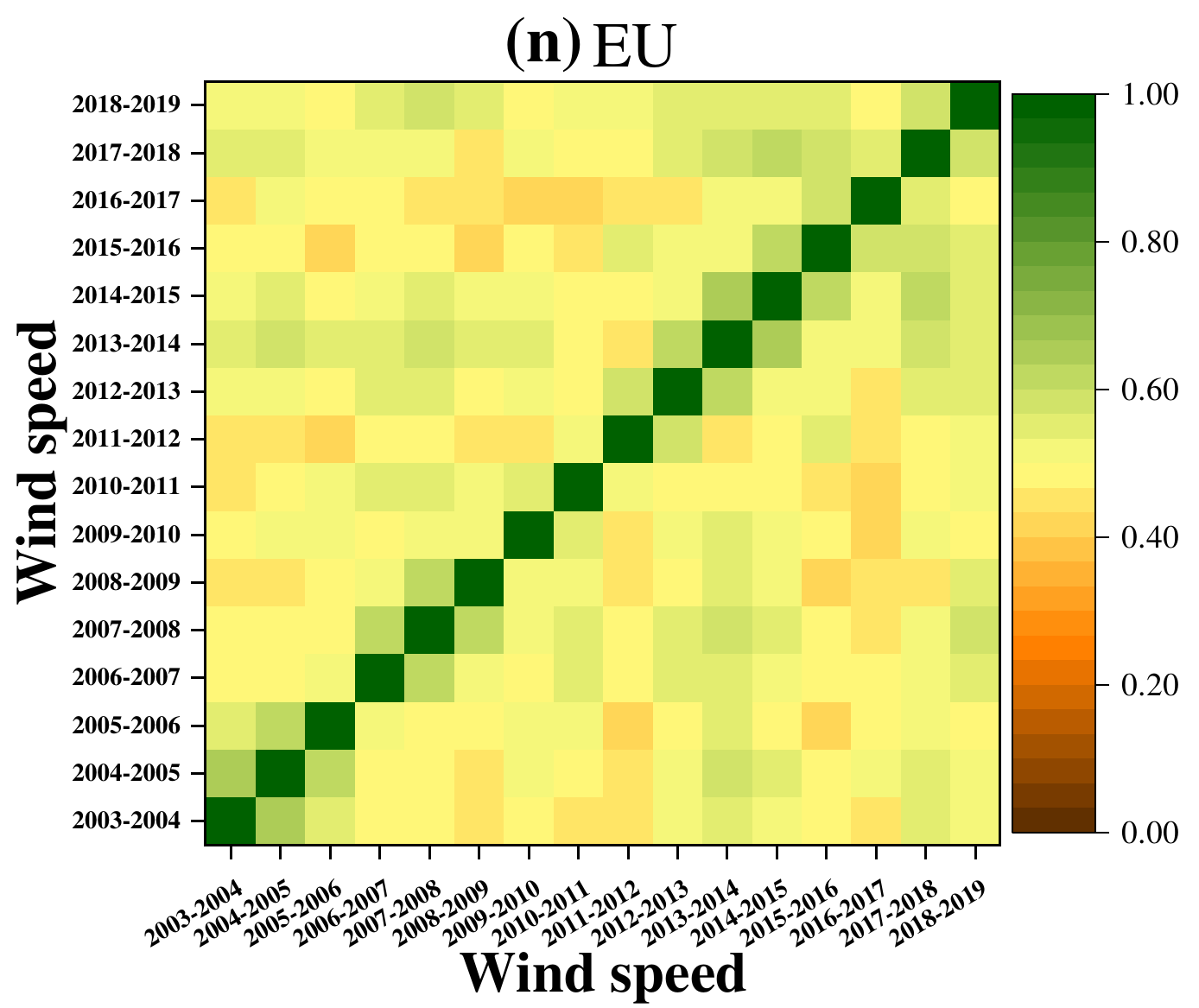}
\includegraphics[width=8em, height=7em]{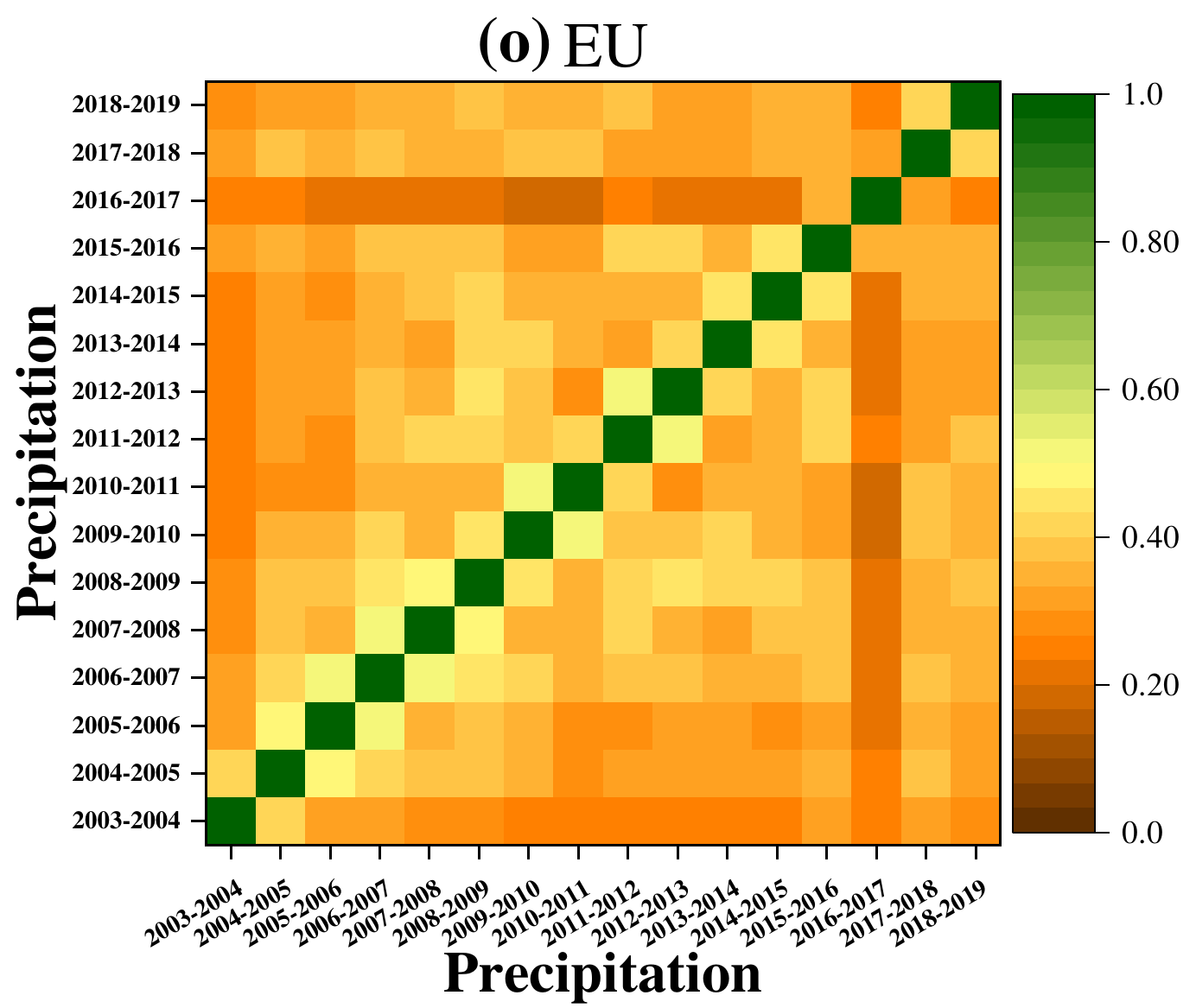}
\end{center}

\begin{center}
\noindent {\small {\bf Fig. S26} The actual Jaccard similarity coefficient matrix of links of lengths above $1000km$ in two networks of different years for each of the climate variables.}
\end{center}

\begin{center}
\includegraphics[width=8em, height=7em]{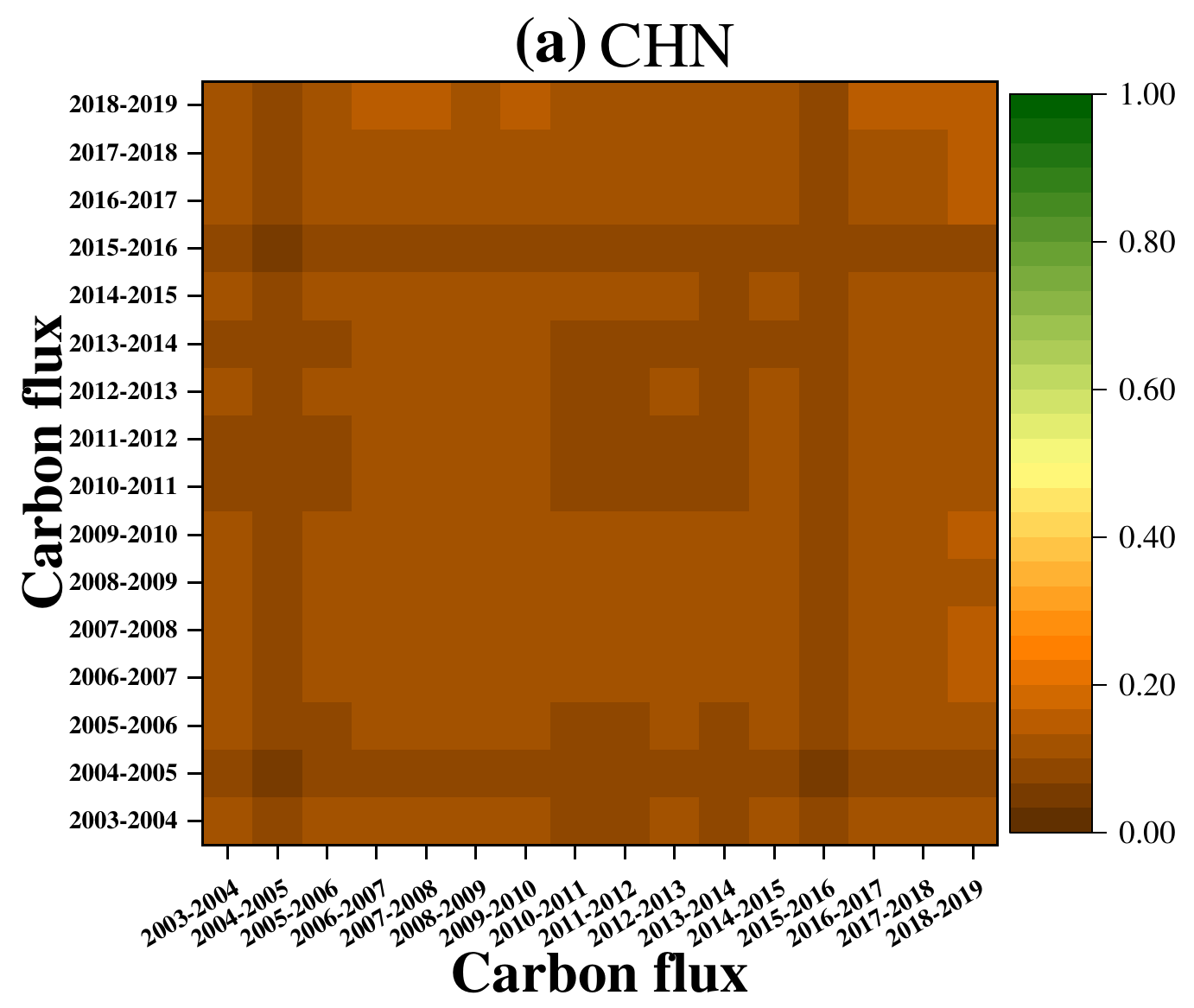}
\includegraphics[width=8em, height=7em]{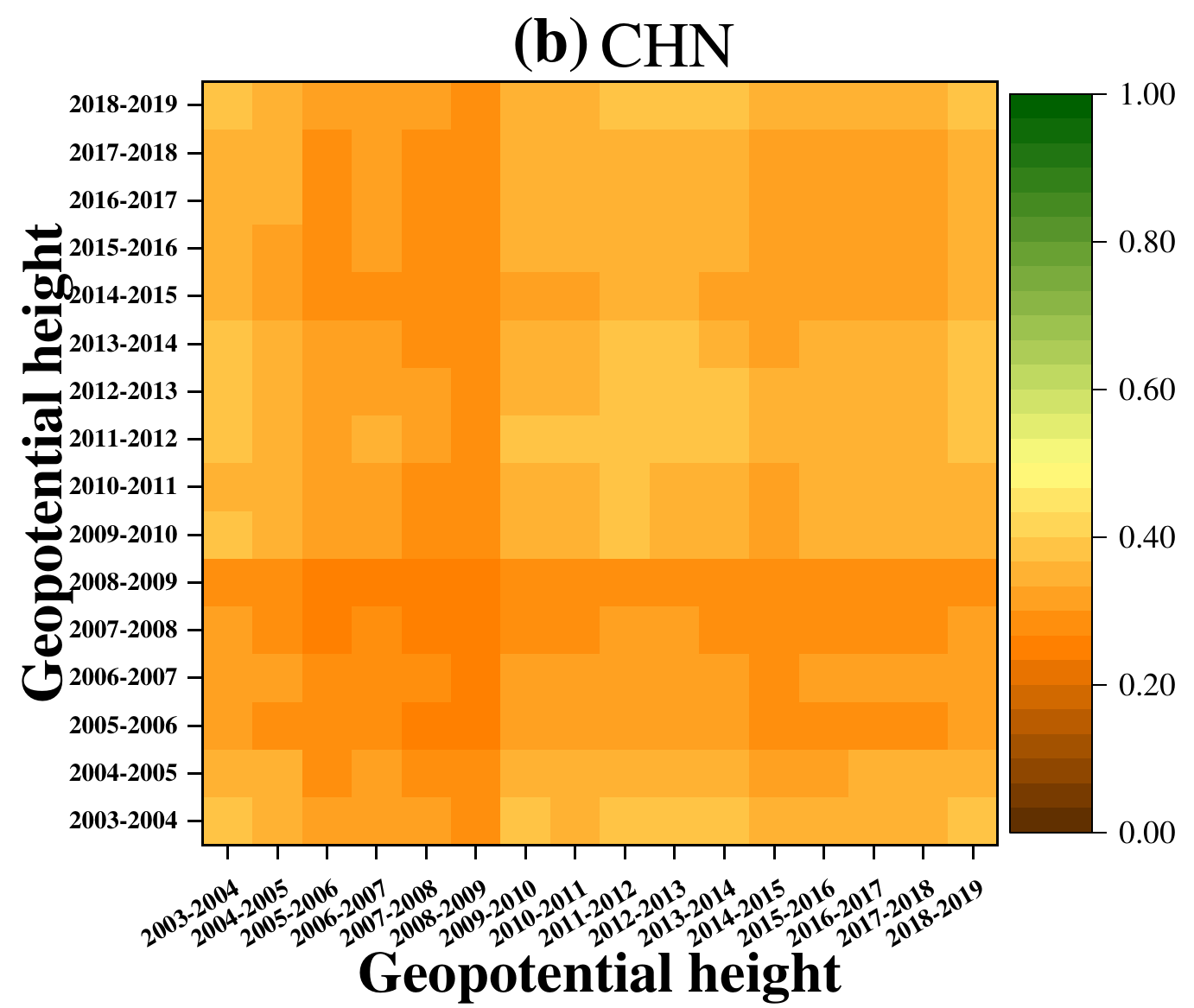}
\includegraphics[width=8em, height=7em]{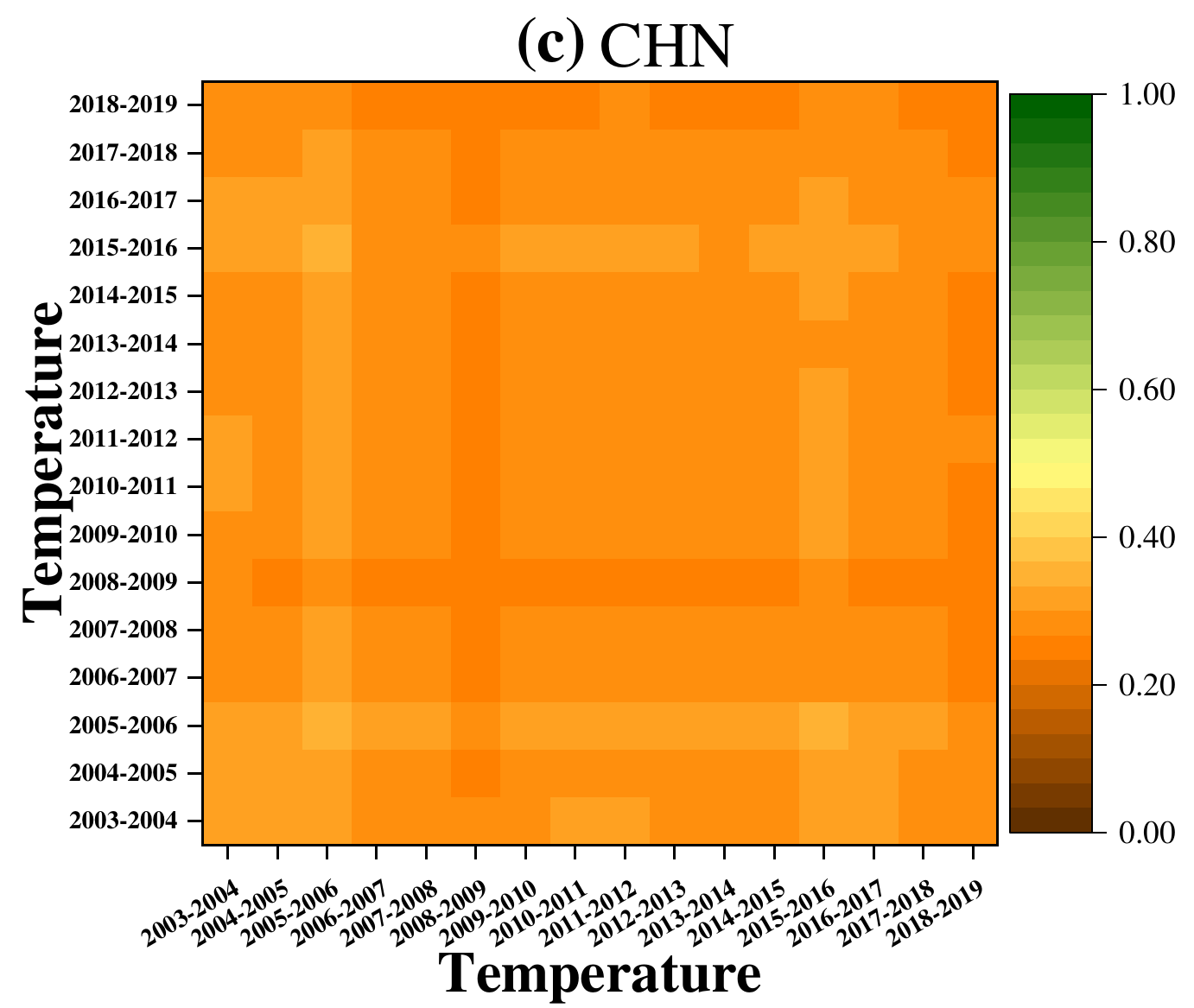}
\includegraphics[width=8em, height=7em]{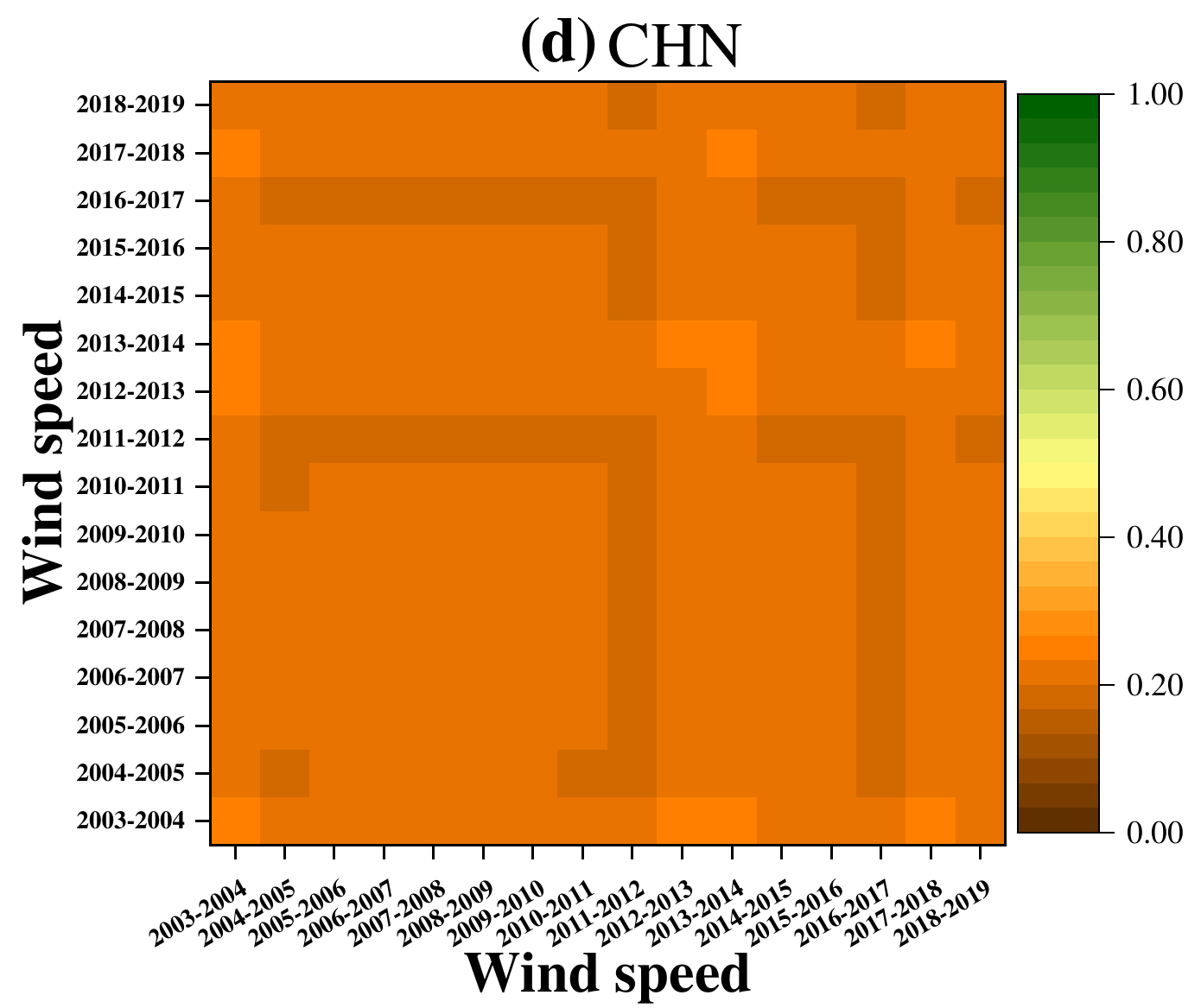}
\includegraphics[width=8em, height=7em]{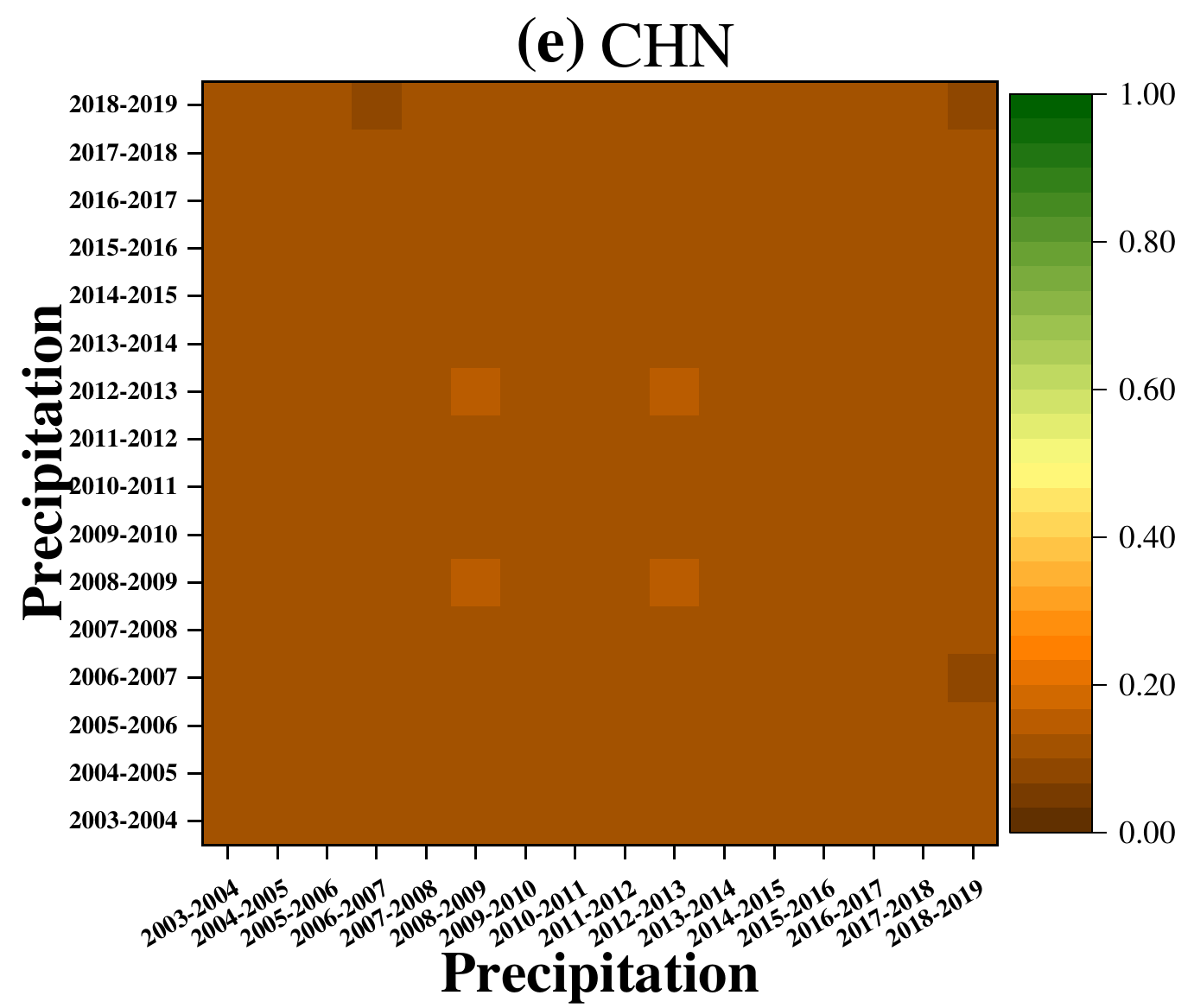}
\includegraphics[width=8em, height=7em]{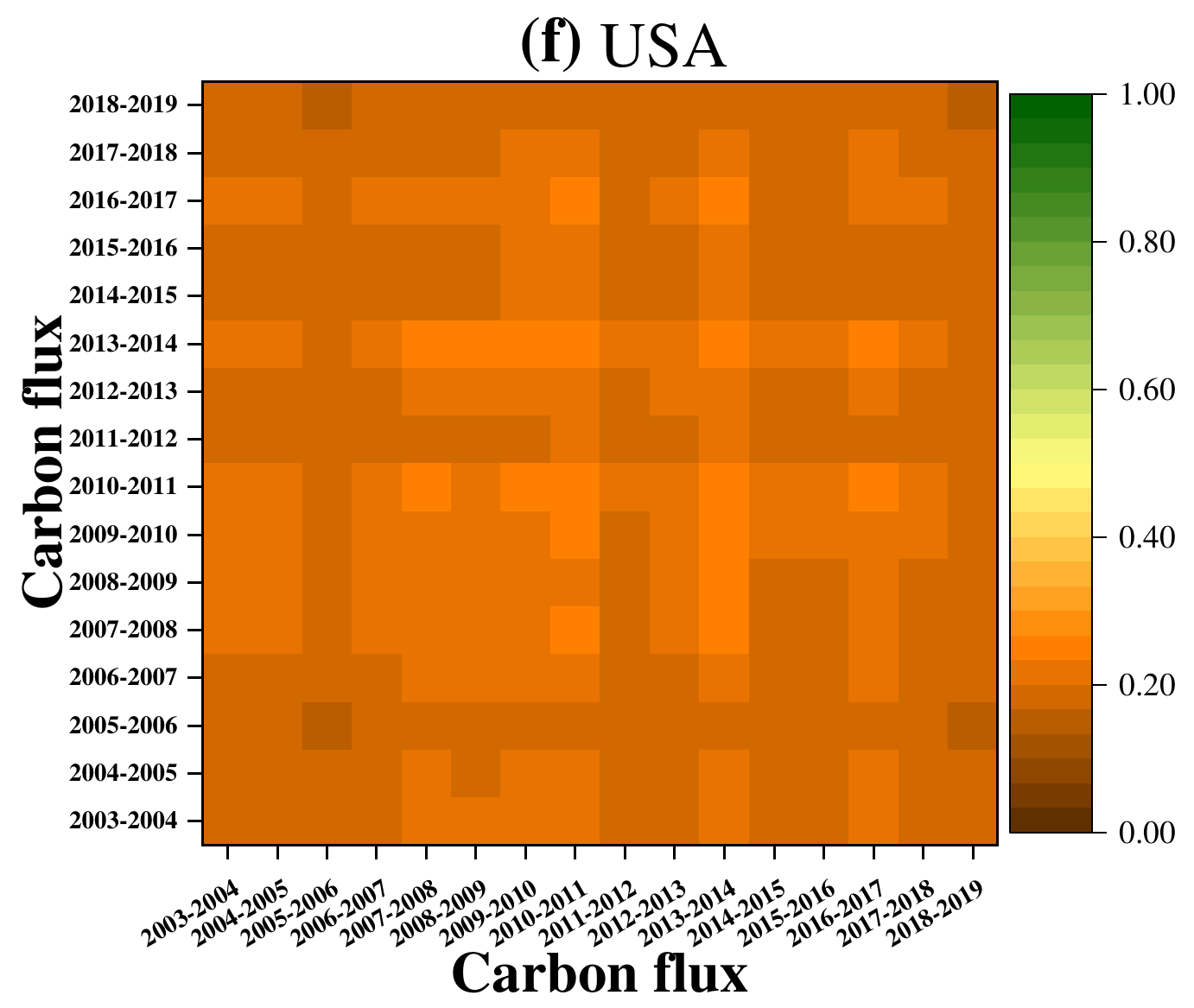}
\includegraphics[width=8em, height=7em]{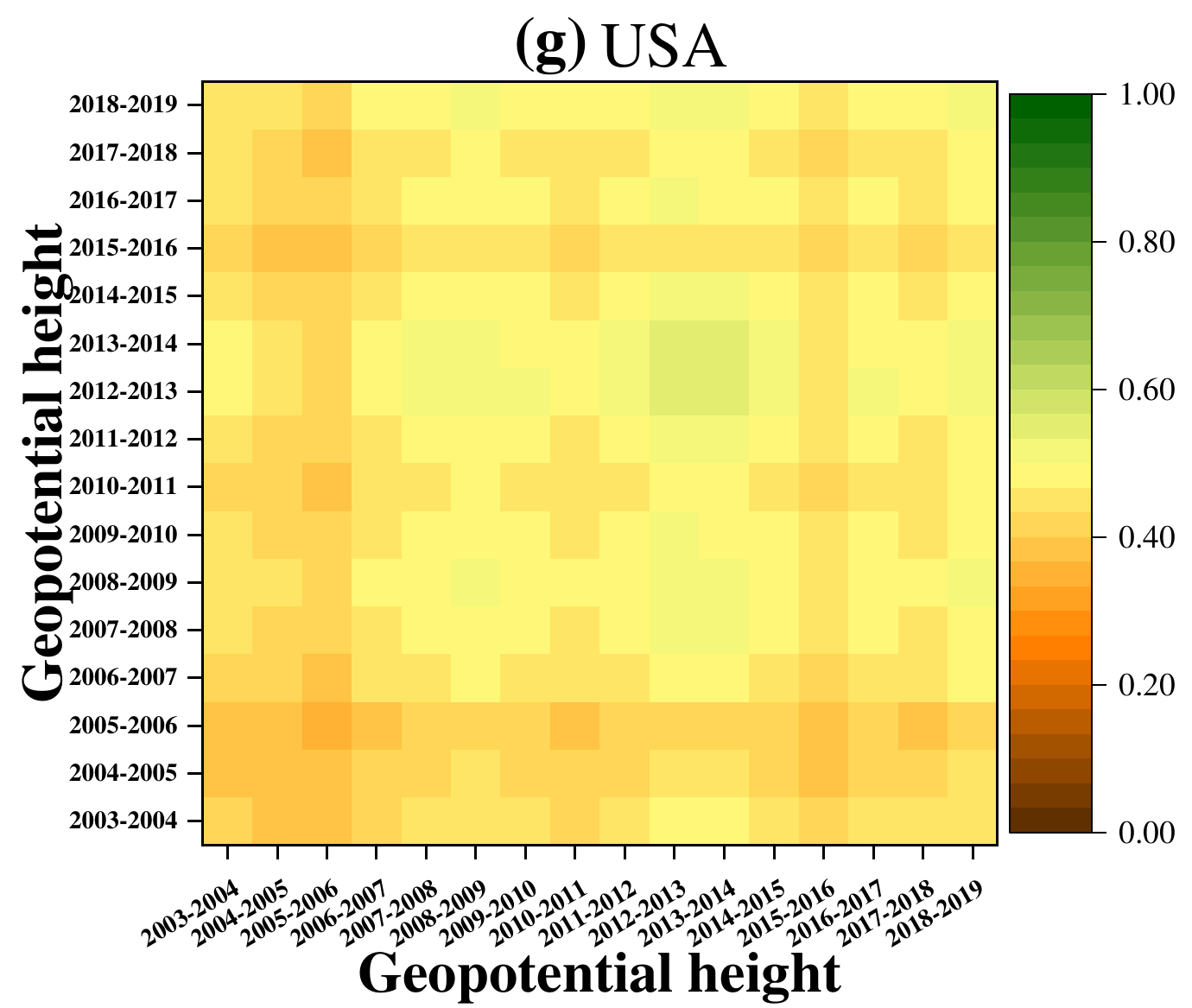}
\includegraphics[width=8em, height=7em]{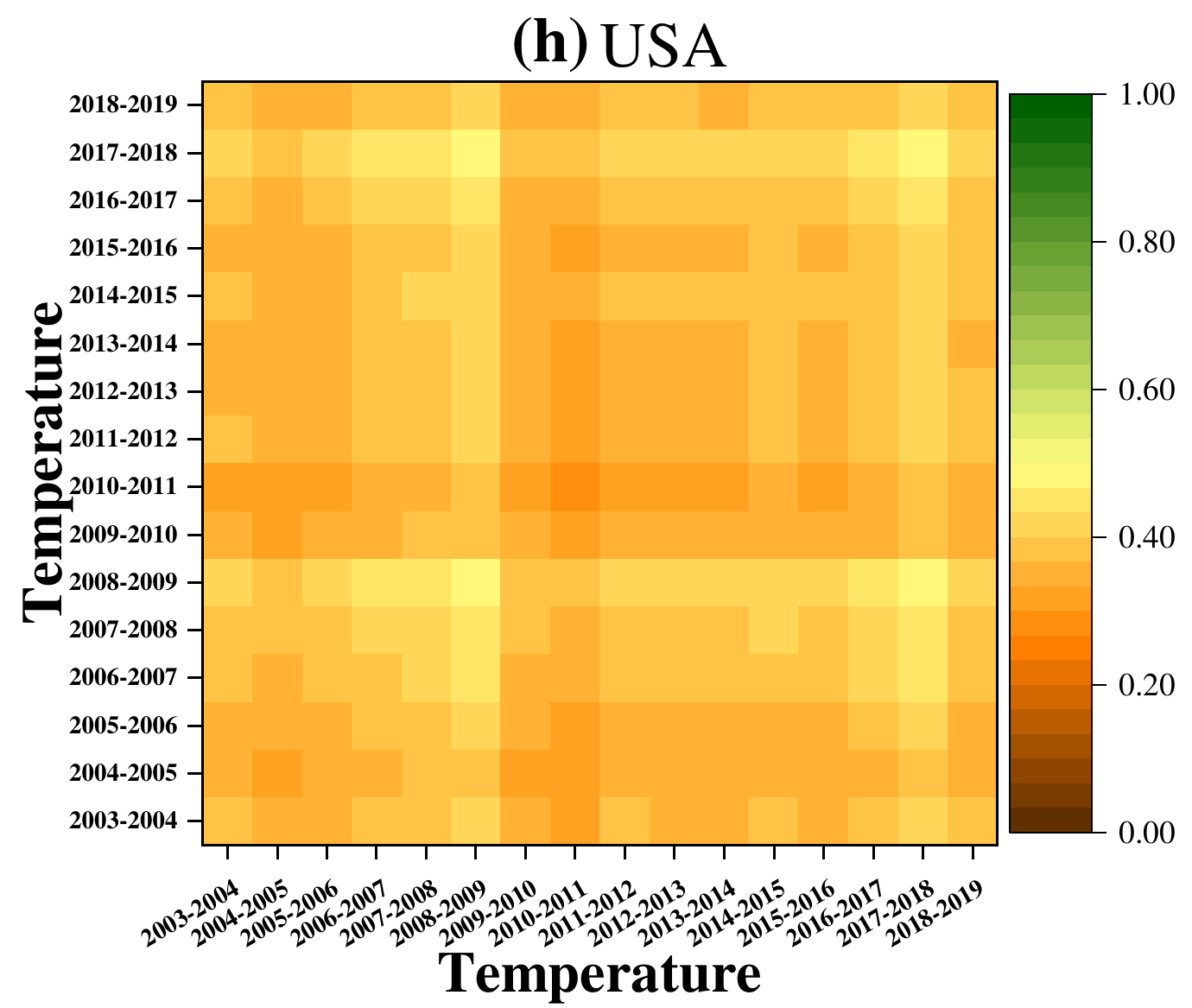}
\includegraphics[width=8em, height=7em]{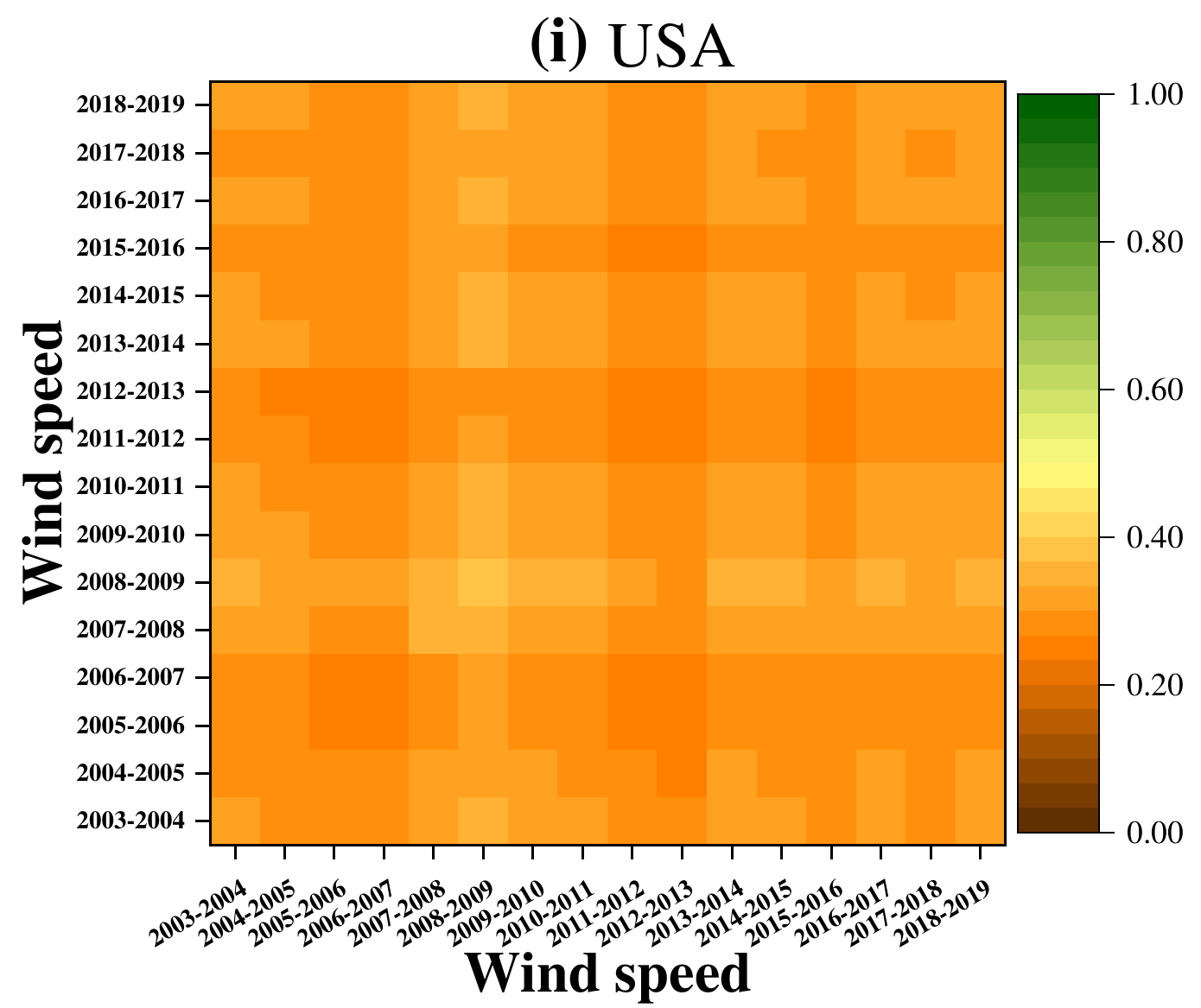}
\includegraphics[width=8em, height=7em]{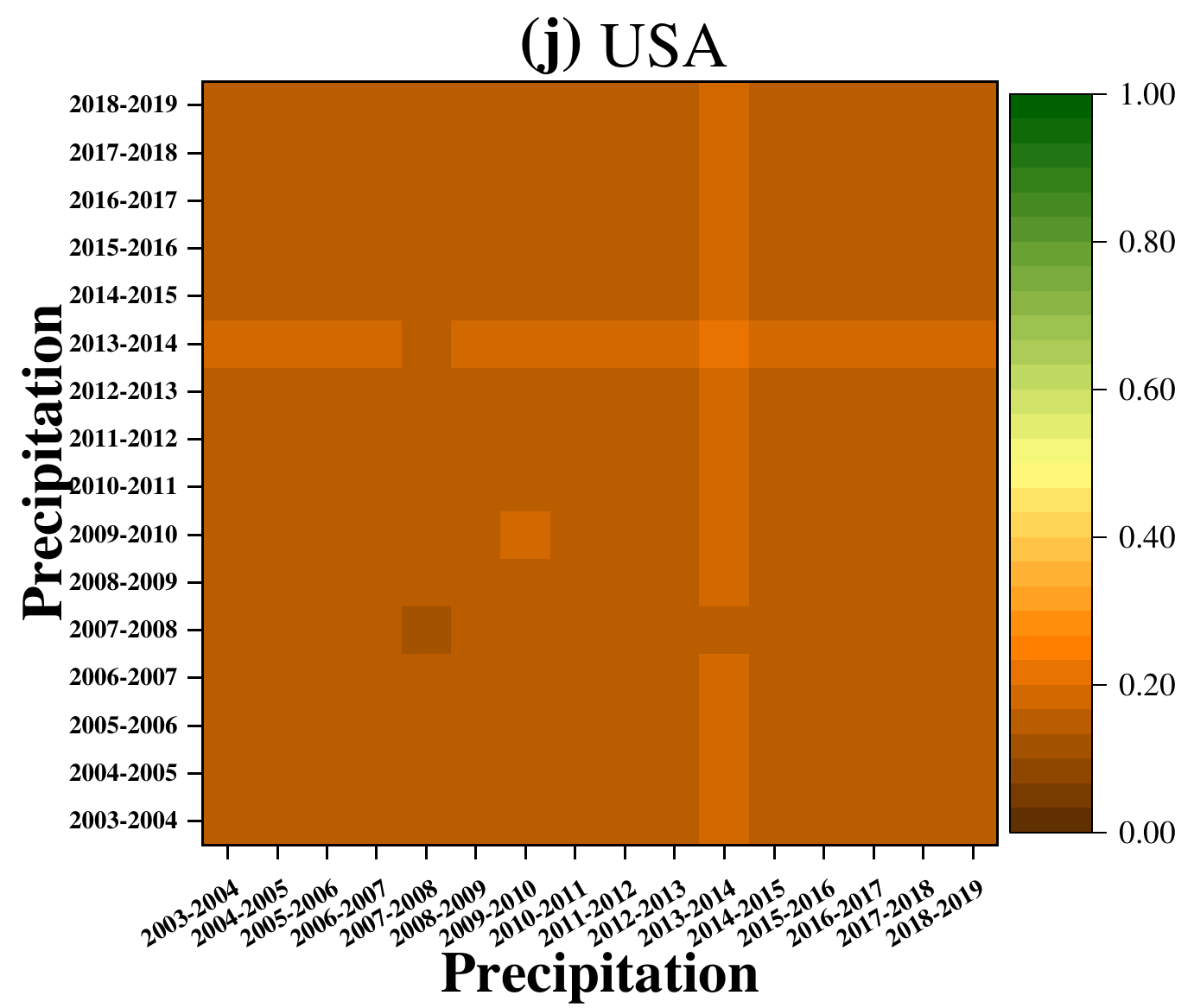}
\includegraphics[width=8em, height=7em]{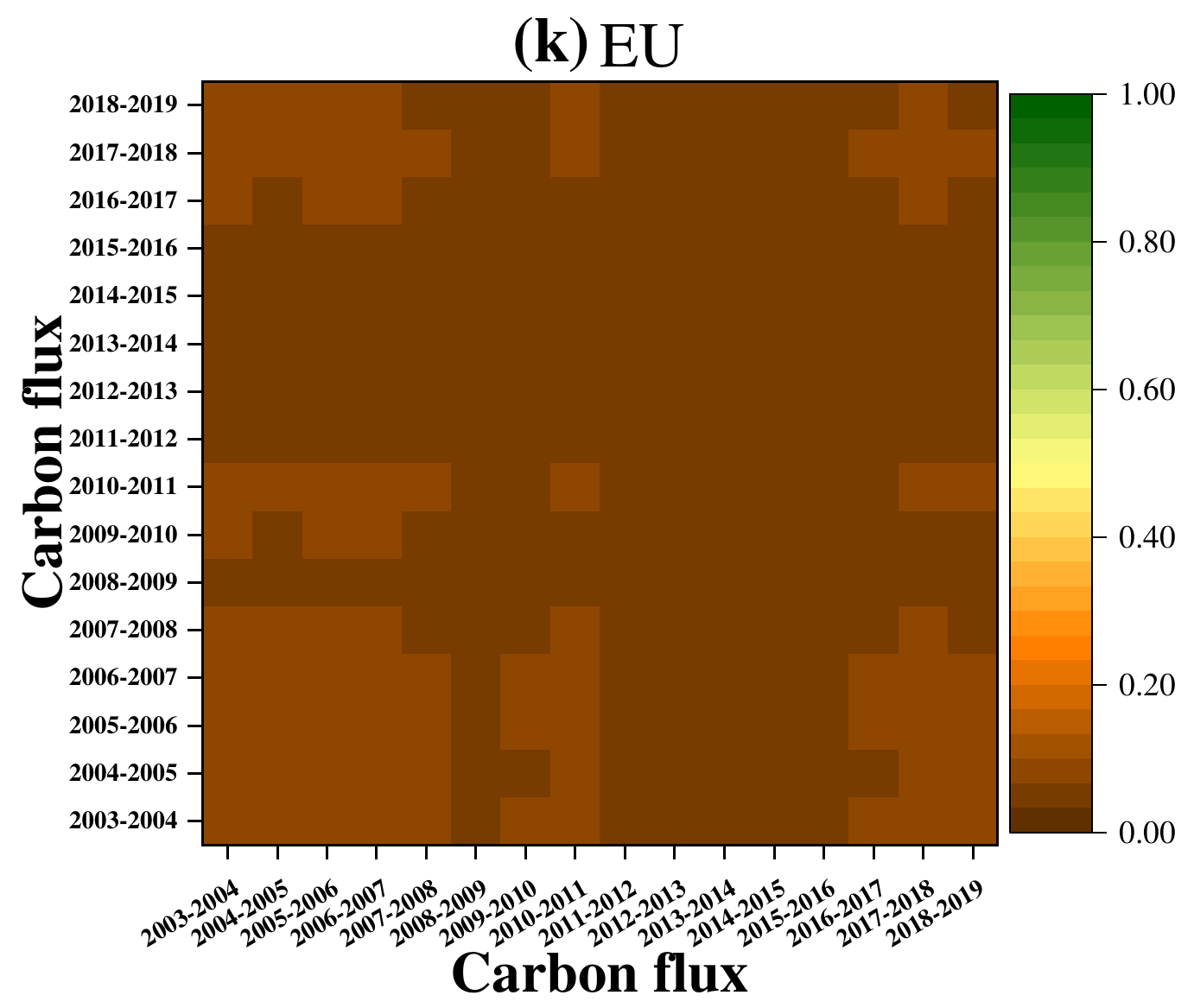}
\includegraphics[width=8em, height=7em]{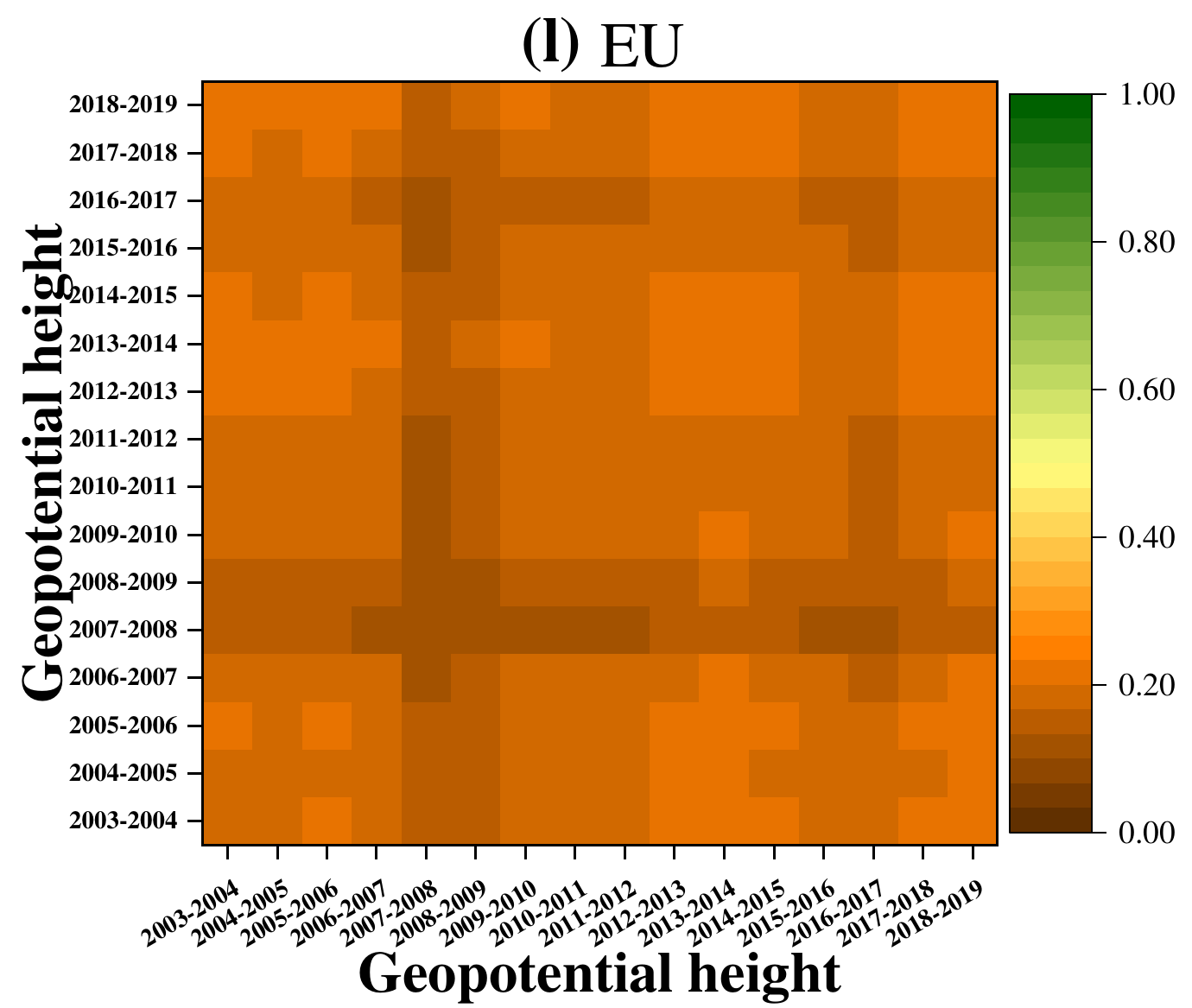}
\includegraphics[width=8em, height=7em]{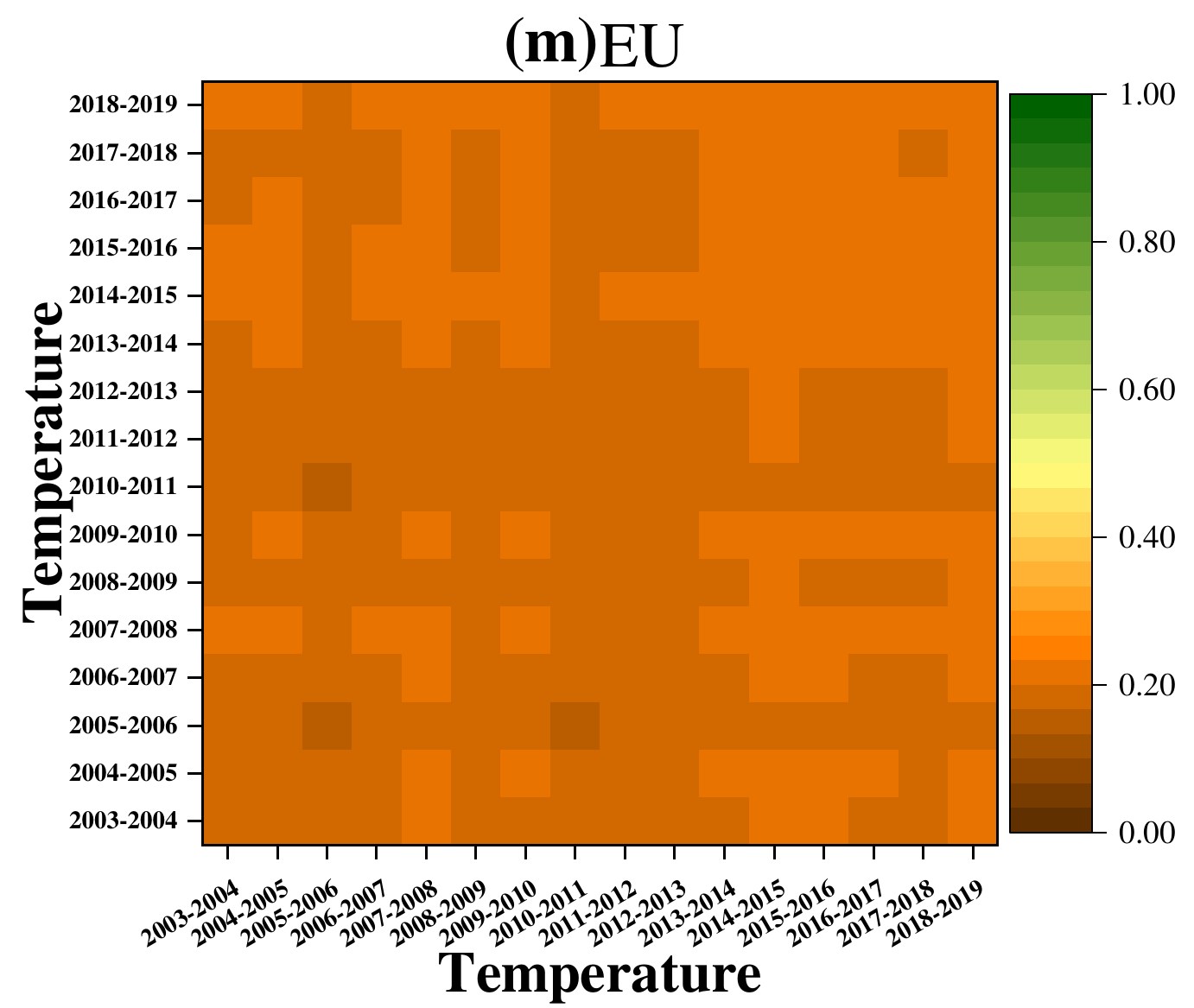}
\includegraphics[width=8em, height=7em]{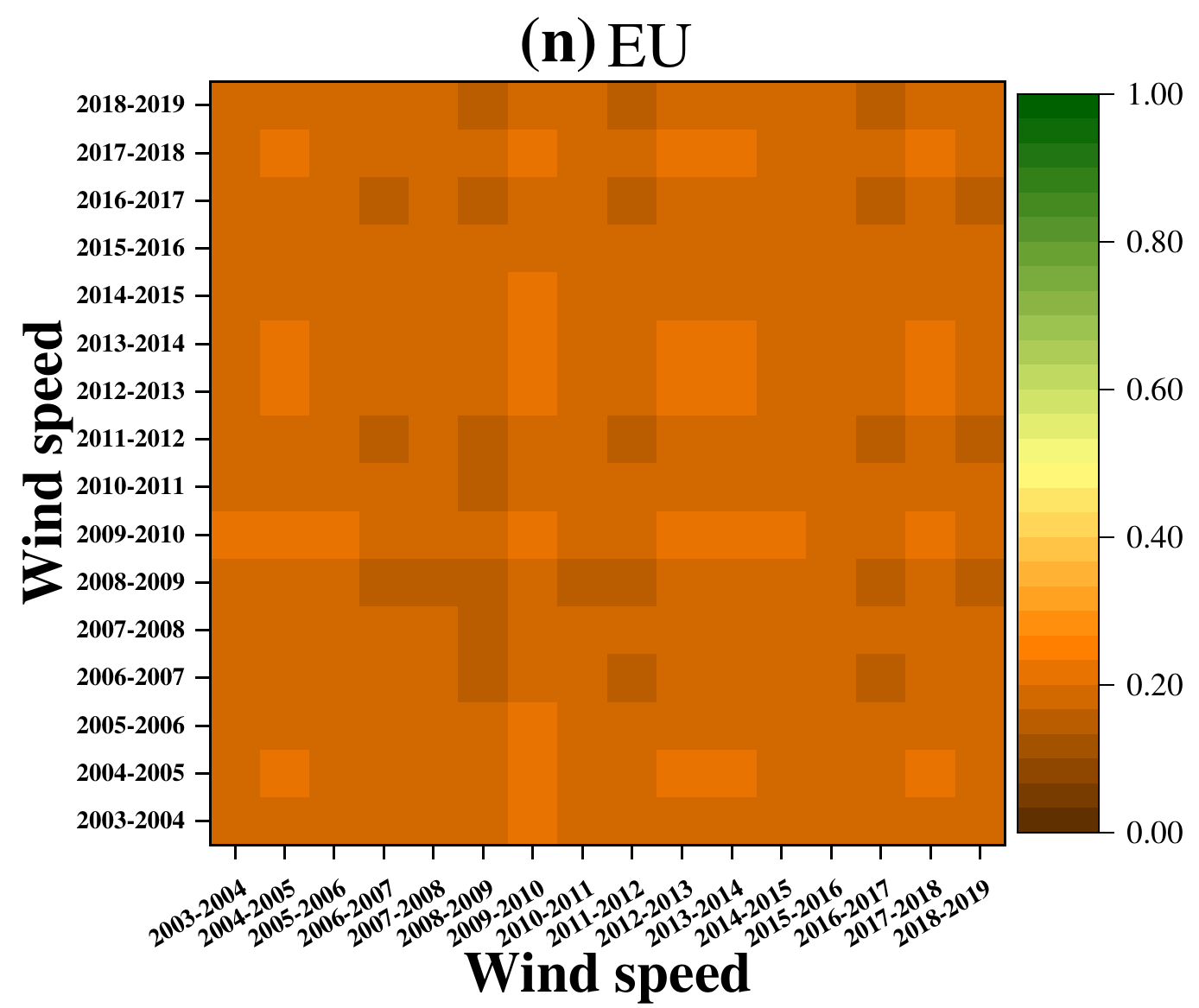}
\includegraphics[width=8em, height=7em]{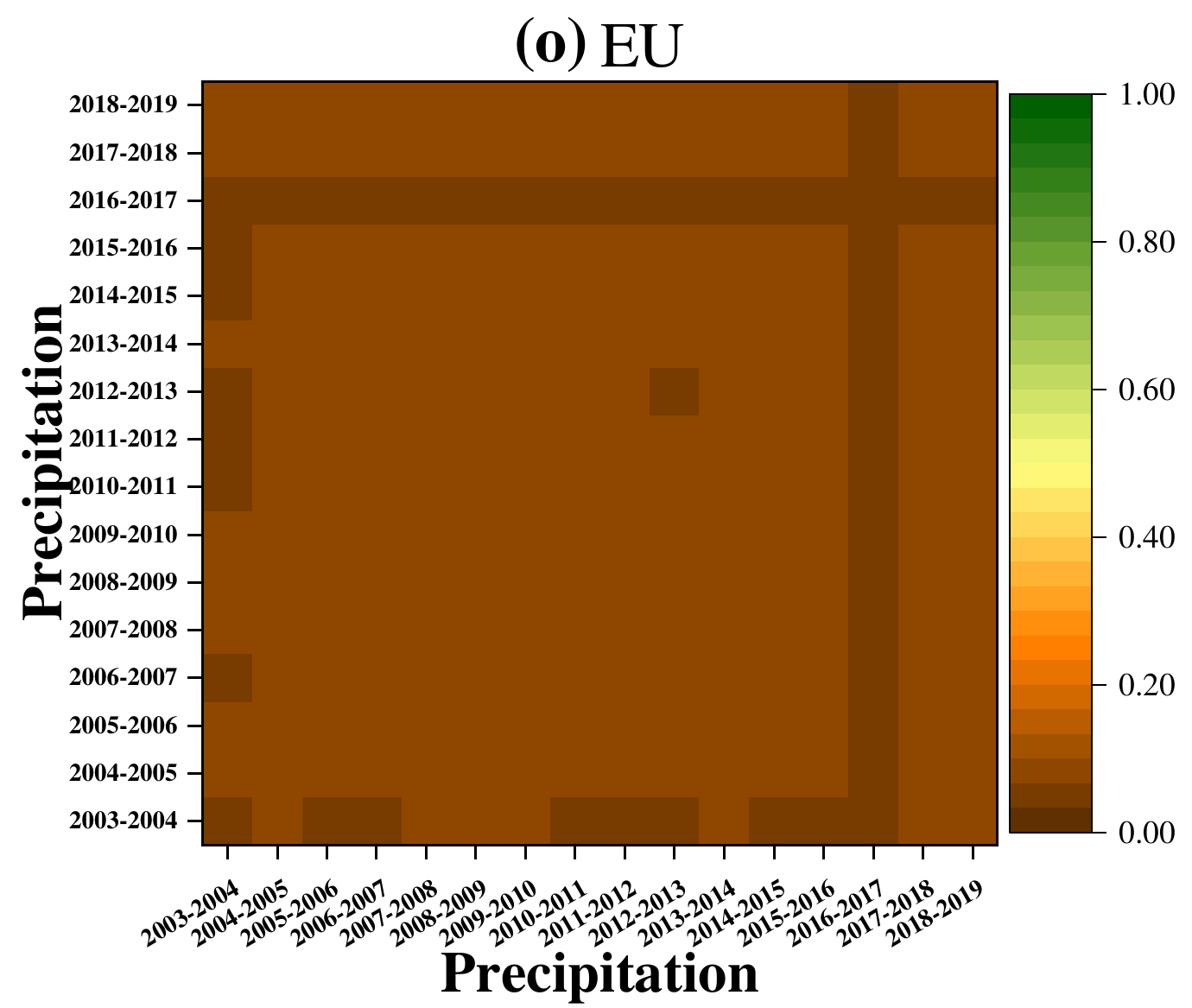}
\end{center}

\begin{center}
\noindent {\small {\bf Fig. S27} For the controlled case, Jaccard similarity coefficient matrix of links in two networks of different years for each of the climate variables.}
\end{center}

\begin{center}
\includegraphics[width=8em, height=7em]{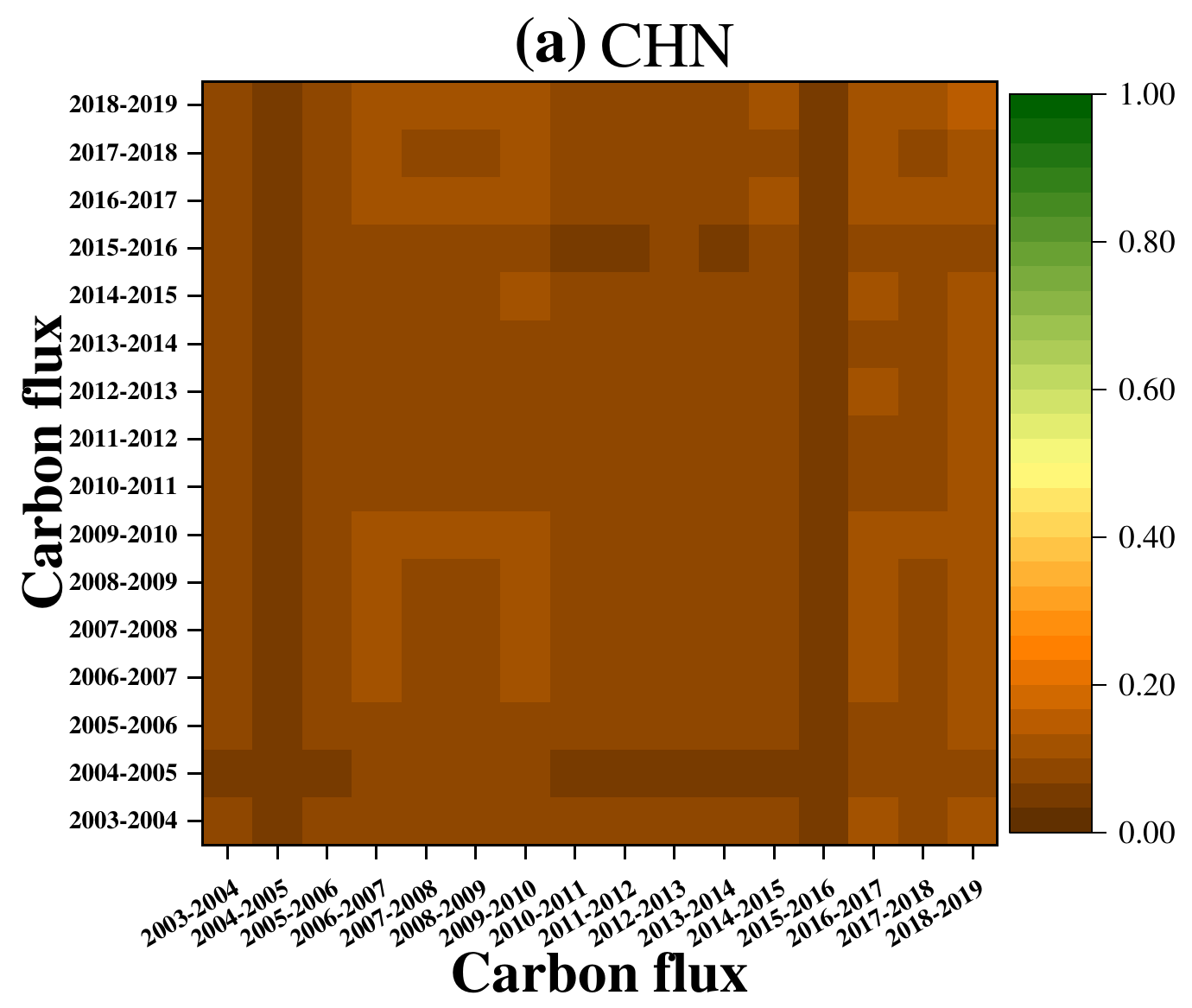}
\includegraphics[width=8em, height=7em]{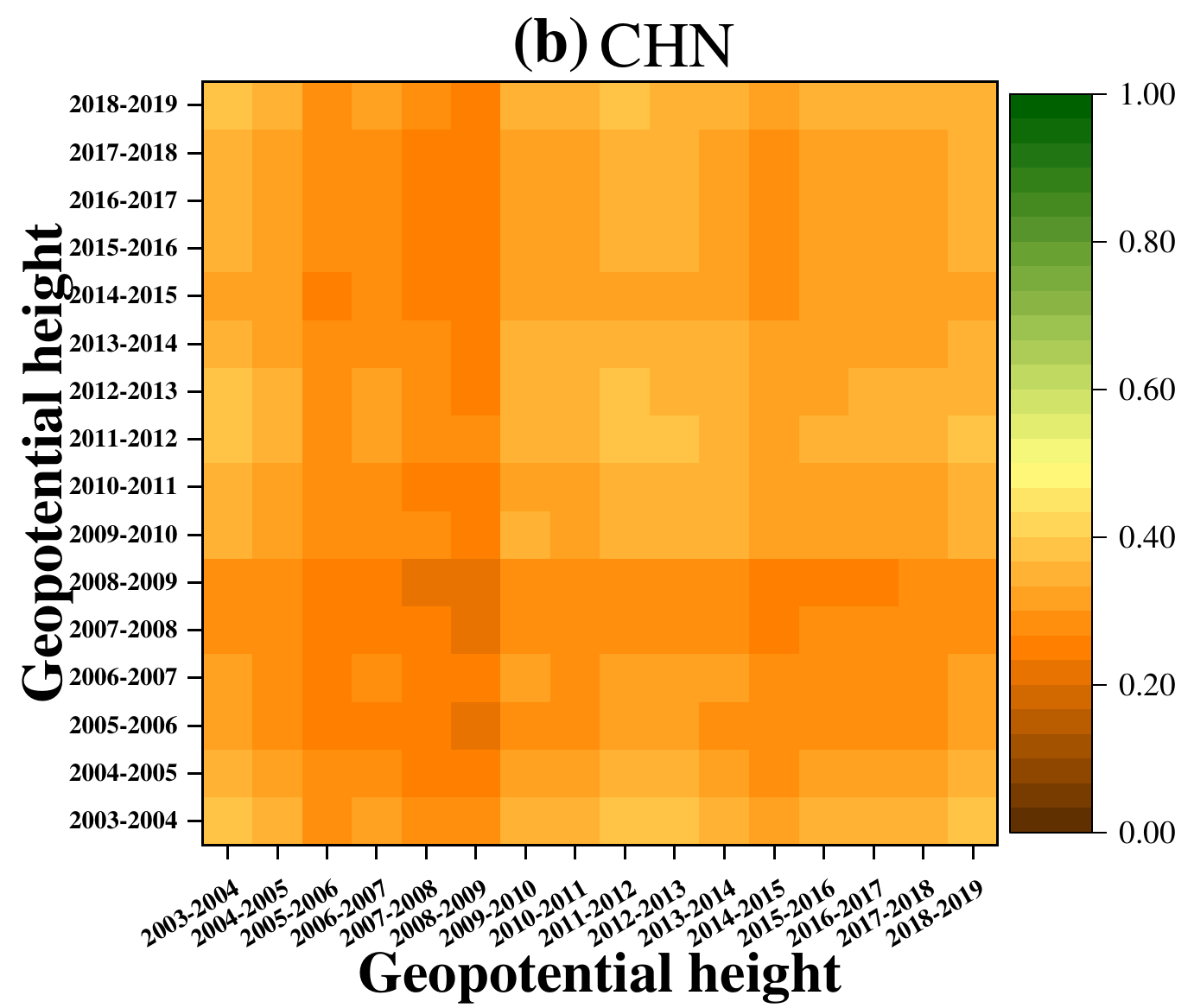}
\includegraphics[width=8em, height=7em]{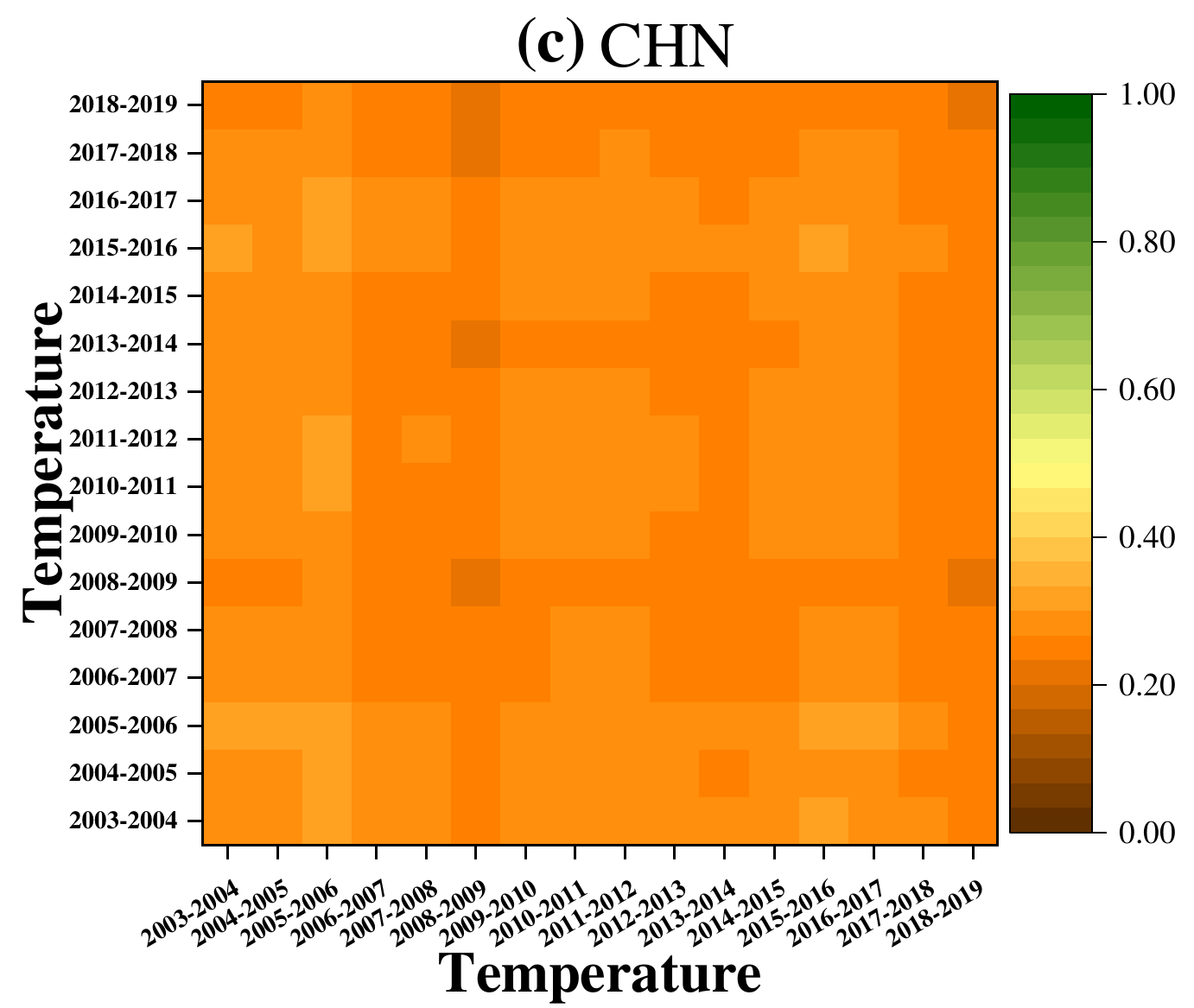}
\includegraphics[width=8em, height=7em]{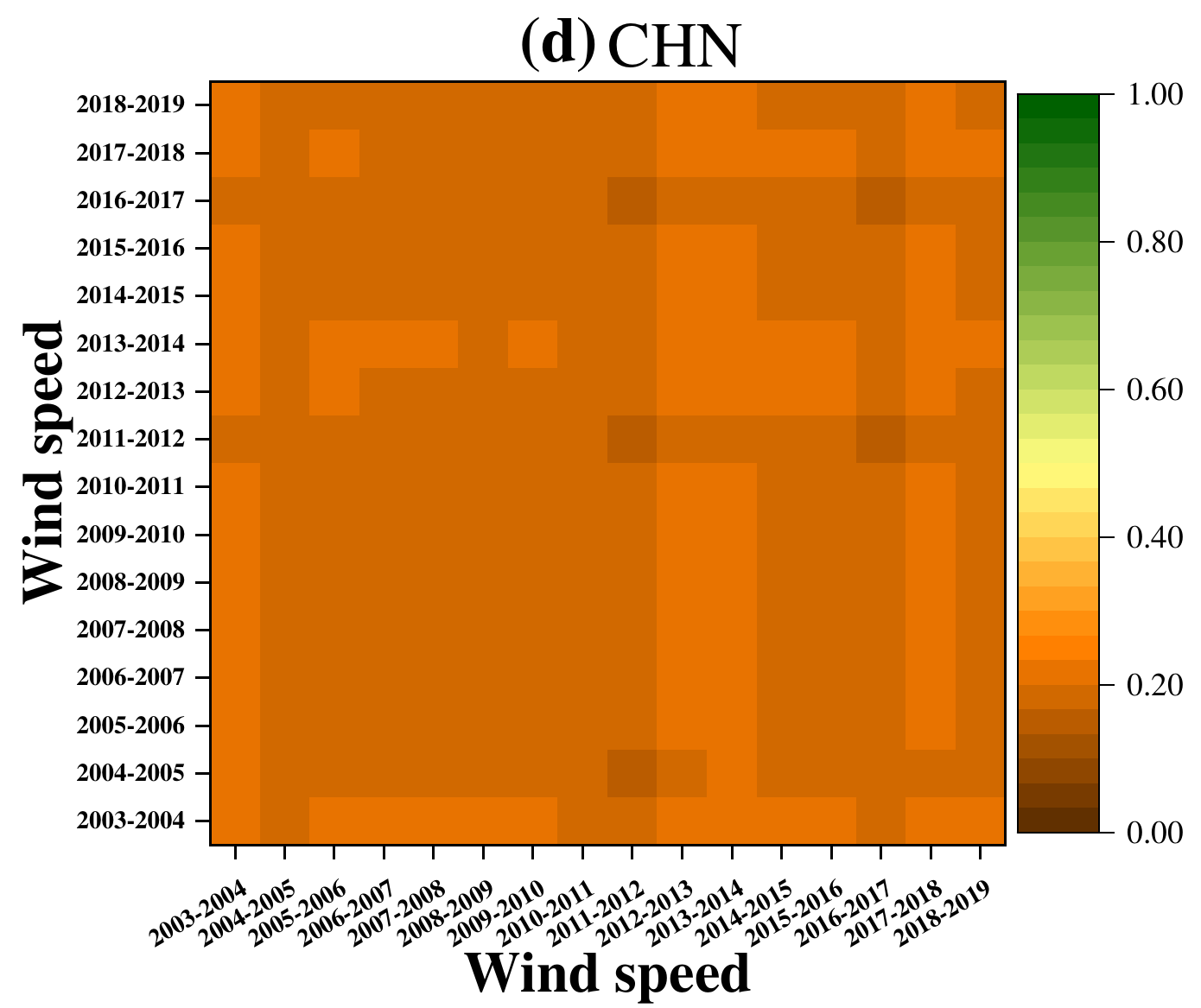}
\includegraphics[width=8em, height=7em]{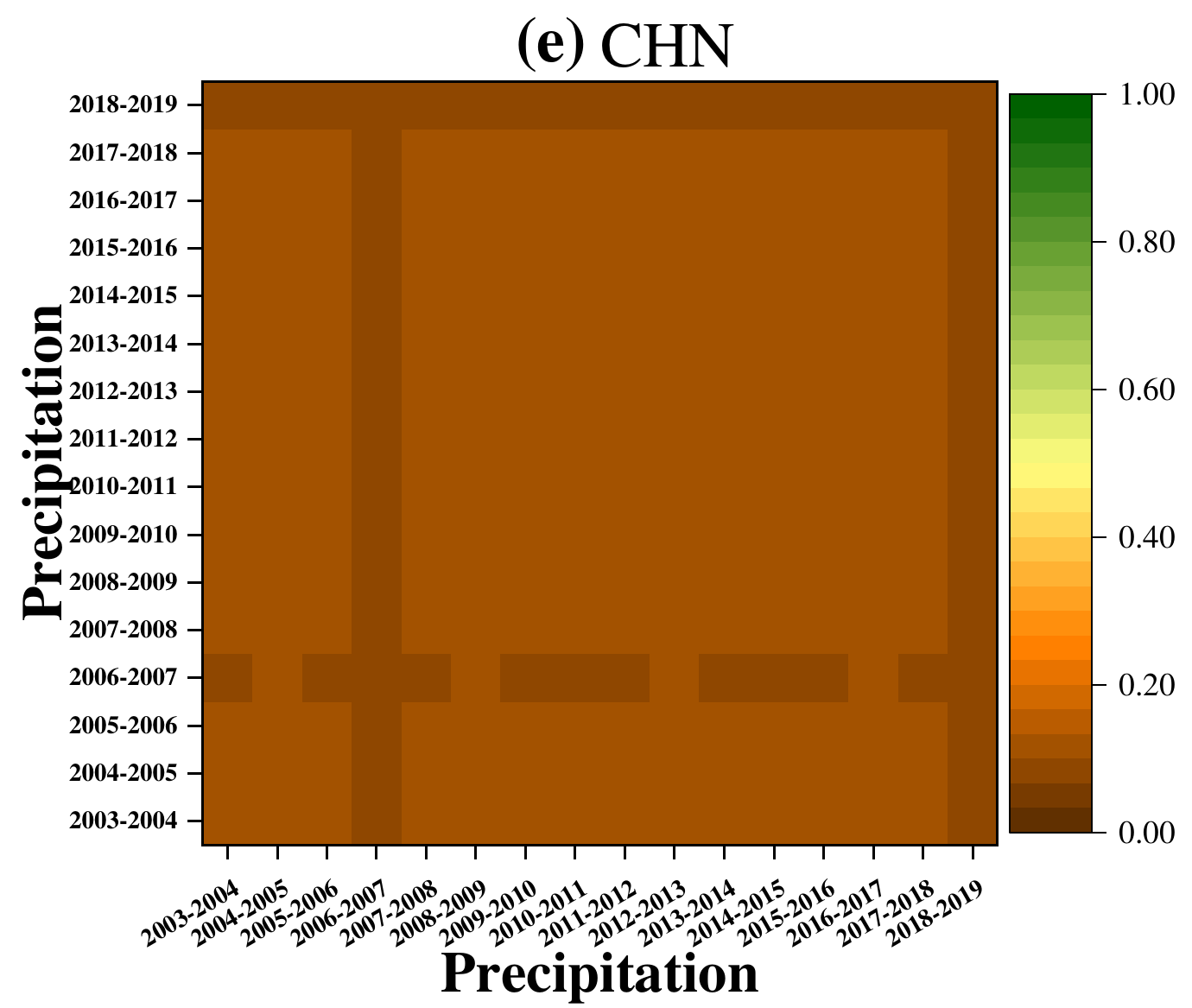}
\includegraphics[width=8em, height=7em]{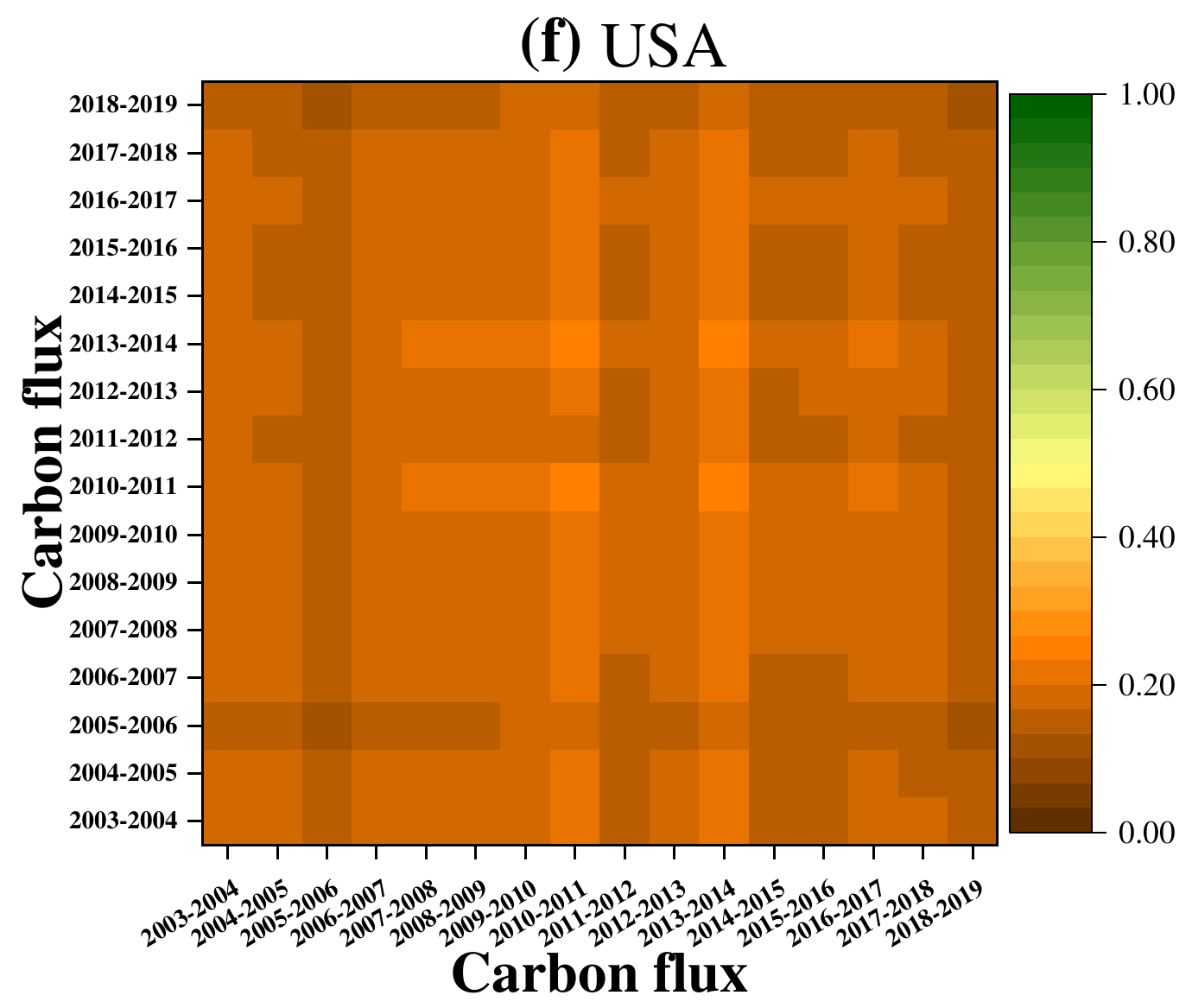}
\includegraphics[width=8em, height=7em]{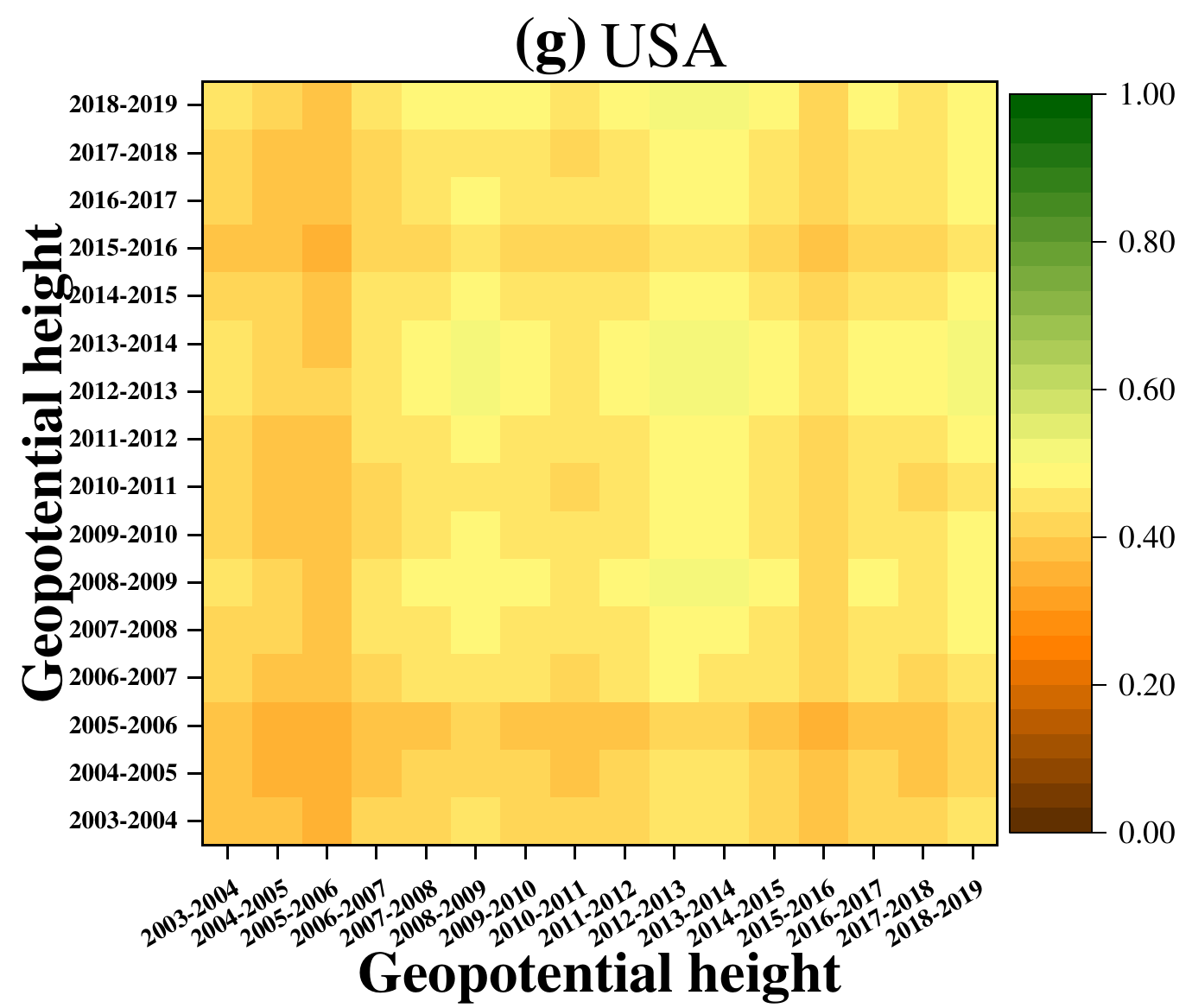}
\includegraphics[width=8em, height=7em]{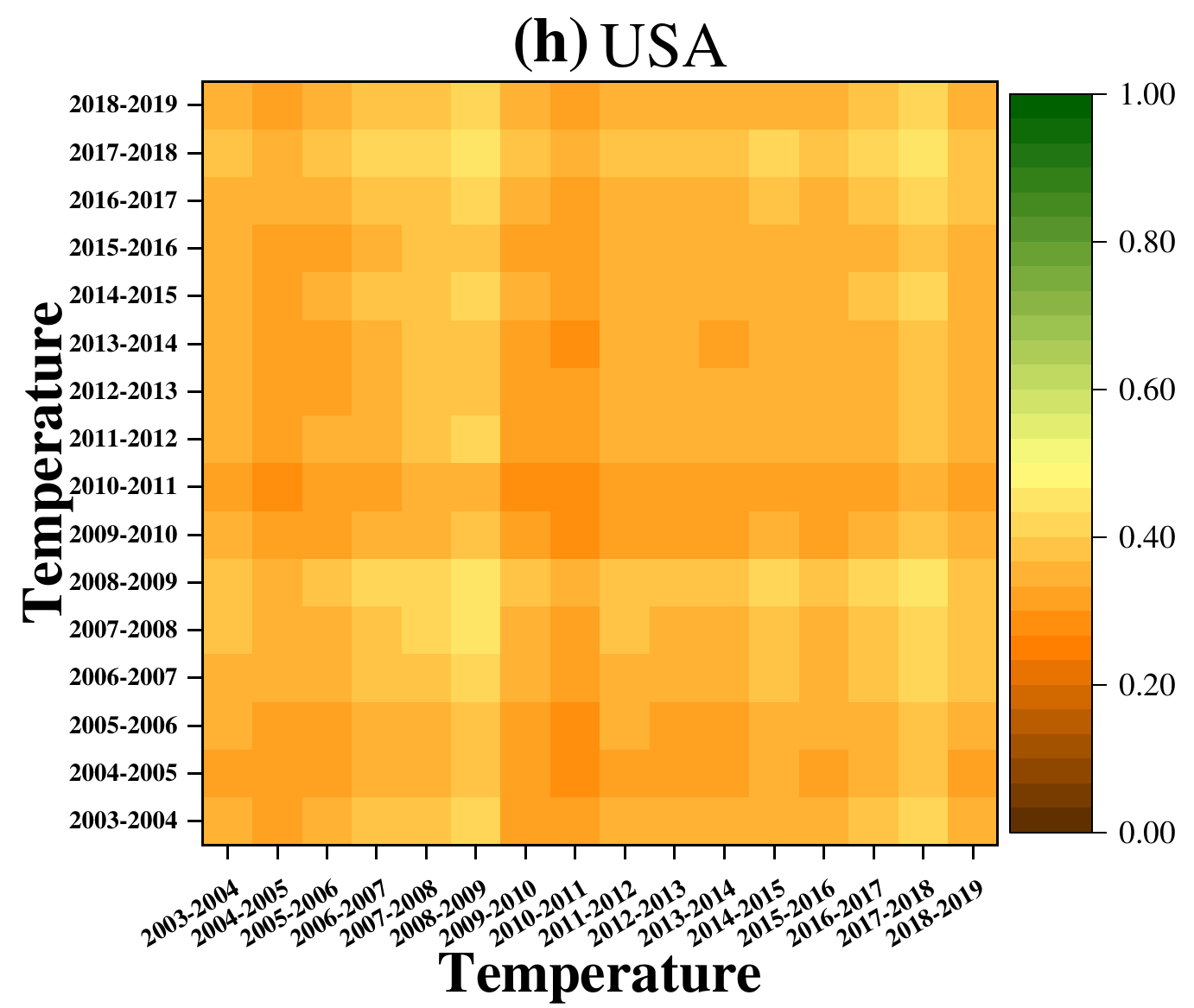}
\includegraphics[width=8em, height=7em]{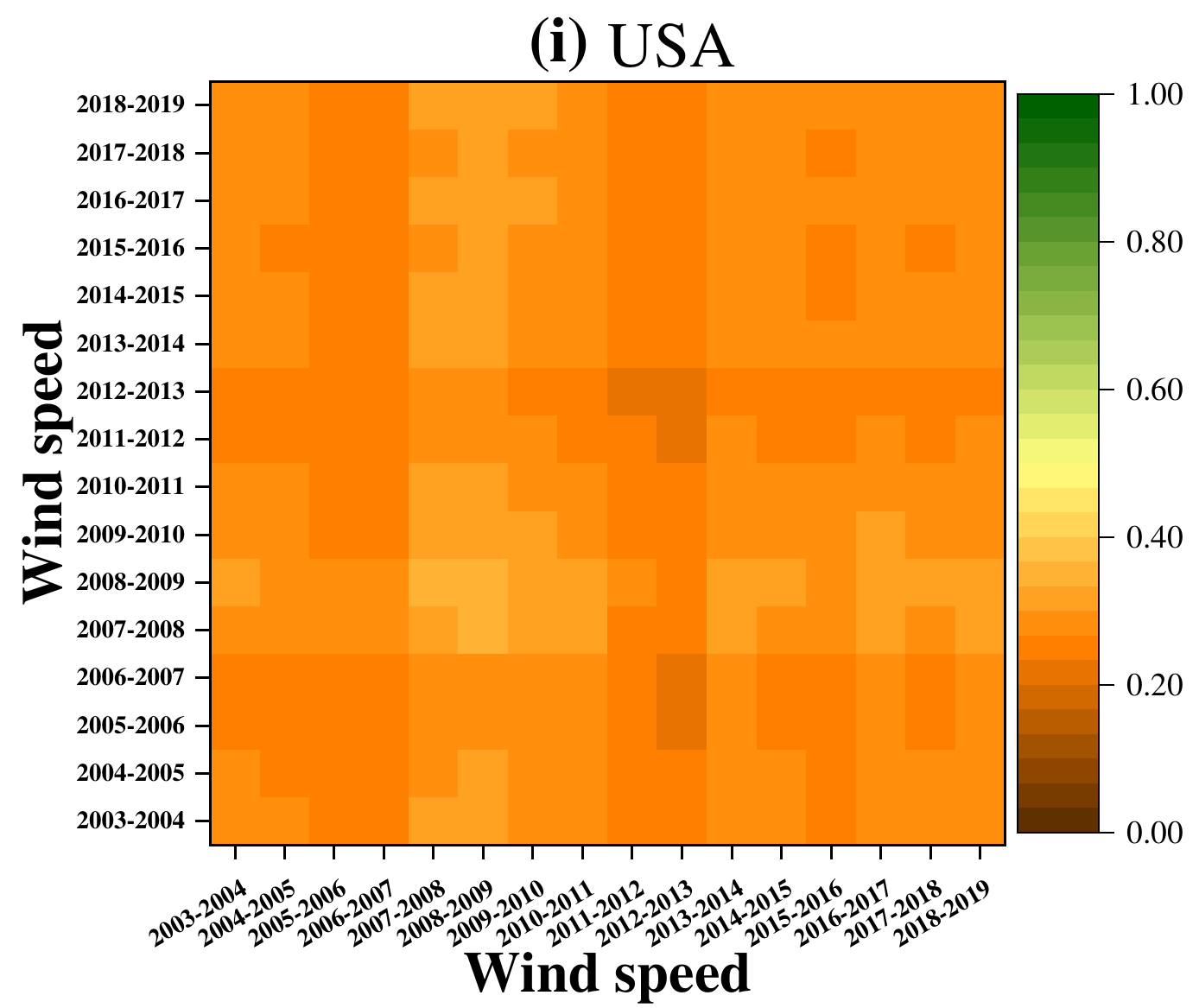}
\includegraphics[width=8em, height=7em]{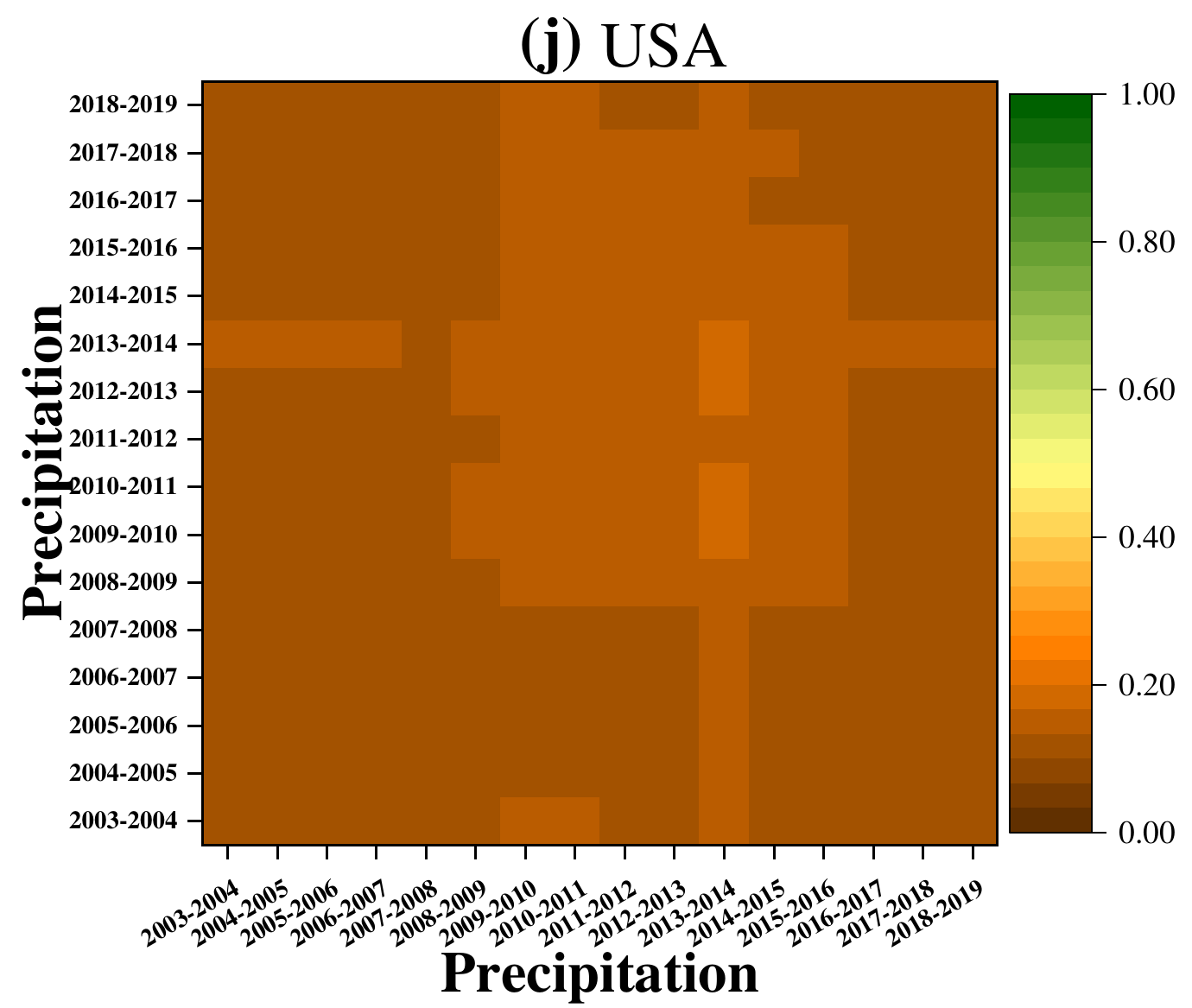}
\includegraphics[width=8em, height=7em]{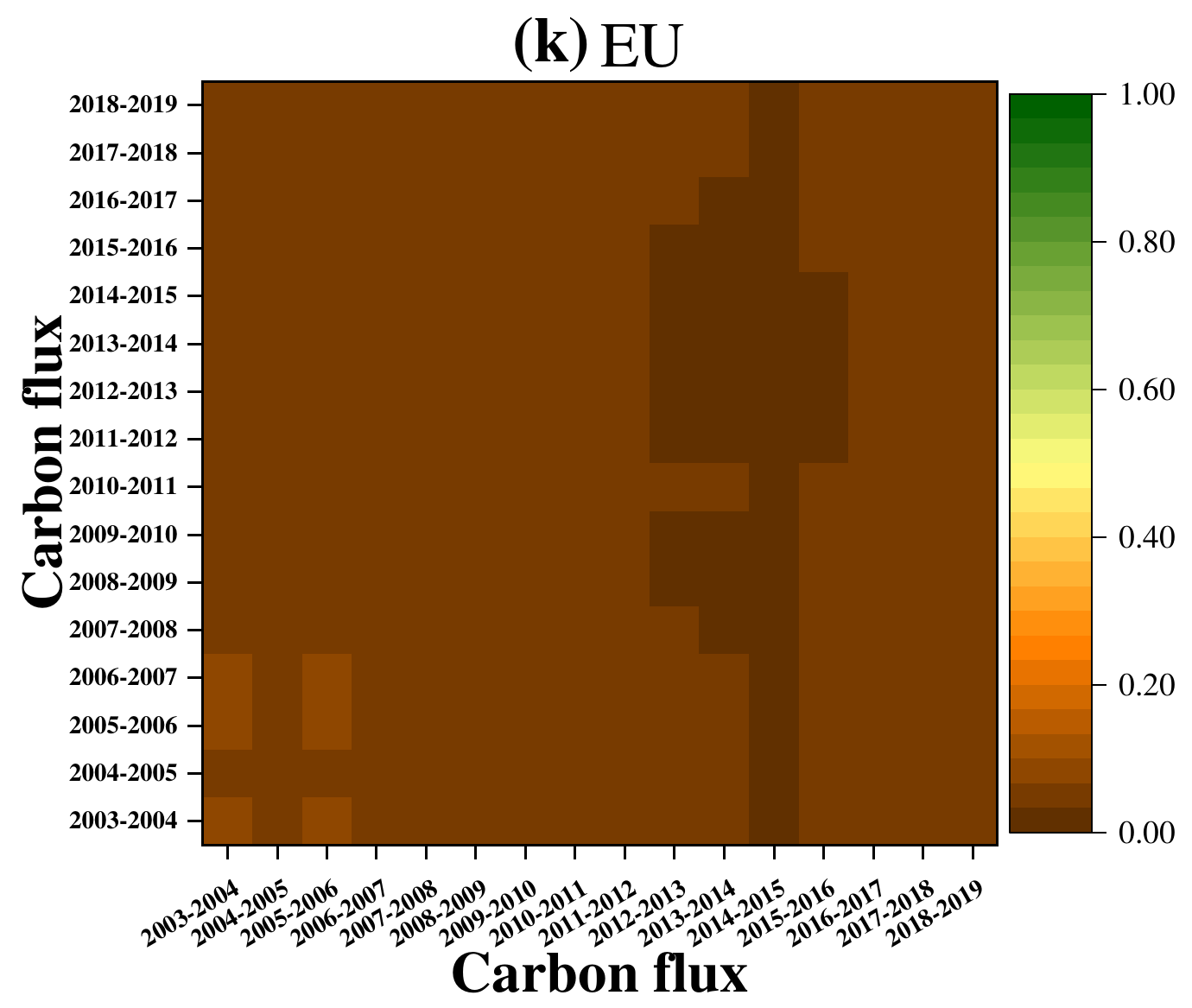}
\includegraphics[width=8em, height=7em]{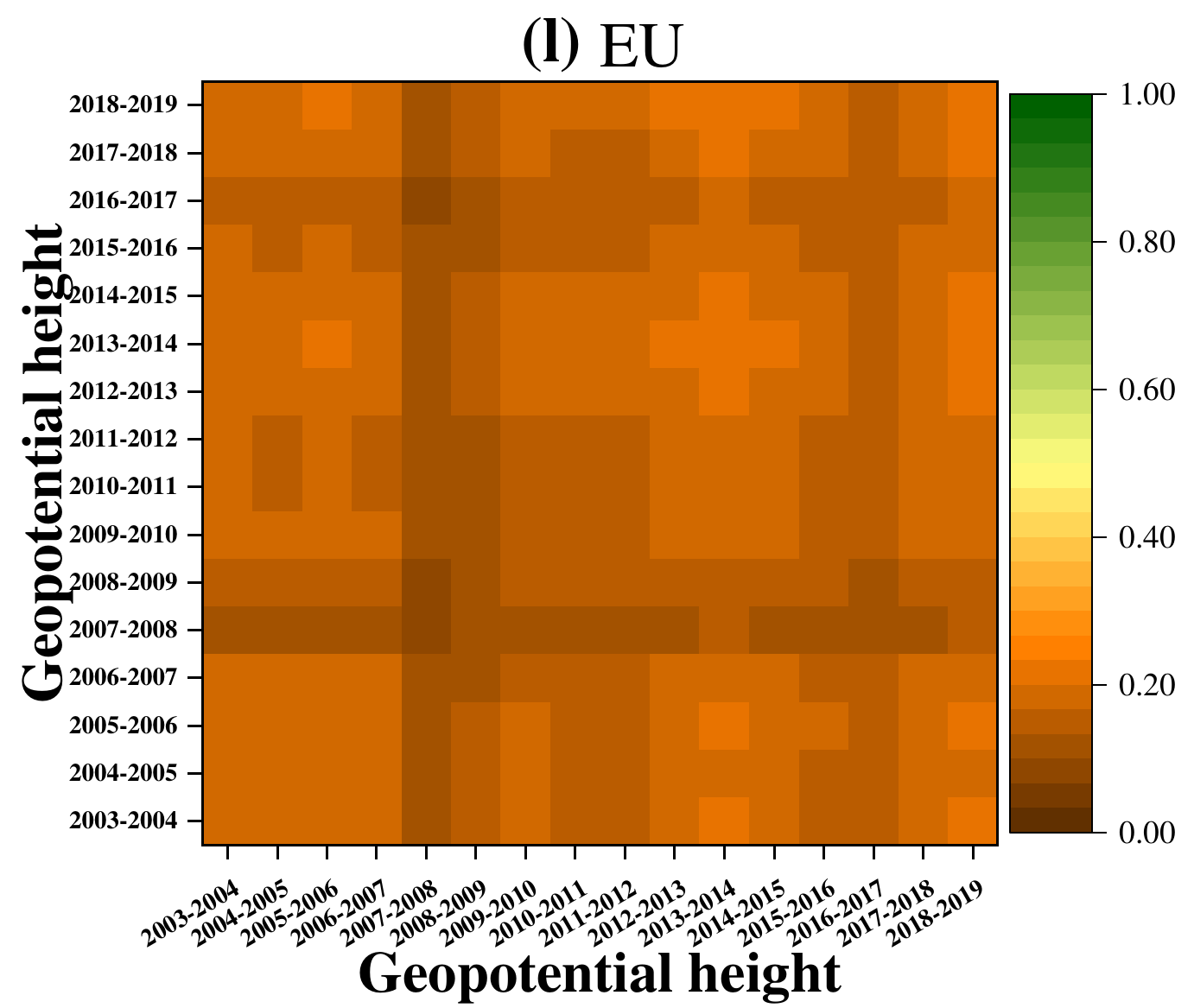}
\includegraphics[width=8em, height=7em]{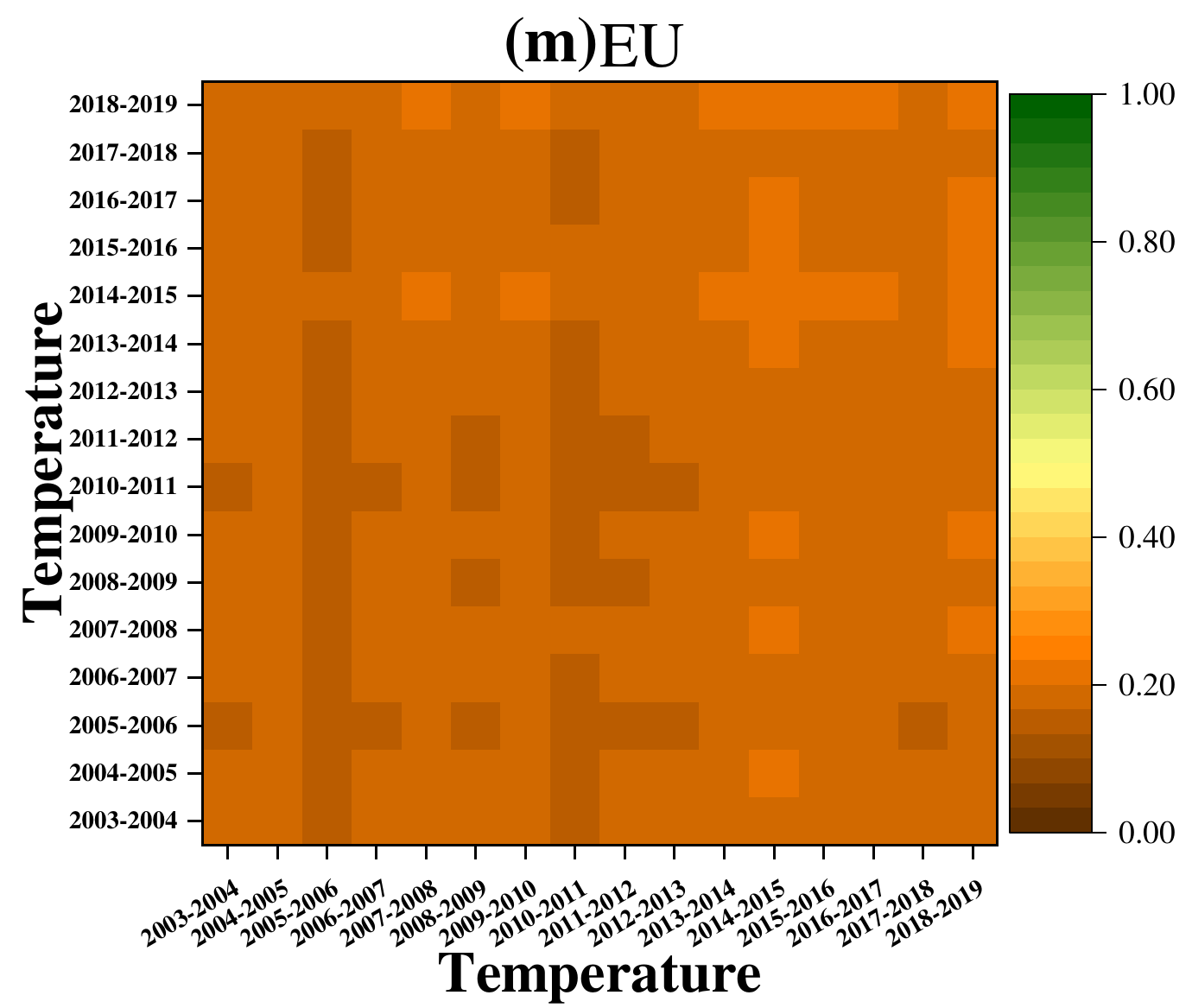}
\includegraphics[width=8em, height=7em]{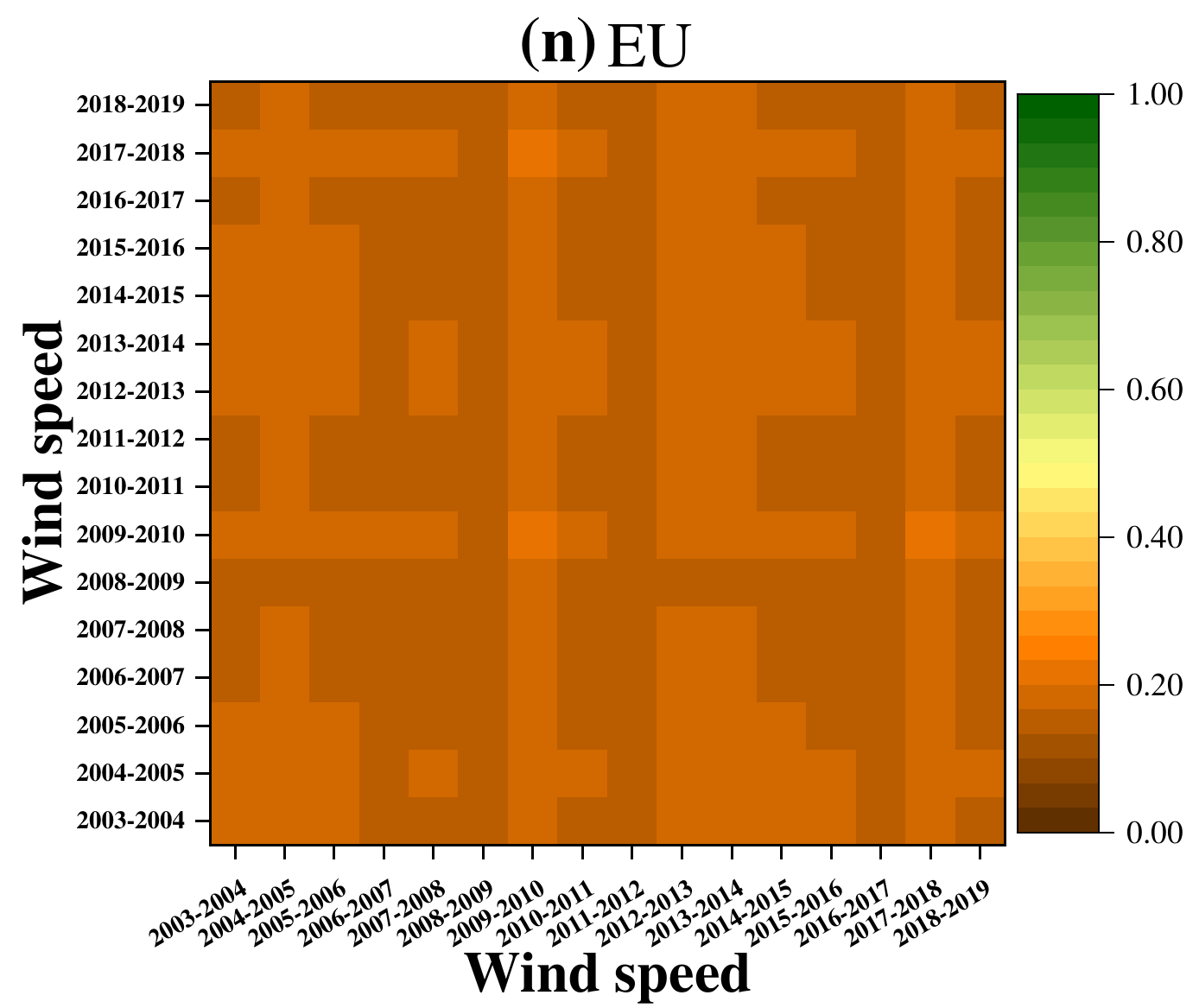}
\includegraphics[width=8em, height=7em]{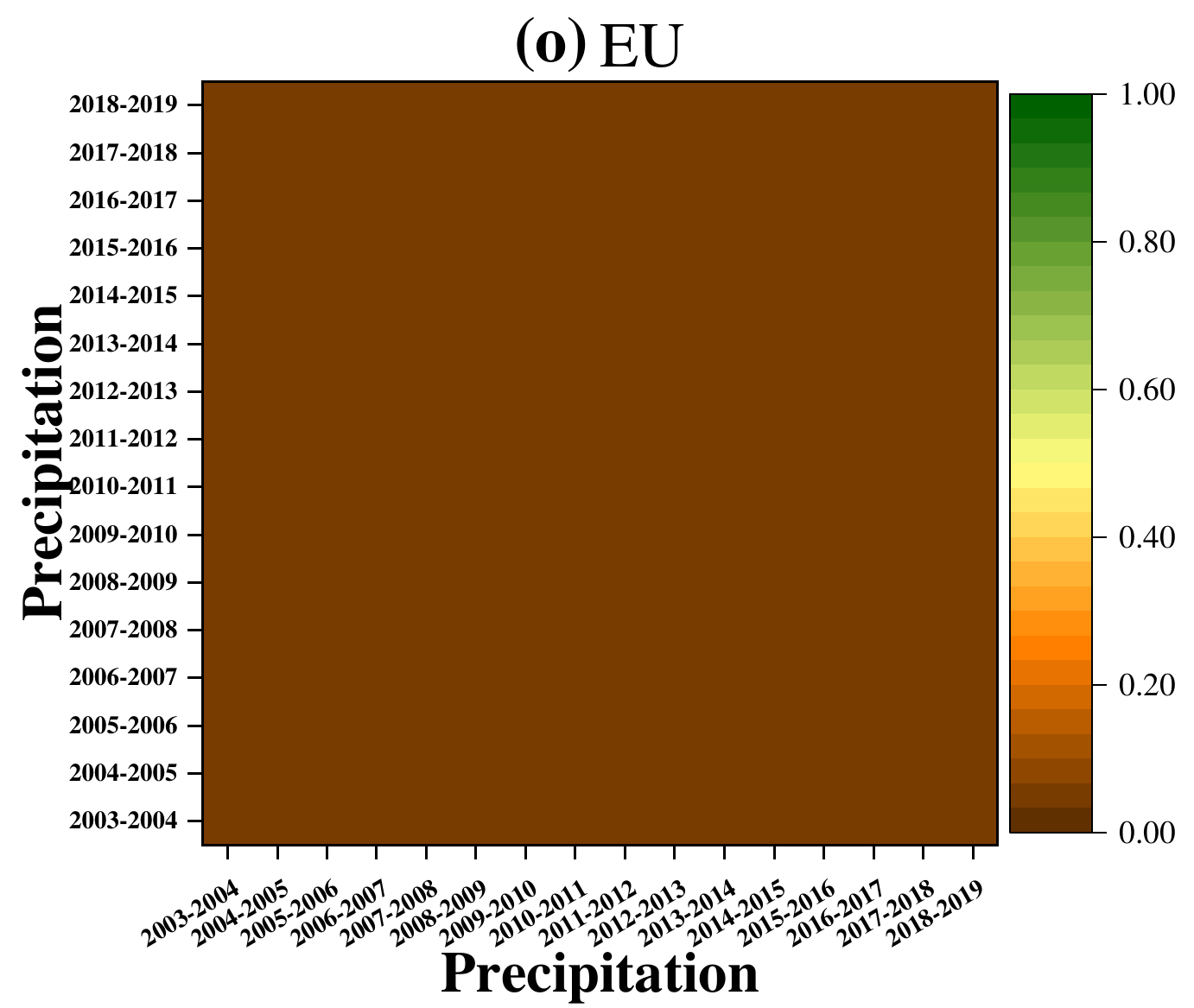}
\end{center}

\begin{center}
\noindent {\small {\bf Fig. S28} For the controlled case for lengths above $500km$, the Jaccard similarity coefficient matrix of links in two networks of different years for each of the climate variables.}
\end{center}

\begin{center}
\includegraphics[width=8em, height=7em]{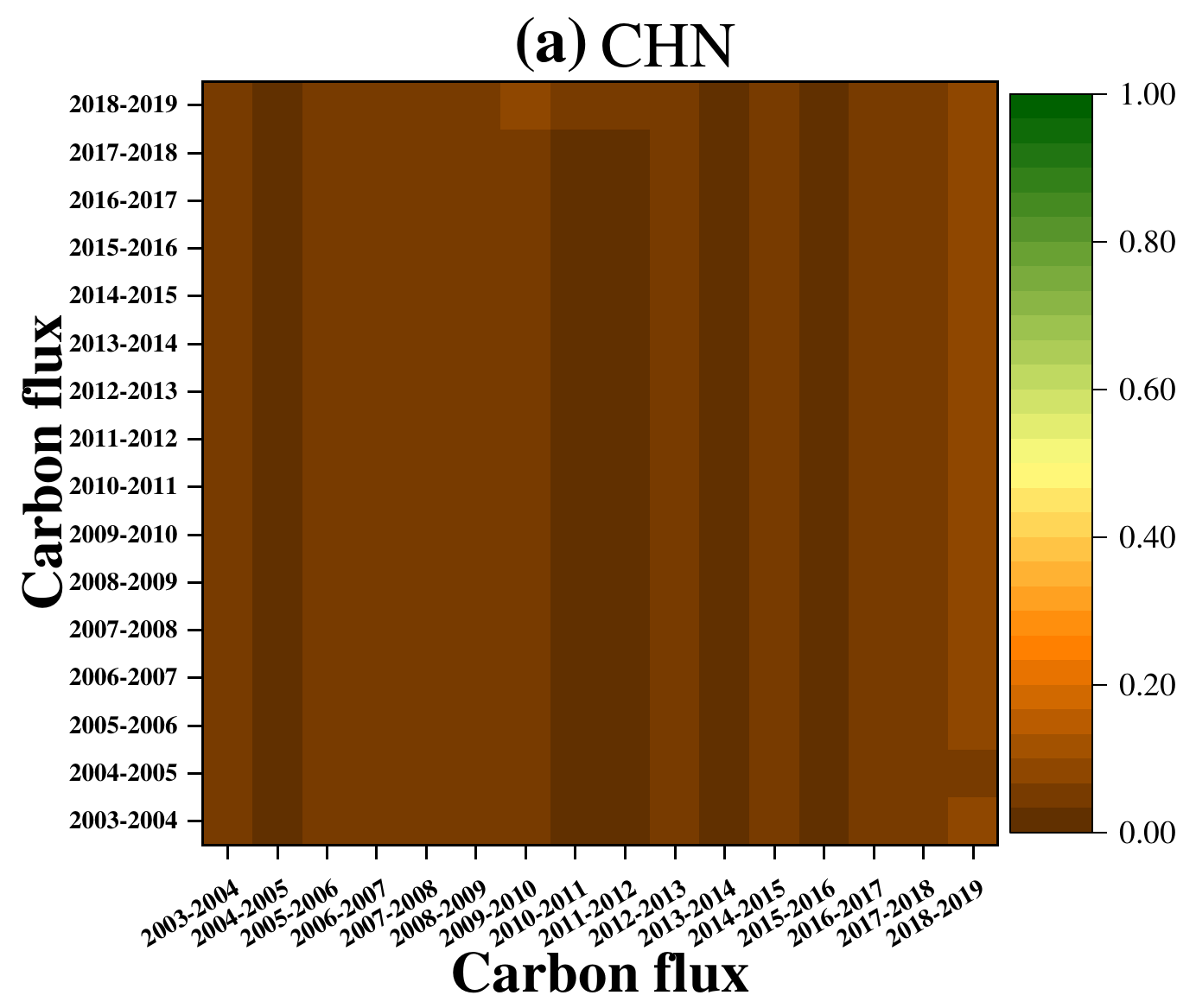}
\includegraphics[width=8em, height=7em]{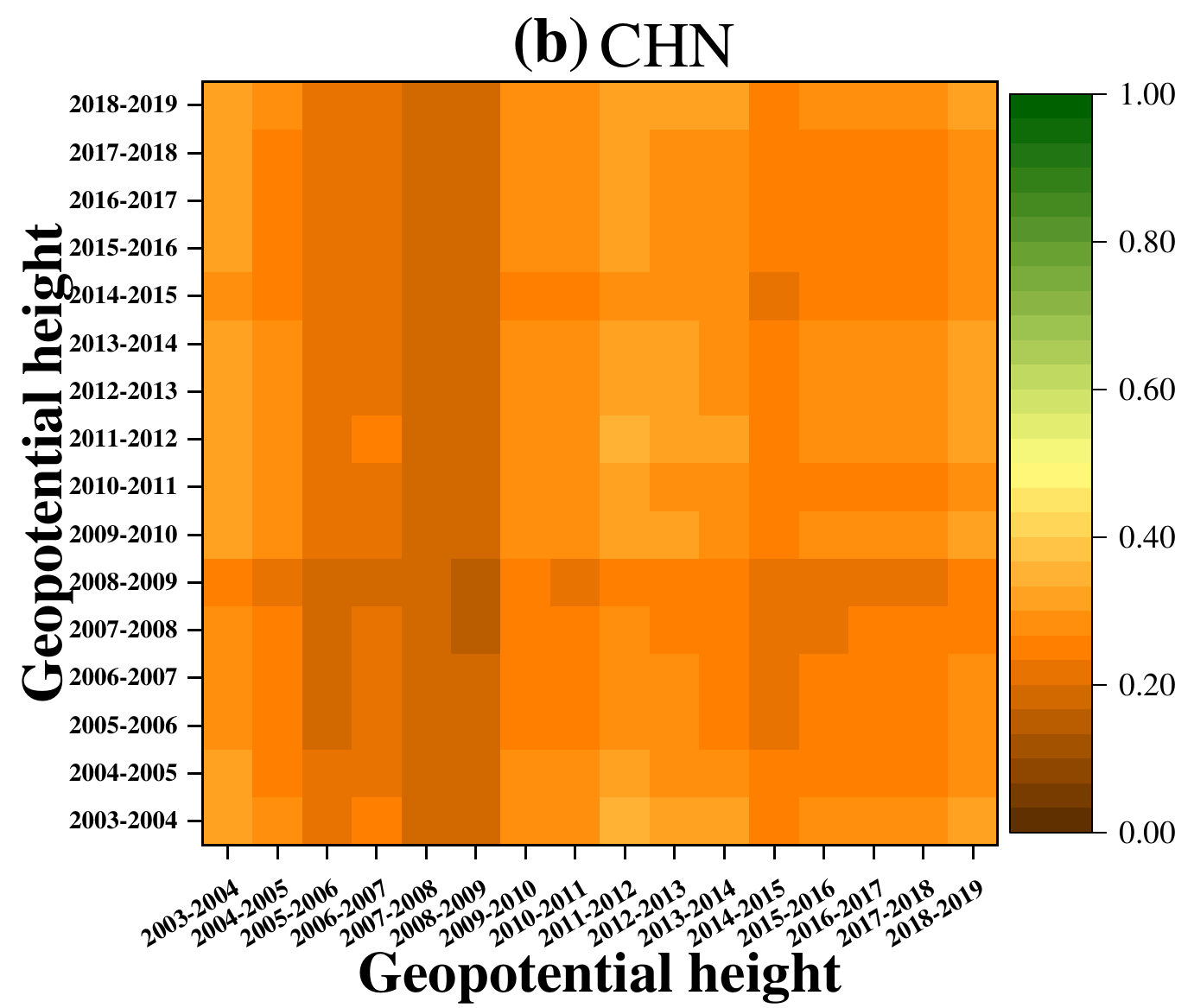}
\includegraphics[width=8em, height=7em]{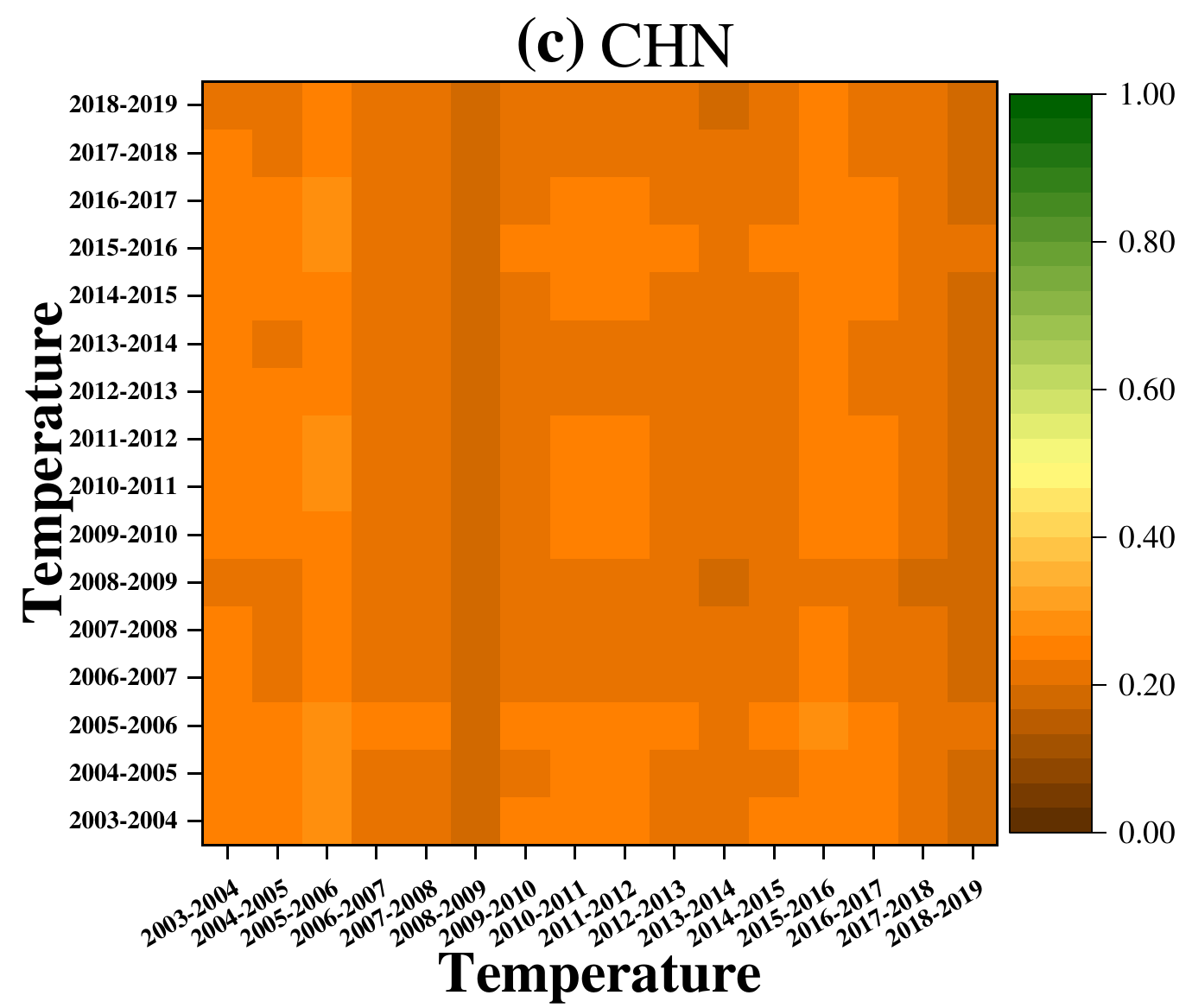}
\includegraphics[width=8em, height=7em]{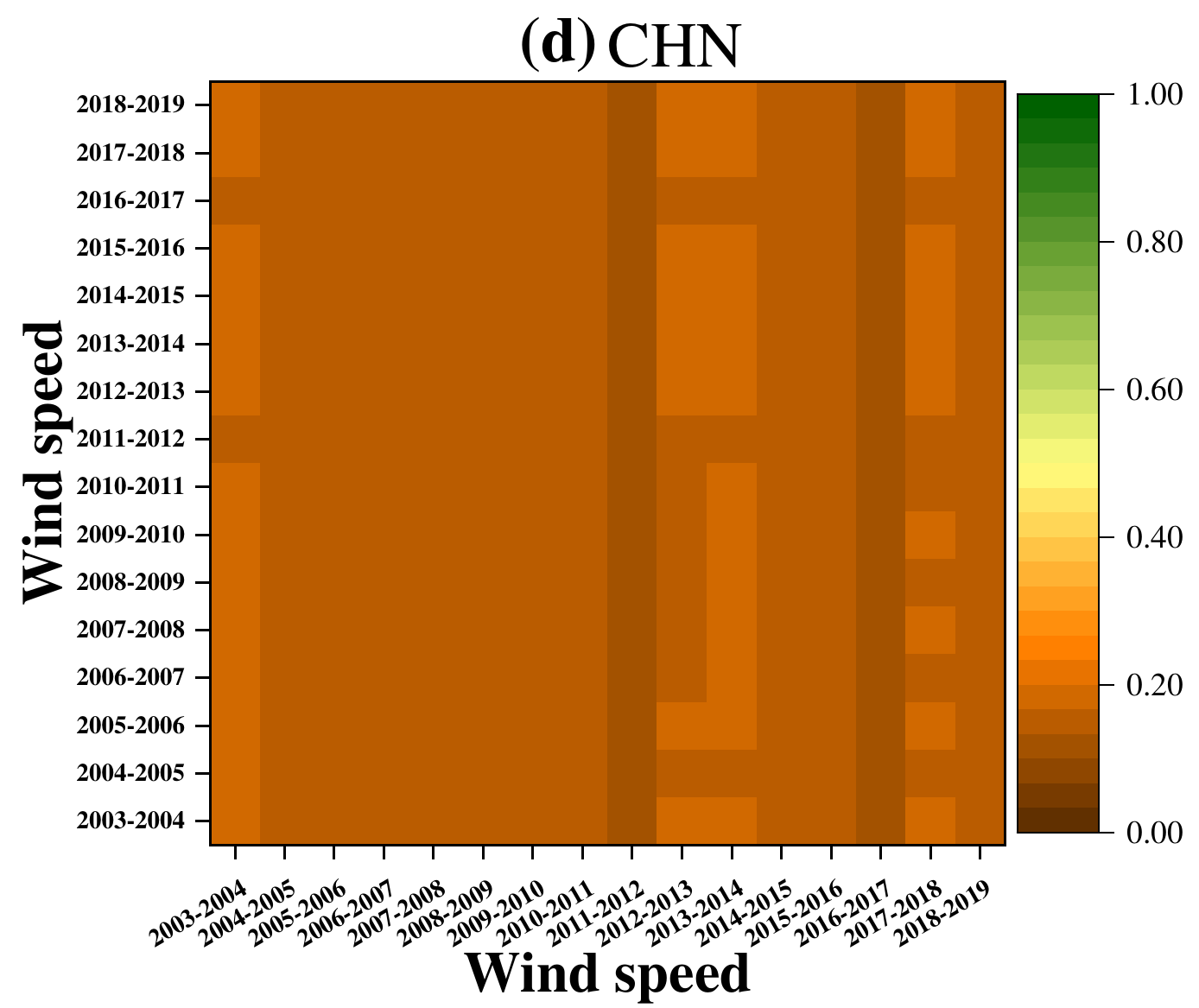}
\includegraphics[width=8em, height=7em]{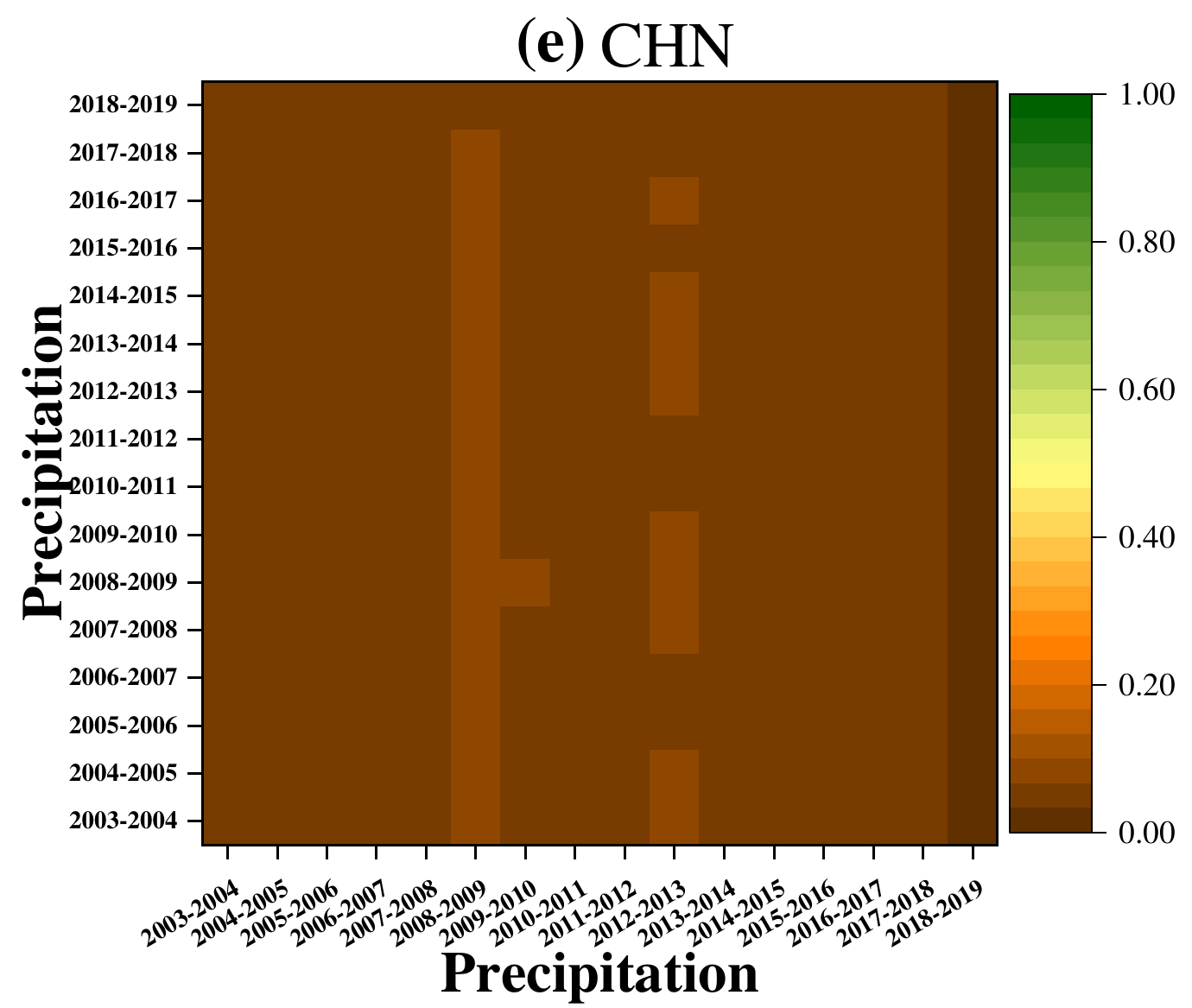}
\includegraphics[width=8em, height=7em]{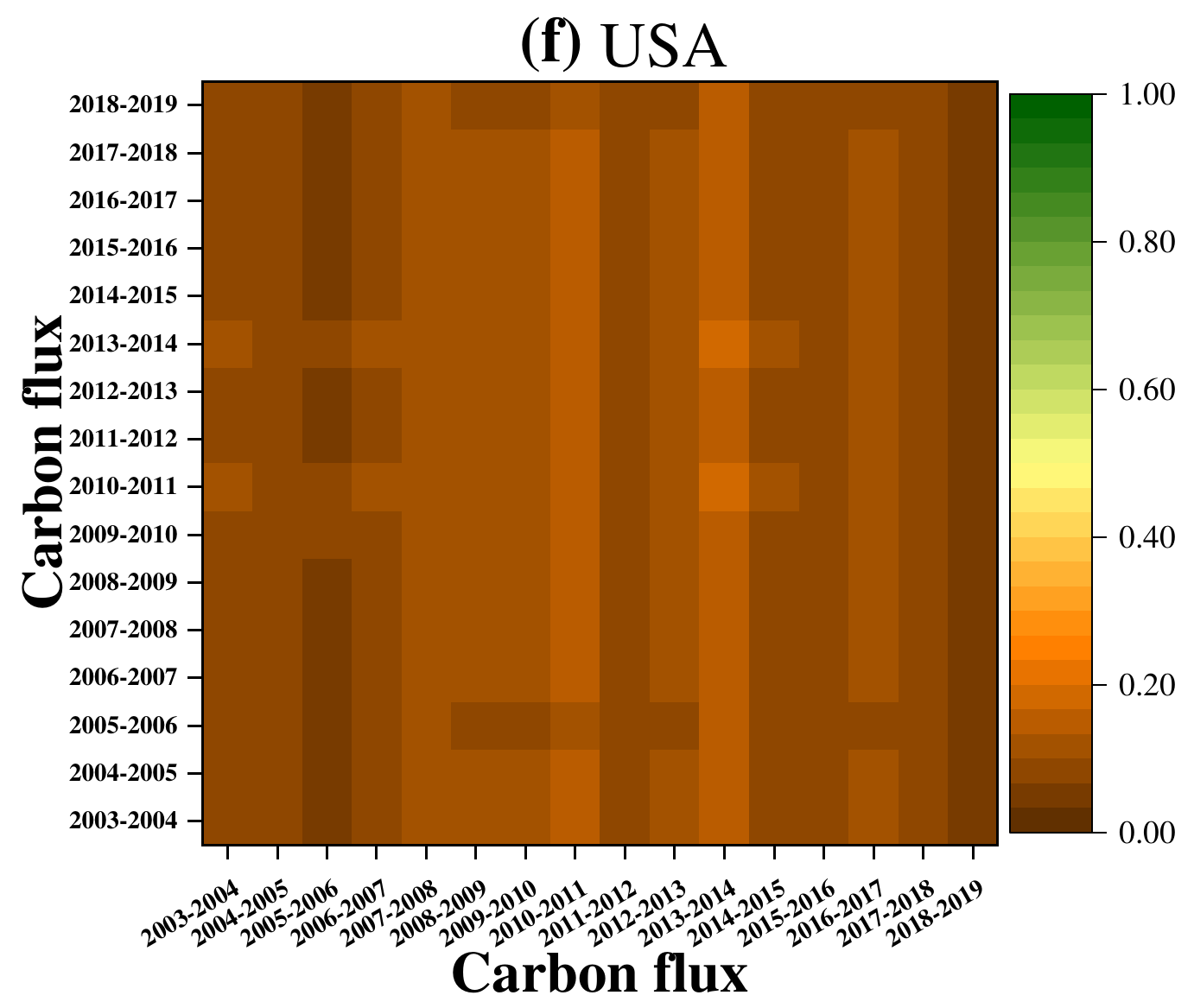}
\includegraphics[width=8em, height=7em]{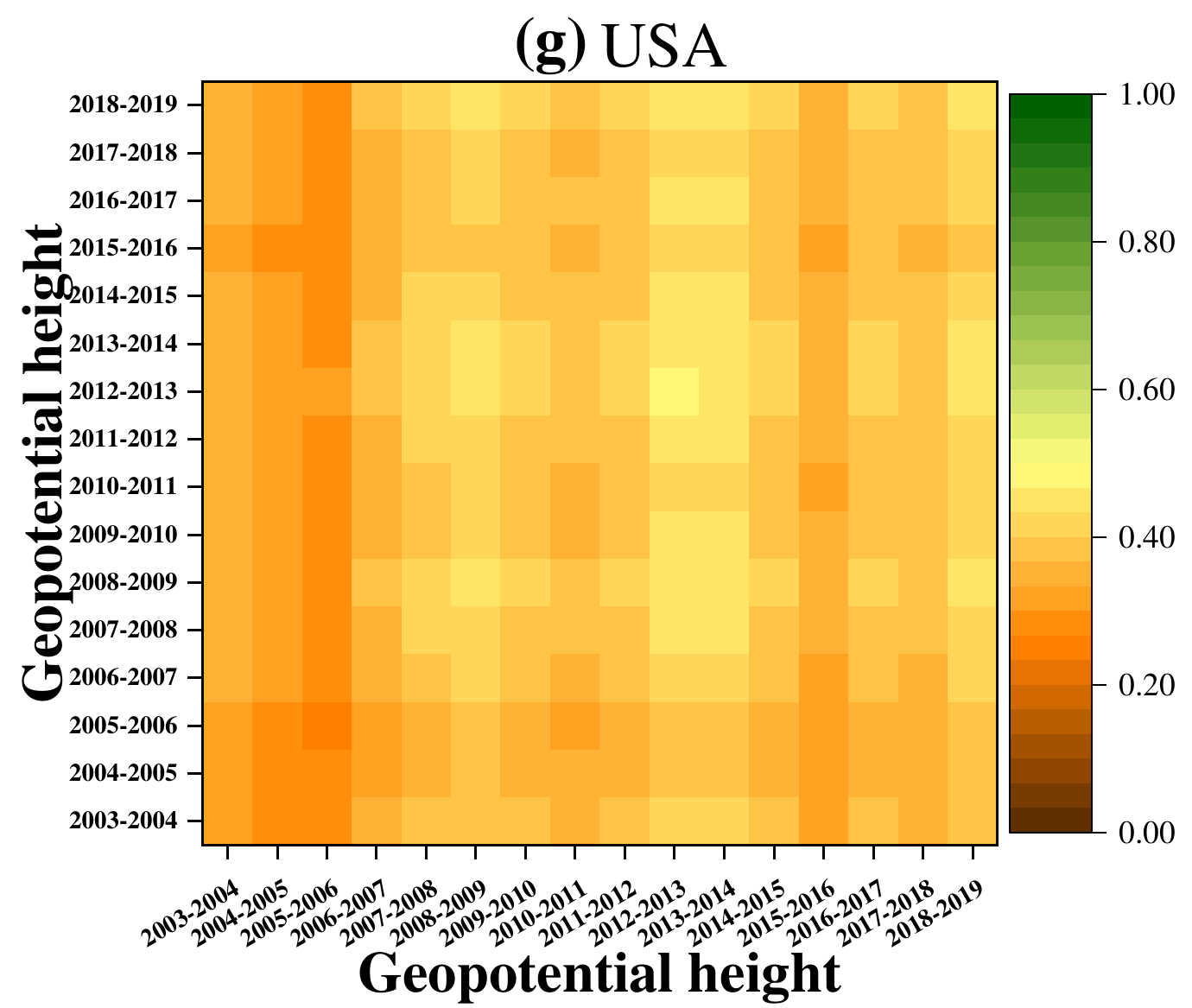}
\includegraphics[width=8em, height=7em]{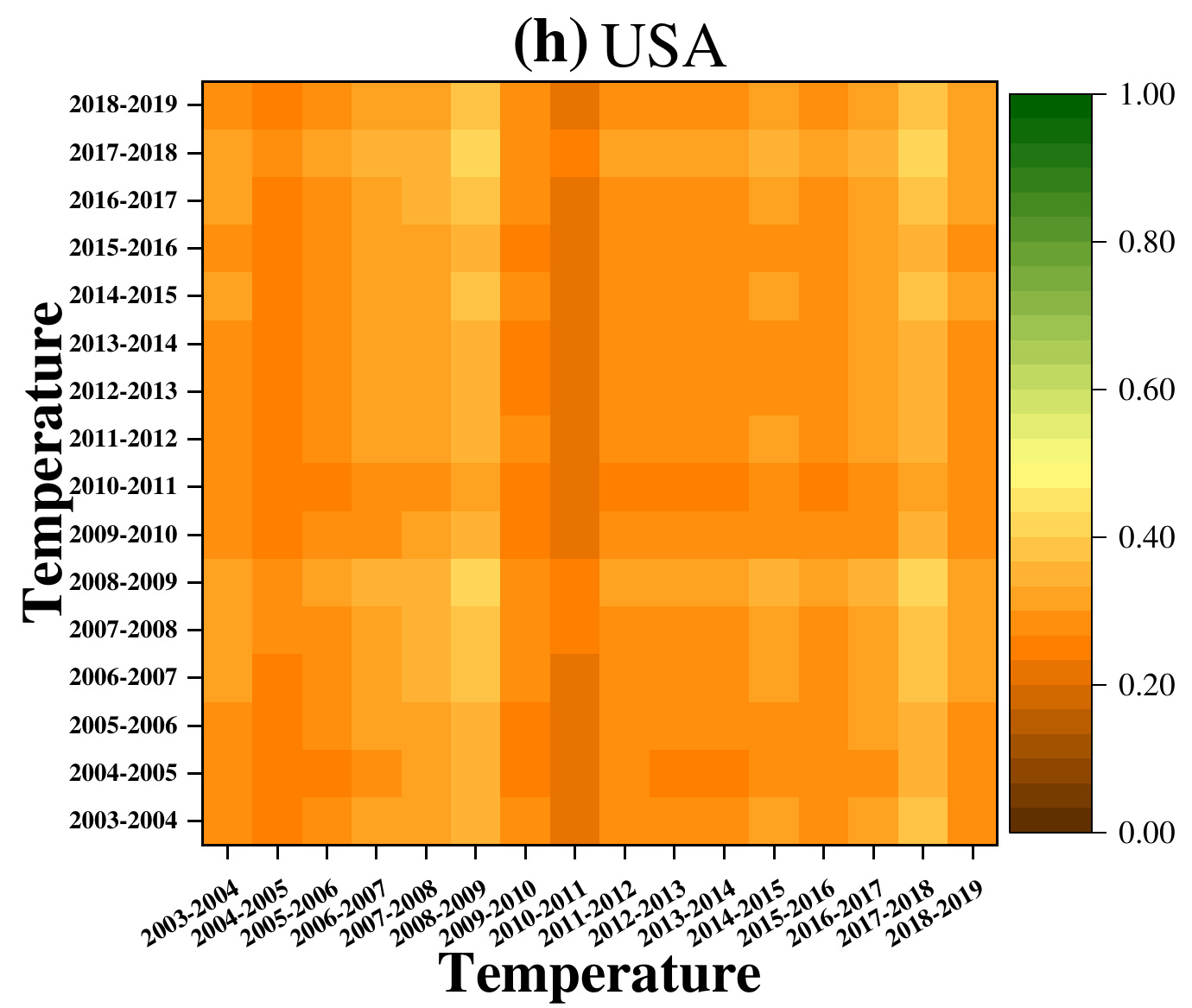}
\includegraphics[width=8em, height=7em]{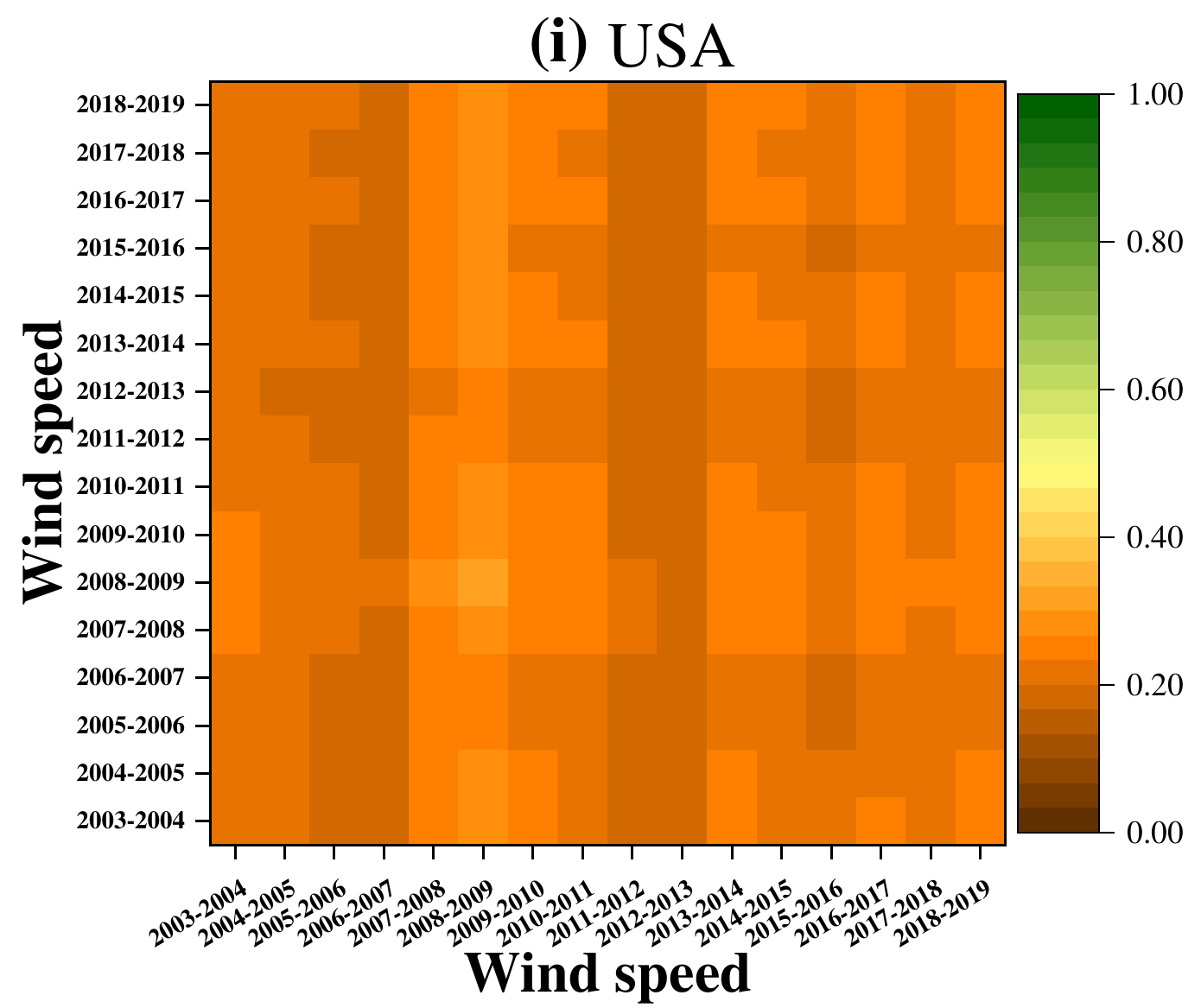}
\includegraphics[width=8em, height=7em]{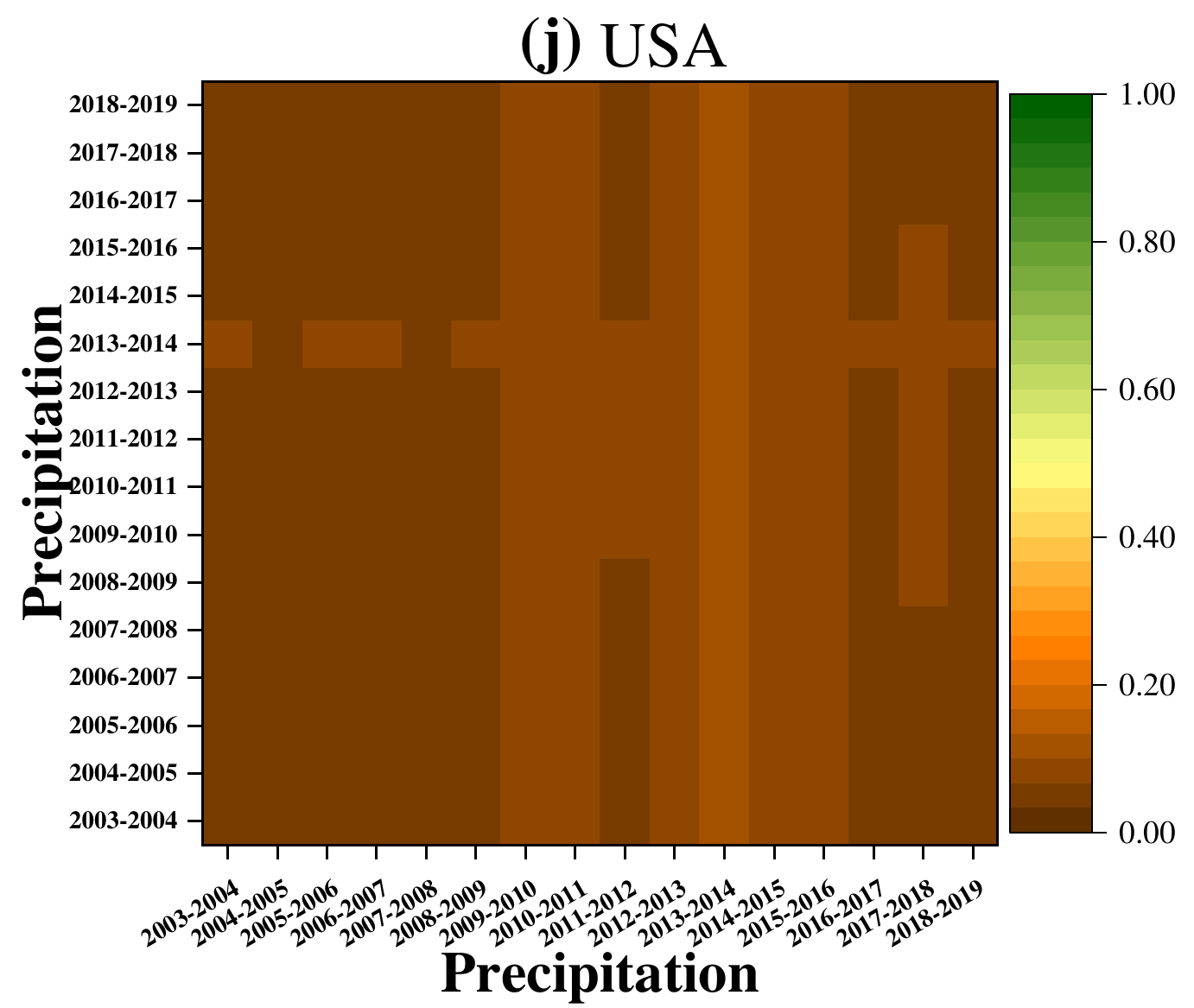}
\includegraphics[width=8em, height=7em]{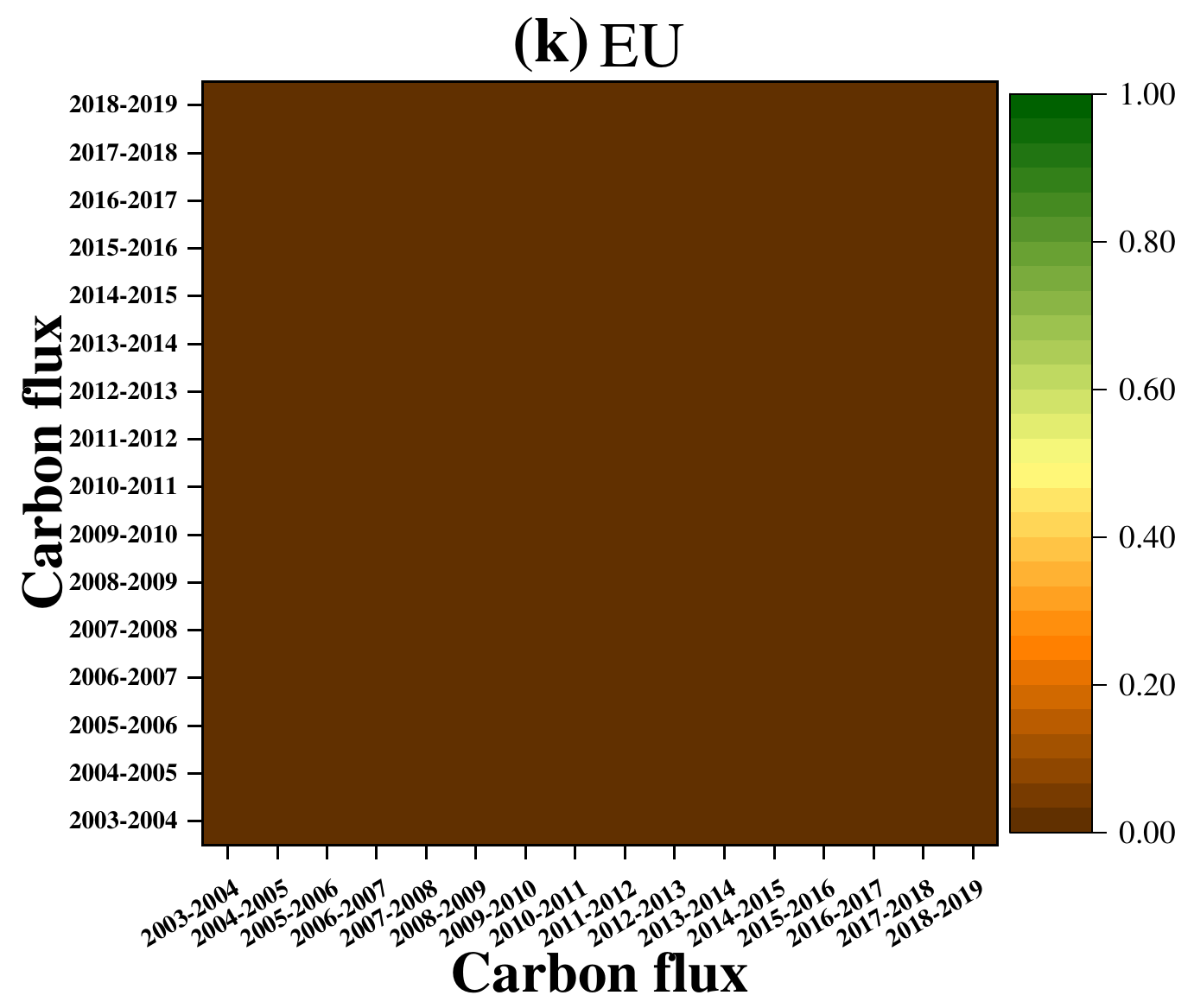}
\includegraphics[width=8em, height=7em]{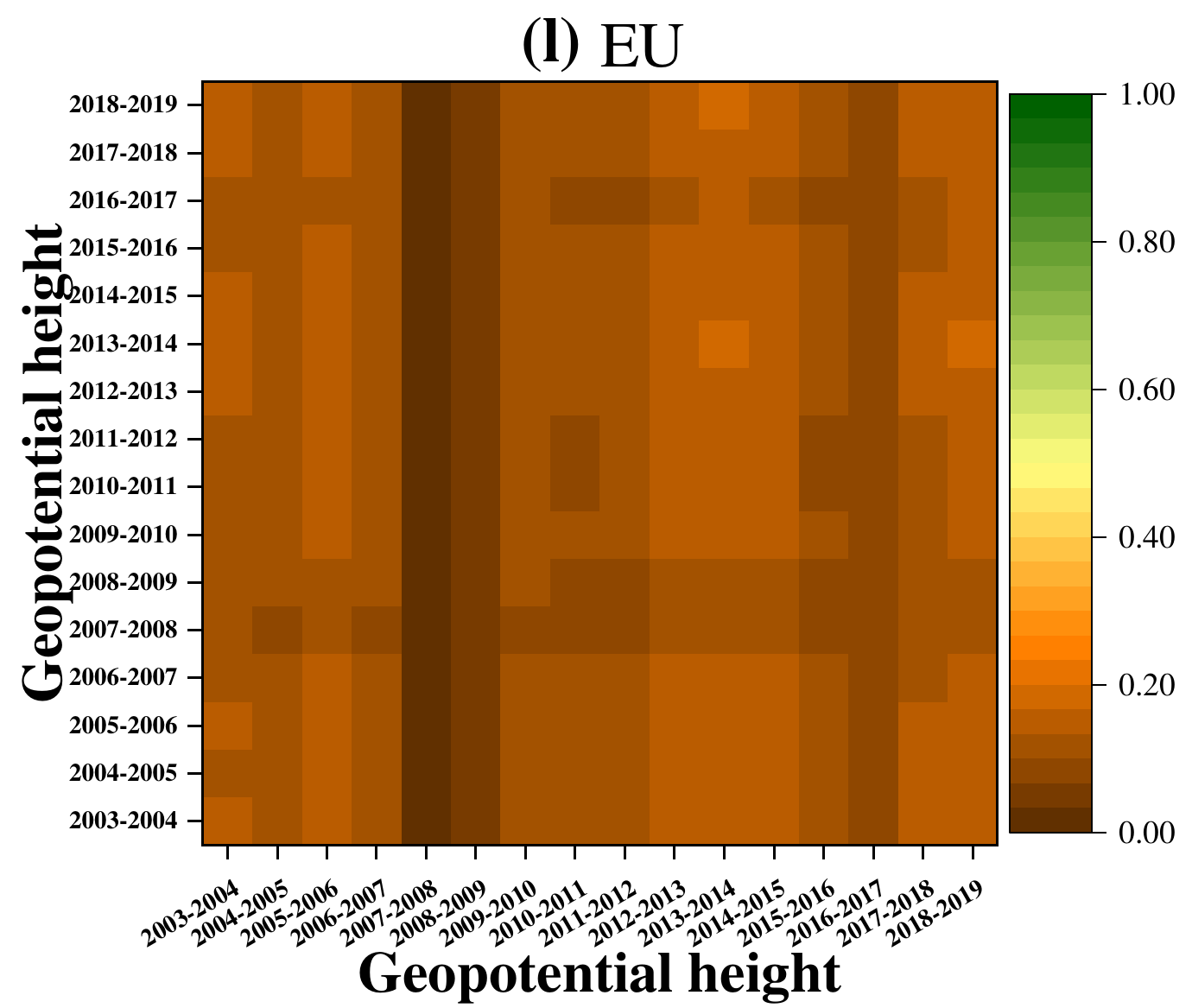}
\includegraphics[width=8em, height=7em]{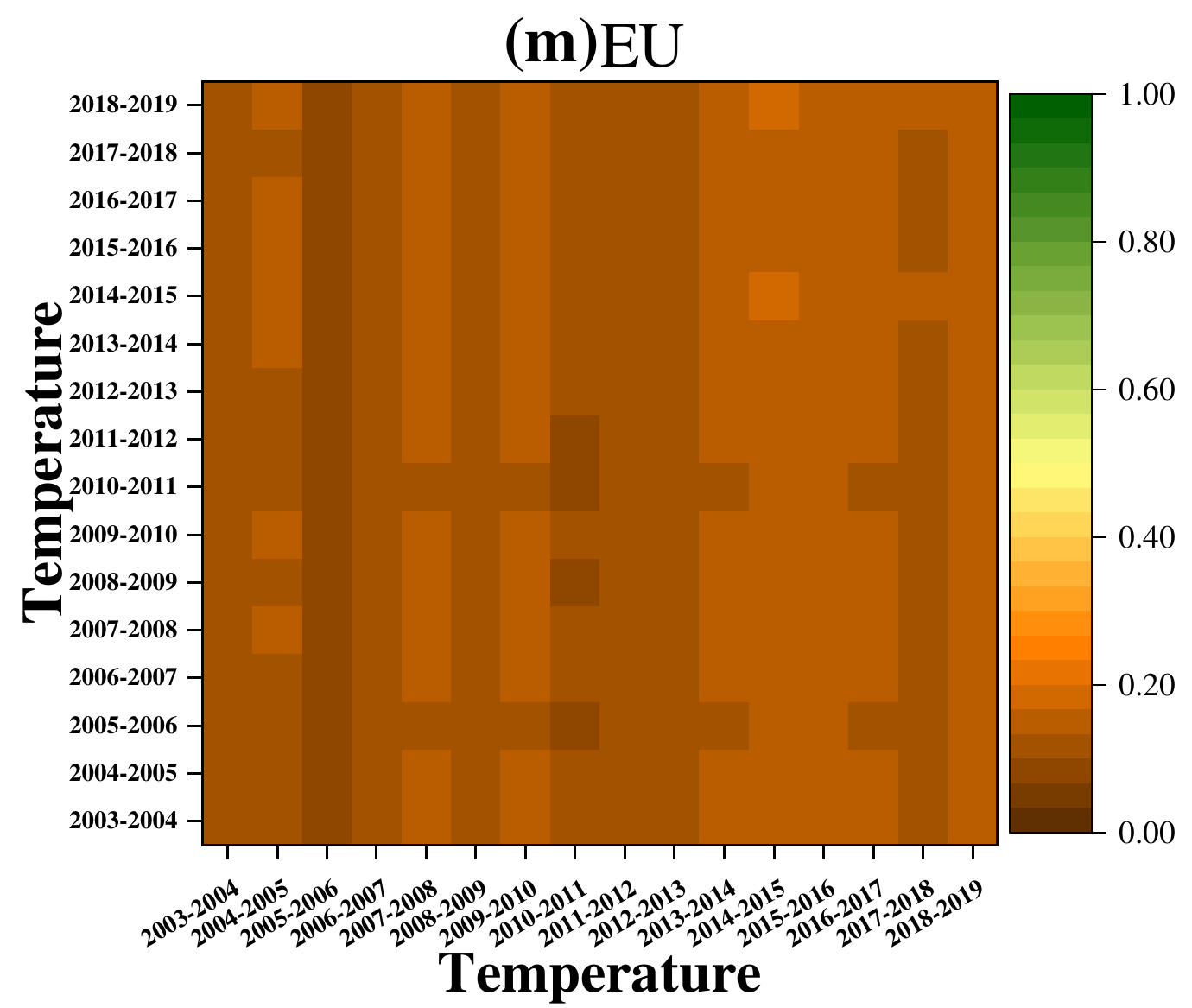}
\includegraphics[width=8em, height=7em]{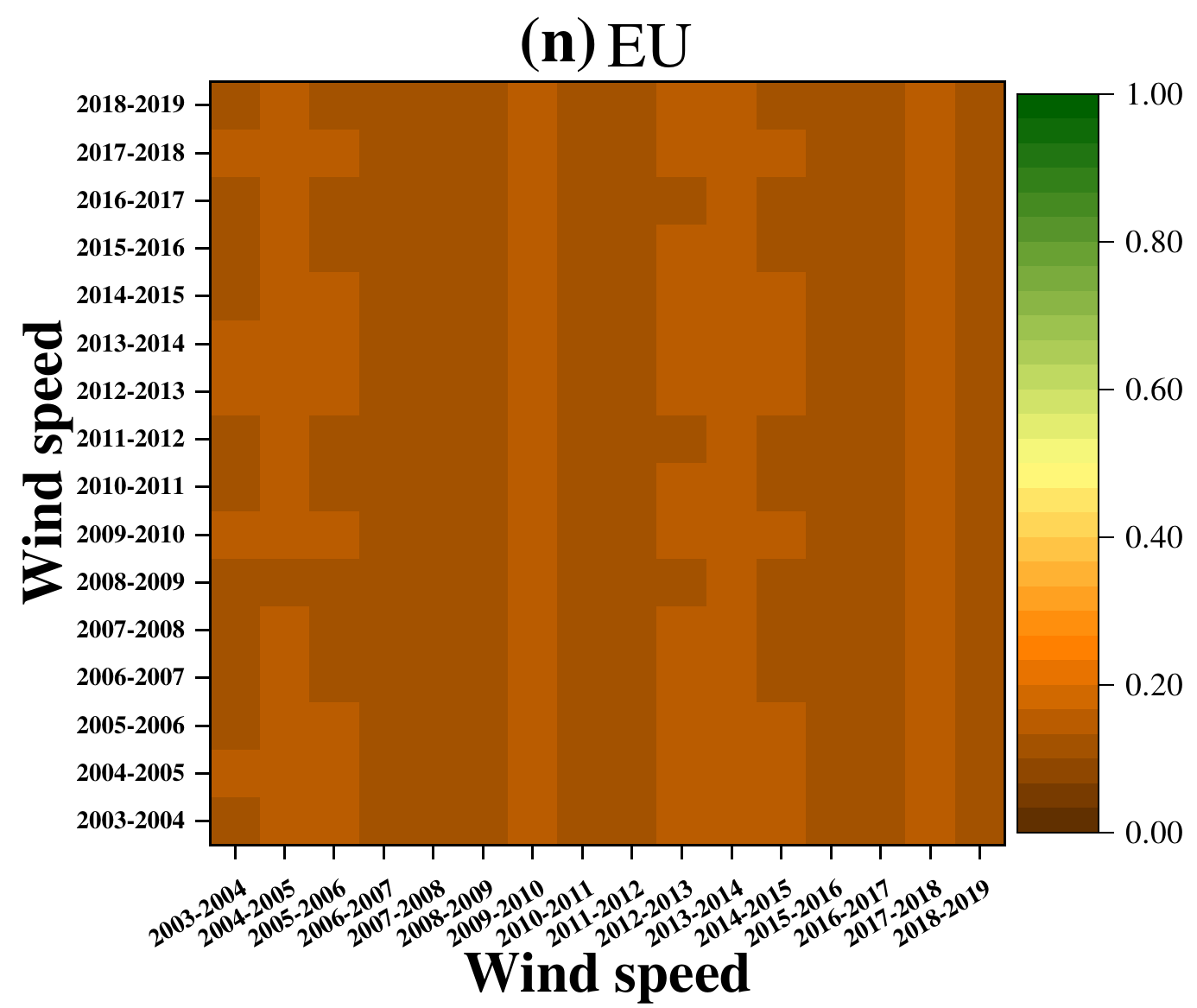}
\includegraphics[width=8em, height=7em]{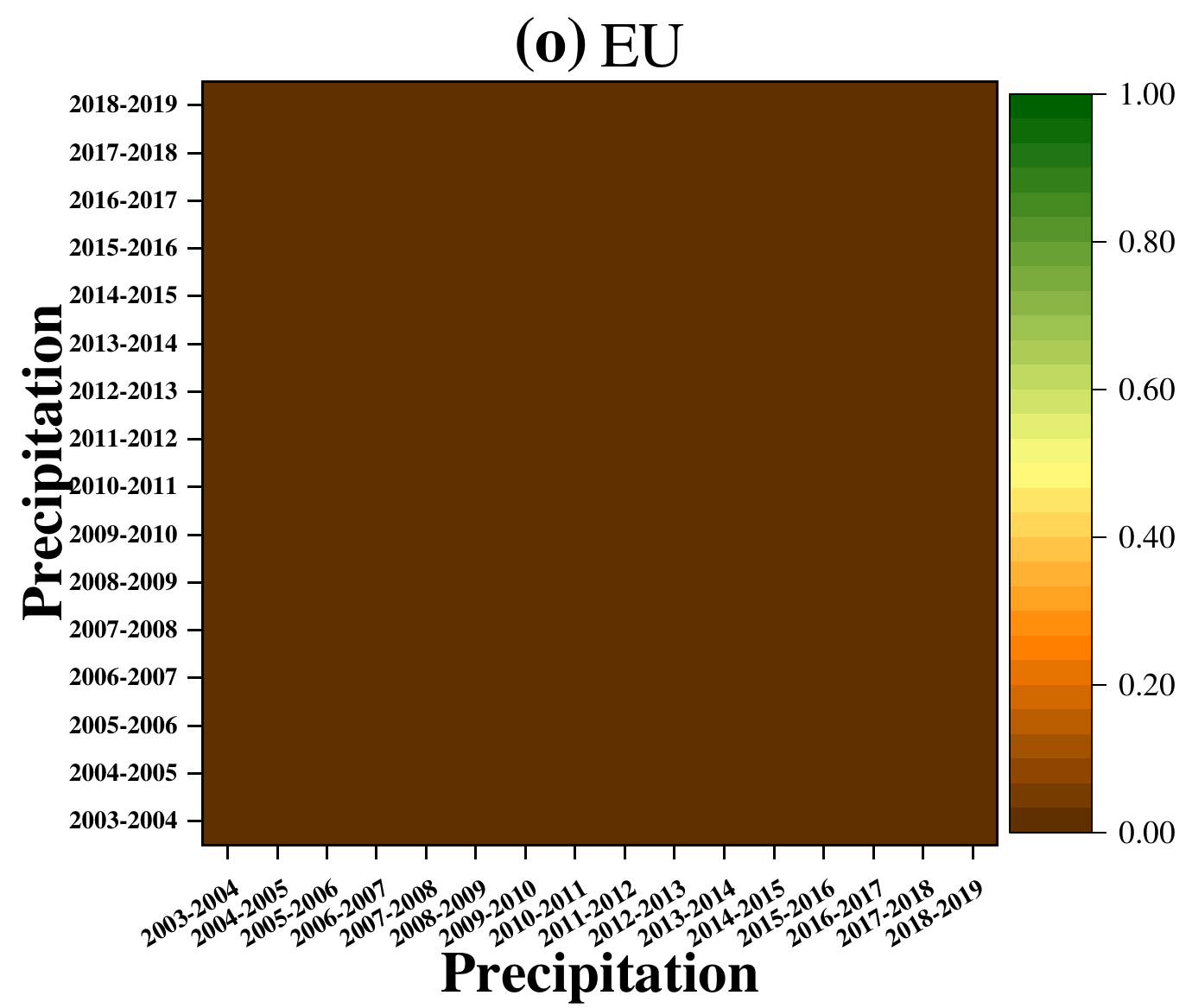}
\end{center}

\begin{center}
\noindent {\small {\bf Fig. S29} For the controlled case for lengths above $1000km$, the Jaccard similarity coefficient matrix of links in two networks of different years for each of the climate variables.}
\end{center}

\begin{center}
\includegraphics[width=8em, height=7em]{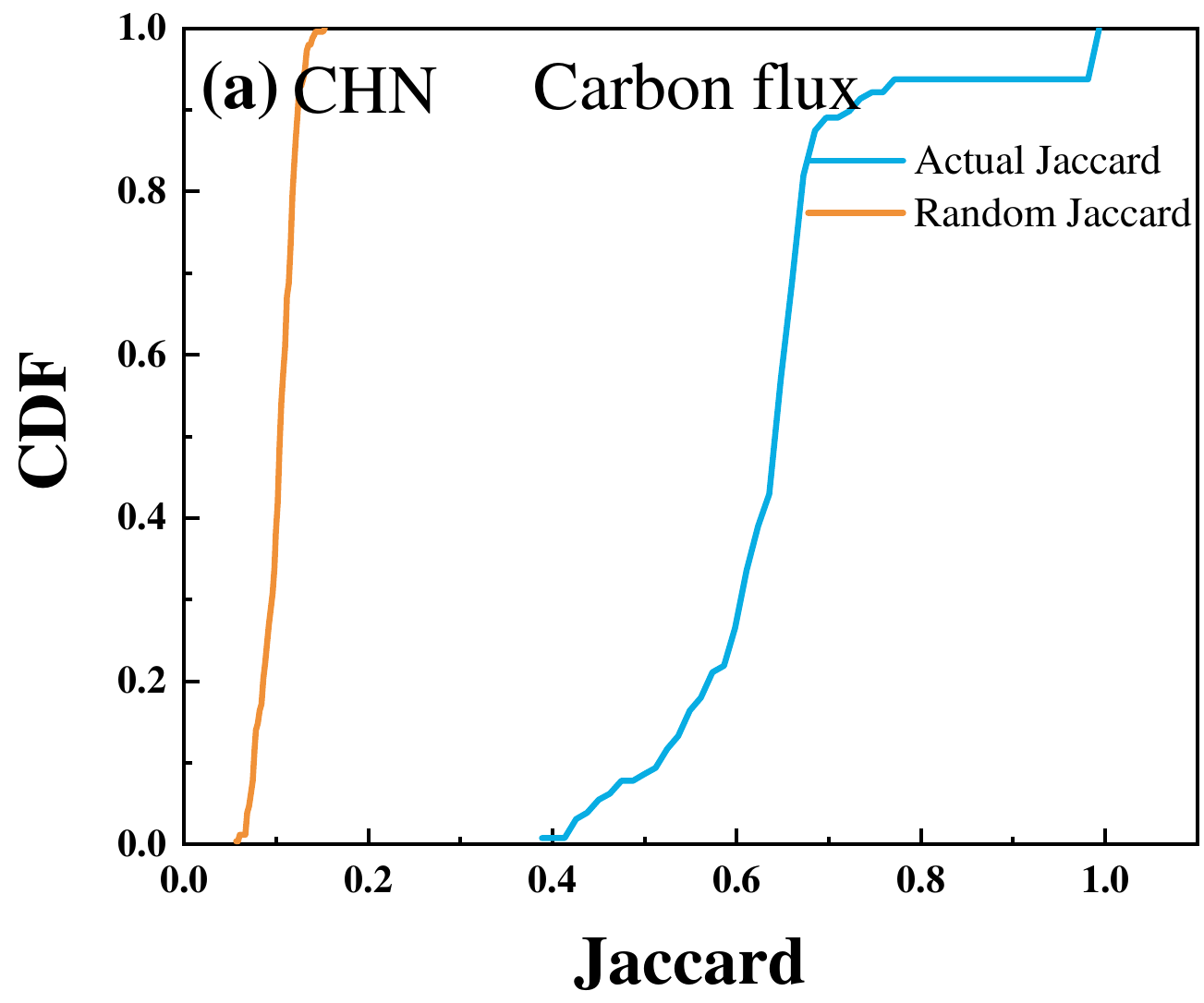}
\includegraphics[width=8em, height=7em]{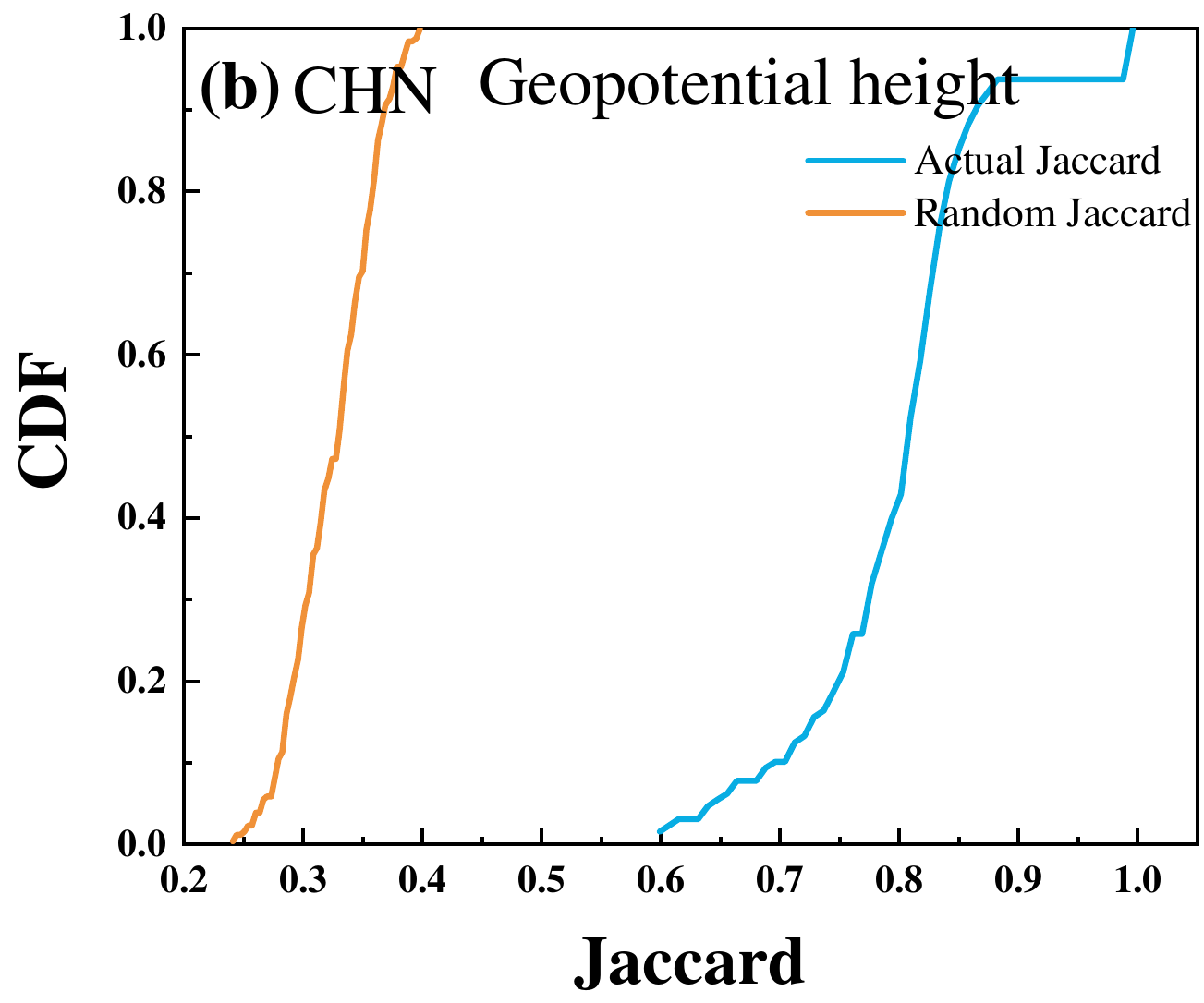}
\includegraphics[width=8em, height=7em]{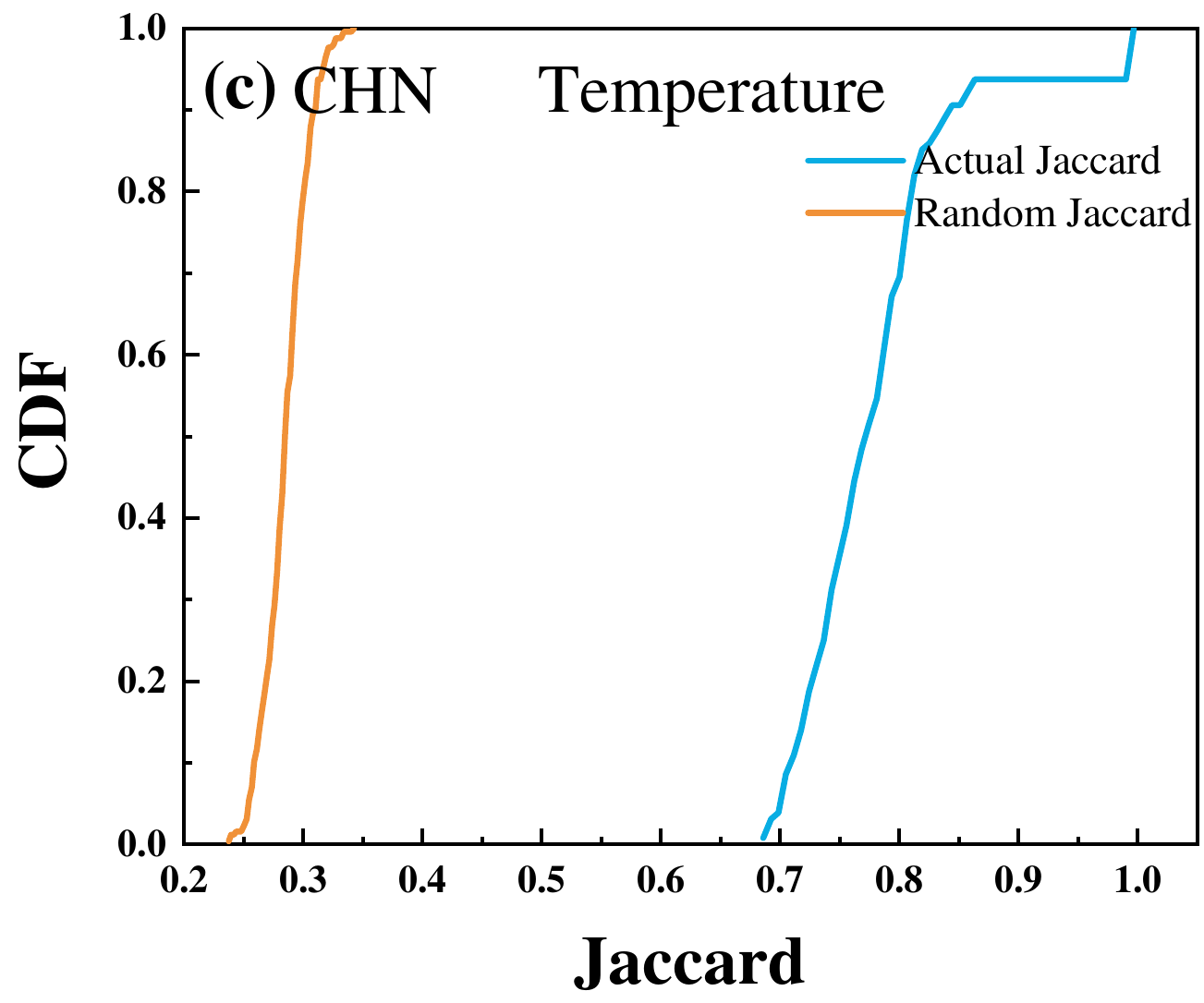}
\includegraphics[width=8em, height=7em]{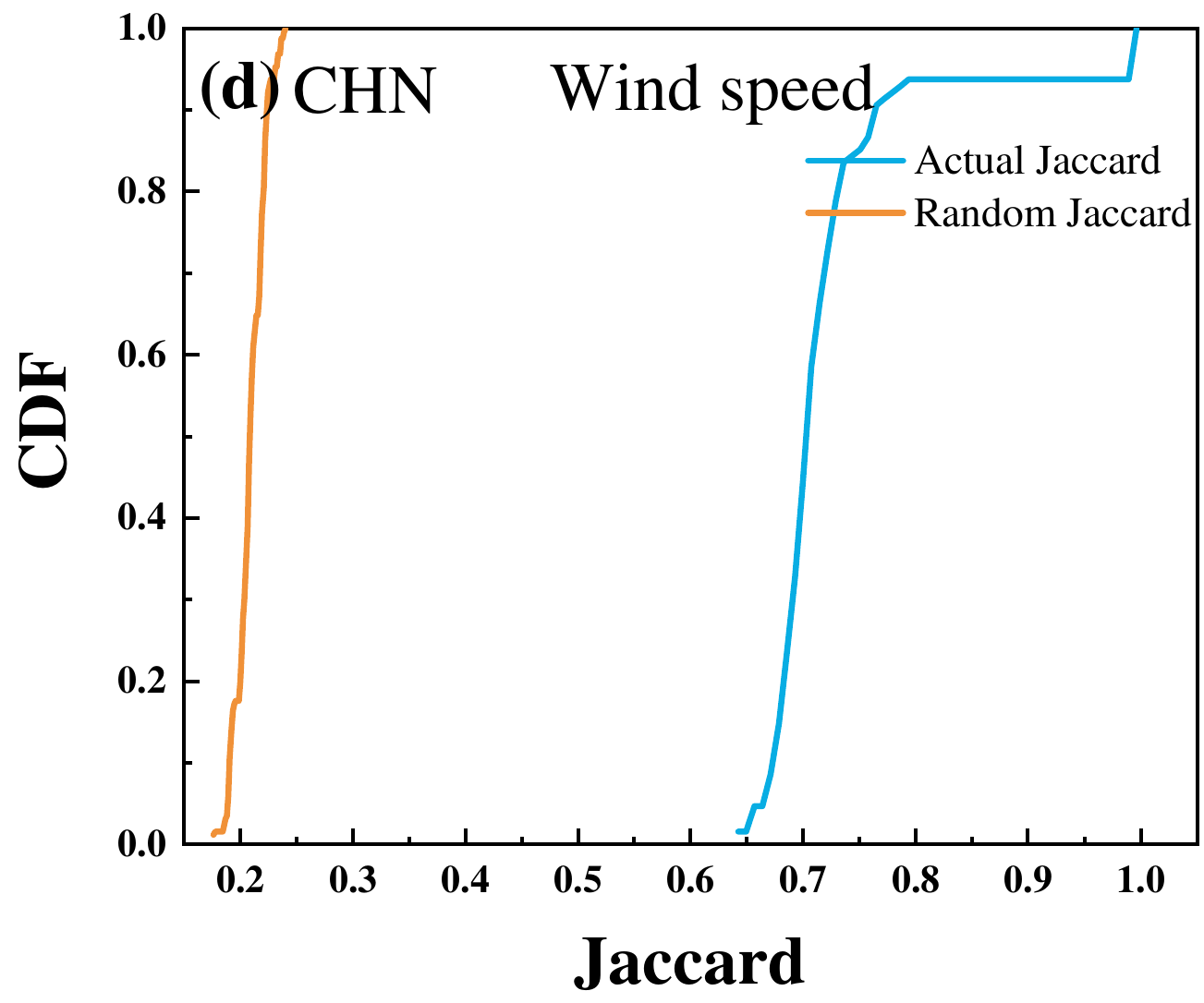}
\includegraphics[width=8em, height=7em]{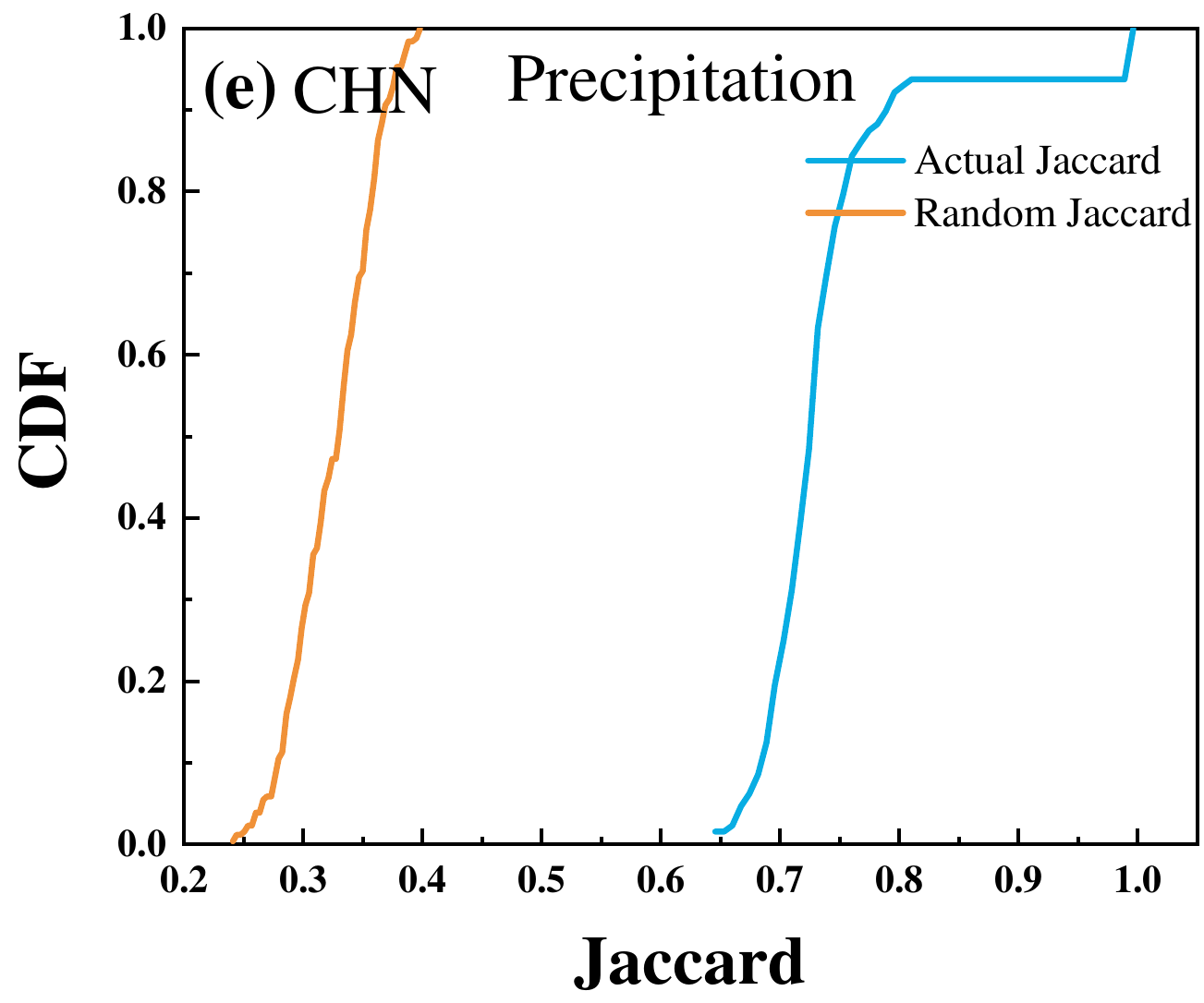}
\includegraphics[width=8em, height=7em]{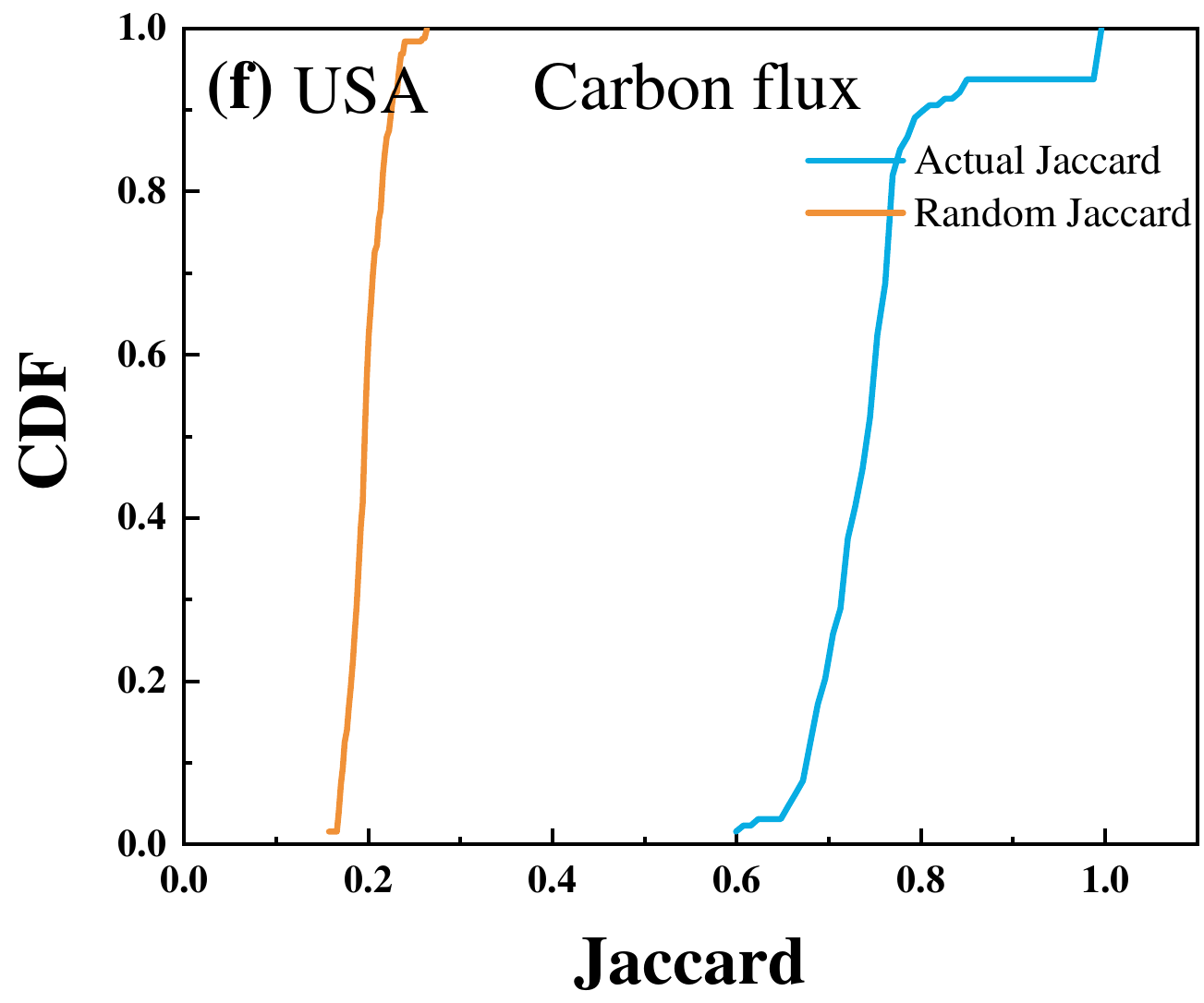}
\includegraphics[width=8em, height=7em]{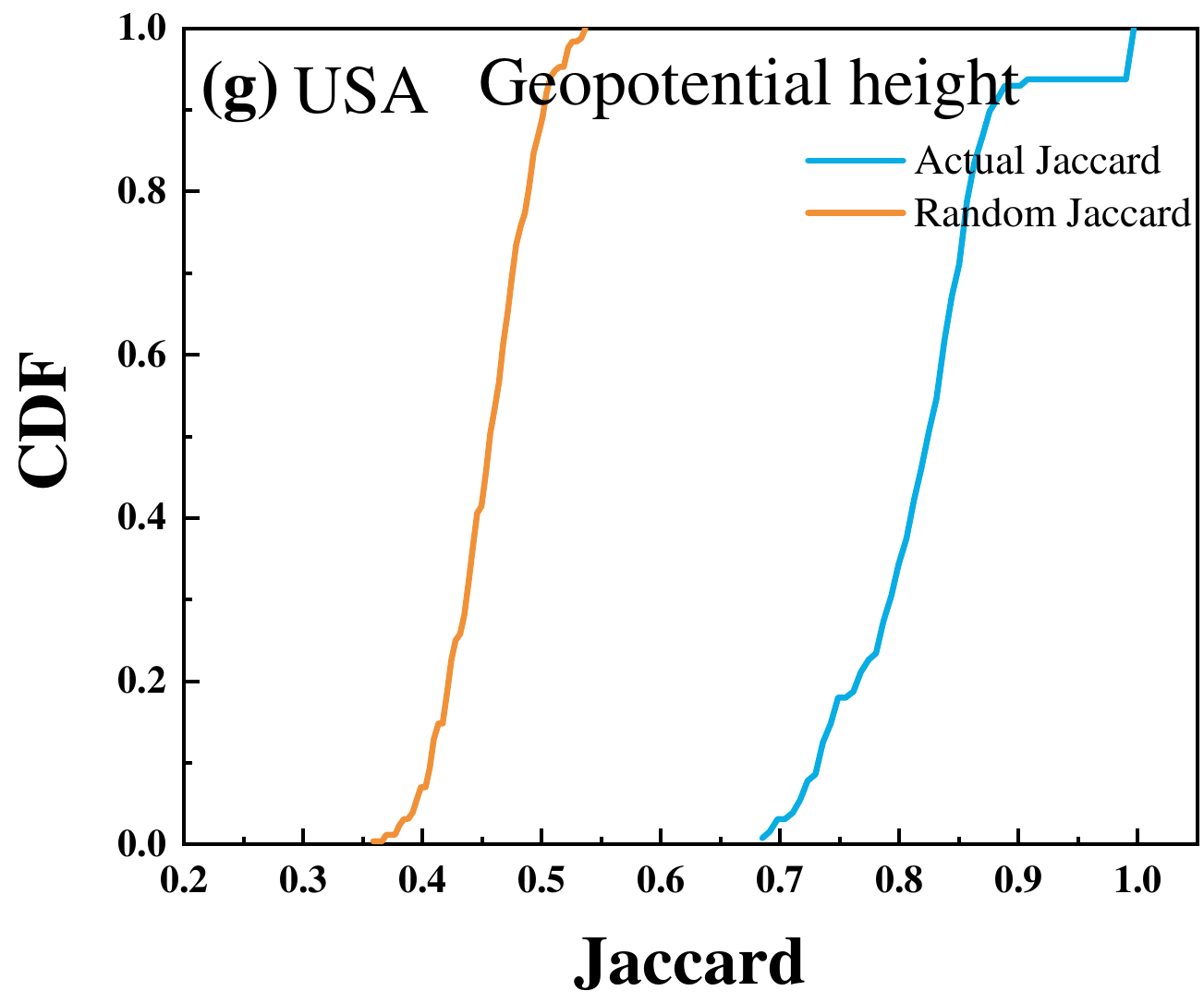}
\includegraphics[width=8em, height=7em]{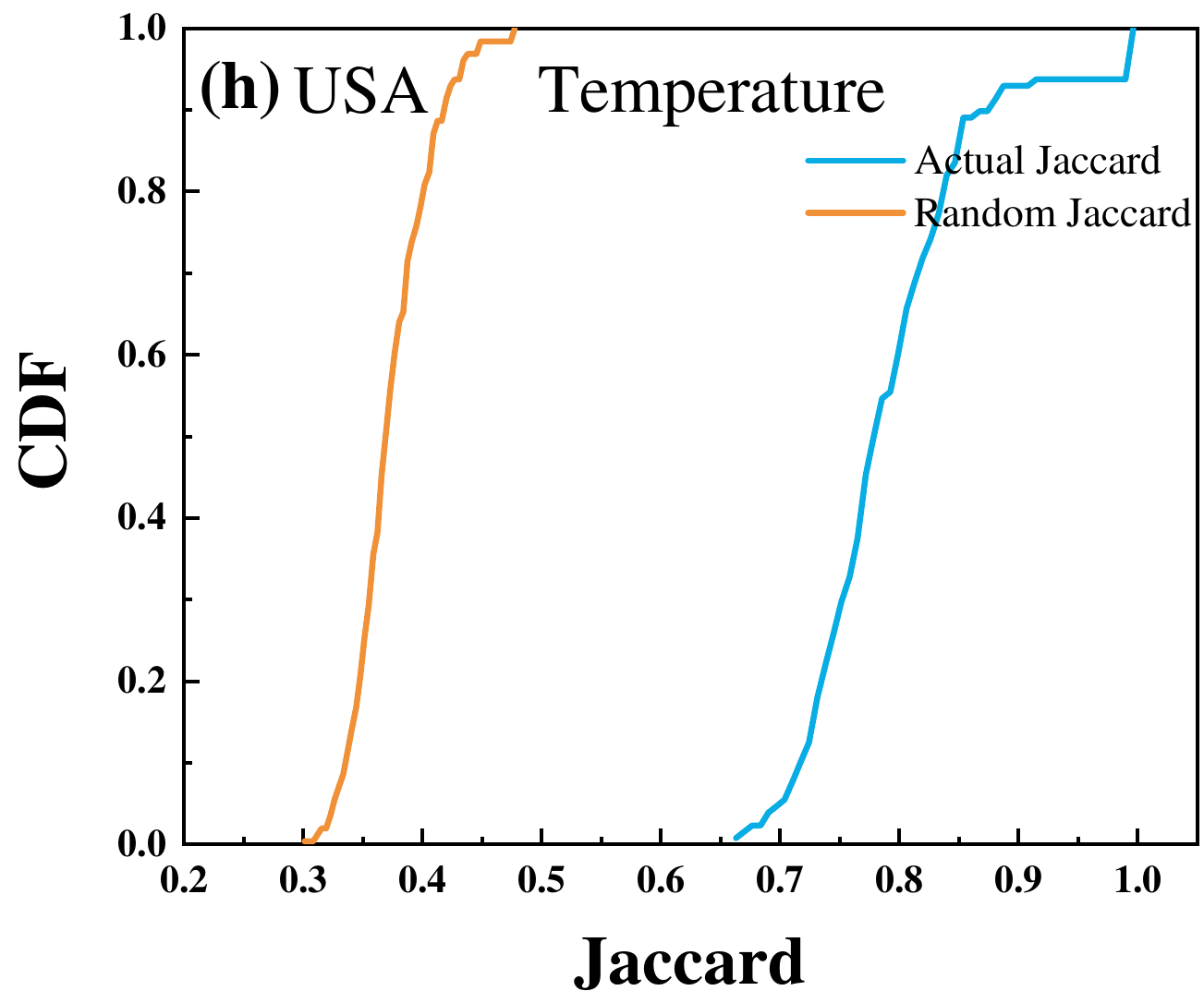}
\includegraphics[width=8em, height=7em]{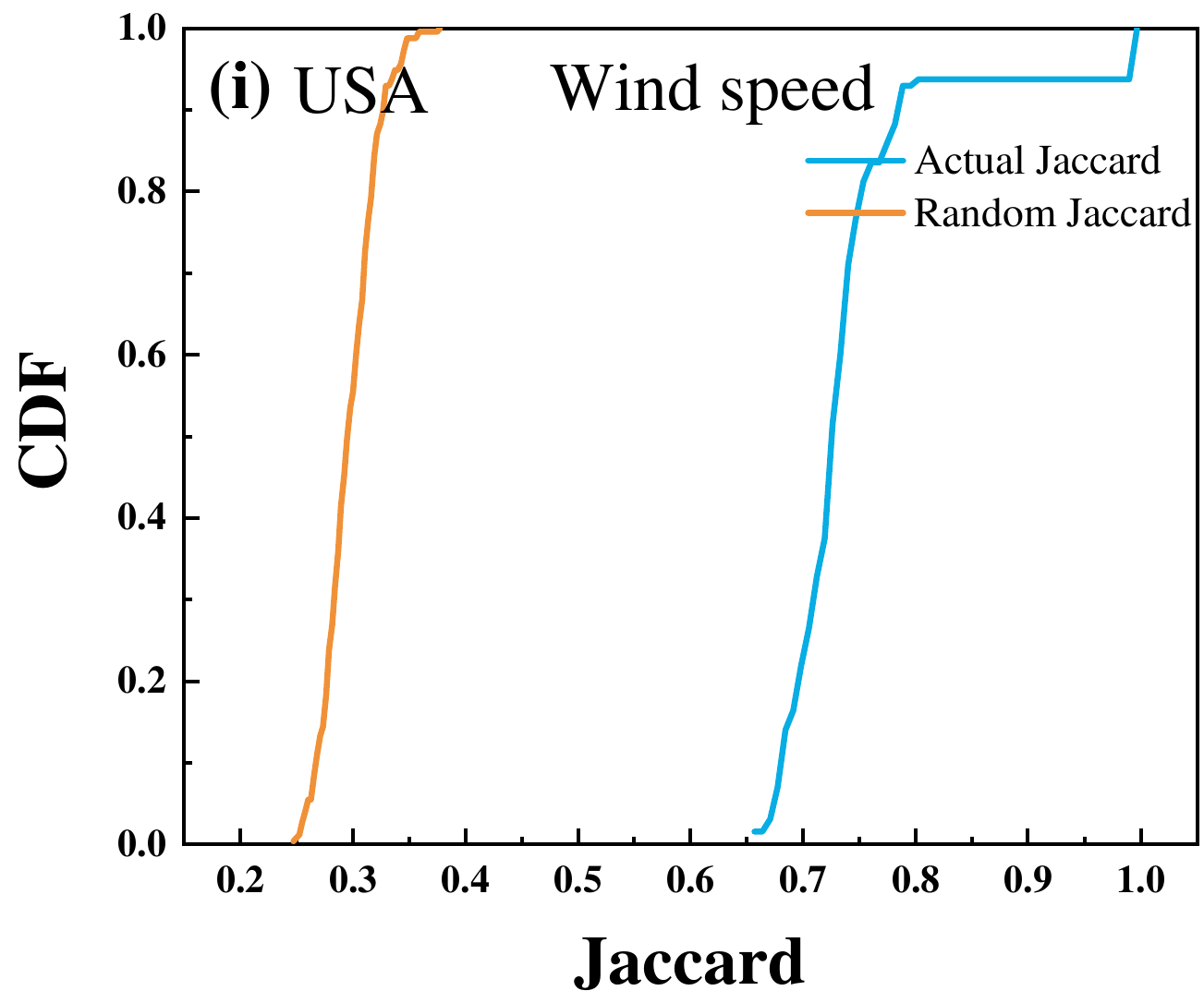}
\includegraphics[width=8em, height=7em]{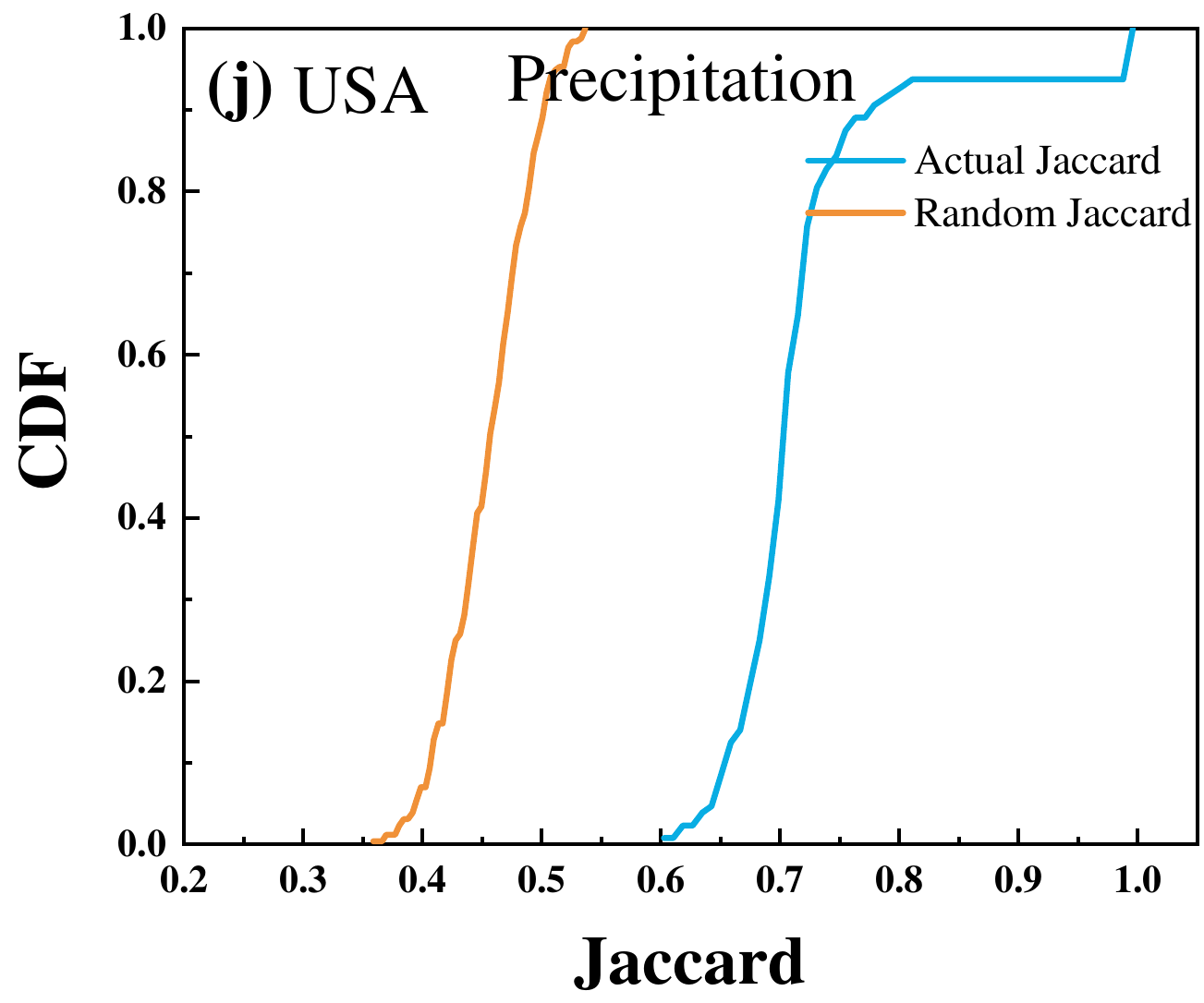}
\includegraphics[width=8em, height=7em]{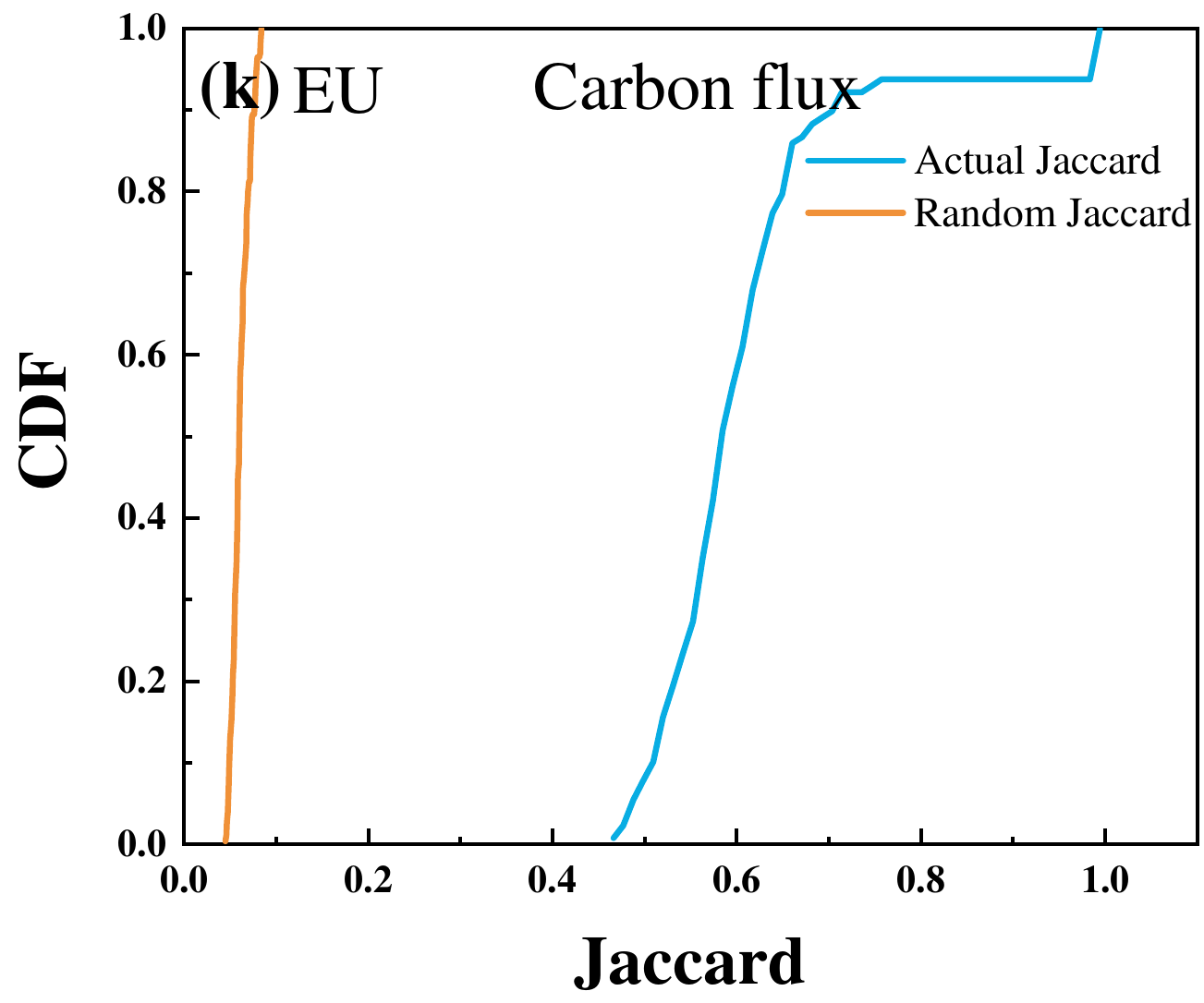}
\includegraphics[width=8em, height=7em]{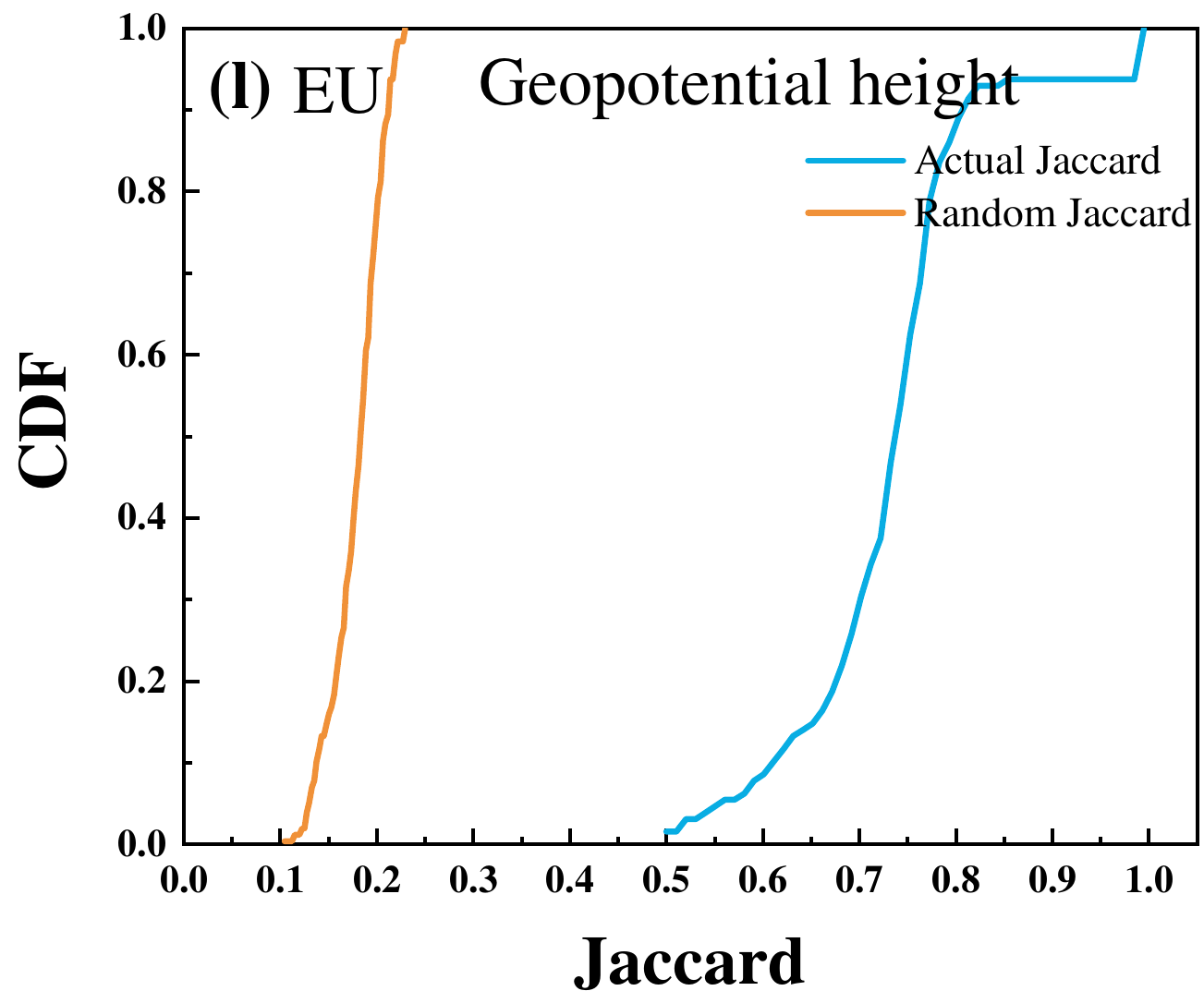}
\includegraphics[width=8em, height=7em]{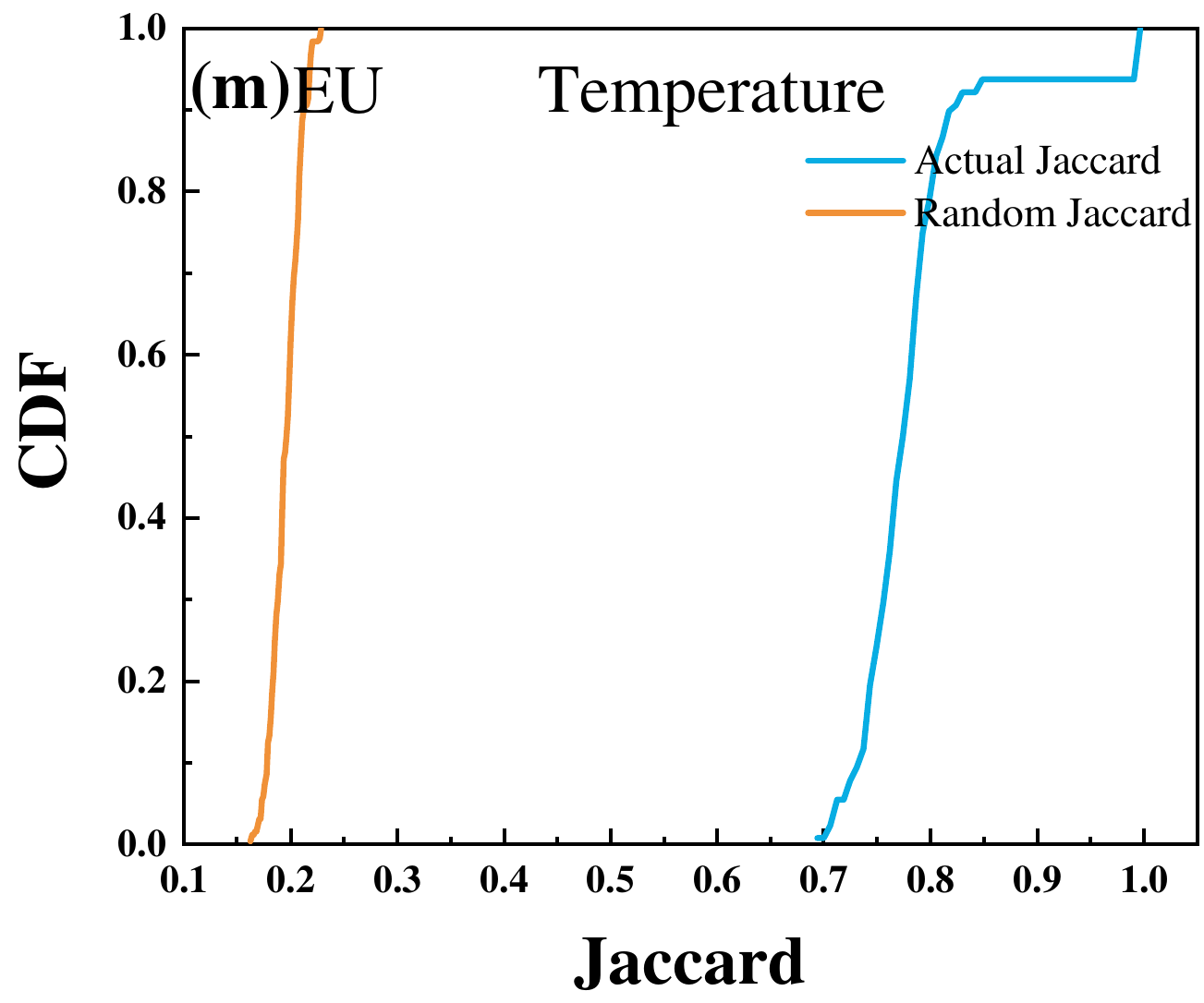}
\includegraphics[width=8em, height=7em]{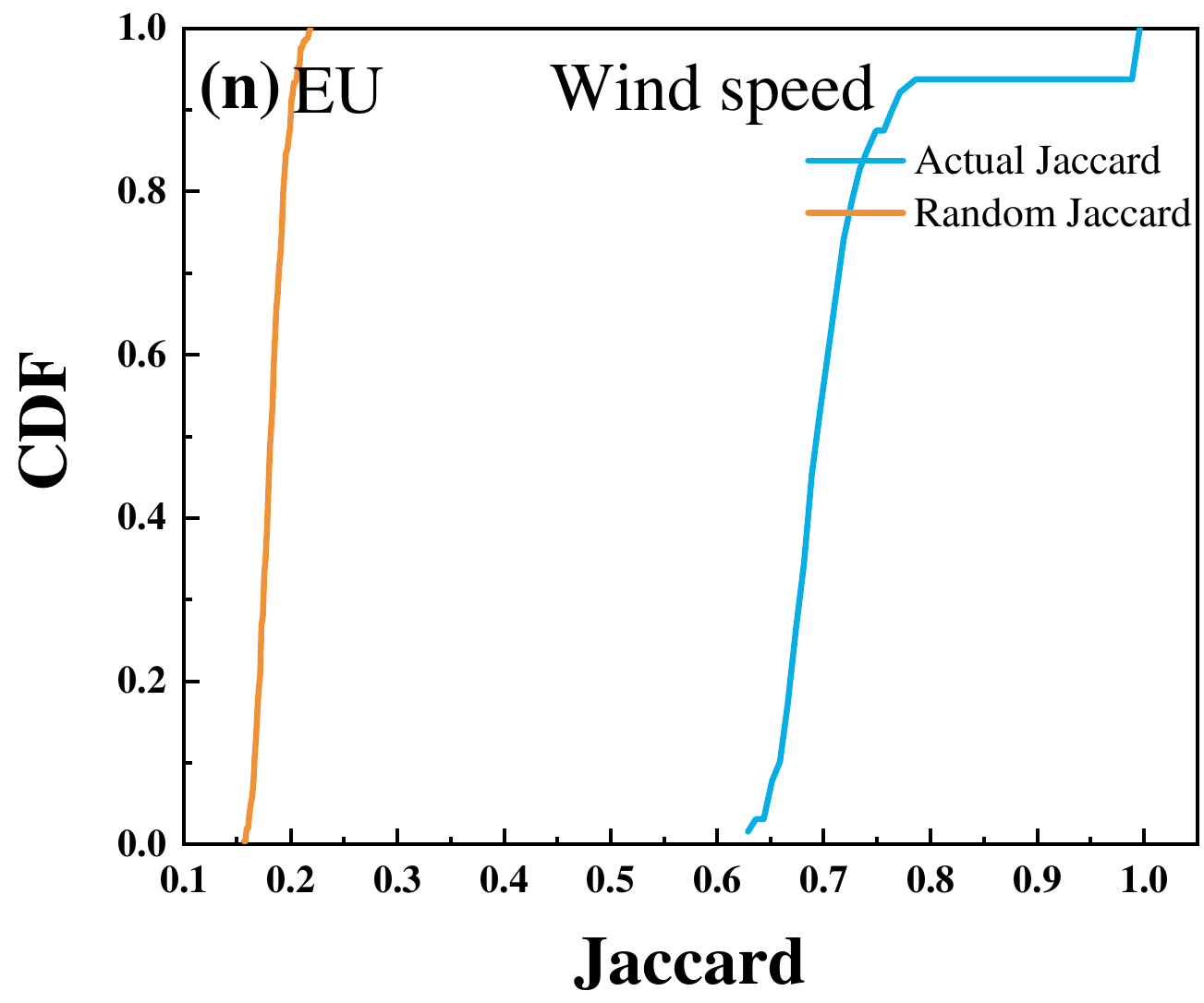}
\includegraphics[width=8em, height=7em]{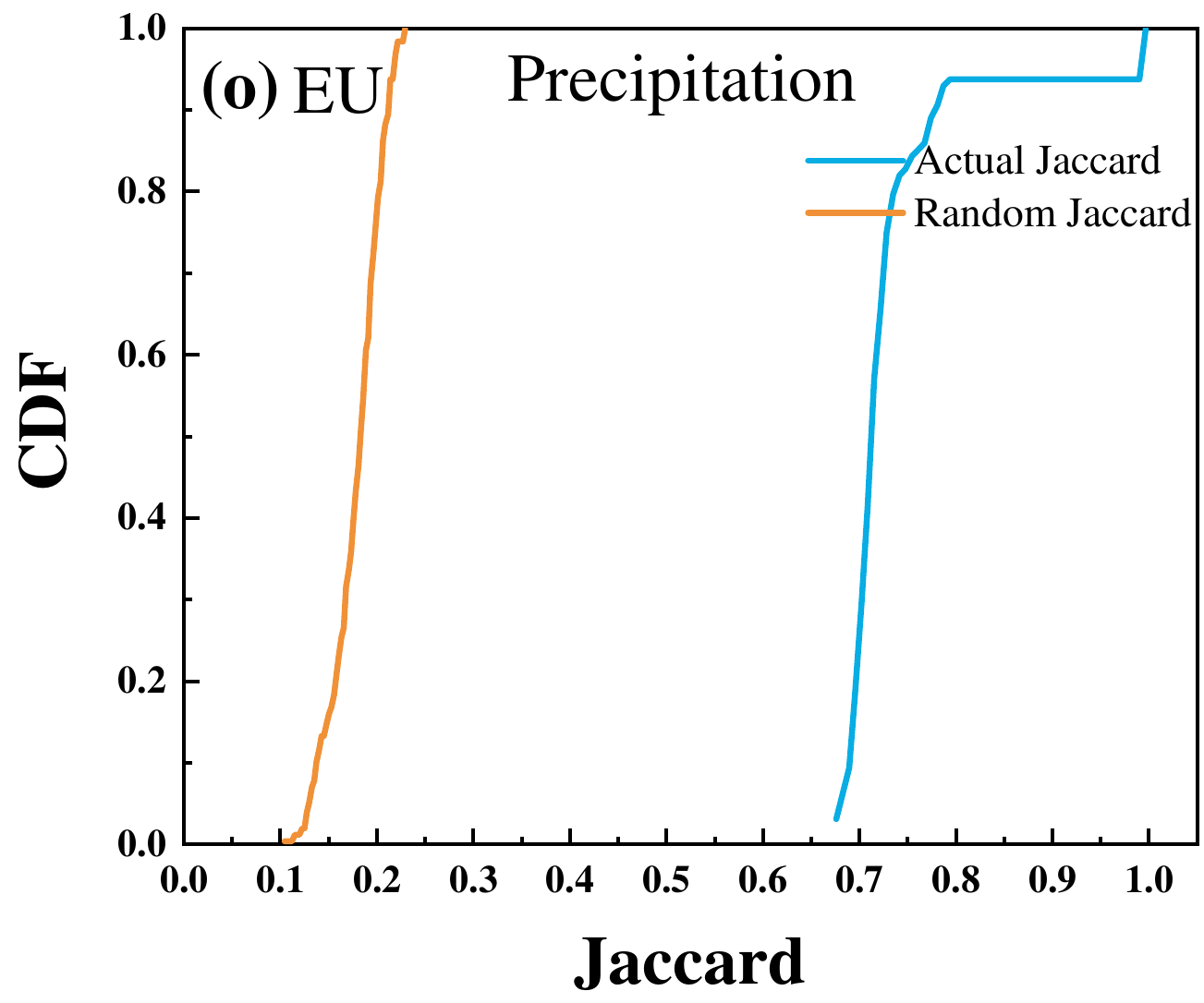}
\end{center}

\begin{center}
\noindent {\small {\bf Fig. S30} The cumulative distribution function (CDF) of the actual Jaccard coefficient (blue line) and Jaccard coefficient in the controlled case (orange line) for different climate variables.}
\end{center}

\begin{center}
\includegraphics[width=8em, height=7em]{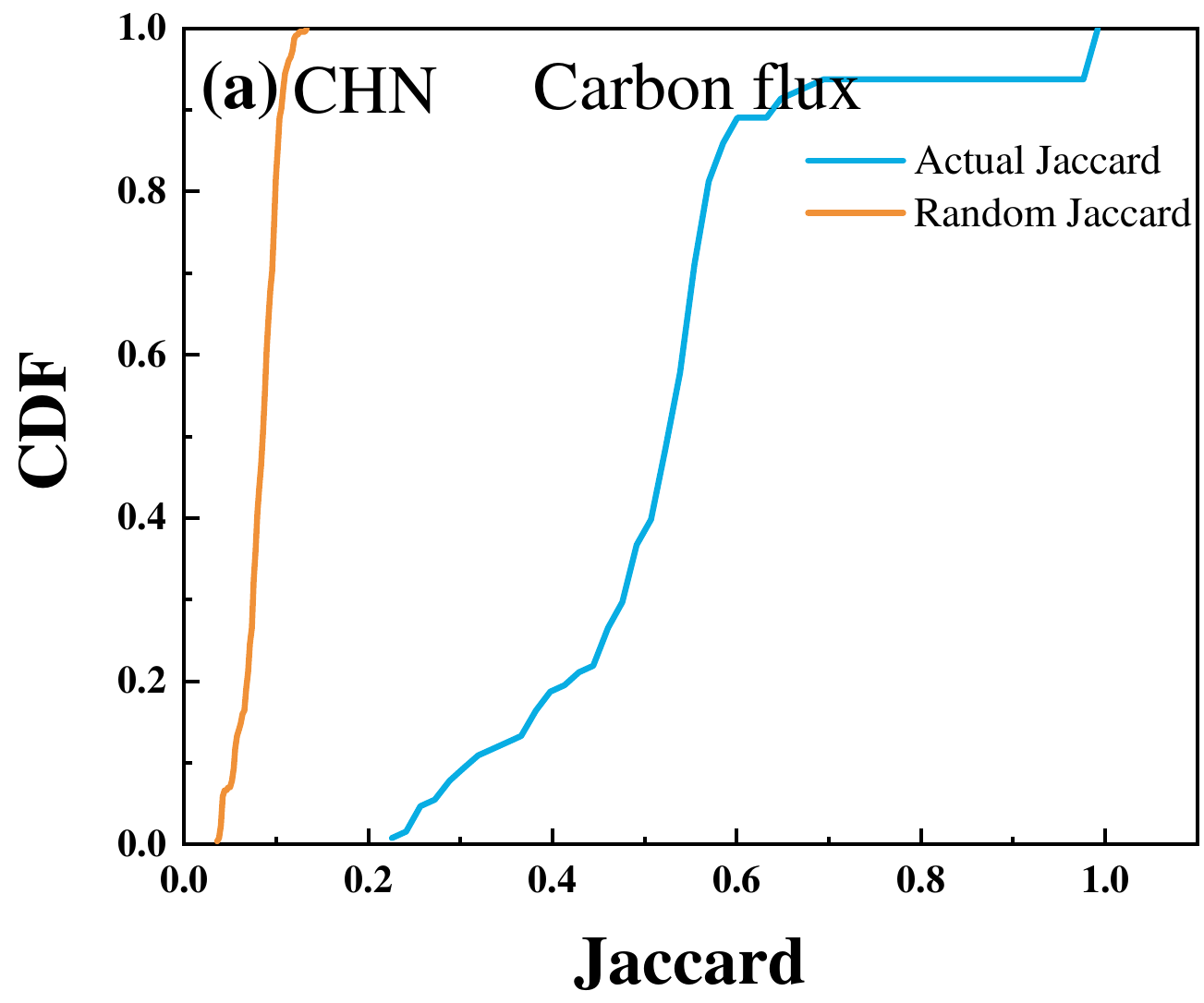}
\includegraphics[width=8em, height=7em]{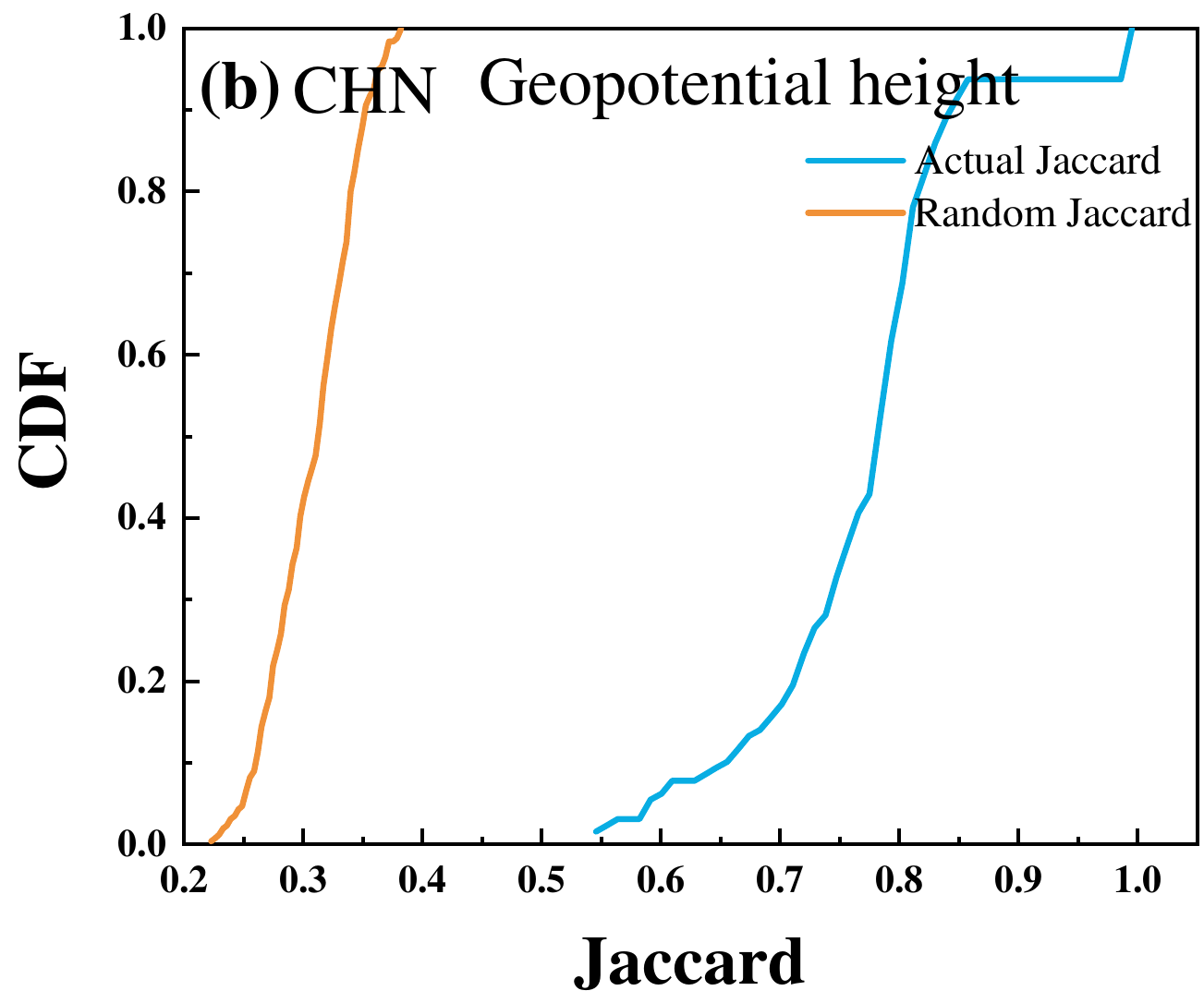}
\includegraphics[width=8em, height=7em]{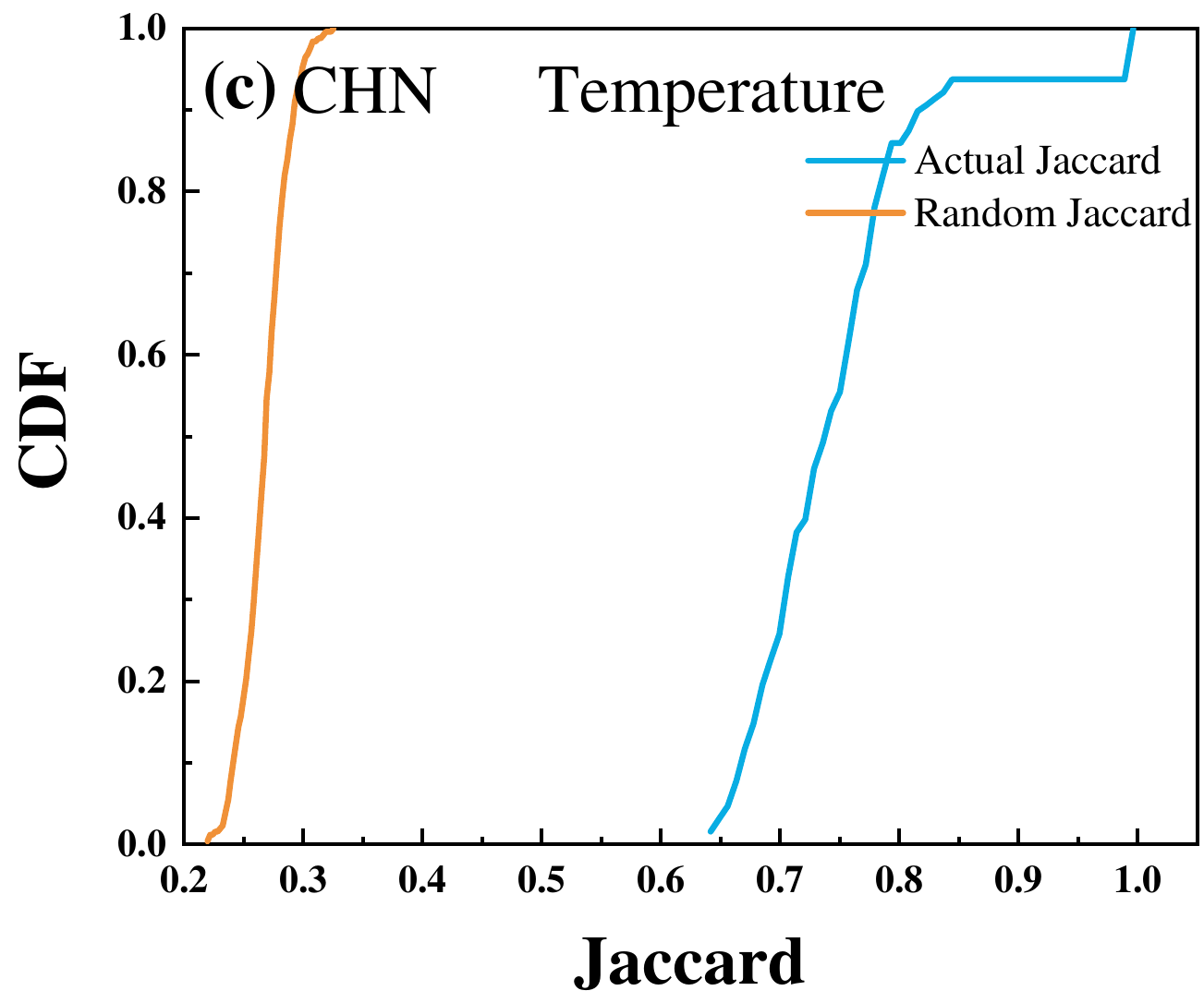}
\includegraphics[width=8em, height=7em]{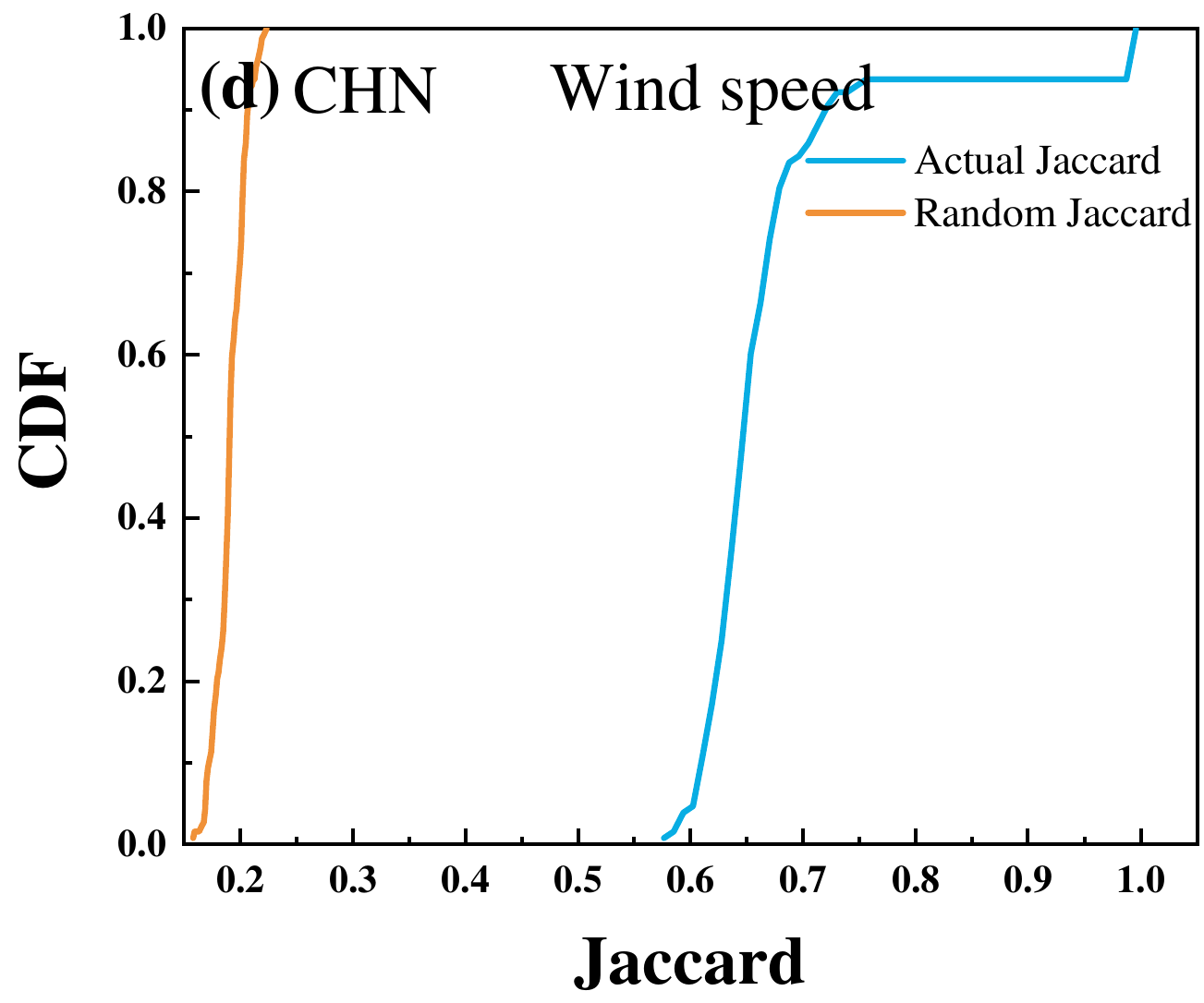}
\includegraphics[width=8em, height=7em]{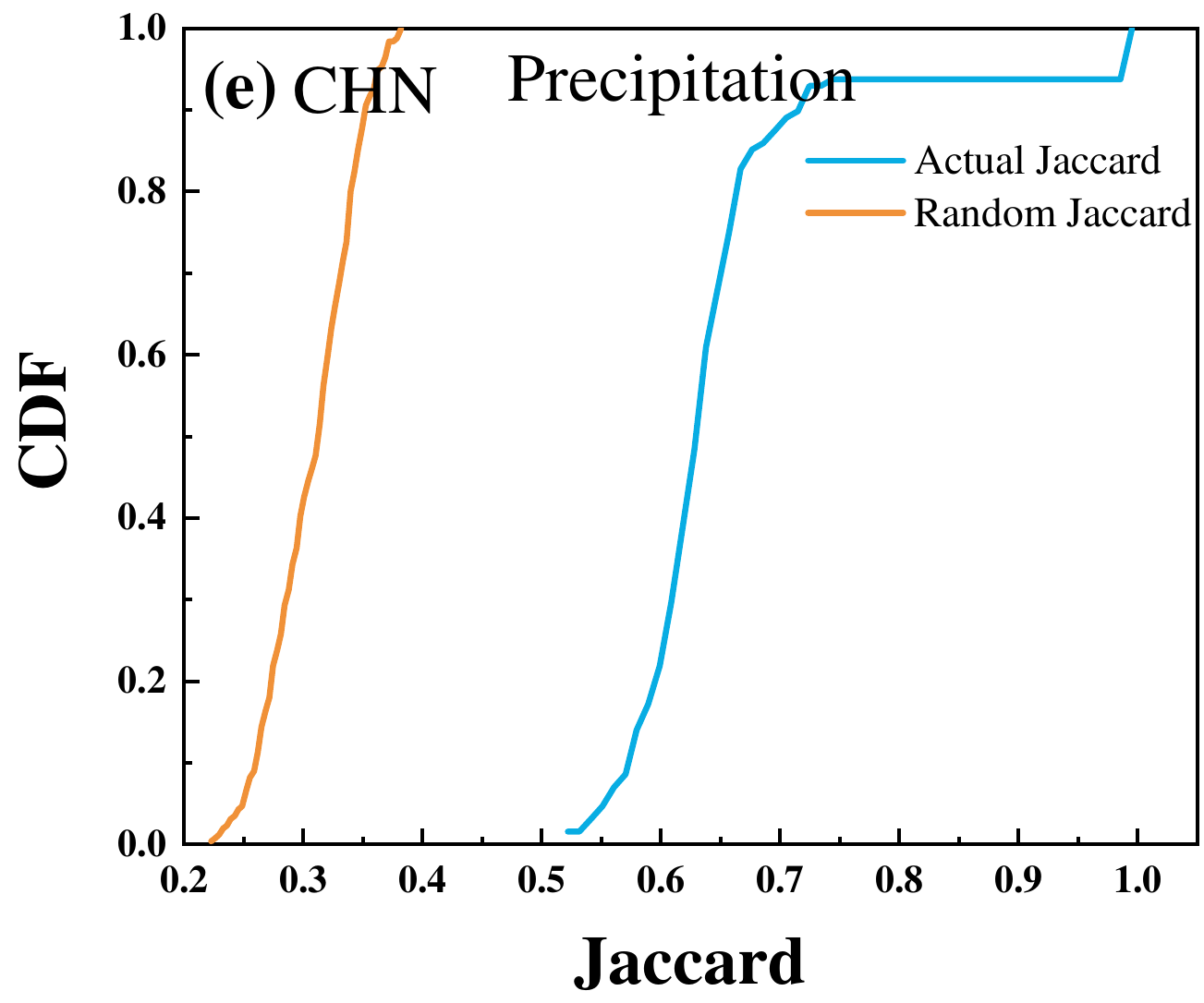}
\includegraphics[width=8em, height=7em]{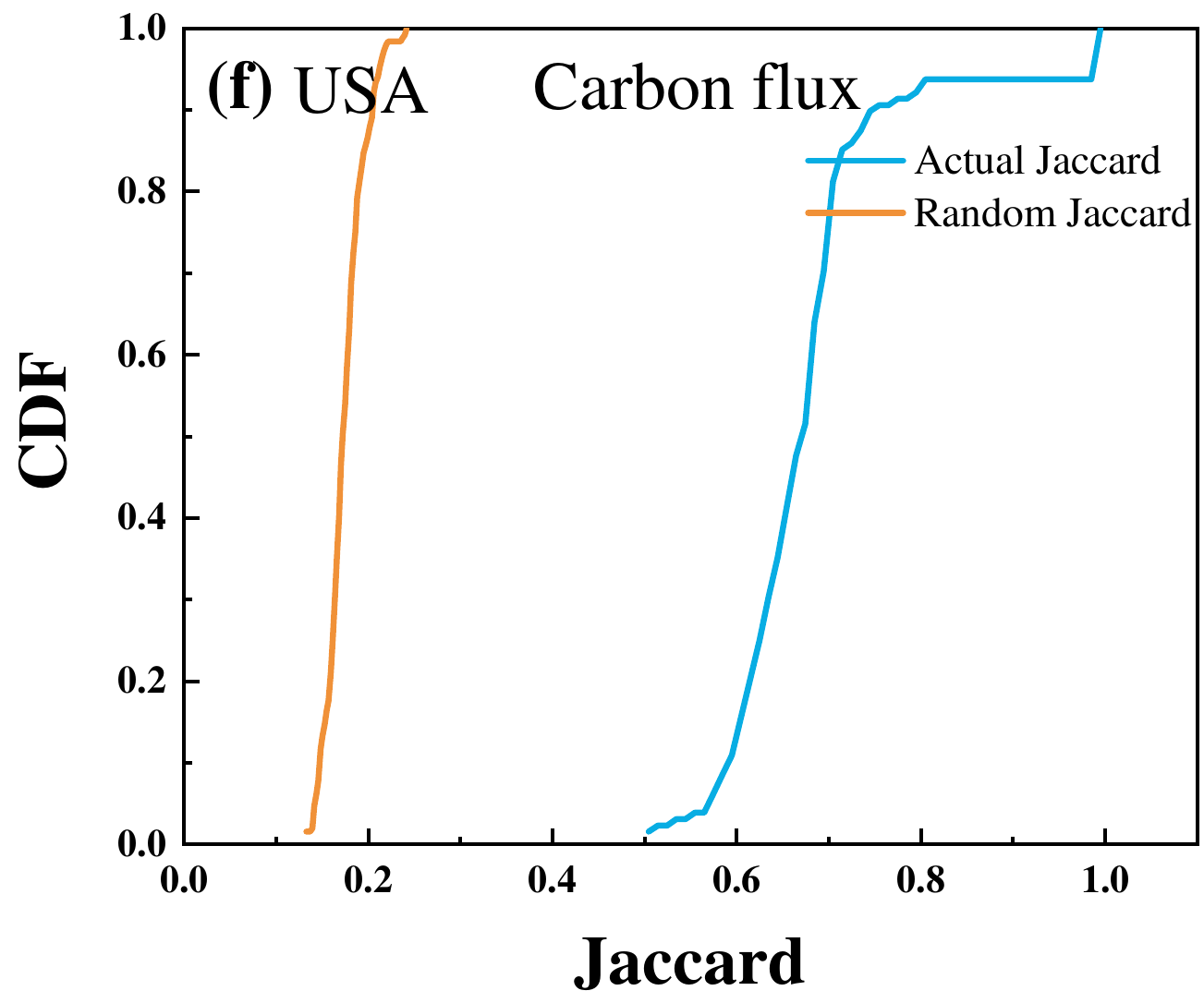}
\includegraphics[width=8em, height=7em]{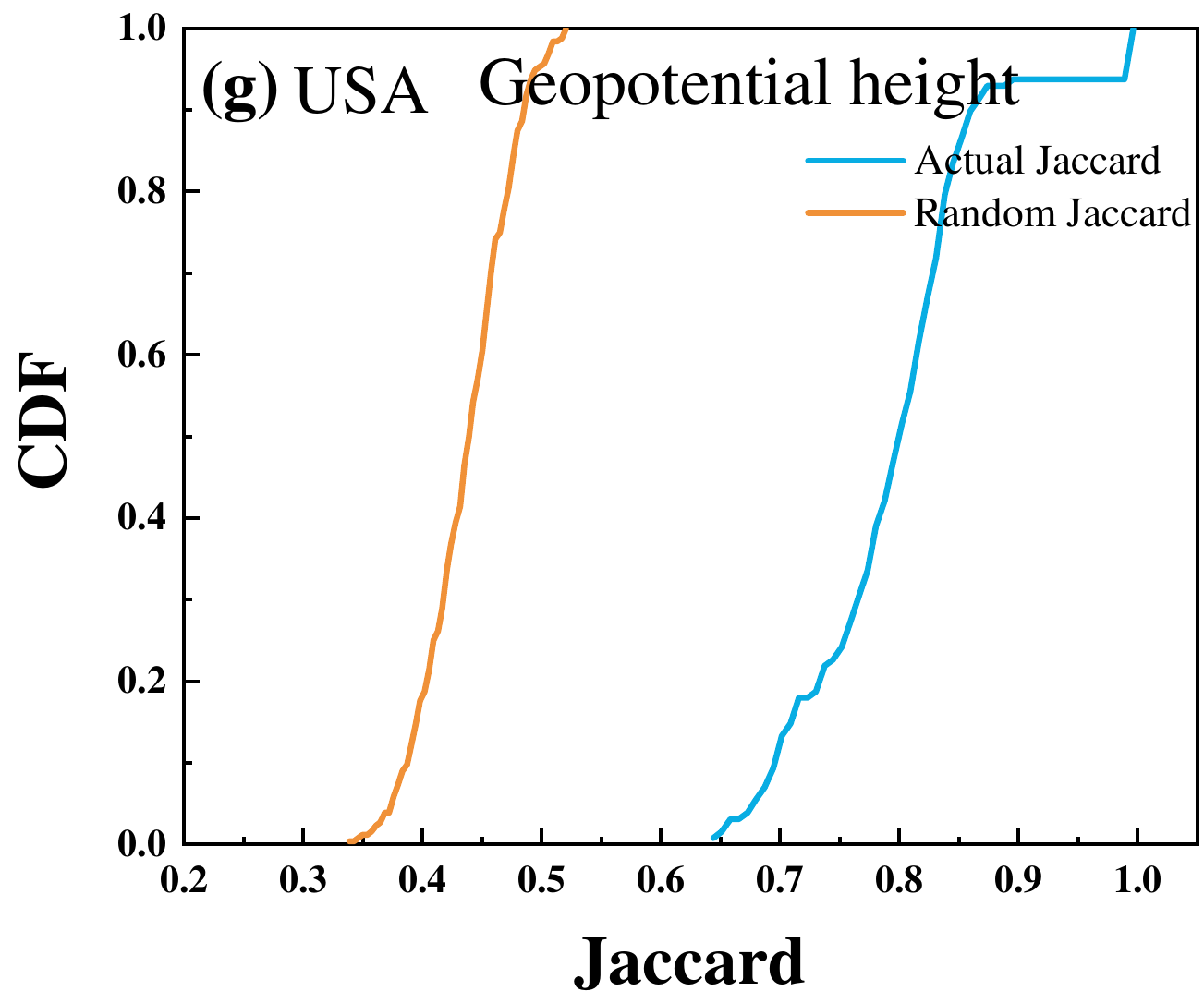}
\includegraphics[width=8em, height=7em]{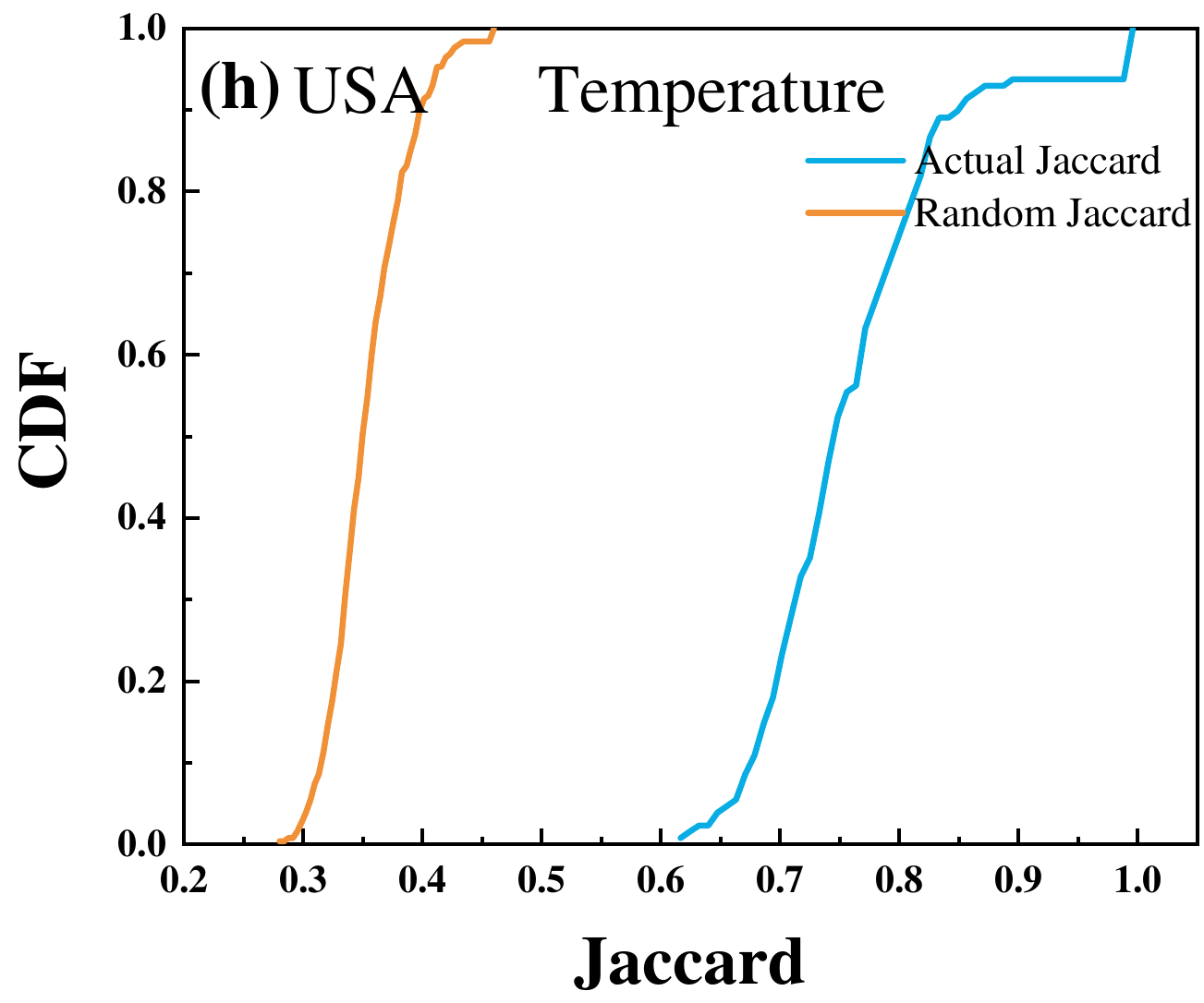}
\includegraphics[width=8em, height=7em]{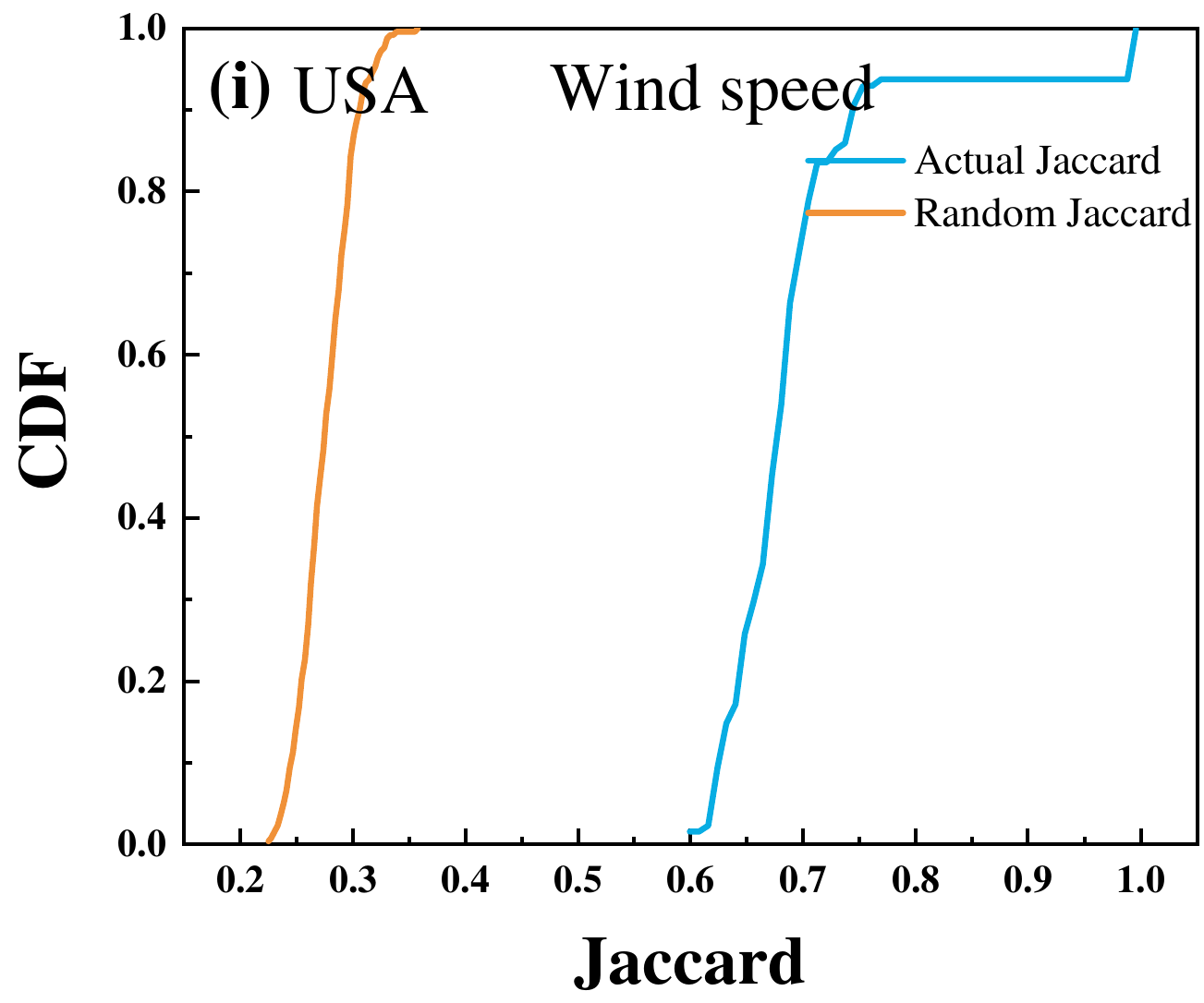}
\includegraphics[width=8em, height=7em]{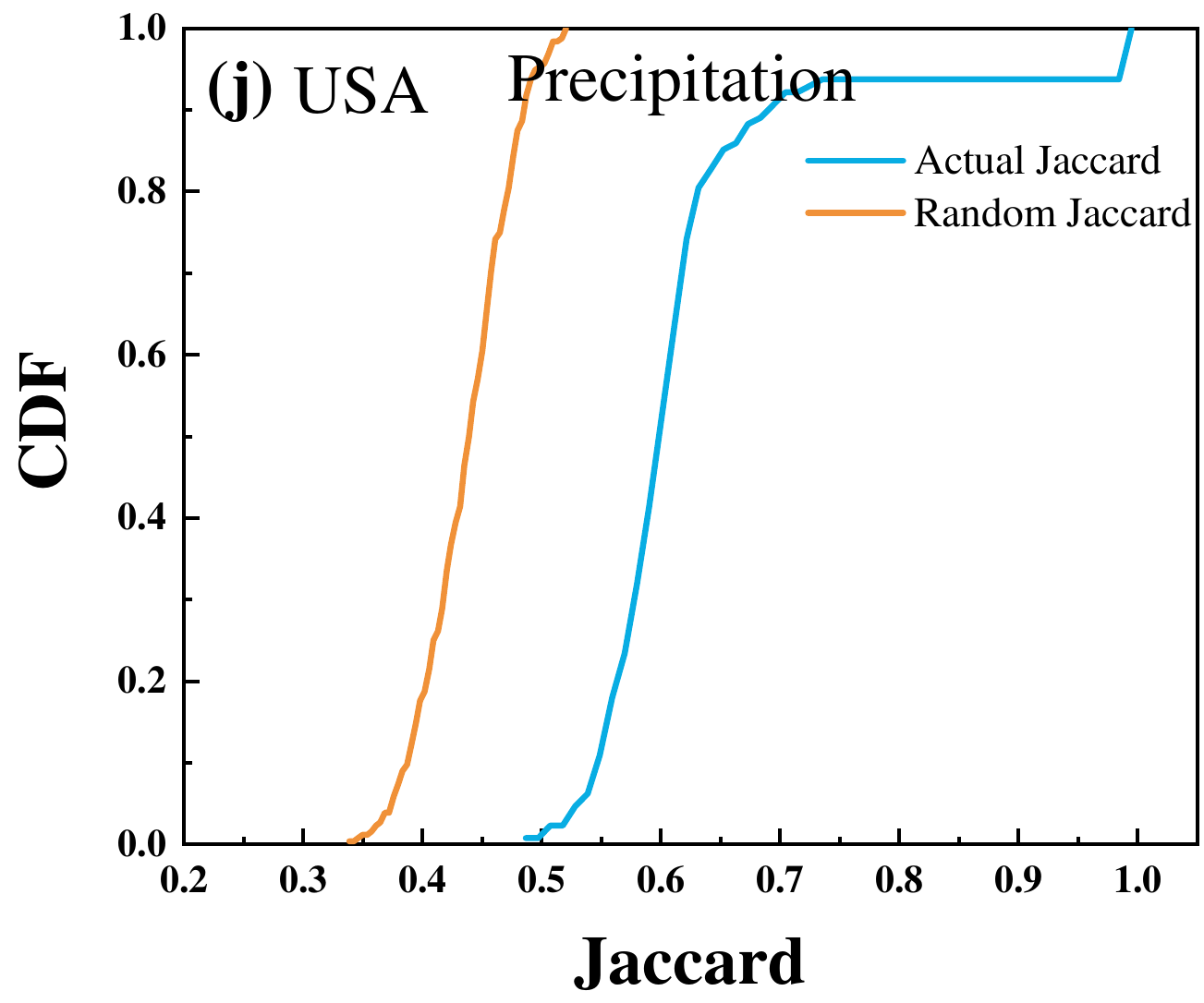}
\includegraphics[width=8em, height=7em]{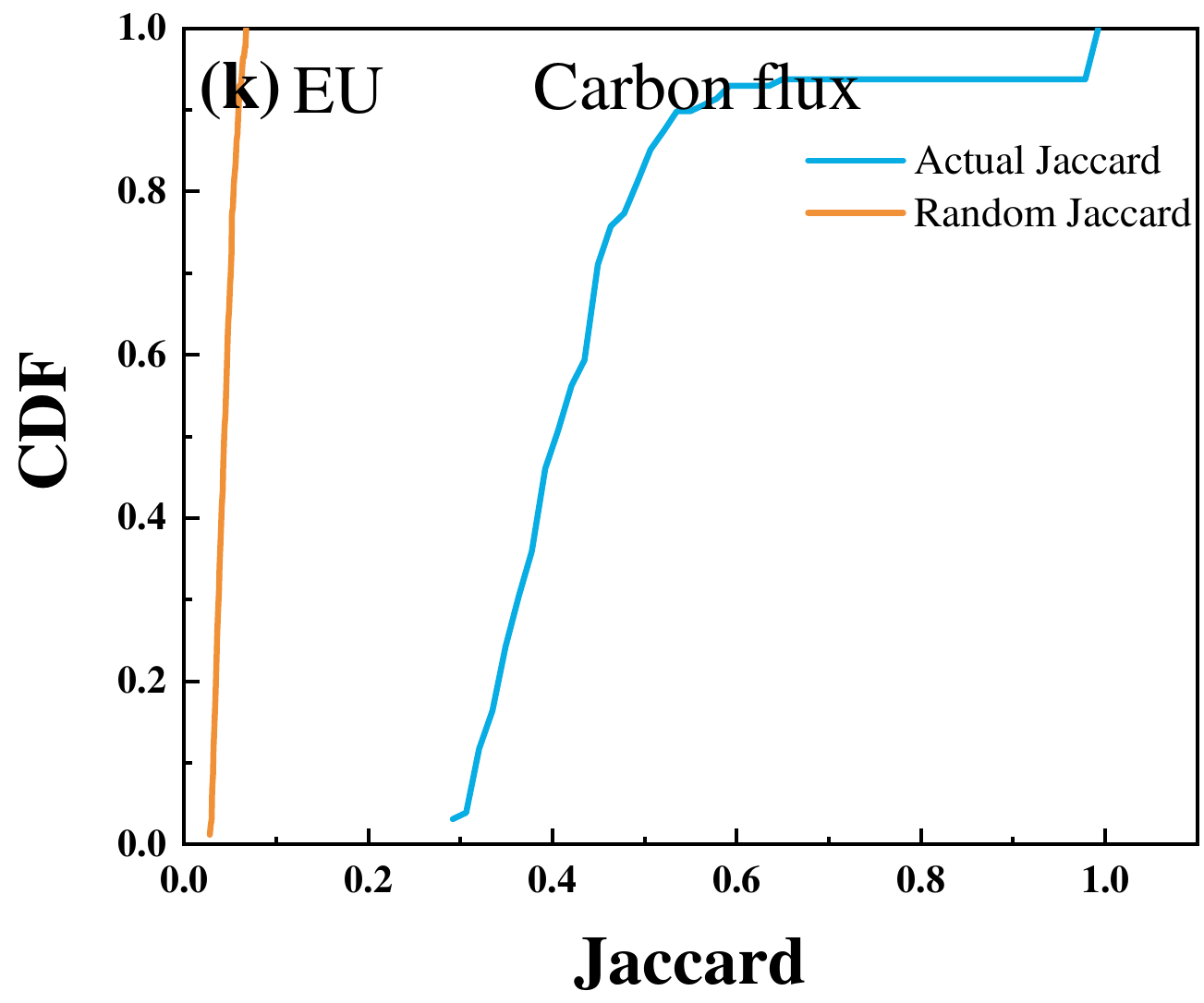}
\includegraphics[width=8em, height=7em]{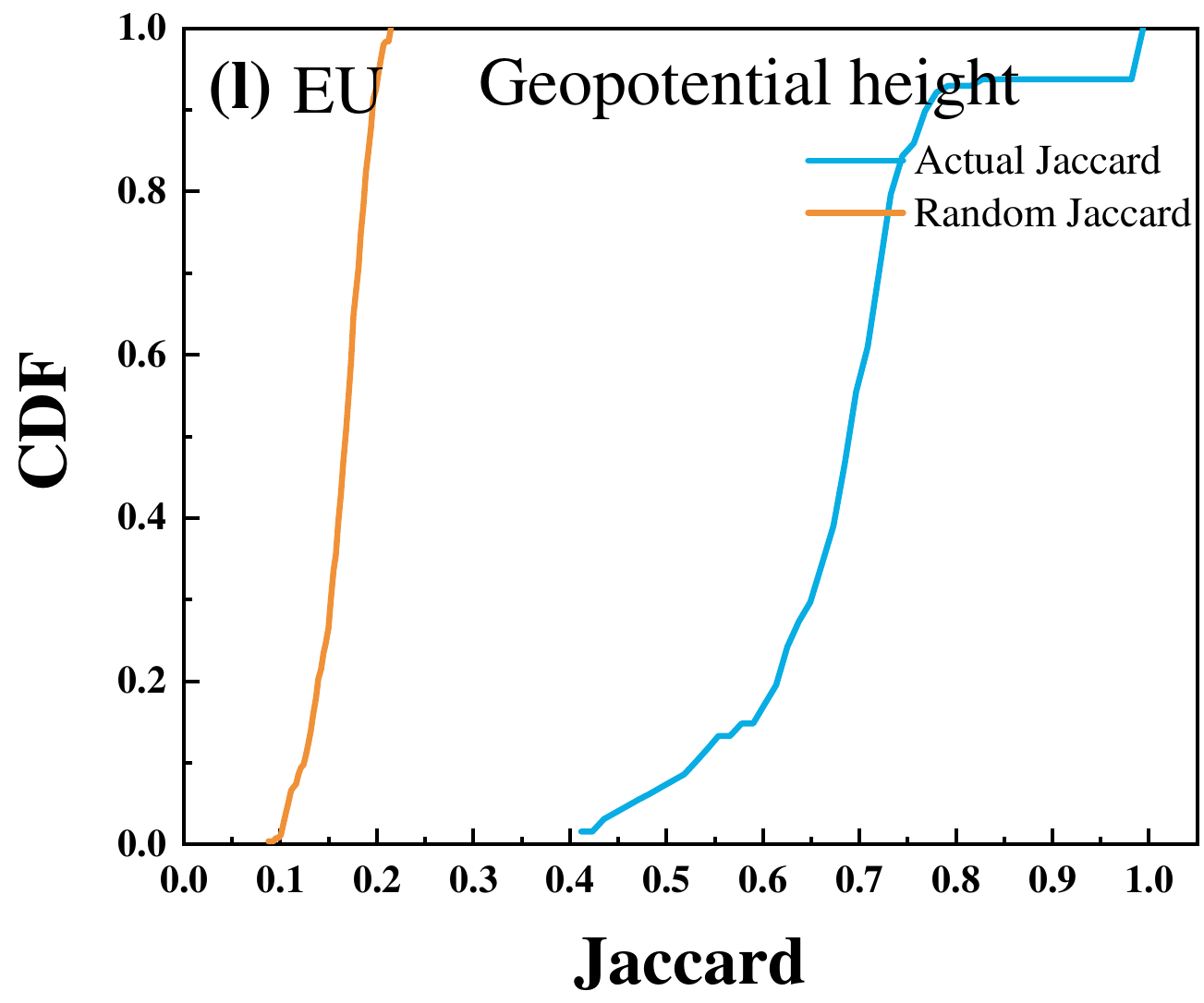}
\includegraphics[width=8em, height=7em]{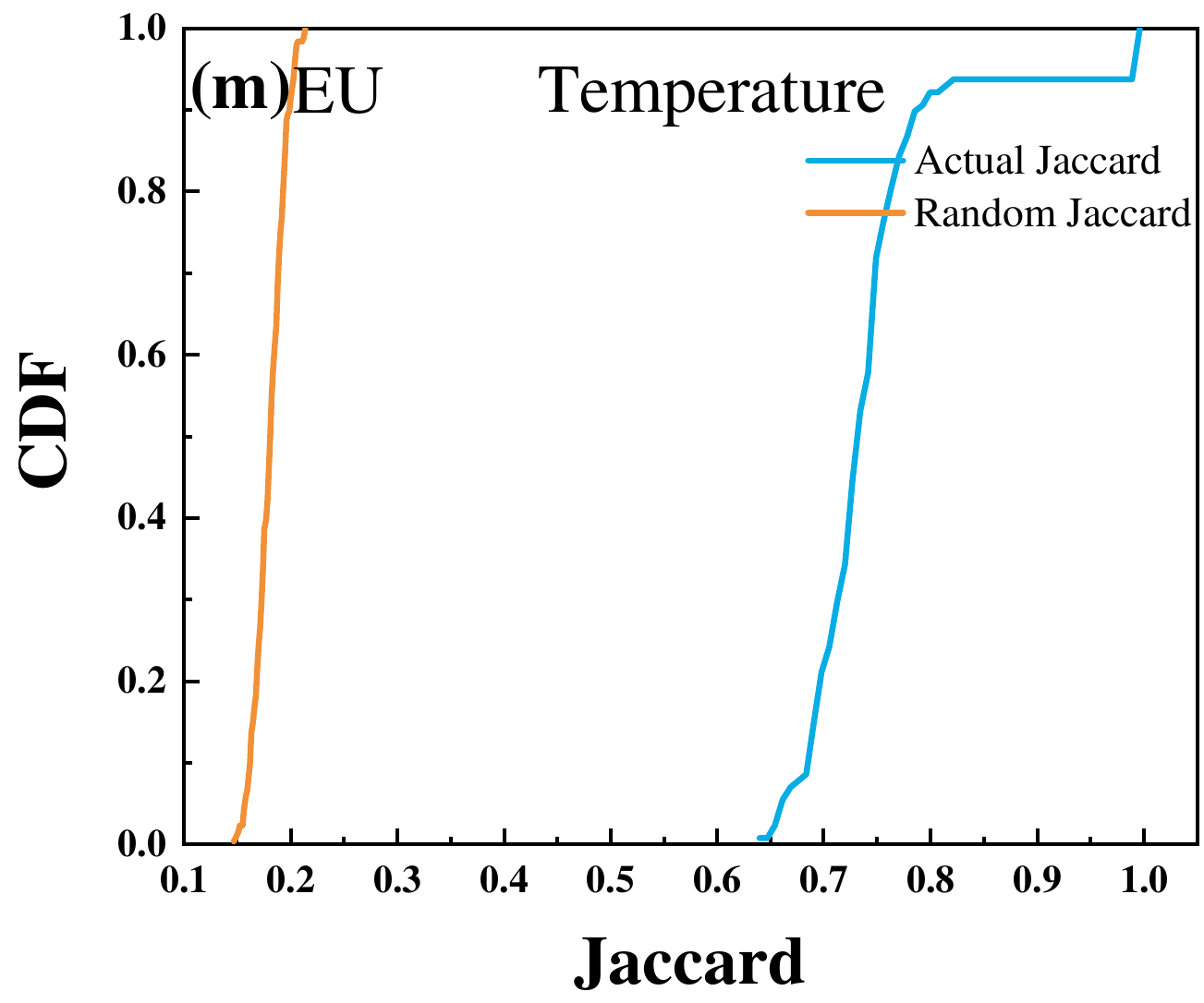}
\includegraphics[width=8em, height=7em]{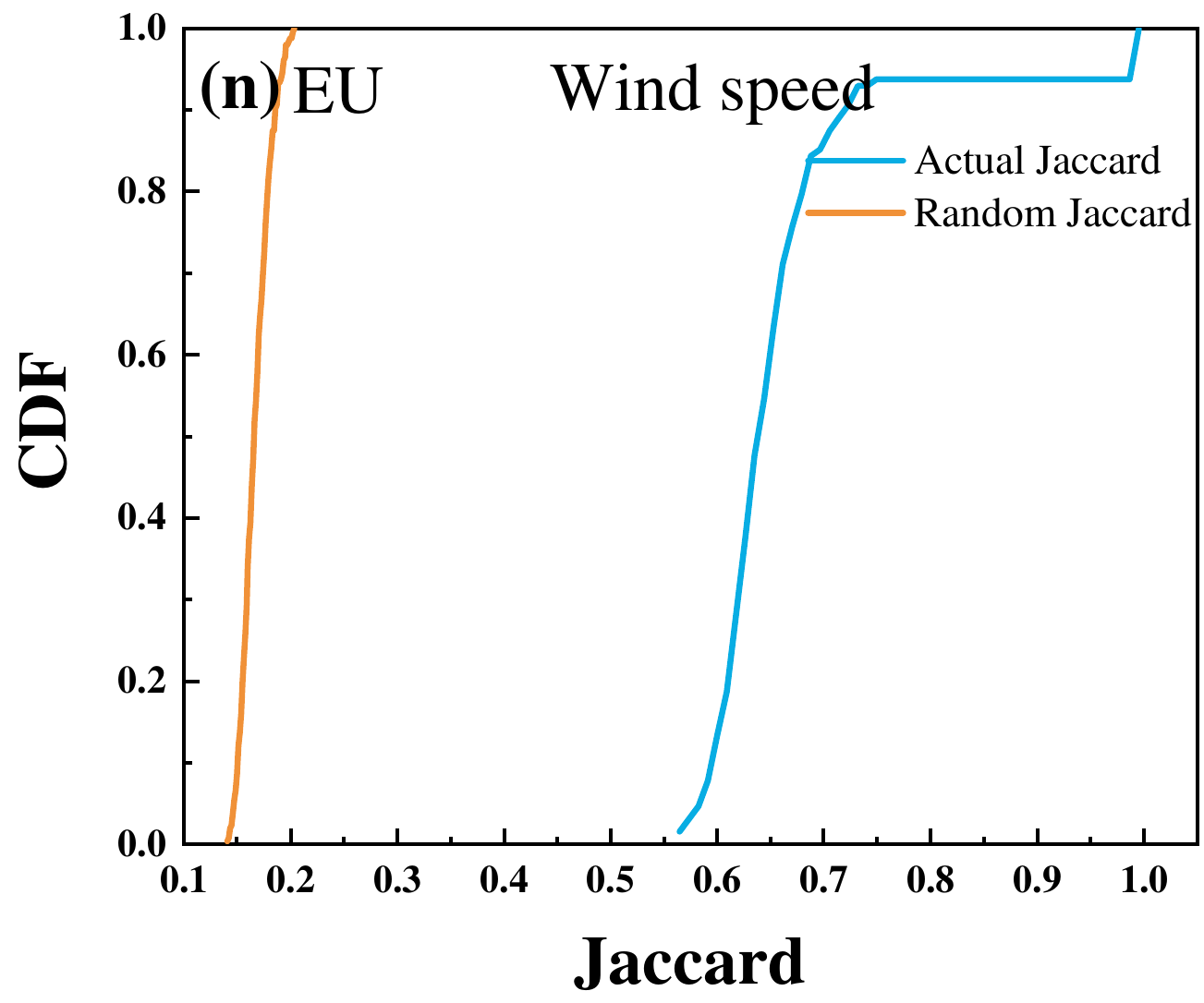}
\includegraphics[width=8em, height=7em]{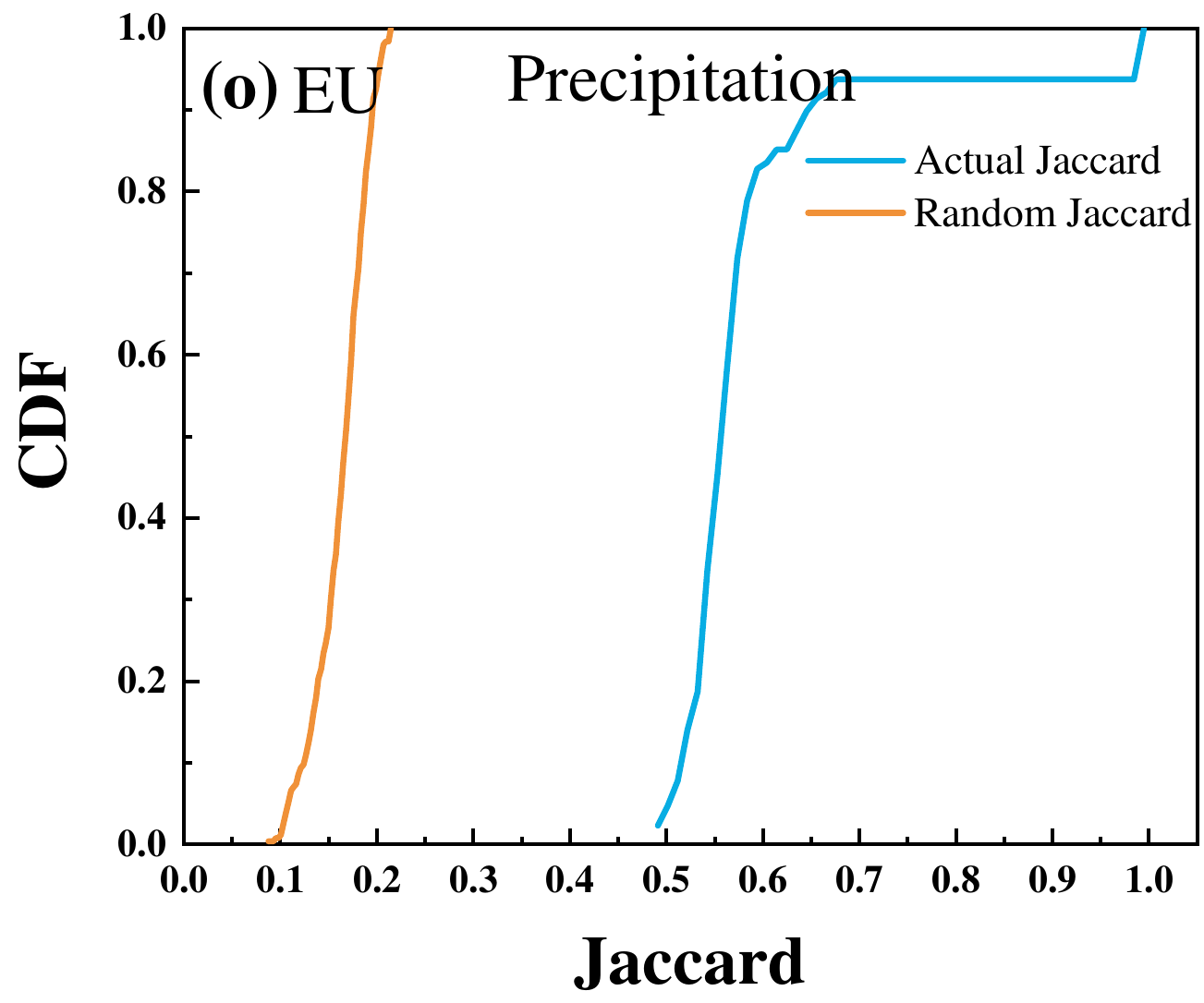}
\end{center}

\begin{center}
\noindent {\small {\bf Fig. S31} For lengths above 500$km$, the cumulative distribution function (CDF) of the actual Jaccard coefficient (blue line) and Jaccard coefficient in the controlled case (orange line) for different climate variables.}
\end{center}

\begin{center}
\includegraphics[width=8em, height=7em]{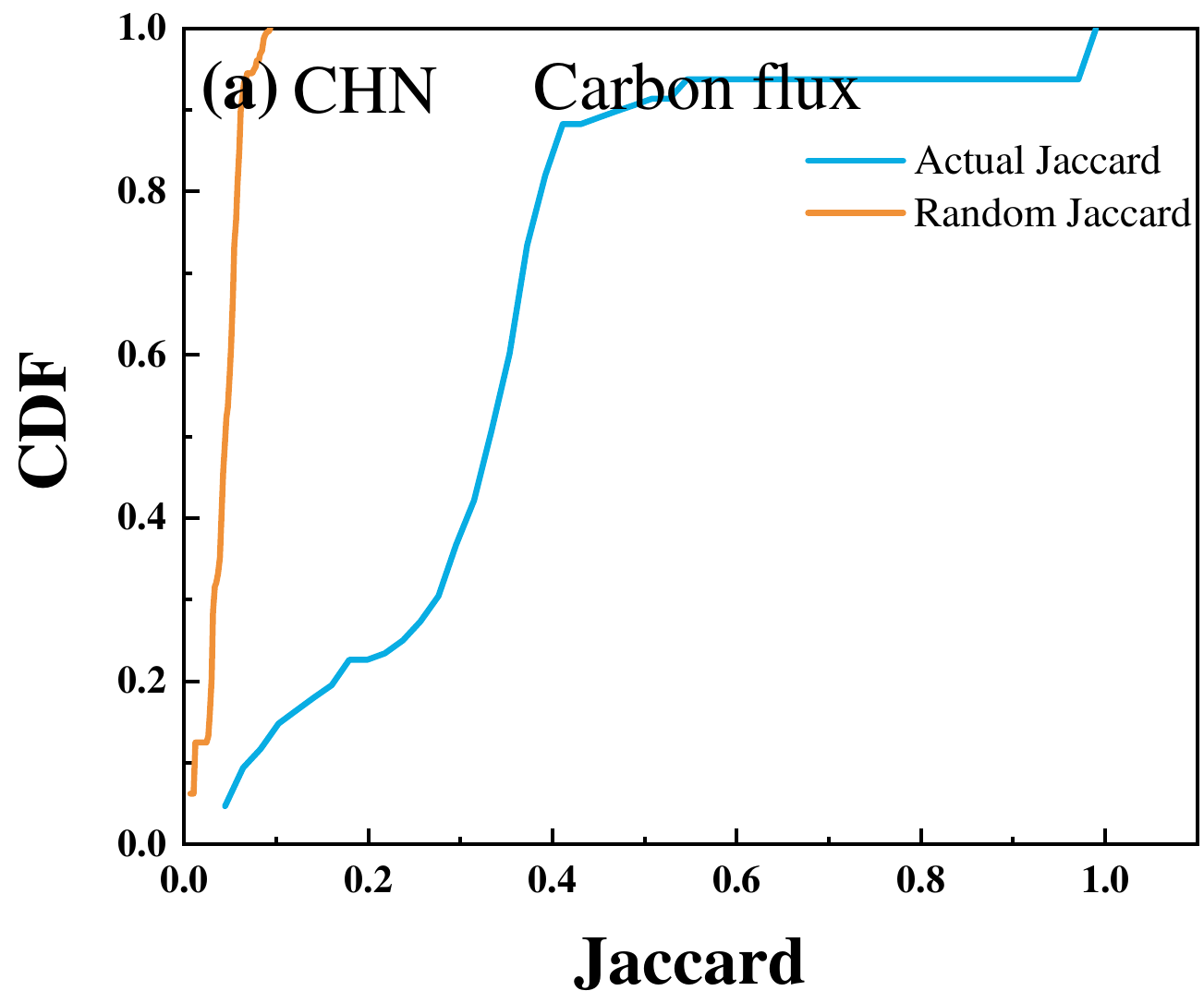}
\includegraphics[width=8em, height=7em]{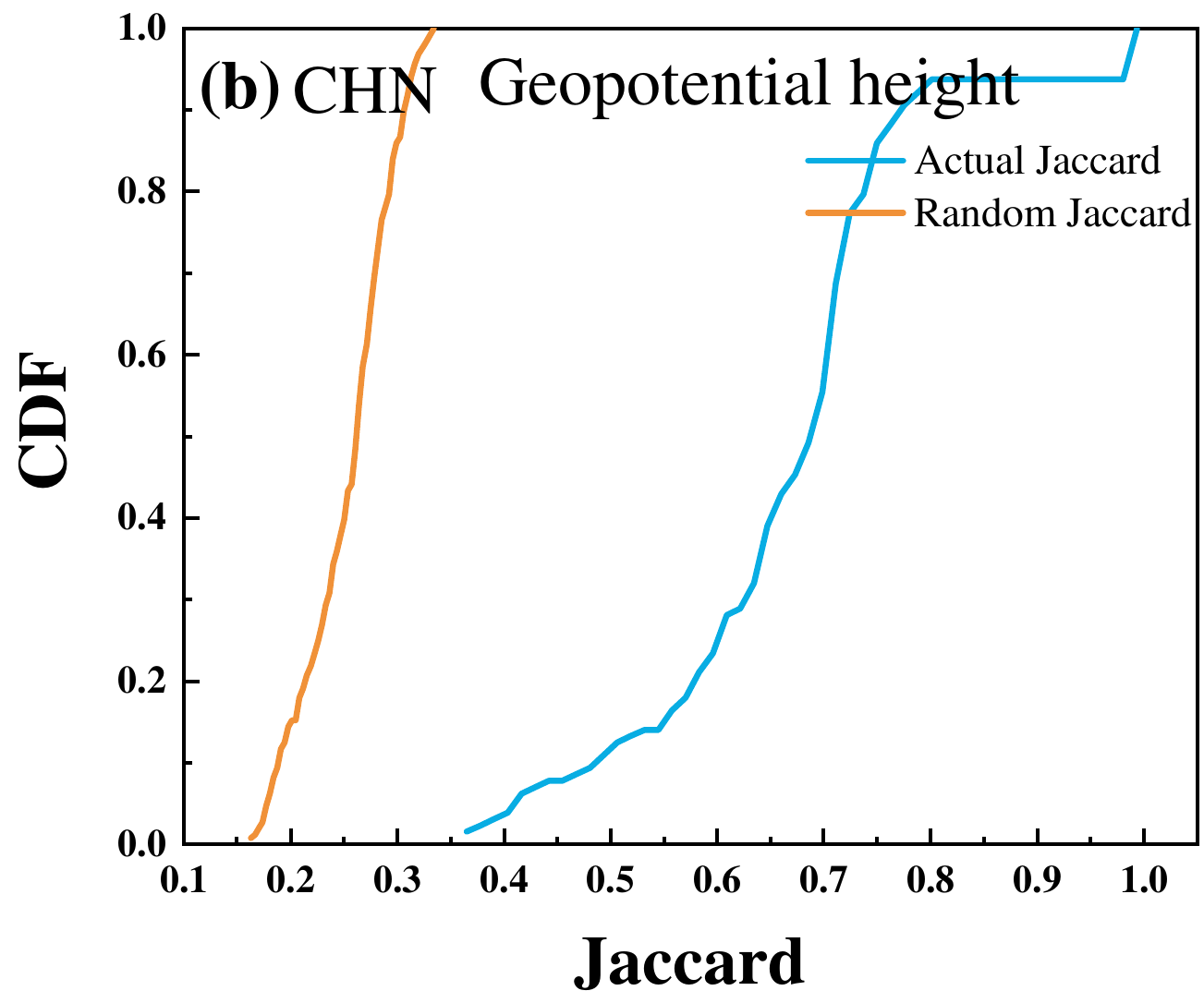}
\includegraphics[width=8em, height=7em]{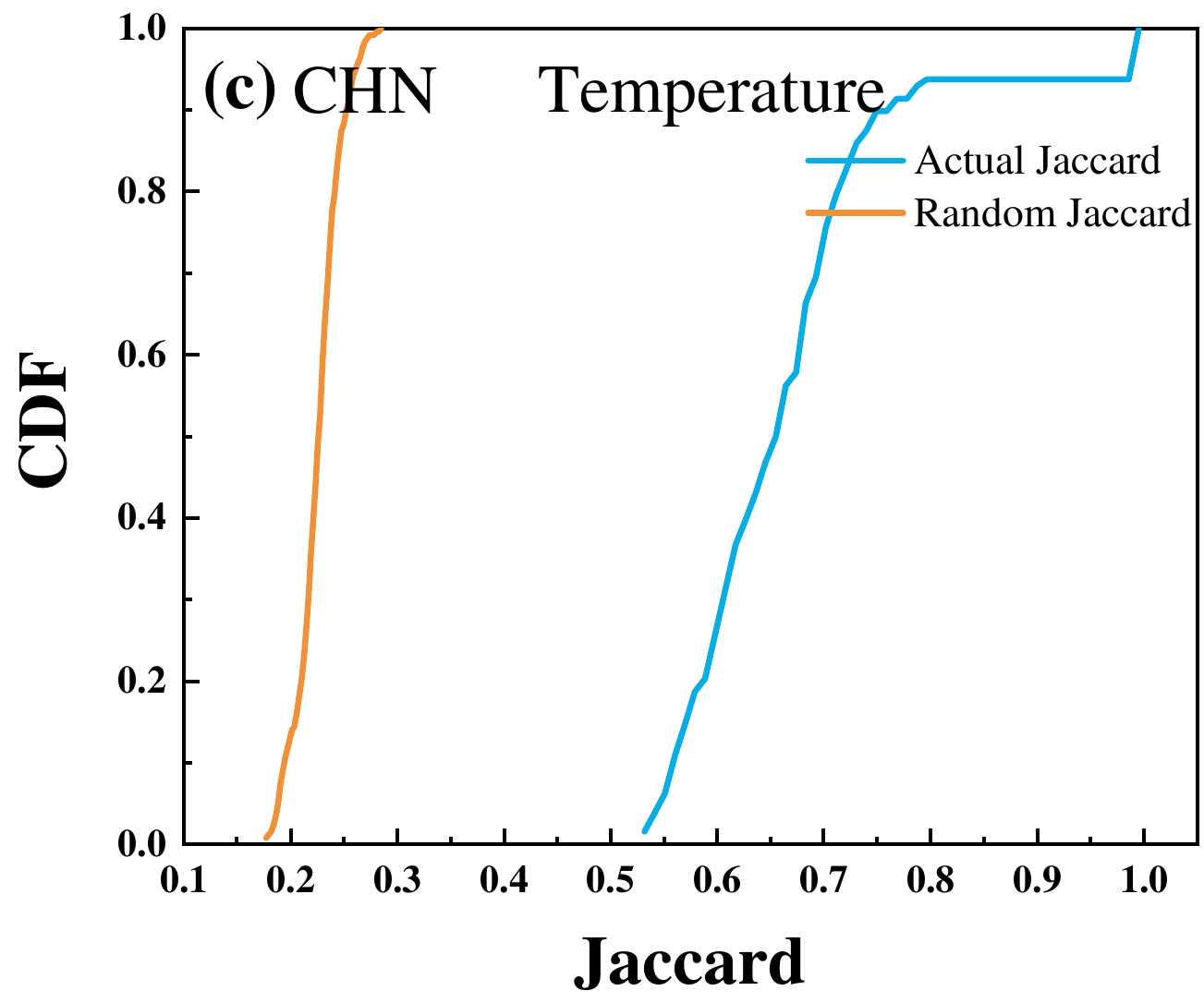}
\includegraphics[width=8em, height=7em]{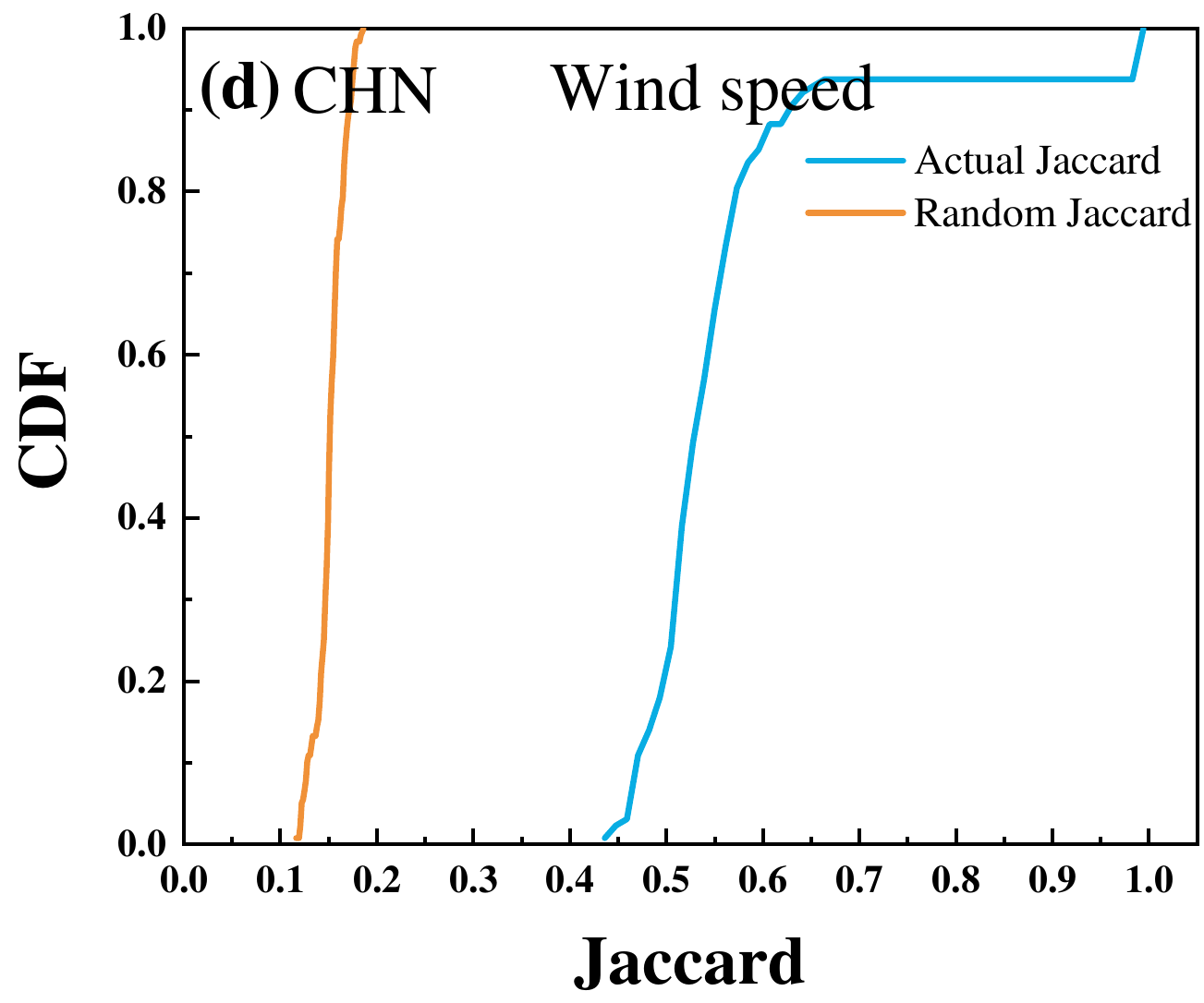}
\includegraphics[width=8em, height=7em]{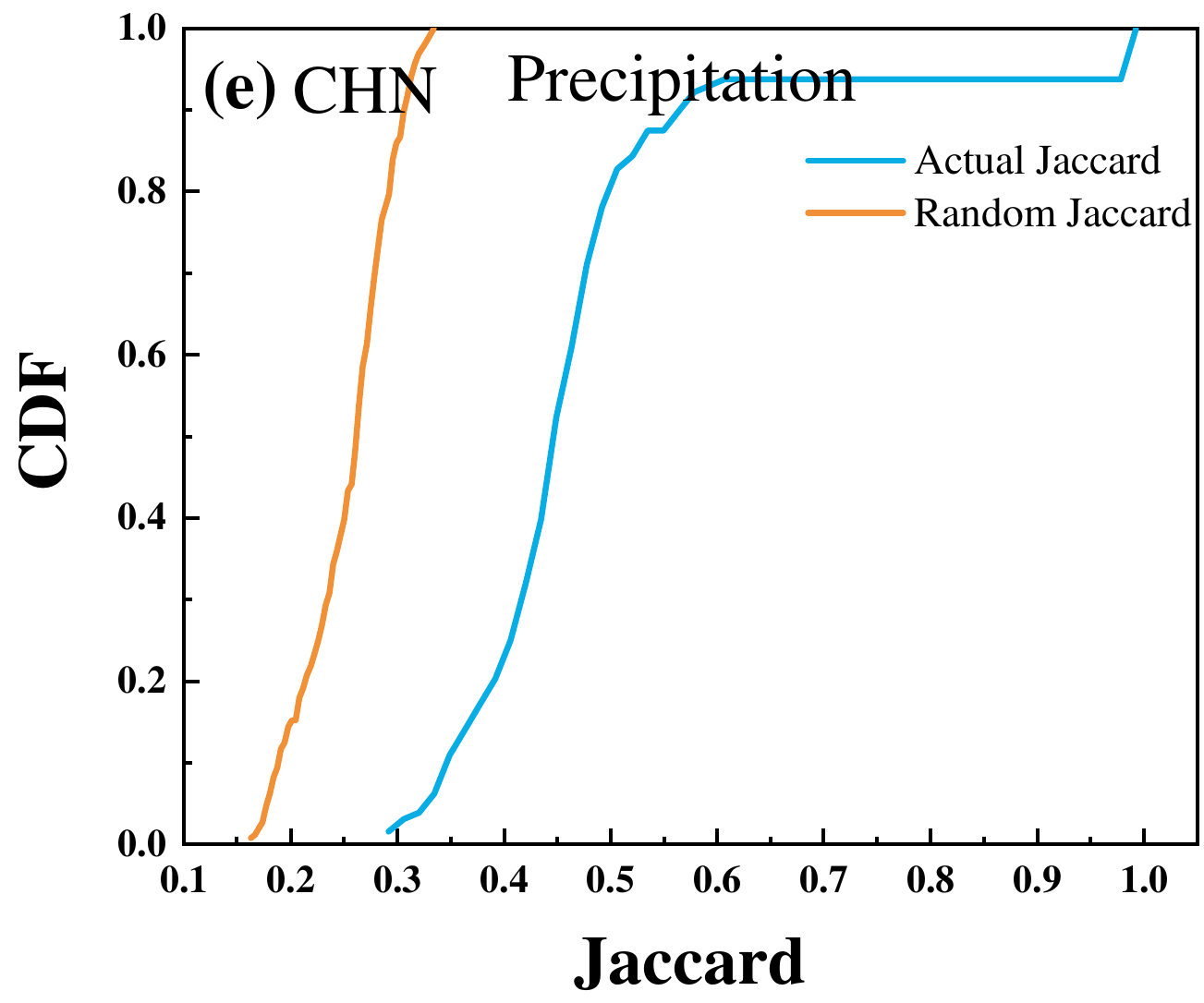}
\includegraphics[width=8em, height=7em]{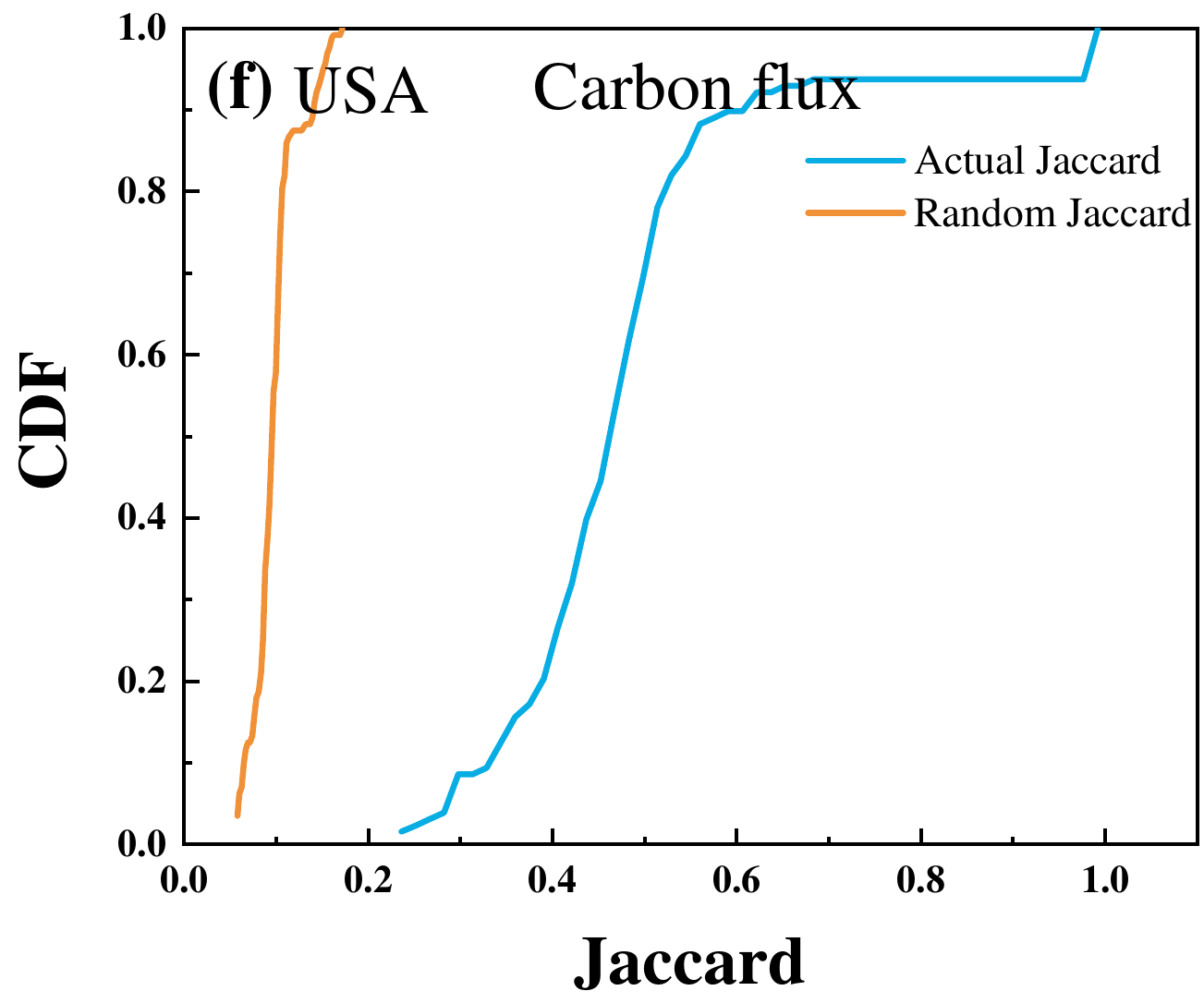}
\includegraphics[width=8em, height=7em]{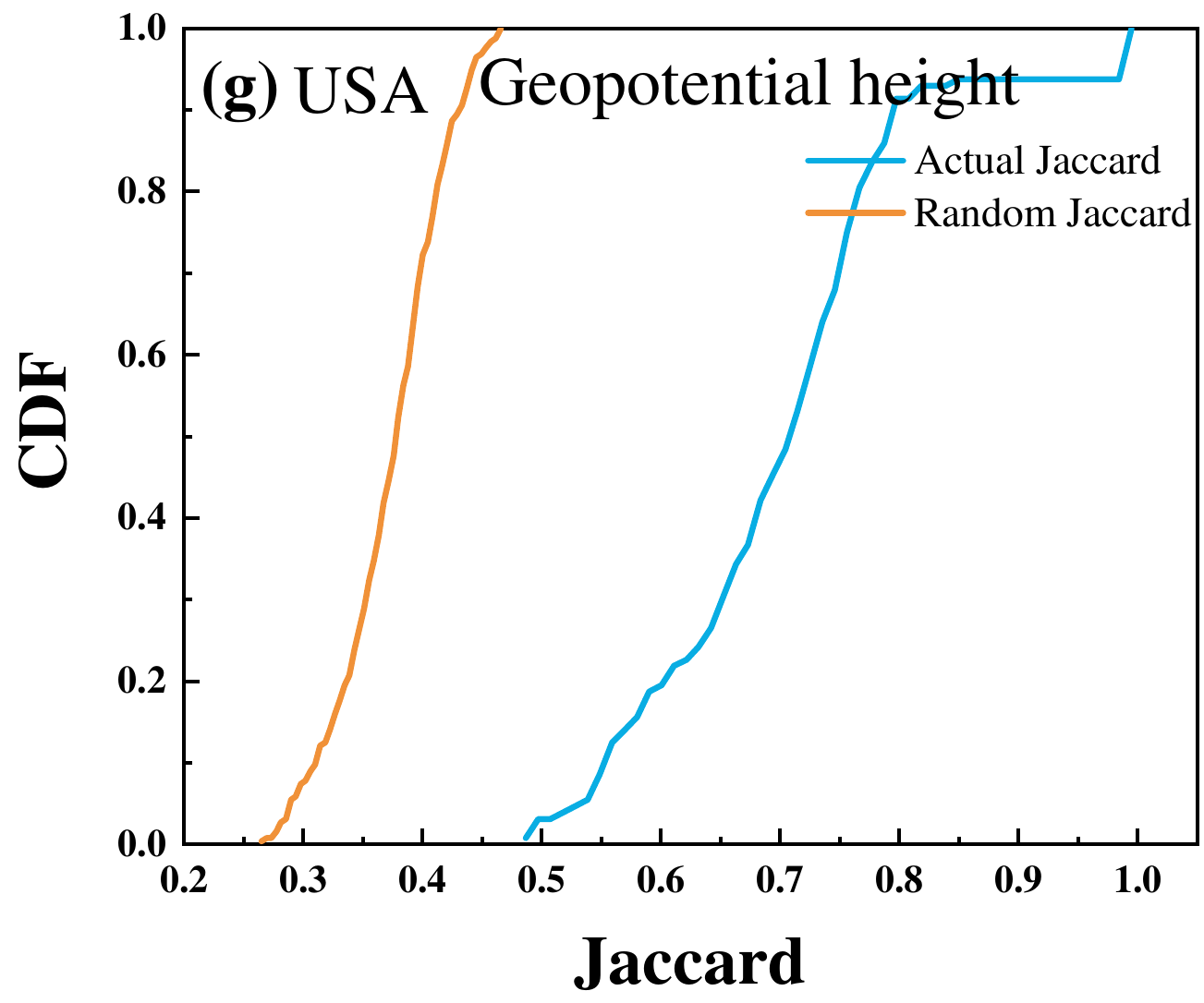}
\includegraphics[width=8em, height=7em]{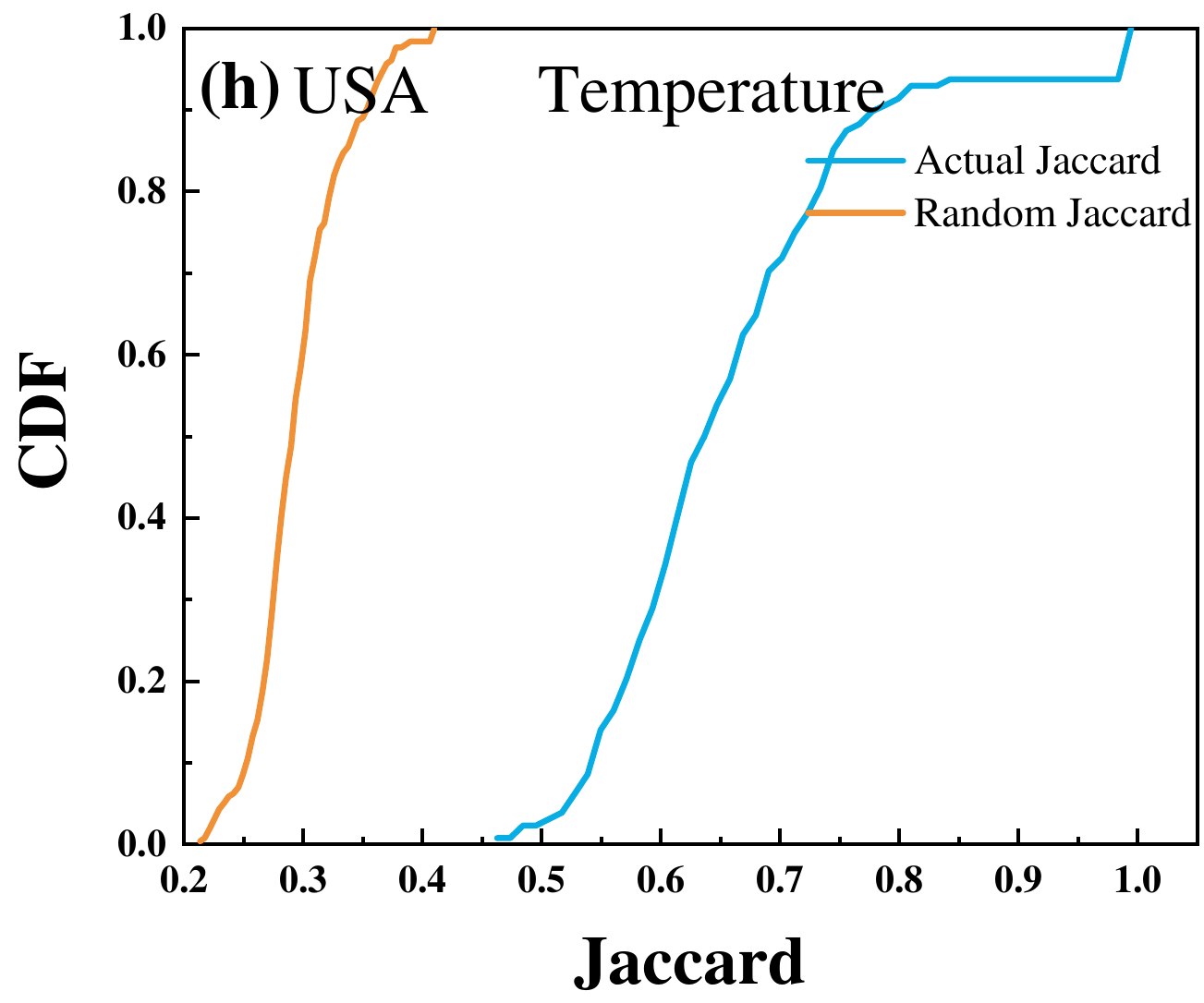}
\includegraphics[width=8em, height=7em]{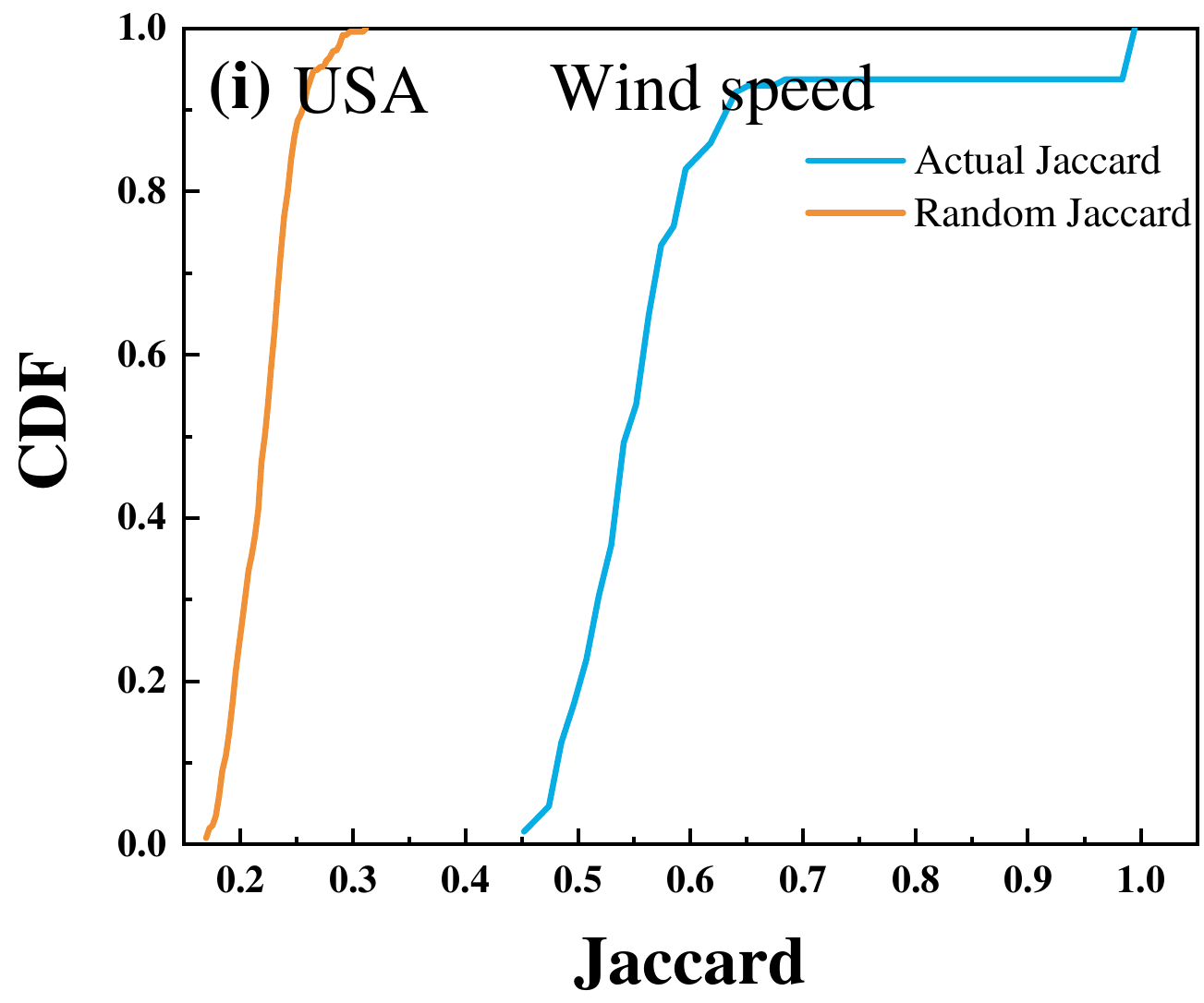}
\includegraphics[width=8em, height=7em]{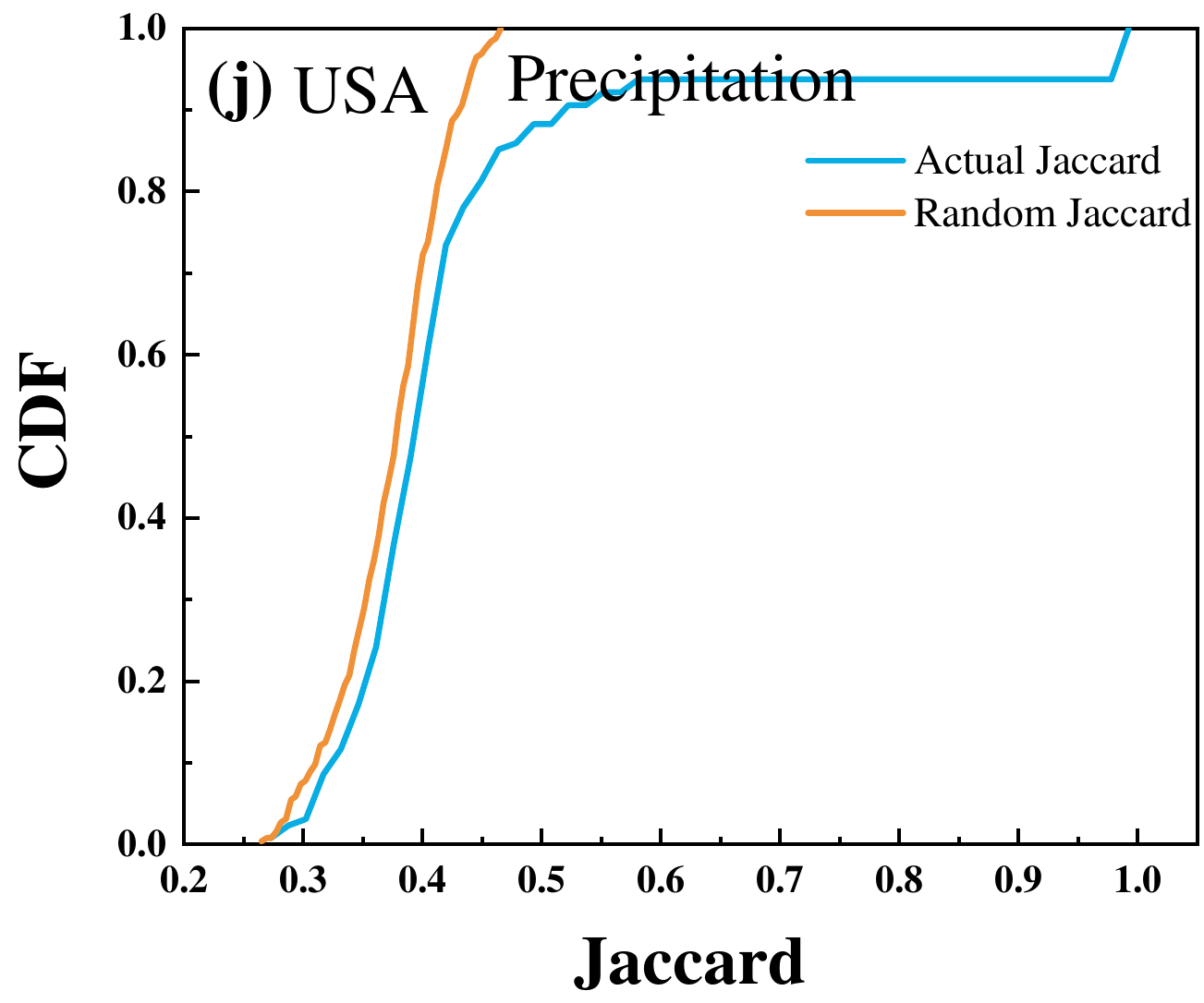}
\includegraphics[width=8em, height=7em]{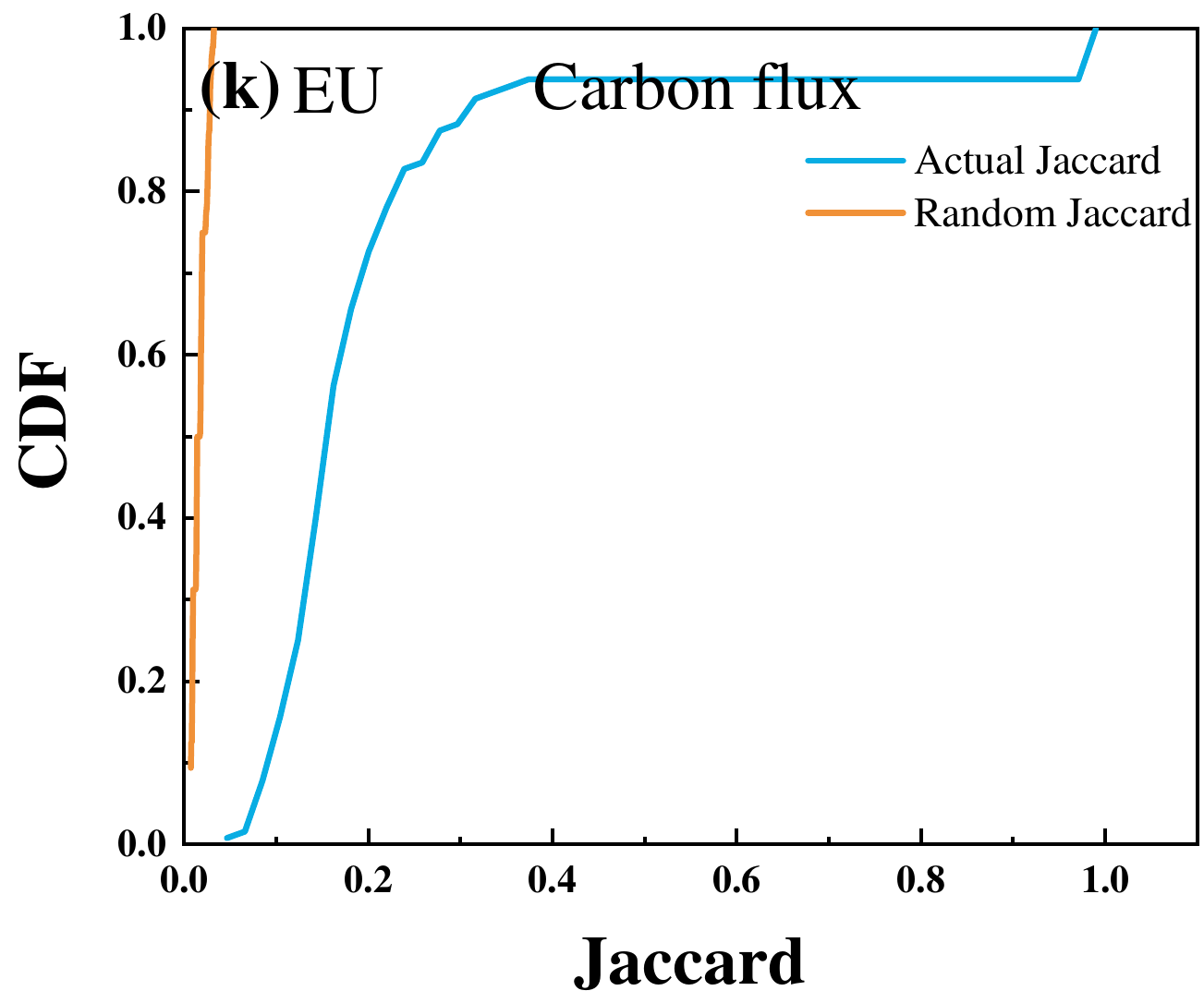}
\includegraphics[width=8em, height=7em]{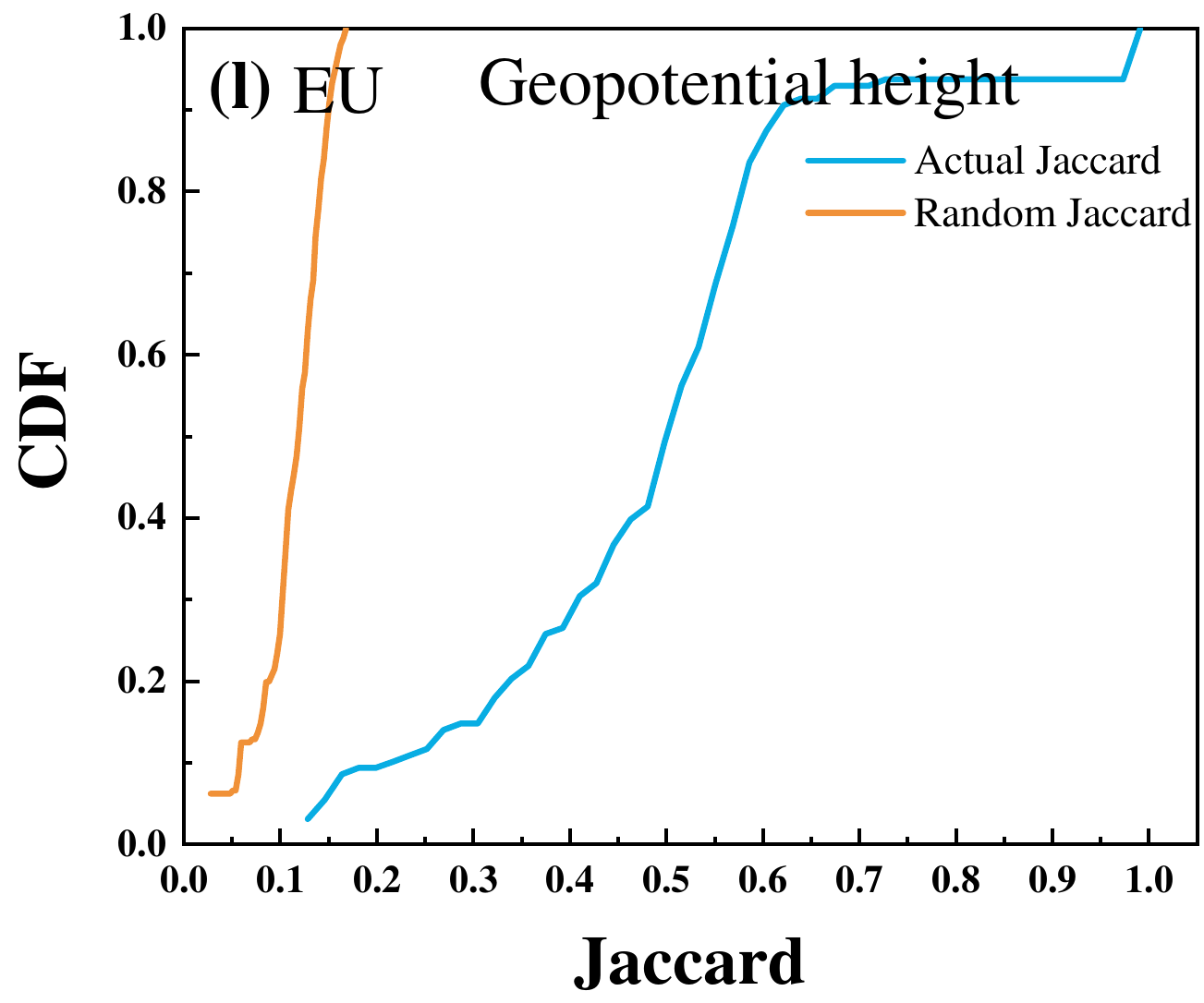}
\includegraphics[width=8em, height=7em]{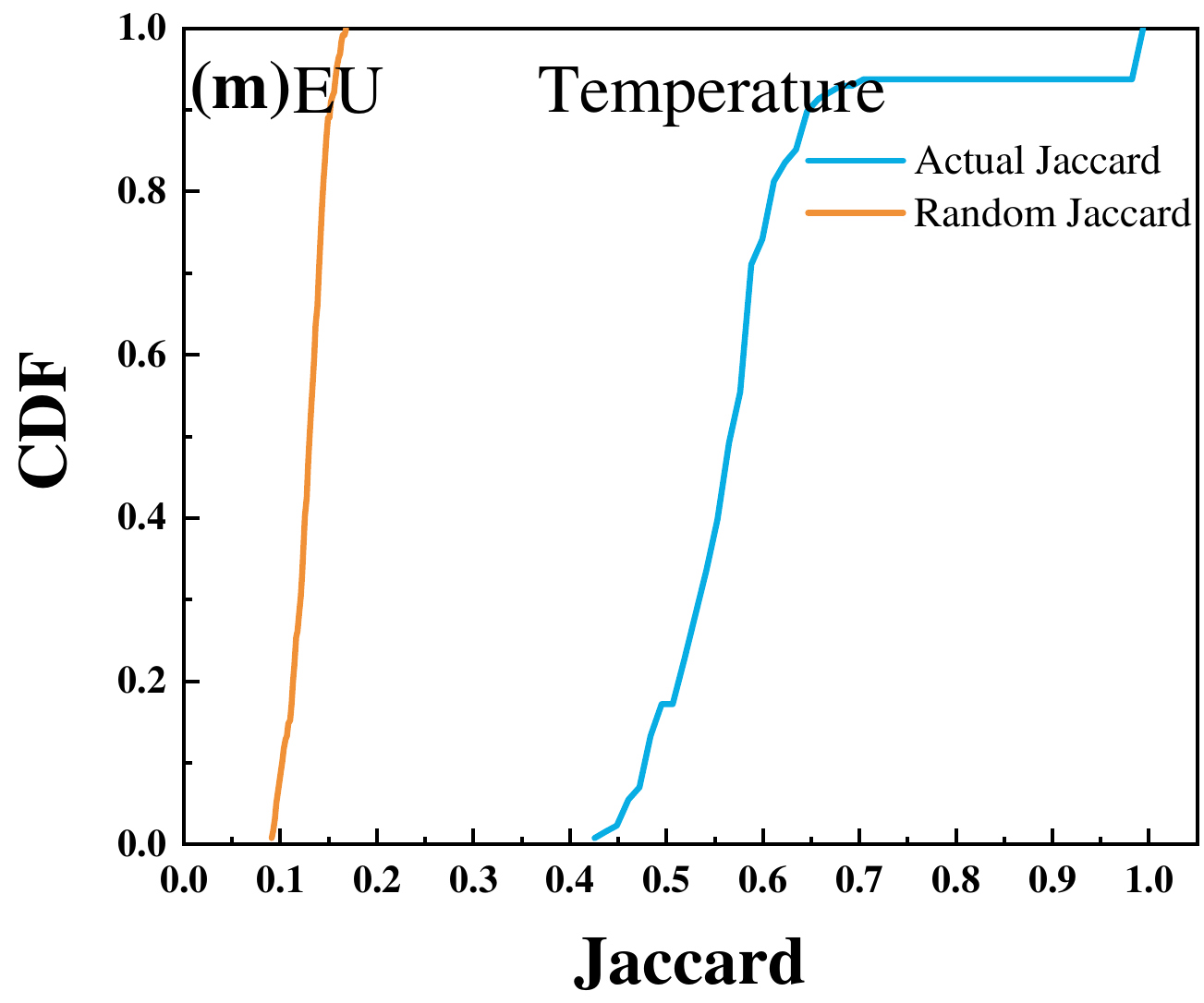}
\includegraphics[width=8em, height=7em]{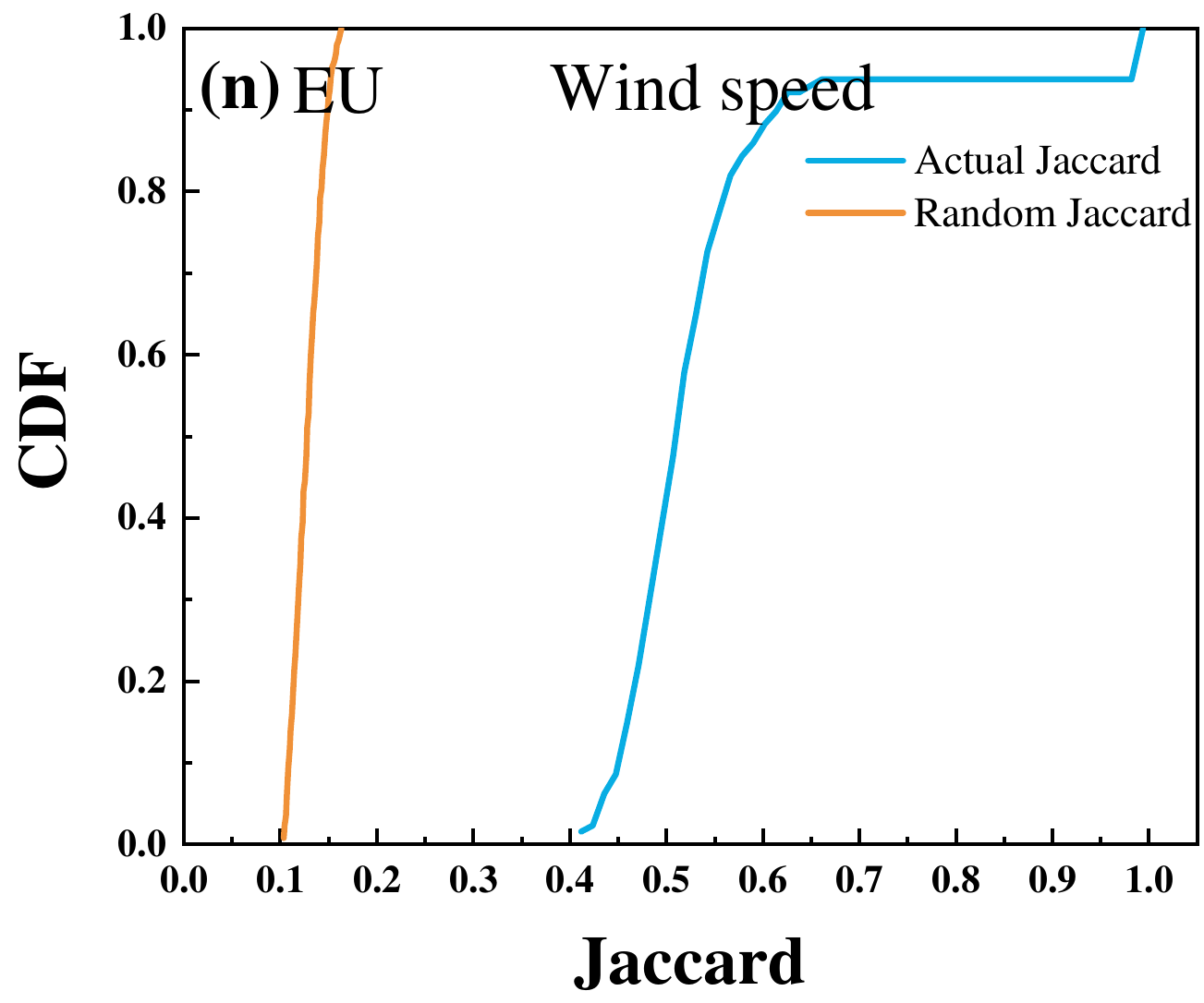}
\includegraphics[width=8em, height=7em]{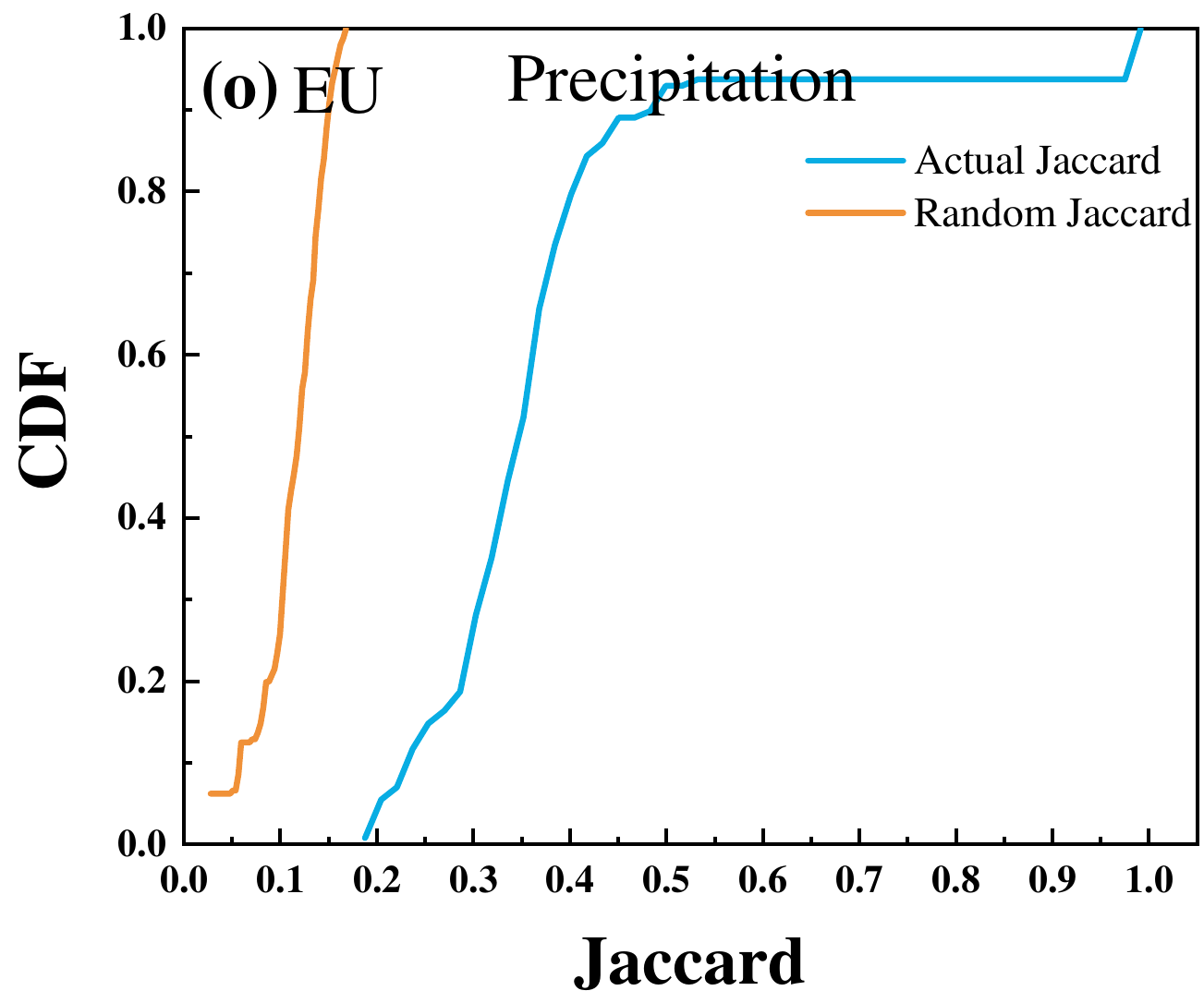}
\end{center}

\begin{center}
\noindent {\small {\bf Fig. S32} For lengths above 1000$km$, the cumulative distribution function (CDF) of the actual Jaccard coefficient (blue line) and Jaccard coefficient in the controlled case (orange line) for different climate variables.}
\end{center}

\begin{table}
\centering
\caption{Average and standard deviation of actual and controlled Jaccard matrices for climate and carbon flux networks ($p<0.05$).}
    \begin{tabular}{|c|c|c|c|c|c|c|c|}
    
    \hline
\multicolumn{3}{|c|}{} & Carbon flux & Geopotential height & Temperature &  Wind speed  & Precipitation \\
\hline
\multirow{6}[12]{*}{CHN} & \multirow{2}[4]{*}{dis$>0$} & Actual & 0.62$\pm$0.07 & 0.79$\pm$0.06 & 0.77$\pm$0.04 & 0.71$\pm$0.03 & 0.73$\pm$0.03 \\
\cline{3-8}          &       & Random & 0.10$\pm$0.02 & 0.33$\pm$0.03 & 0.29$\pm$0.02 & 0.21$\pm$0.01 & 0.12$\pm$0.01 \\
    
\cline{2-8}          & \multirow{2}[4]{*}{dis$>500km$} & Actual & 0.50$\pm$0.1 & 0.76$\pm$0.07 & 0.74$\pm$0.05 & 0.65$\pm$0.03 & 0.63$\pm$0.04 \\

\cline{3-8}          &       & Random & 0.08$\pm$0.02 & 0.31$\pm$0.04 & 0.27$\pm$0.02 & 0.19$\pm$0.01 & 0.10$\pm$0.01 \\
\cline{2-8}          & \multirow{2}[4]{*}{dis$>1000km$} & Actual & 0.29$\pm$0.13 & 0.65$\pm$0.11 & 0.65$\pm$0.06 & 0.54$\pm$0.05 & 0.45$\pm$0.07 \\
\cline{3-8}          &       & Random & 0.04$\pm$0.02 & 0.26$\pm$0.04 & 0.23$\pm$0.02 & 0.15$\pm$0.01 & 0.06$\pm$0.01 \\
    \hline
    \multirow{6}[12]{*}{USA} & \multirow{2}[4]{*}{dis$>0$} & Actual & 0.73$\pm$0.05 & 0.81$\pm$0.05 & 0.78$\pm$0.05 & 0.73$\pm$0.03 & 0.70$\pm$0.04 \\
\cline{3-8}          &       & Random & 0.20$\pm$0.02 & 0.46$\pm$0.04 & 0.38$\pm$0.03 & 0.30$\pm$0.02 & 0.16$\pm$0.01 \\
\cline{2-8}          & \multirow{2}*{dis$>500km$} & Actual & 0.66$\pm$0.06 & 0.79$\pm$0.06 & 0.75$\pm$0.06 & 0.68$\pm$0.04 & 0.60$\pm$0.04 \\
\cline{3-8}          &       & Random & 0.18$\pm$0.02 & 0.44$\pm$0.04 & 0.35$\pm$0.03 & 0.28$\pm$0.02 & 0.13$\pm$0.01 \\
\cline{2-8}          & \multirow{2}*{dis$>1000km$} & Actual & 0.45$\pm$0.09 & 0.69$\pm$0.09 & 0.64$\pm$0.08 & 0.55$\pm$0.05 & 0.40$\pm$0.06 \\
\cline{3-8}          &       & Random & 0.10$\pm$0.02 & 0.38$\pm$0.04 & 0.30$\pm$0.04 & 0.22$\pm$0.03 & 0.07$\pm$0.01 \\
    \hline
    \multirow{6}[12]{*}{EU} & \multirow{2}[4]{*}{dis$>0$} & Actual & 0.59$\pm$0.06 & 0.72$\pm$0.07 & 0.77$\pm$0.03 & 0.70$\pm$0.03 & 0.72$\pm$0.03 \\
\cline{3-8}          &       & Random & 0.06$\pm$0.01 & 0.18$\pm$0.03 & 0.20$\pm$0.01 & 0.18$\pm$0.01 & 0.07$\pm$0.002 \\
\cline{2-8}          & \multirow{2}*{dis$>500km$} & Actual & 0.41$\pm$0.07 & 0.67$\pm$0.09 & 0.73$\pm$0.04 & 0.64$\pm$0.04 & 0.56$\pm$0.04 \\
\cline{3-8}          &       & Random & 0.05$\pm$0.01 & 0.16$\pm$0.03 & 0.18$\pm$0.01 & 0.17$\pm$0.01 & 0.05$\pm$0.002 \\
\cline{2-8}          & \multirow{2}*{dis$>1000km$} & Actual & 0.17$\pm$0.06 & 0.46$\pm$0.14 & 0.56$\pm$0.06 & 0.52$\pm$0.05 & 0.34$\pm$0.07 \\
\cline{3-8}          &       & Random & 0.02$\pm$0.01 & 0.11$\pm$0.04 & 0.13$\pm$0.02 & 0.13$\pm$0.01 & 0.02$\pm$0.002 \\
    \hline
  
	\end{tabular}
 \label{tab:1}%
\end{table}

\begin{center}
\includegraphics[width=8em, height=7em]{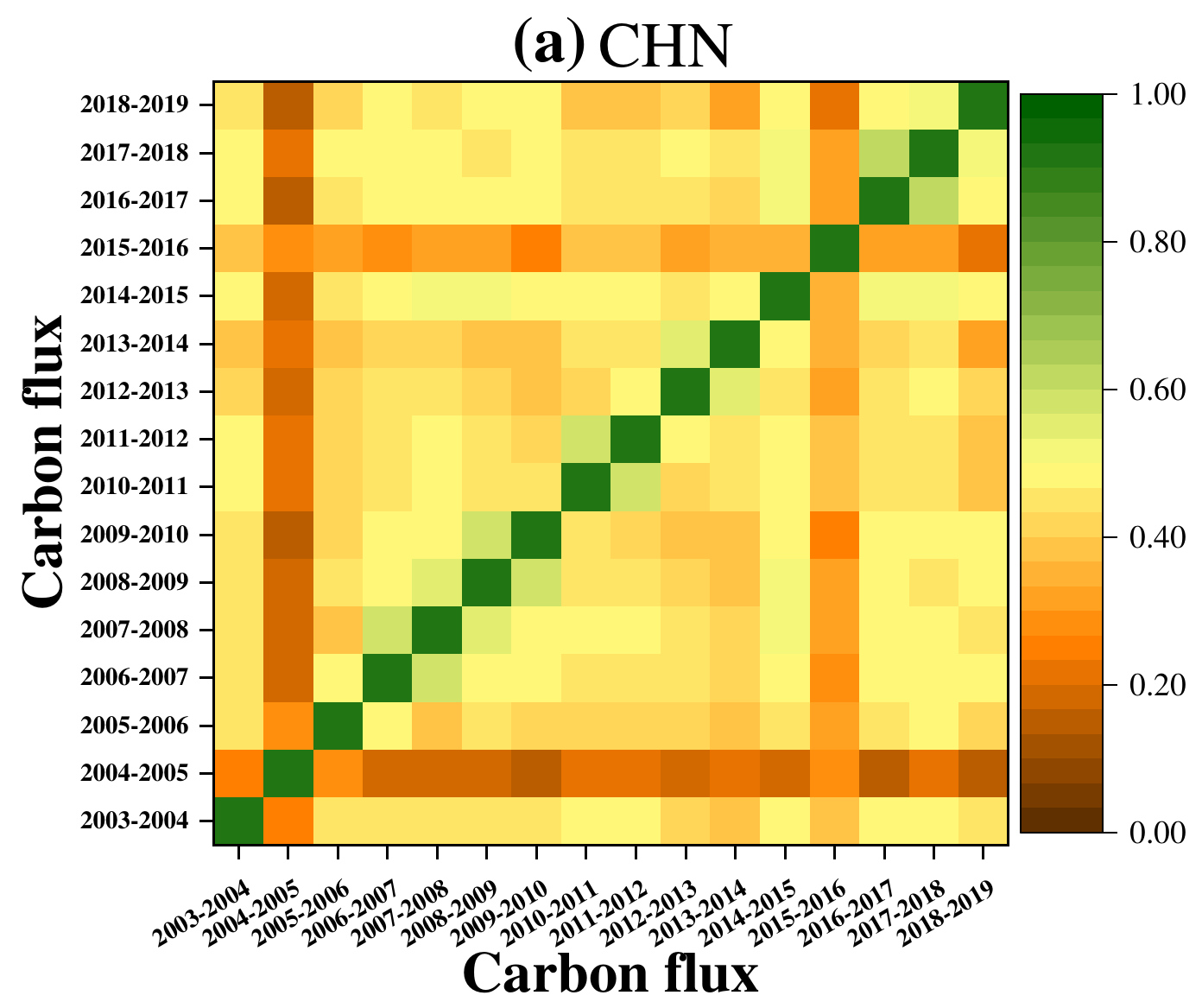}
\includegraphics[width=8em, height=7em]{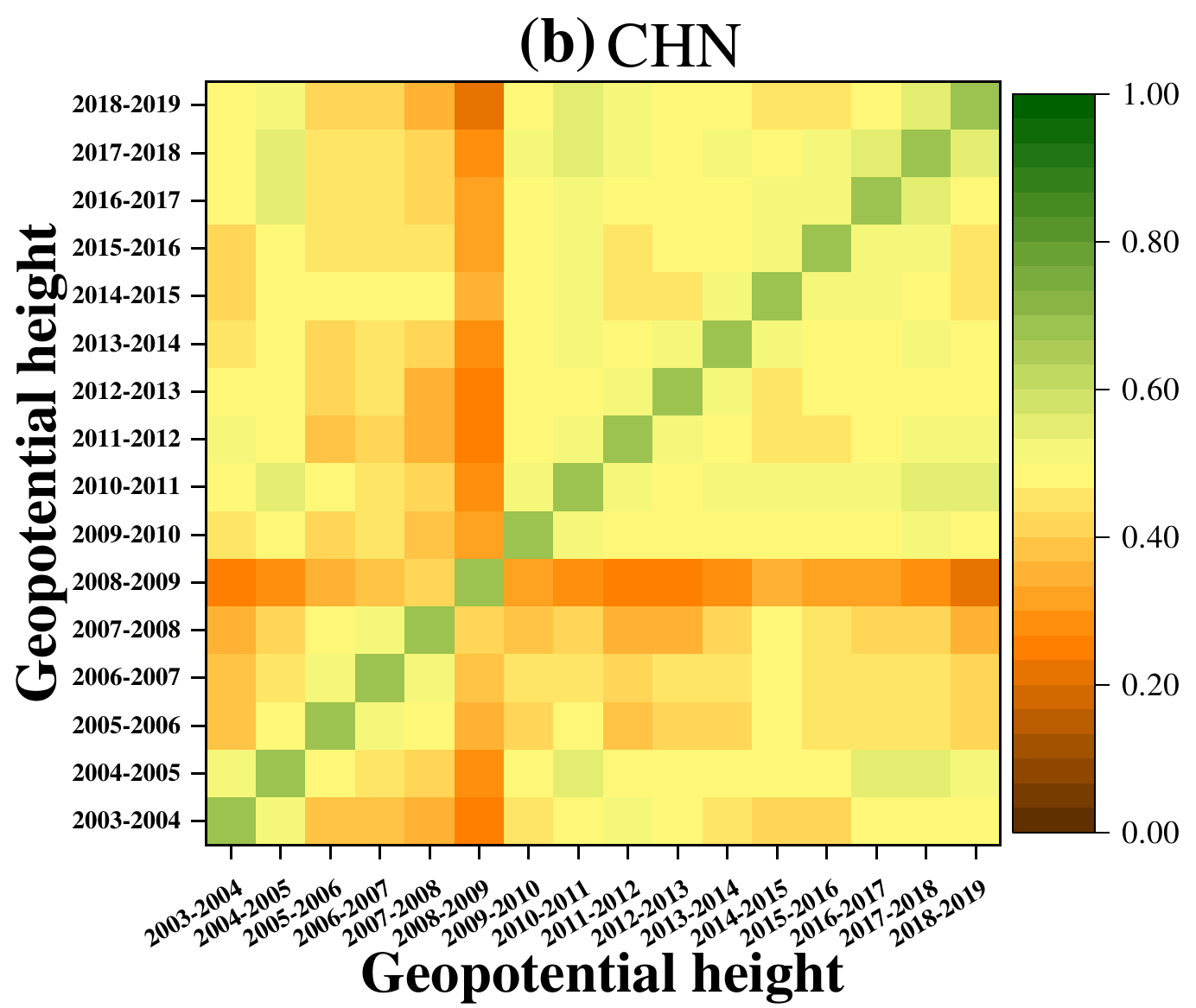}
\includegraphics[width=8em, height=7em]{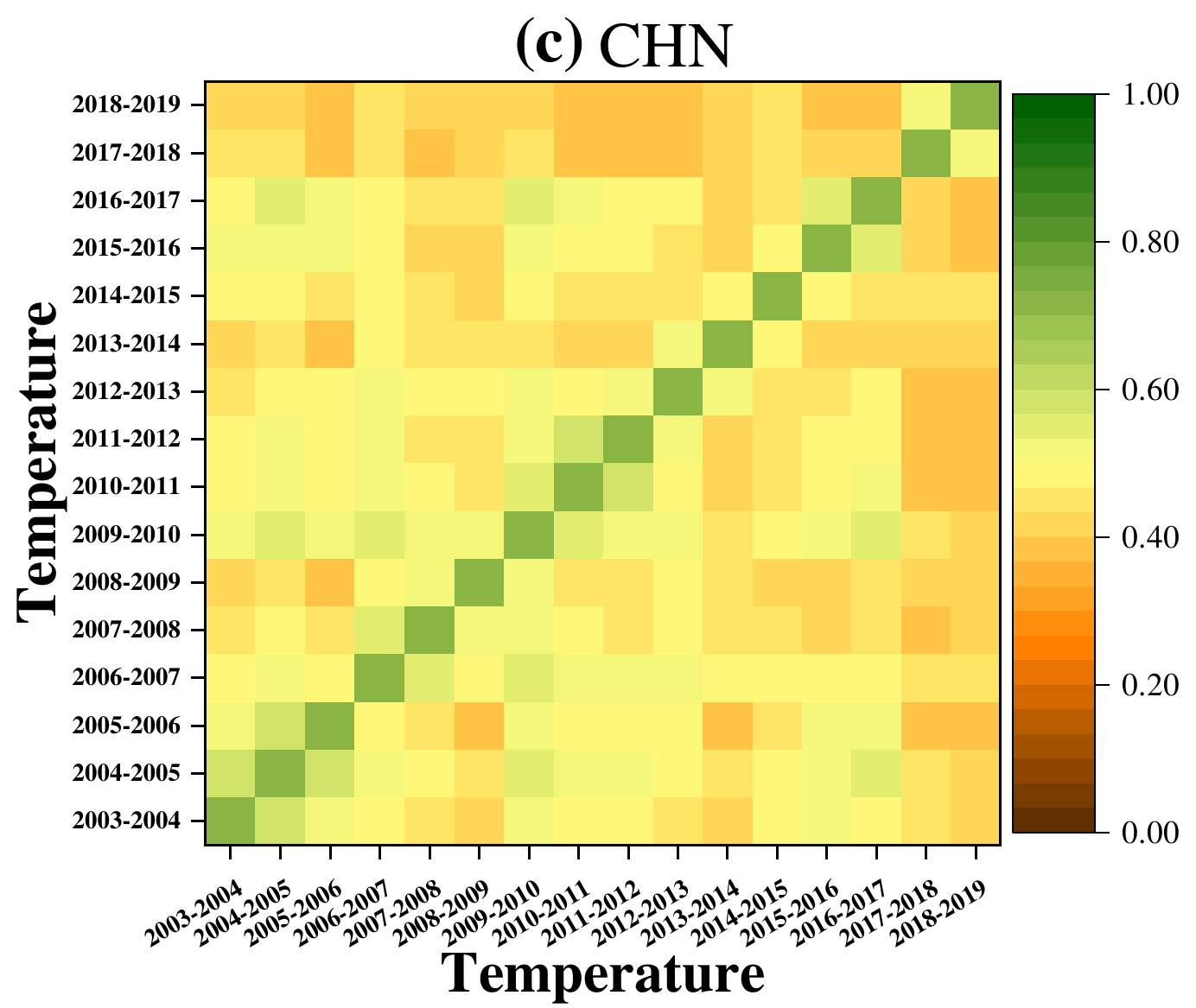}
\includegraphics[width=8em, height=7em]{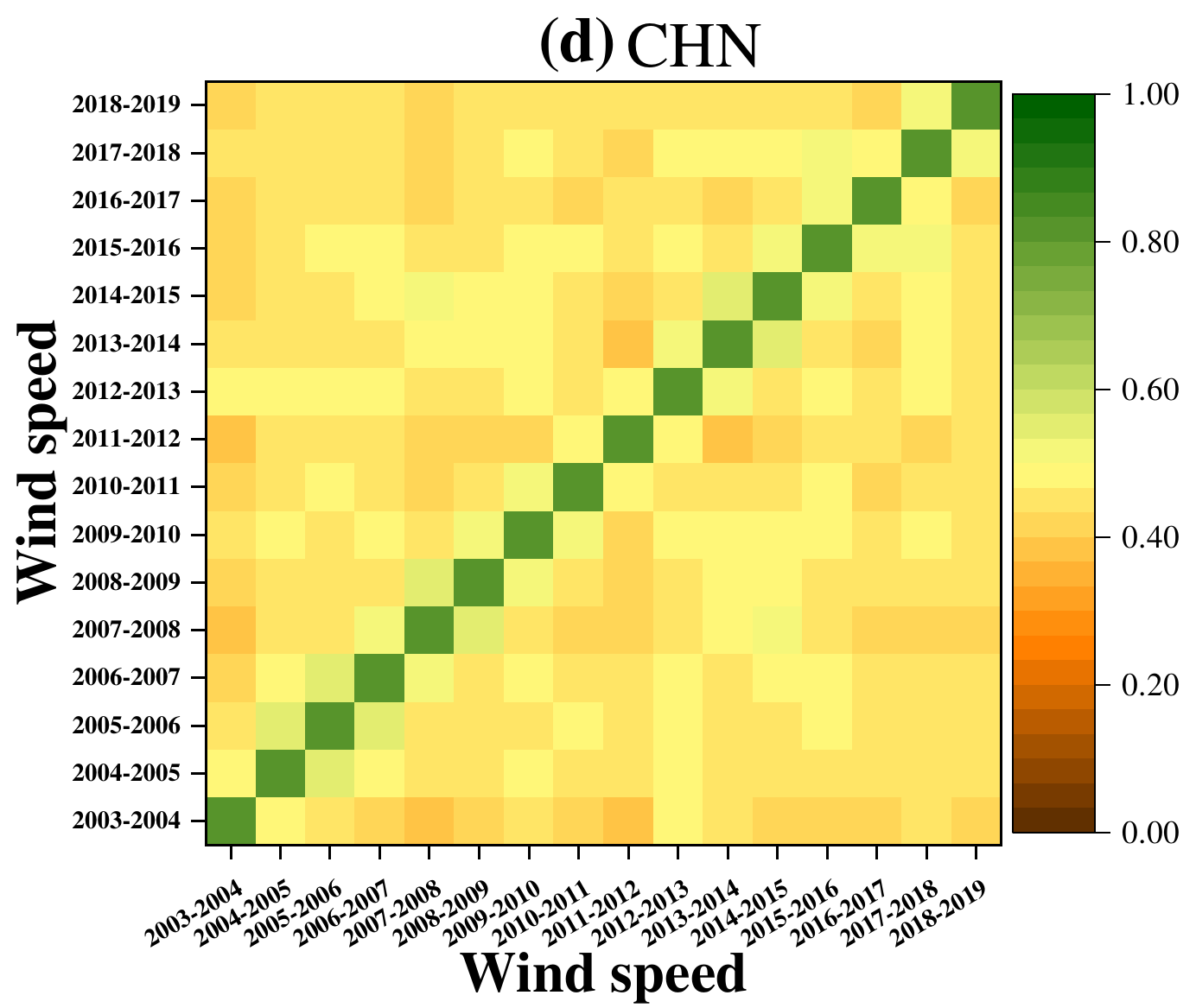}
\includegraphics[width=8em, height=7em]{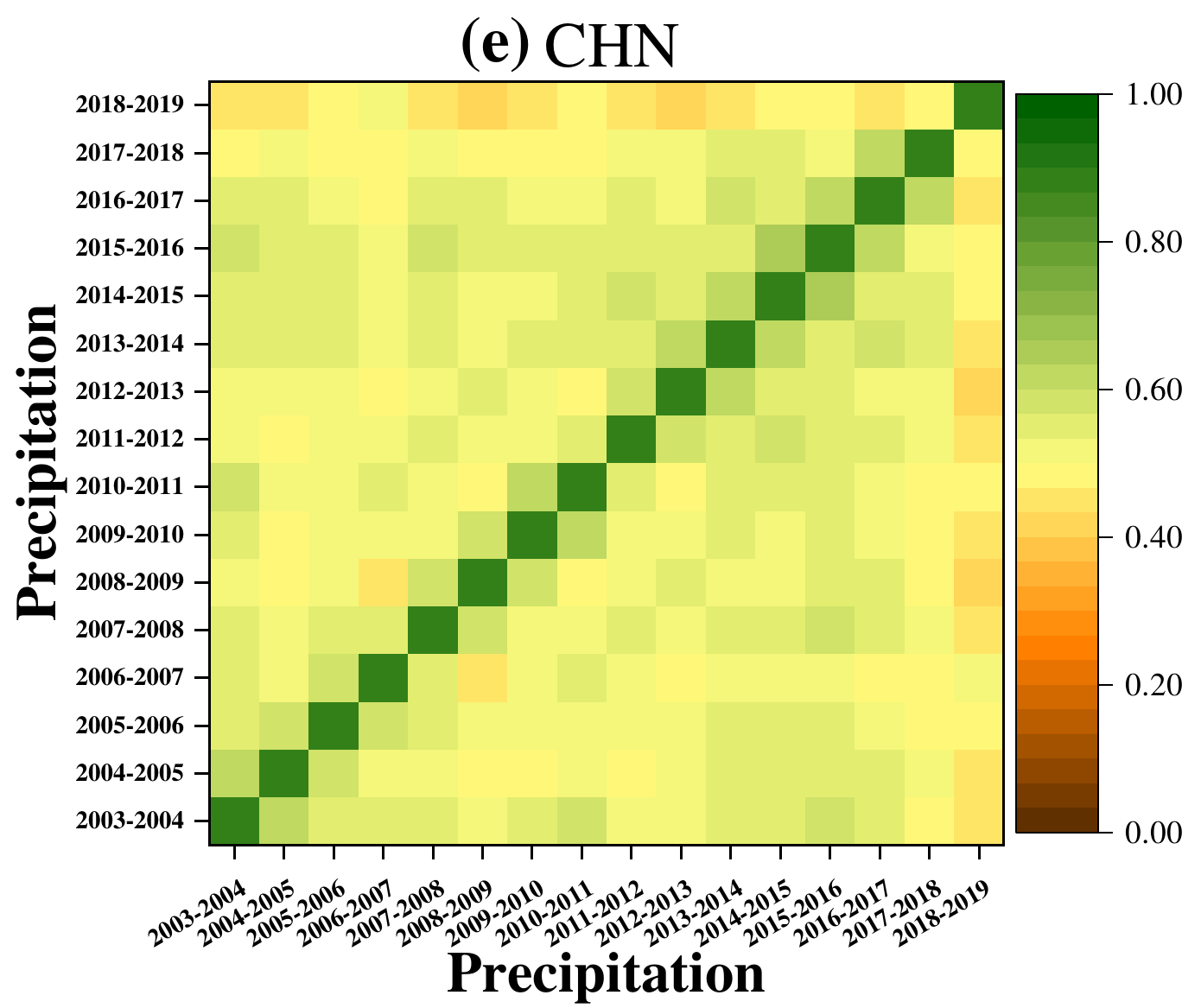}
\includegraphics[width=8em, height=7em]{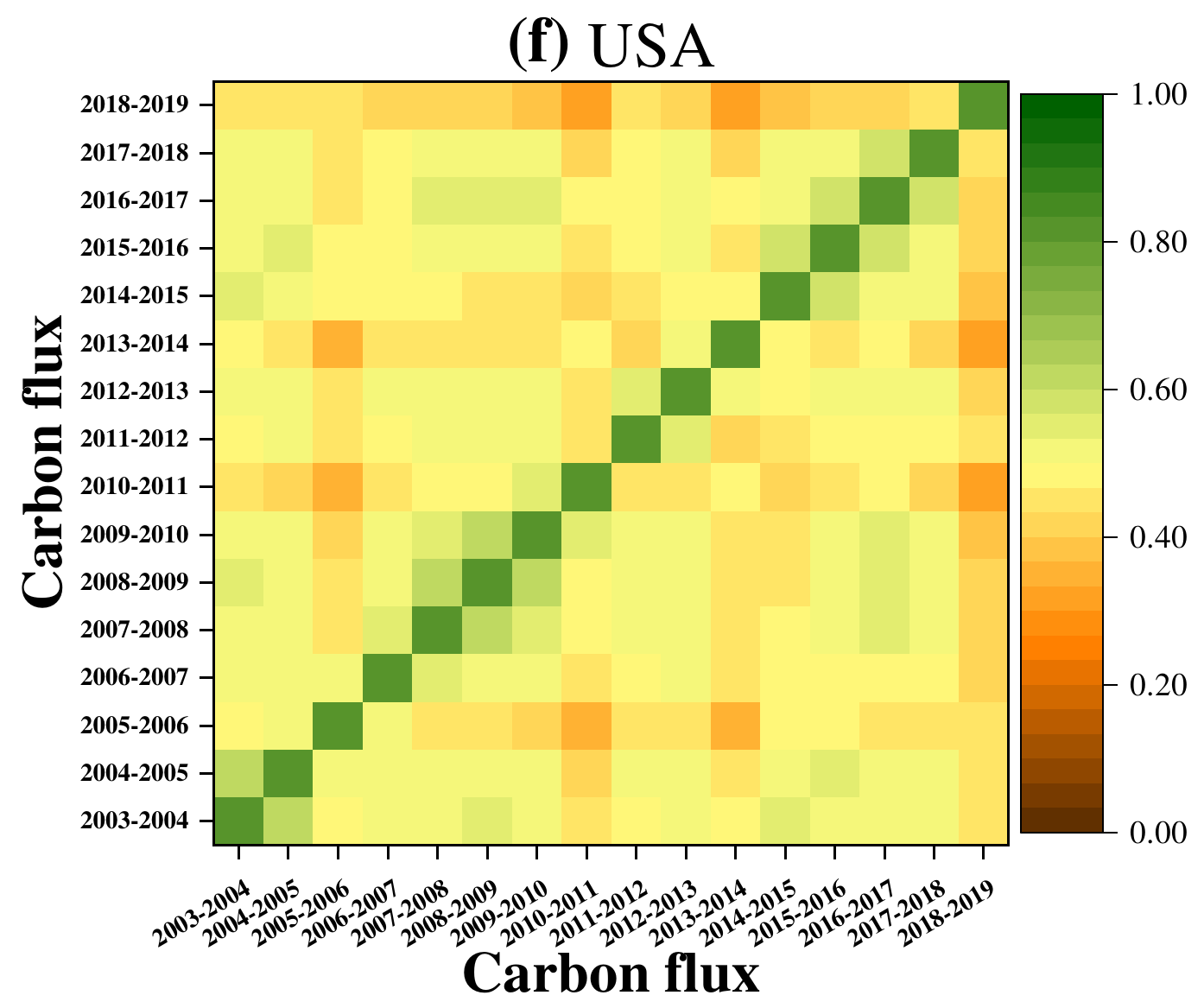}
\includegraphics[width=8em, height=7em]{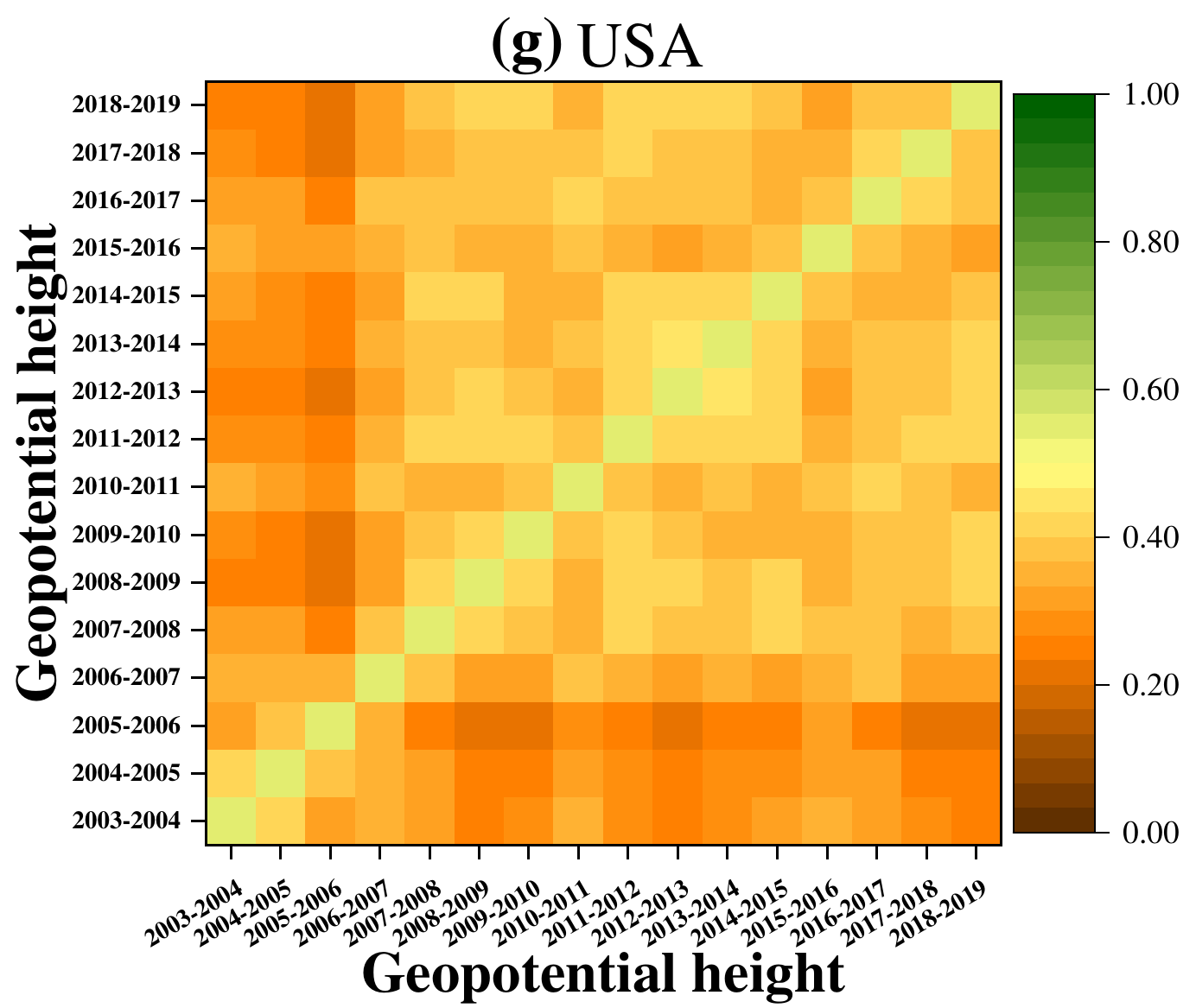}
\includegraphics[width=8em, height=7em]{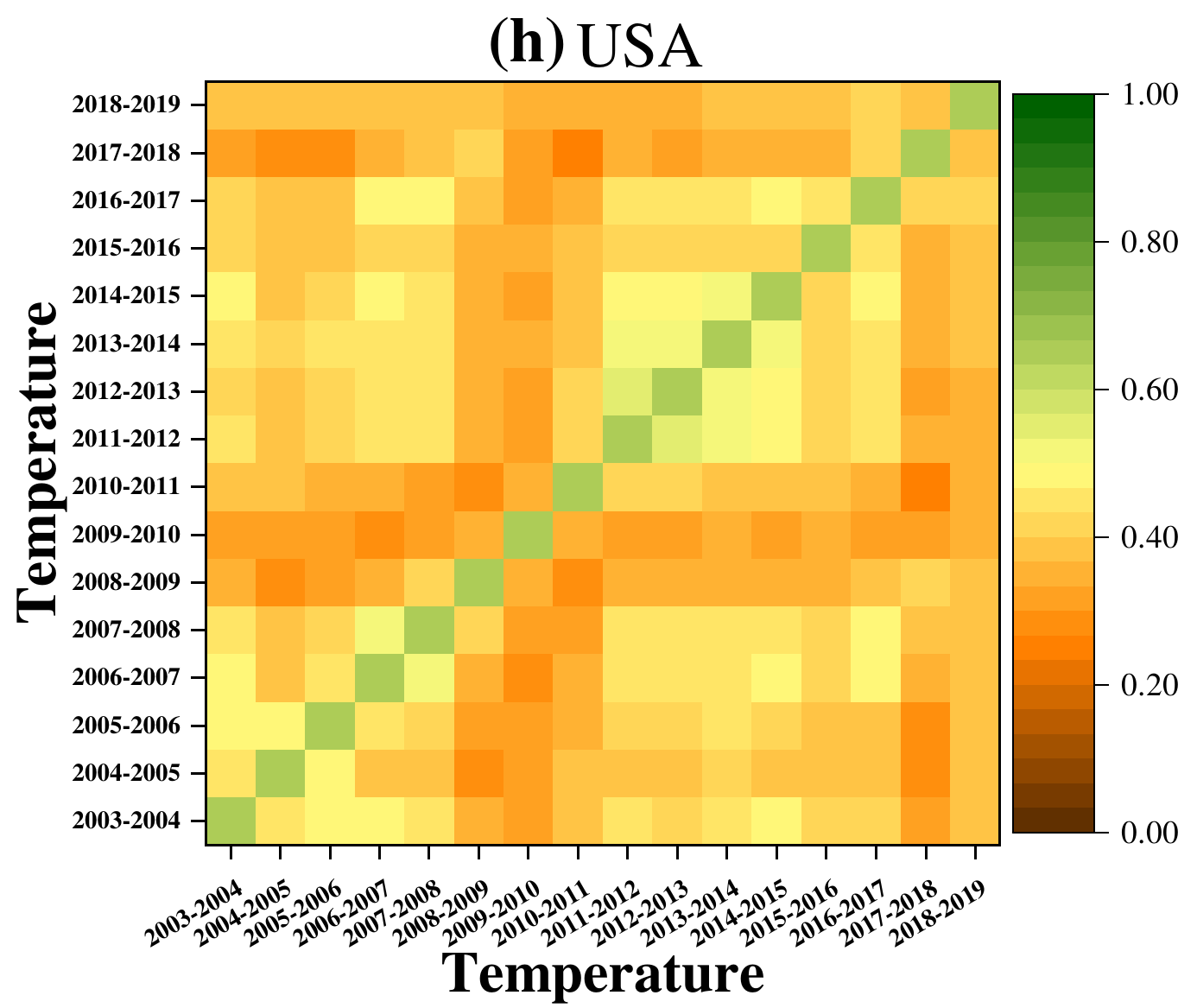}
\includegraphics[width=8em, height=7em]{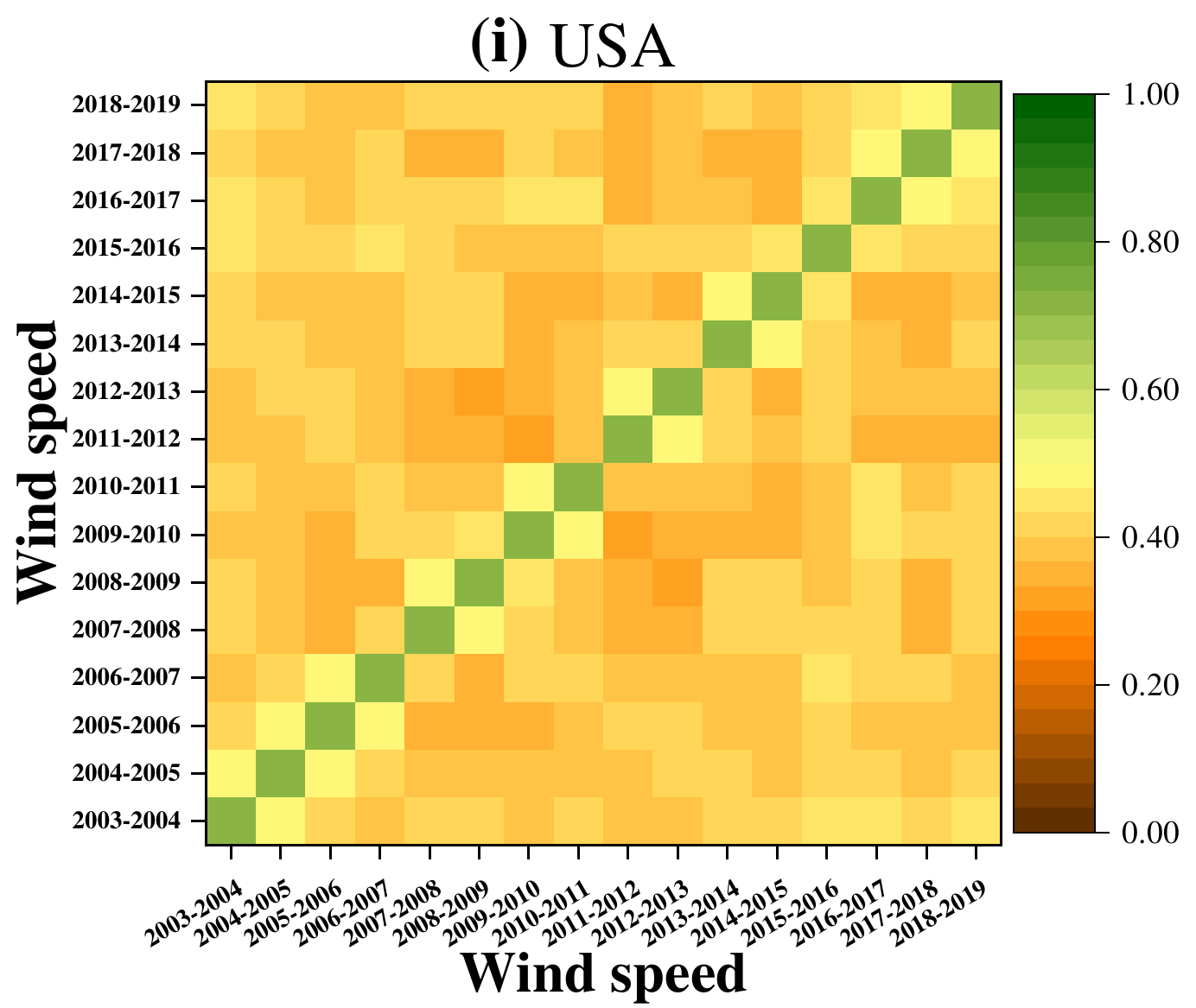}
\includegraphics[width=8em, height=7em]{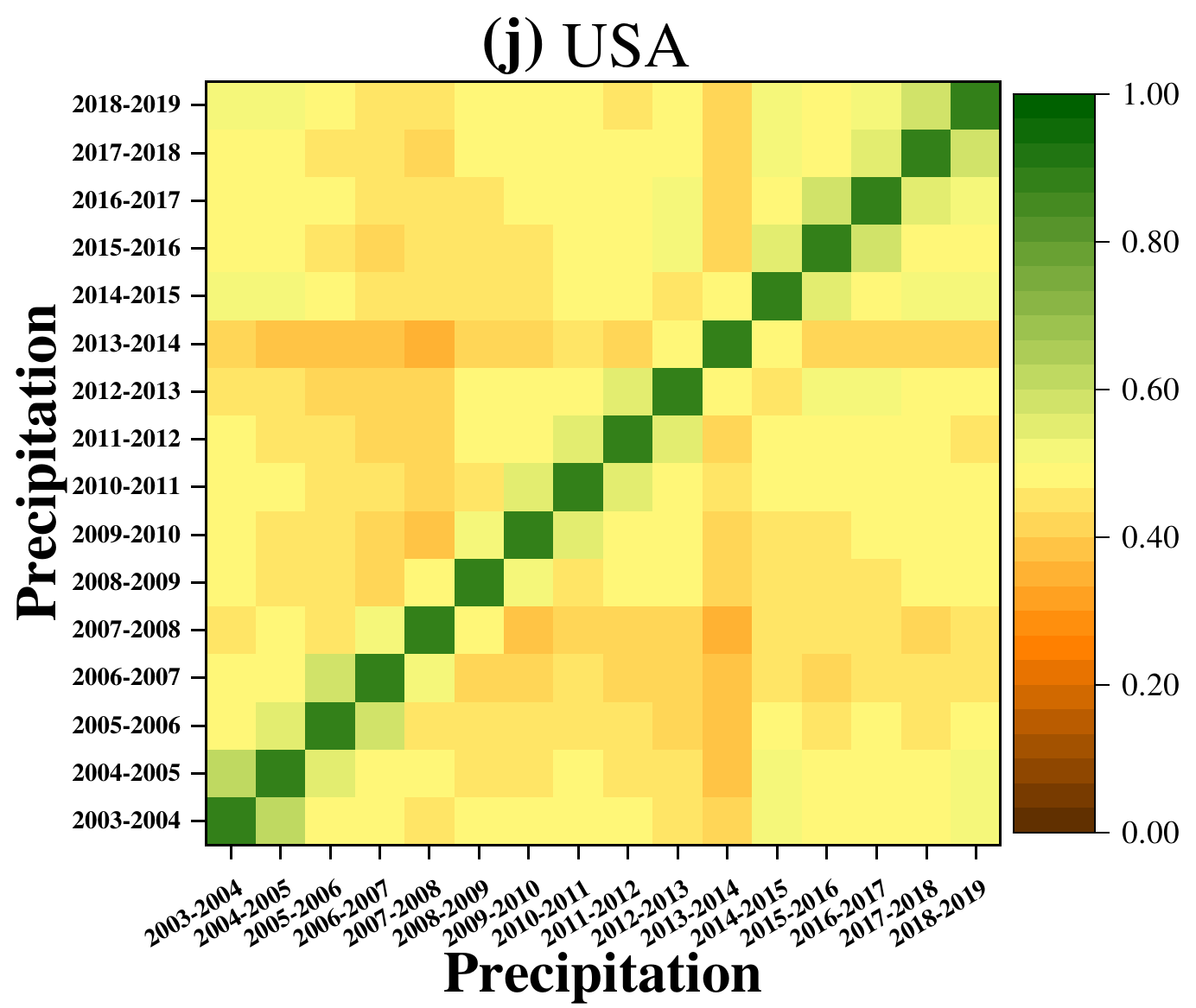}
\includegraphics[width=8em, height=7em]{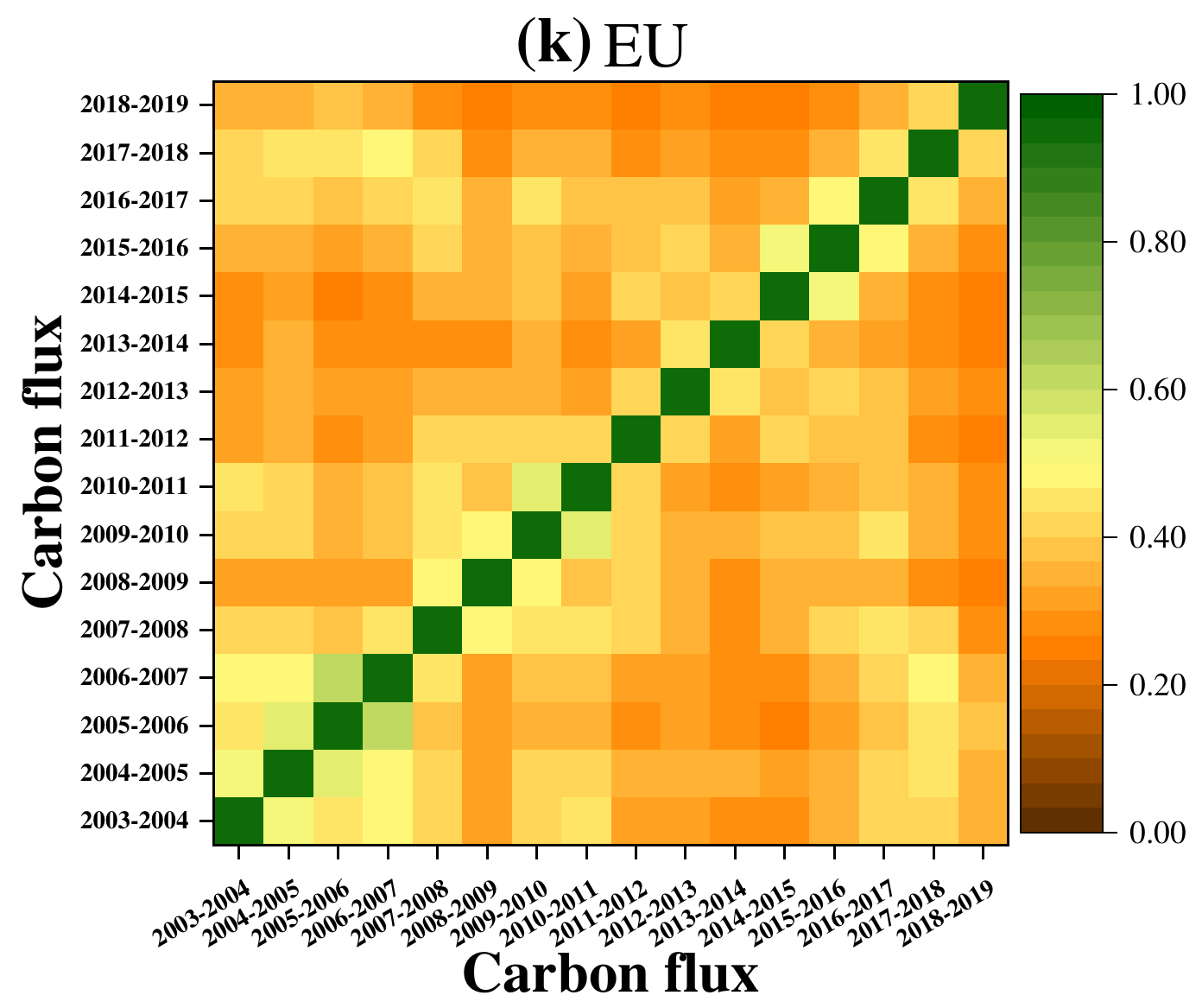}
\includegraphics[width=8em, height=7em]{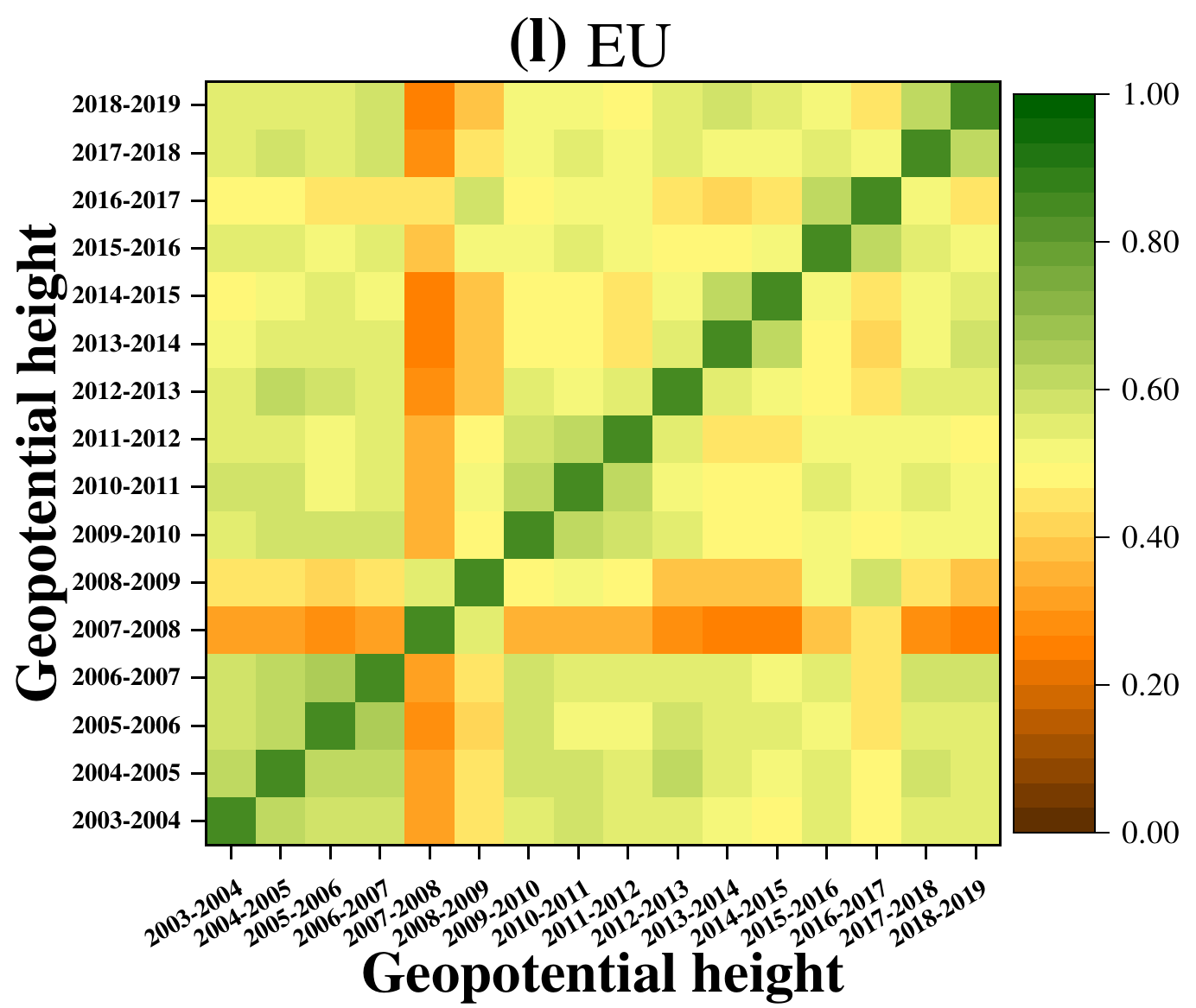}
\includegraphics[width=8em, height=7em]{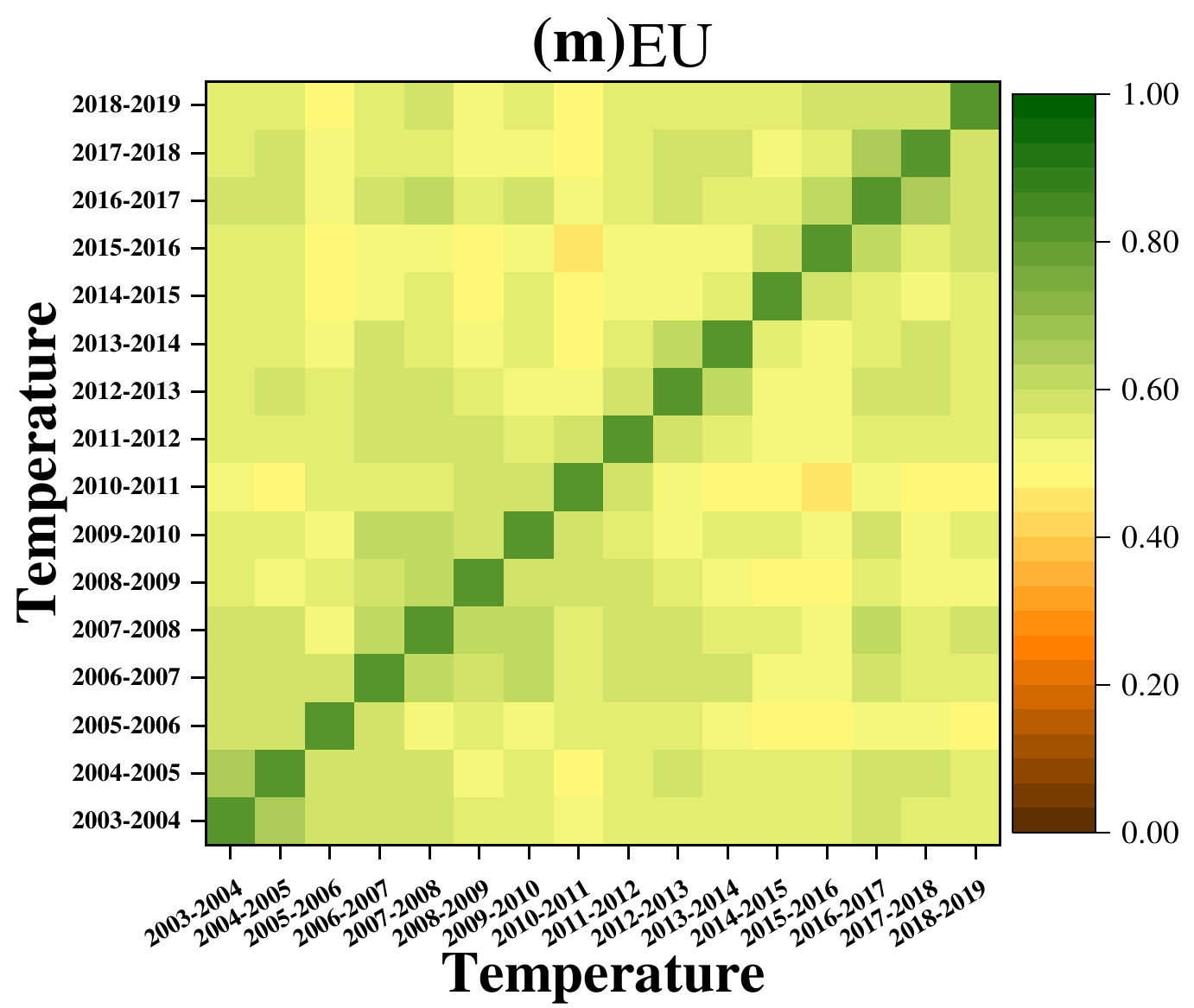}
\includegraphics[width=8em, height=7em]{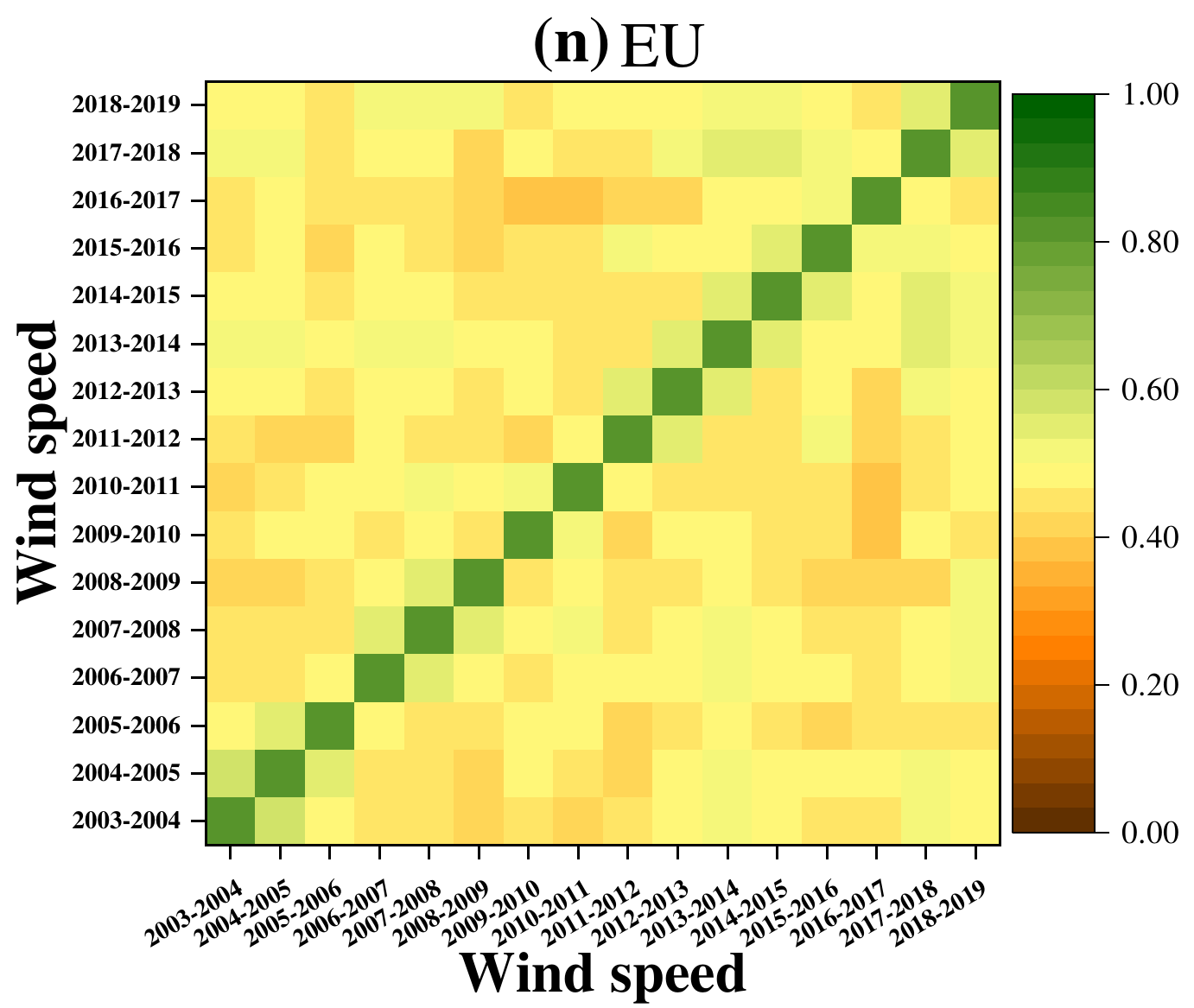}
\includegraphics[width=8em, height=7em]{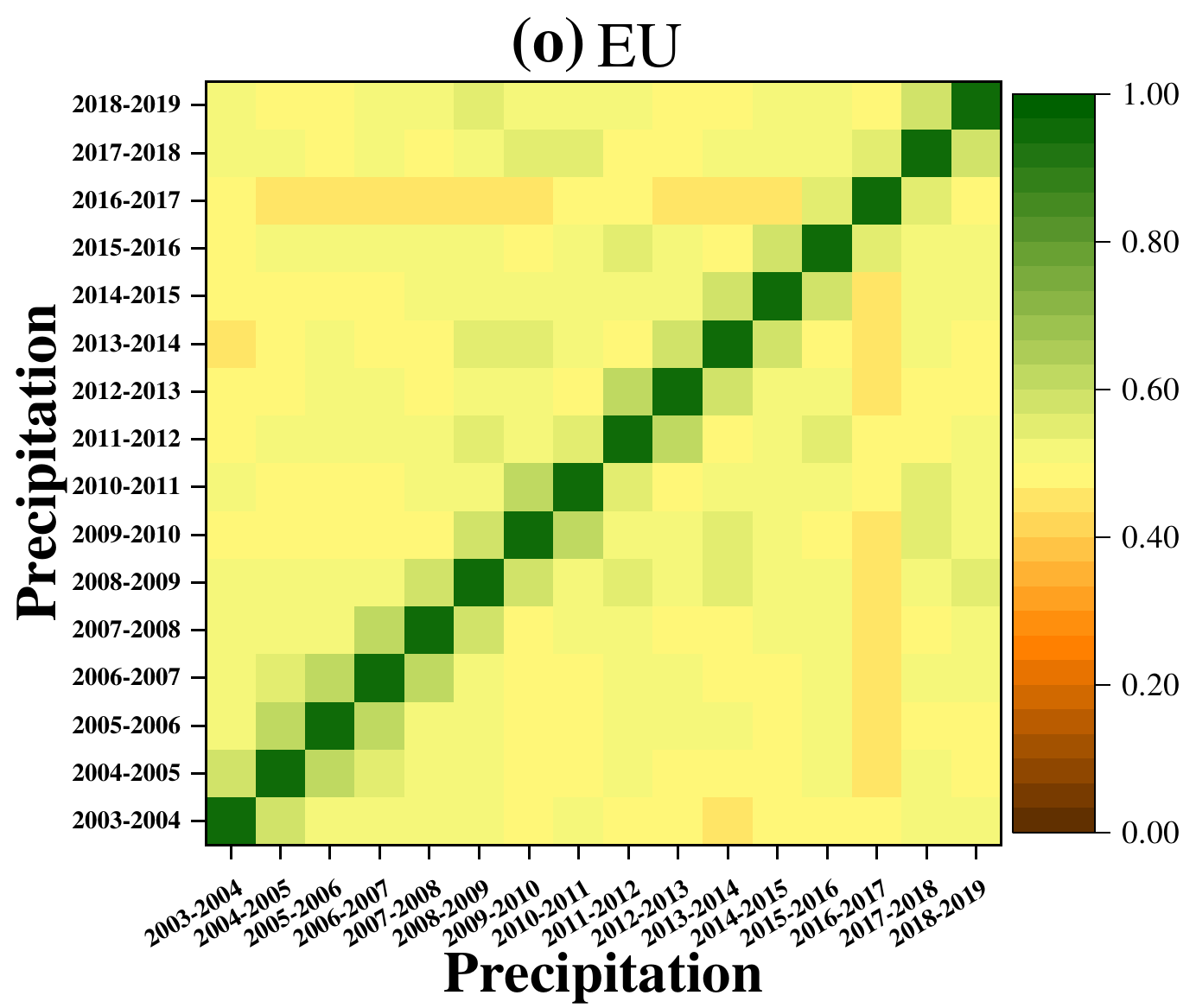}
\end{center}

\begin{center}
\noindent {\small {\bf Fig. S33} The effective Jaccard similarity coefficient matrix for links of lengths above $500km$ in two networks of different years for each of the climate variables. Each matrix element represents the difference between the actual Jaccard similarity coefficient and the corresponding average values obtained from the random controlled case.}
\end{center}

\begin{center}
\includegraphics[width=8em, height=7em]{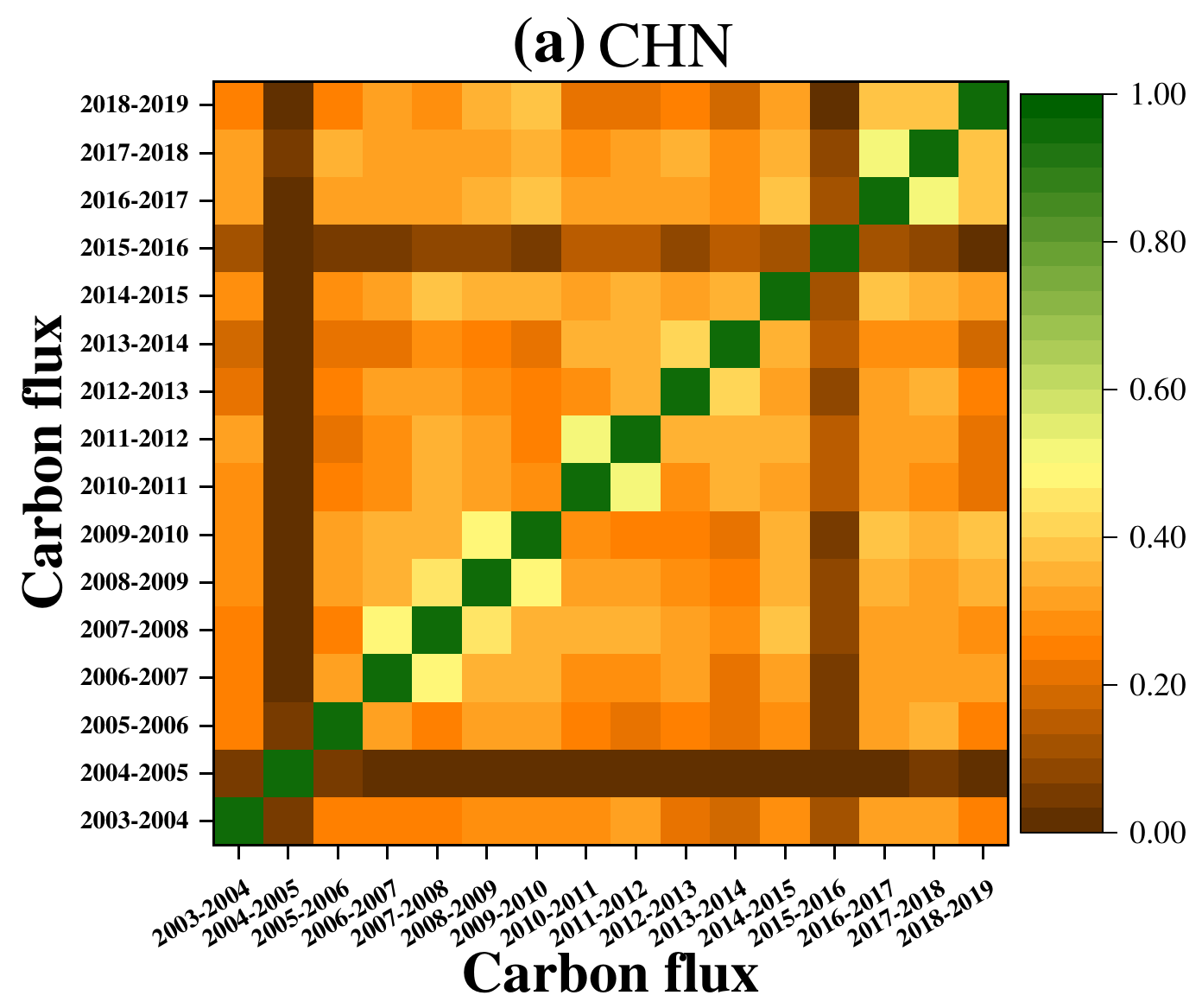}
\includegraphics[width=8em, height=7em]{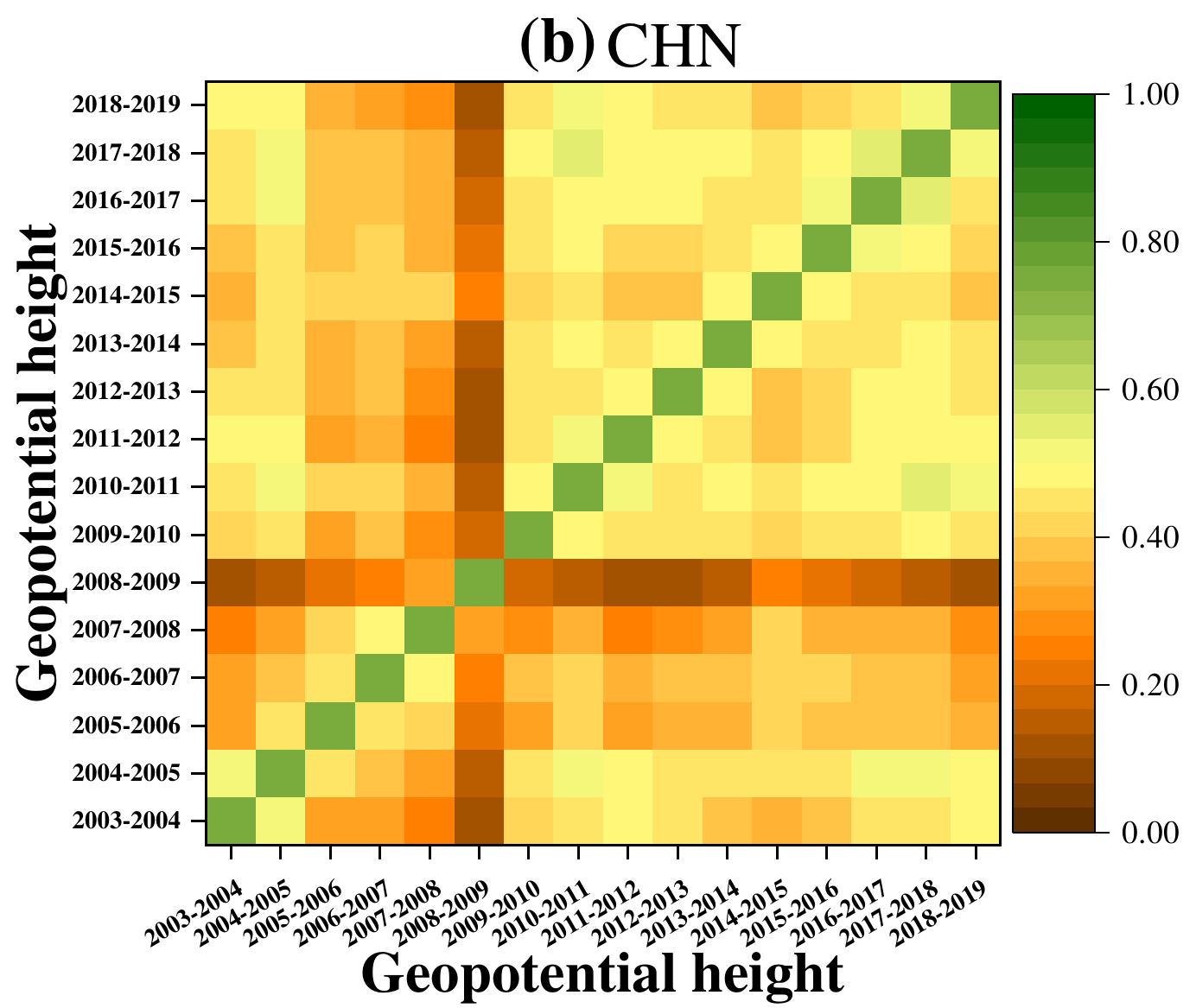}
\includegraphics[width=8em, height=7em]{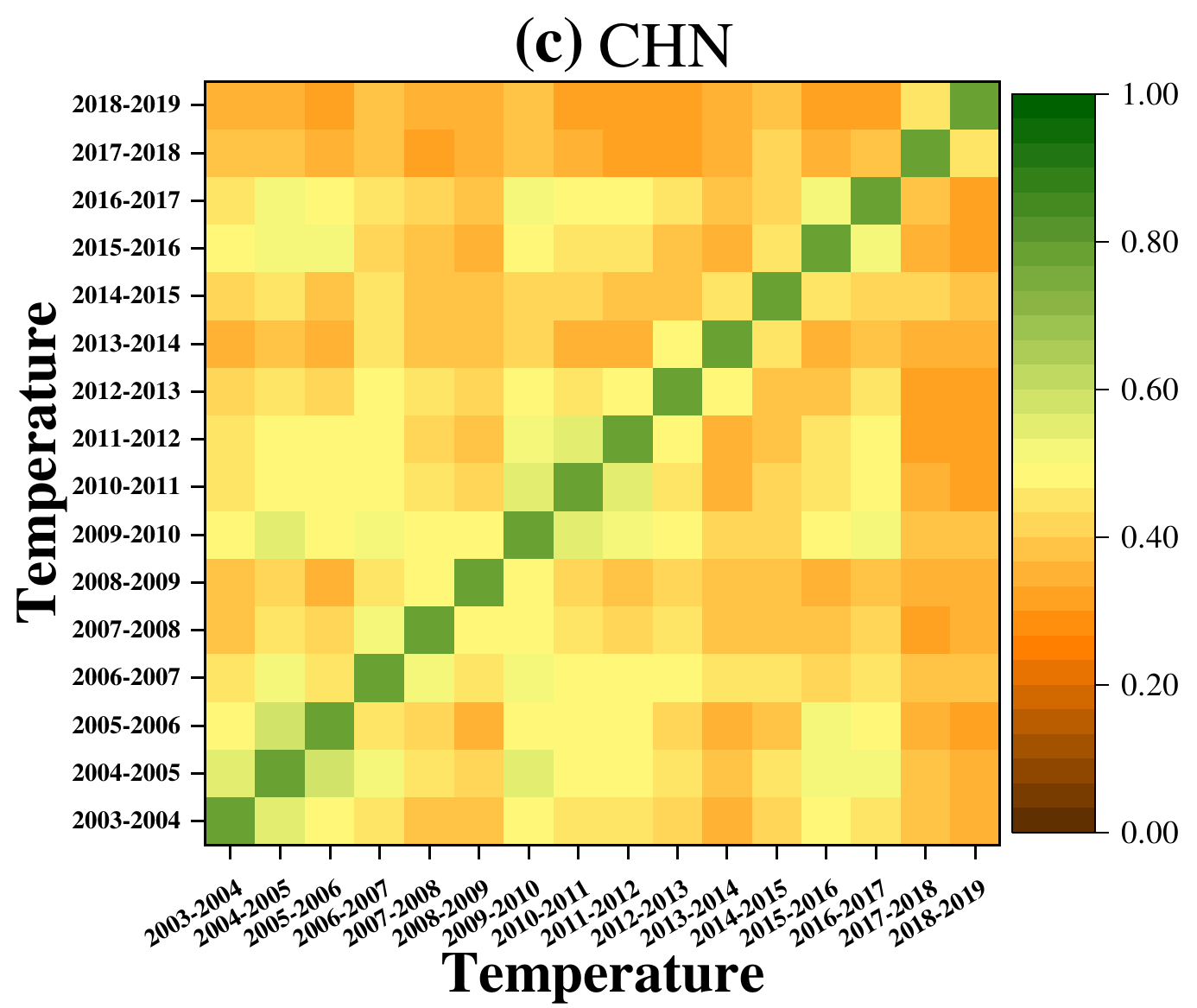}
\includegraphics[width=8em, height=7em]{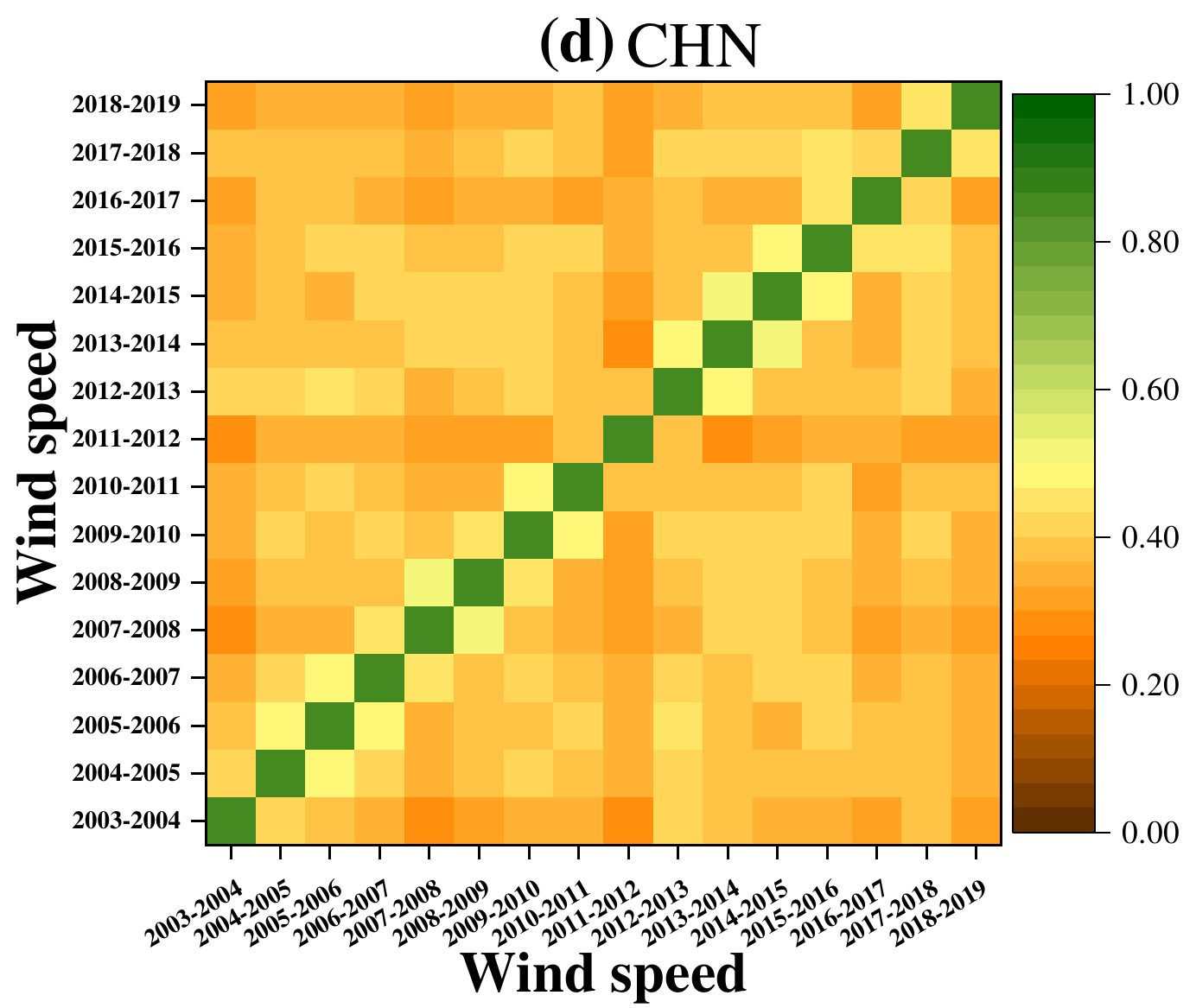}
\includegraphics[width=8em, height=7em]{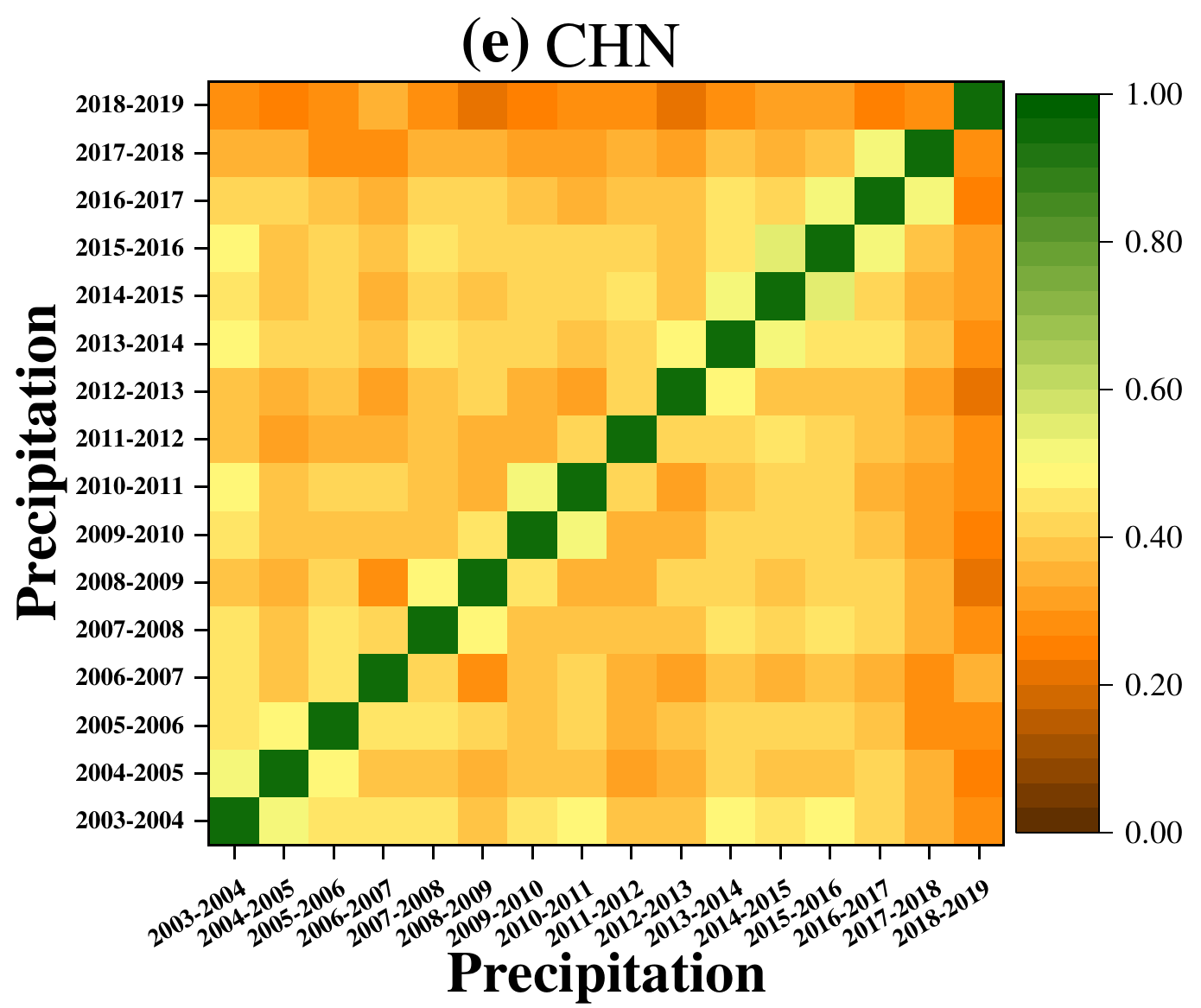}
\includegraphics[width=8em, height=7em]{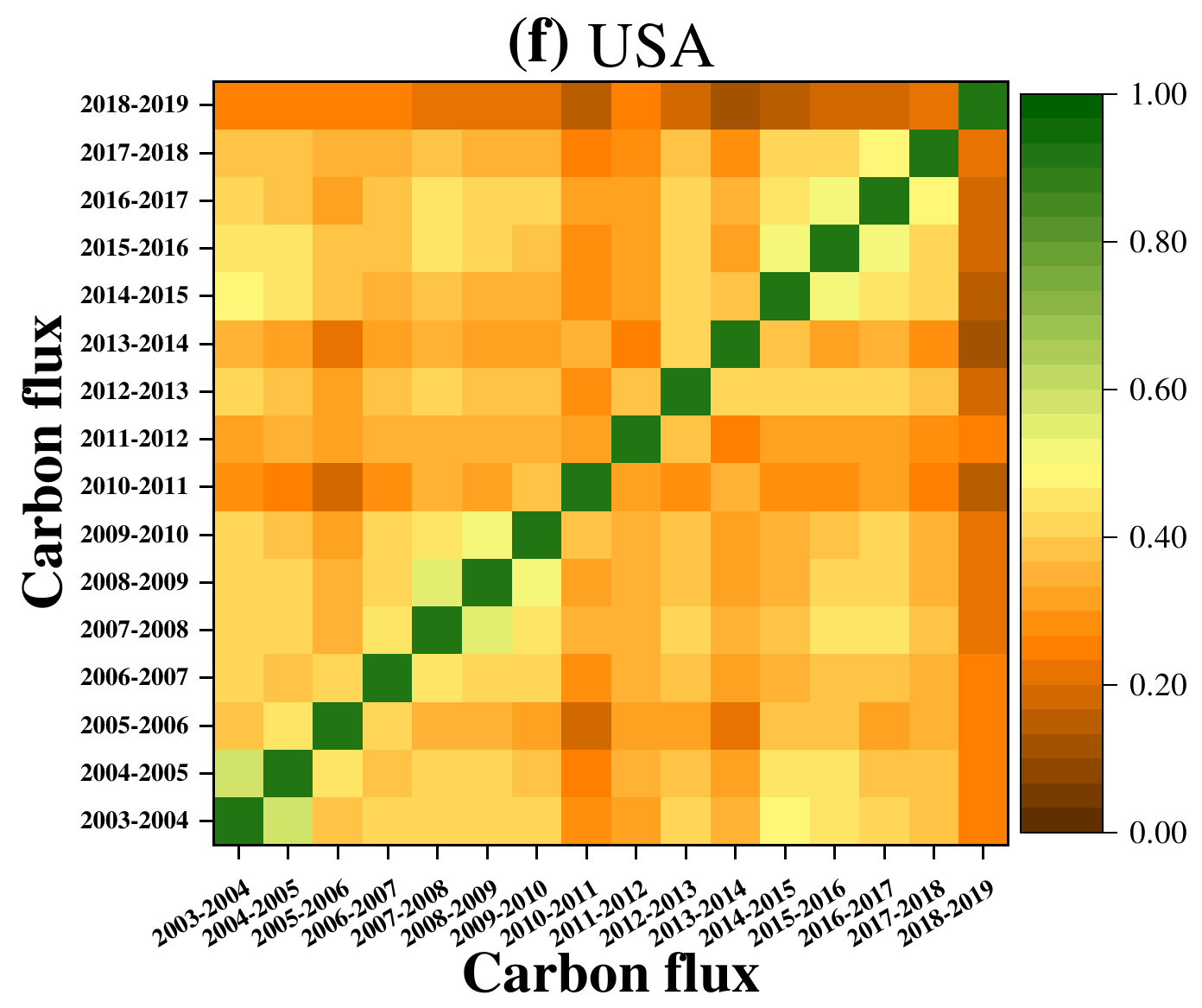}
\includegraphics[width=8em, height=7em]{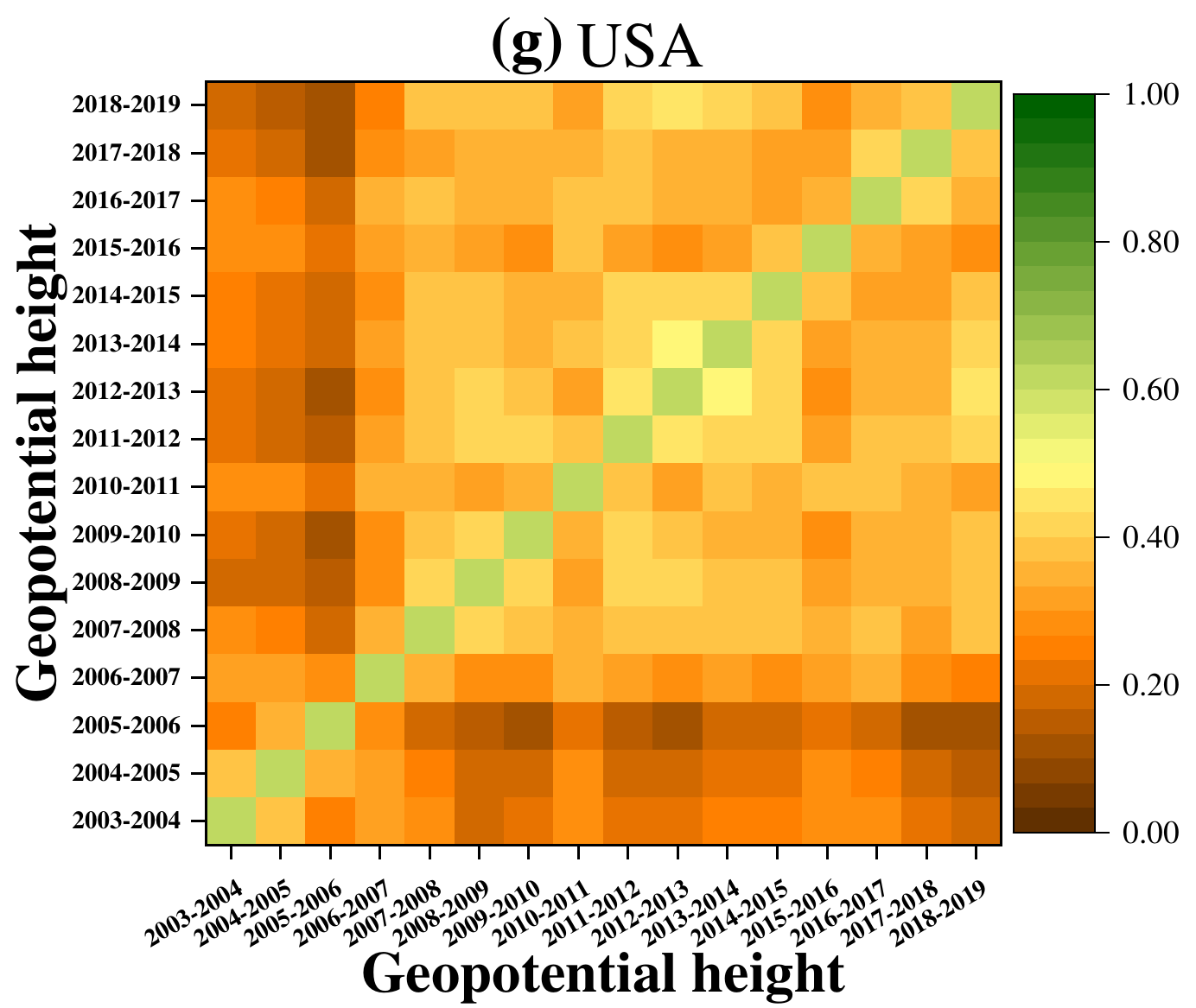}
\includegraphics[width=8em, height=7em]{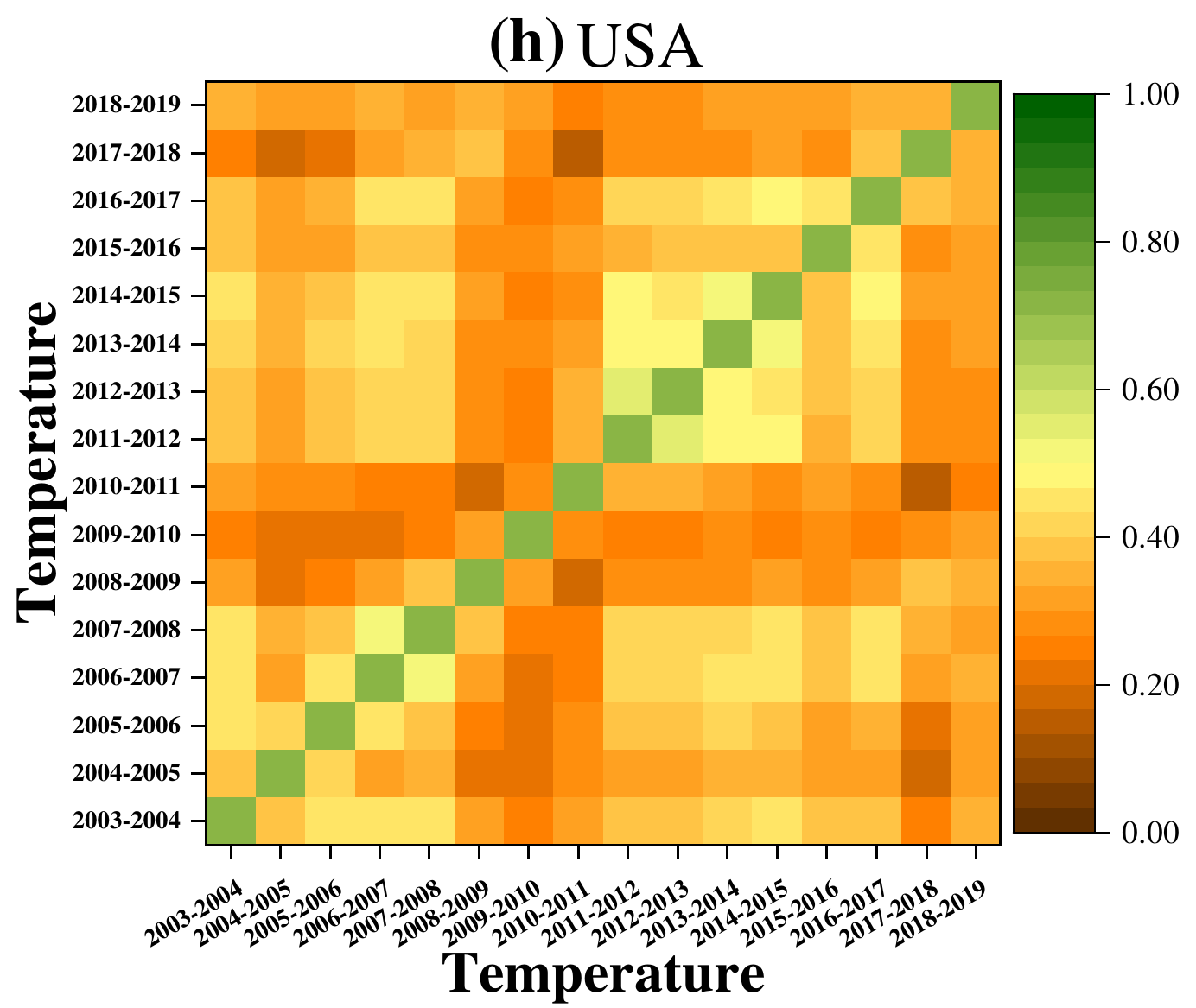}
\includegraphics[width=8em, height=7em]{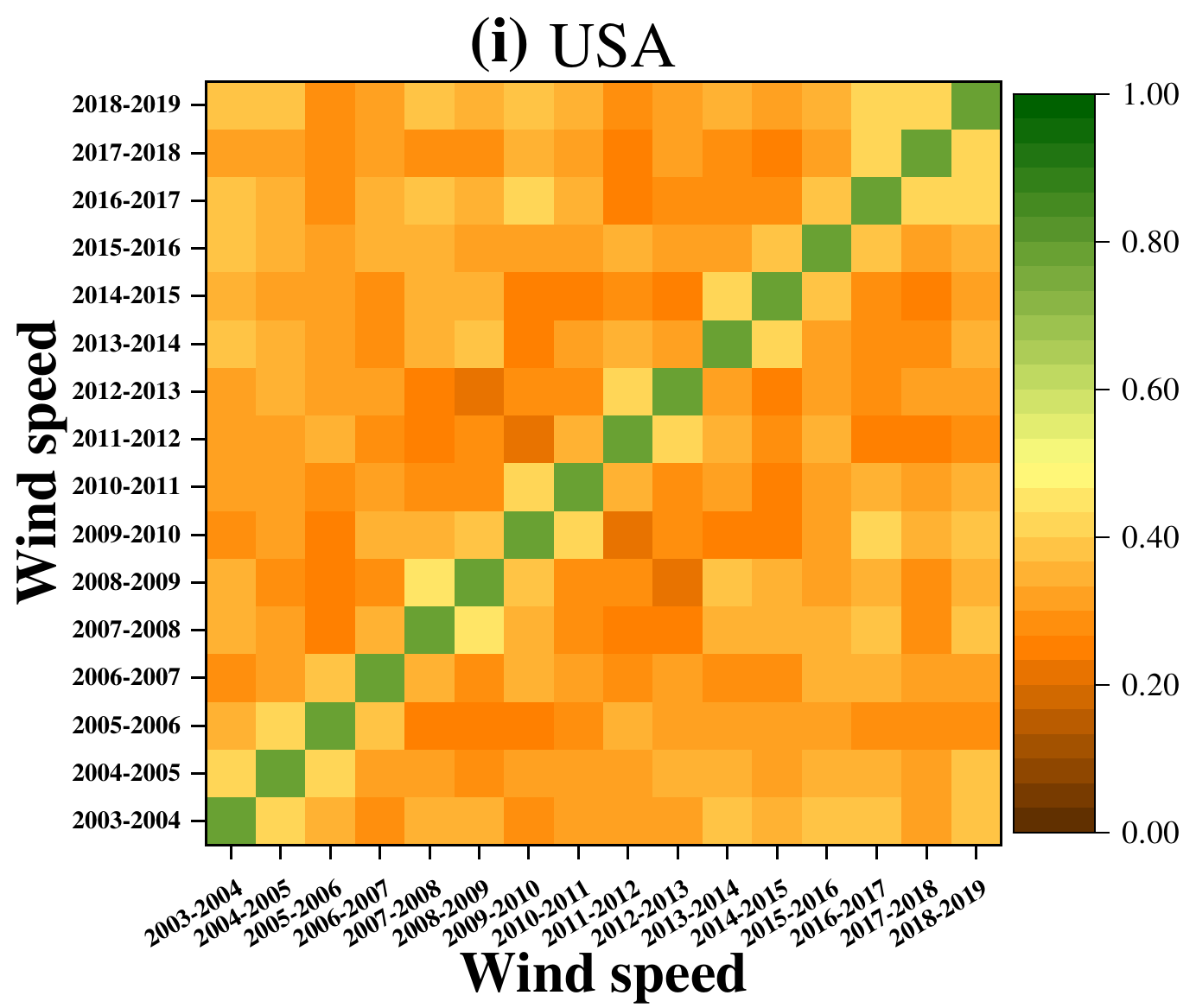}
\includegraphics[width=8em, height=7em]{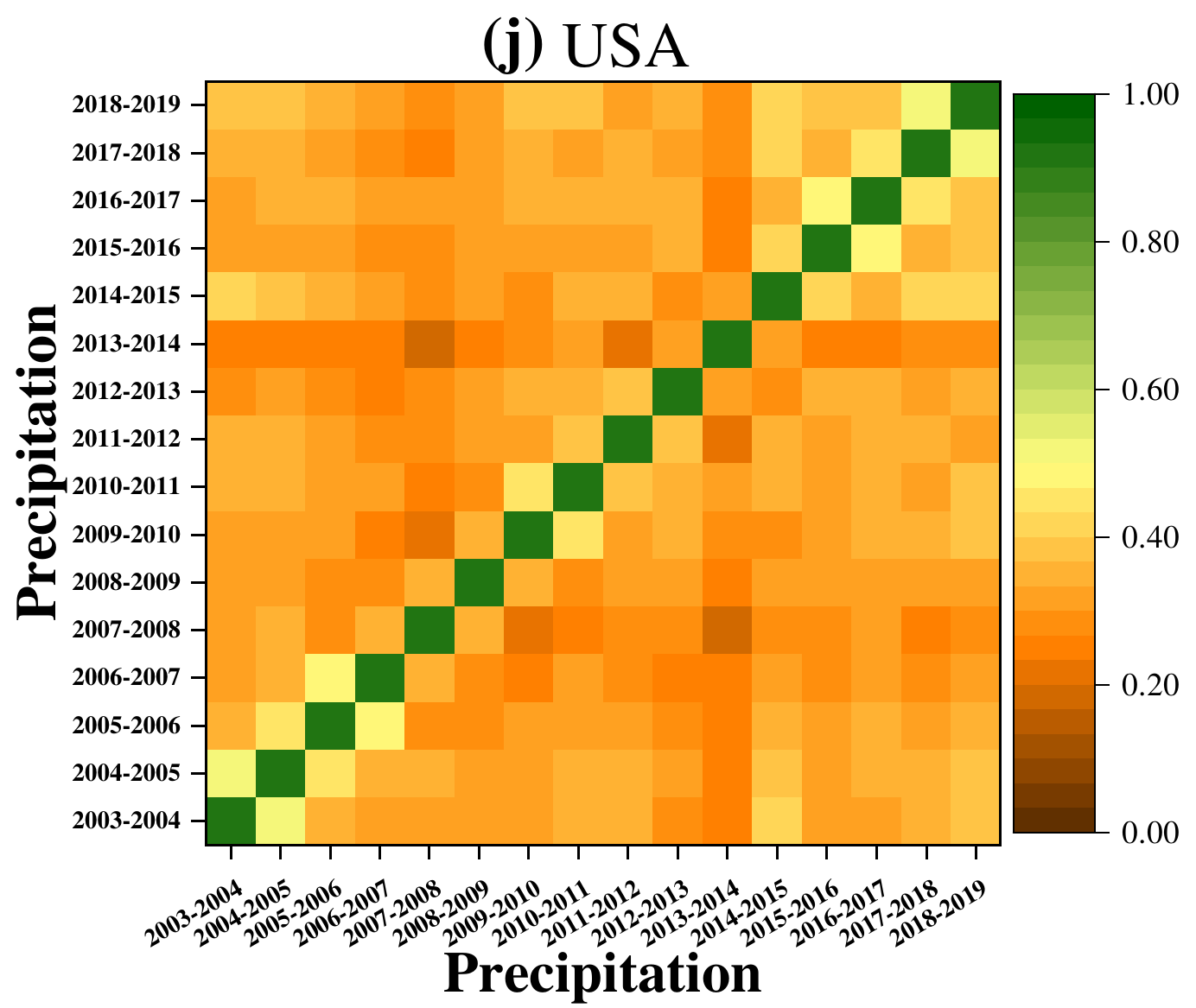}
\includegraphics[width=8em, height=7em]{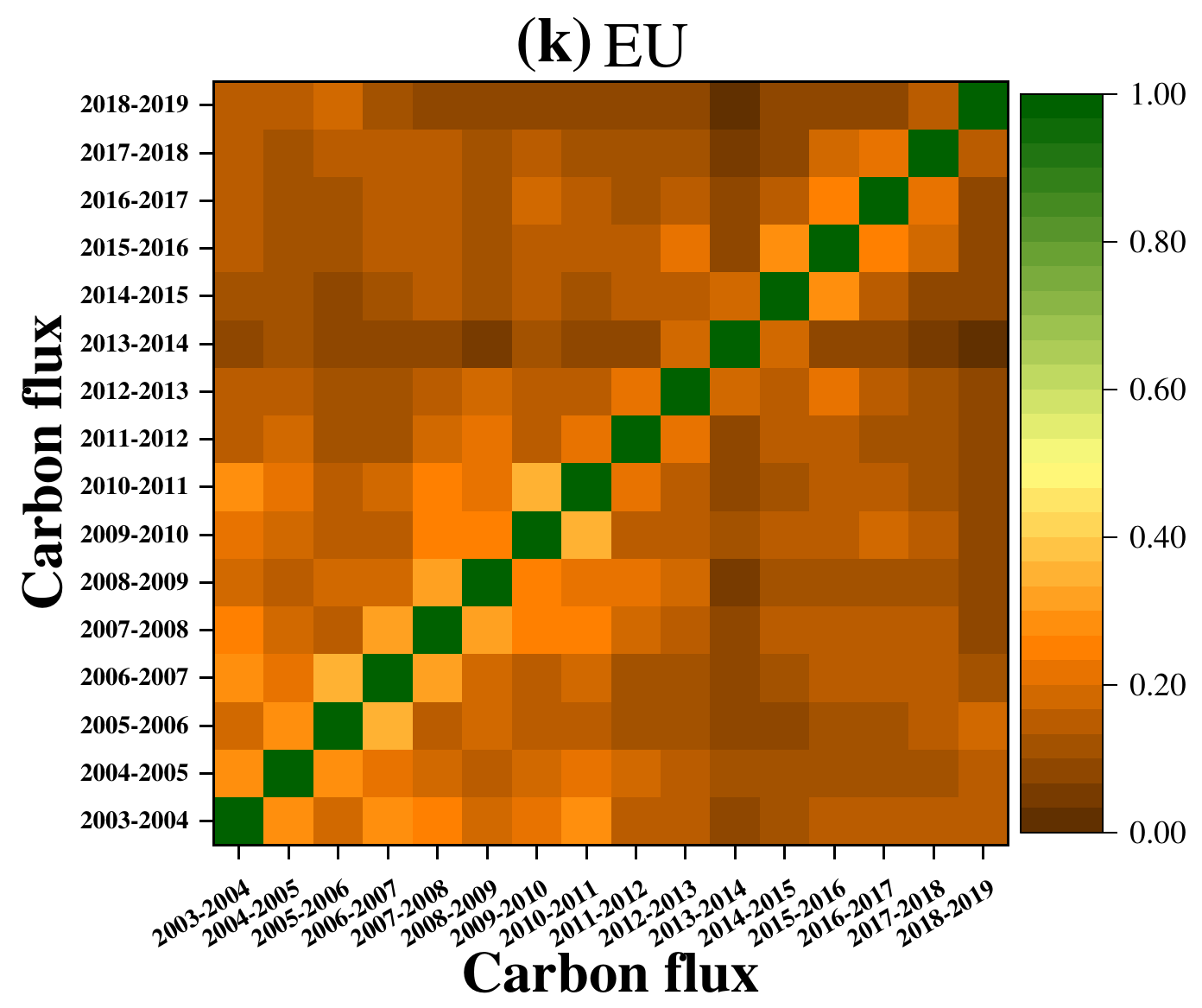}
\includegraphics[width=8em, height=7em]{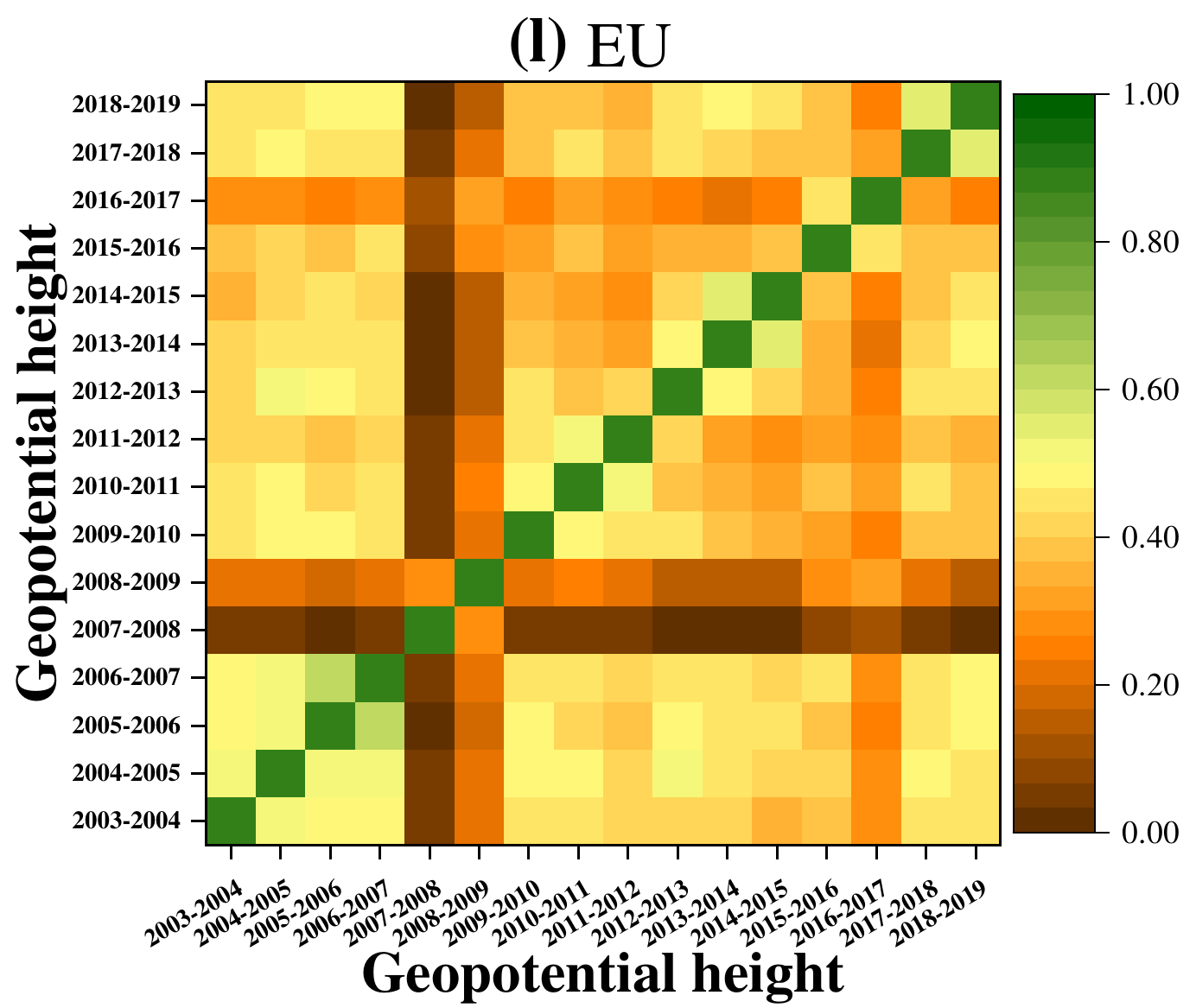}
\includegraphics[width=8em, height=7em]{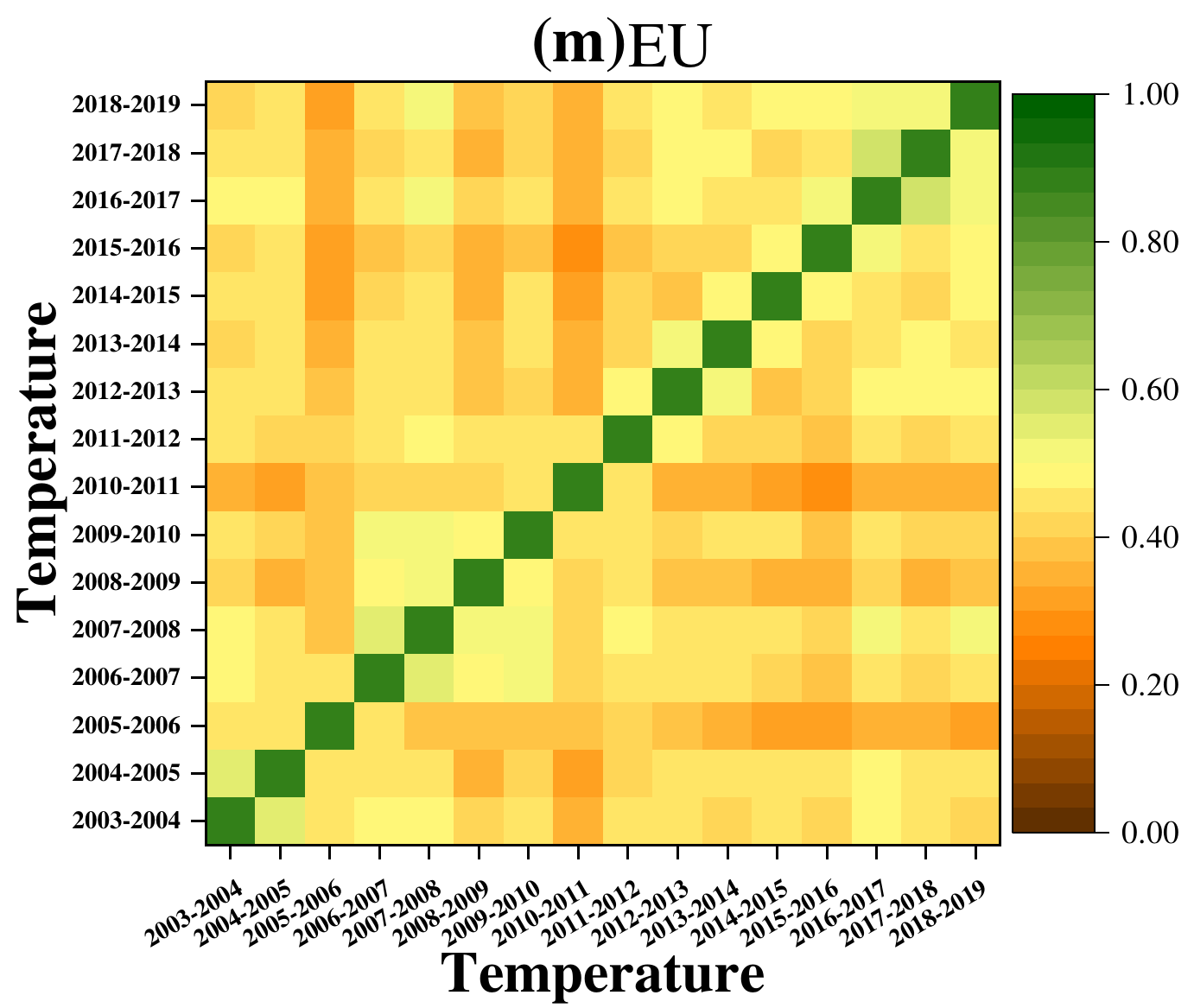}
\includegraphics[width=8em, height=7em]{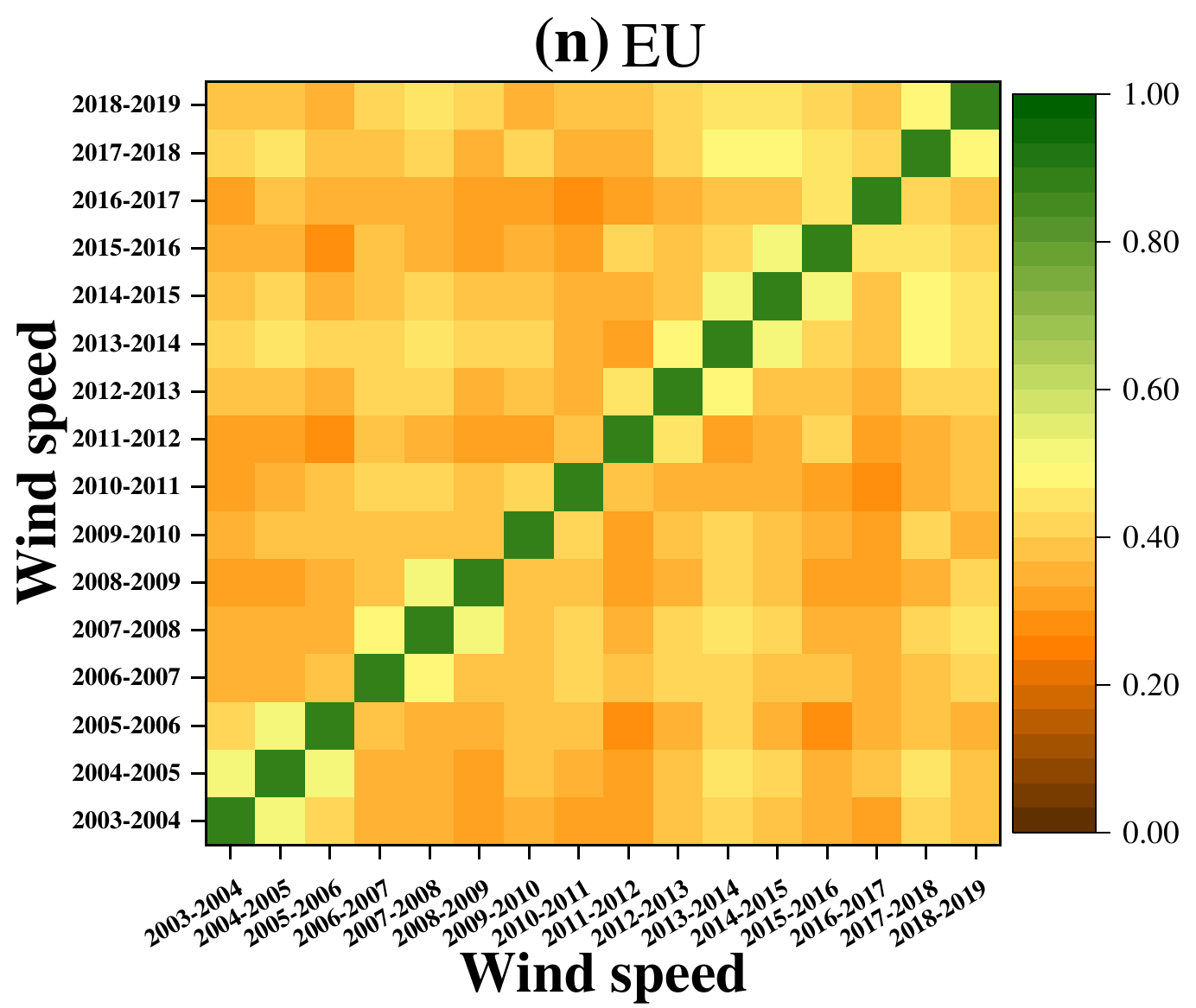}
\includegraphics[width=8em, height=7em]{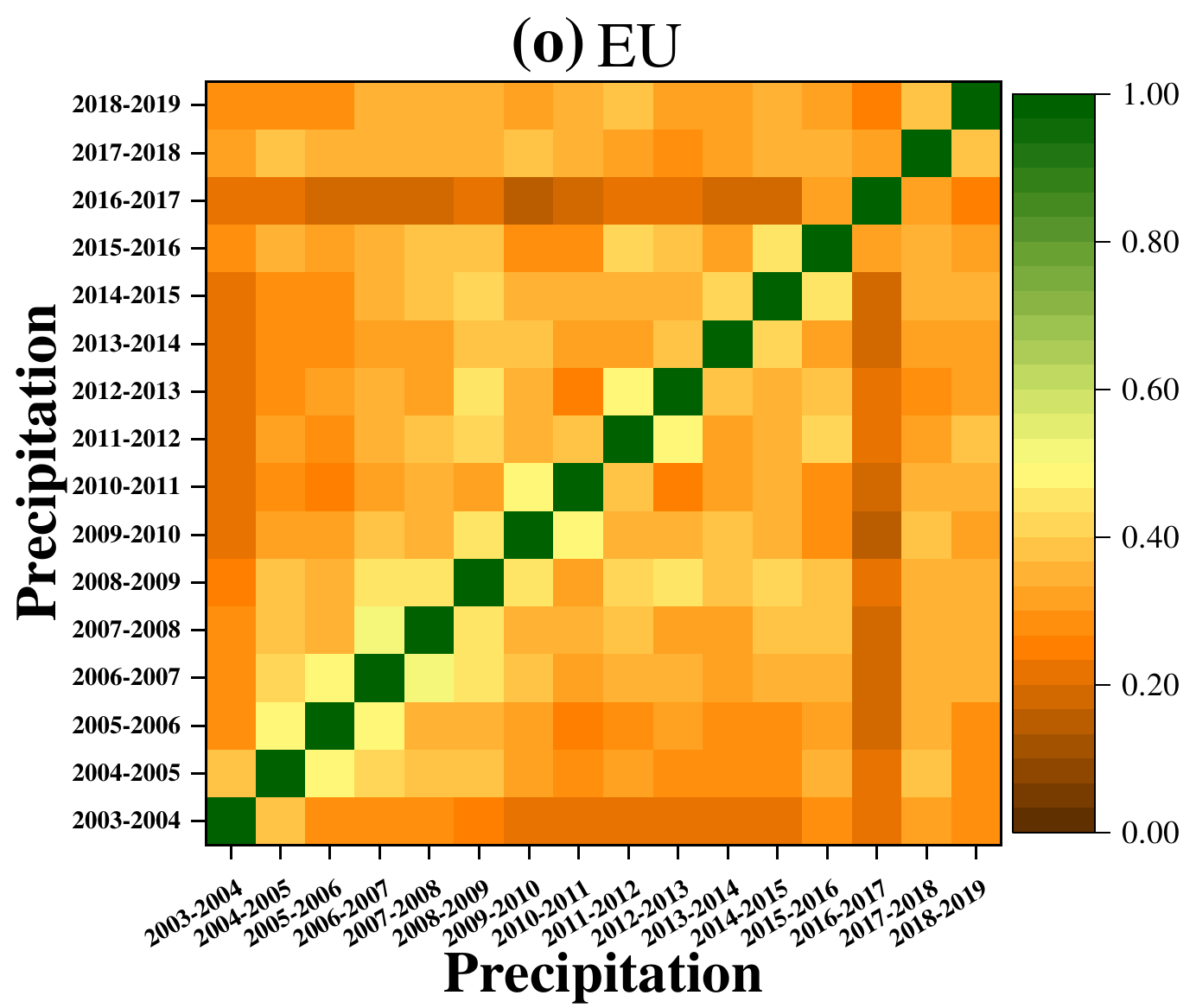}
\end{center}

\begin{center}
\noindent {\small {\bf Fig. S34} The effective Jaccard similarity coefficient matrix for links of lengths above $1000km$ in two networks of different years for each of the climate variables. Each matrix element represents the difference between the actual Jaccard similarity coefficient and the corresponding average values obtained from the random controlled case.}
\end{center}

\begin{center}
\includegraphics[width=8em, height=7em]{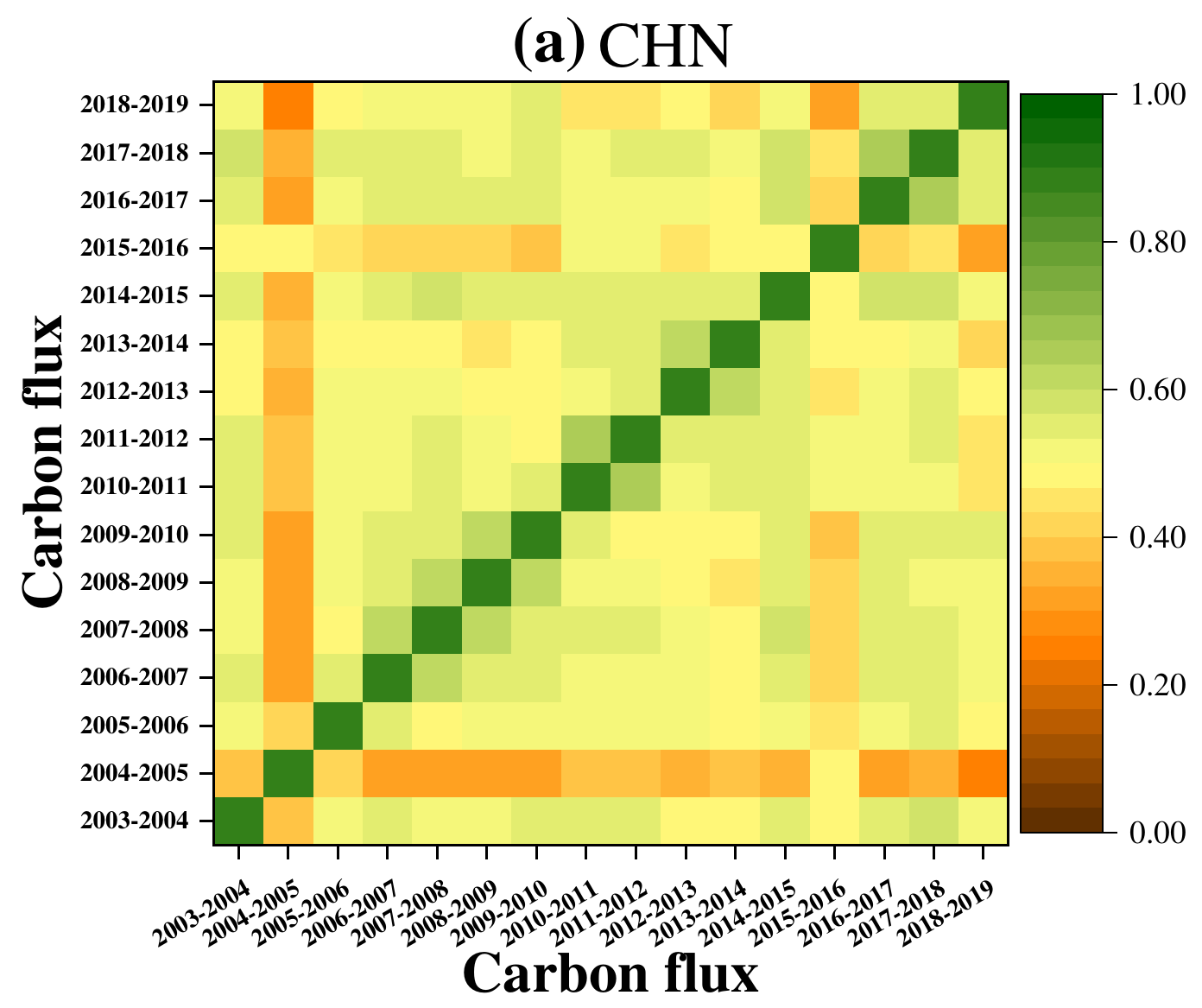}
\includegraphics[width=8em, height=7em]{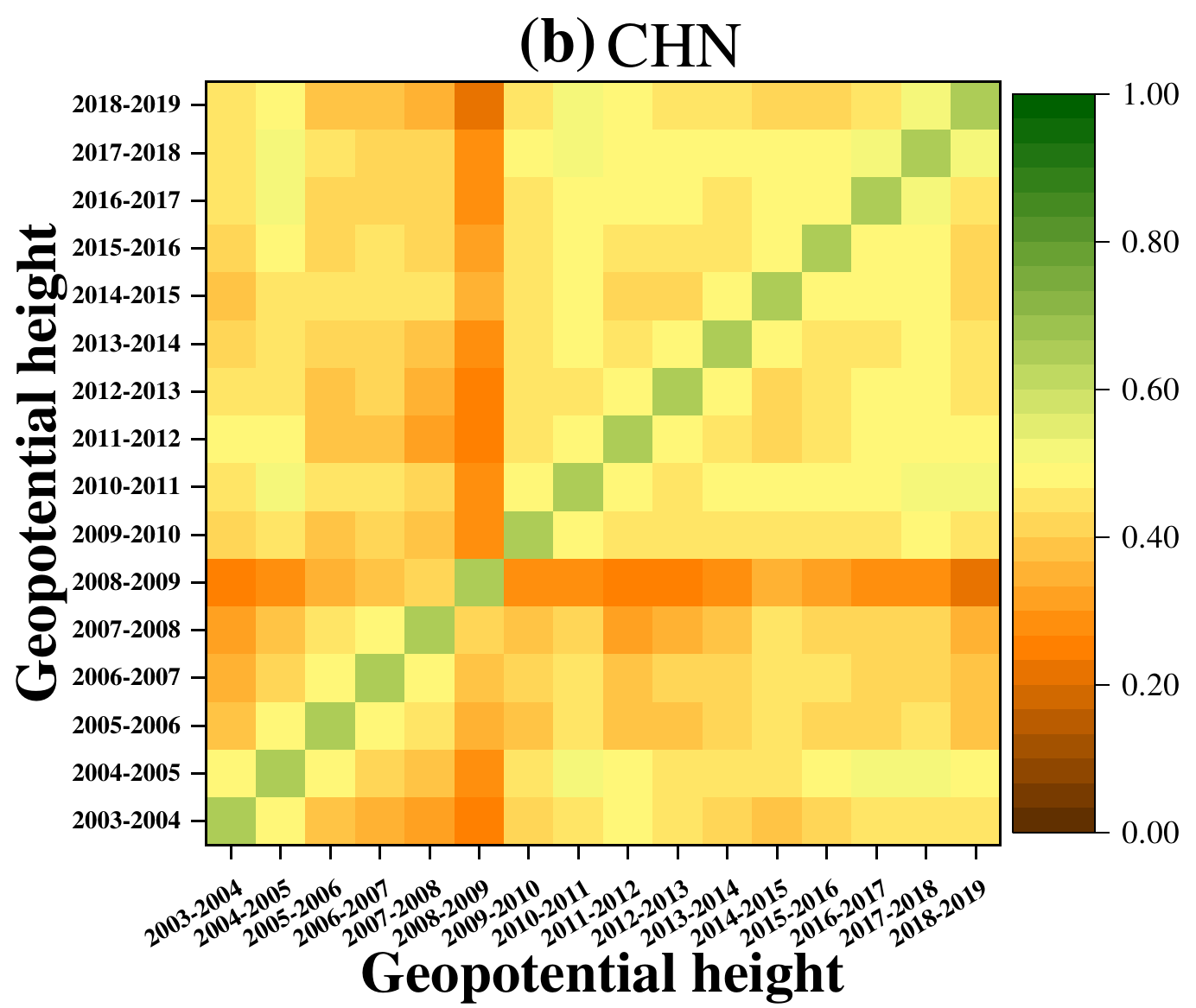}
\includegraphics[width=8em, height=7em]{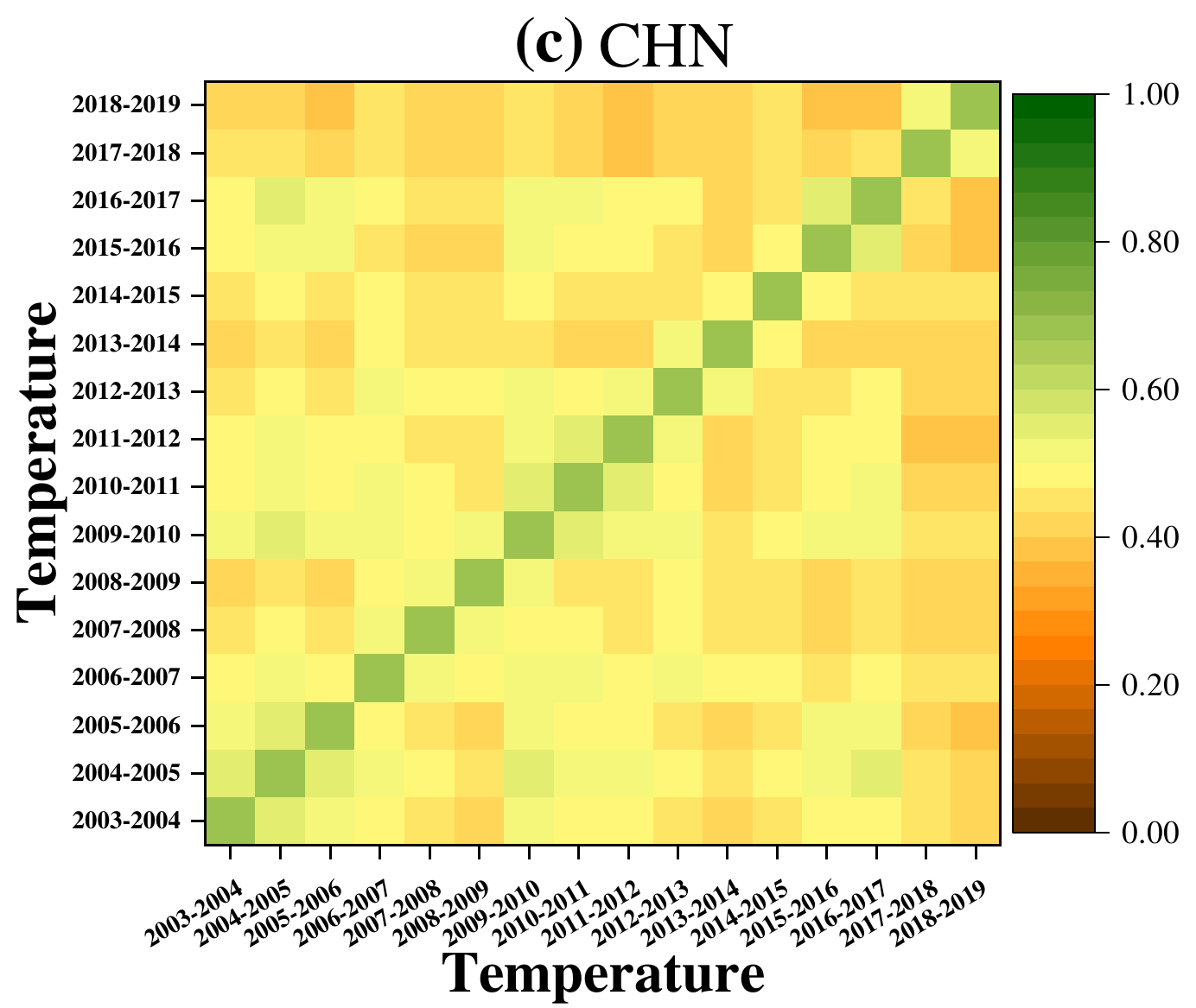}
\includegraphics[width=8em, height=7em]{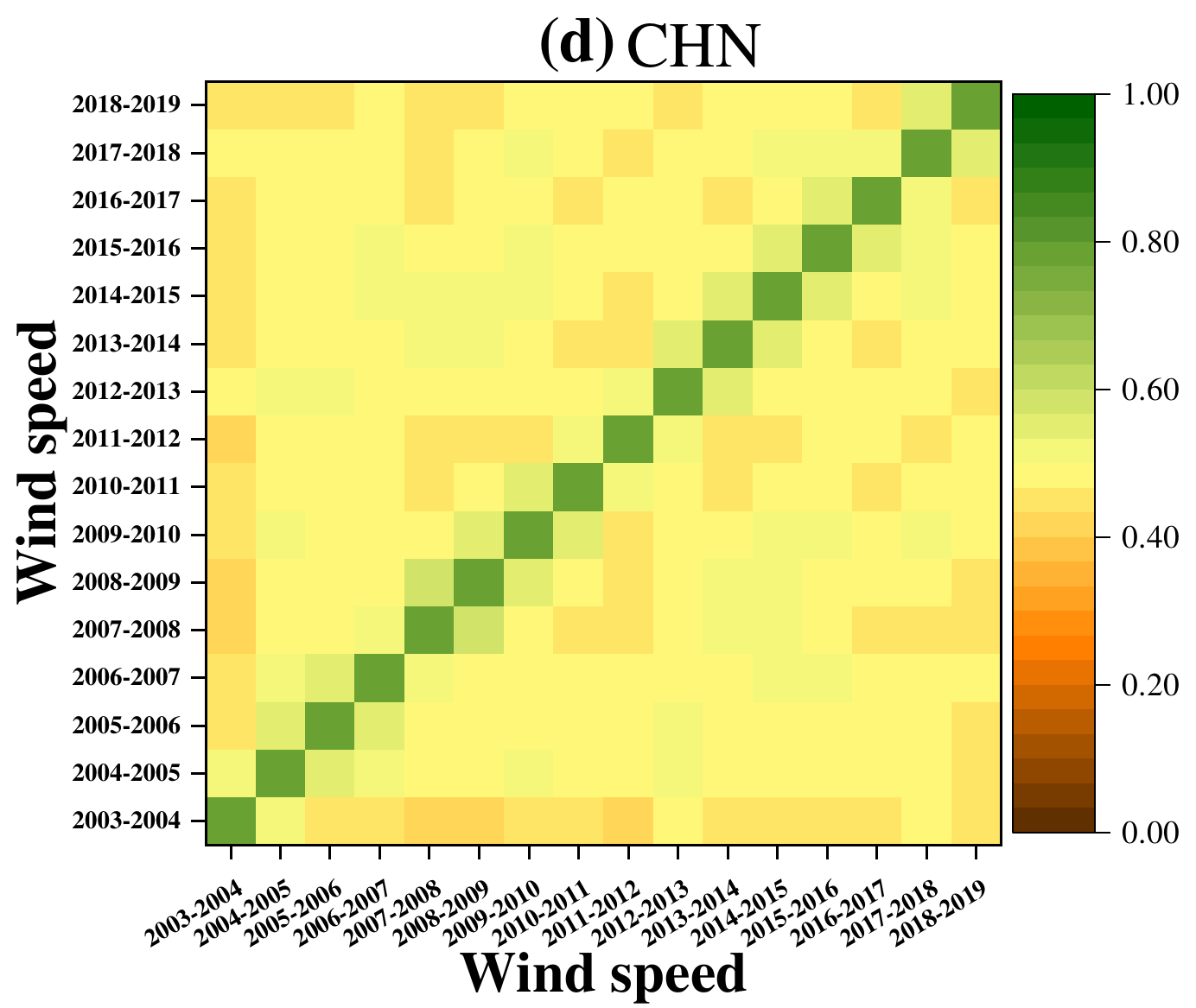}
\includegraphics[width=8em, height=7em]{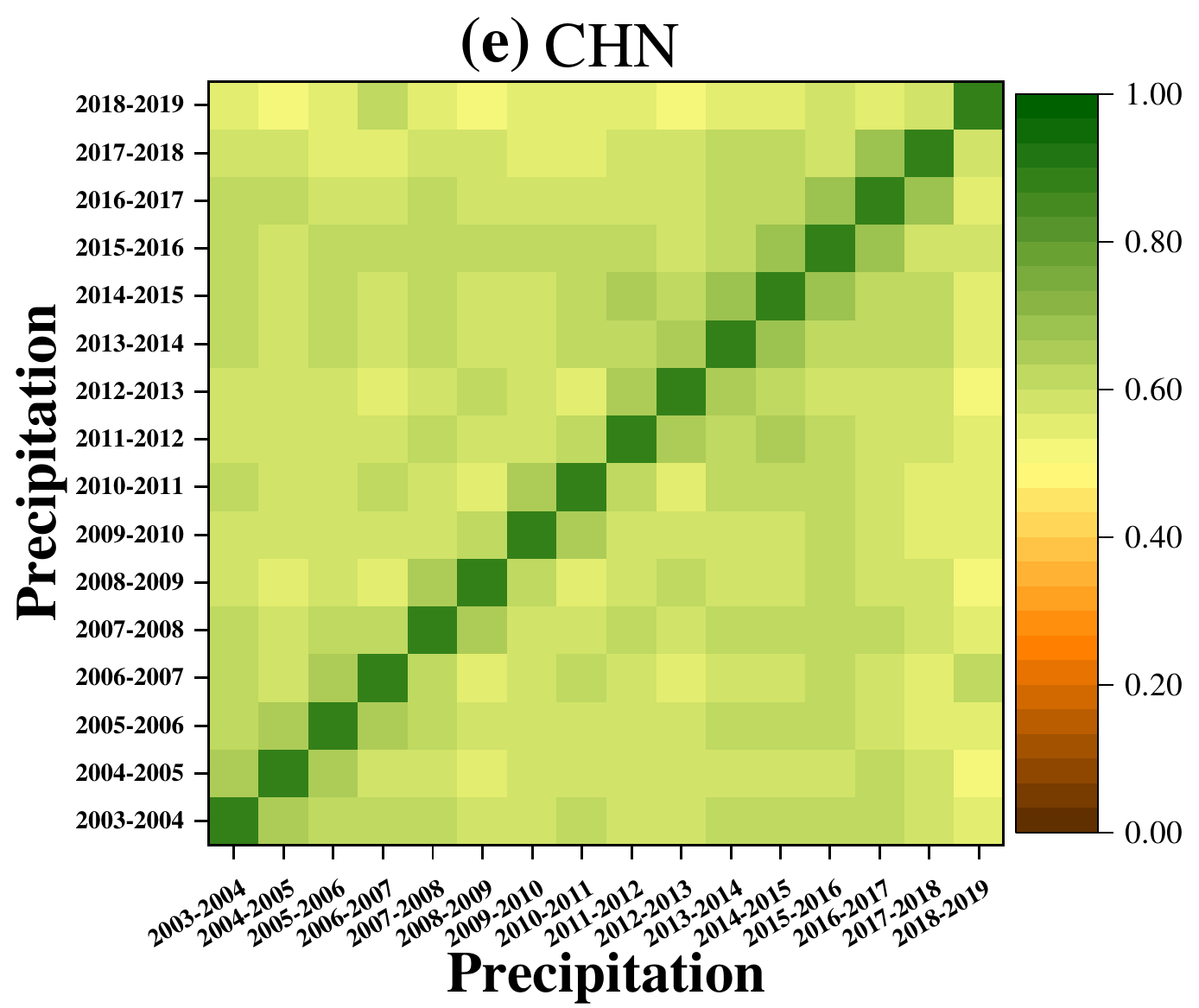}
\includegraphics[width=8em, height=7em]{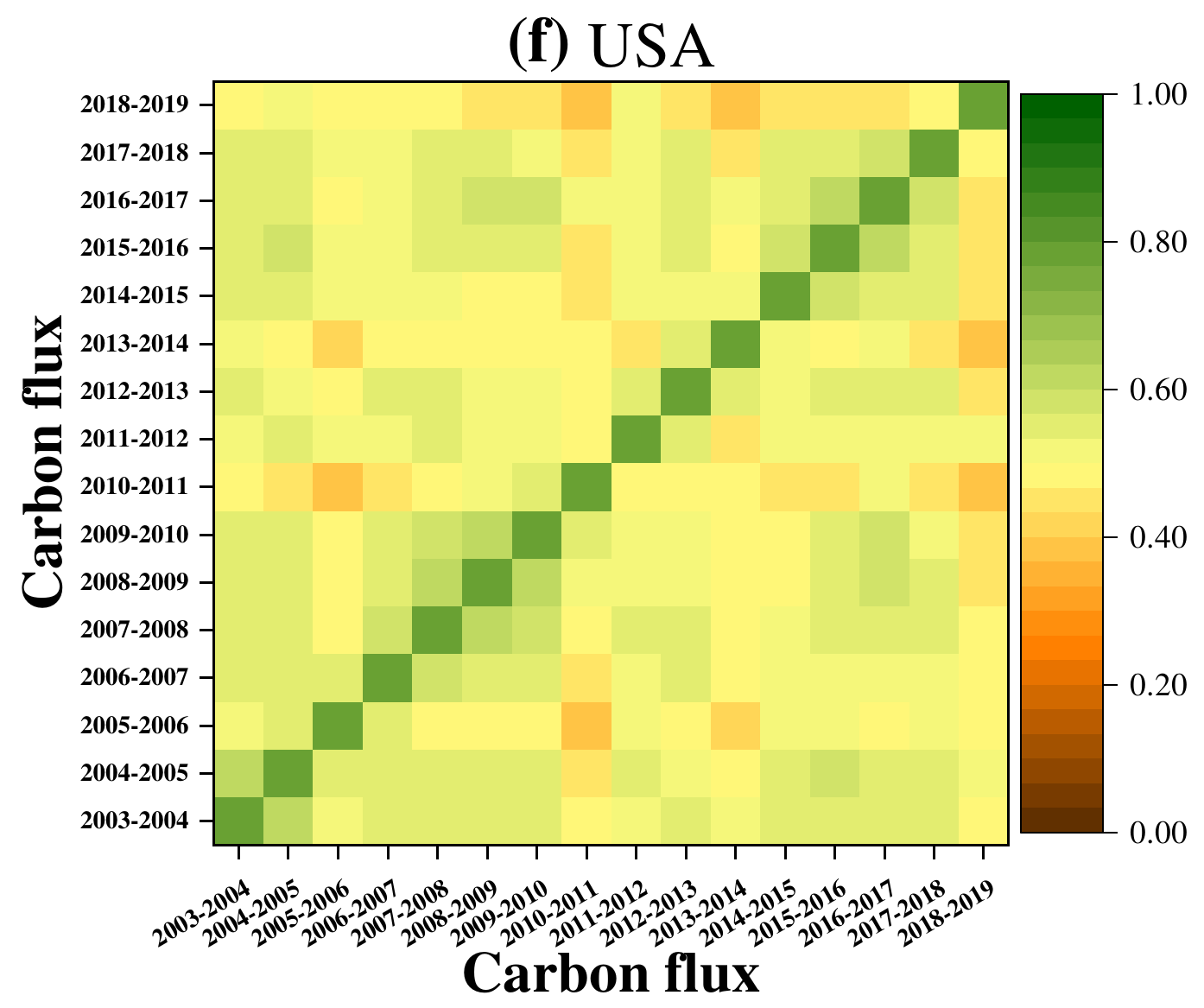}
\includegraphics[width=8em, height=7em]{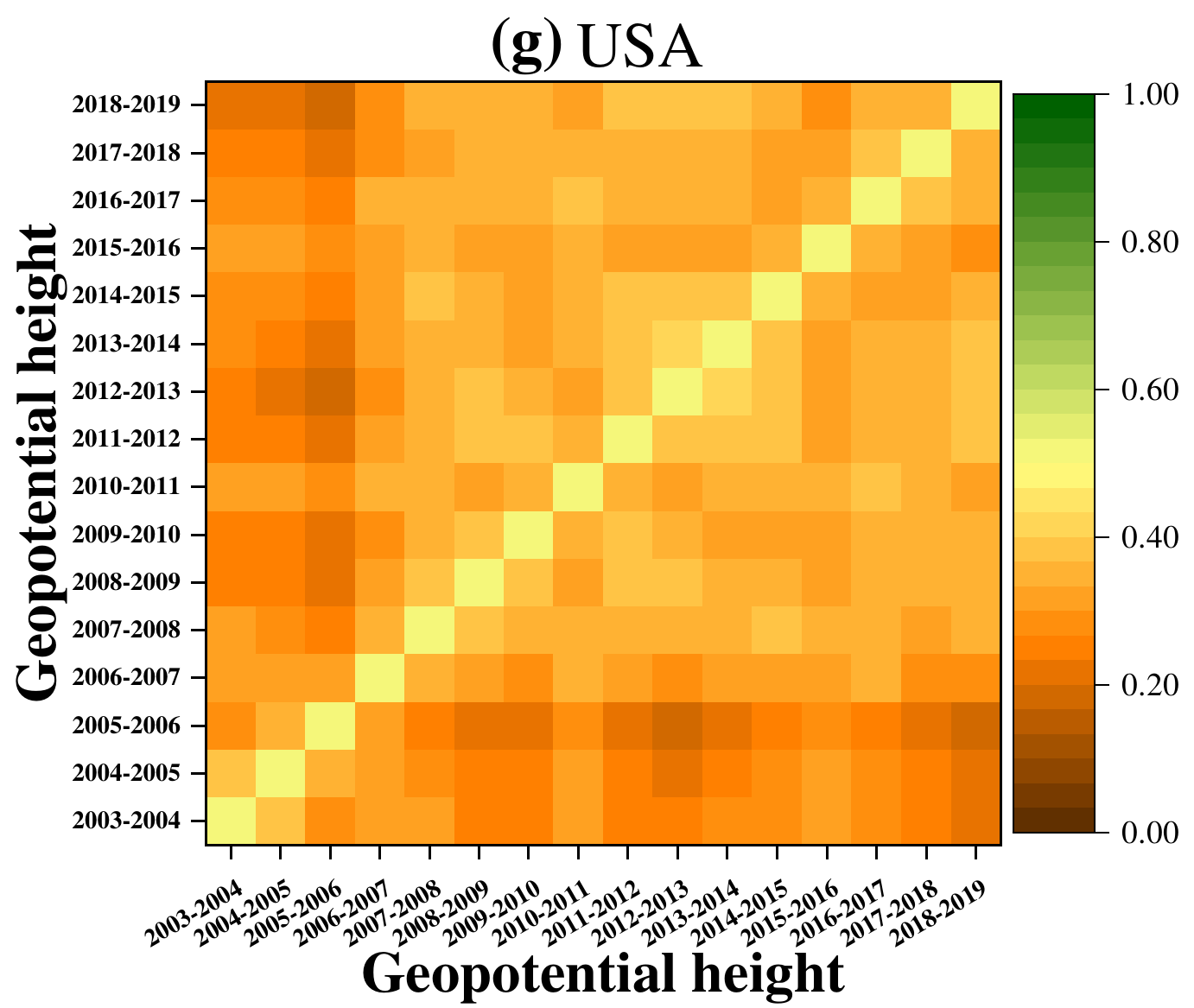}
\includegraphics[width=8em, height=7em]{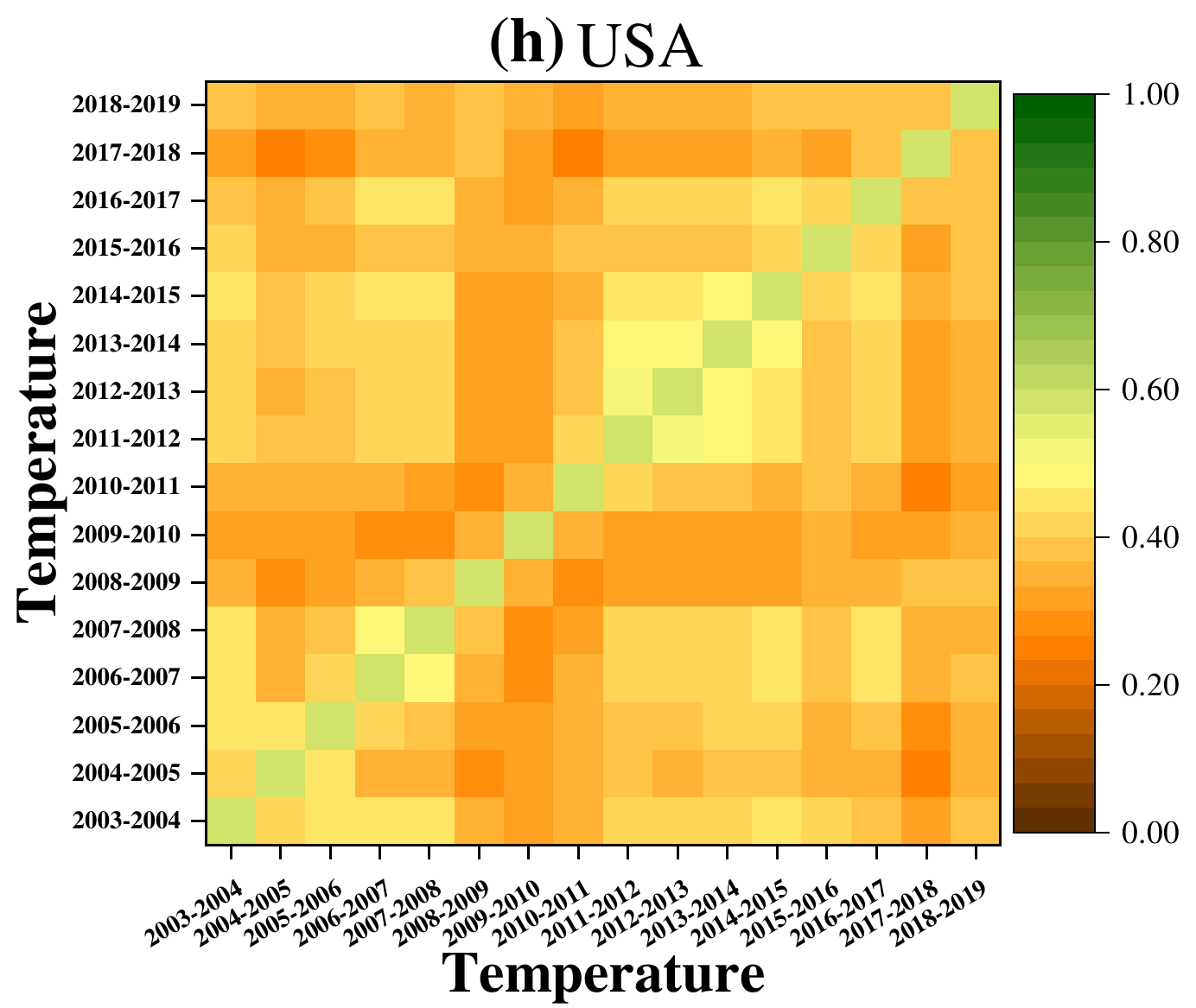}
\includegraphics[width=8em, height=7em]{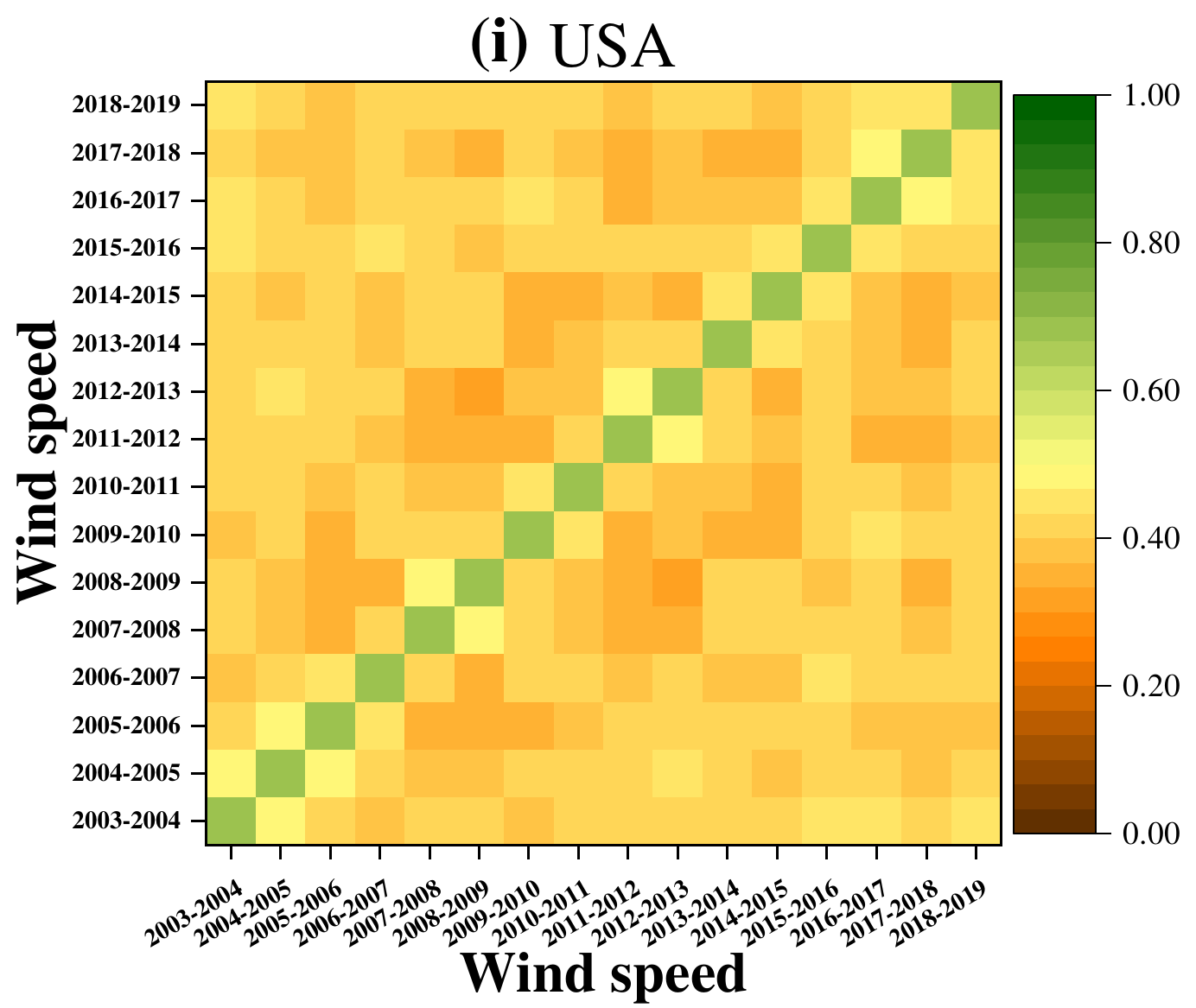}
\includegraphics[width=8em, height=7em]{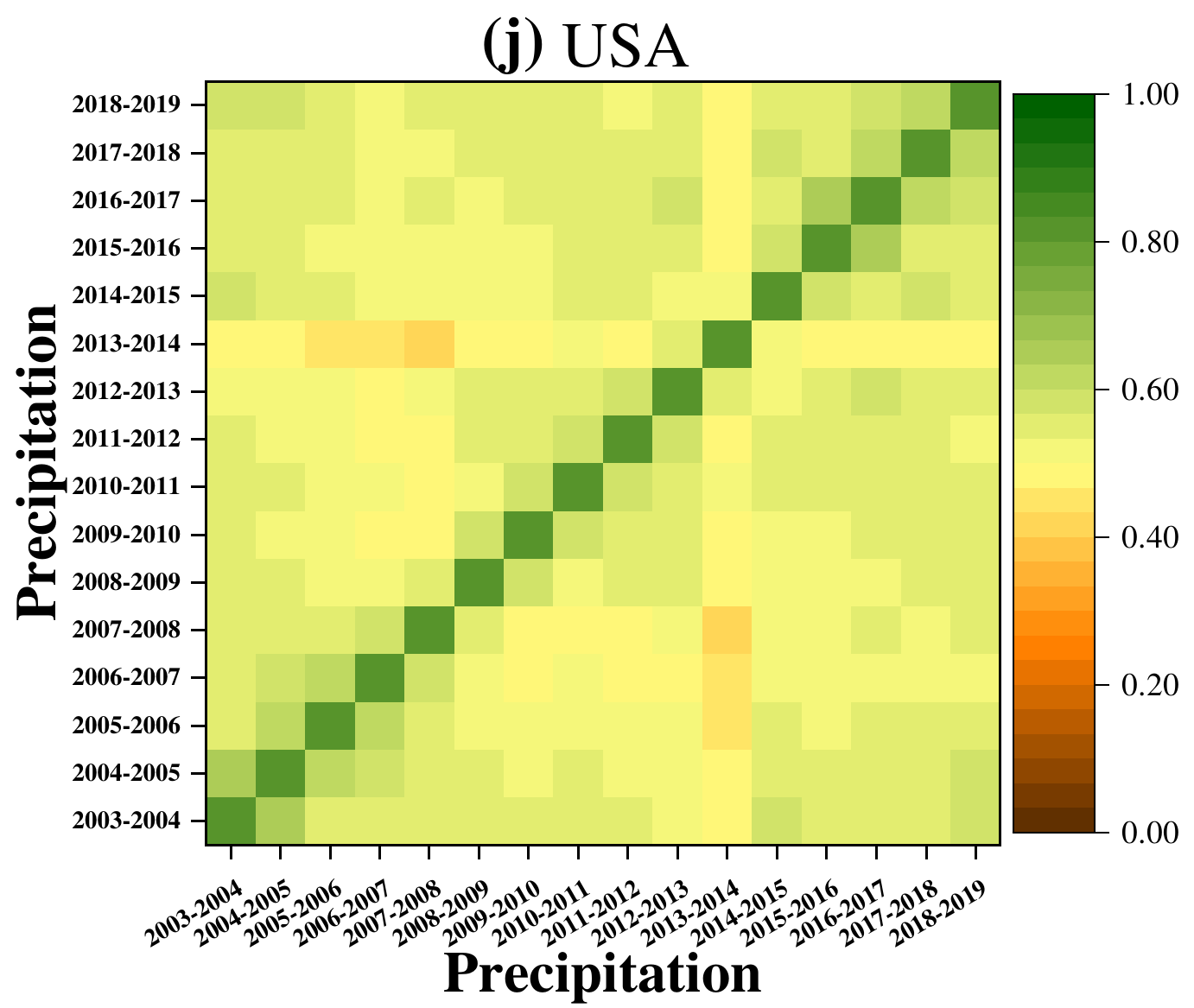}
\includegraphics[width=8em, height=7em]{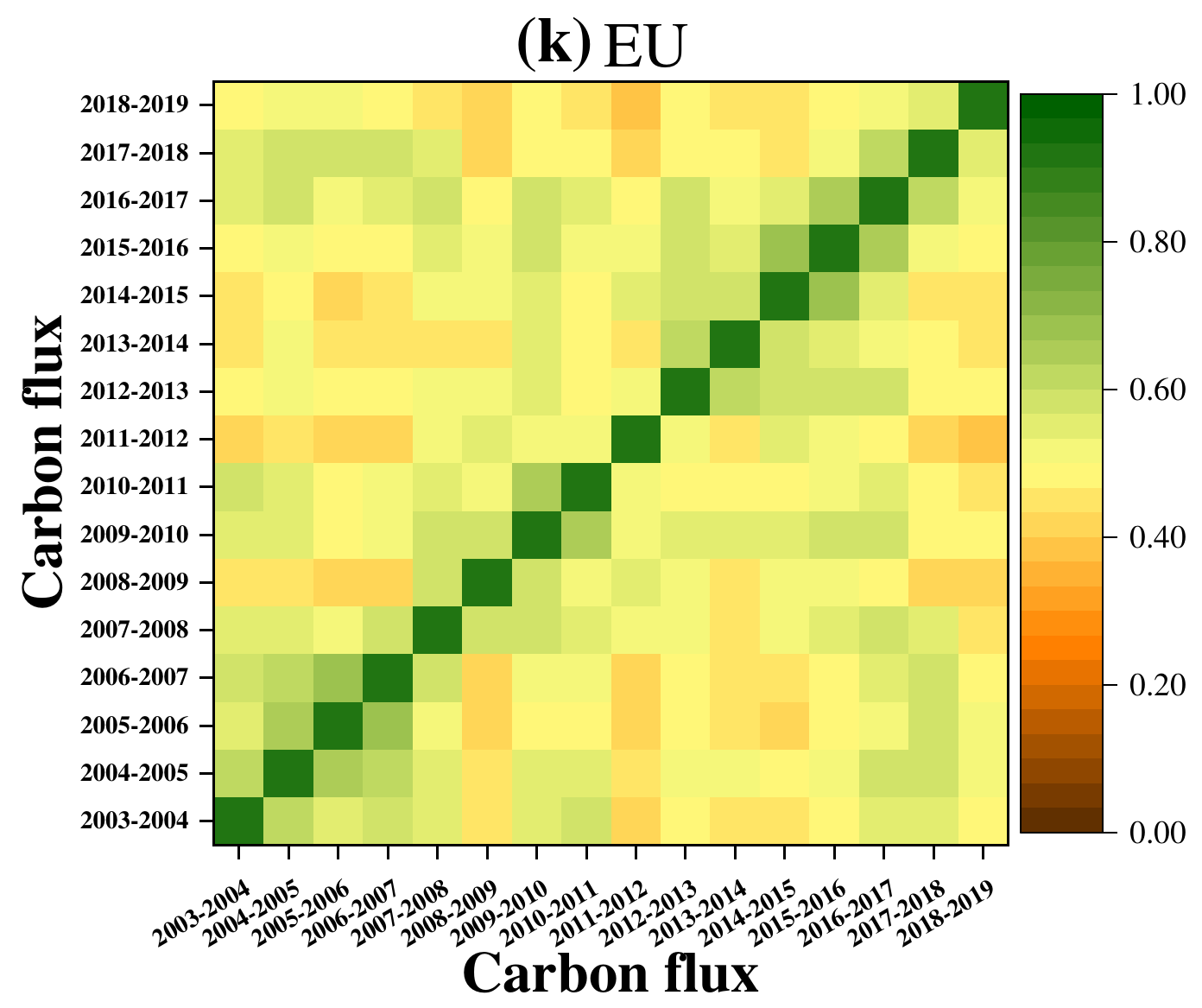}
\includegraphics[width=8em, height=7em]{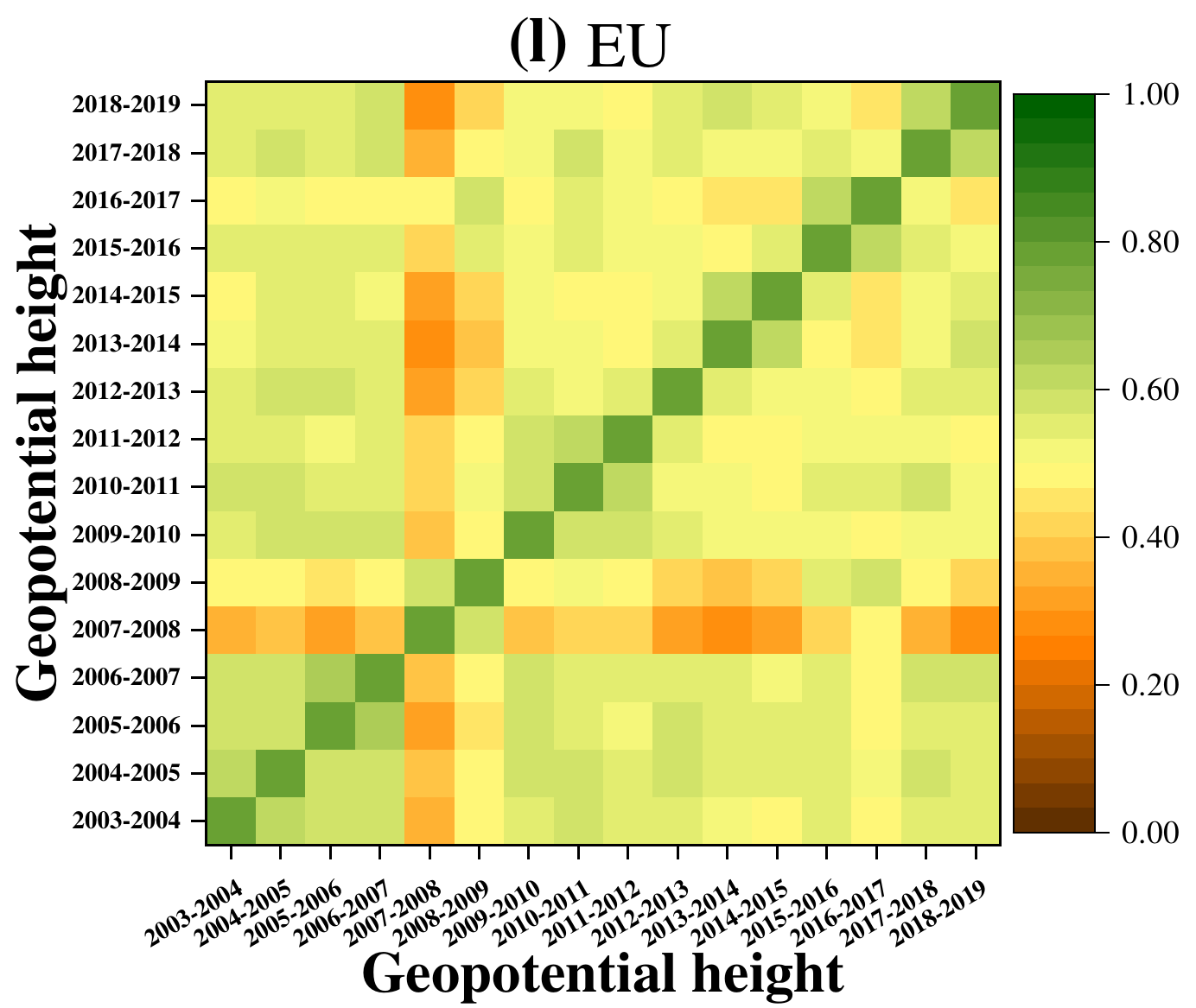}
\includegraphics[width=8em, height=7em]{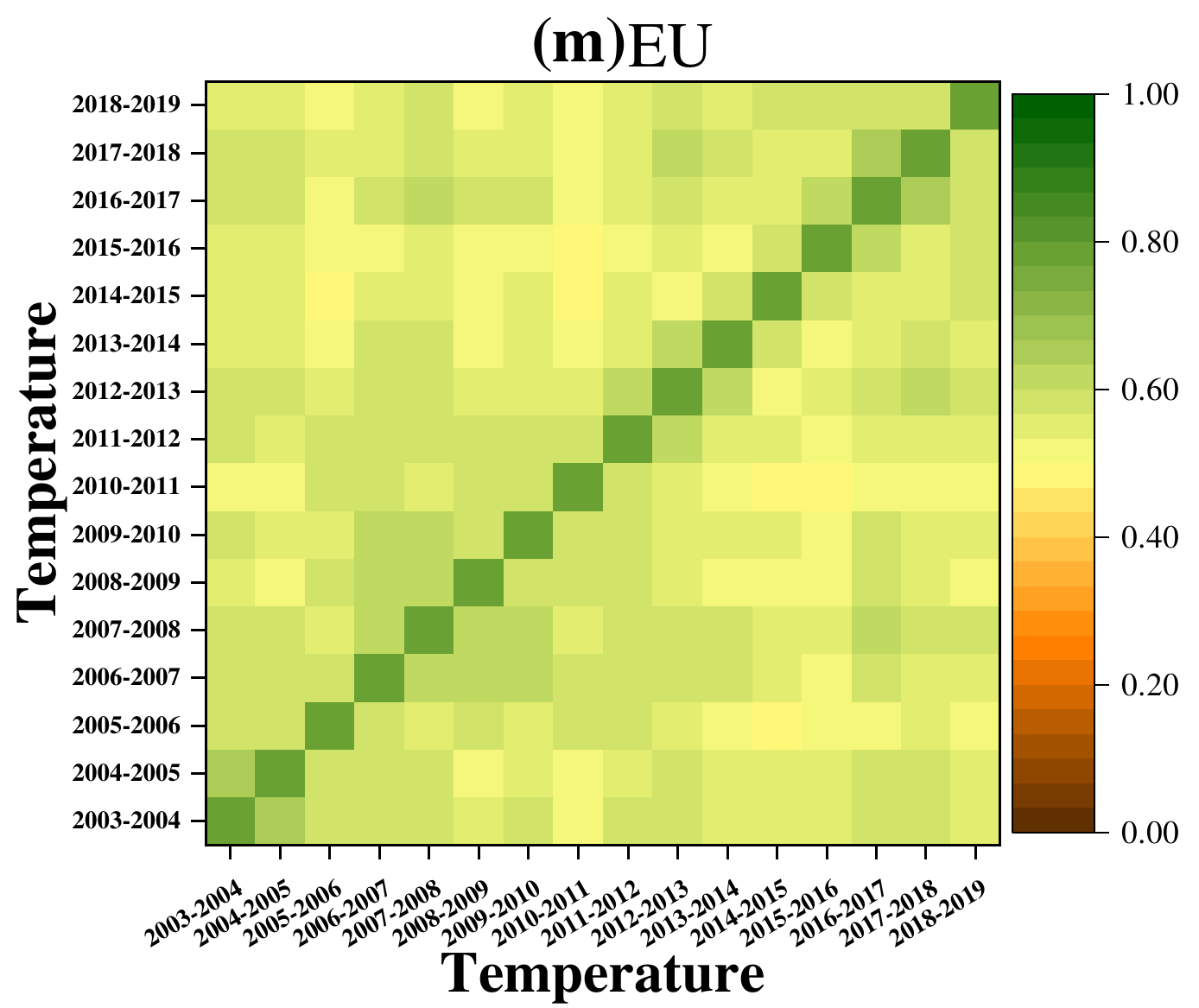}
\includegraphics[width=8em, height=7em]{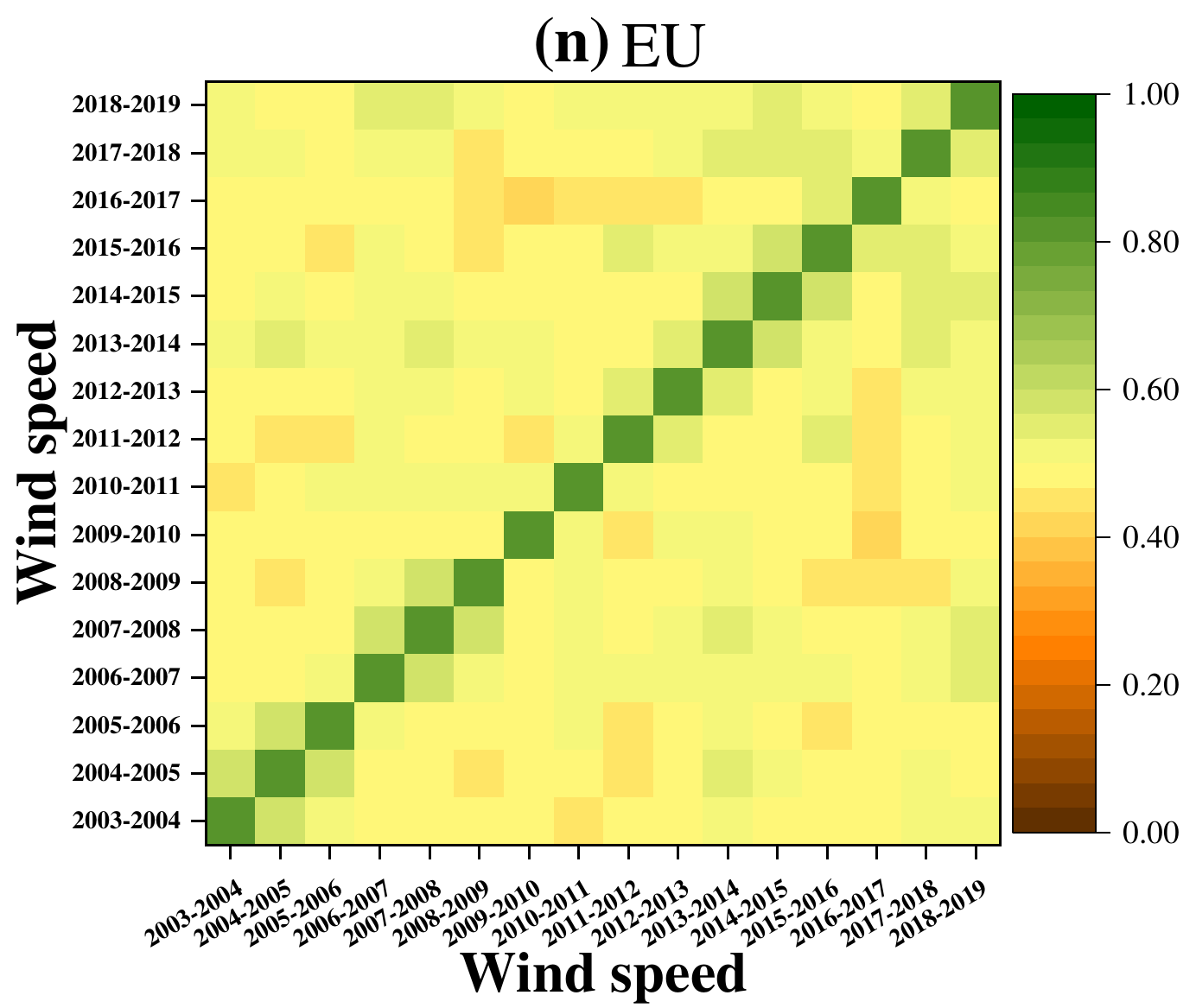}
\includegraphics[width=8em, height=7em]{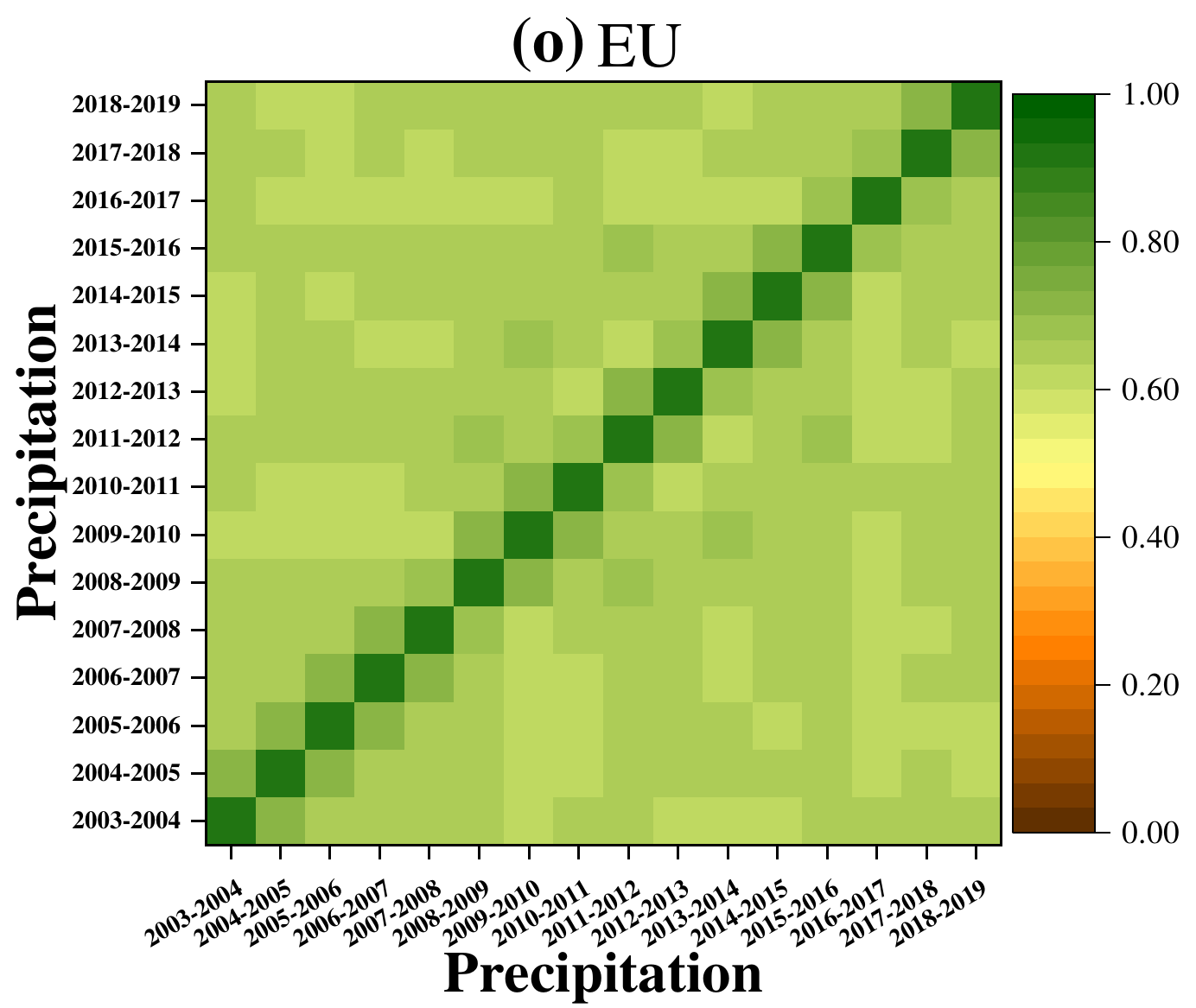}
\end{center}

\begin{center}
\noindent {\small {\bf Fig. S35} The effective Jaccard similarity coefficient matrix for links in two networks of different years for each of the climate variables. Each matrix element represents the difference between the actual Jaccard similarity coefficient and the corresponding average and standard deviation values obtained from the controlled case.}
\end{center}

\begin{center}
\includegraphics[width=8em, height=7em]{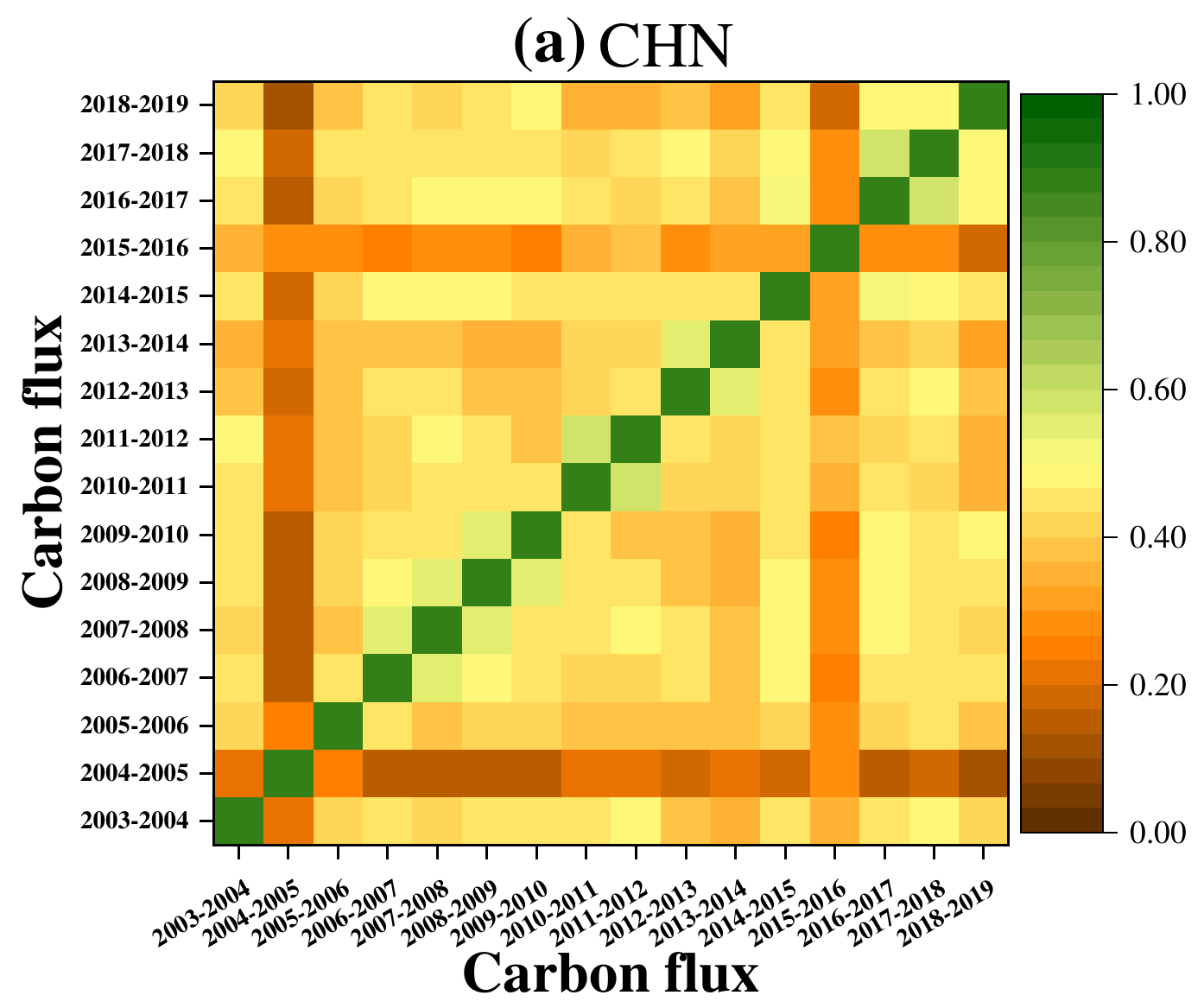}
\includegraphics[width=8em, height=7em]{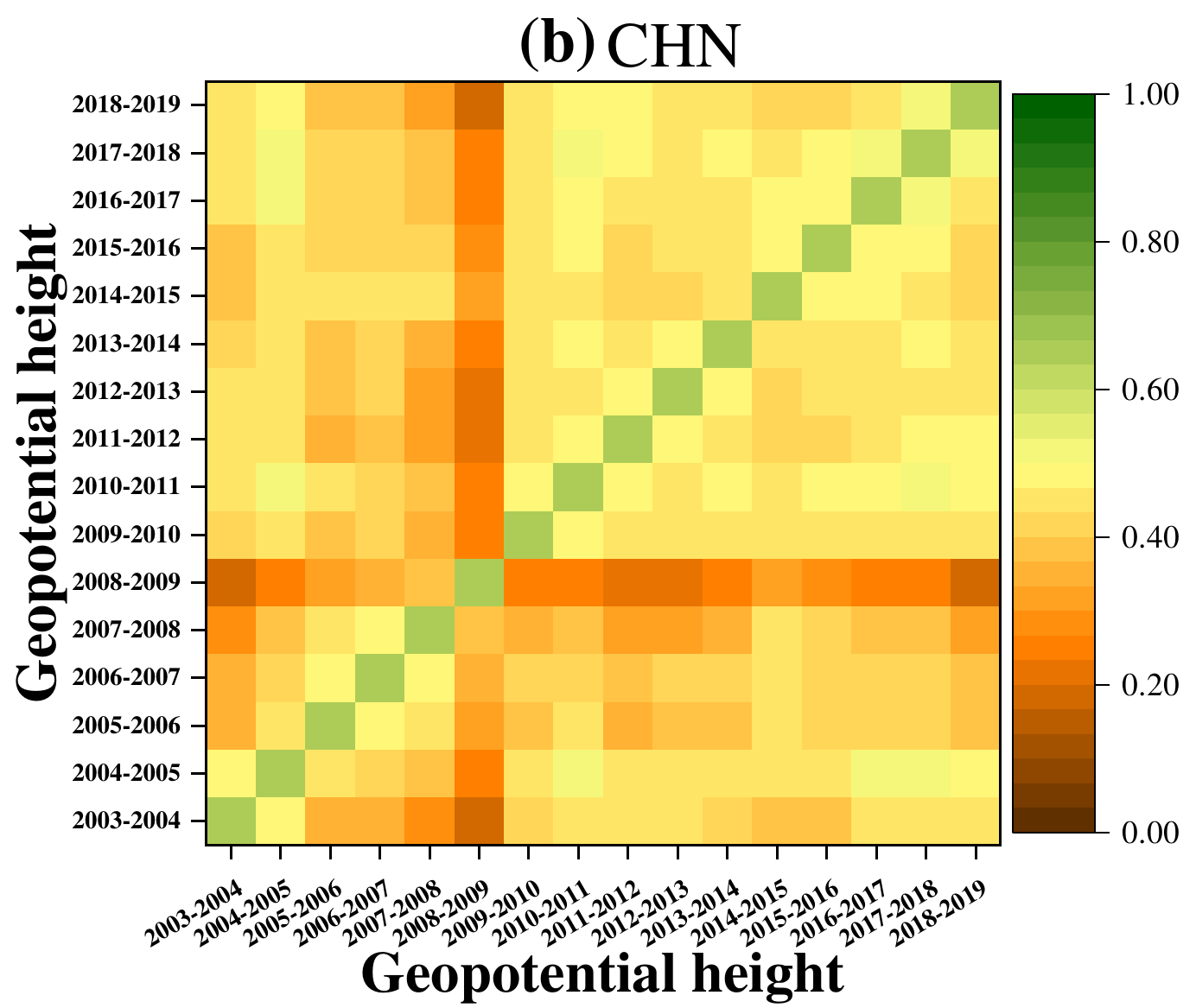}
\includegraphics[width=8em, height=7em]{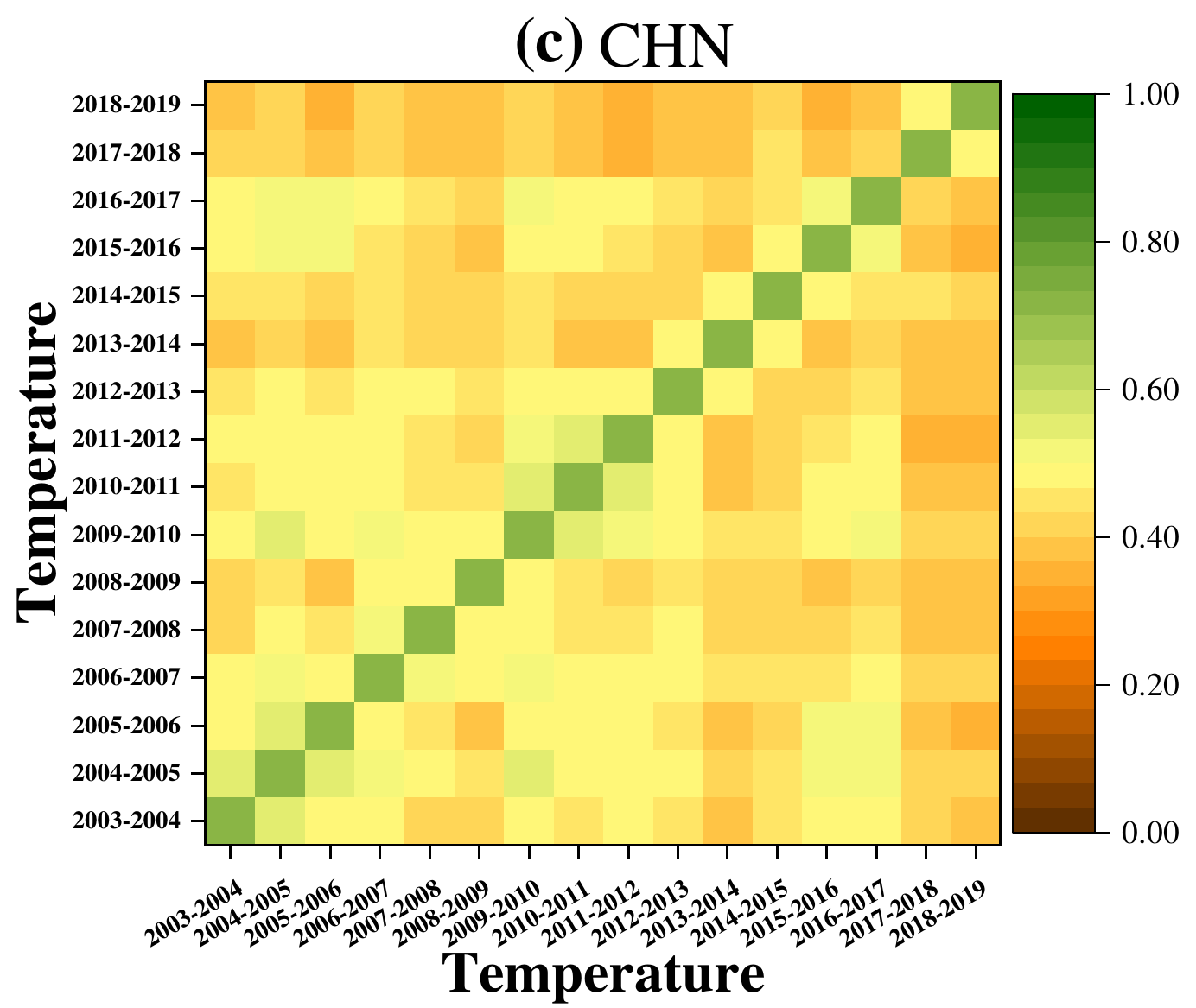}
\includegraphics[width=8em, height=7em]{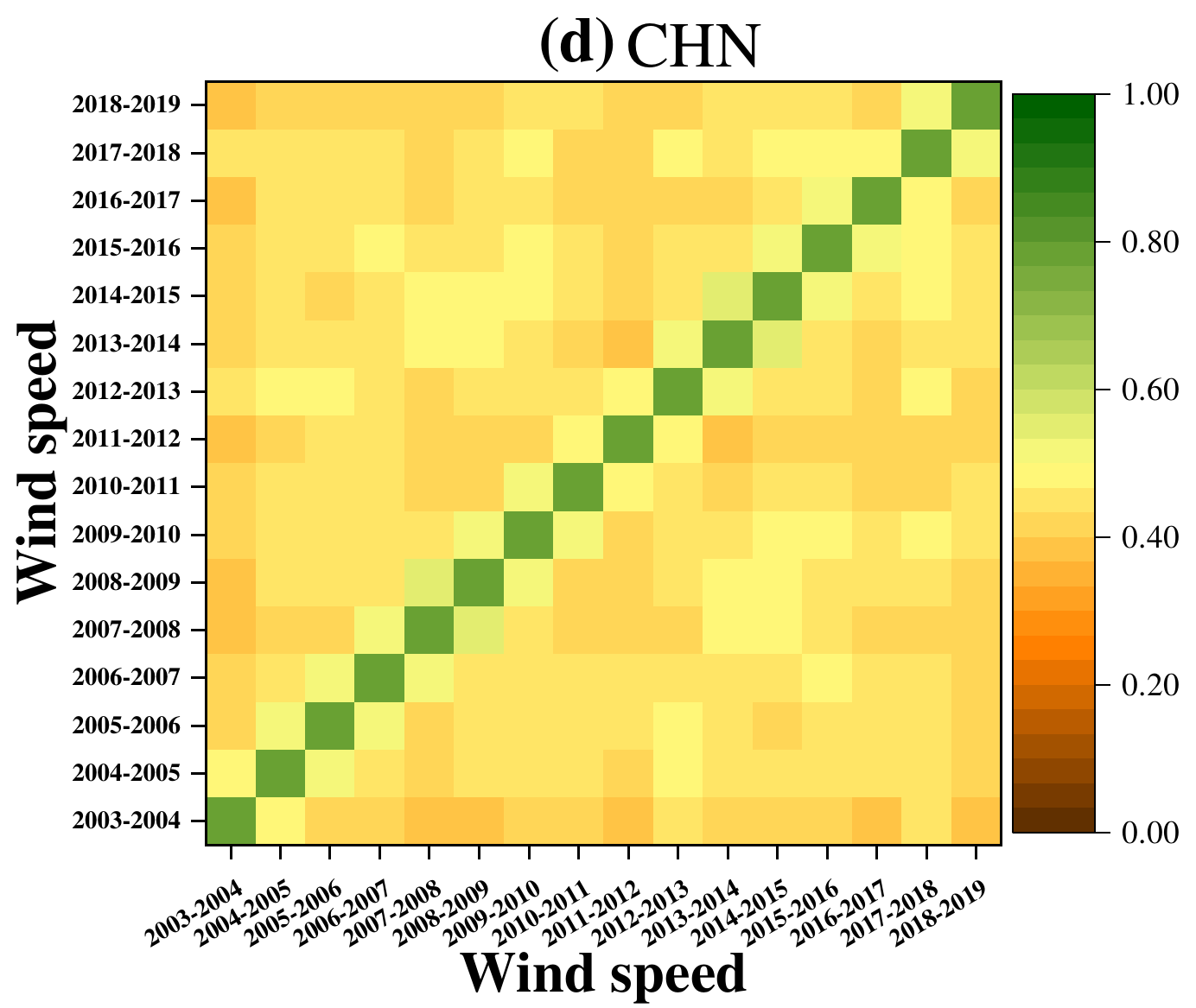}
\includegraphics[width=8em, height=7em]{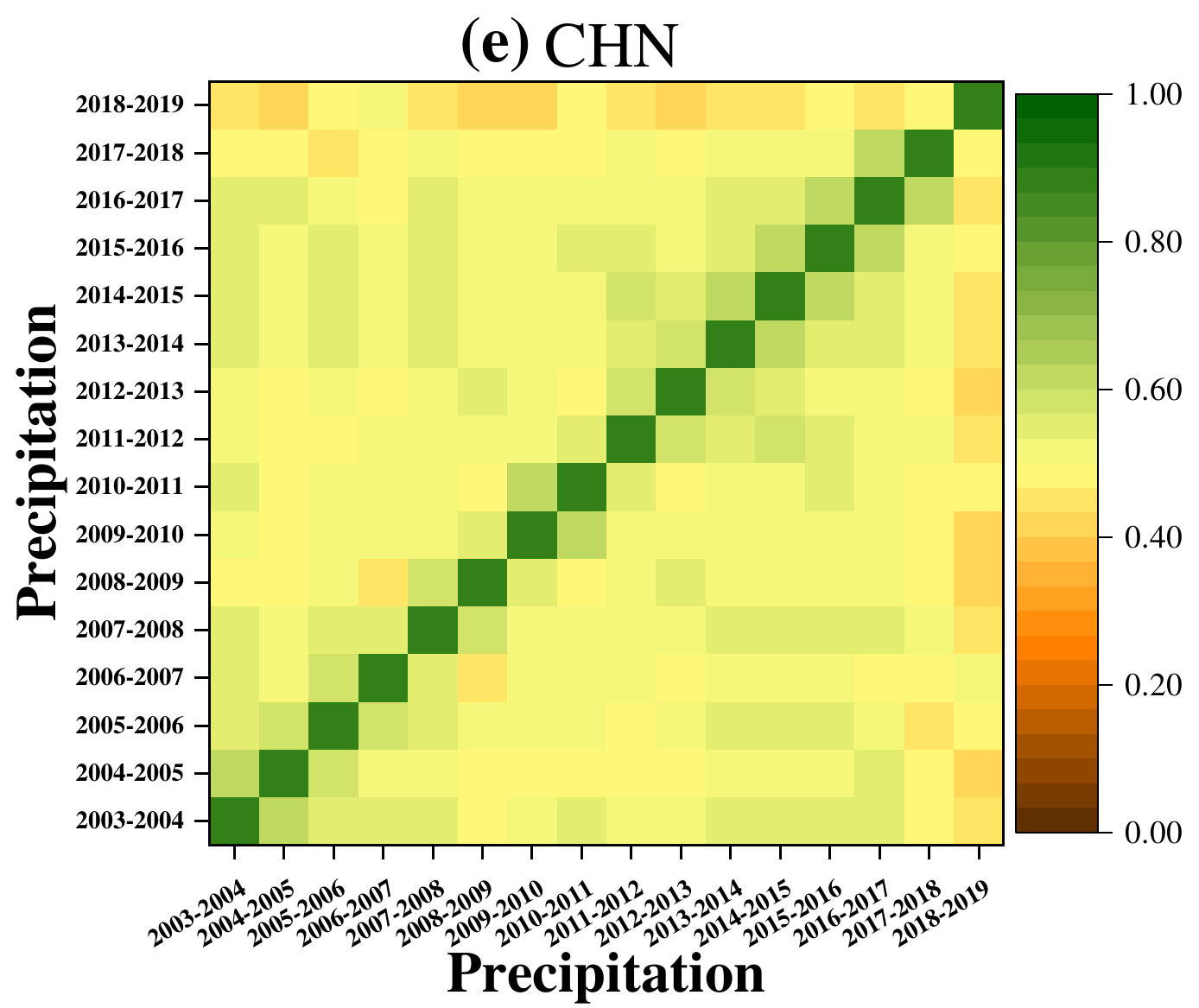}
\includegraphics[width=8em, height=7em]{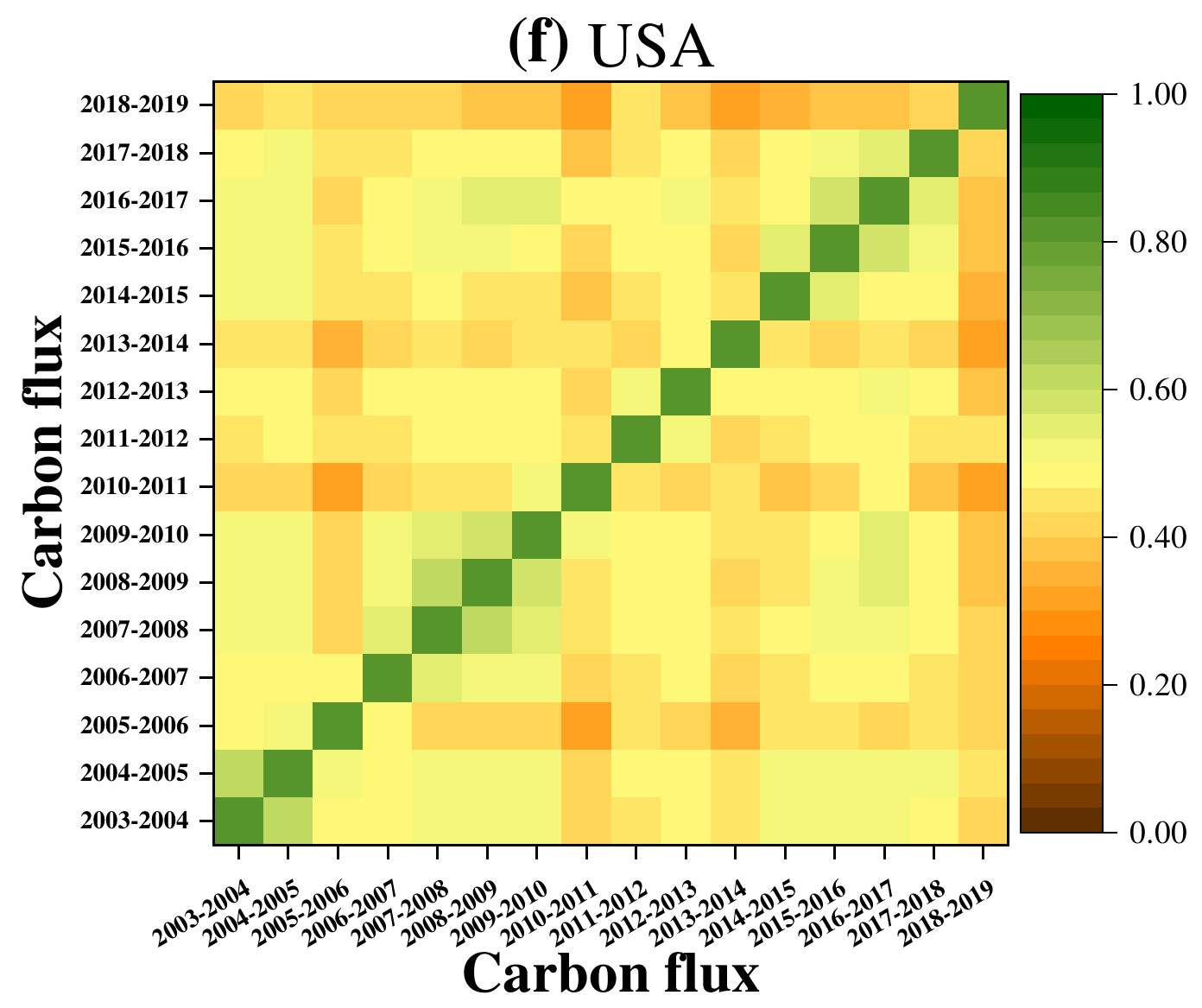}
\includegraphics[width=8em, height=7em]{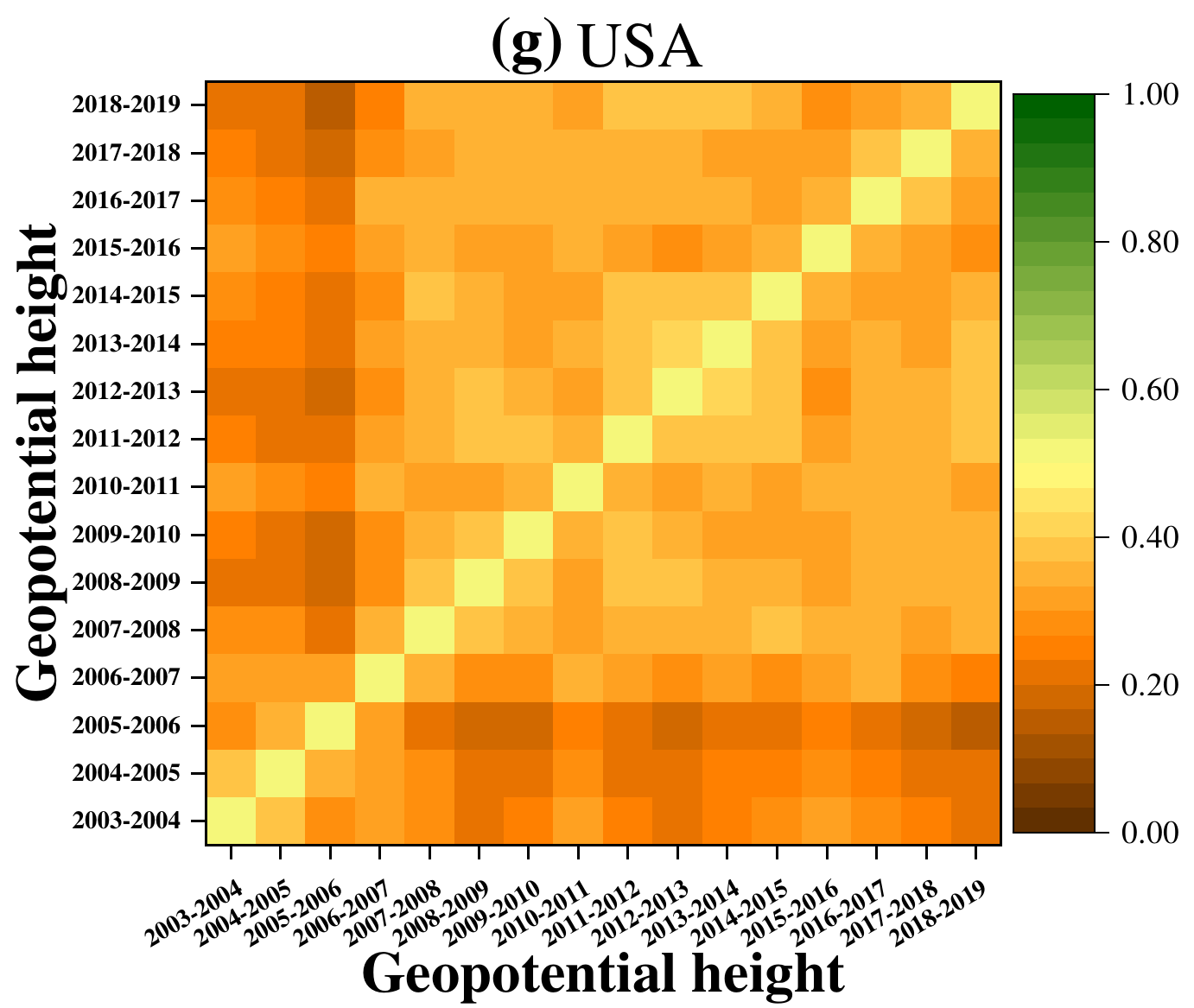}
\includegraphics[width=8em, height=7em]{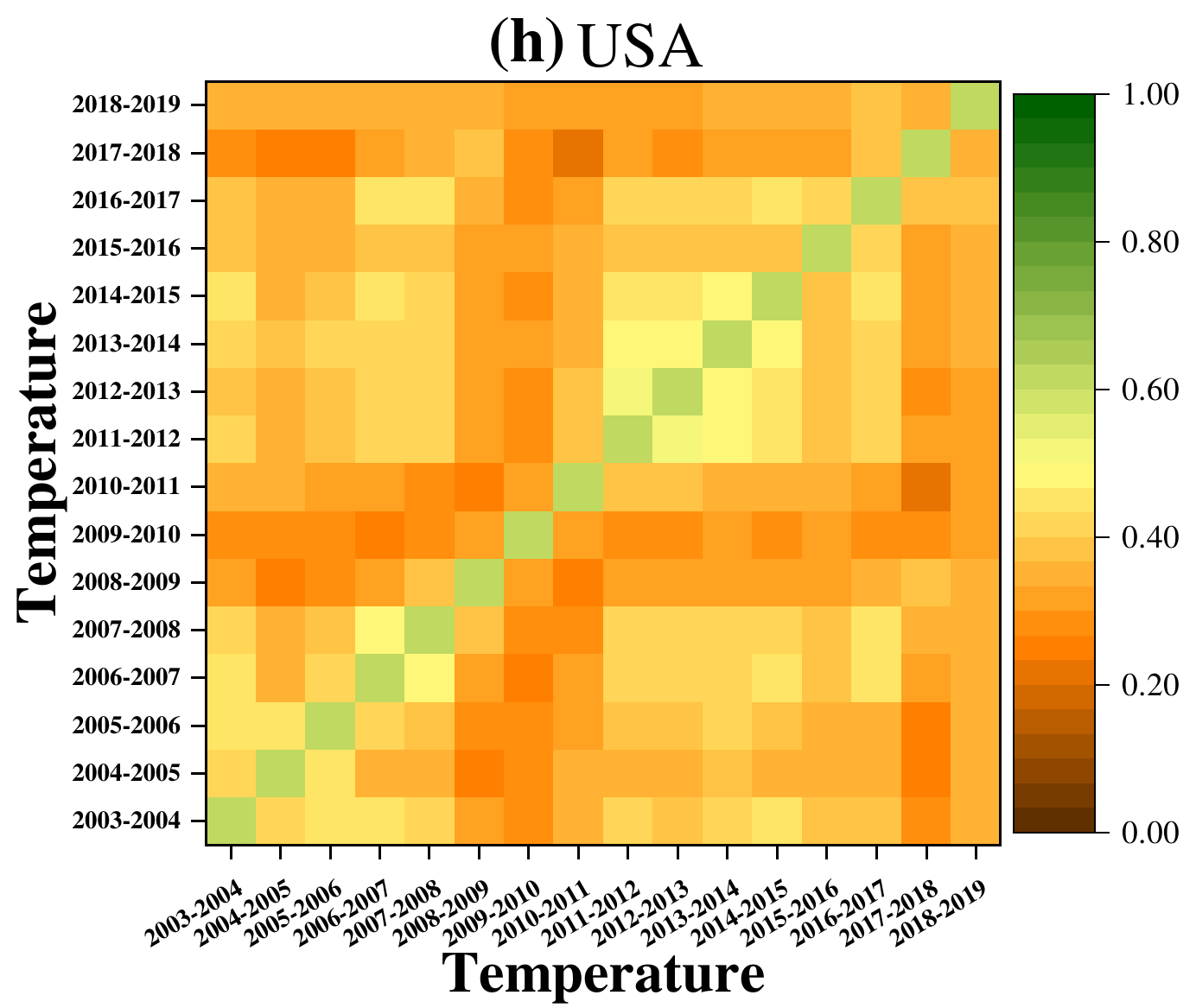}
\includegraphics[width=8em, height=7em]{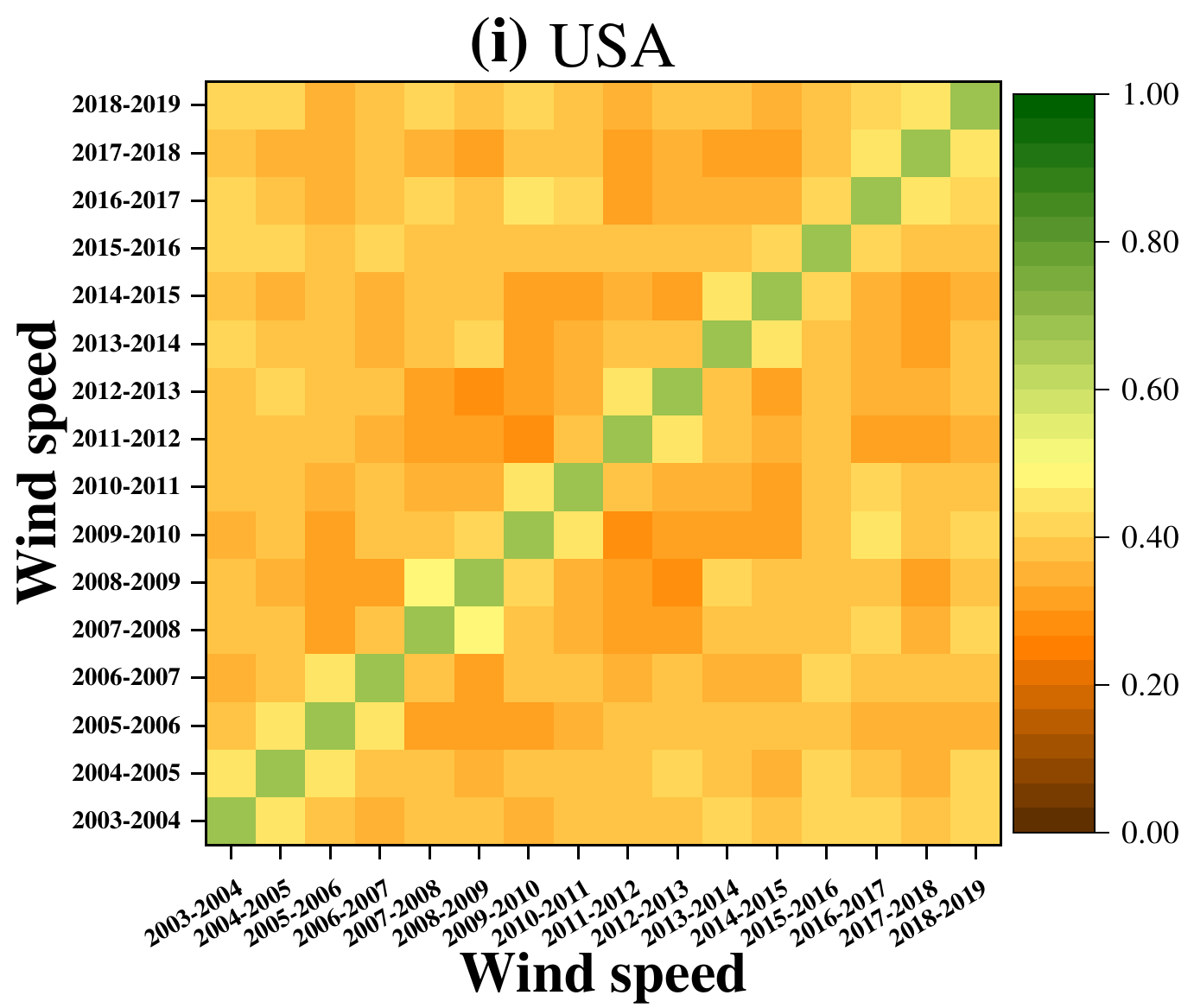}
\includegraphics[width=8em, height=7em]{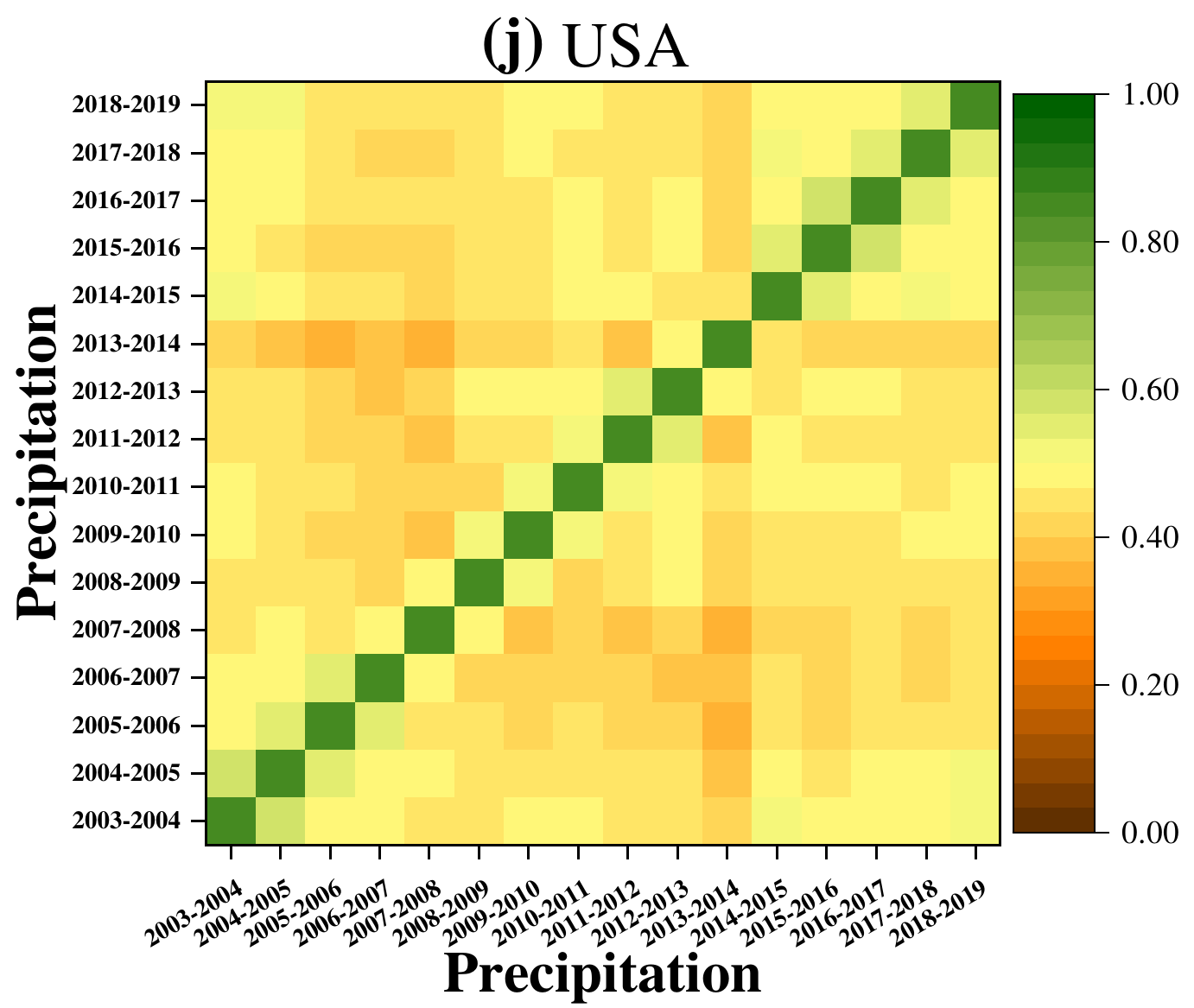}
\includegraphics[width=8em, height=7em]{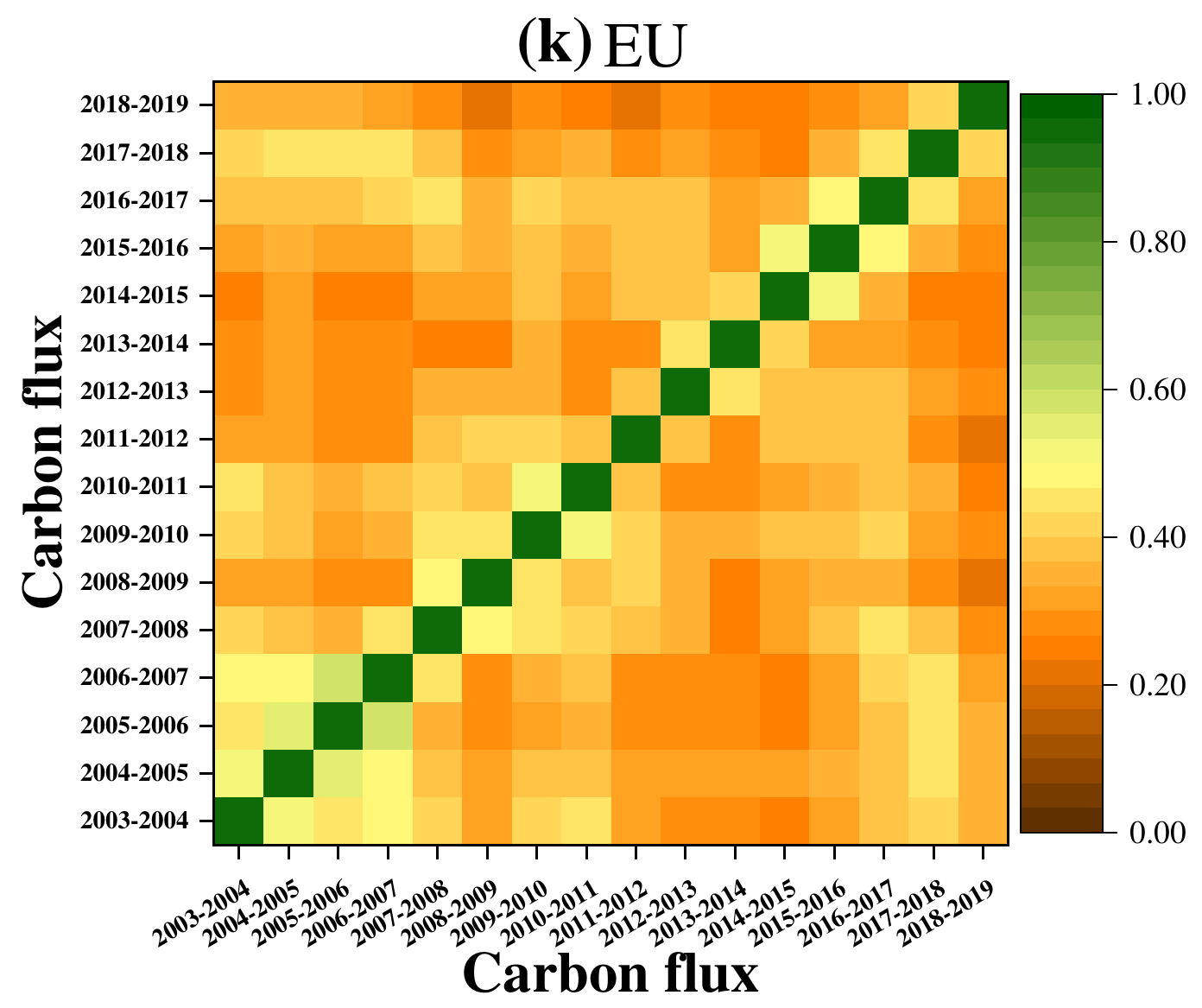}
\includegraphics[width=8em, height=7em]{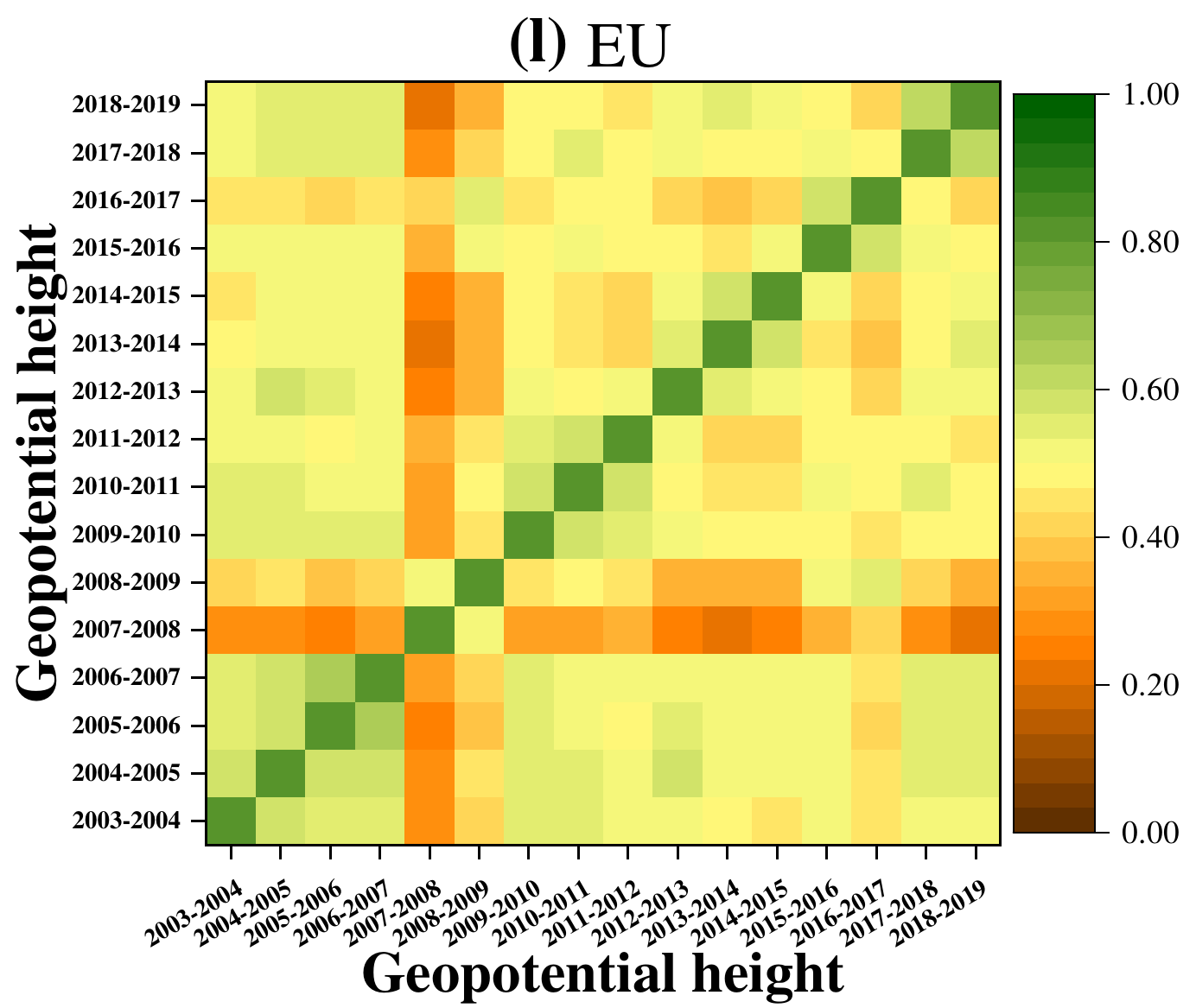}
\includegraphics[width=8em, height=7em]{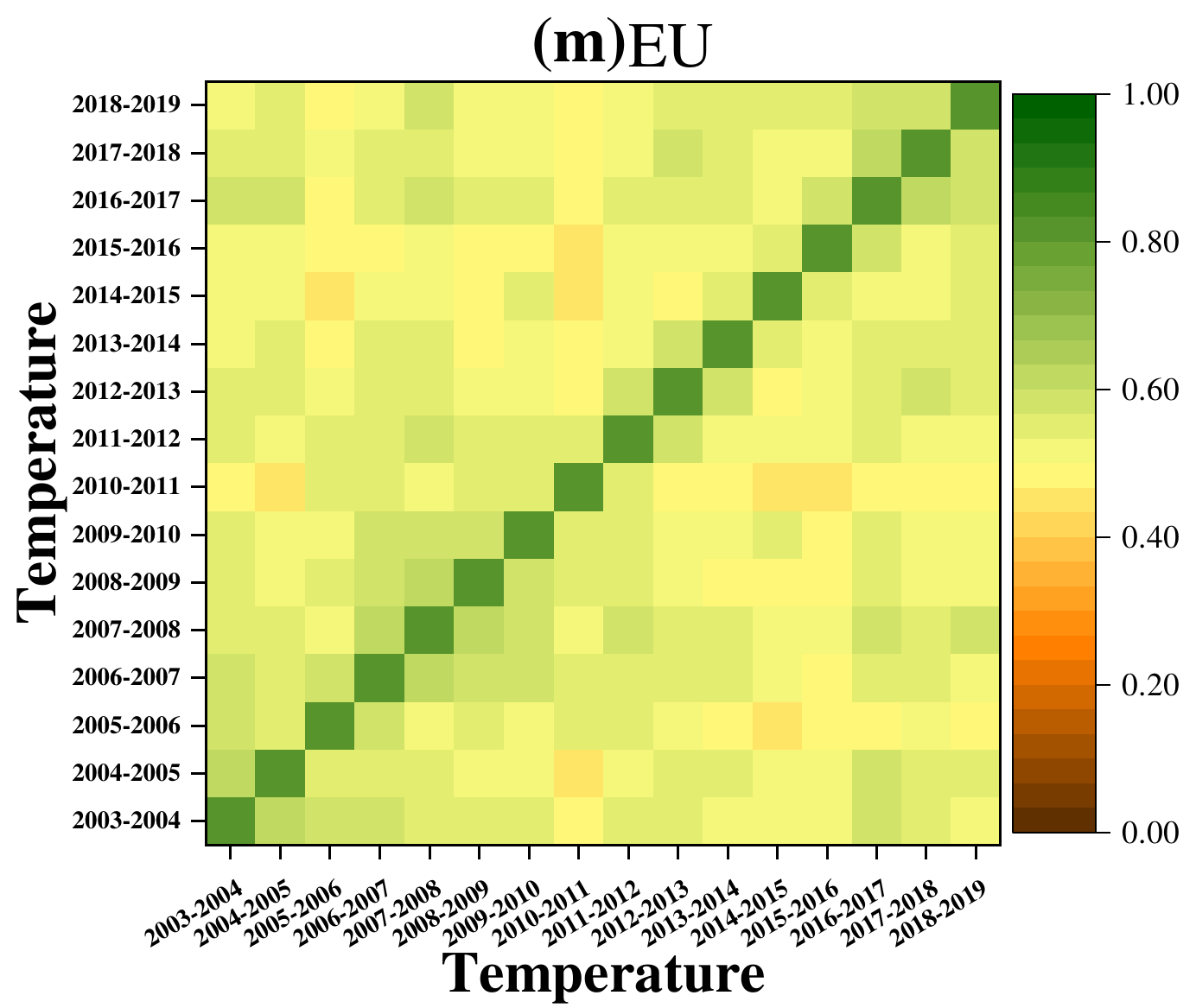}
\includegraphics[width=8em, height=7em]{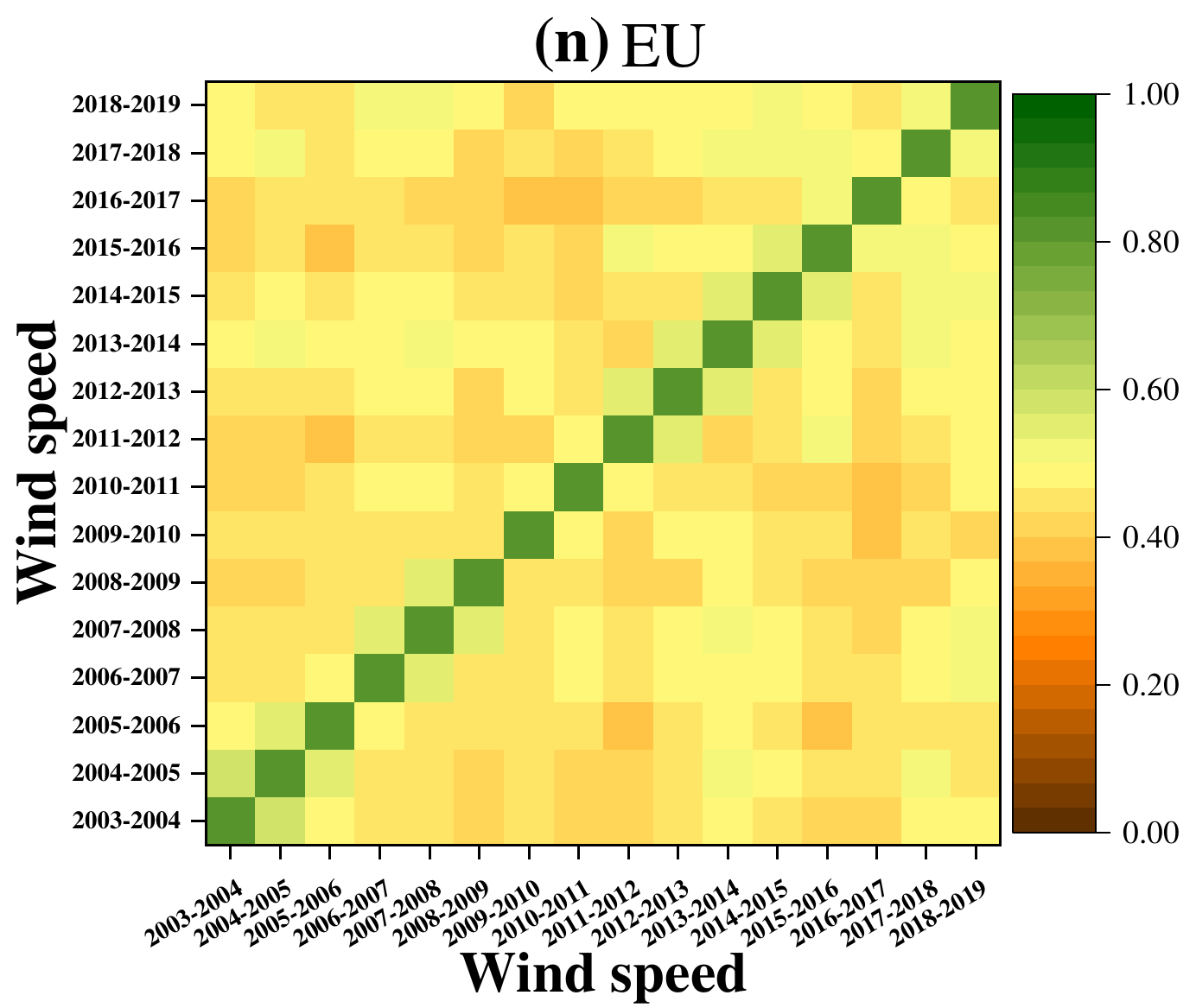}
\includegraphics[width=8em, height=7em]{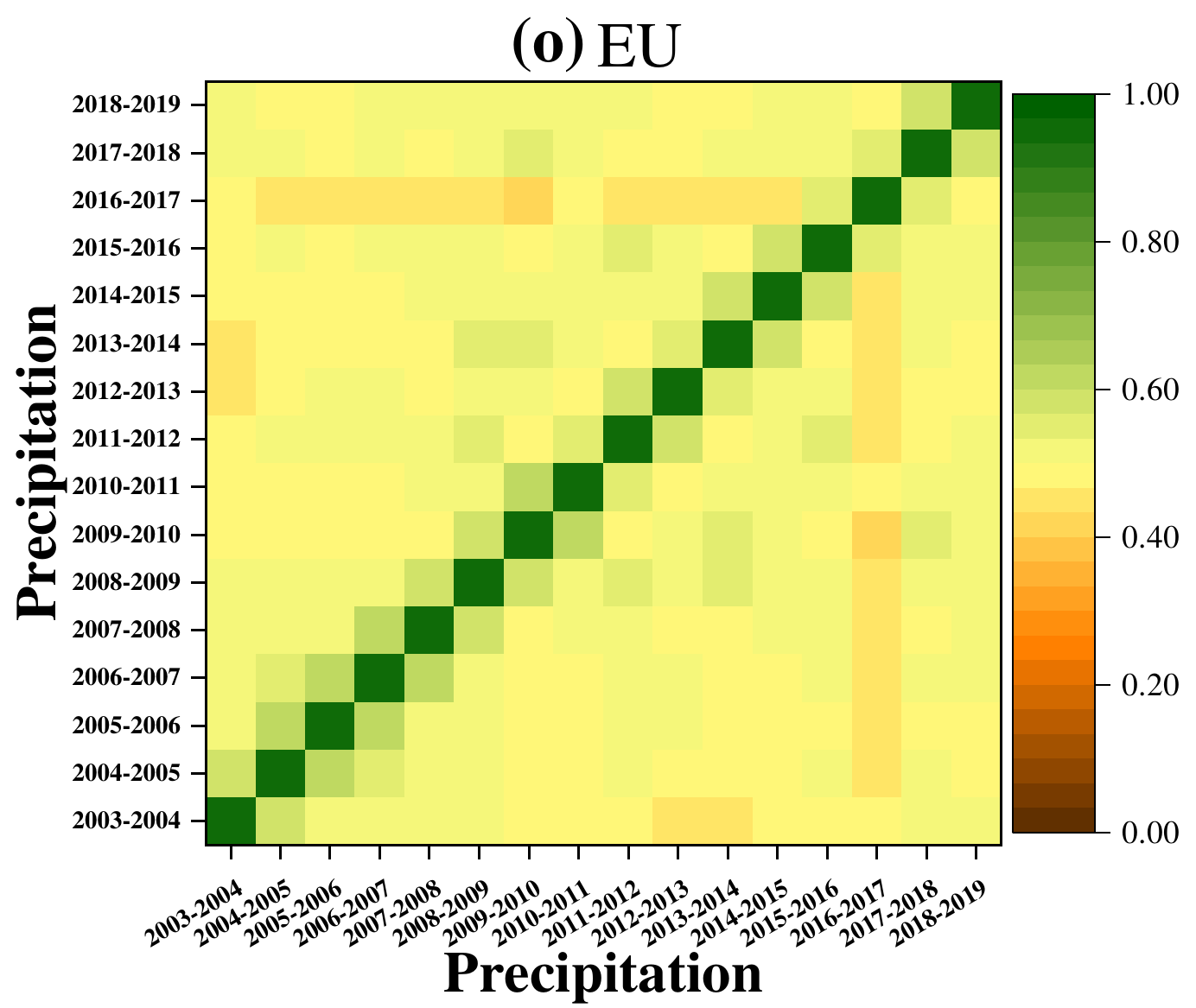}
\end{center}

\begin{center}
\noindent {\small {\bf Fig. S36} The effective Jaccard similarity coefficient matrix for links of lengths above $500km$ in two networks of different years for each of the climate variables. Each matrix element represents the difference between the actual Jaccard similarity coefficient and the corresponding average and standard deviation values obtained from the controlled case.}
\end{center}

\begin{center}
\includegraphics[width=8em, height=7em]{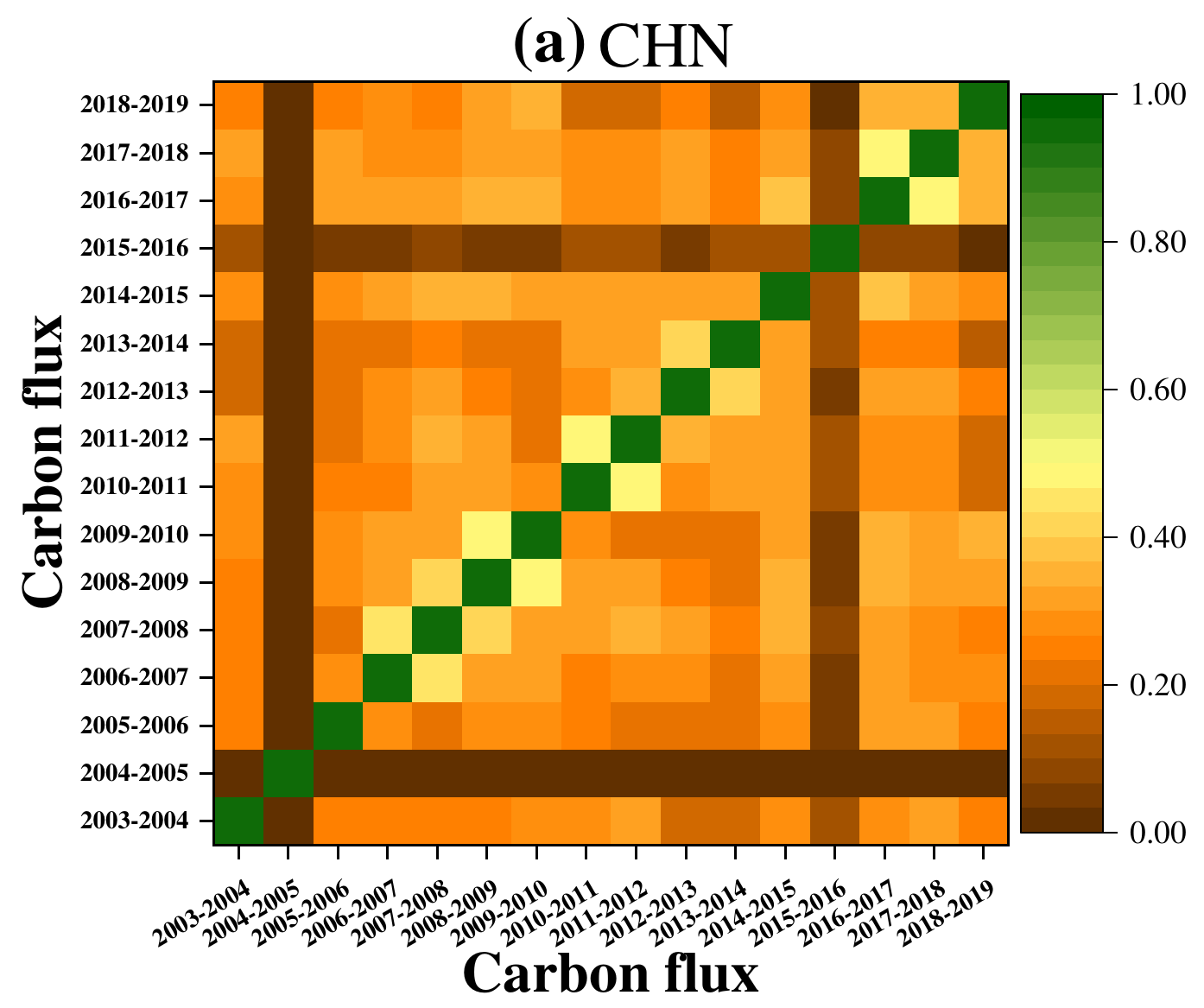}
\includegraphics[width=8em, height=7em]{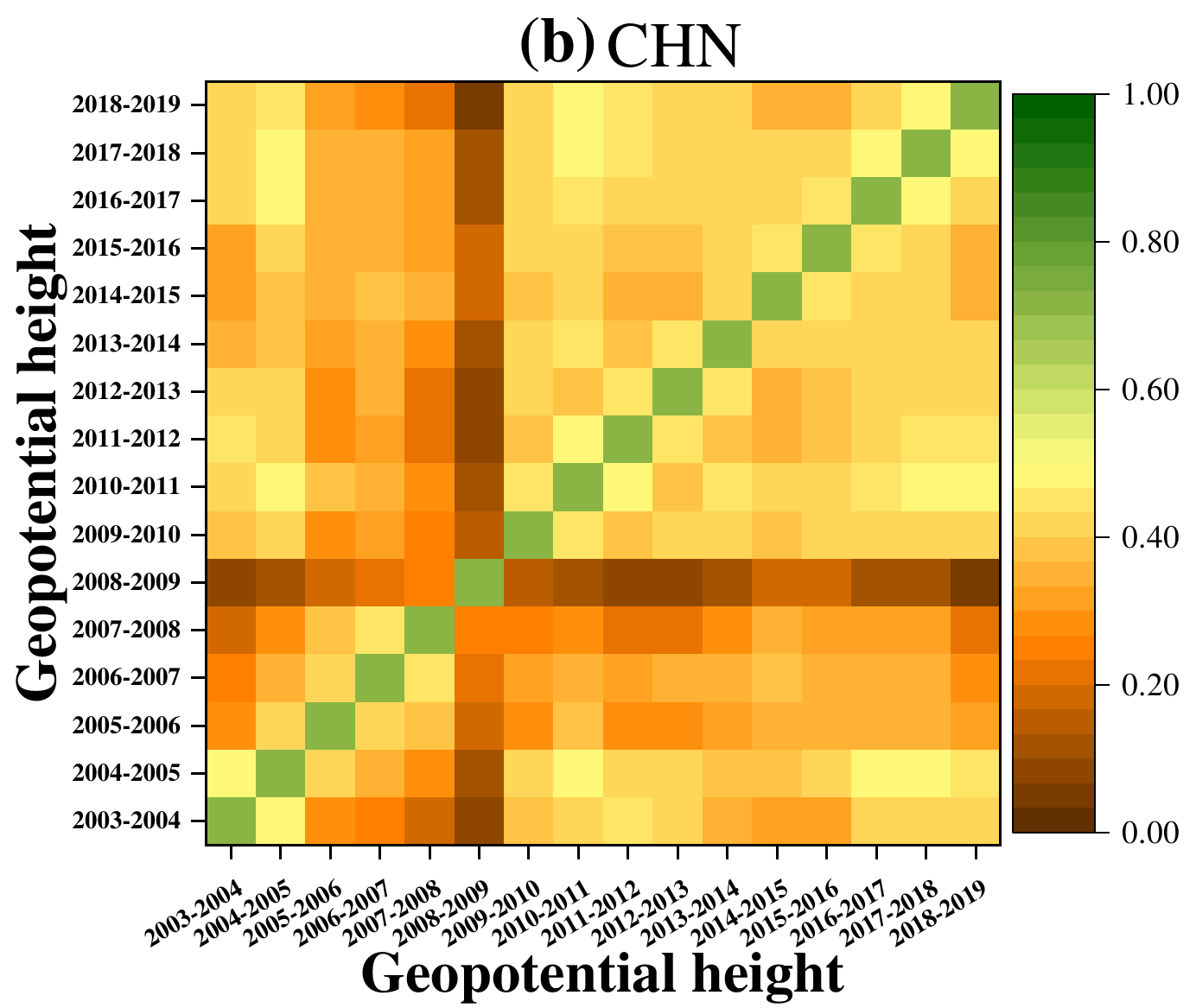}
\includegraphics[width=8em, height=7em]{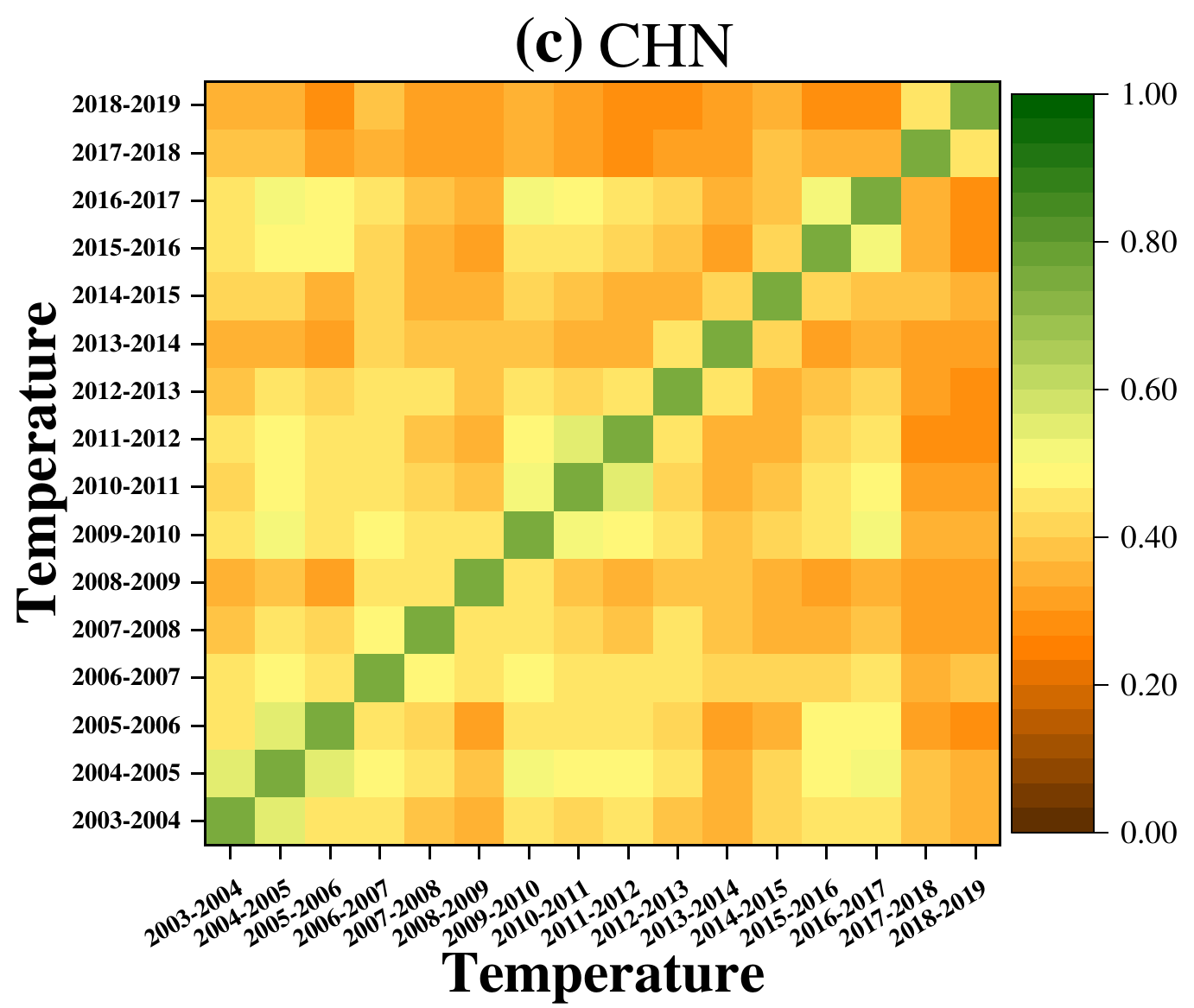}
\includegraphics[width=8em, height=7em]{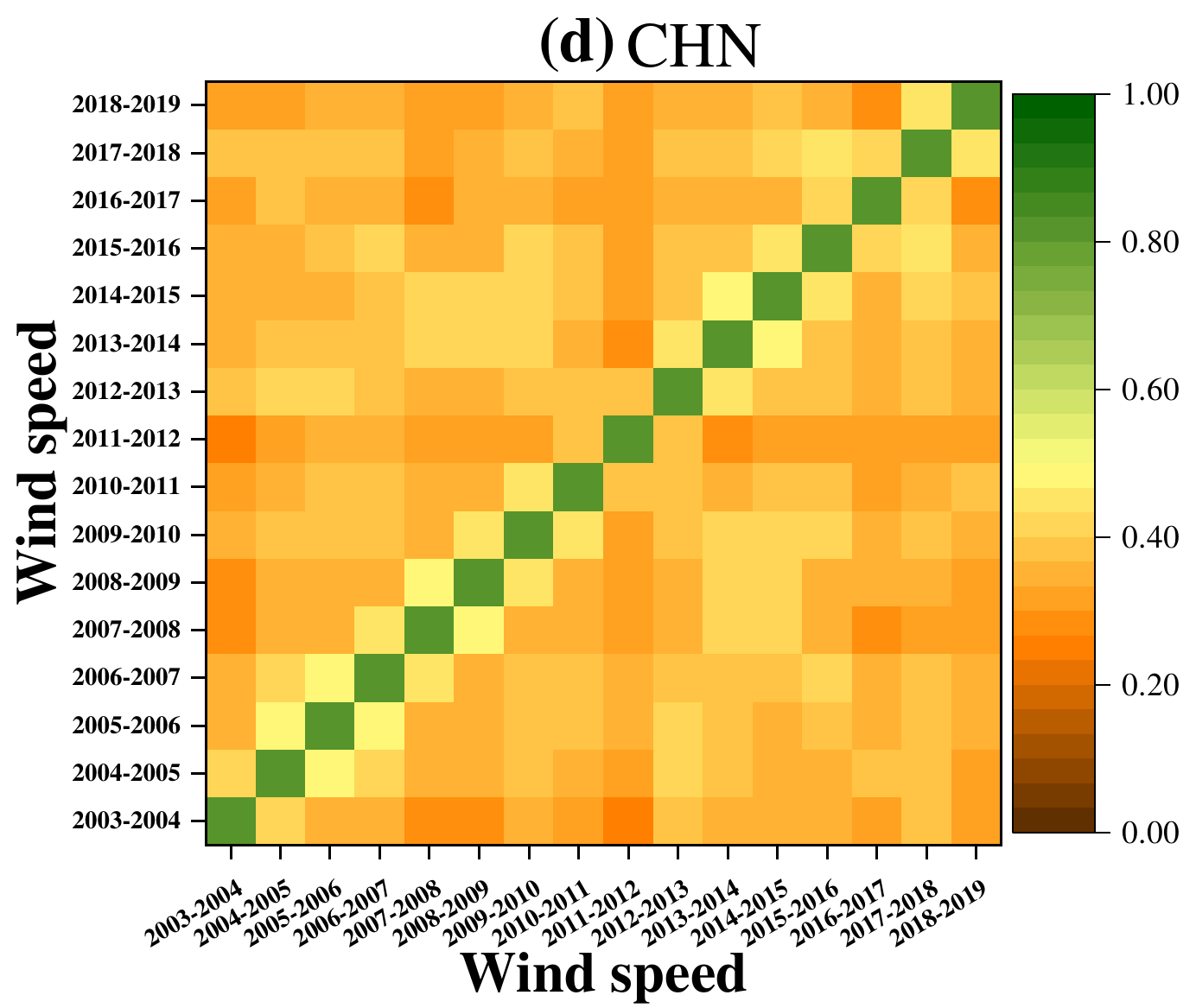}
\includegraphics[width=8em, height=7em]{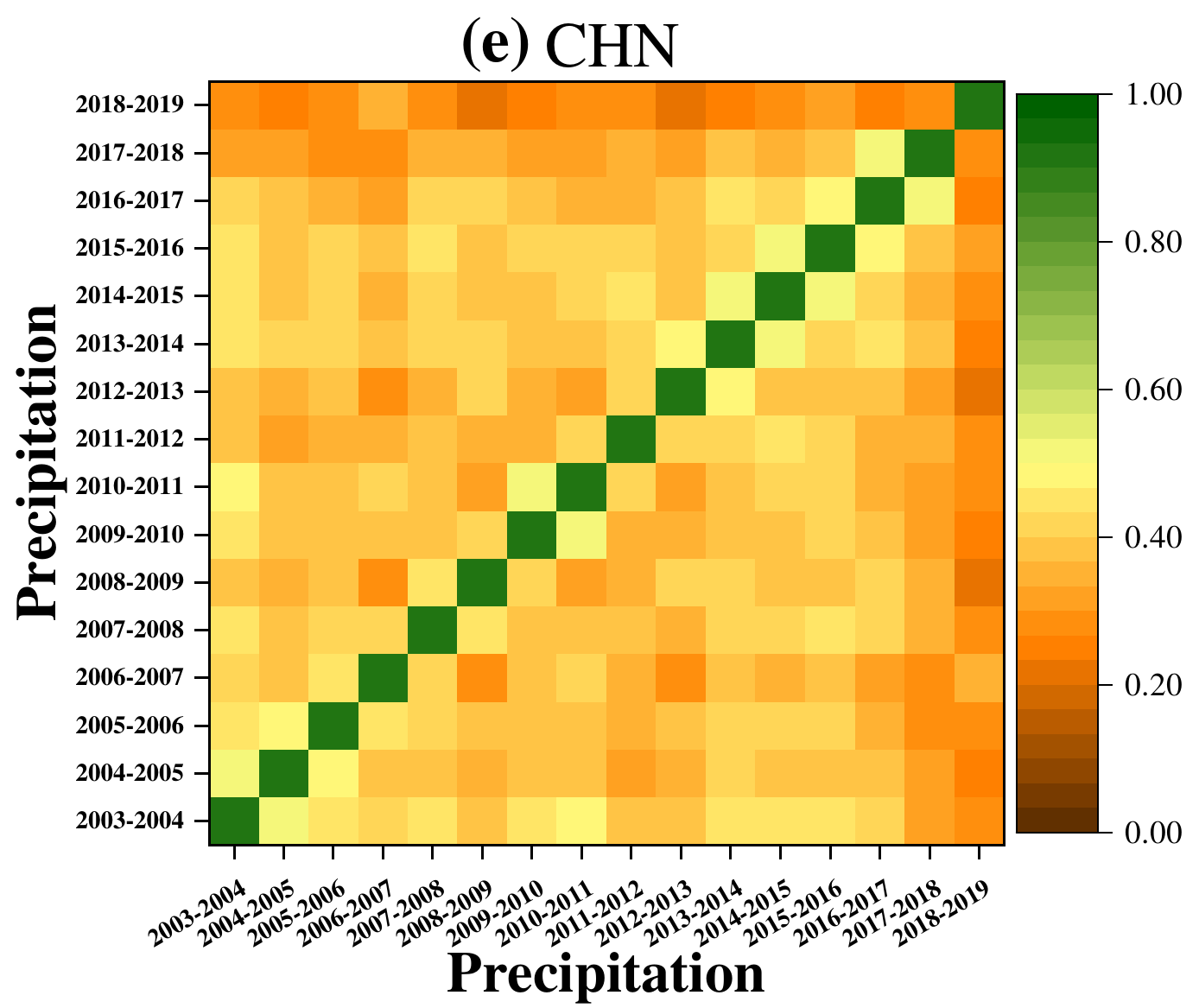}
\includegraphics[width=8em, height=7em]{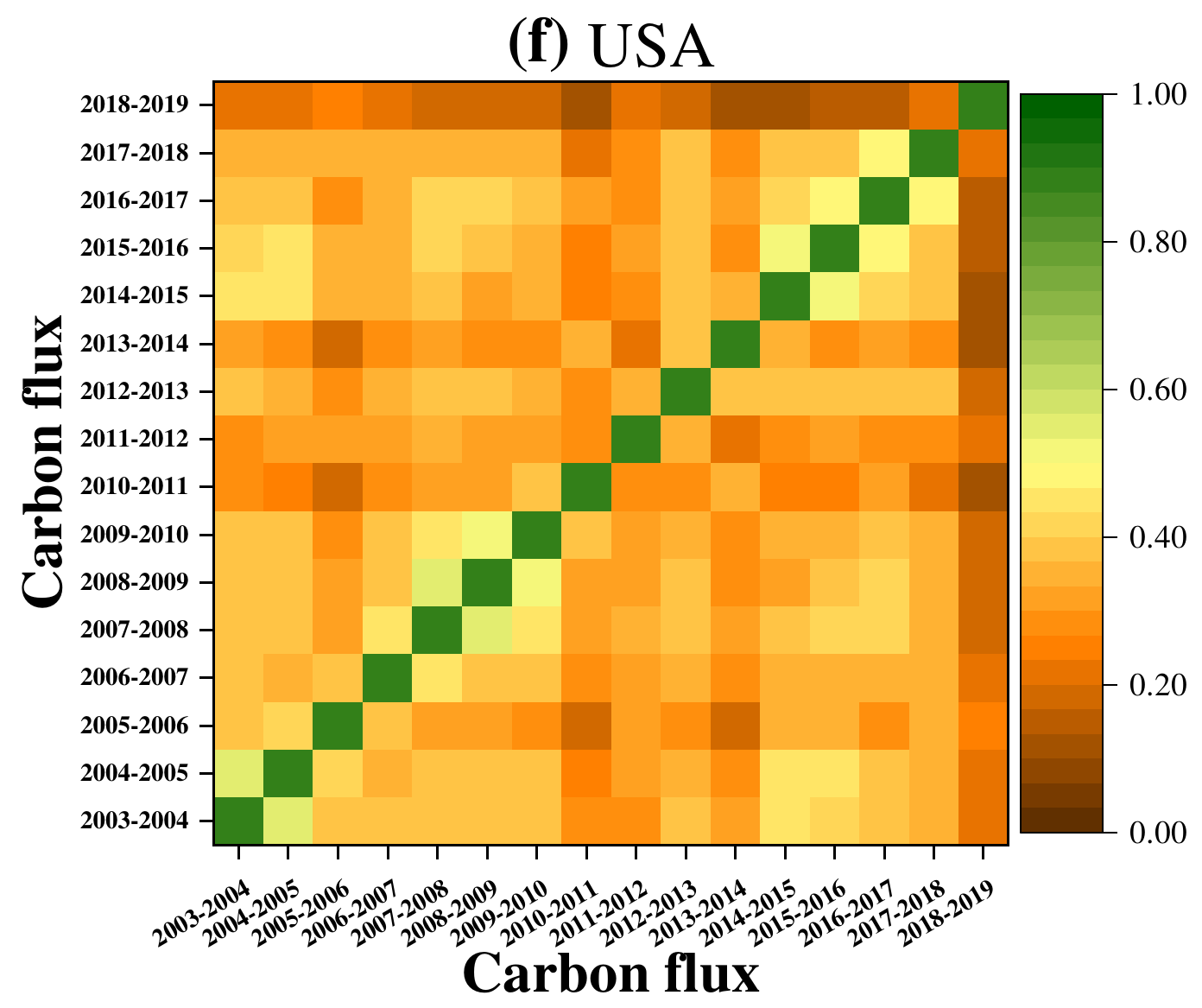}
\includegraphics[width=8em, height=7em]{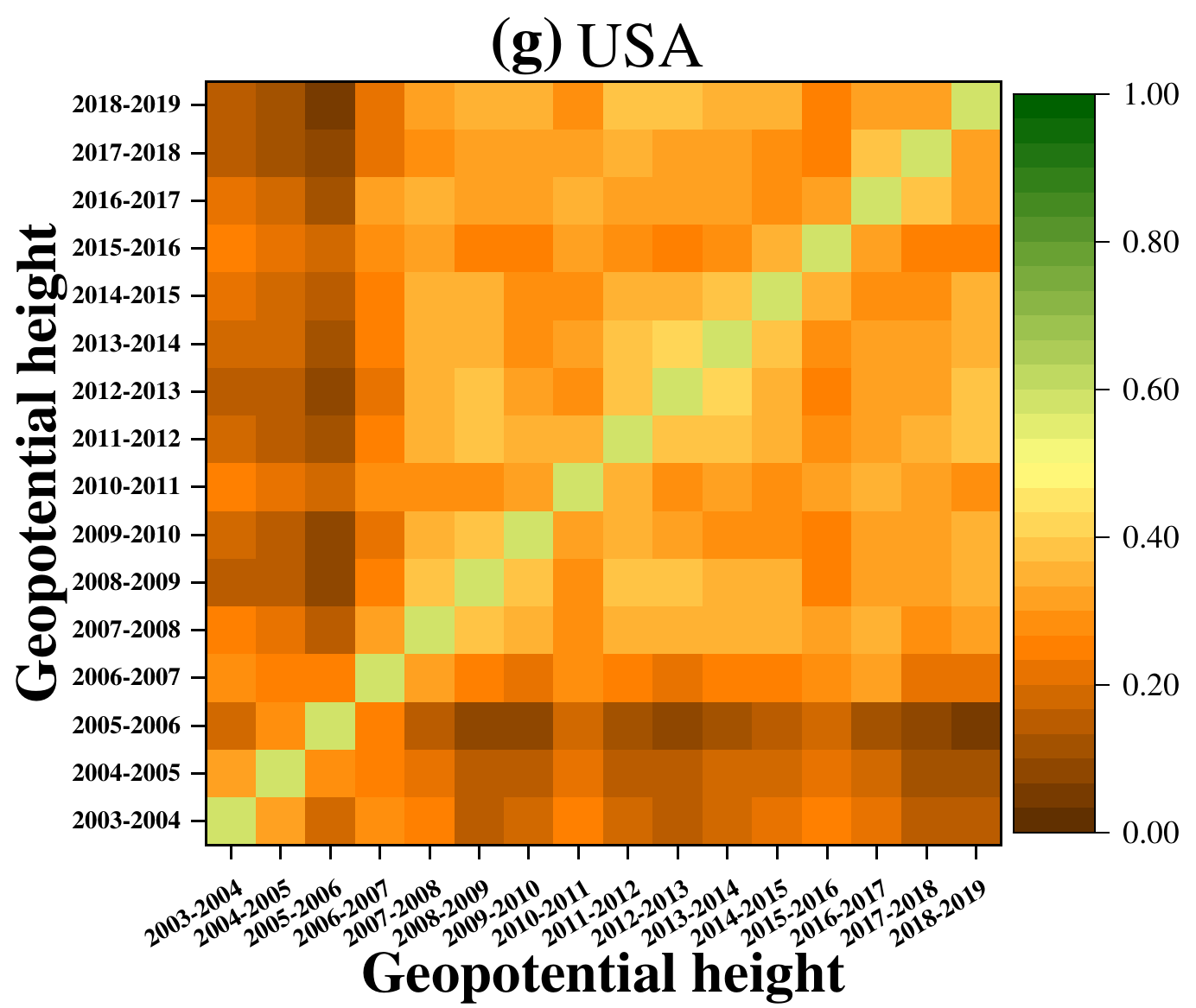}
\includegraphics[width=8em, height=7em]{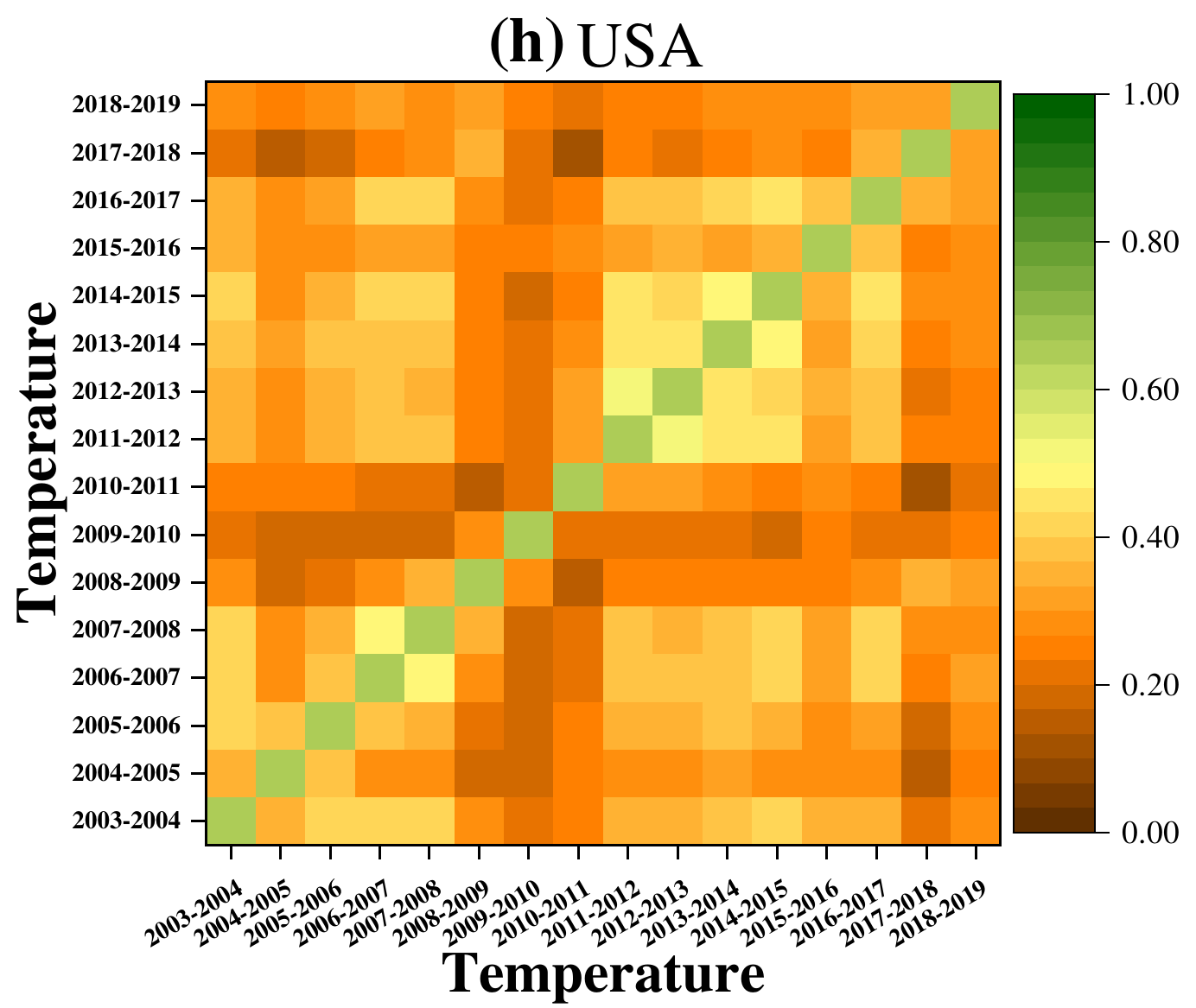}
\includegraphics[width=8em, height=7em]{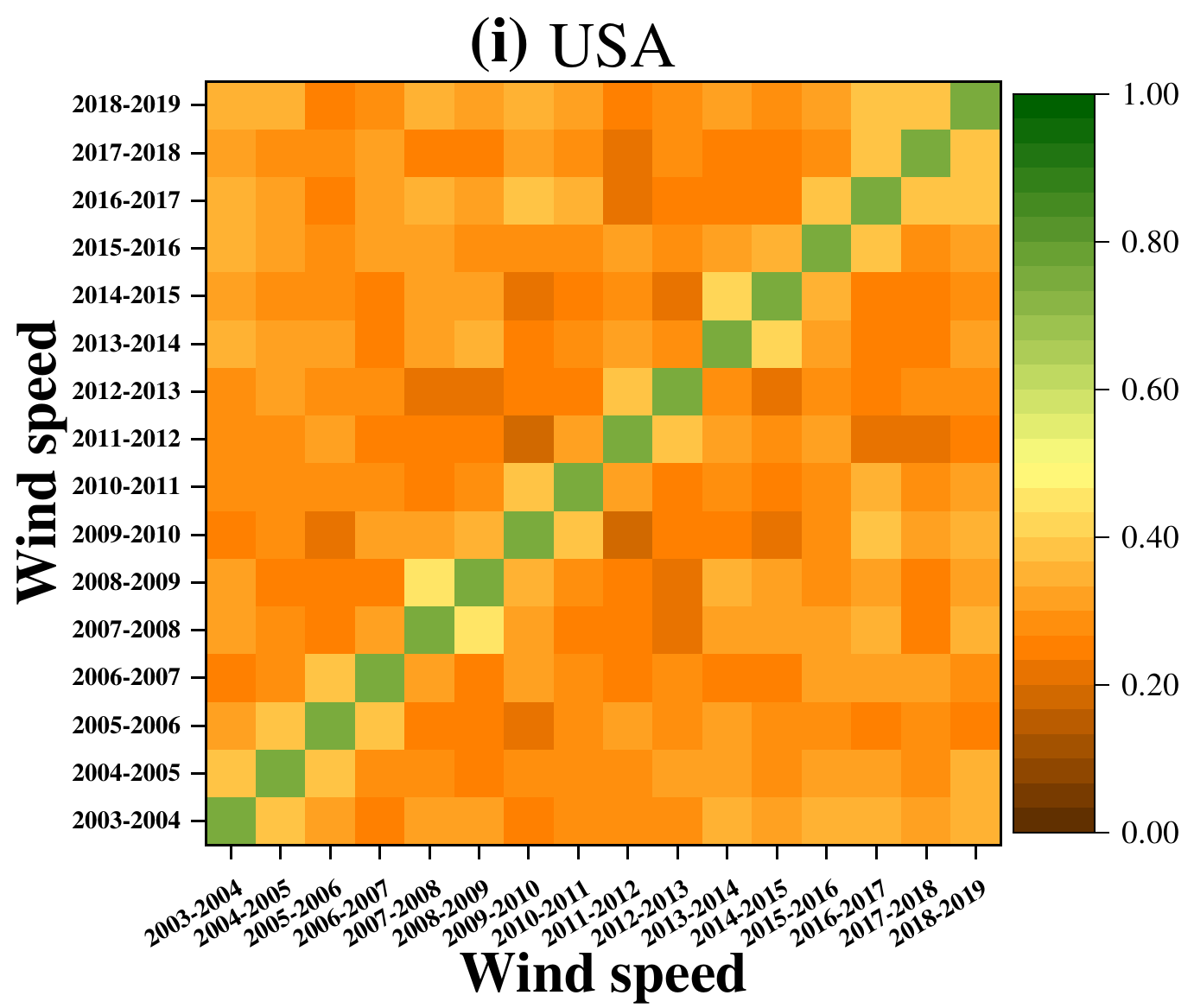}
\includegraphics[width=8em, height=7em]{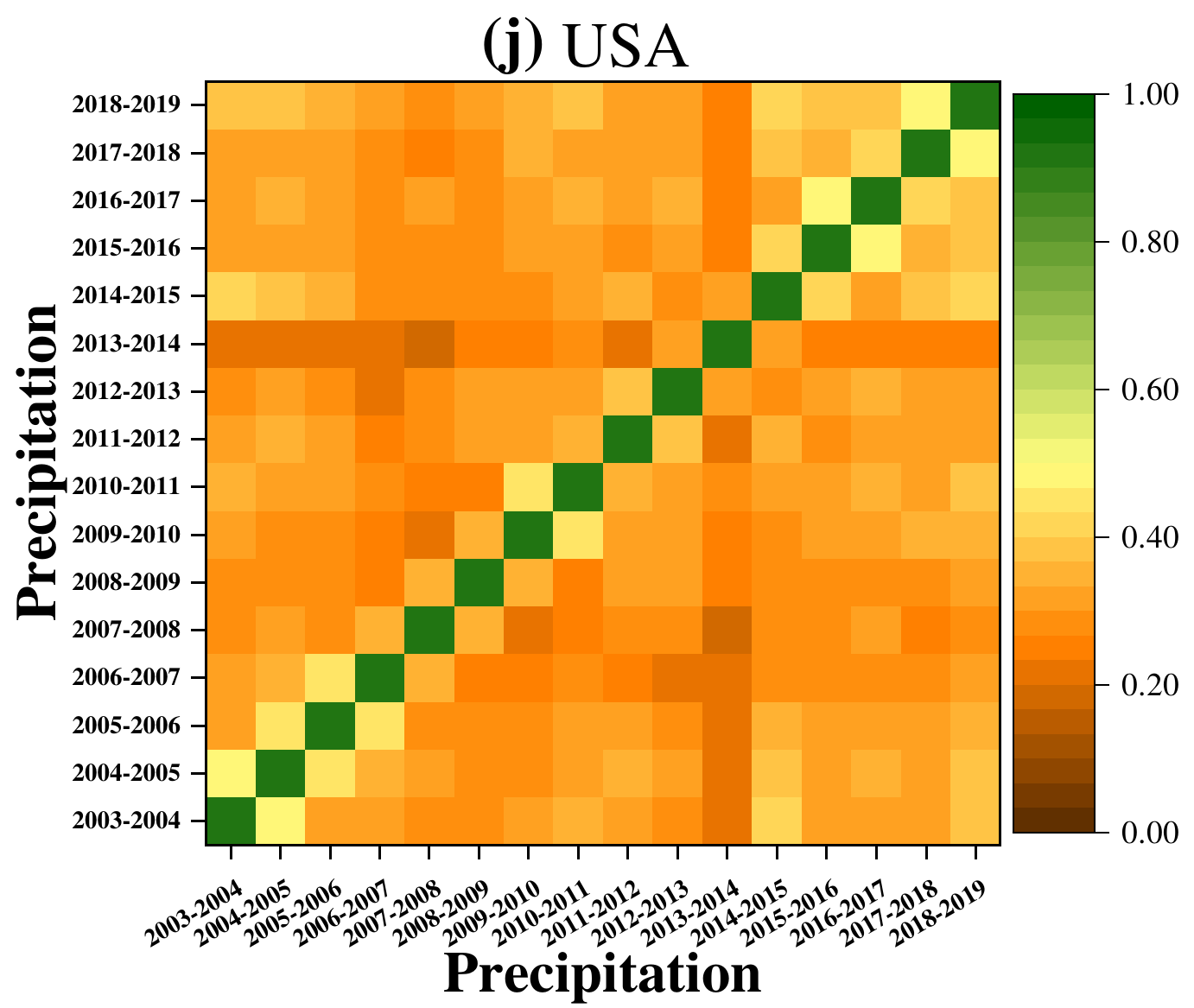}
\includegraphics[width=8em, height=7em]{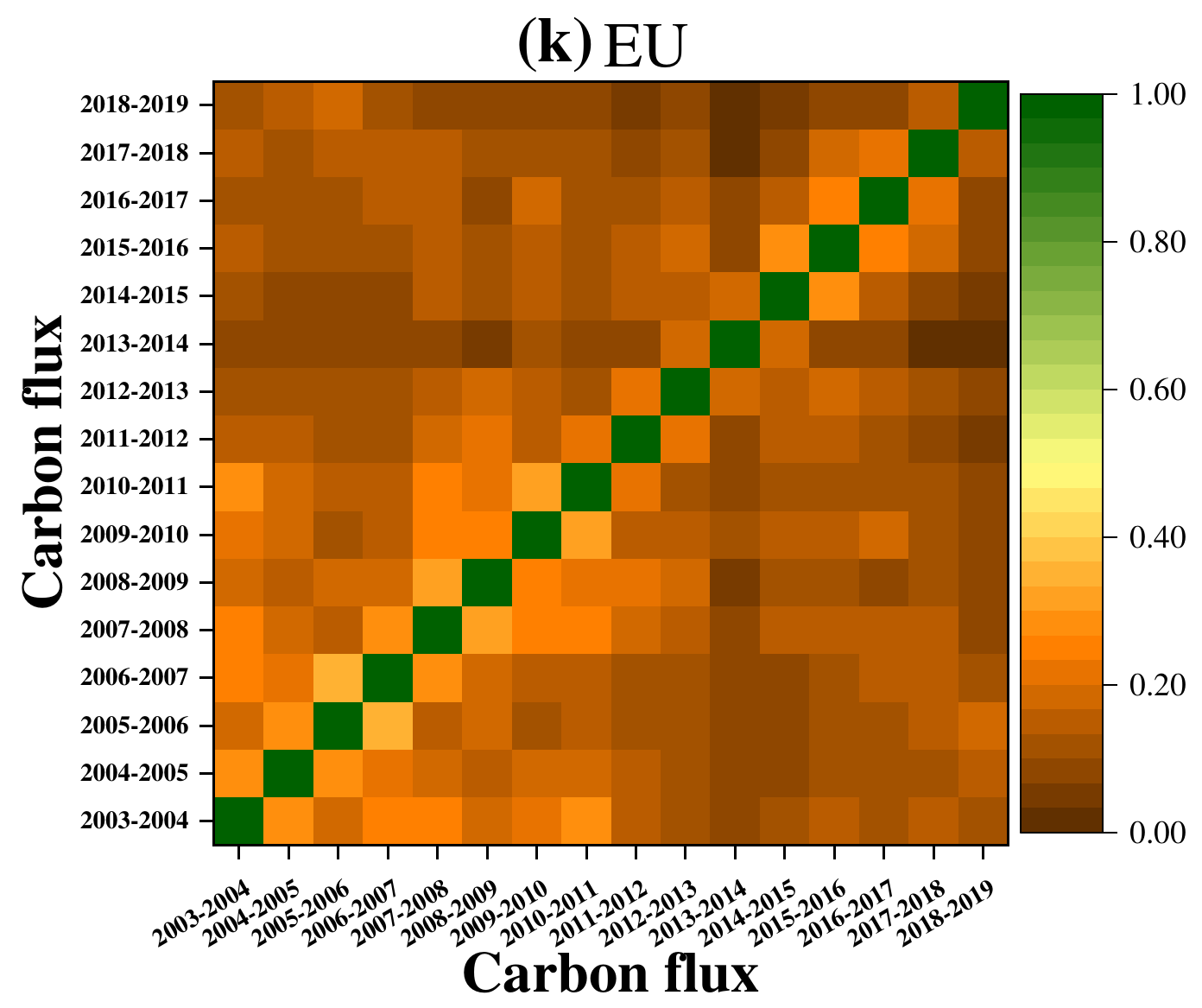}
\includegraphics[width=8em, height=7em]{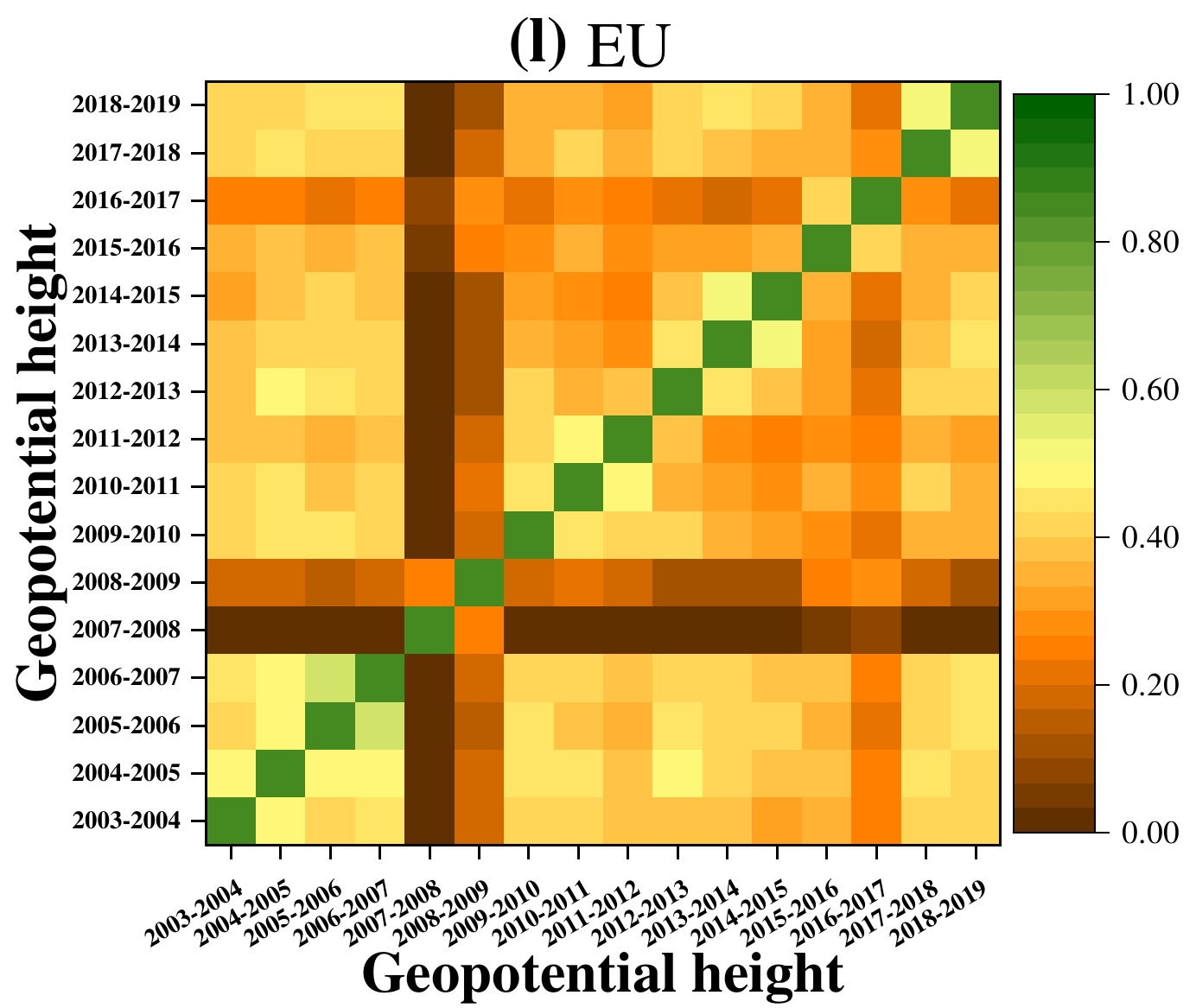}
\includegraphics[width=8em, height=7em]{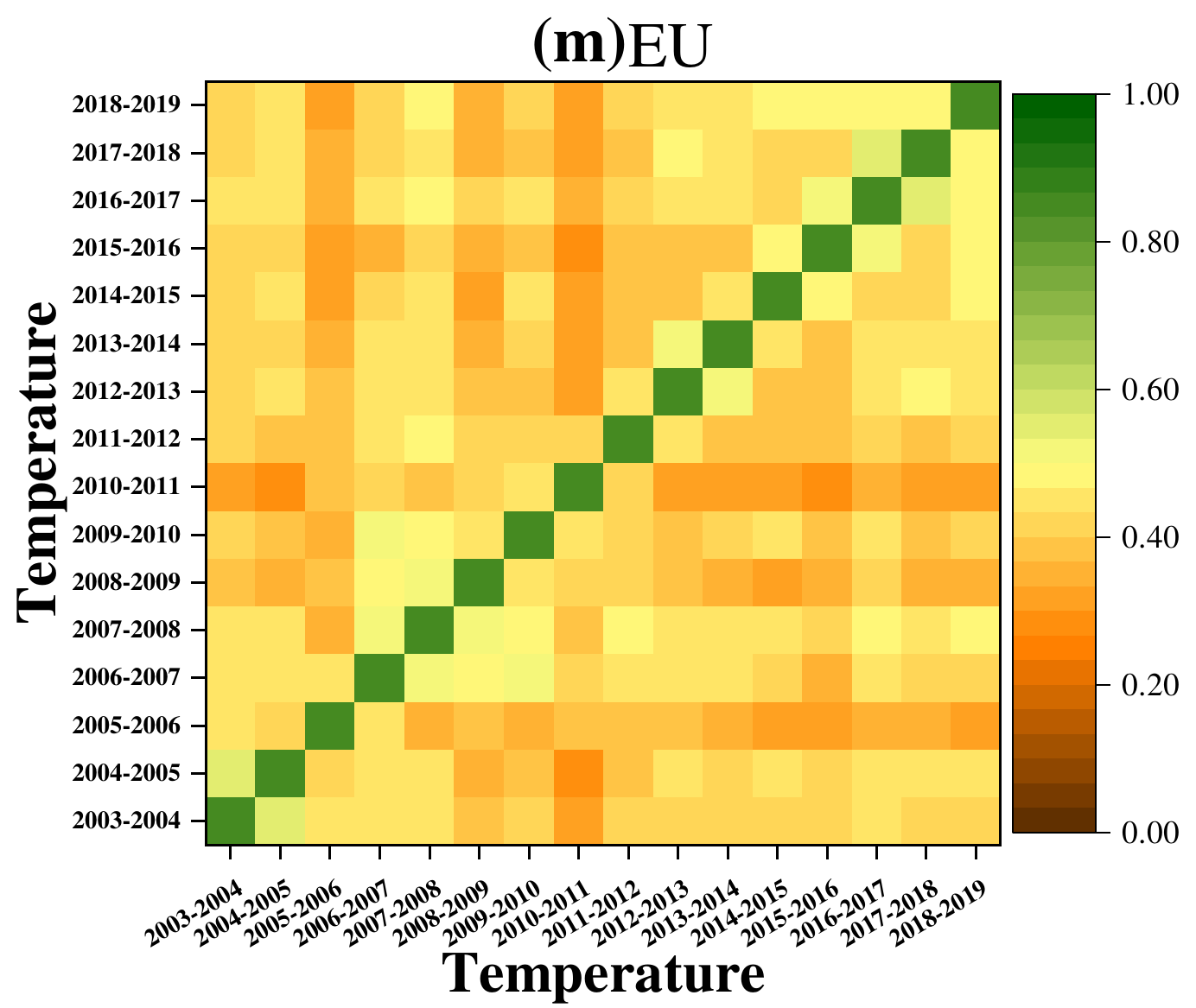}
\includegraphics[width=8em, height=7em]{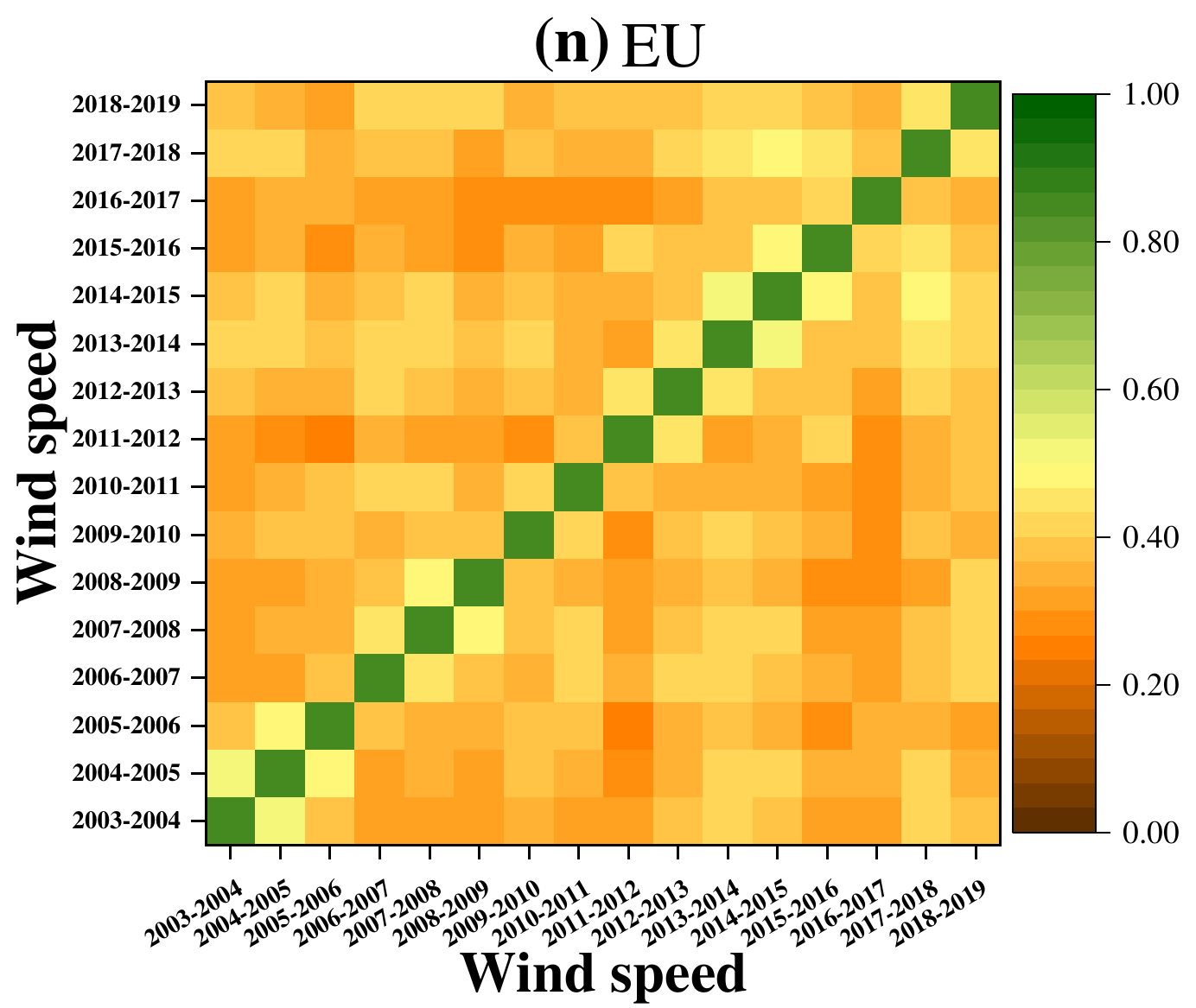}
\includegraphics[width=8em, height=7em]{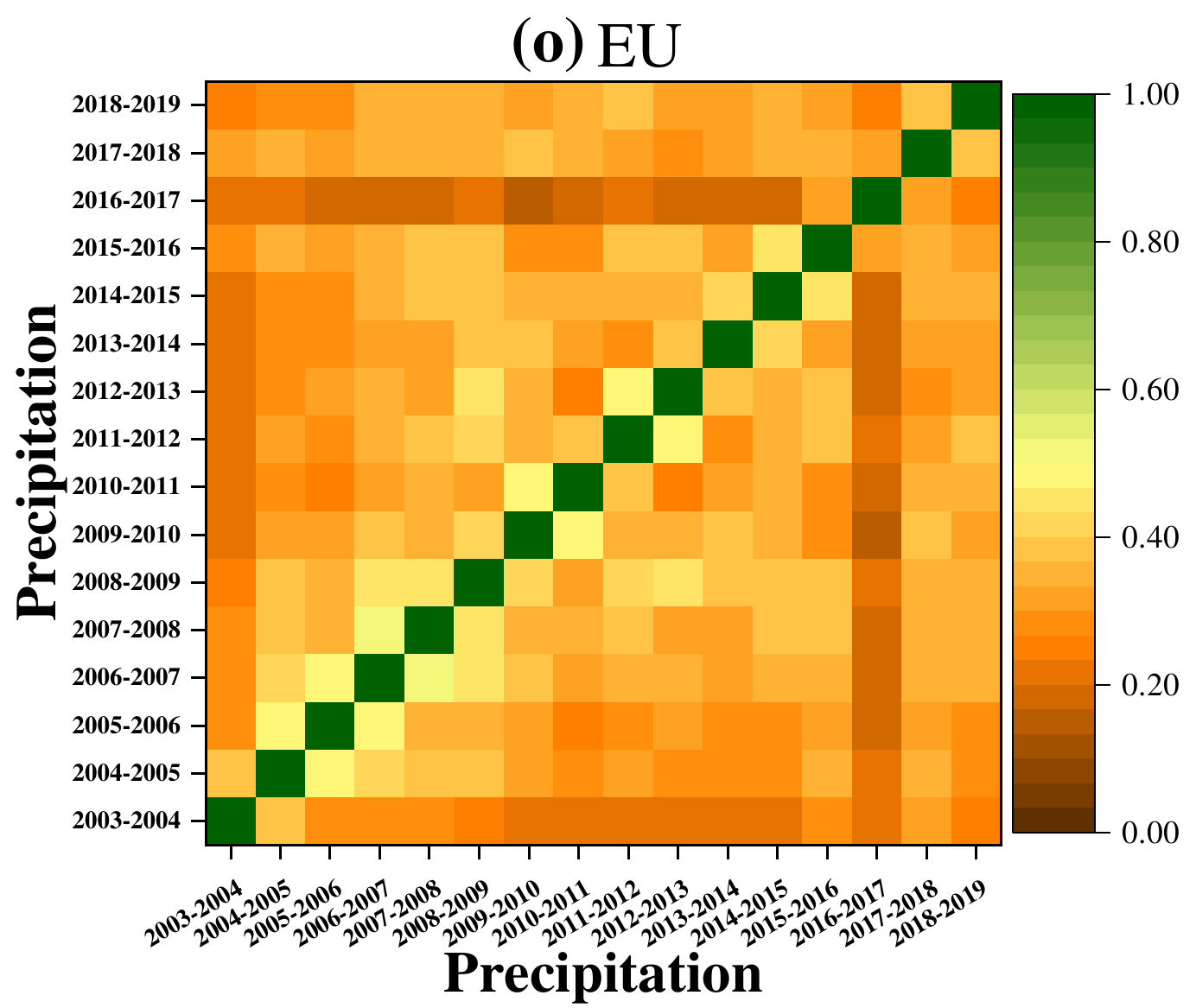}
\end{center}

\begin{center}
\noindent {\small {\bf Fig. S37} The effective Jaccard similarity coefficient matrix for links of lengths above $1000km$ in two networks of different years for each of the climate variables. Each matrix element represents the difference between the actual Jaccard similarity coefficient and the corresponding average and standard deviation values obtained from the controlled case.}
\end{center}

\begin{center}
\includegraphics[width=40em, height=21em]{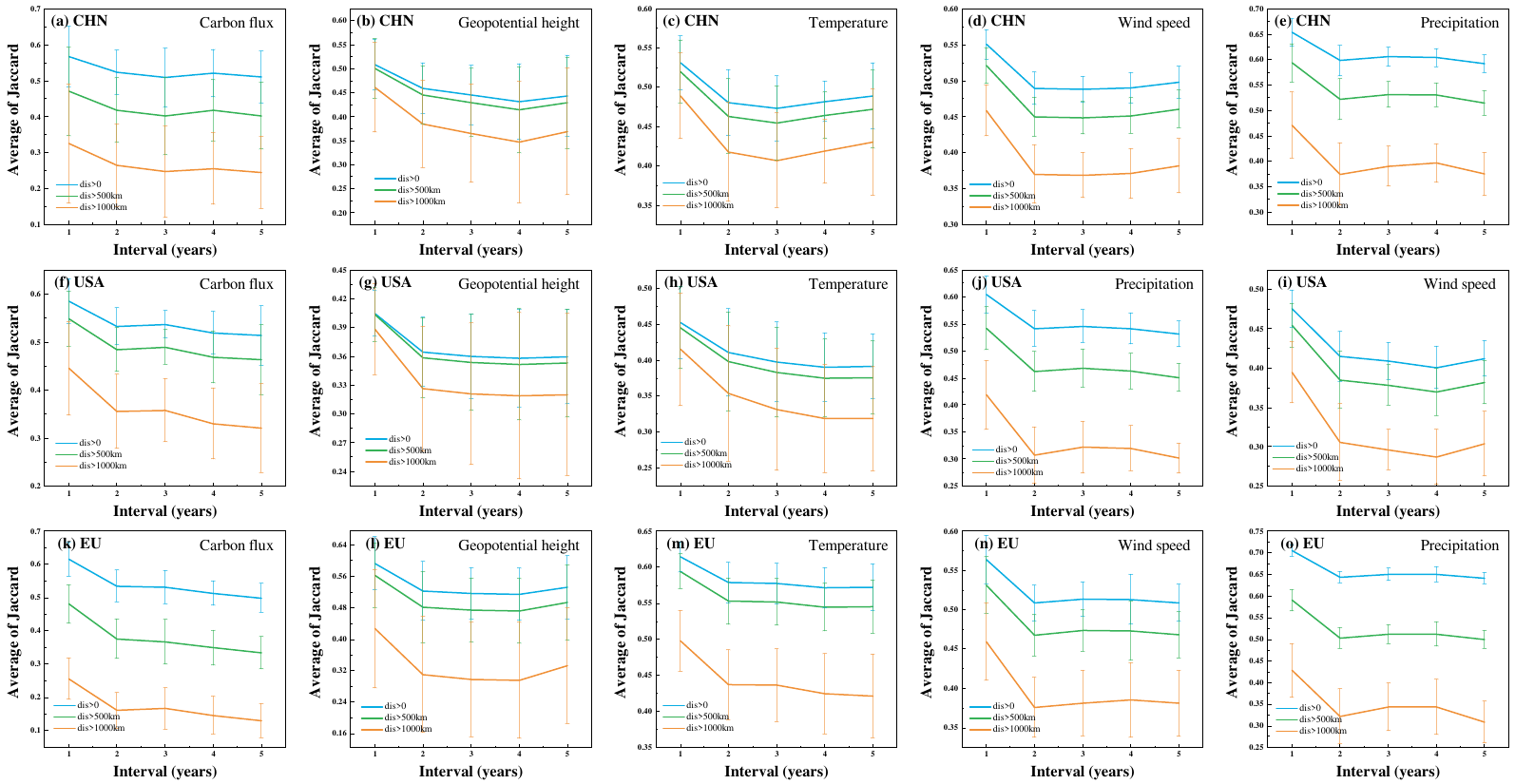}
\end{center}

\begin{center}
\noindent {\small {\bf Fig. S38} The averaged Jaccard values for intervals of one to five years between the years (i.e., averaging over each of the first five diagonal columns below the central dark green column in Fig. 5 of the main text) as a function of years interval. It can be seen generally as a weak decay in similarity with the years interval. The error bars are the standard deviation of Jaccard similarity coefficients.}
\end{center}

\bibliographystyle{elsarticle-num}
  \bibliography{rec}